\documentclass[11pt,twoside]{book}
\usepackage{rotating}
\usepackage{graphicx}
\usepackage{tabularx}
\usepackage{blindtext}
\usepackage[english]{babel}
\usepackage{makeidx}
\usepackage{amsmath}
\usepackage{natbib}
\usepackage{longtable}
\usepackage[normalem]{ulem}
\usepackage{subfigure}
\usepackage{color}
\usepackage{lscape}
\usepackage{colortbl}
\usepackage{rotating}

\def\sun{\hbox{$\odot$}}
\def\earth{\hbox{$\oplus$}} 

\makeindex
\begin{document}

\bibpunct{(}{)}{;}{a}{,}{,}
\bibliographystyle{aa}

\pagestyle{empty} \thispagestyle{empty}

\vspace*{\fill}
\begin{center}
\includegraphics[width=.30\textwidth]{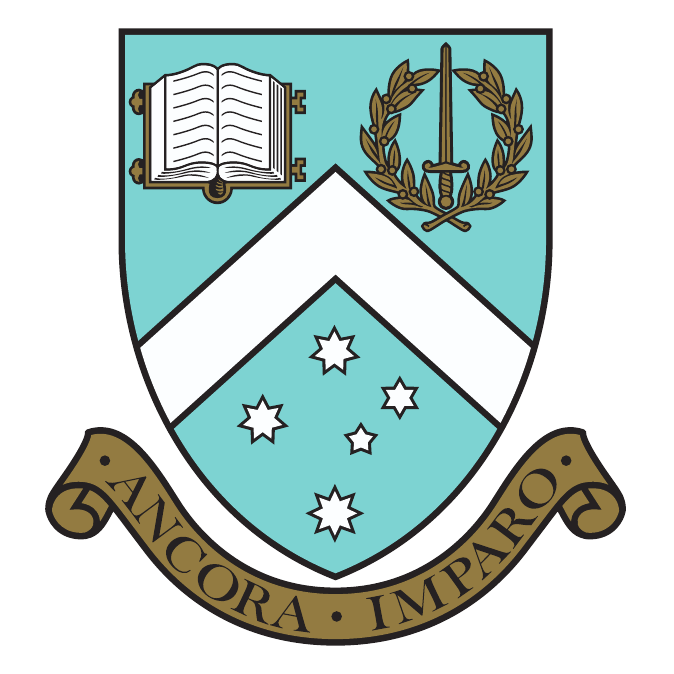}\\[1cm]
\LARGE
\textbf{The Detectability of Moons \\[2mm]of Extra-Solar Planets}\\
\normalsize
\vspace{2cm}
a thesis submitted for the degree of\\
\vspace{3mm}
\large
Doctor of Philosophy\\
\normalsize
\vspace{1cm}
by\\
\vspace{1cm}
\large
\textbf{Karen Michelle Lewis}\\
\normalsize
B.Sc. (Hons)\\
\vspace{2cm}
February, 2011
\end{center}
\vspace*{\fill}

\newpage \thispagestyle{empty}

\cleardoublepage \pagestyle{headings} \pagenumbering{roman}
\markboth{Contents}{Contents} \tableofcontents
\listoffigures
\listoftables

\cleardoublepage \pagestyle{empty} 
\chapter*{Abstract}

The detectability of moons of extra-solar planets is investigated, focussing on the time-of-arrival perturbation technique, a method for detecting moons of pulsar planets, and the photometric transit timing technique, a method for detecting moons of transiting planets.  Realistic thresholds are derived and analysed in the in the context of the types of moons that are likely to form and be orbitally stable for the lifetime of the system.

For the case of the time-of-arrival perturbation technique, the analysis is conducted in two stages.  First, a preliminary investigation is conducted assuming that planet and moon's orbit are circular and coplanar.  This analysis is then applied to the case of the pulsar planet PSR B1620-26 b, and used to conclude that a stable moon orbiting this pulsar planet could be detected, if its mass was $>5$\% of its planet's mass (2.5 $M_{Jup}$), and if the planet-moon distance was $\sim 2$\% of the planet-pulsar separation (23 AU).   Time-of-arrival expressions are then derived for mutually inclined as well as non-circular orbits.

For the case of the photometric transit timing technique, a different approach is adopted.  First, analytic expressions for the timing perturbation due to the moon are derived for the case where the orbit of the moon is circular and coplanar with that of the planet and where the planet's orbit is circular and aligned to the line-of-sight, circular and inclined with respect to the line-of-sight or eccentric and aligned to the line-of-sight.  It is found that when the velocity of the moon is small with respect to the velocity at which the planet-moon barycenter transits the star, that the timing perturbation could be well approximated by a sinusoid.  Second, the timing noise is investigated analytically, for the case of white photometric noise, and numerically, using SOHO lightcurves, for the case of realistic and filtered realistic photometric noise.  It is found the timing noise is normally distributed and uncorrelated for planets likely to host large moons.  In addition it is found that realistic stellar photometric noise results in a dramatic increase in the standard deviation of the timing noise, which is not entirely reversed by filtering.   Finally, using the method of generalised likelihood ratio testing, the work on the form of the timing perturbation due to a moon, and the behaviour of the timing noise are combined to derive both approximate analytic, and exact numerical thresholds.  In particular, a Monte Carlo simulation is run which investigates thresholds for the cases of aligned, inclined and eccentric planet orbits for white, filtered and realistic photometric noise for a range of planet masses ($10 M_{Jup}$, $1M_{Jup}$, $1M_{Ura}$ and $1M_{\earth}$) and semi-major axes (0.2AU, 0.4AU and 0.6AU).  Assuming Kepler quality data, it is found that for the case where the photometric noise is white, physically realistic moons could be detected for gas giant host planets, while for the case where the photometric noise is dominated by intrinsic stellar noise, filtering allows the detection of physically realistic moons for planets with mass $10 M_{Jup}$.

\cleardoublepage \pagestyle{empty} 
\chapter*{Statement}

This thesis contains no material which has been accepted for the award of any other degree or diploma in any university or other institution, and, to the best of my knowledge, contains no material previously published or written by another person, except where otherwise stated in the text.

\vspace{2cm}

\begin{flushright}
Karen M. Lewis
\end{flushright}


\cleardoublepage \pagestyle{empty} 
\chapter*{Acknowledgements}

As any PhD student knows, a PhD is not conducted in isolation, but within the context of a supervisor, institution, research group, collegues, family and friends.  Consequently, I would like to take advantage of the opportunity to acknowledge and thank some of the people who have helped me on this journey.

While I came to Monash to work with Dr. Mardling, it was my fellow PhD students who clinched the deal.  I would particularly like to thank Carolyn ``Linky"  Doherty, Gareth ``GK" Kennedy, Diana Ionescu, Marie Newington and John ``Mc" McCloghlan.  Special thanks go to GK for randomly putting me up in Barcelona and acting like a senpai and to Linky for introducing me to the civilized institution of ``cafe afuera" and letting me invade her office on a near daily basis.  In addition, for the case of the students at ANU, I would like to thank each and every member of the MSO bus committee.  Your help and support made a seemingly impossible task possible.

In addition to my student compatriots, I would like to acknowledge the assistance and friendship given by the range of postdocs I have met over the course of my PhD.  In particular I would like to thank Dr. Jenny McSaveney (and Zane) for introducing me to the Medieval club at ANU, Dr. Daniel Price and Dr. Jesse Andries for the experience of helping organise Monash Astrophysics Day, Dr. Richard ``Ricardo" Stancliff, for listening, advice and for providing temporary emergency accommodation and Dr. Allie ``Elouise" Ford, for the generous application of vegan cake and parrots to any problem (or maybe I should thank Magic and Phoenix for the generous application of Allie to any problem ... either way).  Given that neither of the groups that I was associated with had postdocs, I am very grateful to be ``adopted" and looked after by such kind people.

The help and support of permanent academic staff from other groups and institutions has also been invaluable.  In particular I would like to thank Dr. Kais Hamza and Dr. Aiden Sudbury for their help with the statistics required for my thesis, and for their approachability, and Dr. Pilar Gil-Pons for her tortuga-powered enthusiasm.  The help of such people has not only improved the quality of my thesis, but also the quality of my experience as a PhD student.

I would also like to acknowledge the kindness and help provided by administration and support staff.  For the case of ANU, I would in particular like to thank Pete Walsh and Graeme Blackman for their help, especially with respect the bus, Rebecca Noble for her kindness, and the entire computing section at MSO for just being generally awesome.  For the case of Monash, I would like to thank Gertrude Nayak, Linda Mayer, Doris Herft, Sonia Francis, Melissa Swindle and Rosemary Frigo for their cheerful sympathetic attitude and for help with scholarships, forms, keys and pretty much anything else.  In addition, I would like to thank Dave, Trent and Michael from Monash ITS for their help with  network access and printing, for such things as helping me break into my office when I locked my keys in it and for working out how to trick the coffee machine into thinking it had more beans than it did.  Without your help, this would have been a much rougher ride.

On a more practical note, I would like to acknowledge the financial support given by the Australian Government (Australian Postgraduate Award), the Australian National University (RSAA Supplementary Scholarship), Ms. Joan Duffield (Joan Duffield Research Scholarship) and Monash University (Faculty of Science Dean's Research Scholarship) over the course of my thesis.  I would also like to thank Monash University for its generous travel assistance which allowed me to attend both domestic and international conferences, and the Geneva group for letting me stay at their house.

In addition, over the course of my PhD, I was lucky enough to go observing at the Siding Springs Observatory on three different occasions.  I would like to thank Daniel Bayliss for instructing me on how to use the 40" telescope and Dr. John Wisniewski for observing with me.  I would also like to thank Prof. Penny Sackett (my ANU supervisor), Dr. Leslie Hebb, and Prof. Ken Freeman whose grants allowed me to go.  I really enjoyed observing and was grateful for the opportunity.

On a more personal level, during my candidature I have lived with 17 different housemates in a total of six different households (and this isn't counting the temporary housemates, friends and neighbours that you meet as a result of such a situation).  I have learnt something, or been helped, or been cheered up by each one of my housemates and am deeply grateful for it.  PhD students aren't always the polite, kind, neat, emotionally-centered people that we would like to be and your patience and acceptance was very welcome during these times.  Finally, I would like to acknowledge the effect of a good accommodation environment on my PhD and would particularly like to thank MRS for its culture of kindness towards students in general and me in particular.

In addition, I would like to thank my family, both nuclear and extended.  Their attempts to help me were always taken in the kind spirit in which they were offered.  In addition, the fact that I was studying on the other side of the country resulted in a lot stuffing about on their part (moving from Canberra to Melbourne in a Hyundai Excel for example), but they bore it with good grace and humor.  In particular, I would like to thank my brother for letting me practically stay at his house in 2007 and make a good attempt at watching the entirety of his TV show collection.  I probably wasn't a very good sister at this point, but he was a damn fine brother.

Finally, there are two people without whom this PhD would not have happened, Dr. Rosemary Mardling and Prof. Ken Freeman.  Prof. Freeman believed in me at a time when few people did, a ``baton" he passed onto my PhD supervisor, Rosemary, who ``ran" with it for three years.  For this, these two academics have my heartfelt thanks.  In addition to her kindness, I would also particularly like to thank Dr. Mardling for her time, insight and expertise.  Her well-thought out suggestions and advice have rescued me from a number of blunders and subsequently resulted in a clearer thesis containing much cleaner, neater, more transparent mathematics.  Thank you! 

\cleardoublepage \pagestyle{empty} 

\pagenumbering{arabic}


\cleardoublepage \pagestyle{empty} 
\part{Introduction and Literature Review}
\pagestyle{plain} 
\chapter{Introduction}

\section{Introduction}

Since the announcement of the first two extra-solar planets orbiting the pulsar PSR~B1257+12 \citep{Wolszczanetal1992}, over 500 extra-solar planets have been discovered.\footnote{See e.g http://exoplanet.eu/catalogue.php}  These discoveries have been made using a variety of methods including the transit, radial velocity, microlensing and pulsar time-of-arrival techniques, and have consequently resulted in a broad variety of detected planets.  This wealth of data allows tests to be conducted on individual and ensemble groups of planets to investigate such things as the mode by which gas giant planets formed \citep[e.g.][]{Batyginetal2009,Mardling2010}, the conditions under which they formed and their subsequent orbital evolution \citep[e.g.][]{FabryckyWinn2009,Triaudetal2010}.   This is done through measurement and analysis of planetary properties such as mass, radius, orbital eccentricity and spin orbit misalignment, and by comparing limits on the size of extant moons with limits such as those proposed by \citet{Canupetal2006}.  It is this last issue, in particular, the detection of extrasolar-moons, or moons of extra-solar planets to which this thesis addresses itself.

\section{Structure of this thesis}

In this thesis the detection of extra-solar moons, in particular, using the methods of pulse time-of-arrival perturbation and photometric transit timing is investigated in the context of stability and formation models, as well as other moon detection methods presented in the literature.  As such, this thesis is divided into three main sections.  The first section introduces the preliminary material required to understand this thesis, namely the notation, and literature results on formation, stability and detection of extra-solar moons.  From this context, the second and third parts move on to analyse moon detection using the pulse time-of-arrival perturbation and photometric transit timing methods respectively.  These parts are discussed in turn.

\subsection{Preliminaries and literature review}

We begin in chapter \ref{Intro_Moons_Note} with the task of discussing and defining the notation used in this thesis.  Then, the type of moons that extra-solar planets are expected to have is then summarised in chapter \ref{Intro_Moons_Const}.  In particular, work presented in the literature on moon formation mechanisms and subsequent moon orbital evolution is collated to give a set of likely constraints on the physical and orbital properties of moons of extra-solar planets.  Then, in chapter \ref{Intro_Dect}, the set of methods proposed in the literature for detecting moons of extra-solar planets is summarised, along with the types of moon each method is optimised to detect.  This is done in two main stages.  First, the methods for detecting extra-solar planets are briefly summarised.  Then, within this context, each of the moon detection methods presented in the literature is summarised with particular reference to the two methods investigated in this thesis.

\subsection{Detection of moons of pulsar planets}

We begin our investigation of moon detection in Part \ref{PulsarPart} by focussing on the pulse time-of-arrival perturbation technique.  This investigation is conducted in two stages.  First, a preliminary investigation is conducted into the detectability of moons of pulsar planets.  This work is presented in chapter \ref{Pulsar_Paper} and is published as \citet{Lewisetal2008}.  In this analysis, an expression for the timing perturbation due to planet-moon binarity is derived for the case where the orbit of the planet and the orbit of the moon are both circular, and in the same plane.  This analysis is then used to constrain orbitally stable moons of the pulsar planet PSR~B1620-26~b.  Second, as an extension to this analysis, the effect of mutual inclination and mild eccentricity in the orbit of the planet or the moon is investigated in chapter \ref{Pulsar_Extension}.  In particular, this investigation is conducted using a three-body formalism developed by my PhD supervisor, Dr. Rosemary Mardling, as it allows easy description of hierarchical three-body systems with arbitrary values of inclination and eccentricity.  It was found that mutual inclination or eccentricity introduced additional harmonics into the perturbation, which are summarised in figure~\ref{TOAFreqSplit}.

\subsection{Detection of moons of transiting planets}

In Part \ref{TransitPart}, the detectability of moons of transiting planets using the method of photometric transit timing is addressed.  In chapter \ref{Trans_Intro} the transit technique is reintroduced and expressions for the transit duration and shape of the transit light curve are derived as they are required later in the thesis.  Then, in this context the photometric transit timing statistic $\tau$ is introduced and defined, noting that for this thesis, a slightly more general definition of this statistic is used than that used by \citet{Szaboetal2006}. Then, by algebraically manipulating this definition, expressions for $\Delta \tau$, the timing perturbation due to the moon, and $\epsilon_j$ the timing perturbation due to photometric noise, are derived.  In particular, this formulation allows the effect of the moon and the noise to be investigated separately and then combined to yield thresholds.  

Following on from this, the form of $\Delta \tau$ is investigated in chapter \ref{Transit_Signal}.  In particular, it is investigated for the case where the moon's orbit is circular and in the same plane as the planet's orbit and the planet's orbit is circular and aligned to the line-of-sight (section~\ref{Trans_TTV_Signal_CC}), is circular and slightly inclined with respect to the line-of-sight (section~\ref{Trans_TTV_Signal_Inc}) and eccentric and aligned to the line-of-sight (section~\ref{Trans_TTV_Signal_EccO}).  In addition, the case where the moon's orbit is slightly eccentric is also investigated (appendix \ref{EccMoon_App}).  For the case where the moon's orbit is circular, the motion of the moon is roughly uniform during transit and the velocity of the moon on its orbit is substantially less than the velocity of the planet-moon pair across the face of the star, $\Delta \tau$ is given by a sinusoid with coefficients given in table~\ref{DelTauCoeffTab}.

In chapter \ref{Trans_TTV_Noise}, the quantity $\epsilon_j$, the timing noise on $\tau$, is examined for three realistic photometric noise sources, white noise, intrinsic stellar photometric noise and filtered intrinsic stellar photometric noise.  First, the case of white noise is investigated analytically and compared with the qualitative results given in \citet{Szaboetal2006}.  Using this approach it is found that the size of $\epsilon_j$ does not necessarily decrease with decreasing exposure time as suggested by \citet{Szaboetal2006}, but also depends on the origin of the photometric noise, for example, whether the dominant noise source is shot noise or read noise.  Following on from this, using a method developed in this thesis for deriving the distribution of $\epsilon_j$ using out-of-transit data, the cases of realistic stellar noise and filtered stellar noise are investigated using SOHO light curves.  It is found that $\epsilon_j$ is approximately normal and uncorrelated for all planets likely to host large moons.  In addition, it is found that for the case where the photometric noise is dominated solar-like photometric variability, the amplitude of the timing noise is much larger than that for the equivalent amplitude white noise, and that this effect is only partially negated by filtering.

Finally in chapter \ref{Trans_Thresholds}, the work on the form of $\Delta \tau$ and behaviour of the timing noise is combined to produce preliminary detection thresholds.  In particular, this is done using the method of generalised likelihood ratio testing, which involves comparing the probability that an observed sequence of $\tau$ values was produced by a system containing a planet and a moon as opposed to a system containing only a planet.  Using this method, analytic expressions for thresholds were generated for the case where the number of transits is large.  It is found that the thresholds have a lopsided U-shape with minima defined by the type of photometric noise and inclination, and depth defined by the type of photometric noise and the eccentricity.  In addition to this general trend, the threshold also shows a number of non-detection spikes corresponding to the cases where the moon orbits its host an integer number of times per planetary year, and consequently produces no transit to transit timing variations.  To investigate the more realistic case where the number of observed transits is small, a Monte Carlo simulation is also run.  The thresholds produced showed the same general behaviour as predicted by the analytic analysis.  In addition, it is shown that it may be possible to place limits on physically realistic moons for gas giants hosts using this technique, for planets in the Kepler data set.  

We begin this process by discussing and selecting notation to be used for this thesis.

\chapter{Notation used in this thesis}\label{Intro_Moons_Note}

\section{Introduction}\label{Intro_Moons_Note_Intro}

Before moon detection methods can be investigated, or even literature results summarised, a notation set must be defined and described.  As notation used varies across the literature, the selection of notation is not necessarily straight-forward (for an extreme example of this please see section~\ref{Intro_Moons_Note_Trans}).  Consequently, this chapter is dedicated to discussing and motivating the selection of the notation used in this thesis in three particular contexts.  First, general notation describing the physical properties of the star, planet and moon will be discussed.  Then, the notation required to describe the orbital properties of the star, planet and moon will be motivated and discussed in the context of two-body and three-body theory.  Finally, the discussion will move to the notation required to describe transit light curves.  For reference, the notation selected is presented and summarised in appendix~\ref{VarDef_App}.  We begin with a discussion of general notation, in particular, describing the physical properties of the star, planet and moon.

\section{Notation used for the physical properties of the star, planet and moon}\label{Intro_Moons_Note_Phys}

For both of the two detection methods investigated in this thesis, there are three bodies which need to be described, namely, the star, the planet and the moon. These three bodies have a number of inherent physical properties which can effect moon detection, in particular, their mass and radius.  For this thesis, the notation $M_s$, $M_p$ and $M_m$ will be used for the masses where it is noted that the subscripts $s$, $p$ and $m$ denote properties of the star, planet and moon respectively.  In addition, the radii of the three bodies will be written as $R_s$, $R_p$ and $R_m$ (please see section~\ref{Intro_Moons_Note_Trans} for additional discussion about this choice).

\section{Notation used for the orbital properties of the star, planet and moon}

The selection of notation for the orbital parameters of the star, planet and moon is a little more involved as they are part of a three-body system.  Consequently, to provide a context for this discussion, the motion of two bodies in their mutual gravitational field will be discussed, followed by a general discussion of the motion of three bodies.  Then, the understandings developed from this discussion will then will be applied to the specific case of a star-planet-moon system, and used to define and select intuitively reasonable notation.  We begin with a discussion of general two body motion.

\subsection{General two-body motion}

\begin{figure}[tb]
\begin{center}
\includegraphics[width=.35\textwidth]{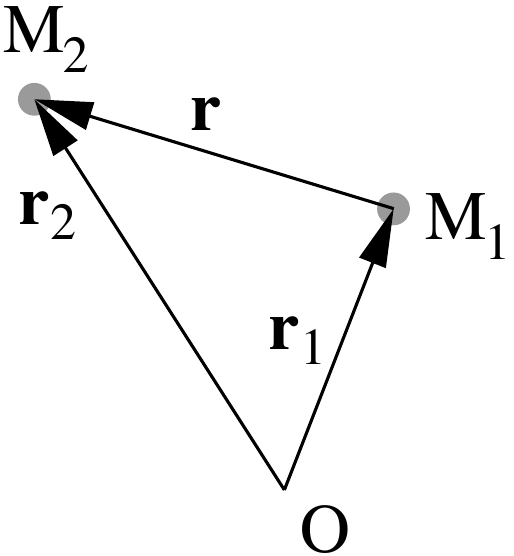}
\caption[Schematic diagram of a general 2-body system.]{Schematic diagram of two masses (grey circles) along with the position vectors to these masses from the origin (O).  The vector from $M_1$ to $M_2$, $\mathbf{r}$, is also shown.}
\label{TwoBodySchematic}
\end{center}
\end{figure}

\begin{figure}[tb]
\begin{center}
\includegraphics[width=.70\textwidth]{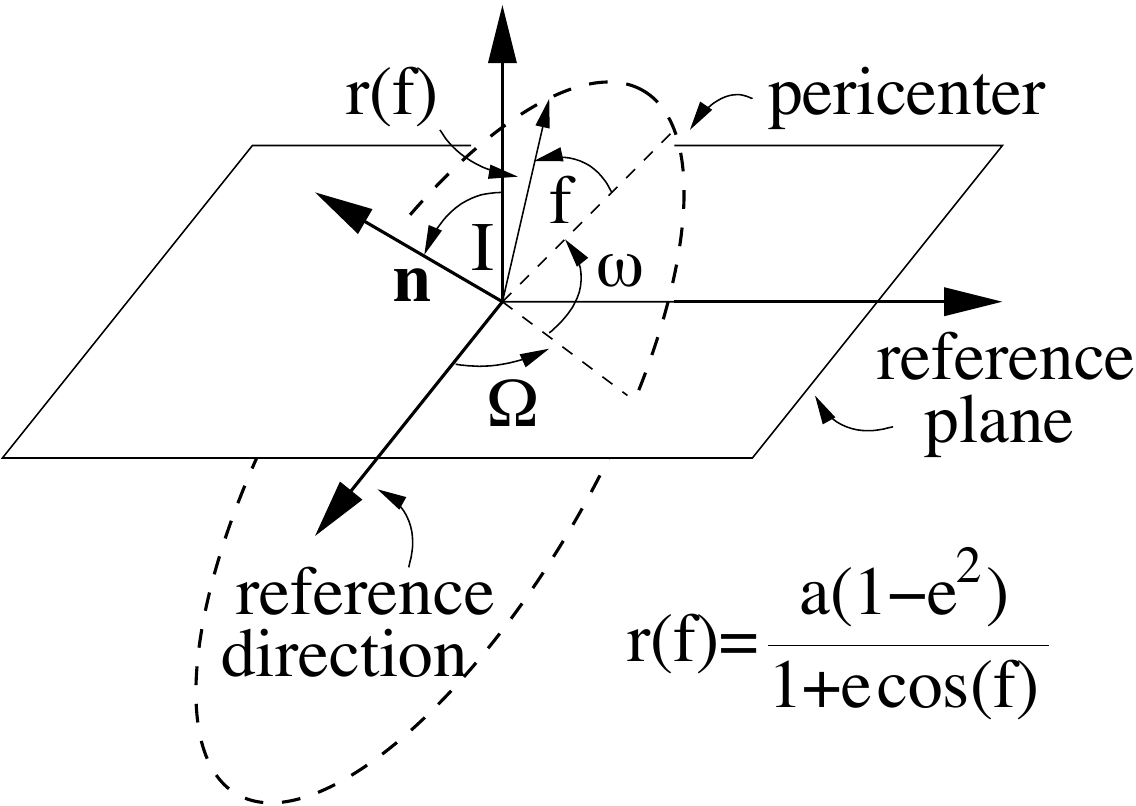}
\caption{Schematic diagram showing the relationship between an elliptical orbit (dashed line), its normal ($\mathbf{n}$), the reference direction and reference plane, and the angles $I$, $\omega$ and $\Omega$.}
\label{IntroOrbitalElements}
\end{center}
\end{figure}

Consider the motion of two bodies, of mass $M_1$ and $M_2$, moving under their mutual gravitational field (see figure~ref{TwoBodySchematic}).  Using Newton's force law, the forces acting on each of the bodies shown in figure~\ref{TwoBodySchematic} can be expressed as
\begin{align}
M_1 \frac{d^2 \mathbf{r}_1}{d t^2} &= \frac{G M_1 M_2}{\left|\mathbf{r}_2 - \mathbf{r}_1\right|^3}(\mathbf{r}_2 - \mathbf{r}_1),\\
M_2 \frac{d^2 \mathbf{r}_2}{d t^2} &= \frac{G M_2 M_1}{\left|\mathbf{r}_1 - \mathbf{r}_2\right|^3}(\mathbf{r}_1 - \mathbf{r}_2),
\end{align}
where $G$ is the universal gravitational constant.

Putting $\mathbf{r} = \mathbf{r}_2 - \mathbf{r}_1$, these two equations combine to give 
\begin{equation}
\frac{d^2 \mathbf{r}}{d t^2} + \frac{G  (M_1 + M_2)}{r^2}\mathbf {\hat{r}} = 0,
\label{intro_not_2bod_govEq}
\end{equation}
where $\mathbf{r}$ is also shown in figure~\ref{TwoBodySchematic}, $\mathbf {\hat{r}}$ is a unit vector in the direction of $\mathbf{r}$, and one of the equations has dropped out as a result of conservation of momentum.

Equation~\eqref{intro_not_2bod_govEq} can be directly solved to give the canonical conic section solutions for two-body motion \citep[see e.g.][]{Murrayetal1999}.  In particular, the orbit is described by six parameters which correspond to the six integration constants from the above second order vector differential equation.  Following the notation of \citet{Murrayetal1999}, the six orbital elements which uniquely define the ellipse of a two body orbit are $a$, the semi-major axis, $e$, the eccentricity, $I$, the inclination, $\omega$, the argument of pericenter, $\Omega$, the longitude of the ascending node, and $f(0)$,\footnote{In order to determine where the body is along its orbit, an initial condition is required.  \citet{Murrayetal1999} use $\tau$, the time of pericenter passage to link $t$ and $f$.  However, as $\tau$ will be used for the photometric transit timing statistic in this thesis, we use $f(0)$ instead.}  the value of the true anomaly at $t=0$ (see figure~\ref{IntroOrbitalElements}). 

Now that two body motion has been introduced, we are in a position to discuss motion of three bodies in their mutual gravitational field, again from a general perspective.  Consequently, we will repeat the above procedure for the case where an additional body has been added.

\subsection{General three-body motion}

\begin{figure}[tb]
\begin{center}
\includegraphics[width=.65\textwidth]{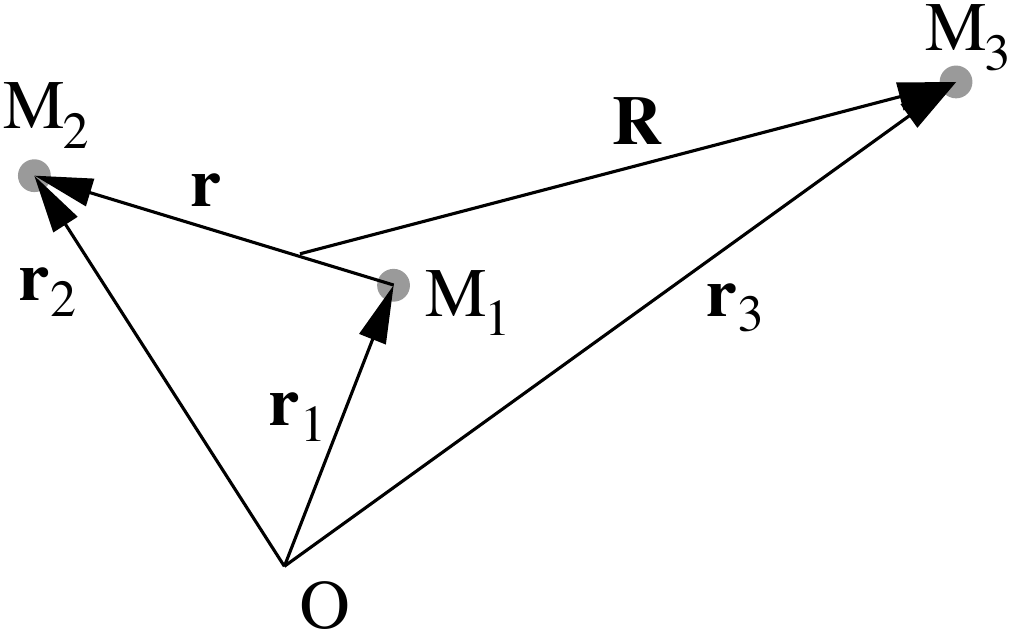}
\caption[Schematic diagram of a general 3-body system.]{Schematic diagram of three masses (grey circles) along with the position vectors to these masses from the origin (O).  In addition, the two Jacobian coordinates $\mathbf{r}$ and $\mathbf{R}$ are shown.}
\label{ThreeBodySchematic}
\end{center}
\end{figure}

We begin by deriving the equations of motion for three massive bodies.  Consider three bodes of mass $M_1$, $M_2$ and $M_3$ moving under the action of their mutual gravitational fields (see figure~\ref{ThreeBodySchematic}).  Again, using Newton's force law, the forces acting on each of the bodies shown in figure~\ref{ThreeBodySchematic} can be expressed as
\begin{align}
M_1 \frac{d^2 \mathbf{r}_1}{d t^2} &= \frac{G M_1 M_2}{\left|\mathbf{r}_2 - \mathbf{r}_1\right|^3}(\mathbf{r}_2 - \mathbf{r}_1) + \frac{G M_1M_3}{\left|\mathbf{r}_3 - \mathbf{r}_1\right|^3}(\mathbf{r}_3 - \mathbf{r}_1),\label{intro_not_3bod_M1Eq}\\
M_2 \frac{d^2 \mathbf{r}_2}{d t^2} &= \frac{G M_2 M_1}{\left|\mathbf{r}_1 - \mathbf{r}_2\right|^3}(\mathbf{r}_1 - \mathbf{r}_2) + \frac{G M_2M_3}{\left|\mathbf{r}_3 - \mathbf{r}_2\right|^3}(\mathbf{r}_3 - \mathbf{r}_2),\label{intro_not_3bod_M2Eq}\\
M_3 \frac{d^2 \mathbf{r}_3}{d t^2} &= \frac{G M_3 M_1}{\left|\mathbf{r}_1 - \mathbf{r}_3\right|^3}(\mathbf{r}_1 - \mathbf{r}_3) + \frac{G M_3M_2}{\left|\mathbf{r}_2 - \mathbf{r}_3\right|^3}(\mathbf{r}_2 - \mathbf{r}_3).
\label{intro_not_3bod_M3Eq}
\end{align}

Equations~\eqref{intro_not_3bod_M1Eq} to \eqref{intro_not_3bod_M3Eq} can be rewritten in terms of Jacobian coordinates, that is, $\mathbf{r}$, the vector from $M_1$ to $M_2$, and $\mathbf{R}$, the vector from the center of mass of $M_1$ and $M_2$ to $M_3$, defined as
\begin{align}
\mathbf{r} &= \mathbf{r}_2 - \mathbf{r}_1, \\
\mathbf{R} &= -\frac{M_1}{M_{1} + M_{2}}\mathbf{r}_1 - \frac{M_2}{M_{1} + M_{1}}\mathbf{r}_2 + \mathbf{r}_3,
\end{align}
where $\mathbf{r}$ and $\mathbf{R}$ are shown in figure~\ref{ThreeBodySchematic}.

Rewriting equations~\eqref{intro_not_3bod_M1Eq} to \eqref{intro_not_3bod_M3Eq} in terms of $\mathbf{r}$ and $\mathbf{R}$ gives
\begin{multline}
\frac{d^2 \mathbf{r}}{d t^2} + \frac{G M_{12} }{ r^3} \mathbf{r} =  - \frac{G M_3}{\left|\mathbf{R}  +  \frac{M_2}{M_{12}}\mathbf{r}\right|^3}\left(\mathbf{R}  +  \frac{M_2}{M_{12}}\mathbf{r}\right) \\+ \frac{G M_3}{\left|\mathbf{R} - \frac{M_1}{M_{12}}\mathbf{r}\right|^3} \left(\mathbf{R} - \frac{M_1}{M_{12}}\mathbf{r}\right), \label{intro_not_3bod_rEq}
\end{multline}
\begin{multline}
\frac{d^2 \mathbf{R}}{d t^2} + \frac{G M_{123} }{R^3} \mathbf{R} = \frac{G M_{123} }{R^3} \mathbf{R} - \frac{G M_{123}}{\left|\mathbf{R} + \frac{M_2}{M_{12}}\mathbf{r} \right|^3}\left(\mathbf{R} + \frac{M_2}{M_{12}}\mathbf{r} \right) \\
- \frac{G M_{123}}{\left|\mathbf{R}  - \frac{M_1}{M_{12}}\mathbf{r} \right|^3} \left(\mathbf{R}  - \frac{M_1}{M_{12}}\mathbf{r}\right),  \label{intro_not_3bod_REq}
\end{multline}
where $M_{12}$ is defined as $M_1 + M_2$, $M_{123}$ is defined as $M_1 + M_2 + M_3$ and where, again, one of the equations has dropped out as a consequence of conservation of momentum.

Equations~\eqref{intro_not_3bod_rEq} and \eqref{intro_not_3bod_REq} can be simplified further by writing the right hand sides in terms of a function known as the disturbing function.  Following \citet{Mardling2008} we define the disturbing function, $\mathcal{R}$, to be
\begin{equation}
\mathcal{R} = -\frac{G M_{12} M_3}{R} + \frac{G M_2
M_3}{| \mathbf{R} - \frac{M_1}{M_{12}}\mathbf{r} |} + \frac{G M_1M_3}{| \mathbf{R} + \frac{M_2}{M_{12}}\mathbf{r} |}
\label{intro_not_3bod_DistFunDef}
\end{equation}
where we note that this definition of the disturbing function has units of energy.  Simplifying equations~\eqref{intro_not_3bod_rEq} and \eqref{intro_not_3bod_REq} using equation~\eqref{intro_not_3bod_DistFunDef} gives
\begin{align}
\frac{M_1 M_2}{M_{12}} \frac{d^2 \mathbf{r}}{d t^2} + \frac{G M_1 M_2}{r^2}\mathbf
{\hat{r}} =&\frac{\partial \mathcal{R}}{\partial \mathbf{r}},\label{Int-Rev-req}\\
\frac{M_{12} M_3}{M_{123}} \frac{d^2 \mathbf{R}}{d t^2}+ \frac{G \left(M_1 + M_2\right)
M_3}{R^2}\mathbf{\hat{R}} &=\frac{\partial \mathcal{R}}{\partial
\mathbf{R}},\label{Int-Rev-Req}
\end{align}
where $r = |\mathbf{r}|$ and $R = |\mathbf{R}|$, where $\mathbf{\hat{r}}$ and $\mathbf {\hat{R}}$ are unit vectors in the directions of $\mathbf{r}$ and $\mathbf{R}$ respectively.

In addition, $\mathbf{r}=\mathbf{r}(x,y,z)$ where the $xyz$ coordinate system has its origin at $M_1$ and $\mathbf{R}=\mathbf{R}(X,Y,Z)$ where the $XYZ$ coordinate system has its origin at the center of mass of $M_1$ and $M_2$, and thus
\begin{align*}
\frac{\partial}{\partial \mathbf{r}} =& \mathbf{i}
\frac{\partial}{\partial x} + \mathbf{j} \frac{\partial}{\partial y}
+ \mathbf{k} \frac{\partial}{\partial z},\\
\frac{\partial}{\partial \mathbf{R}} =& \mathbf{i}
\frac{\partial}{\partial X} + \mathbf{j} \frac{\partial}{\partial Y}
+ \mathbf{k} \frac{\partial}{\partial Z}.
\end{align*}

For the case where the three-body system is hierarchical, that is, the orbit of $M_1$ and $M_2$ about their common barycenter and the orbit of $M_3$ and the $M_1$-$M_2$ barycenter about the system barycenter are described by perturbed two-body motion, these equations have physical meaning.  This can be seen in the structure of the equations.  Conceptually, equations~\eqref{Int-Rev-req} and \eqref{Int-Rev-Req} have two components. The first component consists of the left hand sides of equations~\eqref{Int-Rev-req} and \eqref{Int-Rev-Req}. If the right hand side of the equations were neglected, then both equations would be mathematically equivalent to equation~\eqref{intro_not_2bod_govEq}, the equation for two-body motion.  Consequently, for this case, the motion of $M_1$ and $M_2$ around the $M_1$-$M_2$ barycenter, the ``inner" orbit, would be described by conic sections.  In addition, the motion of $M_3$ and the $M_1$-$M_2$ barycenter around the system barycenter, the ``outer" orbit, would be described by different conic sections.  The second component consists of the terms on the right hand sides of equations~\eqref{Int-Rev-req} and \eqref{Int-Rev-Req}.  These terms allow the ``inner" and ``outer" orbits to interact.  

Now that general-three body theory has been introduced, we are finally in a position to apply it to the case of a star-planet-moon system.

\subsection{Three-body motion for the case of a star, planet and moon}

\begin{figure}[tb]
\begin{center}
\includegraphics[width=.70\textwidth]{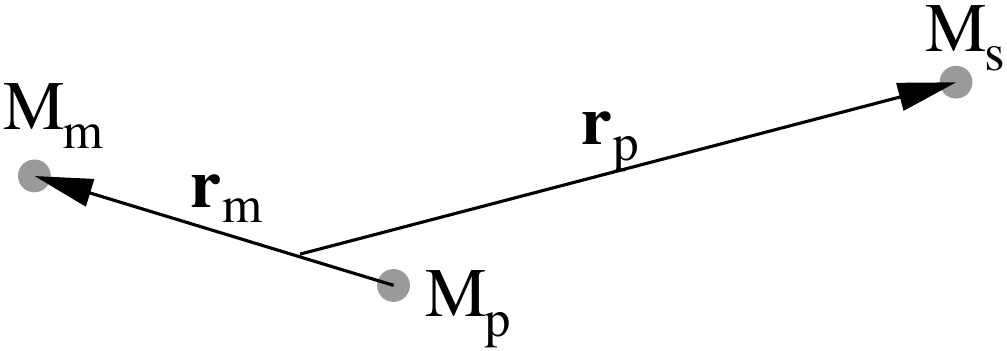}
\caption[Schematic diagram of a 3-body system consisting of a star, planet and moon.]{Schematic diagram of a star-planet-moon three-body system.  As in figures~\ref{TwoBodySchematic} and \ref{ThreeBodySchematic}, the masses are shown as grey circles and labeled by subscripts.  In addition, the two Jacobian coordinates are shown and written as $\mathbf{r}_m$ and $\mathbf{r}_p$ (see text).}
\label{StarPlanetMoonThreeBodySchematic}
\end{center}
\end{figure}

For the case of a planet-moon pair in orbit around a star, we can associate $M_1$ with the planet, $M_2$ with the moon and $M_3$ with the star.  For this case, the ``inner orbit" would describe the elliptical orbit of the planet and moon about their common barycenter while the ``outer orbit" would describe the elliptical orbit of the planet-moon pair, and the star, about the system barycenter.  While the terms ``inner" and ``outer" are general, they are not very intuitive, and it is not immediately obvious how the orbital elements of these orbits should be labeled.  Consequently, for this thesis we will call the ``inner" orbit, the ``moon's" orbit and label the vector and orbital elements associated with it with a subscripted $m$, and call the ``outer" orbit, the ``planet's" orbit and label the vector and the orbital elements associated with it with a subscripted $p$.  Consequently, using this notation, the two governing equations are given by
\begin{align}
\frac{M_m M_p}{M_p + M_m} \frac{d^2 \mathbf{r}_m}{d t^2} + \frac{G M_m M_p}{r_m^2}\mathbf
{\hat{r}}_m =&\frac{\partial \mathcal{R}}{\partial \mathbf{r}_m},\label{Int-Rev-rmeq}\\
\frac{(M_p + M_m) M_s}{M_s + M_p + M_m} \frac{d^2 \mathbf{r}_p}{d t^2}+ \frac{G \left(M_p + M_m\right)
M_s}{r_p^2}\mathbf{\hat{r}}_p &=\frac{\partial \mathcal{R}}{\partial
\mathbf{r}_p},\label{Int-Rev-rpeq}
\end{align}
where $\mathbf{r}_m$ and $\mathbf{r}_p$ are shown in figure~\ref{StarPlanetMoonThreeBodySchematic} and where
\begin{equation}
\mathcal{R} = -\frac{G (M_{p} + M_{m}) M_s}{r_p} + \frac{G M_m
M_s}{| \mathbf{r}_p - \frac{M_p}{M_{p} + M_{m}}\mathbf{r}_m |} + \frac{G M_pM_s}{| \mathbf{r}_p + \frac{M_m}{M_{p} + M_{m}}\mathbf{r_m} |}.
\label{intro_not_spm_DistFunDef}
\end{equation}

While this notation is being used, it should be pointed out that the results of this work are still entirely general, in that the ratio of the moon's mass to the planet's mass can freely range from zero to one.  For example, if the semi-major axis of the orbit of the moon about the planet-moon barycenter is required, it will be written as $(M_p/(M_p + M_m))a_m$.  Similarly, the semi-major axis of the orbit of the planet about the planet-moon barycenter is given by $(M_m/(M_p + M_m))a_m$.  Recalling that the six orbital elements which uniquely define the ellipse of a two body orbit are $a$, $e$, $I$, $\omega$, $\Omega$, and $f(0)$, we have that, the orbital elements of the planet's orbit are given by $a_p$, $e_p$, $I_p$, $\omega_p$, $\Omega_p$ and $f_p(0)$, while the orbital elements of the moon's orbit are given by $a_m$, $e_m$, $I_m$, $\omega_m$, $\Omega_m$ and $f_m(0)$.

In addition to general notation defining the physical and orbital properties of the star, planet and moon, there is one addition situation where additional notation is required, namely, the description of transit light curves.

\section{Notation specific to transit light curves}\label{Intro_Moons_Note_Trans}

\begin{figure}[tb]
\begin{center}
\includegraphics[width=.70\textwidth]{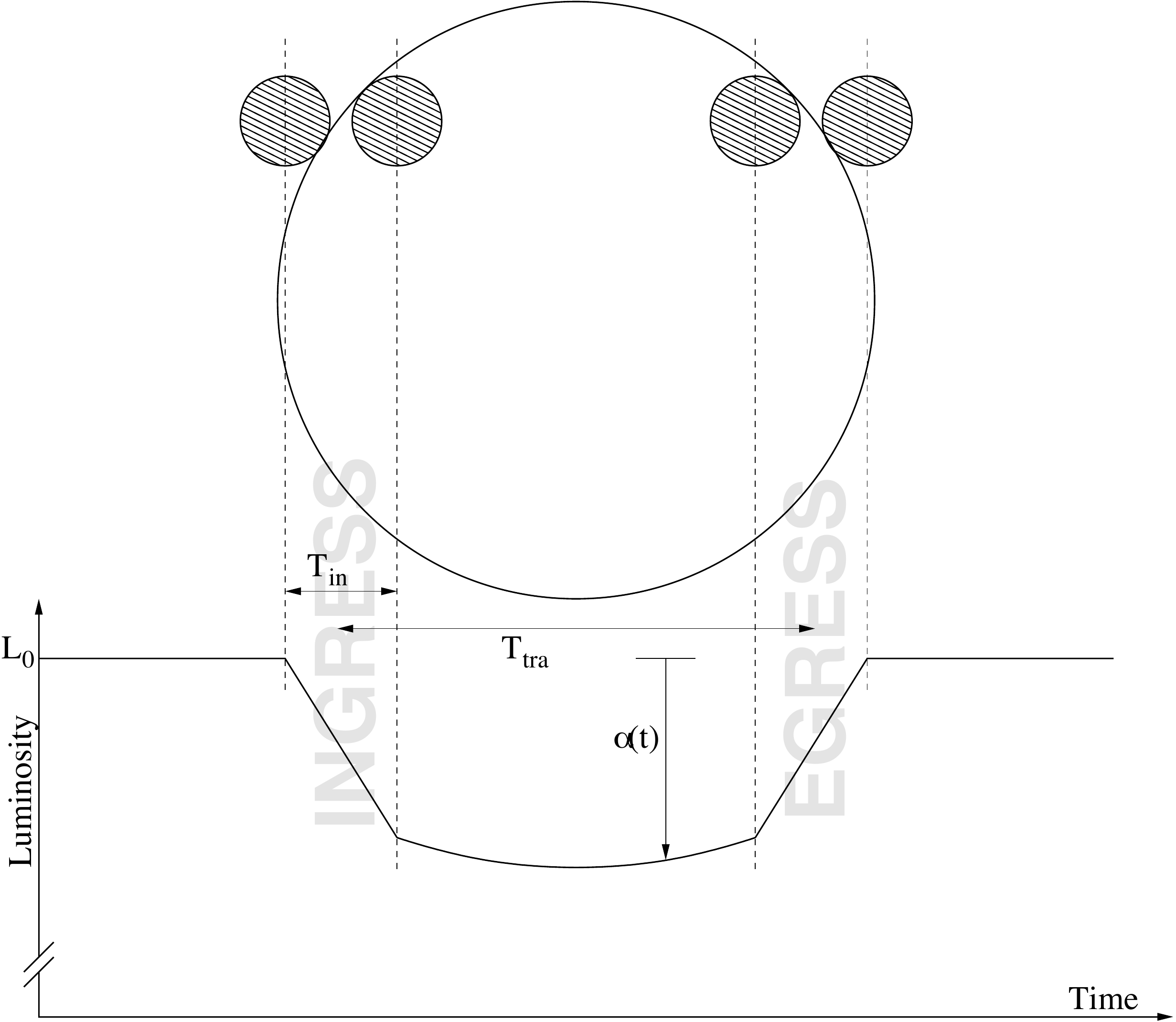}
\caption[Schematic diagram of a transit light curve for the case of a lone planet.]{Diagram showing the different portions of the transit light curve.  The four shaded circles show the planet's position across the face of the star at the beginning and end of ingress, and the beginning and end of egress.  As the position of the planet along the chord of the star is a linear function of time, it can be used as a proxy for time.  Consequently the position of the planet and the value of the light curve resulting from that position are linked by dashed lines.}
\label{IntroTransSch}
\end{center}
\end{figure}

 \begin{sidewaystable}[tbp]
 
   \begin{tabular}{l@{}c@{}c@{}c@{}c@{}c@{}cc@{}c@{}c@{}c}
   \hline
   		& \multicolumn{6}{c}{Transit Light Curve Parameters} & \multicolumn{4}{c}{Geometric Parameters} \\
 Paper & \includegraphics[width=.07\textwidth]{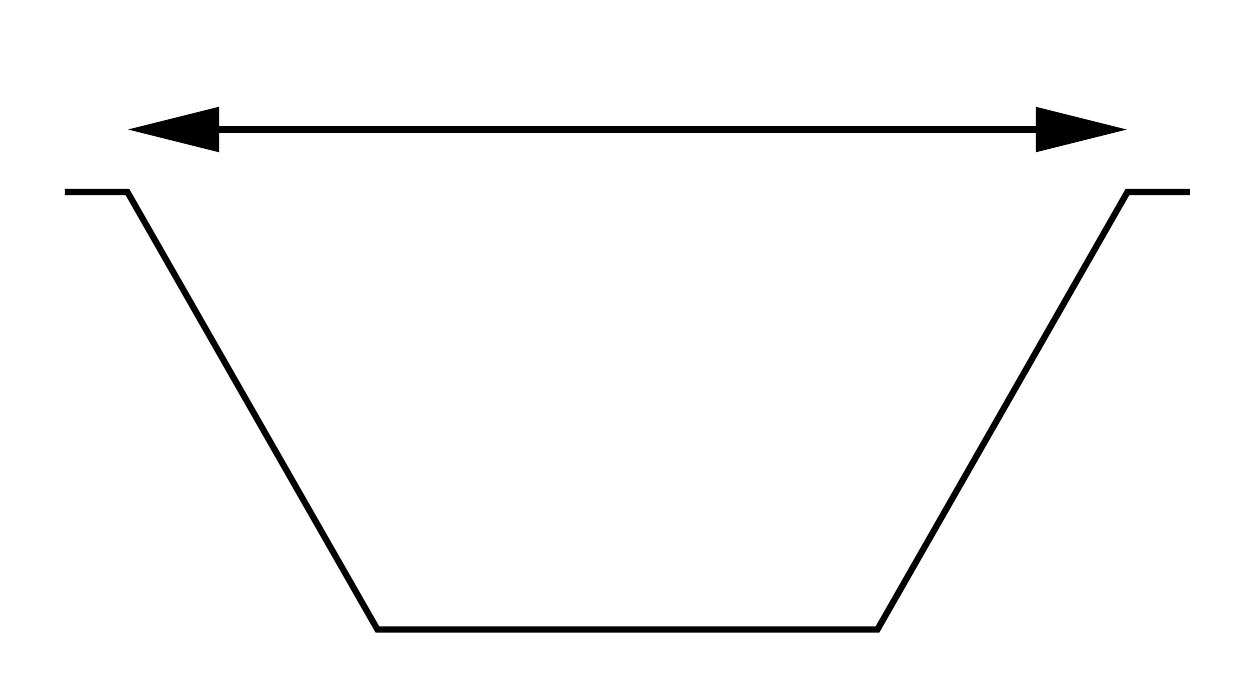} & \includegraphics[width=.07\textwidth]{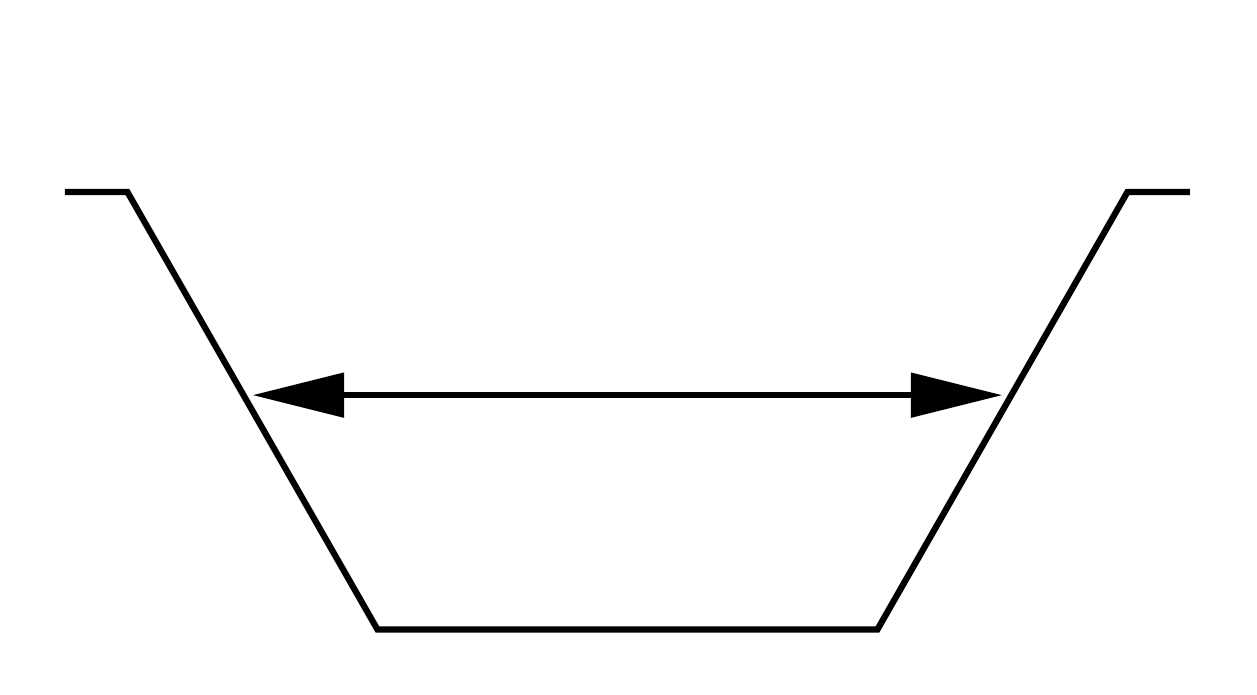} &\includegraphics[width=.07\textwidth]{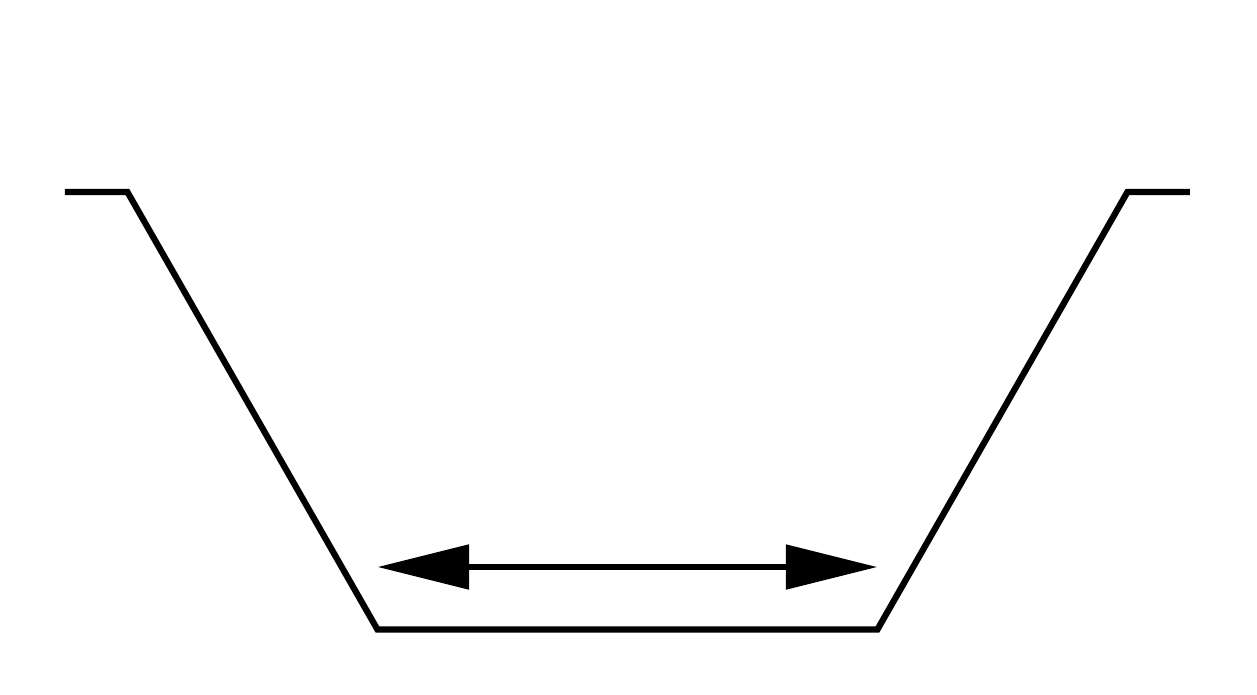} & \includegraphics[width=.07\textwidth]{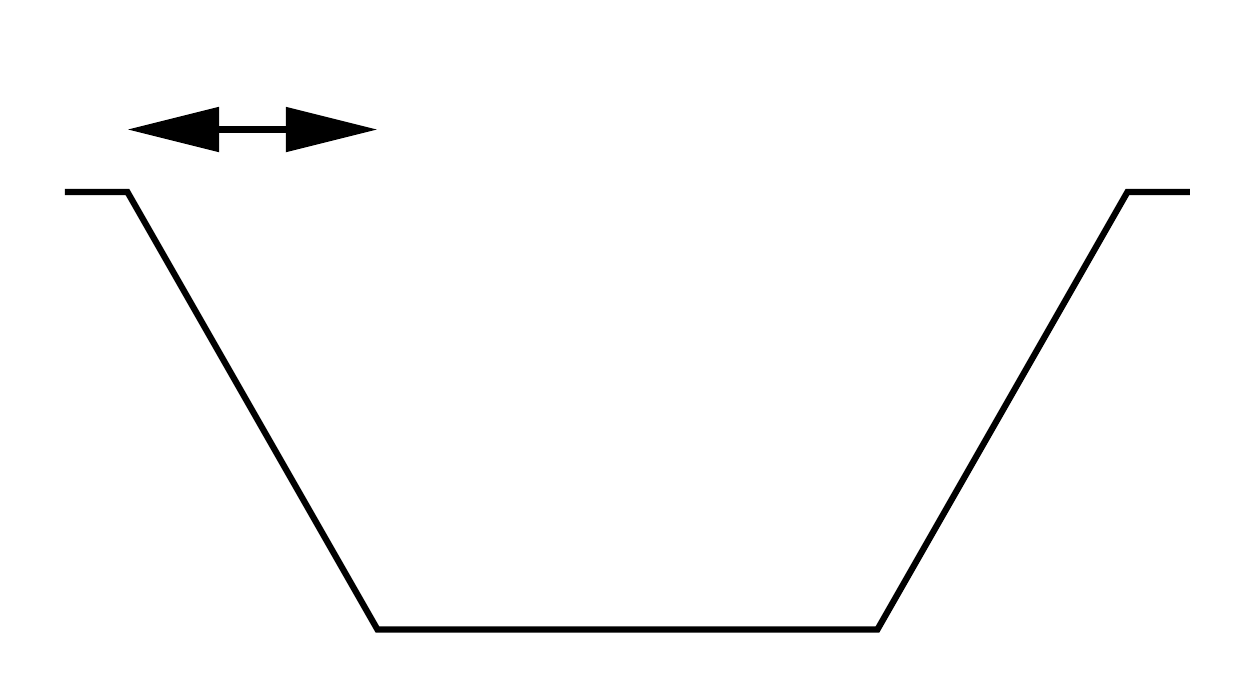} & \includegraphics[width=.07\textwidth]{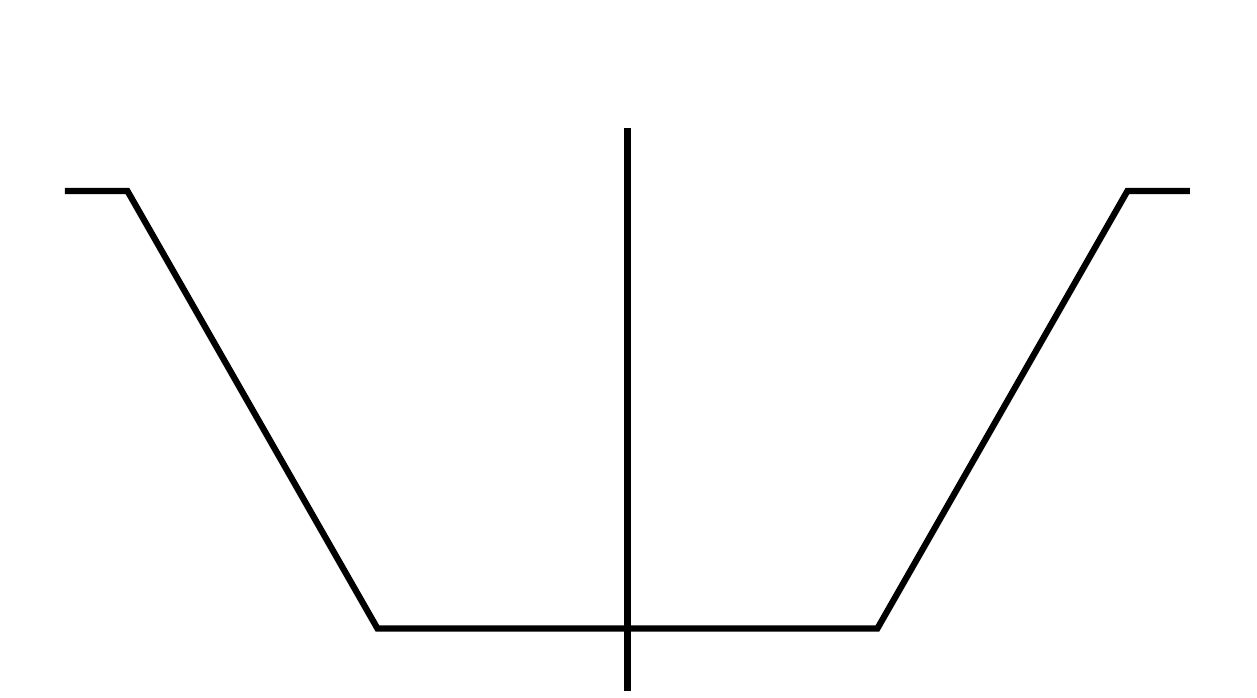} & \includegraphics[width=.07\textwidth]{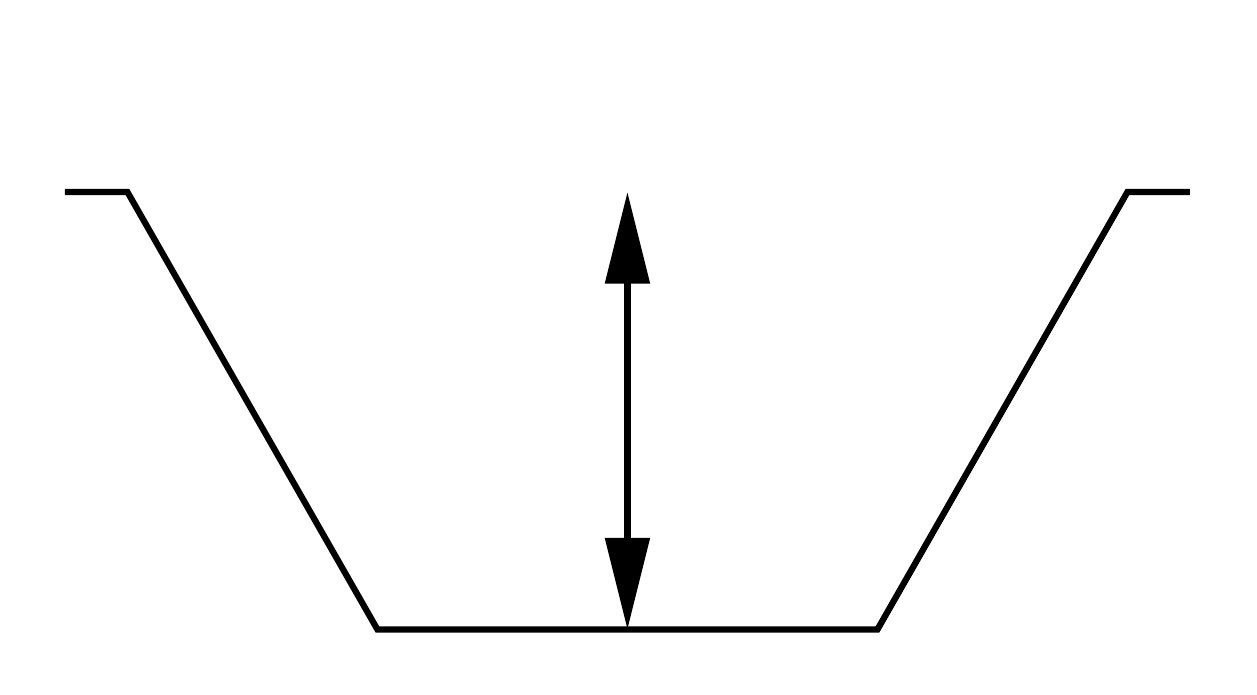} & \includegraphics[width=.07\textwidth]{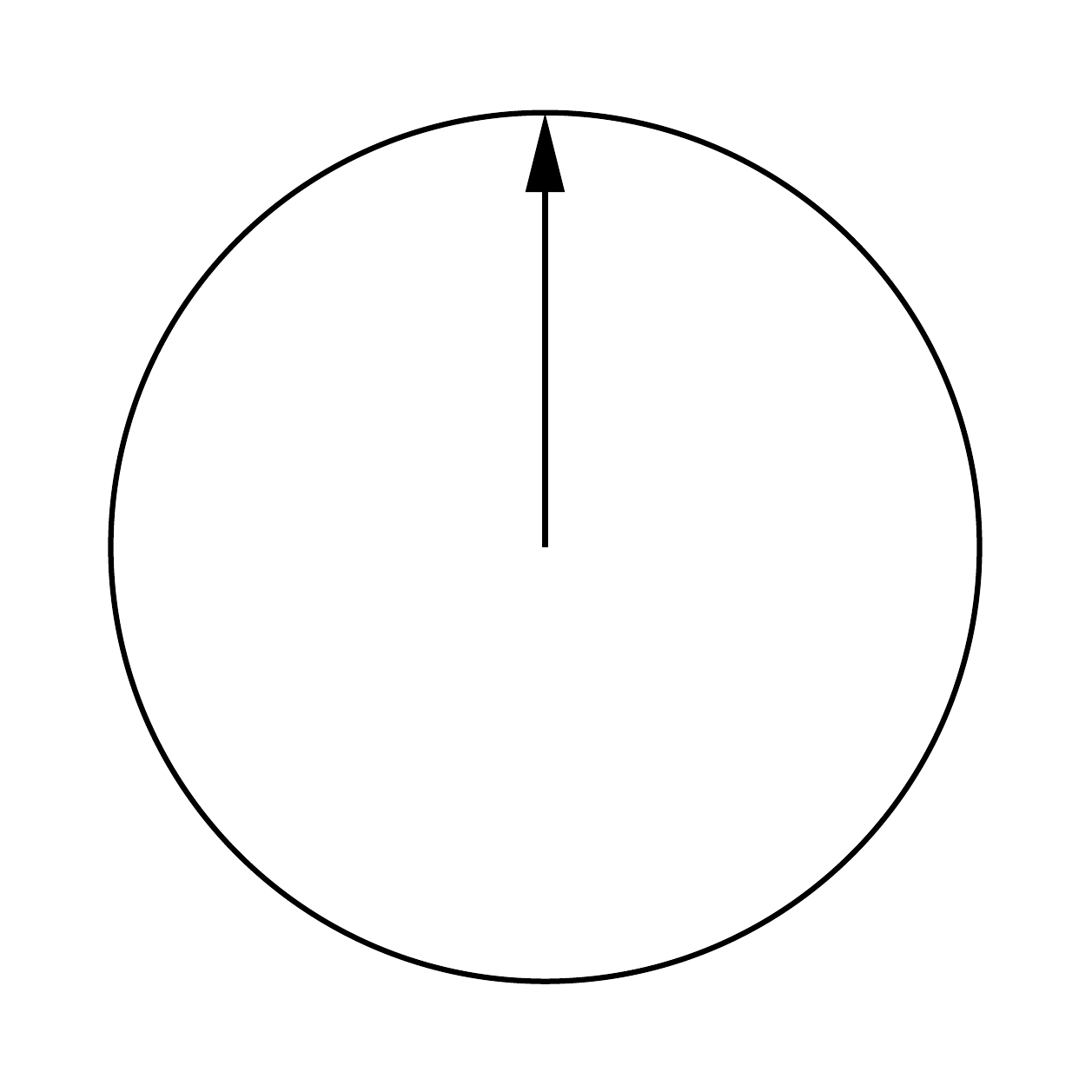} & \includegraphics[width=.07\textwidth]{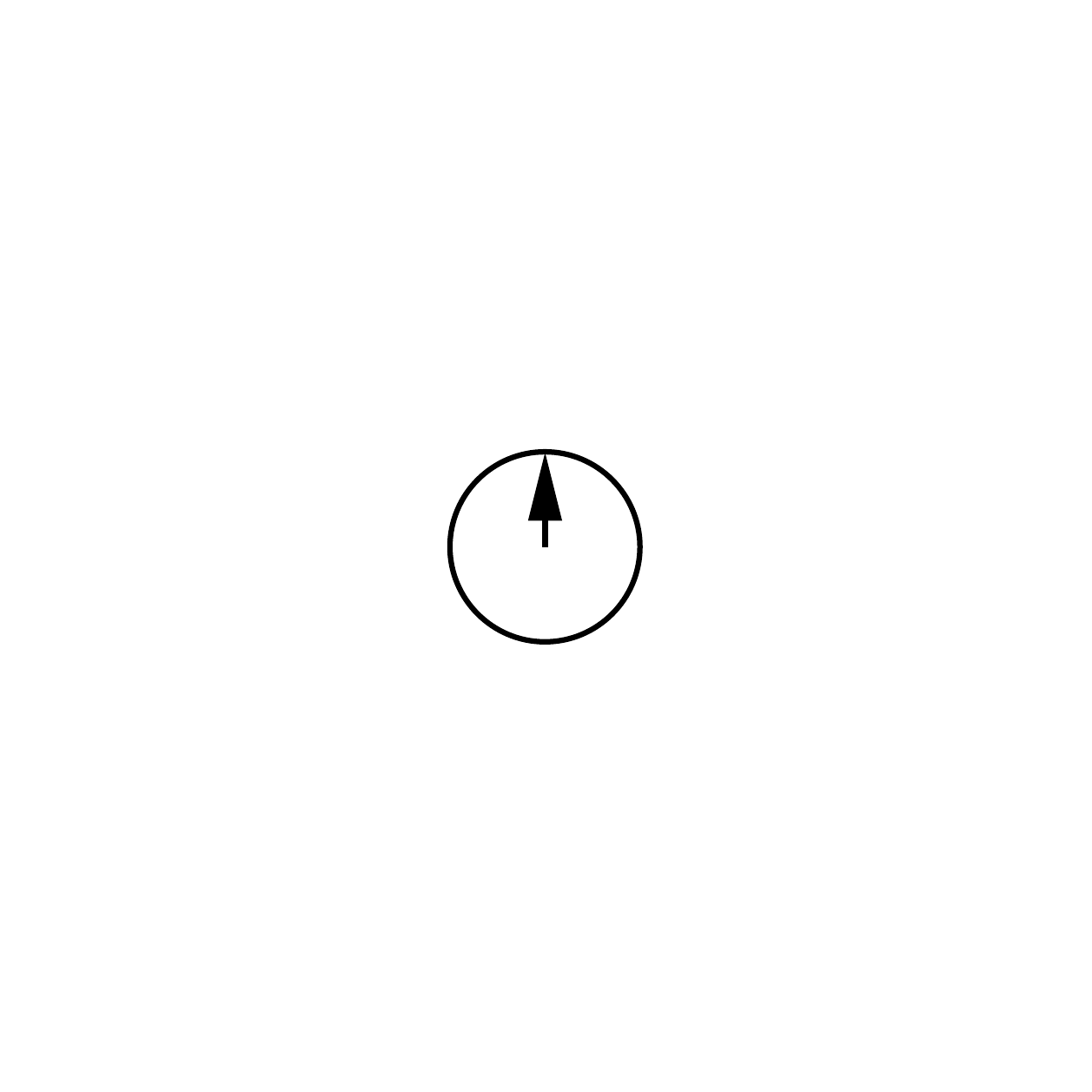} &\includegraphics[width=.07\textwidth]{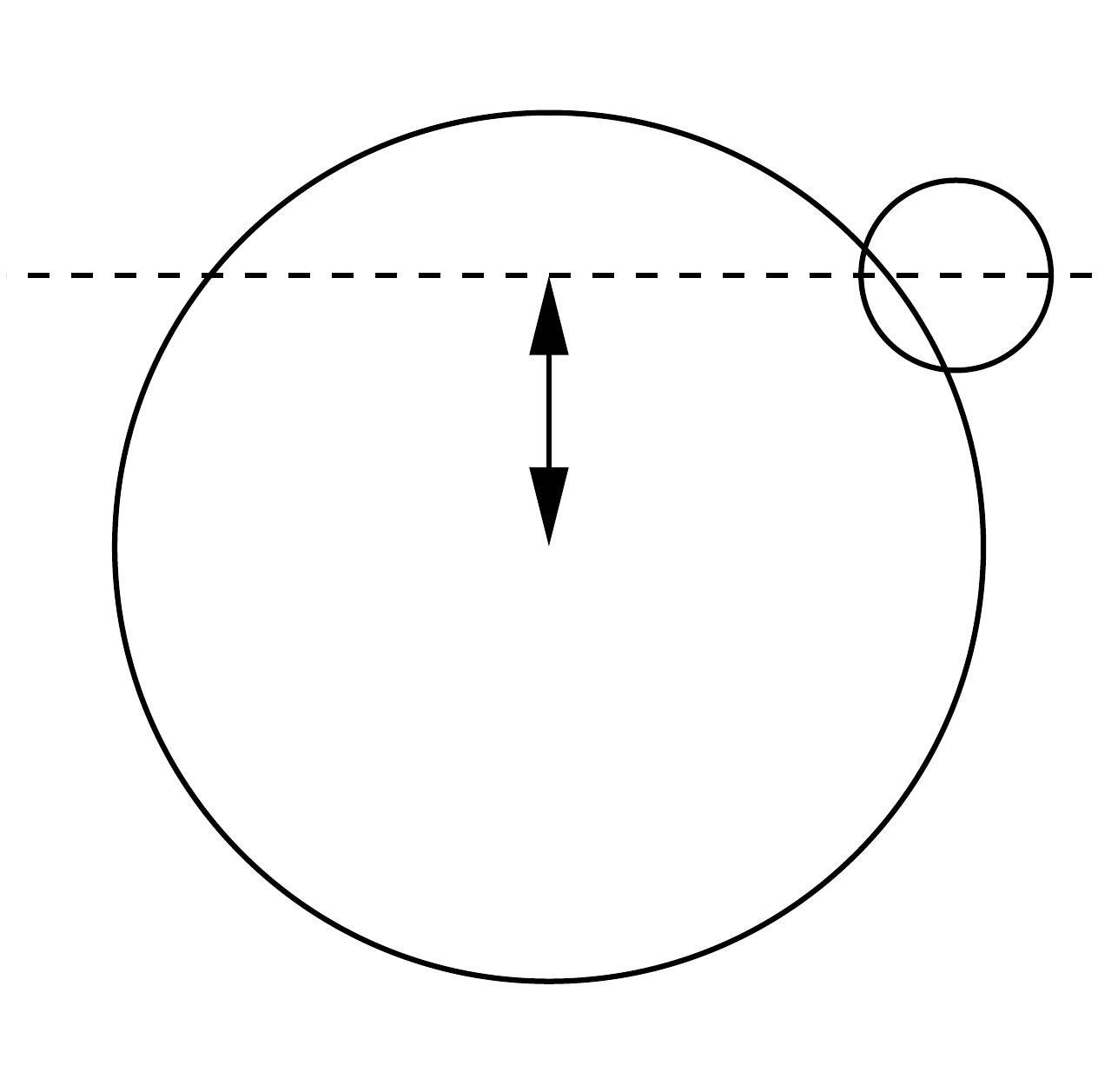} & \includegraphics[width=.07\textwidth]{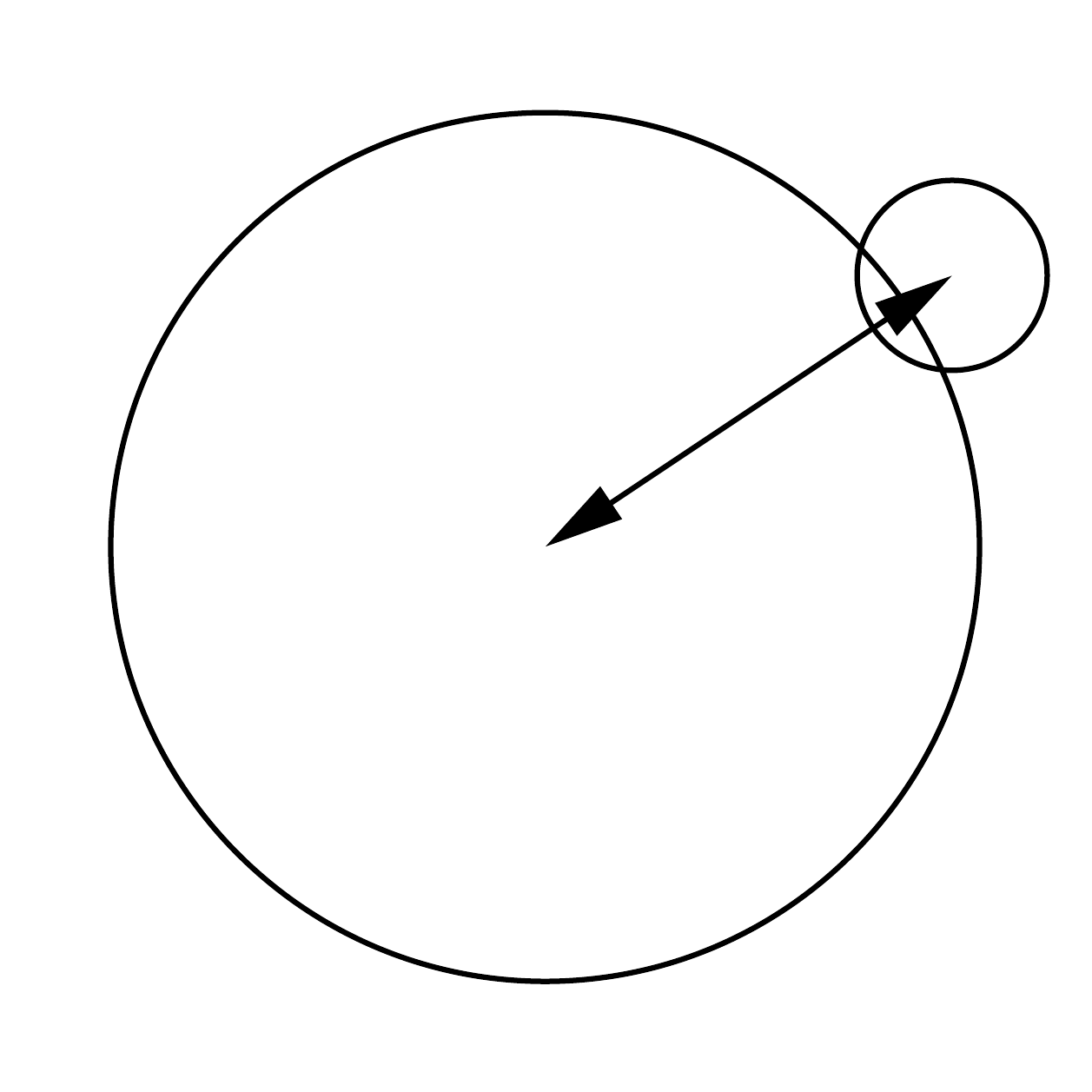}  \\
   \hline
\citet{Deegetal2001}  		& $T_{14}$ 	& --         	& $T_{12}$? 	& $T_{23}$? & $T_c$ & -- & $R^*$ 	& $R_{pl}$         	& $h$ 	&  --\\
\citet{MandelAgol2002}  	& --		 	& --         	& --		 	& -- & -- & $1-F(t)$& $r_*$		 	& $r_p$         	& --		 	& $d(t)$ \\
\citet{Seageretal2003}  	& $t_T$		& --         	& $t_F$		 	& -- & -- & $\Delta F$& $R_*$		& --         	& $b$		 	& -- \\
  \citet{TingleySackett2005}  & $D$ 		& --      	&    --  		& -- & -- & -- & $R_1$ 		& $R_2$     	&    --  		& --  \\
  \citet{Gimenez2006} & --		 	& --         	& --		 	& -- & -- & $\alpha(t)$ & $r_s$		 	& $r_p$        	& --		 	& $\delta(t)$ \\
  \citet{Carteretal2008}		& --		 	& T         	& --		 	& $\tau$ & $t_c$ & $\delta$ & $R_*$		 	& $R_p$         	& --		 	& --  \\
 \citet{Sartorettietal1999}	& -- 		& --         	& -- 			& -- & -- & $\Delta F_*$ & $r_{*}$ 		& $r_0$         	& -- & $r(t)$  \\
  \citet{Deeg2002}  		& $t_{tr}$ 		& --         	& -- 			& -- & $t_0$ & $\Delta L$ & $R^{*}$ 		& $R_{pl}$         	& $R^* \cos \delta$ 			& --  \\
  \citet{Szaboetal2006}  & -- 		& --         	& -- 			& -- & $\tau_0$ & $\Delta m(t_i)$ & -- 		& --         	& -- 			& --  \\
\citet{Kippingetal2009}  & $t_T$ 		& --         	& -- 			& -- & $T_{MID}$ & -- & $R_*$ 		& $R_p$         	& -- 			& $p(t)$  \\
 \hline
 This work			& -- 		& $T_{tra}$         	& -- 		& $T_{in}$ & $t_{mid}$ & $\alpha(t)$ & $R_s$ 		& $R_p$         	& $\delta_{min}$ 			& $\delta(t)$   \\
 \hline
  \end{tabular}\\
 \caption[Table showing the range of transit light curve notation in use for a representative selection of works.]{Table showing the range of transit light curve notation in use for a representative selection of works.  As parameters, for example, the transit duration, have different definitions in different works, their definition is indicated by a cartoon schematic of the light curve, for the case of light curve parameters, or of the star and planet, for the case of geometric parameters.  If a quantity is not explicitly defined in a given work, this is shown in two different ways.  For the case where the notation is sufficiently logical such that the variable that would have been used to represent this quantity can be guessed, the notation is given, followed by a question mark.  For the case where no such guess can be made, the lack is indicated by a dash.}
 \label{TransNotStyles}
 \end{sidewaystable}

A planetary transit occurs when a planet passes between the observer and the face of its host star, and consequently blocks some of that star's light, and a transit light curve is the measured luminosity of a star undergoing a planetary transit.  A schematic of a sample transit light curve is shown in figure~\ref{IntroTransSch}.  As can be seen in figure~\ref{IntroTransSch}, the transit itself consists of three main stages, first the ingress, where the disk of the planet is passing onto the face of the star, second, the main part of the transit, where disk of the planet fully overlaps with the disk of the star, and third, the egress, where the disk of the planet is passing off the face of the star.

In particular, the shape of the light curve contains information about the planet and its orbit, for example, the duration of ingress and egress contains information about the size of the planet and the inclination of the orbit, and the dip depth tells us about the relative size of the planet compared to the star.  In addition to information about the planet, the shape of this light curve also determines the effectiveness of detection of moons using both the method analysed in this thesis, as well as other methods in the literature.   So, in order to investigate these methods, we need to be able to describe the shape of the transit light curve.  

Consequently, the question arises of what notation to use to describe this transit light curve.  In particular, we would like a notation set which is in general use, self-consistent and optimised for describing moon detection.  These issues will be addressed in turn, and then used to decide on a notation system.
 
From a practical perspective we would like a notation set which is easy to understand.  This can be partially ensured if it is already currently in general use in the transiting planet literature.  When considering this goal, two questions naturally arise, 	``What notation is currently in general use in the transiting planet community?" and ``What do we mean by transiting planet community?".  To address the first question, a literature review was conducted, focussing on notation styles.  For reference, table~\ref{TransNotStyles} shows a representative, but by no means a complete list of notation styles in use.  As can be seen from table~\ref{TransNotStyles}, while there are some trends, for example, transit duration is generally represented by a $t$ or a $T$ while radii are represented by a $R$ or $r$ with a subscript,\footnote{Recall that for this thesis it was decided to use $R_s$, $R_p$ and $R_m$ for the radius of the star, planet and moon respectively.} there is no universally accepted notation style.  The origin of this broad range of notation styles, can partially be understood in terms of the answer to the second question.  Within the transiting planet community (and the list of works in table~\ref{TransNotStyles}), there are authors who are interested in the shape of transit light curves, to help streamline the detection process \citep[e.g.][]{TingleySackett2005}, to determine properties of the planet and star \citep[e.g.][]{Seageretal2003} or because they are related to eclipsing binary light curves \citep[e.g.][]{Gimenez2006}.  In addition,  a number of authors are not interested in the shape of the light curve per se, but more in perturbations in the mid-time or duration of transits as they can be used to detect additional bodies such as planets \citep[e.g.][]{HolmanMurray2005} or even moons \citep[e.g.][]{Kippingetal2009}.  From within these groups, notation clusters start to emerge, for example, \citet{Kippingetal2009} and \citet{Carteretal2008} use the similar notation because \citet{Kippingetal2009} adopted some of their notation from \citet{Carteretal2008}.  In addition, work from researchers from other fields carries with it the notation used in those fields, for example, the notation of \citet{Gimenez2006} is inherited from the work of \citet{Kopal1979} on binary star eclipse light curves.
  
In addition to selecting notation which is generally understood, it must also be self-consistent and also consistent with the notation already in use for this thesis.  We will consider the issues of self-consistency and being consistent with the notation already in use, in turn.  As none of the notation sets contain all the variables required for this work, a hybrid, self-consistent notation, combining two or more of these sets must be constructed.  As can be seen from table~\ref{TransNotStyles}, if some sets of notation were combined, it would lead to confusion.  For example, consider a combination of the light curve notation of \citet{Carteretal2008} with the geometric notation of \citet{Gimenez2006}. As \citet{Carteretal2008} use $\delta$ for the depth of the dip and \citet{Gimenez2006} use $\delta(t)$ for the distance between the planet and the star on the plane of the sky, the variable $\delta$ is not uniquely defined.  In addition to selecting a self-consistent set of notation, we would also like a notation set which is compatible with the notation discussed in the previous section and with that used in the literature with respect to photometric transit timing.  For example, the notation used by \citet{Carteretal2008} cannot be used in this work as the symbol they selected for the duration of transit ingress, $\tau$, is the same symbol used by \citet{Szaboetal2006} to describe their timing statistic.  
    
In addition to selecting comprehensible self-consistent notation, the notation selected should be able to easily describe the transiting moon detection method we are focussing on in this thesis, photometric transit timing, but also be able to describe moon detection methods in general for comparison purposes.  The first issue to address is the different ways that different works define light curve parameters, for example \citet{Seageretal2003} define the transit depth to be constant, while \citet{MandelAgol2002} allow it to be a function of time, and \citet{Carteretal2008} define the transit duration as lasting from middle of ingress to the middle of egress, while \citet{TingleySackett2005} define the transit duration as lasting from the beginning of ingress to the end of egress.  Consequently, we need to select the set of light curve parameter definitions which most easily describe the quantities that we want.  

For the case of photometric transit timing we will need to sum over the transit depth for each exposure in the transit, as well as other exposures outside the planetary transit (see equation~\eqref{TraM-TTV-taudef}).  As we will see in section~\ref{Intro_Dect_Moons_Transit}, moons lead to additional dips in the transit light curve, with the dips caused by the moons stretched or compressed and translated from that of their host planet.  To describe these lumpy light curves we need a notation which allows the transit depth to be a function of time.  In particular it was decided to describe the geometry of the transit and the transit depth using the notation of \citet{Gimenez2006}, partially as it is one of the notation styles to describe the dip depth as a function of time, and partially as it allows easy access to a range of mathematical methods, for example, describing dip depth as a surface integral.  Consequently for this thesis, the transit depth is given by $\alpha(t)$, with $\alpha_p(t)$  and $\alpha_m(t)$ representing the portion of the dip resulting from the transit of the planet and moon respectively.  In addition $\delta_p(t)$ and $\delta_m(t)$ represent the projected distance between the center of the star and the planet and moon respectively, while $\delta_{min}$ represents the impact parameter for the transit of both the planet and the moon, as both values are approximately the same for all cases considered in this thesis (see section~\ref{Transit_Signal_Coord_Orient}).  In addition to describing the transit depth, the duration of the transit of the planet and moon must also be described.

Recalling the range of definitions of transit duration given in table~\ref{TransNotStyles}, it can be seen that in order to describe this quantity, both a definition and appropriate notation must be selected.  For this thesis it was decided to define the transit as beginning and ending when the center of the body passes onto and off the stellar limb, and the transit duration as the difference between these two times, that is, the definition shown in column 2 of table~\ref{TransNotStyles}.  This particular definition was selected to reduce the algebraic complexity of the expressions derived in chapter~\ref{Transit_Signal}, by ensuring that transit duration did not depend on the radius of the transiting body.  Informed by the notation styles presented in table~\ref{TransNotStyles}, it was decided to represent the transit duration by $T_{tra}$, with $T_{tra}$ describing the transit duration of a planet with no moon, and with $T_{tra,p}$ and $T_{tra,m}$ denoting the duration of the planet's and moon's transit for the case of a planet with a moon.

In addition, for the case of moon detection using other methods, other light curve parameters also need to be described.  In addition to being able to describe the dip due to the moon and the transit duration, quantities for which notation has already been selected, it would also be useful to be able to describe the duration of ingress and the transit mid-time.  Unfortunately, the only work which explicitly describes the duration of ingress and egress, uses the notation $\tau$, which we cannot use as it is the test statistic for photometric transit timing.  Consequently, the notation selected above was extended such that $T_{in}$ describes the duration of ingress.  Again $T_{in}$ is understood to describe the duration of ingress for the case of a planet with no moon, and with $T_{in,p}$ and $T_{in,m}$ denote the duration of ingress for the case of the transit of the planet and moon the case of a planet with a moon.  Finally, the mid-time of the transit is given by $t_{mid}$, where again $t_{mid}$ represents the transit mid-time for the case of a planet with no moon, while $t_{mid,p}$ and $t_{mid,m}$ represent the transit mid-times for the transit of the planet and moon respectively for the case of a planet and a moon.  In addition, for the 0$^{th}$ transit, $t_{mid}$ may also be written as $t_0$.  For reference each of these decisions is summarised in table~\ref{TransNotStyles} and also given in appendix \ref{VarDef_App}.
   
\section{Conclusion}

The notation that will be used for this thesis has been discussed in the context of three broad areas.  First notation for the physical properties of the star, planet and moon was discussed.  Second, the notation required to describe the orbital motion of the star, planet and moon was discussed in the context of three-body theory.  Finally, the notation required for the description of the transit light curve was selected.  In addition, for reference, the notation selected is summarised in appendix \ref{VarDef_App}.  Now that the issue of notation has been discussed and a framework decided, we can move on to start to discuss the literature, in particular, the types of moons that are likely to exist.
\chapter{Constraints on extra-solar moons}\label{Intro_Moons_Const}

\section{Introduction}

Before considering the types of moons able to be detected, and the mechanics of moon detection, it would be useful to have an understanding of the types of moons extra-solar planets are likely to possess, and in particular, the properties of any large (and consequently detectable moons).  First, we will begin this investigation by looking at the census of moons present in the Solar System.  Second, guided by these results, the characteristics of moons predicted to form will be summarised.  Third, the ways in which tidal and three body effects can modify moon orbits will be discussed.  Finally, these sources of information will be combined to provide an indication of the types of moons that extra-solar planets are likely to possess.

\section{Characteristics of moons in the Solar System}

We begin our discussion of likely moon properties by summarising the properties of the moons that we know about, the moons in the Solar System.  As planets in the Solar System are divided into two distinct types, terrestrial and gas giant, according to their planet-Sun distance, composition and formation history, the properties of moons of terrestrial planets and gas giants will be discussed separately.  We begin with a discussion of the moons of terrestrial planets.

\begin{longtable}{l@{}l@{}llll@{}ll}\small

\tabcolsep 1.8pt

\\
\\
\hline
   \multicolumn{1}{l}{``Planet"} &
   \multicolumn{1}{@{}l}{``Moon"} &
   \multicolumn{1}{@{}l}{$R_m$} &
   \multicolumn{2}{c}{$M_m$} &
   \multicolumn{1}{l}{$a_m$} &
   \multicolumn{1}{@{}l}{$e_m$} &
   \multicolumn{1}{l}{$I_m$}\\
   &  & (km) & ($10^{20}$ kg) & ($M_p$) & ($10^3$ km)  & & ($^{\circ}$)\\
   \hline
   \\[-1.8ex]
\endfirsthead

\multicolumn{3}{c}{{\tablename} \thetable{} -- Continued} \\[0.5ex]
  \hline 
   \multicolumn{1}{@{}l}{``Planet"} &
   \multicolumn{1}{@{}l}{``Moon"} &
   \multicolumn{1}{@{}l}{$R_m$} &
   \multicolumn{2}{c}{$M_m$} &
   \multicolumn{1}{l}{$a_m$} &
   \multicolumn{1}{@{}l}{$e_m$} &
   \multicolumn{1}{l@{}}{$i_m$}\\
  &  & (km) & ($10^{20}$ kg) & ($10^{-4}M_p$) & ($10^3$ km) & & ($^{\circ}$)\\
   \hline
  \\[-1.8ex]
\endhead

  \multicolumn{4}{l}{{Continued on next page \ldots}} \\
\endfoot

\endlastfoot

\rowcolor[gray]{.9} 
   Earth  & Moon      & 1738 & 734.9 & 0.0123 & 384.4 & 0.0554 & 5.16 \\
   Mars   & \small Phobos  & 11* & $1.1 \times 10^{-4}$ & $1.7 \times 10^{-8}$ & 9.376      & 0.0151 & 1.08  \\
               & Deimos  & 6* & $1.8 \times 10^{-5}$ & $2.8 \times 10^{-9}$ & 23.458    & 0.0002 & 1.79 \\
   \rowcolor[gray]{.9} 
   Pluto   & Charon  & 593 & 15 & 0.118 & 17.536    & 0.0022	& 0.001 \\
               & Nix          & -- & $5.8 \times 10^{-3}$ & $4.6 \times 10^{-5}$ & 48.708    & 0.0030	& 0.20 \\
               & Hydra     & -- & $3.2 \times 10^{-3}$ & $2.5 \times 10^{-5}$ & 64.749    & 0.0051	& 0.21 \\
               \hline
 \footnotetext[1]{http://ssd.jpl.nasa.gov/?sat\_elem.} \\
\caption[Physical and orbital properties of the satellites of the Earth, Mars and Pluto. ]{Physical and orbital properties of the satellites of the Earth, Mars and Pluto.  ``Large" satellites are shaded grey.  Radii and masses for the Moon, Phobos, Deimos and Charon are taken from \citep{Murrayetal1999}, while the masses for Nix and Hydra were taken from \citep{Tholenetal2008}.  Orbital parameters were taken from the JPL website.\footnotemark[1]  The inclinations are measured relative to the local Laplace plane.}  \label{TerMoonsTable}
\end{longtable}

\setcounter{footnote}{1}

\subsection{Moons of terrestrial planets}

\begin{figure}
     \centering
     \subfigure[Earth's satellite system.]{
          \label{fig:dl2858}
          \includegraphics[width=.98\textwidth]{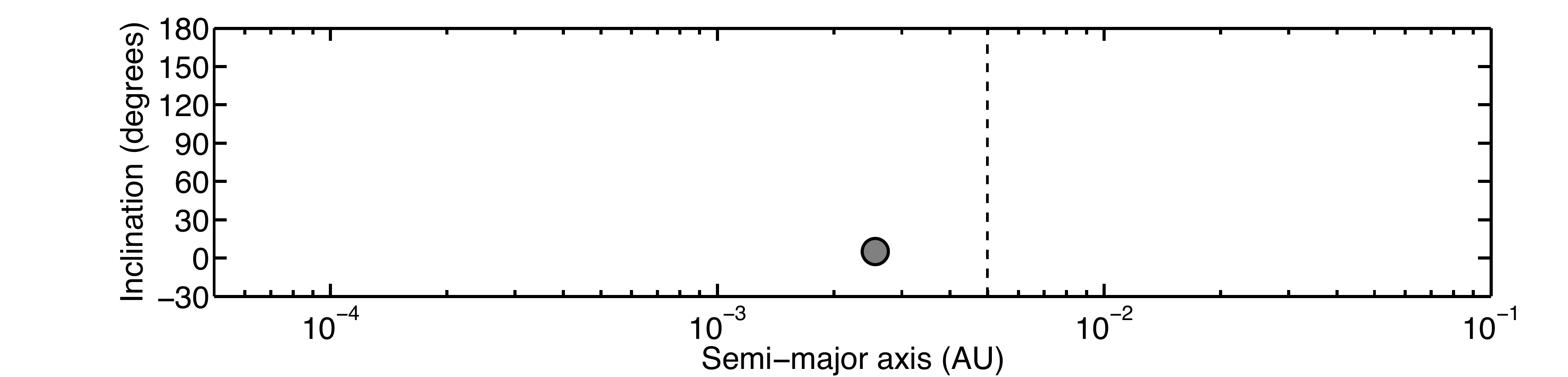}}\\
     \subfigure[Mars' satellite system.]{
          \label{fig:er2858}
          \includegraphics[width=.98\textwidth]{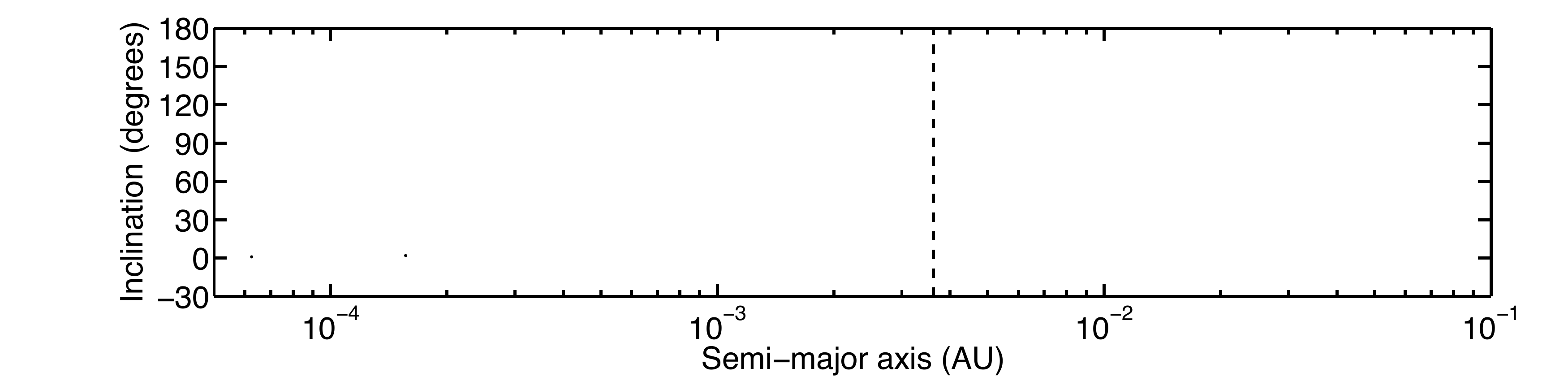}}\\
          \subfigure[Pluto's satellite system.]{
          \label{fig:dl2858}
          \includegraphics[width=.98\textwidth]{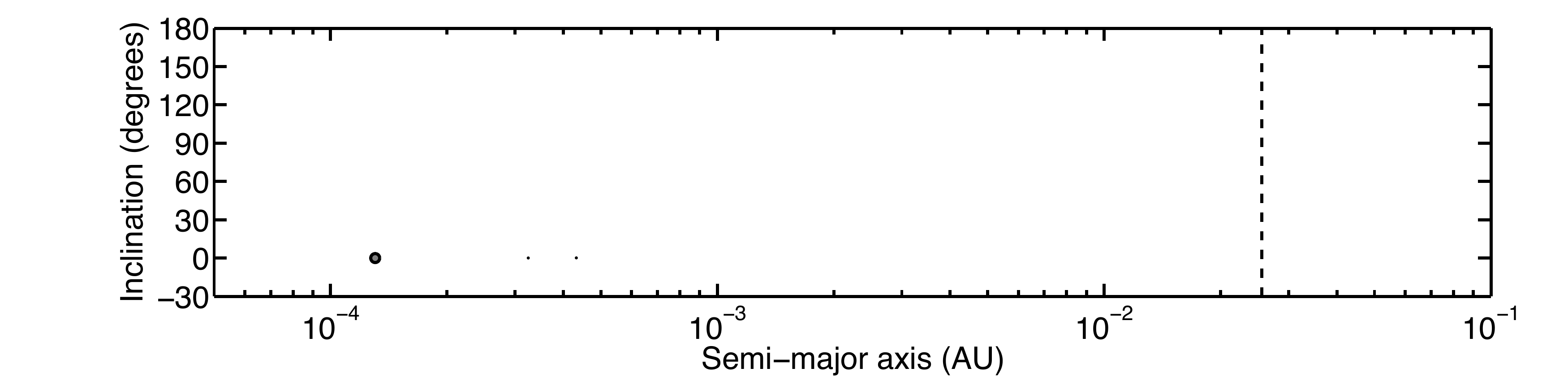}}\\
     \caption[Schematics of the satellite systems of the Earth, Mars and Pluto.]{Schematics of the satellite systems of the Earth, Mars and Pluto.  Large satellites, in particular, the Moon and Charon, are shown as large filled dark grey circles with radii proportional to the radius of the corresponding satellite.  For reference, the Moon, has a radius of 1737km.  Satellites too small to have a spherical shape, are shown as small dots.  The dashed line on the right denotes half a planetary Hill radius.}
     \label{TerPlanMoons}
\end{figure}

Of the four terrestrial planets in our Solar System, only the two most distant planets, Earth and Mars, host moons, and out of these two planets only the Earth hosts a relatively large moon.  In addition, a number of planet-like objects, in particular Pluto, also host ``moons".  For the case of Pluto, this includes its large moon Charon, and its two much smaller moons, Nix and Hydra.  For reference and comparison, the satellite systems of Earth, Mars and Pluto are shown in figure~\ref{TerPlanMoons}, and their properties are summarised in table~\ref{TerMoonsTable}.  As we are discussing moon properties within the context of moon detection, we will focus our attention on large (and consequently detectable) satellites such as the Moon and Charon, and neglect the irregular moons of Mars and Pluto's small moons Nix and Hydra.

Focussing our attention on the Earth-Moon system and the Pluto-Charon system, it can be seen that they share a number of properties.  First, in both of these cases there is only one large moon per satellite system.  Second, this single large moon contains a fair percentage of the mass in the planet-moon system, $\sim$1\% for the case of the Moon and $\sim$10\% for the case of Charon.  Third, the orbit of this moon about their host planet is relatively close and relatively circular.  Finally, in both cases, the moon's orbit is approximately aligned with the planet's equator e.g. the Moon's orbit is tilted by only 5 degrees.  These shared properties suggest a common formation mechanism for these large moons, an issue that will be discussed in section~\ref{Intro_Moons_Form_Impact}.

\subsection{Moons of gas giant planets}

\begin{figure}
     \centering
     \subfigure[Jupiter's satellite system.]{
          \label{fig:dl2858}
          \includegraphics[width=.98\textwidth]{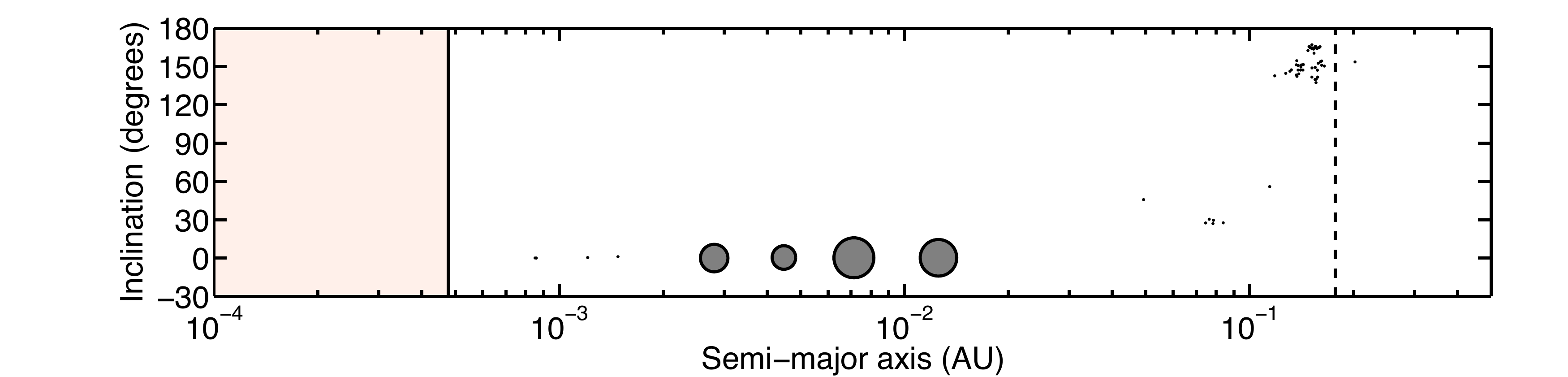}}\\
          \vspace{-0.1cm}
     \subfigure[Saturn's satellite system.]{
          \label{fig:er2858}
          \includegraphics[width=.98\textwidth]{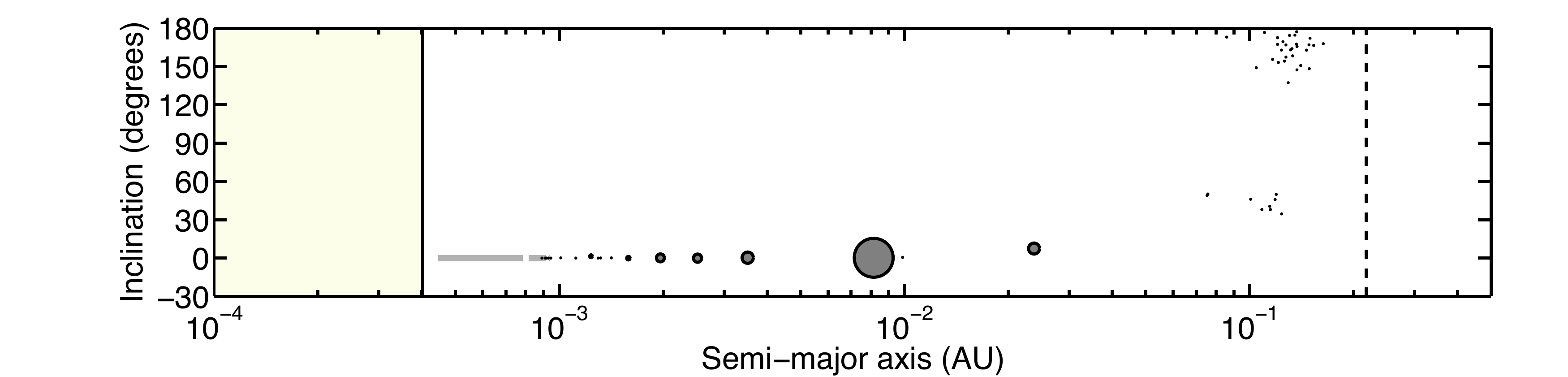}}\\
          \vspace{-0.1cm}
          \subfigure[Uranus' satellite system.]{
          \label{fig:dl2858}
          \includegraphics[width=.98\textwidth]{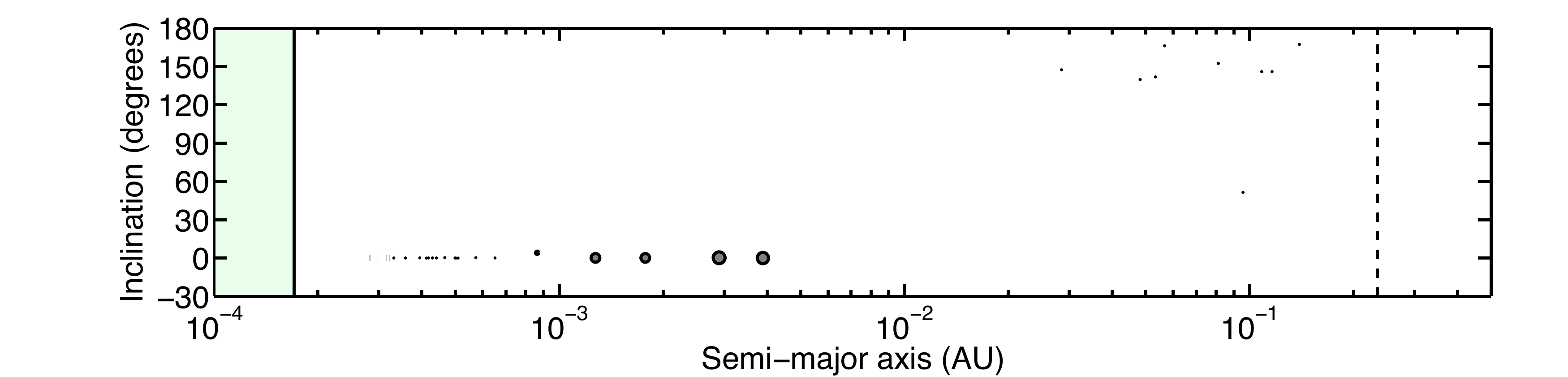}}\\
          \vspace{-0.1cm}
     \subfigure[Neptune's satellite system.]{
          \label{fig:er2858}
          \includegraphics[width=.98\textwidth]{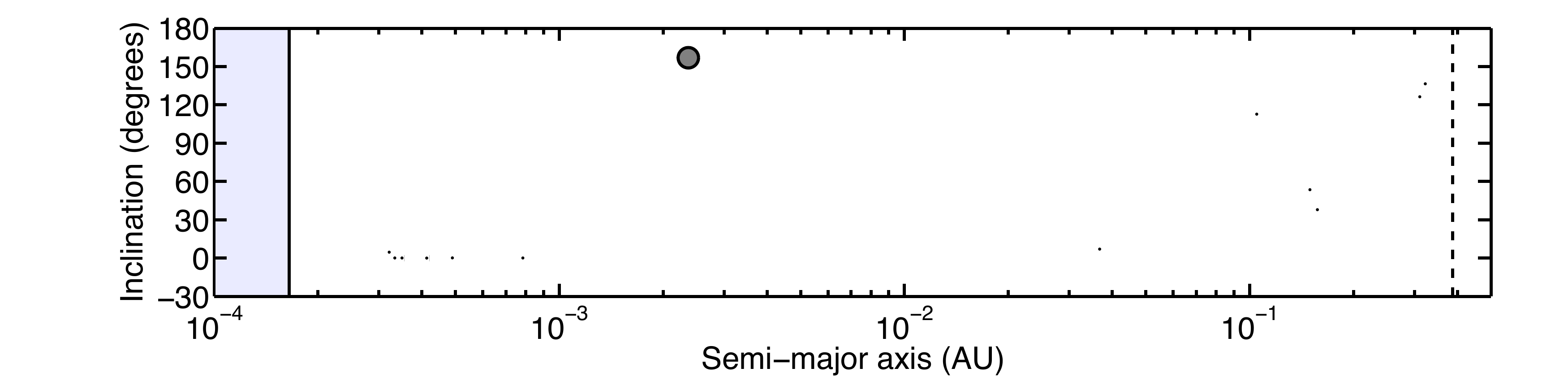}}
          \vspace{-0.1cm}
     \caption[Schematics of the satellite systems of the four gas giants in the Solar System.]{Schematics of the satellite systems of the four gas giants in the Solar System.  The regular satellites are shown as large filled dark grey circles with radii proportional to the radius of the corresponding satellite.  For reference, Titan, the largest satellite of Saturn, has a radius of 2575km.  Satellites too small to have a spherical shape, that is, inner satellites and irregular satellites, are shown as small dots while planetary rings are denoted by a thick light grey line.  The surface of the planet is represented by a thick vertical line on the left, while on the right a dashed line denotes half a planetary Hill radius.}
     \label{GasGiantMoons}
\end{figure}

Unlike the terrestrial planets, each of the four gas giant planets in our Solar System has a plethora of attendant moons with a grand total of 62, 59, 27 and 13 moons for Jupiter, Saturn, Uranus and Neptune respectively, as of the 12$^{th}$ of February 2010.  For reference, the satellite systems of Jupiter, Saturn, Uranus and Neptune are shown in figure~\ref{GasGiantMoons}, while their properties are summarised in tables~\ref{JupMoonsTable},  \ref{SatMoonsTable},  \ref{UraMoonsTable} and  \ref{NepMoonsTable}.  What is immediately obvious from figure~\ref{GasGiantMoons} is that there are three main classes of moon. First, there are small inner moons with orbits very closely aligned to their host planet's equator.  These moons seem to be generally associated with, or help shape the ring systems of their host planet.  Second, outside this set of inner moons each of the gas giant planets has a number of larger, regular moons.  Finally, further away still, are irregular moons with orbits which become increasingly retrograde the further they are from their host planet.  Again noting that the aim of this investigation is gain an intuitive understanding of the the types of satellites that are likely to exist (and consequently be detected), we will focus this discussion on the regular satellites, as they are the largest, and most detectable.

From figure~\ref{GasGiantMoons} it can be seen that the regular satellites of the gas giant planets all share a number of features, with one main exception, Triton, Neptune's only large moon.  Consequently, the satellites of Jupiter, Saturn and Uranus will first be discussed, followed by a separate discussion of the properties and peculiarities of Triton.

The regular satellites of Jupiter, Saturn and Uranus share a number of features.  First, these large moons only occur within a minimum distance from their planet, compared to the irregular moons for example, which occupy orbits reaching to the edge of the stability region, estimated by half a Hill radius.  Second, a planet can have multiple large moons e.g. the four Galilean satellites of Jupiter.  This is in stark comparison with case for terrestrial planets where one large moon (if any) per planet seems to be the norm.  Third, while the mass of these satellites is comparable to that of the Moon, it is small in comparison to the mass of the host planet, in particular, the proportion of the planet-moon system's mass that is in regular satellites is $2.1 \times 10^{-4}$, $2.5 \times 10^{-4}$ and $1.1 \times 10^{-4}$ for Jupiter, Saturn and Uranus respectively.  Finally, the orbits of these moons are circular and aligned with the equator of their host planet.\footnote{Uranus has an orbital obliquity of approximately $98\,^{\circ}$, and consequently its equatorial and orbital planes are very different.  The inclinations shown in figure~\ref{GasGiantMoons} and presented in table~\ref{UraMoonsTable} are measured relative to the local Laplace plane, the plane which defines the axis about which the moon's orbit normal precesses.  This plane is approximately parallel with the equatorial plane of the planet for close in satellites, e.g. the regular satellites, and is approximately aligned with the orbital plane of the planet for more distant satellites, e.g. the irregular satellites.}  A number of processes proposed which will produce such systems will be explained in section~\ref{Intro_Moons_Form_Disk}.

Compared to the satellite systems of Jupiter, Saturn and Uranus, Neptune's satellite system is distinctly odd.  First, it only has one large moon, Triton, and second, this moon's orbit is inclined and retrograde.  However, similar to the moons of Jupiter, Saturn and Uranus, Triton is also relatively close to its host planet and not very massive compared to it.  As a result of these properties it has been proposed that it probably did not form in situ, but was captured.  This process will be discussed further in section~\ref{Intro_Moons_Form_Capture}.

\begin{longtable}{llllllll}

\\
\\
\hline
   \multicolumn{1}{l}{Moon} &
   \multicolumn{1}{l}{$R_m$} &
   \multicolumn{2}{c}{$M_m$} &
   \multicolumn{2}{c}{$a_m$} &
   \multicolumn{1}{l}{$e_m$} &
   \multicolumn{1}{l}{$I_m$}\\
   & (km) & ($10^{20}$ kg) & ($10^{-4}M_p$) & ($10^3$ km) & ($R_c$)  & & ($^{\circ}$) \\
   \hline
   \\[-1.8ex]
\endfirsthead

\multicolumn{3}{c}{{\tablename} \thetable{} -- Continued} \\[0.5ex]
  \hline 
   \multicolumn{1}{l}{Moon} &
   \multicolumn{1}{l}{$R_m$} &
   \multicolumn{2}{c}{$M_m$} &
   \multicolumn{2}{c}{$a_m$} &
   \multicolumn{1}{l}{$e_m$} &
   \multicolumn{1}{l}{$I_m$}\\
  & (km) & ($10^{20}$ kg) & ($10^{-4}M_p$) & ($10^3$ km) & ($R_c$)  & & ($^{\circ}$) \\
   \hline
  \\[-1.8ex]
\endhead

  \multicolumn{4}{l}{{Continued on next page \ldots}} \\
\endfoot

\endlastfoot

Metis	& 20		& -- 		& -- 		&128	& 0.116 	& 0.001 	& 0.019  \\
Adrastea	& 10 		& -- 		& -- 		& 129	& 0.116 	&0.002	& 0.054  \\
Amalthea	& 86* 	& -- 		& -- 		&181.4	& 0.164 	&0.003 	& 0.380  \\
Thebe	& 50 		& -- 		& -- 		& 221.9	& 0.200 	&0.018	& 1.080  \\
\rowcolor[gray]{.9} 
Io		& 1821 	& 893.3 	& 0.4705 	& 421.8	& 0.381 	&0.004 	& 0.036  \\		
\rowcolor[gray]{.9}	
Europa	& 1565 	& 479.7 	& 0.2527 	& 671.1	& 0.606 	& 0.009 	& 0.466  \\
\rowcolor[gray]{.9}
Ganymede& 2634 	& 1482 	& 0.7806 	& 1070.4	& 0.966 	&0.001 	& 0.177  \\
\rowcolor[gray]{.9}
Callisto	& 2403	& 1076	& 0.5667	&1882.7	& 1.699 	& 0.007 	& 0.192  \\ 
Themisto	& -- 		& -- 		& -- 		& 7507 	& 6.775 	& 0.242  	& 43.07	\\
Leda		& 5 		& -- 		& -- 		& 11165 	& 10.08 	& 0.164 	& 27.46 \\
Himalia	& 85 		& -- 		& -- 		& 11461  	& 10.34 	& 0.162  	& 1.438  \\
Lysithea	& 12 		& -- 		& -- 		& 11717  	& 10.57 	& 0.112  	& 28.30 \\
Elara	& 40 		& -- 		& -- 		& 11741  	& 10.60 	& 0.217  	& 143.6	 \\
Carpo   	& -- 		& -- 		& -- 		& 17078 	& 15.41	& 0.444 	& 51.16	\\
S/2003J12 & -- 		& -- 		& -- 		& 17835 	& 16.10 	& 0.488 	& 150.8	\\
Euporie	& -- 		& -- 		& -- 		& 19339 	& 17.45 	& 0.144  	& 145.5	\\	
S/2003J3	& -- 		& -- 		& -- 		& 20230 	& 18.26 	& 0.203 	& 147.8	\\
S/2003J18& -- 		& -- 		& -- 		& 20494 	& 18.50 	& 0.102 	& 146.0	\\
S/2003J16& -- 		& -- 		& -- 		& 20948 	& 18.91 	& 0.231 	& 148.6	\\
Mneme	& -- 		& -- 		& -- 		& 21036 	& 18.99 	& 0.227  	& 148.6	\\
Euanthe	& -- 		& -- 		& -- 		& 21038 	& 18.99 	& 0.231 	& 149.0	\\
Helike   	& -- 		& -- 		& -- 		& 21064 	& 19.01 	& 0.147 	& 154.6	\\
Harpalyke	& -- 		& -- 		& -- 		& 21104 	& 19.05 	& 0.226  	& 148.6	\\
Praxidike	& -- 		& -- 		& -- 		& 21148 	& 19.09 	& 0.230  	& 149.0	\\
Orthosie   & -- 		& -- 		& -- 		& 21164 	& 19.10 	& 0.278  	& 145.9	\\
Thelxinoe	& -- 		& -- 		& -- 		& 21165 	& 19.10 	& 0.219  	& 151.3	\\
Thyone   	& -- 		& -- 		& -- 		& 21192 	& 19.13 	& 0.238 	& 148.8	\\
Iocaste   	& -- 		& -- 		& -- 		& 21272 	& 19.20 	& 0.215  	& 149.4	\\
Ananke	& 10 		& -- 		& -- 		& 21276 	& 19.20 	& 0.244 	&148.9 \\
Hermippe	& -- 		& -- 		& -- 		& 21300 	& 19.22 	& 0.212  	& 150.9	\\
S/2003J15& -- 		& -- 		& -- 		& 22622 	& 20.42 	&0.187 	& 146.4	\\
S/2003J10& -- 		& -- 		& -- 		& 23042 	& 20.80 	&0.428  	& 165.2\\	
Pasithee   & -- 		& -- 		& -- 		& 23090 	& 20.84 	&0.267 	& 165.0\\	
Eurydome	& -- 		& -- 		& -- 		& 23148 	& 20.89 	&0.276 	& 150.2\\
Chaldene	& -- 		& -- 		& -- 		& 23179 	& 20.92 	&0.251  	& 165.2	\\
Isonoe   	& -- 		& -- 		& -- 		& 23231 	& 20.97 	&0.247 	& 165.3	\\
Kallichore	& -- 		& -- 		& -- 		& 23273 	& 21.00 	&0.242  	& 165.1	\\
Erinome	& -- 		& -- 		& -- 		& 23283 	& 21.01 	&0.266 	& 164.9	\\
Kale   	& -- 		& -- 		& -- 		& 23302 	& 21.03 	&0.252 	& 165.1	\\
Aitne   	& -- 		& -- 		& -- 		& 23315 	& 21.04 	&0.266 	& 165.1	\\
Eukelade	& -- 		& -- 		& -- 		& 23322 	& 21.05 	&0.267 	& 165.2	\\
Arche   	& -- 		& -- 		& -- 		& 23355 	& 21.08 	&0.256 	& 164.9	\\
Taygete   	& -- 		& -- 		& -- 		& 23363 	& 21.09	&0.252 	& 165.2	\\
S/2003J9	& -- 		& -- 		& -- 		& 23385 	& 21.11 	&0.264 	& 165.2	\\
Carme	& 15 		& -- 		& -- 		& 23404 	& 21.12 	&0.253 	& 164.9 \\
Herse   	& -- 		& -- 		& -- 		& 23405 	& 21.12	&0.249 	& 164.8\\
S/2003J5	& -- 		& -- 		& -- 		& 23493 	& 21.20 	&0.246 	& 165.3	\\
S/2003J19& -- 		& -- 		& -- 		& 23532 	& 21.24 	&0.262 	& 165.2	\\
S/2003J23& -- 		& -- 		& -- 		& 23549 	& 21.25 	&0.270 	& 146.3	\\
Kalyke   	& -- 		& -- 		& -- 		& 23564 	& 21.27 	&0.246 	& 165.2	\\
Hegemone& -- 		& -- 		& -- 		& 23566 	& 21.27 	&0.344 	& 154.0\\	
Pasiphae	& 18 		& -- 		& -- 		& 23624   & 21.32 	&0.409 	&170.5	 \\
Cyllene   	& -- 		& -- 		& -- 		& 23787 	& 21.47 	&0.418 	& 150.2 	\\
Sponde   	& -- 		& -- 		& -- 		& 23790 	& 21.47 	&0.313 	& 151.2	\\
Magaclite	& -- 		& -- 		& -- 		& 23808 	& 21.49 	&0.421 	& 152.8 \\	
S/2003J4	& -- 		& -- 		& -- 		& 23928 	& 21.60 	&0.356 	& 149.3 	\\
Sinope	& 14 		& -- 		& -- 		& 23939 	& 21.61 	&0.250 	& 158.1	  \\
Aoede   	& -- 		& -- 		& -- 		& 23969 	& 21.63 	&0.432 	& 158.3	\\
Autonoe   & -- 		& -- 		& -- 		& 24033 	& 21.69 	&0.317 	& 152.3	\\
Callirrhoe & -- 		& -- 		& -- 		& 24102 	& 21.75 	&0.283 	& 147.2\\	
Kore   	& -- 		& -- 		& -- 		& 24486 	& 22.10 	&0.332 	& 145.0	\\
S/2003J2	& -- 		& -- 		& -- 		& 28332 	& 25.57 	&0.411 	& 157.1	\\
\hline
\\
\caption[Physical and orbital properties of the satellites of Jupiter.]{Physical and orbital properties of the satellites of Jupiter.  The regular satellites are highlighted in grey.  Radii and mass measurements are taken from \citep{Murrayetal1999}, while orbital parameters are taken from the JPL website.  The inclinations are measured relative to the local Laplace plane.}  \label{JupMoonsTable}
\end{longtable}

\begin{center}
\begin{longtable}{llllllll}

\hline
   \multicolumn{1}{l}{Moon} &
   \multicolumn{1}{l}{$R_m$} &
   \multicolumn{2}{c}{$M_m$} &
   \multicolumn{2}{c}{$a_m$} &
   \multicolumn{1}{l}{$e_m$} &
   \multicolumn{1}{l}{$I_m$}\\
   & (km) & ($10^{20}$ kg) & ($10^{-4}M_p$) & ($10^3$ km) & ($R_c$)  & & ($^{\circ}$) \\
   \hline
   \\[-1.8ex]
\endfirsthead

\multicolumn{3}{c}{{\tablename} \thetable{} -- Continued} \\[0.5ex]
  \hline 
   \multicolumn{1}{l}{Moon} &
   \multicolumn{1}{l}{$R_m$} &
   \multicolumn{2}{c}{$M_m$} &
   \multicolumn{2}{c}{$a_m$} &
   \multicolumn{1}{l}{$e_m$} &
   \multicolumn{1}{l}{$I_m$}\\
  & (km) & ($10^{20}$ kg) & ($10^{-4}M_p$) & ($10^3$ km) & ($R_c$)  & & ($^{\circ}$) \\
   \hline
  \\[-1.8ex]
\endhead

  \multicolumn{8}{l}{{Continued on next page \ldots}} \\
\endfoot

\endlastfoot

Pan			& 10		& --		& -- 			& 133.58			& 0.098 		& 0.000	& 0.001\\
Daphnis		& --		& --		& -- 			&136.5			& 0.100 		& 0.000	& 0.000\\
Atlas			& 16* 	& --		& -- 			&137.67			& 0.101 		& 0.001	& 0.003\\
Prometheus	& 50*	& 0.0014	& $2.5 \times 10^{-6}$ &139.38		& 0.103 		& 0.002	& 0.008\\
Pandora		& 42* 	& 0.0013	& $2.3 \times 10^{-6}$ & 141.72	& 0.104 		& 0.004	& 0.050\\
Epimetheus	& 59* 	& 0.0055	& $9.7 \times 10^{-5}$ &151.41		& 0.111 		& 0.010	& 0.351\\
Janus		& 89* 	& 0.0198	& $3.5 \times 10^{-5}$ &151.46	& 0.111 		& 0.007	& 0.163\\
\rowcolor[gray]{.9} 	
Mimas		& 199	& 0.385	& $6.8 \times 10^{-5}$& 185.54	& 0.137 		& 0.020	& 1.574\\
Methone		& -- 		& --		& -- 			&194.44			& 0.143 		& 0.000	& 0.007\\
Pallene		& --		& --		& -- 			& 212.28			& 0.156 		& 0.004	& 0.181\\	
\rowcolor[gray]{.9} 
Enceladus	& 249	& 0.73	& 0.0013		& 238.04			& 0.175 		& 0.005	& 0.009\\
\rowcolor[gray]{.9} 
Tethys		& 530	& 6.22 	& 0.0109		& 294.67			& 0.217 		& 0.000	& 1.091\\
Telesto		& 11*	& --		& -- 			& 294.71			& 0.217 		& 0.000	& 1.180\\
Calypso		& 10*	& --		& -- 			& 294.71			& 0.217 		& 0.001	& 1.499\\
Polydeuces	& --		& --		& -- 			& 377.2			& 0.278 		& 0.019	& 0.177\\
Helene		& 16*	& --		& -- 			&377.42			& 0.278 		& 0.007	& 0.213\\
\rowcolor[gray]{.9} 
Dione		& 560	& 10.52	& 0.0185		& 377.42			& 0.278 		& 0.002	& 0.028\\
\rowcolor[gray]{.9} 
Rhea		& 764	& 23.1	& 0.0406		& 527.07			& 0.388 		& 0.001	& 0.333\\
\rowcolor[gray]{.9} 
Titan			& 2575	&1345.5	& 2.3669		& 1221.9		& 0.899 		& 0.029	& 0.312\\
Hyperion		& 143*	& -- 		& -- 			& 1500.9 		& 1.104		& 0.023	& 0.615\\
\rowcolor[gray]{.9} 
Iapetus		& 718	& 15.9	& 0.0280		& 3560.9		& 2.620 		& 0.029	& 8.313\\
Kiviuq		& --		& --		& -- 			& 11311			& 8.323 		& 0.164	& 48.53	\\
Ijiraq		& --		& --		& -- 		& 11367		& 8.365 		& 0.458	&47.12\\
Phoebe	& 110 	& --		& -- 		& 12947		& 9.528 		& 0.163	&175.2\\
Paaliaq	& --		& --		& -- 		& 15024		& 11.06 		& 0.540	&41.77	\\
Skathi	& --		& --		& -- 		& 15614		& 11.49 		& 0.294	& 150.8	\\
Albiorix	& --		& --		& -- 		& 16401 		& 12.07 		& 0.484	& 35.51	\\
S/2007\_S2& --		& --		& -- 		& 16723		& 12.31 		& 0.178	& 175.6\\
Bebhionn	& --		& --		& -- 		& 17117 		& 12.60 		& 0.484	& 34.56	\\
Erriapus	& --		& --		& -- 		& 17611 		& 12.96 		& 0.468	& 38.65	\\
Skoll		& --		& --		& -- 		& 17663 		& 13.00 		& 0.470	& 160.2	\\
Tarqeq	& --		& --		& -- 		& 17909 		& 13.18 		& 0.119	& 49.57	\\
Siarnaq	& --		& --		& -- 		& 18015 		& 13.26 		& 0.405	& 44.51	\\
Tarvos	& --		& --		& -- 		& 18263 		& 13.44 		& 0.531	&35.95	\\
S/2004\_S13& --	& --		& -- 		&18408 		& 13.55 		& 0.260	& 169.1	\\
Hyrokkin	& --		& --		& -- 		& 18437 		& 13.57 		& 0.329	& 151.2\\
Greip	& --		& --		& -- 		& 18442 		& 13.57 		& 0.316	& 173.3	\\	
Mundilfari	& --		& --		& -- 		& 18667 		& 13.74 		& 0.205	& 169.2	\\
S/2006\_S1& --		& --		& -- 		&18797 		& 13.83 		& 0.118	& 155.0\\
S/2007\_S3& --		& --		& -- 		&18981  		& 13.97 		& 0.185	& 175.7\\
Bergelmir	& --		& --		& -- 		& 19338 		& 14.23 		& 0.142	& 158.9	\\
Jarnsaxa	& --		& --		& -- 		& 19356 		& 14.24 		& 0.217	& 163.3	\\
Narvi	& --		& --		& -- 		& 19417 		& 14.29 		& 0.426	& 143.2	\\
S/2004\_S17& --	& --		& -- 		&19449 		& 14.31 		& 0.181	& 168.0\\
Suttungr	& --		& --		& -- 		& 19476 		& 14.33 		& 0.114	& 173.9	\\
Hati		& --		& --		& -- 		& 19775		& 14.55 		& 0.373	& 165.0	\\
S/2004\_S12& --	& --		& -- 		&19867		& 14.62 		& 0.323	& 163.3	\\
Bestla	& --		& --		& -- 		& 20278		& 14.92 		& 0.474	& 141.7	\\
Farbaut	& --		& --		& -- 		& 20387		& 15.00 		& 0.245	& 158.0	\\
Thrymr	& --		& --		& -- 		& 20439 		& 15.04 		& 0.466	& 173.7	\\
Aegir	& --		& --		& -- 		& 20749 		& 15.27 		& 0.252	& 167.1	\\
S/2004\_S7& --		& --		& -- 		& 21005 		& 15.46 		& 0.530	& 164.9\\
Kari		& --		& --		& -- 		& 22077 		& 16.25 		& 0.484	& 155.9	\\
S/2006\_S3& --		& --		& -- 		&22100 		& 16.26 		& 0.404	& 158.9\\
Fenrir	& --		& --		& -- 		& 22454 		& 16.52 		& 0.133	& 164.4	\\
Surtur	& --		& --		& -- 		& 22920 		& 16.87 		& 0.447	& 169.1\\
Loge		& --		& --		& -- 		& 23065 		& 16.97 		& 0.188	& 167.2	\\
Ymir		& --		& --		& -- 		& 23140 		& 17.03 		& 0.334	&171.7	\\
Fornjot	& --		& --		& -- 		& 25151 		& 18.51 		& 0.210	& 169.7	\\
\hline
\\
\caption[Physical and orbital properties of the satellites of Saturn.]{Physical and orbital properties of the satellites of Saturn.  The regular satellites are highlighted in grey.  Radii and mass measurements are taken from \citep{Murrayetal1999}, while orbital parameters are taken from the JPL website.  The inclinations are measured relative to the local Laplace plane.}\label{SatMoonsTable} \\
\end{longtable}
\end{center}

\begin{center}
\begin{longtable}{llllllll}

\hline
   \multicolumn{1}{l}{Moon} &
   \multicolumn{1}{l}{$R_m$} &
   \multicolumn{2}{c}{$M_m$} &
   \multicolumn{2}{c}{$a_m$} &
   \multicolumn{1}{l}{$e_m$} &
   \multicolumn{1}{l}{$I_m$}\\
   & (km) & ($10^{20}$ kg) & ($10^{-4}M_p$) & ($10^3$ km) & ($R_c$)  & & ($^{\circ}$) \\
   \hline
   \\[-1.8ex]
\endfirsthead

\multicolumn{3}{c}{{\tablename} \thetable{} -- Continued} \\[0.5ex]
  \hline 
   \multicolumn{1}{l}{Moon} &
   \multicolumn{1}{l}{$R_m$} &
   \multicolumn{2}{c}{$M_m$} &
   \multicolumn{2}{c}{$a_m$} &
   \multicolumn{1}{l}{$e_m$} &
   \multicolumn{1}{l}{$I_m$}\\
  & (km) & ($10^{20}$ kg) & ($10^{-4}M_p$) & ($10^3$ km) & ($R_c$)  & & ($^{\circ}$) \\
   \hline
  \\[-1.8ex]
\endhead

  \multicolumn{4}{l}{{Continued on next page \ldots}} \\
\endfoot

\endlastfoot

Cordelia		& 13		& -- 		& -- 		& 49.8	& 0.034 		& 0.000		& 0.085	\\
Ophelia		& 16		& -- 		& -- 		& 53.8	& 0.037 		& 0.010		& 0.104	\\
Bianca		& 22		& -- 		& -- 		& 59.2	& 0.041 		& 0.001		& 0.193	\\
Cressida		& 33		& -- 		& -- 		& 61.8	& 0.042 		& 0.000		& 0.006	\\
Desdemona	& 29		& -- 		& -- 		& 62.7	& 0.043 		& 0.000		& 0.113\\	
Juliet		& 42		& -- 		& -- 		& 64.4	& 0.044 		& 0.001		& 0.065	\\
Portia		& 55		& -- 		& -- 		& 66.1	& 0.045 		& 0.000		& 0.059	\\
Rosalind		& 29		& -- 		& -- 		& 69.9	& 0.048 		& 0.000		& 0.279	\\
Cupid		& -- 		& -- 		& -- 		& 74.392	& 0.051 		& 0.001		& 0.099\\	
Belinda		& 34		& -- 		& -- 		& 75.3	& 0.052 		& 0.000		& 0.031	\\
Perdita		& -- 		& -- 		& -- 		& 76.417	& 0.052 		& 0.012		& 0.470\\	
Puck			& 77		& -- 		& -- 		& 86		& 0.059 		& 0.000		& 0.319	\\
Mab			& -- 		& -- 		& -- 		& 97.736	& 0.067 		& 0.003		& 0.134	\\
\rowcolor[gray]{.9} 
Miranda		& 761	& 30.14	& 0.3471 	& 129.9	& 0.089 		& 0.001		& 4.338	\\
\rowcolor[gray]{.9} 		
Ariel			& 235*	& 0.659	& 0.00759 &190.9 	& 0.131 		& 0.001		& 0.041	\\
\rowcolor[gray]{.9} 
Umbriel		& 579*	& 13.53	& 0.1558 	& 266	& 0.182 		& 0.004		& 0.128	\\
\rowcolor[gray]{.9} 
Titania		& 585	& 11.72	& 0.1350 	&436.3	& 0.299 		& 0.001		& 0.079	\\
\rowcolor[gray]{.9} 
Oberon		& 789	& 35.27	& 0.4062 	& 583.5	& 0.399 		& 0.001		& 0.068	\\
Francisco		& -- 		& -- 		& -- 		& 4282.9	& 2.932 		& 0.132		& 147.3	\\
Caliban		& -- 		& -- 		& -- 		& 7231.1	& 4.949 		& 0.181		& 141.5	\\
Stephano		& -- 		& -- 		& -- 		& 8007.4	& 5.481 		& 0.225		& 143.8	\\
Trinculo		& -- 		& -- 		& -- 		& 8505.2	& 5.822 		& 0.219		& 167.0	\\
Sycorax		& -- 		& -- 		& -- 		& 12179 	& 8.336 		& 0.522		& 159.4   \\	
Margaret		& -- 		& -- 		& -- 		& 14146 	& 9.683 		& 0.677		& 57.37\\	
Prospero		& -- 		& -- 		& -- 		& 16276 	& 11.14 		& 0.445		& 151.8	\\
Setebos		& -- 		& -- 		& -- 		& 17420 	& 11.92 		& 0.591		& 158.2	\\
Ferdinand		& -- 		& -- 		& -- 		& 20430	& 13.98 		& 0.399		& 169.8	\\
\hline
\\
\caption[Physical and orbital properties of the satellites of Uranus.]{Physical and orbital properties of the satellites of Uranus.  The regular satellites are highlighted in grey.  Radii and mass measurements are taken from \citep{Murrayetal1999}, while orbital parameters are taken from the JPL website.  The inclinations are measured relative to the local Laplace plane.}  \label{UraMoonsTable} \\
\end{longtable}
\end{center}

\begin{center}
\begin{longtable}{llllllll}

\hline
   \multicolumn{1}{l}{Moon} &
   \multicolumn{1}{l}{$R_m$} &
   \multicolumn{2}{c}{$M_m$} &
   \multicolumn{2}{c}{$a_m$} &
   \multicolumn{1}{l}{$e_m$} &
   \multicolumn{1}{l}{$I_m$}\\
   & (km) & ($10^{20}$ kg) & ($10^{-4}M_p$) & ($10^3$ km) & ($R_c$)  & & ($^{\circ}$) \\
   \hline
   \\[-1.8ex]
\endfirsthead

\multicolumn{3}{c}{{\tablename} \thetable{} -- Continued} \\[0.5ex]
  \hline 
   \multicolumn{1}{l}{Moon} &
   \multicolumn{1}{l}{$R_m$} &
   \multicolumn{2}{c}{$M_m$} &
   \multicolumn{2}{c}{$a_m$} &
   \multicolumn{1}{l}{$e_m$} &
   \multicolumn{1}{l}{$I_m$}\\
  & (km) & ($10^{20}$ kg) & ($10^{-4}M_p$) & ($10^3$ km) & ($R_c$)  & & ($^{\circ}$) \\
   \hline
  \\[-1.8ex]
\endhead

  \multicolumn{4}{l}{{Continued on next page \ldots}} \\
\endfoot

\endlastfoot

Naiad	& 29		& --		& -- 		& 48.227	& 0.004 		& 0.000	& 4.691\\
Thalassa	& 40		& --		& -- 		& 50.074	& 0.004 		& 0.000	& 0.135\\
Despina	& 74		& --		& -- 		& 52.526	& 0.005 		& 0.000	& 0.068\\
Galatea	& 79		& --		& -- 		& 61.953	& 0.005 		& 0.000	& 0.034\\
Larissa	& 94*	& --		& -- 		& 73.548	& 0.006 		& 0.001	& 0.205\\
Proteus	& 209*	& --		& -- 		& 117.65	& 0.010 		& 0.000	& 0.075\\
\rowcolor[gray]{.9} 	
Triton	& 1353 	& 215	& 2.10	& 354.76	& 0.031 		& 0.000	& 156.9\\
Nereid	& 170	& --		& -- 		& 5513.8	& 0.475 	& 0.751	& 7.090\\
Halimede	& -- 		& --		& -- 		& 16611	& 1.430 		& 0.265	& 112.7	\\
Sao		& -- 		& --		& -- 		& 22228	& 1.914 		& 0.137	& 53.48\\	
Laomedeia& -- 		& --		& -- 		& 23567	& 2.029 		& 0.397	& 37.87\\	
Psamathe& -- 		& --		& -- 		& 48096	& 4.141 		& 0.381	& 126.3	\\
Neso	& -- 		& --		& -- 		& 49285	& 4.244 		& 0.571	& 136.4	\\
\hline
\\
\caption[Physical and orbital properties of the satellites of Neptune. ]{Physical and orbital properties of the satellites of Neptune.  The regular satellites are highlighted in grey.  Radii and mass measurements are taken from \citep{Murrayetal1999}, while orbital parameters are taken from the JPL website.  The inclinations are measured relative to the local Laplace plane.}  \label{NepMoonsTable}\\
\end{longtable}
\end{center}

\section{Formation models}\label{Intro_Moons_Form}

Planetary formation is currently a vigorous area of research due partially to the exponential increase in  computing power with time and partially to the high rate of discovery of extra-solar planets.  Recent simulations suggest that formation mechanisms can place physical limits on the mass, number and orbital parameters of moons.  As theories of moon formation are built on underlying theories of planet formation the limits on moon mass and orbital parameters will be discussed within the context of the method by which the host planet formed.   

According to current theories, planets generally produce/acquire large moons in the final stages of planet building.  For the case of terrestrial planets, it is proposed that moon formation occurs during the chaotic growth phase of planetary formation.  During this phase, it is believed that moon-sized embryos on eccentric orbits perturb and impact with each other.  For the case of gas giants, it is believed that moon formation occurs during runaway growth, that is, when the proto gas giant becomes large enough to accrete gas directly from the protoplanetary nebula.  The resulting moons are believed to form within the resulting circumplanetary accretion disk.  Finally, for the case of captured moons, the eventual moon properties depend less on the mechanics of planetary formation and more on the population of objects capable of being tidally captured.

\subsection{Impact generated moons}\label{Intro_Moons_Form_Impact}

Simulations of the period of chaotic growth, the phase when it is thought that large impact-generated moons formed, shows that giant impacts which are able to produce moons are common \citep[e.g.][]{Agnoretal1999}.  Simulations of the impact process indicate that impacts between terrestrial mass proto-planets ($M_p < 2.5M_{\earth}$) produce a disk of orbiting debris \citep{Wadaetal2006} which can coalesce into a moon a couple of planetary radii from its host an orbit which can have any value of inclination.  In addition, for standard sized terrestrial planets (0.5 - 1$M_{\earth}$), the dynamics of the impact and interactions between the debris during post impact evolution generally result in a single \citep{CanupLevison1999}, large moon containing up to 4\% of the planets mass \citep[e.g.][]{Canupetal2001}.  As a result of the random nature of the impact, this moon can have any initial orbital inclination.  However, depending on this inclination, the dynamical evolution of the planet-moon system can result in re-impact, a moon on a close, inclined orbit, or a moon on a distant coplanar orbit \citep{AtobeIda2007}.

While only one of the four terrestrial planets in our Solar System has an impact generated moon, giant impacts have been invoked to explain Mercury's high density \citep{Benzetal1998} and Venus' retrograde rotation \citep{Alemietal2006}.  In addition, giant impacts have been proposed to explain the satellite system of Pluto \citep{McKinnon1989,Sternetal2006} and the high obliquity and satellite system of Uranus \citep{Korycanskyetal1990,Slatteryetal1992}.

\subsection{Disk generated moons}\label{Intro_Moons_Form_Disk}

It is believed that the regular satellites of Jupiter, Saturn and Uranus formed within a circumplanetary disk.  This disk may have been the accretion disk of its host planet as it accreted gas and solids from the protoplanetary nebula or, for the case of Uranus, the disk could possibly have been one that was stochastically generated, by a giant impact \citep{Korycanskyetal1990,Slatteryetal1992}.  Independent of the source of the disk, satellite growth within a disk explains the circular orbits of the regular satellites and their low inclination with respect to planetary rotation.  However, the specifics of the method by which a disk of gas and solids is processed into a small number of large satellites is not fully understood.  In addition to the moon properties naturally resulting from accretion from a disk, any proposed model must also be able to explain the masses, distribution of semi-major axes and formation timescales of each of the three sets of regular satellites.  Currently there are two main models for this process, that of \citet{Canupetal2006} and \citet{MosqueiraEstrada2003a,MosqueiraEstrada2003b},  presented in the literature.  These two models, along withe their associated moon formation predictions will be discussed in turn.

\citet{Canupetal2006} use a time-dependant, single component circumplanetary disk model to investigate regular satellite formation.  They suggest that the properties, in particular, the mass, of regular satellites within this disk is determined by the balance between the rate of accretion of material onto the protomoons, and orbital decay of these protomoons within the accretion disk onto the growing gas giant.  This process results in an ordered set of approximately 4 large moons within 60 planetary radii of the planet, with total mass approximately one ten thousandth of their host planet.  This model addresses the issue of formation timescales by proposing that  undifferentiated moons e.g. Callisto, started forming later than their comrades, and remained undifferentiated as a result of cooler disk conditions \citep{BarrCanup2008}.

In comparison, \citet{MosqueiraEstrada2003a,MosqueiraEstrada2003b} propose a two-component disk model with a dense inner sub-disk surrounded by a less dense outer disk.  The edge of this inner disk was set at the centrifugal radius, $R_c$, the radius of the orbit of a gas parcel around the planet, such that the gravitational force from the planet and centrifugal force on that parcel balance, defined as
\begin{equation}
R_c = \frac{j^2}{GM_p},
\end{equation}
where $j$ is the angular momemtum of the gas parcel, $G$ is the universal gravitational constant and $M_p$ is the mass of the planet.  This model was in part proposed to explain the much longer formation timescale measured for Callisto \citep{Andersonetal1998} than its neighbour Ganymede \citep{Schubertetal1996}.  Unlike the model of \citet{Canupetal2006}, this model predicts that the migration timescale of moons is much longer than their formation timescale, mainly as a result of gap opening.  While the model qualitatively describes the ratio of moon mass to planet mass, it does not provide a firm limit, however it does predicts that at most one large satellite should be able to form outside the centrifugal radius.\footnote{\citet{MosqueiraEstrada2003a} use an analytic approximation for the centrifugal radius ($R_c \approx R_H/48$), which was derived for the case of distant gas giants.  More accurate approximations based on simulations are also available \citep{Machidaetal2008,Machida2009}, which are also derived for the case of distant gas giants.  However, for the case of planets which are close to their parent star, where the planet may take up a non-negligible fraction of the centrifugal radius, it is unclear whether these expressions are still valid.  Consequently, for the case of close in gas giant planets, the location of large moons may still be determined by the centrifugal radius, except that the position of $R_c$ for such planets is currently unknown.  For this thesis we use the approximate formula, $R_c \approx R_H/48$ \citep{CassenPettibone1976,Stevensonetal1986}, as this gives the largest centrifugal radius for close in planets.}

\subsection{Captured moons}\label{Intro_Moons_Form_Capture}

Planets can also obtain large satellites through tidal capture, e.g. Neptune's moon Triton.  Currently,\footnote{During the final stages of preparation of this thesis \citet{Podsiadlowskietal2010} proposed a new tidal capture model.  This model is capable of producing gas giant-gas giant binary planets separated from each other by a couple of solar radii.} the only model that can reliably produce large captured moons is the tidal capture model of \citet{AgnorHamilton2006}.  Consequently we will discuss this model in the context of producing large moons of extrasolar planets.

\citet{AgnorHamilton2006} suggest that if a binary system (similar to Pluto-Charon) passed sufficiently close to a host planet, the orbits of the binary pair could be perturbed enough such that one member of the binary gained energy and was ejected while the other lost energy and remained orbiting the planet.  In particular, the binary is likely to be disrupted if $a_b$, the semi-major axis of the binary is approximately equal to its Hill radius, that is,
\begin{equation}
a_b = r_{td} \left(\frac{M_1 + M_2}{3M_p}\right)^{1/3},\label{intro_limits_form_capt_abdef}
\end{equation}
where $r_{td}$ is distance of closest approach and where $M_1$ and $M_2$ are the masses of the two components in the binary respectively.  Also, while \citet{AgnorHamilton2006} found that it was possible for either component to be captured, they found that there was a preference for capturing the lowest mass component, and for this captured moon to be in a retrograde orbit.  

Once this capture has taken place, the new moon will be on a highly elliptical, probably retrograde orbit, with pericenter distance approximately equal to $r_{td}$.  As this moon crosses the region where regular satellites are likely to have formed, either they, or the new moon are likely to be destroyed or ejected.  Consequently, if a planet has a large captured moon, it should be the only large moon.  The new moon's orbit will then tidally circularise, such that $a_m \approx 2r_{td}$.  We can use this to work out the orbital elements of the new moon in terms of the orbital elements of the original binary.  To maximise captured moon mass, we assume that the primordial binary had two equal mass components ($M_m = M_1 = M_2$).  Using equation~\eqref{intro_limits_form_capt_abdef} to substitute for $r_{td}$ gives
\begin{equation}
a_m =  2a_b\left(\frac{3M_p}{2M_m}\right)^{1/3},\label{intro_limits_form_capt_amdef}
\end{equation}
or
\begin{equation}
M_m =  12M_p\frac{a_b^3}{a_m^3}.\label{intro_limits_form_capt_Mmdef}
\end{equation}
As can be seen from equation~\eqref{intro_limits_form_capt_Mmdef}, the maximum mass of a tidally captured moon is inversely proportional to the cube of the final semi-major axis of that moon.  Also, if the rotation of the planet is prograde (a likely consequence of planet formation) and the captured moon's orbit is retrograde (a likely consequence of the capture process), the moon will tidally evolve inwards towards its host planet, reducing its semi-major axis still further.  Consequently this formation mechanism is only capable of producing large moons close to their host planet.

In addition, the efficacy of this moon formation channel also depends on the population of objects, particularly binary objects, available to be captured.  As the formation models proposed for binary trans-neptunian objects require that both objects have large Hill spheres, that is, they are distant from their host star \citep[e.g.][]{Goldreichetal2002}, it follows that a sufficiently large population of large binary objects can only form far from their host star.  Consequently, the only planets capable of capturing such moons must also be distant (so, it is no coincidence that Neptune, the most distant of the gas giants is  the only gas giant to host a large captured moon). 

As the two detection methods investigated in this thesis require that the moon be distant from its host planet and massive (pulsar timing) or orbit a planet close to its host star and be large (transit technique), this formation method does not seem a promising way of producing large detectable moons.  Consequently, within the context of this thesis, the issue of captured moons will not be focussed on.

\section{Stability constraints}

The properties, most particularly the orbital properties of the moons of a given planet depend not only on how and where they formed, but also on their subsequent evolution.  As first pointed out by \citet{Barnesetal2002}, this evolution is governed by two main factors, the slow secular change of the moon's orbital parameters resulting from orbital perturbation, and the more rapid irreversible loss of moons due to tidal disruption, planetary impact or three body instability.  These factors will be discussed in turn and then combined to produce a summary of the mass and distance limits presented in the literature.

\subsection{Moon orbital evolution}\label{Intro_Moons_Stab_Evol}

The main method presented in the literature by which a moon's orbit slowly evolves is though the perturbation on the moon's orbit caused by the tidal bulge raised on the planet by the moon.  For the case where angular velocity associated with the planet's rotation and the angular velocity associated with the moon's orbit are equal, the tidal bulge induced on the planet by the moon is symmetric across the line joining the planet and moon, and no angular momentum is exchanged between the rotation of the planet and the orbit of the moon.  However, for the case where the planet rotates faster/slower than the moon orbits it (i.e. the planet is not tidally locked to the moon), the planet's tidal bulge is ``dragged" ahead of/behind the line joining the planet and moon by an angle $\delta$, which is defined in terms of the tidal dissipation parameter $Q_p$, via $\tan(2\delta) = 1/Q_p$.  The gravitational interaction between this asymmetric bulge and the moon allows an exchange of angular momentum between the planet's rotation and the moon's orbit, leading to a concomitant increase or decrease in semi-major axis depending on whether the planet is rotating faster or slower than the moon.  In particular, from \citet[p. 164]{Murrayetal1999}, we have that the torque on the moon due to the tidal bulge of the planet is given by 
\begin{equation}
\tau_{p-m} = \frac{3}{2} \frac{k_{2p}G M_m^2 R_p^5}{Q_p a_m^6} \mathrm{sgn}(n_{p,rot} - n_m)\label{intro_limits_stab_evol_taupmdef}
\end{equation}
where $G$ is the universal gravitational constant, $k_{2p}$ is the tidal Love number of the planet, and $n_{p,rot}$ and $n_m$ are the angular velocities associated with the planet's rotation and the moon's orbit respectively.  In addition, we note that that $\mathrm{sgn}$, the signum function returns 1 if its argument is positive, 0 if its argument is zero, and -1 if its argument is negative.

Following \citet{Barnesetal2002}, an equation for the evolution of $a_m$ can be determined as a function of the physical parameters of the planet and moon by noting that the torque on the moon is equal to the change in angular momentum, that is
\begin{equation}
\tau_{p-m} = \frac{d}{dt}\left(a_m^2 M_m n_m\right).\label{intro_limits_stab_evol_tauparel}
\end{equation}
From Kepler's law we have that $n_m^2a_m^3 = GM_p$.  Combining this expression with equations~\eqref{intro_limits_stab_evol_taupmdef} and \eqref{intro_limits_stab_evol_tauparel} and simplifying gives
\begin{equation}
\frac{da_m}{dt} = 3\frac{k_{2p}G M_m R_p^5}{\sqrt{GM_p} Q_p a_m^{11/2}} \mathrm{sgn}(n_{p,rot} - n_m).\label{intro_limits_stab_evol_damdef}
\end{equation}

Assuming that $n_{p,rot} - n_m$ does not change sign over the course of the evolution,\footnote{This is a reasonable assumption assuming that the rotation rate of the planet does not change over the lifetime of the system.  If a moon is migrating outward, it is because the rotational period of the planet is shorter than the orbital period of the moon.  As outward migration only increases the orbital period of the moon, the moon will continue to migrate outwards.  Conversely, if a moon is migrating inward, it is because the rotational period of the planet is longer than the orbital period of the moon.  Again, as inward migration results in a decrease in the orbital period of the moon, the moon will continue to migrate inward.  However, the rotation rate of the planet may change if the moon is large enough to modify it, or if the rotation of the planet is influenced by an external factor, for example torque from the host star on the tidal bulges raised on the planet by the host star.  We do not deal with this case in this thesis.} this equation can be integrated to give an explicit form for $a_m$ as a function of time.  For outward migrating moons ($n_{p,rot} > n_m$), we obtain 
\begin{equation}
a_m(t)   = \left(a_m(t_0)^{13/2} + (t - t_0) \frac{33}{2}\frac{k_{2p}G M_m R_p^5}{\sqrt{GM_p} Q_p} \right)^{2/13},\label{intro_limits_stab_evol_amout}
\end{equation}
while for inward migrating moons ($n_{p,rot} < n_m$) we obtain
\begin{equation}
a_m(t)   = \left(a_m(t_0)^{13/2} - (t - t_0) \frac{33}{2}\frac{k_{2p}G M_m R_p^5}{\sqrt{GM_p} Q_p} \right)^{2/13},\label{intro_limits_stab_evol_amin}
\end{equation}
where $t$ is the current time and $t_0$ is the time at which the moon formed.  This equation governs the evolution of the moon's semi-major axis up until the moment it is destroyed or lost to the planet.

\subsection{Processes resulting in moon loss}\label{Intro_Moons_Stab_Dest}

Planets can loose moons through two main processes.  First, if moons are too close to their host planet, they will be destroyed, either through tidal disruption or by impacting with the planet's surface.  Second, if the moon is too distant, it's orbit may become unbound from the planet as a result of  the effect of the periodic perturbation on the moon's orbit by the gravitational field of the host star.  These two processes will be discussed in more detail in turn.

The most dramatic way in which moons can be lost to a host planet is through tidal disruption or impact with that host planet.  To begin, we consider the distance from a planet at which a body held together by self gravity (i.e. a rubble pile) will disrupt due to tidal forces resulting from the planet's non-uniform gravitational field.  This distance is called the Roche limit and is given by
\begin{equation}
R_R = R_p \left(2 \frac{\rho_p}{\rho_m}\right)^{1/3},\label{intro_limits_stab_RRdef}
\end{equation}
where $R_p$ is the radius of the planet and $\rho_p$ and $\rho_m$ are the densities of the planet and moon respectively.  Depending on the ratio of densities of the planet and moon, this radius can range from a couple of planetary radii to within the planet.  In addition, the moon will impact with the planet when the moon's orbit intersects with the surface of the planet.  For the case of circular orbits,\footnote{There are dynamical reasons why close in moons should be in circular orbits around their host.  To see why, note that the timescale for the decay of the moon's orbital eccentricity, $\tau_e$, is proportional to $a_m^5$, resulting in a very rapid decay in orbital eccentricity for moons with small semi-major axes.  In addition, as the timescale for the decay of the semi-major axis is equal to $\tau_e/e_m^2$, the semi-major axis always decays on a timescale longer than that of the eccentricity, giving moons a chance to circularise their orbits.} this corresponds to $a_m \approx R_p$.  

In addition to moon destruction, moons can also be lost from a planet via three-body instability if the  semi-major axis of the moon becomes too large.  The motion of three bodies under their mutual gravitational fields is not a simple issue and can result in  a range of complex behaviour.  In particular, the boundary between stable and unstable orbits is complicated, and most likely fractal.  Fortunately, simple analytic \citep[e.g.][]{Mardling2008} and numerical \citep[e.g.][]{Barnesetal2002,Domingosetal2006} approximations for this boundary are available.  These approximations can generally be expressed in terms of the $R_H$, planetary Hill radius, the distance from the planet where the planet's gravitational force and the tidal force from the star are equal, which is given by
\begin{equation}
R_H = a_p \left(\frac{M_p}{3 M_s}\right)^{1/3},\label{intro_limits_stab_RHdef}
\end{equation}
where $a_p$ is the semi-major axis of the planet's orbit, $M_p$ is the mass of the planet and $M_s$ is the mass of the star.\footnote{See \citet[p. 116]{Murrayetal1999} for a derivation.}  In particular, the approximations used by \citet{Barnesetal2002} and \citet{Domingosetal2006} in their studies of moon stability were 
\begin{equation}
a_{m,max} = 0.36 R_H,
\end{equation}
and 
\begin{equation}
a_{m,max} = 0.4895 R_H (1.0000-1.0305 e_p-0.2738 e_m ),\label{intro_limits_stab_RHDomingos}
\end{equation}
for the case of prograde satellites, and
 \begin{equation}
a_{m,max} = 0.50 R_H,
\end{equation}
and 
\begin{equation}
a_{m,max} = 0.9309 R_H (1.0000-1.0764 e_p-0.9812 e_m ),
\end{equation}
for the case of retrograde satellites respectively, where $a_{m,max}$ is the limiting semi-major axis of the moon's orbit and where $e_p$ and $e_m$ are the eccentricities of the orbit of the planet and moon respectively.  As the effect of planetary eccentricity on moon detection will be investigated in chapter~\ref{Trans_Thresholds}, the approximation of \citet{Domingosetal2006} will be used in this thesis.
 
 \subsection{Limits on moon mass}\label{Intro_Moons_Stab_Lims}
 
The information presented in section~\ref{Intro_Moons_Stab_Evol}, on semi-major axis evolution, and the information presented in section~\ref{Intro_Moons_Stab_Dest} on the region of parameter space where moons are retained can now be combined to provide a lower mass limit on moons as a function of its initial semi-major axis using the method pioneered by \citet{Barnesetal2002}.  Again, as the properties, most particularly the mass and the $Q_p$ value, of terrestrial planets and gas giants differ by a number of orders of magnitude, mass limits for these two cases will be discussed separately.

\subsubsection{Mass limits for moons of terrestrial planets}

As described in section~\ref{Intro_Moons_Form_Impact}, impact generated moons of terrestrial planets form a few planetary radii from their host, on a circular orbit which may be arbitrarily inclined with respect to the plane of the planetary orbit.  In addition, they may contain a sizable fraction of their host planet's mass.

While the moon is formed at this position, it will not remain there long.  If the moon's orbit is highly inclined, simulations \citep{AtobeIda2007} show that while the moon's orbit initially evolves outward, the direction of evolution will reverse, resulting in either the moon being deposited a few planetary radii from the planet or the moon reimpacting with the planet.  In addition, if the spin axis of the planet is anti-aligned with that of the orbit of the moon, the moon's orbit will shrink, again leading to re-impact.  Alternatively, if the moon's orbit is aligned with the planet's orbit, it will survive and rapidly migrate outwards as a result of the low $Q_p$ value (and consequent high rate of tidal energy dissipation) of terrestrial planets.  Thus, large, distant, moons of terrestrial planets, will have orbits which are roughly coplanar with the orbit of their host planet.
  
For the case where the rotational evolution of the planet is not dominated by the moon (that is, the moon is small or the rotational evolution of the planet is dominated by stellar tides) we can use equation~\eqref{intro_limits_stab_evol_amout} to describe semi-major axis evolution as the sign of the $(n_{p,rot} - n_m)$ term in equation~\eqref{intro_limits_stab_evol_damdef} will remain constant.  From equation~\eqref{intro_limits_stab_evol_amout} we have that the location of the moon after a period of time, $T$, is given by 
\begin{equation}
a_m   = \left(R_R^{13/2} + T \frac{33}{2}\frac{k_{2p}G M_m R_p^5}{\sqrt{GM_p} Q_p} \right)^{2/13}.
\label{intro_limits_stab_tp_Mlim}
\end{equation}

However, as impact generated moons can contain a sizable fraction of their host planet's mass, they can, and do noticeably modify their host planet's rotation rate as they evolve.  For the case where the rotational evolution of the planet is dominated by the moon, the planet and moon may tidally lock to the planet (that is, the moon completes one full orbit per planetary revolution).  In this case the $\mathrm{sgn}$ term in equation~\eqref{intro_limits_stab_evol_damdef} is equal to zero, and consequently, the semi-major axis of the moon's orbit will no longer evolve as a result of this mechanism.  Consequently, distant terrestrial planets may retain large moons that they otherwise would have lost.  For example, a system in which tidal locking has occurred is the Pluto-Charon system.

\subsubsection{Mass limits for moons of gas giants}

In contrast to the satellites of terrestrial planets, regular satellites of gas giants do not form at a location, they form within a region.  In addition, compared to the mass of their host planets, the regular satellites of gas giants are very small, and consequently unable to modify their host's rotation.  Consequently, we can use equations~\eqref{intro_limits_stab_evol_amout} and \eqref{intro_limits_stab_evol_amin}, but, as a result of the range of initial formation locations, need to consider both inward and outward orbital evolution.

For the case of terrestrial planets, we were able to predict the location of an impact-generated satellite after a given period of time, as a function of its mass.  However, as moons of gas giants form within a region, we cannot perform a similar extrapolation for the case of moons of gas giants.  However, we can ask what satellites (in particular what mass satellites) will still be extant, albeit in a different position, after a given period of time.

To answer this question we consider equations~\eqref{intro_limits_stab_evol_amout} and \eqref{intro_limits_stab_evol_amin} the equations governing $a_m$ for the case of outwardly and inwardly evolving moons.  For the outward evolution case, the moon is lost to the system when its orbit becomes three body unstable, that is when $a_m > a_{m,max}$.  Setting $a_m(t)$ to $a_{m,max}$ and rearranging equation~\eqref{intro_limits_stab_evol_amout} to get $M_m$ gives
\begin{equation}
M_m  \le \frac{2}{33}\left(a_{m,max}^{13/2} - a_m(t_0)^{13/2}\right) \frac{\sqrt{GM_p} Q_p}{T k_{2p}G   R_p^5},
\label{intro_limits_stab_gg_Mlimout}
\end{equation}
for the case of outwardly evolving moons.  For the inward evolution case, the moon is lost to the system when in impacts with the planet or when it is tidally disrupted.  And as tidal evolution occurs rapidly for $a_m \approx R_p$ ($\tau \propto (R_p/a_m)^5$), $R_R$ and $R_p$ can be considered equivalent with respect to deriving moon mass limits.  Setting $a_m(t)$ to $R_p$ and rearranging to equation~\eqref{intro_limits_stab_evol_amin} to get $M_m$ gives
\begin{equation}
M_m  \le \frac{2}{33}\left(a_m(t_0)^{13/2} - R_p^{13/2}\right) \frac{\sqrt{GM_p} Q_p}{T k_{2p}G   R_p^5},
\label{intro_limits_stab_gg_Mlimin}
\end{equation}
for the case of inwardly evolving moons.

Physically, equations~\eqref{intro_limits_stab_gg_Mlimout} and \eqref{intro_limits_stab_gg_Mlimin} place limits on the maximum mass moon that can exist around a given gas giant.  For example, using the stability criterion of \citet{Barnesetal2002} for a Jupiter-like planet\footnote{$M_p = 1.8986\times 10^27$kg, $R_p = 71398$km, $k_{2p} = 0.51$ and $Q_p = 10^5$.} on a circular orbit with $a_p = 0.2$AU, we have that extant moons must have mass less than $0.88M_{\earth}$.  This limit rises to $6.6M_{\earth}$ for the case where the more generous stability criterion of \citet{Domingosetal2006} is used.

\section{Summary of literature moon limits}

Informed by the census of the large moons present in the solar system, the formation mechanisms and stability properties of large moons have been discussed.  As discussed at the beginning of this chapter the aim of this analysis is to provide a context in which to view moon detection thresholds.  Consequently, the limits for terrestrial and gas giant planets will be summarised with this in mind.

\subsection{Limits for moons of terrestrial planets}

\begin{figure}
     \centering
     \subfigure[$a_p = 0.2$AU]{
          \label{fig:dl2858}
          \includegraphics[width=.80\textwidth]{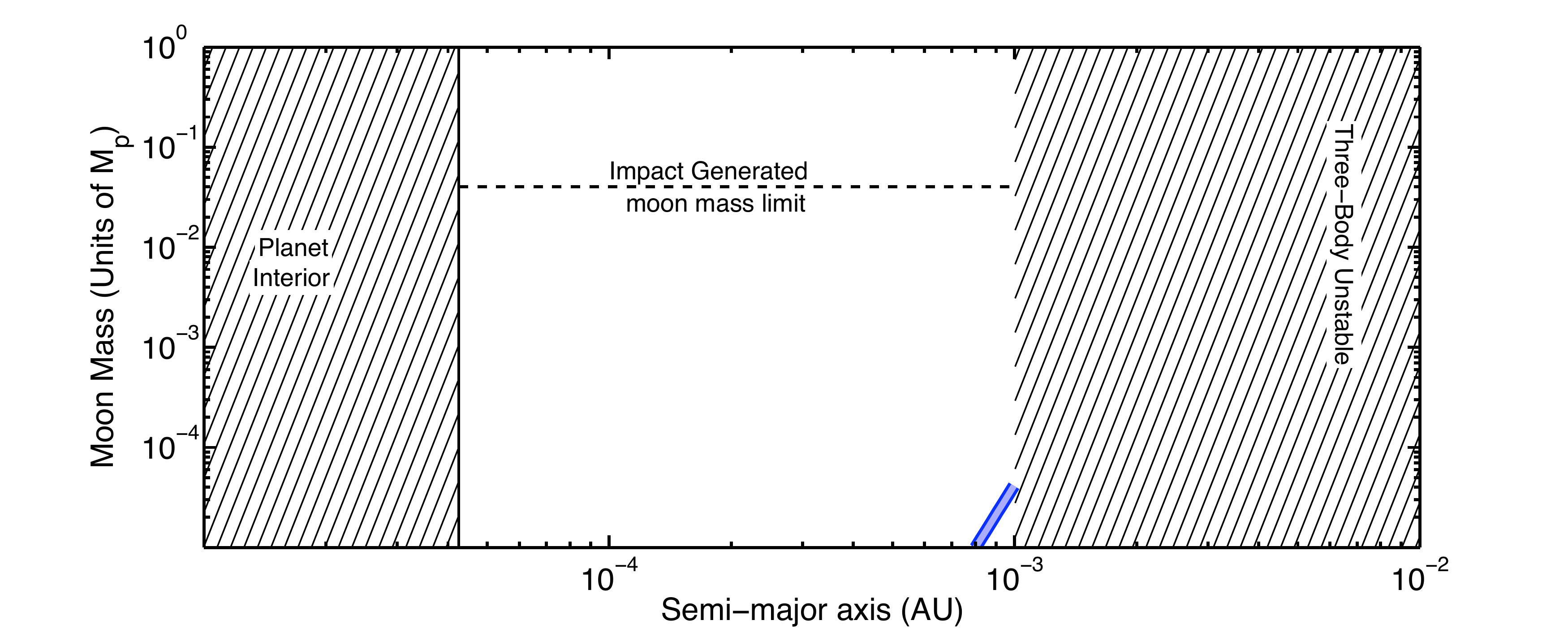}}\\
     \subfigure[$a_p = 0.4$AU]{
          \label{fig:er2858}
          \includegraphics[width=.80\textwidth]{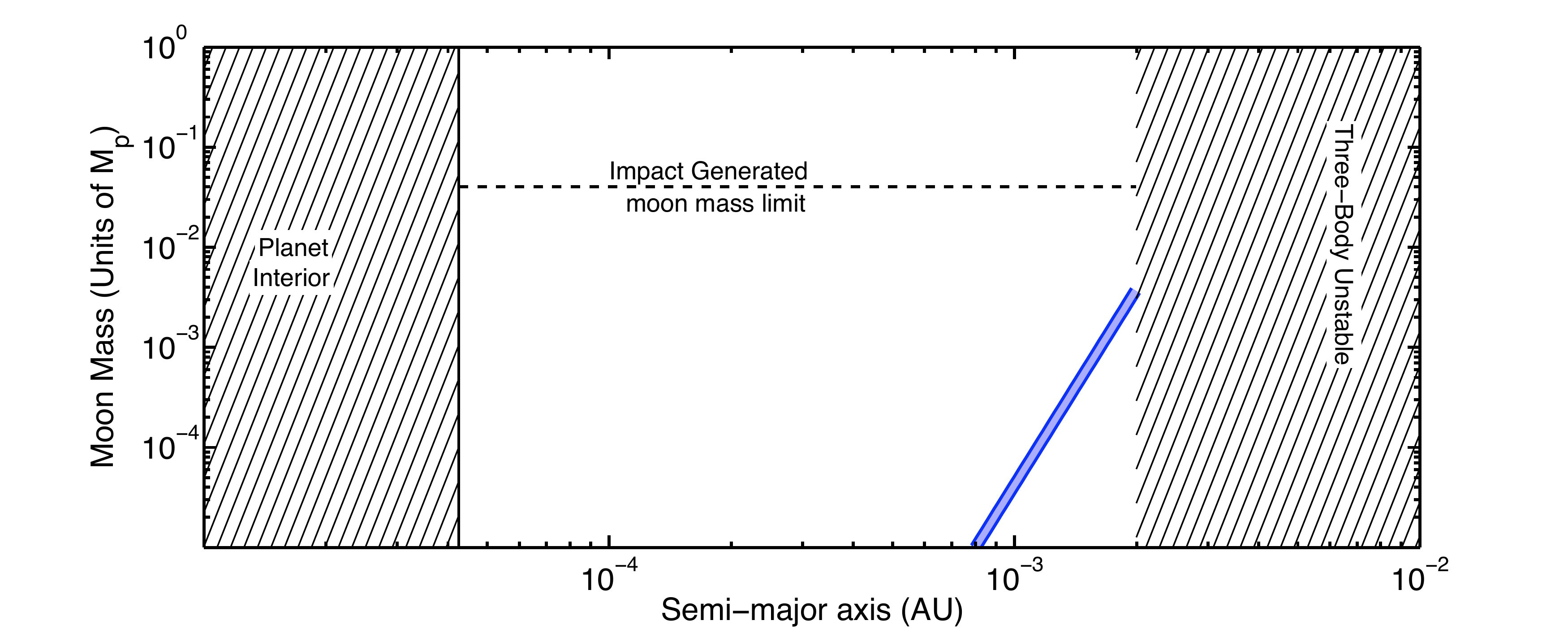}}\\
          \subfigure[$a_p = 1$AU]{
          \label{fig:dl2858}
          \includegraphics[width=.80\textwidth]{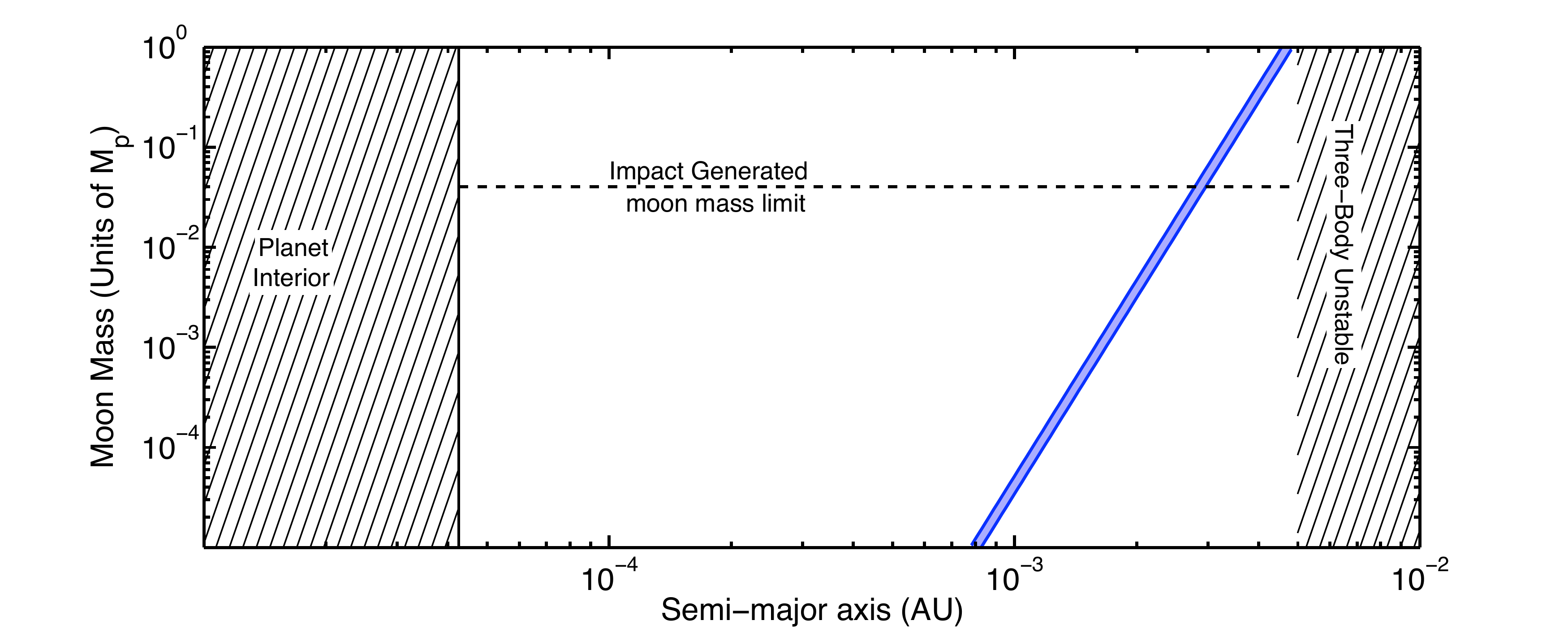}}\\
     \caption[Plot showing the constraints placed on moons of an Earth-analog terrestrial planet as a result of their formation and consequent orbital evolution.]{Plot showing the constraints placed on moons of an Earth-analog terrestrial planet ($M_p = 5.9736 \times 10^{24}$kg, $R_p = 6378$km, $k_{2p} = 0.299$, $Q_p = 12$) as a result of their formation and consequent orbital evolution.  The regions where the moons can never exist (inside the planet and outside the stability region) are cross hatched.  The upper mass limit is denoted by a dashed line.  Finally, for comparison, the semi-major axis an outwardly migrating moon of given mass would evolve to is shown by a blue line, for the case of a system age, $T$, of 5 Gyr.}
     \label{TerPlanMoonLims}
\end{figure}

Constraints can be placed on the physical and orbital characteristics of impact-generated moons as a result of limits imposed by the physics of the formation process, moon orbital evolution and stability.  As discussed in section~\ref{Intro_Moons_Form_Impact}, impact generated moons form from the disk of debris resulting from a collision between an planet-sized impactor and a planet.  For Earth-like terrestrial planets, the impact process is capable of lifting a maximum of 4\% of the total mass contained within of the planet and impactor into orbit, while for larger mass planets, shocks may form in the debris field, resulting in most of the mass being ejected.  Consequently, the mass available to form a moon is limited to at most 4\% of the total mass contained within the two bodies for Earth-sized terrestrial planets, and substantially less for larger planets.  As a result of post-impact interactions, this disk coalesces into a single moon.  While this moon can be formed with any initial orbit orientation, only moons with orbits that are approximately aligned with the planetary orbit will survive and undergo outward orbital evolution. In particular, the semi-major axis of the moon will evolve in accordance with equation~\eqref{intro_limits_stab_tp_Mlim}.  To summarise these results, these limits on moon mass and semi-major axis are shown in figure~\ref{TerPlanMoonLims} for the case of a Earth-like host planet.   
  
\subsection{Limits for moons of gas giants}

\begin{figure}
     \centering
     \subfigure[$a_p = 0.2$AU]{
          \label{fig:dl2858}
          \includegraphics[width=.80\textwidth]{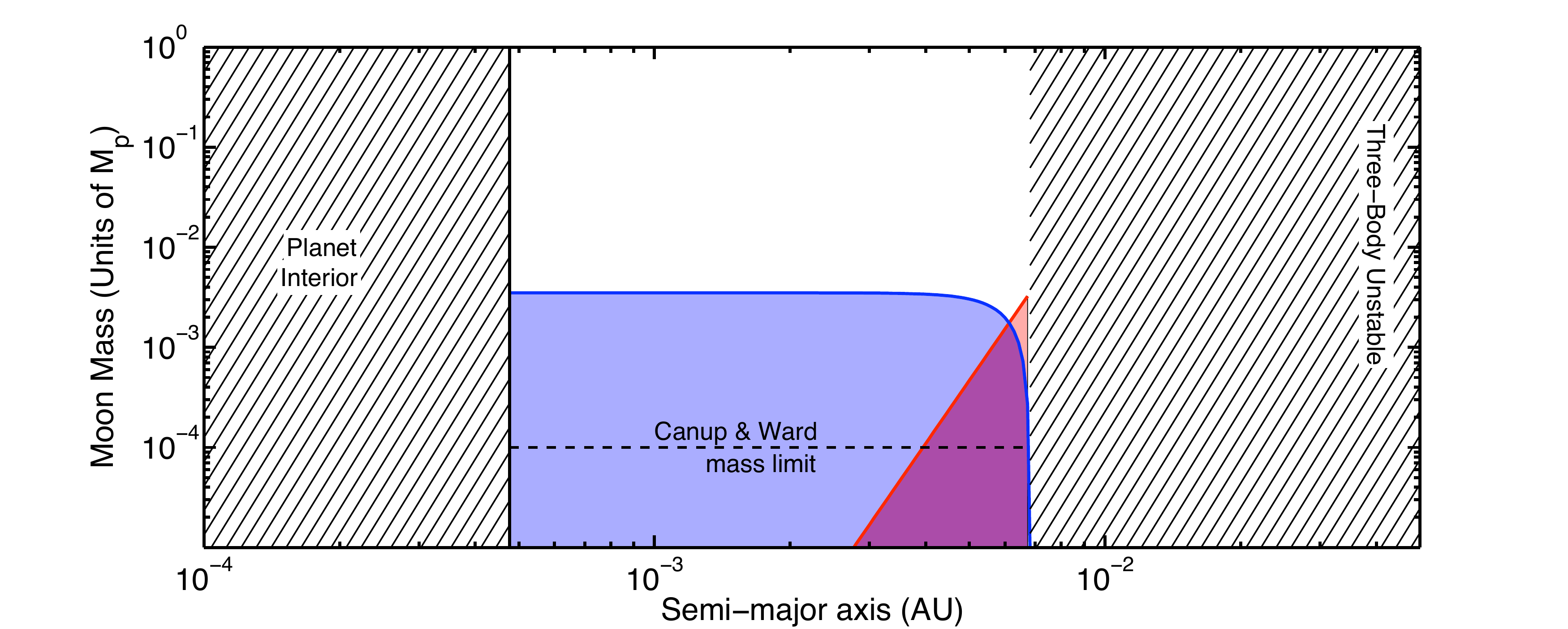}}\\
     \subfigure[$a_p = 0.4$AU]{
          \label{fig:er2858}
          \includegraphics[width=.80\textwidth]{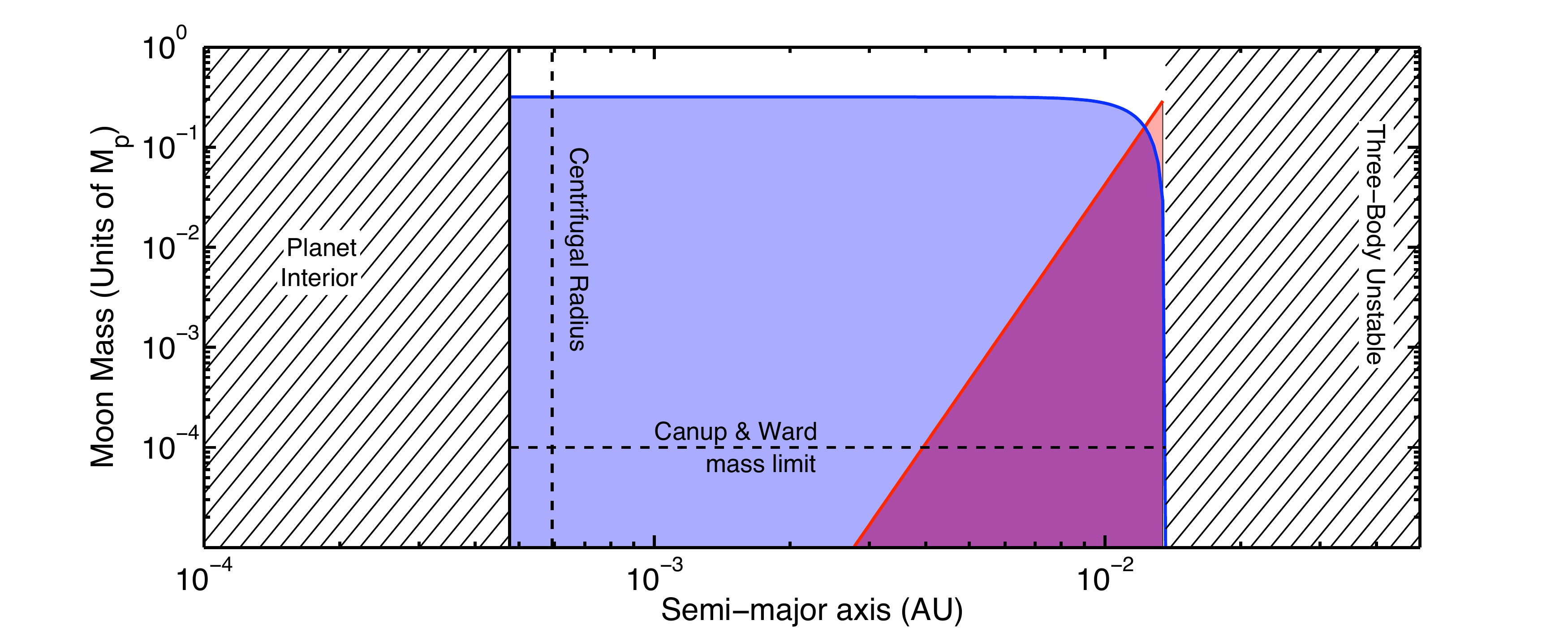}}\\
          \subfigure[$a_p = 1$AU]{
          \label{fig:dl2858}
          \includegraphics[width=.80\textwidth]{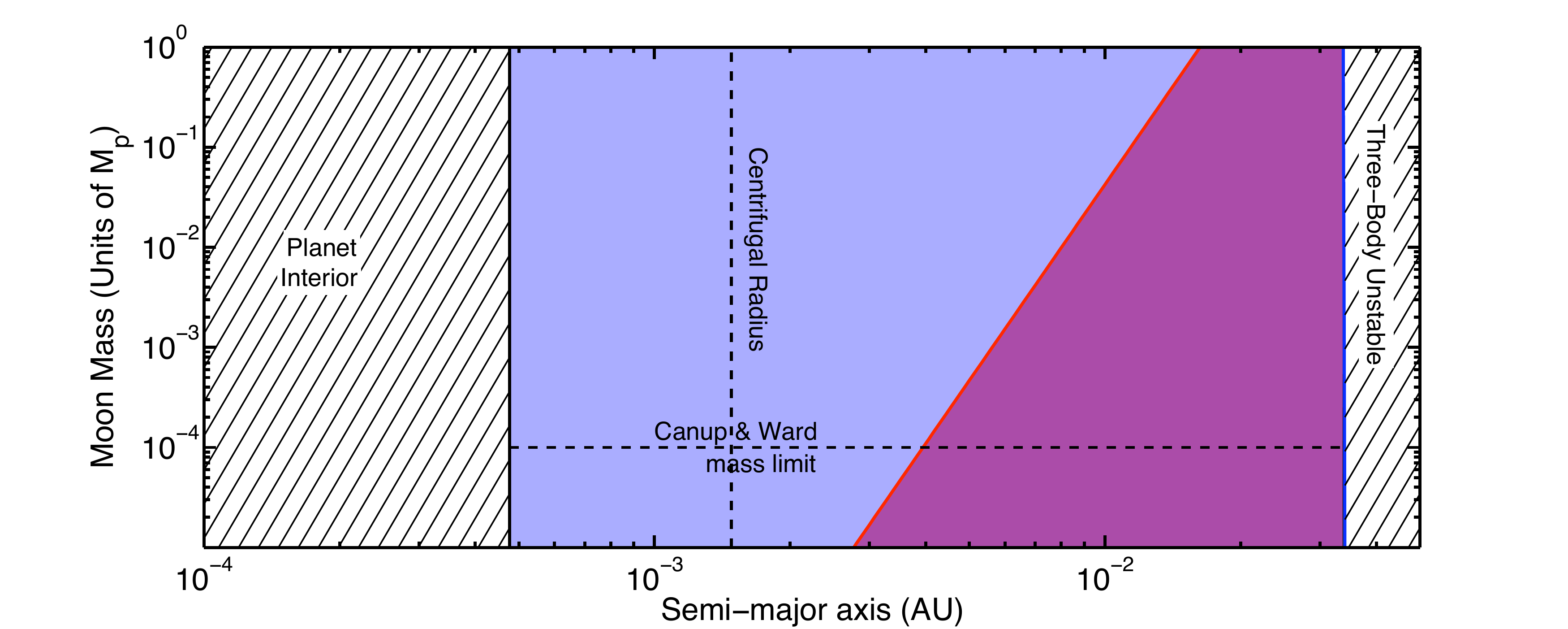}}\\
     \caption[Plot showing the constraints placed on moons of a Jupiter-analog gas giant as a result of their formation and consequent orbital evolution.]{Plot showing the constraints placed on moons of a Jupiter-analog gas giant as a result of their formation and consequent orbital evolution.  The regions where the moons can never exist (inside the planet and outside the stability region) are cross hatched.  The mass limit proposed by \citet{Canupetal2006} and the radius limit proposed by \citet{MosqueiraEstrada2003a} are denoted by dashed lines, and labeled.  Finally, for comparison, the set of moons that, once formed would survive inward and outward migration are denoted by blue and red regions respectively, for the case of a system age, $T$, of 5 Gyr.}
     \label{GasGiantMoonLims}
\end{figure}

Similarly, an understanding of the processes related to moon formation and stability can constrain the set of moons of gas giants that are predicted to form.  As discussed in section~\ref{Intro_Moons_Form_Disk}, it is believed that the regular satellites of gas giants formed within a circumplanetary disk, resulting in a set of moons with orbits which are nearly exactly aligned with the equatorial plane of their host planet.  As the equator of a gas giant should show a tendency to be aligned with the orbital plane, this corresponds to a tendency for the orbit of moons of gas giants to also be aligned with the orbit of their host planet.  In addition, models for this process suggest that the total mass contained within large moons of a gas giant should be approximately $2.5 \times 10^{-4}$ of that gas giant's mass, and the majority of large moons of gas giants should form within a certain distance (be it 60 $R_p$ or $R_c$) of their host planet.  Once formed, the moons will migrate according to equation~\eqref{intro_limits_stab_evol_amout} or \eqref{intro_limits_stab_evol_amin}, and consequently, the maximum mass of a moon formed at a semi-major axis of $a_m(t_0)$ and still retained within the system is given by equations~\eqref{intro_limits_stab_gg_Mlimout} and \eqref{intro_limits_stab_gg_Mlimin} for the case of outward and intward migration respectively.  These limits are summarised for the case of a Jupiter-like\footnote{The term Jupiter-like is a little misleading as some of the properties of Jupiter, in particular, its $Q$-value, are not well constrained.  Recent result suggest that the $Q$-value of a gas giant may depend on the forcing frequency, in this case, on the orbital frequency of the moon.  While Jupiter's $Q$-value has been measured to be $10^5 - 10^6$ by \citep{GoldreichSoter1966}, and $(3.56 \pm 0.66) \times10^4$ by \citep{Laineyetal2009}, there are good theoretical reasons for it to be as high as $10^{12}$ \citep{GoldreichNicholson1977,Wu2005}.  Recent theoretical work \citep{OgilvieLin2004,Wu2005} suggests that the naturally high ($10^{12}$) $Q$-value of gas giants could be suppressed ($10^5-10^9$) for the case where the forcing frequency is less than twice the spin frequency of the planet, neatly explaining both results.  For this work we use a constant $Q$-value of $10^5$, but note that the adoption of a $Q$-value of $10^{12}$ allows moons of even hot Jupiters to be dynamically stable \citep{Cassidyetal2009}.}
 planet in figure~\ref{GasGiantMoonLims}.

\section{Conclusion}

By considering the census of moons within the Solar System along with the current understandings of moon formation and orbital evolution, we have summarised the types of large moons that we expect extra solar planets to host.  Terrestrial planets are expected to host a single large moon (if any) containing up to 4\% of the planet's mass.  In addition, the semi-major axis of this moon will be defined by the system lifetime and the orbital evolution timescale of the moon.  In comparison, gas giant planets are expected each to possess a small number of large moons, where the total mass contained within these moons is approximately equal to $2.5 \times 10^{-4} M_p$.  In addition, these moons should form relatively close to the planet (either $R_c$ or $60R_p$) and then slowly evolve inward or outward depending on whether the rotational period of the planet is longer or shorter than the orbital period of the moon.  From this context we can start to look at the set of proposed moon detection mechanisms.
\chapter{Review of planet and moon detection techniques}\label{Intro_Dect}

\section{Introduction}

To provide a context for the two moon detection methods analysed in this thesis, it would be instructive to review the set of moon detection methods presented in the literature, and the types of moons they are capable of detecting.  As moon detection techniques are in part inspired by planet detection techniques, and a planet must be detected using one of these techniques before a moon of that planet can be detected, we will begin our discussion with a brief overview of planet detection.  In particular, we will look at planet detection techniques in terms of the physics they are based on, how effective they are at finding planets, and the types of planets likely to be discovered using them.  Building on this framework, moon detection will then be discussed.  In particular it will be discussed with respect to physics of the detection method, the type of host planet required, and the types of moons that can be discovered.

\section{Planet detection techniques}

\begin{figure}[tb]
\begin{center}
\includegraphics[width=.98\textwidth]{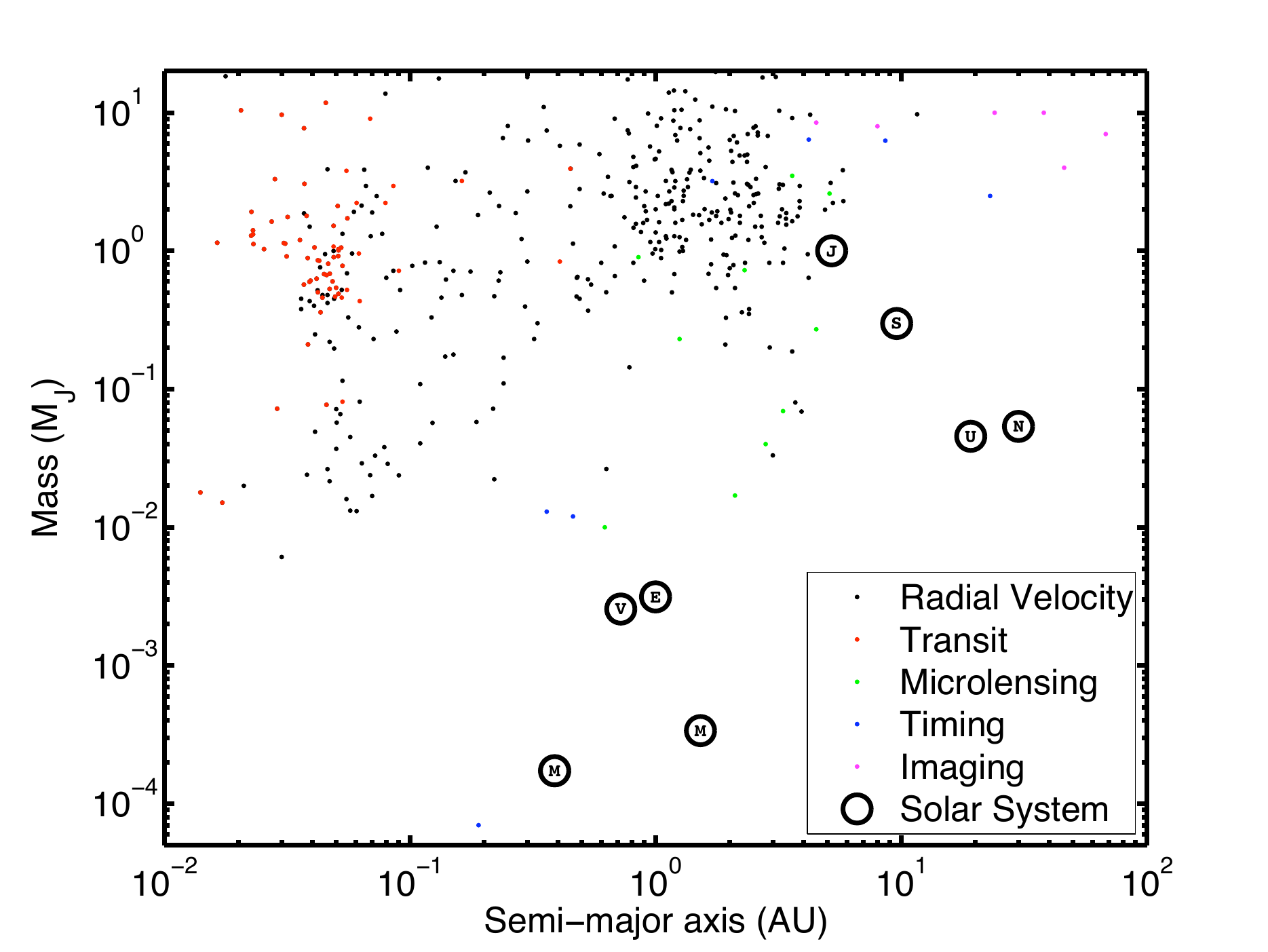}
\caption[Plot of all known extrasolar planets (small dots) as of the 5$^{th}$ of April 2010, colour-coded by detection method. ]{Plot of all known extrasolar planets (small dots) as of the 5$^{th}$ of April 2010, colour-coded by detection method.  For the cases where $M_p \sin I_p$ is known, while $M_p$ is not, for example, the planets detected by radial velocity, $M_p \sin I_p$ has been used in place of $M_p$, as it indicates the lowest possible value of $M_p$.  For comparison, the planets of our Solar System are also shown.}
\label{SetOfAllXOPlanets}
\end{center}
\end{figure}

Currently, over 500 extrasolar planets have been discovered.\footnote{See e.g
http://exoplanet.eu/catalogue.php}  These discoveries have been made using a variety of methods, including, the radial velocity, transit, microlensing, timing and direct imaging techniques.  In addition, other techniques, such as astrometry, have been proposed, but have yet resulted in no successful detections of planets.  To provide a context for the upcoming discussion on moon detection, these techniques will be described in terms of the physical basis for the detection technique, and the type and number of planets detected.  For reference and comparison, a diagram showing all the planets as of 5$^{th}$ of April 2010 is given in figure~\ref{SetOfAllXOPlanets}.  We will begin with the radial velocity technique, the most successful planet detection technique to date.

\subsection{The radial velocity technique}

The radial velocity technique is a planetary detection technique which involves measuring the reflex ``wobble" of the
planet's parent star due to the planet's motion.  Practically, this is done by measuring the periodic shift of the absorption lines in the star's spectrum and converting this to a line-of-sight velocity.  The amplitude of this velocity for a planet on a circular orbit is given by
\begin{align}
max(v_{p,los}) &= \frac{M_p}{M_s} \sqrt{\frac{GM_s}{a_p}} \sin I_p,\\
 &= 28.4 \frac{M_p \sin I_p}{M_J} \left(\frac{M_{\sun}}{M_s}\right)^{1/2} \left(\frac{1 AU}{a_p}\right)^{1/2} \text{ms}^{-1},
 \label{intro_det_planet_rvamp}
\end{align}
where $M_p$ and $M_s$ are the mass of the planet and star respectively, and, $a_p$ and $I_p$ are the semi-major axis and line-of-sight inclination of the planet's orbit.  As can be seen in equation~\eqref{intro_det_planet_rvamp} and figure~\ref{SetOfAllXOPlanets}, this technique is more sensitive to planets in close orbits, but is still capable of detecting more distant planets.  While this technique is capable of detecting planets with any orbital inclination, it does have the drawback of only yielding $M_p\sin I_p$ where $M_p$ is the true mass of the planet and $I_p$ is the inclination of the orbit. Despite requiring high resolution, high signal-to-noise spectra that are expensive in terms of telescope time, it is currently the most successful planetary detection technique, with nearly 500 planets detected.

\subsection{The transit technique}


For the case where a planet's orbit is aligned such that it transits across the face of its host star as viewed from the Earth, the presence of the planet can be inferred by the corresponding dip of the intensity of its host star during this passage.  Such a transit will have a depth of approximately $(R_p^2/R_s^2)L_0$,\footnote{Note that $R_p^2/R_s^2$ is equal to the ratio of the projected area of the planet to that of the star.} where $R_p$ and $R_s$ are the radius of the planet and star respectively, and $L_0$ is the unoccluded luminosity, and for a planet on a circular orbit, can last up to approximately
\begin{align}
T_{tra} &= \frac{2 R_s}{v_{tr}} = 2R_s\left(\frac{GM_s}{a_p}\right)^{-1/2} \\
 &= 13 \left(\frac{R_s}{R_{\sun}}\right) \left(\frac{M_{\sun}}{M_s}\right)^{1/2}\left(\frac{a_p}{1 AU}\right)^{1/2} \text{hrs},
 \label{intro_det_planet_transDur}
\end{align}
depending on the planet's orbital inclination.  As planets can only be detected if they transit (probability of transiting is $\propto a_p^{-1}$) and during transit (time between transits is proportional to $a_p^{-3/2}$), this technique is strongly biased toward detecting short period planets (see figure~\ref{SetOfAllXOPlanets}).  While not all planets will be detected, this technique is photometric as opposed to spectroscopic, which means that  each star in a crowded field can be monitored simultaneously, so many more stars (and many fainter stars) can be investigated for planets.  So far, 115 transiting planets are known.\footnote{The number of planets discovered using the transit technique is not an easily definable quantity.  This is partially because, in order to be confirmed, a planet detected via the transit technique must also be detected using radial velocity.  In addition, planets initially detected using radial velocity may later be discovered to transit.}  This method is discussed in greater detail in section~\ref{Trans_Intro_Transtech}.

\subsection{Microlensing}

Planets can also be detected by the perturbations caused by their gravitational field.  To see how, consider a relatively nearby star (the lens star) moving against a background of more distant stars.  If the projected distance on the sky between this lens star and a background star (the source star) becomes small enough, the gravitational field of the lens star can perturb the path of photons leading to image the image of the source star being magnified or demagnified.  In addition, if this lens star has a planet, the planet can lead to additional spikes in the light curve of the source star.  This technique is sensitive to planets with a projected distance from their host star of approximately one stellar Einstein radius, where the stellar Einstein radius is defined as
\begin{align}
R_E &= D_L\theta_E,\\
 &= D_L \left(\frac{4GM_L}{c^2}\right)^{1/2} \left(\frac{1}{D_L} - \frac{1}{D_S}\right)^{1/2},\\
 &= 6AU \left(\frac{D_L}{6\text{kpc}}\right) \left(\frac{M_L}{M_{\sun}}\right)^{1/2} \left(\frac{8\text{kpc}}{D_S}\right)^{1/2}  \left(\frac{D_S}{D_L} - 1\right)^{1/2}, 
\end{align}
where $\theta_E$ is the angular einstein radius, $G$ is the universal gravitational constant, and $D_L$ and $D_S$ are the distances between the observer and the lens and the observer and the source respectively, and where the scale distance for $D_S$ of $~$8kpc (the Milky Way bulge) could just have equally been $~$50kpc (Large Magellanic Cloud), or $~$60kpc (Small Magellanic Cloud).  This distance ranges from 1AU to 5AU for typical microlensing systems, which corresponds well with the 1AU to 5AU band in which microlensing planets have been discovered (see figure~\ref{SetOfAllXOPlanets}).  Currently, twelve planets, in eleven systems have been detected using this technique.

\subsection{Timing}

The timing technique involves measuring perturbations in the arrival time of periodic events associated with a host star, resulting from the reflex motion caused by orbiting planets. Example periodic events include radio pulses associated with millisecond pulsars, pulsations associated with giant stars or with white dwarf stars and eclipses of binary stars.  The timing amplitude for a planet in a circular orbit about a host with mass $M_s$ is given by 
\begin{align}
max(t_{pert}) &= \frac{1}{c}\frac{M_p}{M_s} a_p \sin I_p,\\
 &= 0.5 \frac{M_p \sin I_p}{M_J} \frac{a_p}{1 AU} \frac{M_{\sun}}{M_s} \text{s},
 \label{intro_det_planet_timeamp}
\end{align}
where $M_p$ and $M_s$ are the mass of the planet and host, $a_p$ is the semi-major axis of the planet's orbit and $c$ is the speed of light.  Similar to the case for the radial velocity technique, the planetary mass appears only in the term $M_p \sin I_p$.  Consequently, only $M_p \sin I_p$ can be directly measured.  However, for the case of millisecond pulsar host stars, this technique is so sensitive, that second order effects, such as resonance effects, can be used to define limits to the orbital inclination and consequently provide an estimate of the planet's mass \citep{KonackiWolszczan2003}.  Currently ten planets have been discovered using this technique, four in two millisecond pulsar systems \citep{Wolszczanetal1992,Wolszczan1994,Backeretal1993}, one around a pulsating horizontal branch star \citep{Silvottietal2007} and and five planets in three circumbinary systems \citep{Leeetal2009,Qianetal2009,Qianetal2010}.  This technique is discussed further with respect to millisecond pulsars, in Part~\ref{PulsarPart}.

\subsection{Direct imaging}

As its name suggests, the technique of direct imaging involves detecting planets through an image.  Unfortunately, making an image of an extra-solar planet is challenging for two main reasons.  First, host stars are many orders of magnitude brighter than their attendant planets ($10^{9}$ for the case of a Jupiter analog around a Sun-like star).  Second, the planets are very close, generally in the wings of the stellar point-spread function.  Consequently, special techniques such as coronography or adaptive optics need to be employed to implement this method, and even then, it is most sensitive to hot distant planets.  While distant hot planets are uncommon \citep{Nielsenetal2008}, they do exist.  Currently, using this technique, seven planets have been detected around four stars.\footnote{While other candidate planets have been detected, the errors in their masses are so large that their planetary status is uncertain, or they are orbiting brown dwarfs.}  Despite the technical challenges imposed by this method, it can potentially offer high returns in terms of planetary followup, for example, detection of the planetary spectrum.  As a result, the capability to directly detect extra-solar planets is a strong science goal in a number of proposals for the next generation of telescopes, for example, the Terrestrial Planet Finder (TPF) and the Giant Magellan Telescope (GMT).
 
\subsection{Astrometry}

Astrometry is a planet detection technique that involves measuring the motion of a star about the planet moon barycenter via the perturbation of the star's position on the plane of the sky.  Unlike the radial velocity technique, this technique can reveal a planet's true orbital inclination and, combined with an estimate of the star's mass, the planet's true mass.  For circular orbits, $\alpha$, the maximum amplitude of such an angular displacement is given by 
\begin{equation}
\alpha = \frac{M_p}{M_s} \frac{a_p}{D} = 0.3 \frac{M_{\sun}}{M_s} \frac{M_p}{M_{\earth}} \frac{a_p}{1\text{AU}} \frac{1\text{pc}}{D} \mu\text{as},
\end{equation}
where $M_p$ is the mass of the planet, $M_s$ is the mass of the star, $a_p$ is the semi-major axis of the planet's orbit and $D$ is the distance between the host star and the Earth.  As the amplitude of the signal is inversely proportional to $D$, and proportional to $a_p$, this method is optimised to detect distant planets of nearby low-mass stars.  As of the 24$^{th}$ of March 2010, no known extra-solar planets has also been successfully detected astrometrically \citep{PravdoShaklan2009,Beanetal2010}.  However, future space missions, most notably SIM,\footnote{See for example, http://sim.jpl.nasa.gov/keyPubPapers/SIMLiteBook/SIM-Book-Full-Book-LR.pdf.} are planned to use this technique to find low mass distant planets.

\section{Moon detection techniques}

\begin{figure}[tb]
\begin{center}
\includegraphics[width=.91\textwidth]{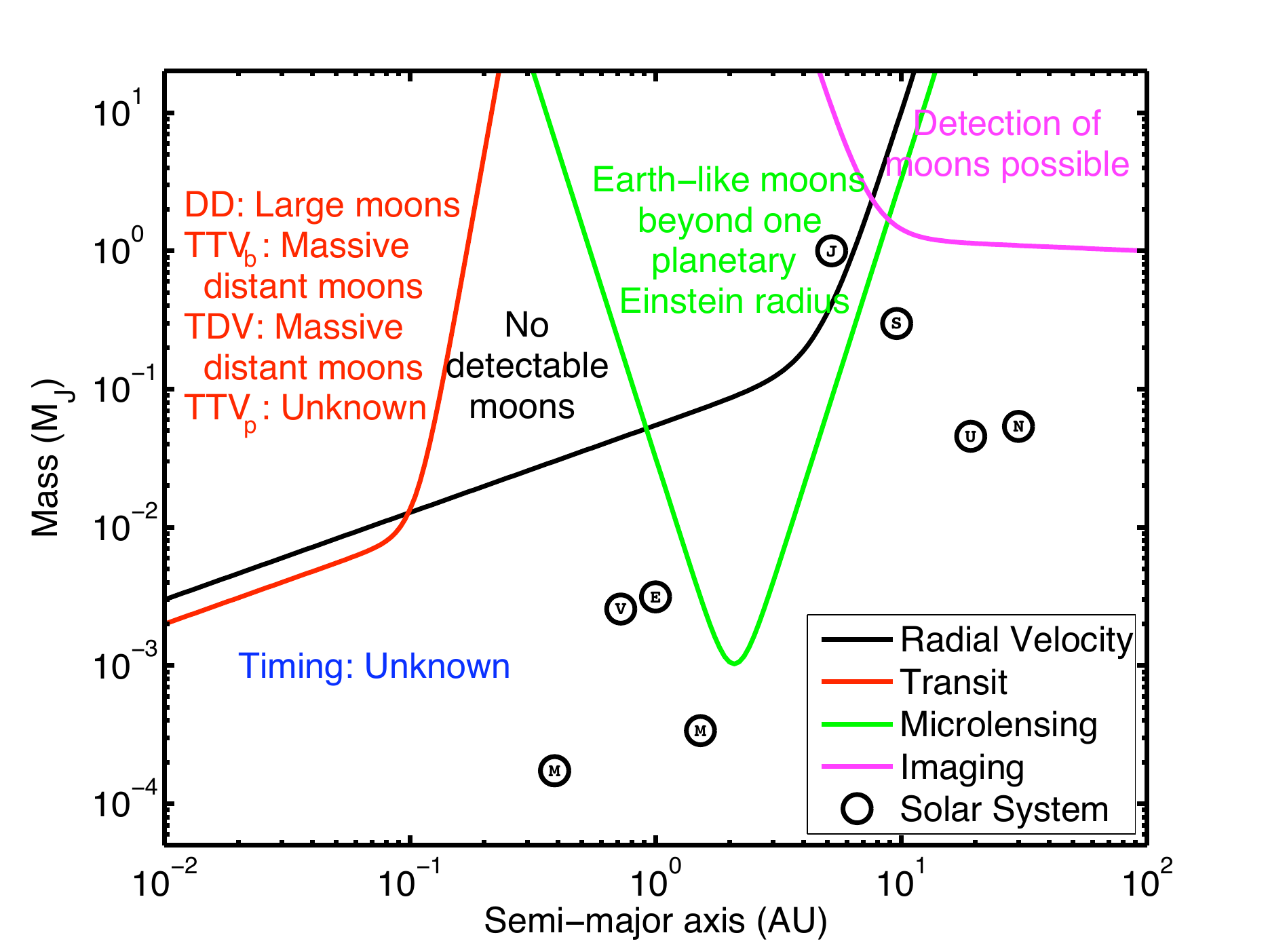}
\caption[A diagram using the same scale and colour scheme as figure~\ref{SetOfAllXOPlanets}, summarising the types of moons which can be detected as a function of planetary detection method.]{A diagram using the same scale and colour scheme as figure~\ref{SetOfAllXOPlanets}, summarising the types of moons which can be detected as a function of planetary detection method.  Information on the type of moons which can be detected, is provided inside the cartoon representation of the region in which planets have currently been discovered using a given method.  For the case of planets detected by timing, no cartoon was included as the detectability of a given planet depends more on the properties of its host that its properties.  Again, for comparison, the Solar System planets are overlaid.}
\label{XOMoonProps}
\end{center}
\end{figure}

To date, no extra-solar moons have been discovered, with only upper limits placed on moon radii and mass of the planets orbiting HD~209458 \citep{Brownetal2001} and OGLE-TR-113b \citep{Gillonetal2006}. Despite this, a number of investigations have been conducted into possible methods for detecting extrasolar moons.  As most of these methods are extensions of planet detection techniques, the corresponding moon detection techniques will be summarised in the same order used in the preceding section.  In addition, the properties of these detectable extra-solar moons are summarised in figure~\ref{XOMoonProps}, according to the type of host planet, that is, the type of method used to detect the host, and the moon detection technique used.

\subsection{Radial velocity perturbation}

As the majority of planets have been discovered using the radial velocity technique and approximately half of these planets are distant enough from their stars to host sizable moons (see section~\ref{Intro_Moons_Stab_Lims}), it seems intuitively sensible to try and extend the radial velocity technique to search for moons of these radial velocity planets.  While the possibility of using this technique has been suggested in the literature \citep[e.g.][]{Szaboetal2006}, it was qualified by the statement that success would be ``unlikely" due to the small signals produced.

Currently, no analysis of the radial velocity perturbation specific to planet-moon systems has been performed.  However, analogous systems have been studied. \citet{Schneideretal2006} studied the radial velocity perturbation caused by a pair of equal mass binary stars on a companion and found the amplitude of the perturbation to be
\begin{equation}
max(v_{pert}) = 9 \sqrt{\frac{G M_A}{a_A^3}} \frac{a_B^5}{a_A^4},
\label{intro_det_moon_rvstaramp}
\end{equation}
where $G$ is the universal gravitational constant, $M_A$ is the mass of the companion star, $a_A$ is the distance between the companion star and the center of mass of the binary, and $a_B$ is the distance between one of the binary stars and the center of mass of the binary pair.  Designating one of the binary pair to be the ``planet", and the other the ``moon", and setting $a_m = 2A_B$,\footnote{\citet{Schneideretal2006} defined $a_B$ to be the distance between the center of mass of the binary pair and one of the components.  For this thesis $a_m$ was defined to be the semi-major axis of the orbit of the moon relative to the planet.  From the perspective of the ``planet" star, the ``moon" star orbits at a distance of $2a_B$ ($a_B$ to get from the ``planet" star to the center of mass, and $a_B$ to get tot he ``moon" star).  Consequently $a_m = 2a_B$.} then setting the mass of the companion to the mass of the host star and $a_p = a_A$, equation~\eqref{intro_det_moon_rvstaramp} becomes
\begin{align}
\text{max}(v_{pert}) &= \frac{9}{2^5} \sqrt{\frac{G M_s}{a_p^3}} \frac{a_m^5}{a_p^4}, \label{intro_det_moon_rvmoonamp1}\\
 &= 0.013 \left(\frac{1 AU}{a_p}\right)^{1/2} \left( \frac{M_{\sun}}{M_s} \right)^{7/6} \left( \frac{M_p}{M_J} \right)^{5/3} \left(\frac{a_m}{R_H}\right)^{5} \text{ms}^{-1}.\label{intro_det_moon_rvmoonamp2}
\end{align}

From equation~\eqref{intro_det_moon_rvmoonamp2} we have that the perturbation is maximised for planet moon pairs that are massive, distant from each other, but close to their host star.  For such large moons, the minimum size of $a_p$ is limited by the timescale for orbital decay to approximately 0.6AU \citep{Barnesetal2002}, and the maximum size of $a_m$ is limited by three body stability to $\sim 0.5 R_H$ for prograde orbits and $\sim R_H$ for retrograde orbits (see section~\ref{Intro_Moons_Stab_Dest}).  Setting the mass of the host star to $0.1M_{\sun}$, setting $a_p = 0.6AU$, setting $a_m = R_H$ and setting the mass of the planet and moon to $5M_J$ gives a perturbation signal of amplitude 3.7ms$^{-1}$.  As that these are very extreme conditions, and radial velocity measurements are currently limited to a little below 1ms$^{-1}$ due to stellar surface motion, detection is indeed ``unlikely".  Consequently, this method is unlikely to be of practical use.

\subsection{Perturbation to transit light curve}\label{Intro_Dect_Moons_Transit}

\begin{figure}[tb]
\begin{center}
\includegraphics[width=.70\textwidth]{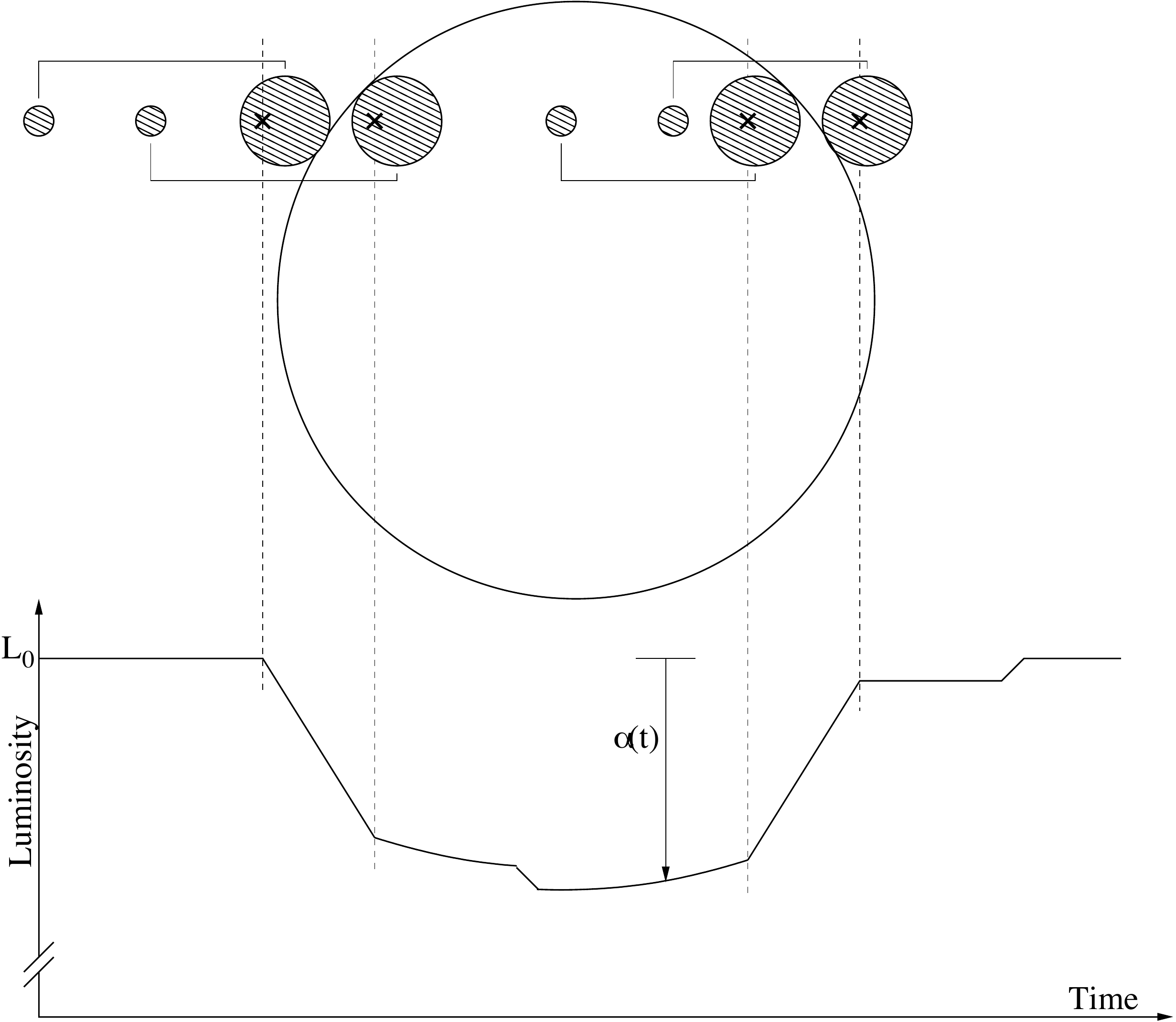}
\caption[Schematic diagram of a transit light curve for the case of a planet and a moon]{Diagram showing the different portions of the transit light curve for the case where both a planet and moon transit.  Four silhouettes of the planet and moon are shown, corresponding to the the beginning and end of planetary ingress, and the beginning and end of planetary egress.  Planet-moon pairs which correspond to a single silhouette are joined by a solid line, while the location of the planet-moon barycenter is indicated by cross.  As the position of the planet-moon barycenter is a linear function of time it can be used as a proxy for time.  Consequently the position of the barycenter and the value of the light curve resulting from that position are linked by dashed lines.  }
\label{PlanetMoonTransitSch}
\end{center}
\end{figure}

\begin{figure}
     \centering
     \subfigure[$R_p$=$R_J$, $a_p=1$AU, $\delta_{min} = 0$.]{
          \label{TransitSignalCoordSysDirectCC}
          \includegraphics[width=.485\textwidth]{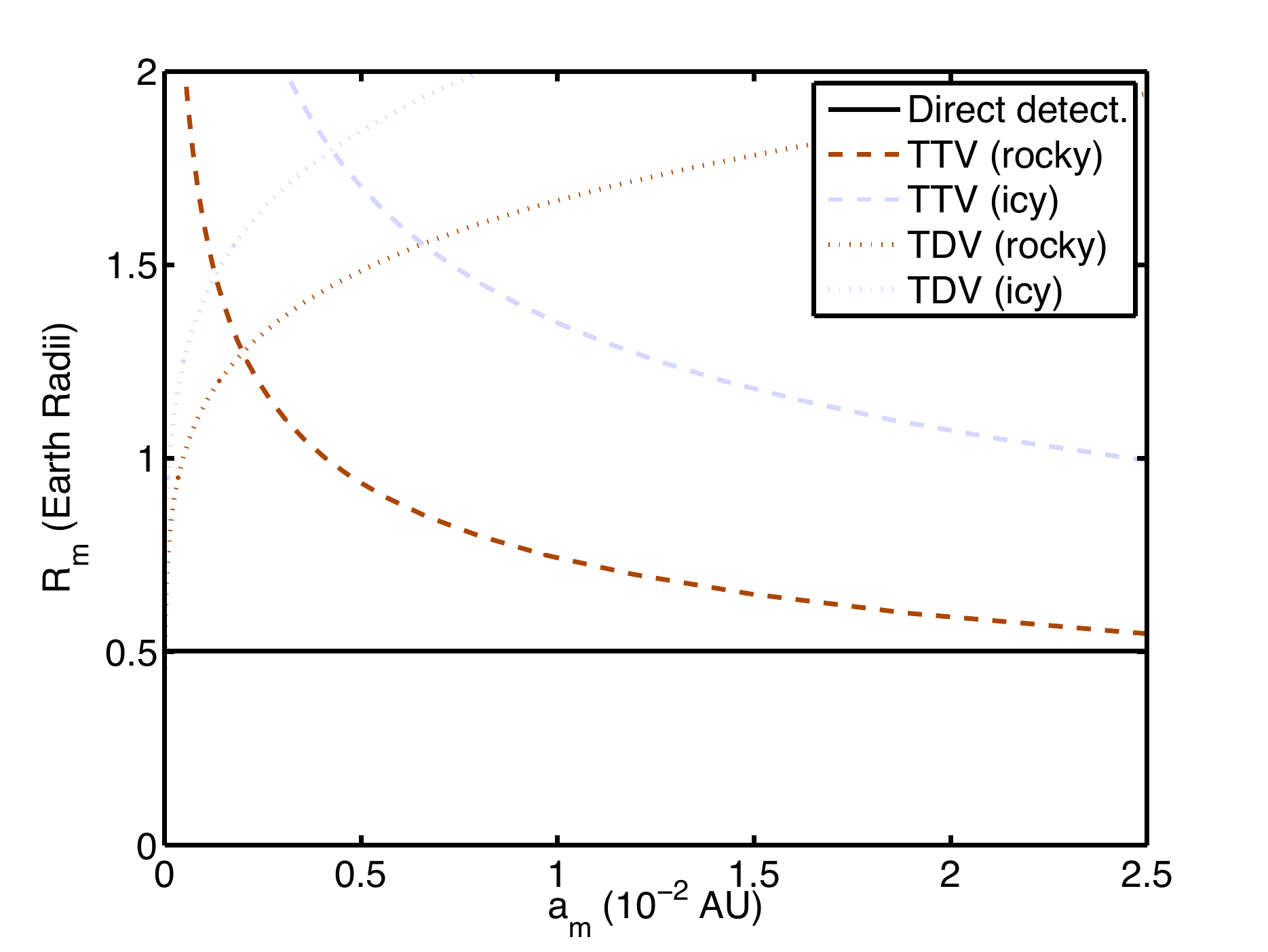}}
     \subfigure[$R_p$=$R_{\earth}$, $a_p=1$AU, $\delta_{min} = 0$.]{
          \label{TransitSignalCoordSysDirectLOS}
          \includegraphics[width=.485\textwidth]{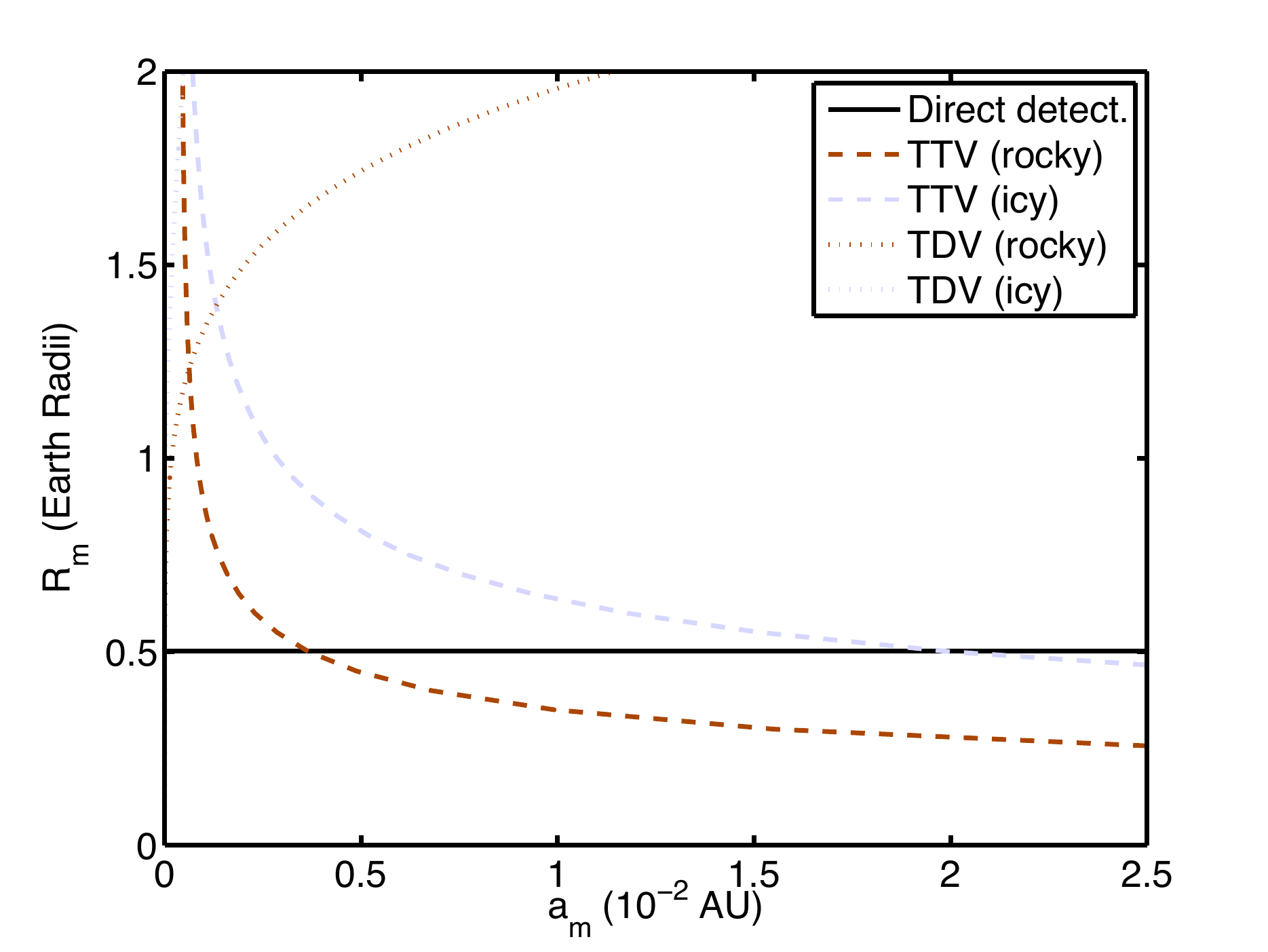}}\\
     \subfigure[$R_p$=$R_J$, $a_p=0.2$AU, $\delta_{min} = 0$.]{
          \label{TransitSignalCoordSysDirectLOS}
          \includegraphics[width=.485\textwidth]{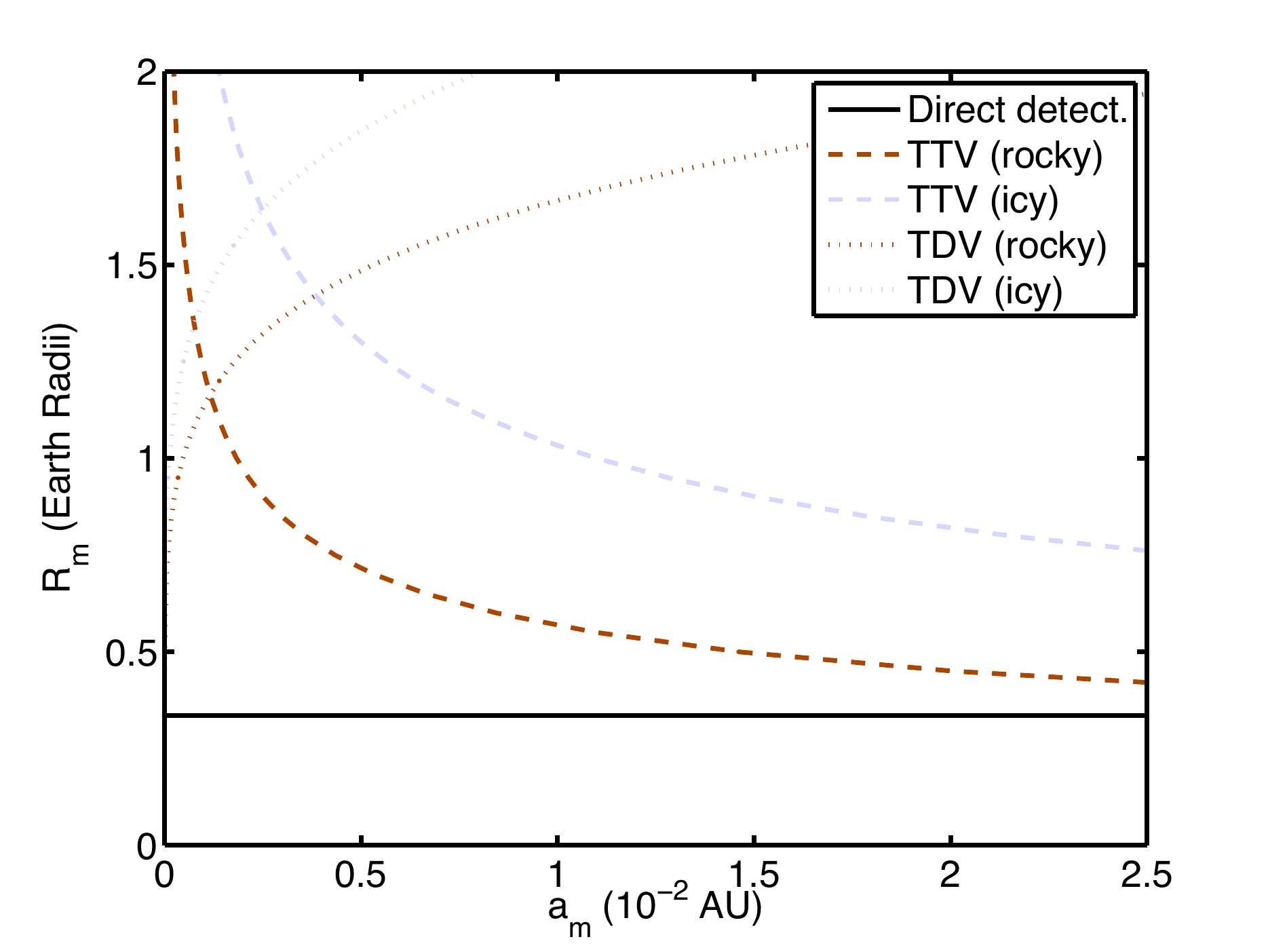}}
     \subfigure[$R_p$=$R_J$, $a_p=1$AU, $\delta_{min} = 0.5 R_{\sun}$.]{
          \label{TransitSignalCoordSysDirectLOS}
          \includegraphics[width=.485\textwidth]{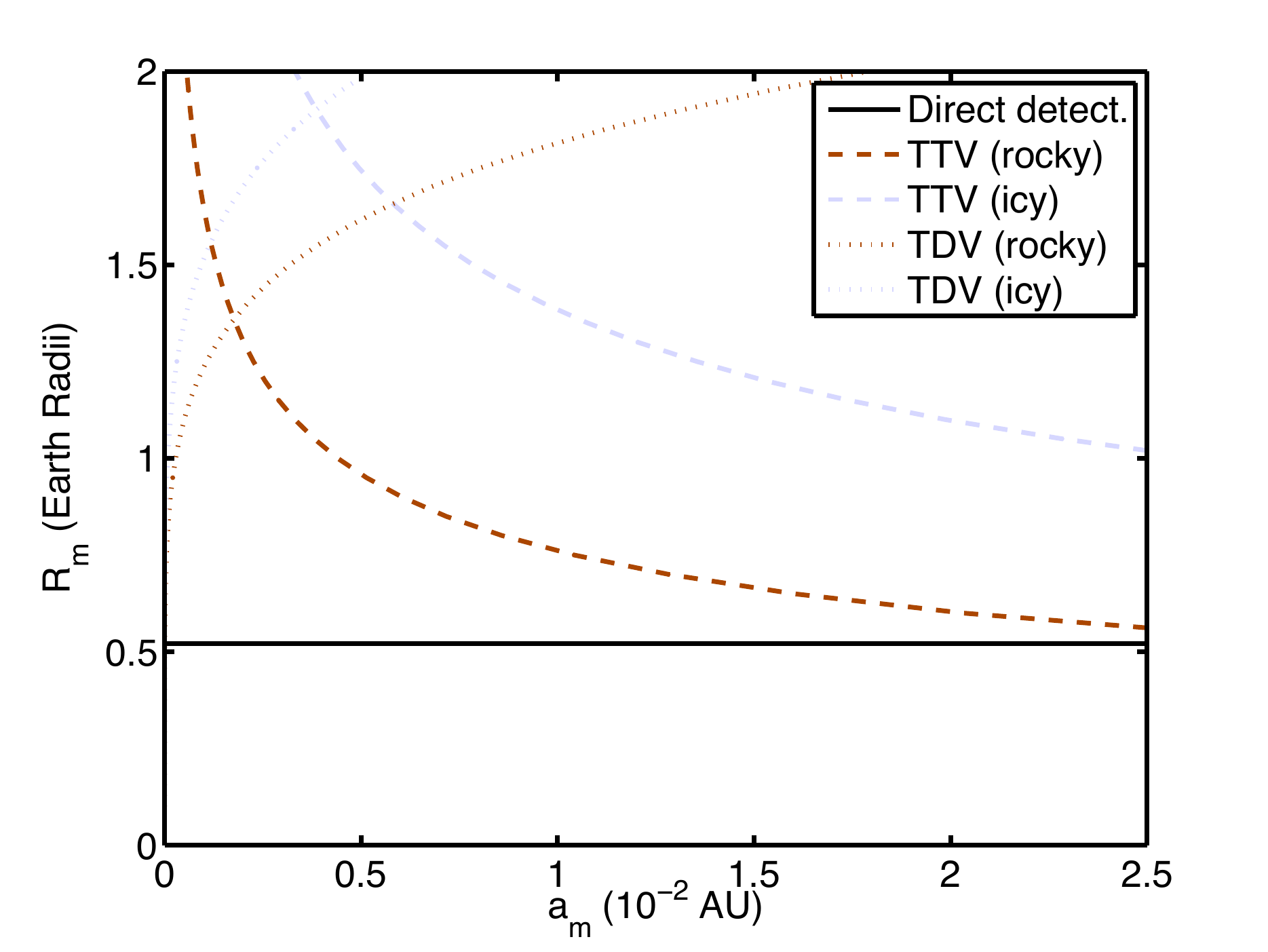}}
 \caption[Approximate moon detection thresholds for the direct detection, barycentric transit timing and transit duration variation methods calculated using equations~\eqref{intro_det_moon_DDtransthresh}, \eqref{intro_det_moon_TTVbtransthresh} and \eqref{intro_det_moon_TDVtransthresh}.]{Approximate moon detection thresholds for the direct detection, barycentric transit timing and transit duration variation methods calculated using equations~\eqref{intro_det_moon_DDtransthresh}, \eqref{intro_det_moon_TTVbtransthresh} and \eqref{intro_det_moon_TDVtransthresh}.  It is assumed that the length of time over which observations were recorded was four years, corresponding to $N=4$ for the case of $a_p = 1$AU and $N=44$ for the case of $a_p = 0.2$AU.  For the case of the barycentric transit timing and transit duration variation methods, the thresholds for rocky ($\rho_m = 5515$kgm$^{-3}$) and icy ($\rho_m = 916.7$kgm$^{-3}$) moons are shown in brown and blue respectively.  These thresholds are shown for four different cases, for a standard (a), and three comparison cases (b) - (d).  The comparison between (a) and (b) shows the effect of changing from a Jupiter-like host planet to an Earth-like host planet.  The comparison between (a) and (c) shows the effect of reducing the semi-major axis of the host planet from 1AU to 0.2AU (and consequently reducing the transit duration, but increasing the number of transits).  The comparison between (a) and (d) shows the effect of decreasing the length of the chord the planet makes across the face star (and consequently increasing the duration of ingress while decreasing the transit duration).}
     \label{ApproximateMoonDetThresh}
\end{figure}

Moons of a transiting planet can alter the light curve in a number of ways, each corresponding to different physical processes.  Consequently a number of different methods, and corresponding test statistics have been proposed and investigated in the literature.  These include:
\begin{itemize}
\item Direct Detection:  Detecting the extra dip in the light curve due to the moon.
\item Barycentric Transit Timing:  Detecting motion of the planet around the planet-moon barycenter through the possible lead or lag in the planetary transit mid-time.
\item Transit Duration Variation:  Detecting motion of the planet about the planet-moon barycenter through variations in planet transit duration.
\item Photometric Transit Timing: A hybrid test statistic which measures the distortion of the light curve due to the transit of the moon.
\end{itemize}
To highlight the physics underlying each of the methods, formulae will be derived, and the corresponding moon detection thresholds for each method compared.  

Unfortunately, as different authors use different underlying assumptions, the thresholds cannot simply be copied from the respective papers.  In particular, the issues of the size and orbital parameters of the moon, the number of observed transits, the detection threshold and the type of photometric noise will be discussed.  For simplicity and ease of comparison it was decided to investigate moons on circular orbits which are aligned to the line-of-sight.\footnote{For a discussion of why moon orbits which are coplanar to the orbits of their host planets and moon orbits which are aligned to the line-of-sight are equivalent, see section~\ref{Transit_Signal_Coord_Orient}.}  In addition, guided by the properties of Solar System moons, it was decided to investigate the case where the moon is small compared to the planet (i.e. $R_m \ll R_p$ and $M_m \ll M_p$).  To simplify the mathematics, it was decided to consider the case where $N$, the number of transits is large (i.e $\gg 4$).  As a detection threshold needs to be selected, it was decided to use the three sigma (99.7\%) detection threshold as this is the threshold that will be used in chapter~\ref{Trans_Thresholds}.  Finally, while correlated noise can strongly affect planet detection \citep{Pontetal2006}, any correlated noise encountered will be specific to that system (combination of star, telescope, and observing conditions).  Consequently, for ease of comparison, the formulae were derived under the assumption of normal uncorrelated noise.   Within this context, each of the four methods introduced, will be discussed and analysed, in turn.

\subsubsection{Direct detection of the moon's transit}

Similar to the case for transiting planets, the passage of a moon across the face of the star can also block some of the star's light, resulting in an additional dip in the light curve.  For the case of the planet-moon pair shown in figure~\ref{PlanetMoonTransitSch}, the additional dip due to the moon is translated to the right of the dip due to the planet.  This effect has been investigated in the literature in terms of the probability that a given moon will transit \citep{Sartorettietal1999}, the depth of the dip produced \citep{Sartorettietal1999} and the effect of mutual events, such as the moon eclipsing or being eclipsed by the planet \citep{Sartorettietal1999,CabreraSchneider2007} on the shape of this dip.   In addition to theoretical investigations, this method has been used to place a limit of 1.2 Earth radii on the radius of moons orbiting HD~209458 \citep{Brownetal2001}.\footnote{Unfortunately, such large moons are not tidally stable (see section~\ref{Intro_Moons_Stab_Lims}) so this limit is more a statement on the quality of the data than on the size of moons of HD~209458.}

By analogy with the planetary detection case, directly detecting a moon in a transit light curve involves comparing the average photon deficit due to the moon with the error in the average photon deficit due to the photometric noise.  The degree to which the deficit due to the moon is larger than the error due to photometric noise determines whether or not a moon could be detected.  From geometry we have that the average depth of the dip caused by the moon is given by
\begin{equation}
\alpha_m \approx \left(R_m^2/R_s^2\right) L_0,
\end{equation}
where $R_m$ is the radius of the moon, $L_0$ is the unoccluded luminosity of the star, and where the effect of mutual events has been ignored.

Following \citet{Pontetal2006}, we can estimate the error in the deficit due to photometric noise ($\sigma_{L}$ per exposure) over the $N_{exp}$ exposures found in the region of the light curve where the moon is transiting.\footnote{The question of how we know which sections of light curve correspond to when the moon is transiting is one that will not be addressed in this work.}  Assuming that the noise in one exposure is normally distributed and independent, the Central Limit Theorem gives
\begin{equation}
\sigma_\alpha = \frac{\sigma_{L}}{\sqrt{N_{exp}}},
\end{equation}
where $\sigma_\alpha$ is the standard deviation of the error in the measured depth and $\sigma_{L}$ is the absolute photometric noise.

To evaluate this expression, we require information about $N_{exp}$.  Consider, that as a result of orbital motion about the planet moon barycenter, the duration of the transit of the planet \citep[e.g.][]{Kipping2008b} and the moon will vary from transit to transit.  However, as the time-averaged velocity of a moon about its host planet is equal to zero (if it wasn't, then it would drift away from its host planet), we can approximate the duration of a typical moon transit with that of a typical planetary transit.  Consequently we can approximate $N_{exp}$ as $N T_{tra}/\Delta t$, where $T_{tra}$ is the duration of the planetary transit for the case of no moon, $\Delta t$ is the exposure time, and $N$ is the number of transits.  Thus
\begin{equation}
\sigma_{\alpha} =  \frac{\sigma_{L}}{\sqrt{ N T_{tra}/ \Delta t }}.
\end{equation}

Recalling that we are using an three sigma detection threshold, we have that, in order to be detected, the average depth of the dip due to the moon must be three times the average error in this depth, that is
\begin{equation}
\frac{R_m^2}{R_s^2} L_0 =  3 \frac{\sigma_{L}}{\sqrt{ N T_{tra}/ \Delta t }},
\end{equation}
or
\begin{equation}
R_m = 0.0065 R_s  \frac{1}{N^{1/4}} \left[\frac{\sigma_{L}/L_0}{3.95 \times 10^{-4}} \left(\frac{\Delta t}{1\text{min}} \right)^{1/2}\right]^{1/2} \left(\frac{13\text{hrs}}{T_{tra}}\right)^{1/4}.
\label{intro_det_moon_DDtransthresh}
\end{equation}

While equation~\eqref{intro_det_moon_DDtransthresh} indicates the detection threshold for the case of direct detection, it is not necessarily indicative of the true detection threshold for moons.  For example, equation~\eqref{intro_det_moon_DDtransthresh} neglects effects due to fitting the planetary light curve  and the effects of correlated noise.  In particular, \citet{Pontetal2006} showed that for planet detection, equation~\eqref{intro_det_moon_DDtransthresh} became increasingly inaccurate as the transit duration and proportion of low frequency red noise increased.  However, as Kepler is billed as being able to detect Earth-like planets at 1AU in the presence of realistic photometric noise, it seems reasonable that it should be able to detect Earth-like moons of planets at 1 AU.

\subsubsection{Barycentric transit timing}

Moons of transiting planets can also be detected through the motion of their host planet about the planet-moon barycenter \citep{Sartorettietal1999}.  The technique of barycentric transit timing (TTV$_b$) aims to detect this motion through deviations in the transit mid-time from strict periodicity (caused by the planet's physical displacement from the planet-moon barycenter).  For example, in figure~\ref{PlanetMoonTransitSch} the transit of the planet occurs earlier than would have happened had there been no moon.  Using this technique, \citet{Brownetal2001} and \citet{Gillonetal2006} placed upper limits on the masses of moons orbiting HD~209458 and OGLE-TR-113b of 3 Earth masses and 7 Earth masses respectively.  

As will be derived in section~\ref{Trans_Thresholds_ExpBehav}, the three sigma detection threshold for a sinusoidal signal for the case where $N$, the number of samples is large, is given by
\begin{equation}
13.95 = \frac{A^2}{2\sigma^2}\label{intro_det_moon_3sigdef}
\end{equation}
where $A$ is the amplitude of the signal and $\sigma$ is the standard deviation of the noise.

From \citep{Sartorettietal1999} we that the amplitude of the timing signal caused by the moon is given by
\begin{align}
TTV_{b} &= a_m \frac{M_m}{M_{p}} \frac{T_p}{2\pi a_p},\\
&= a_m \frac{M_m}{M_{p}} \sqrt{\frac{a_p}{GM_s}},
\end{align}
where $T_p$ is the period of the planetary orbit, and where we note that \citet{Kipping2009} investigated and extended this expression for the case of eccentric moon orbits.

From \citet{Deeg2002} we have that $\sigma_{t_{mid,p}}$, the standard deviation of the timing error on this transit mid-time for a single transit is given by
\begin{equation}
\sigma_{t_{mid,p}} =  \sigma_{L}\left[\sum_i \left(\frac{\partial L(t_i,t_{mid,p})}{\partial t_{mid,p}} \right)^2\right]^{-1/2}
\label{intro_det_moon_TTVbnoisedef}
\end{equation}
where the sum runs over the length of the planetary transit, where $L(t_i,t_{mid,p})$ is the expression for a transit light curve centered at $t_{mid,p}$ at time $t_i$, and where it is assumed that only the transit mid-time varies from transit to transit.  Noticing that the sum is dominated by the regions of the transit light curve with the largest gradient,  that is, the ingress and egress, equation~\eqref{intro_det_moon_TTVbnoisedef} can be simplified.  Following \citet{Carteretal2008} we approximate the transit light curve by three straight line segments, one each for the ingress, flat bottom and egress of the transit.\footnote{For a discussion of the effect of neglecting limb darkening, see \citet{Carteretal2008}.}  We also assume that the transit has depth $\alpha_p$ and $T_{in}$ is the duration of both the ingress and egress.  Consequently, for both the ingress and egress the square of the partial derivative in equation~\eqref{intro_det_moon_TTVbnoisedef} is given by $(\alpha_p /T_{in})^2$.  Noting that there are $T_{in}/\Delta t$ exposures during both the ingress and egress, and substituting the above approximation into equation~\eqref{intro_det_moon_TTVbnoisedef} gives
\begin{align}
\sigma_{t_{mid,p}} &=  \sigma_{L}\left[2 \frac{T_{in}}{\Delta t} \left(\frac{\alpha_p}{T_{in}} \right)^2\right]^{-1/2},\\
 &= \frac{ \sigma_{L}}{\sqrt{2}}  \frac{\sqrt{\Delta t T_{in}}}{\alpha_p}.  
\end{align}

Substituting these expressions into equation~\eqref{intro_det_moon_3sigdef} and simplifying gives
\begin{multline}
 R_m  = 0.0168R_s \frac{1}{N^{1/6}} \left[\frac{ \sigma_{L}/L_0}{3.95\times 10^{-4}} \left(\frac{\Delta t}{1 \text{min}}\right)^{1/2} \right]^{1/3} \left(\frac{13\text{hrs}}{T_{tra}}\right)^{1/6} \\
 \times \left(\frac{R_s}{a_m}\right)^{1/3} \left(\frac{\rho_p}{\rho_m}\right)^{1/3} \left(\frac{R_p}{0.1R_s}\right)^{1/2},
 \label{intro_det_moon_TTVbtransthresh}
\end{multline}
where we have used equation \eqref{AppIngDurDtraApprox} for $T_{in}$ and the expressions $\alpha_p = L_0 R_p^2/R_s^2$, $M_m = \rho_m 4/3\pi R_m^3$ and $M_p = \rho_p 4/3\pi R_p^3$, where $\rho_m$ and $\rho_p$ are the densities of the moon and planet respectively.  As can be seen by the dependence of $R_m$ on $a_m$ and $\rho_m$, this statistic is good for detecting massive, distant moons (see figure~\ref{ApproximateMoonDetThresh}).

Unfortunately, other physical systems can also cause periodicities in transit timing e.g. the presence of additional planet \citep{MiraldaEscude2002,SteffenAgol2005,Agoletal2005}.   In addition, as yet, no analysis has been performed on the effect of realistic stellar noise on this statistic.  However, as the ingress and egress of the transit are relatively short compared to the transit duration, it seems intuitively reasonable that this technique would be more robust to the effects of red photometric noise than direct detection.

\subsubsection{Transit duration variation}

The transit duration variation technique (TDV) is an alternative way to detect motion of the planet about the planet-moon barycenter.  This involves detecting transit to transit variation of the planetary transit duration to measure any perturbation in the velocity of the transiting planet  as it moves around the planet-moon barycenter \citep{Kipping2009}.  For example in figure~\ref{PlanetMoonTransitSch}, the motion of the planet around the planet-moon barycenter is in the opposite direction as the bulk motion of the planet-moon pair, resulting in a longer planetary transit duration than if there had been no moon.  This method was introduced by \citet{Kipping2009}, and extended by \citet{Kipping2009b} to include inclined moon orbits.  While it has not been used to set limits on moons of currently known planets, it is predicted to be able to detect moons as small as $0.2M_{\earth}$ in Kepler data \citep{Kippingetal2009}.

From \citet{Kipping2009} we have that the amplitude of the timing perturbation in the transit duration caused by a planet and moon on circular coplanar orbits is given by 
\begin{equation}
TDV_{rms} = \sqrt{\frac{a_p}{a_m}}\sqrt{\frac{M_m^2}{M_p M_s}} T_{tra}.
\label{intro_det_moon_TDVamp}
\end{equation}
Where we have transformed from the root mean squared amplitude presented in \citet{Kipping2009} to an amplitude by multiplying by $\sqrt{2}$ and where it is assumed that $M_p \ll M_s$.  In addition, from \citet{Carteretal2008} we have that the expression for the error in transit duration due to photometric noise can be approximated by
\begin{equation}
\sigma_{T_{tra}} =   \frac{\sigma_{L}}{\alpha} \sqrt{\frac{\Delta t}{T_{tra}}} T_{tra} \sqrt{2\frac{T_{in}}{T_{tra}}}.
\label{intro_det_moon_TDVnoise}
\end{equation}

Using the method described in the previous section, equation~\eqref{intro_det_moon_TDVamp} and \eqref{intro_det_moon_TDVnoise} can be combined to give the to give the three sigma detection threshold, which is
\begin{multline}
R_m = 0.0197 R_s  \frac{1}{N^{1/6}} \left[\frac{\sigma_{L}/L_0}{3.95\times10^{-4}} \left(\frac{\Delta t}{1\text{min}}\right)^{1/2} \right]^{1/3} \left(\frac{13 \text{hrs}}{T_{tra}}\right)^{1/2} \\
\times \left(\frac{a_m}{R_s}\right)^{1/6} \left(\frac{\rho_p}{\rho_{ice}}\right)^{1/6} \left(\frac{\rho_{ice}}{\rho_m}\right)^{1/3},
\label{intro_det_moon_TDVtransthresh}
\end{multline}
where $\rho_{ice}$, the density of ice is taken to be $916.7$kgm$^{-3}$, and only the highest order terms in $R_p/R_s$ have been retained. As $R_m$ decreases as $a_m$ decreases, this statistic is optimised for detecting close-in moons (see figure~\ref{ApproximateMoonDetThresh}).

From the work of \citet{Kippingetal2009} it seems that this technique is relatively robust in terms of correlated stellar noise and instrumental variability.  They investigated the shape of the distribution of errors in transit duration for the case of white and synthetic red noise and found that the distributions agreed and were both normal.  This result could be again due to the fact that the size of the timing error is dominated by the ingress and egress, and is not strongly affected by red noise as a result of the short time over which red noise has to act.

\subsubsection{Photometric transit timing}

The photometric transit timing technique ($TTV_p$) also uses the timing of transits to search for moons, but in this case, the time used is not the midpoint of the transit, but $\tau$, the first moment of the photon deficit caused by the planet-moon pair.  Qualitatively, this statistic measures the position of the center of the transit, which is altered both by bulk motion of the transit signal and asymmetry of the transit light curve.  The method was proposed by \citet{Szaboetal2006} as being equivalent to that of barycentric transit timing, but in a later paper \citep{Simonetal2007} the differences between these two methods were discussed.  Again, while this method has not been used to place limits on moons of known planets, \citet{Szaboetal2006} used this updated method to investigate the number of moons expected to be detected by missions such as COROT and Kepler.  They investigated this in terms of both giant and
terrestrial planets, and suggested that Earth-Moon type systems could be detected. 

While photometric transit timing has been investigated using a Monte Carlo simulation \citep{Szaboetal2006} and from a more theoretical standpoint \citep{Simonetal2007}, it is not possible to derive the signal form using literature results.  In particular, the only literature result relating to the signal is given in \citet{Simonetal2007}, which states that the maximum size of this perturbation is given by 
\begin{equation}
\text{max}\left(TTV_p\right) = \frac{a_m}{v_{tr}} \left| \left(\frac{R_m}{R_p}\right)^2 - \frac{M_m}{M_p} \right|.
\end{equation}

As we do not have a signal form,\footnote{The form of the TTV$_p$ signal can be guessed using equation~(15) and Fig. 2 of \citet{Simonetal2007}.  Assuming that the silhouette of the planet and moon shown in this correspond to a planet and moon not at maximum separation, equation~(15) can be used to show that the TTV$_p$ signal should be approximately a sinusoidal function of time.  However, equation~(15) was derived assuming that there is no transverse motion of the planet and moon, which is nearly true at maximum separation, but may not be true at other times, especially if the planet and moon are close.  Consequently, it is not apparent how this effect will modify the sinusoidal signal.} this cannot be converted into a threshold similar to those given.  In addition, there is no expression given in the literature for the error in $\tau$ assuming any type of noise, uncorrelated or otherwise.

Consequently, in Part~\ref{TransitPart} of this thesis, these gaps will be addressed. First, in chapter~\ref{Transit_Signal}, the form of $\tau$ as a function of time will be derived for the case of circular and coplanar planet and moon orbits.  This analysis will then be extended to the case of eccentric planet orbits.  Then, in chapter~\ref{Trans_TTV_Noise}, the functional form and standard deviation of the probability distribution of the error in $\tau$ caused by photometric noise will be derived and investigated for the of white noise and more realistic correlated photometric noise.  These results will then be combined in chapter~\ref{Trans_Thresholds}, with the aim of first, deriving a simplified equation for the detection threshold similar to equations~\eqref{intro_det_moon_DDtransthresh}, \eqref{intro_det_moon_TTVbtransthresh} and \eqref{intro_det_moon_TDVtransthresh}, so that the four methods can be compared on an equal footing, and second, generating realistic moon detection threshold maps for the case of photometric transit timing.

\subsection{Microlensing}

The possibility of detecting moons of planets detected by microlensing through their perturbations on the microlensing light curve has started to be explored in the literature.  An initial investigation into whether or not moons of planets detected by microlensing could be detected suggested that detection was unlikely due to the finite source effect \citep{Hanetal2002}.  However, a more in depth investigation found that Earth-sized moons may be detectable if the distance from their host planet is similar to or greater than its Einstein radius \citep{Han2008}.

\subsection{Timing}

For the case of planets detected by timing, the possibility exists to detect the additional timing perturbation due to planet-moon binarity, given sufficient timing sensitivity of the host.  Currently the only host stars to display the required timing sensitivity are the millisecond pulsar hosts.  Unfortunately, pulsar planets are rare, with only four discovered to date.   Prior to this thesis, no analysis has been conducted on time-of-arrival perturbation due to planet-moon pairs.  However, many other second order signals, such as the effect of the 2:3 resonance in the PSR B1257+12 system \citep{Wolszczan1994}, free precession \citep{Link2003} and Shapiro delay \citep{Ordetal2006} have been detected.  Consequently, moon detection using this technique may be possible.  This technique will be analysed in detail in Part~\ref{PulsarPart} of this thesis, with the aim of determining which moons (if any) of the four pulsar planets may be detectable.

\subsection{Detecting moons of imaged planets}\label{Intro_Dect_Moons_Image}

For planets which have been directly imaged we have a a dot, attributed to the ``planet", which is physically separated from a majority of the light from the star.  However, as we do not yet have the technical capability to spatially separate this dot into an image of a  planet and a moon for the case of Solar System planet-moon analogs, this dot is really a combination of light from a planet, and light from any moons of that planet.  While moons cannot be directly detected, a number of techniques have been proposed that use light from this ``planetary" dot to search for companion moons.  These proposed techniques include using photometry \citep{Moskovitzetal2009,CabreraSchneider2007} astrometry \citep{CabreraSchneider2007} or spectral information \citep{Williamsetal2004,CabreraSchneider2007} to infer the presence of moons.

The photometric effect of a moon on the light curve from a directly detected planet has been investigated in two different ways.  First, the effect of moon-like satellites on the infra-red light curve of Earth-like planets was investigated by \citet{Moskovitzetal2009}.  They found that, as a result of a degeneracy between the effect of a moon and the effect of an inclined planetary spin axis, a TPF-like mission could only detect large (Mars-sized) satellites of terrestrial planets.  In addition to the effect of the bulk motion of the planet-moon system about the star on the light curve, \citet{CabreraSchneider2007} have also investigated the effect on both optical and infra-red light curves of discrete mutual events such as the moon eclipsing or casting a shadow on the planet and visa versa.  They suggest that such events could allow for the detection of lunar-sized moons of Earth-analogs by missions similar to TPF-C.

In addition to using ``planet" light curves to detect moons, moons may also be detectable from perturbations in the position of the photocenter of the dot attributed to the planet.  This method was also investigated by \citet{CabreraSchneider2007}, and the possibility of detecting Earth-like moons of gas giant planets using this method was addressed in the science case for the ELT \citep{Hooketal2005}.

Finally, moons of directly detected planets may also be detected using spectra.  This can involve measuring the Doppler shift of the planet's spectrum due to its motion about the planet-moon barycenter \citep{CabreraSchneider2007} or by noticing that Earth-like moons are much brighter in the 1-4$\mu$m CH$_4$ hole, than host gas giants in the habitable zone \citep{Williamsetal2004}.

As yet, none of these techniques have been used to place limits on moon sizes for any of the seven directly detected planets.  However, \citep{Kalasetal2009} report that the optical emission from the planet Fomalhaut b is consistent with an extended circum-planetary disk, the size of the Galilean satellite system.  This lends support to the idea that moon systems may exist about such planets, and may, in future, be detectable.

\section{Conclusion}

To provide a context for extra-solar moon detection, the techniques of extra-solar planet detection were discussed in terms of the physics behind the technique, their effectiveness, and the types of planets discovered using them.  This discussion was then extended to include moon detection, with particular emphasis on moons that can be detected through the pulsar timing and transit techniques.  From this position we can now begin to discuss the first of the two moon detection techniques that will be investigated in this thesis, pulsar timing.



\cleardoublepage \pagestyle{empty} 
\part{Detecting Moons of Pulsar Planets}\label{PulsarPart}
\pagestyle{plain} 
\chapter{Possibility of detecting moons of pulsar planets through time-of-arrival analysis}\label{Pulsar_Paper}

\Large

\textbf{Authored by:  Karen M. Lewis, Penny D. Sackett and Rosemary A. Mardling}

\normalsize

\vspace{1cm}

\noindent
This chapter is a reformatted and expanded version of the paper:

\noindent
K. M. Lewis, P. D. Sackett, R. A. Mardling, 2008, ``Possibility of Detecting Moons of Pulsar Planets Through Time-of-Arrival Analysis", Astrophysical Journal Letters, \textbf{685}, L153-L156.

\vspace{2cm}

\section{Abstract}\label{paper_abstract}

The perturbation caused by planet-moon binarity on the
time-of-arrival signal of a pulsar with an orbiting planet is
derived for the case in which the orbits of the moon and the planet-moon
barycenter are both circular and coplanar. The signal consists of
two sinusoids with frequency $(2n_m - 3n_p )$ and $(2n_m - n_p )$,
where $n_m$ and $n_p$ are the mean motions of the planet and
moon around their barycenter, and the planet-moon system around the host, respectively.
The amplitude of the signal is the fraction $\sin I_p[9(M_p M_m)/16(M_p + M_m)^2] [r_m/r_p]^5$
of the system crossing time $r_p/c$, where $M_p$ and $M_m$ are the the masses of the planet and moon, 
$r_m$ is their orbital separation, $r_p$ is the distance between the host pulsar and planet-moon barycenter, $I_p$ is the inclination of the orbital plane of the planet, and $c$ is the speed of light.   The analysis is
applied to the case of PSR~B1620-26~b, a pulsar planet, to constrain
the orbital separation and mass of any possible moons.  We find that a stable moon orbiting this pulsar planet could be detected, if its mass was $>$5\% of its planet's mass, and if the planet-moon distance was $\sim$2\% of the planet-pulsar separation.

\section{Introduction to extra-solar moons}

In the past decade and a half, over three hundred extra-solar planets have been discovered.\footnote{See, for example,
http://exoplanet.eu/catalogue.php} With the data expected to be
produced by satellites such as COROT \citep{Auvergneetal2003} and
Kepler \citep*{Basrietal2005}, it will not only be possible to find
smaller planets, but moons of those planets as well
\citep{Szaboetal2006}. As a result, the detectability of extra-solar
moons is starting to be explored in terms of their effect on
planetary microlensing \citep{Hanetal2002} and transit lightcurves
\citep{Sartorettietal1999, Szaboetal2006,Simonetal2007}. Upper limits
have already been placed on the mass and radius of putative moons of
the planets HD~209458~b \citep{Brownetal2001}, OGLE-TR-113 b \citep{Gillonetal2006} and
HD~189733~b \citep{Pontetal2007}.

While the limitations of microlensing and the transit technique for
detecting moons have been discussed and used in the literature, the
limitations of other techniques such as the time-of-arrival (TOA)
technique have not. This technique involves determining the
variations in line-of-sight position to the host star, usually a
pulsar, using the observed time of periodic events associated with
that host. The aim of this analysis is to explore what the
TOA signal of a planet-moon pair is, and relate it to the planetary
systems that can give the most precise timing information, those around millisecond pulsars.

\section[Review of detection of pulsar planets]{Review of planetary detection around millisecond pulsars}

The first planetary system outside the Solar System was detected
around the millisecond pulsar PSR~1257+12 \citep{Wolszczanetal1992}.
This detection was made by investigating periodic variations in the
time of arrival of its radio pulses using a timing model.  An example timing model for the case in which the planet's orbit around the pulsar is circular is
\begin{multline}\label{TOA_example}
\left(t_N - t_0\right) = \left(T_N - T_0\right) + \Delta T_C +
\Delta T_R + \mathbf{R_e.\hat{n}}/c + \Delta T_S + \Delta T_{n_e}
\\+ TOA_{pert,p}(M_s,M_p,r_p,I_p,f_{p}(0)),
\end{multline}
\citep[for example,][]{Backer1993} where $t_0$ and $t_N$ are the
times the initial and $N^{th}$ pulses are emitted in the pulsar's
frame, $T_0$ and $T_N$ are the times the initial and $N^{th}$ pulses
are received in the observatory's frame, and where the terms $\Delta
T_C$, $\Delta T_R$, $\mathbf{R_e.\hat{n}}/c$, $\Delta T_S$ and
$\Delta T_{n_e}$ act to change the frame of reference from the
observatory on Earth to the barycenter of the pulsar system. The
terms $\Delta T_C$ and $\Delta T_R$ are clock correction terms.
$\Delta T_C$ converts the time recorded by the observatory atomic
clock to terrestrial proper time while $\Delta T_R$ contains time
dilation corrections due to the transverse doppler effect and
gravitational red-shift due to the Earth's motion through the
gravitational potential of the Solar System. The term
$\mathbf{R_e.\hat{n}}/c$, corrects for the annual motion of the
earth. $\Delta T_S$ and $\Delta T_{n_e}$ correct for propagation
effects, namely, variations in the amount of ray bending due to
gravitational fields both inside and outside our solar system, and
varying electron density along the line of sight to the pulsar
respectively.  The final term represents the effect of a planet on the motion of the pulsar, where $r_p$ is the planet-pulsar distance, $I_p$ is the angle between the normal of the planet-pulsar orbit and the line-of-sight, $M_s$ and $M_p$ are the mass of the pulsar and the planet respectively, and $f_p(0)$ is the initial angular position of the planet measured from the $x$-axis, about the system barycenter.

Currently, four planets around two millisecond pulsars have been
discovered, three around PSR~1257+12
\citep{Wolszczanetal1992,Wolszczan1994} and one around PSR~B1620-26
\citep*{Backeretal1993}. These four planets include one with mass
$~0.02$ Earth masses, the lowest mass extra-solar planet known. This high timing precision of millisecond pulsars indicates that they are optimal targets for planet, and consequent moon searches.

\section[The TOA perturbation caused by a moon]{What is the TOA perturbation caused by a moon?}

\begin{figure}
 \includegraphics[width=.90\textwidth]{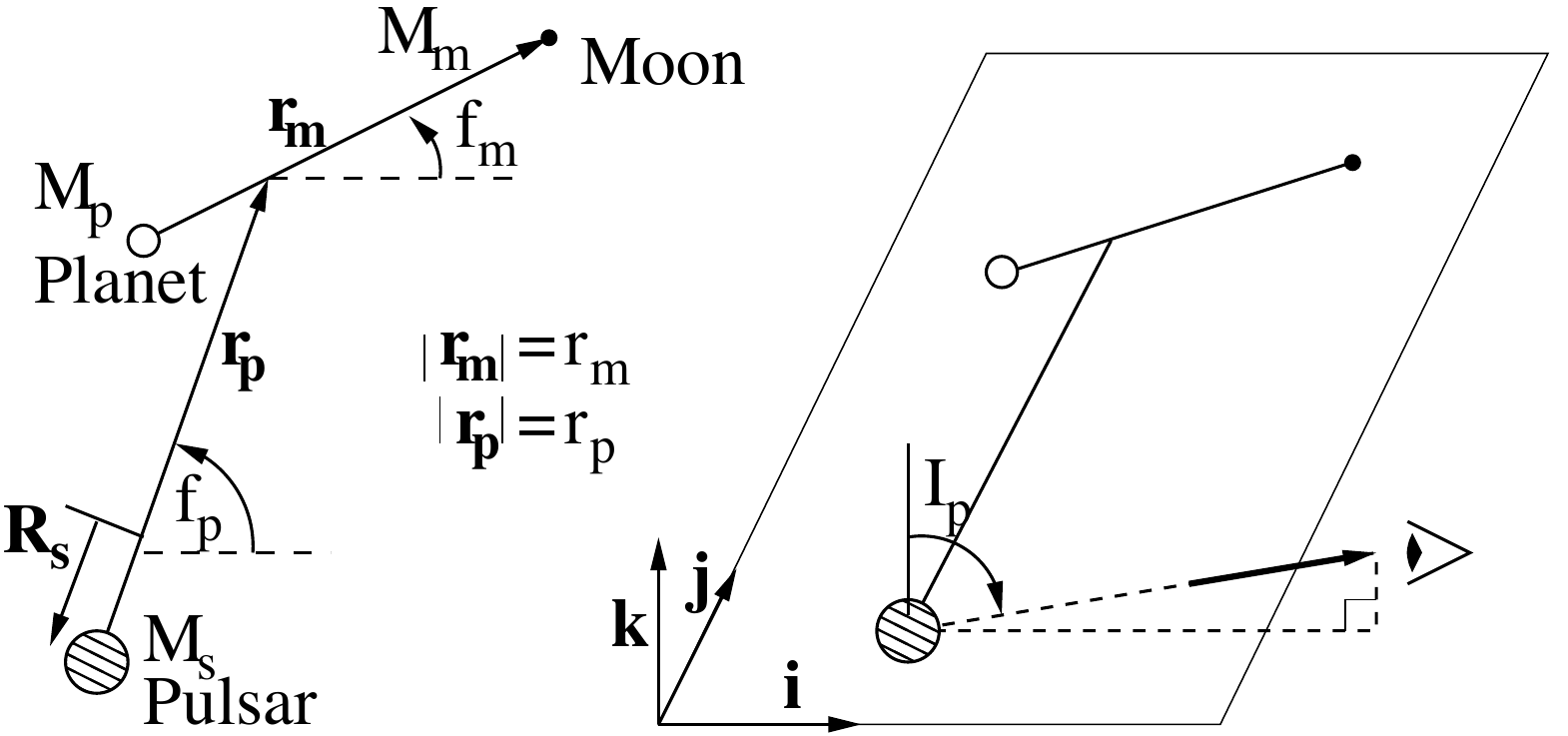}
\caption[Coordinate system used in the analysis of the TOA perturbation
caused by a moon.]{Coordinate system used in the analysis of the TOA perturbation
caused by a moon. The left diagram shows the quantities used
to describe the position of the three bodies in their mutual orbital
plane, while the right diagram shows the relationship between
this orbital plane and the observer.}
\label{CoordSystem}
\end{figure}

In order to investigate the perturbation caused by planet-moon binarity, the
timing model presented in equation~(\ref{TOA_example}) must be
updated to include effects due to the presence of the moon. For simplicity, we consider here only systems in which both the orbit of the planet and moon around their common barycenter, and the orbit of the planet-moon barycenter around the pulsar, are circular and lie in the same plane.
The resulting updated model is
\begin{multline}\label{TOA_example_m}
\left(t_N - t_0\right) = \left(T_N - T_0\right) + \Delta T_C +
\Delta T_R + \mathbf{R_e.\hat{n}}/c + \Delta T_S + \Delta T_{n_e}
\\+ TOA_{pert,p}(M_s,M_p,r_p,I_p,f_{p}(0)) 
\\+ TOA_{pert,pm}(M_s, M_p, M_m, r_p, r_m, I_p, f_{p}(0),f_{m}(0)).
\end{multline}
We have explicitly modified $TOA_{pert,p}$ to indicate that it
depends on the combined planet-moon mass, and included another term,
$TOA_{pert,pm}$, to account for planet-moon binarity.  Here $M_m$ is
the mass of the moon, $r_m$ is the distance between the planet and the moon, and $f_m(0)$ is the initial angular position of the moon measured from the $x$-axis, about the planet-moon barycenter (see figure~\ref{CoordSystem}).  $TOA_{pert,pm}$ can be derived from $\mathbf{R}_s$, the vector between the system barycenter and the pulsar, using
\begin{equation}
\frac{1}{c}\int_0^t\int_0^{t'} \mathbf{\ddot{R}}_s \cdot \mathbf{n} dt'' dt' =
TOA_{pert,p} + TOA_{pert,pm},\label{TOA-TOApertpmdef}
\end{equation}
where $c$ is the speed of light and $\mathbf{n}$ is a unit vector pointing along the line of sight.  From figure~\ref{CoordSystem} we have that
\begin{equation}
\mathbf{n} = \sin I_p \mathbf{i} + \cos I_p \mathbf{k},\label{TOA-ndef}
\end{equation}
where the vectors $\mathbf{i}$, $\mathbf{j}$ and $\mathbf{k}$ are defined in figure~\ref{CoordSystem}.

The governing equation for $\mathbf{R_s}$ can be written as the sum of the zeroth order term, which describes $TOA_{pert,p}$ and the tidal terms, which describe $TOA_{pert,pm}$,
\begin{equation}
\frac{d^2\mathbf{R_s}}{dt^2} = \frac{G (M_p +
M_m)}{r_p^3}\mathbf{r}_p +  \left[-\frac{M_m + M_p + M_s}{M_m + M_p} \nabla_{\mathbf{r}_p} \mathcal{R}\right],\label{TOA-d2Rs}
\end{equation}
where the second term is the tidal perturbation to the orbit due to
the presence of the moon, which has been written in terms of the disturbing function $\mathcal{R}$ \citep[for example][p. 226]{Murrayetal1999}, where $\mathbf{r}_p = X\mathbf{i} + Y\mathbf{j} + Z\mathbf{k}$,
\begin{equation}
\nabla_{\mathbf{r}_p} = \mathbf{i}\frac{\partial}{\partial X} + \mathbf{j}\frac{\partial}{\partial Y} + \mathbf{k}\frac{\partial}{\partial Z},
\end{equation}
and where
\begin{multline}
\mathcal{R} = -\frac{M_m + M_p}{M_m + M_p + M_s}\left[\frac{G\left(M_m + M_p\right)}{r_p}  - \frac{GM_p}{\left|\mathbf{r}_p - \frac{M_m}{M_m+M_p} \mathbf{r}_m\right|} \right. \\
\left. - \frac{GM_m}{\left|\mathbf{r}_p + \frac{M_p}{M_m+M_p} \mathbf{r}_m\right|} \right].
\end{multline}
The disturbing function can be expanded using multipole analysis \citep[for example][p. 92]{Jackson1975} in terms of  Legendre polynomials.  Assuming $r_m \ll r_p$, the expansion can be truncated to order $r_m^2/ r_p^2$ giving
\begin{equation}
\mathcal{R} = \frac{M_m + M_p}{M_m + M_p + M_s} \frac{G M_m M_p}{M_m + M_p} \frac{
r_m^2}{r_p^3}\frac{1}{2}(3\cos^2\left(f_m -f_p\right)-1)\label{TOA-distfundef}.
\end{equation}
As the orbits are both circular and coplanar, $r_m$ and $r_p$ are both constant, $f_p(t) = n_p t + f_p(0)$, and $f_m(t) = n_m t + f_m(0)$ where $n_p$ and $n_m$ are the mean motions of the two respective orbits and are both constants, and the inclination, $I_p$, describes the planes of both orbits.
Using polar coordinates in the plane of the orbit to evaluate $\nabla_{\mathbf{r}_p}\mathcal{R}$ we have,
\begin{multline}
\frac{d^2\mathbf{R_s}}{dt^2}\cdot \mathbf{n} = \sin I_p\frac{G (M_p +
M_m)}{r_p^3}r_p \cos f_p + \sin I_p \frac{G M_p M_m}{(M_m +
M_p)}\frac{r_m^2}{r_p^4} \\\times \left[
\frac{6}{4}\sin(2(f_p - f_m))\sin f_p + \left( \frac{3}{4} + \frac{9}{4}
\cos(2(f_p - f_m))\right) \cos f_p\right].\label{TOA-d2Rs2}
\end{multline}
So that from equation \eqref{TOA-TOApertpmdef},
\begin{multline}
TOA_{pert,pm} =  \frac{- \sin I_p G M_p M_m}{c (M_m +
M_p)}\frac{r_m^2}{r_p^4}\left[ \frac{3}{4 n_p^2}\cos f_p  \right. \\
\left.+
\frac{3}{8(n_p - 2n_m)^2} \cos(f_p - 2f_m) + \frac{15}{8(3n_p - 2n_m)^2} \cos(3f_p - 2f_m) \right]
.\label{TOA-pert1}
\end{multline}
The $\cos f_p$ term in equation~\eqref{TOA-pert1} has the same
frequency as the signal of a lone planet and it acts to increase the measured value of $M_p +M_m$ derived from $TOA_{pert,p}$ by $(3/4)(r_m^2/r_p^2)(M_pM_m/(M_p+M_m))$.  Consequently, this term can be neglected as it will be undetectable as a separate signal. Also, the
stability region for a prograde satellite of the low-mass component
of a high-mass ratio binary extends from $r_R$, the Roche radius, to $0.36r_H$ for the case
of circular orbits, where $r_H =
r_p\left[(M_p)/(3M_s)\right]^{1/3}$ is the secondary's Hill
radius \citep{Holmanetal1999}.  As moon detectability increases as $r_m^5$, and $r_R$ is equal to only a few planetary radii, this limit can be safely ignored. When $r_m$ is equal to $0.36r_H$, $n_m \approx 8n_p$.  As $n_p \ll n_m$ is likely, we have that the denominators of the $\cos(f_p - 2f_m)$ and $\cos(3f_p - 2f_m)$ terms will never approach zero.  This, in addition to the assumption of zero eccentricities, means that resonance effects can be neglected. Consequently, equation~(\ref{TOA-pert1}) can be simplified
by neglecting $n_p$ in the denominators, giving
\begin{multline}
TOA_{pert,pm} =  \frac{- \sin I_p G M_p M_m}{c (M_m +
M_p)}\frac{r_m^2}{r_p^4}\left[\frac{3}{32n_m^2} \cos(f_p -
2f_m) \right. \\
\left. + \frac{15}{32n_m^2} \cos(3f_p - 2f_m) \right]
.\label{TOA-pert2}
\end{multline}
Writing $n_p$ in terms of $r_m$, using Kepler's law, gives
\begin{multline}
TOA_{pert,pm} =  - \sin I_p\frac{M_p M_m}{(M_m +
M_p)^2}\frac{r_p}{c}\left(\frac{r_m}{r_p}\right)^5  \left[\frac{3}{32}
\cos(f_p - 2f_m)  \right. \\
\left. + \frac{15}{32} \cos(3f_p - 2f_m)
\right],\label{TOA-pert}
\end{multline}
where we are making no assumptions about the size of $M_m/M_p$.

A similar study was conducted by \citet{Schneideretal2006}, investigating the radial velocity perturbation due to an equal-mass pair of binary stars on a distant companion.  Converting their radial velocity perturbation to a timing perturbation, setting $M_p = M_m$, and noting that their $a_A$ is equivalent to $r_m/2$, our results agree.

\section[Detecting moons of pulsar planets]{Is it possible to detect moons of planets orbiting millisecond pulsars?}

To investigate whether or not it is possible to detect moons of
pulsar planets, we simplify equation~(\ref{TOA-pert}) by summing the
amplitudes of the sinusoids, giving the maximum possible amplitude
\begin{equation}\label{ApproxTOAPert}
max\left(TOA_{pert,pm}\right) = \frac{9\sin I_p}{16}\frac{M_m M_p}{(M_m +
M_p)^2}\frac{r_p}{c}\left(\frac{r_m}{r_p}\right)^5.
\end{equation}
Thus, the size of the
perturbation varies as $[M_mM_p/(M_m + M_p)^2][r_m/r_p]^5$ times the system crossing time, $r_p/c$. So, the best hope of a
detectable signal occurs when the planet-moon pair widely are separated from each other, both quite massive, and very accurate
timing data is available.  For example, a stable system such as a 0.1AU
Jupiter-Jupiter binary located 5.2AU from a host pulsar would
produce a  $TOA_{pert,pm}$ of amplitude 960ns, which compares well
with the 130ns residuals obtained from one of the most stable
millisecond pulsars, PSR J0437-4715 \citep{vanStratenetal2001}.

\begin{figure}[htp]   
 \includegraphics[width=.90\textwidth]{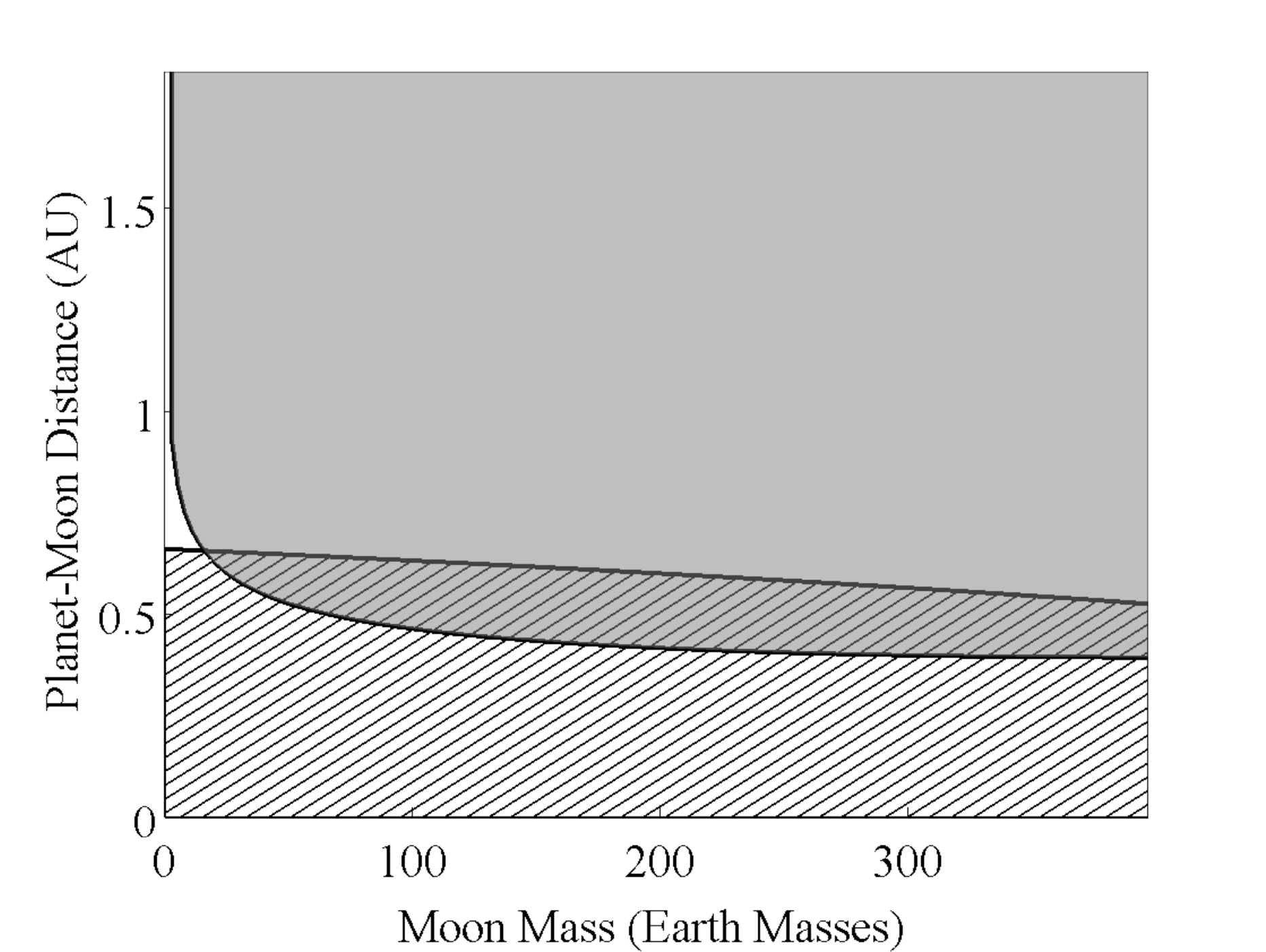}
\caption[The regions of parameter space containing detectable and stable moons of the planet PSR~B1620-26 b are shown as a function of planet-moon separation and moon mass.]{The regions of parameter space containing detectable (shaded) and stable
(cross-hatched) moons of the planet PSR~B1620-26 b are shown as a function of planet-moon separation and moon mass.  The total mass and the
distance of the planet-moon pair from the parent pulsar were assumed
to be 2.5 Jupiter masses and 23AU respectively
\citep{Sigurdssonetal2003}, while it was assumed that $\sin I_p = 1$.
The mass of the host was set at 1.7 solar masses (the sum of the
mass of the pulsar and its white dwarf companion). The $3 \sigma$
detection threshold was calculated assuming the $\sim 40\mu s$
timing residuals given in \citet{Thorsettetal1999} are uncorrelated
and that similar accuracy TOA measurements of PSR~B1620-26 continue
to the present day.}
\label{PulsarPlanetMoon}
\end{figure}

To demonstrate this method, the expected maximum signals from a moon
orbiting each of the four known pulsar planets were explored. It was
found that in the case of PSR~B1620-26~b, signals that are in
principle detectable could confirm or rule out certain
configurations of moon mass and orbital parameters (see
figure~\ref{PulsarPlanetMoon}).

In the particular case of PSR~B1620-26~b, the perturbation signal
will not match the signal shown in equation~(\ref{TOA-pert})
due to the effect of its white dwarf companion.  As a side project, this companion's effect was investigated and found to be the introduction of additional
perturbations on top of the $TOA_{pert,p}$ and $TOA_{pert,pm}$
calculated.  Consequently, the detection threshold represents an
upper limit to the minimum detectable signal and the analysis
is still valid.

Unfortunately, there are practical limits to the applicability of
this method.  They include discounting other systems that could
produce similar signals, sensitivity limits due to intrinsic pulsar
timing noise, and limits imposed by moon formation and stability.

First, other systems that could produce similar signals need to be
investigated. Possible processes include pulsar precession
\citep*[e.g.,][]{Akgunetal2006}, periodic variation in the
ISM \citep{Schereretal1997}, gravitational waves
\citep{Detweiler1979}, unmodelled interactions between planets
\citep{Laughlinetal2001} and other small planets. To help
investigate the last two options, we plan on completing a more
in-depth analysis of the perturbation signal of an extra-solar moon,
including the effects of inclination and 
eccentricity.\footnote{See chapter \ref{Pulsar_Extension}.}

\begin{figure}[htp]   
 \includegraphics[width=.90\textwidth]{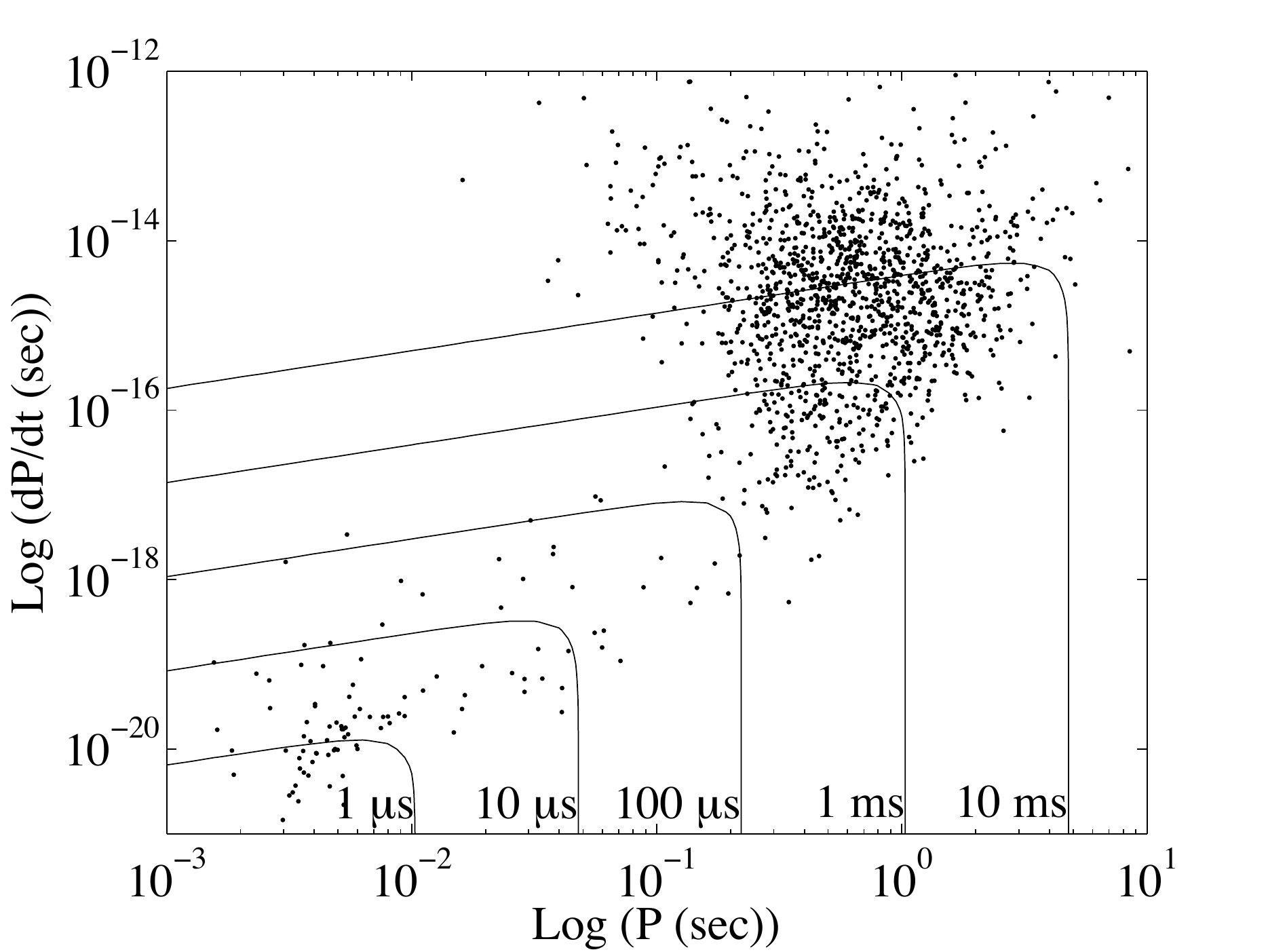}
\caption[Contour plot of
predicted timing noise as a function of pulsar rotation period and
period derivative.]{Contour plot of
predicted timing noise as a function of pulsar rotation period and
period derivative. This plot is based on figure 9 from
\citet{Cordes1993}. The functions and assumptions used
to generate the contours are the same as given in
\citet{Cordes1993}, noting that the TOA integrations are 1000
seconds long. Note that correlated timing noise measured for individual
pulsars can vary from the predicted values by two orders of
magnitude \citep[e.g.,][]{Arzoumanianetal1994}. }
\label{PulsarPlot}
\end{figure}

Second, the noise floor of the system needs to be examined. The
suitability of pulsars for signal detection is limited by two main
noise sources, phase jitter and red timing noise \citep[e.g.,][]{Cordes1993}. Phase jitter is error due to pulse-to-pulse
variations and leads to statistically independent errors for each TOA
measurement.  Phase jitter decreases with increasing rotation rate
(decreasing $P$) due to the increase in the number of
pulses sampled each integration. Red timing noise refers to noise
for which neigbouring TOA residuals are correlated. Red timing noise has
been historically modeled as a random walk in phase, frequency or
frequency derivative \citep[e.g.,][]{Boyntonetal1972,Cordes1980,Kopeikin1997}. Red
noise is strongly dependent on $\dot{P}$.  It has been proposed that red noise is due to
non-homogeneous angular momentum transport either between components
within the pulsar \citep[e.g.][]{Jones1990} or between it
and its environment \citep[e.g.][]{Cheng1987}. To illustrate
the effect of these two noise sources, an estimate
of the resulting TOA residuals as a function of $P$ and $\dot{P}$ is
shown in figure~\ref{PulsarPlot}.  For comparison, the values of $P$
and $\dot{P}$ of every pulsar listed in the The ATNF Pulsar
Catalogue\footnote{http://www.atnf.csiro.au/research/pulsar/psrcat/.}
\citep{Manchesteretal2005} are also included.

Third, whether or not moons will be discovered depends on whether or
not they \emph{exist} in certain configurations, which depends on
their formation history and orbital stability. Recent research
suggests that there are physical mass limits for satellites of both
gas giants \citep{Canupetal2006} and terrestrial planets
\citep{Wadaetal2006}. Also, tidal and three-body effects can
strongly affect the longevity of moons
\citep*{Barnesetal2002,Domingosetal2006,AtobeIda2007}.

Finally, while this method was investigated for the specific case of
a pulsar host, this technique could also be applied to planets
orbiting other clock-like hosts such as pulsating giant stars
\citep{Silvottietal2007} and white dwarfs \citep*{Mullallyetal2006}.

\chapter{Effect of mutual inclination and eccentricity on the time-of-arrival perturbation}\label{Pulsar_Extension}

\section{Introduction}

As mentioned in \citet{Lewisetal2008}, in addition to considering the case where the planet and moon's orbits were circular and coplanar, it would be of use to determine the effect of mutual inclination and eccentricity in the orbit of the planet and moon, on the time-of-arrival perturbation due to planet-moon binarity.  As the method used in chapter~\ref{Pulsar_Paper} to investigate the time-of-arrival perturbation for the case of circular coplanar orbits cannot be easily extended to deal with these cases, a more general method will be used.  To begin, a set of expansions developed by my PhD supervisor, Dr. Rosemary Mardling, will be introduced, which allow the disturbing function to be written in terms of the semi-major axis, eccentricity, inclination, longitude of the ascending node, argument of periastron and the mean anomaly corresponding to the planet and moon orbits.  Then, the selection of reference plane and direction required for the definition of Euler angles is motivated and discussed.  Using this coordinate system, the equations defining this perturbation are then reformatted such that the expansions can be easily applied on a case by case basis.  First, the case of circular coplanar orbits will be re-investigated, to demonstrate the use of the method and to show that the expressions produced in chapter \ref{Pulsar_Paper} and using this method are equivalent.  Then, building on this foundation, the effect of mutual inclination, low eccentricity in the moon's orbit and low eccentricity in the planet's orbit on the time of arrival perturbation due to planet-moon binarity will be investigated in turn.  We begin by introducing the expansions that will be used to write the disturbing function in terms of the orbital elements of the planet and moon's orbits.

\section{Writing the disturbing function in terms of the orbital elements of the planet and moon}\label{Pulsar_Ext_ExpansionIntro}

As can be seen from equations~\eqref{TOA-TOApertpmdef} and \eqref{TOA-d2Rs}, the form of the time-of-arrival perturbation due to planet-moon binarity is entirely specified by $\mathcal{R}$, the disturbing function, and modified by $\mathbf{n}$, the unit vector directed along the line-of-sight.  Consequently, the ability to write the disturbing function in terms of time and the orbital elements of the planet's and moon's orbits, corresponds to the ability to determine the time-of-arrival perturbation due to planet-moon binarity for any orbital configuration as a function of time.  For this work I will be using a method pioneered by my PhD supervisor, Dr. Rosemary Mardling, which allows the disturbing function to be written in terms of these orbital elements and is valid for all values of eccentricity and inclination.  While aspects of this method are presented in the literature \citep{Mardling2008,Mardling2008b}, for completeness the fundamental mathematics required for this chapter will be summarised.

\begin{figure}[tb]
\begin{center}
\includegraphics[width=.80\textwidth]{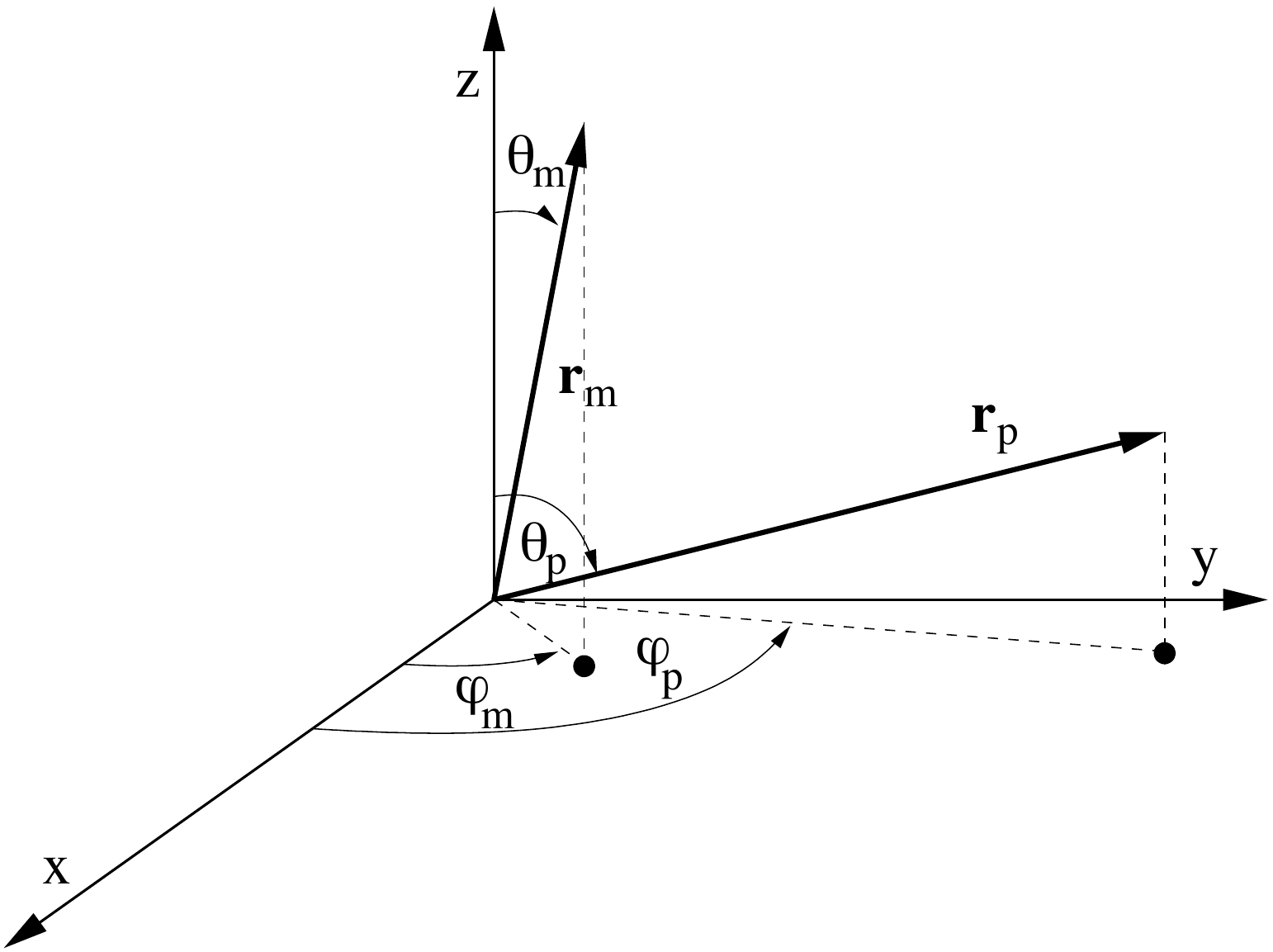}
\caption{Coordinate system used to describe $\mathbf{r}_p$ and $\mathbf{r}_m$ in terms of the spherical polar angles $\theta_p$, $\psi_p$, $\theta_m$ and $\psi_m$.} \label{RpRmOrientDia}
\end{center}
\end{figure}

To begin, recall that, for this work, the disturbing function is given by 
\begin{equation}
\mathcal{R} = - \frac{G (M_{p} + M_{m}) M_s}{r_p} + \frac{G M_m
M_s}{| \mathbf{r}_p - \frac{M_p}{M_{p} + M_{m}}\mathbf{r}_m |} + \frac{G M_pM_s}{| \mathbf{r}_p + \frac{M_m}{M_{p} + M_{m}}\mathbf{r_m} |}.
\label{pulsar_signal_DistFunDef}
\end{equation}
While this function can be expanded in terms of Legendre polynomials as was done in chapter \ref{Pulsar_Paper}, it can also be expanded in terms of spherical harmonics.  To do this we use the identity
\begin{equation}
\frac{1}{|\mathbf{b}-\mathbf{a}|} = \sum_{l=0}^\infty \sum_{m=-l}^l
\frac{4 \pi}{2l+1}
\frac{a^l}{b^{l+1}}Y_{lm}(\theta_a,\psi_a)Y_{lm}^*(\theta_b,\psi_b),
\end{equation}
where $|\mathbf{a}| < |\mathbf{b}|$, $a = |\mathbf{a}|$ and $b = |\mathbf{b}|$, and where $\theta_a$ and $\psi_a$, and $\theta_b$ and $\psi_b$ represent the orientation of $\mathbf{a}$ and $\mathbf{b}$ expressed in spherical polar coordinates.  In addition, we have that
\begin{equation}
Y_{lm}(\theta,\psi) = \sqrt{\frac{2l+1}{4
\pi}\frac{(l-m)!}{(l+m)!}}P_l^m(\cos \theta)e^{im\psi},\label{Int-Rev-YlmDef}
\end{equation}
where $P_l^m(\cos \theta)$ is the associated Legendre polynomial of degree $l$ and order $m$ given by
\begin{equation}
P_l^m(x) = \frac{(-1)^m}{2^l l!}(1-x^2)^{m/2}
\frac{d^{l+m}}{dx^{l+m}}(x^2-1)^l,\label{Int-Rev-PlmDef}
\end{equation}
where
\begin{equation}
P_l^{-m}(x) = (-1)^m \frac{(l-m)!}{(l+m)!}P_l^{m}(x).\label{Int-Rev-PlmProp}
\end{equation}
Using this expansion on $\mathcal{R}$ gives
\begin{equation}
\mathcal{R} = - \frac{G M_m M_p M_s}{M_m + M_p}\sum_{l=2}^\infty
\sum_{m=-l}^l \frac{4 \pi}{2l+1} M_l \frac{
r_m^l}{r_p^{l+1}}Y_{lm}(\theta_m,\psi_m)Y_{lm}^*(\theta_p,\psi_p),\label{pulsar_signal_DistFunExpan}
\end{equation}
where
\begin{equation*}
M_l = \frac{M_m^{l-1}-\left(-M_p\right)^{l-1}}{\left(M_m +
M_p\right)^{l-1}},
\end{equation*}
and where $\theta_m$, $\psi_m$, $\theta_p$ and $\psi_p$ describe the angular orientation of $\mathbf{r}_m$ and $\mathbf{r}_p$ in spherical polar coordinates (see figure~\ref{RpRmOrientDia}), and where the monopole ($l=0$) and dipole ($l=1$) terms are exactly zero.

Writing the disturbing function in this way, it can be seen that it consists of a sum of terms, each comprised of five factors, which are, a constant, the distance between the planet and moon raised to the power $l$, the distance between the star and the planet-moon barycenter raised to the power $-(l+1)$, a spherical harmonic depending on the orientation of $\mathbf{r}_m$, and a spherical harmonic depending on the orientation of $\mathbf{r}_p$.  Consequently, we need a way to express each of these factors in terms of time and the orbital parameters of the system.  We begin with the spherical harmonics.

\subsection{Writing $Y_{lm}(\theta,\psi)$ in terms of $I$, $\omega$, $\Omega$ and $f$}

Following Mardling (private communication), $Y_{lm}(\theta,\psi)$ can be written in terms of the orbital elements $I$, $\omega$, $\Omega$ and $f$, such that, 
\begin{equation}
Y_{lm}(\theta,\psi) = \sum^l_{m' = -l,2}
D_{lmm'}\left(I,\omega,\Omega\right)e^{im'f},
\label{Int-Rev-RotDefn}
\end{equation}
where $I$, $\omega$ and $\Omega$ are the Euler
angles specifying the orientation of the orbit and where
$D_{lmm'}(I,\omega,\Omega)$ are Wigner D-functions, such that
\begin{equation}
D_{lmm'}\left(I,\omega,\Omega \right) = (-i)^{2l + m + m'}Y_{lm}\left(\frac{\pi}{2},0\right)
\gamma_{lmm'}\left(I \right)e^{i(m'\omega + m \Omega)},
\label{Int-Rev-DfunDefn}
\end{equation}
where, for completeness, the inclination functions $\gamma_{lmm'}(I)$ are tabulated in Appendix~\ref{Incl_App}. Applying this expansion to both spherical harmonics in equation~\eqref{pulsar_signal_DistFunExpan} and rearranging gives
\begin{multline}
\mathcal{R} = \frac{G M_m M_p M_s}{M_m + M_p}\sum_{l=2}^\infty
\sum_{m=-l}^l \sum^l_{m' = -l,2} \sum^l_{m'' = -l,2} \frac{4 \pi}{2l+1} \frac{a_m^1}{a_p^{l+1}} M_l \left(Y_{lm}\left(\frac{\pi}{2},0\right)\right)^2  \\ 
\times (-i)^{2m + m' + m''} \gamma_{lmm'}\left(I_m \right) \gamma_{lmm''}\left(I_p \right)  e^{i(m'\omega_m + m \Omega_m -m''\omega_p - m \Omega_p)} \\
\times \left[ \frac{r_m^l}{a_m^l} e^{im'f_m}\right]  \left[\frac{a_p^{l+1}}{r_p^{l+1}} e^{-im''f_p} \right].\label{pulsar_signal_DistFunExpanIncl}
\end{multline}
As can be seen, using this transformation functionally replaces each term with a set of new terms each consisting of three factors, which are a constant which depends on $a_m$, $I_m$, $\omega_m$, $\Omega_m$, $a_p$, $I_p$, $\omega_p$ and $\Omega_p$, a term which depends on $r_m$ and $f_m$ and a term which depends on $r_p$ and $f_p$.  These last two terms can also be described in terms of $a_m$, $e_m$, $M_m(t)$, $a_p$, $e_p$ and $M_p(t)$ using a different expansion, where we note that the time dependance differentiates between the mean anomalies $M_m(t)$ and $M_p(t)$, and the masses $M_m$ and $M_p$.

\subsection{Writing terms of the form $(r/a)^l e^{imf}$ in terms of $a$, $e$ and $M(t)$}

To begin, we take a closer look at the quantities in square brackets in equation~\eqref{pulsar_signal_DistFunExpanIncl}.  While these terms are not simple functions of time, they are approximately periodic, which means that they can be described as Fourier series, that is
\begin{equation}
\left(\frac{r_m}{a_m}\right)^l e^{i m f_m} = \sum_{n=-\infty}^{\infty}s^{(lm)}_n(e_m)e^{inM_m(t)} \label{PulM_Form_EccFTMoon}
\end{equation}
and
\begin{equation}
\left(\frac{a_p}{r_p}\right)^{l+1} e^{-i m f_p} = \sum_{n=-\infty}^{\infty}F^{(lm)}_n(e_p)e^{inM_p(t)} \label{PulM_Form_EccFTPlanet}
\end{equation}
where the coefficients $s^{(lm)}_n(e_m)$ and $F^{(lm)}_n(e_p)$ are functions of the eccentricity and are defined by
\begin{equation}
s^{(lm)}_n(e_m) = \frac{1}{2\pi}\int_0^{2 \pi} \frac{r_m^l}{a_m^l} e^{i m f_m} e^{-inM_m(t)}dM_m(t) \label{PulM-Ecc-Inn-Eq2}
\end{equation}
and
\begin{equation}
F^{(lm)}_n(e_p) = \frac{1}{2\pi}\int_0^{2 \pi} \frac{a_p^{l+1}}{r_p^{l+1}} e^{-i m f_p} e^{inM_p(t)}dM_p(t) \label{PulM-Ecc-Inn-Eq2}
\end{equation}
The properties of these coefficients were investigated in \citet{Mardling2008} and are summarised in appendix~\ref{App_Ecc_Fun}.  Also, the dependence of the coefficients $s^{(lm)}_n(e_m)$ and $F^{(lm)}_n(e_p)$ on $e_m$ and $e_p$ to third order is presented in tables \ref{App-Ecc-Inn-sTab} and \ref{App-Ecc-Out-FTab}.  

\section{Selection of the reference plane and direction}\label{Pulsar_Ext_Reference}

\begin{figure}
     \centering
     \subfigure[Coordinate system for planet's orbit.]{
          \label{PulsarCoordSystemPlanet}
          \includegraphics[width=.48\textwidth]{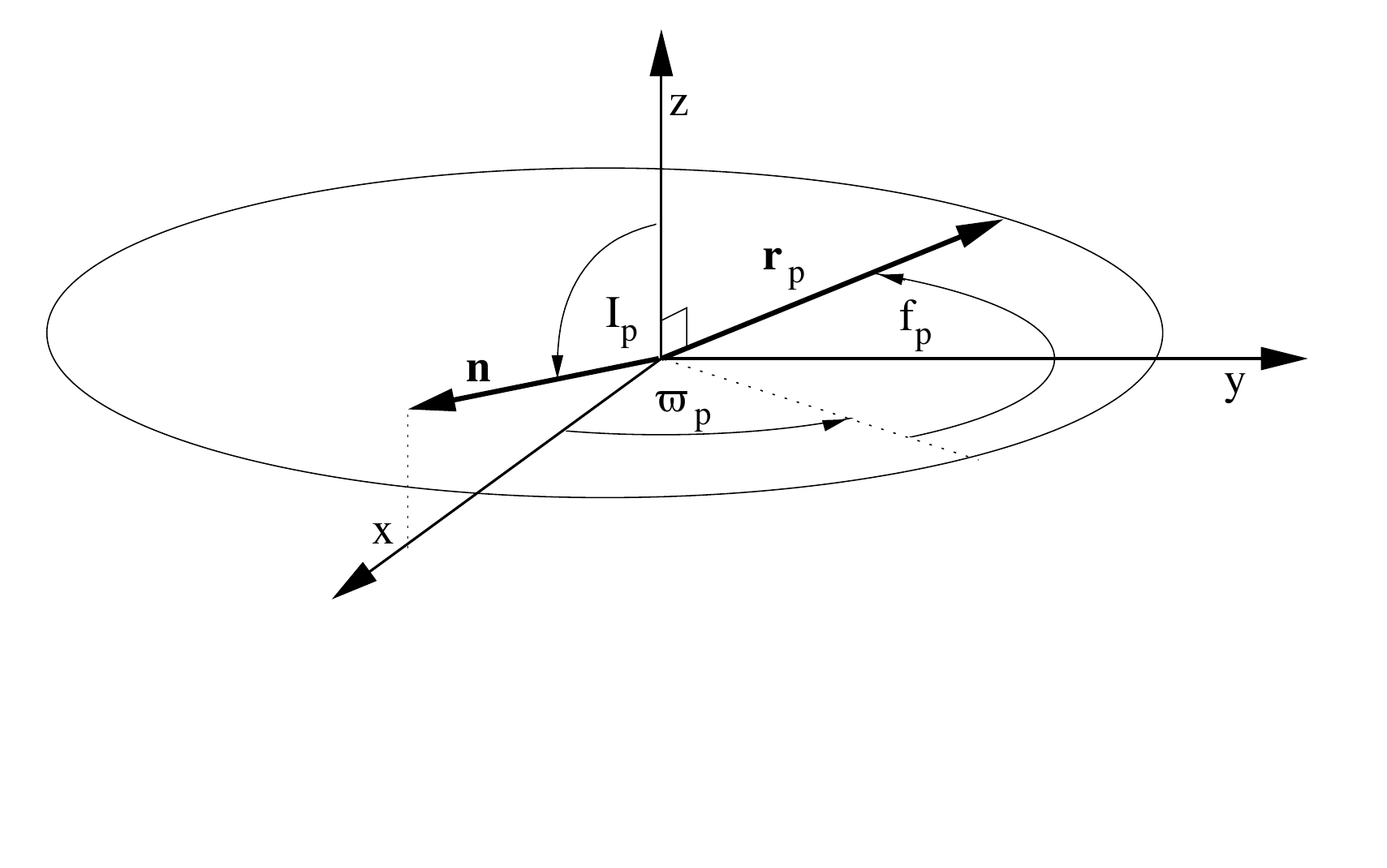}}
     \subfigure[Coordinate system for moon's orbit. ]{
          \label{PulsarCoordSystemMoon}
          \includegraphics[width=.48\textwidth]{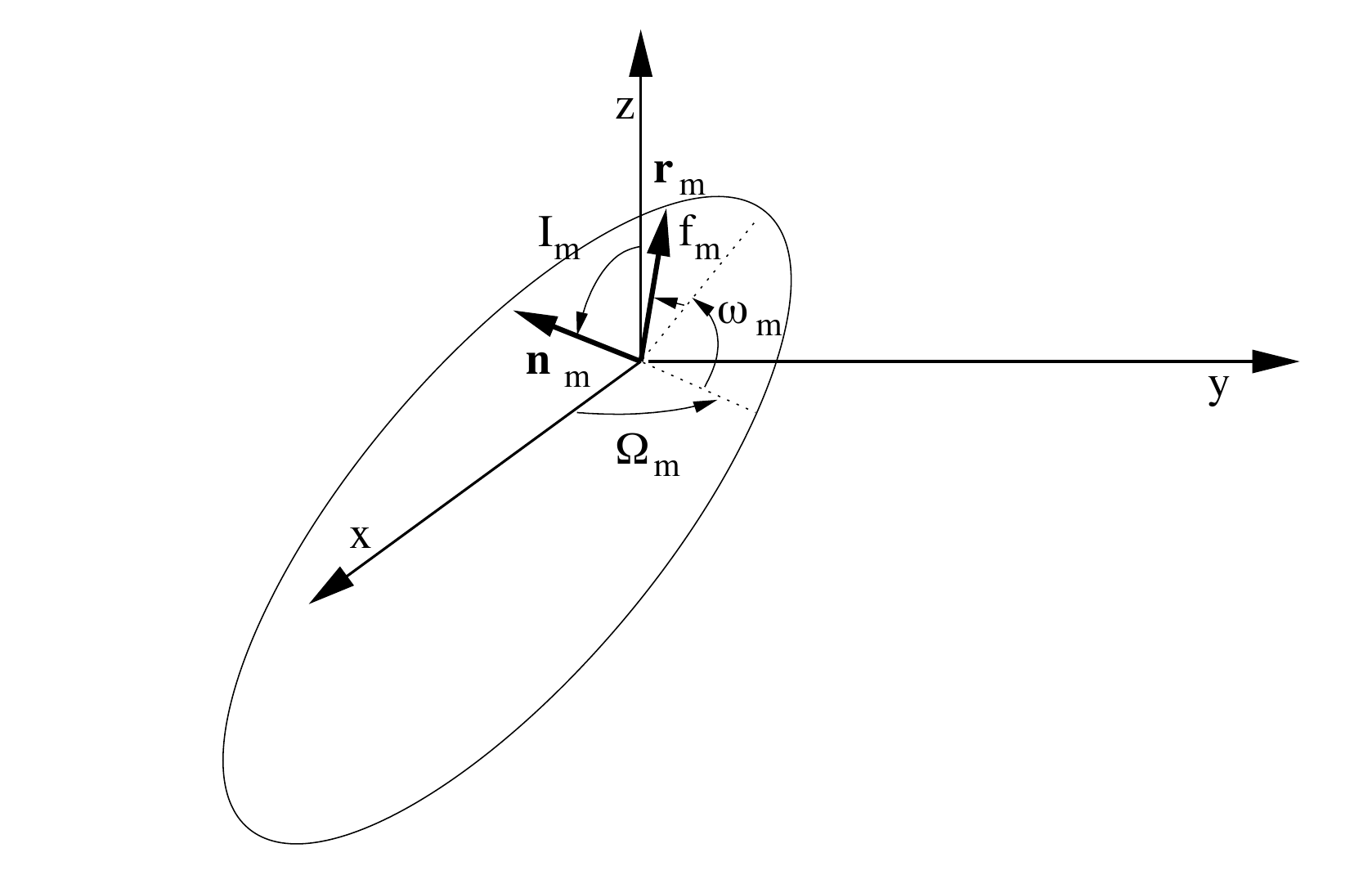}}
     \caption[Diagram showing the definition of the Euler angles $I_p$, $\Omega_p$, $\omega_p$, $I_m$, $\Omega_m$ and $\omega_m$ used in this chapter.]{Diagram showing the definition of the Euler angles $I_p$, $\Omega_p$, $\omega_p$, $I_m$, $\Omega_m$ and $\omega_m$ used in this chapter, where we define $\varpi_p = \omega_p + \Omega_p$.  In particular note that $\mathbf{n}_m$ is the orbit normal to the moon's orbit while $\mathbf{n}$ is directed along the line of sight}
     \label{PulsarCoordSystem}
\end{figure}

In order to derive expressions for the time-of-arrival perturbation due to planet-moon binarity, a reference plane and direction required for the definition of the Euler angles $I$, $\omega$ and $\Omega$, needs to be selected.  

To simplify the expression for $\mathcal{R}$ we choose the coordinate system such that the $x$-$y$ plane coincides with the planet's orbit.  For this choice we do not need to expand the spherical harmonic corresponding to the planet's orbit as $\theta_p = \pi/2$ and $\psi_p$ is equal to the sum of $f_p$, $\omega_p$, $\Omega_p$ and an additive constant which depends on the reference direction.  Unfortunately, an unwanted side effect of this choice is that the coordinate system is no longer inertial, and any process which acts to alter the orientation of the planet's orbit acts to alter the orientation of the coordinate system.  This issue is beyond the scope of this thesis, but will be briefly discussed in chapter~\ref{Conclusion} in context of directions for future research.  We now move onto the selection of reference direction.

For this work, the $x$-axis was selected as the reference direction, such that, for our choice of reference plane $\psi_p = f_p + \omega_p + \Omega_p$.  In addition, we choose the $x$-axis to correspond with the projection of the line between the observer and the system barycenter onto the plane of the planet's orbit.  This decision was made to reduce the number of non-zero components in $\mathbf{n}$.  For reference, these definitions are summarised in figure~\ref{PulsarCoordSystem}.

These two decisions have a number of ramifications, especially for the definition of the inclinations $I_m$ and $I_p$.  These will be briefly highlighted.  First, as the reference plane is the planet's orbit, the Euler angles for the moon's orbit are measured relative to the planet's orbit (see figure~\ref{PulsarCoordSystemMoon}).  In particular $I_m$ represents the mutual inclination between the orbit of the planet and that of the moon, such that $I_m=0$ implies that the orbits are coplanar and $I_m \ne 0$ implies that they are not.  Second, for the planet's orbit, the situation is a little more involved.  As the orbital plane of the planet is the reference plane, using the standard definition of inclination, the planet's orbit would have zero inclination by definition.  Consequently, following \citet{Lewisetal2008}, for this chapter we define $I_p$ to be the angle between $\mathbf{n}$, the vector along the line of sight and the normal to the planet's orbital plane (see figure~\ref{PulsarCoordSystemPlanet}).  Now that a coordinate system has been selected and discussed, we can use the methods introduced in section~\ref{Pulsar_Ext_ExpansionIntro} to produce a more useful form for the expression for the time-of-arrival perturbation.

\section{Derivation of the general equation}

We begin by deriving a general expression for the time-of-arrival perturbation, to which the expansions described in section~\ref{Pulsar_Ext_ExpansionIntro} can be applied where appropriate.  To start consider equation~\eqref{TOA-TOApertpmdef},
\begin{equation*}
\frac{1}{c}\int_0^t\int_0^{t'} \mathbf{\ddot{R}}_s \cdot \mathbf{n} dt'' dt' =
TOA_{pert,p} + TOA_{pert,pm},
\end{equation*}
where we recall that $TOA_{pert,p}$ is the time-of-arrival perturbation due to orbit of the the planet-moon system about the pulsar, $TOA_{pert,pm}$ is the timing perturbation due to planet-moon binarity, and $\mathbf{R}_s$ is the vector from the system barycenter to the star.  From chapter \ref{Intro_Moons_Note}, we have that $\mathbf{r}_p$ is the vector from the planet-moon barycenter to the star, and thus $\mathbf{R}_s = [(M_p + M_m)/(M_p + M_m + M_s)]\mathbf{r}_p$.  Using this expression and equation~\eqref{Int-Rev-rpeq} we then have that
\begin{equation}
TOA_{pert,pm} = -\frac{1}{c} \frac{1}{M_s} \int_0^t \int_0^{t'}
\frac{\partial \mathcal{R}}{\partial \mathbf{r}_p} 
\cdot \mathbf{n} dt' dt,\label{PulM-GovEq-GovEq2}
\end{equation}
where we note that a different definition of $\mathcal{R}$ is used in \citet{Lewisetal2008} than used in this chapter.  Expanding the disturbing function using  equation~\eqref{pulsar_signal_DistFunExpan} gives
\begin{align}
TOA_{pert,pm} &= -\frac{1}{c} \frac{1}{M_s} \int_0^t \int_0^{t'}
\frac{\partial}{\partial \mathbf{r}_p} \left(\frac{G M_m M_p
M_s}{M_m + M_p}\sum_{l=2}^\infty \sum_{m=-l}^l \frac{4 \pi}
{2l+1} M_l \right. \notag \\
&\left.  \times
\frac{r_m^l}{r_p^{l+1}}Y_{lm}(\theta_m,\psi_m)Y_{lm}^*(\theta_m,\psi_m)\right)
\cdot \mathbf{n} dt' dt, \\ 
 &= - \frac{1}{c} \frac{G M_m
M_p}{M_m + M_p} \int_0^t \int_0^{t'} \left(\sum_{l=2}^\infty \sum_{m=-l}^l \frac{4
\pi}{2l+1} M_l \frac{ r_m^l}{r_p^{l+2}}\right.
\notag \\
\times Y_{lm}(\theta_m,\psi_m) & \left.\Bigg[ -(l+1) Y_{lm}^*(\theta_p,\psi_p)\mathbf{e}_{r_p} + \left.\frac{\partial Y_{2m}^*(\theta_p,\psi_p)}{\partial \theta_p}\right|_{\theta_p = \frac{\pi}{2}} \mathbf{e}_{\theta_p}
\right. 
\notag \\
& \left. - im Y_{2m}^*(\theta_p,\psi_p) \mathbf{e}_{\psi_p}\Bigg] \right)\cdot \mathbf{n} dt' dt,
\label{PulM-GovEq-GovEq3}
\end{align}
where we note that the expression $\theta_p = \pi/2$ has been used.  Using the fact that $r_m/r_p \ll 1$, we retain the  $l=2$ terms only,\footnote{For the case of the four known pulsar planets PSR~1257+12~b, PSR~1257+12~c, PSR~1257+12~d and PSR B1620-26 b, the highest values of $r_m/r_p$ allowed by orbital stability are 0.0012, 0.0034, 0.0033, and 0.039, where we note that the ratio for PSR~B1620-26~b was calculated under the unrealistic \citep{Fordetal2000,Sigurdssonetal2004} assumption that its orbit is not eccentric.  Consequently, for these four cases the $l=3$ terms will be at least 0.0012, 0.0034, 0.0033, and 0.039 times smaller than the $l=2$ terms respectively, and can thus be neglected.} equation~\eqref{PulM-GovEq-GovEq3} simplifies to
\begin{multline}
TOA_{pert,pm} = - \frac{1}{c} \frac{G M_m M_p}{M_m + M_p} \int_0^t \int_0^{t'} \left( \sum_{m=-2}^2 \frac{4 \pi}{5} \frac{ r_m^2}{r_p^4} Y_{lm}(\theta_m,\psi_m) \right.\\
\left.\times \Bigg[ -3 Y_{2m}^*(\theta_p,\psi_p)\mathbf{e}_{r_p} + \left.\frac{\partial
Y_{2m}^*(\theta_p,\psi_p)}{\partial \theta_p}\right|_{\theta_p =
\frac{\pi}{2}} \mathbf{e}_{\theta_p}
\right. \\
\left.  - im Y_{2m}^*(\theta_p,\psi_p)
\mathbf{e}_{\psi_p}\Bigg] \right)\cdot \mathbf{n} dt'
dt.\label{PulM-GovEq-GovEq4}
\end{multline}
where we note that $M_2 = 1$.  We now consider the definition of $\mathbf{n}$.  As the direction of the unit vectors in spherical polar coordinates depends on position, it would be useful to convert $\mathbf{n}$ into Cartesian coordinates. Using
the fact that $\theta_p = \pi/2$ we have that
\begin{align}
\mathbf{e}_{x_p} &= \cos \psi_p \mathbf{e}_{r_p} - \sin \psi_p \mathbf{e}_{\psi_p},
\label{PulM-GovEq-eX}
\\
\mathbf{e}_{y_p} &= \sin \psi_p \mathbf{e}_{r_p} + \cos \psi_p \mathbf{e}_{\psi_p},
\label{PulM-GovEq-eY}
\\
\mathbf{e}_{z_p} &= - \mathbf{e}_{\theta_p}. \label{PulM-GovEq-eZ}
\end{align}
We now write the three components of $\mathbf{n}$, $n_{x_p}$, $n_{y_p}$ and $n_{z_p}$, in terms of the quantities of the system.  Comparing with figure~\ref{PulsarCoordSystemPlanet} and noting that $\mathbf{n}$ lies in the $x$-$z$ plane gives
\begin{align}
n_{x_p} &= \sin I_p, \label{PulM-GovEq-nX}
\\
n_{y_p} &= 0, \label{PulM-GovEq-nY}
\\
n_{x_p} &= \cos I_p, \label{PulM-GovEq-nZ}
\end{align}
where $I_p$ is the angle between the line-of-sight and the vector normal to the planet's orbit.

So, using this notation we have that
\begin{equation}
\mathbf{n} = \left[\cos \psi_p \mathbf{e}_{r_p} - \sin \psi_p \mathbf{e}_{\psi_p} \right]\sin I_p + 0 + \left[ -
\mathbf{e}_{\theta_p} \right]\cos I_p, \label{PulM-GovEq-ndef1}
\end{equation}
which is equivalent to
\begin{equation}
\mathbf{n} = \sin I_p \cos \psi_p \mathbf{e}_{r_p}
 - \cos I_p \mathbf{e}_{\theta_p} -\sin I_p \sin \psi_p
 \mathbf{e}_{\psi_p}. \label{PulM-GovEq-ndef2}
\end{equation}
The two equations we will require are equations~\eqref{PulM-GovEq-GovEq4} and
\eqref{PulM-GovEq-ndef2}. These equations will form the basis for the following analyses into the perturbation in the case of circular coplanar, inclined and eccentric orbits.

\section{Circular coplanar orbits}
\label{Sec_PulM_CC}

To begin the investigation into the form of the time-of-arrival perturbation due to a moon, we revisit the case of circular and coplanar planet and moon orbits.  First, we must write the spherical polar angles representing the position of the planet ($\theta_p$ and $\psi_p$) and the moon ($\theta_m$ and $\psi_m$) in terms of the angles $f_p$, $\omega_p$ and $\Omega_p$ and $f_m$, $\omega_m$ and $\Omega_m$.  As mentioned in section~\ref{Pulsar_Ext_Reference}, the orbit of the planet is defined to lie in the horizontal plane defined by $\theta_p = \pi/2$.  As the orbit of the moon is coplanar with that of the planet, we also have that $\theta_m = \pi/2$.  Finally, recalling that the $x$-axis is the reference direction for both the spherical polar coordinate system and the definition of longitudes, we have that $\psi_p = f_p + \omega_p + \Omega_p$ and $\psi_m = f_m + \omega_m + \Omega_m$. Substituting these values into equation~\eqref{PulM-GovEq-GovEq4} and noting that $P_l^m(0)=0$ for $m+l$ odd and $\left.\partial P_l^m(\cos \theta_p)/\partial \theta_p \right|_{\cos\theta_p = 0}=0$ for $m+l$ even, this simplifies to
\begin{multline}
TOA_{pert,pm} = - \frac{1}{c} \frac{G M_m M_p}{M_m + M_p} \int_0^t \int_0^{t'} \left( \sum_{m=-2,2}^2 \frac{(2-m)!}{(2+m)!} \frac{ r_m^2}{r_p^4}\right.\\
\left.\times \left(P_2^m(0)\right)^2 e^{im(f_m + \varpi_m - f_p - \varpi_p)}\left[-3 \mathbf{e}_{r_p} -im \mathbf{e}_{\psi_p} \right] \right)\cdot \mathbf{n} dt' dt,\label{PulM-CC-Eq1}
\end{multline}
where the expressions $\varpi_p = \omega_p + \Omega_p$ and $\varpi_m = \omega_m + \Omega_m$ have been used.  Expanding the sum and rewriting the complex exponentials in terms of sinusoids then gives
\begin{multline}
TOA_{pert,pm} = - \frac{1}{c} \frac{G M_m M_p}{M_m + M_p} \int_0^t \int_0^{t'} \left( \frac{ r_m^2}{r_p^{4}} \right. \\
\left. \times \left( \left[ \frac{-9}{4} \cos(2f_m + \varpi_m - f_p - \varpi_p) - \frac{3}{4} \right]\mathbf{e}_{r_p} \right.\right.\\
\left.\left. + \frac{3}{2} \sin(2f_m + \varpi_m - f_p - \varpi_p) \mathbf{e}_{\psi_p} \right) \right)\cdot \mathbf{n} dt' dt.\label{PulM-CC-Eq3}
\end{multline}
Writing $\mathbf{n}$ out in full using equation~\eqref{PulM-GovEq-ndef2} and combining like terms gives
\begin{multline}
TOA_{pert,pm} = - \frac{1}{c}\frac{G M_m M_p\sin I_p}{M_m + M_p} \int_0^t \int_0^{t'} \frac{r_m^2}{r_p^{4}} \left(- \frac{3}{4}\cos
\left(f_p + \varpi_p\right)\right. \\
\left. - \frac{15}{8} \cos(2f_m + 2\varpi_m - 3f_p - 3\varpi_p) \right. \\
\left.- \frac{3}{8} \cos(2f_m + 2\varpi_m - f_p - \varpi_p) \right) dt' dt.\label{PulM-CC-Eq4}
\end{multline}
As both the planet and moon orbits are circular, $r_m = a_m$ and $r_p = a_p$, and thus these terms can be moved outside the integral as they are constant.  Also, we have that, $\frac{df_p}{dt}=n_p$ and $\frac{df_m}{dt}=n_m$, where $n_p$ and $n_m$ are the mean motions of the planet and moon respectively, are also constant. Taking advantage of these simplifications and performing the two integrals gives
\begin{multline}
TOA_{pert,pm} = -\frac{1}{c}\frac{G M_m M_p\sin I_p}{M_m + M_p}\frac{
r_m^2}{r_p^{4}} \left(\frac{3}{4n_p^2}\cos \left(f_p + \varpi_p\right)\right.
\\
\left. + \frac{15}{8(2n_m-3n_p)^2} \cos(2f_m + 2\varpi_m - 3f_p - 3\varpi_p) \right.
\\
\left.+ \frac{3}{8(2n_m-n_p)^2} \cos(2f_m + 2\varpi_m - f_p - \varpi_p)
\right).\label{PulM-CC-Eq5}
\end{multline}
Noting that  $f_p + \varpi_p$ and $f_m + \varpi_m$ in this work is equivalent to $f_p$ and $f_m$ in chapter \ref{Pulsar_Paper}, this is exactly the same expression as given in equation~\eqref{TOA-pert1}.  Also, as pointed out in chapter \ref{Pulsar_Paper}, the first term of equation~\eqref{PulM-CC-Eq5} cannot be distinguished from $TOA_{pert,p}$, the the time-of-arrival signal due to the orbit of the the planet-moon system about the pulsar, as they have the same angular frequency, and consequently it can be neglected. 

\begin{figure}[tb]
\begin{center}
\includegraphics[width=.83\textwidth]{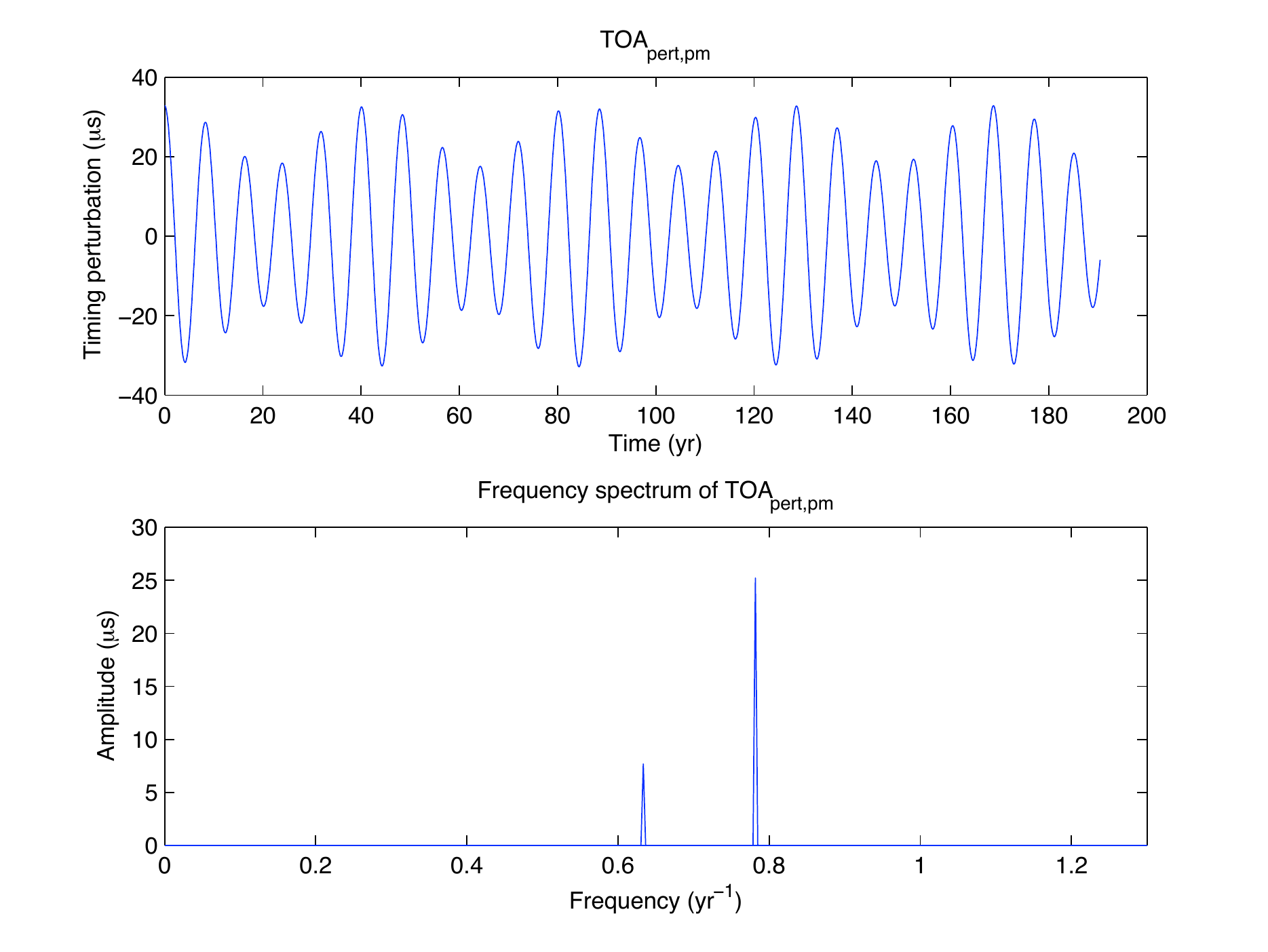}
\caption[Functional form and frequency composition of the time-of-arrival perturbation due to planet-moon binarity for the case where the planet and moon's orbits are circular and coplanar.]{Functional form and frequency composition of the time-of-arrival perturbation due to planet-moon binarity for the case where the planet and moon's orbits are circular and  coplanar.  This curve was calculated for the case of a PSR~B1620-26~b analog, in particular, it was assumed that $M_p =2.3M_J$, $a_p =23$AU, $M_m = 0.2M_J$ and $a_m = 0.8$AU.} \label{TOACoplanarSig}
\end{center}
\end{figure}

For reference, a realisation of $TOA_{pert,pm}$ calculated for an example planet-moon pair corresponding to PSR~B1620-26~b is presented in figure~\ref{TOACoplanarSig}.  While data corresponding to a full orbit of PSR~B1620-26~b is not available as the pulsar has only been observed for a little over 20 years, and the orbital period is of the order of a century, the orbital elements can be constrained by measuring the period derivatives and measuring perturbations on the orbit of the white dwarf companion \citep[e.g.][]{JoshiRasio1997}.  Conversely, the perturbation due to planet-moon binarity causes timing variations over much shorter timescales, and recalling from chapter~\ref{Pulsar_Paper} that the timing errors are of the order of 40$\mu$s, this perturbation is potentially detectable.

In particular, as can be seen in figure~\ref{TOACoplanarSig}, for the case of circular and coplanar planet and moon orbits, $TOA_{pert,pm}$ looks like a high frequency sinusoid which has been multiplied by an envelope function.  While this is not strictly mathematically the case, in that $TOA_{pert,pm}$ is given by the sum of a sinusoid and a beat function, this analogy is intuitively useful.  Noting that the sinusoids with frequency $f_1 = 2n_m - n_p$ and $f_2 = 2n_m - 3n_p$ are causing the beating in $TOA_{pert,pm}$, and recalling that the frequency of the envelope function for a beat is given by $(f_1 - f_2)/2 = n_p$ while the frequency of the high frequency sinusoid that it modifies is given by $(f_1 + f_2)/2 = n_m - n_p$, it can be surmised that the envelope function is defined by the effect of the planet's orbit and the high frequency sinusoid defines the effect of the moon's orbit.  We now move to the first of the more complex cases analysed in this chapter, the case where the orbits of the planet and moon are mutually inclined.

\section{Mutually inclined orbits}

As discussed in chapter~\ref{Intro_Moons_Const}, moons are more likely to form or be captured into orbits which are roughly coplanar with that of their host planet. However, highly inclined moon systems do form, for example, the satellite system of Uranus.  Consequently, it is of scientific interest to investigate the effect of mutual inclination in two regimes.  First, the effect of slight orbital misalignment will be investigated by deriving an expression for the time-of-arrival perturbation correct to order $\sin I_p$.  Then the case of arbitrary orbital misalignment will be considered.  However, before either of these cases can be investigated, equation~\eqref{PulM-GovEq-GovEq4} must be written in terms of the orbital elements.

As in the previous case, we begin by writing the angles $\theta_p$, $\psi_p$, $\theta_m$ and $\psi_m$ in terms of the orbital elements $f_p$, $\omega_p$, $\Omega_p$, $f_m$, $\omega_m$ and $\Omega_m$.  As for the case where the orbits were coplanar, we have that $\theta_p = \pi/2$ and $\psi_p = f_p + \omega_p + \Omega_p$, however, as the moon's orbit is now inclined, writing $\theta_m$ and $\psi_m$ in terms of $f_m$, $\omega_m$ and $\Omega_m$ is a little more challenging. This leads to two issues that must be resolved before analytic progress can be made:
\begin{enumerate}
\item The inclusion of terms of the form $\left. \partial
Y_{lm}(\theta_p,\psi_p)/\partial \theta_p \right|_{\theta_p = \pi/2}$ in the sum.
\item The time-of-arrival perturbation is no longer independent of $\theta_m$.
\end{enumerate}
The solution to both of these issues is to expand the term in question, however, the way this is done is slightly different for each of the two cases.

\subsection{Recasting terms of the form $\left.\frac{\partial Y_{lm}(\theta_p,\psi_p)}{\partial
\theta_p}\right|_{\theta_p = \pi/2}$}

To begin, we note that there are a number of recurrence identities which relate Legendre polynomials of different orders.  One such identity is
\begin{equation}
\sin \theta_p \frac{d P^m_l(\cos \theta_p)}{d \theta_p} = l \cos \theta_p
P^m_l(\cos \theta_p) - (l+m)P^m_{l-1}(\cos
\theta_p).\label{PulM-Incl-ThetaDer1}
\end{equation}
This identity is valid for both positive and negative $m$.
Substituting in $\theta_p = \pi/2$, gives
\begin{equation}
\left.\frac{d P^m_l(\cos \theta_p)}{d \theta_p}\right|_{\theta_p = \pi/2}
= -(l+m)P^m_{l-1}(0). \label{PulM-Incl-ThetaDer3}
\end{equation}
Comparing this with equation~\eqref{Int-Rev-YlmDef}, it can be seen
that
\begin{equation}
\left.\frac{\partial Y_{lm}(\theta_p,\psi_p)}{\partial
\theta_p}\right|_{\theta_p = \pi/2} =
-\sqrt{\frac{(2l+1)(l-m)}{(2l-1)(l+m)}}(l+m)Y_{(l-1)m}
\left(\frac{\pi}{2},\psi_p\right). \label{PulM-Incl-ThetaDer4}
\end{equation}
Substituting equation~\eqref{PulM-Incl-ThetaDer4} into equation~\eqref{PulM-GovEq-GovEq4} and noting that $l=2$ gives
\begin{multline}
TOA_{pert,pm} = - \frac{1}{c} \frac{G M_m
M_p}{M_m + M_p} \int_0^t \int_0^{t'} \left( \sum_{m=-2}^2 \frac{4 \pi}{5} \frac{ r_m^2}{r_p^{4}} Y_{2m} (\theta_m,\psi_m )\right.
\\
\left.\times \left[-3 Y_{2m}^*\left(\frac{\pi}{2},\psi_p \right)\mathbf{e}_{r_p} -\sqrt{\frac{5(2-m)}{3(2+m)}}(2+m)
Y^*_{1m}\left(\frac{\pi}{2},\psi_p \right) \mathbf{e}_{\theta_p}
\right.\right. 
\\
\left. \left. -im Y_{2m}^*\left( \frac{\pi}{2},\psi_p \right)\mathbf{e}_{\psi_p} \right] \right)\cdot \mathbf{n} dt' dt.\label{PulM-Incl-ThetaDer5}
\end{multline}

\subsection{Describing a rotated moon orbit}

As the moon's orbit no longer lies in the same plane as the planet's orbit, its orbital plane is no longer given by $\theta_m = \pi/2$.  To deal with this inclined orbit, the expansion described in section~\ref{Pulsar_Ext_ExpansionIntro} will be employed. Using equations~\eqref{Int-Rev-RotDefn} and \eqref{Int-Rev-DfunDefn} to expand equation~\eqref{PulM-Incl-ThetaDer5} gives
\begin{multline}
TOA_{pert,pm} = \frac{1}{c} \frac{G M_m M_p}{M_m + M_p} \int_0^t \int_0^{t'} \left.\Bigg( \sum_{m=-2}^2 \sum_{m'=-2,2}^2 \frac{ r_m^2}{r_p^4}(-i)^{m+m'}\right.\\
\left.\times
\gamma_{2mm'}(I_m)\sqrt{\frac{(2-m')!}{(2+m')!}}P^{m'}_{2}(0)e^{i(m'f_m + m'\omega_m
- m f_p - m \varpi_p)}e^{im\Omega_m}\right.
\\
\left.\left[-3 \sqrt{\frac{(2-m)!}{(2+m)!}}P^{m}_{2}(0)\mathbf{e}_{r_p}
+ \sqrt{\frac{(2-m)!}{(2+m)!}}(2+m)P^m_1(0) \mathbf{e}_{\theta_p}
\right.\right. \\
\left. \left. -im
\sqrt{\frac{(2-m)!}{(2+m)!}}P^{m}_{2}(0)\mathbf{e}_{\psi_p}\right]
 \right)\cdot \mathbf{n} dt' dt.\label{PulM-Incl-Rot2}
\end{multline}
where, for easy reference, a table of $\gamma_{lmm'}(I_m)$ functions is given in appendix~\ref{Incl_App}.  Armed with equation~\eqref{PulM-Incl-Rot2}, we are now in a position to calculate the form of the perturbation for the cases of small and arbitrary amounts of misalignment.

\subsection{Solution in the case of circular orbits and small mutual
inclination}

\begin{figure}[tb]
\begin{center}
\includegraphics[height=2in,width=3.5in]{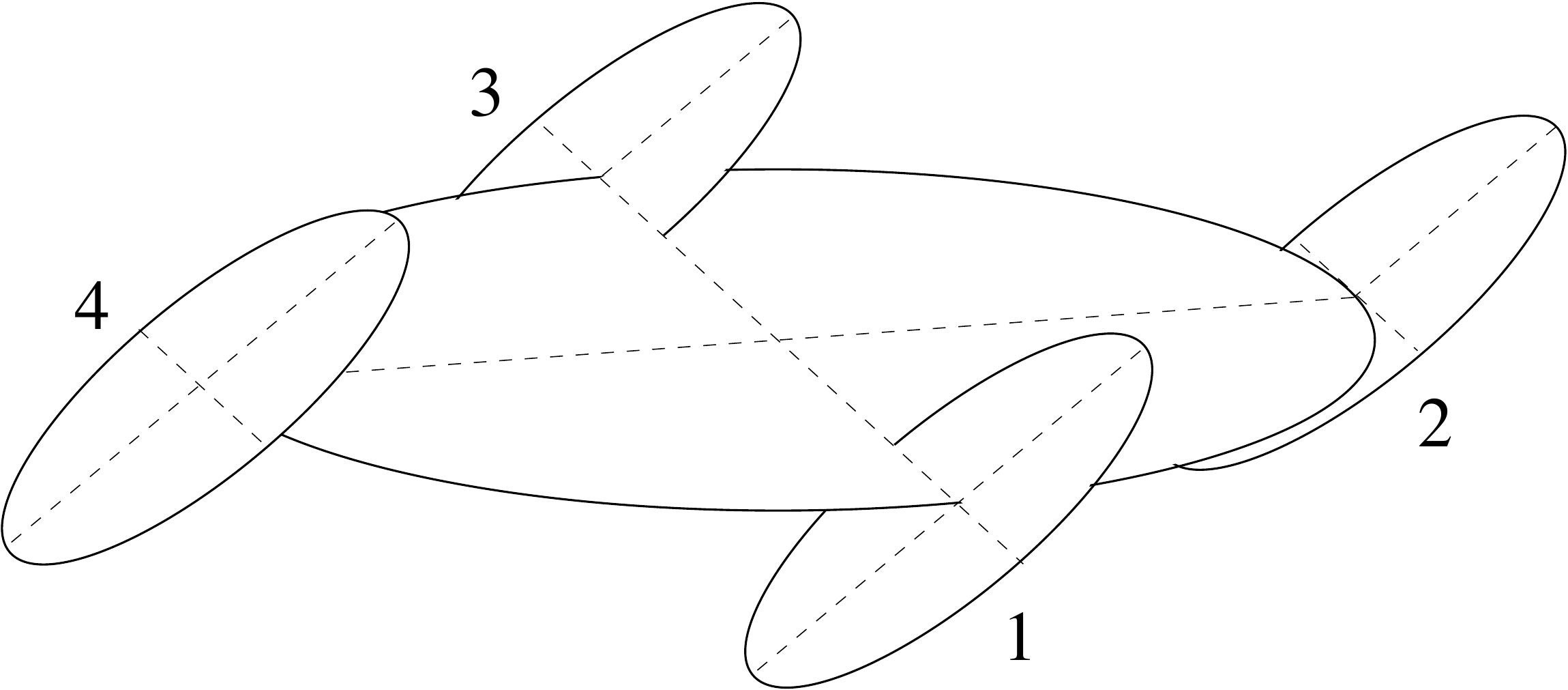}
\caption[Schematic diagram of the orbital orientations of the moon
and planet orbits at four stages of the ``year".]{Schematic diagram of the orbital orientations of the moon
and planet orbits at four stages of the ``year".} \label{LowIncPos}
\end{center}
\end{figure}

For the case where $\sin I_m$, the sine of the relative inclination is near zero, we have that
\begin{equation}
\cos I_m \approx 1-\frac{1}{2}\sin^2 I_m.
\end{equation}
As a result of the prediction that low inclinations should be
common, it would be interesting to only include the terms of order
$\sin I_m$ and investigate the signal form.  As $\gamma_{lmm'}(I_m)$ are the
terms which contain $I_m$, a quick investigation of the table in
Appendix~\ref{Incl_App} shows that only $\gamma_{21m'}(I_m)$ and
$\gamma_{2-1m'}(I_m)$ have terms that are first order in $\sin I_m$.  Also, to
first order in $\sin I_m$, all other terms are exactly equal to the
values found in the circular coplanar case, which allows us to
write
\begin{multline}
TOA_{pert,pm} = TOA_{pert,cc} - \frac{1}{c} \int_0^t \int_0^{t'} \frac{G M_m M_p}{M_m + M_p} \sum_{m=-1,2}^1 \sum_{m'=-2,2}^2 \frac{ r_m^2}{r_p^4}\\
\times \gamma_{2mm'}(I_m)\sqrt{\frac{(2-m')!(2-m)!}{(2+m')!(2+m)!}}(2+m) P^{m'}_{2}(0)P^m_1(0)
\\
\times (-i)^{m+m'} e^{i(m'(f_m + \omega_m) - m (f_p + \varpi_p))}e^{im\Omega_m} \mathbf{e}_{\theta_p} \cdot \mathbf{n} dt' dt.\label{PulM-Incl-SmallI1}
\end{multline}
where $TOA_{pert,cc}$ is the perturbation $TOA_{pert,pm}$ for the case of circular coplanar orbits. Expanding the sum, substituting in the values of $\gamma_{2mm'}(I_m)$ to order $\sin I_m$ from Appendix \ref{Incl_App}, the expression for $\mathbf{n}$ from equation~\eqref{PulM-GovEq-ndef2} and combining the complex exponentials into sinusoids gives
\begin{multline}
TOA_{pert,pm} = TOA_{pert,cc} - \frac{1}{c} \int_0^t \int_0^{t'} \frac{G M_m M_p\cos I_p}{M_m + M_p} \frac{ r_m^2}{r_p^4}
\\
\times \left(\frac{3}{4} \sin I_m \sin(2f_m + 2\omega_m -  f_p - \varpi_p + \Omega_m) \right.
\\
\left.+ \frac{3}{4} \sin I_m \sin(2f_m  + 2\omega_m + f_p + \varpi_p - \Omega_m) \right.
\\
\left. - \frac{3}{2} \sin I_m \sin(f_p + \varpi_p - \Omega_m) \right) dt''
dt'.\label{PulM-Incl-SmallI3}
\end{multline}
Noting that as the orbits are circular, $r_p = a_p$, $r_m = a_m$, $\frac{df_p}{dt} = n_p$ and
$\frac{df_m}{dt} = n_m$, the integration can be performed to give
\begin{multline}
TOA_{pert,pm} = TOA_{pert,cc} - \frac{1}{c} \frac{G M_m M_p \sin I_m \cos
I_p}{M_m + M_p} \frac{ a_m^2}{a_p^4}
\\
\times \left(\frac{-3}{4(2n_m - n_p)^2} \sin(2f_m + 2\omega_m -  f_p - \varpi_p + \Omega_m)
\right.
\\
\left. -  \frac{3}{4(2n_m + n_p)^2} \sin(2f_m + 2\omega_m + f_p + \varpi_p - \Omega_m)
\right.
\\
\left. + \frac{3}{2n_p^2} \sin(f_p + \varpi_p - \Omega_m) \right).
 \label{PulM-Incl-SmallI4}
\end{multline}

As can be seen above, low inclination results in the inclusion of one\footnote{Terms with frequency $2n_m - n_p$ and $n_p$ are already present in the expression corresponding to the circular coplanar case.  Consequently, the term with frequency $2n_m + n_p$ corresponds to the only ``new" frequency.} additional frequency in the solution.  It is interesting to note that this additional signal looks like a beat function.  

To see why, consider how the position of the planet and moon relative to the pulsar changes over a moon orbit as a function of the position of the planet-moon pair about the planet's orbit (see figure~\ref{LowIncPos}).  During the two sections of the planet's year when the moon's orbital plane is aligned with the vector pointing to the pulsar (stage 1 and 3 in figure~\ref{LowIncPos}), the moon moves from being $a_p + (M_p/(M_m + M_p))a_m$ away from the pulsar to being $a_p - (M_p/(M_m + M_p))a_m$ away from the pulsar, while the planet moves from being $a_p + (M_m/(M_m + M_p))a_m$ away from the pulsar to being $a_p - (M_m/(M_m + M_p))a_m$ away from the pulsar during a single moon orbit.  As these are the same values as for circular coplanar orbits it should be unsurprising that these two times correspond to the zero of the envelope function of the beat.  Conversely, when the moon orbit is more face on to the pulsar (stage 2 and 4 in figure~\ref{LowIncPos}), the moon moves from being $(a_p^2 + (M_p/(M_m + M_p))^2a_m^2 + 2a_p(M_p/(M_m + M_p))a_m\cos I_m)^{1/2}$ away from the pulsar to being $(a_p^2 + (M_p/(M_m + M_p))^2a_m^2 - 2a_p(M_p/(M_m + M_p))a_m\cos I_m)^{1/2}$ away from the pulsar while the planet moves from being $(a_p^2 + (M_m/(M_m + M_p))^2a_m^2 + 2a_p(M_m/(M_m + M_p))a_m\cos I_m)^{1/2}$ away from the pulsar to being $(a_p^2 + (M_m/(M_m + M_p))^2a_m^2 - 2a_p(M_m/(M_m + M_p))a_m\cos I_m)^{1/2}$ away from the pulsar during one moon orbit. As these values are the most different from those for circular coplanar orbits, it should be unsurprising that it is at these times that the perturbation to the circular coplanar signal is greatest.  We now explore the effect of arbitrary mutual inclination.

\subsection{Solution in the case of circular orbits and arbitrary mutual
inclination}

For the case of arbitrary mutual inclination we no longer neglect the higher order terms in $\sin I_m$.  Taking equation~\eqref{PulM-Incl-Rot2}, expanding the sum, collecting the complex exponentials into sine and cosine functions and substituting in the expressions for $\gamma_{lmm'}(I_m)$, where we note that that $\gamma_{lmm'}(I_m) = (-1)^{m+m'}\gamma_{l-m-m'}(I_m)$ (see Appendix \ref{Incl_App}), gives
\begin{multline}
TOA_{pert,pm} = - \frac{1}{c}\frac{G M_m M_p}{M_m + M_p} 
\int_0^t \int_0^{t'} \frac{ r_m^2}{r_p^4} \left( - \frac{3}{8} (3\cos^2 I_m - 1) \mathbf{e}_{r_p} \right.
\\
\left.- \frac{9}{16} (1 + \cos I_m)^2 \cos(2f_m + 2\varpi_m - 2 f_p - 2 \varpi_p) \mathbf{e}_{r_p} \right.
\\
\left.- \frac{3}{8} (1 + \cos I_m)^2 \sin(2f_m + 2\varpi_m - 2 f_p  - 2 \varpi_p) \mathbf{e}_{\psi_p} \right.
\\
\left. + \frac{3}{4} \sin I_m (1 + \cos I_m) \sin(2f_m + 2\omega_m - f_p - \varpi_p + \Omega_m)
\mathbf{e}_{\theta_p}   \right.
\\
\left. +  \frac{3}{4} \sin I_m (1 - \cos I_m) \sin(2f_m + 2\omega_m + f_p + \varpi_p -\Omega_m)
\mathbf{e}_{\theta_p}\right.
\\ 
\left. - \frac{9}{8} \sin^2 I_m \cos(2 f_p +2 \varpi_p - 2\Omega_m) \mathbf{e}_{r_p} \right.
\\ \displaybreak
\left.+ \frac{3}{8} \sin^2 I_m \sin(2 f_p + 2 \varpi_p - 2\Omega_m) \mathbf{e}_{\psi_p} \right.
\\
\left. - \frac{9}{16} (1 - \cos I_m)^2 \cos(2f_m + 2\omega_m +2 f_p +2 \varpi_p -2\Omega_m)\mathbf{e}_{r_p} \right.
\\
\left.- \frac{9}{8} \sin^2 I_m \cos(2f_m + 2\omega_m) \mathbf{e}_{r_p} \right.
\\
\left.+ \frac{3}{8} (1 - \cos I_m)^2 \sin(2f_m + 2\omega_m +2 f_p + 2 \varpi_p - 2 \Omega_m)\mathbf{e}_{\psi_p} \right.
\\
\left. + \frac{3}{4} \sin (2I_m) \sin(f_p +  \varpi_p - \Omega_m) \mathbf{e}_{\theta_p}  \right)\cdot \mathbf{n} dt''
dt'.\label{PulM-Incl-GenI2}
\end{multline}
Substituting in equation~\eqref{PulM-GovEq-ndef2} and collecting
like coefficients gives
\begin{multline}
TOA_{pert,pm} = -\frac{1}{c}\frac{G M_m M_p}{M_m + M_p} \int_0^t \int_0^{t'} \frac{ r_m^2}{r_p^4}\\
\times \left( - \frac{3}{8} (3\cos^2 I_m - 1) \sin I_p \cos (f_p + \varpi_p) \right.
\\
\left.+ \frac{3}{32} (1 + \cos I_m)^2 \sin I_p\cos(2f_m + 2\varpi_m - 3 f_p - 3 \varpi_p) \right.
\\
\left.- \frac{21}{32}(1 + \cos I_m)^2 \sin I_p \cos(2f_m + 2\varpi_m - f_p - \varpi_p)  \right.
\\
\left.- \frac{3}{4} \sin I_m (1 - \cos I_m) \cos I_p\sin(2f_m + 2\omega_m - f_p  - \varpi_p + \Omega_m)
\right.
\\
\left.
- \frac{3}{4} \sin I_m (1 - \cos I_m) \cos I_p \sin(2f_m + 2\omega_m  + f_p + \varpi_p -\Omega_m)\right.
\\
\left.- \frac{15}{16} \sin^2 I_m \sin I_p \cos(f_p + \varpi_p - 2\Omega_m) \right.
\\
\left. - \frac{3}{16} \sin^2 I_m  \sin I_p \cos(3 f_p + 3 \varpi_p - 2\Omega_m)\right.
\\
\left. - \frac{9}{16} \sin^2 I_m \sin I_p \cos(2f_m + 2\omega_m + f_p  + \omega_p) \right.
\\
\left. - \frac{9}{16} \sin^2 I_m \sin I_p \cos(2f_m + 2\omega_m - f_p - \varpi_p) \right.
\\
\left.- \frac{15}{32} (1 - \cos I_m)^2 \sin I_p \cos(2f_m + 2\omega_m + f_p + \varpi_p -2\Omega_m) \right.
\\
\left.- \frac{3}{32} (1 - \cos I_m)^2 \sin I_p \cos(2f_m + 2\omega_m +3 f_p +3 \varpi_p -2\Omega_m) \right.
\\ 
\left. - \frac{3}{4} \sin(2I_m) \cos I_p  \sin(f_p +\omega_p - \Omega_m) \right) dt' dt.\label{PulM-Incl-GenI3}
\end{multline}
As both orbits are circular, $r_m = a_m$, $r_p = a_p$, $f_m = n_m t + f_m(0)$ and $f_p = n_p t + f_p(0)$.  Using these simplifications and performing the integrations gives
\begin{multline}
TOA_{pert,pm} = -\frac{1}{c}\frac{G M_m M_p}{M_m + M_p} \frac{ a_m^2}{a_p^4}\\
\times \left( \frac{3}{8} \frac{(3\cos^2 I_m - 1) \sin I_p}{n_p^2} \cos (f_p + \varpi_p) \right.
\\
\left.- \frac{3}{32} \frac{(1 + \cos I_m)^2 \sin I_p}{(2n_m - 3n_p)^2} \cos(2f_m + 2\varpi_m - 3 f_p - 3 \varpi_p) \right.
\\
\left.+ \frac{21}{32} \frac{(1 + \cos I_m)^2 \sin I_p}{(2n_m - n_p)^2} \cos(2f_m + 2\varpi_m - f_p - \varpi_p)  \right.
\\
\left.+ \frac{3}{4} \frac{\sin I_m (1 - \cos I_m) \cos I_p}{(2n_m - n_p)^2}\sin(2f_m + 2\omega_m - f_p  - \varpi_p + \Omega_m)
\right.
\\
\left.
+ \frac{3}{4} \frac{\sin I_m (1 - \cos I_m) \cos I_p}{(2n_m + n_p)^2} \sin(2f_m + 2\omega_m  + f_p + \varpi_p -\Omega_m)\right.
\\
\left.+ \frac{15}{16} \frac{\sin^2 I_m \sin I_p}{n_p^2} \cos(f_p + \varpi_p - 2\Omega_m) \right.
\\
\left. + \frac{3}{16} \frac{\sin^2 I_m  \sin I_p}{(3n_p)^2} \cos(3 f_p + 3 \varpi_p - 2\Omega_m)\right.
\\
\left. + \frac{9}{16} \frac{\sin^2 I_m \sin I_p}{(2n_m + n_p)^2} \cos(2f_m + 2\omega_m + f_p  + \omega_p) \right.
\\
\left. + \frac{9}{16} \frac{\sin^2 I_m \sin I_p}{(2n_m - n_p)^2} \cos(2f_m + 2\omega_m - f_p - \varpi_p) \right.
\\
\left.+ \frac{15}{32} \frac{(1 - \cos I_m)^2 \sin I_p}{(2n_m + n_p)^2} \cos(2f_m + 2\omega_m + f_p + \varpi_p -2\Omega_m) \right.
\\
\left.+ \frac{3}{32} \frac{(1 - \cos I_m)^2 \sin I_p}{(2n_m + 3n_p)^2} \cos(2f_m + 2\omega_m +3 f_p +3 \varpi_p -2\Omega_m) \right.
\\ 
\left. + \frac{3}{4} \frac{\sin(2I_m) \cos I_p}{n_p^2}  \sin(f_p +\omega_p - \Omega_m) \right). \label{PulM-Incl-GenI3}
\end{multline}
Consequently, higher values of inclination modify the amplitude of the sinusoids with frequency $2n_m - n_p$ and $2n_m - 3n_p$ (the frequencies corresponding to the circular coplanar case), with frequency $2n_m + n_p$ (the frequency corresponding to the slightly inclined case) and introduce sinusoids of frequency $3n_p$ and $2n_m + 3n_p$.  

\begin{figure}[tb]
\begin{center}
\includegraphics[width=.83\textwidth]{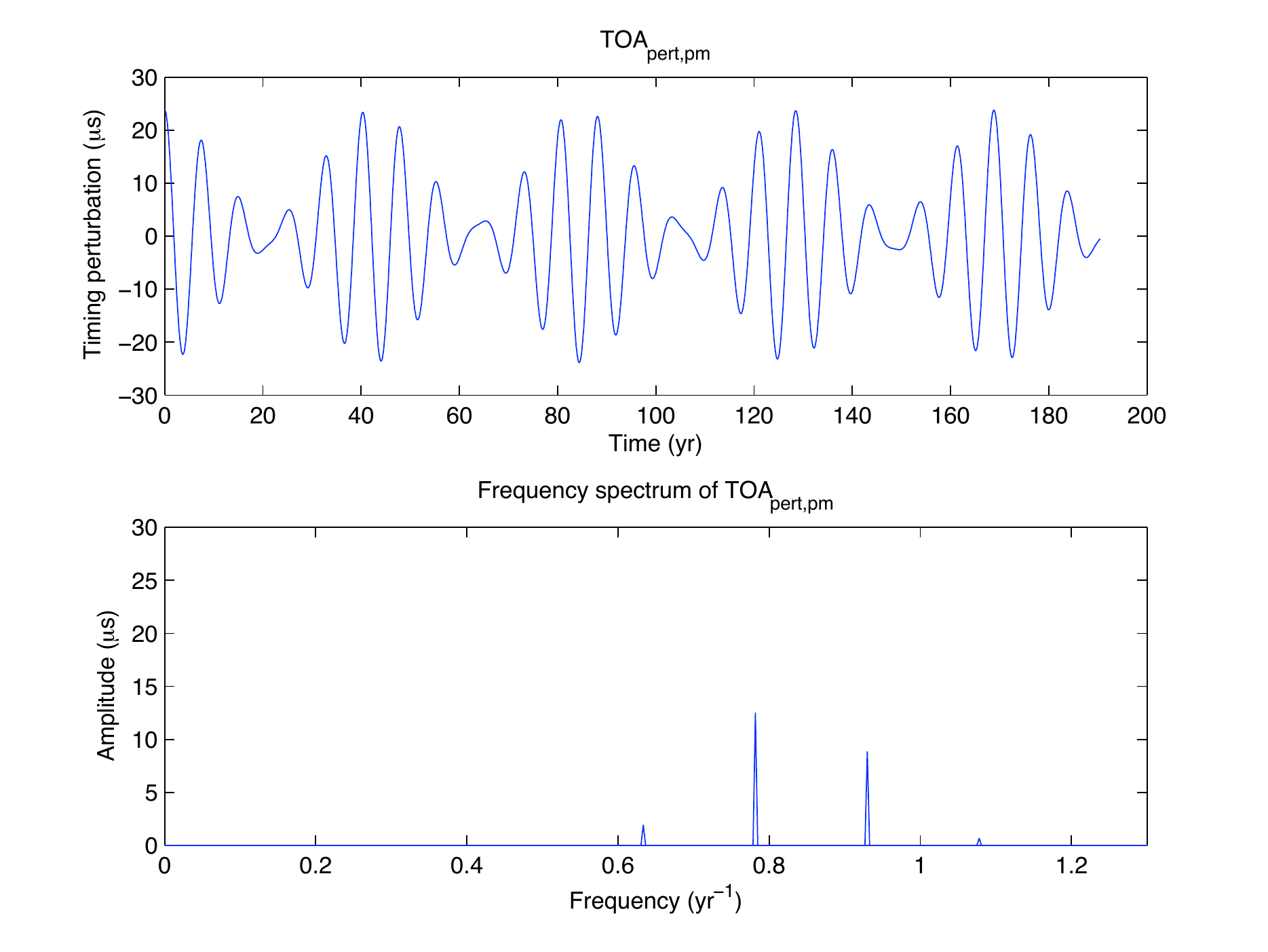}
\caption[Functional form and frequency composition of the time-of-arrival perturbation due to planet-moon binarity for the case where the planet and moon's orbits are circular and the orbital planes of the planet and moon are  perpendicular ($I_m = \pi/2$).]{Functional form and frequency composition of the time-of-arrival perturbation due to planet-moon binarity for the case where the planet and moon's orbits are circular and the orbital planes of the planet and moon are perpendicular ($I_m = \pi/2$).  These curves were calculated for the case of a PSR B1620-26 b analog, in particular, it was assumed that $M_p =2.3M_J$, $a_p =23$AU, $M_m = 0.2M_J$ and $a_m = 0.8$AU.} \label{TOAInclinedSig}
\end{center}
\end{figure}

In addition to adding new frequencies to the perturbation, high mutual inclination can also change the form of the perturbation (compare figures~\ref{TOACoplanarSig} and \ref{TOAInclinedSig}).  In particular, mutual inclination is capable of changing $TOA_{pert,pm}$ from something that looks like a sinusoid to something that looks like a beat (see figure~\ref{TOAInclinedSig}).  As can be seen, while mutual inclination does not substantially alter the maximum amplitude of the perturbation over a full planetary orbital period, it can reduce the amplitude for lengths of time much smaller than an orbital period, where we recall that the period of the envelope function is half a planetary orbital period.  Consequently, taking mutual inclination into account is very important for placing limits on moons of pulsar planets such PSR~B1620-26~b which have orbital periods much longer than length of time over which they have been observed.  Now that the effect of mutual inclination on the time-of-arrival perturbation has been investigated, we move on to investigate the effect of eccentricity, in particular, the effect of eccentricity in the moon's orbit.

\section{Slightly eccentric moon orbits}

While most moons are expected to form on circular orbits (see chapter~\ref{Intro_Moons_Const}), captured moons begin with very  elliptical orbits.  In addition, even for the case of a moon on an initially circular orbit, subsequent orbital evolution can increase the orbit's eccentricity \citep[e.g.][]{Hut1981}.  Consequently, it is of interest to investigate the case where the moon's orbit is eccentric.  In this section the case of slightly eccentric moon orbits will be investigated by considering the expression for the time-of-arrival perturbation correct to first order in $e_m$.  This case was selected as first, it indicates the types of effects eccentricity in the moon orbit can have on the perturbation and second as it is substantially simpler than the general case.

For the case where the planet and moon's orbits are coplanar, but the moon's orbit is eccentric, $\theta_p = \pi/2$, $\psi_p = f_p + \omega_p + \Omega_p$, $\theta_m = \pi/2$ and $\psi_m = f_m + \omega_m + \Omega_m$.  However, as the moon's orbit is eccentric, terms involving $r_m$ and $f_m$ are no longer simple functions of time.  Using these expressions, equation~\eqref{PulM-GovEq-GovEq4} becomes
\begin{multline}
TOA_{pert,pm} = -\frac{1}{c} \frac{G M_m M_p}{M_m + M_p} \int_0^t \int_0^{t'} \left( \sum_{m=-2,2}^2 \frac{(2-m)!}{(2+m)!} \left[\frac{r_m^2}{a_m^2} e^{i m f_m}\right]\right.
\\
\left. \times e^{i(m \varpi_m -i m f_p - m \varpi_p)}  \frac{a_m^2}{r_p^4} \left(P_2^m(0)\right)^2\left[-3 \mathbf{e}_{r_p} -im \mathbf{e}_{\psi_p} \right] \right)\cdot \mathbf{n} dt''
dt',\label{PulM-Ecc-Eq2}
\end{multline}
where the terms corresponding to the moon's orbit have been grouped into one factor using square brackets.  Applying the expansion presented in equation~\eqref{PulM_Form_EccFTMoon} to the term in square brackets, where we note that the Fourier coefficients $s^{(2m)}_n(e_m)$ are given to order $e_m^2$ in appendix~\ref{App_Ecc_Fun}, gives 
\begin{multline}
TOA_{pert,pm} = - \frac{1}{c} \frac{G M_m M_p}{M_m + M_p}  \int_0^t \int_0^{t'} \left(\sum_{m=-2,2}^{2} \sum_{n=-\infty}^{\infty} \frac{(2-m)!}{(2+m)!}
P^{m}_{2}(0)^2
\frac{a_m^2}{r_p^4}\right.\\
\left. s^{(2m)}_n(e_m)e^{inM_m(t)} \times e^{i(m \varpi_m - m f_p - m \varpi_p)} \left[-3
\mathbf{e}_{r_p} -im \mathbf{e}_{\psi_p} \right] \right)\cdot \mathbf{n}
dt' dt.\label{PulM-Ecc-Inn-Eq3}
\end{multline}
This equation describes the time-of-arrival perturbation for the case of eccentric moon orbits. 

In the case of low eccentricity, the terms of order $e_m^2$ and above
can be neglected.  Truncating the expansions in table~\ref{App-Ecc-Inn-sTab} to order $e_m$ gives
\begin{align}
s^{(22)}_{2}(e_m) &= 1,\\
s^{(22)}_1(e_m) &= -3e_m,\\
s^{(22)}_3(e_m) &= e_m,\\
s^{(20)}_0(e_m) &= 1,\\
s^{(20)}_1(e_m) &= -e_m.
\end{align}
Only including the above terms in equation~\eqref{PulM-Ecc-Inn-Eq3}, noting that $s^{(lm)}_n =
s^{(l-m)*}_{-n}$, and combining the complex exponentials into sinusoids gives
\begin{multline}
TOA_{pert,pm} = - \frac{1}{c} \frac{G M_m M_p}{M_m + M_p}  \int_0^t \int_0^{t'} \frac{a_m^2}{r_p^4} \left(- \frac{3}{4} \mathbf{e}_{r_p} \right.\\
\left. - \frac{9}{4} \cos(2M_m(t) + 2 \varpi_m - 2 f_p - 2 \varpi_p) \mathbf{e}_{r_p} \right.
\\
\left. - \frac{6}{4} \sin(2M_m(t) + 2 \varpi_m - 2 f_p - 2 \varpi_p) \mathbf{e}_{\psi_p} \right.
\\
\left. - \frac{9e_m}{4} \cos(3M_m(t) + 2 \varpi_m - 2 f_p - 2 \varpi_p) \mathbf{e}_{r_p}\right.
\\
\left. - \frac{6e_m}{4}  \sin(3M_m(t) + 2 \varpi_m - 2 f_p - 2 \varpi_p) \mathbf{e}_{\psi_p}  \right.
\\
\left. + \frac{27e_m}{4} \cos(M_m(t) + 2 \varpi_m - 2 f_p - 2 \varpi_p) \mathbf{e}_{r_p} \right.
\\
\left. + \frac{18e_m}{4} \sin(M_m(t) + 2 \varpi_m - 2 f_p - 2 \varpi_p) \mathbf{e}_{\psi_p} \right.
\\
\left. + \frac{3e_m}{2} \cos(M_m(t)) \mathbf{e}_{r_p}  \right)\cdot \mathbf{n}
dt' dt.\label{PulM-Ecc-Inn-Small3}
\end{multline}
Substituting in equation~\eqref{PulM-GovEq-ndef2} and expanding the trigonometric products gives
\begin{multline}
TOA_{pert,pm} = - \frac{1}{c} \frac{G M_m M_p}{M_m + M_p} \sin I_p \int_0^t \int_0^{t'} \frac{a_m^2}{r_p^4} \left(- \frac{3}{4} \cos (f_p + \varpi_p) \right.
\\
\left. - \frac{15}{8} \cos(2M_m(t) + 2 \varpi_m - f_p - \varpi_p) \right.
\\
\left. - \frac{3}{8} \cos(2M_m(t) + 2 \varpi_m - 3 f_p - 3 \varpi_p) \right.
\\
\left. -  \frac{15e_m}{8} \cos(3M_m(t) + 2 \varpi_m -  f_p -  \varpi_p) \right.
\\
\left. - \frac{3e_m}{8} \cos(3M_m(t) + 2 \varpi_m - 3 f_p - 3 \varpi_p) \right.
\\
\left. + \frac{45e_m}{8}  \cos(M_m(t) + 2 \varpi_m - f_p - \varpi_p) \right.
\\
\left. + \frac{9e_m}{8}  \sin(M_m(t) + 2 \varpi_m - 3 f_p - 3 \varpi_p) \right.
\\
\left. + \frac{3e_m}{4} \cos(M_m(t) - f_p - \varpi_p) \right.
\\
\left. + \frac{3e_m}{4} \cos(M_m(t) + f_p + \varpi_p)   \right)
dt' dt.\label{PulM-Ecc-Inn-Small4}
\end{multline}
Noting that $r_p = a_p$, $\frac{df_p}{dt} = n_p$ and $\frac{dM_m(t)}{dt} = n_m$, performing the integrals and simplifying then gives
\begin{multline}
TOA_{pert,pm} = -\frac{1}{c} \frac{G M_m M_p}{M_m + M_p} \sin I_p \frac{a_m^2}{a_p^4} \left(\frac{3}{4 n_p^2} \cos (f_p + \varpi_p) \right.
\\
\left. + \frac{15}{8(2n_m - n_p)^2} \cos(2M_m(t) + 2 \varpi_m - f_p - \varpi_p) \right.
\\ 
\left. + \frac{3}{8(2n_m - 3n_p)^2} \cos(2M_m(t) + 2 \varpi_m - 3 f_p - 3 \varpi_p) \right.
\\
\left.+ \frac{15e_m}{8(3n_m - n_p)^2} \cos(3M_m(t) + 2 \varpi_m -  f_p -  \varpi_p) \right.
\\
\left. + \frac{3e_m}{8(3n_m - 3n_p)^2} \cos(3M_m(t) + 2 \varpi_m - 3 f_p - 3 \varpi_p) \right.
\\
\left. - \frac{45e_m}{8(n_m - n_p)^2}  \cos(M_m(t) + 2 \varpi_m - f_p - \varpi_p) \right.
\\ \displaybreak
\left. - \frac{9e_m}{8(n_m - 3n_p)^2}  \sin(M_m(t) + 2 \varpi_m - 3 f_p - 3 \varpi_p) \right.
\\
\left. - \frac{3e_m}{4(n_m - n_p)^2} \cos(M_m(t) - f_p - \varpi_p) \right.
\\
\left. - \frac{3e_m}{4(n_m + n_p)^2} \cos(M_m(t) + f_p + \varpi_p)   \right).\label{PulM-Ecc-Inn-Small5}
\end{multline}

As seen by comparing equation~\eqref{PulM-Ecc-Inn-Small5} with equation~\eqref{PulM-CC-Eq5},
the perturbation in the case of circular coplanar orbits, the effect
of a small amount of eccentricity in the moon's orbit, is to split each of the frequencies corresponding to the circular coplanar case into three frequencies (see figure~\ref{TOAFreqSplit}).  The origin of this splitting can be seen by
considering that the only non-zero coefficients to
appear in the low eccentricity case correspond to terms with $n = m+1$, $m$ or $m-1$.  For orbits which are eccentric enough such that the assumption of low eccentricity is no longer valid, more coefficients will be relevant, and thus more than three frequencies will be produced by the splitting.  

\begin{figure}[tb]
\begin{center}
\includegraphics[width=.83\textwidth]{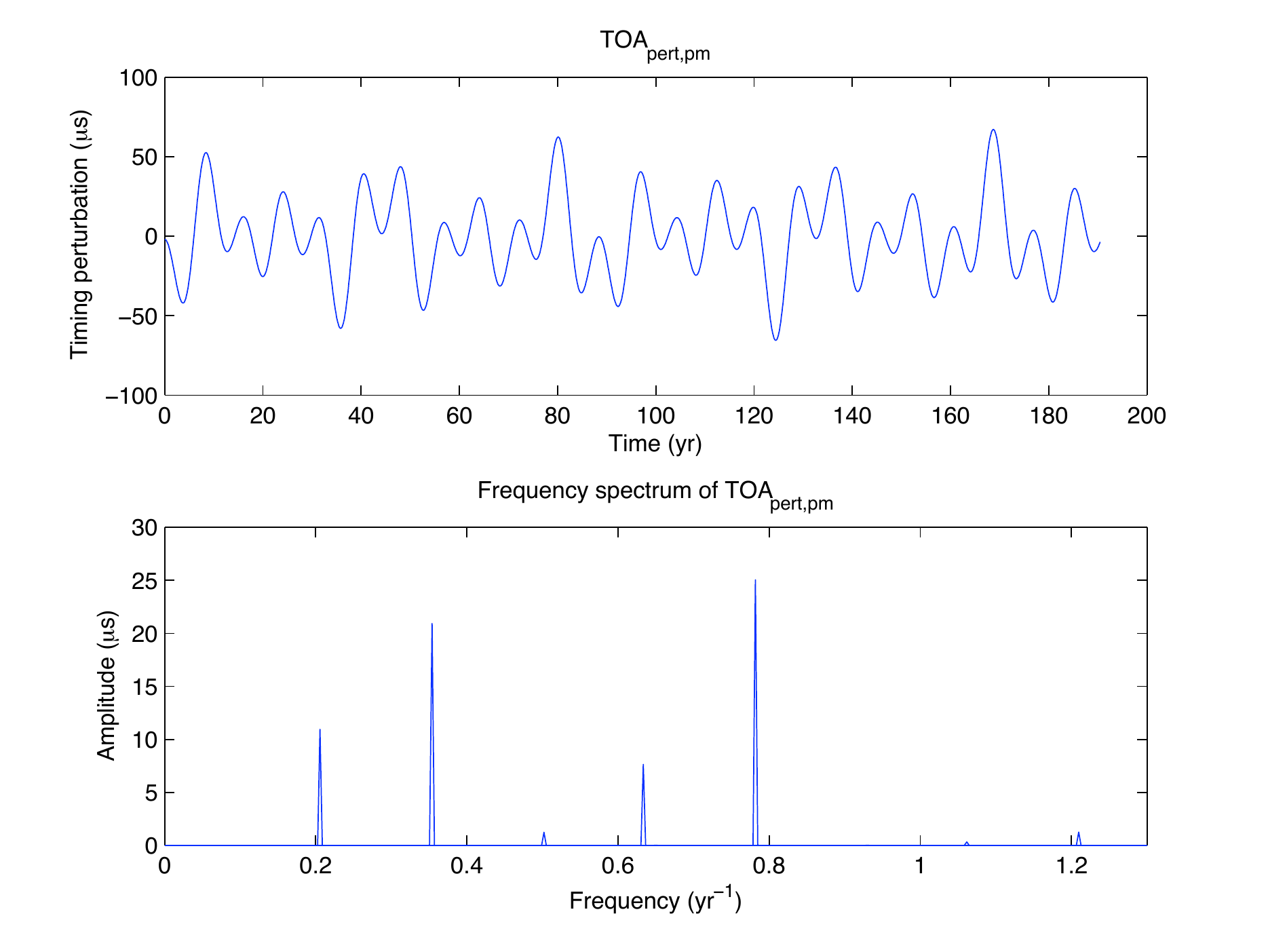}
\caption[Functional form and frequency composition of the time-of-arrival perturbation due to planet-moon binarity for the case of an eccentric moon orbit.]{Functional form and frequency composition of the time-of-arrival perturbation due to planet-moon binarity for the case where the planet and moon's orbits are coplanar, the planet's orbit is circular, and the moon's orbit is eccentric ($e_m = 0.05$).  These curves were calculated for the case of a PSR B1620-26 b analog, in particular, it was assumed that $M_p =2.3M_J$, $a_p =23$AU, $M_m = 0.2M_J$ and $a_m = 0.8$AU.} \label{TOAMoonEccSig}
\end{center}
\end{figure}

As for the case of mutually inclined orbits these additional frequencies act to modify the form of $TOA_{pert,pm}$ (see figure~\ref{TOAMoonEccSig}).  Again considering the beat analogy and attributing the envelope function to the motion of the planet-moon pair about the pulsar and the high frequency ``sinusoid" which the envelope function modifies to the motion of the planet and moon about their common barycenter, we expect that eccentricity in the moon's orbit would modify this ``sinusoid".  From figure~\ref{TOAMoonEccSig} we can see that this is truly the case.  We now move on to investigating the effect of eccentricity in the planet's orbit.

\subsection{Slightly eccentric planet orbits}

An investigation of the effect of eccentricity in the planet's orbit is scientifically interesting for two main reasons.  First, while planets in the Solar System have orbits that are well approximated by circles, many extra-solar planets do not, for example, nearly half of the planets presented in the extrasolar planet encyclopedia have eccentricities larger than 0.1.  In addition, for the particular case of pulsar planets, the two outer planets in the PSR~B1257+12 system have orbits with low, but non-zero eccentricities of 0.0186 and 0.0252, and it is thought that the orbit of PSR~B1620-26~b is eccentric \citep{Fordetal2000,Sigurdssonetal2004}.  Second, recall from section~\ref{paper_abstract}, that for the case of circular coplanar orbits, the amplitude of $TOA_{pert,pm}$ is approximately $\sin I_p[9(M_p M_m)/16(M_p + M_m)^2] [r_m/r_p]^5$ times the system crossing time $r_p/c$.  As a result of the dependance on $r_p$, it can be intuitively seen that variation in $r_p$ over a full planetary orbit is likely to have a marked effect on the size and structure of the perturbing signal.  As for the case of moon eccentricity, we investigate the effect of planetary eccentricity in the low eccentricity regime by deriving an expression for the time-of-arrival perturbation due to planet-moon binarity correct to first order in $e_p$.

Assuming the moon's orbit is circular and coplanar with that of the planet, equation~\eqref{PulM-GovEq-GovEq4} becomes
\begin{multline}
TOA_{pert,pm} =- \frac{1}{c} \frac{G M_m M_p}{M_m + M_p} \int_0^t \int_0^{t'} \left( \sum_{m=-2,2}^2 \frac{(2-m)!}{(2+m)!} \left[\frac{a_p^4}{r_p^4} e^{- i m f_p}\right]\right.
\\
\left. \times e^{i(m f_m + m \varpi_m - m \varpi_p)}  \frac{r_m^2}{a_p^4} \left(P_2^m(0)\right)^2\left[-3 \mathbf{e}_{r_p} -im \mathbf{e}_{\psi_p} \right] \right)\cdot \mathbf{n} dt''
dt',\label{PulM-Ecc-Eq2}
\end{multline}
where
\begin{equation}
\mathbf{n} = \sin I_p \cos (f_p + \varpi_p) \mathbf{e}_{r_p}
 - \cos I_p \mathbf{e}_{\theta_p} -\sin I_p \sin (f_p + \varpi_p)
 \mathbf{e}_{\phi_p}.
\end{equation}\label{PulM-Ecc-Out-ndef}
It can be seen that in the case of eccentric planet orbits some extra
work must be done before progress can be made.  This is because
$f_p$ is no longer a linear function of time.  Consequently, the
$\cos(f_p + \varpi_p)$ and $\sin (f_p + \varpi_p)$ terms within
$\mathbf{n}$ must also be included in the Fourier expansion. For the
$\cos(f_p + \varpi_p)$ term we have
\begin{equation}
\frac{a_p^4}{r_p^4}e^{-imf_p}\cos(f_p + \varpi_p) =
e^{i\varpi_p}\frac{1}{2}\frac{a_p^4}{r_p^4}e^{-i(m-1)f_p} +
e^{-i\varpi_p}\frac{1}{2}\frac{a_p^4}{r_p^4}e^{-i(m+1)f_p},
\label{PulM-Ecc-Out-cos_prep}
\end{equation}
which can be written as 
\begin{multline}
\frac{a_p^4}{r_p^4}e^{-imf_p}\cos(f_p + \varpi_p) = e^{i\varpi_p}\frac{1}{2}\sum_{n=-\infty}^\infty
F_n^{(3,m-1)}(e_p)e^{-inM_p(t)} \\
+ e^{-i\varpi_p}\frac{1}{2}\sum_{n=-\infty}^\infty
 F_n^{(3,m+1)}(e_p)e^{-inM_p(t)}. \label{PulM-Ecc-Out-cos}
\end{multline}
Similarly for the $\sin(f_p + \varpi_p)$ term we have that
\begin{multline}
\frac{a_p^4}{r_p^4}e^{-imf_p}\sin(f_p + \varpi_p) = e^{i\varpi_p} \frac{1}{2i} \sum_{n=0}^\infty F_n^{(3,m-1)}(e_p)e^{-in M_p(t)} \\
- e^{i\varpi_p} \frac{1}{2i}\sum_{n=0}^\infty
 F_n^{(3,m+1)}(e_p)e^{-i n M_p(t)},\label{PulM-Ecc-Out-sin}
\end{multline}
where it is now written in terms of $F^{(3,m-1)}$ and $F^{(3,m+1)}. $\footnote{Both
$F^{(3,m-1)}$ and $F^{(3,m+1)}$ can correspond to parent spherical
harmonics where the absolute value of ``$m$'' is larger than $l$. As
$Y_l^{m} \equiv 0$ for $|m|>l$ (see equation~\eqref{Int-Rev-PlmDef}) the term
$Y_l^{m}F^{(l,m)}$ in equation~\eqref{PulM-GovEq-GovEq4} is equal to zero independent
of the value of $F^{(l,m)}$. As $F^{(l,m)}$ are Fourier
coefficients of functions of the form
$(a_p^{l+1}/r_p^{l+1})e^{-im f_p}$, there is no physical limit on
the values of $l$ and $m$. So, while coefficients of the form
$F^{(3,4)}$ and $F^{(3,-4)}$ have meaning, they are not
naturally occurring in the expansion of the disturbing function as
usually they would be premultiplied by zero.}

Using equations~\eqref{PulM-Ecc-Out-cos} and \eqref{PulM-Ecc-Out-sin},
equation~\eqref{PulM-Ecc-Eq2} can now be written as
\begin{multline}
TOA_{pert,pm} = - \frac{1}{c} \frac{G M_m M_p}{M_m + M_p} \sin I_p \int_0^t \int_0^{t'} \left( \sum_{m=-2,2}^2 \frac{(2-m)!}{(2+m)!} \right.
\\
\left. e^{- im \varpi_p} \frac{r_m^2}{a_p^4} e^{i(m f_m + m \varpi_m)}
\left(P_2^m(0)\right)^2 \left[\frac{-3 + m}{2}
e^{i\varpi_p}\sum_{n=-\infty}^\infty F_n^{(3,m-1)}(e_p)e^{-inM_p(t)} \right.\right.\\
\left.\left. - \frac{3 + m}{2}e^{-i\varpi_p}\sum_{n=-\infty}^\infty
 F_n^{(3,m+1)}(e_p)e^{-inM_p(t)}\right] \right) dt' dt.\label{PulM-Ecc-Out-Eq3}
\end{multline}
This is the governing equation for eccentric planet orbits.  In the limit of low eccentricity, terms
of order $e_p^2$ and above can be safely neglected.  The only terms
in table \ref{App-Ecc-Out-FTab} which are non-zero once the $e_p^2$ terms have
been neglected are:
\begin{align}
F_2^{(33)}(e_p) &= -e_p, \\
F_3^{(33)}(e_p) &= 1, \\
F_4^{(33)}(e_p) &= 5e_p, \\
F_2^{(32)}(e_p) &= 1, \\
F_3^{(32)}(e_p) &= 4e_p, \\
F_0^{(31)}(e_p) &= e_p, \\
F_1^{(31)}(e_p) &= 1, \\
F_2^{(31)}(e_p) &= 2e_p, \\
F_0^{(30)}(e_p) &= 1, \\
F_1^{(30)}(e_p) &= 2e_p. 
\end{align}

Expanding equation~\eqref{PulM-Ecc-Out-Eq3}, only retaining these terms and combining them into sinusoids gives
\begin{multline}
TOA_{pert,pm} = -\frac{1}{c} \frac{G M_m M_p \sin I_p}{M_m + M_p} \int_0^t \int_0^{t'}  \frac{r_m^2}{a_p^4} \left(- \frac{3}{4} \cos(M_p(t) + \varpi_p)  \right. 
\\
\left.
- \frac{3}{8} \cos(2 f_m + 2 \varpi_m -M_p(t) -  \varpi_p)  - \frac{15}{8} \cos(2 f_m +2 \varpi_m  -3M_p(t) - 3 \omega_p)
\right. 
\\
\left.
- \frac{3e_p}{8} \cos(2 f_m + 2 \varpi_m -  \varpi_p) - \frac{9e_p}{8} \cos(2 f_m + 2 \varpi_m -2M_p(t) -  \varpi_p) \right. \\
\left. + \frac{15e_p}{8} \cos(2 f_m + \varpi_m - 2M_p(t) - 3 \varpi_p) - \frac{9e_p}{4} \cos(2M_p(t) + \omega_p)\right.
\\
\left. - \frac{75e_p}{8} \cos(2 f_m + 2 \varpi_m - 4M_p(t) - 3 \varpi_p)  - \frac{3e_p}{4} \cos(\varpi_p)  \right) dt' dt.\label{PulM-Ecc-Out-Eq5}
\end{multline}
Noting that $r_m = a_m$ and that $f_m = n_m t + f_m(0)$ and $M_p(t) = n_p t + M_p(0)$, and performing the double integration gives
\begin{multline}
TOA_{pert,pm} = -\frac{1}{c} \frac{G M_m M_p \sin I_p}{M_m + M_p}  \frac{r_m^2}{a_p^4} \left( \frac{3}{4n_p^2} \cos(M_p(t) + \varpi_p) \right. 
\\
\left.+ \frac{15}{8(2n_m - 3n_p)^2} \cos(2 f_m +2 \varpi_m  -3M_p(t) - 3 \omega_p) \right.
\\
\left.
+ \frac{3}{8(2n_m - n_p)^2} \cos(2 f_m + 2 \varpi_m -M_p(t) -  \varpi_p) \right. 
\\
\left.+ \frac{9e_p}{8(2n_m - 2n_p)^2} \cos(2 f_m + 2 \varpi_m -2M_p(t) -  \varpi_p) \right. \\
\left. - \frac{15e_p}{8(2n_m - 2n_p)^2} \cos(2 f_m + \varpi_m - 2M_p(t) - 3 \varpi_p) \right. 
\\
\left. + \frac{75e_p}{8(2n_m - 4n_p)^2} \cos(2 f_m + 2 \varpi_m - 4M_p(t) - 3 \varpi_p)  \right. 
\\
\left.+ \frac{9e_p}{4(2n_p)^2} \cos(2M_p(t) + \varpi_p)  + \frac{3e_p}{8(2n_m)^2} \cos(2 f_m + 2 \varpi_m -  \varpi_p) \right), \label{PulM-Ecc-Out-Eq5}
\end{multline}
where the constant term has been neglected.  As can be seen from equation~\eqref{PulM-Ecc-Out-Eq5}, eccentricity in the planet's orbit leads to the inclusion of terms with frequency $2n_m$, $2n_m - 2n_p$ and $2n_m - 4n_p$ in the perturbation.\footnote{Eccentricity in the planet's orbit also leads to the inclusion terms with frequency $n_p$ and $2n_p$ in the perturbation.  However, as the planet is on an eccentric orbit, $TOA_{pert,p}$ already contains terms with these frequencies. Consequently these terms will be undetectable as a separate signal.}

\begin{figure}[tb]
\begin{center}
\includegraphics[width=.83\textwidth]{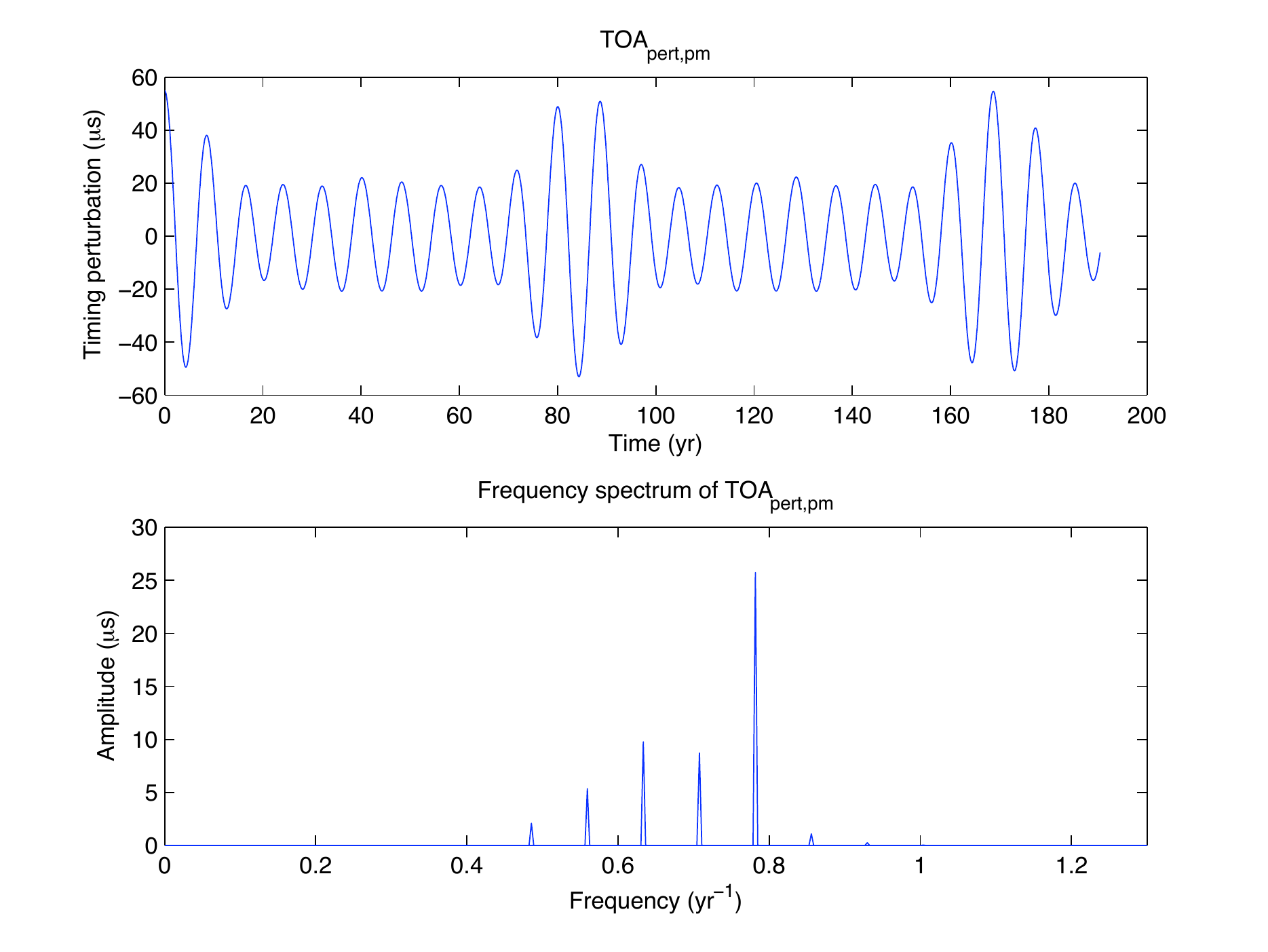}
\caption[Functional form and frequency composition of the time-of-arrival perturbation due to planet-moon binarity for the case of an eccentric planet orbit.]{Functional form and frequency composition of the time-of-arrival perturbation due to planet-moon binarity for the case where the planet and moon's orbits are coplanar, the moon's orbit is circular, and the planet's orbit is eccentric ($e_p = 0.1$).  These curves were calculated for the case of a PSR B1620-26 b analog, in particular, it was assumed that $M_p =2.3M_J$, $a_p =23$AU, $M_m = 0.2M_J$ and $a_m = 0.8$AU.} \label{TOAPlanetEccSig}
\end{center}
\end{figure}

\begin{figure}[tbp]
\begin{center}
\includegraphics[height=4.8in,width=4.8in]{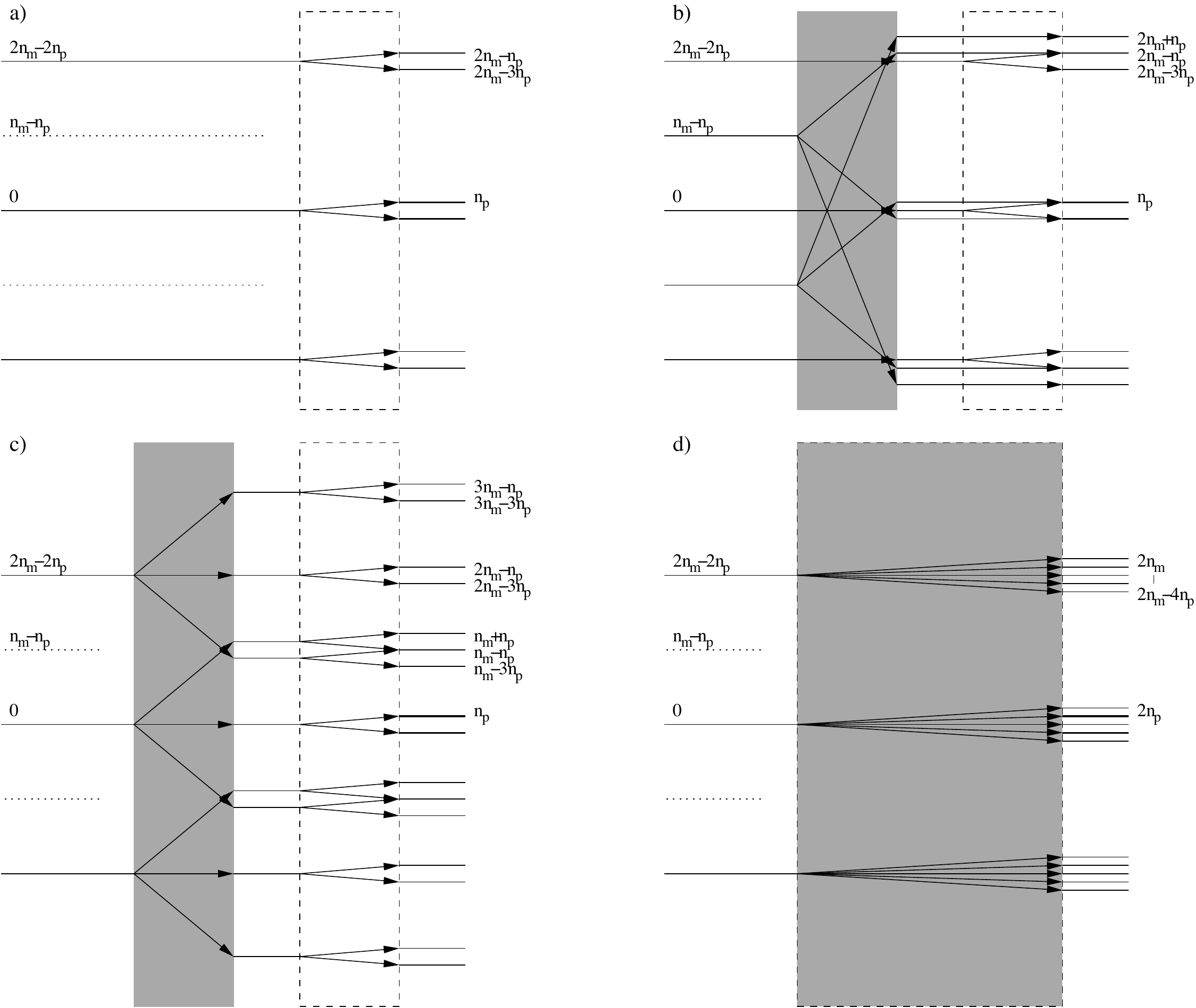}
\caption[Diagram showing the effect on the time-of-arrival perturbation frequencies due to constant viewing angle and symmetry breaking processes such as mutual inclination and eccentricity for a) circular coplanar orbits b) orbits with low mutual inclination c) moon orbits low eccentricity and d) planet orbits with low  eccentricity]{Diagram showing the effect on the time-of-arrival perturbation frequencies due to constant viewing angle and symmetry breaking processes such as mutual inclination and eccentricity for a) circular coplanar orbits b) orbits with low mutual inclination c) moon orbits with low eccentricity and d) planet orbits with low  eccentricity.  The splitting induced by inclination or eccentricity is shown in the gray box, while the splitting due to constant viewing angle is shown in the dashed box. In the case of an elliptical planet orbit, the splitting from both of these sources are inseparable and are consequently shown in a grey box with a dashed border.  The initial and resulting frequencies are labeled, except in the case where neighbouring frequencies differ by $n_p$. In this case, the lower and upper limits are given, separated by a vertical line. Finally, the fundamental frequency of the zeroth order time-of-arrival
signal is denoted by a bold line.}\label{TOAFreqSplit}
\end{center}
\end{figure}

The effect of these extra frequencies is to modify the shape of the perturbation.  In particular eccentricity in the planet's orbit modulates the envelope function over a planetary period (see figure~\ref{TOAPlanetEccSig}).  This result is physically sensible in that $\mathbf{r}_p$ is periodic over a planetary orbital period.  Recalling that for the circular-coplanar case, the timing perturbation due to planet-moon binarity was proportional to $1/r_p^4$, we would expect $TOA_{pert,pm}$ to be large at periastron (when $r_p$ is small), and small at apastron (when $r_p$ is large). Noticing that the orbit shown in figure~\ref{TOAPlanetEccSig} is at periastron at $t = 0$, this is exactly what is obtained.  Such an understanding is practically useful as it indicates that for the case of a pulsar planet on an eccentric orbit, observations aimed at detecting if it has a moon should be scheduled during or near periastron where it is expected that the perturbation is largest.

\section{Conclusion}

Expressions for the timing perturbation due to planet-moon binarity have been derived using a three-body formalism developed by my PhD supervisor, Dr. Rosemary Mardling.  Using this formalism, the cases where the planet and moon's orbit were circular and coplanar, circular and mutually inclined, coplanar with an eccentric moon orbit and coplanar with an eccentric planet orbit were investigated.  For the case of circular coplanar orbits, the expressions derived using this more general method exactly matched those produced in chapter~\ref{Pulsar_Paper}.  Then, building on this analysis, the cases of mutually inclined planet and moon orbits and slightly eccentric planet and moon orbits were investigated.  For the case of mutually inclined orbits, it was found that slight misalignment resulted in additional terms with frequency $2n_m - n_p$ and $2n_m + n_p$, and with amplitude proportional to the degree of the misalignment being added to the base circular coplanar signal form, while larger values of mutual inclination altered the perturbation from something which looked like a sinusoid to something that looked like a beat function (compare figures~\ref{TOACoplanarSig} and \ref{TOAInclinedSig}).  In addition,  for the case of slightly eccentric orbits, it was found that, as for the case of slight mutual inclination, the expression for the perturbation was given by the sum of the perturbation for the case of circular coplanar orbits and a perturbation term proportional to the relevant eccentricity.  In particular, for the case of slightly eccentric moon orbits the perturbation term contained sinusoids of frequency $3n_m - n_p$, $3n_m - 3n_p$, $2n_m - n_p$, $2n_m - 3n_p$, $n_m + n_p$, $n_m - n_p$ and $n_m - 3n_p$, while for the case of slightly eccentric planet orbits the perturbation term contained sinusoids of frequency $2n_m$, $2n_m - n_p$, $2n_m - 2n_p$, $2n_m - 3n_p$ and $2n_m - 4n_p$.  From a more qualitative perspective these additional frequencies resulted in a change in the shape of the high frequency oscillations in $TOA_{pert,pm}$ for the case of eccentricity in the moon's orbit, and a modulation of the envelope function of $TOA_{pert,pm}$ over an planetary orbital period for the case of eccentricity in the planet's orbit.  These results are summarised in figure~\ref{TOAFreqSplit}.  In line with the motivation of this chapter, these expressions, along with the transparent way in which they were derived, allow an understanding of the physical origin of form of the perturbation signal as a function of the orbital elements of the moon's orbit and allow this method to be extended to include pulsar planets on inclined or eccentric orbits.  Now that moon detection around pulsar planets has been investigated, used to place limits on moons of a real pulsar planet and extended, we shift our focus to the second moon detection technique analysed in this thesis, photometric transit timing.

\cleardoublepage \pagestyle{empty} 
\part{Detecting Moons of Transiting Planets} \label{TransitPart}
\pagestyle{plain} 
\chapter{Introduction}\label{Trans_Intro}

\section{Introduction}

Before investigating which moons of transiting planets are detectable using the photometric transit timing technique (TTV$_p$), it would be instructive to summarize the mathematics and main results associated with the transit technique, detection of moons of transiting planets in general, and the TTV$_p$ technique in particular.  This will be done in three main stages.  First, pertinent results from the transiting planet literature will be summarised, in particular, the way in which the transit duration and the shape of the transit light curve depend on the system parameters will be discussed and the corresponding formulae for these quantities introduced.  Second, the set of methods proposed in the literature to find moons of transiting planets using the transit light curves will be reviewed.  Finally, the TTV$_p$ technique will be focussed on, with the aim of summarising the results presented in previously published work, defining where my work fits in that context, and also providing a more mathematically useful description of $\Delta \tau$, the TTV$_p$ test statistic.  We begin with a discussion of the transit technique.

\section{Description of the transit technique}\label{Trans_Intro_Transtech}

\begin{figure}[tb]
\begin{center}
\includegraphics[height=2.83in,width=3.25in]{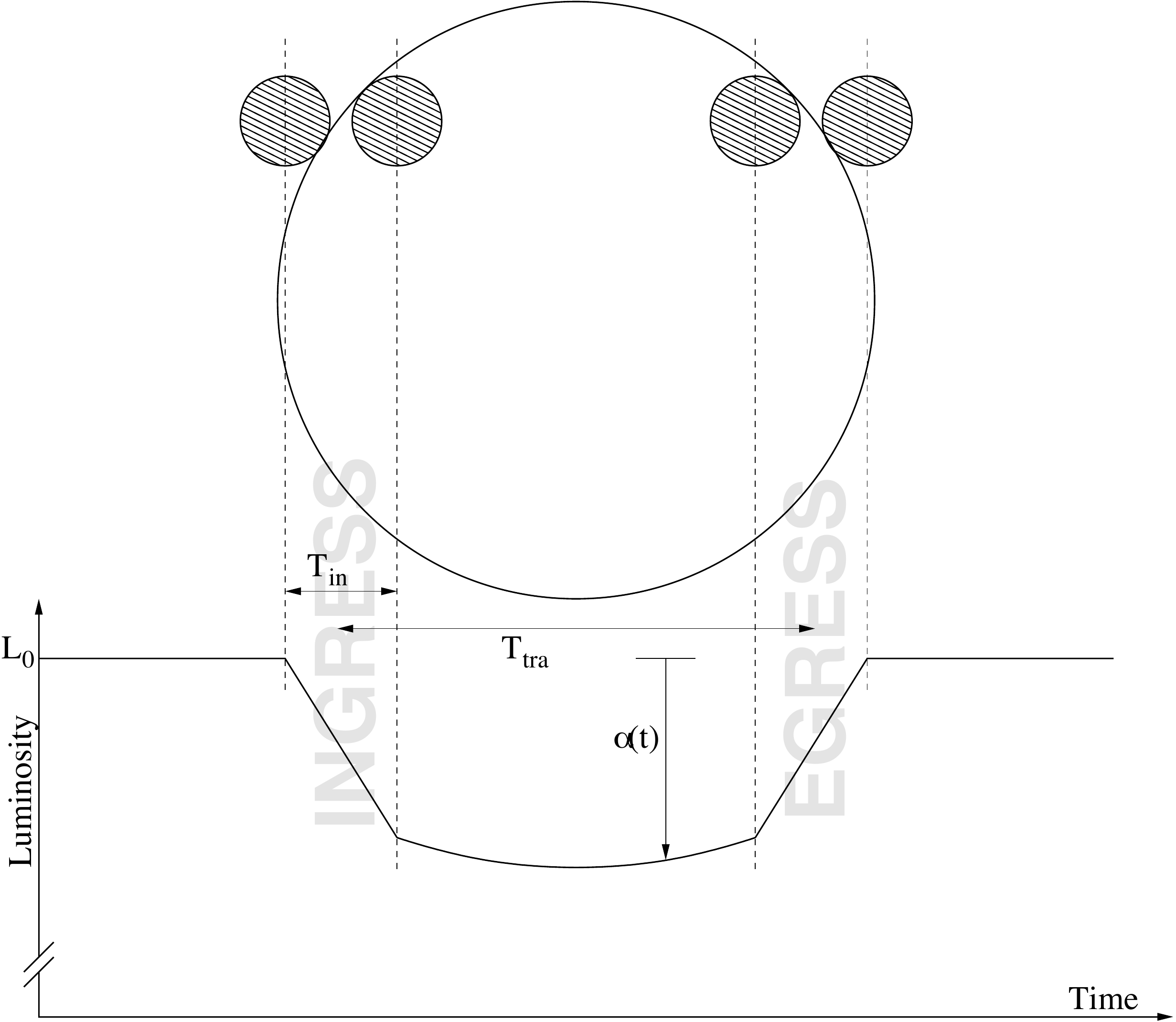}
\caption[Schematic diagram of a transit light curve for the case of a lone planet.]{Diagram showing the different portions of the transit light curve.  The four shaded circles show the planet's position across the face of the star at the beginning and end of ingress, and the beginning and end of egress.  As the position of the planet along the chord of the star is a linear function of time, it can be used as a proxy for time.  Consequently the position of the planet and the value of the light curve resulting from that position are linked by dashed lines.}
\label{PlanetTransitSchematic2}
\end{center}
\end{figure}

The transit technique is a planetary detection technique where the presence of the planet is deduced by the dip in received intensity of its host star as the planet passes in front of it.  This technique was first proposed by \citet{Struve1952}, who used images taken using photographic plates to search for transiting planets.  While the technique was periodically revisited \citep[e.g.][]{Rosenblatt1971,Boruckietal1984, Boruckietal1985}, it wasn't until the advent of CCD technology which made wide-field surveys plausible, that the disadvantages of this technique started to be outweighed by its advantages \citep{Kjeldsenetal1992}, and could start to produce results.

This method's main disadvantage is that in order to be detected, the orbit of the planet must be such that it passes in front of its parent star.  As the probability of a given planet transiting is $\propto a_p^{-1}$ \citep{Boruckietal1984, Barnes2007}, where $a_p$ is the semi-major axis of the planet's orbit,  it can be seen that a given planet is more likely to transit, and thus to be discovered, if its semi-major axis is small than if its semi-major is large.  However, as the transit technique uses the star's total intensity, as opposed to the radial velocity technique, where the light must be split up to give high resolution, high signal-to-noise spectra, it can be seen that fainter stars can be targeted.  Consequently, the disadvantages of this technique as a result of selection effects can be partially rectified as many hundreds or even thousands of stars can be monitored at once.

With the advent of wide field CCD surveys, this planet detection technique has come of age with over 100 planets discovered using this technique.\footnote{See, for example, http://exoplanet.eu/catalogue.php.}  Not only does the transit technique allow for planetary detection, it also allows for the measurement of planetary radius, orbital inclination as well as observables such as orbital orientation relative to the star's spin axis \citep[e.g.][]{Quelozetal2000,Naritaetal2007}, planetary oblateness \citep{Huietal2002,Barnesetal2003}, atmospheric composition \citep[e.g.][]{Charbonneauetal2002,VidalMadjaretal2004,Richardsonetal2007, Tinettietal2007} and even the presence or absence of moons \citep{Sartorettietal1999}.

As with the concept, the mathematical techniques required to analyse these light curves are well established in the literature \citep{Gimenez2006}.  This is because the light curves from transiting planets are related to the light curves of eclipsing binary stars, which have been extensively studied \citep[e.g.][]{Kopal1979}.  An example transit light curve is shown in figure~\ref{PlanetTransitSchematic2}.  The duration and the shape of this light curve  depend on the shape and inclination of the planet's orbit, the relative sizes of the planet and the star and the degree of limb darkening exhibited by the star.  To aide in further derivations, the effect of these variables on the duration and shape of a transit will be summarised.

To begin the investigation on transit duration, we recall from chapter~\ref{Intro_Moons_Note}, that for this thesis, the transit duration is defined as the time between the center of ingress ($t_{in,p}$) and the center of egress ($t_{eg,p}$).  That is,
\begin{equation}
T_{tra} = t_{eg,p} - t_{in,p}. \label{transit_intro_dur_durdef}
\end{equation}
As the position of a given planet is generally written in terms of the true anomaly, $f_p$, as opposed to the time, $t$, an expression will be constructed for the transit duration in terms of $f_p$.  Following \citet{Kipping2008b} and using Kepler's second law (see equations~(2.10) and (2.26) of \citet{Murrayetal1999}), we have that
\begin{equation}
dt = \frac{r_p^2}{n_pa_p^2 \sqrt{1-e_p^2}} df_p, \label{transit_intro_dur_kepeq1}
\end{equation}
where $t$ is time, $r_p$ is the distance between the planet and the star, $a_p$ is the semi-major axis of the orbit, $e_p$ is the eccentricity of the orbit and $f_p$ is the true anomaly.  When integrated between $t_{in,p}$ and $t_{eg,p}$, this equation gives
\begin{align}
\int_{t_{in,p}}^{t_{eg,p}} dt  &= \int_{f_{in,p}}^{f_{eg,p}} \frac{r_p^2}{n_pa_p^2 \sqrt{1-e_p^2}} df_p,\\
  t_{eg,p} - t_{in,p} &= \int_{f_{in,p}}^{f_{eg,p}} \frac{r_p^2}{n_pa_p^2 \sqrt{1-e_p^2}} df_p,
\end{align}
thus
\begin{equation}
T_{tra} = \int_{f_{in,p}}^{f_{eg,p}} \frac{r_p^2}{n_pa_p^2 \sqrt(1-e_p^2)} df_p.\label{transit_intro_dur_durdefint}
\end{equation}
So, in order to determine the transit duration, $f_{in,p}$ and $f_{eg,p}$, the values of $f_p$ corresponding to the middle of ingress and middle of egress, are required.

\begin{figure}[tb]
\begin{center}
\includegraphics[height=0.85in,width=4.7in]{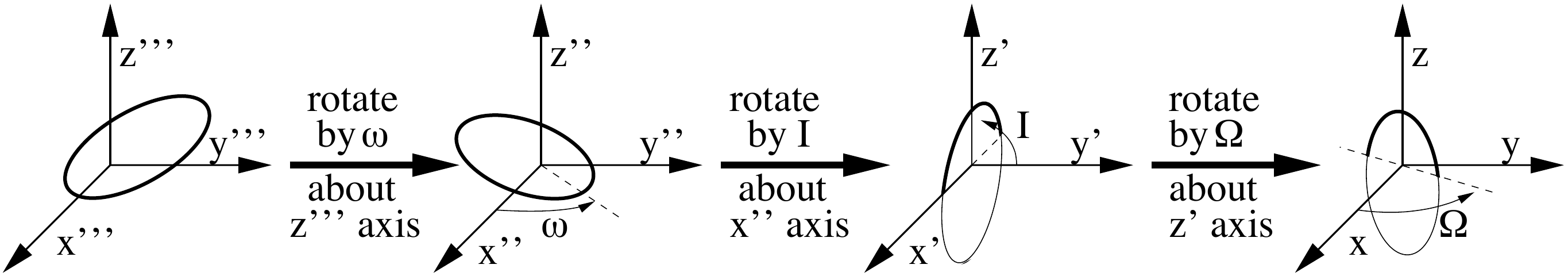}
\caption[Schematic diagram of the method used to describe an orbit of orientation given by the Euler angles $\omega$, $I$ and $\Omega$.]{Schematic diagram of the method used to describe an orbit of arbitrary orientation.  The $x'''$, $y'''$, $z'''$ coordinate frame is fixed to the orbit, such that the pericenter points along the positive $x'''$ axis, and the orbit lies in the $x'''$-$y'''$ plane.  This orbit is then rotated sequentially through the three Euler angles, $\omega$, $I$ and $\Omega$ to give a description of the orbit in the $x$, $y$, $z$ coordinate frame.  This unprimed coordinate system describes the orbit with respect to an inertial reference frame.  In particular, for this application, the $x$ and $y$ coordinate axes lie in the plane of the sky, while the $z$-axis point along the line-of-sight.}
\label{Transit_Intro_CoordFrame}
\end{center}
\end{figure}

To obtain these values we begin by considering a keplerian planet orbit given by 
\begin{equation}
r_p(t) = \frac{a_p(1-e_p^2)}{1+e_p\cos f_p(t)},\label{transit_intro_dur_rdef}
\end{equation}
where $r_p$ is the distance between the planet and the star, $a_p$ is the semi-major axis of the orbit, $e_p$ is the eccentricity of the orbit, and $f_p$ is the true anomaly.  Rewriting this expression using Cartesian coordinates and rotating this orbit by the three Euler angles (see figure~\ref{Transit_Intro_CoordFrame}) gives
\begin{align}
x_p &= r_p\cos \Omega_p \cos(f_p + \omega_p) - r_p\sin \Omega_p \cos I_p \sin(f_p + \omega_p),\label{transit_intro_dur_xpdef} \\
y_p &= r_p\sin \Omega_p \cos(f_p + \omega_p) + r_p\cos \Omega_p \cos I_p \sin(f_p + \omega_p), \label{transit_intro_dur_ypdef}
\end{align}
where the three Euler angles, $\Omega_p$, $\omega_p$ and $I_p$, represent the longitude of the ascending node, the argument of periastron and the inclination, respectively.

Now, the center of transit ingress and the center of transit egress occur when the center of the silhouette of the planet and just touches the limb of the star, that is, when the center of the planet is $R_s$ from the center of the star, where $R_s$ is the radius of the star.  Mathematically this occurs when
\begin{equation}
R_s^2 = x_p^2 + y_p^2.\label{transit_intro_dur_edgedef}
\end{equation}
Substituting in equation~\eqref{transit_intro_dur_xpdef} and \eqref{transit_intro_dur_ypdef} for $x_p$ and $y_p$ and simplifying gives
\begin{equation}
R_s^2 = r_p^2[\cos^2(f_p + \omega_p) + \cos^2 I_p \sin^2(f_p + \omega_p) ].\label{transit_intro_dur_edgedefgen}
\end{equation}
This equation describes the values of $f_p$ corresponding to the beginning and end of the primary transit.\footnote{This equation may also describe the beginning and end of the secondary transit as well.}  In addition, we have that $f_p$ depends on $R_s$, $r_p$, $\omega_p$ and $I_p$, but not on $\Omega_p$, as it does not appear in this equation.  This is reasonable as altering $\Omega_p$ only alters the orientation of the path taken by the planet on the face of the star, and not the intensity along it.

Continuing, to determine the transit duration, equation~\eqref{transit_intro_dur_edgedefgen} needs to be solved in terms of $f_p$ for the times of ingress and egress.  As this is a high order equation in $f_p$, this is not trivial (see \citet{Kipping2008b} for a derivation of general expressions for $T_{tra}$).  For this thesis, the full general expression is not required, so, we will look at three specific cases, partially to highlight the physics and partially as expressions for these quantities will be required in later chapters.  These cases correspond to cases where the planet's orbit is circular and aligned to the line-of-sight, is circular, but slightly inclined to the line-of-sight and eccentric and aligned to the line-of-sight.  Expressions for $f_{in,p}$ and $f_{eg,p}$ and thus $T_{tra}$ will be derived in turn for these three cases.

For the case where the planet's orbit is circular and aligned to the line-of-sight, we have that $r_p = a_p$ and $I_p = \pi/2$.  Consequently, equation~\eqref{transit_intro_dur_edgedefgen} becomes 
\begin{equation}
R_s^2 = a_p^2[\cos^2(f_p + \omega_p)], \label{transit_intro_dur_cc_edgedef}
\end{equation}
and thus
\begin{equation}
R_s = \pm a_p \cos (f_p + \omega_p).\label{transit_intro_dur_cc_edgedef2}
\end{equation}
where, assuming $n_p$, the mean motion, is positive, the plus and minus represent the egress and ingress respectively.  Rearranging equation~\eqref{transit_intro_dur_cc_edgedef2} to give an explicit expression for $f_p$ gives
\begin{align}
f_p &= \cos^{-1}\left(\pm \frac{R_s}{a_p} \right) - \omega_p, \\
 &=  \frac{\pi}{2} \pm (\sin^{-1}\left( \frac{R_s}{a_p} \right) - \omega_p, \label{transit_intro_dur_cc_edgedef3}
\end{align}
where we have used the identity that $\sin (A \pm \pi/2) = \pm \cos (A)$ and we have kept the solutions relevant to the primary transit.

For the case of the planets of interest, we have that $R_s/a_p \ll 1$ as only distant planets are likely to keep their moons \citep{Barnesetal2002}.  To first order in $R_s/a_p$, equation~\eqref{transit_intro_dur_cc_edgedef3} becomes
\begin{equation}
f_p = \frac{\pi}{2} \pm \frac{R_s}{a_p} - \omega_p.\label{transit_intro_dur_cc_fdef}
\end{equation}

Substituting this expression into equation~\eqref{transit_intro_dur_durdefint}, setting $r_p = a_p$ and $e_p = 0$, we have that
\begin{align}
T_{tra} &= \int_{\frac{\pi}{2} - \frac{R_s }{a_p} - \omega_p}^{\frac{\pi}{2} + \frac{R_s}{a_p} - \omega_p} \frac{a_p^2}{n_pa_p^2} df_p,\\
 &= \frac{1}{n_p} \int_{\frac{\pi}{2} - \frac{R_s}{a_p} - \omega_p}^{\frac{\pi}{2} + \frac{R_s}{a_p} - \omega_p} df_p\\
 &= \frac{2 R_s}{a_p n_p}.\label{transit_intro_dur_cc_Ddef}
\end{align}
This is exactly the result that one would expect.  Consider the numerator and the denominator of equation~\eqref{transit_intro_dur_cc_Ddef}.  The numerator is exactly the distance that the planet must travel to cross from one side of the star to the other, while the denominator is the velocity of a planet on a circular orbit.

For the case where the planet's orbit is still circular, but slightly inclined, we have that $r_p = a_p$, $I_p \ne \pi/2$ and $e_p = 0$.  Thus equation~\eqref{transit_intro_dur_edgedefgen} becomes
\begin{equation}
R_s^2 = a_p^2[\cos^2(f_p + \omega_p) + \cos^2 I_p \sin^2(f_p + \omega_p) ].\label{transit_intro_dur_inc_edgedef}
\end{equation}
Using the identity that $\sin^2 A + \cos^2 A = 1$ we have that
\begin{equation}
R_s^2 - a_p^2 \cos^2 I_p  = a_p^2 \sin^2 I_p \cos^2(f_p + \omega_p),\label{transit_intro_dur_inc_edgedef2}
\end{equation}
which simplifies to 
\begin{equation}
\sqrt{R_s^2 - a_p^2 \cos^2 I_p } = \pm a_p \sin I_p \cos(f_p + \omega_p).\label{transit_intro_dur_inc_edgedef3}
\end{equation}
Following the same method as used above, this can be written as
\begin{align}
f_p &= \cos^{-1}\left(\pm \frac{\sqrt{R_s^2 - a_p^2 \cos^2 I_p}}{a_p \sin I_p} \right) - \omega_p,\\
       & = \frac{\pi}{2} \pm \sin^{-1}\left(\frac{\sqrt{R_s^2 - a_p^2 \cos^2 I_p}}{a_p \sin I_p} \right) - \omega_p.
\end{align}
Noting that $\cos I_p$ is of order $R_s/a_p$,\footnote{Recall that $\cos I_p = \delta_{min}/a_p$ where $\delta_{min}$ is the impact parameter.  In addition, in order for the planet to transit, we must have that $\delta_{min} < R_s + R_p \approx R_s$.  Thus $\cos I_p$ is of order $R_s/a_p$.} and again only retaining terms up to first order in $R_s/a_p$, we have that
\begin{equation}
f_p  = \frac{\pi}{2} \pm \sqrt{(R_s/a_p)^2 -  \cos^2 I_p}  - \omega_p. \label{transit_intro_dur_inc_fdef}
\end{equation}
Again conducting the integral we obtain 
\begin{equation}
T_{tra} = \frac{2\sqrt{R_s^2 - a_p^2 \cos^2 I_p}}{a_p n_p}.\label{transit_intro_dur_inc_Ddef}
\end{equation}
Again, this is what we would expect.  The distance travelled by the planet is exactly given by the numerator while the velocity is given by the denominator.  In addition, from equation~\eqref{transit_intro_dur_inc_Ddef} we can see that the reduction in the transit duration with the increase in $|I_p - \pi/2|$, results from the shorter length chord over which the planet transits, and not a change in velocity.

Finally, for the case where the orbit is aligned to the line-of-sight, but eccentric, we have that $I_p = \pi/2$ but that $r_p$ is given by equation~\eqref{transit_intro_dur_rdef}.  Consequently equation~\eqref{transit_intro_dur_edgedefgen} becomes 
\begin{equation}
R_s^2 = \left[\frac{a_p(1-e_p^2)}{1+e_p\cos f_p(t)}\right]^2 \cos^2(f_p + \omega_p),\label{transit_intro_dur_ecc_edgedef}
\end{equation}
which simplifies to
\begin{equation}
R_s = \pm \frac{a_p(1-e_p^2)}{1+e_p\cos f_p(t)} \cos(f_p + \omega_p).
\label{transit_intro_dur_ecc_edgedef2}
\end{equation}

As in the previous sections, we would like an expression for $f_p$ correct to first order in $R_s/a_p$, but in this case we will use a perturbation expansion to obtain it.  To simplify the expression we multiply both sides by $(1+e_p\cos f_p(t))$, and to ensure that the small term, $R_s/a_p$, is clearly identified we divide by $a_p$, giving
\begin{equation}
\frac{R_s}{a_p}(1+e_p\cos f_p(t)) = \pm (1-e_p^2) \cos(f_p + \omega_p).
\label{transit_intro_dur_ecc_edgedef3}
\end{equation}

We begin by writing $f_p$ as a perturbation expansion
\begin{equation}
f_p = f_{p,0} + \epsilon f_{p,1} + ...
\end{equation}
where the small parameter $\epsilon$ is equal to $R_s/a_p$.  Substituting into equation~\eqref{transit_intro_dur_ecc_edgedef3} and grouping terms of like orders we find that the zeroth and first order equations are
\begin{equation}
0 = \pm (1-e_p^2) \cos(f_{p,0} + \omega_p),
\label{transit_intro_dur_ecc_0thorder}
\end{equation}
and 
\begin{equation}
1+e_p\cos f_{p,0} = \pm (1-e_p^2) ( - f_{p,1} \sin(f_{p,0} + \omega_p)).
\label{transit_intro_dur_ecc_1storder}
\end{equation}
 
The zeroth order equation can be solved to give 
\begin{equation}
f_{p,0} = \frac{\pi}{2} - \omega_p.
\end{equation}
This result is unsurprising as it agrees with the expressions for $f_{p}$ for the two previous cases to zeroth order (see equations~\eqref{transit_intro_dur_cc_fdef} and \eqref{transit_intro_dur_inc_fdef}).  Substituting this into equation~\eqref{transit_intro_dur_ecc_1storder} gives
 \begin{equation}
f_{p,1}  = \pm\frac{1+e_p\sin \omega}{1-e_p^2}.
\end{equation}
Thus,
\begin{equation}
f_p = \frac{\pi}{2} \pm \frac{R_s}{a_p}\frac{1+e_p\sin \omega}{1-e_p^2} - \omega_p,
\label{transit_intro_dur_ecc_fdef}
\end{equation} 
to first order  in $R_s/a_p$.
 
Consider the equation for the transit duration,
\begin{align}
T_{tra} &= \int_{f_{p,0} - \epsilon f_{p,1}}^{f_{p,0} + \epsilon f_{p,1}} \frac{1}{n_pa_p^2 \sqrt{1-e_p^2}} \left[\frac{a_p(1-e_p^2)}{1+e_p\cos f_p(t)}\right]^2 df_p,\label{transit_intro_dur_ecc_Dint1}\\
&= \frac{1}{n_p \sqrt{1-e_p^2}} \int_{f_{p,0} - \epsilon f_{p,1}}^{f_{p,0} + \epsilon f_{p,1}} \left[\frac{1-e_p^2}{1+e_p\cos f_p(t)}\right]^2 df_p.\label{transit_intro_dur_ecc_Dint2}
\end{align}
Following \citet{Kipping2008b} (see equation~(A36)), we have that
 \begin{multline}
 \int \left[\frac{1-e_p^2}{1+e_p\cos f_p(t)}\right]^2 df_p = 2\sqrt{1 - e_p^2} \tan^{-1}\left( \sqrt{\frac{1-e_p}{1+e_p}} \tan \frac{f_p}{2} \right) \\- \frac{e_p(1 - e_p^2)\sin f_p}{1 + e_p \cos f_p}.\label{transit_intro_dur_ecc_Kipint}
 \end{multline}
 
Converting equation~\eqref{transit_intro_dur_ecc_Kipint} into a definite integral with integration limits $f_{p,0} - \epsilon f_{p,1}$ and $f_{p,0} + \epsilon f_{p,1}$, taking the Taylor expansion about $f_p = f_{p,0}$ and retaining terms of order $\epsilon$ gives
\begin{multline}
 \int_{f_{p,0} - \epsilon f_{p,1}}^{f_{p,0} + \epsilon f_{p,1}} \left[\frac{1-e_p^2}{1+e_p\cos f_p(t)}\right]^2 df_p = \\
 \left[(1 - e_p) \frac{\cos^2 \left(\tan^{-1}\left( \sqrt{\frac{1-e_p}{1+e_p}} \tan \frac{f_{p,0}}{2} \right) \right)}{\cos^2 \frac{f_{p,0}}{2}}\right. \\- \left.\frac{e_p(1 - e_p^2) (\cos f_{p,0} + e_p)}{(1 + e_p \cos f_{p,0})^2}\right] 2 \epsilon f_{p,1}. \label{transit_intro_dur_ecc_Kipint2}
\end{multline}

Comparing this with equation~\eqref{transit_intro_dur_ecc_Dint2}, and substituting in the expressions for $f_{p,0}$, $f_{p,1}$ and $\epsilon$ gives
\begin{equation}
T_{tra} =  \frac{2R_s}{n_p a_p \left(F(e_p, \omega_p)\right)^{-1}} ,\label{transit_intro_dur_ecc_Ddef}
\end{equation}
where $F(e_p, \omega_p)$ is given by
\begin{multline}
F(e_p, \omega_p) =  \left[ \frac{(1+e_p\sin \omega)}{(1 + e_p) \sqrt{1-e_p^2}} \frac{\cos^2 \left(\tan^{-1}\left( \sqrt{\frac{1-e_p}{1+e_p}} \tan \left(\frac{\pi}{4} - \frac{\omega_p}{2}\right) \right) \right)}{ \cos^2 \left(\frac{\pi}{4} - \frac{\omega_p}{2}\right)}\right. \\- \left.\frac{e_p (\sin \omega_p + e_p)}{\sqrt{1-e_p^2} (1 + e_p \sin \omega_p)}\right].\label{transit_intro_dur_ecc_Fdef}
\end{multline}
Again, we have from equation~\eqref{transit_intro_dur_ecc_Ddef} that the transit duration is dictated to by two factors.  Again the numerator represents the distance travelled by the planet while the denominator (along with $F(e_p, \omega_p)$) represents the velocity of the planet.  Consequently eccentricity modifies the transit duration by modifying the velocity of the host planet during transit.

So, as discussed, the transit duration is determined by two factors, the distance travelled across the face of the star (which is modified by the inclination) and the velocity at which it is travelled (which is modified by the eccentricity).  To give a feel for these values some example transit durations and transit velocities are shown in table~\ref{EgTransDurVel}.  In addition to determining the transit duration, the shape of the transit light curve is determined by the orbital and physical properties of the planet and star. 

\begin{figure}[tb]
\begin{center}
\includegraphics[height=3.2in,width=5in]{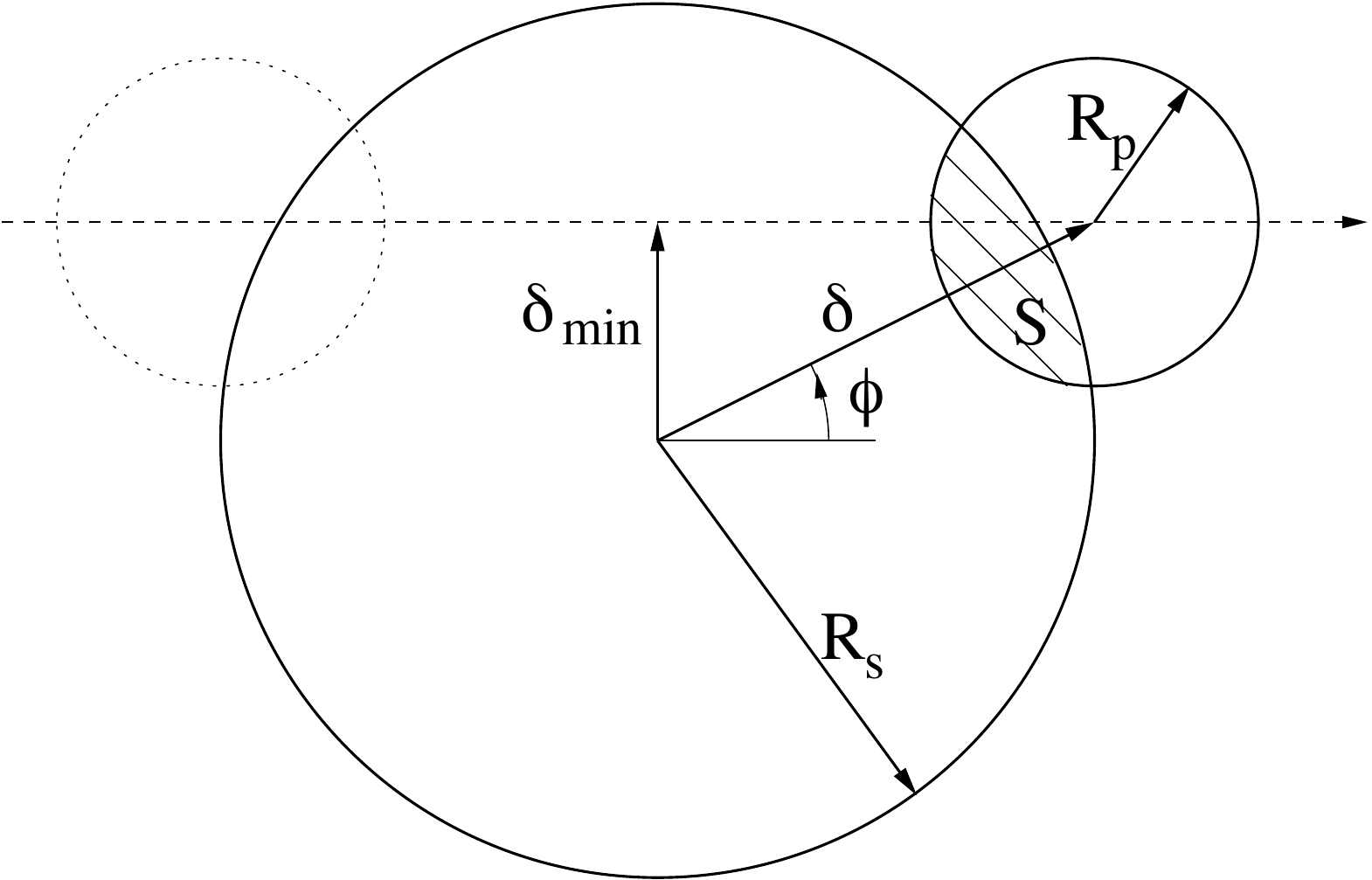}
\caption[Schematic diagram of the coordinates used to determine the shape of the transit light curve.]{Schematic diagram of the coordinates used to determine the shape of the transit light curve.  The path of the center of the planet across the star's face is indicated with a dashed line.  Also, one of the previous locations of the planet is shown using a dotted line.}
\label{PlanetTransitSchematic}
\end{center}
\end{figure}

\begin{table}[tb]
\begin{center}
  \begin{tabular}{llcll}
   \hline
 $a_p$ &  $e_p$  &  Orbit  			& $T_{tra}$ & $v_{tr}$   \\
 (AU)     &               &  Orientation                	& (hr) & (kms$^{-1}$)   \\
   \hline
   0.2  & 0     & --                       & 5.83 & 66.31 \\
           & 0.5 & P  & 3.36        & 114.85 \\
           & 0.5 & A & 10.09        & 38.28 \\
   0.3  & 0     & --  & 7.14 & 54.15 \\
           & 0.5 & P  & 4.12        & 93.79\\
           & 0.5 & A & 12.36       & 31.26 \\
   0.4  & 0     & --  & 8.24 & 46.89 \\
          & 0.5 & P  & 4.76 & 81.21 \\
           & 0.5 & A & 14.27 & 27.07 \\
    0.7  & 0     & --  & 10.90 & 35.45 \\
           & 0.5 & P  & 6.29 & 61.40 \\
           & 0.5 & A & 18.88 & 20.47 \\
      1  & 0     & --  & 13.03 & 29.66\\
           & 0.5 & P  & 7.52 & 51.37 \\
           & 0.5 & A & 22.56 & 17.12 \\
  \end{tabular}\\
 \caption[Example transit durations ($T_{tra}$) and mid-transit planetary velocities ($v_{tr}$) for a planet which transits the central chord of its Sun-like star, for a range of different values of $a_p$, $e_p$ and orbital orientation. ]{Example transit durations ($T_{tra}$) and mid-transit planetary velocities ($v_{tr}$) for a planet which transits the central chord of its Sun-like star, for a range of different values of $a_p$, $e_p$ and orbital orientation.  Note that the letters P and A correspond to a transit occurring at periastron and apastron respectively and the symbol ``--", corresponds to the case for a circular orbit.}
 \label{EgTransDurVel}
  \end{center}
 \end{table}

\begin{figure}[tb]
\begin{center}
\includegraphics[height=2.5in,width=5in]{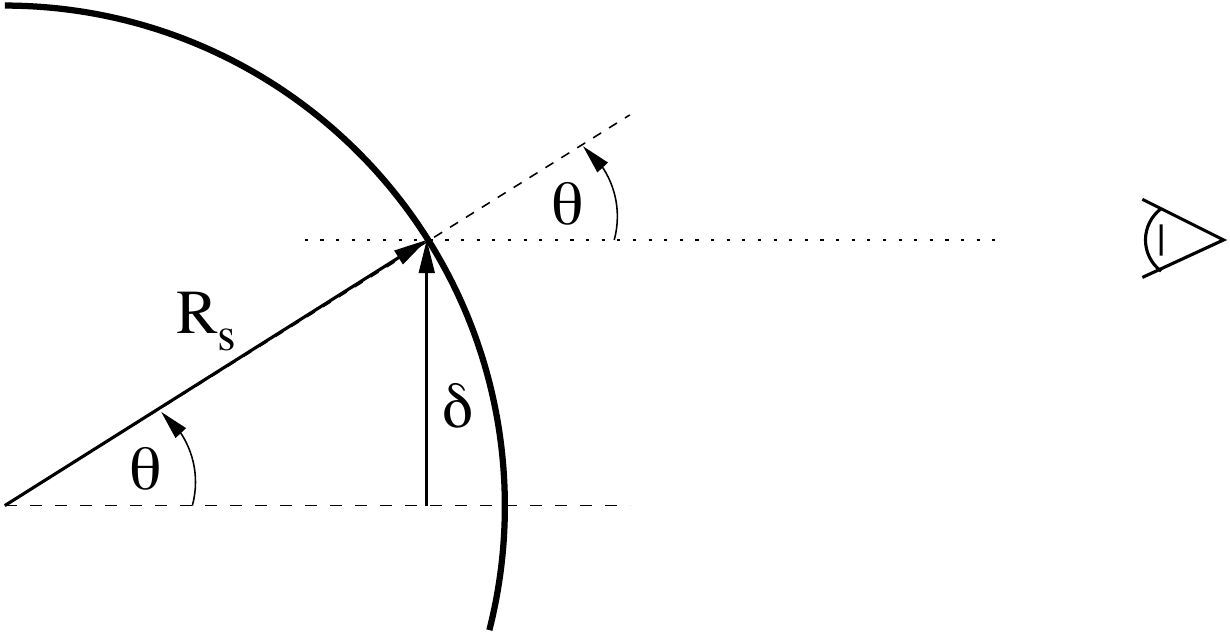}
\caption[Diagram showing the relationship between $\delta$ the distance between the center of the star's face and the position of interest and the angle $\theta$, the angle between the surface normal and the line-of-sight.]{Diagram showing the relationship between $\delta$ the distance between the center of the star's face and the position of interest and the angle $\theta$, the angle between the surface normal and the line-of-sight.  The surface of the star is shown using a thick line, the direction along the line-of-sight is shown using a dotted line, while lines showing the surface normal are dashed.}
\label{Thetadia}
\end{center}
\end{figure}

To determine the shape of a transit light curve produced by a given planet as it travels across a given star we consider the method of \citep{Gimenez2006}.  We can write the luminosity of the star, $L(t)$ as
\begin{equation}
L(t) = L_0 - \alpha_p(\delta(t)),\label{TraM-DescT-Ldef}
\end{equation}
where $L_0$ is the luminosity of the star out-of-transit and $\alpha_p$ is the amount of light occulted by the planet.  In addition, we note that $\alpha_p$ depends on $\delta$, the distance between the centers of the planet and star, but does not depend on $\phi$, the angular position of the planet on the face of the star (see figure~\ref{PlanetTransitSchematic} for definitions of $\delta$ and $\phi$).  Continuing, we have that $\alpha_p$ is defined as
\begin{equation}
\alpha_p(\delta(t)) = \int_S I(\mu)\mu dA,\label{TraM-DescT-alphadef}
\end{equation}
where $S$ is the region of the star which is occulted by the planet, $\mu$ is the cosine of $\theta$, the angle between the surface normal and the line-of-sight (see figure~\ref{Thetadia}), dA is an infinitesimally small area element and $I(\mu)$ is the intensity.  While both theoretical \citep[e.g.][]{Claret2000} and observational \citep[e.g.][]{Sing2010} constraints on $I(\mu)$ exist, as will be shown in chapter~\ref{Transit_Signal}, the size and form of the $TTV_p$ signal does not strongly depend on the form of $I(\mu)$.

Assuming that the velocity of the planet during transit can be considered to be constant, $\delta(t)$ can be written as
\begin{equation}
\delta(t) = \sqrt{\delta_{min}^2 + ((t - t_0)v_{tr})^2}.\label{TraM-DescT-deltadef}
\end{equation}
Consequently, the transit light curve is given by
\begin{equation}
L(t) = L_0 - \alpha\left(\sqrt{\delta_{min}^2 + ((t - t_{mid})v_{tr})^2}\right),\label{TraM-DescT-Ldef2}
\end{equation}
where $\delta_{min}$ is the distance of closest approach between the center of the star and the planet, $t_{mid}$ is the time at which this closest approach occurs and $v_{tr}$ is the projected velocity of the planet across the star's face during transit.  While equation~\eqref{TraM-DescT-Ldef2} can be written explicitly in terms of Jacobi polynomials \citep{Gimenez2006}, that level of detail will not be required for this analysis. 

As can be deduced from equation~\eqref{TraM-DescT-Ldef2}, and figure~\ref{PlanetTransitSchematic}, the light curve $L$ has a number of properties.  The first property, symmetry about $t_{mid}$, can be seen by noting that replacing $t - t_{mid}$ by $t_{mid} - t$ does not alter the equation.  This symmetry is a direct consequence of the intensity $I$, being a function of the angle between the line-of-sight and the surface normal only, and that the velocity of the planet during transit remains constant.  The second property is that the exact shape of the light curve depends on the chord it makes across the star and how fast it travels along the chord, which is determined by the planet's orbit, the relative sizes of the planet and star, and the mid-time of the transit.\footnote{See appendix \ref{App_Omega_Dep} for a proof that the shape of the light curve does not depend on $\Omega_p$.}  As a result, formally $L$ should be written as
\begin{equation}
L(t) = L(R_s,R_p,a_p,e_p,\omega_p,I_p,n_p,t_{mid};t),
\label{TraM-DescT-Ldeffull}
\end{equation}
where $R_s$ and $R_p$ are the radius of the star and planet, $a_p$, $e_p$, $\omega_p$, $I_p$ and $n_p$ are the semi-major axis, eccentricity, argument of periastron, inclination of the planet's orbit around the star and the mean motion, and $t_{mid}$ is the time at which the projected distance between the center of the planet and that of the star is smallest.

\section[Techniques to find transiting moons]{Extending the transit technique to find moons of transiting planets}

\begin{figure}[tb]
\begin{center}
\includegraphics[height=2.83in,width=3.25in]{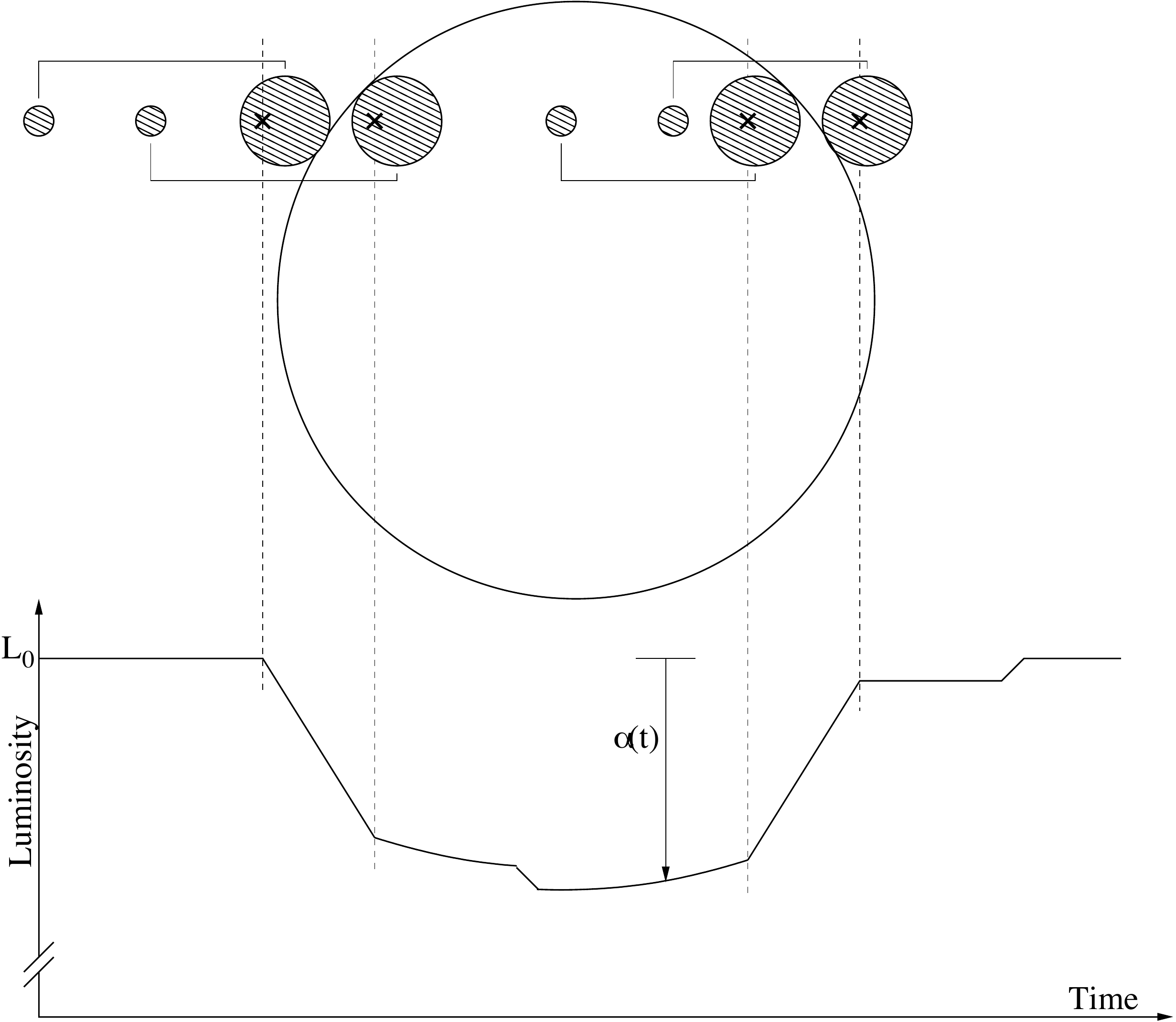}
\caption[Schematic diagram of a transit light curve for the case of a planet and moon.]{Diagram showing the different portions of the transit light curve for the case where both a planet and moon transit.  Four silhouettes of the planet and moon are shown, corresponding to the the beginning and end of planetary ingress, and the beginning and end of planetary egress.  Planet-moon pairs which correspond to a single silhouette are joined by a solid line, while the location of the planet-moon barycenter is indicated by cross.  As the position of the planet-moon barycenter is a linear function of time it can be used as a proxy for time.  Consequently the position of the barycenter and the value of the light curve resulting from that position are linked by dashed lines.  }
\label{PlanetTransitSchematic3}
\end{center}
\end{figure}

As discussed in chapter~\ref{Intro_Dect}, there are many ways to extend the transit technique to look for moons of transiting planets.   While this chapter focusses on the photometric transit timing technique, four methods have been proposed to extend the the transit technique to search for moons.  These methods are direct detection, barycentric transit timing, photometric transit timing and transit duration variation. Consequently, to provide a context for the following investigation, the rationale for each of these four methods will be briefly summarised using the transit light curve shown in figure~\ref{PlanetTransitSchematic3}, before concentrating our investigation on the photometric transit timing technique.

\subsection{Direct detection}

The process of direct detection involves searching the region of the light curve near the planetary transit for any extra dips due to putative moons.  For example, for the case of the light curve shown in figure~\ref{PlanetTransitSchematic3}, the additional dip caused by the moon can be seen translated to the right of the dip caused by the planetary transit.

\subsection{Barycentric transit timing}

The barycentric transit timing technique (TTV$_b$) involves searching for transit timing variations (TTV) where the time of transit is defined by the center of the planetary transit ($t_{mid,p}$).  Departures of consecutive transit times from strict periodicity, a result of  motion of the planet around the planet-moon barycenter, could indicate the presence of a moon. For example, for the case shown in figure~\ref{PlanetTransitSchematic3}, the mid-time of the planetary transit occurs earlier than would be expected due to the presence of the moon.

\subsection{Photometric transit timing}

The photometric transit timing (TTV$_p$) also involves searching for aperiodicity in transit times.  However in this case the times used are no longer the center of each planetary transit, but the mean time during the transit, weighted by the photon deficit (see equation~\eqref{TraM-TTV-taudef} for a definition).  This particular formulation is interesting as it is affected both by the extra dip due to the moon as well as by any lead or lag in the planet transit time caused by the presence of the moon.

\subsection{Transit duration variation}

Finally transit duration variation (TDV) a technique proposed by \citet{Kipping2009,Kipping2009b} also uses timing to search for moons of a given planet, but instead of measuring the ``mid-time" of the transits, the duration of the transit is used.  Instead of focussing on timing deviations due to the changing position of the planet about the planet-moon barycenter as the barycentric and photometric transit timing methods do, this method endeavours to measure perturbations to the planet's velocity across the face of the star due to the moon.  For the case shown in figure~\ref{PlanetTransitSchematic3}, the motion of the planet and moon about their common barycenter during transit result in a longer planetary transit duration than would have occurred had there been no moon.

As this chapter concentrates on the photometric transit timing method, it would be useful to expand upon the short description given above.  Consequently, the published results and limitations of this method will be discussed in greater detail.

\section{The $TTV_p$ method}\label{Trans_TTV}

\subsection{Introduction}

To provide a context for the work presented in this thesis, we begin by summarising the current state of the field with respect to $TTV_p$.  In particular, this will involve a statement of the definition of the $TTV_p$ test statistic followed by a summary of all previous work currently presented in the literature, with particular emphasis on the results and the gaps.  Informed by this summary, a more general definition of the $TTV_p$ test statistic will be proposed.  Using this definition, expressions for the timing perturbation caused by the moon (named $\Delta \tau$) and the error on this time (named $\epsilon_j$) are constructed.  We begin with the definition of the $TTV_p$ test statistic $\tau$.

\subsection{Literature definition of $\tau$, the $TTV_p$ test statistic}

\begin{figure}[tb]
\begin{center}
\includegraphics[width=.9\textwidth]{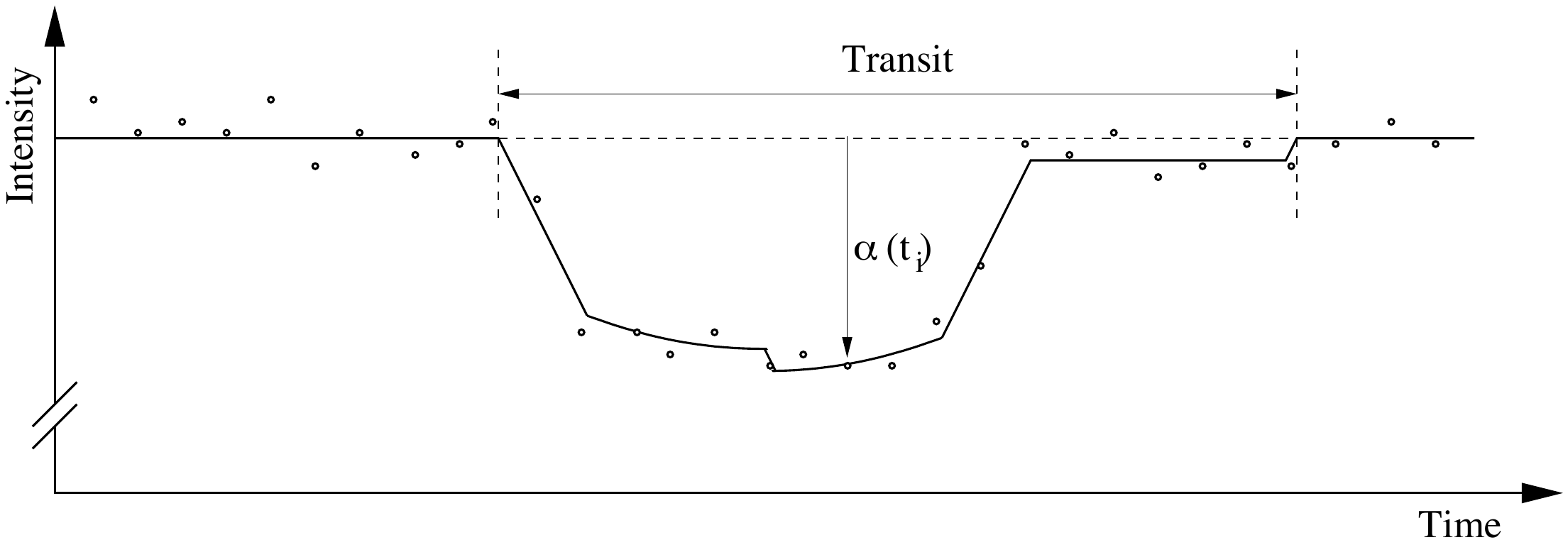}
\caption[Schematic of the transit light curve with sample experimental data points.]{Schematic of the transit light curve.  The experimental data points are represented by dots and the theoretical light curve is represented by a thick line.}
\label{SzaboSchematic}
\end{center}
\end{figure}

The photometric transit timing method (TTV$_p$) was proposed by \citet{Szaboetal2006}, and involves the statistic $\tau$, the first moment of the dip in the light curve, to search for timing perturbations due to moons.  Following \citep{Szaboetal2006}, $\tau$ is defined as
\begin{equation}
\tau = \frac{\sum_i t_i \alpha(t_i)}{\sum_i \alpha(t_i)},\label{TraM-TTV-taudef}
\end{equation}
where $t_i$ and $\alpha(t_i)$ are the times and observed absolute photon deficits for the $i^{th}$ exposure, and where the sum is carried out only over the region marked ``transit" in figure~\ref{SzaboSchematic}.  

\subsection{Summary and discussion of previously published $TTV_p$ results}

While some work has been presented on whether or not moons could be detected \citep{Szaboetal2006} and which physical properties of these moons could be determined \citep{Simonetal2007}, these analyses are by no means a complete description of the capabilities of this technique.  In particular, the work presented in this chapter extends these analyses in three important ways.

First, the analysis of \citet{Szaboetal2006} used two unrealistic assumptions.  The first assumption was that the ingress and egress times of the moon's transit is known, so that the sum could be carried out only over the planet and moon transits.  The second assumption was that the total unoccluded luminosity of the star was known for the in-transit period so that the difference between this unoccluded luminosity and the measured luminosity, could be used to calculate $\alpha(t_i)$.  Unfortunately, both these quantities will not be known for real transiting systems.  Consequently, the effect of relaxing these assumptions will be investigated within the context of the physical limits inherent to the system, for example, constraints to the time between planet and moon transit resulting from the requirement that the system is three-body stable.

Second, the way that the moon detectability depended on the physical parameters of the system is unknown.  The Monte Carlo analysis used by \citet{Szaboetal2006} involved producing 500 realisations for a range of randomly selected planet moon systems including both terrestrial and gas giant planet-moon pairs.  In addition to determining that realistic moons could be detected, \citet{Szaboetal2006} also used the results of their simulations to propose a number of factors which increased moon detectability.  These were:
\begin{itemize}
\item Shorter exposure time
\item Increased planet semi-major axis
\item Increased moon semi-major axis
\item Decreased relative photometric noise.
\end{itemize}
While this approach indicated that, realistic moons could be detected using this technique, and identified a number of factors which increased moon detectability, they did not give the functional dependence on these factors.  Consequently, for the case where these factors may be related, for example, for the case where the photometric noise is shot noise dominated, decreasing the exposure time increases the relative photometric noise, the result of altering a variable such as the exposure time is unknown.  Consequently, in this Part, approximate analytic relationships will be derived which relate the detectability of a given moon to physical parameters of the star, host planet and moon.

Finally, the approach adopted by \citet{Szaboetal2006} is of limited use to observers as it does not provide a simple way to determine the statistical significance of a detection or calculate a TTV$_p$ detection threshold.  To do this using the Monte Carlo method of \citet{Szaboetal2006} would involve constructing many realisations of the light curves and determining the percentage of these virtual moons which would have been detected.  As this would involve creating a set of models spanning the range of possible physical ($R_m$ and $M_m$) and orbital parameters ($a_m$, $e_m$, $I_m$, $\omega_m$, $\Omega_m$ and $f_m(0)$) of the putative moon, and requires a large number of realistic realisations of the photometric noise, this is not a trivial procedure.  Alternatively, in this Part, the issue of determining statistical significance, and generating thresholds, is addressed in three main ways.  First a method for determining the significance level of a detection in terms of measured variables will be presented.  Second, using this method along with analytic expressions describing the timing perturbation and the behaviour timing noise, expressions which approximately describe the detection threshold will be derived.  Third, as will be seen, for the cases where a Monte Carlo simulation must be run, the insight gained from the derivation of these approximate detection thresholds can be used to determine which variables are important and which variables are not.  By concentrating on these physically important variables, the computational load of calculating a threshold is dramatically reduced.

With these three aims it mind, it can be seen that the definition of $\tau$ needs to be investigated in two important ways before analytic expressions for moon detectability can be derived.  First, the definition of $\tau$ needs to be expanded such that knowledge of of the moon's position during transit and the unoccluded intensity of the star is not required.  Second, expressions for the mean value and error in $\tau$ for a given transit need to be determined in terms of the parameters of the system.

\subsection{Generalising the definition of $\tau$}\label{Transit_Intro_Taudef}

\begin{figure}[tb]
\begin{center}
\includegraphics[width=.9\textwidth]{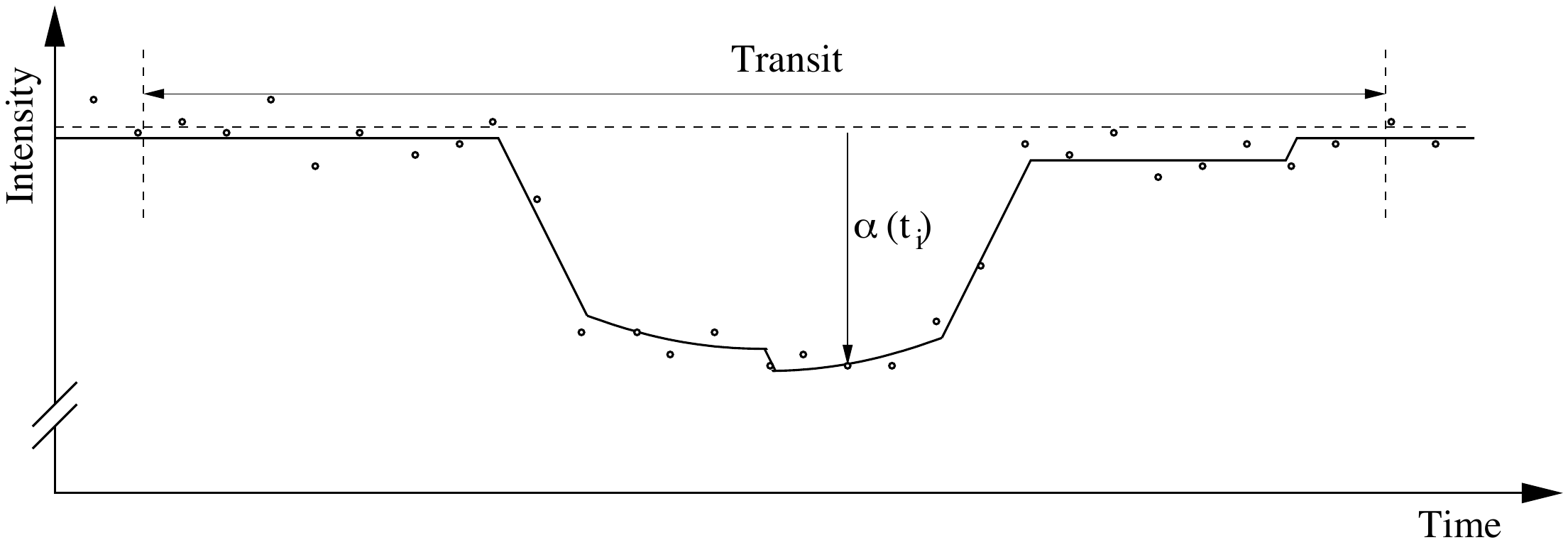}
\caption[Updated schematic of the transit light curve with sample experimental data points.]{Updated schematic of the transit light curve. The experimental 
data points are represented by dots and the theoretical light curve is represented 
by a thick line. A possible ``transit" region is also indicated. }
\label{LewisSchematic}
\end{center}
\end{figure}

In order to generalise the definition of $\tau$, the assumptions made by \citet{Szaboetal2006} need to be relaxed.  In both cases, this process can be informed by the physics of the system.

In order to relax the assumption that the unoccluded luminosity of the star is known, we are confronted with two main issues.  First, the unoccluded luminosity of the star is not necessarily constant across the length of the transit, for example, the luminosity of the star may drop as a starspot passes onto its face.  Second, the out-of-transit portions of the light curve cannot be used to form an estimated unoccluded luminosity as these portions may be contaminated by the transit of the moon.  Consequently it was decided to treat the unoccluded intensity of the star as if it were constant and absorb the variability of the stellar intensity into the photometric variability.  Fortunately, it can be shown that a small amount of error in the exact value of the unoccluded luminosity selected has little to no effect on the value or variability of any given $\tau$ value (see appendix~\ref{FluxErr_App}).  Thus, the practical effect of this error can be ignored.
 
The definition of $\tau$ given by equation~\eqref{TraM-TTV-taudef} assumes that the position of the moon during each planetary transit, and consequently the location of the dip caused by the moon in the light curve, are known before the moon is detected.  Unfortunately, this information is not known a priori, and thus the definition of $\tau$ must be modified.  Consequently, it was decided that instead of evaluating the sum over an a priori unknown interval, the sum would be evaluated over a region of length $T_{obs}$ centered on the planetary transit.  For simplicity, the region selected for this thesis was the smallest region always ensured to include the moon's transit (see figure~\ref{LewisSchematic}).  In particular, this region includes the planetary transit, along with a margin of length $a_m (1 + e_m)/v_{tr}$ either side of the planetary transit, where we note that $a_m (1 + e_m)$ is the distance between the planet and moon at apocenter and that $v_{tr}$ is the velocity at which the planet-moon barycenter transits the star.  While this assumption does require that something must be known about the moon before it is detected, it is useful for two reasons.  First, it is an improvement, in that this work only assumes knowledge of one variable, $a_m (1 + e_m)$, as opposed to three,  $a_m$, $e_m$ and $f_m(0) + \omega_m$.  Second, it results in a significant reduction in complexity when determining detection thresholds (this issue is further discussed in section~\ref{Trans_Thresholds_Method_Window}).  Finally, limits on the size of this margin (in particular limits relating to $a_m$) can be constrained using limits from our understanding of moon formation and orbital evolution.

 As discussed in chapter~\ref{Intro_Moons_Const}, limits can be placed on the properties of moons of extra-solar planets as a result of their formation and consequent evolution. In particular, limits on their semi-major axis (and consequently the time delay between their transit and that of their host planet). These limits come in two main varieties.  First, an educated guess on where moons are likely to be (based on the current understanding of moon formation).  Second, a more broad description of where moons could possibly exist without being rapidly destroyed or lost from the planet.  These two cases will be discussed in turn.
 
We begin by considering the places where moons are likely to be found.  As formation processes and evolution timescales differ for terrestrial and gas giant planets, these cases will be discussed separately.  

For the case of terrestrial planets, impact generated moons form very close to their host planet, and then, their orbits evolve outward.  Consequently, the mass of these moons is determined by the impact process and their final semi-major axis depends on the orbital evolution of the moon, with the orbit of more massive moons evolving more quickly than that of less massive moons (see section~\ref{Intro_Moons_Stab_Evol}).  Consequently, given an upper mass limit (0.04$M_p$), a model of the evolution process, an estimate of the Love number ($k_{2p}$), the $Q$-value ($Q_p$) and the age of the system, the semi-major axis of such a moon should be able to be predicted (e.g. using equation~\eqref{intro_limits_stab_tp_Mlim}).  Assuming that the physics used to calculate the orbital evolution is correct, this semi-major axis can be treated as an outer limit as other processes, such as tidal locking can halt orbital evolution.  For the case of an Earth-like\footnote{$k_{2p} = 0.299$, $Q_p = 12$, $T = 5$Gyr, $M_p = M_{\earth}$ and $R_p = R_{\earth}$.} host planet at 0.2AU and 1AU, the largest region allowed by the equation~\eqref{intro_limits_stab_tp_Mlim} constraint will begin and end 3.4 hours and 7.7 hours, before and after the planetary transit respectively.  

\begin{sidewaystable}[p] 
\centering       
\begin{tabular}{llllllll}    
\hline                  
Planet      	&  Moon 	& $R_m$        	&  $a_m$ 		&  $v_m$ 		& $e_m$ & $I_m$ & $T_m$\\
	 	&     		& ($10^{-2}R_{\sun}$)	& ($R_{\sun}$)	& (kms$^{-1}$) &                &  & (days) \\
\hline
Earth  	& Moon      & 0.250  	& 0.552 		& 1.02 &	0.0554 & 5.16 &  27.32\\
Jupiter  	& Io		& 0.262 		& 0.606		& 17.33 		&0.0041	& 0.036  & 1.77\\		
		& Europa	& 0.225 		& 0.964		& 13.74 	      	& 0.0094  & 0.466  & 3.55\\
		& Ganymede& 0.378	& 1.538		& 10.88           	&0.001      & 0.177  & 7.15\\
		& Callisto	& 0.345		& 2.705		& 8.21 	       	& 0.0074   & 0.192  & 16.69\\
Saturn 	& Mimas	& 0.029		& 0.267	& 14.32  	         & 0.0196   & 1.574 &  0.94\\
		& Enceladus& 0.036		& 0.342	& 12.63	& 0.0047 & 0.009 & 1.37\\
		& Tethys	& 0.076		& 0.423	&11.35	& 0.0001 & 1.091 & 1.89\\
		&Dione	& 0.081		& 0.542	& 10.03 	& 0.0022 & 0.028 & 2.74\\ 
		& Rhea	& 0.110	 & 0.757		&  8.48	& 0.0010 & 0.333 & 4.52\\
		& Titan	& 0.370	& 1.756	& 5.57	& 0.0288 & 0.312 & 15.95\\
		& Iapetus	& 0.103	& 5.116	&  3.26		& 0.0293 & 8.313 & 79.32\\
Uranus	& Miranda& 0.109	& 0.187		& 6.68 	& 0.0013 & 4.338 & 1.41\\
		& Ariel	& 0.034	&0.274 		& 5.52 & 0.0012   & 0.041 & 2.52\\
		& Umbriel	& 0.083	& 0.382	& 4.67 & 0.0039		& 0.128 & 4.14\\
		& Titania	& 0.084 	& 0.627	& 3.64 & 0.0011		& 0.079 & 8.71\\
		& Oberon	& 0.113	& 0.838	& 3.15 & 0.0014		& 0.068 & 13.46\\
Neptune	& Triton	& 0.194 	& 0.510	& 4.39 & 0.0000	& 156.865 & 5.88\\
\hline
\end{tabular} 
\caption[Physical and orbital data for the regular satellites presented in tables~\ref{TerMoonsTable}, \ref{JupMoonsTable}, \ref{SatMoonsTable}, \ref{UraMoonsTable} and \ref{NepMoonsTable}. ]{Physical and orbital data for the regular satellites presented in tables~\ref{TerMoonsTable}, \ref{JupMoonsTable}, \ref{SatMoonsTable}, \ref{UraMoonsTable} and \ref{NepMoonsTable}.  $v_m$, the orbital velocity was calculated assuming that the moon's orbit was circular.  In addition, note that $10^{-2}R_{\sun}$ is approximately an Earth radius.}  
\label{SSMoonsTable} 
\end{sidewaystable}

For the case of gas giant planets, large moons generally form in an extended region which is relatively close to their host planet.  Unfortunately, as the way in which the size of this region in which regular satellites form scales with planetary parameters is not fully understood,\footnote{For example, \citet{MosqueiraEstrada2003a,MosqueiraEstrada2003b} suggest that the size of this region scales with $R_H$ while \citet{Canupetal2006} suggest that it scales with $R_p$.} it is difficult to use it to place firm limits on the location of regular satellites.  However, using Solar System data we can suggest qualitative limits on where large moons of gas giant planets are likely to be found.  From table~\ref{SSMoonsTable}, we have that all the large regular satellites of gas giants lie within three solar radii of their host.  Assuming that the host-star is Sun-like, for the case of a host planet at 0.2AU and 1AU, the largest region allowed by this constraint will begin and end 8.7 hours and 19.5 hours, before and after the planetary transit respectively.

In addition to the window in which moons are likely to be found, there is also the window in which moons can possibly exist, that is, not be instantly destroyed or lost by tidal disruption, impact with the host planet or, by three body instability.  In particular, moons are destroyed if their orbits are too close to their host planet, and are three-body unstable if their orbits are too distant.  As we would like an upper limit to the size of the window that we need to search, we will consider three-body instability only.  From equation~\eqref{intro_limits_stab_RHDomingos} we have that, for a prograde moon to be orbitally stable, the semi-major axis of the moon must be less than approximately 0.5 Hill Radi, where $R_H$, the Hill radius is defined as
\begin{equation*}
R_H = a_p\left(\frac{M_p}{3M_s}\right)^{1/3}.\label{TraM-TTV-rH}
\end{equation*}
This corresponds to a constraint that the moon can lead or lag the planetary transit by a factor of $0.5R_H/2\pi a_p$ times the orbital period of the planet.  For example, using this approximation, for the case of a Earth-like planet orbiting a Sun-like star at 0.2AU and 1AU, this requirement means that the transit of a moon can only lead or lag the planetary transit at most by 0.6 hours and 7.0 hours respectively.   Similarly, for the case of a Jupiter-like planet orbiting a Sun-like star at 0.2AU and 1AU, this requirement means that the transit of a moon can only lead or lag the planetary transit at most by 3.4 hours and 48.3 hours respectively.

\subsection{Analytic groundwork}\label{Transit_Intro_Deriv}

Now that $\tau$ has been redefined, the mechanics required for searching for a signal must be constructed.  In order to search the sequence of $\tau$ values, we need to be able to write $\tau$ as a function of transit number, moon and planet orbital parameters as well as the influence of any photometric noise.  Assuming that the planet has only one moon,\footnote{For the case where additional moons are suspected, equation~\eqref{TraM-TTV-taufitdef} can be modified by including additional $\Delta \tau$ terms.} $\tau_j$, the $\tau$ value calculated from the $j^{th}$ transit where the numbering begins at zero, can be written as
\begin{multline}
\tau_j = t_0 + jT_p \\
+ \Delta \tau(j,a_m,e_m,f_m(t_0),\omega_m,I_m,I_p,e_p,\omega_p,\Omega_m - \Omega_p,T_p,T_m) + \epsilon_j, \label{TraM-TTV-taufitdef}
\end{multline}
where $\Delta \tau$ is a function representing the effect of the presence of the moon on $\tau_j$, $\epsilon_j$ represents the timing error due to photometric noise, $t_0$ would have been the mid-time of the zeroth transit for the case where the planet did not have a moon, and where $T_p$ and $T_m$ are the orbital periods of the planet and moon orbits respectively.  In addition, $\Delta \tau$ is a function of $a_m$, $e_m$, $f_m(t_0)$, $I_m$ and $\omega_m$, which are the semi-major axis, eccentricity, true anomaly at time $t_0$, the inclination and argument of pericenter of the moon's orbit, $I_p$, $e_p$ and $\omega_p$, which are  the inclination, eccentricity and argument of periastron of the planet's orbit, and $\Omega_m - \Omega_p$, the difference between the longitudes of the ascending node for the planet and moon's orbit.\footnote{See section~\ref{Transit_Signal_Coord_Pos}.}  We begin by writing 
\begin{equation}
\alpha(t_i) = \alpha_p(t_i) + \alpha_m(t_i) + \alpha_n(t_i),\label{TraM-TTV-alphasplit}
\end{equation}
where $\alpha_p$ is the absolute dip due to the transit of the planet, $\alpha_m$ is the absolute dip due to the transit of the moon\footnote{For the case where the moon is eclipsed by or eclipses its host planet, the moon will not cause an additional dip, i.e. $\alpha_m = 0$.  However, for both these cases $\Delta \tau \approx 0$, that is, the true value of $\Delta \tau$ is approximately equal to the value of $\Delta \tau$  if there were no moon.  Consequently the situation where  $\alpha_m = 0$,  when the moon is in front of or behind the planet, will be neglected in this thesis.} and $\alpha_n$ is the noise on the light curve due to photometric variability.  Note that $\alpha_n$ is defined such that it has a mean of zero and consequently it can have positive or negative values.  In particular, using equation~\eqref{TraM-DescT-alphadef}, we can write $\alpha_p$ and $\alpha_m$ as
\begin{align}
\alpha_p(\delta_p(t)) &= \int_{S_p} I(\mu_p)\mu_p dA,\label{TraM-DescT-alphapdef}\\
\alpha_m(\delta_m(t)) &= \int_{S_m} I(\mu_m)\mu_m dA,\label{TraM-DescT-alphamdef}
\end{align}
where $S_p$ and $S_m$ represent the region of stellar surface occulted by the planet and moon respectively and $\mu_p$ and $\mu_m$ are the $\mu$ values, again corresponding to the position of the planet and moon respectively. Also, as we will investigate the effects of different types of photometric noise on $\tau$, we will keep the definition of $\alpha_n$ as general as possible.

Substituting equation~\eqref{TraM-TTV-alphasplit} into equation~\eqref{TraM-TTV-taudef} gives
\begin{equation}
\tau = \frac{\sum_i t_i (\alpha_p(t_i) + \alpha_m(t_i) + \alpha_n(t_i))}{\sum_i \alpha_p(t_i) + \alpha_m(t_i) + \alpha_n(t_i)}.\label{TraM-TTV-tauderiv1}
\end{equation}
As $\alpha_n$ is present in both the numerator and the denominator, the numerator and denominator are correlated.  

To see how this correlation manifests practically, we consider an example.  First, for clarity, equation~\eqref{TraM-TTV-tauderiv1} is reformatted by writing $t_i = t_i - t_{mid,p} + t_{mid,p}$, where $t_{mid,p}$ is the mid-time of the window, and simplifying, giving
\begin{equation}
\tau = t_{mid,p} + \frac{\sum_i (t_i - t_{mid,p}) (\alpha_p(t_i) + \alpha_m(t_i) + \alpha_n(t_i))}{\sum_i \alpha_p(t_i) + \alpha_m(t_i) + \alpha_n(t_i)}.
\end{equation}
Now, consider the case where one value of $\alpha_n$, lets say $\alpha_n(t_k)$, is large (and positive) and where $t_k > t_{mid,p}$ (i.e a data point near the egress of the transit).  As $(t_k - t_{mid,p})$ is positive, $\alpha_n(t_k)$ acts to make both the denominator and numerator of the fraction more positive.  These two effects partially cancel, thus, the error in $\tau$ resulting from the effect of $\alpha_n(t_k)$ will be small.  Conversely, for the case where $\alpha_n(t_k)$ is still large and positive, but where $t_k < t_{mid,p}$ (i.e a data point near the ingress of the transit) this is no longer the case.  As $(t_k - t_{mid,p})$ is now negative $\alpha_n(t_k)$ now acts to make the numerator smaller but the denominator larger.  Consequently, for this case the error in $\tau$ resulting from the effect of $\alpha_n(t_k)$ will be amplified.  For the case where $\alpha_n(t_k)$ is large and negative, these effects are reversed.  As a result of these behaviours, we cannot investigate the denominator and numerator of equation~\eqref{TraM-TTV-tauderiv1} in isolation, and then combine the results.  

Consequently, to make analytic progress, equation~\eqref{TraM-TTV-tauderiv1} must be reformatted to remove this correlation.  Assuming that $\sum_i \alpha_n(t_i) \ll \sum_i (\alpha_p(t_i) + \alpha_m(t_i))$, the binomial expansion can be used to expand the denominator of equation~\eqref{TraM-TTV-tauderiv1}, giving
\begin{equation}
\tau = \frac{\sum_i t_i (\alpha_p(t_i) + \alpha_m(t_i) + \alpha_n(t_i))}{\sum_i \alpha_p(t_i) + \alpha_m(t_i)}\left(1 - \frac{\sum_i\alpha_n(t_i)}{\sum_i \alpha_p(t_i) + \alpha_m(t_i) }\right).\label{TraM-TTV-tauderiv2}
\end{equation}
Expanding equation~\eqref{TraM-TTV-tauderiv2}, neglecting terms of order $(\sum \alpha_n/\sum (\alpha_p + \alpha_m))^2$ and gathering terms linear in $\alpha_n$ under the same sum sign gives
\begin{multline}
\tau = \frac{\sum_i t_i (\alpha_p(t_i) + \alpha_m(t_i))}{\sum_i \alpha_p(t_i) + \alpha_m(t_i)}  \\+\frac{1}{\sum_i \alpha_p(t_i) + \alpha_m(t_i)}\sum_i\left[t_i -  \frac{\sum_i t_i (\alpha_p(t_i) + \alpha_m(t_i))}{\sum_i \alpha_p(t_i) + \alpha_m(t_i)} \right]\alpha_n(t_i).\label{TraM-TTV-tauderiv3}
\end{multline}
Note that as this first term does not contain $\alpha_n$, it is exact and consequently only contributes to $jT_p + t_0 + \Delta \tau$.  Expanding and then contracting the first term of equation~\eqref{TraM-TTV-tauderiv3} gives
\begin{multline}
\tau = \frac{\sum_i \alpha_p(t_i)}{\sum_i \alpha_p(t_i) + \alpha_m(t_i)}\frac{\sum_i t_i \alpha_p(t_i)}{\sum_i \alpha_p(t_i)} + \frac{\sum_i \alpha_m(t_i)}{\sum_i \alpha_p(t_i) + \alpha_m(t_i)}\frac{\sum_i t_i \alpha_m(t_i)}{\sum_i \alpha_m(t_i)}   \\+\frac{1}{\sum_i \alpha_p(t_i) + \alpha_m(t_i)}\sum_i\left[t_i -  \frac{\sum_i t_i (\alpha_p(t_i) + \alpha_m(t_i))}{\sum_i \alpha_p(t_i) + \alpha_m(t_i)} \right]\alpha_n(t_i),\label{TraM-TTV-tauderiv4}
\end{multline}
or
\begin{multline}
\tau = \frac{A_p\tau_p + A_m\tau_m}{A_p + A_m}  \\+\frac{1}{\sum_i \alpha_p(t_i) + \alpha_m(t_i)}\sum_i\left[t_i -  \frac{\sum_i t_i (\alpha_p(t_i) + \alpha_m(t_i))}{\sum_i \alpha_p(t_i) + \alpha_m(t_i)} \right]\alpha_n(t_i),\label{TraM-TTV-tauderiv5}
\end{multline}
where $A_p = \sum_i \alpha_p$, $A_m = \sum_i \alpha_m$, the area of the dips caused by the planet and moon respectively and where $\tau_p$ and $\tau_m$ are defined by $\tau_p = (\sum_i t_i \alpha_p(t_i))/(\sum_i \alpha_p(t_i))$ and $\tau_m = (\sum_i t_i \alpha_m(t_i))/(\sum_i \alpha_m(t_i))$.  $A_m$ and $A_p$ can be written explicitly in terms of the position of the planet and the moon respectively using equations~\eqref{TraM-DescT-alphapdef} and \eqref{TraM-DescT-alphamdef}.  Noting that $\int \alpha dt \approx \Delta t \sum \alpha$, we have that
\begin{align}
A_p &= \frac{1}{\Delta t} \int \alpha_p dt = \int \int_{S_p} I(\mu_p)\mu_p dA dt,\\
A_m & = \frac{1}{\Delta t} \int \alpha_m dt = \int \int_{S_m} I(\mu_m)\mu_m dA dt.
\end{align}
where $\Delta t$ is the exposure time and the integrals are conducted over the same time period as the sum (see figure~\ref{LewisSchematic}).  Similar expressions could be presented for $\tau_p$ and $\tau_m$.  However, these will not be required as we will be assuming that the planet and moon move with a constant velocity during transit.  For this case the transit light curves become symmetric and $\tau_p$ and $\tau_m$ are given by the mid-times of their respective transits.

Returning to the derivation of expressions for $\Delta \tau$ and $\epsilon_j$, and examining the second term of equation~\eqref{TraM-TTV-tauderiv3}, it can be seen that it consists of a sum of random variables with zero mean.  Consequently, the second term of equation~\eqref{TraM-TTV-tauderiv3}  can only contribute to $\epsilon_j$.  As the first term of equation~\eqref{TraM-TTV-tauderiv3} contributes only to $jT_p + t_0 + \Delta \tau$ and the second term of equation~\eqref{TraM-TTV-tauderiv3} only contributes to $\epsilon_j$, we have that
\begin{align}
jT_p + t_0 + \Delta \tau &= \frac{A_p\tau_p + A_m\tau_m}{A_p + A_m}, \label{transit_intro_ground_deltaudef}\\
\epsilon_j &= \frac{1}{A_p + A_m}\sum_i\left[t_i -  (jT_p + t_0 +\Delta \tau) \right]\alpha_n(t_i),\label{transit_intro_ground_noisedef}
\end{align}
where the definitions of $A_p$, $A_m$ and equation~\eqref{transit_intro_ground_deltaudef} have been used to simplify the coefficient of $\alpha_n$.  Thus, the process of determining if a given moon is detectable is the process of comparing equation~\eqref{transit_intro_ground_deltaudef}, the `signal', and equation~\eqref{transit_intro_ground_noisedef}, the `noise', for the set of available transits.  Consequently, these two equations will each be investigated in turn in chapters~\ref{Transit_Signal} and \ref{Trans_TTV_Noise} and combined to produce detection thresholds in chapter~\ref{Trans_Thresholds}.

\section{Conclusion}

A general introduction to transiting planets and the detection of moons of transiting planets has been presented, with the aim of introducing the material required for the work conducted in this Part.  This was done in three broad stages.    First the transit technique was discussed, and mathematically investigated, with particular emphasis on deriving expressions for the transit duration, and the shape of the transit light curve.  Then the literature relating to the detection of moons of transiting planets was summarised.  Finally, the TTV$_p$ technique was focussed on in terms of describing the work previously presented in the literature, discussing how my work fits within that context and using the definition of $\tau$ to derive equations for the the timing perturbation due to the moon $\Delta \tau$ and the timing noise $\epsilon_j$.  Now that the background work has been summarised and the equations defining the problem have been introduced,   the form of $\Delta \tau$, the TTV$_p$ signal, can be investigated.

\chapter[$TTV_p$ Signal]{$TTV_p$ Signal caused by an Extra-solar Moon}\label{Transit_Signal}

\section{Introduction}\label{Transit_Signal_Intro}

The first step in determining the attributes of detectable moons using the photometric transit timing method is to investigate the timing signal, $\Delta \tau$, defined in section~\ref{Trans_TTV}.  As observing time is limited, it would be of use to know the way in which the detectability of a given moon depends on the properties of its host planet, and thus be able to focus follow-up on the set of planets most likely to have detectable moons.  Consequently, in this chapter, this signal is investigated in terms of the physical parameters of the system, that is, the masses and radii of the host star, planet and moon and the orbital elements of both the planet and moon orbits,\footnote{Recall that ``planet orbit" refers to the orbit of the planet-moon barycenter about the star and the ``moon orbit" refers to the orbit of the moon about the planet.} but with particular emphasis on the physical and orbital properties of the planet.  As a result of processes such as motion of the planet and moon about their common barycenter during transit, analytically deriving the general form of $\Delta \tau$ in terms of these elements is not a trivial problem.  Consequently, three representative cases, which highlight the types of planet orbits likely to be encountered, were selected.  These are:
\begin{enumerate}
\item A circular planet orbit aligned to the line-of-sight,
\item A circular planet orbit slightly inclined to the line-of-sight and
\item An eccentric planet orbit aligned to the line-of-sight.
\end{enumerate}
In addition, for simplicity, it is assumed that the moon's orbit is circular\footnote{The effect of a small amount of eccentricity in the moon orbit will be investigated in appendix~\ref{EccMoon_App}.} and coplanar with the planet's orbit.  These cases will be investigated in turn, in terms of the form and associated properties of $\Delta \tau$ resulting from these configurations.  However, in order to perform this investigation, an appropriate coordinate system and method needs to be selected.

\section{Definition of the coordinate system}\label{Transit_Signal_Coord}

In order to calculate $\Delta \tau$, the positions of both the planet and moon on the face of the star need to be known as a function of time.  Consequently, to provide a framework for this description, a coordinate system must be selected. It would be optimal if this coordinate system could be used to simply describe the systems most likely to form, easily provide pertinent information, such as, the position of the planet and moon on the face of the star, while also relating to coordinate systems used in the literature.  To select a coordinate system with these properties, two issues must be addressed.  First, the orientation of the coordinate system needs to be decided, with respect to ``natural" standards such as the plane of the sky and the line-of-sight.  Second, using this coordinate system, two reference directions, required for the definition of the Euler angles (which define the orientation of the planet and moon orbits), need to be selected.  For example, two of the coordinate axes could be selected for use as the reference directions.  As the orientation of the coordinate axes informs the choice of the reference directions, the choice of coordinate axes will be discussed first.

\subsection{Orientation of coordinate system}\label{Transit_Signal_Coord_Orient}

The selection of the orientation of the Cartesian coordinate system to be used in this analysis should be informed by both the physics and mathematics of the system.  However, there are two main choices for the orientation of the three axes.

First, one of the axes could be chosen such that it is parallel to the projection of the line-of-sight onto the plane of the planet's orbit (see figure~\ref{TransitSignalCoordSysDirectCC}).  This coordinate system is physically motivated in that it has the advantage that it can easily describe the case where the planet's and moon's orbits are coplanar.  As a majority of formation mechanism relevant to large moons, have a preference for producing moons in circular coplanar orbits (see section~\ref{Intro_Moons_Form}), a simple description of these cases is clearly an advantage.  Unfortunately, as this coordinate system is tied to the planet's orbit, it is non-inertial, that is, if processes such as orbital precession act on the planet's orbit, they will also act to change the orientation of the coordinate system.

Second, the coordinate system could be defined such that one axis of the coordinate system is oriented along the line-of-sight (see figure~\ref{TransitSignalCoordSysDirectLOS}).  This choice of coordinate system allows easy description of the most mathematically simple configuration, the case where the planet and moon traverse the same chord of the star when they transit.  In order for this to occur, the moon's orbit would have to be slightly inclined with respect to the planet's orbit.  While this configuration is the most simple, there is no good physical reason why the moon orbital plane would have exactly the right inclination and orientation such that it would be specifically aligned with our line-of-sight.  Despite this, this coordinate system has the added advantages of being inertial and that its two remaining axes lie on the plane of the sky.

\begin{figure}
     \centering
     \subfigure[Moon orbit coplanar with planet orbit.]{
          \label{TransitSignalCoordSysDirectCC}
          \includegraphics[width=.48\textwidth]{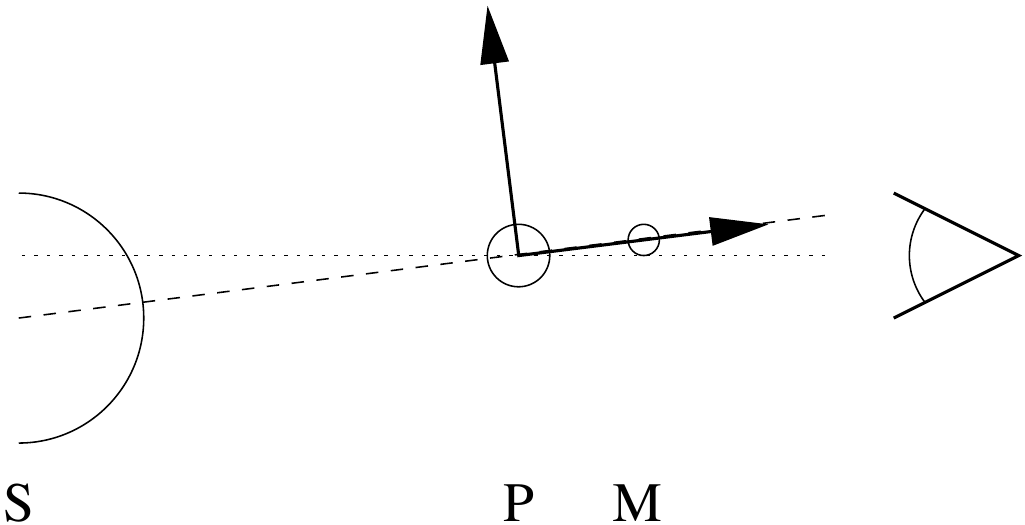}}
     \subfigure[Moon orbit aligned to the line-of-sight.]{
          \label{TransitSignalCoordSysDirectLOS}
          \includegraphics[width=.48\textwidth]{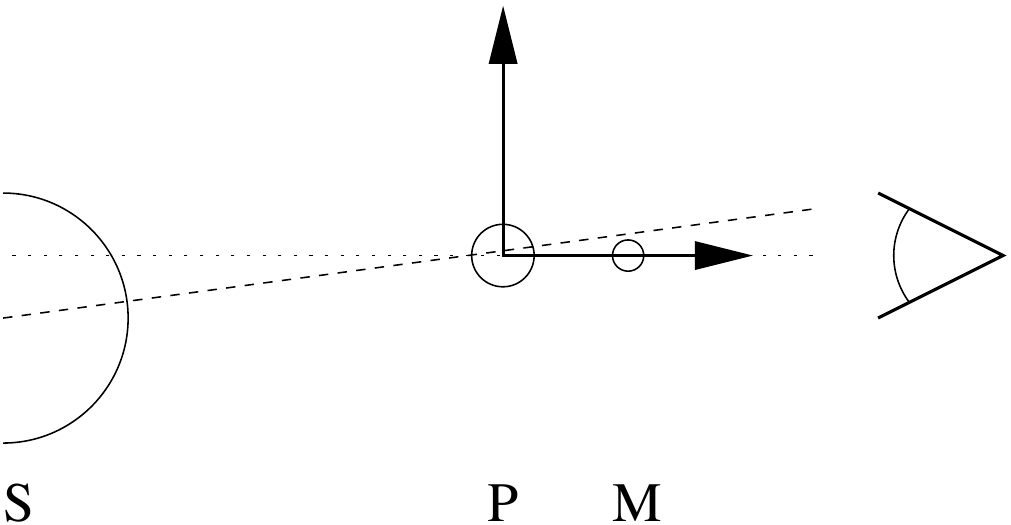}}
     \caption[Schematic diagram of the two proposed coordinate systems overlaid by the planet-moon systems they are optimised to describe.]{Schematic diagram of the two proposed coordinate systems overlaid by the planet-moon systems they are optimised to describe (see text).  For each diagram, the star (S), planet (P), moon (M) and observer (represented by an eye) are arranged from left to right.  In addition, in both diagrams, a dashed line and a dotted line is used to show the alignment of the moon's orbital plane with respect to the planet's orbital plane and the observer respectively.}
     \label{TransitSignalCoordSysDirect}
\end{figure}

\begin{figure}[tb]
\begin{center}
\includegraphics[width=.95\textwidth]{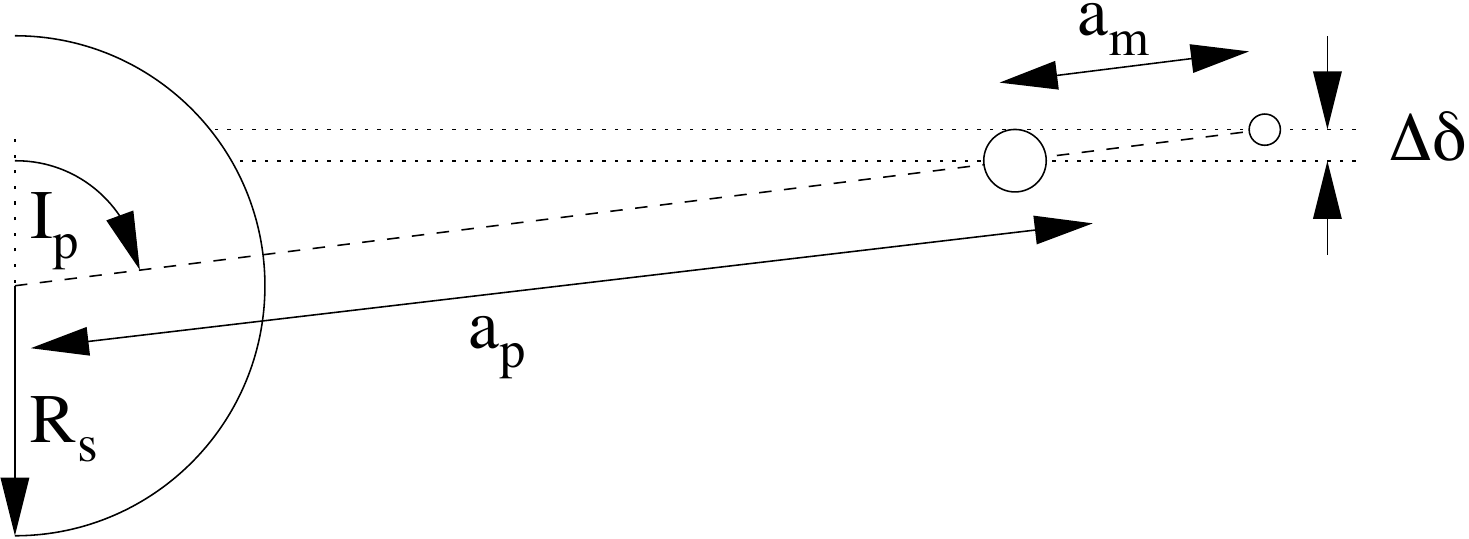}
\caption[An enlarged version of figure~\ref{TransitSignalCoordSysDirectCC} showing the orbital elements $I_p$, $a_p$ and $a_m$.]{An enlarged version of figure~\ref{TransitSignalCoordSysDirectCC} showing the orbital elements $I_p$, $a_p$ and $a_m$.  The star, planet and moon are assumed to be collinear, as this is the configuration which leads to the maximum vertical displacement between the planet and the moon on the face of the star.}
\label{TransitSignalCoordSysDeltadelta}
\end{center}
\end{figure}

Fortunately, to the accuracy required for this work, these two coordinate systems are essentially equivalent, in that the error incurred by approximating the planet-moon system shown in figure~\ref{TransitSignalCoordSysDirectCC} with the one shown in figure~\ref{TransitSignalCoordSysDirectLOS} is not measurable.  This is demonstrated below.

To see this equivalence, consider a planet-moon pair with circular coplanar orbits.\footnote{The increase in $\Delta \delta$ corresponding to a small rotation, is maximised for orbits initially aligned to the line-of-sight.}  As can be seen from figure~\ref{TransitSignalCoordSysDeltadelta}, $\Delta \delta$, the maximum distance between the planet and moon chord across the face of the star is
\begin{equation}
\Delta \delta =  a_m \cos I_p.\label{transit_signal_coord_eqn1}
\end{equation}
Using the most extreme value of $I_p$ such that the planet still transits, that is,
\begin{equation}
\cos I_p =  \frac{R_s}{a_p},\label{transit_signal_coord_eqn2}
\end{equation}
we have that
\begin{equation}
\Delta \delta =  R_s \frac{a_m}{a_p}.\label{transit_signal_coord_eqn3}
\end{equation}
Taking $a_m$ to be that of the most distant stable moon orbit, and taking this to be 0.5 Hill radii, gives
\begin{equation}
\Delta \delta =  R \times 0.5\left(\frac{M_p}{3M_s}\right)^{1/3}.\label{transit_signal_coord_eqn4}
\end{equation}
Consequently, the maximum deviation possible for the cases where $M_p/M_s$ is equal to $10^{-2}$, $10^{-3}$ and $10^{-4}$, is equal to 7.5\%, 3.5\% and 1.6\% of a stellar radius respectively.  As moons are likely to form close to their host planet and not near the boundary for orbital stability (see section~\ref{Intro_Moons_Form}), the deviations observed for real moons are likely to be much smaller.  Consequently a moon which is on a circular coplanar orbit can be treated as if its orbit is aligned to the line-of-sight and visa versa.  Thus, formally, we can select one of these coordinate systems while still retaining the benefits of the other.  To allow ease of mathematical description, it was decided to select a Cartesian coordinate system with one axis pointed along the line-of-sight and the two remaining axes in the plane of the sky.  In addition, in order to take advantage of the mathematical intuition associated with the Cartesian coordinate system, it was decided to define the $x$ and $y$-axes to be in the plane of the sky, aligned such that the $y$-axis points north, such that the position of the planet and moon on the face of the star are given by their respective $x$ and $y$-coordinates, and the $z$-axis to lie along the line-of-sight.

\subsection{Selection of reference directions}\label{Transit_Signal_Coord_References}

In order to describe the orientation of the moon's and planet's orbits, a reference direction, required for the definition of the inclinations, $I_p$ and $I_m$, and a second reference direction required for the definition of $\omega_p$ and $\omega_m$, the periapse arguments, and $\Omega_p$ and $\Omega_m$, the longitudes of the ascending node, needs to be selected.  With this in mind, the selection of the two reference directions will be discussed in turn.  

The selection of the reference direction for the definition of inclinations was informed by the conventions present in the literature.  In the transiting planet literature, the line-of-sight is generally used as the reference direction, for example, planetary orbital inclinations are given as the angle between the planetary orbit normal, and the line-of-sight.  Consequently, for this work, the line-of-sight was used as the reference direction for both the planet and moon orbits.

For this work, the $x$-axis is selected as the reference direction for longitudes (recall that the $y$-axis is defined to be north).  While this selection was informed by the coordinate systems used in the literature \citep[e.g.][p. 48]{Murrayetal1999}, the choice is slightly arbitrary as the coordinate system will be rotated about the $z$ axis by $- \Omega_p$ in section~\ref{Transit_Signal_Coord_Pos} for mathematical convenience.

These two choices and the consequent definitions of $I_p$, $I_m$, $\omega_p$, $\omega_m$, $\Omega_p$ and $\Omega_m$ are summarised in figure~\ref{TransitSignalGenCoordSys}.  Now that the coordinate system has been defined, it can be used to describe the path of the planet and moon across the face of the star during transit.

\subsection{Position of the planet and moon on the face of the star}\label{Transit_Signal_Coord_Pos}

\begin{figure}[tb]
\begin{center}
\includegraphics[width=.95\textwidth]{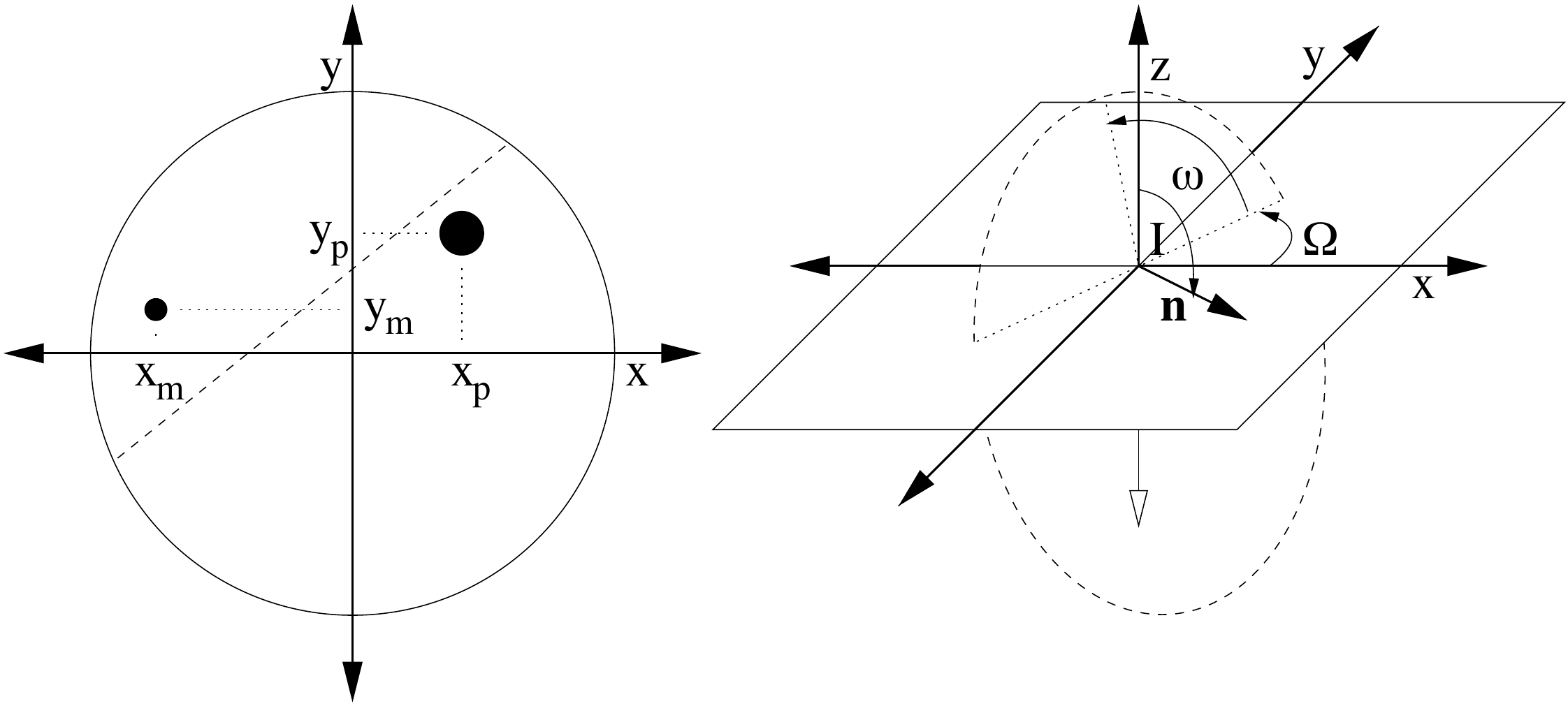}
\caption[Schematic diagram of the coordinate system.  The diagram on the left shows the position of the planet and moon on the face of the star during transit.  The diagram on the right shows how the orientation of a general orbit is related to its three Euler angles ($I$, $\Omega$ and $\omega$).]{Schematic diagram of the coordinate system.  The diagram on the left shows the position of the planet (large circle) and moon (small circle) on the face of the star during transit.  The dashed line indicates the path of the planet-moon barycenter across the face of the star.  The diagram on the right shows how the orientation of a general orbit is related to its three Euler angles ($I$, $\Omega$ and $\omega$), noting that $\mathbf{n}$ is a unit vector normal to the plane of the orbit.  This is the system used to describe the orientation of the planet's orbit in terms of the Euler angles $I_p$, $\Omega_p$ and $\omega_p$, and the orientation of moons orbit in terms of the Euler angles $I_m$, $\Omega_m$ and $\omega_m$.
}
\label{TransitSignalGenCoordSys}
\end{center}
\end{figure}

Now that a coordinate system has been selected, the positions of the planet and moon on the face of the star can be specified as a function of time.  Using the coordinate system shown in figure~\ref{TransitSignalGenCoordSys}, rotating the planet and moon orbits by their associated Euler angles gives
\begin{multline}
x_p = r_p\cos \Omega_p \cos(f_p + \omega_p) - r_p\sin \Omega_p \cos I_p \sin(f_p + \omega_p) \\
- \frac{M_m}{M_{pm}} r_m\left[\cos\Omega_m \cos(f_m + \omega_m) - \sin \Omega_m \cos I_m \sin(f_m + \omega_m)\right], \label{transit_signal_coord_xp}
\end{multline}
\begin{multline}
x_m = r_p\cos \Omega_p \cos(f_p + \omega_p) - r_p\sin \Omega_p \cos I_p \sin(f_p + \omega_p) \\
+ \frac{M_p}{M_{pm}} r_m \left[\cos\Omega_m \cos(f_m + \omega_m) - \sin \Omega_m \cos I_m \sin(f_m + \omega_m)\right], \label{transit_signal_coord_xm}
\end{multline}
\begin{multline}
y_p = r_p\sin \Omega_p \cos(f_p + \omega_p) + r_p\cos \Omega_p \cos I_p \sin(f_p + \omega_p) \\
- \frac{M_m}{M_{pm}} r_m \left[\sin\Omega_m \cos(f_m + \omega_m) + \cos \Omega_m \cos I_m \sin(f_m + \omega_m)\right], \label{transit_signal_coord_yp}
\end{multline}
\begin{multline}
y_m = r_p\sin \Omega_p \cos(f_p + \omega_p) + r_p\cos \Omega_p \cos I_p \sin(f_p + \omega_p) \\
+ \frac{M_p}{M_{pm}} r_m \left[\sin\Omega_m \cos(f_m + \omega_m) + \cos \Omega_m \cos I_m \sin(f_m + \omega_m)\right], \label{transit_signal_coord_ym}
\end{multline}
where $M_{pm} = M_p + M_m$.  In addition, for these equations $r_p$ and $r_m$ are given by
\begin{align}
r_p &= \frac{a_p(1-e_p^2)}{1+e_p\cos f_p},\label{transit_signal_coord_Rdef}\\
r_m &= \frac{a_m(1-e_m^2)}{1+e_m\cos f_m},\label{transit_signal_coord_rdef}
\end{align}
and $e_p$ and $e_m$ are the eccentricities of the planet's and moon's orbit respectively.  The  true anomalies of the planet and moon orbits, $f_p$ and $f_m$, are related to time through
\begin{align}
\cos f_p &= \frac{\cos E_p - e_p}{1-e_p\cos E_p},\label{transit_signal_coord_fodef}\\
\cos f_m &= \frac{\cos E_m - e_m}{1-e_m\cos E_m},\label{transit_signal_coord_fidef}
\end{align}
and
\begin{align}
n_p (t - t_0) &= E_p - e_p\sin E_p,\label{transit_signal_coord_Eodef}\\
n_m (t - t_0) &= E - e_m\sin E_m,\label{transit_signal_coord_Eidef}
\end{align}
where $n_p$ and $n_m$ are the mean motions of the planet's and moon's orbit respectively.

\begin{figure}[tb]
\begin{center}
\includegraphics[width=.95\textwidth]{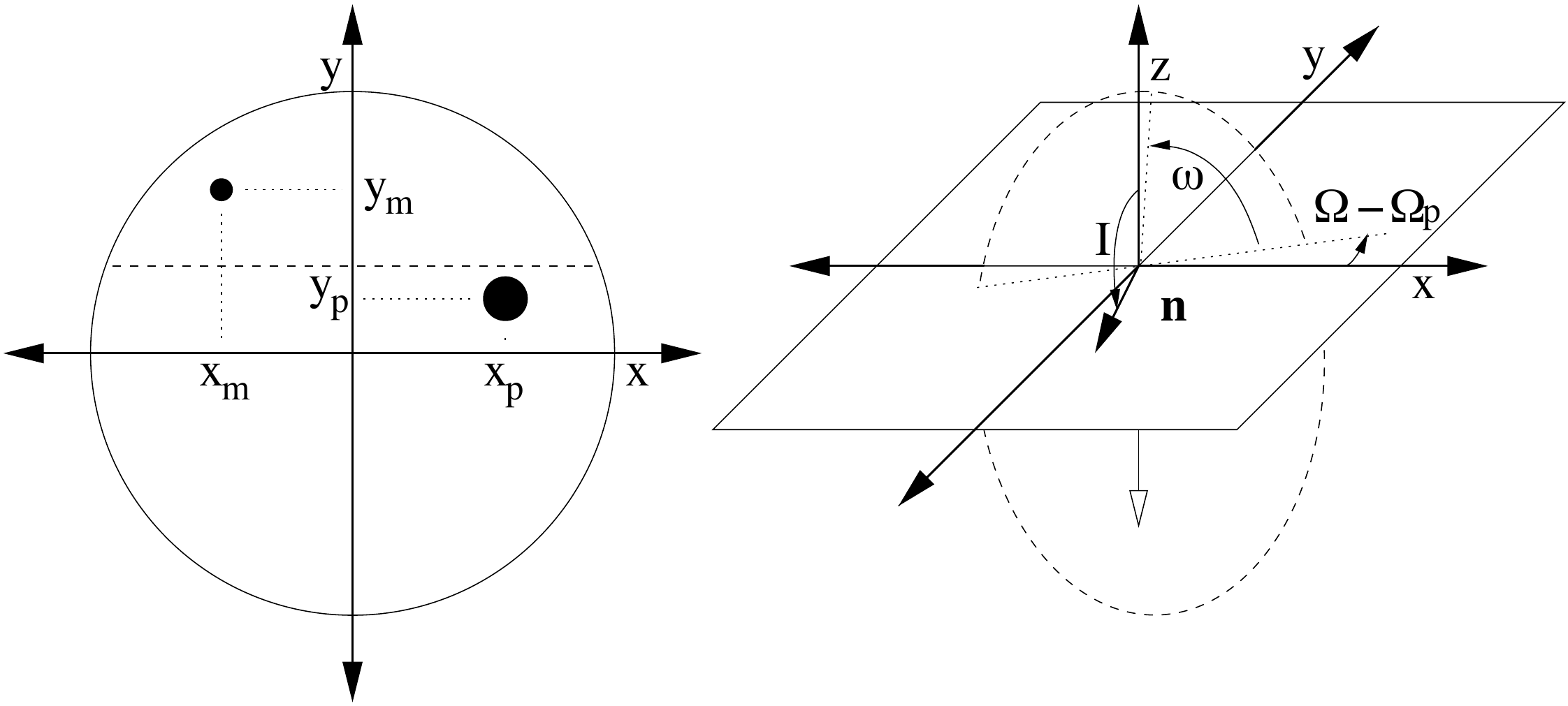}
\caption[Schematic diagram of the positions of the planet and moon during transit, after rotation of the coordinate system by $-\Omega_p$.]{Schematic diagram of the positions of the planet and moon during transit, after rotation of the coordinate system by $-\Omega_p$.  Note that the $x$-axis is now parallel to the chord made by the planet-moon barycenter and the positive $y$-axis now bisects this chord.}
\label{TransitSignalGenCoordSysRot}
\end{center}
\end{figure}

Equations~\eqref{transit_signal_coord_xp} to \eqref{transit_signal_coord_ym} can be simplified further.  These equations describe a transit where the path of the planet-moon barycenter across the star makes an angle of $\Omega_p$ with the $x$-axis.  For purely mathematical reasons, it would be useful if this path was parallel to either the $x$ or $y$-axes.  As the shape of the transit light curve does not depend on $\Omega_p$ (see appendix~\ref{App_Omega_Dep}), the coordinate system can by rotated about the $z$-axis by $-\Omega_p$, effectively setting $\Omega_p$ to zero and making the path of the planet-moon barycenter across the face of the star horizontal.  Performing this rotation (see figure~\ref{TransitSignalGenCoordSysRot}), and simplifying gives
\begin{multline}
x_p = r_p \cos(f_p + \omega_p) - \frac{M_m}{M_{pm}}r_m\left[\cos (\Omega_m - \Omega_p) \cos(f_m + \omega_m)\right. \\ \left.
+  \sin(\Omega_m - \Omega_p) \cos I_m \sin(f_m + \omega_m)\right],\label{transit_signal_coord_xpdef}
\end{multline}
\begin{multline}
x_m = r_p \cos(f_p + \omega_p) + \frac{M_p}{M_{pm}}r_m\left[\cos(\Omega_m - \Omega_p) \cos(f_m + \omega_m)\right. \\ \left.
+  \sin(\Omega_m - \Omega_p) \cos I_m \sin(f_m + \omega_m)\right],\label{transit_signal_coord_xmdef}
\end{multline}
\begin{multline}
y_p = r_p\cos \Omega_p \cos I_p \sin(f_p + \omega_p) - \frac{M_m}{M_{pm}}r_m \left[\sin(\Omega_m - \Omega_p) \cos(f_m + \omega_m) \right. \\ \left.
+ \cos(\Omega_m - \Omega_p) \cos I_m \sin(f_m + \omega_m)\right],\label{transit_signal_coord_ypdef}
\end{multline}
\begin{multline}
y_m = r_p\cos \Omega_p \cos I_p \sin(f_p + \omega_p) + \frac{M_p}{M_{pm}}r_m \left[\sin(\Omega_m - \Omega_p) \cos(f_m + \omega_m) \right. \\ \left.
+ \cos(\Omega_m - \Omega_p) \cos I_m \sin(f_m + \omega_m)\right].\label{transit_signal_coord_ymdef}
\end{multline}

Now that equations describing the position of the planet and moon on the face of the star have been derived, we can begin to investigate what method would be most useful to determine $\Delta \tau$ using these equations in terms of the three special cases under consideration.

\section{Discussion of method} \label{Transit_Signal_Method}

\subsection{Introduction}

Now that we have expressions for the location of the planet and moon on the face of the star as a function of their orbital elements and time (equations~\eqref{transit_signal_coord_xpdef} to \eqref{transit_signal_coord_ymdef}), we can combine these with expressions for the amount of light blocked by these objects, $\alpha_p$ and $\alpha_m$, as a function of their position (equations~\eqref{TraM-DescT-alphapdef} and \eqref{TraM-DescT-alphamdef}) and consequently calculate $\Delta \tau$ via $\tau_p$, $\tau_m$, $A_p$ and $A_m$ using equation~\eqref{transit_intro_ground_deltaudef}.  

Unfortunately, the case where $\alpha_p$ and $\alpha_m$ are dominated by non-uniform motion across a limb darkened star is not mathematically simple.  For example, consider the calculation of $A_p$ via $\alpha_p$.  As can be seen from equation~\eqref{TraM-DescT-alphapdef}, $\alpha_p$ is defined in terms of a surface integral with domain $S_p$, where the domain represents the region of the face of the star physically occluded by the planet. Consequently, the shape of this domain $S_p$ depends on whether the planet is on the face of the star ($S_p$ is circular) whether the planet is in ingress or egress ($S_p$ is lens-shaped) or whether the planet is off the face of the star ($S_p$ is non-existent).  In order to perform the time integral required to evaluate equation~\eqref{TraM-DescT-alphapdef}, the equation which defines $A_p$ in terms of $\alpha_p$, the shape and location of $S_p$, a function of planet position, needs to be determined as a function of time.  As equations~\eqref{transit_signal_coord_xpdef} to \eqref{transit_signal_coord_ymdef}, the equations which relate planetary position to time, are transcendental equations in time, this is not a trivial problem.  So, while expressions for $A_p$ and similarly $A_m$, can be derived, they are so complex that they do not give much physical insight into this system.  Consequently, we need to use physically appropriate approximations to simplify the equations, especially equations~\eqref{TraM-DescT-alphapdef} and \eqref{TraM-DescT-alphamdef}, in order to highlight the underlying physics.

As a result, it was assumed that the planet and moon move across the face of the star with constant, but not necessarily equal, velocities.  This approximation was selected as first, the characteristic timescale over which the orbital velocity of the moon changes, e.g. its orbital period, is generally much larger than a transit duration, (see tables~\ref{EgTransDurVel} and \ref{SSMoonsTable}) and second, it results in substantial mathematical simplification.  Recall from section~\ref{Trans_Intro_Transtech}, that for the case of uniform motion, transit light curves become symmetric about their midpoint.  This simplification results in two important properties, which will be stated here, but derived in section~\ref{Transit_Signal_Method_Implementation}.  First, $\tau_p$ and $\tau_m$ become the time-coordinates of the lines of symmetry of $\alpha_p(t)$ and $\alpha_m(t)$ respectively.  Second, the equations for $A_p$ and $A_m$ reduce to the product of the transit duration with a geometric term, which depends on the transit geometry, planetary and stellar radii and stellar limb darkening parameters.  To take advantage of these properties, a practical method for implementing this assumption needs to be investigated.

\subsection{Implementation of method} \label{Transit_Signal_Method_Implementation}

The simplest way of implementing this approximation is to determine $t_{in,p}$, $t_{eg,p}$, $t_{in,m}$ and $t_{eg,m}$, the time of ingress and egress of the planet and moon respectively, and use them to calculate $\tau_p$, $\tau_m$, $A_p$ and $A_m$ and thus $\Delta \tau$.  

\subsubsection{Evaluating $\tau_p$ and $\tau_m$  in terms of $t_{in,p}$, $t_{eg,p}$, $t_{in,m}$ and $t_{eg,m}$}

For the case of $\tau_p$ and $\tau_m$, writing these quantities in terms of $t_{in,p}$, $t_{eg,p}$, $t_{in,m}$ and $t_{eg,m}$ is simple as $\alpha_p(t)$ and $\alpha_m(t)$ are symmetric, as a result of the physics and geometry of the system (see section~\ref{Trans_Intro_Transtech}).  Thus, $\tau_p$ and $\tau_m$ should correspond to the geometric mean of the time of egress and ingress of the planet and moon respectively. Consequently
\begin{align}
\tau_p &= \frac{t_{eg,p} + t_{in,p}}{2}\label{transit_signal_method_taupdef},\\
\tau_m &= \frac{t_{eg,m} + t_{in,m}}{2}\label{transit_signal_method_taumdef}.
\end{align}

\subsubsection{Evaluating $A_p$ and $A_m$  in terms of $t_{in,p}$, $t_{eg,p}$, $t_{in,m}$ and $t_{eg,m}$}

For the case of $A_p$ and $A_m$ we have that 
\begin{align}
A_p &= \frac{1}{\Delta t} \int_{t_{in,p} - 1/2T_{in}}^{t_{eg,p} + 1/2T_{in}} \alpha_p(t) dt \label{transit_signal_method_Apeq1}\\
A_m &= \frac{1}{\Delta t} \int_{t_{in,p} - 1/2T_{in}}^{t_{eg,p} + 1/2T_{in}} \alpha_m(t) dt \label{transit_signal_method_Ameq1}
\end{align}
As can be seen, finding a simple way of writing $A_p$ and $A_m$ in terms of of $t_{in,p}$, $t_{eg,p}$, $t_{in,m}$ and $t_{eg,m}$, will take a little more work.  Consider equation~\eqref{transit_signal_method_Apeq1}.  $A_p$ is defined as the integral of a complicated function of $t$, with the limits on the integral being given by a function of the variables of interest, $t_{in,p}$ and $t_{eg,p}$ \ldots not an optimal format!  To separate the dependance of $A_p$ on the times $t_{in,p}$ and $t_{eg,p}$, from its dependance on the brightness profile of the star, equation~\eqref{transit_signal_method_Apeq1} was rewritten using $\delta_p$, the projected distance between the planet and the center of the star, given by 
\begin{equation}
\delta_p(t) = \sqrt{\delta_{min}^2 + ((t - \tau_p)v_{tr,p})^2 },\label{transit_signal_method_deltadef}
\end{equation}
where 
\begin{equation}
v_{tr,p} = \frac{2\sqrt{R_s^2 - \delta_{min}^2}}{t_{eg,p} - t_{in,p}}.\label{transit_signal_method_vtrpdef}
\end{equation}
where $\delta_{min}$ is the smallest projected distance between the center of the planet and the star during transit and where $\delta$, which for this application is $\delta_p$, and $\delta_{min}$ are both shown in figure~\ref{PlanetTransitSchematic}.

Dividing the integral into two halves to account for the different behaviour of $\delta_p$ for the cases $t < \tau_p$ and $t > \tau_p$, using equation~\eqref{transit_signal_method_deltadef} to substitute for $t$, and using $\delta_p$ as the integration variable, equation~\eqref{transit_signal_method_Apeq1} becomes
\begin{align}
A_p =&  \frac{1}{\Delta t} \int_{R_s + R_p}^{\delta_{min}}  \alpha_p(\delta_p) \frac{dt}{d \delta_p} d\delta_p +  \frac{1}{\Delta t} \int_{\delta_{min}}^{R_s + R_p}  \alpha_p(\delta_p) \frac{dt}{d \delta_p} d\delta_p, \label{transit_signal_method_Apeq2}\\
 =&  \frac{1}{\Delta t} \int_{R_s + R_p}^{\delta_{min}} \frac{\alpha_p(\delta_p) \times -\delta_p}{v_{tr+p} \sqrt{\delta_p^2 - \delta_{min}^2}} d\delta_p \notag \\
  &+  \frac{1}{\Delta t} \int_{\delta_{min}}^{R_s + R_p} \frac{\alpha_p(\delta_p) \times \delta_p}{v_{tr+p}\sqrt{\delta_p^2 - \delta_{min}^2} } d\delta_p, \label{transit_signal_method_Apeq2}\\
 =&   \frac{1}{v_{tr+p}}2 \frac{1}{\Delta t}\int_{\delta_{min}}^{R_s + R_p}  \frac{\alpha_p(\delta_p) \delta_p}{\sqrt{\delta_p^2 - \delta_{min}^2} } d\delta_p. \label{transit_signal_method_Apeq3}
\end{align}

Now, consider the case where the planet has no moon.  In this case the transit velocity will be given by $v_{tr}$, so, from equation~\eqref{transit_signal_method_Apeq3} we have that
\begin{equation}
\hat{A}_p = \frac{1}{v_{tr}}2 \frac{1}{\Delta t} \int_{\delta_{min}}^{R_s + R_p}  \frac{\alpha_p(\delta_p) \delta_p}{\sqrt{\delta_p^2 - \delta_{min}^2} } d\delta_p, \label{transit_signal_method_Aphatdef}
\end{equation}
where the hat has been added to show that this is a comparison case.  Assuming that $\delta_{min}$ is the same for both cases (true for the three special cases investigated in this chapter), we can write $A_p$ in terms of $\hat{A}_p$
 using equation~\eqref{transit_signal_method_Aphatdef}, and thus equation~\eqref{transit_signal_method_Apeq3} can be written as
\begin{align}
A_p &= \frac{v_{tr}}{v_{tr+p}} \hat{A}_p, \\
 &= \frac{t_{eg,p} - t_{in,p}}{2\sqrt{R_s^2 - \delta_{min}^2}} v_{tr} \hat{A}_p. \label{transit_signal_method_Apdef}
\end{align}
Note that the necessity of performing the surface integral in equation~\eqref{TraM-DescT-alphapdef} is now entirely avoided.  Similarly, we have that
\begin{equation}
A_m = \frac{t_{eg,m} - t_{in,m}}{2\sqrt{R_s^2 - \delta_{min}^2}} v_{tr} \hat{A}_m, \label{transit_signal_method_Amdef}
\end{equation}
where 
\begin{equation}
\hat{A}_m = \frac{1}{v_{tr}}2 \frac{1}{\Delta t} \int_{\delta_{min}}^{R_s + R_m}  \frac{\delta_m \alpha_m(\delta_m)}{\sqrt{\delta_m^2 - \delta_{min}^2} } d\delta_m. \label{transit_signal_method_Amhatdef}
\end{equation}

To demonstrate that the expressions derived for $\tau_p$, $\tau_m$, $A_p$ and $A_m$ are valid for the systems of interest, the quantities given by equations~\eqref{transit_signal_method_taupdef}, \eqref{transit_signal_method_taumdef}, \eqref{transit_signal_method_Apdef} and \eqref{transit_signal_method_Amdef} were compared to the full expressions using a simulation.

\begin{figure}
     \centering
     \noindent\makebox[\textwidth]{%
     \subfigure[$a_m = 2R_s$, $v_m/v_{tr} = 0.66$.]{
          \label{fig:dl2858}
          \includegraphics[width=.65\textwidth]{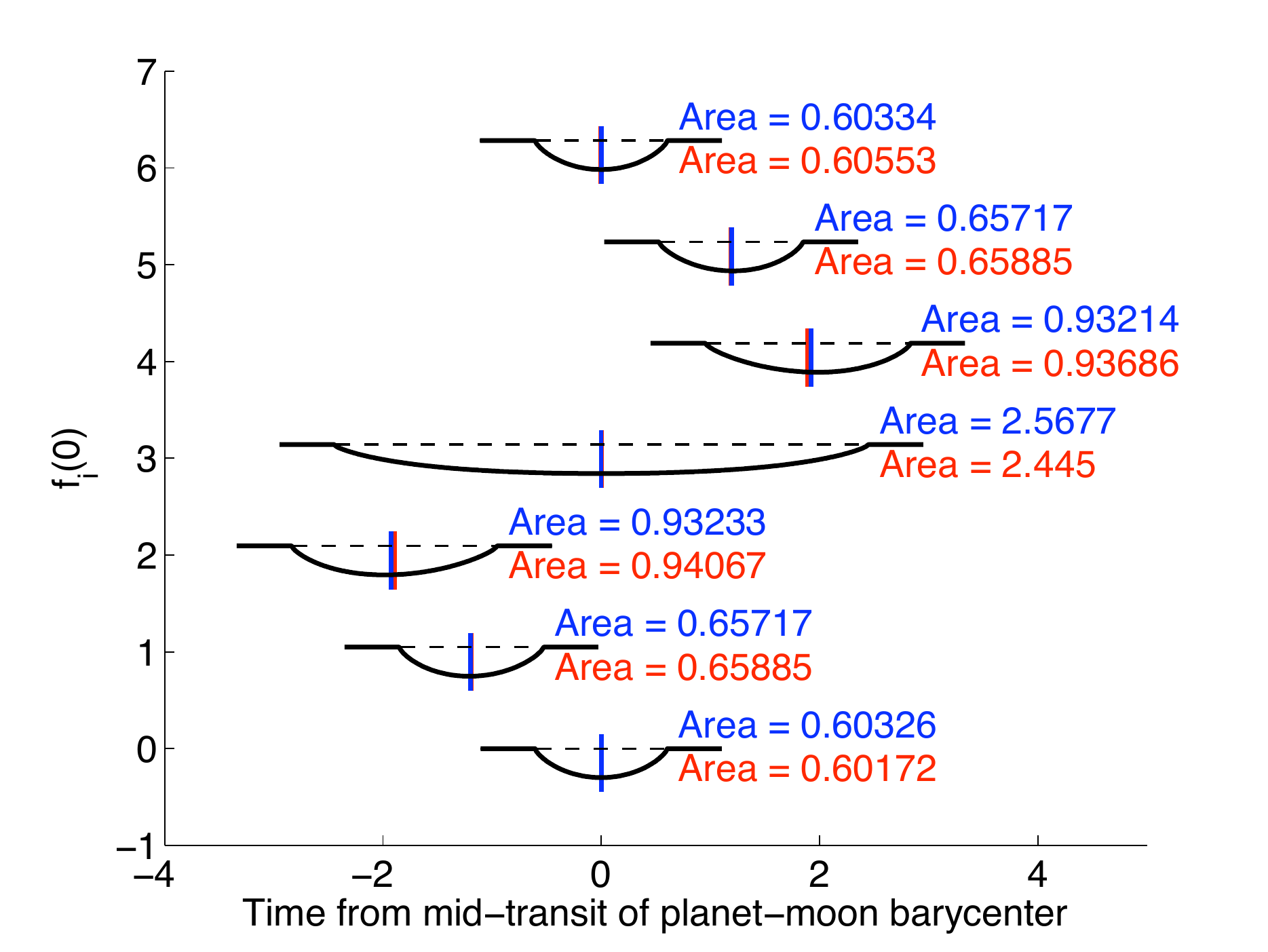}}
          \hspace{-0.8cm}
     \subfigure[$a_m = R_s$, $v_m/v_{tr} = 0.66$.]{
          \label{fig:er2858}
          \includegraphics[width=.65\textwidth]{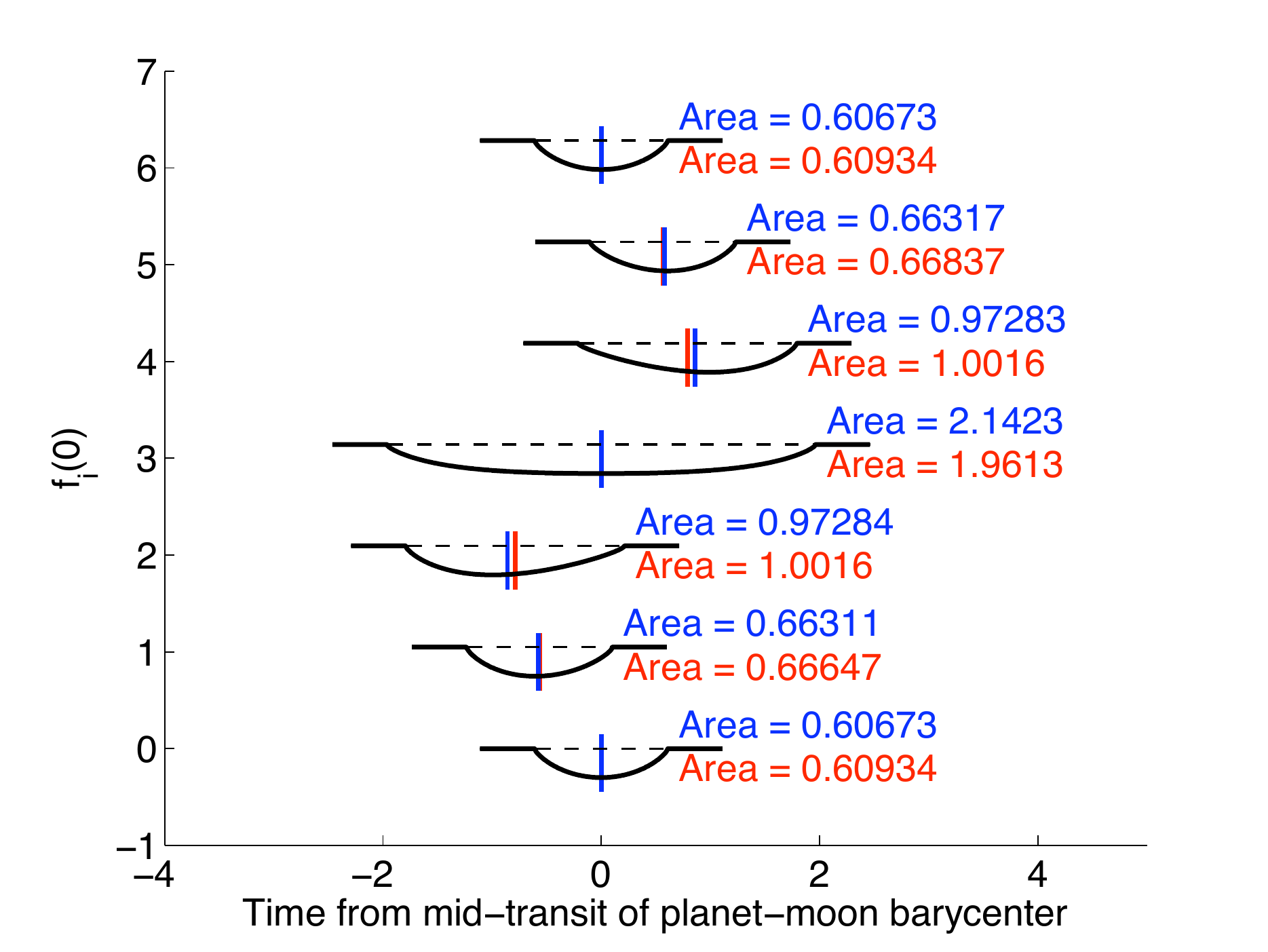}}}\\
          \noindent\makebox[\textwidth]{%
     \subfigure[$a_m = 0.5R_s$, $v_m/v_{tr} = 0.66$.]{
           \label{MoonTransitShape05B66}
           \includegraphics[width=.65\textwidth]{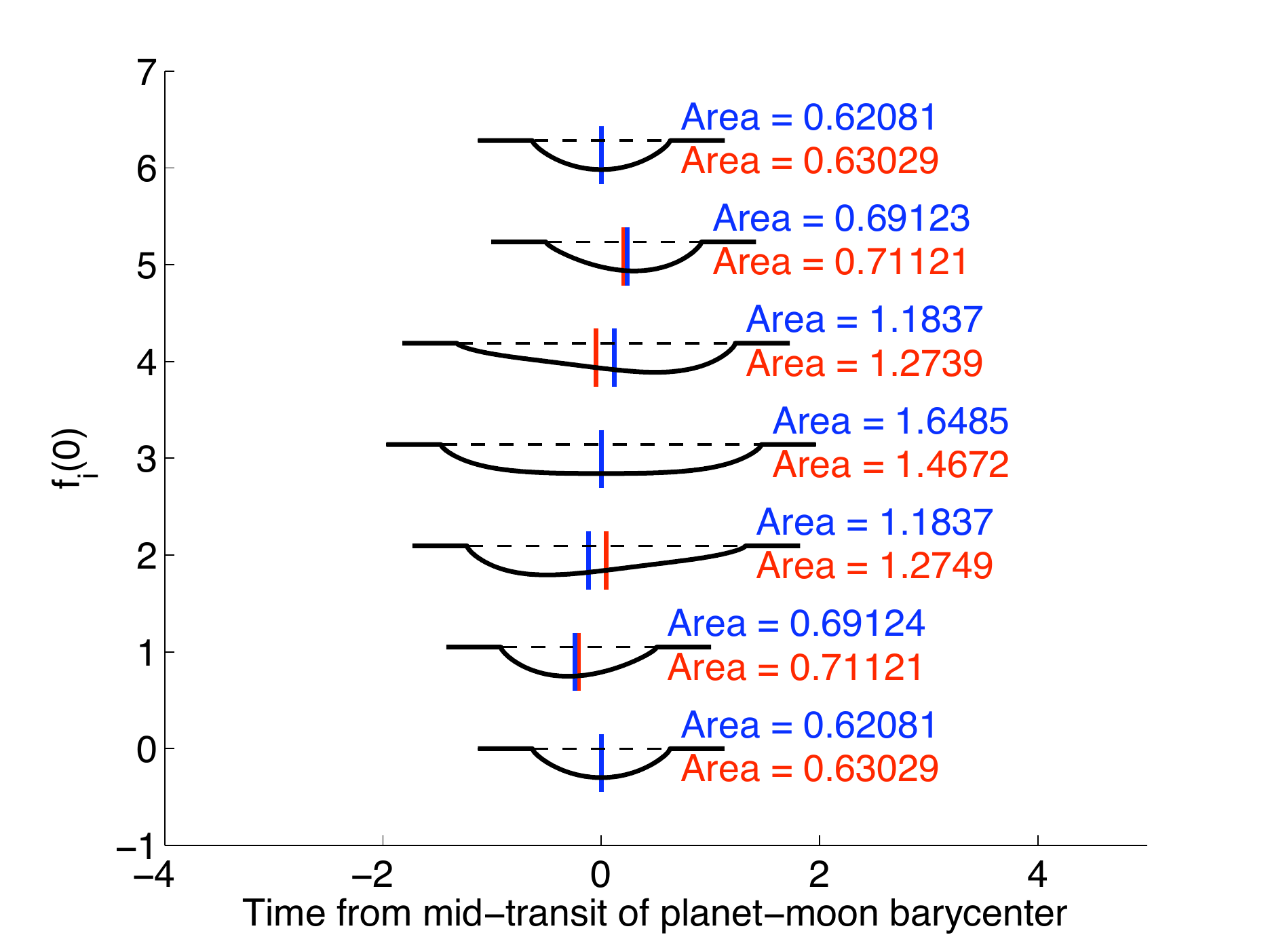}}
           \hspace{-0.8cm}
           \subfigure[$a_m = R_s$, $v_m/v_{tr} = 0.33$.]{
           \label{fig:cminusscalar2858}
           \includegraphics[width=.65\textwidth]{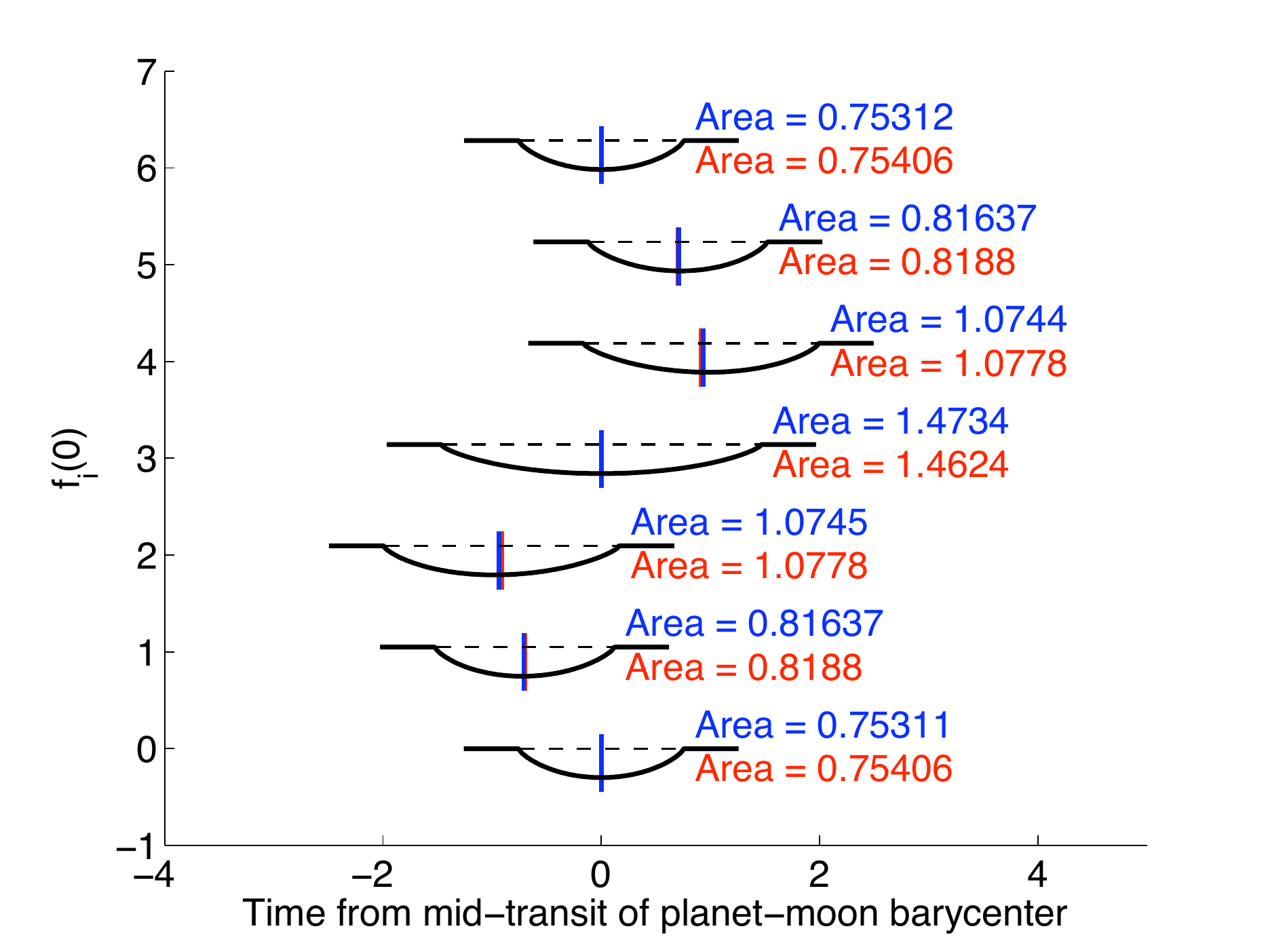}}}
     \caption[Diagram showing the shape of the transit light curve of the moon, along with the true and predicted values of $\tau_m$ and $A_m$ as a function of orbital phase $f_m(0)$ for four different combinations of moon orbital radius and velocity ratio.]{Diagram showing the shape of the transit light curve of the moon (black line) as a function of orbital phase $f_m(0)$ for four different combinations of moon orbital radius and velocity ratio.  The time axis is defined such that the planet-moon barycenter takes two time units to cross the face of the star, consequently on average the dip caused by the moon should be two time units long.  In addition, the depth of the dip is scaled such that the area of dip caused by a moon moving at the same velocity as the planet-moon barycenter would be equal to one.  The true and predicted values of $\tau_m$ are shown as blue and red bars on the transit light curves, while the true and predicted areas of the transit light curves are displayed to the right of each light curve, again in blue and red.}
     \label{MoonTransitShape}
\end{figure}

\subsection{Validation of approximation of uniform velocities}\label{Transit_Signal_Method_Validation}

To investigate the validity of the assumption that the planet and moon move with uniform velocity during transit, a set of simulated light curves were constructed with the aim of comparing the values of $\tau_m$ and $A_m$ calculated directly from the light curve, with the approximate values of $\tau_m$ and $A_m$ calculated using equations~\eqref{transit_signal_method_taumdef} and \eqref{transit_signal_method_Amdef}.  This simulation focussed specifically on the dip caused by the moon for a range of reasons which will be described in the next paragraph.  The set of conditions under which the assumption of uniform velocity during transit breaks down was investigated by considering an approximate expression for the maximum change in the velocity of a moon during transit.  In particular it was found that high values of $v_m/v_{tr}$ or $R_s/a_m$ could result in non-uniform moon (or planet) motion.  Consequently four simulations were run looking at the transit of the moon for a range of different values of $v_m/v_{tr}$ and $R_s/a_m$.  These issues, and the results of the simulations will be discussed in turn.

The dip due to the moon was chosen as the target for these simulations for three main reasons.  First, as was shown by \citep{Szaboetal2006} and as will be shown in section~\ref{Trans_TTV_Signal_CC_Form}, the value of $\Delta \tau$ is dominated by the effect of the moon.  Second, the orbital motion of the moon around the planet-moon barycenter is much more pronounced than the orbital motion of the planet around the planet-moon barycenter, and consequently any asymmetry in the transit light curve caused by this motion will also be more pronounced for the light curve corresponding to the moon as opposed to that corresponding to the planet.  Third, for the case where the moon is small with respect to the star ($R_m \ll R_s$) and the planet ($M_m \ll M_p$), the orbit of the moon becomes independent of it's mass\footnote{Recall that the orbital mean motion of the moon is given by $(G(M_p + M_m)/a_m^3)^{1/2}$.  Thus for $M_m \ll M_p$, $n_m \approx (G M_p/a_m^3)^{1/2}$.} and the shape of the light curve becomes independent of the radius while its depth becomes proportional to the cross-sectional area of the moon.\footnote{To see this, consider a generic transit light curve consisting of ingress, eclipse and egress.  The duration of ingress and egress is proportional to the radius of the moon (see appendix~\ref{IngressDur_App}), while the duration of the transit is of the order of $R_s$ (see section~\ref{Trans_Intro_Transtech}).  Consequently, in the limit that the moon is small, the transit light curve is dominated by the eclipse portion and the ingress and egress can be neglected.  In addition, as the moon becomes smaller, the region of star that it is blocking becomes more homogeneous.  As a result, in this regime changing the size of the moon does not affect the shape of the light curve, only the relative depth.}  In addition, for simplicity, it was decided to simulate the case of circular coplanar orbits.  Also, it was assumed that the planet and moon transited the central chord of the star as these systems suffer from the most extreme transit distortions, for example, due to the associated longer transit durations.  The light curve caused by a moon was modeled by using equations~\eqref{transit_signal_coord_xmdef} and \eqref{transit_signal_coord_ymdef} to determine the position of the moon as a function of time, and equation~\eqref{TraM-DescT-alphamdef} to determine the corresponding value of $\alpha_m$.  

In order to focus our investigation on the region of parameter space where the assumption of uniform velocities breaks down, and consequently select representative scenarios for the numerical simulations, we begin by estimating  the degree to which velocity of the moon across the plane of the sky changes during transit.  In particular, the maximum change in velocity which can occur during transit is equal to
\begin{equation}
\text{max}(\Delta v) \approx T_{tra} \times \text{max}(\dot{v}_m),
\end{equation}
where $T_{tra}$ is the transit duration, and $\text{max}(\dot{v}_m)$ is the maximum acceleration along the plane of the sky.  For an moon on a circular orbit, the maximum acceleration, and thus the maximum possible acceleration along the plane of the sky is given by
\begin{equation}
\text{max}(\dot{v}_m) \le \frac{v_m^2}{a_m}.
\end{equation}
In addition, the maximum value of $T_{tra}$ can be approximated using equation~\eqref{transit_intro_dur_cc_Ddef}.  Consequently, 
\begin{equation}
\text{max}(\Delta v) \le \frac{2R_s}{a_m} \frac{v_m}{v_{tr}} v_m
\end{equation}
Noting that $v_m$ is proportional to $(R_s/a_m)^{1/2}$ we have that for a given star (constant $R_s$)
\begin{equation}
\text{max}(\Delta v) \propto \left(\frac{R_s}{a_m}\right)^{3/2} \frac{v_m}{v_{tr}}.
\end{equation}
Consequently, the change in velocity for a given star depends on two ratios, $v_m/v_{tr}$ and $R_s/a_m$.  As a result of this dependance, moon light curves were simulated for a range of values of $v_m/v_{tr}$ and $R_s/a_m$.  As the analysis in this chapter is limited to $v_m/v_{tr} < 0.66$ as a result of the expansion that will be used to derive ingress and egress times, it was decided to investigate moon light curves using two different values of $v_m/v_{tr}$.  First, the case of $v_m/v_{tr} = 0.66$ was investigated as it corresponds to the worst case scenario that can be described by this analysis.  Second, the case of $v_m/v_{tr} = 0.33$ was investigated as a comparison case.  The selection of appropriate values of $a_m$ was informed by the regular satellites in the Solar System.  As these satellites are found between 0.27 (Ariel) and 2.7 (Callisto) solar radii from their hosts (see table~\ref{SSMoonsTable} ) it was decided to investigate the cases where $a_m$ is equal to $0.5R_s$, $R_s$ and $2R_s$.  These simulations were performed, and the resulting light curves are shown in figure~\ref{MoonTransitShape}.

While distortion of the light curves is evident (see figure~\ref{MoonTransitShape}), the position of the true first moment, compared to the value calculated from equation~\eqref{transit_signal_method_taumdef}, agree well for the majority of the orbital period of the moon for $v_m/v_{tr}$ up to 0.66 and $a_m/R_s \ge 1$.  For moons with $a_m/R_s < 1$, there is some disagreement.  However, for $a_m/R_s < 1$ the signal caused by the moon is small and so it is likely that moons for which the assumption of uniform motion does not apply, will also not be detectable.  In addition, as can be seen from figure~\ref{MoonTransitShape}, for the case of $a_m = 0.5R_s$, the magnitude of the predicted value of $\tau_m$ is always less than the magnitude of the true value.  Consequently, using these assumptions, we will still be able to place limits, albeit generous, on the population of these inner moons.  Finally, values of $A_m$ were calculated using equation~\eqref{transit_signal_method_Amdef} and all agree well with the values predicted for $A_m$ given by the simulation.  Consequently the effect of the non-uniform motion of the moon on the value of $\tau$ can be safely neglected, and equations~\eqref{transit_signal_method_taupdef}, \eqref{transit_signal_method_taumdef}, \eqref{transit_signal_method_Apdef} and \eqref{transit_signal_method_Amdef} used.

\section[Circular orbit aligned to the line-of-sight]{Circular planet orbit aligned to the line-of-sight}\label{Trans_TTV_Signal_CC}

\begin{figure}[tb]
\begin{center}
\includegraphics[height=2.55in,width=4.5in]{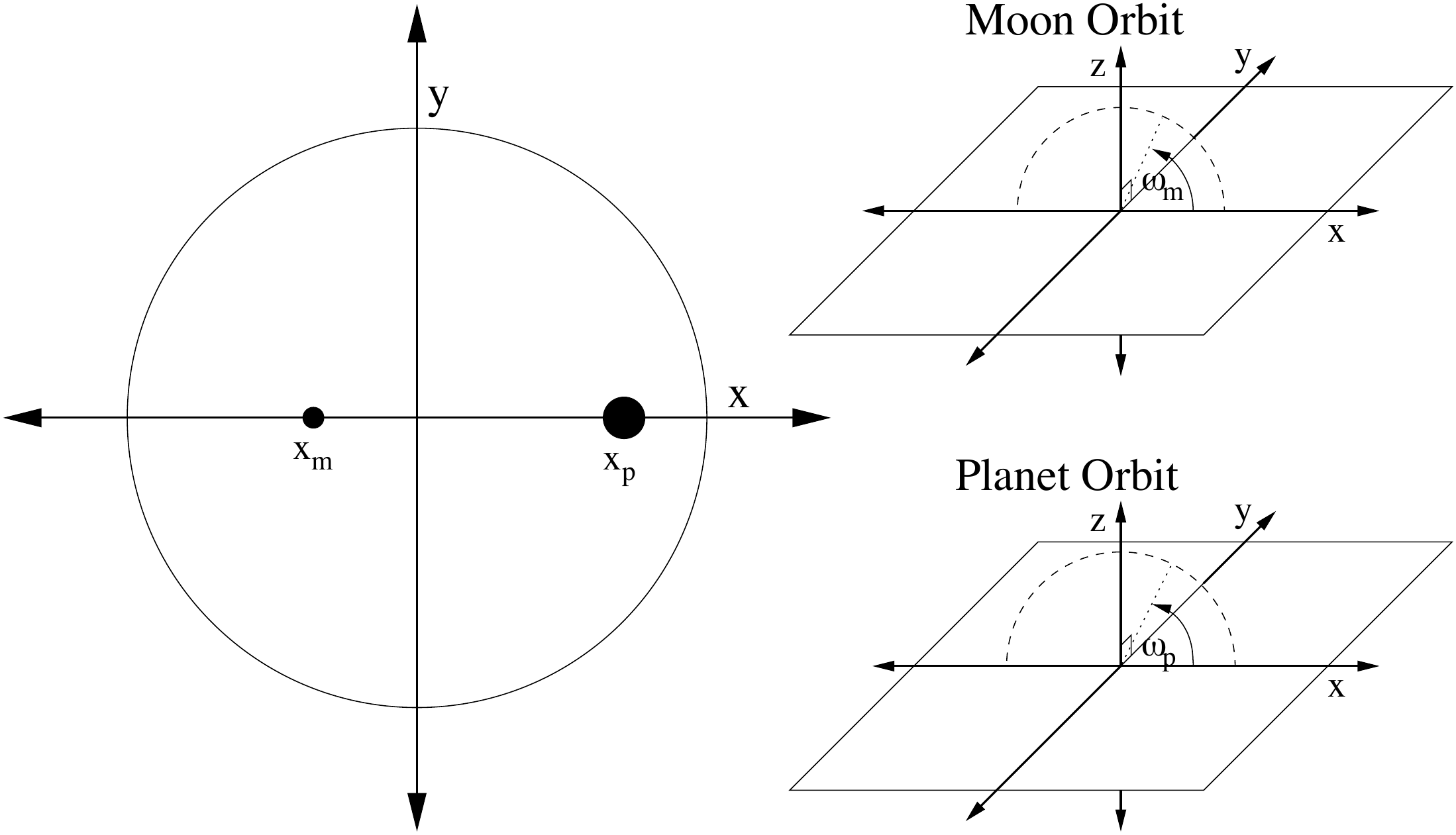}
\caption[Schematic diagram of the same form as figure~\ref{TransitSignalGenCoordSysRot} of the coordinate system for the case of circular coplanar orbits.]{Schematic diagram of the same form as figure~\ref{TransitSignalGenCoordSysRot} of the coordinate system for the case of circular coplanar orbits. In particular, it is assumed that $I_p = \pi/2$, $I_m = \pi/2$ and $\Omega_m = \Omega_p$. }
\label{TransitSignalCoordSysCC}
\end{center}
\end{figure}

Now that the preliminary work in describing a coordinate system and determining a method for deriving $\Delta \tau$ is complete, we can concentrate on determining the effect of the physical parameters of a given planet-moon system on $\Delta \tau$. We begin with the first, and simplest, of our three special cases, the case where both orbits are circular and coplanar, and both the planet and moon transit the central chord of the star.  These assumptions result in substantial simplification.  As both orbits are circular ($e_m = 0$ and $e_p = 0$), equations~\eqref{transit_signal_coord_Rdef} and \eqref{transit_signal_coord_rdef} simplify to $r_m = a_m$ and $r_p = a_p$, that is, both $r_m$ and $r_p$ are constant.  Similarly, equations~\eqref{transit_signal_coord_fidef} and \eqref{transit_signal_coord_Eidef} also simplify to $f_m  = n_m t + f_m(0)$, that is, the planet-moon pair progresses around its orbit with constant angular velocity.  Also, we assume that the planet and the moon's orbit are aligned with the line-of-sight, that is, $I_p = \pi/2$, $I_m = \pi/2$ and $\Omega_m = \Omega_p$ (see figure~\ref{TransitSignalCoordSysCC}).  Now we are in a position to use the coordinate system discussed in section~\ref{Transit_Signal_Coord} and the method described in section~\ref{Transit_Signal_Method} to investigate $\Delta \tau$ for this special case.  

The first stage in deriving $\Delta \tau$ is to determine $t_{in,p}$, $t_{in,m}$, $t_{eg,p}$ and $t_{eg,m}$, the times of ingress and egress for the transit of the planet and moon.  We begin with the equations describing $x_p$, $x_m$, $y_p$ and $y_m$, equations~\eqref{transit_signal_coord_xpdef} to \eqref{transit_signal_coord_ymdef}.  Using the simplifications described above, and ignoring the $y$ components as they are identically zero, the position of the planet and moon along the chord on which they are transiting is given by
\begin{equation}
x_p = a_p \cos(f_p + \omega_p) - \frac{M_m}{M_m + M_p}a_m \cos(n_m t + f_m(0) + \omega_m),\label{transit_signal_cc_xp}
\end{equation}
\begin{equation}
x_m = a_p \cos(f_p + \omega_p) + \frac{M_p}{M_m + M_p} a_m \cos(n_m t + f_m(0) + \omega_m), \label{transit_signal_cc_xm}
\end{equation}
where $x_p$, $x_m$, $\omega_p$ and $\omega_m$ are defined in figure~\ref{TransitSignalCoordSysCC} and where the first term represents the motion across the face of the star due to the motion of the planet-moon barycenter, and the second term represents the motion of the planet and moon about the planet-moon barycenter.  For the case where the orbital period of the planet is much longer than the transit, the motion of the planet can be accurately approximated by uniform motion.  Recalling that we do not expect short period planets to host large moons and expanding the first term of equations~\eqref{transit_signal_cc_xp} and \eqref{transit_signal_cc_xm} about $t=jT_p+t_0$, the central time the $j^{th}$ planetary transit would have occurred if there were no moon, gives
\begin{align}
x_p &= v_{tr} (t - (jT_p + t_0)) - \frac{M_m}{M_m + M_p}a_{m} \cos(n_m t + f_m(0) + \omega_m),\label{transit_signal_cc_xpexp}\\
x_m &= v_{tr} (t - (jT_p + t_0)) + \frac{M_p}{M_m + M_p}a_{m} \cos(n_m t + f_m(0) + \omega_m),\label{transit_signal_cc_xmexp}
\end{align}
where $v_{tr} = a_p n_p$ is the velocity of the planet-moon barycenter across the face of the star.  

An alternative way of viewing equations~\eqref{transit_signal_cc_xpexp} and \eqref{transit_signal_cc_xmexp} is that they implicitly define $t$ for a given $x_p$ or $x_m$.  In particular, this equation defines the ingress and egress times of the planet and moon when the values of $x_p$ and $x_m$ on the left hand side of equations~\eqref{transit_signal_cc_xpexp} and \eqref{transit_signal_cc_xmexp}, correspond to the limb of the star.  

The position of the limb of the star for this transit geometry is given by
\begin{align}
x_p &= \pm R_s,\\
x_m &= \pm R_s.
\end{align}
Consequently, the equations describing the ingress and egress times of the planet and moon's transit can be defined implicitly through
\begin{multline}
- R_s = v_{tr} (t_{in,p}- jT_p - t_0) \\- \frac{a_{m} M_m}{M_m + M_p} \cos(n_m t_{in,p} + f_m(0) +\omega_m),\label{transit_signal_cc_tinp}
\end{multline}
\begin{multline}
- R_s = v_{tr} (t_{in,m}- jT_p - t_0) \\+ \frac{a_{m} M_p}{M_m + M_p} \cos(n_m t_{in,m} + f_m(0) +\omega_m),\label{transit_signal_cc_tinm}
\end{multline}
\begin{multline}
R_s = v_{tr} (t_{eg,p}- jT_p - t_0) \\- \frac{a_{m} M_m}{M_m + M_p} \cos(n_m t_{eg,p} + f_m(0) +\omega_m),\label{transit_signal_cc_tegp}
\end{multline}
\begin{multline}
R_s = v_{tr} (t_{eg,m}- jT_p - t_0) \\+ \frac{a_{m} M_p}{M_m + M_p} \cos(n_m t_{eg,m} + f_m(0) +\omega_m),\label{transit_signal_cc_tegm}
\end{multline}
where $t_{in,p}$ and $t_{in,m}$ are the ingress times for the planet and moon and where $t_{eg,p}$ and $t_{eg,m}$ are the corresponding egress times.

The argument of the cosine function in equations~\eqref{transit_signal_cc_tinp}, \eqref{transit_signal_cc_tinm}, \eqref{transit_signal_cc_tegp} and \eqref{transit_signal_cc_tegm} is a measure of the position of the moon around its orbit during planetary ingress, moon ingress, planetary egress and moon egress respectively.  Setting $\theta = f_m + \omega_m + \pi/2$, we have that $\theta_{in,p} = n_m t_{in,p} + f_m(0) +\omega_m + \pi/2$, $\theta_{eg,p} = n_m t_{eg,p} + f_m(0)+\omega_m + \pi/2$, $\theta_{in,m} = n_m t_{in,m} + f_m(0)+\omega_m + \pi/2$ and $\theta_{eg,m} = n_m t_{eg,m} + f_m(0)+\omega_m + \pi/2$.  Substituting these expressions into the above equations and rearranging gives
\begin{align}
f_m(0) &+\omega_m + \frac{\pi}{2}+ n_m (jT_p + t_0) -\frac{n_m R_{s}}{v_{tr}} = \theta_{in,p}  - \frac{v_p}{v_{tr}} \sin(\theta_{in,p}), \label{transit_signal_cc_tinp2}\\
f_m(0) &+ \omega_m + \frac{\pi}{2}+  n_m (jT_p + t_0) -\frac{n_m R_{s}}{v_{tr}} = \theta_{in,m} + \frac{v_m}{v_{tr}} \sin(\theta_{in,m}), \label{transit_signal_cc_tinm2}\\
f_m(0) &+ \omega_m + \frac{\pi}{2}+  n_m (jT_p + t_0) + \frac{n_m R_{s} }{v_{tr}} = \theta_{eg,p}  - \frac{v_p}{v_{tr}} \sin(\theta_{eg,p}), \label{transit_signal_cc_tegp2}\\
f_m(0) &+ \omega_m + \frac{\pi}{2}+  n_m (jT_p + t_0) + \frac{n_m R_{s}}{v_{tr}} = \theta_{eg,m} + \frac{v_m}{v_{tr}} \sin(\theta_{eg,m}), \label{transit_signal_cc_tegm2}
\end{align}
where $v_p$ and $v_m$, the velocity of the planet and moon about the planet-moon barycenter are defined as
\begin{align}
v_p &= n_m a_{m}\frac{M_m}{M_{pm}},\\
v_m &= n_m a_{m}\frac{M_p}{M_{pm}},
\end{align}
where $M_{pm} = M_p + M_m$.

Equations~\eqref{transit_signal_cc_tinp2}, \eqref{transit_signal_cc_tegp2}, \eqref{transit_signal_cc_tinm2}, and \eqref{transit_signal_cc_tegm2} can be written as
\begin{equation}
\Phi = \theta_{cc} + B \sin(\theta_{cc}),\label{transit_signal_cc_AB}
\end{equation}
where the subscript ``cc" implies that the orbits are circular and coplanar, where $\Phi$ and $B$ are known constants, and where $B$ is the ratio of the velocity of the planet or moon around their barycenter to the velocity of the barycenter around the star.  For reference, the $\Phi$ and $B$ corresponding to each of the four equations are given in table~\ref{ABTable}.

Now is a good time to pause and take stock.  Equation~\eqref{transit_signal_cc_AB} exactly describes $\theta_{cc}$ in terms of $\Phi$ and $B$ for all values of $\Phi$ and $B$.  While we could solve equation~\eqref{transit_signal_cc_AB} numerically for a grid of $\Phi$ and $B$ values representing all the values of  $R_s$, $n_m$, $f_m(0) +\omega_m$ and $v_{tr}$ of interest, this approach is not optimal for three reasons.  First, a numerical approach may lead to canceling errors which would not occur if a more analytic approach were employed.  For example, consider the dependance of the detection threshold on $\hat{A}_p + \hat{A}_m$, a quantity dominated by the size of the planet.  As will be found in sections~\ref{Trans_TTV_Signal_CC_Form_smallB} and \ref{transit_noise_method_method} both $\Delta \tau$ and $\epsilon_j$ are inversely proportional to $\hat{A}_p + \hat{A}_m$.  Consequently, when the ratio of the amplitude of $\Delta \tau$ and $\epsilon_j$ is formed to determine the detection threshold, as will be done in chapter~\ref{Trans_Thresholds}, it should be independent of $\hat{A}_p + \hat{A}_m$, and thus planetary radius.  However, if a numerical approach was employed, we would find that $\Delta \tau$ was approximately proportional to $\hat{A}_p + \hat{A}_m$ and consequently that the ratio between $\Delta \tau$ and $\epsilon_j$ may not be independent of $\hat{A}_p + \hat{A}_m$.  As the difference between a slight dependance of $\Delta \tau$ on $\hat{A}_p + \hat{A}_m$ and no dependance is scientifically interesting, the analytic approach is preferred.  Second, the expressions derived to evaluate $\Delta \tau$ assume that either $v_m/v_{tr}$ is small, or $R_s/a_m$ is small.  There is little point obtaining precise numerical values for $\theta_{cc}$ if they are to be used in an approximate expression, and consequently will not yield more precise values of $\Delta \tau$.  Third, for the case where the moon orbit is no longer circular and coplanar, the form of equation~\eqref{transit_signal_cc_AB} and the form of $\Phi$ and $B$ will change.  Consequently a new numerical grid will have to be evaluated for each new moon orbit.  Alternatively, as demonstrated in appendix~\ref{EccMoon_App}, an analytic method can be extended to investigate different types of moon orbits without evaluating numerical grids of $\Phi$ and $B$ values.  As a result of these reasons, the approach that is used, is to approximate the solution of equation~\eqref{transit_signal_cc_AB} using an analytic expansion.  In particular, the expansion that was been selected is most accurate for small values of $B$, and can be extended to other types of moon orbits.

\begin{table}[tb]
	\begin{center}
  \begin{tabular}{l|c|c}
 $X$ & $\Phi_X$  & $B_X$ \\
  \hline
  ${in,p}$  & $f_m(0) + \omega_m + \frac{\pi}{2} +  n_m(jT_p + t_0) -\frac{n_m R_s}{v_{tr}}$ & $ - \frac{v_p}{v_{tr}}$ \\
  ${in,m}$  & $f_m(0) + \omega_m + \frac{\pi}{2}+  n_m(jT_p + t_0) -\frac{n_m R_s}{v_{tr}}$ & $\frac{v_m}{v_{tr}}$ \\
  ${eg,p}$  & $f_m(0) + \omega_m + \frac{\pi}{2}+  n_m(jT_p + t_0) + \frac{n_m R_s}{v_{tr}}$ & $- \frac{v_p}{v_{tr}}$ \\
  ${eg,m}$  & $f_m(0) + \omega_m + \frac{\pi}{2}+  n_m(jT_p + t_0) + \frac{n_m R_s}{v_{tr}}$ & $\frac{v_m}{v_{tr}}$\\
  \end{tabular}\\
 \caption{The values of $\Phi$ and $B$ corresponding to equations~\eqref{transit_signal_cc_tinp2} to \eqref{transit_signal_cc_tegm2}.}
 \label{ABTable}
 \end{center}
 \end{table}
 
We begin by noting that equation~\eqref{transit_signal_cc_AB} is mathematically equivalent to Kepler's equation
\begin{equation}
M = E + e\sin E, \label{transit_signal_cc_KepEq}
\end{equation}
where M, the mean anomaly, is equivalent to $\Phi$, and $E$, the eccentric anomaly, is equivalent to $\theta_{cc}$ and $e$, the orbital eccentricity, is equivalent to $B$.  As equation~\eqref{transit_signal_cc_KepEq} can be expanded to give an explicit expression for $E$ in terms of $M$ and $e$, equation~\eqref{transit_signal_cc_AB} can be expanded to give an expression for $\theta_{cc}$ in terms of $\Phi$ and $B$.

Following \citet[][p39]{Murrayetal1999} and writing $\sin E$ as a Fourier series in $M$, equation~\eqref{transit_signal_cc_KepEq} can be written in terms of Bessel functions, as follows
\begin{equation}
E = M + \sum_{k =1}^\infty \frac{2}{k} J_k(ke) \sin(nM),\label{transit_signal_cc_Edef}
\end{equation}
where $J_k(x)$ is a Bessel function of the first kind \citep[e.g. ][p. 150]{Spiegeletal1999}, defined as
\begin{equation}
J_k(x) = \sum_{j=0}^\infty \frac{(-1)^j (x/2)^{k+2j}}{j! \Gamma(k+j+1)},\label{transit_signal_cc_Besseldef}
\end{equation}
where $\Gamma$ is the gamma function.  For reference, expansions to order $x^5$ for Bessel functions with $k = 1\ldots5$ are presented in table~\ref{BesselTable}.  While the solution of equation~\eqref{transit_signal_cc_KepEq} is true for all values of $M$ and $e$, the expansion given by equation~\eqref{transit_signal_cc_Edef} only converges for $e<0.6627$ \citep[][p. 510]{Hagihara1970} and can consequently only be used when $e<0.6627$.  By analogy, we also have that
\begin{equation}
\theta_{cc} = \Phi + \sum_{k =1}^\infty \frac{2}{k} J_k(kB) \sin(k\Phi),\label{transit_signal_cc_tccdef}
\end{equation}
for cases where $|B| < 0.6627$.

\begin{table}[tb]
\begin{center}
  \begin{tabular}{ccr}
  Bessel Function   & Expansion to $\mathcal{O}(x^5)$   \\
  \hline
  $J_1(x)$  & $\frac{1}{2}x -\frac{1}{16}x^3 +\frac{1}{384}x^5 + \mathcal{O}(x^7)$  \\
  $J_2(x)$  & $\frac{1}{8}x^2 - \frac{1}{96}x^4 + \mathcal{O}(x^6)$  \\
  $J_3(x)$  & $\frac{1}{48}x^3 -\frac{1}{768}x^5 + \mathcal{O}(x^7)$ \\
  $J_4(x)$  & $\frac{1}{384}x^4 + \mathcal{O}(x^6)$ \\
  $J_5(x)$  & $\frac{1}{3840}x^5 +  \mathcal{O}(x^7)$ \\
  \end{tabular}\\
 \caption{List of the Taylor expansion of the first five Bessel functions up to order $x^5$.}
 \label{BesselTable}
 \end{center}
 \end{table}

\begin{figure}
     \centering
     \subfigure[$M_p/M_s = 10^{-2}$]{
          \label{fig:dl2858}
          \includegraphics[width=.48\textwidth]{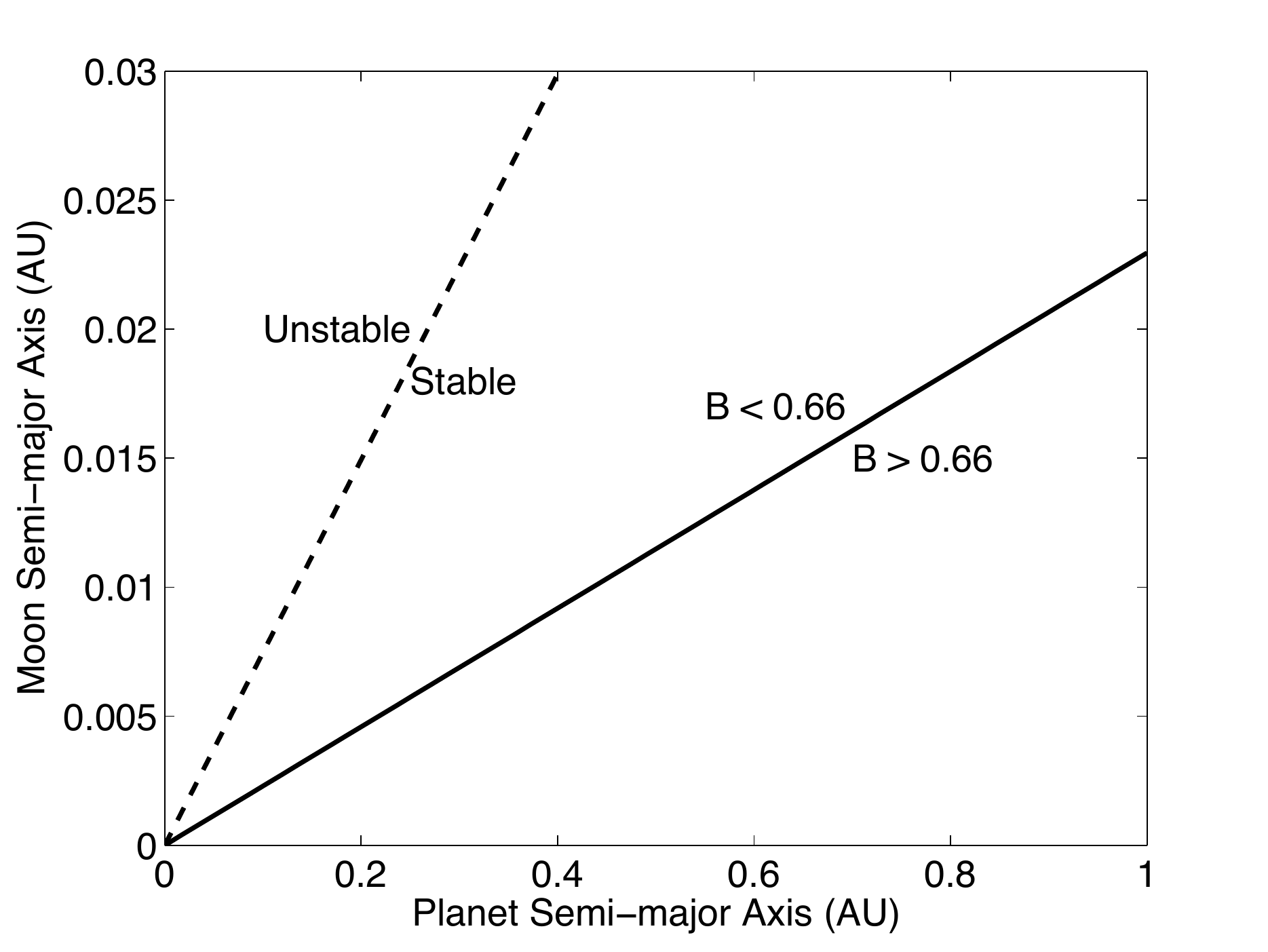}}
     \subfigure[$M_p/M_s = 10^{-3}$]{
          \label{fig:er2858}
          \includegraphics[width=.48\textwidth]{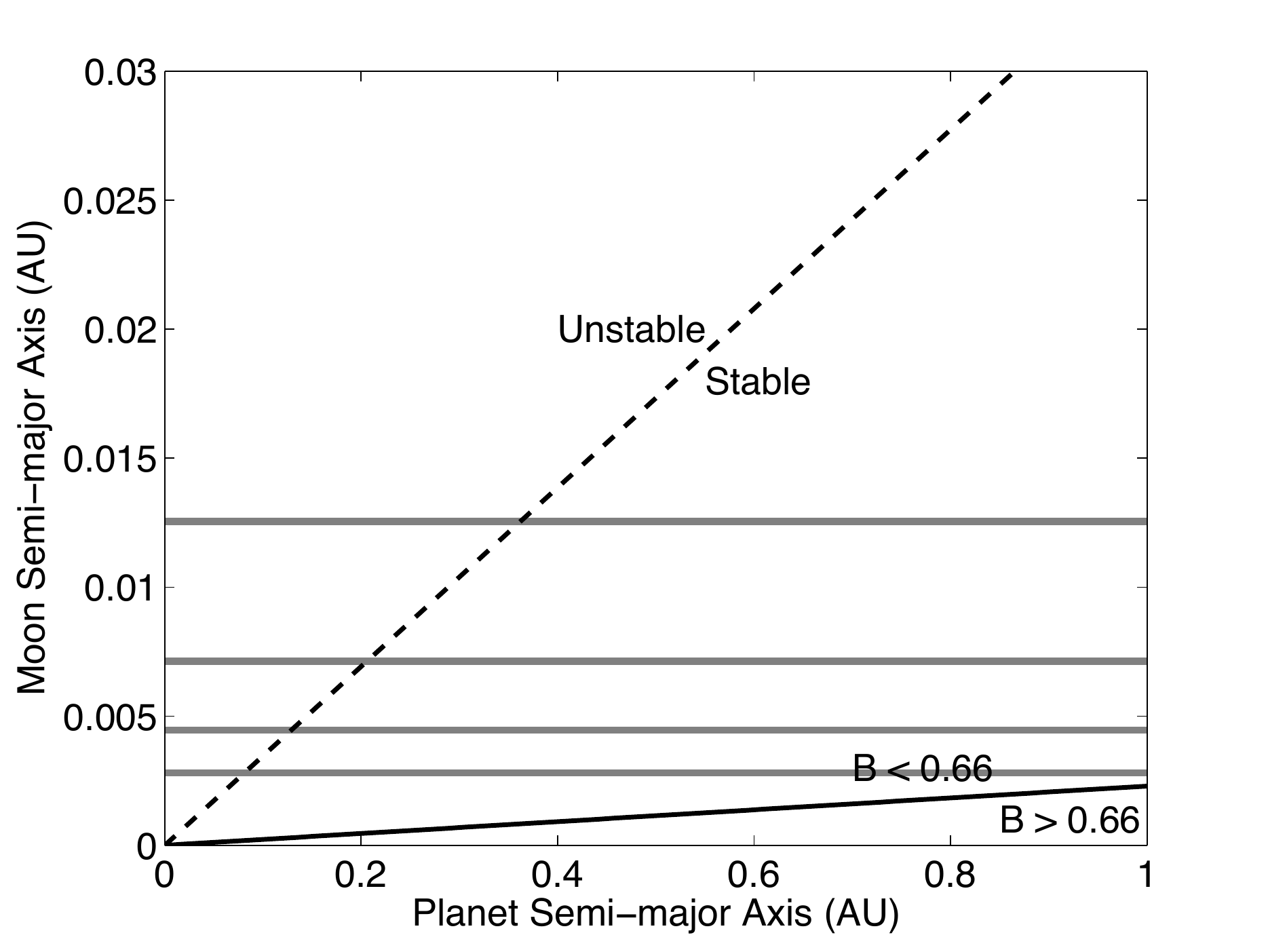}}\\
     \subfigure[$M_p/M_s = 10^{-4}$]{
           \label{fig:cminusscalar2858}
           \includegraphics[width=.48\textwidth]
                {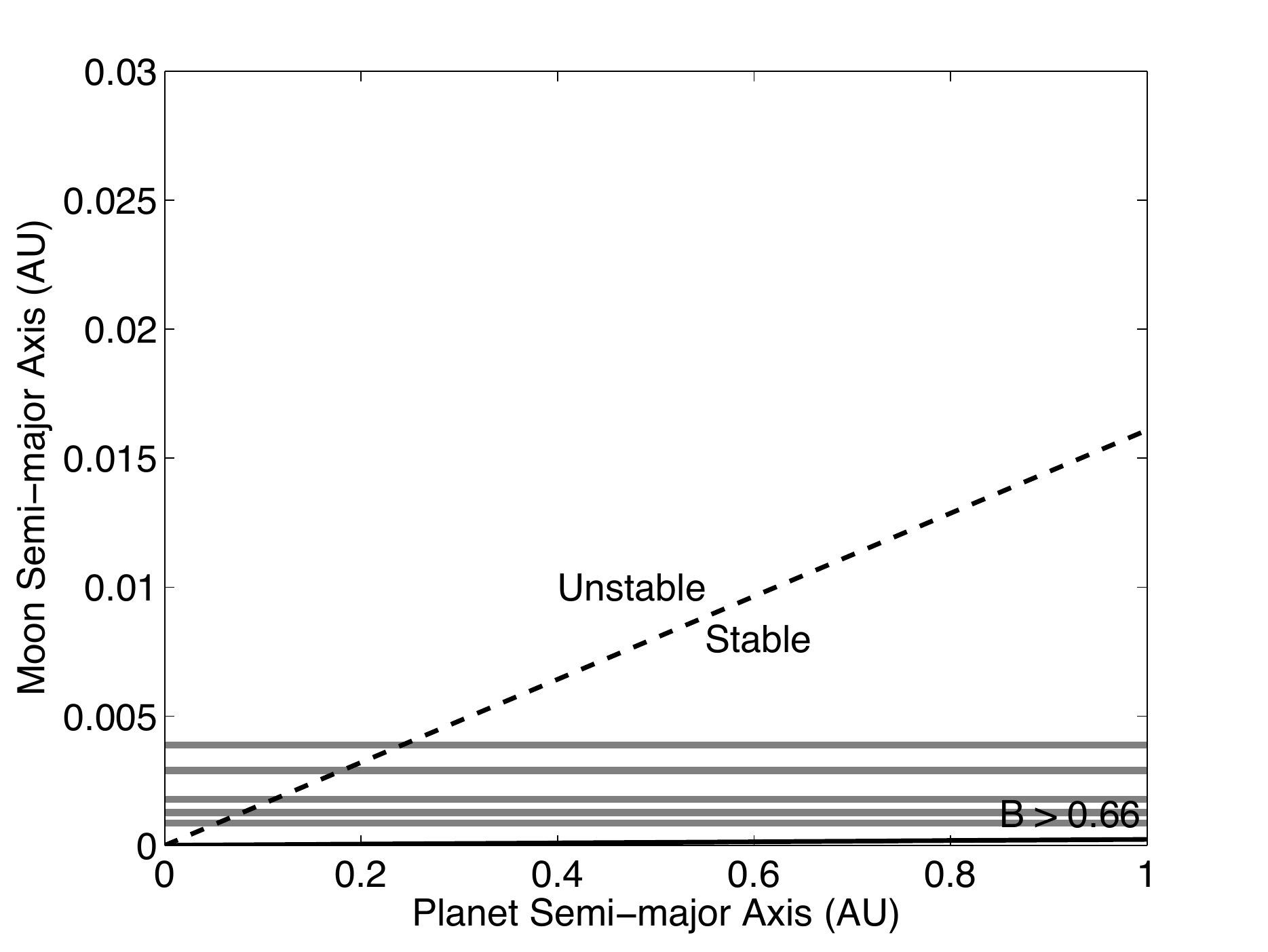}}
     \caption[Diagram showing the position of the three body instability boundary (dashed line) and the boundary of the region where the expansion given by equation~\eqref{transit_signal_cc_tccdef} fails (solid line), for three values of the planet star mass ratio.]{Diagram showing the position of the three body instability boundary (dashed line) and the boundary of the region where the expansion given by equation~\eqref{transit_signal_cc_tccdef} fails (solid line), for three values of the planet star mass ratio.  For comparison, for the case of $M_p/M_s = 10^{-3}$ and $M_p/M_s = 10^{-4}$, the semi-major axis of the regular satellites of Jupiter and Uranus respectively are also shown (grey lines).}
     \label{ExpFail}
\end{figure}

There are two possible issues with using equation~\eqref{transit_signal_cc_tccdef} in its present form to derive $t_{in,p}$, $t_{eg,p}$, $t_{in,m}$ and $t_{eg,m}$ and thus $\Delta \tau$.  The first issue is that the expansion may fail, that is, that $|B| > 0.6627$.  The second issue is that the expression, in particular, the infinite sum, is functionally complex.  Fortunately, both these problems are simply resolved as, first, excluding the region in which the expansion fails does not exclude many physically realistic moons, and second that most of the terms in the infinite sum can (and will) be neglected as they do not substantially increase the accuracy of the calculated value of $\Delta \tau$.  These issues will be discussed in turn.

As $|B|$ can correspond to both $v_m/v_{tr}$ and $v_p/v_{tr}$, and $v_m \ge v_p$ (as $M_m \le M_p$), the requirement that $|B| < 0.6627$ corresponds to the limit of $v_m/v_{tr} < 0.6627$.  Fortunately, this limit does not strongly restrict the range of detectable moons, for example, for the Jupiter-Callisto moon system were at 1AU and transiting a Sun-like star, the velocity ratio would be approximately 0.28.  The regions where this expansion fails are shown in figure~\ref{ExpFail} for the cases where $M_p/M_s=10^{-2}$, $M_p/M_s=10^{-3}$ and $M_p/M_s=10^{-4}$.  As can be seen, the motion of the moon is well described by this expansion for most of the range of planet masses and radii which are likely to be detected. 

In addition to the region where the expansion fails, there is also the region where the assumption that the planet and moon have constant velocities during transit leads to substantial inaccuracies in the calculated value of $\Delta \tau$.  As mentioned in section~\ref{Transit_Signal_Method_Validation}, the assumption of constant velocities is most accurate for planet moon pairs with small values of $v_m/v_{tr}$ (that is, $B$) or $a_m/R_s$.  In addition, as also mentioned in section~\ref{Transit_Signal_Method_Validation}, as it is difficult to detect moons with small values of $a_m/R_s$, $B$ is a good indicator of when the assumption of uniform velocities is likely to fail. So for the case where $B$ is small, only the lowest order terms in $B$ are required (as $B$ is small), and for the case where $B$ is not small, additional terms do not substantially increase the accuracy in the final calculated value of $\Delta \tau$ as the approximation of constant velocities becomes increasingly inaccurate (see figure~\ref{TauAgreement} in section~\ref{Trans_TTV_Signal_CC_Form_largeB}).  As a result, the infinite sum can be approximated by the first couple of terms, and the Bessel functions can be approximated by the first couple of terms in their Tailor expansion.\footnote{For example, to first order in $B$, only the $k=1$ term in equation~\eqref{transit_signal_cc_tccdef} contributes, so only it has be be retained.  For the case where equation~\eqref{transit_signal_cc_tccdef} is taken to order $B^2$ only the $k=1$ and $k=2$ terms contribute.}  In particular, in section~\ref{Trans_TTV_Signal_CC_Form}, the section where expressions for $\Delta \tau$ are calculated, $\Delta \tau$ will be calculated to first order (section~\ref{Trans_TTV_Signal_CC_Form_smallB}) and second order (section~\ref{Trans_TTV_Signal_CC_Form_largeB}) in $B$.  

Continuing, in order to derive expressions for $t_{in,p}$, $t_{eg,p}$, $t_{in,m}$ and $t_{eg,m}$, and consequently $\Delta \tau$, equation~\eqref{transit_signal_cc_tccdef}, the expression for $\theta_{cc}$ in terms of $\Phi$ and $B$ must be reformatted to give expressions for $t_{in,p}$, $t_{eg,p}$, $t_{in,m}$ and $t_{eg,m}$.  Using equation~\eqref{transit_signal_cc_tccdef}, and the values in table~\ref{ABTable}, the ingress and egress times of the moon can be written in terms of the planet and moon masses and orbital parameters.  This gives
\begin{align}
t_{in,p} &=  jT_p + t_0 -\frac{R_s}{v_{tr}} + \frac{1}{n_m }\sum_{k =1}^\infty \frac{2}{k} J_k(kB_{in,p}) \sin(k\Phi_{in,p}),\label{transit_signal_cc_Besstinp}\\
t_{in,m} &=  jT_p + t_0 - \frac{R_s}{v_{tr}} + \frac{1}{n_m }\sum_{k =1}^\infty \frac{2}{k} J_k(kB_{in,m}) \sin(k\Phi_{in,m}),\label{transit_signal_cc_Besstinm}\\
t_{eg,p} &= jT_p + t_0 + \frac{R_s}{v_{tr}} + \frac{1}{n_m } \sum_{k =1}^\infty \frac{2}{k} J_k(kB_{eg,p}) \sin(k\Phi_{eg,p}),\label{transit_signal_cc_Besstegp}\\
t_{eg,m} &=  jT_p + t_0 + \frac{R_s }{v_{tr}} + \frac{1}{n_m }\sum_{k =1}^\infty \frac{2}{k} J_k(kB_{eg,m}) \sin(k\Phi_{eg,m}),\label{transit_signal_cc_Besstegm}
\end{align}
where we have left the $\Phi$ and $B$ terms in the sum for readability.  The form of equations~\eqref{transit_signal_cc_Besstinp} to \eqref{transit_signal_cc_Besstegm} is reassuring in that it is exactly what we would expect.  For example, equation~\eqref{transit_signal_cc_Besstegp} indicates that the planet's time of egress is the sum of the the time we would expect the planet to reach the middle of the stellar disk ($jT_p + t_0$), the time it would take a lone planet to travel from the center to the limb of the star ($R_s/v_{tr}$), and an additional modifying term (the infinite sum), indicating the effect of the moon.

Now that expressions for $t_{in,p}$, $t_{eg,p}$, $t_{in,m}$ and $t_{eg,m}$ have been determined, equations~\eqref{transit_signal_method_taupdef}, \eqref{transit_signal_method_taumdef}, \eqref{transit_signal_method_Apdef} and \eqref{transit_signal_method_Amdef} from section~\ref{Transit_Signal_Method_Implementation} can be used to write expressions for $\tau_p$, $\tau_m$, $A_p$ and $A_m$, and consequently derive $\Delta \tau$.  For simplicity, these expressions will be calculated in the following section, after the equations have been reduced to the appropriate order.

\subsection{Form of $\Delta \tau$}\label{Trans_TTV_Signal_CC_Form}

Now that expressions for $\tau_p$, $\tau_m$, $A_p$ and $A_m$ can be derived, $\Delta \tau$ will be investigated.  As equations~\eqref{transit_signal_cc_Besstinp} to \eqref{transit_signal_cc_Besstegm} are quite complex, in order to build mathematical intuition, $\Delta \tau$ will be investigated for two cases. First, $\Delta \tau$ will be investigated for the case where motion during transit is negligible.  Second, $\Delta \tau$ will be investigated for the case where motion during transit is non-negligible, but where the largest ratio of the velocities, $v_m/v_{tr}$ is less that 0.66.  These results will then be combined to provide a qualitative description of the behaviour produced.

\subsubsection{Case where $v_m/v_{tr} \ll 1$}\label{Trans_TTV_Signal_CC_Form_smallB}

Expanding equations~\eqref{transit_signal_cc_Besstinp}, \eqref{transit_signal_cc_Besstinm}, \eqref{transit_signal_cc_Besstegp} and \eqref{transit_signal_cc_Besstegm} to first order in $v_p/v_{tr}$ or $v_m/v_{tr}$ gives
\begin{multline}
t_{in,p} =  jT_p + t_0 -\frac{R_s}{v_{tr}} \\- \frac{1}{n_m } \frac{v_p}{v_{tr}} \sin\left(f_m(0) + \omega_m + \frac{\pi}{2} +  n_m(jT_p + t_0) -\frac{n_m R_s}{v_{tr}}\right),\label{transit_signal_cc_sB_tinp}
\end{multline}
\begin{multline}
t_{in,m} =  jT_p + t_0 - \frac{R_s}{v_{tr}} \\+ \frac{1}{n_m }\frac{v_m}{v_{tr}} \sin\left(f_m(0) + \omega_m + \frac{\pi}{2}+  n_m(jT_p + t_0) -\frac{n_m R_s}{v_{tr}}\right),\label{transit_signal_cc_sB_tinm}
\end{multline}
\begin{multline}
t_{eg,p} = jT_p + t_0 + \frac{R_s }{v_{tr}} \\- \frac{1}{n_m }\frac{v_p}{v_{tr}} \sin\left(f_m(0) + \omega_m + \frac{\pi}{2}+  n_m(jT_p + t_0) + \frac{n_m R_s}{v_{tr}}\right),\label{transit_signal_cc_sB_tegp}\\
\end{multline}
\begin{multline}
t_{eg,m} =  jT_p + t_0 + \frac{R_s }{v_{tr}} \\+ \frac{1}{n_m }\frac{v_m}{v_{tr}} \sin\left(f_m(0) + \omega_m + \frac{\pi}{2}+  n_m(jT_p + t_0) + \frac{n_m R_s}{v_{tr}}\right).\label{transit_signal_cc_sB_tegm}
\end{multline}
Then, substituting equations~\eqref{transit_signal_cc_Besstinp} to \eqref{transit_signal_cc_Besstegm} into equations~\eqref{transit_signal_method_taupdef}, \eqref{transit_signal_method_taumdef}, \eqref{transit_signal_method_Apdef} and \eqref{transit_signal_method_Amdef} and retaining terms to order  $v_p/v_{tr}$ or $v_m/v_{tr}$ gives
\begin{equation}
\tau_p =  jT_p + t_0 - \frac{1}{n_m } \frac{v_p}{v_{tr}} \cos\left(\frac{n_m R_s}{v_{tr}}\right) \\ \times \cos\left(f_m(t_0) + \omega_m +  n_m jT_p \right),  
\end{equation}
\begin{equation}
\tau_m =  jT_p + t_0 + \frac{1}{n_m }\frac{v_m}{v_{tr}} \cos \left(\frac{n_m R_s}{v_{tr}}\right) \cos\left(f_m(t_0) + \omega_m +  n_m jT_p \right),
\end{equation}
\begin{equation}
A_p = \hat{A}_p - \frac{1}{n_m }\frac{v_p}{v_{tr}}\frac{v_{tr}}{R_s}\hat{A}_p \sin \left(\frac{n_m R_s}{v_{tr}}\right) \sin\left(f_m(t_0) + \omega_m +  n_m jT_p\right),\label{transit_signal_cc_BessAp}
\end{equation}
\begin{equation}
A_m = \hat{A}_m + \frac{1}{n_m }\frac{v_m}{v_{tr}}  \frac{v_{tr}}{R_s} \hat{A}_m \sin \left( \frac{n_m R_s}{v_{tr}}\right) \sin\left(f_m(t_0) + \omega_m +  n_m jT_p \right),\label{transit_signal_cc_BessAm}
\end{equation}
where we have used the identity that $f_m(t_0) = f_m(0) + n_m t_0$.

Substituting this into equation~\eqref{transit_intro_ground_deltaudef} and neglecting all terms of order $v_p^2/v_{tr}^2$, $v_mv_p/v_{tr}^2$ or $v_m^2/v_{tr}^2$ and above gives
\begin {multline}
\Delta \tau = \frac{\hat{A}_m M_p -\hat{A}_p M_m}{\hat{A}_{pm} M_{pm}}  \frac{a_m}{v_{tr}} \cos \left(\frac{n_m R_s}{v_{tr}}\right) \\
\times \cos\left(f_m(t_0) + \omega_m +  n_m jT_p \right),\label{transit_signal_cc_form_lB}
\end{multline}
where we define $M_{pm} = M_p + M_m$ and $\hat{A}_{pm} = \hat{A}_p + \hat{A}_m$.

Consequently, for the case of negligible motion of the planet and moon during the transit, $\Delta \tau$ is given by equation~\eqref{transit_signal_cc_form_lB}.  This equation can be further simplified by investigating the quantity $(\hat{A}_p M_m - \hat{A}_m M_p)/\hat{A}_{pm} M_{pm}$.  We begin by noting that $\hat{A}_m/\hat{A}_p \approx R_m^2/R_p^2$, $M_p = 4\pi/3 R_p^3 \rho_p$ and $M_m = 4\pi/3 R_m^3 \rho_m$.  Consequently the ratio of the size of the first term in the numerator to the second is equal to
\begin{equation}
\frac{\hat{A}_m M_p}{\hat{A}_p M_m} \approx \frac{R_p}{R_m} \frac{\rho_p}{\rho_m}.
\end{equation}
For the case where the planet is dominated by solids, for example the terrestrial planets, the ratio $\rho_p/\rho_m$ is likely to be close to one and the ratio $R_p/R_m$ is likely to be large.  Consequently the term $\hat{A}_m M_p$ will be much larger than $\hat{A}_p M_m$.  For the case where the planet is dominated by gas, for example, the gas giants, the ratio $R_p/R_m$ will be very large, so while the ratio of the densities could be less than one, $\hat{A}_m M_p$ will again dominate $\hat{A}_p M_m$.

As a result, the term $\hat{A}_p M_m$ can be neglected in the numerator, to give the approximation
\begin{equation}
\Delta \tau  \approx  \frac{\hat{A}_m}{\hat{A}_{pm}}\frac{M_p}{M_{pm}} \frac{a_m}{v_{tr}} \cos\left(\frac{n_mR_{s}}{v_{tr}}\right) \cos (f_m(t_0) + \omega_m + jn_mT_p)).\label{transit_signal_cc_form_lBsimp}
\end{equation}
As an aside, for the case where $\hat{A}_{pm} \approx \hat{A}_{p}$ the neglected term corresponds to $\Delta t_p$, the barycentric transit timing perturbation.  Consequently, we have that $\Delta \tau > \Delta t_p$.

\subsubsection{Case where $v_m/v_{tr} \le 0.66$}\label{Trans_TTV_Signal_CC_Form_largeB}

The expressions for $\Delta \tau$ derived above are only correct to first order in $v_m/v_{tr}$ and $v_p/v_{tr}$.  To explore and quantify the error caused by truncating to this order, $\Delta \tau$ will be calculated to second order in $v_m/v_{tr}$ and $v_p/v_{tr}$ and compared to a full simulation.

\begin{figure}
     \centering
     \subfigure[$a_m$=$2 R_s$, $v_m/v_{tr}$=$0.66$.]{
          \label{TauAgreement2B066}
          \includegraphics[width=.31\textwidth]{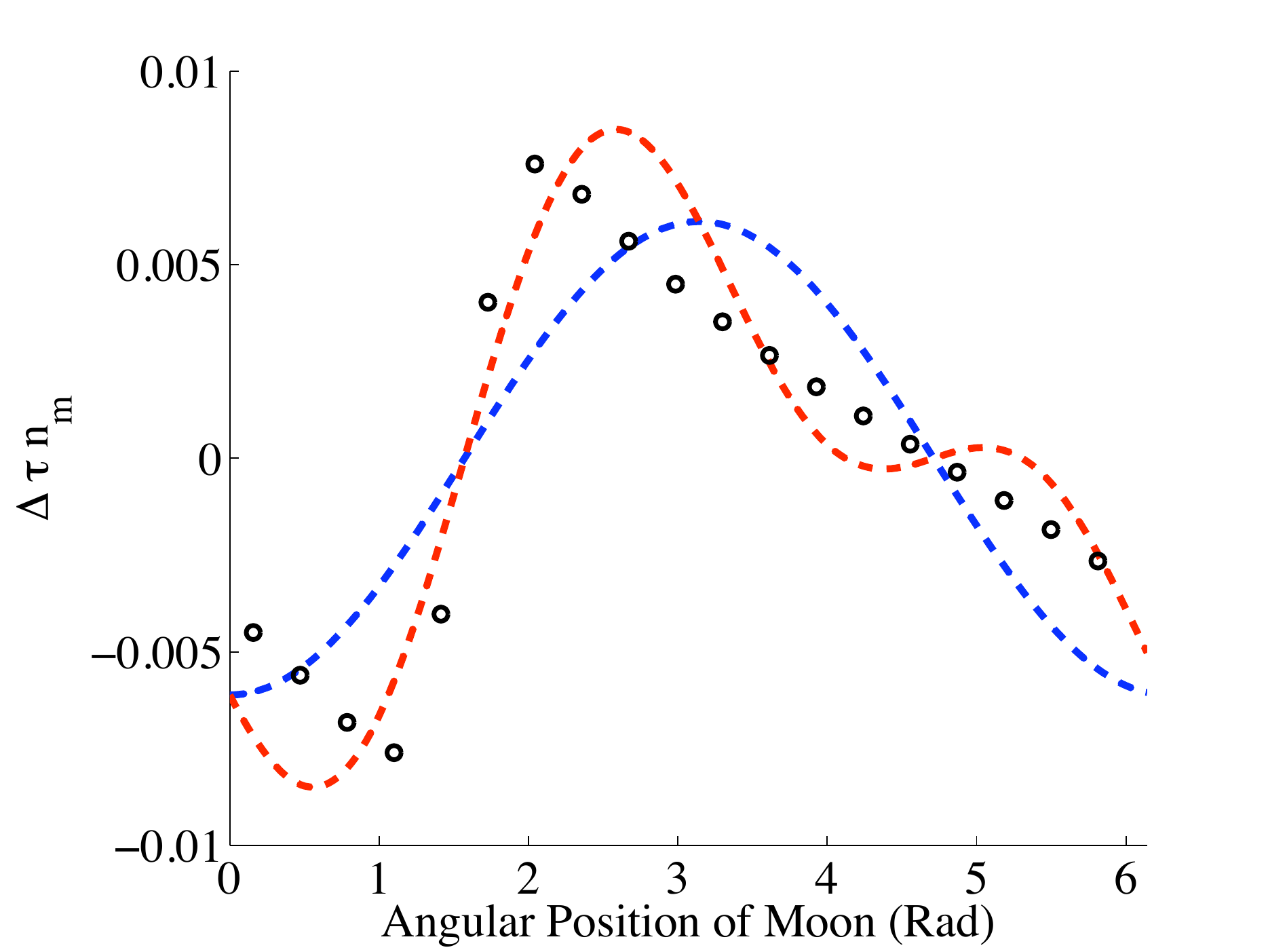}}
     \subfigure[$a_m$=$R_s$, $v_m/v_{tr}$=$0.66$.]{
          \label{TauAgreement1B066}
          \includegraphics[width=.31\textwidth]{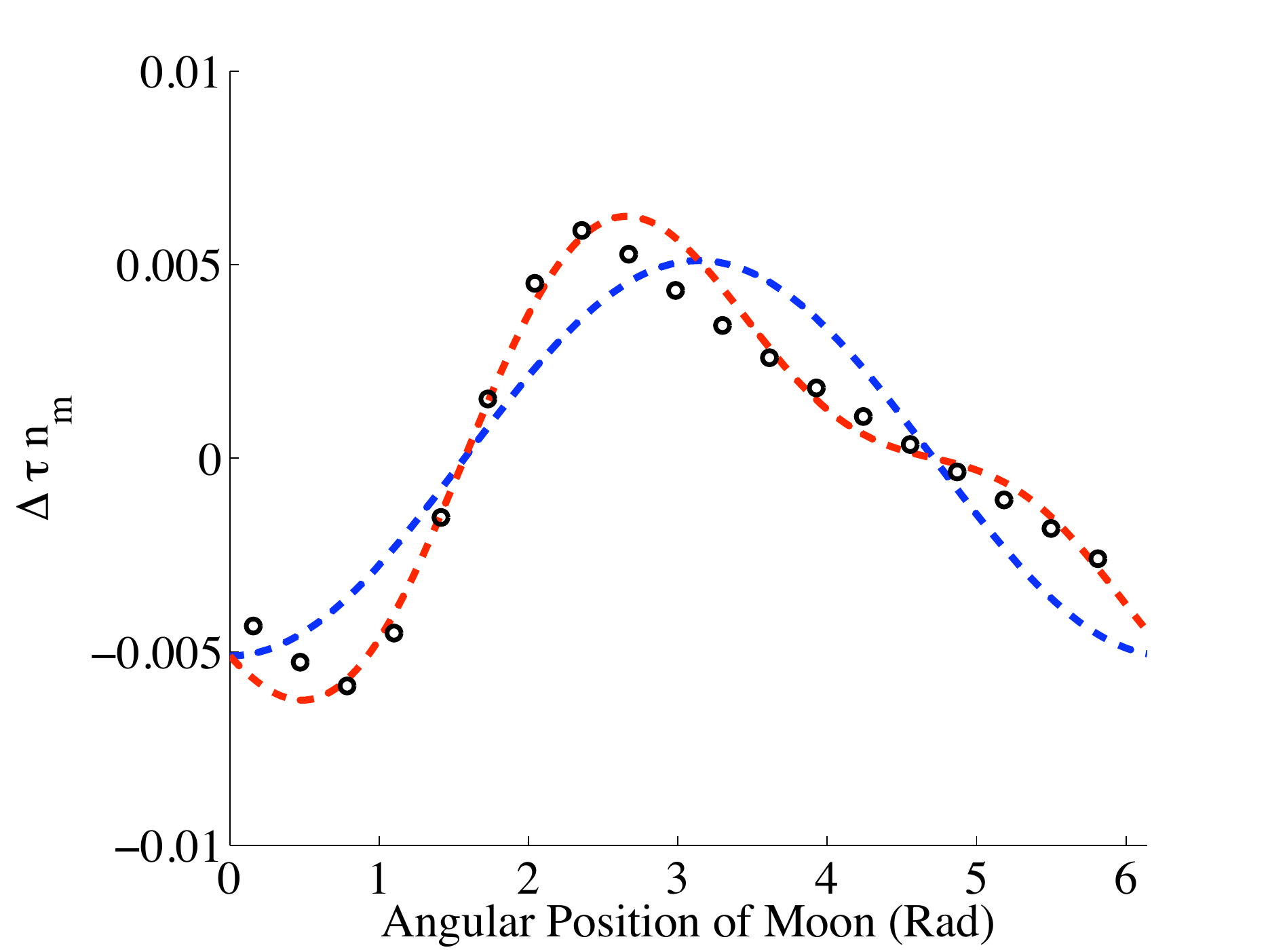}}
     \subfigure[$a_m$=$0.5 R_s$, $v_m/v_{tr}$=$0.66$]{
           \label{TauAgreement05B066}
           \includegraphics[width=.31\textwidth]{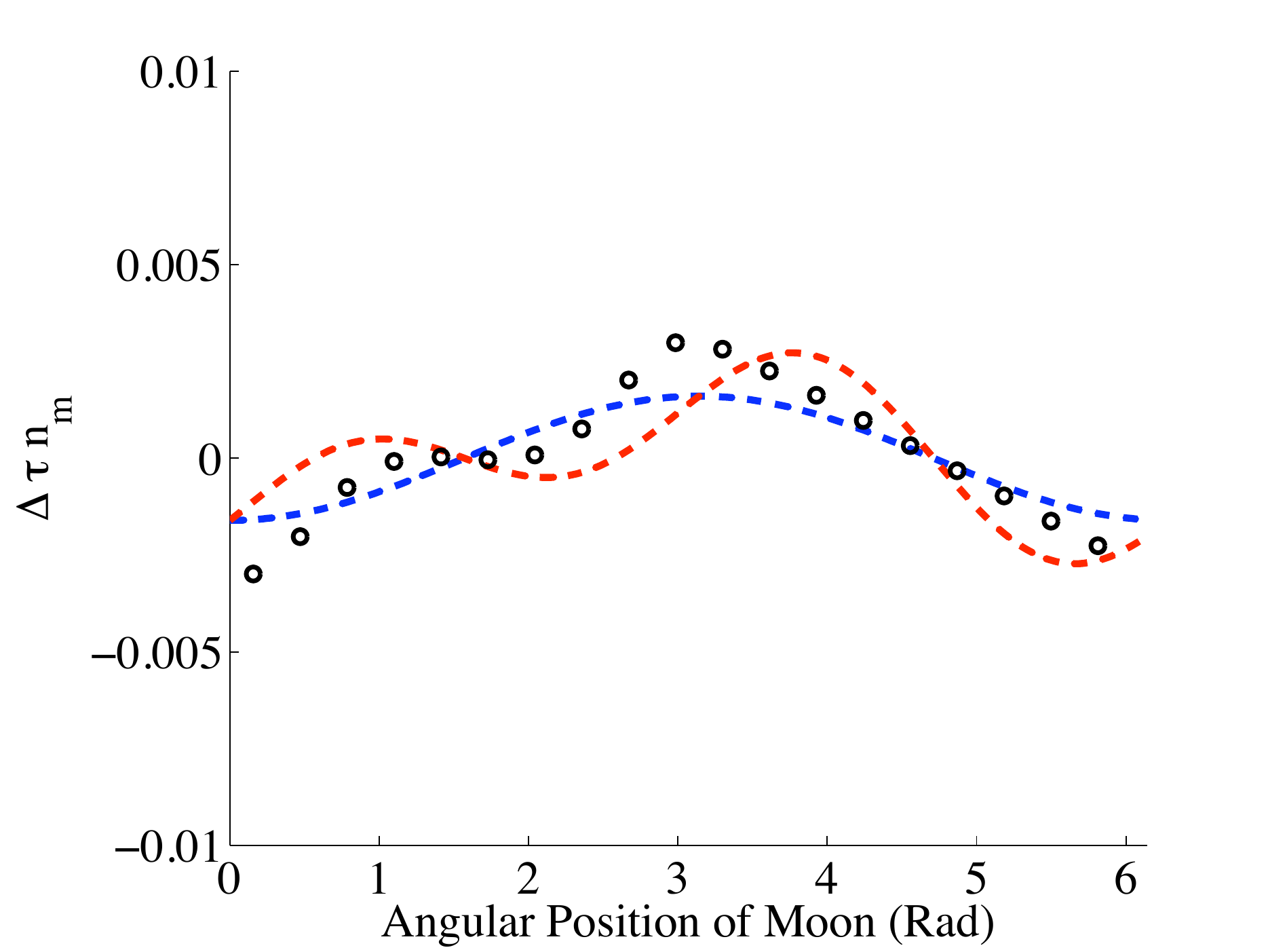}}\\
           \subfigure[$a_m$=$2 R_s$, $v_m/v_{tr}$=$0.49$.]{
          \label{TauAgreement2B049}
          \includegraphics[width=.31\textwidth]{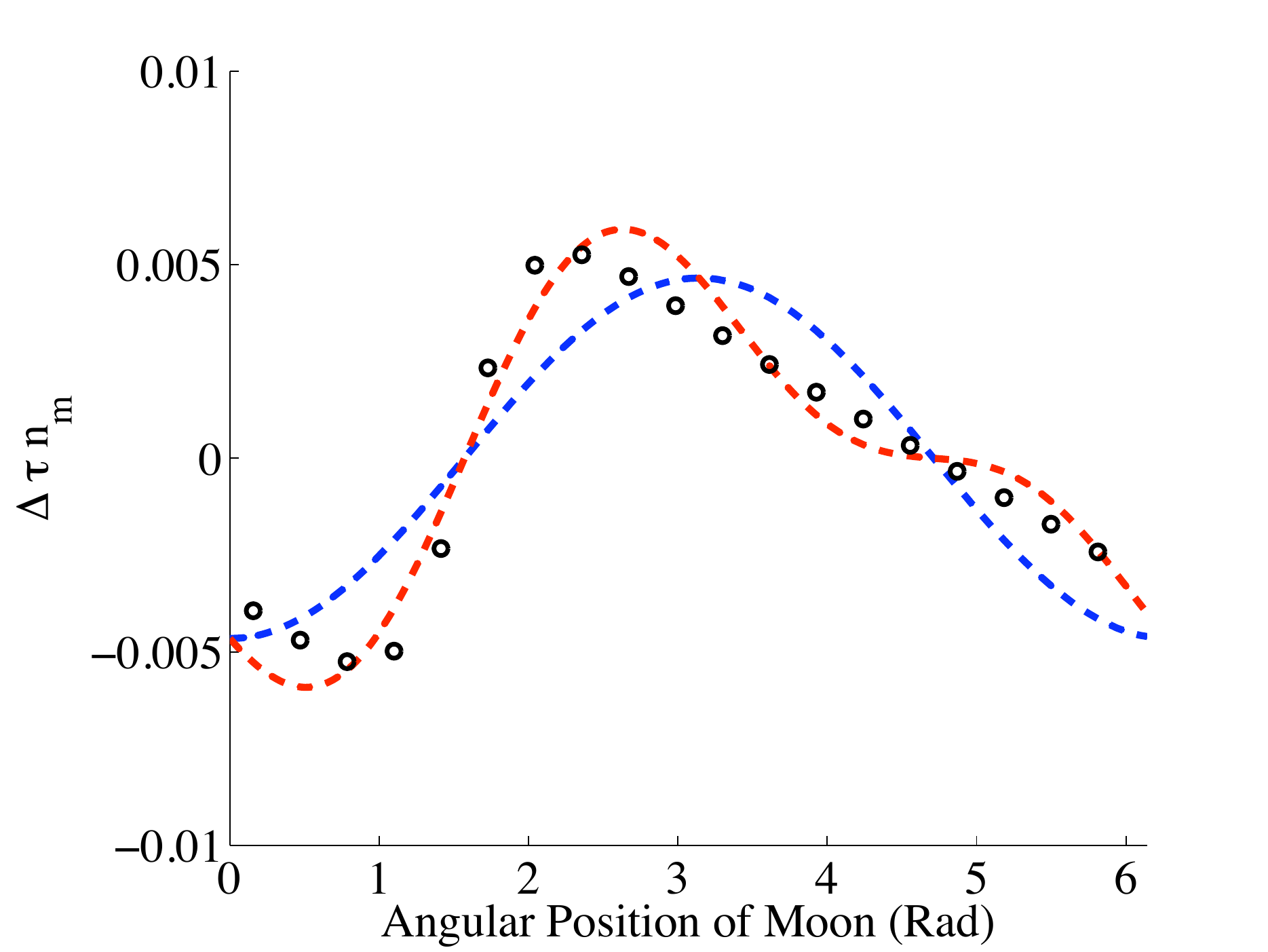}}
     \subfigure[$a_m$=$R_s$, $v_m/v_{tr}$=$0.49$.]{
          \label{TauAgreement1B049}
          \includegraphics[width=.31\textwidth]{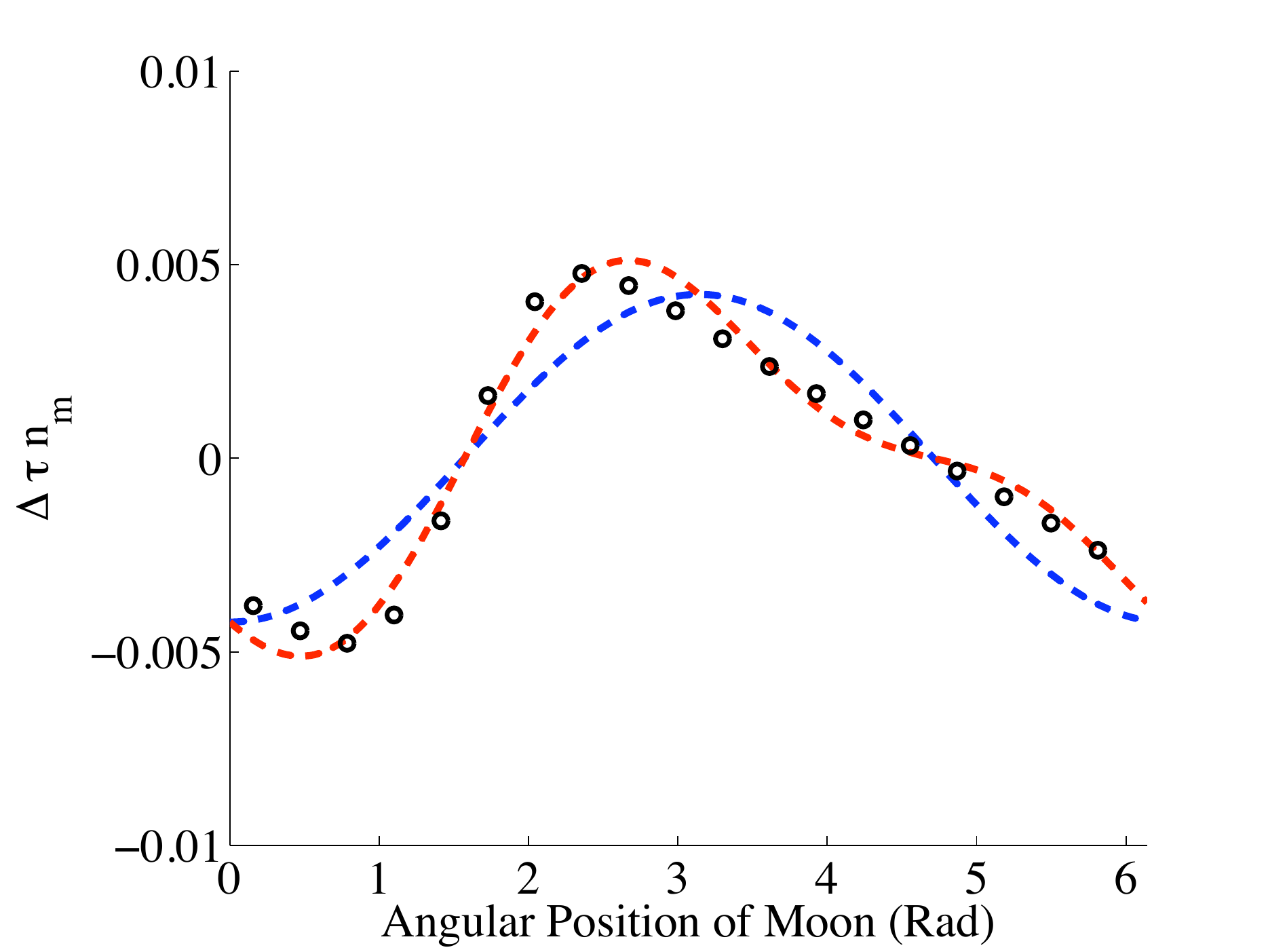}}
     \subfigure[$a_m$=$0.5 R_s$, $v_m/v_{tr}$=$0.49$.]{
           \label{TauAgreement05B049}
           \includegraphics[width=.31\textwidth]{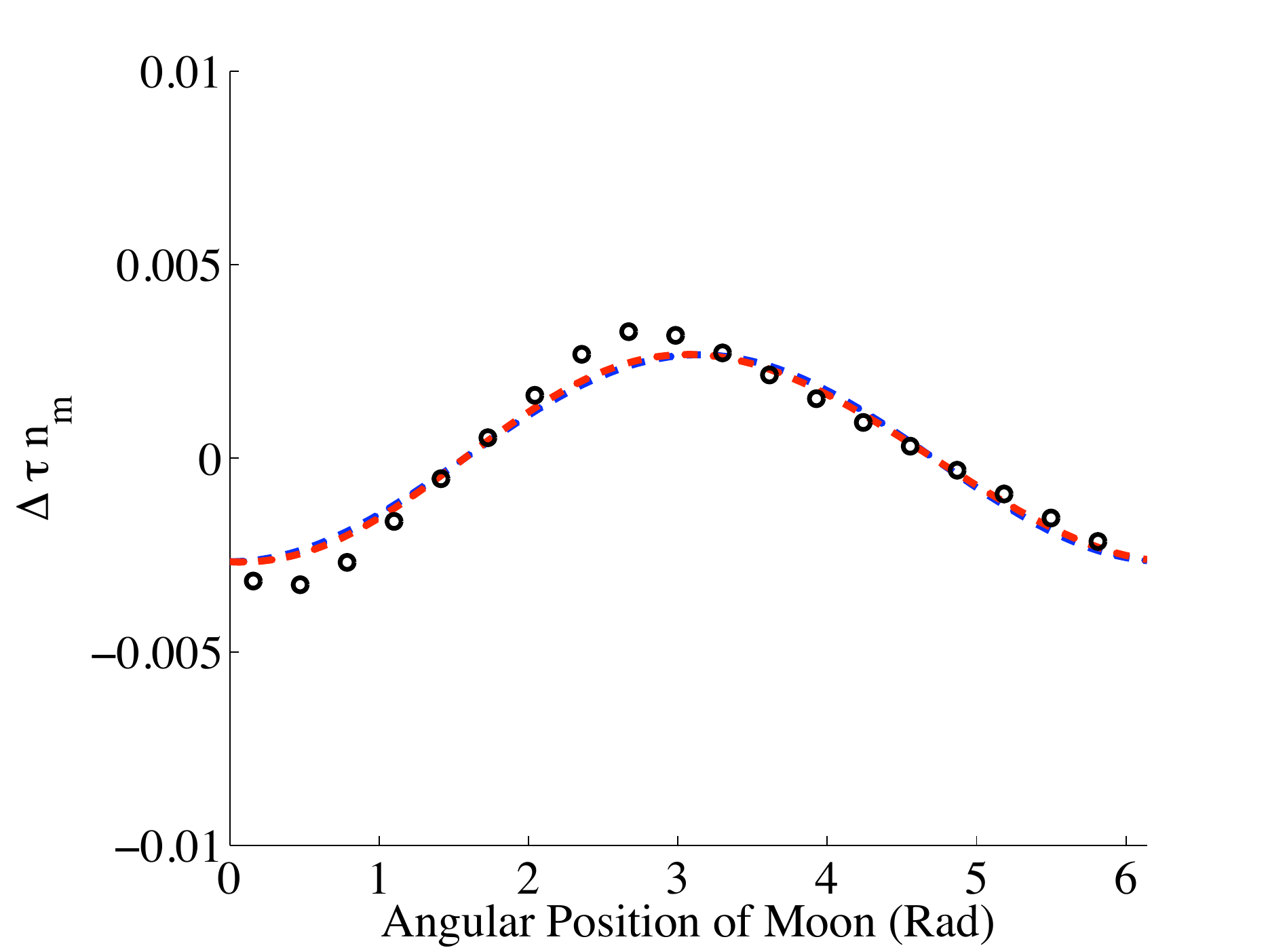}}\\
           \subfigure[$a_m$=$2 R_s$, $v_m/v_{tr}$=$0.33$.]{
          \label{TauAgreement2B033}
          \includegraphics[width=.31\textwidth]{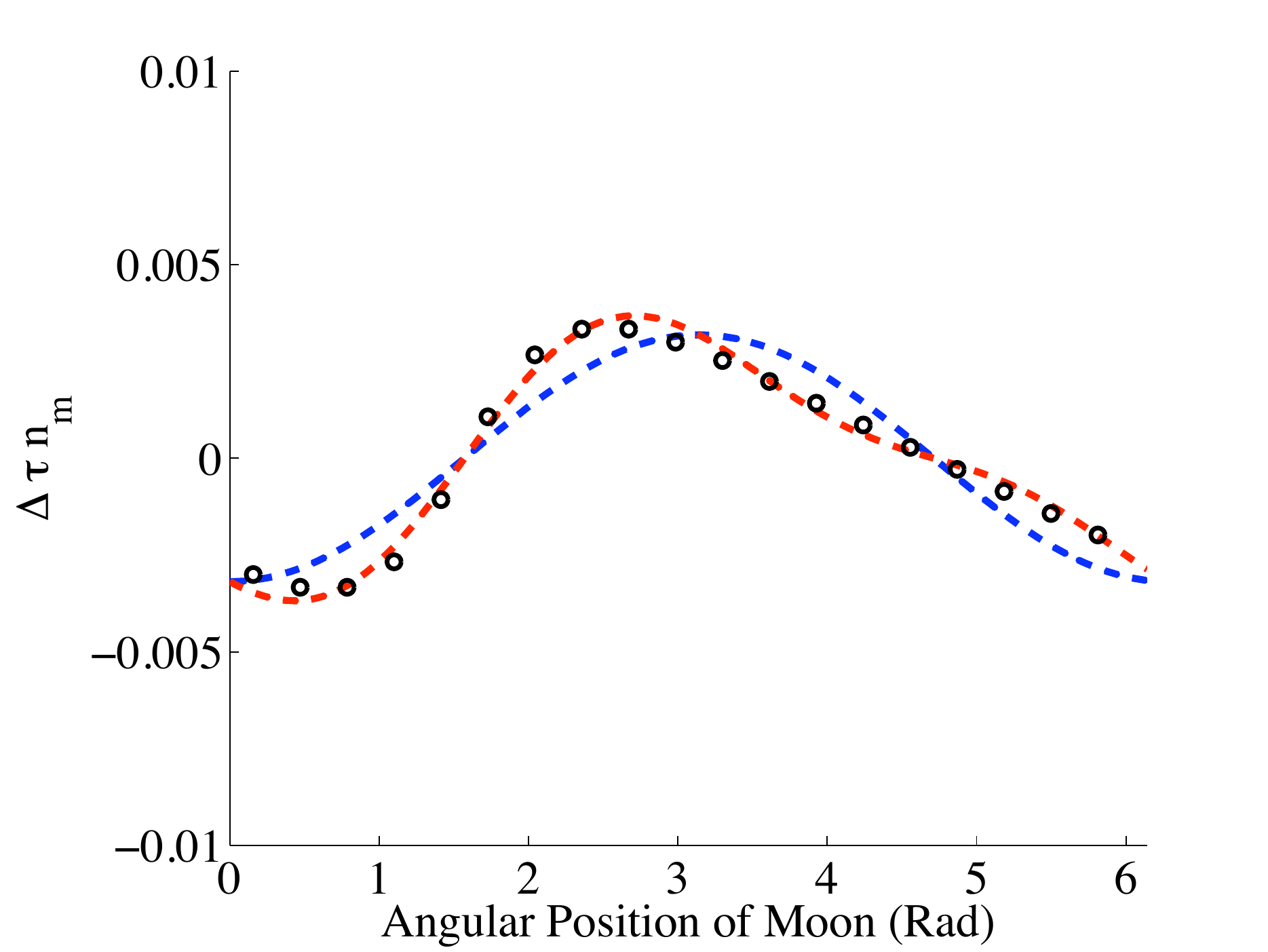}}
     \subfigure[$a_m$=$R_s$, $v_m/v_{tr}$=$0.33$.]{
          \label{TauAgreement1B033}
          \includegraphics[width=.31\textwidth]{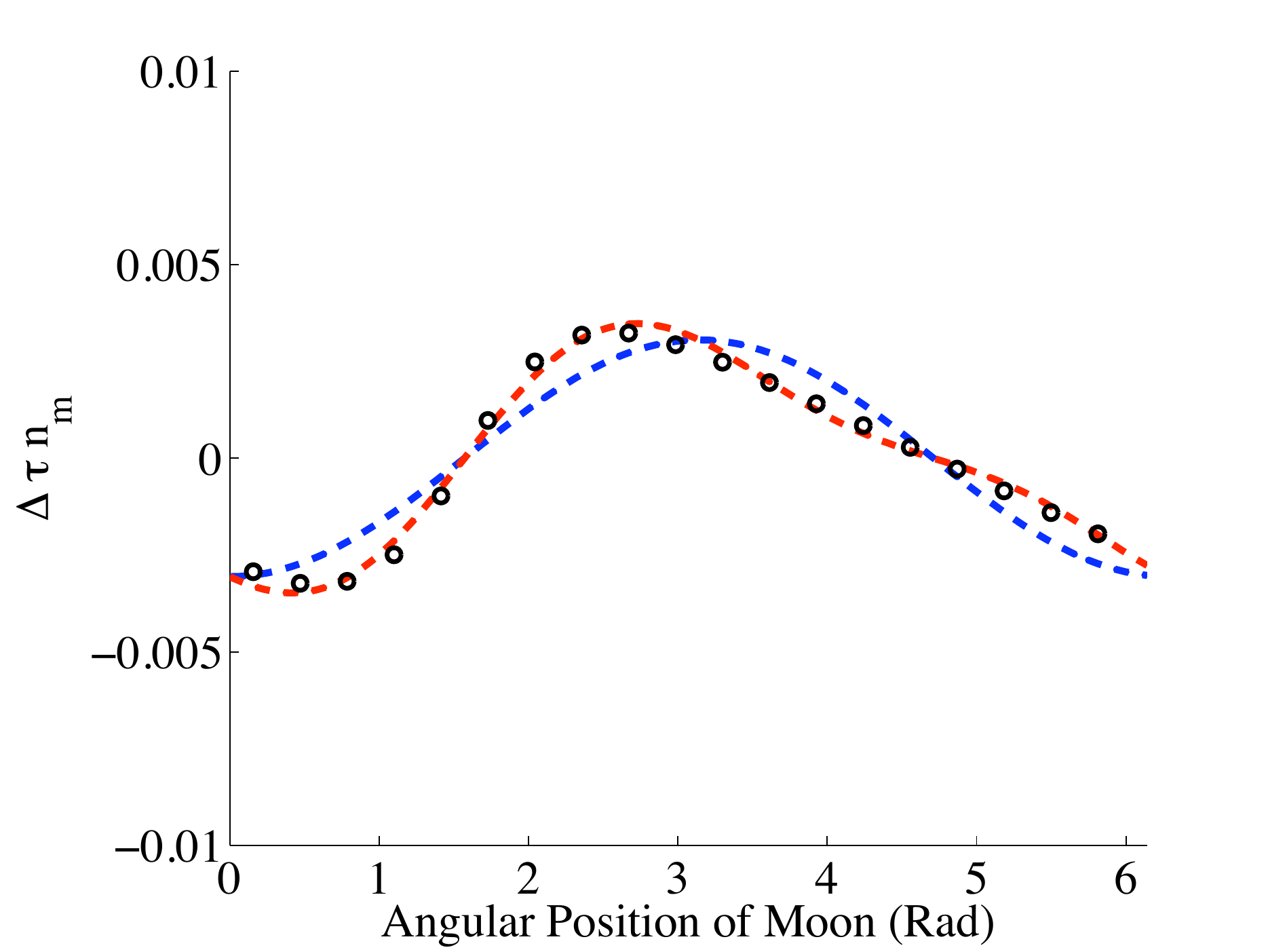}}
     \subfigure[$a_m$=$0.5 R_s$, $v_m/v_{tr}$=$0.33$.]{
           \label{TauAgreement05B033}
           \includegraphics[width=.31\textwidth]{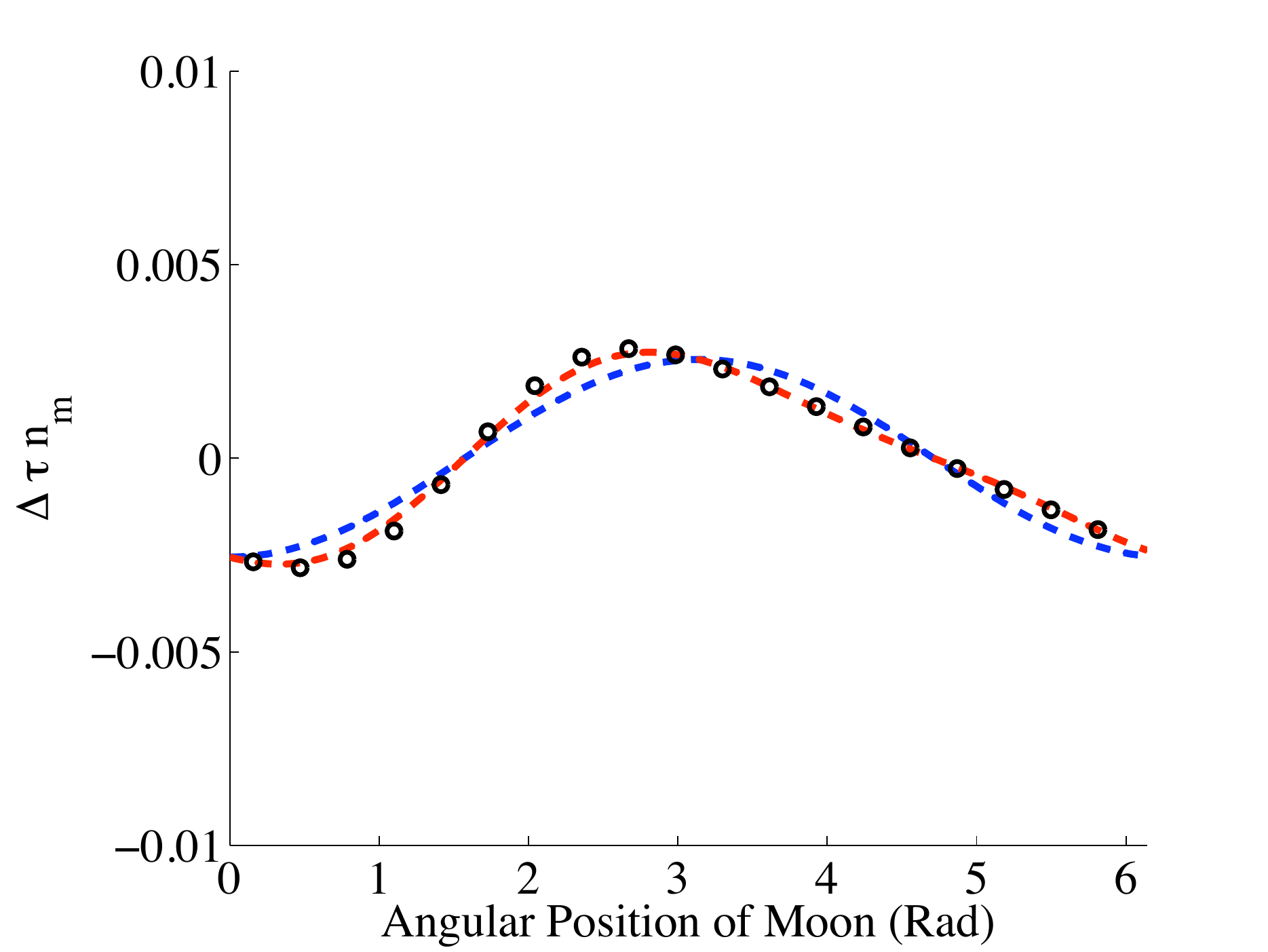}}\\
           \subfigure[$a_m$=$2 R_s$, $v_m/v_{tr}$=$0.16$.]{
          \label{TauAgreement2B016}
          \includegraphics[width=.31\textwidth]{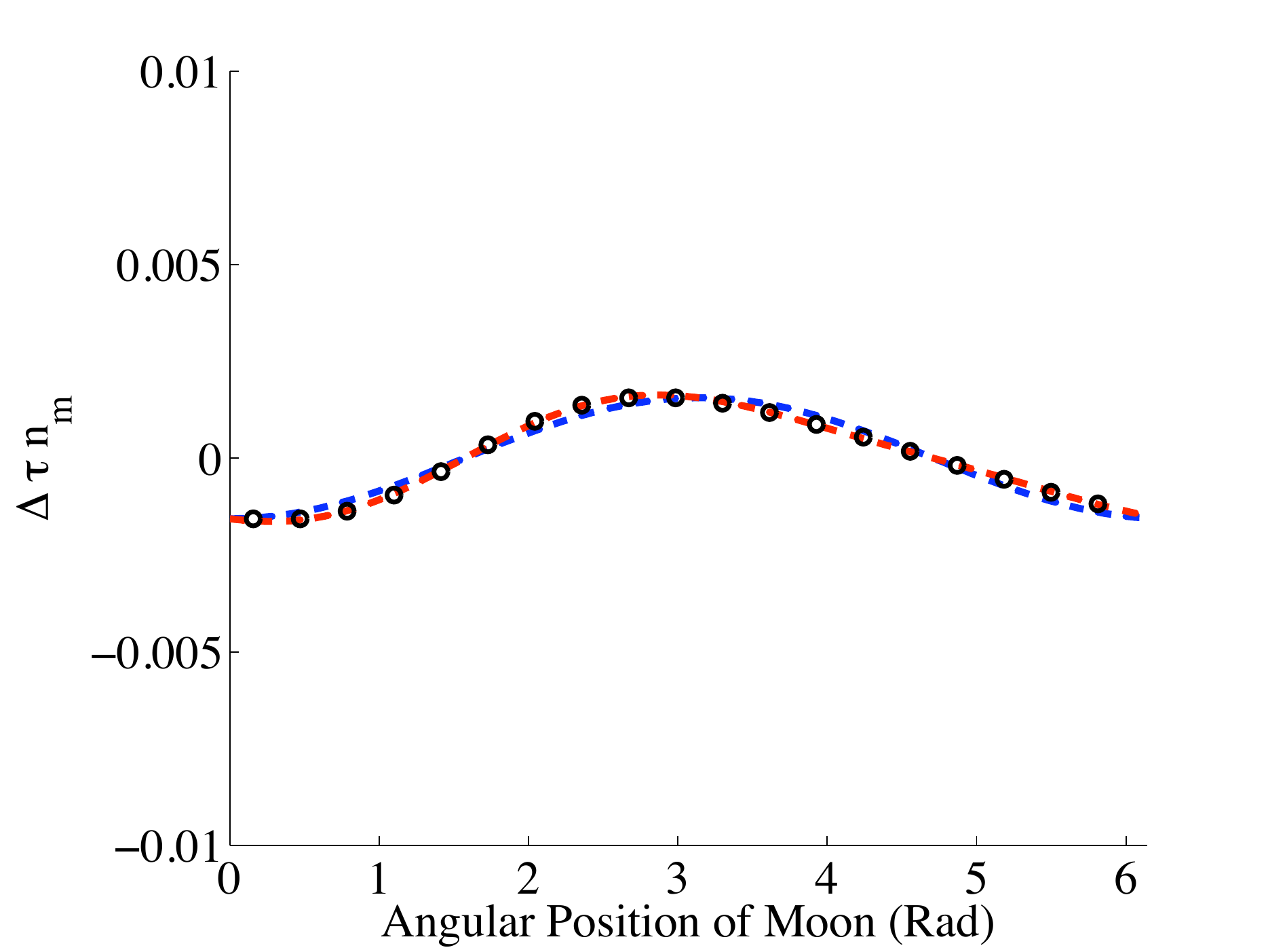}}
     \subfigure[$a_m$=$R_s$, $v_m/v_{tr}$=$0.16$.]{
          \label{TauAgreement1B016}
          \includegraphics[width=.31\textwidth]{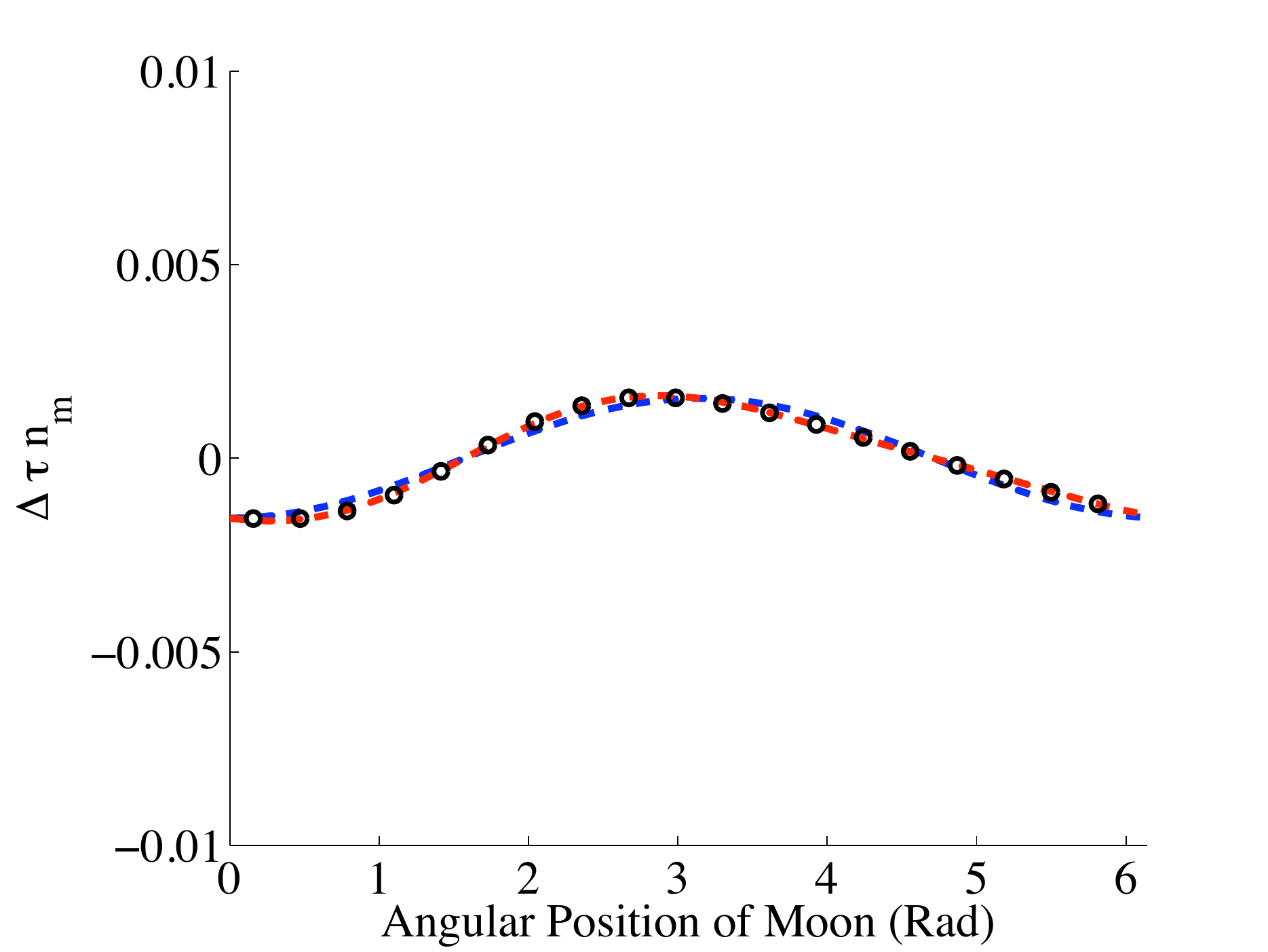}}
     \subfigure[$a_m$=$0.5 R_s$, $v_m/v_{tr}$=$0.16$.]{
           \label{TauAgreement05B016}
           \includegraphics[width=.31\textwidth]{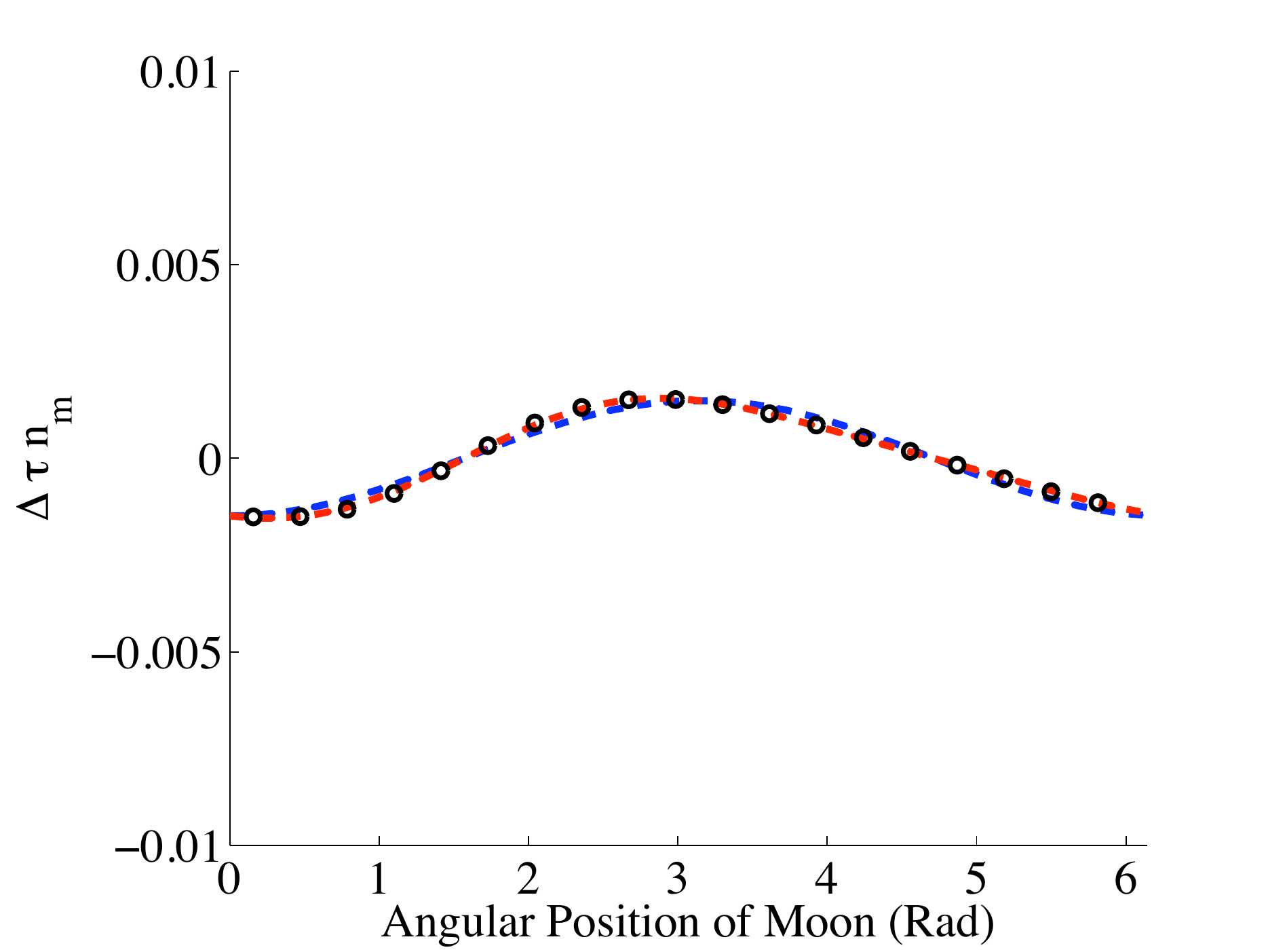}}
     \caption[Comparison of the value of $\Delta \tau$ calculated directly from simulated transit light curves (black), with that of equations~\eqref{transit_signal_cc_form_lB} and \eqref{transit_signal_cc_form_hB}, the analytic approximations to $\Delta \tau$ accurate to first (blue) and second (red) order in velocity ratio.]{Comparison of the value of $\Delta \tau$ calculated directly from simulated transit light curves (black), with that of equations~\eqref{transit_signal_cc_form_lB} and \eqref{transit_signal_cc_form_hB}, the analytic approximations to $\Delta \tau$ accurate to first (blue) and second (red) order in velocity ratio.  As the degree of agreement of the curves is more important then their exact value, and to reduce the number of independent variables, $\Delta \tau n_m$ is plotted against the angle $(f_m(t_0)+\omega_m + jn_mT_p)$.  These plots were constructed for the case of a large gas giant moon, in particular, it was assumed that $R_p = 0.1 R_s$ and $R_m = 0.01 R_s$.}
     \label{TauAgreement}
\end{figure}

Expanding equations~\eqref{transit_signal_cc_Besstinp}, \eqref{transit_signal_cc_Besstinm}, constructing expressions for $\tau_p$, $\tau_m$, $A_p$ and $A_m$ and consequently $\Delta \tau$, and only retaining terms of order $(v_m/v_{tr})^2$, $v_m v_p/v_{tr}^2$ and $(v_p/v_{tr})^2$ gives
\begin{multline}
\Delta \tau = \frac{\hat{A}_m M_p - \hat{A}_p M_m}{ \hat{A}_{pm} M_{pm} } \frac{a_m}{v_{tr}}  \cos\left(\frac{n_mR_{s}}{v_{tr}}\right) \cos \left(  f_m(t_0) + \omega_m + jn_mT_p \right) \\
- \frac{\hat{A}_p M_m^2 + \hat{A}_m M_p^2}{ \hat{A}_{pm} M_{pm} ^2} \frac{a_m n_m}{v_{tr}} \frac{a_m}{v_{tr}} \cos\left(2\frac{n_mR_{s}}{v_{tr}}\right) \sin \left( 2 \left( f_m(t_0) + \omega_m + jn_mT_p\right)\right)\\
 + \frac{1}{4 } \frac{\hat{A}_m\hat{A}_p}{\hat{A}_{pm}^2} \frac{a_m}{v_{tr}} \frac{a_m}{R_{s}} \sin\left(2\frac{n_mR_{s}}{v_{tr}}\right) \sin \left(2(f_m(t_0) + \omega_m + jn_mT_p)\right) . \label{transit_signal_cc_form_hB}
 \end{multline}
 As can be seen, the addition of the higher order terms results in higher order harmonics being included in the expression for $\Delta \tau$.  Physically these additional terms make $\Delta \tau$ asymmetric (see figure~\ref{TauAgreement}).
 
 To determine the increase in accuracy in $\Delta \tau$, equations~\eqref{transit_signal_cc_form_lB} and \eqref{transit_signal_cc_form_hB} were compared to $\Delta \tau$ values directly calculated from a set of simulated light curves.  For comparison purposes, it was decided to select scenarios which included those shown in figure~\ref{MoonTransitShape}.  Consequently the cases where $v_m/v_{tr}$ was equal to 0.16, 0.33, 0.49 and 0.66 and $a_m$ was equal to $0.5 R_s$, $R_s$ and $2R_s$ were selected.  These curves were calculated assuming that $R_p = 0.1 R_s$ and $R_m = 0.01 R_s$ and are shown in figure~\ref{TauAgreement}.  This figure shows the degree of agreement between the two analytic approximations for $\Delta \tau$ (blue and red) and the true value (black) as a function of the angle $(f_m(t_0)+\omega_m + jn_mT_p)$.  As can be seen, equation~\eqref{transit_signal_cc_form_lB}, the first order approximation, reproduces the broad behaviour of $\Delta \tau$, in particular its amplitude, while equation~\eqref{transit_signal_cc_form_hB}, the second order equation is more successful at reproducing the finer detail (see in particular figure~\ref{TauAgreement1B033}).  Reassuringly, for the regions where the assumption of uniform velocities is an acceptable one (figures~\ref{TauAgreement2B066}, \ref{TauAgreement1B066} and \ref{TauAgreement1B033}), both approximations perform well, and for the scenario where the assumption of uniform velocities is not as effective (see figures~\ref{TauAgreement05B066} and \ref{MoonTransitShape05B66}) the two expressions do not agree very well with the exact waveform.  Consequently, for the cases where the assumptions that the planet and moon travel with constant velocity during transit and that $v_m/v_{tr} < 0.66$ hold, equation~\eqref{transit_signal_cc_form_lB}, the first order equation, gives a good qualitative description of $\Delta \tau$ and equation~\eqref{transit_signal_cc_form_hB} gives a more accurate description.

Now that we have two physically realistic approximations for $\Delta \tau$, the properties of $\Delta \tau$ can now be investigated.  In particular, we can begin to consider how the form of these equations (and the physics of the system) can affect the amount of perturbation timing signal a given moon can produce and the amount of this signal which can be detected.


\subsection{Properties of $\Delta \tau$}\label{Sec-TraM-TTV-CC-prop}

Now that we have expressions for $\Delta \tau$, it is a good time to take a step back and consider what these expressions tell us about the system.  Recall from section~\ref{Transit_Intro_Deriv} that that an observer cannot measure $\Delta \tau$ directly, and can only measure a sequence of $\tau$ values
\begin{equation}
\tau_1, \tau_2, \ldots, \tau_N, \label{transit_signal_cc_prop_tausequence}
\end{equation} 
corresponding to the $N$ measured transits.  Consequently we will discuss the form of $\Delta \tau$ for the case of circular coplanar orbits in two different contexts.  First we will look at the properties of $\Delta \tau$ in isolation, with particular reference to how the amplitude depends on the physical parameters of the system, and how the form of $\Delta \tau$ does not allow differentiation between prograde and retrograde orbits.  Then we will discuss $\Delta \tau$ within the context of being part of a signal train such as the one given by equation~\eqref{transit_signal_cc_prop_tausequence}.

\subsubsection{Properties of the amplitude of $\Delta \tau$}\label{Trans_TTV_Signal_CC_PropAmp}

Intuitively it can be seen that the larger the amplitude of $\Delta \tau$, the more detectable the perturbation in a sequence of $\tau$ values.  As we now have a number of approximations to $\Delta \tau$ given by equations~\eqref{transit_signal_cc_form_lB}, \eqref{transit_signal_cc_form_lBsimp} and \eqref{transit_signal_cc_form_hB}, we consequently can start to look at the relationship between this amplitude, and the size, mass and semi-major axis of the moon.  To provide the maximum mathematical intuition with the least amount of mathematical complexity, equation~\eqref{transit_signal_cc_form_lBsimp}, the simplest equation for $\Delta \tau$, will be investigated.

Recasting this equation into physical variables and grouping like terms in square brackets gives
\begin{multline}
\Delta \tau  \approx 0.01 \left[ \frac{2 R_{s}}{v_{tr}} \right] \left[ \frac{a_m}{2R_{s}} \right] \left[ \frac{100 \hat{A}_m}{\hat{A}_p + \hat{A}_m} \right] \\ \times \left[ \frac{M_p}{M_p + M_m} \right] \left[\cos\left(\frac{v_m}{v_{tr}} \frac{R_{s}}{a_m}\right)\right] \cos (f_m(t_0) + \omega_m + jn_mT_p)).
\end{multline}
There are three terms in this equation which substantially affect the amplitude of $\Delta \tau$ (on the first line) and two that don't as they are approximately unity (on the second line).  These five terms in the amplitude will be discussed in turn.

The first term in square brackets is the length of the transit duration (see equation~\eqref{transit_intro_dur_cc_Ddef}). This indicates that the longer the transit duration (i.e. the more distant the planet-moon pair is from the star), the larger the amplitude of $\Delta \tau$.  Also, as all other terms in the amplitude are either fractions of like quantities, or functions of fractions of like quantities, it is this term that give $\Delta \tau$ the units of time.

The second term in the square brackets is the ratio between the size of the moon's orbital semi-major axis and a characteristic scale size of the star, in this case, the diameter of the star.  As can be seen, $\Delta \tau$ scales with $a_m$ and consequently more distant ($a_m \ge R_{s}$) moons have larger $\Delta \tau$ amplitudes than similar moons with a smaller semi-major axis.

The third term is a ratio between the area of the dip caused by a moon and the total dip area corresponding to both planet and moon, and has been scaled such that a Earth-like moon of a Jupiter-like planet would render this term approximately equal to one.  As dip depth is approximately proportional to radius of the body squared, and as planets are generally much larger than moons, this term can be though of a comparison between $R_m^2$ and $R_p^2$.

As mentioned previously, the remaining two terms do not substantially effect $\Delta \tau$ as they are both approximately equal to one.  However, they will be discussed for completeness.

The fraction $M_p/(M_p + M_m)$ describes the proportion of the mass in the planet-moon system that is taken up by the planet.  As this proportion will differ from unity by at most 0.01\% and 4\% for the case of disk generated and impact generated moons respectively (see section~\ref{Intro_Moons_Form}) it can be safely neglected.

The final amplitude term is approximately equal to one for the case where moons are detectable, and the expansion is accurate.  To see this, recall from section~\ref{Transit_Signal_Method_Validation}, that in order for the assumption of uniform velocities to be accurate, either $v_m/v_{tr}$ or $R_s/a_m$ must be small.  In addition, the amplitude of $\Delta \tau$ is proportional to $a_m/R_s$, so detectable moons should have appreciable values of $a_m/R_s$ (and thus small values of $R_s/a_m$).  As a result of these two two factors, the argument of the cosine function is likely to be small.  Thus, as $\cos \theta \approx 1$  for $\theta$ small, this term is approximately 1.

As discussed, the amplitude of $\Delta \tau$ depends linearly on the transit duration and the semi-major axis of the moon, suggesting that more distant planets with more distant moons have higher $\Delta \tau$ amplitudes.  In addition $\Delta \tau$ is also proportional to $\hat{A}_m/(\hat{A}_p + \hat{A}_m) \approx R_m^2/R_p^2$, but not $M_m$ or $M_p$.  This suggests that it is the physical size of the planet and the moon which affects the amplitude of $\Delta \tau$, and not their masses.  Now that the properties of the amplitude of $\Delta \tau$ signal have been discussed some more properties of the signal will be described, in particular the degeneracy in $\Delta \tau$ with respect to prograde and retrograde orbits.

\subsubsection{Properties of the form of $\Delta \tau$}\label{Trans_TTV_Signal_CC_PropForm}

\begin{figure}[tb]
\begin{center}
\includegraphics[width=.40\textwidth]{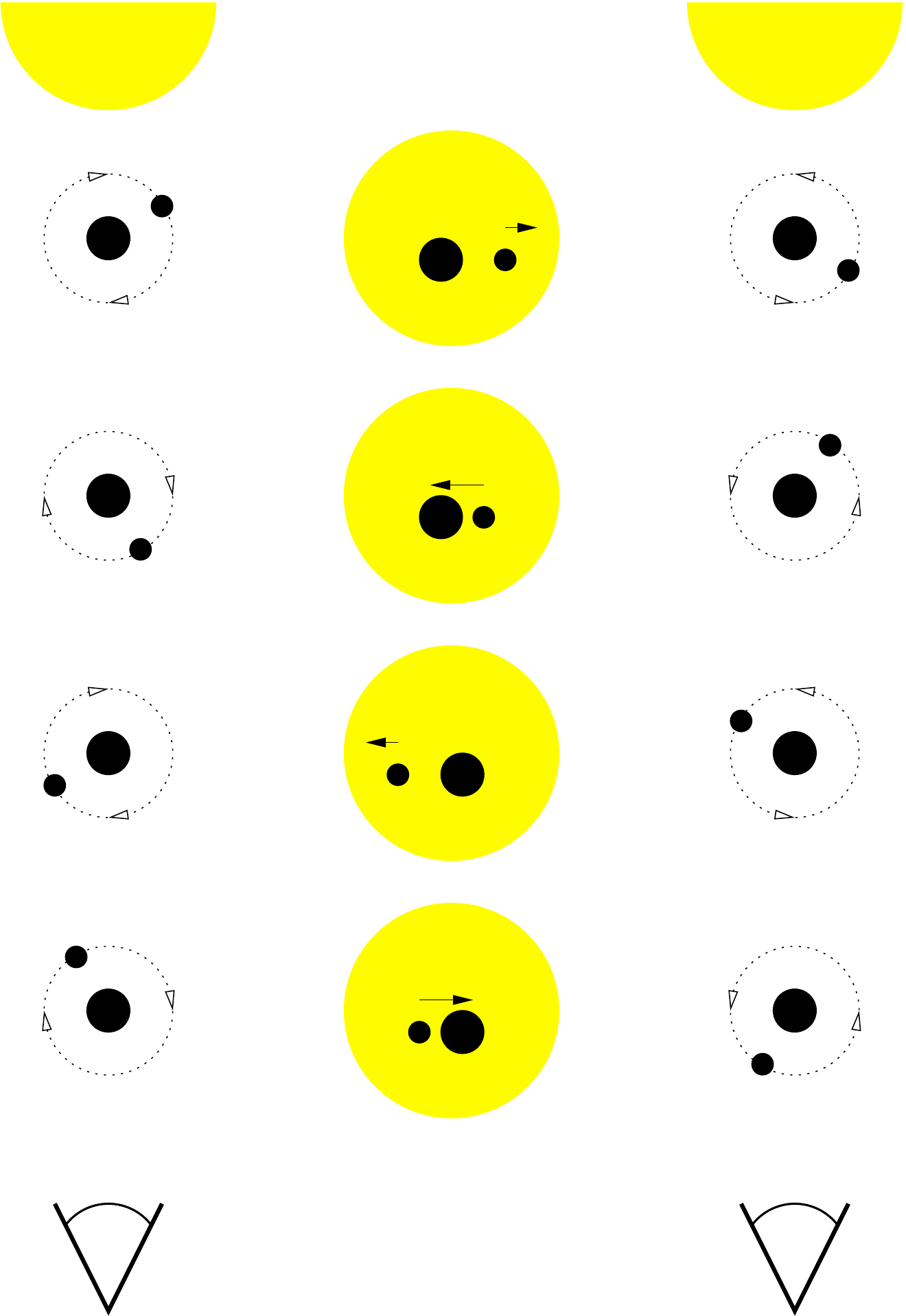}
\caption[An example of a prograde (left) and retrograde (right) system that will show the same silhouette (middle) during transit, and thus have the same $\Delta \tau$ values.]{An example of a prograde (left) and retrograde (right) system that will show the same silhouette (middle) during transit, and thus have the same $\Delta \tau$ values.  The two systems are shown from above, such that the yellow hemisphere and the eye represent the star and the observer, and the vertical sequence of diagrams show the relative planet-moon orientation corresponding to a sequence of four consecutive transits.  The silhouettes are shown from the point of view of the observer, where the large and small black dots mark the mid-transit position of the planet and moon respectively and an arrow is used to show the direction and magnitude of the transverse orbital velocity of the moon in its orbit about the planet.}
\label{ProgRetSym}
\end{center}
\end{figure}

$\Delta \tau$ is a function of the projected position of the planet and moon on the plane of the sky, and to a lesser extent their velocities in that plane.  However, there are two orbits with the same positions and velocities across the plane of the sky as a function of time (see figure~\ref{ProgRetSym}).  One corresponds to a prograde orbit with an initial position angle of $f_m(t_0)$ and argument of perihelion of $\omega_m$, while the second corresponds to a retrograde orbit with initial position angle $-f_m(t_0)$ and argument of perihelion of $-\omega_m$.  To see this mathematically, replace $f_m(t_0)$ with $-f_m(t_0)$, $\omega_m$ with $-\omega_m$ and $n_m$ with $-n_m$ in any of equations~\eqref{transit_signal_cc_form_lB}, \eqref{transit_signal_cc_form_lBsimp} and \eqref{transit_signal_cc_form_hB} and note that the form of $\Delta \tau$ remains the same.  Consequently $\Delta \tau$ cannot be used to distinguish between prograde and retrograde orbits.  This is unfortunate as different formation mechanisms predict different types of orbital structures, e.g. captured moons are likely to be retrograde, while regular moons are likely to be prograde (see section~\ref{Intro_Moons_Form}).  Thus TTV$_p$ (and all other transit-based moon detection techniques) cannot be used to investigate moon properties with respect to whether orbits are prograde or retrograde.

This result is very robust in that inclined or eccentric orbits suffer from the same issue\footnote{In these cases prograde orbits pair up with the retrograde orbits which have been reflected across the plane of the sky.} and that mutual events (moon passing in front or behind planet) do not help differentiate between the cases.  Only a direct measure the effect of the moon on the planet's light curve or spectra (see section~\ref{Intro_Dect_Moons_Image}), can break the symmetry and differentiate between prograde and retrograde orbits.

This symmetry leads to one more interesting property, $\Delta \tau$ depends on $a_m$ alone, and not the sign of $n_m$.  Consequently any detection plot produced with $a_m$ as one of the axes can represent both prograde and retrograde orbits.

\subsubsection{Properties of the signal containing $\Delta \tau$}\label{Trans_TTV_Signal_CC_PropSig}

Now that the properties of $\Delta \tau$ have been discussed, we can look at the ``size" of the detectable component of $\Delta \tau$ within the context of a signal, such as the one given by equation~\eqref{transit_signal_cc_prop_tausequence}.  As the $TTV_{p}$ detection process involves fitting a linear ($t_0 + jT_p$) and a quasi-sinusoidal function ($\Delta \tau$) of $j$, the transit number, to the $\tau$ values, the pertinent amplitude is not actually the amplitude of $\Delta \tau$, but the amplitude of $\Delta \tau$  once linear trends have been removed, that is, the amplitude of the perturbation that would be seen in the corresponding O-C diagram.  This process leads to a number of behaviours, and in particular, we will concentrate on two of these.  First, as $\Delta \tau$ is aliased, moons with the same aliased orbital frequency will display similar behaviour in terms of the size of the detectable component of $\Delta \tau$ in their corresponding sequence of $\tau$ values.  Second, moons which complete an integer number of orbits each planetary year will not be detectable.  These ``non-detection spikes" are due to the fact that the position of the moon relative to the planet during transit will be the same for every transit, consequently $\Delta \tau$ will be the same for each transit, and thus be absorbed into the fitting parameter $t_0$.  These aspects will be discussed in turn.

As a result of stability constraints,\footnote{For the case where $a_m = 0.5R_H$, an estimate for the orbit with the lowest value of $n_m$, it can be shown that $n_m = \sqrt{24}n_p$.  Consequently, $n_m > n_p$, and $T_m$ the orbital period of the moon is always smaller than $T_p$, the orbital period of the planet.} moons complete many orbits of their host planet during a single planetary year (Kipping, 2009a).  Consequently, $\Delta \tau$ also goes through many cycles between one transit and the next, in other words it is aliased, and thus there are a quantised set of angular frequencies which will produce the same series of snapshots and consequently produce similar values of $\Delta \tau$.  To see this, consider the parameter that defines the angle that a moon has progressed around its orbit from one transit to the next, the number of months per planetary year, $n_m/n_p$, and in particular, the fractional part of this quantity.  Consequently, if $n_m$ were increased by $n_p$, such that the fraction $n_m/n_p$ increased by one, then the snapshot of the position of the planet and moon observed at each transit would still be the same.  Mathematically, this can be seen by replacing $n_m$ with $n_m + kn_p$, where $k$ is an integer, in equation~\eqref{transit_signal_cc_form_lB}, the equation for the case where $v_m/v_{tr}$ is small, and noting that the equation remains unchanged.\footnote{For larger values of $v_m/v_{tr}$, where motion of the planet and moon during transit is non-negligible, such as the situations described by equation~\eqref{transit_signal_cc_form_hB}, $\Delta \tau$ will change when $n_m$ is incremented by $n_p$, but only slightly.}

As we will investigate how moon detectability changes with semi-major axis $a_m$ in sections~\ref{Trans_Thresholds_ExpBehav} and \ref{Trans_Thresholds_MC}, it would be useful to recast this result in terms of $a_m$ as opposed to $n_m$.  The angular frequency $n_m$ is related to the semi-major axis through
\begin{equation}
n_m^2 = \frac{G(M_m + M_p)}{a_m^3}.  \label{transit_signal_cc_prop_ndef}
\end{equation}
Using implicit differentiation to take the derivative of this equation with respect to $n_m$ gives
\begin{equation}
2n_m = -3\frac{G(M_m + M_p)}{a_m^4}\frac{d a_m}{d n_m}.
\end{equation}

Now, consider $\Delta a_m$, the change in moon semi-major axis, which occurs when $n_m$ is increased by $n_p$.  As the relative change in $a_m$ and $n_m$ moving from an orbit with mean motion $n_m$ to one with mean motion $n_m + n_p$ is small (i.e $n_p \ll n_m$ and $\Delta a_m \ll a_m$), we can replace the derivatives with $\Delta$s.  Doing this, noting that $\Delta n_m$, the change in mean motion between similar orbits is given by $n_p$, and rearranging to give $\Delta a_m$ as a function of $n_m$ and $n_p$ gives
\begin{equation}
\Delta a_m = - n_m a_m^4 n_p \frac{2}{3 G(M_m + M_p)}. \label{transit_signal_cc_prop_Deltaadef}
\end{equation}
From equation~\eqref{transit_signal_cc_prop_ndef} we have that $n_m \propto a_m^{-3/2}$. Combining this with equation~\eqref{transit_signal_cc_prop_Deltaadef} and noting that $n_p$ is constant, we have that $\Delta a_m \propto a_m^{5/2}$.  Now that the general case has been discussed, we will look at a special set of orbits which result in undetectable moons even though $\Delta \tau \ne 0$.

\begin{figure}[p]
\begin{center}
\includegraphics[width=.95\textwidth]{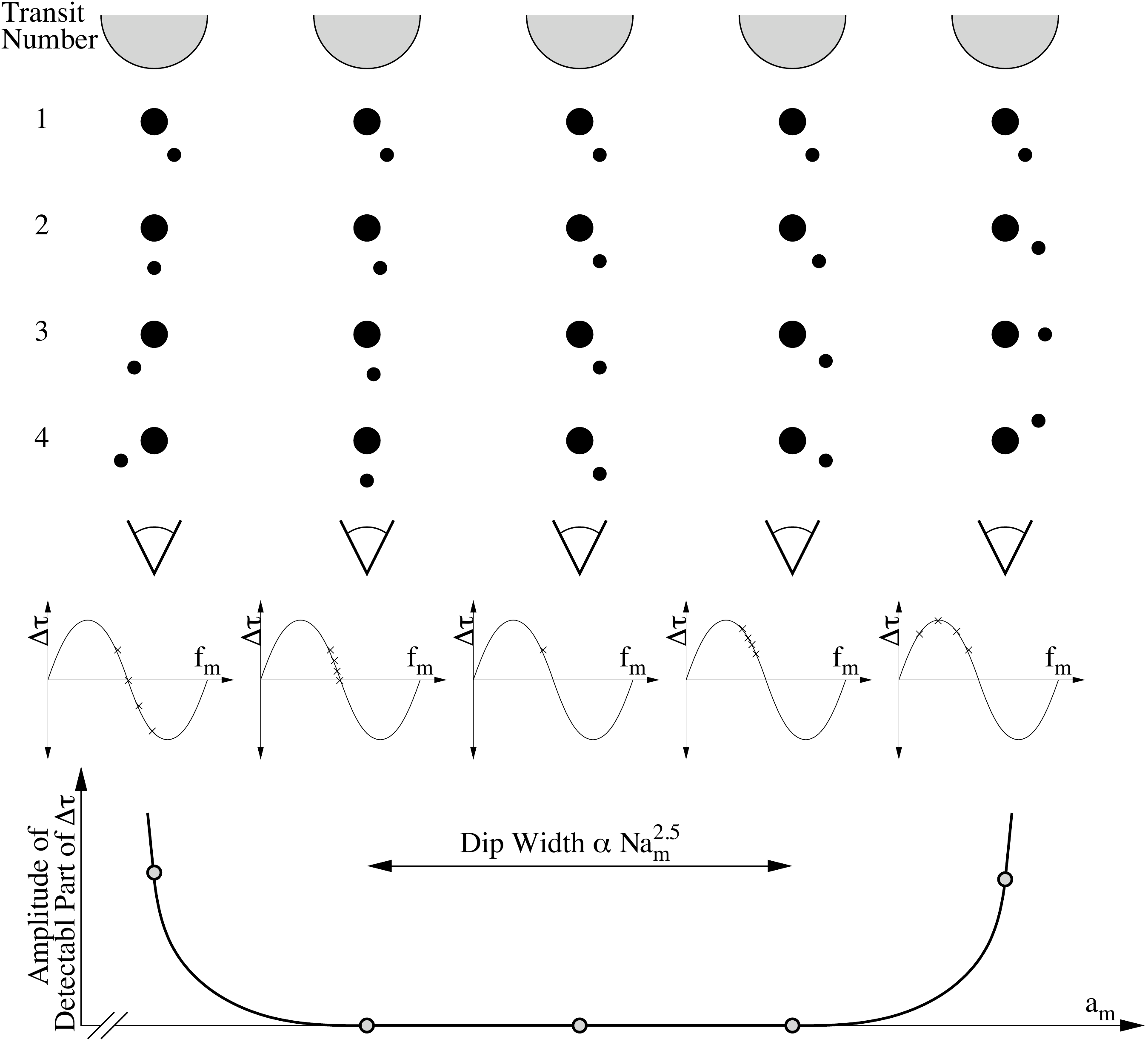}
\caption[Schematic diagram showing four transits for five planet-moon systems with slightly different moon semi-major axes.  The corresponding phase-wrapped sequence of $\Delta \tau$ values (crosses) is then shown in the small plot below.  Finally, the amplitude of $\Delta \tau$ (once linear trends have been removed) is shown in the lower plot by a thick black line.]{Schematic diagram showing four transits for five planet-moon systems with slightly different moon semi-major axes. The different planet-moon systems are arranged from left to right.  For each planet moon system the relative orientation of the planet and moon during transit is shown in the vertical column at the top of the diagram with the first transit being at the top and the last transit at the bottom, where the position of the star and the observer is given by the grey semi-circle and the eye respectively.  The corresponding phase-wrapped sequence of $\Delta \tau$ values (crosses) is then shown in the small plot below.  Finally, the amplitude of $\Delta \tau$ (once linear trends have been removed) is shown in the lower plot by a thick black line, while the values for  each of the five systems are shown by grey dots.}
\label{DipSch}
\end{center}
\end{figure}

For the case where $n_m = k2\pi/T_p$, where $k$ is an integer, that is, the moon completes an integer number of orbits each planetary year, the moon will not be detectable.  This is because the same segment of the moon's orbit will be sampled each transit, and consequently no periodic perturbation in $\tau$ will be observed.  The resulting structure on a plot of the amplitude of the detectable portion of $\Delta \tau$ vs. $a_m$ is a non-detection spike, i.e. the function will go to zero for these particular orbits.  In particular, the relative width of these dips can be estimated.  

To do this, consider the case where a moon has initial position angle $f_m(t_0)$, and orbital angular frequency $n_m = k2\pi/T_p + \Delta n_m$, where $\Delta n_m$ is very much smaller than $2\pi$.  The moon will progresses around its orbit by an angle of $\Delta n_m$ from transit to transit and the corresponding orbit is $\Delta n_m$ away from the center of a spike.  If the in-transit position angle of the moon only samples a region of $\Delta \tau$ which is well approximated by a straight line, it doesn't matter how many measurements are taken, or the accuracy of these measurements, the moon will not be detectable, and the orbit will correspond to a position inside the spike (see figure~\ref{DipSch}).  

For the inner three systems shown in figure~\ref{DipSch}, the region of $\Delta \tau$ sampled by the four observed transits can either be well described by a straight line or a by point, resulting in the perturbation $\Delta \tau$ being undetectable for this small, but non-zero, range of $a_m$.  To begin, let the angle over which $\Delta \tau$ is well approximated by a straight line near the angle $f_m(t_0)$ be $\Delta f_m$ where we note that $\Delta f_m$ is a function of $f_m(t_0)$.  If the difference between the position angle of the moon during the first transit and the position angle of the moon during the last transit is larger than this threshold value, then the moon will be detectable, while if it is smaller than the threshold value, then the moon will be undetectable and lie in the spike.

Mathematically, the $\Delta n_n$ value which corresponds to the edge of the spike is given by
\begin{equation}
(f_m(t_0) + (N-1)\Delta n_m T_p) - f_m(t_0) = (N-1)\Delta n_m T_p = \Delta f_m.\label{transit_signal_cc_prop_spwid}
\end{equation}
Rearranging equation~\eqref{transit_signal_cc_prop_spwid}, in terms of $\Delta n_m$ gives
\begin{equation}
\Delta n_m  =  \frac{\Delta f_m}{T_p(N-1)}.
\end{equation}
It can be seen that the width of these regions depends on $f_m(t_0)$ through the $\Delta f_m$ term, and that it is  inversely proportional to $(N - 1)$ and independent of $n_m$.  As discussed in the previous section, small intervals with width that are independent of $n_m$ have width proportional to $a_m^{2.5}$ when written in terms of $a_m$.  Consequently, for large $N$, the core of these dips have width proportional to $N^{-1}a_m^{2.5}$.

\subsubsection{Summary of properties}

\begin{figure}[tb]
\begin{center}
\includegraphics[width=.95\textwidth]{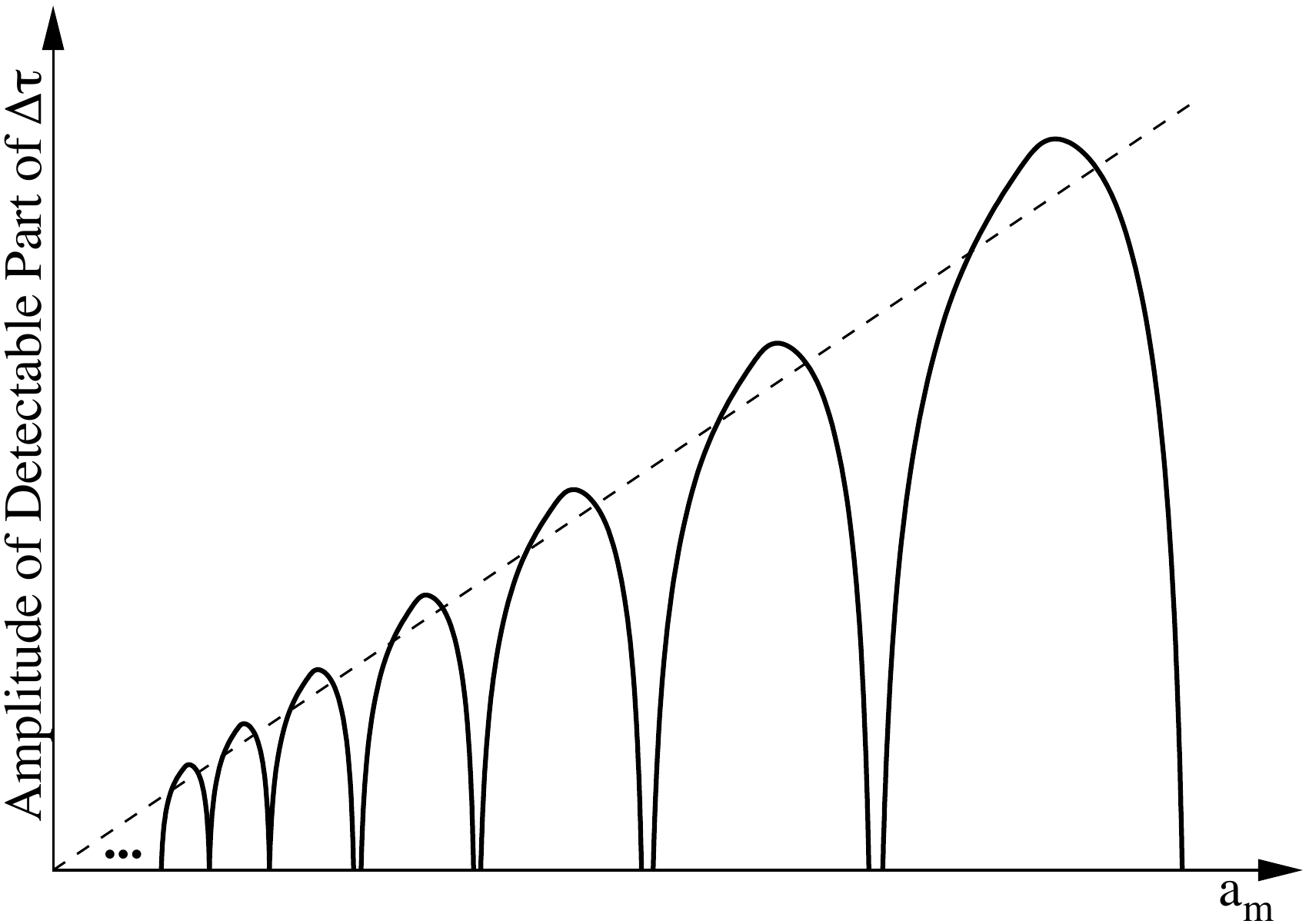}
\caption[Sketch of the size of $\Delta \tau$, once linear trends have been removed, as a function of moon semi-major axis.]{Sketch of the size of $\Delta \tau$, once linear trends have been removed, as a function of moon semi-major axis.  For reference, a dashed line proportional to $a_m$ is also shown.}
\label{EgAmpMap}
\end{center}
\end{figure}

Bringing all this work together, we can now summarise the expected properties of the detectable portion of $\Delta \tau$ as a function of $a_m$.  From section~\ref{Trans_TTV_Signal_CC_PropAmp} we have that gross behaviour of this function is linear in $a_m$, however from section~\ref{Trans_TTV_Signal_CC_PropSig} we know that fine level structure is also present in this function.  Due to aliasing the function is comprised of discrete, repeating blocks, with length proportional to $a_m^{2.5}$.  Also, at the start and end of these blocks are non-detection spikes with width proportional to $N^{-1}a_m^{2.5}$, where the proportionality constant depends on $f_m(t_0)$.  To represent this behaviour, a model diagram was constructed showing the ``amount" of $\Delta \tau$ that could be detected, for the case of a moon with known (and constant) mass and size but with varying  semi-major axis (see figure~\ref{EgAmpMap}).   Now that the case of circular coplanar orbits has been fully explored, let us expand our analysis to the case of inclined orbits.   

\section[Circular orbit inclined to the line-of-sight]{Circular planet orbit inclined to the line-of-sight}\label{Trans_TTV_Signal_Inc}

\begin{figure}[tb]
\begin{center}
\includegraphics[height=2.55in,width=4.5in]{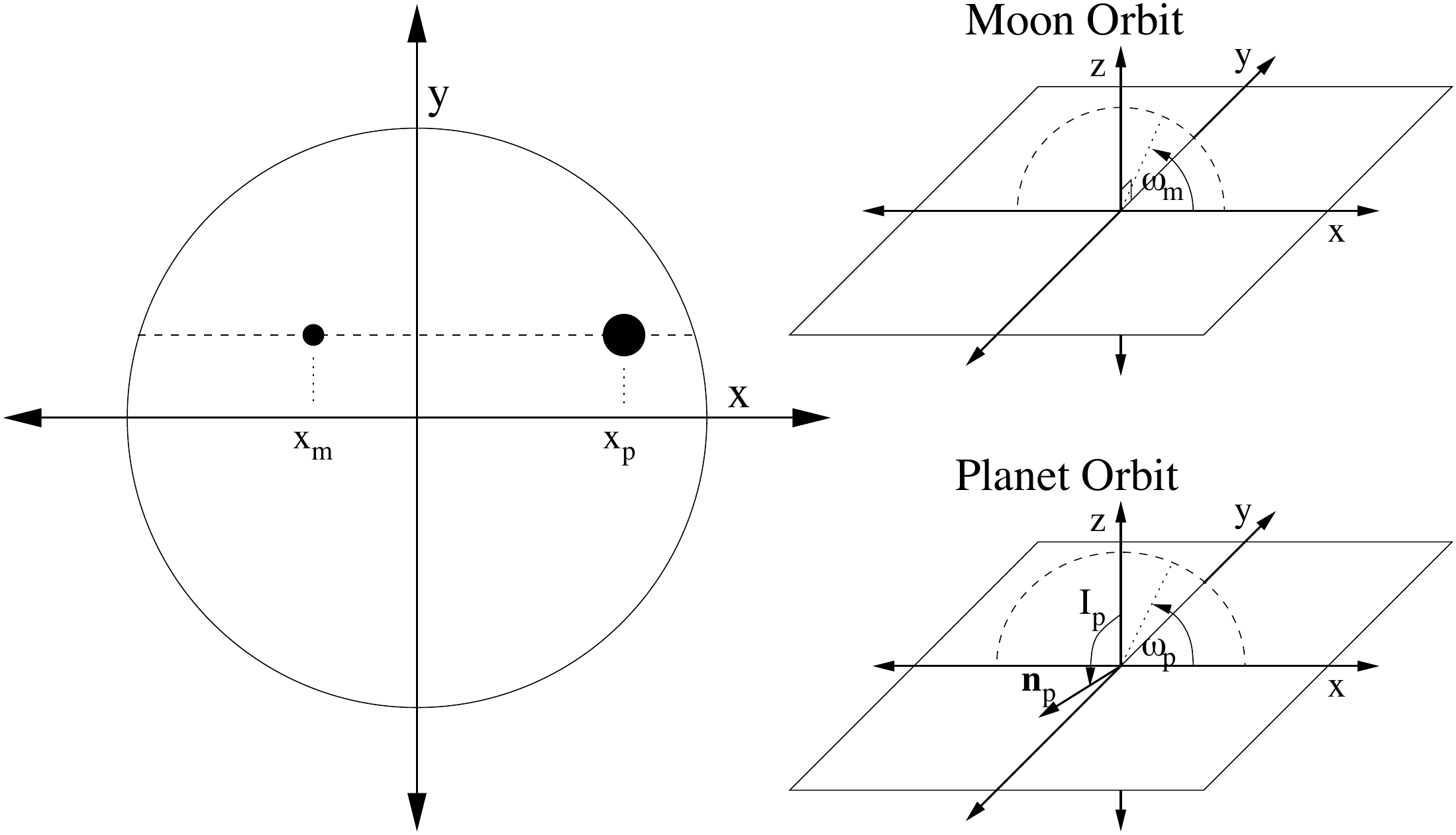}
\caption[Schematic diagram of the same form as figure~\ref{TransitSignalGenCoordSysRot} of the coordinate system for the case of a circular but slightly inclined planet orbit.]{Schematic diagram of the same form as figure~\ref{TransitSignalGenCoordSysRot} of the coordinate system for the case of a circular but slightly inclined planet orbit. In particular, it is assumed that $I_m = \pi/2$ and $\Omega_m = \Omega_p$. }
\label{TransitSignalCoordSysInc}
\end{center}
\end{figure}

For the case where the orbit of the planet is still circular, but slightly inclined with respect to the line-of-sight, the planet and moon no longer transit the central chord of their host star (see figure~\ref{TransitSignalCoordSysInc}).  As for the case of circular and coplanar orbits, we begin the process of deriving $\Delta \tau$ by considering and simplifying the equations of motion of the planet and moon.  For the case of a circular, slightly inclined planet orbit we have that $e_p = 0$ and $r_p = a_p$.  Similarly, for a circular moon orbit aligned to the line-of-sight\footnote{Recall from section~\ref{Transit_Signal_Coord_Orient} that the case where the orbit of the moon is aligned with the line-of-sight is indistinguishable from the case of circular coplanar orbits.} we have that $I_m = \pi/2$, $\Omega_p = \Omega_m$, $e_m = 0$, $r_m = a_m$ and $f_m = n_m t + f_m(0)$.  Using the expressions given above, equations~\eqref{transit_signal_coord_xpdef} to \eqref{transit_signal_coord_ymdef} simplify to 
\begin{align}
x_p &= a_p \cos(f_p + \omega_p) - \frac{M_m}{M_{pm}}a_m \cos(n_m t + f_m(0) + \omega_m),\label{transit_signal_inc_xp}\\
x_m &= a_p \cos(f_p + \omega_p) + \frac{M_p}{M_{pm}}a_m \cos(n_m t + f_m(0) + \omega_m),\label{transit_signal_inc_xm}\\
y_p &= a_p\cos \Omega_p \cos I_p \sin(f_p + \omega_p),\label{transit_signal_inc_yp}\\
y_m &= a_p\cos \Omega_p \cos I_p \sin(f_p + \omega_p), \label{transit_signal_inc_ym}
\end{align}
where we note that the $y$-components are constant to order $R_s/a_p$ and can be safely subsequently neglected.\footnote{Recall from section~\ref{Trans_Intro_Transtech} that the value of $f_p$ corresponding to the planetary transit mid-time is approximately given by $\pi/2 - \omega_p$.  Performing a Taylor expansion of equations~\eqref{transit_signal_inc_yp} and \eqref{transit_signal_inc_ym} about $f_p = \pi/2 - \omega_p$, gives $y_p = y_m = a_p\cos \Omega_p \cos I_p(1 - 1/2(f_p - (\pi/2 - \omega_p))^2)$.  Again recalling from section~\eqref{Trans_Intro_Transtech} that the change in $f_p$ during transit is of order $R_s/a_p$, we have that $y_p = y_m = a_p\cos \Omega_p \cos I_p$ to first order in $R_s/a_p$.}

Again performing a Taylor expansion of the first terms in equations~\eqref{transit_signal_inc_xp} and \eqref{transit_signal_inc_xm} about the time that the $j^{th}$ transit would have occurred if there had been no moon, gives
\begin{align}
x_p &= v_{tr} (t - (jT_p + t_0)) - \frac{M_m}{M_{pm}}a_m \cos(n_m t + f_m(0) + \omega_m),\\
x_m &= v_{tr} (t - (jT_p + t_0)) + \frac{M_p}{M_{pm}}a_m \cos(n_m t + f_m(0) + \omega_m).
\end{align}
These expressions for the position of the planet and moon on the face of the star can again be used to determine the time of ingress and egress by equating the left hand side of the equations to the $x$-coordinate of the limb of the star.

For the case of an inclined planetary orbit, the $x$-position of the limb of the star is given by
\begin{align}
x_p &= \pm \sqrt{R_s^2 - \delta_{min}^2},\\
x_m &= \pm \sqrt{R_s^2 - \delta_{min}^2},
\end{align}
for the case of the planet and moon respectively, where $\delta_{min}$ is the minimum projected distance between the center of the planet and the star and is equal to $(R_s^2 - a_p^2 \cos^2 I_p)^{1/2}$ for this type of orbit (see section~\ref{Trans_Intro_Transtech}).  Consequently, the equations describing the ingress and egress times of the planet and moon's transit can be defined implicitly using
\begin{multline}
-\sqrt{R_s^2 - \delta_{min}^2} = v_{tr} (t_{in,p} - (jT_p + t_0)) \\- \frac{M_m}{M_{pm}}a_m \cos(n_m t_{in,p} + f_m(0) + \omega_m),\label{transit_signal_inc_pin}
\end{multline}
\begin{multline}
- \sqrt{R_s^2 - \delta_{min}^2} = v_{tr} (t_{in,m} - (jT_p + t_0)) \\+ \frac{M_p}{M_{pm}}a_m \cos(n_m t_{in,m} + f_m(0) + \omega_m),\label{transit_signal_inc_min}
\end{multline}
\begin{multline}
\sqrt{R_s^2 - \delta_{min}^2} = v_{tr} (t_{eg,p} - (jT_p + t_0)) \\- \frac{M_m}{M_{pm}}a_m \cos(n_m t_{eg,p} + f_m(0) + \omega_m),\label{transit_signal_inc_peg}
\end{multline}
\begin{multline}
\sqrt{R_s^2 - \delta_{min}^2} = v_{tr} (t_{eg,m} - (jT_p + t_0)) \\+ \frac{M_p}{M_{pm}}a_m \cos(n_m t_{eg,m} + f_m(0) + \omega_m).\label{transit_signal_inc_meg}
\end{multline}

Again defining $\theta_{in,p} = n_m t_{in,p} + f_m(0) +\omega_m + \pi/2$, $\theta_{eg,p} = n_m t_{eg,p} + f_m(0)+\omega_m + \pi/2$, $\theta_{in,m} = n_m t_{in,m} + f_m(0)+\omega_m + \pi/2$ and $\theta_{eg,m} = n_m t_{eg,m} + f_m(0)+\omega_m + \pi/2$, and substituting these expressions into equations~\eqref{transit_signal_inc_pin} to \eqref{transit_signal_inc_meg} we obtain
\begin{multline}
-\frac{n_m\sqrt{R_s^2 - \delta_{min}^2}}{v_{tr}} + \frac{\pi}{2} + \omega_m + f_m(0) + n_m(jT_p + t_0) \\=  \theta_{in,p}  - \frac{v_m}{v_{tr}} \sin(\theta_{in,p}),\label{transit_signal_inc_tinp2}
\end{multline}
\begin{multline}
- \frac{n_m\sqrt{R_s^2 - \delta_{min}^2}}{v_{tr}} + \frac{\pi}{2} +\omega_m  + f_m(0) + n_m(jT_p + t_0)\\=  \theta_{in,m} + \frac{v_m}{v_{tr}} \sin(\theta_{in,m}),\label{transit_signal_inc_tinm2}
\end{multline}
\begin{multline}
\frac{n_m\sqrt{R_s^2 - \delta_{min}^2}}{v_{tr}}+ \frac{\pi}{2} +\omega_m + f_m(0) + n_m(jT_p + t_0)\\= \theta_{in,m} - \frac{v_p}{v_{tr}} \sin(\theta_{in,m}),\label{transit_signal_inc_tegp2}
\end{multline}
\begin{multline}
\frac{n_m\sqrt{R_s^2 - \delta_{min}^2}}{v_{tr}} + \frac{\pi}{2} +\omega_m + f_m(0) + n_m(jT_p + t_0) \\= \theta_{eg,m} + \frac{v_m}{v_{tr}} \sin(\theta_{eg,m}).\label{transit_signal_inc_tegm2}
\end{multline}
Comparing these equations to equations~\eqref{transit_signal_cc_tinp2} to \eqref{transit_signal_cc_tegm2}, the equivalent set for the case of circular coplanar orbits, it can be seen that they are of the same form.  In particular, they can also be represented by equation~\eqref{transit_signal_cc_AB},
\begin{equation*}
\Phi = \theta + B \sin(\theta).
\end{equation*}
However, in this case, the expressions for $\Phi$ are no longer given by those in table~\ref{ABTable}, but instead by those in table~\ref{ABTableInc}.

To solve this equation, we note that in section~\ref{Trans_TTV_Signal_CC} an expression for $\theta$ was derived for the case where $B < 0.6627$, which was accurate for all values of $\Phi$.  As the values of $B$ for this case are the same as those for the case of circular coplanar orbits, and only the values of $\Phi$ are altered, the solution derived for the case of circular coplanar orbits can be directly used to determine $\theta$ and thus $\Delta \tau$ for the case of inclined orbits.

\begin{table}[tb]
	\begin{center}
  \begin{tabular}{l|c|c}
 $X$ & $\Phi_X$  & $B_X$ \\
  \hline
  ${in,p}$  & $f_m(0) + \omega_m + \frac{\pi}{2} +  n_m(jT_p + t_0) -\frac{n_m \sqrt{R_s^2 - \delta_{min}^2}}{v_{tr}}$ & $ - \frac{v_p}{v_{tr}}$ \\
  ${in,m}$  & $f_m(0) + \omega_m + \frac{\pi}{2}+  n_m(jT_p + t_0) -\frac{n_m \sqrt{R_s^2 - \delta_{min}^2}}{v_{tr}}$ & $\frac{v_m}{v_{tr}}$ \\
  ${eg,p}$  & $f_m(0) + \omega_m + \frac{\pi}{2}+  n_m(jT_p + t_0) + \frac{n_m \sqrt{R_s^2 - \delta_{min}^2}}{v_{tr}}$ & $- \frac{v_p}{v_{tr}}$ \\
  ${eg,m}$  & $f_m(0) + \omega_m + \frac{\pi}{2}+  n_m(jT_p + t_0) + \frac{n_m \sqrt{R_s^2 - \delta_{min}^2}}{v_{tr}}$ & $\frac{v_m}{v_{tr}}$\\
  \end{tabular}\\
 \caption{The values of $\Phi$ and $B$ corresponding to equations~\eqref{transit_signal_inc_tinp2} to \eqref{transit_signal_inc_tegm2}.}
 \label{ABTableInc}
 \end{center}
 \end{table}

\subsection{Form of $\Delta \tau$}\label{Trans_TTV_Signal_inc_Form}

To begin the process of deriving expressions for $\Delta \tau$, we note that equations~\eqref{transit_signal_cc_tinp2} to \eqref{transit_signal_cc_tegm2} and \eqref{transit_signal_inc_tinp2} to \eqref{transit_signal_inc_tegm2} only differ in their $\Phi$ terms.  In addition, as the expressions for $\Phi$ for the case of circular and coplanar orbits can be transformed to those for inclined orbits by replacing $R_s$ with $(R_s^2 - \delta_{min}^2)^{1/2}$, and $R_s$ and $(R_s^2 - \delta_{min}^2)^{1/2}$ do not occur anywhere else in the equations, the expressions for $\Delta \tau$ can be modified to give equivalent expressions for the case of inclined orbits by simply replacing $R_s$ with $(R_s^2 - \delta_{min}^2)^{1/2}$.  This will be done for the two cases considered in section~\ref{Trans_TTV_Signal_CC_Form}, the case where $v_m/v_{tr} \ll 1$, and the case where $v_m/v_{tr} < 0.66$.

\subsubsection{Case where $v_m/v_{tr} \ll 1$}\label{Trans_TTV_Signal_inc_Form_smallB}

In section~\ref{Trans_TTV_Signal_CC_Form_smallB}, two expressions were derived for the case where $v_m/v_{tr} \ll 1$, equation~\eqref{transit_signal_cc_form_lB}, a general equation, and equation~\eqref{transit_signal_cc_form_lBsimp}, a simplified version of this equation.  The equivalent expressions for the case of inclined planetary orbits are given by replacing all instances of $R_s$ with $(R_s^2 - \delta_{min}^2)^{1/2}$ in these two equations. Doing this, the following is obtained
\begin {multline}
\Delta \tau = \frac{\hat{A}_m M_p -\hat{A}_p M_m}{\hat{A}_{pm} M_{pm}}  \frac{a_m}{v_{tr}} \cos \left(\frac{n_m \sqrt{R_s^2 - (a_p\cos I_p)^2}}{v_{tr}}\right) \\
\times \cos\left(f_m(t_0) + \omega_m +  n_m jT_p \right),\label{transit_signal_inc_form_lB}
\end{multline}
and
\begin{multline}
\Delta \tau  \approx  \frac{\hat{A}_m}{\hat{A}_{pm}}\frac{M_p}{M_{pm}} \frac{a_m}{v_{tr}} \cos\left(\frac{n_m\sqrt{R_s^2 - (a_p\cos I_p)^2}}{v_{tr}}\right) \\ \times \cos (f_m(t_0) + \omega_m + jn_mT_p)),\label{transit_signal_inc_form_lBsimp}
\end{multline}
where the expression $\delta_{min} = a_p\cos I_p$ has been used, and where equation~\eqref{transit_signal_inc_form_lB} represents the general solution and equation~\eqref{transit_signal_inc_form_lBsimp} represents the simplified solution.  In addition, the higher order approximation to $\Delta \tau$ for the case of circular coplanar orbits can also be modified to give the associated expression for inclined orbits.

\subsubsection{Case where $v_m/v_{tr} < 0.66$}

\begin{figure}
     \centering
     \subfigure[$a_m$=$2 R_s$, $v_m/v_{tr}$=$0.66$.]{
          \label{TauAgreementInc2B066}
          \includegraphics[width=.31\textwidth]{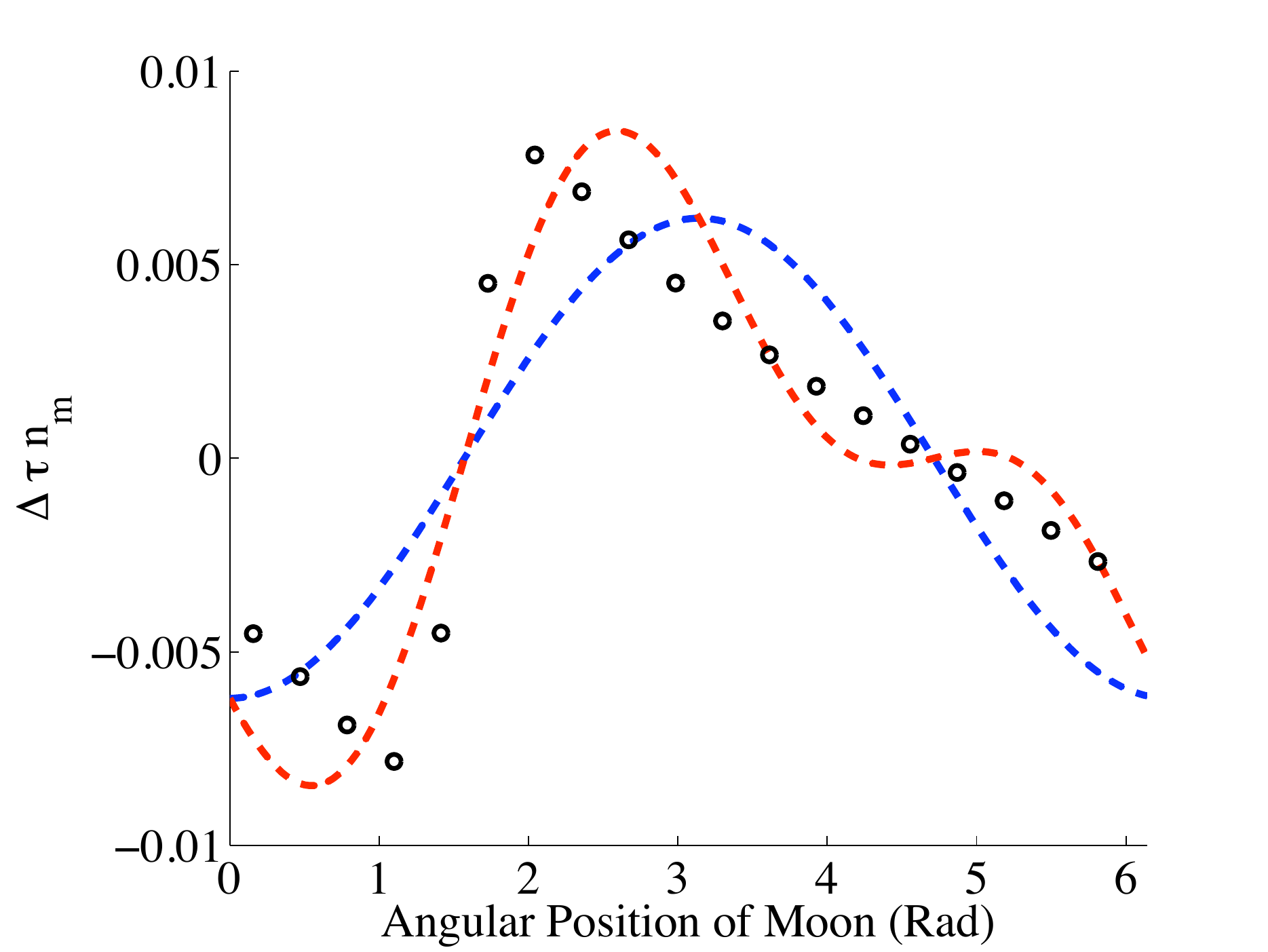}}
     \subfigure[$a_m$=$R_s$, $v_m/v_{tr}$=$0.66$.]{
          \label{TauAgreementInc1B066}
          \includegraphics[width=.31\textwidth]{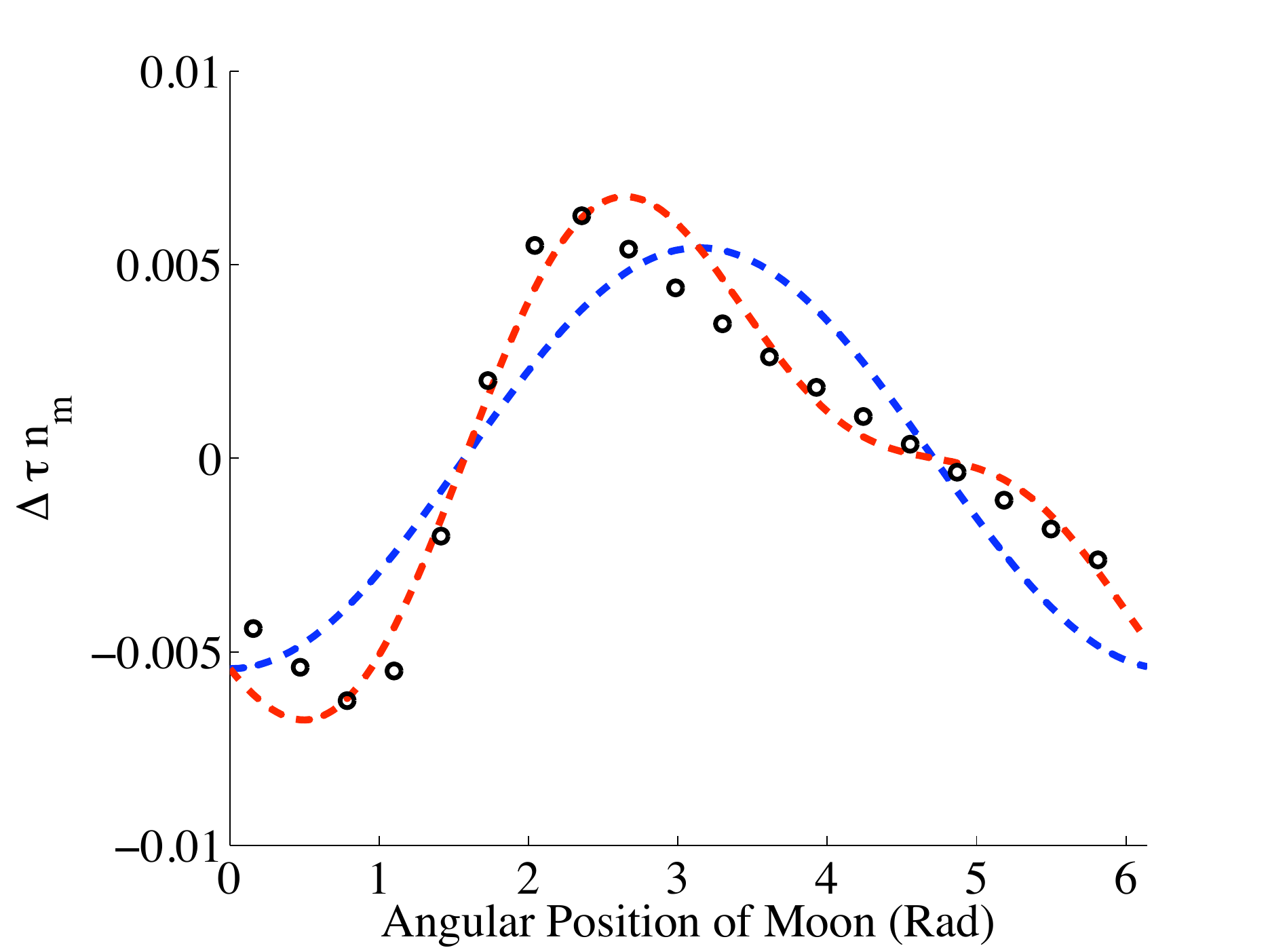}}
     \subfigure[$a_m$=$0.5 R_s$, $v_m/v_{tr}$=$0.66$]{
           \label{TauAgreementInc05B066}
           \includegraphics[width=.31\textwidth]{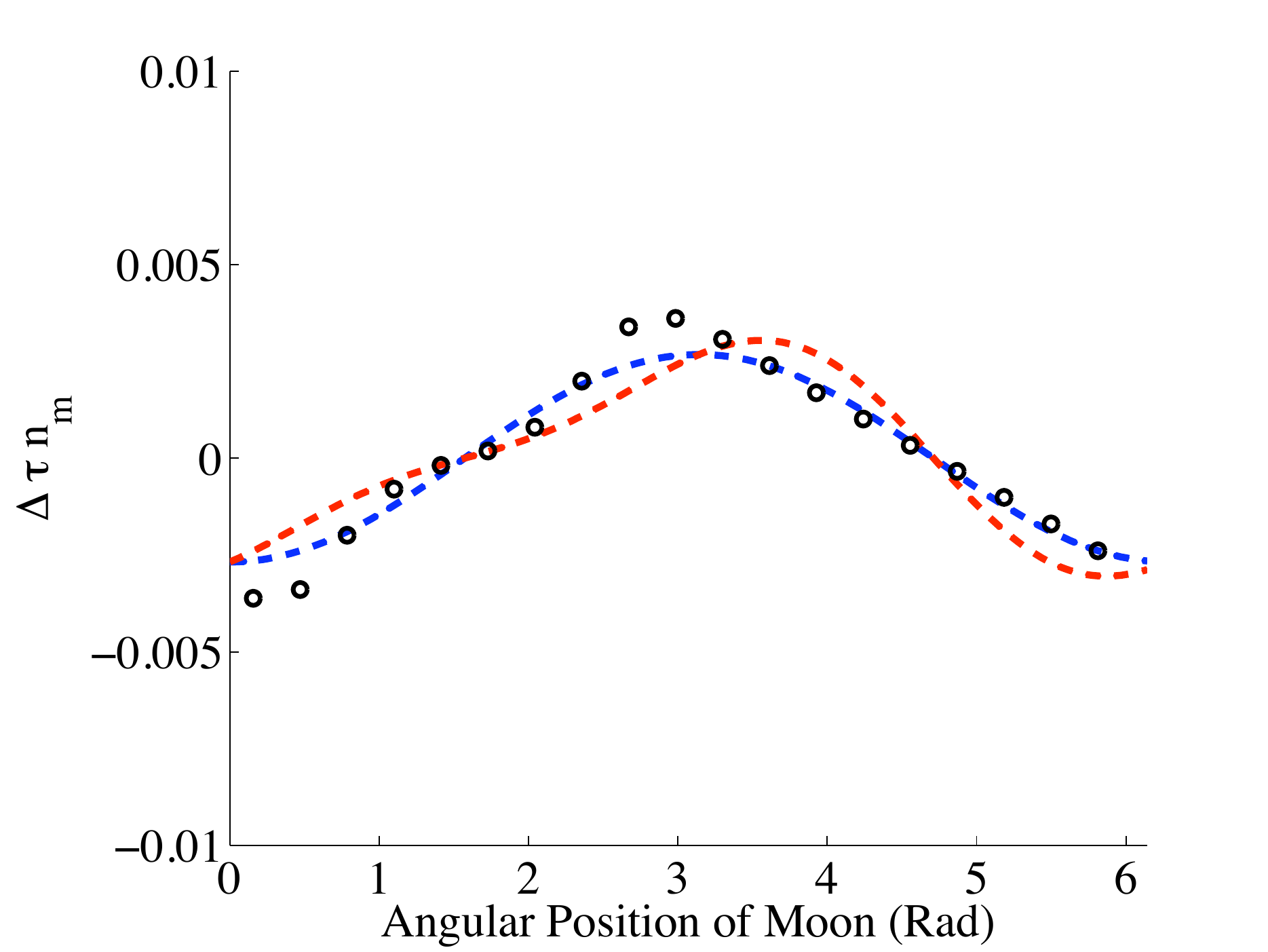}}\\
           \subfigure[$a_m$=$2 R_s$, $v_m/v_{tr}$=$0.49$.]{
          \label{TauAgreementInc2B049}
          \includegraphics[width=.31\textwidth]{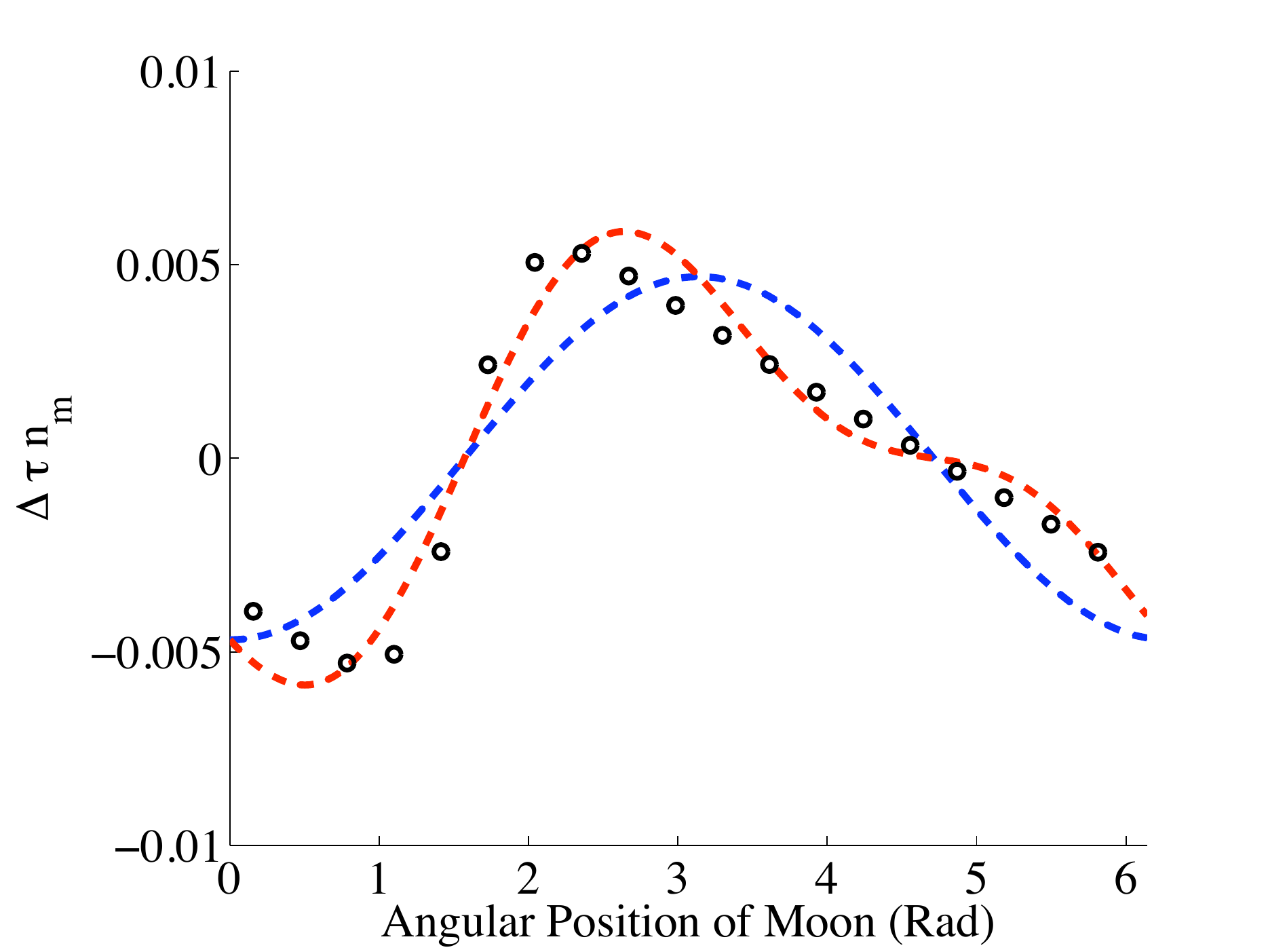}}
     \subfigure[$a_m$=$R_s$, $v_m/v_{tr}$=$0.49$.]{
          \label{TauAgreementInc1B049}
          \includegraphics[width=.31\textwidth]{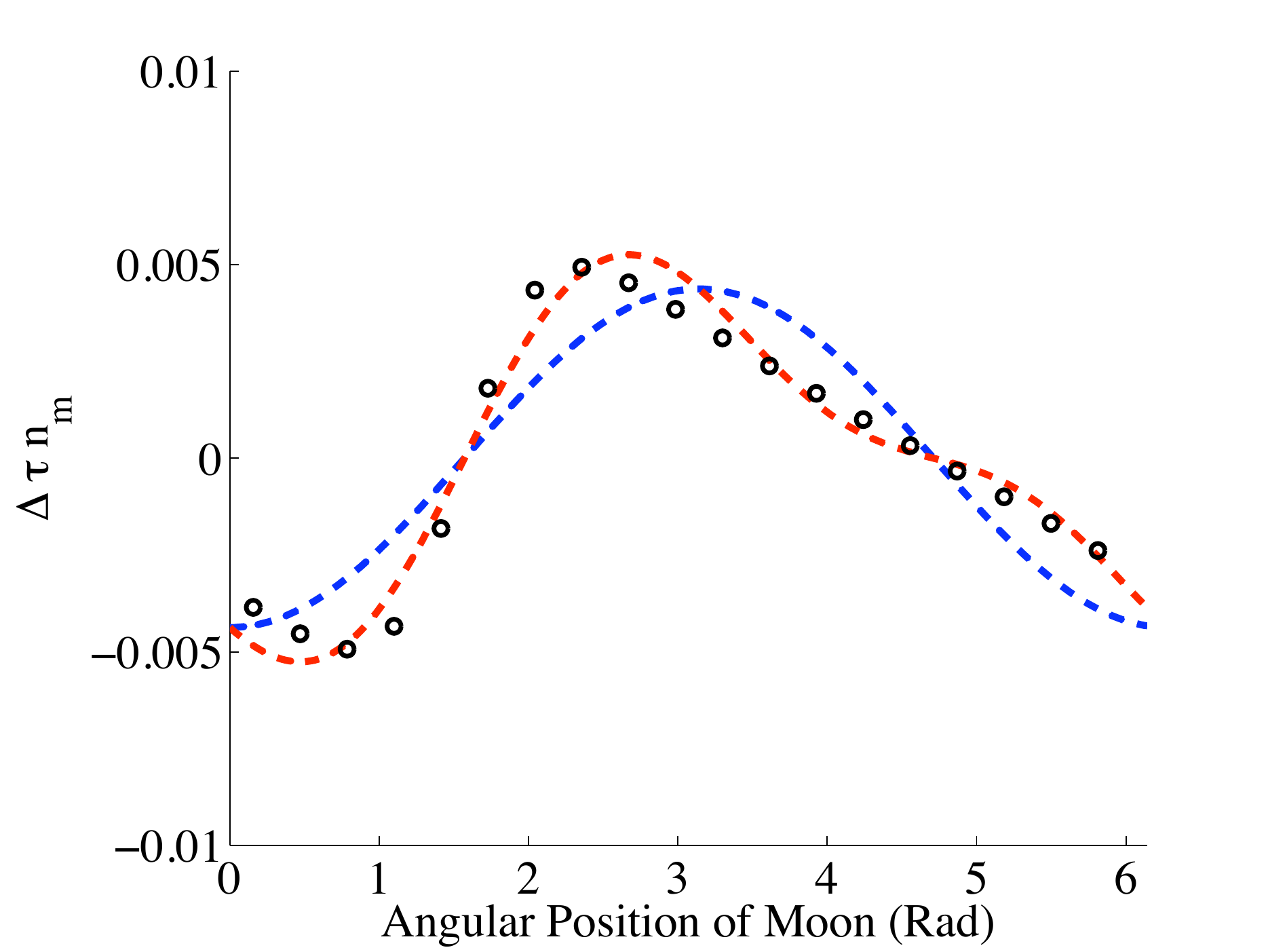}}
     \subfigure[$a_m$=$0.5 R_s$, $v_m/v_{tr}$=$0.49$.]{
           \label{TauAgreementInc05B049}
           \includegraphics[width=.31\textwidth]{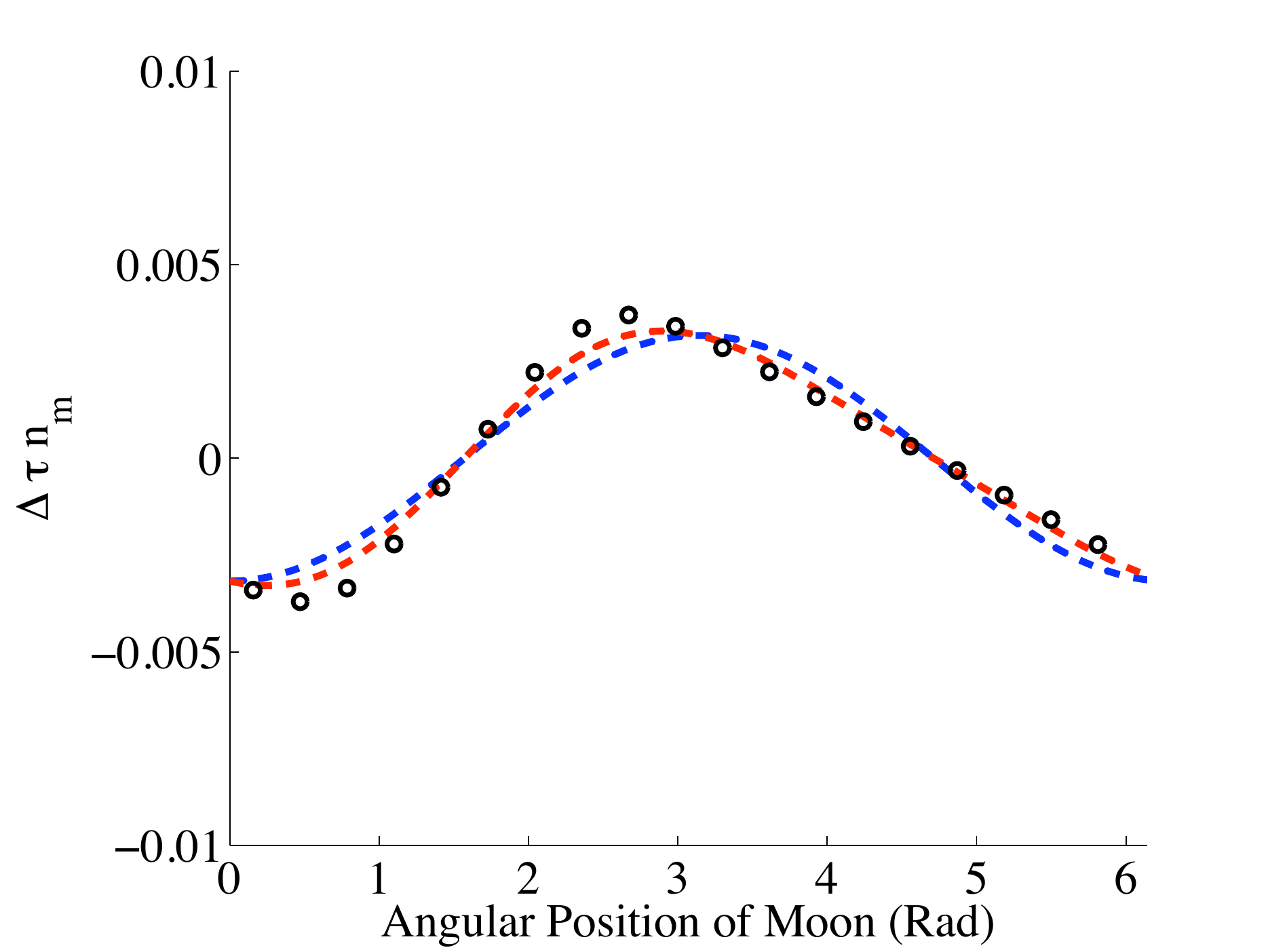}}\\
           \subfigure[$a_m$=$2 R_s$, $v_m/v_{tr}$=$0.33$.]{
          \label{TauAgreementInc2B033}
          \includegraphics[width=.31\textwidth]{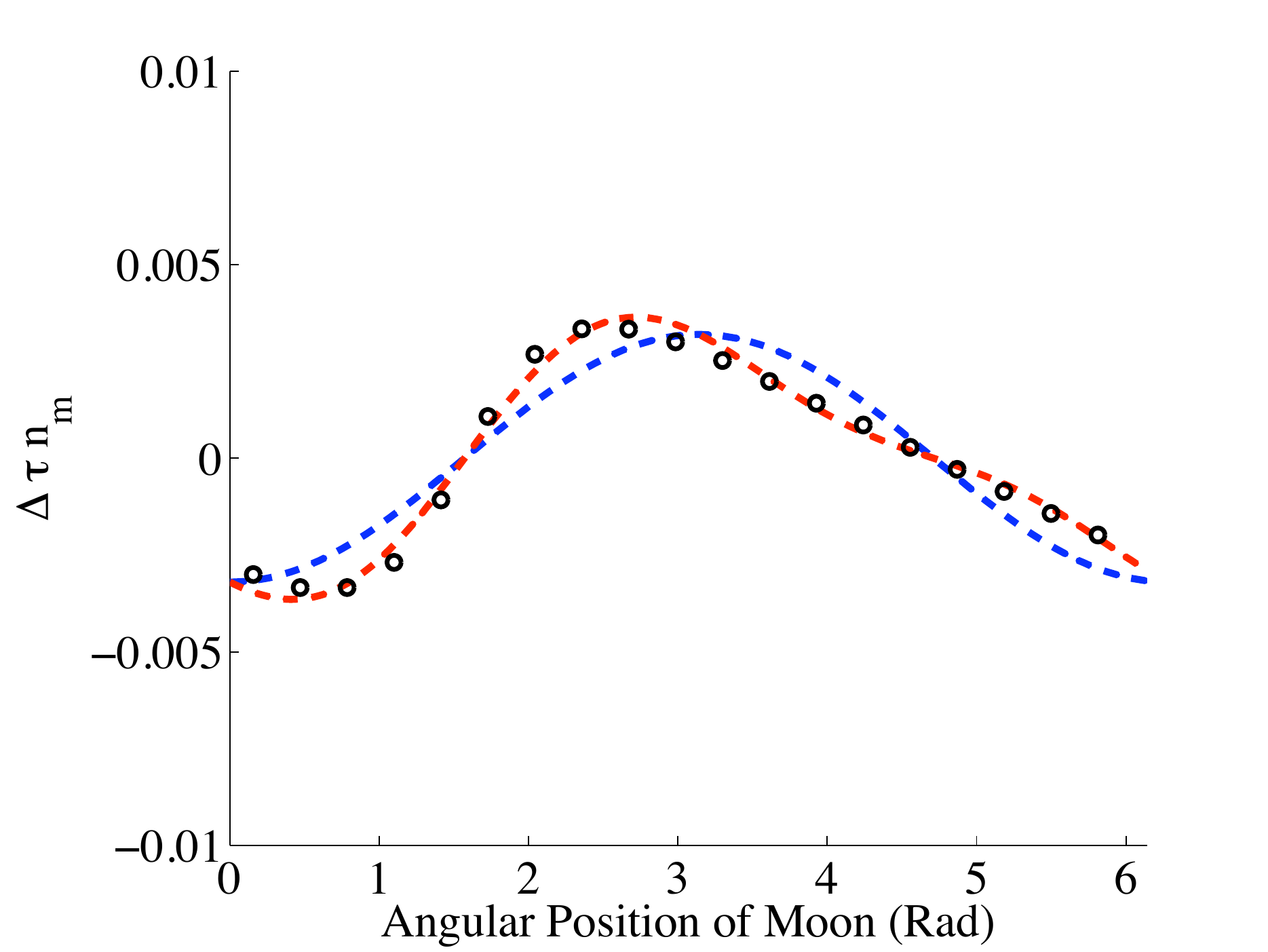}}
     \subfigure[$a_m$=$R_s$, $v_m/v_{tr}$=$0.33$.]{
          \label{TauAgreementInc1B033}
          \includegraphics[width=.31\textwidth]{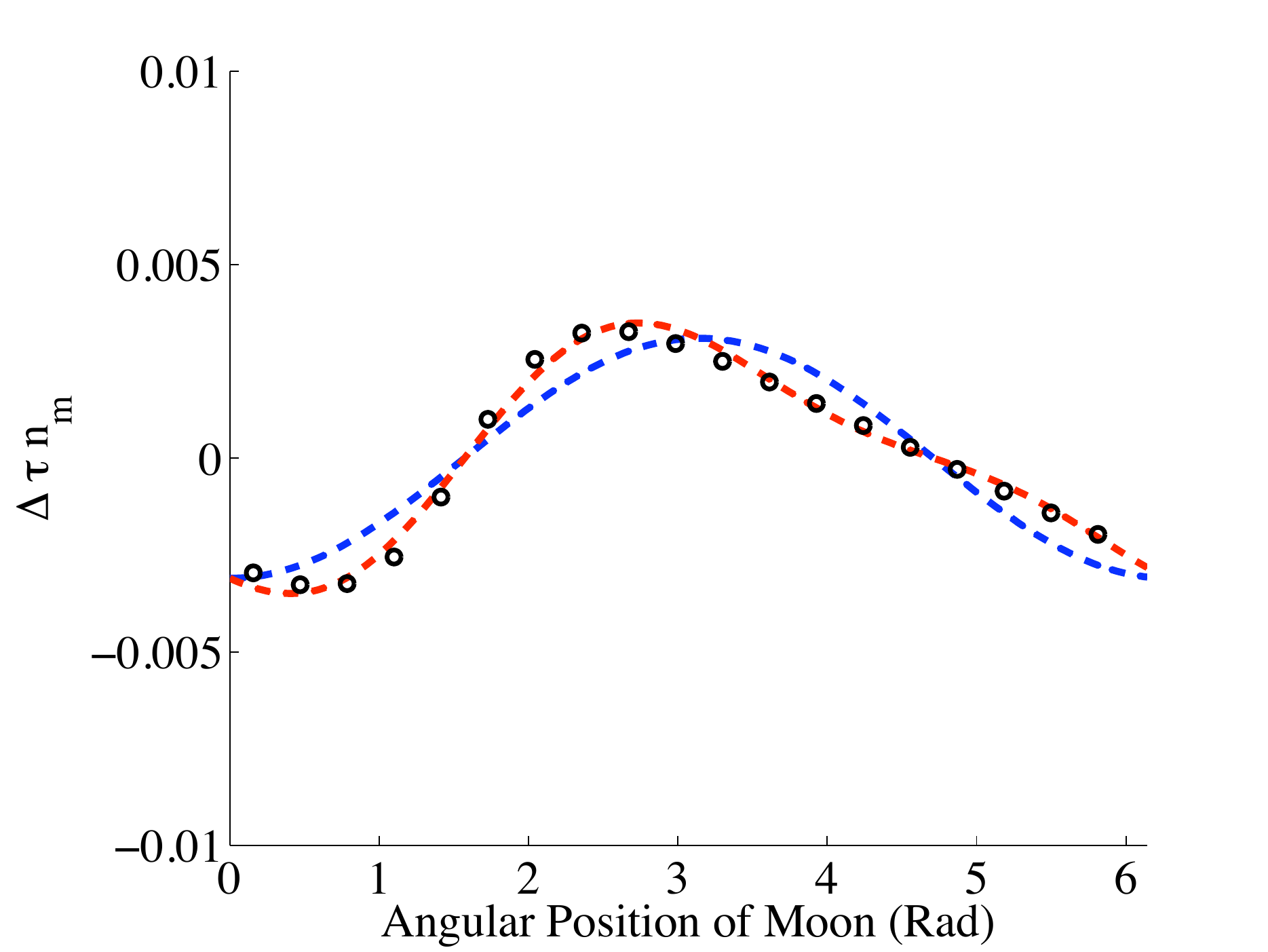}}
     \subfigure[$a_m$=$0.5 R_s$, $v_m/v_{tr}$=$0.33$.]{
           \label{TauAgreementInc05B033}
           \includegraphics[width=.31\textwidth]{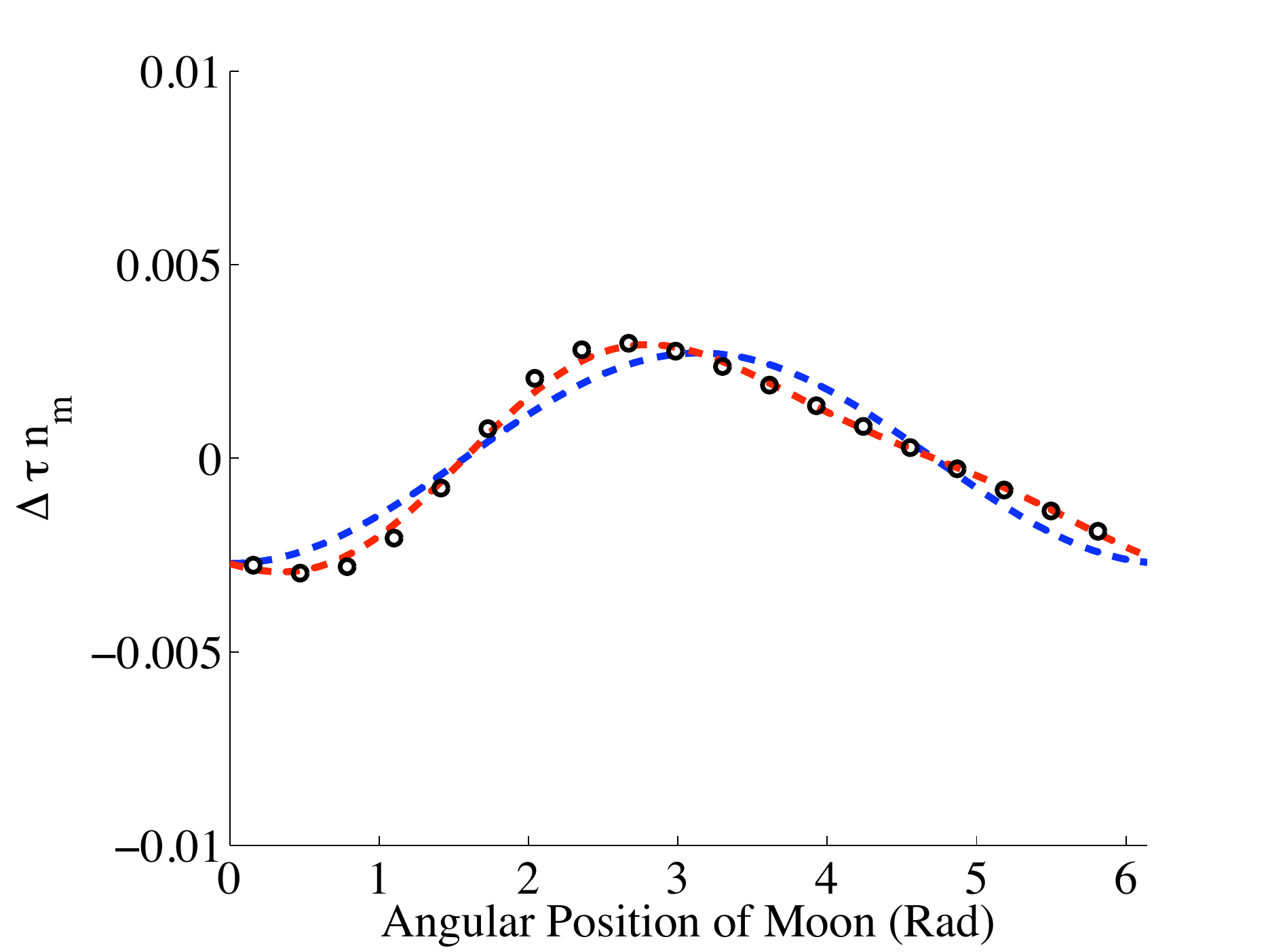}}\\
           \subfigure[$a_m$=$2 R_s$, $v_m/v_{tr}$=$0.16$.]{
          \label{TauAgreementInc2B016}
          \includegraphics[width=.31\textwidth]{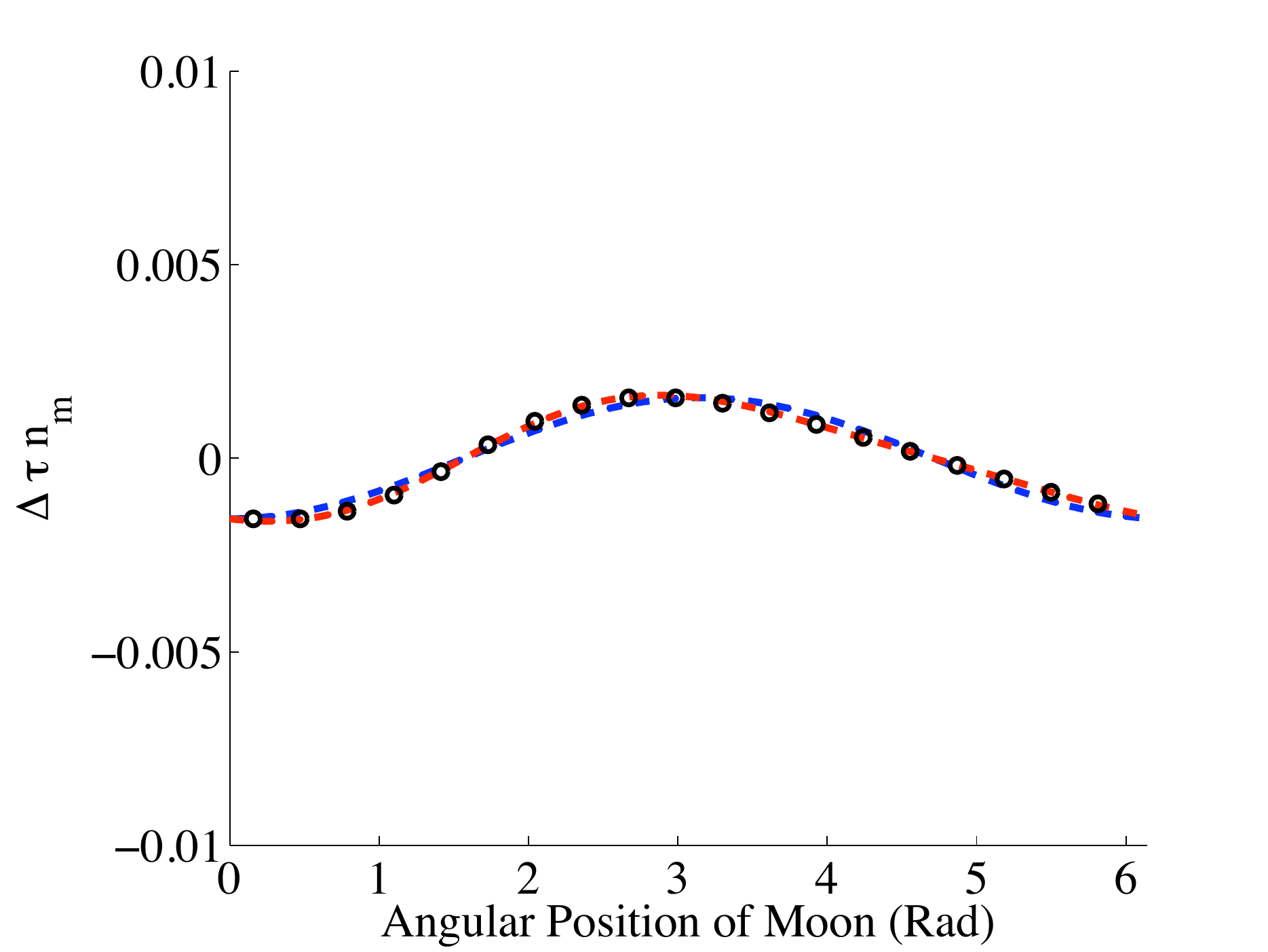}}
     \subfigure[$a_m$=$R_s$, $v_m/v_{tr}$=$0.16$.]{
          \label{TauAgreementInc1B016}
          \includegraphics[width=.31\textwidth]{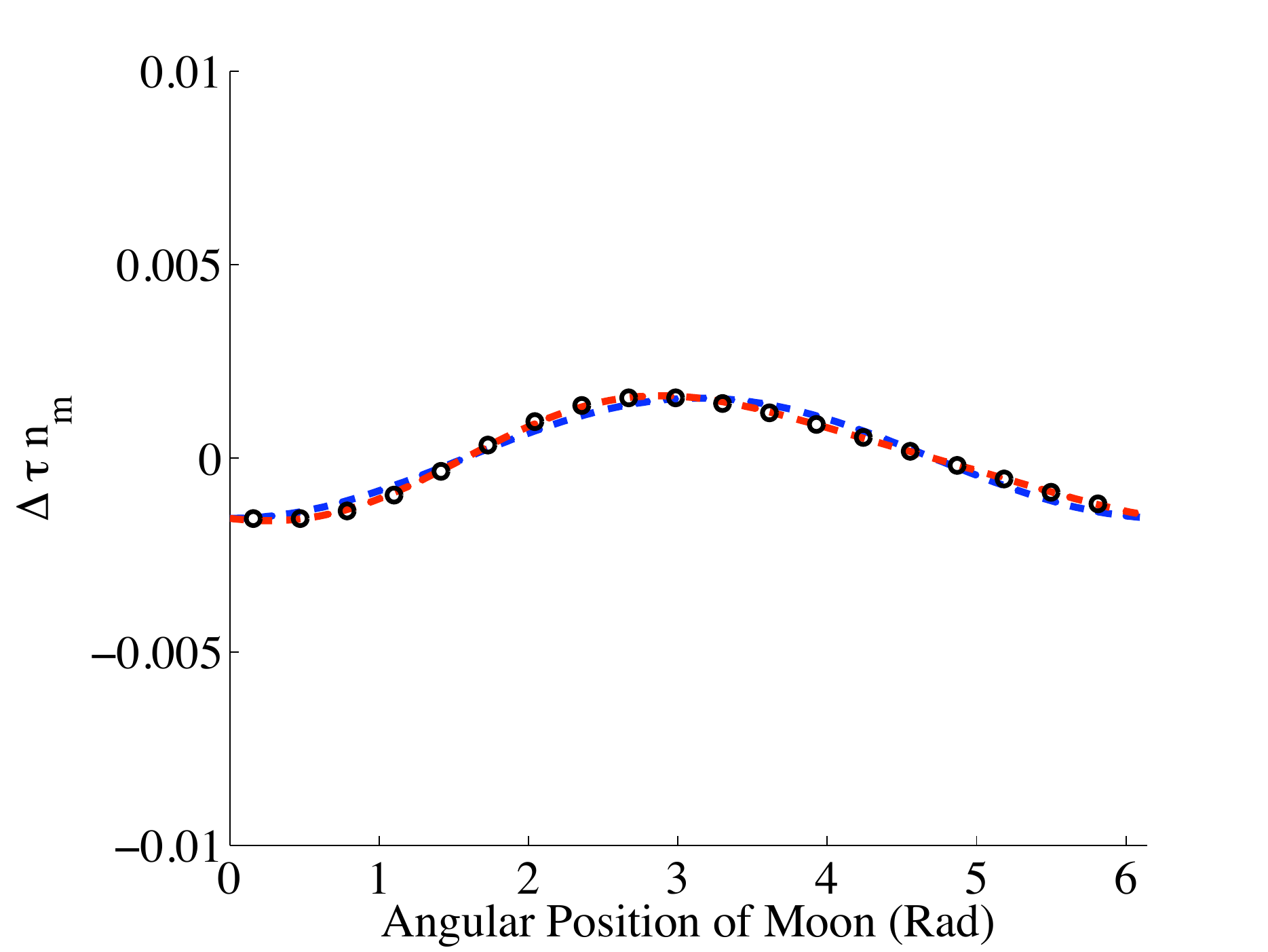}}
     \subfigure[$a_m$=$0.5 R_s$, $v_m/v_{tr}$=$0.16$.]{
           \label{TauAgreementInc05B016}
           \includegraphics[width=.31\textwidth]{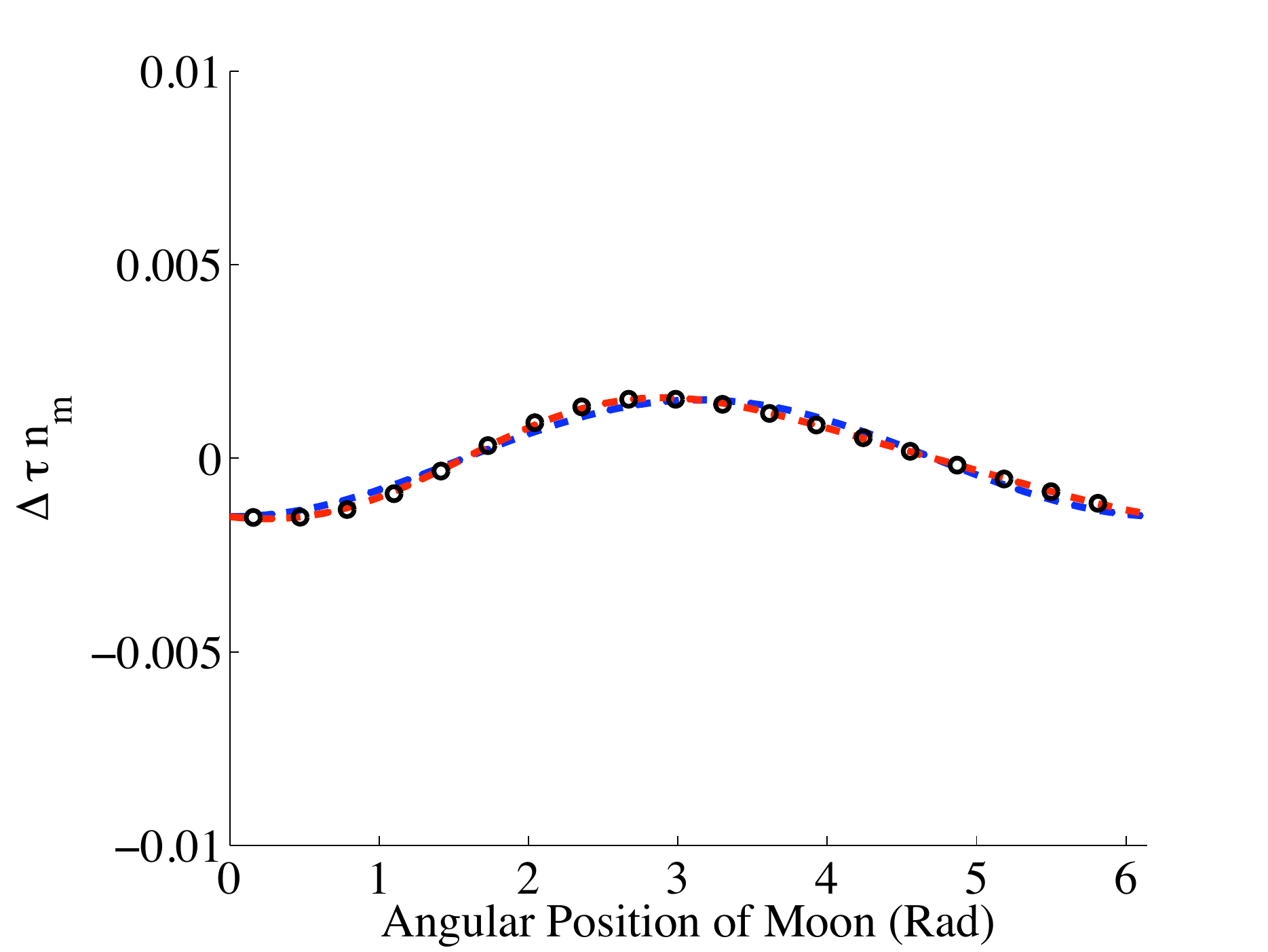}}
     \caption{Figure of the same form as figure~\ref{TauAgreement}, but calculated assuming $\delta_{min} = 0.5 R_s$.}
     \label{TauAgreementInc}
\end{figure}

Again, expressions for $\Delta \tau$ can be investigated for larger values of $B$ by looking at the expression correct to second order in $B$.  Replacing $R_s$ with $(R_s^2 - \delta_{min}^2)^{1/2}$ in equation~\eqref{transit_signal_cc_form_hB}, the equivalent equation for the case of circular coplanar orbits, gives
\begin{multline}
\Delta \tau = \frac{\hat{A}_m M_p - \hat{A}_p M_m}{ \hat{A}_{pm} M_{pm} } \frac{a_m}{v_{tr}}  \cos\left(\frac{n_m\sqrt{R_s^2 - (a_p\cos I_p)^2}}{v_{tr}}\right) \\ \times \cos \left(  f_m(t_0) + \omega_m + jn_mT_p \right) \\
- \frac{\hat{A}_p M_m^2 + \hat{A}_m M_p^2}{ \hat{A}_{pm} M_{pm} ^2} \frac{a_m n_m}{v_{tr}} \frac{a_m}{v_{tr}} \cos\left(2\frac{n_m\sqrt{R_s^2 - (a_p\cos I_p)^2}}{v_{tr}}\right) \\ \times \sin \left( 2 \left( f_m(t_0) + \omega_m + jn_mT_p\right)\right)\\
 + \frac{1}{4 } \frac{\hat{A}_m\hat{A}_p}{\hat{A}_{pm}^2} \frac{a_m}{v_{tr}} \frac{a_m}{R_{s}} \sin\left(2\frac{n_m\sqrt{R_s^2 - (a_p\cos I_p)^2}}{v_{tr}}\right) \\ \times \sin \left(2(f_m(t_0) + \omega_m + jn_mT_p)\right) . \label{transit_signal_inc_form_hB}
 \end{multline}

Again, to determine the effect of truncating the expansion at first order in $B$, equations~\eqref{transit_signal_inc_form_lB} and \eqref{transit_signal_inc_form_hB} were compared with the value of $\Delta \tau$ calculated from a simulated light curve.  These results are presented in figure~\ref{TauAgreementInc}.  As was the case for circular and coplanar orbits, the values of $\Delta \tau$ given by equations \eqref{transit_signal_inc_form_lB} and \eqref{transit_signal_inc_form_hB} qualitatively agree with the numerically calculated values, even for extreme values of $v_m/v_{tr}$.  Now that the form of $\Delta \tau$ has been derived for the case of slightly inclined planetary orbits we can explore the effect of inclination on the properties of the signal.

\subsection{Effect of inclination on $\Delta \tau$}

Inclination of the planetary orbit has two main effects on $\Delta \tau$.  First, the amplitude of the signal is slightly increased.  Second, it results in an increase in the accuracy of $\Delta \tau$ through an increase in the accuracy of  approximation of uniform velocities during transit.  These two effects will be discussed in turn.

As can be seen by comparing equation~\eqref{transit_signal_cc_form_lB}, the expression for $\Delta \tau$ for the case of circular coplanar orbits with equation~\eqref{transit_signal_inc_form_lB}, the expression for $\Delta \tau$ for the case of slightly inclined orbits, the only term which is modified is the cosine term in the amplitude.  As discussed in section~\ref{Trans_TTV_Signal_CC_PropAmp}, this term is likely to be approximately unity as the the argument of the cosine function is likely to be small.  In addition, as $(R_s^2 - \delta_{min}^2)^{1/2} < R_s$ by definition, any inclination of the planetary orbit will act to reduce the argument of the cosine term further and consequently it the cosine term will tend to increase toward one.  As a result, inclination of the planetary orbit causes a slight increase in the amplitude of $\Delta \tau$.

In addition to modifying the amplitude of $\Delta \tau$, changing the inclination of the planetary orbit also increases the accuracy of our approximation to $\Delta \tau$ by increasing the accuracy of the assumption that the planet and moon travel with uniform velocities during transit.  Physically this is a result of the reduced transit duration (and thus the shorter amount of time in which the planet or moon has to accelerate).  Consequently for the case of inclined orbits, the expression derived for $\Delta \tau$ is likely to be more accurate than the equivalent expression for circular coplanar orbits (in particular compare figure~\ref{TauAgreement05B066} and figure~\ref{TauAgreementInc05B066}).

Now that the form and properties of $\Delta \tau$ for the case of inclined orbits have been investigated, we will look at our third and final special case, the case of eccentric planetary orbits.

\section[Eccentric orbit aligned to the line-of-sight]{Eccentric planet orbit aligned to the line-of-sight}\label{Trans_TTV_Signal_EccO}

\begin{figure}[tb]
\begin{center}
\includegraphics[height=2.55in,width=4.5in]{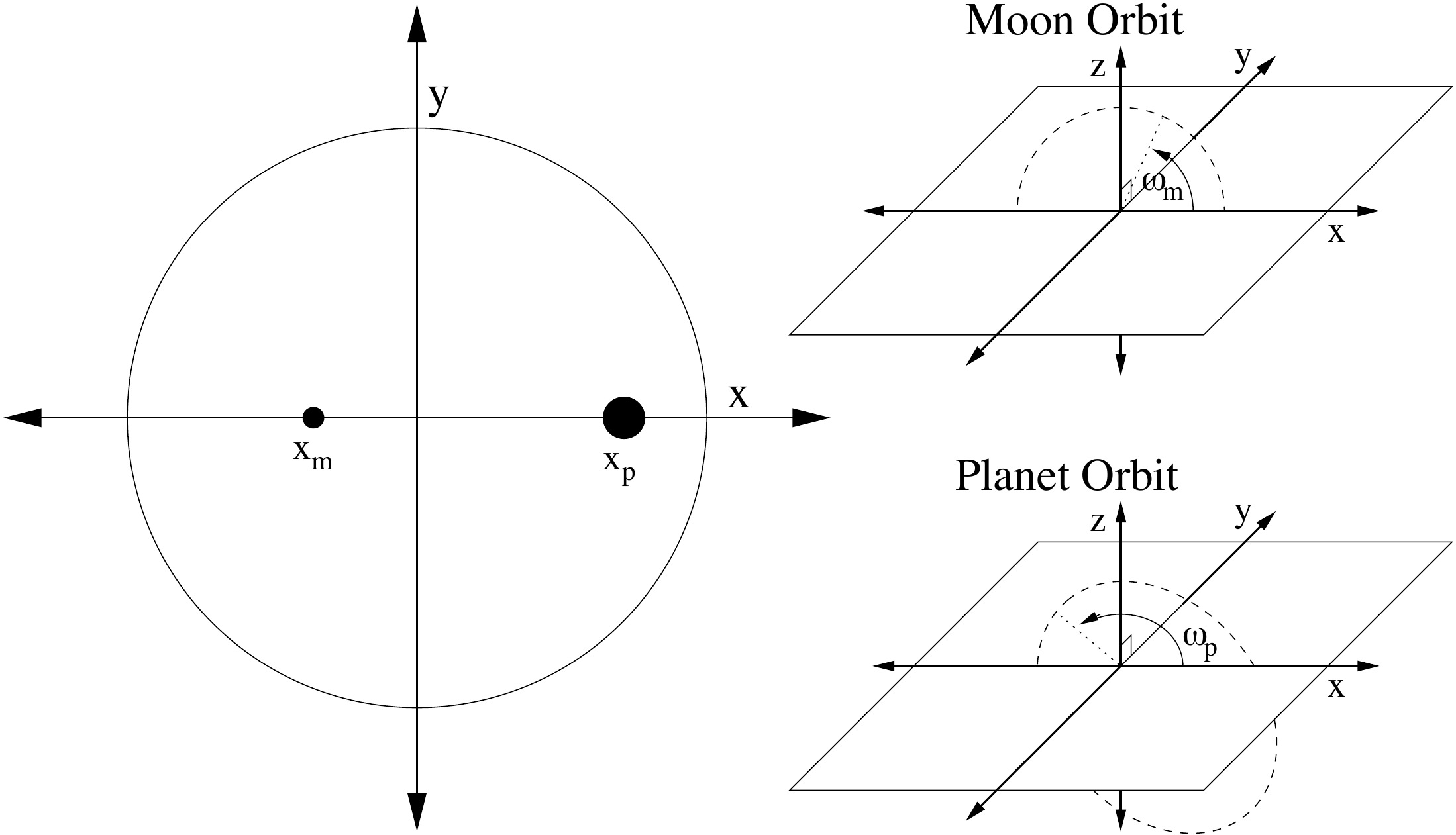}
\caption[Schematic diagram of the same form as figure~\ref{TransitSignalGenCoordSysRot} of the coordinate system for the case of an eccentric planet orbit aligned to the line-of-sight.]{Schematic diagram of the same form as figure~\ref{TransitSignalGenCoordSysRot} of the coordinate system for the case of an eccentric planet orbit aligned to the line-of-sight. In particular, it is assumed that $I_p = \pi/2$, $I_m = \pi/2$ and $\Omega_m = \Omega_p$. }
\label{TransitSignalCoordSysOe}
\end{center}
\end{figure}

Eccentricity in the planet's orbit can also alter the behaviour of $\Delta \tau$.  To isolate the effect of non-zero eccentricity in the planet's orbit, it was assumed that the orbits of the planet and the moon were coplanar, and aligned to the line-of-sight, that is, $I_p = \pi/2$, $I_m = \pi/2$ and $\Omega_p = \Omega_m$, and that the moon's orbit was circular, that is, $r_m = a_m$ and $f_m = n_m t + f_m(0)$.

Again, the dependence of $\Delta \tau$ on the orbital parameters of the planet and moon can be determined from an examination of the time of ingress and egress of the planet and moon.  Rewriting equations~\eqref{transit_signal_coord_xpdef} and \eqref{transit_signal_coord_xmdef}, under these assumptions gives
\begin{align}
x_p &= r_p\cos(f_p + \omega_p) - \frac{M_m}{M_p + M_m}a_p\cos(n_m t + f_m(0) + \omega_m),\label{TraM-TTV-Oecc-xp}\\
x_m &= r_p\cos(f_p + \omega_p) + \frac{M_m}{M_p + M_m}a_p\cos(n_m t + f_m(0) + \omega_m),\label{TraM-TTV-Oecc-xm}
\end{align}
where $r_p$ is defined by equation \eqref{transit_signal_coord_Rdef} and $f_p$ is defined by equations \eqref{transit_signal_coord_fodef}and \eqref{transit_signal_coord_Eodef}.  We can again Taylor expand the first terms in equations~\eqref{TraM-TTV-Oecc-xp} and \eqref{TraM-TTV-Oecc-xm} about the the time that the mid-time of the transit would have occurred had there been no moon, giving
\begin{multline}
x_p = \frac{n_p a_p}{F(e_p,\omega_p)}(t - (jT_p + t_0)) \\- \frac{M_m}{M_p + M_m}a_p\cos(n_m t + f_m(0) + \omega_m)\label{TraM-TTV-Oecc-TExp},
\end{multline}
\begin{multline}
x_m = \frac{n_p a_p}{F(e_p,\omega_p)}(t - (jT_p + t_0)) \\+ \frac{M_m}{M_p + M_m}a_p\cos(n_m t + f_m(0) + \omega_m)\label{TraM-TTV-Oecc-TExm},
\end{multline}
where we recall that the transit velocity for a circular orbit is given by $n_p a_p$, and that for the case of eccentric orbits, the transit velocity is modified by a factor $F^{-1}$, given by equation~\eqref{transit_intro_dur_ecc_Fdef}, and restated below for convenience
\begin{multline*}
F(e_p, \omega_p) =  \left[ \frac{(1+e_p\sin \omega)}{(1 + e_p) \sqrt{1-e_p^2}} \frac{\cos^2 \left(\tan^{-1}\left( \sqrt{\frac{1-e_p}{1+e_p}} \tan \left(\frac{\pi}{4} - \frac{\omega_p}{2}\right) \right) \right)}{ \cos^2 \left(\frac{\pi}{4} - \frac{\omega_p}{2}\right)}\right. \\- \left.\frac{e_p (\sin \omega_p + e_p)}{\sqrt{1-e_p^2} (1 + e_p \sin \omega_p)}\right].
\end{multline*}

Now, equations~\eqref{TraM-TTV-Oecc-TExp} and \eqref{TraM-TTV-Oecc-TExm} are mathematically equivalent to equations~\eqref{transit_signal_cc_xpexp} and \eqref{transit_signal_cc_xmexp}, the equations describing circular coplanar orbits.  In addition, as the $x$-coordinates describing the start and end of the transit ($x = \pm R_s$) are also the same, the expressions derived for $\Delta \tau$ for the case of circular coplanar orbits can be directly modified to give $\Delta \tau$ for the case of eccentric outer orbits, by replacing $v_{tr}$ with $n_pa_p/F(e_p,\omega_p)$.

\subsection{Form of $\Delta \tau$}

Using this method of substituting $n_pa_p/F(e_p,\omega_p)$ for $v_{tr}$ into equations~\eqref{transit_signal_cc_form_lB}, \eqref{transit_signal_cc_form_lBsimp} and \eqref{transit_signal_cc_form_hB}, expressions for $\Delta \tau$ for the case of eccentric planet orbits can be derived.  Again, these expressions will be investigated for the case where $v_m/v_{tr} \ll 1$ and the case where $v_m/v_{tr} < 0.66$.

\subsubsection{Case where $v_m/v_{tr} \ll 1$}

For the case where $v_m/v_{tr} \ll 1$, this substitution results in
\begin {multline}
\Delta \tau = \frac{\hat{A}_m M_p -\hat{A}_p M_m}{\hat{A}_{pm} M_{pm}}  F(e_p,\omega_p) \frac{a_m}{a_p n_p} \cos\left(\frac{n_mR_{s}F(e_p,\omega_p)}{a_p n_p}\right) \\
\times \cos\left(f_m(t_0) + \omega_m +  n_m jT_p \right),\label{transit_signal_oe_form_lB}
\end{multline}
and 
\begin{multline}
\Delta \tau  \approx  \frac{\hat{A}_m}{\hat{A}_{pm}}\frac{M_p}{M_{pm}} F(e_p,\omega_p) \frac{a_m}{a_p n_p} \cos\left(\frac{n_mR_{s}F(e_p,\omega_p)}{a_p n_p}\right) \\ \times \cos (f_m(t_0) + \omega_m + jn_mT_p)),\label{transit_signal_oe_form_lBsimp}
\end{multline}
where the first term in the quotient has again been neglected.

\subsubsection{Case where $v_m/v_{tr} \le 0.66$}

Similarly, for the case where $v_m/v_{tr} \le 0.66$, this substitution results in
\begin{multline}
\Delta \tau = \frac{\hat{A}_m M_p - \hat{A}_p M_m}{ \hat{A}_{pm} M_{pm} } \frac{a_m}{a_p n_p} F(e_p,\omega_p) \cos\left(\frac{n_mR_{s}F(e_p,\omega_p)}{a_p n_p}\right) \\
\times \cos \left(  f_m(t_0) + \omega_m + jn_mT_p \right) - \frac{\hat{A}_p M_m^2 + \hat{A}_m M_p^2}{ \hat{A}_{pm} M_{pm} ^2} \frac{a_m n_m}{a_p n_p} \frac{a_m}{a_p n_p} \\
\times F(e_p,\omega_p)^2 \cos\left(2\frac{n_mR_{s}F(e_p,\omega_p)}{a_p n_p}\right) \sin \left( 2 \left( f_m(t_0) + \omega_m + jn_mT_p\right)\right)\\
 + \frac{1}{4 } \frac{\hat{A}_m\hat{A}_p}{\hat{A}_{pm}^2} \frac{a_m}{a_p n_p} \frac{a_m}{R_{s}} F(e_p,\omega_p) \sin\left(2\frac{n_mR_{s}F(e_p,\omega_p)}{a_p n_p}\right) \\ \times \sin \left(2(f_m(t_0) + \omega_m + jn_mT_p)\right) . \label{transit_signal_oe_form_hB}
 \end{multline}
 
\subsection{Effect of eccentricity on $\Delta \tau$}

As the signal in the case where the planet's orbit is eccentric is mathematically equivalent to the case for circular coplanar orbits, all the properties discussed in section~\ref{Sec-TraM-TTV-CC-prop} still hold.  However, eccentricity in the planet's orbit alters the transit velocity, which means that it alters the amplitude of $\Delta \tau$ and 
also alters the point at which the expansion used to derive the expressions for $\Delta \tau$ breaks down ($v_m/v_{tr} < 0.66$).  These two issues will be discussed in turn.

For the case where the planet's orbit is eccentric, its transit velocity may no longer be equal to $a_p n_p$.  In particular, the value of $v_{tr}$ depends on both $e_p$ and $\omega_p$ and can include values from $n_pa_p\sqrt{1+e_p}/\sqrt{1-e_p}$ at periastron to $n_pa_p\sqrt{1-e_p}/\sqrt{1+e_p}$ at apastron \citep[e.g.][p. 31]{Murrayetal1999}. As the amplitude of $\Delta \tau$ is proportional to $1/v_{tr}$, eccentricity in the planet's orbit can affect $\Delta \tau$, and does so in three different regimes:
\begin{enumerate}
\item For the case where the planet's orbit is highly eccentric, and the transit occurs near pericenter, the transit velocity is increased and the amplitude of $\Delta \tau$ is consequently reduced from the equivalent value for circular orbits.
\item For the case where the planet's orbit is not very eccentric, or has just the correct orientation, the transit velocity of the planet can be approximated to be equal to that for the equivalent planet on a circular orbit.  In this case, unsurprisingly, $\Delta \tau$ is unchanged.
\item Finally, for the case where the planet's orbit is highly eccentric, but the transit occurs near apocenter, the transit velocity is decreased and the amplitude of $\Delta \tau$ is consequently increased from the equivalent value for circular orbits.
\end{enumerate}
While it is possible that a planet on an eccentric orbit could transit near apocenter, it is not likely (see appendix~\ref{EccProb_App}).  Consequently, for a majority of transiting planets, eccentricity in the planet's orbit will either decrease, or leave unchanged the amplitude of $\Delta \tau$.

In addition to affecting its amplitude, the value of $v_{tr}$ also determines whether or not the expansions used to derive the expressions for $\Delta \tau$ are accurate.  In particular, for the case where $v_{tr}$ is high (case 1 above) the expansions is more accurate, for the case where $v_{tr}$ is approximately equal to that for the equivalent circular  orbit (case 2 above), the expressions are as good as for the circular case and for the case where $v_{tr}$ is low (case 3 above) the expansions is less accurate.  Fortunately, it is more probable that a given planet will transit while its velocity is greater than $a_p n_p$ than for it to transit while less than than $a_p n_p$ (see appendix~\ref{EccProb_App}), especially for highly eccentric orbits.  Consequently, for the case of eccentric planetary orbits, the expansion used to derive $\Delta \tau$ is likely to be more accurate than for the case of a circular orbit.

For the case of planets with eccentric orbits, eccentricity has the following two effects on $\Delta \tau$.  As planets are more likely to transit when their velocities are higher, the amplitude of $\Delta \tau$ is likely to be reduced, but the accuracy of the analytic expression for $\Delta \tau$ will increase.  

\section{Conclusion}

The form and properties of $\Delta \tau$ have been investigated for three cases for which the moon's orbit was circular and coplanar with the planet's orbit (sections~\ref{Trans_TTV_Signal_CC}, \ref{Trans_TTV_Signal_Inc} and \ref{Trans_TTV_Signal_EccO}), and one case for which it was not (appendix~\ref{EccMoon_App}).  Analytic expressions for $\Delta \tau$ were derived assuming that the velocity at which the planet-moon pair orbited each other was much smaller than the velocity at which they transited their host star.  In addition, a number of properties of the signal arising from its form were derived for each of these cases.   The form of these expressions, and their properties will be discussed in turn.

\subsection{The form of $\Delta \tau$}

\begin{sidewaystable}[p] 
\centering       
 \begin{tabular}{cccccc}
\hline
   \multicolumn{2}{c}{Planet Orbit} &
   \multicolumn{1}{c}{Equation} &
   \multicolumn{1}{c}{$A$} &
   \multicolumn{1}{c}{$\omega$} &
   \multicolumn{1}{c}{$\phi$} \\
 $I_p$  & $e_p$ & No. & 		 & 		 & 		 \\
   \hline 
  $\frac{\pi}{2}$  & 0      & \eqref{transit_signal_cc_form_lB} & $\frac{\hat{A}_m M_p -\hat{A}_p M_m}{\hat{A}_{pm} M_{pm}}  \frac{a_m}{v_{tr}} \cos \left(\frac{n_m R_s}{v_{tr}}\right)$ & $n_mT_p$ & $f_m(t_0) + \omega_m$  \\
  $\ne \frac{\pi}{2}$ & 0 & \eqref{transit_signal_inc_form_lB} & $\frac{\hat{A}_m M_p -\hat{A}_p M_m}{\hat{A}_{pm} M_{pm}}  \frac{a_m}{v_{tr}} \cos \left(\frac{n_m \sqrt{R_s^2 - (a_p\cos I_p)^2}}{v_{tr}}\right)$ & $n_mT_p$ & $f_m(t_0) + \omega_m$        \\
$\frac{\pi}{2}$ & $\ne 0$ & \eqref{transit_signal_oe_form_lB} & $\frac{\hat{A}_m M_p -\hat{A}_p M_m}{\hat{A}_{pm} M_{pm}}  F(e_p,\omega_p) \frac{a_m}{a_p n_p} \cos\left(\frac{n_mR_{s}F(e_p,\omega_p)}{a_p n_p}\right)$ & $n_mT_p$ & $f_m(t_0) + \omega_m$     \\
               \hline
 \end{tabular}

\caption{Expressions for $A$, $\omega$ and $\phi$ in terms of physical variables for the three cases investigated in this chapter.}  
\label{DelTauCoeffTab}
\end{sidewaystable}

For the case where the moon's orbit was circular and coplanar with the planet's orbit, the form of $\Delta \tau$ was investigated for planet orbits which were circular and aligned to the line-of-sight (section~\ref{Trans_TTV_Signal_CC}), circular and slightly inclined to the line-of-sight (section~\ref{Trans_TTV_Signal_Inc}) and eccentric and aligned to the line-of-sight (section~\ref{Trans_TTV_Signal_EccO}).  In addition, the case where the moon's orbit was slightly eccentric and coplanar with the planet's orbit, for the case where the planet's orbit was circular and aligned to the line-of-sight was investigated in appendix~\ref{EccMoon_App}.  From the work in sections~\ref{Trans_TTV_Signal_CC}, \ref{Trans_TTV_Signal_Inc} and \ref{Trans_TTV_Signal_EccO}, the cases where the moon's orbit was circular and coplanar, it can be seen (compare equations~\eqref{transit_signal_cc_form_lB}, \eqref{transit_signal_inc_form_lB} and \eqref{transit_signal_oe_form_lB}) that $\Delta \tau$ is given by a function of the form
\begin{equation}
\Delta \tau = A \cos (\omega j + \phi).\label{transit_signal_conc_form}
\end{equation}
For reference the values of the coefficients $A$, $\omega$ and $\phi$ for the three cases are given in table~\ref{DelTauCoeffTab}.  To provide physical intuition the amplitudes that would have been produced by a Jupiter-Ganymede and Earth-Moon planet-moon pair for a range of different orbital configurations are given in table~\ref{DelTauEgs}.  As a result of its general applicability, equation~\eqref{transit_signal_conc_form} is the equation which will be used in the analysis in chapter~\ref{Trans_Thresholds}.  However, while this is the equation that will be used,  it should be noted that it will not hold for all cases.  In particular, as shown in appendix~\ref{EccMoon_App} and section~\ref{Trans_TTV_Signal_CC_Form_largeB}, non-negligible values of $e_m$ and $v_m/v_{tr}$ can break the symmetry and introduce higher order sinusoid components.
  
\subsection{The properties of $\Delta \tau$}

In addition to its form, the $\Delta \tau$ signal has two main types of properties.  First it has a characteristic size, parameterised by the amplitude.  Second, owing to the way in which $\tau$ and thus $\Delta \tau$ are measured, only part of this amplitude is detectable.  This detectable component shows a range of interesting and relevant behaviour as a function of moon semi-major axis.  These aspects will be summarised in turn.

\subsubsection{The properties of the amplitude of $\Delta \tau$}

 \begin{table} 
  \begin{tabular}{llrcc}
   \hline
 $a_p$ (AU) &  $e_p$  &  $\omega_p$  & Earth-Moon & Jupiter-Ganymede   \\
   \hline
   0.2  & 0     & 0                       & \sout{428 s} 	& \sout{21.7 s} \\
           & 0.5 & $\frac{\pi}{2}$  & \sout{247 s}  	& \sout{12.5 s} \\
           & 0.5 & -$\frac{\pi}{2}$ & \sout{740 s}  	& \sout{37.5 s} \\
   0.3  & 0     & 0                       & \sout{524 s} 	& 26.6 s \\
           & 0.5 & $\frac{\pi}{2}$  & \sout{303 s}  	& \sout{15.4 s} \\
           & 0.5 & -$\frac{\pi}{2}$ & \sout{907 s} 	& \sout{46.0 s} \\
   0.4  & 0     & 0                       & \sout{605 s} 	& 30.7 s \\
          & 0.5 & $\frac{\pi}{2}$  & \sout{350 s} 	& \sout{17.8 s} \\
           & 0.5 & $-\frac{\pi}{2}$ & \sout{1049 s} 	& \sout{53.3 s} \\
    0.7  & 0     & 0                       & 801 s 		& 40.7 s \\
           & 0.5 & $\frac{\pi}{2}$  & \sout{462 s} 	& 23.5 s \\
           & 0.5 & $-\frac{\pi}{2}$ & \sout{1387 s} 	& 70.5 s \\
      1  & 0     & 0                       & 957 s 			& 48.6 s \\
           & 0.5 & $\frac{\pi}{2}$  & \sout{553 s} 	& 28.1 s \\
           & 0.5 & $-\frac{\pi}{2}$ & \sout{1658 s} 	& 84.2 s \\
  \end{tabular}\\
 \caption[$\Delta \tau$ amplitudes calculated for a range of values of planetary semi-major axis, eccentricity and orbital orientation for the case of an Earth-Moon and Jupiter-Ganymede planet-moon pair assuming they transit the central chord of a Sun-like star.]{$\Delta \tau$ amplitudes calculated for a range of values of planetary semi-major axis, eccentricity and orbital orientation for the case of an Earth-Moon and Jupiter-Ganymede planet-moon pair assuming they transit the central chord of a Sun-like star.  For reference $\omega_p = \pi/2$ and $\omega_p = -\pi/2$ correspond to a transit which occurs during periastron and apastron respectively.  The values for planet-moon pairs which are three-body unstable are crossed out.}
 \label{DelTauEgs}
 \end{table}
 
One way to consider $\Delta \tau$ is to consider the average size of the perturbation caused by the moon.  As can be seen from equation~\eqref{transit_signal_conc_form}, $\Delta \tau$ can be approximated by a sinusoid and thus the degree to which a given moon perturbs transit timings from strict periodicity can be estimated by inspecting the amplitude.  Mathematically, as can be seen from table~\ref{DelTauCoeffTab}, the $\Delta \tau$ amplitude mainly depends on three factors $\hat{A}_m/\hat{A}_p$, $a_m$ and $v_{tr}$.  These variables will be discussed in turn.

First, the amplitude of $\Delta \tau$ is proportional to $\hat{A}_m/\hat{A}_p$ to first order in $v_m/v_{tr}$.  This implies that it is the physical size of the moon and planet\footnote{As will be shown in the next chapter, the amplitude of $\epsilon_j$, the timing noise, is proportional to $(\hat{A}_s/\hat{A}_p)\times((R_s+a_m)/R_s)$.  Consequently, the detectability of a given moon, determined by comparing the sizes of $\Delta \tau$ and $\epsilon_j$, will not depend on $\hat{A}_p$, only $\hat{A}_m$.} which affects the amplitude of $\Delta \tau$, not their masses.  This is in striking comparison with the two other timing methods presented in the literature, barycentric transit timing and transit duration variation, which have timing amplitudes proportional to $M_m/M_p$ and $M_m/M_p^{1/2}$ respectively.  

Second, the moon's semi-major axis also affects the amplitude of $\Delta \tau$.  In particular, $\Delta \tau \propto a_m$, such that moons which are distant from their host planet will have larger $\Delta \tau$ amplitudes than moons which are close to their host planet.  Recall from section~\ref{Intro_Dect_Moons_Transit} that the size of the signal for direct detection does not depend on $a_m$, while the timing amplitude for the case of barycentric transit timing and transit duration variation is proportional to $a_m$ and $a_m^{-1/2} $respectively.  These differences again indicate that there may be regions of parameter space which the TTV$_p$ method could be optimised to probe.

Third and finally, the amplitude of the $\Delta \tau$ signal depends on the transit velocity of the planet.  As discussed during this chapter, the transit velocity of the planet depends on the semi-major axis, eccentricity and orientation of the planet's orbit, aspects which will be discussed in turn.  The semi-major axis determines the size of the orbit as well as the velocity of the planet along that orbit.  In particular, distant planets will have low transit velocities, long transit durations and large $\Delta \tau$ amplitudes, while closer planets will have high transit velocities, short transit durations and low $\Delta \tau$ amplitudes.  In addition to depending on the size of the orbit, the transit velocity also depends on the eccentricity and orientation of the orbit.  While it is possible to have transit velocities ranging from $n_pa_p\sqrt{1+e_p}/\sqrt{1-e_p}$ at perihelion to $n_pa_p\sqrt{1-e_p}/\sqrt{1+e_p}$, (recall that for a circular orbit $v_{tr} = a_p n_p$), it is more likely (see appendix~\ref{EccProb_App}) that a planet on an eccentric orbit will transit while its velocity is higher.  Consequently planets on eccentric orbits are likely to have smaller $\Delta \tau$ values than those for planets  on a circular orbit with the same semi-major axis.

To summarise, the amplitude of $\Delta \tau$ depends on the physical size of the planet and moon, the semi-major axis of the moon and on the semi-major axis, eccentricity and orientation of the planet's orbit, through the transit velocity.  However, the detectability of a given moon is not just determined by the amplitude of $\Delta \tau$, it also depends on the amount of this amplitude which can practically be measured, which, in turn depends on the properties of $\Delta \tau$.

\subsection{The properties of the signal $\Delta \tau$}

In particular to gain a full understanding of $\Delta \tau$ we not only need an understanding of its amplitude, but also on the proportion of its amplitude which can be detected, which, is defined by how $\Delta \tau$ is measured.  As discussed in section~\ref{Sec-TraM-TTV-CC-prop}, $\Delta \tau$ is measured as a perturbation to a linear signal and can only be measured once per orbit.  These issues will be discussed in turn.

First, consider the time that is measured, $\tau_j$.  From the definition of $\tau_j$,
\begin{equation*}
\tau_j = t_0 + jT_p + \Delta \tau + \epsilon_j
\end{equation*}
it can be seen that it is comprised of three terms, a linear term ($t_0 + jT_p$), $\Delta \tau$, a sinusoidal term and a noise term, $\epsilon_j$.  However, for the case where the moon orbits an integer number of times around its planet per planetary orbit or a near integer number of times, $\Delta \tau$ can be constant or approximately linear.  As a result, for these cases, the perturbation due to the moon will be included in the linear fit, and the moon will be effectively undetectable.  As a result, when the detectable portion of $\Delta \tau$ is plotted against moon semi-major axis, a moon which fulfills the above criterion will reside in a non-detection spike with width proportional to $N^{-1}a_m^{2.5}$, where $N$ is the number of transits recorded, and where the proportionality constant depends on $\phi$.

Second, $\Delta \tau$ can only be measured once per orbit, in particular only during transit.  As a result, it is the configuration of the planet and moon during transit which determines $\Delta \tau$.  Consequently the form of the TTV$_p$ signal of a moon which orbits its planet, for example, $n+1/4$ times per planetary period (i.e the moon advances around its orbit by $1/4 \times 2\pi$ radians from one transit to the next) will be the same as a moon which orbits its planet $(n+1) + 1/4$ times per planetary period (i.e. the moon also advances around its orbit by $1/4\times2\pi$ radians from one transit to the next).  Consequently, when plotted against moon semi-major axis, moon detection thresholds should contain repeating blocks which look the same.

\subsection{Conclusion}

The TTV$_p$ signal was investigated by deriving expressions and and exploring properties of $\Delta \tau$  for a range of realistic planet-moon configurations.  In particular, $\Delta \tau$ was investigated for three cases where the moon orbit was circular and coplanar with the planet orbit, specifically, where the planet's orbit was circular and aligned to the line-of-sight (section~\ref{Trans_TTV_Signal_CC}), circular and slightly inclined to the line-of-sight (section~\ref{Trans_TTV_Signal_Inc}) and eccentric and aligned to the line-of-sight (section~\ref{Trans_TTV_Signal_EccO}).  In addition, to determine the effect of altering properties of the moon's orbit, the case where the planet's orbit was circular and aligned to the line-of-sight, but the moon's orbit was eccentric was also investigated (appendix~\ref{EccMoon_App}).  It was discovered for the case where the moon's orbit was circular and coplanar with the planet's orbit that $\Delta \tau$ could be approximated by a sinusoid, and that high values of $v_m/v_{tr}$ or eccentricity in the moon's orbit introduce higher order terms (see section~\ref{Trans_TTV_Signal_CC_Form_largeB} and appendix~\ref{EccMoon_App}).  In addition, $\Delta \tau$ has a number of properties as a result of the fact that it can only be measured once per transit, and is measured in the context of a linear trend ($t_0 + jT_p$).  In particular it was found that for moons which orbit their host planet an integer number of times per planetary orbit are not detectable, as for this case $\Delta \tau$ becomes a linear function of transit number and cannot be independently determined using the fit.  Armed with these expressions and properties, only information about $\epsilon_j$, the timing noise, is needed to derive and interpret detection thresholds.  With this in mind, the timing noise $\epsilon_j$ will be investigated in the following chapter.

\chapter{Measurement error in $TTV_p$ signal}\label{Trans_TTV_Noise}

\section{Introduction}

Whether the TTV$_p$ signal of a given moon can be detected depends not only on $\Delta \tau$, the timing signal produced by the moon, but also on the characteristic size and behaviour of $\epsilon_j$, the timing noise masking this signal. This timing noise is a result of the perturbations on $\tau$ caused by the sequence of $\alpha_n$ values for a given transit, where $\alpha_n$ is the sum of all sources of photometric error for a particular exposure.  As shown in section~\ref{Trans_TTV}, this error, to first order in $\sum_i \alpha_n/(A_p + A_m)$, is given by equation~\eqref{transit_intro_ground_noisedef}
\begin{equation*}
\epsilon_j = \frac{1}{A_p + A_m}\sum_i\left[t_i -  (jT_p + t_0 +\Delta \tau) \right]\alpha_n(t_i).
\end{equation*}

The photometric error in a light curve, $\alpha_n$, can come from many sources, including detection error,  propagation effects, shot noise (error due to small number statistics) and variability in the signal source, in this case the host star.  As the behaivour of $\epsilon_j$ is dictated by the behavour of $\alpha_n$ through equation~\eqref{transit_intro_ground_noisedef}, $\epsilon_j$ will be investigated within the context of three realistic noise sources.  First, the distribution of $\epsilon_j$ is determined analytically assuming all noise sources produced white noise.  This case is selected as white noise is easy to theoretically manipulate, as shot noise and common types of instrumental noise such as dark noise and read noise are well described by white noise, and as initial investigations into TTV$_p$ detection, assuming white noise, are present in the literature \citep[e.g.][]{Szaboetal2006}.  Second, the distribution of $\epsilon_j$ is numerically investigated for the case of realistic stellar photometric noise. This case is selected as intrinsic photometric variability of stars is not a white noise process, and it has long been known that red photometric noise due to stellar variability, is a limiting factor in transit surveys \citep[e.g.][]{Boruckietal1985}.  While this type of noise cannot be completely avoided, methods have been proposed in the transit detection literature to reduce its effects on planet detection.  Most of these involve some form of filtering.  Consequently, the distribution of $\epsilon_j$ is also determined using realistic stellar noise which had been ``filtered" for trends due to the rotational modulation of starspots using the method of \citet{Lanzaetal2003}.

Before the distribution and behaviour of $\epsilon_j$ can be investigated for these three cases, appropriate methods for determining $\epsilon_j$ using equation~\eqref{transit_intro_ground_noisedef} need to be selected.  Consequently, the properties of equation~\eqref{transit_intro_ground_noisedef} will be explored with the aim of selecting the optimal method for determining $\epsilon_j$ analytically, for the case of white noise, and numerically, for the case of realistic and filtered realistic stellar photometric noise.

\section{Method}

\begin{center}
\begin{figure}
 \includegraphics[width=.95\textwidth]{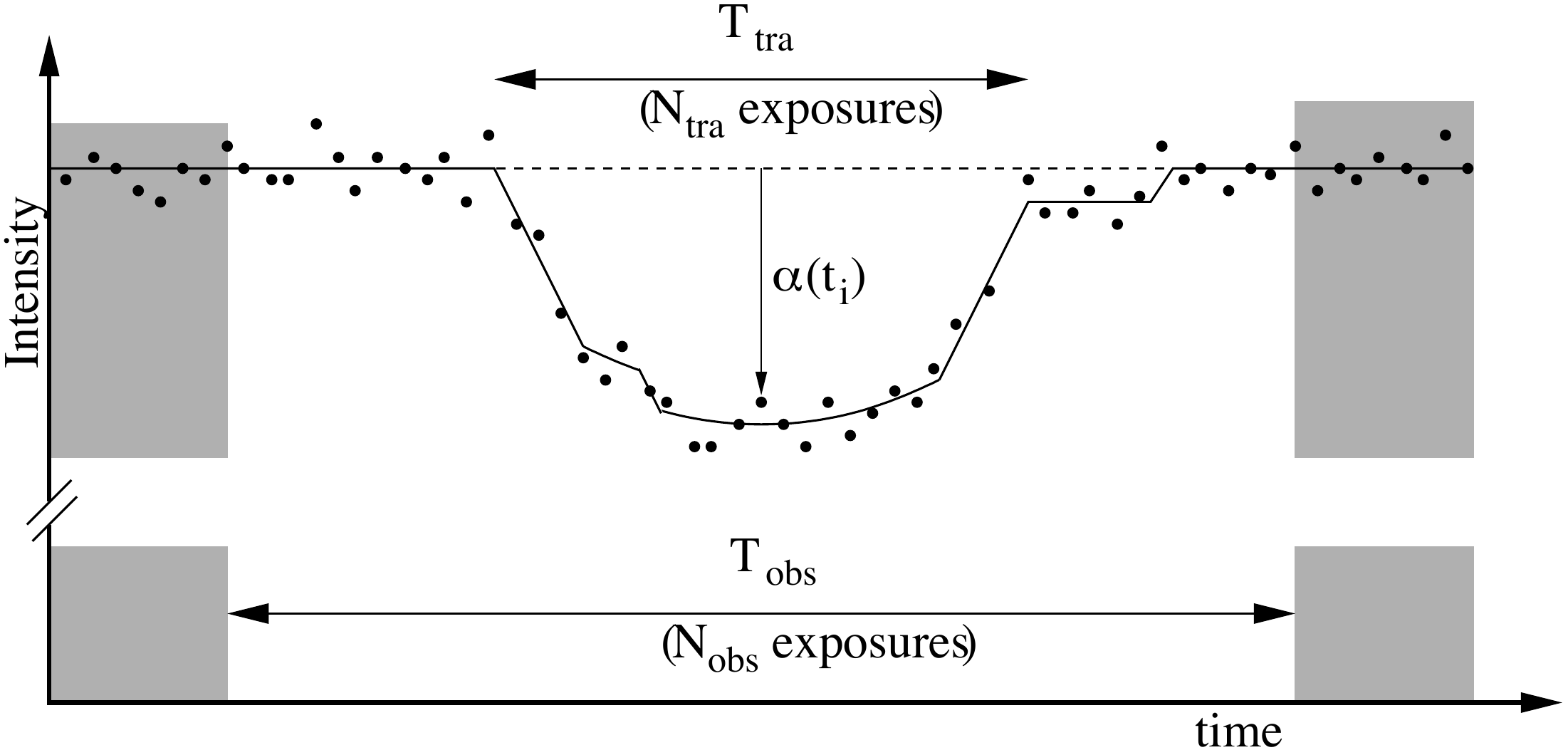}
\caption[An example transit showing measured intensities (dots) and theoretical intensity (line).]{An example transit showing measured intensities (dots) and theoretical intensity (line).  The quantity $\alpha(t_i)$ is shown for one of the exposures and the region which is not counted within the sum is shaded grey.}
\label{EgTransit}
\end{figure}
\end{center}

 Before investigating the effect of white, realistic and filtered stellar photometric noise on $\epsilon_j$, two issues must be addressed.  First, the method for determining the properties of $\epsilon_j$ for a given sequence of $\alpha_n$ needs to be discussed.  Second, for the cases of realistic and filtered photometric noise, photometric noise from real stars is required.  Consequently, the issue of whether solar photometric data is representative of the photometric noise from a typical star will be discussed.
 
 \subsection{Method for calculating $\epsilon_j$}\label{transit_noise_method_method}

We begin our investigation with equation~\eqref{transit_intro_ground_noisedef}, the equation which links $\epsilon_j$ and $\alpha_n$,  restated below for convenience
\begin{equation*}
\epsilon_{j} = \frac{1}{A_p + A_m} \sum_{i}\left[t_i - t_0 - jT_p - \Delta \tau \right] \alpha_n(t_i)\label{theory_eps2}.
\end{equation*}
As described in section~\ref{Transit_Intro_Taudef}, the sum is evaluated over a time interval of length $T_{obs}$ centered on the planetary transit (see figure~\ref{EgTransit}).  While equation~\eqref{transit_intro_ground_noisedef} is exact, it is not in the most useful form.  In this chapter, an approximation to equation~\eqref{transit_intro_ground_noisedef} will be used, mainly as it is physically realistic, and the simplified version has a suite of properties which can be exploited.

To begin, we consider the relative sizes of the terms in square brackets.  First, $t_i$ is the time at which the $i^{th}$ exposure is taken in a given transit, while $t_0 + jT_p$ is approximately the mid-transit time for that transit.  Consequently the quantity $t_i - t_0 - jT_p$ ranges from $-1/2T_{obs}$ to $1/2T_{obs}$.  As the length of the observing window must be at least as long as the transit duration (see figure~\ref{EgTransit}) and transit durations are typically many hours long (see table~\ref{EgTransDurVel}), $t_i - t_0 - jT_p$ is of the order of many hours.  Conversely, as determined in chapter~\ref{Transit_Signal}, $\Delta \tau$ is of the order of minutes for physically realistic moons (see table~\ref{DelTauEgs}).  As equation~\eqref{transit_intro_ground_noisedef} is dominated by large values of $t_i - t_0 - jT_p$, the $\Delta \tau$ term can be neglected, giving
\begin{equation}
\epsilon_{j} = \frac{1}{A_p + A_m} \sum_{i}\left[t_i - t_0 - jT_p \right] \alpha_n(t_i)\label{transit_noise_method_wndef}.
\end{equation}
This is the equation that will be used to analytically investigate the case of white noise.

For the case of realistic and filtered noise, real photometric light curves will be manipulated to give a sample of $\epsilon_j$ values, which will consequently be used to estimate the distribution and behaviour of  $\epsilon_j$.\footnote{As an aside, the assumption that out-of-transit light curves can be used to estimate in-transit photometric variability is not necessarily valid.  For example, models \citep{Silva2003} and observations \citep[e.g.][]{Pontetal2007} show that if a planet passes in front of a finite structure on the surface of its host star, e.g. a starspot, during transit, this will result in additional perturbations to the transit light curve.  The effect of this additional noise source on $\Delta \tau$ will not be investigated in this thesis.}   To begin, consider an explicitly measured light curve $L(t_i)$, given by 
\begin{equation}
L(t_i) = L_0 + L_n(t_i),\label{transit_noise_method_Ldef}
\end{equation}
where $L_0$ is the average value of $L(t_i)$, and $L_n(t_i)$ is the zero-mean photometric noise.  Now, some stars may exhibit more photometric noise than the star from which our real light curves were taken, while others may exhibit less.  To explore the effect that the amplitude of the photometric noise has on $\epsilon_{j}$, we will focus on the case where $L_0$ is the same, but the amplitude of the photometric noise has been scaled  by a factor $\beta$.\footnote{Simple noise amplitude scaling will not fully describe the behaviour of $\epsilon_j$ for all stars.  The particular shape of the noise power spectrum depends on the processes that produced the red noise, for example, granulation \citep{RabelloSoaresetal1997} and rotational modulation of active regions \citep{Leanetal1998} and their associated characteristic timescales.  While methods do exist \citep[e.g.][]{Aigrainetal2004} for predicting red noise characteristics of stars as a function of physical and spectral properties, this extension will not be investigated in this thesis.}  Let us define $L^*$, the luminosity associated with this general light curve, as
\begin{equation}
L^*(t_i) = L_0 + \beta L_n(t_i),\label{transit_noise_method_Lstardef}
\end{equation}
where $\beta$ is a known constant.  

Currently equation~\eqref{transit_noise_method_wndef} is formulated in terms of $\alpha_n(t_i)$.  As we will be using real light curves to estimate $\epsilon_j$, we would like to write equation~\eqref{transit_noise_method_wndef} in terms of $L(t_i)$ as opposed to $L^*(t_i)$ or $\alpha_n(t_i)$.  To do this, we define
\begin{equation}
\tau^* = \frac{\sum_{i} t_i L(t_i) }{\sum_{i} L(t_i)} = \frac{\sum_{i} t_i (L_0 + L_n(t_i)) }{\sum_{i} (L_0 + L_n(t_i))} = \frac{\sum_{i} t_i \beta (L_0 + L_n(t_i)) }{\sum_{i} \beta(L_0 + L_n(t_i))}, \label{transit_noise_method_taustardef}
\end{equation}
where we have multiplied the numerator and denominator by $\beta$ to simplify the algebra later on.

As $\sum_i \beta L_0 \gg \sum_i \beta L_n$, we can expand $\tau^*$ using the same method used in section~\ref{Transit_Intro_Deriv} to expand $\tau$.  Noting that $\sum_i \beta L_n/\sum_i \beta L_0$ is our small parameter, and performing the binomial expansion of the denominator gives
\begin{equation}
\tau^* = \frac{1}{\sum_i \beta L_0} \left(\sum_{i} t_i \beta (L_0 + L_n(t_i)) \right)\left(1 - \frac{\sum_i \beta L_n(t_i)}{\sum_i \beta L_0}\right).\label{transit_noise_method_taustareq1}
\end{equation}
Now, as $L_0$ is constant, $\sum_i L_0 = N_{obs} L_0$, where $N_{obs}$ is the total number of exposures in the observing window.  Using this expression to simplify equation~\eqref{transit_noise_method_taustareq1} and only retaining terms up to first order in $L_n$, gives,
\begin{align}
\tau^* &= \frac{\sum_{i} t_i \beta L_0}{\beta N_{obs} L_0} + \frac{\sum_{i} t_i \beta  L_n(t_i)}{\beta N_{obs} L_0} - \frac{\sum_i \beta L_n(t_i)}{\beta N_{obs} L_0} \frac{\sum_{i} t_i \beta (L_0)}{\beta N_{obs} L_0},\label{transit_noise_method_taustareq2}\\
	&= \frac{\sum_{i} t_i}{N_{obs}} + \frac{\sum_{i} t_i \beta  L_n(t_i)}{\beta N_{obs} L_0} - \frac{\sum_i \beta L_n(t_i)}{\beta N_{obs} L_0} \frac{\sum_{i} t_i}{N_{obs}}.\label{transit_noise_method_taustareq3}
\end{align}
As the exposures are evenly spaced, $\sum_{i} t_i/N_{obs}$, the average of the times at which the exposures are taken, is equal to $t_{mid}$, the time corresponding to the center of the window.  Consequently, 
\begin{equation}
\tau^* = t_{mid} + \frac{1}{\beta N_{obs} L_0}\sum_{i} [t_i - t_{mid}] \beta  L_n(t_i).\label{transit_noise_method_taustareq4}
\end{equation}

We can now compare this equation with equation~\eqref{transit_noise_method_wndef}, the equation describing $\epsilon_{j}$.  First, we note, that for the case described by equation~\eqref{transit_noise_method_wndef}, $t_{mid,p}$ is given by $t_0 + jT_p$.  In addition, for the case where the luminosity of the star is given by equation~\eqref{transit_noise_method_Lstardef}, $\alpha_n(t_i)$ is equal to $-\beta L_n(t_i)$.\footnote{Recall that $\alpha_n$ is defined as a photon deficit, while $L_n$ is defined as a photon surplus.}  Using these two expressions, we can rewrite equation~\eqref{transit_noise_method_wndef} as
\begin{equation}
\epsilon_{j} = -\frac{1}{A_p + A_m} \sum_{i}\left[t_i - t_{mid,p} \right] \beta  L_n(t_i).
\end{equation}
Substituting equation~\eqref{transit_noise_method_taustareq4} into equation~\eqref{transit_noise_method_wndef} gives
\begin{align}
\epsilon_{j} &= -\frac{\beta N_{obs} L_0}{A_p + A_m} \left(\tau^* - t_{mid,p}\right),\label{transit_noise_method_rndef}\\
 &= -\frac{\beta N_{obs} L_0}{A_p + A_m} \epsilon^* \label{transit_noise_method_rndefeps},
\end{align}
where
\begin{equation}
\epsilon^* = \tau^* - t_{mid}\label{transit_noise_method_epsstardef}.
\end{equation}

Consequently, the error $\epsilon_{j}$, recorded for the case where the host star's luminosity is given by equation~\eqref{transit_noise_method_Lstardef}, is given by equation~\eqref{transit_noise_method_rndef}, where $ \tau^*$ is defined by equation~\eqref{transit_noise_method_taustardef} and where $t_{mid,p}$ is the middle of the piece of light curve used to determine the effective value of $\epsilon_{j}$.  Equations~\eqref{transit_noise_method_rndef} and \eqref{transit_noise_method_taustardef} provide a simple method for determining a sequence of representative $\epsilon_j$ values for a transit with a given $A_p + A_m$, using out of transit light curves.  

\subsection{Suitability of solar photometric data for calculating $\epsilon_j$}\label{Transit_Noise_Method_Solar}

In order to determine the effect of intrinsic stellar photometric variability on $\epsilon_j$ using equation~\eqref{transit_noise_method_rndef}, a suitable sample light curve which is dominated by realistic photometric stellar variability corresponding to a typical star is required.  This data series should optimally have a number of properties.  First, it needs to be long enough such that a statistically valid estimate of $\epsilon
_j$ can be constructed.  Second, the data must have high signal to noise, to ensure that the majority of recorded photometric noise is inherent to the star, and not resulting from instrumental or statistical effects.  Third, the data needs to be high cadence, such that the effect of long and short exposure times on $\epsilon_j$ can be investigated.  Finally, this data must be easily available and easy to manipulate.  Fortunately, photometric measurements of the Sun meet all these requirements.

Over twelve years of high quality solar data are available as a result of the Solar and Heliospheric Observatory (SOHO).  This satellite is positioned at the L1 point between Earth and the Sun, which allows it uninterrupted access to the Sun's behaviour.  For this work it is the measurements of Total Solar Irradiance (TSI), the total intensity of the solar face, that are important.  This measurement is made within the Virgo module \citep{Frohlichetal1995}, by comparing the output of the photometers PMO6V and DIARAD \citep{Frohlichetal1997}.   This data was kindly made available by SOHO team and can be freely downloaded from the SOHO archive.\footnote{http://seal.nascom.nasa.gov/cgi-bin/gui\_plop.}

While high quality solar data may be freely available, whether it should be used depends on whether the photometric behaviour of the Sun is representative of the photometric behaviour of Sun-like stars, in particular, the stars to be targeted by COROT and Kepler.  Indeed, it has been found that the Sun shows two to three times less variation on the decadal time scale compared to similar stars \citep{Lockwoodetal1997,Radicketal1998,Lockwoodetal2007}.  While this discrepancy may seem discouraging, it needs to be viewed in the context of two other findings, the effect of star orientation with respect to the observer, and the results of studies of extra solar planet host stars.  First, it has been suggested that the low observed value of solar photometric noise is due to our privileged observing position, that is, in the plane of the Sun's equator.  The maximum predicted magnitude of this effect ranges from an increase in photometric variability of 6 \citep{Schatten1993} to 1.3 \citep{Knaacketal2001} times that observed in the equatorial plane as the viewing angle is altered.  Fortunately, it has been observed for both transiting \citep{FabryckyWinn2009} and non transiting \citep{Bouquinetal2009} planets that there is a preference for the orbital angular momentum vector of the planet and the spin axis of the host star to be aligned.  Consequently, as the planets of interest transit, the equator of the target star should also be preferentially aligned with the line-of-sight.  Second, while the survey conducted by \citep{Lockwoodetal2007} into the photometric variability of Sun-like stars is the most temporally complete, it is not the only survey.  In particular, the survey of \citet{Henryetal2000} has looked at the photometric variability of extra-solar planet host stars detected by the radial velocity technique.  Within this subset, the Sun appears typical.  Unfortunately, as only relatively magnetically inactive stars are chosen as targets for radial velocity searches, and long term photometric variability increases with increasing magnetic activity \citep[e.g][]{Radicketal1998}, this set of stars is statistically biased.  Consequently, the study of \citet{Henryetal2000} cannot be used to argue that the Sun's photometric stability is representative of most stars.  However, while not all planetary hosts will display the same photometric stability as the Sun e.g. CoRoT-2b \citep{Alonsoetal2008}, this study does indicate that there is a substantial subset of planetary host stars which will.  Within this context it was felt that the solar data was a suitable testbed for this analysis.

For the case of the Sun, equation~\eqref{transit_noise_method_rndefeps} becomes
\begin{equation}
\epsilon_{j} = -\frac{\beta N_{obs}L_o}{A_p + A_m} \epsilon_{\sun}, \label{red_epsdef}
\end{equation}
where $\epsilon_{\sun}$ is given by 
\begin{equation}
 \epsilon_{\sun} = \frac{\sum_{i} t_i L(t_i) }{\sum_{i} L(t_i)} - t_{mid,p},
\end{equation}
where the luminosities $L(t_i)$ are given by the total solar intensity measurements taken by SOHO.  By evaluating $\epsilon_{\sun}$ for enough segments of solar light curve, an observed distribution for $\epsilon_{\sun}$, and correspondingly $\epsilon_{j}$, can be constructed.  This formulation will be used to analyse the cases of realistic and filtered stellar noise.

Finally, it should be noted that the specifics of the behaviour of $\epsilon_j$ of course depend on the photometric behaviour of the host star selected and will not necessarily share the the same behaviour as derived for the Sun.  Fortunately, $\epsilon_j$ can be derived for any given host star using the method demonstrated above, but using out of transit light curves as opposed to SOHO data.  To ensure accurate assessments of moon detectability, this procedure would have to be completed for the host star of every transiting planet targeted for followup.

Now that the method and data have been discussed, we can move on to analysing the behaviour of $\epsilon_j$ for the three noise sources under investigation.  First we will look at white noise.

\section{White noise: analytic derivation}\label{Trans_TTV_Noise_White}

\subsection{Introduction to white noise}

White noise is a type of additive noise which occurs in many physical processes such as photon counting.  It is known as white, as the Fourier transform of white noise contains the same power at all frequencies.  Using the analogy between frequency and colour, as there is no dominant colour, the noise is ``white".

In order for a discrete sequence of numbers $x = \{x_1, x_2, \ldots x_j, \dots, x_n\}$ to describe white noise, two conditions must be satisfied.  First, each of the $x_j$ must be drawn from a Gaussian distribution, that is, they are described by the probability density function
\begin{equation}
P(x_j) = \frac{1}{\sigma \sqrt{2 \pi}} e^{-\frac{(x_j - \mu)^2}{2\sigma^2}},\label{white_intro_gaudef}
\end{equation}
where $\mu$ and $\sigma$ are the mean and the standard deviation respectively.  Second, the individual values of $x$ must be uncorrelated, that is, the probability distribution for any $x_{j}$ is unaffected by the values of any of the $x_k$ where $k\ne j$.

Uncorrelated, normally distributed random variables, such as those used to produce white noise, have a number of useful properties.  In particular, if $X_1$ and $X_2$ are uncorrelated normally distributed random variables with mean $\mu_1$ and $\mu_2$ and standard deviation $\sigma_1$ and $\sigma_2$, then a linear combination of these variables, $aX_1 + bX_2$ is also  normally distributed with mean $\mu_{1+2}$ and standard deviation $\sigma_{1+2}$, given by
\begin{align}
\mu_{1+2} &= a\mu_1 + b\mu_2, \label{TraM-Noi-White-Prop-summean}\\
\sigma_{1+2} &= \sqrt{a^2\sigma_1^2 + b^2\sigma_2^2}. \label{TraM-Noi-White-Prop-sumstd}
\end{align}

\subsection{Derivation of $\epsilon_j$}\label{Trans_TTV_Noise_White_epj}

Restating the definition of $\epsilon_j$ given by equation~\eqref{transit_noise_method_wndef}, 
\begin{equation*}
\epsilon_j = \frac{1}{A_p + A_m}\sum_i\left[t_i -  (jT_p + t_0) \right]\alpha_n(t_i),
\end{equation*}
it can be seen that $\epsilon_j$ is defined in terms of a sum of values $\alpha_n$, drawn from a distribution which has first been premultiplied by the value $t_i -  (jT_p + t_0)$.  As $\epsilon_j$ is the sum of a set of normally distributed random variables, $\alpha_n$, it too is normally distributed.  In addition, assuming the distribution of $\alpha_n$ has mean 0 and standard deviation $\sigma_L$ we can use equations~\eqref{TraM-Noi-White-Prop-summean} and \eqref{TraM-Noi-White-Prop-sumstd} to determine $\mu_\epsilon$ and $\sigma_\epsilon$ the mean and standard deviation of the distribution of $\epsilon_j$ respectively.   Applying equations~\eqref{TraM-Noi-White-Prop-summean} and \eqref{TraM-Noi-White-Prop-sumstd} iteratively to equation~\eqref{transit_noise_method_wndef}, we have that
\begin{equation}
\mu_\epsilon = 0\label{TraM-Noi-White-summean}
\end{equation}
and
\begin{equation}
\sigma_\epsilon = \frac{\sigma_L}{A_p + A_m}\sqrt{\sum_i(t_i -  (jT_p + t_0))^2}.\label{TraM-Noi-White-sumstd}
\end{equation}
While equation~\eqref{TraM-Noi-White-summean} is fully evaluated, equation~\eqref{TraM-Noi-White-sumstd} is not.   In order to evaluate this expression, we need to write $t_i$ in terms of physical variables such as the $t_{mid,p}$, the midpoint of the window, $\Delta t$, the exposure length, and $N_{obs}$, the number of exposures used to calculate $\tau$.  Using these variables, $t_i$ can be defined
\begin{equation}
t_i = t_{mid,p} + \left(i - \frac{N_{obs}}{2}\right)\Delta t.
\end{equation}

Noting that the window over which the transit is examined is centered on the planetary transit, we have that the center of the $j^{th}$ transit will occur at
\begin{equation}
t_{mid,p} = jT_p + t_0 + \Delta t_p,
\end{equation}
where $\Delta t_p$ is the time delay in the center of the planetary transit due to the motion of the planet about the planet moon barycenter (i.e. the TTV$_b$ signal).  Consequently $t_i$ is given by
\begin{equation}
t_i = jT_p + t_0 + \Delta t_p + \left(i - \frac{N_{obs}}{2}\right)\Delta t. \label{transit_noise_white_tidef}
\end{equation}
Now, as was shown in section~\ref{Trans_TTV_Signal_CC_Form_smallB}, we know that the TTV$_b$ signal is less than, or at the very least the same order of magnitude as the equivalent TTV$_p$ signal ($\Delta \tau$).  As a result, as we have neglected $\Delta \tau$ term from equation~\eqref{transit_intro_ground_noisedef}, as it was much smaller than $t_i - (jT_p + t_0)$, we can also neglect the $\Delta t_p$ term from equation~\eqref{transit_noise_white_tidef} for the same reason.

Thus, writing out $t_i$ and the sum limits in full, we have that
\begin{equation}
\sigma_\epsilon =  \frac{\sigma_L}{A_p + A_m}\sqrt{\sum_{i = 0}^{N_{obs}-1} \left(i - \frac{N_{obs}}{2}\right)^2\Delta t^2}.\label{TraM-Noi-White-epder1}
\end{equation}

Expanding the term under the square root
\begin{equation}
\sigma_\epsilon =  \frac{\sigma_L}{A_p + A_m}\sqrt{\sum_{i = 0}^{N_{obs}-1}  \Delta t^2i^2 - N_{obs}\Delta t^2 i  +\frac{N_{obs}^2}{4}\Delta t^2 } ,\label{TraM-Noi-White-epder2}
\end{equation}
and using the identities
\begin{align}
\sum_{i = 0}^{n} i &= \frac{n(n+1)}{2}\label{TraM-Noi-White-sumi},\\
\sum_{i = 0}^{n} i^2 &= \frac{n(n+1)(2n + 1)}{6},\label{TraM-Noi-White-sumi2}
\end{align}
equation~\eqref{TraM-Noi-White-epder2} can be expanded to give
\begin{equation}
\sigma_\epsilon =  \frac{\sigma_L}{A_p + A_m} \sqrt{\Delta t^2\frac{N_{obs}^3}{12} + \Delta t^2\frac{N_{obs}}{6}}.\label{TraM-Noi-White-epder3}
\end{equation}

In addition, as the first term is approximately $N_{obs}^2$ times the second term, where $N_{obs}$, the number of exposures used to calculate $\tau$, is a large number, the second term can be neglected giving
\begin{equation}
\sigma_\epsilon =  \sigma_L \frac{N_{obs} \Delta t}{A_p + A_m} \sqrt{\frac{N_{obs}}{12}}.\label{transit_noise_white_sigdef}
\end{equation}
To provide intuitive understanding of this expression, equation~\eqref{transit_noise_white_sigdef} was recast into physical variables,\footnote{In particular the reference system is a Jupiter-like planet about a Sun-like star at an orbit of 1 AU, while the reference instrument is Kepler.  The reference relative photometric accuracy was calculated by assuming that the nominal relative photometric precision of $2 \times 10^{-5}$ in a 6.5 hour exposure for a magnitude 12 star \citep[e.g.][]{Boruckietal2003} is dominated by shot noise, and consequently calculating the corresponding shot noise that would be observed in a one minute exposure.} giving
\begin{multline}
\sigma_\epsilon =  47.9\text{s} \left[\left(\frac{\sigma_L / L_0}{3.95 \times 10^{-4}}\right)\left(\frac{\Delta t}{1 \text{min}}\right)^{1/2}\right] \left[\frac{100(A_p + A_m)}{N_{tra} L_0} \right]^{-1} \\
\times \left[\left(\frac{T_{obs}}{24 \text{hrs}}\right)^{3/2} \left(\frac{T_{tra}}{13 \text{hrs}}\right)^{-1}\right].\label{transit_noise_white_sigdefphys}
\end{multline}

\begin{figure}[tb]
\begin{center}
\includegraphics[width=.95\textwidth]{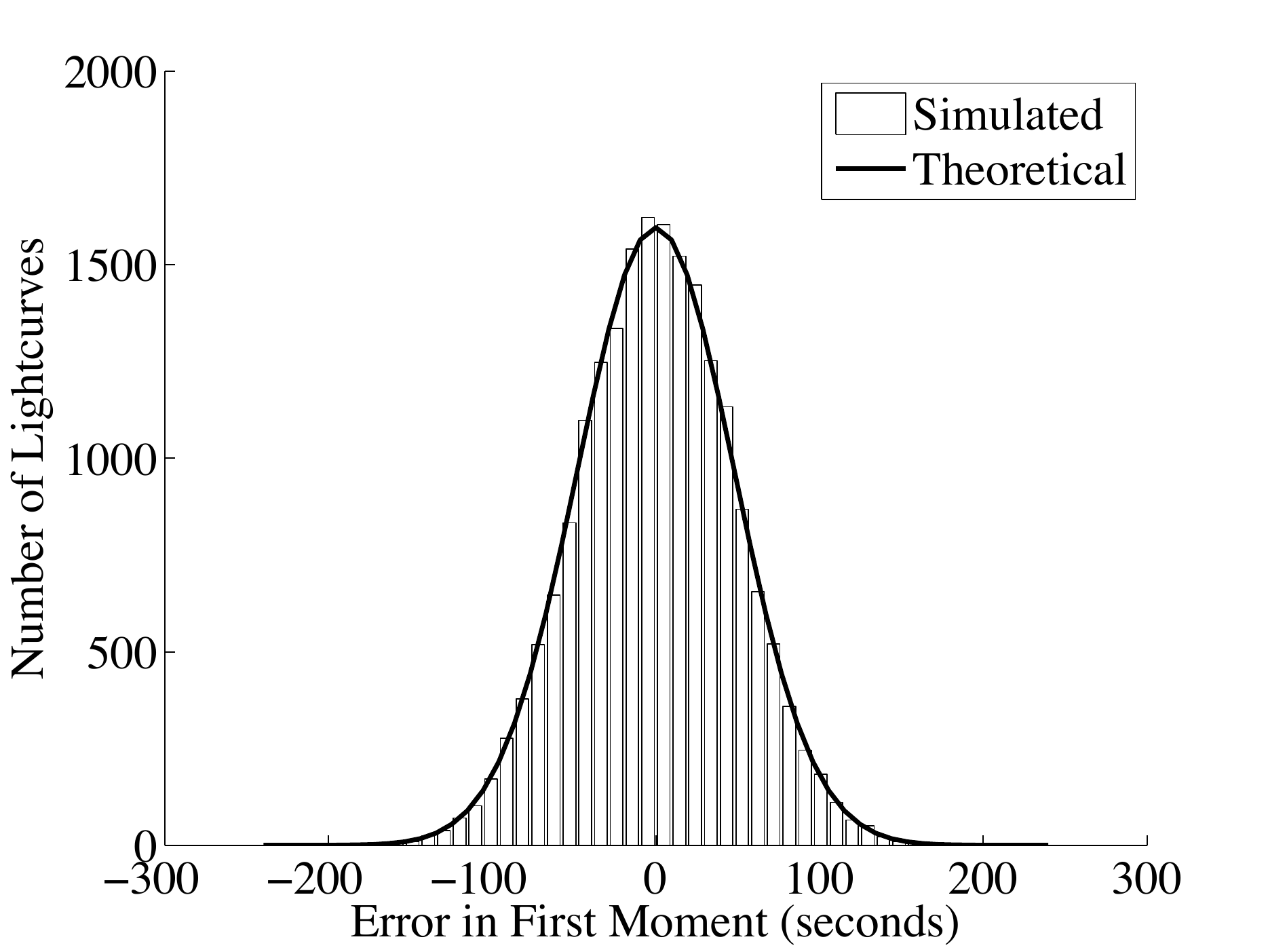}
\caption{Comparison between a Monte Carlo simulation (white bar) and theoretical
prediction (black line) for the distribution of first moments for transit contaminated with white noise.}
\label{MCCompare}
\end{center}
\end{figure}

To check this formula, a Monte Carlo simulation was run.  A mock transit light curve was constructed using three straight line segments corresponding to the ingress, flat bottom and egress.  To this, Gaussian noise with a known standard deviation was added to produce a sequence of model transit light curves. The $\epsilon_j$ corresponding to each of these light curves was then calculated. The histogram of $\epsilon_j$ along with the theoretical prediction are both shown in figure~\ref{MCCompare}.  As can be seen, the agreement is very strong.  

\subsection{Properties of $\epsilon_j$}

As can be seen from equation~\eqref{transit_noise_white_sigdef}, the standard deviation of $\epsilon_j$ for the case of white photometric noise depends on a number of physical quantities, in particular, the photometric accuracy, $\sigma_L$, the exposure time, $\Delta t$, the size of the planet (parameterised by $A_p$) and the length of the transit and observing windows.  As a different physical process is highlighted in each of the three terms of equation~\eqref{transit_noise_white_sigdefphys}, this equation will be used as a scaffold for this discussion.  These three terms will be discussed in turn, in the context of equation~\eqref{transit_noise_white_sigdefphys} and in the context of previous results assuming white photometric noise, in particular, those from \citet{Szaboetal2006}.  In addition $\epsilon_j$ will also be discussed in terms of two physically important quantities that it does not depend on, $\Delta \tau$ and $j$, mainly as this independence leads to a number of important statistical properties that will be used in chapter~\ref{Trans_Thresholds} to construct thresholds.  To begin this exploration of the properties of $\epsilon_j$, we look at the dependance of $\sigma_\epsilon$ on the relative photometric accuracy per exposure and the exposure time.

The first term in equation~\eqref{transit_noise_white_sigdefphys} describes both the dependence of $\sigma_\epsilon$ on the relative photometric accuracy ($\sigma_L / L_0$) and the exposure time ($\Delta t$).  Unsurprisingly, the smaller the relative photometric error, the smaller the value of $\sigma_\epsilon$ and consequently, the smaller the error in $\tau$.  However, while the relative photometric noise can be altered without changing the exposure time, for example, by moving to a larger telescope or by upgrading the instrumentation, changing the exposure time can alter the photometric noise depending on the source of the photometric noise.  For the case of shot noise, error due to small number statistics, the total error is proportional to the square root of the total number of photons collected for that star (i.e. $\propto \sqrt{\Delta t}$) while the total intensity is proportional to the total number of photons (i.e. $\propto \Delta t$), so the relative photometric error is proportional to $\sqrt{\Delta t}/\Delta t = \Delta t^{-1/2}$.\footnote{A similar argument can be constructed for the case of dark noise, the noise caused by electron motion in the CCD chip during an exposure.}  For the case of read noise, error resulting from the process of measuring the number of photons collected at the end of an exposure, the total error does not depend on the exposure time.  Consequently, the relative photometric error is proportional to $(\Delta t)^{-1}$.  So, for a star with white photometric noise that is dominated by shot noise, $\sigma_\epsilon$ is independent of exposure time, while for a star dominated by read noise, $\sigma_\epsilon$ should decrease with increasing exposure time.\footnote{For the case where the relative photometric noise is independent of exposure time, such as the case explored by \citet{Szaboetal2006}, $\sigma_\epsilon$ will decrease with decreasing exposure time, which is in agreement with their findings.}

The second bracketed term in equation~\eqref{transit_noise_white_sigdefphys} represents the effect of planet size on the size of $\sigma_\epsilon$.  This can be seen by noting that $L_0 N_{tra}$ is the total number of photons emitted by a star during a transit duration, while $A_p + A_m$ is the number of photons blocked by the planet during the same time.  Consequently, this fraction can be thought of as a ratio between the projected area of the planet-moon pair and the star on the sky respectively.  While this initially seems to indicate that moons should be more detectable around large transiting planets as a result of their large $A_p + A_m$ values, recall from section~\ref{Trans_TTV_Signal_CC_PropAmp} that the amplitude of $\Delta \tau$ also decreases as $(\hat{A}_p + \hat{A}_m)^{-1}$.  As both these terms can be thought of as representing the cross-sectional area of the planet, the gain in moon detectability caused by decreasing $\sigma_\epsilon$ is matched by the loss in moon detectability caused by the decreasing $\Delta \tau$ amplitude.

The final term in square brackets describes the dependance of $\sigma_\epsilon$ on the transit duration and the length of the observing window.  Physically, this corresponds to investigating  the effect on  $\sigma_\epsilon$ of the distance of the planet-moon pair from the star, and the distance of the planet-moon pair from each other.  For a given planet-moon system, the optimal length of the observing window will scale with the transit duration.  Consequently $\sigma_\epsilon \propto \sqrt{T_{tra}}$.  This result, along with the dependance of $\Delta \tau$ on transit duration,\footnote{Recall from section~\ref{Trans_TTV_Signal_CC_PropAmp} that the amplitude of $\Delta \tau$ is inversely proportional to $v_{tr}$, and thus proportional to $T_{tra}$.} implies that for the case of white photometric noise, the detectability of a given moon increases with increasing transit duration of its host planet.  This is in agreement with the work of \citet{Szaboetal2006} who found that more distant planets (with larger transit durations) had more detectable moons.  The relationship between $\sigma_\epsilon$ and $T_{obs}$ (and consequently the planet-moon separation) is a little more involved as the position of the moon is not known before detection.  From equation~\eqref{transit_noise_white_sigdefphys} it can be seen that as $T_{obs}$ increases, $\sigma_\epsilon$ increases as $T_{obs}^{3/2}$.  Consequently it would be useful to use the smallest observing window possible, while still including the moon's transit.  Consider the best case scenario, where the observing window selected is the shortest window such that the full transit of the moon will always be captured, that is, $T_{obs} = 2a_m/v_{tr} + T_{tra}$.  Consequently for a small semi-major axis ($2a_m/v_{tr} \ll T_{tra}$), $\sigma_\epsilon$ will be independent of $a_m$ and for large semi-major axis ($2a_m/v_{tr} \gg T_{tra}$) is proportional to $a_m^{3/2}$.  Comparing this to the result from section~\ref{Trans_TTV_Signal_CC_PropAmp}, that the amplitude of $\Delta \tau$ is proportional to $a_m$, it can be seen that very close and very distant moons are not particularly detectable.  This is at odds with the result found by \citet{Szaboetal2006} in that they proposed that only close moons were undetectable.

Now that the physical parameters that affect $\sigma_\epsilon$ have been discussed, $\sigma_\epsilon$ will also be discussed in the context of two additional parameters, $\Delta \tau$ and the transit number $j$.  In particular the fact that $\sigma_\epsilon$ does not strongly depend on either $\Delta \tau$ or $j$ will be of great use in constructing thresholds in chapter~\ref{Trans_Thresholds}.  We begin with an investigation of the relationship between $\sigma_\epsilon$ and $\Delta \tau$.

The first parameter of interest is $\Delta \tau$.  As can be seen from equation~\eqref{transit_noise_white_sigdefphys}, $\sigma_\epsilon$ does not depend on the measured value of $\Delta \tau$ for that transit.  This lack of dependance is partially a result of neglecting the $\Delta \tau$ term in equation~\eqref{transit_intro_ground_noisedef} (as it was small) and partially as a result of neglecting higher order terms in the expansion of equation~\eqref{TraM-TTV-tauderiv1} required for the derivation of equation~\eqref{transit_intro_ground_noisedef} (see section~\ref{transit_noise_method_method} and appendix~\ref{SecOrdNoise_App}).  While $\sigma_\epsilon$ does not formally depend on $\Delta \tau$, it does depend on a quantity that changes as $\Delta \tau$ changes, $(A_p + A_m)$.  However, for the purposes of determining the detection threshold it acts as a constant,\footnote{The statistical method that will be used in chapter~\ref{Trans_Thresholds} uses ratios between the signal and the noise to construct thresholds and will only be evaluated to first order in $v_m/v_{tr}$.  As these ratios will only be evaluated to first order in $v_m/v_{tr}$, and $\Delta \tau$ is already first order in $v_m/v_{tr}$, only the zeroth order component of $A_p + A_m$ need be retained i.e. $A_p + A_m = \hat{A}_p + \hat{A}_m$.  The disappearance of these first order terms is heartening as they depend on the orbital parameters of the system in a non-trivial way.  This issue will be further discussed in sections~\ref{Trans_Thresholds_ExpBehav} and \ref{Trans_Thresholds_MC}.} and so can be ignored.  This weak dependence of $\epsilon_j$ on $\Delta \tau$ means that the noise in $\tau$ can be approximated with little to no knowledge of the orbital elements of any putative moons.  This property will simplify the mathematics in the next chapter, when the expressions describing $\Delta \tau$ and $\epsilon_j$ are combined to determine the set of detectable moons.

The second parameter of interest is the transit number, $j$.  To understand the (lack of) dependance of $\epsilon_j$ on the transit number $j$, consider the following.  From the definition of white noise, the error in each exposure is uncorrelated to the error in any other exposure.  Consequently, the error in any quantity calculated from a sequence of exposures (such as $\tau$) should be uncorrelated with the error in that quantity calculated from a different sequence of exposures.  That is, for the case of white noise, $\epsilon_j$ is independent of the transit number $j$.  The fact that the errors in $\tau$ for consecutive transits are uncorrelated with each other is another very useful statistical property in terms of fitting the sequence of transits, and consequently, calculating detection thresholds.

Now that we have explored and discussed the properties of $\epsilon_j$ within the context of white noise, we have an understanding of the types of behaviour that photometric noise can have on $\tau$.  We will use this as a basis for investigating the effect of more complex types of noise on the error in $\tau$, and in particular, move onto the case where realistic (solar-like) noise is the dominant noise source.

\begin{figure}[tb]
\begin{center}
\includegraphics[width=.95\textwidth]{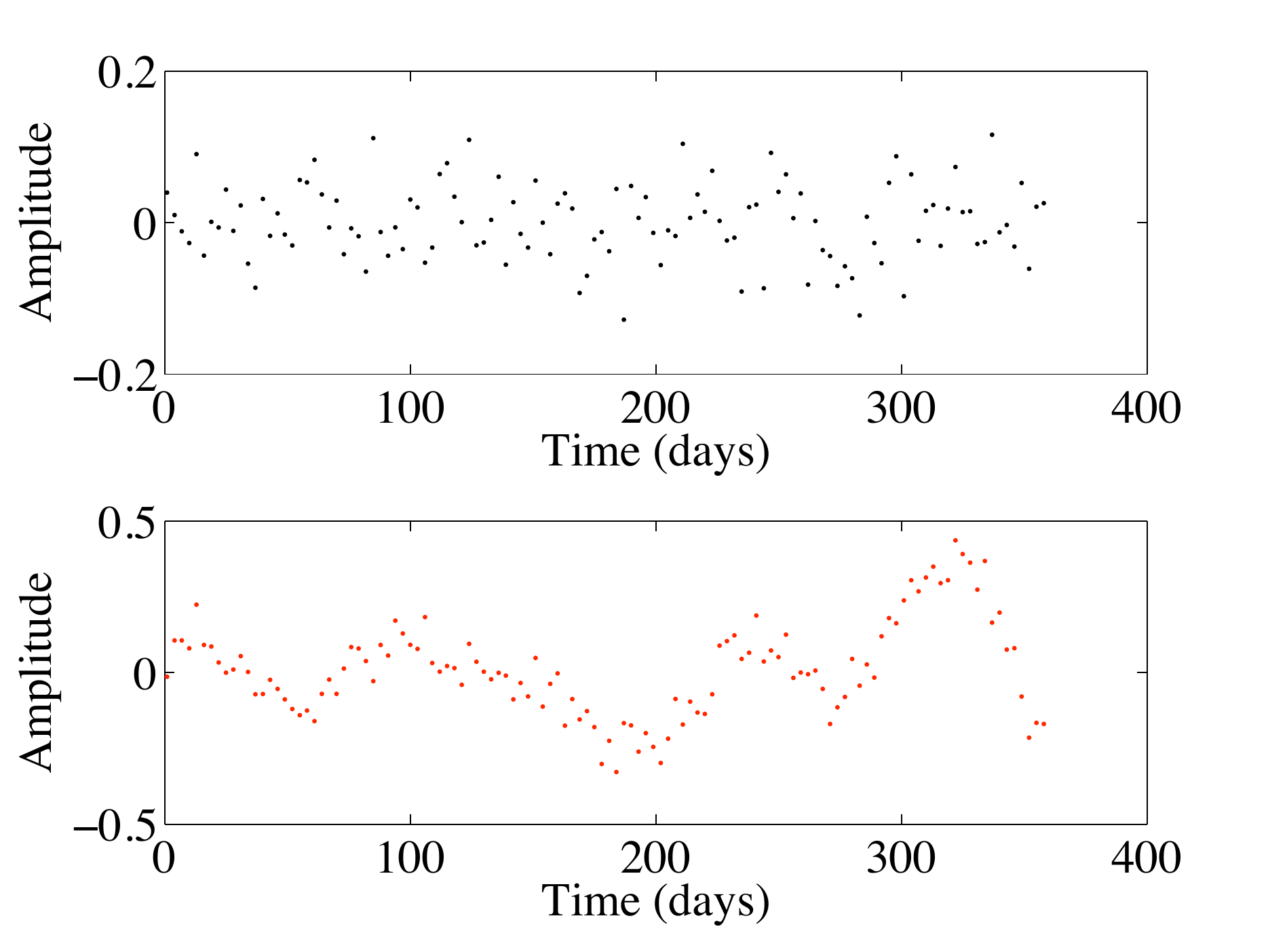}
\caption{Comparison between white noise (top panel) and red noise (bottom panel).}
\label{RedNoiseEg}
\end{center}
\end{figure}
 
\section{Red noise: observational derivation}\label{Transit_Noise_Red}

\subsection{Introduction to red noise}

Red noise is additive noise which contains an excess of power in lower (or ``redder") frequencies.  While in signal processing red noise is a slang term for brownian noise or random walk noise and has spectral density proportional to $f^{-2}$, where $f$ is the frequency, in astrophysics, the term is colloquially used to describe any noise process which produces an excess of low frequencies.  For a more in-depth introduction into correlated noise processes within the context of transits, please see \citet{CarterWinn2009}.

The fact that stars in general \citep[e.g.][]{Radicketal1982,DorrenGuinan1982} and the Sun in particular \citep[e.g.][]{Willsonetal1981} have a red noise component in their light curves has long been known.  Due to this excess of long wavelength components, consecutive data points in a light curve dominated by the intrinsic variability of the host star are correlated, that is, there are long term trends in the data (see figure~\ref{RedNoiseEg}).  Consequently, the statistical methods used in the previous section no longer apply.  As discussed in section~\ref{Transit_Noise_Method_Solar}, for the case of realistic photometric noise, the distribution and behaviour of $\epsilon_j$ will be calculated numerically using solar photometric light curves.

\subsection{Derivation of $\epsilon_j$}

For this thesis, Total Solar Intensity (TSI) data, taken using the the DIARAD instrument on the SOHO satellite was used.  In particular, DIARAD TSI data taken between 30/1/96 to 1/9/07 was downloaded from the SOHO website.\footnote{http://seal.nascom.nasa.gov/cgi-bin/gui\_plop.}  This data was in the form of a FITS compliant text file (see figure~\ref{Examplefile}).  To indicate a NULL reading, the intensity 99.99 and the flag 99 were used. There were a number of issues with this data.  First some of the data was not available for certain dates (presumably as the photometer was not running).  Second, during days where data was available, it was not available for all times.   Finally, numerous formatting inconsistencies, such as the running together of the last intensity value and the NULL flag and the value of the recorded intensities reversing sign, meant that preprocessing was required.

To deal with the issue of missing days and formatting inconsistencies, a Perl program was written to reformat the data.  This program:
\begin{itemize}
\item Read in all the data from the data files.
\item Inserted the correct number of NULL values (99.99 and 99) for each missing day.
\item Separated the intensity value and the 99 from the end of each measured sequence.
\item Checked the status flag for each exposure, and if the code was not valid ($\ne16380$) then set the intensity and the intensity flag to 99.99 and 99 respectively.
\item Reversed the sign of the intensity if it was negative.
\item Corrected the intensity for instrumental variation using the table given on the SOHO website.\footnote{ftp://ftp.pmodwrc.ch/pub/data/irradiance/virgo/TSI/korr\_tot6\_002\_0904.dat.}
\item Output the intensity of each exposure and the time in minutes since the beginning of observation, in this case, since 30/1/1996, into a single text file.  
\end{itemize}
Now that a corrected TSI time series is available the set of variable values for which the distribution of $\epsilon_j$ will be evaluated needs to be considered.

As this is a numerical investigation into $\epsilon_j$ the set of variable values for which the distribution of $\epsilon_j$ will be evaluated needs to be selected.  Informed by the work on white noise, physically important variables include $T_{obs}$, $T_{tra}$, $\Delta t$, $\beta$ and $A_p + A_m$.  However, as the terms $\beta$ and $A_p + A_m$ (and consequently $T_{tra}$) only appear in the premultiplied factor in equation~\eqref{red_epsdef}, not in $\epsilon_{\sun}$ we only need to consider a grid of $T_{obs}$ and $\Delta t$ values.  The grid of values selected for this analysis will be discussed.

 \begin{table}[tb]
   \begin{tabular}{lcccc}
   \hline
                     &   DIARAD        & DIARAD        &  PMOV6V &  PMOV6V\\
 $T_{obs}$ &  3 mins        &  30 mins        &  1 min           & 30 mins  \\
   \hline
   30 mins  & 163039 & --          & 132185 & -- \\
   1 hr         & 78971   & 76956  & 63802   &  61681     \\
   2 hr         & 37208   & 36244  & 29736   &  29604 \\
   4 hr        & 16503    & 16113  & 12657   &  12153\\
   8 hr        & 6471      & 6318     & 4077     & 3679 \\
   12 hr      & 3438     & 3365      & 475       & 462 \\
   16 hr     & 2080     & 2038      & 238        &  228 \\
   24 hr      & 824      & 805         & 73          &  71   \\
   36 hr      & 317      & 307         & 54          &  53   \\
  \end{tabular}\\
 \caption{ Number of complete lengths of data of a given duration, $T_{obs}$, for each of the four possible data sets.}
 \label{DataLengths}
 \end{table}
 
An investigation of the effect of the size of the observation window into the distribution of $\epsilon_j$ is pertinent for two main reasons.  First, to document the effect of the size of the observing window for cases relevant to the set of planets likely to be discovered by COROT and Kepler.  Second, to check the method by comparing the behaviour of $\epsilon_j$ with the behaviour calculated for the case of white noise.  First, the range of observing windows relevant to the COROT and Kepler missions needs to be determined.  As the observation window cannot be smaller than the transit duration, the minimum observation window is limited by the  minimum transit duration.  As discussed in chapter~\ref{Intro_Moons_Const}, it is unlikely that moons will be discovered around planets with semi-major axes smaller than 0.2AU as a result of moon orbital evolution and subsequent loss.  In addition,  as a result of the finite duration of the COROT and Kepler missions, and thus the limited number of transits which can be observed per planet, it is unlikely that moons will be discovered around planets with semi-major axes larger than 1AU.  For a Sun-like star, these two limits correspond to a transit duration of 8 hours and 12 hours respectively, for the case where the planet passes across the central chord of the star.  The actual transit duration may be shorter depending on the inclination of the planet's orbit to the line of sight.  As 86.6\% percent of transiting planets will have transit durations longer than 50 percent of the nominal duration, a conservative estimate of the minimum length observation window is four hours.  Also, as will be discussed in section~\ref{Trans_Thresholds_ExpBehav}, the most detectable moons are located between $1/2R_s$ and $2R_s$ from the host planet, the exact value depending on the behaviour of the noise on the light curve.  For the case of planets at 0.2AU and 1 AU, this corresponds to an observing window of length 16 and 24 hours respectively, for a central transit.  As we would not only like to search for the most detectable moons, it was decided to investigate observing windows with lengths between 4 and 36 hours.  Second, in order to compare the results of this analysis with that of the analysis of white noise, we need to select the set of observing windows for which the data most resembles white noise.  The shorter the observing window, the smaller the effects of the long term trends caused by the red noise component, the more ``white", the noise should be. Consequently it was decided to also investigate observing windows which were shorter than four hours.  As the largest exposure time used by COROT and Kepler is 30 minutes it was decided that this would be the shortest observing window investigated.  As a result of these two factors, it was decided to investigate the cases where the observing window was of length 30 minutes, 1 hour, 2 hours, 4 hours, 8 hours, 12 hours, 16 hours, 24 hours and 36 hours.

An investigation of the relationship between exposure time and the distribution of $\epsilon_j$ may allow optimisation of the selected observing strategy with respect to moon detection.  As the satellite missions COROT and Kepler are the most likely to find the types of planets which could harbour moons and also have the required sensitivity to do the TTV$_p$ followup, it is most useful to investigate the effect of exposure time on $\epsilon_j$ with respect to the capabilities of these satellites.  Both these satellites are capable of long and short exposures.  For the case of COROT, a 512 second exposure 
time is used for its catalogue of approximately 12000 targets, but it is also capable of 32 s readout for a subset of 1000 highlighted sources \citep{Quentinetal2006}.  For the case of Kepler, a thirty minute exposure time is used for its set of 3000 sources, but it is also capable of one minute exposures for a subset of 512 highlighted sources \citep{Boruckietal2007}.  So, ideally, it would be useful to compare the distribution of $\epsilon_j$ for the case of ``short" exposures (32 seconds or 1 minute) and ``long" exposures (16 or 30 minutes).  Unfortunately, PMO6V and DIARAD, the two instruments on the SOHO satellite, have exposure times of 1 minute and 3 minutes respectively, so, solar data with cadence below 1 minute are not available.  In addition, while data with cadence of 1 minute are available, it is quite patchy.  As a result of this patchiness, there are only a small number of long contiguous lengths of data which are 24 or 36 hours long (see table~\ref{DataLengths}).  As we would like to investigate the case of short exposure times for long observing windows, it was decided to use the 3 minute cadence DIARAD data.  To maximise the difference between long and short exposures it was decided to investigate the case of 30 minute exposures.  These exposures were created by first dividing the three minute data into thirty minute intervals.  For the case where all the data within a given interval was valid, the average of the intensities of the constituent three minute exposures was recorded as the intensity for that 30 minute ``exposure".  For the case where some of the data within the interval was not valid, a NULL intensity was recorded for that 30 minute ``exposure".  

\begin{figure}[tb]
  \begin{tabular}{lrl}
   SIMPLE  =                    &T & / file does conform to FITS standard\\
   BITPIX  =                      & 8 & / number of bits per data pixel\\
   NAXIS   =                     & 0 & / number of data axes\\
   \vdots                            &  & \vdots \\
   EXTNAME = `DIARAD LEVEL 1'     & &/ name of this ASCII table extension\\
   TNULL1  = `99.99'           & & / Undefined value SOLAR\_CT\\
   TNULL2  = `99'          & &  / Undefined value STATUS\\
   END\\  
  \end{tabular}
  \\
  \begin{tabular}{llllllll}
   99.99        & 99        &  99.99       & 99       & 99.99        &   99      &  99.99   \\
   99              &99.99   & 99              & 99.99 & 99              & 99.99  &   99       \\
   99.99        & 99        &99.99         & 99       &  99.99       & 99        & 99.99  \\
   99              &  99.99 & 99             & 99.99  & 99              &  99.99 & 99        \\
   1366.747 & 16380 & 1366.952 & 16380 & 1366.827 & 16380 & 1366.856\\
   16380       & 1366.790 & 16380 & 1366.870 & 16380 &1366.781 & 16380\\
                        & \vdots  &                   & \vdots &                    & \vdots  &                 \\
   1366.958 &16380 & 1366.957 & 16380 & 1366.806 & 16380 & 1366.849 \\
   16380 & 1366.795 & 16380 & 1366.861& 1638099.99 & 99            & 99.99\\
    99            & 99.99  &   99            & 99.99  &   99             & 99.99      & 99 \\
                            & \vdots  &                   & \vdots &                    & \vdots  &                 \\
    1366.997 & 16380   & 1366.966 &16380 & 1366.939 &16380 & 1366.757 \\
    16380      &1366.883 & 16380     &1366.875 & 16380 & 1366.921 & 16380 \\
    1366.897 & 16380 & 1366.851& 16380\\ 
  \end{tabular}\\
 \caption{Abridged example DIARAD data file.  In this case, the file corresponds to the measurements taken on the  1/1/2000.}
 \label{Examplefile}
 \end{figure}
 
 \begin{figure}
     \centering
     \subfigure[$T_{obs}$ = 1hr.]{
          \label{fig:dl2858}
          \includegraphics[width=.48\textwidth]{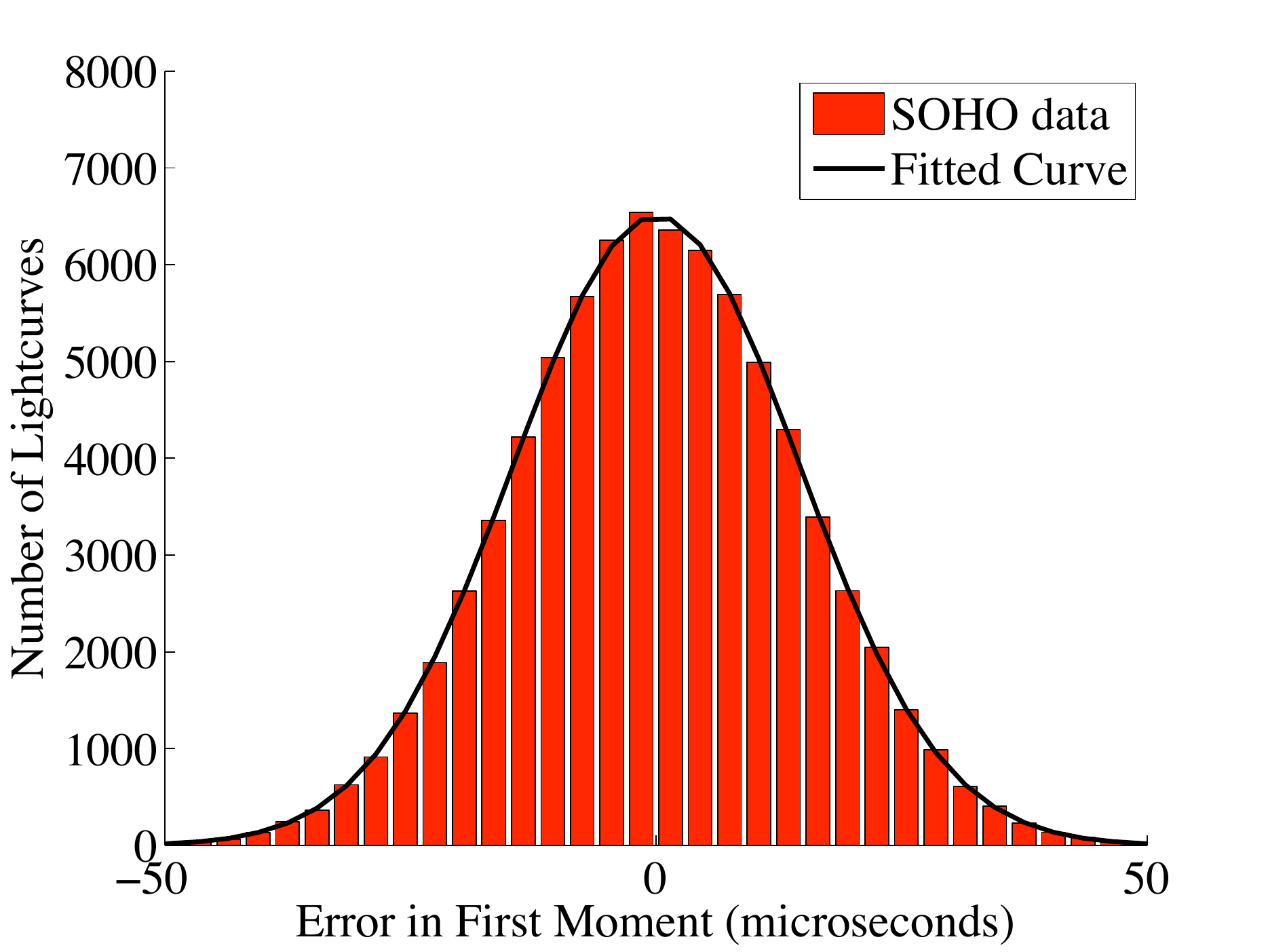}}
     \subfigure[$T_{obs}$ = 2hr.]{
          \label{fig:er2858}
          \includegraphics[width=.48\textwidth]{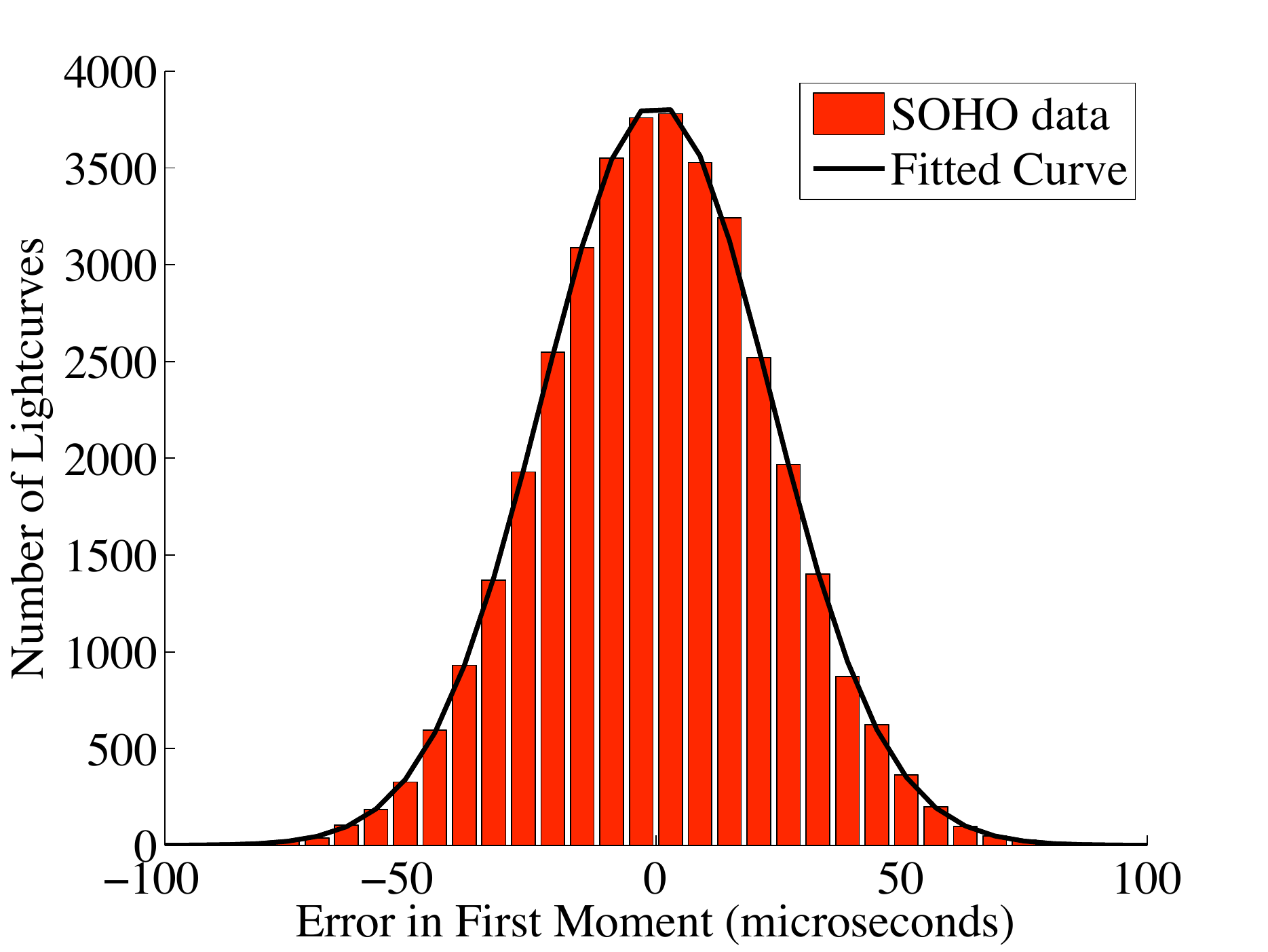}}\\
          \subfigure[$T_{obs}$ = 4hr.]{
          \label{fig:dl2858}
          \includegraphics[width=.48\textwidth]{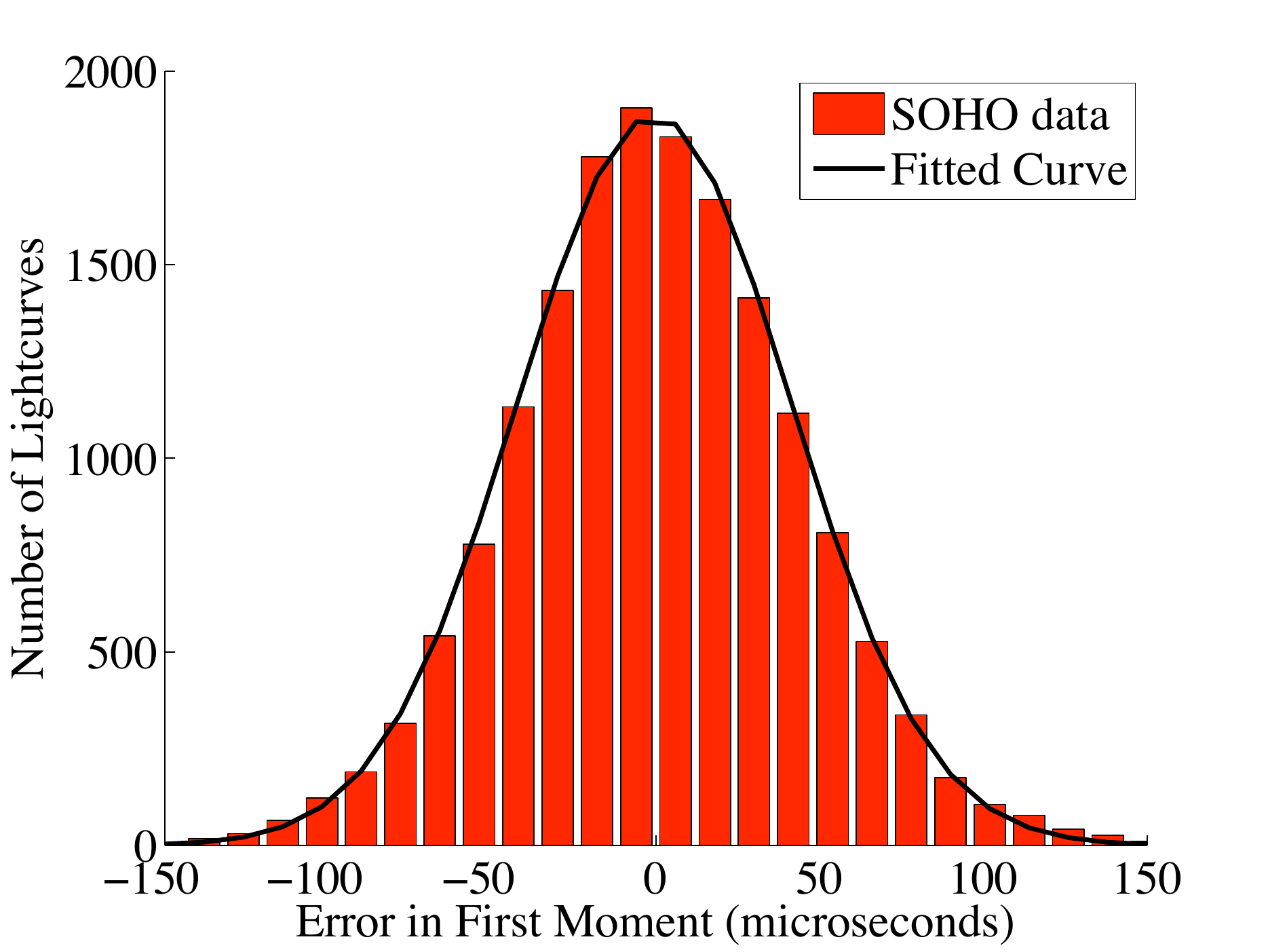}}
     \subfigure[$T_{obs}$ = 8hr.]{
          \label{fig:er2858}
          \includegraphics[width=.48\textwidth]{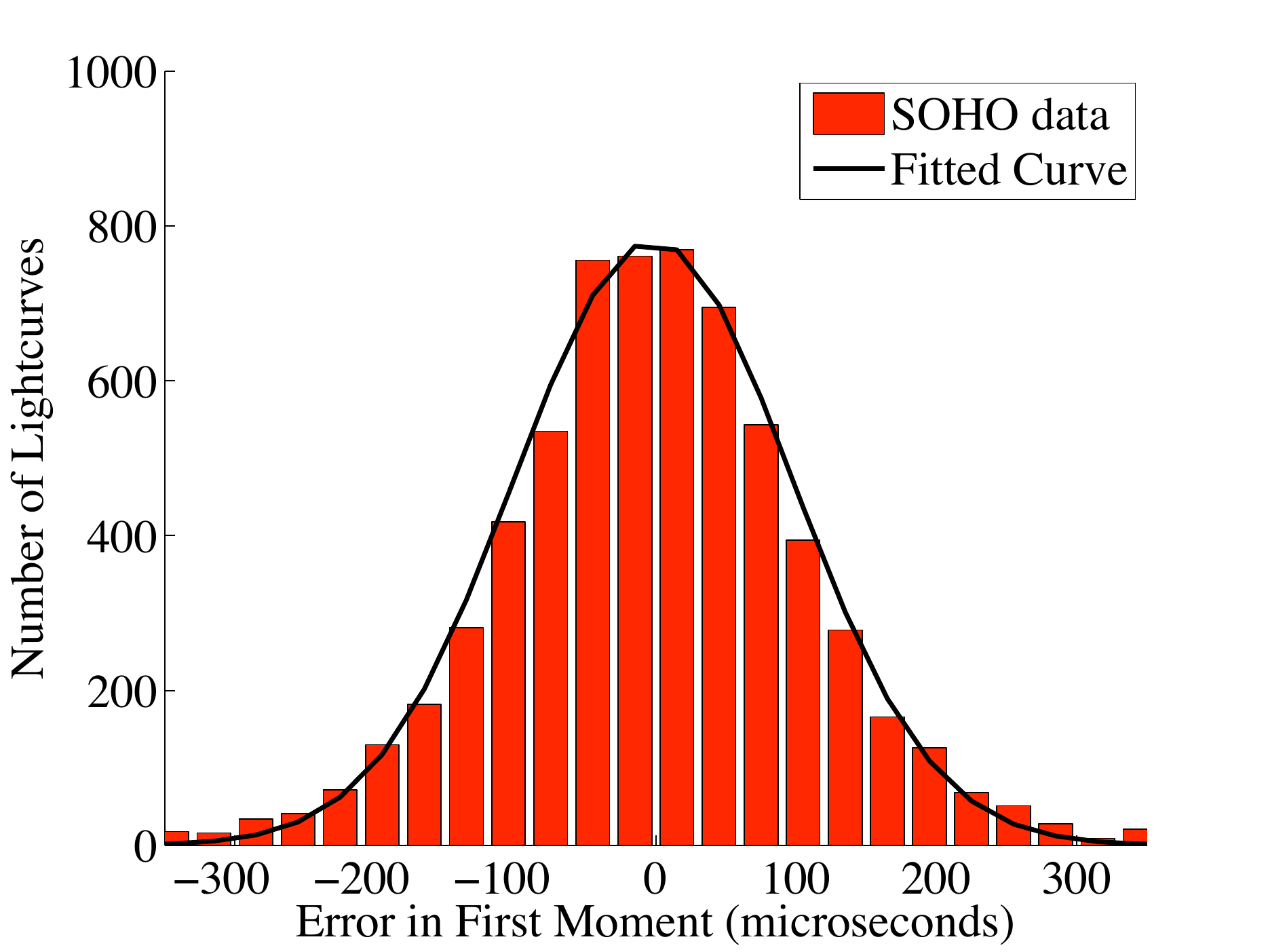}}\\
           \subfigure[$T_{obs}$ = 12hr.]{
           \label{fig:cminusscalar2858}
           \includegraphics[width=.48\textwidth]{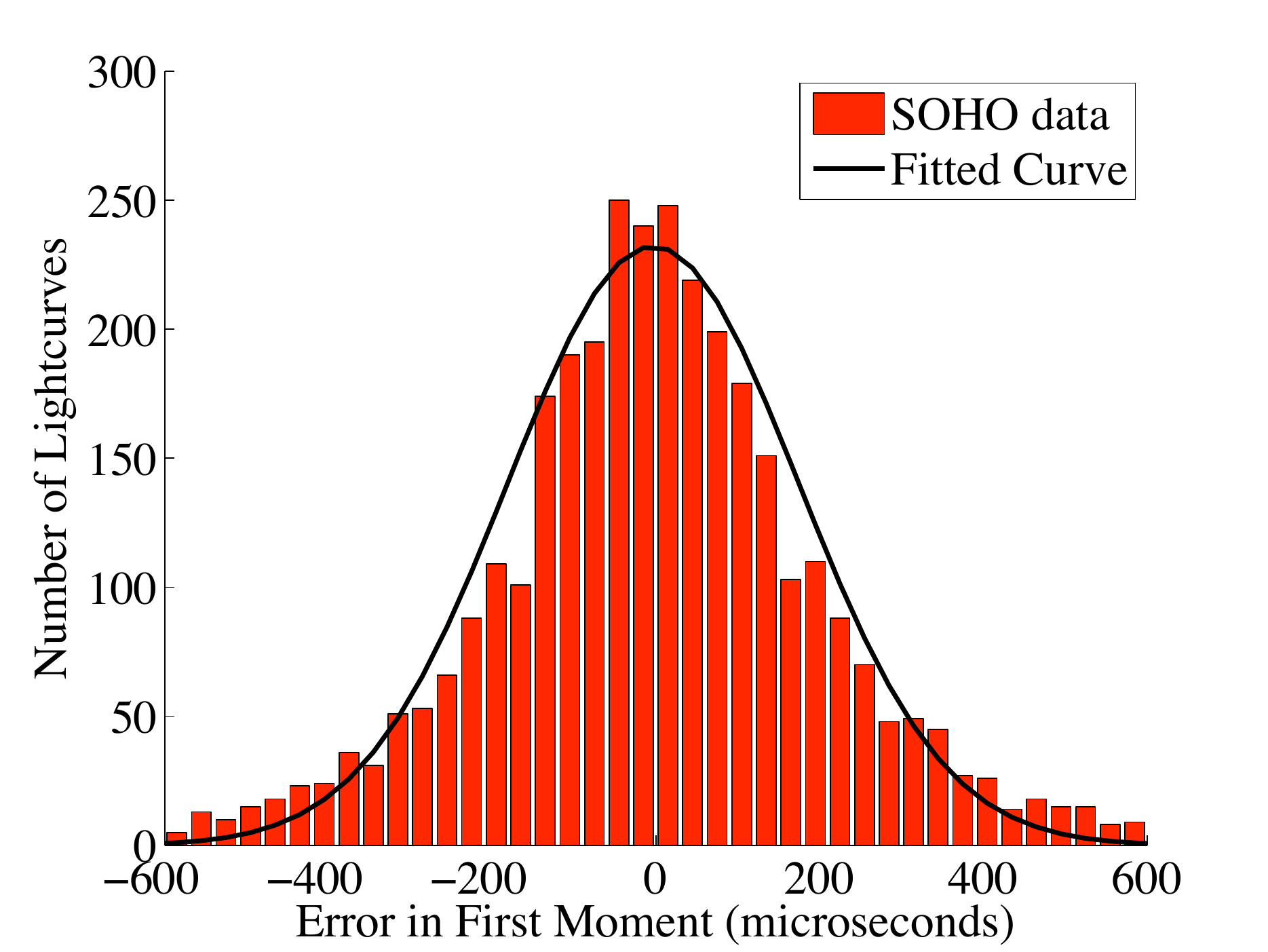}}
           \subfigure[$T_{obs}$ = 24hr.]{
           \label{fig:cminusscalar2858}
           \includegraphics[width=.48\textwidth]{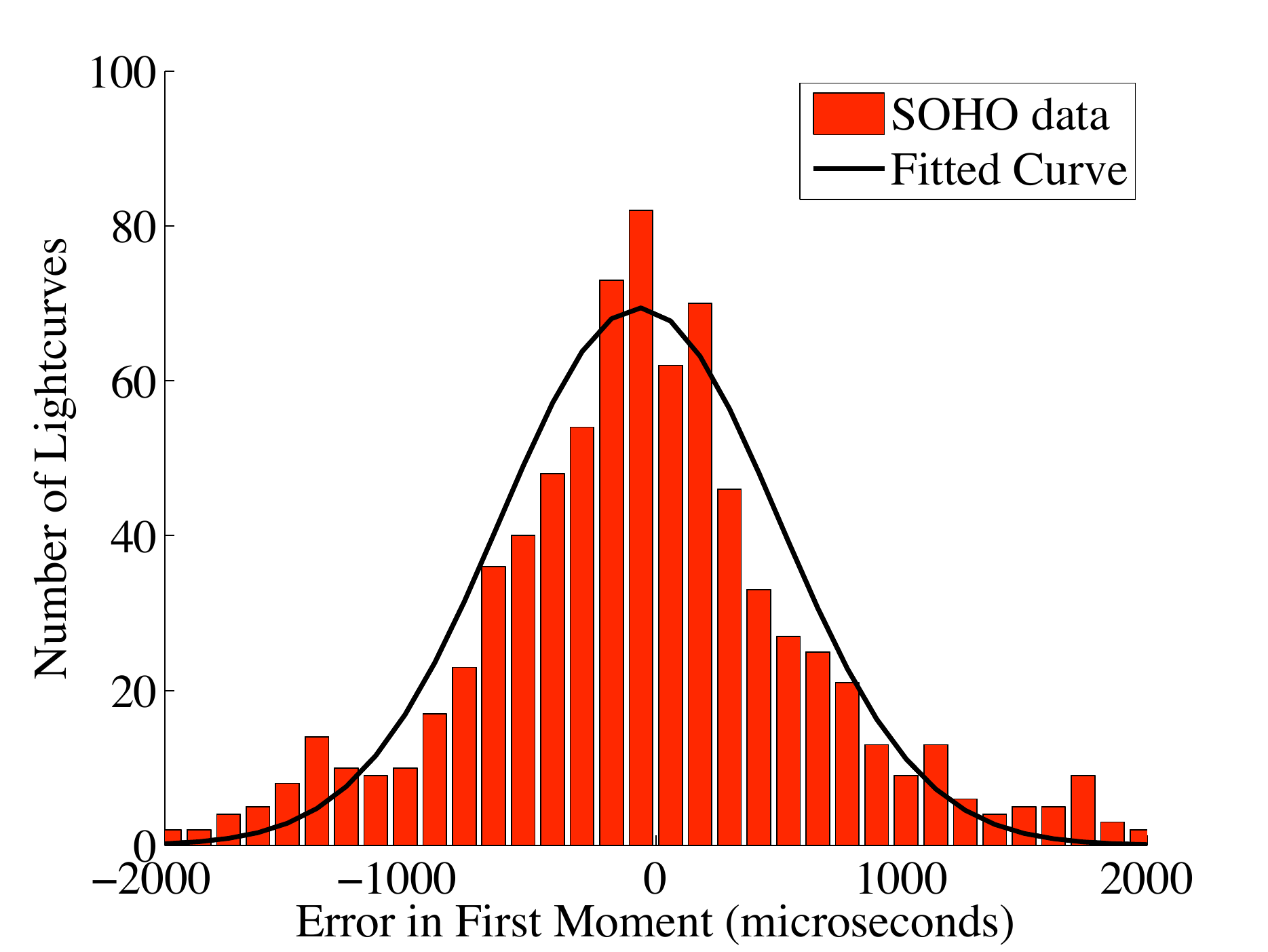}}
     \caption[The observationally determined distribution of $\epsilon_{\sun}$ (red bar) for the case of realistic solar photometric noise for six different length observing windows.]{The observationally determined distribution of $\epsilon_{\sun}$ (red bar) for the case of realistic solar photometric noise for six different length observing windows.  Note that for observing windows less than 12 hours long, the distribution of $\epsilon_{\sun}$ strongly represents a normal curve (black line).}
     \label{RedDist}
\end{figure}

To implement this numerical investigation into $\epsilon_j$ for this grid of $T_{obs}$ and $\Delta t$ values, a second program was written.  It read in either the 3 minute or 30 minute corrected TSI values and divided each section into ``transits" of a given duration $T_{obs}$, ignoring any remainder.  As discussed above, the distribution of $\epsilon_j$ can be calculated from the TSI directly and then scaled by $\beta L_{0}N_{obs}/(A_p + A_m)$ for the transit in question.  Thus, for each of these lengths, the error associated with each of these transits were calculated and then binned to give an experimental distribution of $\epsilon_{\sun}$ (see figure~\ref{RedDist}).  As these distributions were approximately Gaussian for all $T_{obs}$ examined it was decided to model the distribution using the standard deviation.  Using the MATLAB function \texttt{fminsearch} to perform a least squares fit to the histogram, $\mu_{\sun}$ and $\sigma_{\sun}$, and the errors associated with these values were derived.  In all cases the mean was approximately zero, consequently only $\sigma_{\sun}$ was recorded.  

\begin{figure}[tb]
\begin{center}
\includegraphics[width=.90\textwidth]{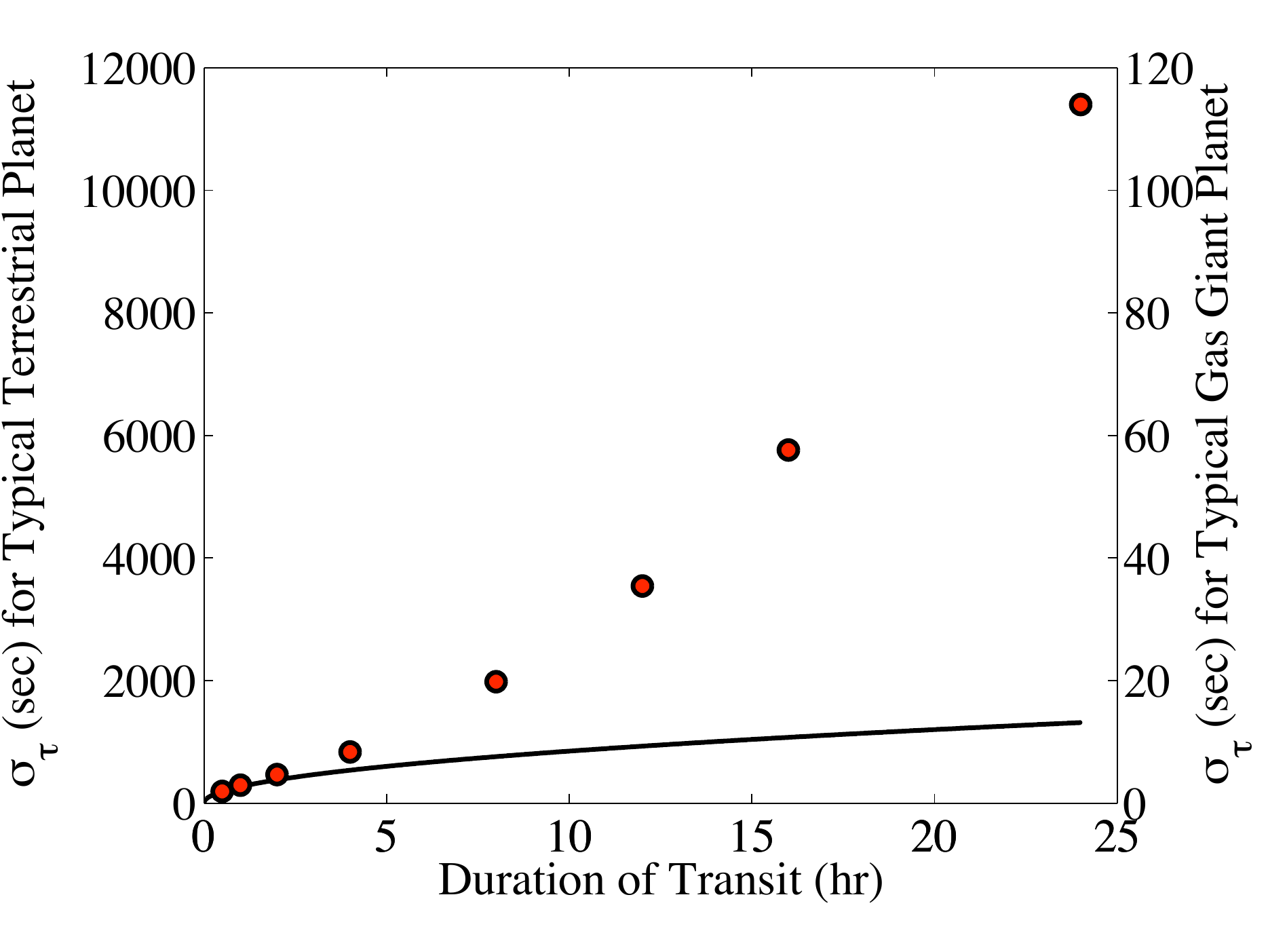}
\caption[Comparison between calculated errors in $\tau$ using solar light curves (red dots) and theoretically predicted errors in $\tau$ using white noise with the same power as the noise in the solar light curves (thick line). ]{Comparison between calculated errors in $\tau$ using solar light curves (red dots) and theoretically predicted errors in $\tau$ using white noise with the same power as the noise in the solar light curves (thick line).  The method used to derive this white noise amplitude is presented in appendix \ref{App_WhiteNoiseAmp}.  As the results for 3 minute and 30 minute exposures are so similar, only the data for 3 minute exposures is shown.  These relations are shown for the case of the transit of a gas giant ($(A_p + A_m)/L_0 N_{tra} = 10^{-2}$) and a terrestrial planet ($(A_p + A_m)/L_0 N_{tra} = 10^{-4}$), assuming $T_{obs} \approx 2 T_{tra}$.}
\label{RedNoiseCompare}
\end{center}
\end{figure}

The results are presented in figure~\ref{RedNoiseCompare} and table~\ref{RedNoiseepssun}.  As can be seen in table~\ref{RedNoiseepssun}, the values of $\sigma_{\sun}$ corresponding to the two different exposure times for $T_{obs}$ greater than 4 hours agree within error bounds.  As the vast majority of realistic values of $T_{obs}$ are likely to be greater than four hours, these two data sets are effectively equal.  Performing a quadratic fit to this data, the following formula for the standard deviation of $\sigma_{\sun}$ is obtained
\begin{equation}
\sigma_{\sun} = 5.32 \times 10^{-3}\text{s} + 6.48 \times 10^{-3} \text{s hr}^{-1} T_{obs} + 6.96 \times 10^{-4} \text{s hr}^{-2} T_{obs}^2 \label{TraM-Noi-Red-Dfit}
\end{equation}
Substituting this into equation~\eqref{red_epsdef} and recasting into physical variables gives
\begin{multline}
\sigma_\epsilon = 103.7\text{s} \left[\beta \right]\left[\frac{L_{0}N_{tra}}{100(A_p + A_m)}\right] \left[ \left(\frac{T_{tra}}{13 \text{hr}}\right)^{-1} \right. \\ \left.
\times \left(0.010 \left(\frac{T_{obs}}{24 \text{hr}}\right) + 0.277 \left(\frac{T_{obs}}{24 \text{hr}}\right)^2 + 0.714 \left(\frac{T_{obs}}{24 \text{hr}}\right)^3\right)\right] \label{transit_noise_red_sigdefphys}
\end{multline}
Using this result, some general comments can now be made about the behaviour of $\epsilon_j$ for the case where the light curve is dominated by realistic stellar photometric noise.

\begin{table}[tb]
\begin{center}
   \begin{tabular}{lll}
   \hline
 $T_{obs}$ &   $\Delta t$ = 3 mins         &  $\Delta t$ =  30 mins   \\
   \hline
   30 mins  & $9.492  \times 10^{-3}$ s    & --  \\
   1 hr         & $1.462 \times 10^{-2}$ s     & $1.261\times10^{-2}$ s   \\
   2 hr        	& $2.326 \times 10^{-2}$ s      & $2.259\times10^{-2}$ s  \\
   4 hr        	& $4.18 \times 10^{-2}$ s        & $4.18\times10^{-2}$ s    \\
   8 hr        & $9.90 \times 10^{-2}$ s        & $9.90\times10^{-2}$ s \\
   12 hr      & $1.77 \times 10^{-1}$ s        & $1.79\times10^{-1}$ s \\
   16 hr      & $2.88 \times 10^{-1}$ s       & $2.88\times10^{-1}$ s  \\
   24 hr      & $5.7 \times 10^{-1}$ s         & $5.7\times10^{-1}$ s         \\
   36 hr      & $1.1$ s                                   & $1.1$ s \\
   \end{tabular}\\
 \caption{The size of $\sigma_{\sun}$ as a function of length of observation window, $T_{obs}$, and exposure time $\Delta t$.  The value of $\sigma_{\sun}$ is recorded to the last significant figure.}
 \label{RedNoiseepssun}
  \end{center}
 \end{table}

 \begin{figure}[tb]
\begin{center}
\includegraphics[height=3.0in,width=4.0in]{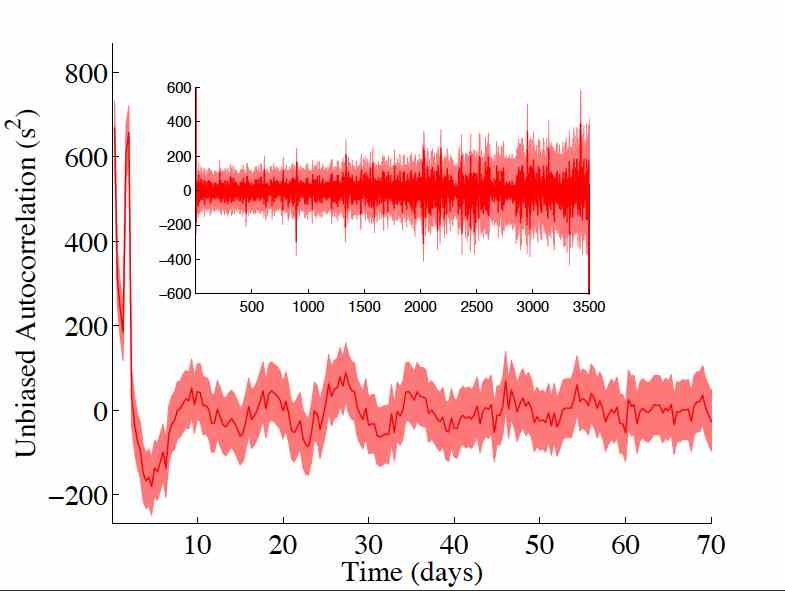}
\caption[Unbiased autocorrelation function of $\epsilon_j$ (red line) calculated using realistic stellar noise.]{Unbiased autocorrelation function of $\epsilon_j$ (red line) calculated using realistic stellar noise for the case where the planet is Jupiter-like ($(A_p + A_m)/L_0 N_{tra} = 10^{-2}$ and $T_{obs} \approx 2 T_{tra}$) and where the observing window is 8 hours.  For reference, the one sigma error bars (pink) are also shown.  As short period planets are more likely to be discovered by the transit technique, the inner section of the autocorrelation function is shown in the main plot.  For completeness, the full autocorrelation function is shown in the inset.}
\label{RedNoiseCorr}
\end{center}
\end{figure}
 
\subsection{Properties of $\epsilon_j$}

As for the case of white noise, the properties of $\epsilon_j$ will be discussed using equation~\eqref{transit_noise_red_sigdefphys} in terms of $\beta$, $\Delta t$, $A_p + A_m$, $T_{tra}$ and $T_{obs}$.  Then the discussion will proceed to a more general investigation in terms of $\Delta \tau$ and $j$.

The first term in equation~\eqref{transit_noise_red_sigdefphys}, describes the effect of the amplitude of the photometric noise relative to the amplitude of the photometric noise of the Sun, through the term $\beta$.  Unsurprisingly, the larger the amplitude of the intrinsic photometric noise of the star, the larger the resulting error in $\tau$.  This is equivalent to the linear dependance of $\epsilon_j$ on $\sigma_L$ for the case of white photometric noise.  Also, as discussed previously, for the case of realistic solar photometric noise, the distribution of $\epsilon_j$ does not seem to depend on the exposure time.  This again is unsurprising as the noise is inherent to the host object, not on the way in which it is measured, so altering the exposure time merely acts to smear out the photometric noise, not alter its form or amplitude (or its effect on $\epsilon_j$).

Similar to the case of white photometric noise, the second term of equation~\eqref{transit_noise_red_sigdefphys} represents the effect of the relative size of the planet compared to the star.  As this dependance is a property of equation~\eqref{transit_noise_red_sigdefphys}, it should not be surprising.

The third term of equation~\eqref{transit_noise_red_sigdefphys} describes the dependance of $\epsilon_j$ on planet-star separation and planet-moon separation through the transit duration $T_{tra}$ and the observing duration $T_{obs}$ respectively.  As mentioned in the section on white noise, the optimal observing duration scales as the transit duration for a given planet-moon pair.  Applying this to equation~\eqref{transit_noise_red_sigdefphys} we find that $\epsilon_j$ is a superlinear function of transit duration.  In addition, it is also a superlinear function of observing duration (a good proxy for planet-moon distance).  Comparing these results to the relations for the case of white noise ($\epsilon_j \propto T_{tra}^{1/2}$  and $\epsilon_j \propto T_{obs}^{3/2}$) and to the amplitude dependance of $\Delta \tau$ ($\Delta \tau \propto T_{tra}$), it can be seen that the addition of realistic photometric noise to a light curve makes a substantial difference to moon detection.  In particular, for the case of realistic photometric noise it would be easier to detect moons which are closer to their host planet and to detect moons of planets with shorter transit times, due to, for example, an inclined orbit.  

As with the case of white noise, equation~\eqref{transit_noise_red_sigdefphys} does not depend on $\Delta \tau$, again as these dependancies were neglected (as a result of their small contribution) in the derivation of equations~
\eqref{transit_intro_ground_noisedef} and \eqref{transit_noise_method_rndef}.  By analogy with the case of white noise, we would expect this discrepancy to become large for large $\Delta \tau$, or for the case where the noise amplitude is large relative to the transit depth (that is, for large $\beta$).

In addition to determining the distribution of $\epsilon_j$, it is also important to determine whether or not consecutive values of $\epsilon_j$ are correlated.  Consequently the autocorrelation\footnote{For this work the autocorrelation is defined as $\sum_j \epsilon_j \epsilon_{j+k}/N_k$ where $\epsilon_j$ and $\epsilon_{j+k}$ represent all $\epsilon_j$ pairs separated by $k \times 8$ hours and $N_k$ is the number of such pairs.  This unbiased definition was selected to allow ease of comparison between values calculated for different lag times.} of the sequence of $\epsilon_j$ values was computed (see figure~\ref{RedNoiseCorr}).  From physical intuition, we would expect that there would be some correlation on timescales comparable with the rotational period of the Sun ($\approx 25$ days) as the gradient of the luminosity is also correlated over those timescales.  As can be seen from figure~\ref{RedNoiseCorr}, $\epsilon_j$ is effectively uncorrelated for all transiting planets with orbital periods above forty days.  As the shortest interval between consecutive transits for a planet with stable moons around a Sun-like star is approximately one month, only the region of the autocorrelation curve corresponding to orbital periods greater than one month need be examined.  Consequently, for nearly all planets likely to host moons, $\epsilon_j$ is uncorrelated.

 Now that the case of raw realistic stellar noise has been investigated, the analysis can be refined by ``filtering" the data to remove long term trends.  In the case to be investigated these trends are due to the advection of active regions across the solar surface.

\section{Filtered red noise: observational derivation}\label{Trans_TTV_Noise_Filtered}

\subsection{Introduction to filtered noise}\label{Trans_TTV_Noise_Filtered}

As the presence of red noise in light curves significantly alters the detection probability of transiting planets \citep[e.g.][]{Boruckietal1985,Pontetal2006}, a range of methods for reducing the effect of this type of noise have been investigated.  These include methods where the filtering is performed in combination with a transit finding algorithm \citep[e.g.][]{Jenkins2002}, methods where the red noise is preprocessed with a whitening filter before transit detection \citep[e.g.][team 3]{Carpanoetal2003, GuisBarge2005, Moutouetal2005}, and methods where the physical processes believed to be underlying the red noise are modeled \citep{Lanzaetal2003}.

Since for this application moon searches will be conducted on a star by star basis, it was decided to use a method which modeled the physical processes appropriate to each star.  In particular it was decided to use the three spot model method of \citet{Lanzaetal2003} as:
\begin{itemize}
\item This method was developed using SOHO data.
\item It was designed to be adapted to model photometric variation for any Sun-like star \citep{Lanzaetal2004}.
\item It has been used to model variability in real stellar data, specifically, the COROT targets CoRoT-Exo-2 \citep{Lanzaetal2009a} and CoRoT-Exo-4 \citep{Lanzaetal2009}.
\end{itemize}
For completeness, the particulars of this method will be discussed.

\subsection{Description of the three spot model}

One source of long term photometric variability of the Sun is rotational modulation of the active regions, that is, regions on the Sun's surface where the magnetic field strengths are high.  Active regions are complex structures comprising of clusters of magnetic phenomenon such as darker sunspots and brighter faculae.  While there is a log normal distribution of active region sizes \citep{Bodganetal1988}, \citet{Lanzaetal2003} determined that the long term photometric variation of the Sun could be effectively modeled by only accounting for three distinct active regions on the face of the star in any given fourteen day window.

To perform the fitting required to implement this ``three spot model" the extent to which the substructure of the active regions needs to be modeled, must be determined.  In particular, the issues of active region composition, evolution and distribution across the solar surface need to be explored.  While the ratio between the area of faculae and starspots changes as a function of active region size and position in the solar cycle \citep{Chapmanetal1997}, \citet{Lanzaetal2003} assumed a constant ratio of sunspot area to facular area of 1:10.  Consequently this assumption was also used for this thesis.  Not only are active regions composed of complex structures, these structures evolve over time.  Fortunately, the timescale for active region evolution is longer than that for solar rotation.  Consequently, the solar intensity data can be divided into blocks, such that it can be assumed that the position and size of an active region doesn't change within the block.  \citet{Lanzaetal2003} found that a good compromise between modeling the effect of solar rotation and minimising active region evolution was 14 days.\footnote{The optimum block length depends on the characteristics of the host star.  For example, for the case of the Sun the optimal length is 14 days, while for the case of CoRoT-Exo-4 a length of 8.2 days \citep{Lanzaetal2009} was used.}  Consequently, for this thesis, data was divided into 14 day segments, each consisting of 6720 data points.  In addition, as active regions are generally much smaller than the radius of the Sun, the entire active region can be modeled as having the same $\mu$ value, where the $\mu$ value is given by the cosine of the angle between the surface normal and the direction of the line-of-sight at that position on the solar surface.  Consequently, three variables are required to model each active region, two describing the position of the region on the face of the Sun, and one variable describing the effective area of the active region.  With these assumptions and simplifications in mind, the fitting model can be introduced.

Following \citet{Lanzaetal2003}, this results in the following model for $L$, the photometric intensity of the Sun
\begin{multline}
L - L_0 = L_r + L_0 \mathcal{C} \sum_{i = 1:\mu_i > 0}^3 A_i \mu_i \left[ a_p + b_p \mu_i + c_p \mu_i^2\right] \\ \times \left[ (c_s-1)+Q(c_f + c_f' \mu_i - 1) \right],
\end{multline}
where the $\mu_i$ correspond to the $\mu$ values of each of the star spots, where $\mu_i$ will be defined in equation~\eqref{transit_noise_filt_mudef}.  For this study the parameters relating to the intrinsic properties of the Sun, $L_0$, $\mathcal{C}$, $a_p$, $b_p$, $c_p$, $Q$, $c_s$, $c_f$ and $c_f'$ were set to the same values used in \citet{Lanzaetal2003}. For completeness the values and the purpose of each of these model parameters will be stated and explained.  The limb darkening parameters, $a_p$, $b_p$, $c_p$ and $\mathcal{C}$ were set to 0.36, 0.84, -0.20 and 4.88 respectively.  In addition, the parameter $c_s$, which describes the dependence of sunspot luminosity on position was set to 0.67, and the parameters $c_f$ and $c_f'$ which describe the dependence of facular luminosity as a function of position were set to the solar values of 1.115 and -0.115 \citep{Foukaletal1991}.  $Q$, the ratio of sunspot area to faculae area was set to 10.  As discussed in \citet{Chapmanetal1997}, this ratio can change throughout the eleven year solar cycle.

In addition to the variables remaining constant for all fits, eleven variables are allowed to change.  First the additional background intensity is described by $L_r$.  Second the relative areas of the three starspots are given by $A_1$, $A_2$ and $A_3$.  Finally, the effect of the position of the star spots is parameterised by their $\mu$ value, defined as
\begin{equation}
\mu_i = \cos i_{sun} \sin \theta_i + \sin i_{sun} \cos \theta_i \cos (\lambda_i + \Omega t - L_0)\label{transit_noise_filt_mudef}
\end{equation}
where $\Omega$ is the angular velocity of the Sun due to its rotation and $\lambda_i$ and $\theta_i$ are the longitude and latitude of the $i^{th}$ starspot.  Consequently, the eleven variables which must be fitted for this model are $A_1$, $\lambda_1$, $\theta_1$, $A_2$, $\lambda_2$, $\theta_2$, $A_3$, $\lambda_3$, $\theta_3$, $L_r$ and $\Omega$ for each 14 day segment of data.  The fitted value of these variables is such that the $\chi^2$ value, defined as
\begin{equation}
\chi^2 = \frac{1}{M}\sum_{i=1}^M \frac{(L(t_i) - L_{fit}(t_i))^2}{\sigma^2}
\end{equation}
where $L$ is the observed luminosity, $L_{fit}$ is the fitted luminosity and where $\sigma = 2\times 10^{-5}L_0$, is minimised.

In addition to fitting these eleven variables, there are a number of constraints imposed by the physics of the problem on the values the  variables can take, in particular with respect to starspot areas and the rotation rate.   The values of the starspot areas were constrained to be positive and below a threshold size.  \citet{Lanzaetal2003} calculated this threshold size to be $8.2\times 10^{-4}$ by assuming that the largest dip in the data was due to the rotational modulation of a single active region. However, as the span of SOHO data used for this thesis is longer than that used by \citet{Lanzaetal2003}, the upper limit derived may not be able to explain all the behaviour in the light curve analysed. In particular there is a region of apparently high sunspot activity occurring between 15/10/2003 and 11/11/2003 containing dips which correspond to spots with a relative area of $1.7\times 10^{-3}$.  Consequently, the upper limit of $8.2\times 10^{-4}$ on starspot area was used for all 14 day blocks except for the two blocks between 15/10/2003 and 11/11/2003, where the limit $1.7\times 10^{-3}$ was used.  In addition to starspot area, solar rotation rate is an observationally constrained quantity.  For this work, the limits of 23.0 days to 33.5 days on the solar rotation period, derived by \citet{Lanzaetal2003}, were used without modification.

For this thesis, the simplex method described by \citet{Lagariasetal1998}, implemented using the multidimensional Matlab fitting function \texttt{fminsearch} was used to complete this constrained minimisation.   This approach required that starting values and a function for calculating $\chi^2$ were provided to the method.  The starting values for each fit were either taken from the previous fit, or manually estimated.  As suggested by \citet{Lanzaetal2003}, the variables,  $A_1$, $\lambda_1$, $\theta_1$, $A_2$, $\lambda_2$, $\theta_2$, $A_3$, $\lambda_3$, $\theta_3$ and $L_r$ were fitted independently to $\Omega$, which was then optimised.  In addition, to ensure that the starspot areas remained within the physical bounds, the function which returned the $\chi^2$ value was modified such that it returned a very large value if active region areas became too large or negative.  Unfortunately, while these inputs resulted in a set of acceptable values for the fitting parameters being returned, issues with the simplex method itself, such as the fact that it may not converge \citep[e.g.][]{Mckinnon1998}, required that a number of tests were conducted to ensure that the true minimum had been found.

To ensure that the fit had converged, three checks were made.  First, each fit was checked by eye.  Second, for each block of data, the fitting procedure was repeated, that is the output of the preceding fit was used as the starting variables for the next fit, until the fitted variables remained approximately the same between two successive fits.  Third, the dependence of $\chi^2$ on each of the eleven fitting parameters was checked for 29 randomly selected blocks.  All curves inspected indicated that for these blocks the simplex method had converged to the $\chi^2$ minimum.

Once the fitting had been performed, the best fit parameters were recorded in a text file.  These parameters were in turn used to produce a model light curve.  By subtracting the model from the data and then adding the average intensity, a detrended light curve was constructed.

\subsection{Derivation of $\epsilon_j$}

This detrended data was then subjected to the same procedure as that of the unfiltered red noise (see section~\ref{Transit_Noise_Red}) to determine $\sigma_\epsilon$ as a function of transit area, transit duration and exposure time.  The results are presented in table~\ref{FilteredNoiseepssun} and figure~\ref{FilteredNoiseCompare}.  Again, as the data for 3 and 30 minutes are effectively equal, only the three minute data was used.  Fitting a line to these data points gives
\begin{equation}
\sigma_{\sun} = -2.16 \times 10^{-3} \text{s} + 1.21 \times 10^{-3} \text{s hr}^{-1} T_{obs}. \label{TraM-Noi-Filt-Dfit}
\end{equation}
Substituting this into equation~\eqref{red_epsdef} and recasting into physical variables gives
\begin{multline}
\sigma_\epsilon = 53.2\text{s} \left[\beta\right] \left[\frac{100(A_p + A_m)}{L_{0}N_{tra}}\right]^{-1} \\ \times\left[ \left(\frac{T_{tra}}{13 \text{hr}}\right)^{-1} \left(-8 \times 10^{-3} \left(\frac{T_{obs}}{24 \text{hr}}\right) + 1.008 \left(\frac{T_{obs}}{24 \text{hr}}\right)^2 \right)\right]. \label{transit_noise_filt_sigdefphys}
\end{multline}

As with the previous two types of noise, the behaviour of $\epsilon_j$ for the case of filtered photometric noise will be discussed in the context of equation~\eqref{transit_noise_filt_sigdefphys}.

\begin{table}[tb]
\begin{center}
   \begin{tabular}{lll}
   \hline
 $T_{obs}$ &   $\Delta t$ = 3 mins         &  $\Delta t$ =  30 mins   \\
   \hline
   30 mins  & $9.4764  \times 10^{-3}$ s    & --  \\
   1 hr         & $1.446 \times 10^{-2}$ s     & $1.246\times10^{-2}$ s   \\
   2 hr        	& $2.255 \times 10^{-2}$ s      & $2.179\times10^{-2}$ s  \\
   4 hr        	& $3.77 \times 10^{-2}$ s        & $3.78\times10^{-2}$ s    \\
   8 hr        & $7.52 \times 10^{-2}$ s        & $7.38\times10^{-2}$ s \\
   12 hr      & $1.16 \times 10^{-1}$ s        & $1.16\times10^{-1}$ s \\
   16 hr      & $1.64 \times 10^{-1}$ s       & $1.64\times10^{-1}$ s  \\
   24 hr      & $2.7 \times 10^{-1}$ s         & $2.7\times10^{-1}$ s         \\
   36 hr      & $4.2 \times 10^{-1}$ s         & $4.2\times10^{-1}$ s \\
   \end{tabular}\\
 \caption{The size of $\sigma_{\sun}$ as a function of length of observation window, $T_{obs}$, and exposure time $\Delta t$.  The value of $\sigma_{\sun}$ is recorded to the last significant figure.}
 \label{FilteredNoiseepssun}
 \end{center}
 \end{table}

\begin{figure}
     \centering
     \subfigure[$T_{obs}$ = 1hr.]{
          \label{fig:dl2858}
          \includegraphics[width=.48\textwidth]{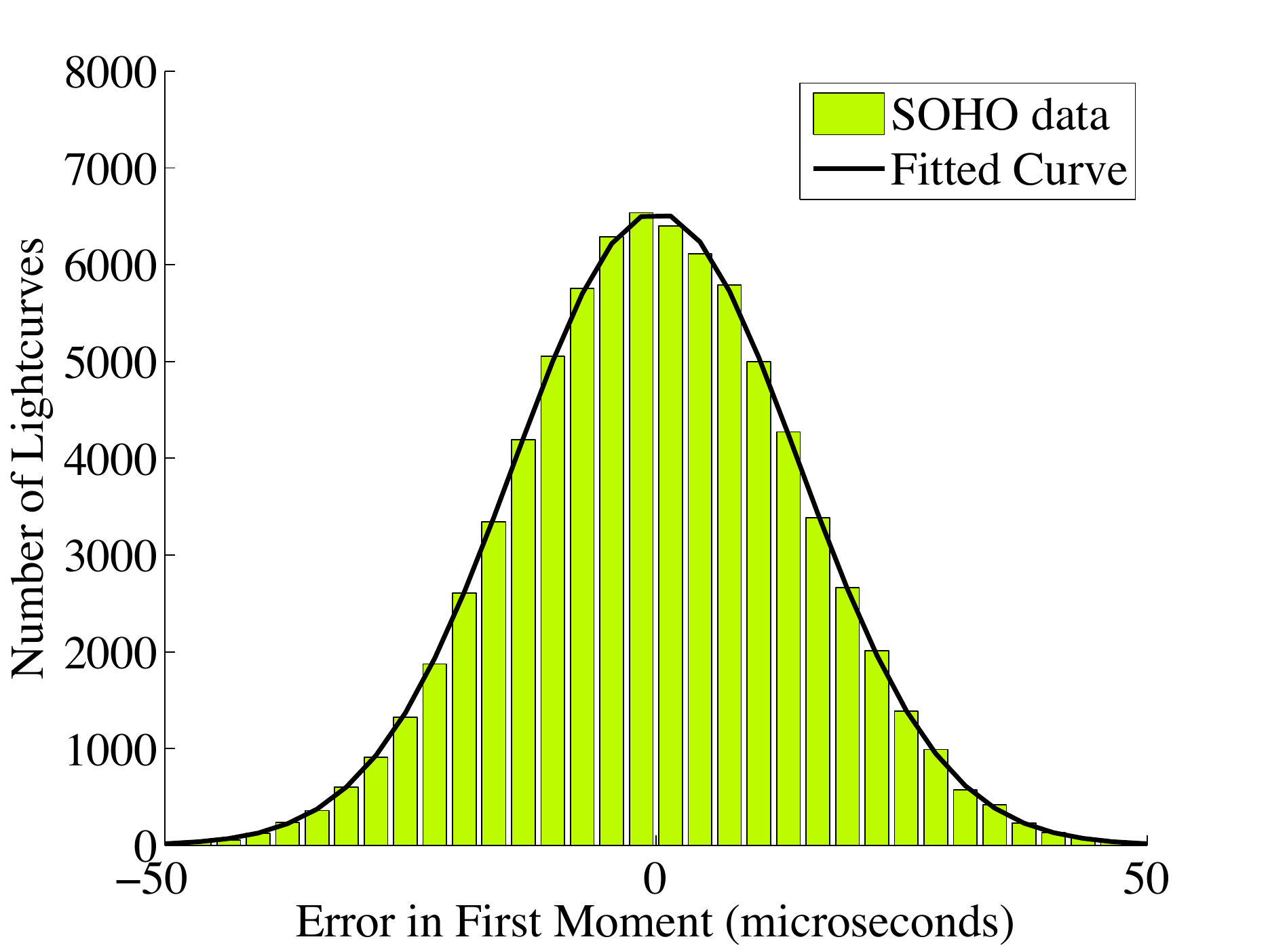}}
     \subfigure[$T_{obs}$ = 2hr.]{
          \label{fig:er2858}
          \includegraphics[width=.48\textwidth]{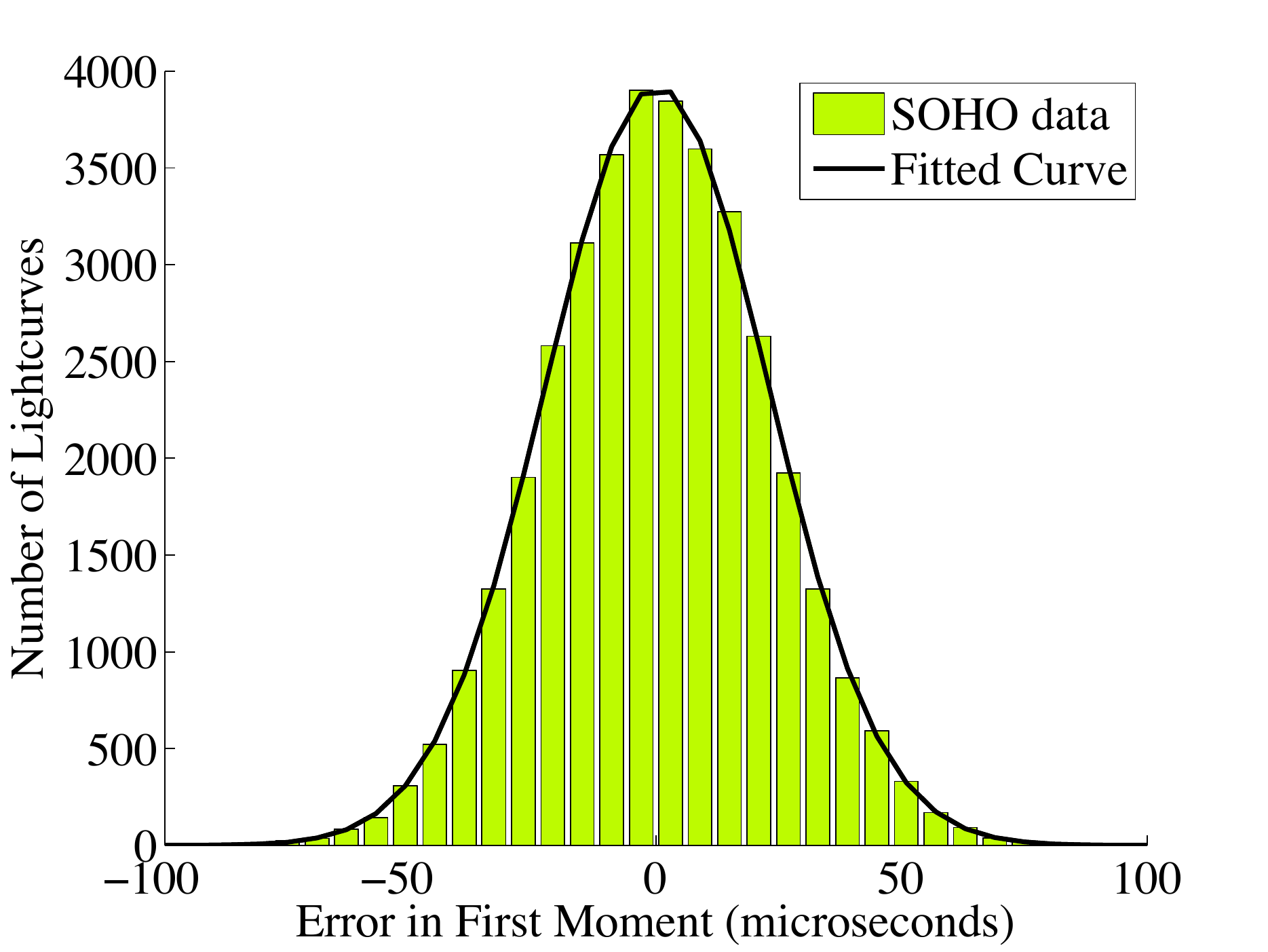}}\\
          \subfigure[$T_{obs}$ = 4hr.]{
          \label{fig:dl2858}
          \includegraphics[width=.48\textwidth]{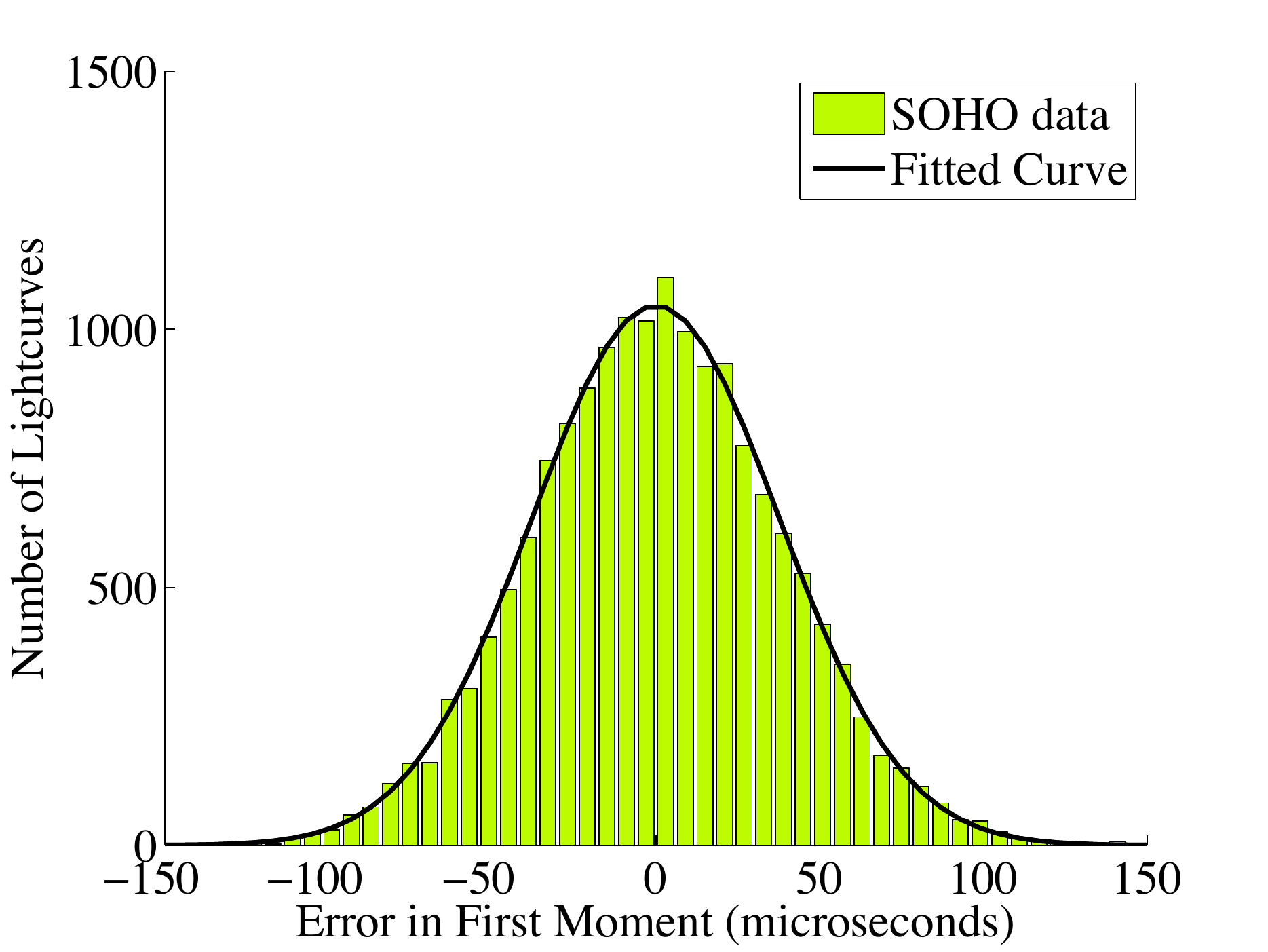}}
     \subfigure[$T_{obs}$ = 8hr.]{
          \label{fig:er2858}
          \includegraphics[width=.48\textwidth]{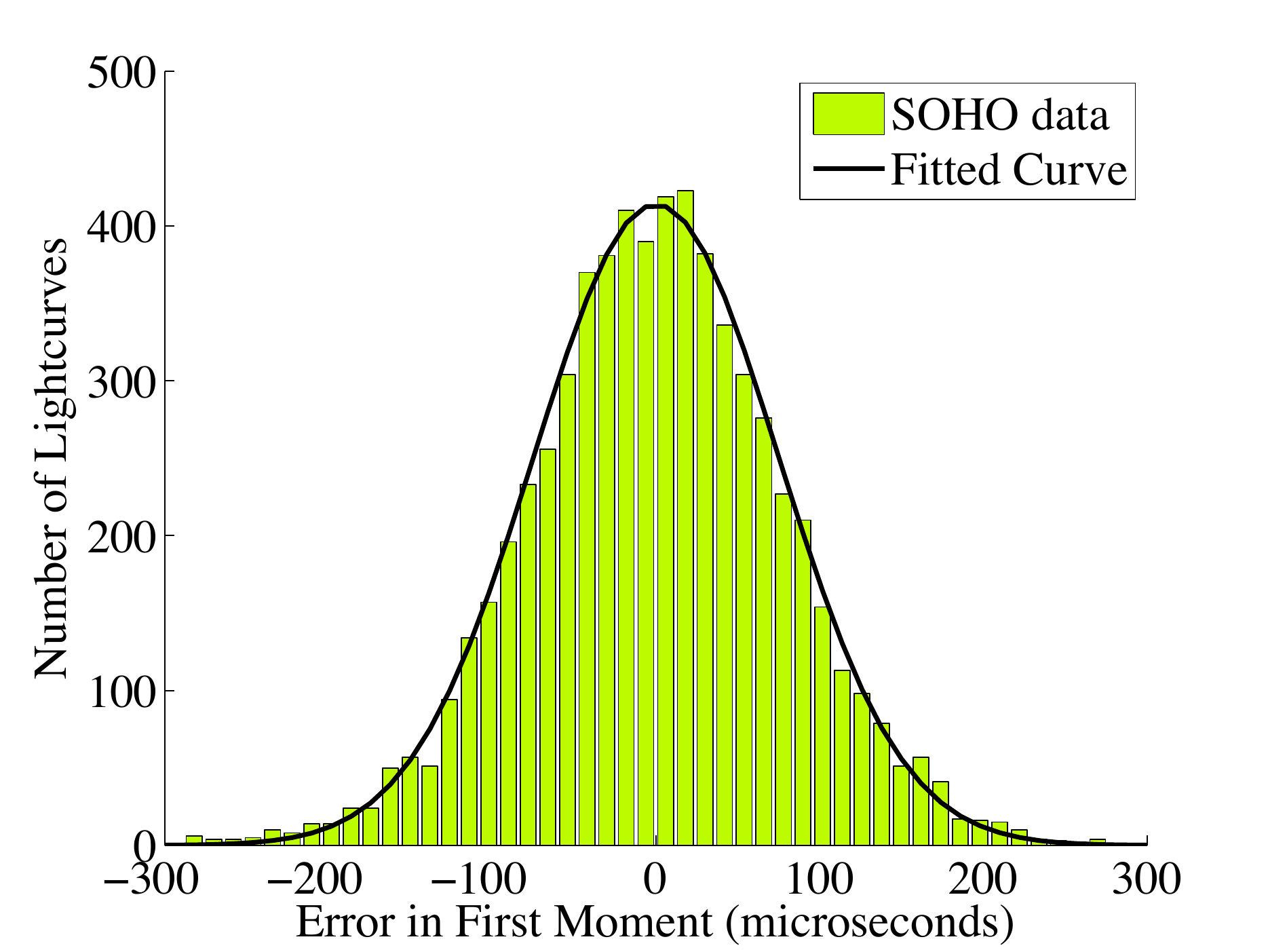}}\\
           \subfigure[$T_{obs}$ = 12hr.]{
           \label{fig:cminusscalar2858}
           \includegraphics[width=.48\textwidth]{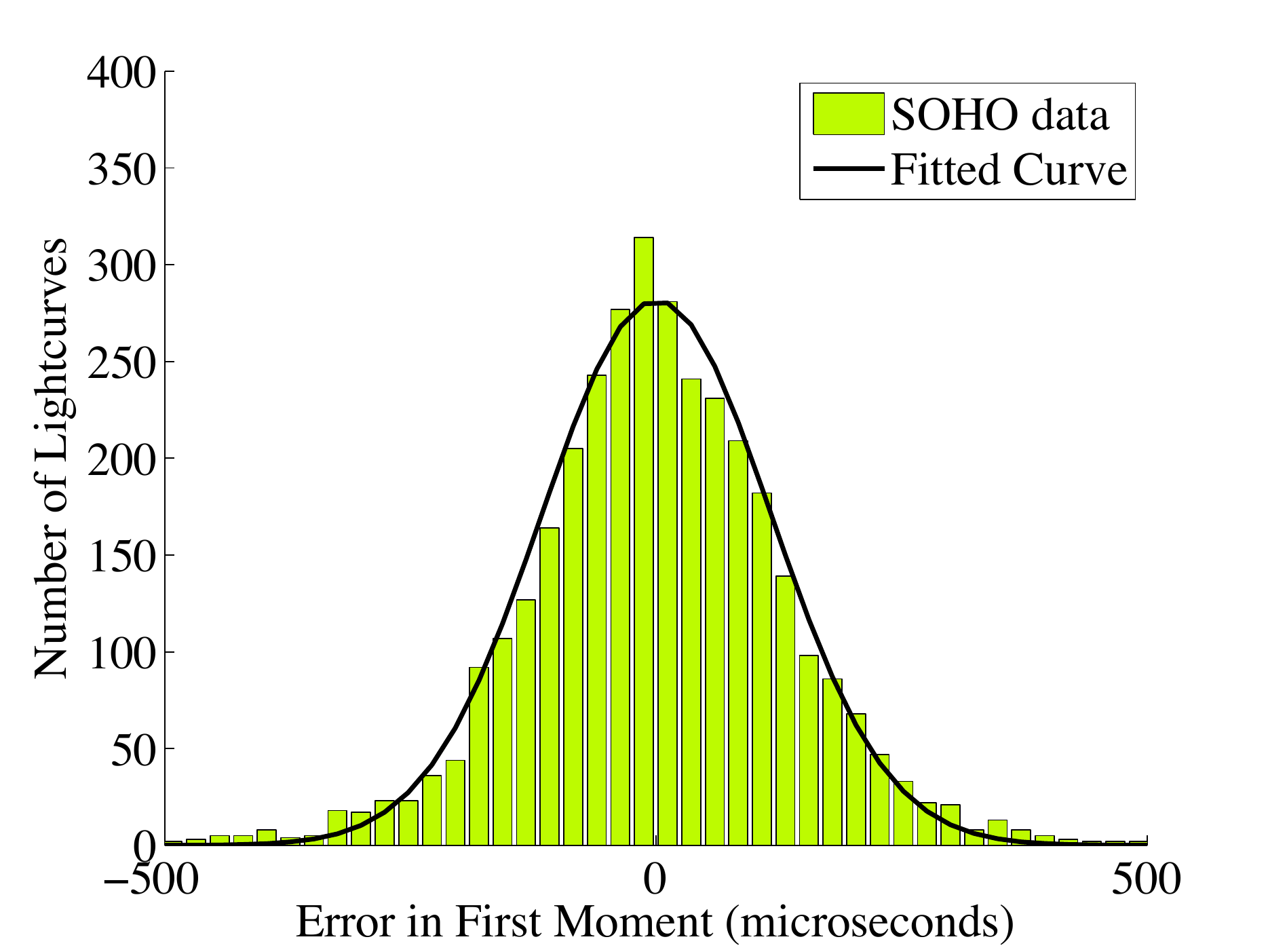}}
           \subfigure[$T_{obs}$ = 24hr.]{
           \label{fig:cminusscalar2858}
           \includegraphics[width=.48\textwidth]{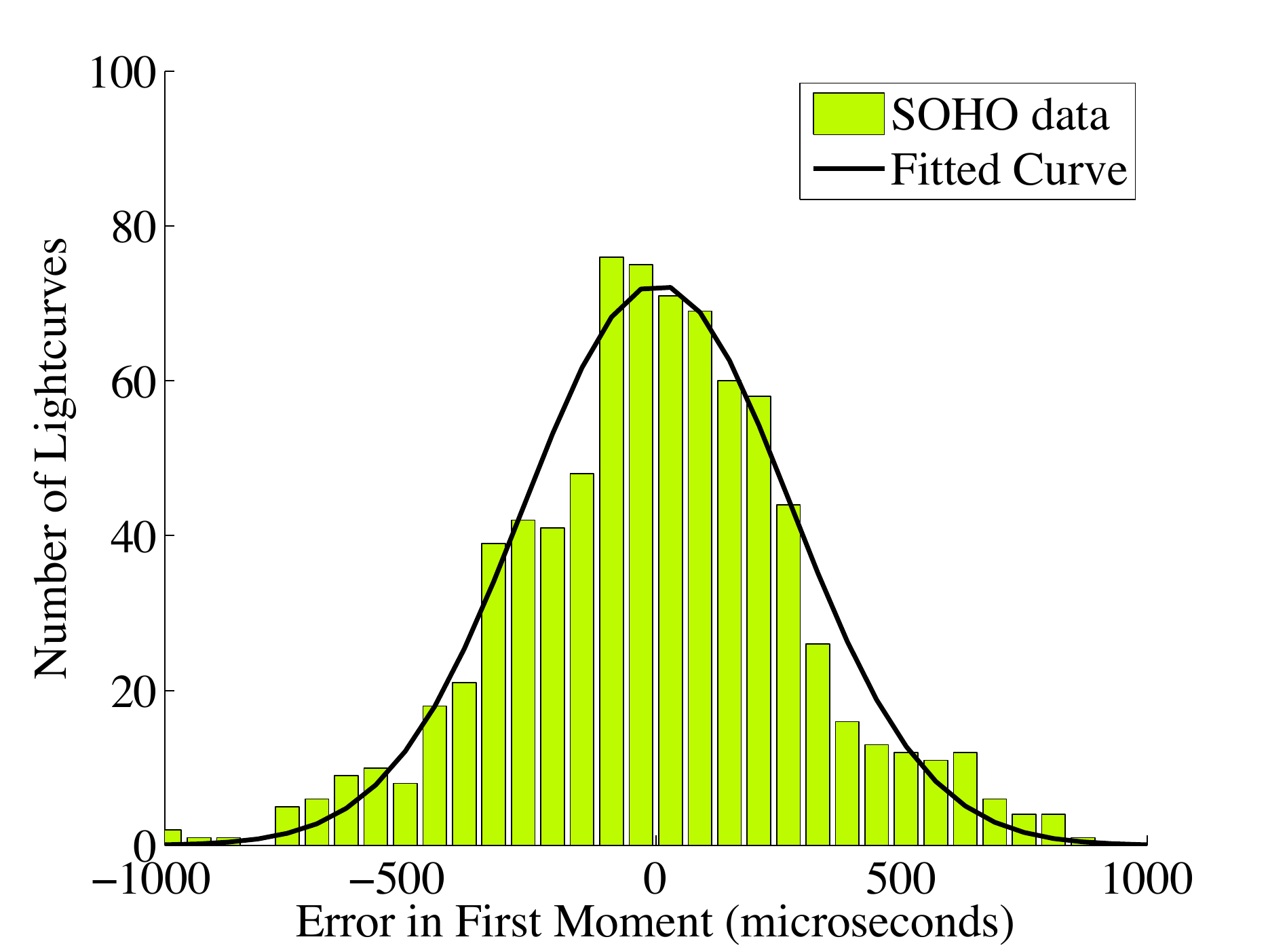}}
     \caption[The observationally determined distribution of $\epsilon_{\sun}$ (green bar) for the case of filtered solar photometric noise for six different length observing windows.]{The observationally determined distribution of $\epsilon_{\sun}$ (green bar) for the case of filtered solar photometric noise for six different length observing windows.  Note that the distribution of $\epsilon_{\sun}$ strongly represents a normal curve (black line).}
     \label{FiltDist}
\end{figure}

\begin{figure}[tb]
\begin{center}
\includegraphics[width=.90\textwidth]{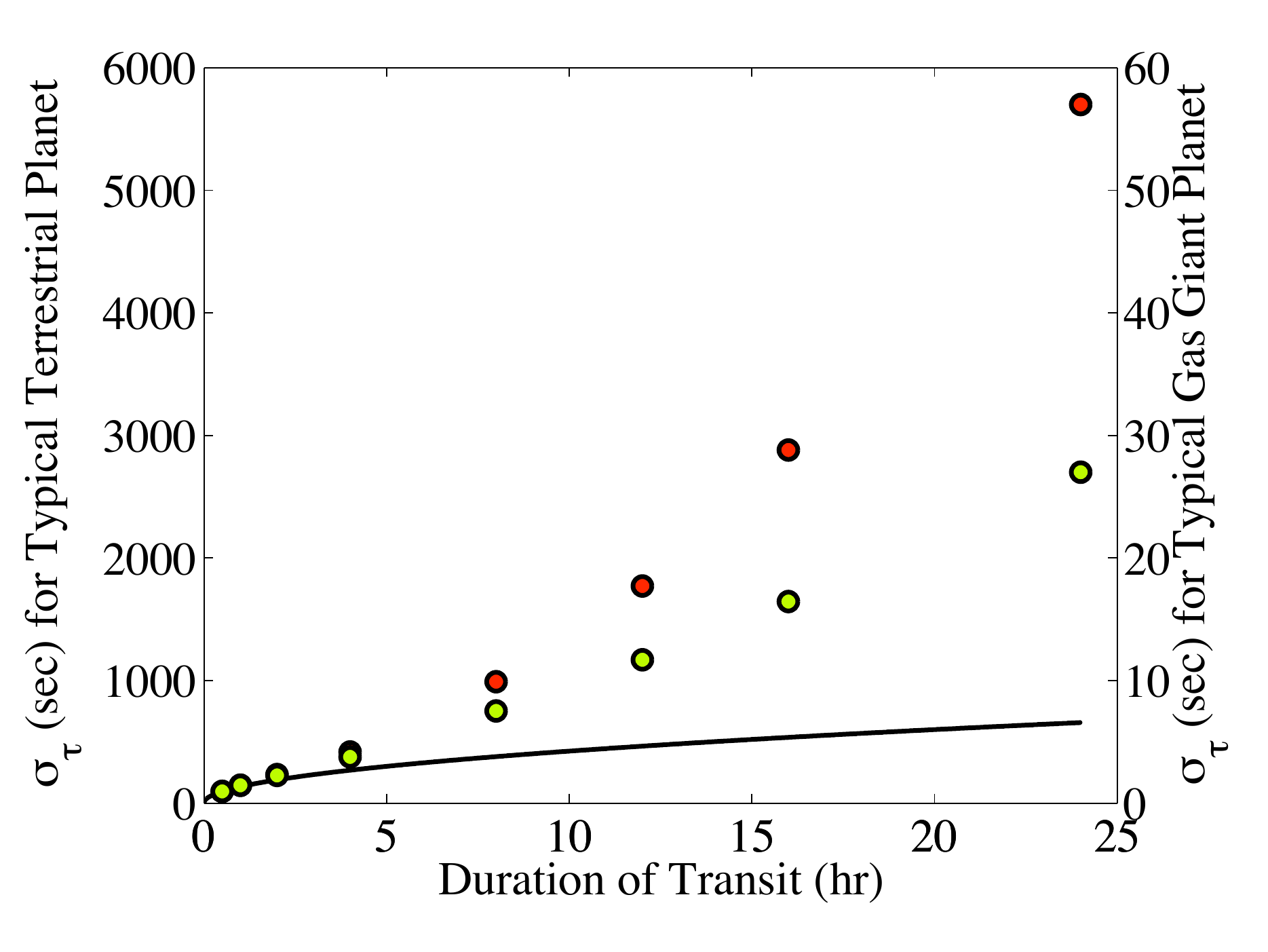}
\caption[Comparison between calculated errors in $\tau$ using solar light curves which have been ``filtered" using the method of \citet{Lanzaetal2003} (green dots) and theoretically predicted errors in $\tau$ using white noise with the same power as the noise in the solar light curves (thick lines).]{Comparison between calculated errors in $\tau$ using solar light curves which have been ``filtered" using the method of \citet{Lanzaetal2003} (green dots) and theoretically predicted errors in $\tau$ using white noise with the same power as the noise in the solar light curves (thick lines).  In addition the errors in $\tau$ for the case of unfiltered noise are also plotted for comparison (red dots).  These relations are shown for the case of the transit of a gas giant ($(A_p + A_m)/L_0 N_{tra} = 10^{-2}$) and a terrestrial planet ($(A_p + A_m)/L_0 N_{tra} = 10^{-4}$), assuming $T_{obs} \approx 2 T_{tra}$.}
\label{FilteredNoiseCompare}
\end{center}
\end{figure}

\subsection{Properties of $\epsilon_j$}

\begin{figure}[tb]
\begin{center}
\includegraphics[height=3.0in,width=4.0in]{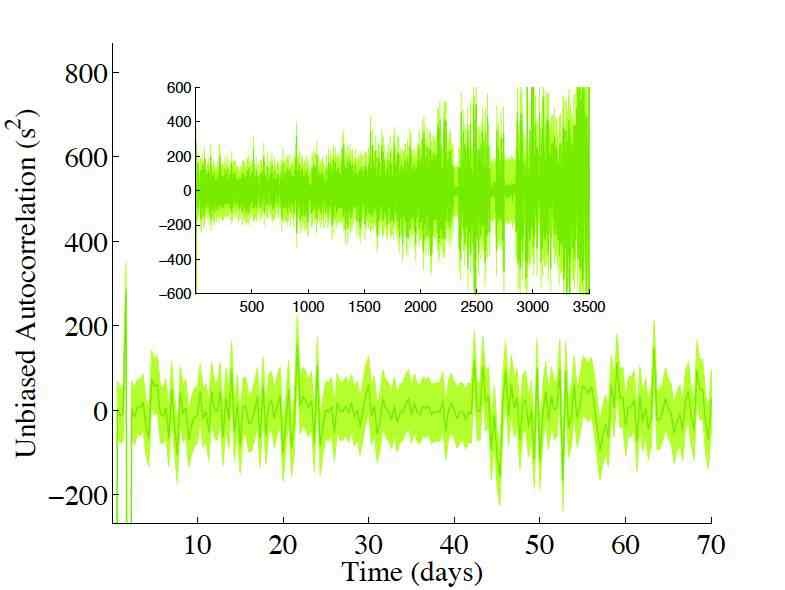}
\caption[Normalised autocorrelation function of $\epsilon_j$ (green line) calculated using filtered stellar noise.]{Normalised autocorrelation function of $\epsilon_j$ (green line) calculated using filtered stellar noise for the case where the observing window is 8 hours.  For reference, the one sigma error bars (light green) are also shown.  As short period planets are more likely to be discovered by the transit technique, the inner section of the autocorrelation function is shown in the main plot.  For completeness, the full autocorrelation function is shown in the inset.}
\label{FilteredNoiseCorr}
\end{center}
\end{figure}

The effect of filtered photometric noise on $\epsilon_j$ will be discussed with respect to the terms in equation~\eqref{transit_noise_filt_sigdefphys} and with respect to $\Delta \tau$ and $j$.  In particular, these results will be compared to the case of unfiltered noise.

Again, from the first two terms, equation~\eqref{transit_noise_filt_sigdefphys} depends linearly on $\beta$ and is inversely proportional to $A_p + A_m$.  As these dependancies have been found for the case of white and realistic stellar noise, they will not be further discussed.  However, the third term shows some difference.

The difference between the cases of filtered and unfiltered realistic photometric noise become apparent in the third term of equation~\eqref{transit_noise_filt_sigdefphys}, which describes the behaviour of $\epsilon_j$ as a function of 
 planet-star and planet-moon distance through $T_{tra}$ and $T_{obs}$.  Noting that the $T_{obs}/24$hr term in the Taylor expansion is dominant, $\sigma_\epsilon$ is proportional to $T_{tra}$.  As the signal amplitude is also proportional to $T_{tra}$, this means to first order, the detection threshold does not depend on the transit duration (and thus the planet-star distance) for the case  where the light curve is contaminated with filtered realistic solar photometric noise.  Similarly, the amplitude of $\epsilon_j$ is proportional to $T_{obs}^2$ for constant transit duration where the observing window is being altered to look for more distance moons of a given planet.  Again this physically corresponds to very close and very distant moons being undetectable.

Finally, as for the cases of white and realistic photometric noise, the dependance of $\epsilon_j$ on $\Delta \tau$ was negligible.  In addition, the autocorrelation (see figure~\ref{FilteredNoiseCorr}) was again calculated, and it was found that $\epsilon_j$ is again effectively uncorrelated from transit to transit.

Now that we have investigated the properties of $\epsilon_j$ for the case of filtered noise, we will summarise the behaviour of $\epsilon_j$ resulting from the different types of noise before continuing with the analysis.

\section{Conclusion}\label{Transit_Noise_Conclusion}

As discussed, $\epsilon_j$ values resulting from white, realistic and filtered realistic photometric noise show quite similar behaviour in some ways, for example, their dependance on $A_p + A_m$, but quite different behaviour in others, for example, their dependance on $T_{tra}$.  These differences in behaviour will be summarised in turn, with the aim of highlighting the results required for later work.

In general, the amplitude, form or behaviour of $\epsilon_j$ does not depend on exposure time.  The only case where it does (among the three noise sources investigated) is for white noise processes which are dependent on the behaviour on the instrument, e.g. read noise.  For these cases, the exposure time should be optimised with respect to instrumental specifications and not the physics of the planet-moon system.

For all three types of noise, the amplitudes of $\epsilon_j$ were found to scale linearly with the amplitude of the noise, while the proportionality constant depended on the specific type of noise (white, realistic or realistic filtered) and the physics of the system. As this is a property of equation~\eqref{red_epsdef}, it should hold true for all types of photometric noise.

The behaviour with respect to $T_{tra}$ and $T_{obs}$ was found to vary significantly with noise type.  For the case where the planet-star distance was altered without affecting the transit geometry, that is, altering $T_{tra}$, the detectability of moons increased with, decreased with and was independent of planet-star distance for the case of white noise, realistic stellar noise and filtered realistic noise respectively.  For the case where the planet-moon distance was altered (that is, $T_{obs}$ changes but $T_{tra}$ does not) all these three  noise types show similar behaviour, in that there is an optimum planet-moon distance for moon detection, and moons that are closer or further away than this are less detectable.  In particular, this optimal distance is largest for white noise, smallest for red noise and intermediate for filtered noise.  The exact position will depend on the star and the transit geometry, an issue that is further investigated in the next chapter. 

Finally, for realistic planet-moon systems, the distribution of $\epsilon_j$ can be well approximated by a normal distribution for the case of white and filtered realistic photometric noise.  For the case of realistic photometric noise, the distribution of $\epsilon_j$ is well approximated by a normal distribution for observing window lengths shorter than about 12 hours, but becomes slightly non-normal for the case of observing windows longer than about 12 hours.  In addition, the values of $\epsilon_j$ corresponding to a sequence of transits are uncorrelated, and only very weakly depend on the values of $\Delta \tau$ corresponding to the that transit.  These three very useful statistical properties will be exploited in the next chapter.

Now that the effect of three physically sensible noise sources have been investigated, we can now combine the results from chapter~\ref{Transit_Signal} with those of this chapter to produce detection thresholds.
\chapter[Thresholds]{Detection Threshold for Moons of Transiting Planets}\label{Trans_Thresholds}

\section{Introduction} 

Now that $\Delta \tau$, the form of the TTV$_p$ perturbation has been derived, and the behaviour of $\epsilon_j$, the timing noise, has been explored, we are finally in a position to determine the set of moons that can be detected using the TTV$_p$ technique.  We begin by first motivating and defining the set of planetary systems that will be explored in this chapter.  Then the method used to determine these thresholds will be discussed and the mathematics required to calculate these thresholds introduced and defined.  Then using these formulae, the general behaviour of the detection threshold will be explored using simplifications that occur when the number of transits is large.  This analysis will then in turn be used to perform a comparison between the detection threshold calculated using the TTV$_p$ technique and the detection thresholds corresponding to the three other transit moon detection methods calculated and discussed in section~\ref{Intro_Dect_Moons_Transit}.  Finally the TTV$_p$ moon detection threshold will be numerically calculated for realistic systems to investigate the effect of inclination, eccentricity and orientation of the planet's orbit for the case where the number of observed transits is finite.  However, to perform such an investigation, the region of parameter space to be investigated, must be defined.  With this in mind, we begin with a discussion of the region of parameter space to be explored.

\section{Region of parameter space considered}\label{Trans_Thresholds_Parameters}

While it would be nice to have a simple two-dimensional plot which describes moon detection in terms of all moon detection behaviour, this is not possible, as the detection thresholds depend on many more than two parameters.  In particular, they depend on the physical and orbital parameters of the star, planet and moon in question.  Thus, the variables to be investigated have to be selected very carefully to make best use of this wealth of information.  Consequently we will discuss the selection of variables relating to moon detection with respect to the star, planet and moon in turn.

Transiting planets have been found around stars with radii ranging from 0.2$R_{\sun}$ to 2.1$R_{\sun}$.  However, while the physical parameters of stars along the main sequence can be easily derived (e.g. $R_s \propto M_s$), the behaviour of the photometric noise of stars as a function of their position on the main sequence is not a simple function of their physical parameters.\footnote{\citet{Aigrainetal2004} proposed a method for predicting the inherent photometric noise spectrum main sequence stars of any spectral type.  However, in order to construct this method, they had to assume that some of the noise parameters of stars in general were the same as that of the Sun (as the parameters were poorly constrained).}  As the analysis conducted in chapter~\ref{Trans_TTV_Noise} into realistic and filtered photometric noise only applies to the Sun, it can only be extended to other solar-like host stars.  Consequently, in this work it was decided to investigate the case of a 12$^{th}$ magnitude Sun-like host star with photometric noise which is either dominated by white, realistic solar or filtered solar photometric noise, with amplitude defined by the Kepler reference case.\footnote{A relative photometric precision of $2\times10^{-5}$ over a 6.5 hour exposure (e.g. Borucki et al., 2003).  For the case of white noise this translates to a relative photometric precision of $7.2 \times 10^{-5}$ for a thirty minute exposure.} In particular, for the case of realistic solar or filtered solar photometric noise, this involves selecting a scaling factor $\beta = 1.9$.

The planet can also influence the moon detection thresholds through inherent properties such as mass and radius, and through properties of its orbit. As discussed in chapter~\ref{Trans_TTV_Noise}, planetary radius does not strongly affect detection thresholds as both the amplitude of $\Delta \tau$ and $\epsilon_j$ are inversely proportional to $\hat{A}_p + \hat{A}_m$.  Consequently, for this work, moon detection will not be investigated with respect to the radius of the planet.  The mass of the planet is a different story.  While planetary mass does not directly affect either the amplitude of $\Delta \tau$ or the characteristic size of $\epsilon_j$, it is important as it alters the  position where the assumption that $v_m \ll v_{tr}$ (required for deriving the expression for $\Delta \tau$) breaks down, and as predictions from moon stability and formation theory depend on planetary mass.  As these quantities are useful for comparison purposes, the effect of the mass of the planet will be investigated.  The orbital parameters of the planet's orbit can also affect the detection threshold, in particular $a_p$, $I_p$ and $e_p$.  $a_p$ alters moon detection both by altering the transit duration through altering $v_{tr}$, but also by altering $N$, the total number of transits that will be observed (as the orbital period of the planet depends on $a_p$).  Consequently, the effect of $a_p$ on the thresholds will be investigated.  In addition, $e_p$ and $I_p$ alter moon detection by altering the transit duration by altering $v_{tr}$ and the chord length respectively, and their effect on the detection threshold will be investigated specifically in sections~\ref{Trans_Thresholds_Thresh_Inclined} and \ref{Trans_Thresholds_Thresh_Eccentric}.  Consequently, planets will be investigated in terms of $M_p$ and $a_p$.  In particular, the values of $M_p$ and $a_p$ that will be investigated will be $10 M_J$, $1 M_J$, $1 M_U$ and $1 M_{\earth}$, to cover the full range of available planets, and $a_p = 0.2$AU (the closest and thus the most detectable planets that can host moons), 0.4AU and 0.6AU (the most distant planets that can host moons that are detectable by TTV$_p$ using Kepler).

Finally, the moon itself can (unsurprisingly) affect its detection threshold through its physical properties and its orbital parameters. The physical properties a moon can possess include its size (parameterised by its radius), and its mass.  As detection thresholds depend predominantly on moon size and not mass (see section~\ref{Trans_TTV_Signal_CC_Form}), moon detection will be investigated in terms of moon radius.  While moon radius cannot be simply determined from $A_m$, it can be approximated.  \citet{Sartorettietal1999} looked at the error in assuming that $\hat{A}_m/L_0 N_{tra} = (R_m/R_s)^2$ and found that it was small for all but the most extreme values of $I_p$.  Consequently this approximation will also be used for this work.  Also, as some of the moon formation and stability limits are formulated in terms of a moon mass as opposed to a radius e.g. the mass limit proposed by \citet{Canupetal2006} (see section~\ref{Intro_Moons_Form_Disk}), it would be useful to have a conversion factor (i.e. a density) so that mass limits can be written in terms of moon radii.  As all models are well inside the snow line, it was decided to use a density of 3000 kgm$^{-3}$.  For comparison, the moon's density is 3346 kgm$^{-3}$ and Callisto's is 1860 kgm$^{-3}$.  The orbital parameters of the moon, in particular, $a_m$, $e_m$, $I_m$ and $f_m(0) + \omega_m$, may also affect moon detectability.  As this is a preliminary investigation, we will not be looking at the effect of $e_m$ and $I_m$ on the TTV$_p$ moon detection threshold, and only consider moons on circular orbits which are aligned to the line-of-of sight.  In addition, the effects\footnote{\label{phifootnote}In addition to the non-detection spikes associated with $a_m$ there are also non-detection spikes associated with $\phi = f_m(0) + \omega_m$.  For example, consider the case where $\omega = \pi$ and $\phi = 0$.  From equation~\eqref{transit_signal_conc_form} we have that $\Delta \tau$ is given by the sequence $A$, $-A$, $A$, $-A$ $\dots$, a signal which could be detected.  Contrast this with the case where $\omega = \pi$ and $\phi = \pi/2$.  For this case $\Delta \tau$ is given by the sequence $0$, $0$, $0$, $0$ $\dots$, a signal which is not detectable.} of $f_m(0) + \omega_m$ on the detection threshold will be side-stepped in this work as first, the shape of the detection threshold becomes independent of $f_m(0) + \omega_m$ as $N\to \infty$ and second, as $\phi = f_m(0) + \omega_m$ will be randomly selected for the Monte Carlo simulation conducted in section~\ref{Trans_Thresholds_MC}.  Consequently we will also investigate moon detectability in terms of $a_m$.

 So in summary, the detection thresholds will be investigated in terms of $M_p$, $a_p$, $R_m$ and $a_m$.  As moons will be detected for a given planet, the thresholds will be displayed in a three by four grid of planetary mass and semi-major axes.  Thus each of the twelve plots represents a host planet with a given value of $M_p$ and $a_p$.  The plots themselves show moon detection thresholds as a function of moon radius and semi-major axis.  Now that the range of parameter space to be investigated has been discussed, the method for constructing these threshold maps needs to be decided.
 
\section{Method}

In order to numerically determine moon detection thresholds using TTV$_p$, three main issues must be addressed.  First, in order to calculate the value of and error in $\tau$ for each transit, a particular observation window must be used.  As discussed in chapter~\ref{Trans_Intro}, for this thesis, we will use the smallest window which will always include the moon's transit.  As we are now in a position to be be able to discuss this choice, the selection of this particular window will be revisited.  Second, a method needs to be selected for determining the detection threshold.  Finally, these two choices must be mathematically implemented in order to provide expressions which can give the detection threshold of a given moon.  We begin with a discussion of the selection of the observation window.

\subsection{Selection of observation window}\label{Trans_Thresholds_Method_Window}

\begin{figure}
\centering   
          \includegraphics[width=.90\textwidth]{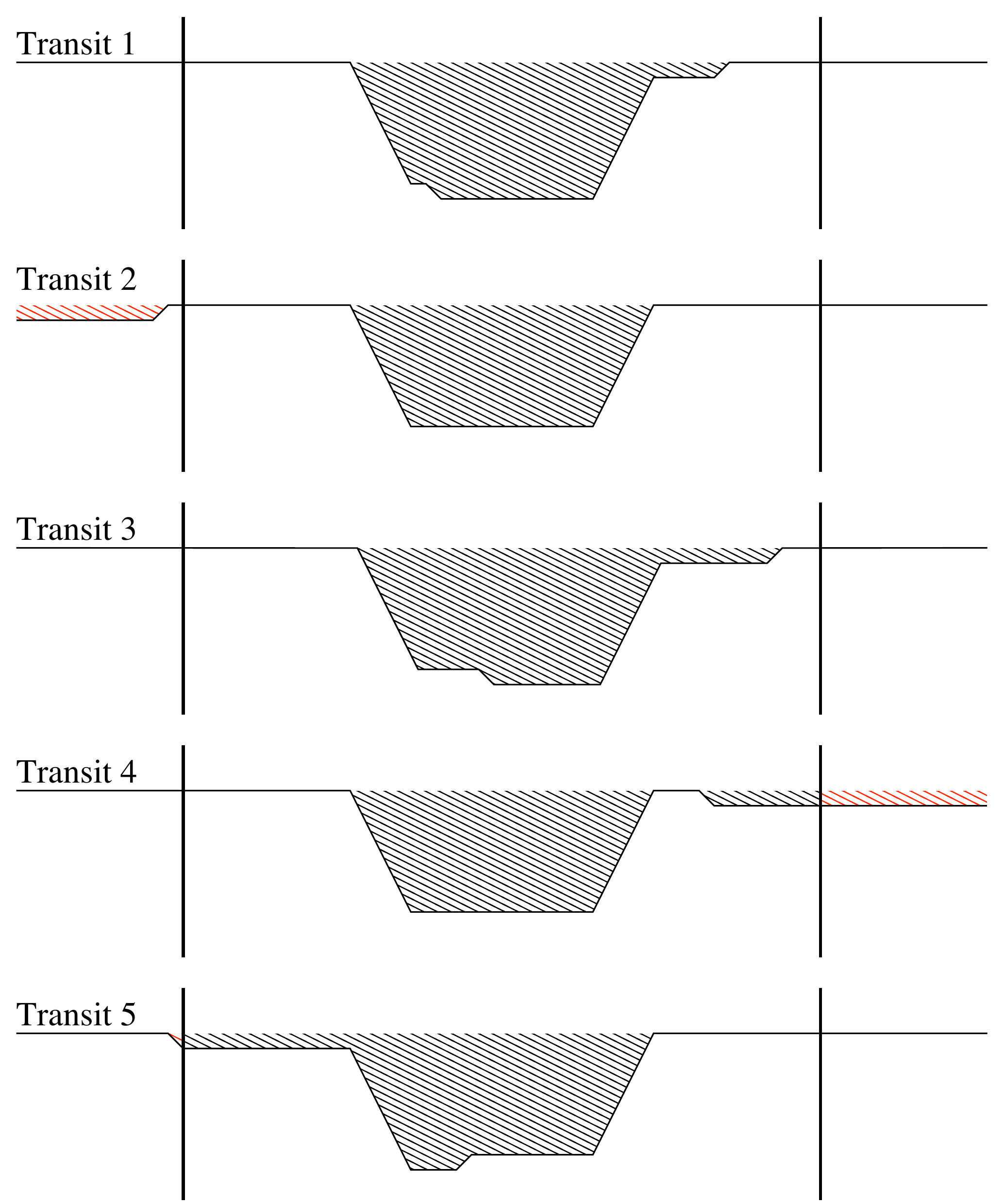}
     \caption[Cartoon of a sequence of five light curves caused by the transit of a planet-moon pair for the case where the observing window is too short to always contain the transit of the moon.]{Cartoon of a sequence of five light curves caused by the transit of a planet-moon pair for the case where the observing window is too short to always contain the transit of the moon.  The beginning and end of the observation window are denoted by thick vertical lines, and each light curve is centered on the planetary transit.  The region of the dip caused by the planet or moon inside the window and and consequently included in the sum used to calculate $\tau$, is crosshatched in black, while the region not included in the sum is crosshatched red.}
     \label{ShortObsWind}
\end{figure}

\begin{figure}
     \centering
     \subfigure[White noise.]{
          \label{WindowNoiseCorrWhite}
          \includegraphics[width=.48\textwidth]{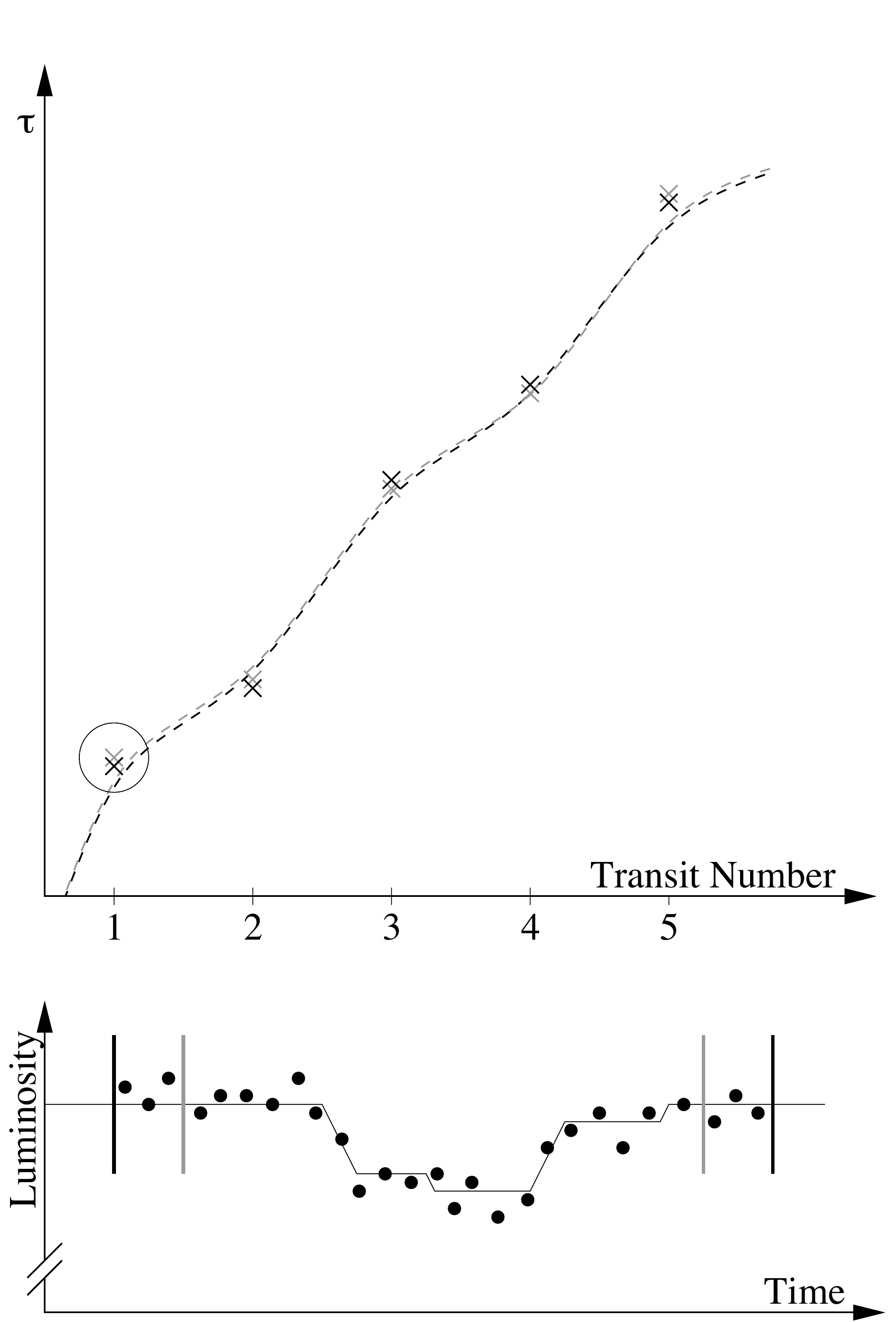}}
     \subfigure[Corellated noise.]{
          \label{WindowNoiseCorrRed}
          \includegraphics[width=.48\textwidth]{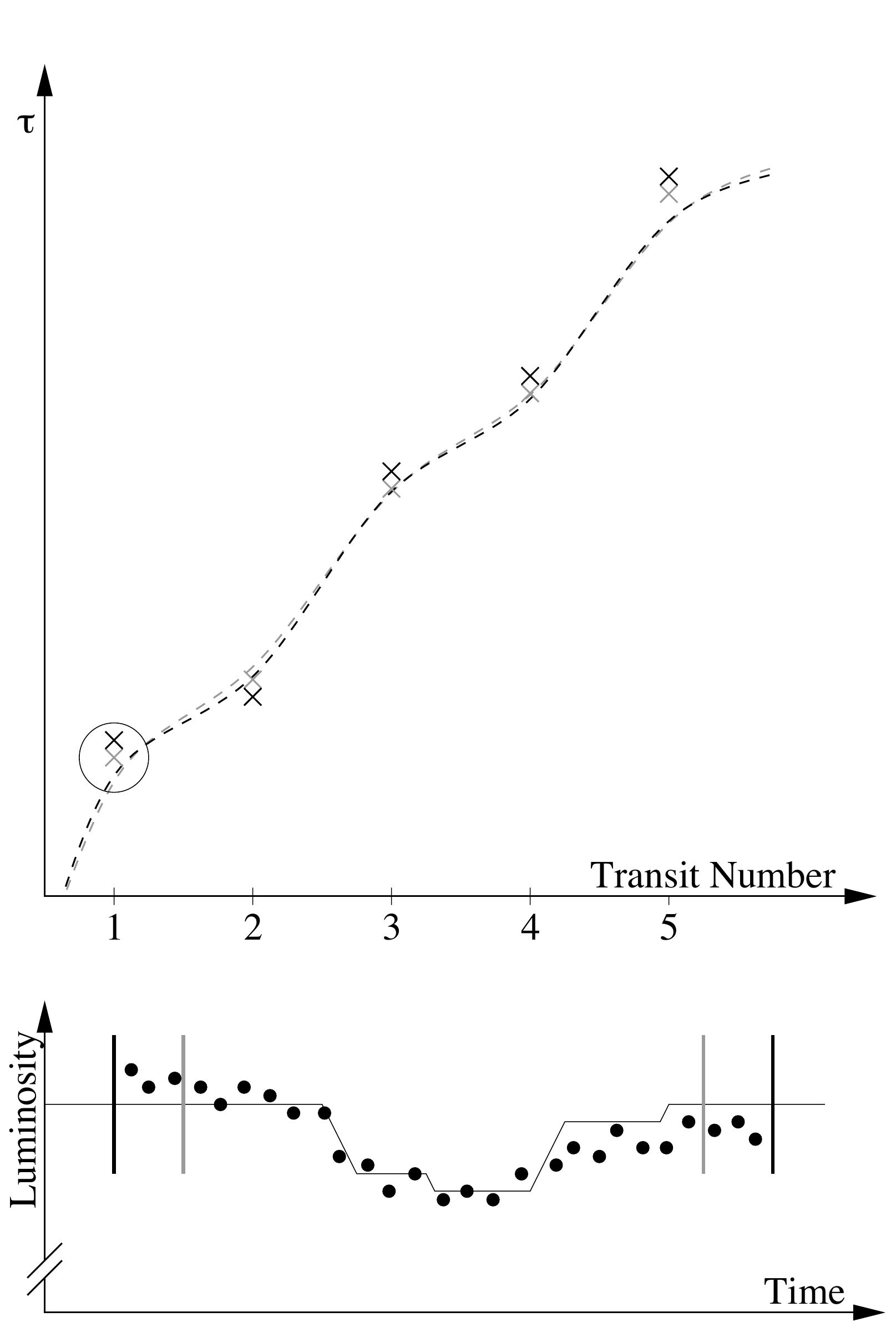}}
     \caption[Diagram demonstrating how the the measured values, best fit function and errors in $\tau$ change with observation window size and the type of photometric noise contaminating the light curve, for the cases of white and correlated photometric noise.]{Diagram demonstrating how the the measured values, best fit function and errors in $\tau$ change with observation window size and the type of photometric noise contaminating the light curve, for the cases of white and correlated photometric noise.  A typical light curve (black dots) is shown at the base of each figure along with a cartoon theoretical light curve for the case of no photometric noise (thin line).  In addition, on this light curve two concentric observing windows are indicated by bold vertical lines, the shorter inner window beginning at the grey line on the left and finishing at the grey line on the right, and a longer outer window beginning at the black line on the left and finishing at the black line on the right.  The diagram above shows measured $\tau$ values (crosses) as a function of transit number along with the best fit $\tau$ function (dashed line) for the case where the grey observing window and black observing window have been used to calculate $\tau$.  In particular, the transit light curve shown corresponds to the circled point on the diagram above, where the $\tau$ values calculated using the grey observing window and the black observing window are indicated by a grey and black cross respectively.}
     \label{WindowNoiseCorr}
\end{figure}

As discussed in the introduction to this Part, for this thesis, $\tau$ will be calculated using a section of light curve centered on the planetary transit and of length 
\begin{equation}
T_{obs} = T_{tra} + \frac{2 a_m(1 + e_m)}{v_{tr}}.  
\label{transit_thresholds_method_DobsDef}
\end{equation}
This particular length window was selected as it is the smallest window which is ensured to always contain the moons transit.  Now that the behaviour of the timing perturbation and the noise have been discussed, we are finally in a position to revisit this assumption, and discuss why it was made.

Realistically, one would like to select a concentric set of windows of differing sizes, check each one for signs of a moon, and then combine the results to give a statistical description of the types of moons that could or could not be detected.  Unfortunately, there are two problems with this.  First, if the window selected is too small, sections of the moon's transit may be neglected for some of the transits.  Second, if multiple windows are selected, the calculated values of $\tau$ are not statistically independent and it is difficult to construct thresholds.  These issues will be discussed in turn.

For the case where the window selected does not always include the all of moon's transit (see figure~\ref{ShortObsWind}), problems can arise.  In particular, these problems occur for transits where a section of the moon's transit is not included e.g. transits 2, 4 and 5 of figure~\ref{ShortObsWind}.  For these cases, the approximate expressions for $\Delta \tau$ derived in chapter~\ref{Transit_Signal} are no longer correct and can no longer be used to determine thresholds.  In addition this problem is fairly insidious as it only acts on the transits with large planet moon separations (and thus large $\Delta \tau$) and always acts to reduce the size of $\Delta \tau$.  Unfortunately, there are also issues with using a concentric set of large windows to try and find a window with the correct size.

To begin, we consider the effect on the measured values of $\tau$, and consequently on the derived thresholds, of moving from an observation window, centered on the planetary transit and long enough to always include the moon's transit to a longer observation window which is also centered on the planetary transit.  To provide a context for this discussion, consider the two parts of figure~\ref{WindowNoiseCorr}.  Figure~\ref{WindowNoiseCorrWhite} shows a typical sequence of $\tau$ values as a function of transit number for the case of two different length observations windows (black and grey).  In addition, the best fit $\tau$ function and an example transit light curve containing white noise are also shown for the two cases.  Figure~\ref{WindowNoiseCorrRed} shows the same thing but for red noise.  The discussion will be conducted using these two examples

Beginning with the case of white noise, consider figure~\ref{WindowNoiseCorrWhite}, and in particular consider the circled point which corresponds to the example transit light curve.   Moving from using the shorter observing window (grey) to the longer window (black) to determine $\tau$, involves using all the points in the grey window, but also including the six extra data points (three from the left and three from the right), along with the additional noise associated with them.  Importantly, as the sum in the black window also includes all the points in the grey window, the error in $\tau$ for both windows will not be independent.\footnote{In particular the noise is random walk with the step-size taken from a Gaussian distribution with standard deviation $\sigma_L \sqrt{ \sum_i(t_0 + jT_p - t_i)^2}$, where the sum is carried out only over the new sections of window.}  In addition, this will be true for each of the transits, and consequently for the case of the black window, the best fit curve and whatever measures of goodness of fit are selected, will depend to an extent on the best fit curve and goodness of fit for the case of the grey window.  As a result of this correlation, it is not easily apparent how to simply and accurately determine the statistical significance of a positive detection in one of the windows.

For the case of correlated noise, the situation is even more complicated, for example consider the example light curve shown in figure~\ref{WindowNoiseCorrRed}.  In this example, the correlated photometric noise can be seen as an downward trend in the light curve.  As the dip is deeper later in the light curve, $\tau$ will be biased toward later times and thus will be overestimated (i.e. $\epsilon_j$ is positive).  In addition, as the noise is correlated, the new pieces of light curve included in the larger window are likely exhibit similar behaviour (resulting in an even more extreme value of $\epsilon_j$).  Thus, it seems reasonable that when moving from a smaller observation window to a larger one, the error in $\tau$ will preferentially increase in magnitude, but stay the same sign.  Consequently the issue of determining statistical significance of a detection in one of a number of windows for the case of correlated noise is even less clear.

While the selection of an appropriate observation window is an important issue, it is outside the scope of this  thesis as a result of the two reasons discussed above.  However, as discussed in section~\ref{Transit_Intro_Taudef}, guesses can be made about the size of physically realistic windows from formation and stability constraints.  As a result, for this work we will use equation~\eqref{transit_thresholds_method_DobsDef} to describe the size of the observing window, continue with the analysis, and leave further discussion of this issue until chapter~\ref{Conclusion}, within the context of future research directions.

\subsection{Selection of technique for calculating thresholds}

Now that we have expressions for $\Delta \tau$, the timing perturbation caused by the moon, a description of $\epsilon_j$ the timing noise, along with a well defined region of parameter space for which we would like moon detection thresholds, we are in a position to introduce and describe the method used in this thesis to calculate detection thresholds. To introduce this method, we first consider what detecting a moon means statistically, and formulate this into a null and alternative hypothesis.  This description is then used to motivate the choice of statistical method used, likelihood radio testing, and in particular generalised likelihood radio testing.  Finally, the expressions required to practically use this method are derived and discussed.

\subsubsection{Formulating a null and alternative hypothesis}

For this work we would like to differentiate between the case of ``detecting" a moon, and not ``detecting" a moon, where the ``detection" depends on the statistical threshold selected.  For the case where there is no moon, we expect the transit timings to be strictly periodic.  Formulating this as the null hypothesis we have that
\begin{equation}
H_0\colon  \tau_j = t_0 + jT_p + \epsilon_j,
\end{equation}
where $\epsilon_j$ is normally distributed with zero mean and known constant standard deviation $\sigma_\epsilon$.  For the case where there is a moon, we expect the transit timings to be the sum of the above linear function with a  low amplitude sinusoidal perturbation.  Formulating this as the alternative hypothesis we have that
\begin{equation}
H_1\colon  \tau_j = t_0 + jT_p + A\cos(\omega j + \phi) + \epsilon_j,
\end{equation}
where $\epsilon_j$ is again, normally distributed with zero mean and known constant\footnote{Near the detection threshold $A$ is of the same order of magnitude as $\sigma_\epsilon$.  As the second harmonic of $\Delta \tau$ is being neglected as it's amplitude is of order $v_m/v_{tr}$ times smaller than $A$, the perturbation in $\sigma_\epsilon$ due to the $(A_p + A_m)$ term must also be neglected as the changes in $A_p$ and $A_m$ are also of order $v_m/v_{tr}$ times smaller than $A_p$ and $A_m$ (see equations~\eqref{transit_signal_cc_BessAp} and \eqref{transit_signal_cc_BessAm}).  Thus $\sigma_\epsilon$ can be considered constant for this application.} standard deviation $\sigma_\epsilon$.  So, in a purely qualitative sense, the process of determining if a moon is detectable in a given sequence of data is the process of determining to the relative probability that the data would occur under $H_0$ or $H_1$.

\subsubsection{Introduction to likelihood ratio testing}

A statistical method exists, likelihood ratio testing, which uses just this approach to determine if $H_1$ is a better description of the data than $H_0$.  In particular, the quantity calculated, $\Lambda$ is the probability that the observed data were generated from the model corresponding to the null hypothesis divided by the probability that the data were generated from the model corresponding to the alternative hypothesis.  In addition, it can be shown that this method is the optimal\footnote{The test that is least likely to accept the null hypothesis when it isn't true.} statistical method for selecting between models for a broad range of model types \citep[e.g.][p. 303]{Rice1995}.  To motivate and provide intuition with respect to this method we consider the case where $t_0$, $T_p$, $A$, $\omega$ and $\phi$ are known a priori (a useful, but unphysical assumption) and where we have a sequence $\tau_1, \tau_2, \ldots, \tau_N$ of recorded data points that we wish to test.

To begin, consider the first data point $\tau_1$.  According to the null hypothesis, $\tau_1$ should be normally distributed with mean $t_0+T_p$ and standard deviation $\sigma_\epsilon$, that is, it should have a probability distribution
\begin{equation}
P(\tau_1) = \frac{1}{\sigma_\epsilon\sqrt{2\pi}} e^{\frac{(\tau_1 - (t_0+T_p))^2}{2 \sigma_\epsilon^2}}.
\end{equation}
Consequently, assuming this model is correct, the probability that a value between $\tau_1$ and $\tau_1 + \delta \tau$ is measured is given by 
\begin{equation}
P(\tau_1)\delta \tau = \frac{1}{\sigma_\epsilon\sqrt{2\pi}} e^{\frac{(\tau_1 - (t_0+T_p))^2}{2 \sigma_\epsilon^2}}\delta \tau.
\end{equation}

Similarly for the case of the alternative hypothesis, $\tau_1$ should be normally distributed with mean $t_0 + T_p + A\cos(\omega + \phi)$ and standard deviation $\sigma_\epsilon$, that is, have a probability distribution\begin{equation}
P(\tau_1) = \frac{1}{\sigma_\epsilon\sqrt{2\pi}} e^{\frac{(\tau_1 - (t_0+T_p + A\cos(\omega + \phi)))^2}{2 \sigma_\epsilon^2}}.
\end{equation}
Again, assuming this model is correct, the probability that a value between $\tau_1$ and $\tau_1 + \delta \tau$ is measured is given by
\begin{equation}
P(\tau_1)\delta \tau = \frac{1}{\sigma_\epsilon\sqrt{2\pi}} e^{\frac{(\tau_1 - (t_0+T_p + A\cos(\omega + \phi)))^2}{2 \sigma_\epsilon^2}}\delta \tau.
\end{equation}

By analogy, it follows that the probability that a value of $\tau_j$ ranging from $\tau_j$ to $\tau_j + \delta \tau$ is measured under the null hypothesis,  is given by
\begin{equation}
P(\tau_j)\delta \tau = \frac{1}{\sigma_\epsilon\sqrt{2\pi}} e^{\frac{(\tau_j - (t_0+jT_p))^2}{2 \sigma_\epsilon^2}}\delta \tau,
\end{equation}
while for the alternative hypothesis it is given by
\begin{equation}
P(\tau_j)\delta \tau = \frac{1}{\sigma_\epsilon\sqrt{2\pi}} e^{\frac{(\tau_j - (t_0+jT_p+ A\cos(\omega j + \phi)))^2}{2 \sigma_\epsilon^2}}\delta \tau.
\end{equation}

Now, as each of the $\epsilon_j$'s are uncorrelated (see section~\ref{Transit_Noise_Conclusion}), the probabilities corresponding to $\tau_1, \tau_2, \dots, \tau_N$ are independent for both the null and alternative hypotheses.  For the case of independent events, the probability that a set of events all occur e.g. each of the $\tau_1$ to $\tau_N$ is produced under the model, it equal to the product of the individual probabilities, that is
\begin{align}
P(\tau_1, \tau_2, \ldots, \tau_N)(\delta \tau)^N &= P(\tau_1)\delta \tau \times P(\tau_2)\delta \tau \times ... \times P(\tau_N)\delta \tau,\\
&= \prod_{j = 1}^N \frac{1}{\sigma_\epsilon\sqrt{2\pi}} e^{\frac{(\tau_j - (t_0+jT_p))^2}{2 \sigma_\epsilon^2}}\delta \tau,
\end{align}
for the case of the null hypothesis and
\begin{align}
P(\tau_1, \tau_2, \ldots, \tau_N)(\delta \tau)^N &= P(\tau_1)\delta \tau \times P(\tau_2)\delta \tau \times ... \times P(\tau_N)\delta \tau,\\
&= \prod_{j = 1}^N \frac{1}{\sigma_\epsilon\sqrt{2\pi}} e^{\frac{(\tau_j - (t_0+jT_p + A\cos(\omega j + \phi)))^2}{2 \sigma_\epsilon^2}}\delta \tau,
\end{align}
for the case of the alternative hypothesis.  Recalling that the test statistic, $\Lambda$, is the ratio of the probability that  the data was produced under the null hypothesis to the probability that the data was produced under the alternative hypothesis, we obtain
\begin{align}
\Lambda &= \frac{\prod_{j = 1}^N \frac{1}{\sigma_\epsilon\sqrt{2\pi}} e^{ \frac{(\tau_j - (t_0+jT_p))^2}{2 \sigma_\epsilon^2}}\delta \tau}{\prod_{j = 1}^N \frac{1}{ \sigma_\epsilon\sqrt{2\pi}} e^{\frac{(\tau_j - (t_0+jT_p+ A\cos(\omega j + \phi)))^2}{2 \sigma_\epsilon^2}}\delta \tau},\\
 &= \frac{ e^{\sum_{j = 1}^N\frac{(\tau_j - (t_0+jT_p))^2}{2 \sigma_\epsilon^2}}}{ e^{\sum_{j = 1}^N\frac{(\tau_j - (t_0+jT_p+ A\cos(\omega j + \phi)))^2}{2 \sigma_\epsilon^2}}}.
\end{align}
where the factors of $\sigma_\epsilon\sqrt{2\pi}$ and $\delta \tau$ have been cancelled.  As $\Lambda$ is the ratio of the probability that the null hypothesis describes the data to the probability that the alternative hypothesis describes the data, high values ($\Lambda \approx 1$) suggest that the null hypothesis is more likely, while low values  ($\Lambda \ll 1$) suggest that the alternative hypothesis is more likely.

However, as previously mentioned, the system parameters, $t_0$, $T_p$, $A$, $\omega$ and $\phi$ are not known prior to, or after detection.  Consequently,  a more general version of this method will be used.

\subsubsection{Introduction to generalised likelihood ratio testing}\label{Trans_Thresholds_Method_GenLikelihood}

For this work, generalised likelihood ratio testing will be used to calculate detection thresholds.  This method has the same principles as likelihood ratio testing, except that instead of using known model parameters, the model parameters used are those most likely to have produced the data under that model for the case of the null and alternative hypotheses respectively.  The cost of this generalisation is that the technique may no longer be optimal.  Consequently, for this case, the null hypothesis is given by
\begin{equation}
H_0\colon  \tau_j = \overline{t}_0 + j\overline{T}_p + \epsilon_j,
\end{equation}
where $\epsilon_j$ is normally distributed with zero mean and known standard deviation $\sigma_\epsilon$, and where $\overline{t}_0$ and $\overline{T}_p$ are the coefficients derived from a linear least squares fit.\footnote{As the errors are normally distributed, the model, $\overline{\tau}$ most likely to have produced the data is the model which is fitted in the least-squares sense, that is, it minimises $\sum_{j = 1}^N (\tau_j - \overline{\tau}_j)^2$, where $\overline{\tau}_j$ are the values of $\tau_j$ predicted by the model \citep[see e.g.][p. 651]{Pressetal1992}.}  Similarly, the alternative hypothesis is given by
\begin{equation}
H_1\colon  \tau_j = \hat{t}_0 + j\hat{T}_p + \hat{A}\cos(\hat{\omega} j + \hat{\phi}) + \epsilon_j,
\end{equation}
where $\epsilon_j$ is again normally distributed with zero mean and known standard deviation $\sigma_\epsilon$, and where $\hat{t}_0$, $\hat{T}_p$, $\hat{A}$, $\hat{\omega}$ and $\hat{\phi}$ are the coefficients derived from a least squares fit for the case of a line plus a sinusoid.  In particular, note that $t_0$, $\overline{t}_0$ and $\hat{t}_0$ are not necessarily equal, and that $T_p$, $\overline{T}_p$ and $\hat{T}_p$ are also not necessarily equal.\footnote{See section~\ref{Trans_TTV_Signal_CC_PropSig} for a discussion of the roll of incorrect fitted values of $T_p$ on the formation of non-detection spikes.}  Consequently, the expression for $\Lambda$ is given by
\begin{equation}
\Lambda = \frac{ e^{\sum_{j = 1}^N\frac{(\tau_j - (\overline{t}_0+j\overline{T}_p))^2}{2 \sigma_\epsilon^2}}}{ e^{\sum_{j = 1}^N\frac{(\tau_j - (\hat{t}_0+j\hat{T}_p+ \hat{A}\cos(\hat{\omega} j + \hat{\phi})))^2}{2 \sigma_\epsilon^2}}}.\label{transit_threshold_method_Lambdadef}
\end{equation}

Now that we have an expression for our test statistic $\Lambda$, we need a way to determine the statistical significance associated with a particular value of $\Lambda$, or as we will see with $2\log(\Lambda)$, as it is more useful in practice.  To determine this statistical significance we consider the distribution of $2\log(\Lambda)$ for the case where the null hypothesis is true (i.e. that the low value of $\Lambda$ occurred by chance).

We begin with equation~\eqref{transit_threshold_method_Lambdadef} by taking the natural logarithm and simplifying to give 
\begin{multline}
2\log(\Lambda) = \sum_{j = 1}^N\frac{(\tau_j - (\overline{t}_0+j\overline{T}_p))^2}{\sigma_\epsilon^2} \\- \sum_{j = 1}^N\frac{(\tau_j - (\hat{t}_0+j\hat{T}_p+ \hat{A}\cos(\hat{\omega} j + \hat{\phi})))^2}{\sigma_\epsilon^2}.\label{transit_thresholds_method_thresholddef}
\end{multline}
As can be seen, this expression is comprised of two sums, which we will consider in turn.  First, an expression of the form of the first sum is exactly described by a $\chi^2$ distribution\footnote{Recall that a $\chi^2$ distribution with $\nu$ degrees of freedom  is generated by summing the square of $\nu$ independent standard normally distributed variables.} with $N-2$ degrees of freedom, where we note that the 2 is due to the two fitting variables ($\overline{t}_0$ and $\overline{T}_p$).  Similarly, an expression of the form of the second sum is approximately\footnote{It is not exactly described by a $\chi^2$ distribution as the fitting formula is non-linear in the fitting parameters $\omega$ and $\phi$.} described by a $\chi^2$ distribution with $N-5$ degrees of freedom, where we again note that the 5 is due to the five fitting variables ($\hat{t}_0$, $\hat{T}_p$, $\hat{A}$, $\hat{\omega}$ and $\hat{\phi}$).  As the $\chi^2$ distribution describes the sum of the square of a sequence of independent normally distributed variables with mean zero and standard deviation one, the sum or difference of two variables which are $\chi^2$ distributed will also be $\chi^2$ distributed.  Consequently, as $2\log(\Lambda)$ is the difference between two variables which are approximately $\chi^2$ distributed, it will also be $\chi^2$ distributed.  In particular it is approximately distributed as a $\chi^2$ variable with three degrees of freedom,\footnote{Corresponding to the three degrees of freedom lost by also fitting the sinusoid.} an approximation which becomes increasingly accurate as $N$ becomes large \citep[e.g.][p. 310]{Rice1995}.  So, for the case where $N$ is large, $2\log(\Lambda)$ should be approximately distributed as a $\chi^2$ variable with three degrees of freedom under the null hypothesis (there is no moon), and thus we can determine the probability that such a large value of $2\log(\Lambda)$ (small value of $\Lambda$) occurred due to random chance.  For example, the 95\% confidence limit for the $\chi^2_3$ distribution is 7.816.  Thus, if a value of $2\log(\Lambda)$ was calculated corresponding to 7.816 we would be 95\% sure that this wasn't just a statistical deviation, and it actually corresponded to a moon detection.  For the case where $N$ is small we can simulate a statistically significant number of realisations of $\tau$, and use these to determine the null distribution of $2\log(\Lambda)$.  For example, for the case where $N = 9$ the 95\% confidence limit is given by $8.45 \pm 0.09$.

Now that we have a method for calculating the detection threshold, we can begin to explore the behaviour of this threshold.  This will be done in three main stages.  First, analytic expressions for the location of the thresholds will be derived for the case where $N \to \infty$.  Then, these will be compared to similar moon detection thresholds derived in section~\ref{Intro_Dect_Moons_Transit} for the three other methods presented in the literature.  Finally, a Monte Carlo simulation will be used to investigate the case of low $N$, and in this context, the effect of $N$, and inclination, eccentricity and orientation of the planet's orbit on moon detection will be discussed.

\section{Expected behaviour of detection thresholds}\label{Trans_Thresholds_ExpBehav}

Armed with the equation determining whether or not a moon is detectable in a given sequence of $\tau$ values, equation~\eqref{transit_thresholds_method_thresholddef}, we now investigate the behaviour of the detection thresholds.  While equation~\eqref{transit_thresholds_method_thresholddef} is analytic (from $\tau_1$, $\tau_2$, $\ldots$, $\tau_N$, the values $\overline{t}_0$, $\overline{T}_p$, $\hat{t}_0$, $\hat{T}_p$, $\hat{A}$, $\hat{\omega}$ and $\hat{\phi}$ can be calculated), it is so complex that is does not provide much in the way of intuitive understanding.

As most of the complexity stems from the finite $N$ nature of the fitting process, we will use a two pronged method.  First, for the sections of the threshold where the finite nature of $N$ is important e.g. in a non-detection spike, we will use the analysis conducted in chapter~\ref{Transit_Signal}, summarised below for convenience.  For the regions of the threshold where the finite nature of $N$ is not important we will use an approximation assuming large $N$ to determine the gross behaviour.

\subsection{Summary of behaviour of non-detection spikes}

For the case where a moon orbits the planet an integer number of times per transit, the moon detection threshold will show a non-detection spike.  Physically, this spike results from the fact that the planet and moon present the same orientation each transit and thus produce no transit to transit timing perturbation.  The distribution and shape of these spikes was discussed and investigated in section~\ref{Sec-TraM-TTV-CC-prop}, and in particular it was found that the moon detection threshold should be decorated with a ``comb" of non-detection spikes, where the spacing between neighbouring spikes is proportional to $a_m^{2.5}$.  In addition, each individual spike should be approximately symmetric and have width proportional to $Na_m^{2.5}$.  Consequently, as $N$, the number of observed transits increases, each spike should remain centered on the same semi-major axis, but decrease in width, such that as $N$ tends to infinity, the spikes become infinitely thin, and can be neglected.  We now concentrate on the shape of the threshold for this case.

\subsection{Limit as $N \to \infty$}

To investigate the shape of the threshold as $N$, the number of measured transits, becomes large, we begin by assuming that our $\tau$ values are described by
\begin{equation}
\tau_j = t_0 + jT_p + A\cos(\omega j + \phi) + \epsilon_j,
\end{equation}
where $\epsilon_j$, the timing noise, is normally distributed with known standard deviation $\sigma_\epsilon$.  As $N$ increases, we expect that our fitted parameters, $\hat{t}_0$, $\hat{T}_p$, $\hat{A}$, $\hat{\omega}$ and $\hat{\phi}$, would tend towards\footnote{The fitted value of $\omega$ may vary from the true value by a multiple of 2$\pi$ as a result of aliasing.} the true values and would asymptotically approach them as $N \to \infty$.  In addition, the fitted values, $\overline{t}_0$ and $\overline{T}_p$, for the linear fit also tend to $t_0$ and $T_p$ as $N \to \infty$ (see appendix~\ref{App_t0Tp}).  Consequently, equation~\eqref{transit_thresholds_method_thresholddef} becomes
\begin{multline}
2\log(\Lambda) = \frac{1}{\sigma_\epsilon^2} \sum_{j = 1}^N (t_0 + jT_p + A\cos(\omega j + \phi) + \epsilon_j - (t_0+j T_p))^2 \\- (t_0 + jT_p + A\cos(\omega j + \phi) + \epsilon_j - (t_0+j\hat{T}_p+ \hat{A}\cos(\hat{\omega} j + \hat{\phi})))^2,\label{transit_thresholds_expbehav_thresholdeq1}
\end{multline}
which simplifies to
\begin{equation}
2\log(\Lambda) = \frac{1}{\sigma_\epsilon^2} \sum_{j = 1}^N (A \cos(\omega j + \phi))^2 + \frac{1}{\sigma_\epsilon^2} \sum_{j = 1}^N \epsilon_j A\cos(\omega j + \phi).\label{transit_thresholds_expbehav_thresholdeq2}
\end{equation}

Consider the first sum on the right hand side of equation~\eqref{transit_thresholds_expbehav_thresholdeq2}.  As all the terms under this sum sign are squares of real numbers, they are positive by definition.  In addition, as each of these terms are of order $A^2$, the characteristic size of the sum is $N \times A^2/\sigma_\epsilon^2$.  In comparison, the second sum consists of terms which are the product of a normally distributed variable $\epsilon_j$, and an expression of characteristic size $A$.  Recalling that the standard deviation of the sum of $N$ independent, normally distributed variables with standard deviation $\sigma$ is $\sqrt{N}\sigma$, we have that the characteristic size of the second sum in equation~\eqref{transit_thresholds_expbehav_thresholdeq2} is $\sqrt{N} \times A/\sigma_\epsilon$.  Consequently, for large $N$ ($N \gg (A/\sigma_\epsilon)^{-2}$), the size of the second sum in equation~\eqref{transit_thresholds_expbehav_thresholdeq2} will be much smaller than the first, and can thus be neglected.

Now, assuming that the orbit of the moon is not in resonance with the orbit of the planet, the sequence of in-transit moon positions corresponding to the sequence of $A\cos(\omega j + \phi)$ values will never repeat.  Consequently, after a sufficiently large number of transits, all parts of the moon's orbit will be sampled equally.  Thus, the sum in equation~\eqref{transit_thresholds_expbehav_thresholdeq2} can be replaced with $N$ multiplied by $\overline{\Delta \tau^2}$, the average value of $A^2\cos^2(\omega j + \phi)$.  Averaging over a full orbit, we obtain
\begin{align}
\overline{\Delta \tau^2} &= \frac{\int_0^{2\pi/\omega} (A\cos(\omega t + \phi))^2 dt}{2\pi/\omega},\\
 &= \frac{2\pi/\omega (A^2)/2 }{2\pi/\omega},\\
& =  \frac{A^2}{2}.
\end{align}

Finally, the distribution of $2\log(\Lambda)$ is also a function of $N$.  In particular, as mentioned previously, as $N \to \infty$, it tends to a $\chi^2$ distribution with three degrees of freedom.  The 99.7\% limit for a $\chi^2_3$ distribution is 13.93.  Consequently, the equation describing the 99.7\% threshold for moon detection would be given by replacing $2\log(\Lambda)$ with 13.93 in equation~\eqref{transit_thresholds_expbehav_thresholdeq2}.

Applying these three simplifications to equation~\eqref{transit_thresholds_expbehav_thresholdeq2}, the following expression for the detection threshold is obtained,
\begin{equation}
13.95 = N \times \frac{A^2}{2 \sigma_\epsilon^2}.\label{transit_thresholds_expbehav_thresholdeq3}
\end{equation}

From equation~\eqref{transit_signal_cc_form_lBsimp} we have that for the case of circular and coplanar orbits, $A$, the amplitude of $\Delta \tau$, is given by
\begin{equation}
A = \cos\left(\frac{n_mR_{ch}}{v_{tr}}\right) \frac{\hat{A}_m}{\hat{A}_p + \hat{A}_m}\frac{M_p}{M_p + M_m} \frac{a_m}{v_{tr}}.\label{transit_thresholds_expbehav_Adef}
\end{equation}
Recalling from section~\ref{Trans_TTV_Signal_CC_PropAmp} that for moons which are detectable and can be described by this analysis, the cosine term in equation~\eqref{transit_thresholds_expbehav_Adef} is approximately equal to one, and can consequently be neglected.  Performing this simplification, substituting this into equation~\eqref{transit_thresholds_expbehav_thresholdeq3} and rearranging such that all the terms involving $\hat{A}_p$ and $\hat{A}_m$ are on the left hand side and all the other terms are on the right gives
\begin{equation}
 \frac{\hat{A}_m}{\hat{A}_p + \hat{A}_m}  = 5.28 \frac{\sigma_\epsilon}{\sqrt{N}} \frac{M_p + M_m}{M_p} \frac{v_{tr}}{a_m}.\label{transit_thresholds_expbehav_thresholdeq4}
\end{equation}
Now, assuming that the planet is much larger than the moon ($\hat{A}_p \gg \hat{A}_m$ and $M_p \gg M_m$) equation~\eqref{transit_thresholds_expbehav_thresholdeq4} simplifies to
\begin{equation}
 \frac{\hat{A}_m}{\hat{A}_p}  = 5.28 \frac{\sigma_\epsilon}{\sqrt{N}} \frac{v_{tr}}{a_m}.\label{transit_thresholds_expbehav_thresholdeq5}
\end{equation}

As discussed in section~\ref{Trans_Thresholds_Parameters}, $\hat{A}_m/L_0 N_{tra}$ can be approximated by $(R_m/R_s)^2$.  Similarly, $\hat{A}_p/L_0 N_{tra}$ can be approximated by $(R_p/R_s)^2$.  Consequently, $\hat{A}_m/\hat{A}_p$ can be approximated by $R_m^2/R_p^2$.  Using this expression gives
\begin{equation}
R_m^2 = 5.28 R_p^2 \frac{\sigma_\epsilon}{\sqrt{N}} \frac{v_{tr}}{a_m}.\label{transit_thresholds_expbehav_thresholddef}
\end{equation}

As $\sigma_\epsilon$ is different for the cases of white, filtered and red noise, equation~\eqref{transit_thresholds_expbehav_thresholddef} will be investigated separately for each of these cases.  In particular, the fact that moons with $a_m \ll R_s$ and $a_m \gg R_s$ are not detectable will be discussed, followed by an investigation of the threshold minimum for the case where $a_m \approx R_s$, corresponding to the detectable moon with the smallest radius.

\subsubsection{Behaviour of threshold in the case of white noise}

For the case of white photometric noise, $\sigma_\epsilon$ is given by equation~\eqref{transit_noise_white_sigdefphys},
\begin{multline*}
\sigma_\epsilon =  47.9s \left[\left(\frac{\sigma_L / L_0}{3.95 \times 10^{-4}}\right)\left(\frac{\Delta t}{1 \text{min}}\right)^{1/2}\right] \left[\frac{100(A_p + A_m)}{N L_0} \right]^{-1} \\
\times \left[\left(\frac{T_{obs}}{24 \text{hrs}}\right)^{3/2} \left(\frac{T_{tra}}{13 \text{hrs}}\right)^{-1}\right].
\end{multline*}

Noting that $N_{tra} L_0/(A_p + A_m)$ is approximately equal to $R_s^2/R_p^2$ and substituting this expression into equation~\eqref{transit_thresholds_expbehav_thresholddef}, and simplifying gives
\begin{multline}
R_m^2 = 2.15\times 10^{-5} R_s^2 \frac{1}{\sqrt{N}} \left[\left(\frac{\sigma / L_0}{3.95 \times 10^{-4}}\right)\left(\frac{\Delta t}{1 \text{min}}\right)^{1/2}\right] \\
\times \left(\frac{T_{tra}}{13 \text{hrs}} \right)^{-1/2} \left[\frac{(2R_s + 2a_m)^{3/2}}{a_m\sqrt{2R_s}} \right],\label{transit_thresholds_expbehav_whitethresh}
\end{multline}
where equations~\eqref{transit_thresholds_method_DobsDef} and \eqref{transit_intro_dur_cc_Ddef} were used to substitute for $T_{obs}$ and $T_{tra}$.  This equation describes the minimum radius of a moon that can be detected to three sigma significance as a function of the moon's semi-major axis.  As discussed in section~\ref{Transit_Noise_Conclusion}, for the case where $a_m \ll R_s$, the $R_s$ term dominates in the $(2R_s + 2a_m)$ term and so $R_m \propto 1/a_m^{1/2}$, and thus such moons are undetectable.  Similarly, for the case where $a_m \gg R_s$, the $a_m$ term dominates the $(2R_s + 2a_m)$ expression, we have that $R_m \propto a_m^{1/4}$, and again, moons are undetectable.  However, for the case where $a_m \approx R_s$, there is a region for which moon detection is possible.  To investigate the shape of this region, the minimum of this threshold curve, that is, the semi-major axis which gives the smallest value of $R_m$, will be determined.

To find the minimum of this function, the derivative is taken with respect to $a_m$ and then $\frac{dR_s}{d a_m}$ is set equal to zero, giving
\begin{align}
0 &= \frac{1}{a_m } \frac{3 (2R_s + 2a_m)^{1/2}}{\sqrt{2R_s} } - \frac{1}{a_m^2 }\frac{(2R_s + 2a_m)^{3/2}}{\sqrt{2R_s} },\\
 &= \frac{(2R_s + 2a_m)^{1/2}}{a_m^2\sqrt{2R_s}} \left[ 3 a_m - (2R_s + 2a_m) \right].
\end{align}
where all the terms which do not depend on $a_m$ have been neglected.  As $a_m$ cannot be infinite, we multiply by $a_m^2\sqrt{2R_s}$, to give
\begin{equation}
0 = (2R_s + 2a_m)^{1/2} \left[ a_m - 2R_s \right].
\end{equation}
Noting that the first term cannot equal zero (as only positive semi-major axes are physical) the equation can be divided through by it, giving
\begin{equation}
0 =  a_m - 2R_s,
\end{equation}
or 
\begin{equation}
a_m =  2R_s.
\end{equation}

Consequently, for the case where the photometric noise is white, we expect that moons with $a_m \ll R_s$ will be undetectable, moons with $a_m \approx R_s$ will be detectable, with the most detectable moons having semi-major axis equal to a stellar diameter and that moons with $a_m \gg R_s$ will also be undetectable.  In addition, for the case where the planet's orbit is inclined ($\delta_{min} \ne 0$), it can be shown using an equivalent derivation that for this case the most detectable moons have semi-major axis equal to the length of the chord they make on the star, that is, $a_m = 2 R_s (1 - \delta_{min}^2)^{1/2}$.  To put this result in context, recall from figure~\ref{TauAgreement} that $\Delta \tau$ is very accurately represented by a sinusoid, for moons with $a_m \approx 2 R_s$, even for large values of $v_m/v_{tr}$.  Consequently, this result is very robust.

\subsubsection{Behaviour of threshold in the case of filtered noise}

For the case where the photometric noise is dominated by filtered realistic noise, $\sigma_\epsilon$ is given by equation~\eqref{transit_noise_filt_sigdefphys},
\begin{multline*}
\sigma_\epsilon = 53.2s \left[\beta\right] \left[\frac{100(A_p + A_m)}{L_{0}N_{tra}}\right]^{-1} \\ \times\left[ \left(\frac{T_{tra}}{13 \text{hrs}}\right)^{-1} \left(-8 \times 10^{-3} \left(\frac{T_{obs}}{24 \text{hrs}}\right) + 1.008 \left(\frac{T_{obs}}{24 \text{hrs}}\right)^2)\right)\right]. 
\end{multline*}
Neglecting the constant term in the large round brackets as it is much smaller than the other term for all transits of interest gives
\begin{equation}
\sigma_\epsilon \approx 53.6s \left[\beta\right] \left[\frac{100(A_p + A_m)}{L_{0}N_{tra}}\right]^{-1}  \left(\frac{T_{tra}}{13 \text{hrs}}\right)^{-1}  \left(\frac{T_{obs}}{24 \text{hrs}}\right)^2. 
\end{equation}
Substituting this into equation~\eqref{transit_thresholds_expbehav_thresholddef} and simplifying gives
\begin{equation}
R_m^2 = 1.77\times 10^{-5} R_s^2 \frac{\beta}{\sqrt{N}} \frac{(2R_s + 2a_m)^2}{2R_sa_m},\label{transit_thresholds_expbehav_filt_thresh}
\end{equation}
where we again note that $N_{tra} L_0/(A_p + A_m) \approx R_s^2/R_p^2$ and where where equations~\eqref{transit_thresholds_method_DobsDef} and \eqref{transit_intro_dur_cc_Ddef} have been used to substitute for $T_{obs}$ and $T_{tra}$
Again, as for the case of white noise, moons with $a_m \ll R_s$ or $a_m \gg R_s$ will be undetectable.  In particular, for the case of $a_m \ll R_s$, $R_m \propto 1/a_m^{1/2}$, and for the case of $a_m \gg R_s$, $R_m \propto a_m^{1/2}$.  While very close or very distant moons are undetectable, moons with $a_m \approx R_s$ may be detectable.  To investigate this case we again consider the semi-major axis corresponding to the smallest detectable moon.

Taking the derivative of equation~\eqref{transit_thresholds_expbehav_filt_thresh} with respect to $a_m$, setting $\frac{d R_m}{d a_m} = 0$ and simplifying gives
\begin{align}
0 &= \frac{1}{a_m} \frac{4 (2R_s + 2a_m)}{2R_s} - \frac{1}{a_m^2} \frac{(2R_s + 2a_m)^2}{2R_s},\\
 &= \frac{(2R_s + 2a_m)}{2R_s a_m^2}\left[ 4a_m -  (2R_s + 2a_m)\right].
\end{align}
Noting that the first factor again cannot be equal to zero, we have that
\begin{equation}
0 = 4a_m -  (2R_s + 2a_m).
\end{equation}
or
\begin{equation}
a_m = R_s
\end{equation}

So, for the case where the light curve is dominated by filtered noise, moons with $a_m \ll R_s$ are undetectable, moons with $a_m \approx R_s$ are possibly detectable, with the most detectable moons having semi-major axis given by $a_m = R_s$ and finally, moons with $a_m \gg R_s$ are again undetectable.  In addition, repeating this analysis for the case of inclined orbits again alters the semi-major axis corresponding to the minimum.  In particular, for this case, the minimum occurs for $a_m = R_s(1 - \delta_{min}^2)^{1/2}$.  Again, recall from figure~\ref{TauAgreement}, that $\Delta \tau$ can be well described by a sinusoid for moons with $a_m = R_s$.  Consequently, this result is also relatively robust.

\subsubsection{Behaviour of threshold in the case of red noise}

For the case where the photometric noise is dominated by red noise, $\sigma_\epsilon$ is given by equation~\eqref{transit_noise_red_sigdefphys},
\begin{multline*}
\sigma_\epsilon = 103.7s \left[\beta \right]\left[\frac{L_{0}N_{tra}}{100(A_p + A_m)}\right] \left[ \left(\frac{T_{tra}}{13 \text{hrs}}\right)^{-1} \right. \\ \left.
\times \left(0.277 \left(\frac{T_{obs}}{24 \text{hrs}}\right)^2 + 0.714 \left(\frac{T_{obs}}{24 \text{hrs}}\right)^3\right)\right] ,
\end{multline*}
where again the constant term has been neglected.  Unlike the case of filtered noise, both remaining terms are dominant, and so an analysis of the type conducted for the case of white and filtered noise will not be possible.  While simple analytic expressions for the semi-major axis corresponding to the threshold minimum are not available, the behaviour of the minimum can be bracketed by investigating the cases of $\sigma_\epsilon \propto T_{obs}^2/T_{tra}$ and $\sigma_\epsilon \propto T_{obs}^3/T_{tra}$.  In addition, only the case of $\sigma_\epsilon \propto T_{obs}^3/T_{tra}$ needs to be analysed as the case of $\sigma_\epsilon \propto T_{obs}^2/T_{tra}$ was investigated in the previous section in the context of filtered noise.

Substituting this into equation~\eqref{transit_thresholds_expbehav_thresholddef} and simplifying gives
\begin{equation}
R_m^2 \propto R_s^2 \frac{\beta}{\sqrt{N}}  \frac{(2R_s + a_m)^3}{2R_sa_m}.
\end{equation}
For the case where $\sigma_\epsilon \propto T_{obs}^3/T_{tra}$, we again have that moons with $a_m \ll R_s$ or $a_m \gg R_s$ are undetectable as $R_m \propto 1/a_m^{1/2}$ and $R_m \propto a_m$ respectively.  However, for $a_m \approx R_s$, there is a chance of detection.  Again, determining the semi-major axis corresponding to the  minimum value of $R_m$, by differentiation gives
\begin{align}
0 &= \frac{1}{a_m}  \frac{6(2R_s + 2a_m)^2}{2R_s} - \frac{1}{a_m^2}  \frac{(2R_s + 2a_m)^3}{2R_s},\\
 &=  \frac{(2R_s + 2a_m)^2}{2R_sa_m^2} \left[6a_m -  (2R_s + 2a_m)\right],
\end{align}
which gives, 
\begin{equation}
0 = 6a_m -  (2R_s + 2a_m),
\end{equation}
or
\begin{equation}
a_m =  \frac{1}{2}R_s.
\end{equation}

For the case where the light curve is dominated by red photometric noise, moons with $a_m \ll R_s$ or $a_m \gg R_s$ are again undetectable, while moons with $a_m \approx R_s$ may again be detectable.  In particular, the semi-major axis corresponding to the smallest detectable moon ranges from $R_s$ to $1/2R_s$ depending on the transit duration, where it tends towards $R_s$ for short transit durations and $1/2 R_s$ for longer transit durations.  Finally, repeating this analysis for the case of inclined planet orbits, it is found that this range changes from one half to a quarter of the diameter of the star to one half to a quarter of the length of the chord made by the path of the planet across the face of the star.  Unlike the case for white and filtered noise, this minima may lie in the region where our approximation for $\Delta \tau$ is no longer accurate (see figure~\ref{TauAgreement05B066}).  Consequently, for this case, the specific behaviour of the threshold will also depend on $v_m/v_{tr}$.

\subsection{Summary of expected properties}

\begin{figure}[tb]
\begin{center}
\includegraphics[width=.98\textwidth]{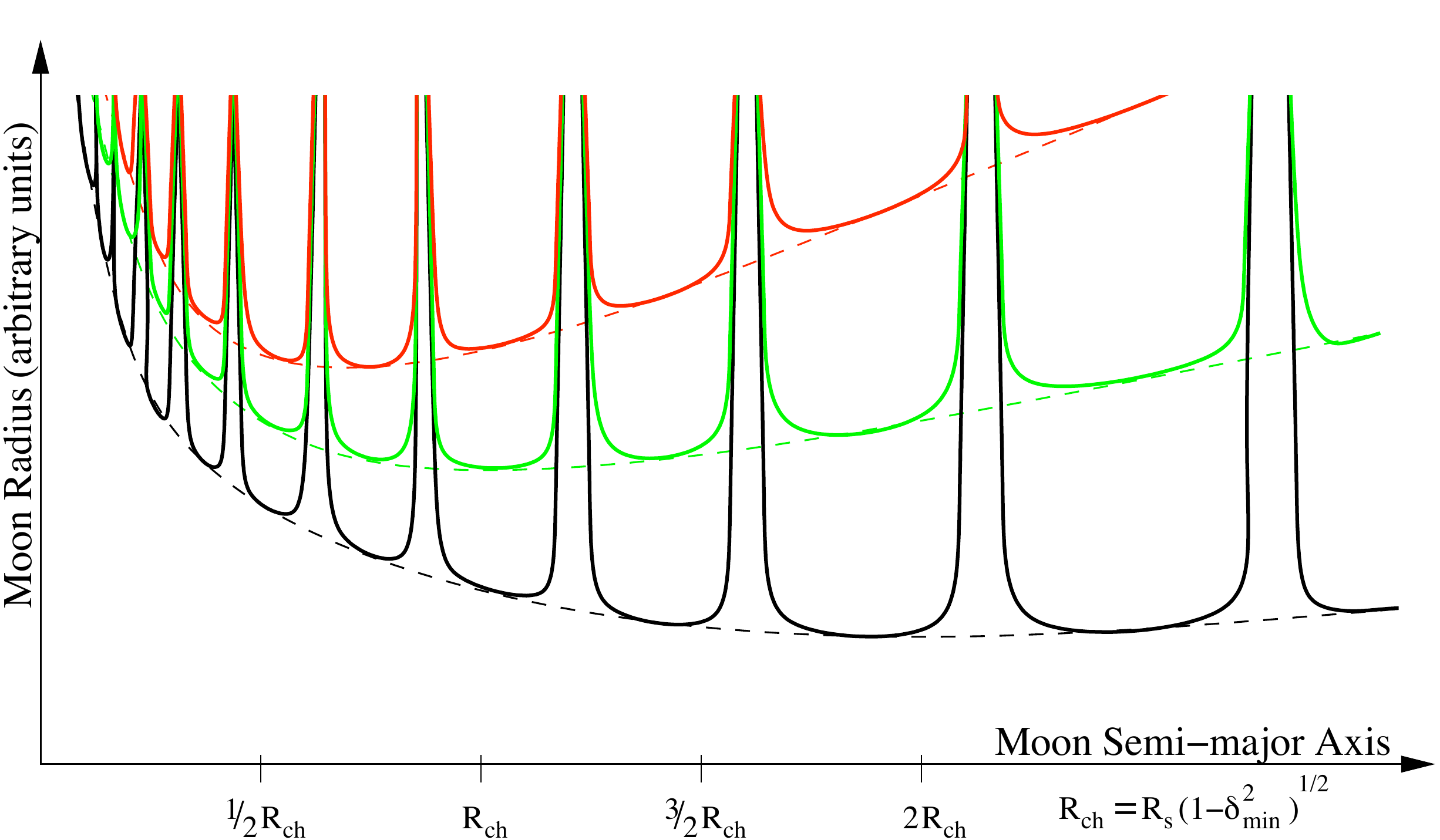}
\caption[Cartoon showing the expected features of TTV$_p$ moon detection thresholds for the case of a light curve contaminated with white (black line), filtered (green line) and red (red line) photometric noise.]{Cartoon showing the expected features of TTV$_p$ moon detection thresholds for the case of a light curve contaminated with white (black line), filtered (green line) and red (red line) photometric noise.  The detection threshold estimated by assuming $N$ is large is denoted by a dashed line, while the full detection threshold is shown as a thicker solid line.}
\label{CartoonDetThresh}
\end{center}
\end{figure}

Combining the summary of the properties of the non-detection spikes with the properties derived for the case where the number of transits becomes large, a comprehensive picture of the type of behaviour that we expect the detection threshold to show can be formed (see figure~\ref{CartoonDetThresh}).  To begin, we expect the thresholds to be in the shape of a distorted ``U", such that the minimum radius of a detectable moon tends to infinity as the moon semi-major axis tends either to 0 or to infinity.  In addition, the minima of this curve should occur at $a_m = 2R_s(1 - \delta_{min}^2)^{1/2}$ for the case of white noise, $a_m = R_s(1 - \delta_{min}^2)^{1/2}$ for the case of filtered noise and between $R_s(1 - \delta_{min}^2)^{1/2}$ and $1/2\,R_s(1 - \delta_{min}^2)^{1/2}$ for the case of red noise, where we note that $2R_s(1 - \delta_{min}^2)^{1/2}$ is the length of the chord that the planet makes across the face of its host star.  In addition to this general behaviour, the threshold should also be decorated with a comb of non-detection spikes with spacing proportional to $a_m^{2.5}$ and width proportional to $Na_m^{2.5}$.  Now that we have an understanding of the shape and behaviour of the TTV$_p$ detection threshold, we can compare it with similar detection thresholds calculated for the three other transiting moon detection techniques.

\section{Comparison with literature thresholds}

\begin{figure}[htp]
     \centering
     \subfigure[$R_p$=$R_J$, $a_p=1$AU, $\delta_{min} = 0$.]{
          \label{UpdatedStdNinfThresh}
          \includegraphics[width=.485\textwidth]{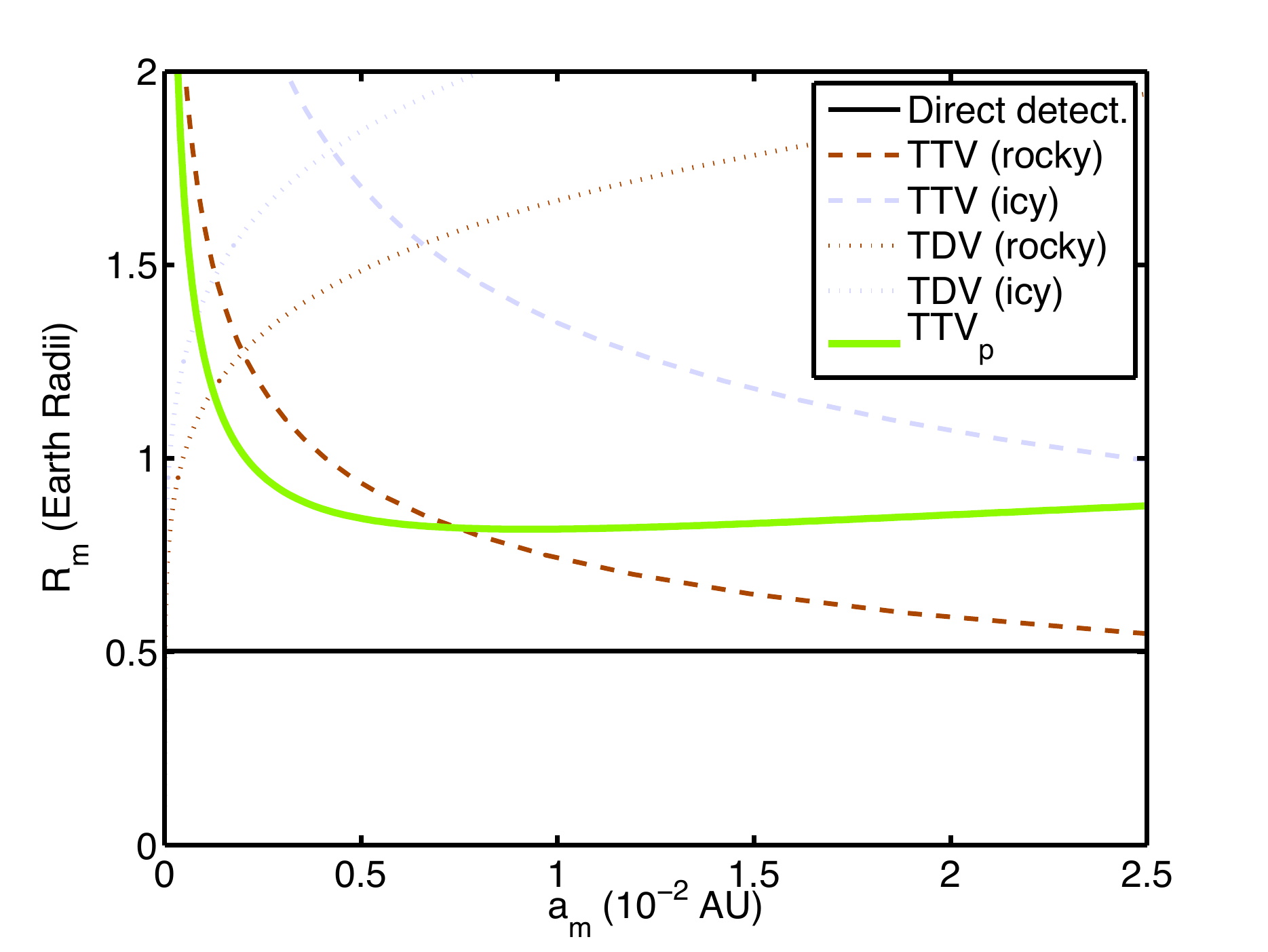}}
     \subfigure[$R_p$=$R_{\earth}$, $a_p=1$AU, $\delta_{min} = 0$.]{
          \label{UpdatedMeNinfThresh}
          \includegraphics[width=.485\textwidth]{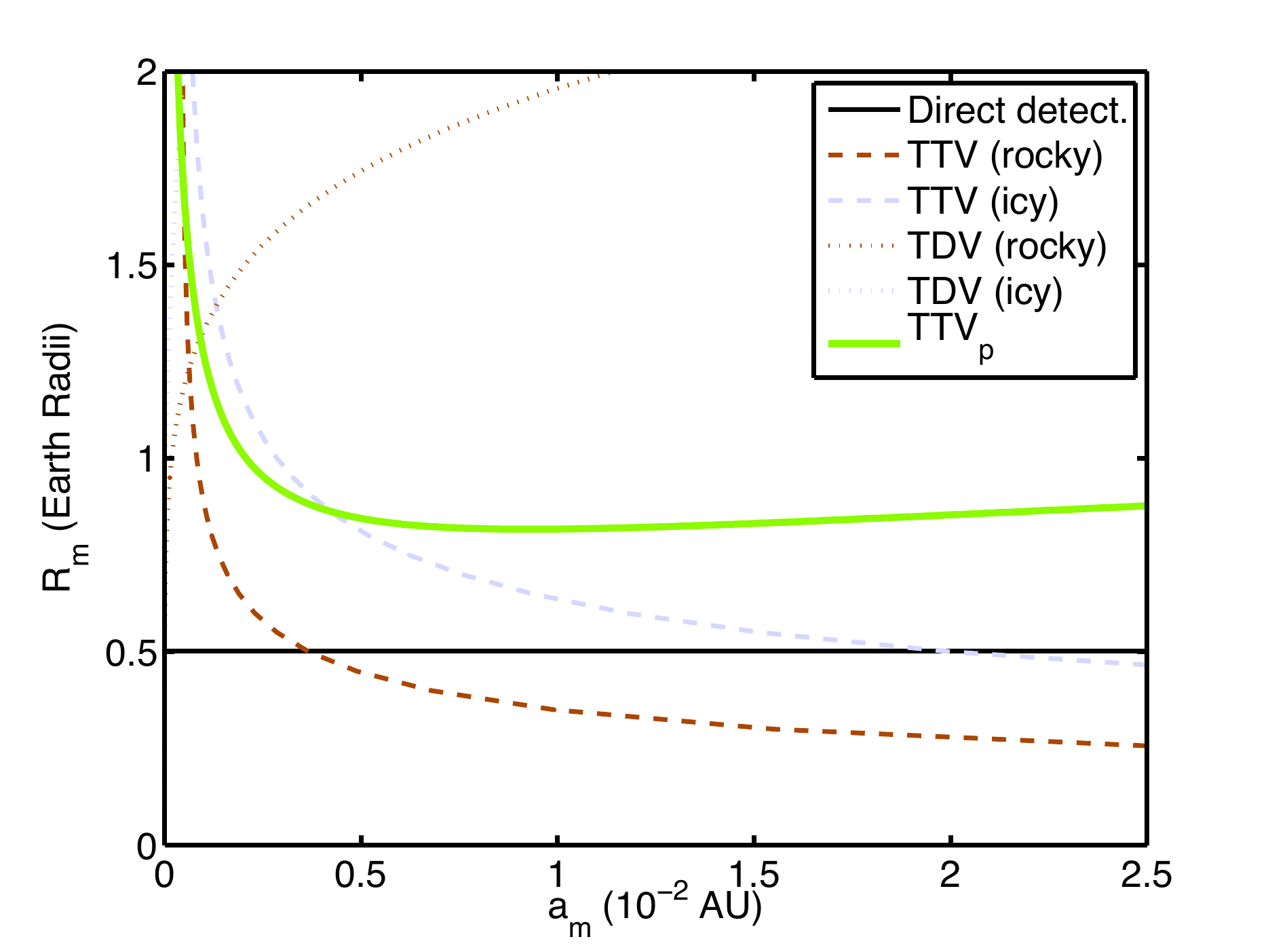}}\\
     \subfigure[$R_p$=$R_J$, $a_p=0.2$AU, $\delta_{min} = 0$.]{
          \label{Updated02NinfThresh}
          \includegraphics[width=.485\textwidth]{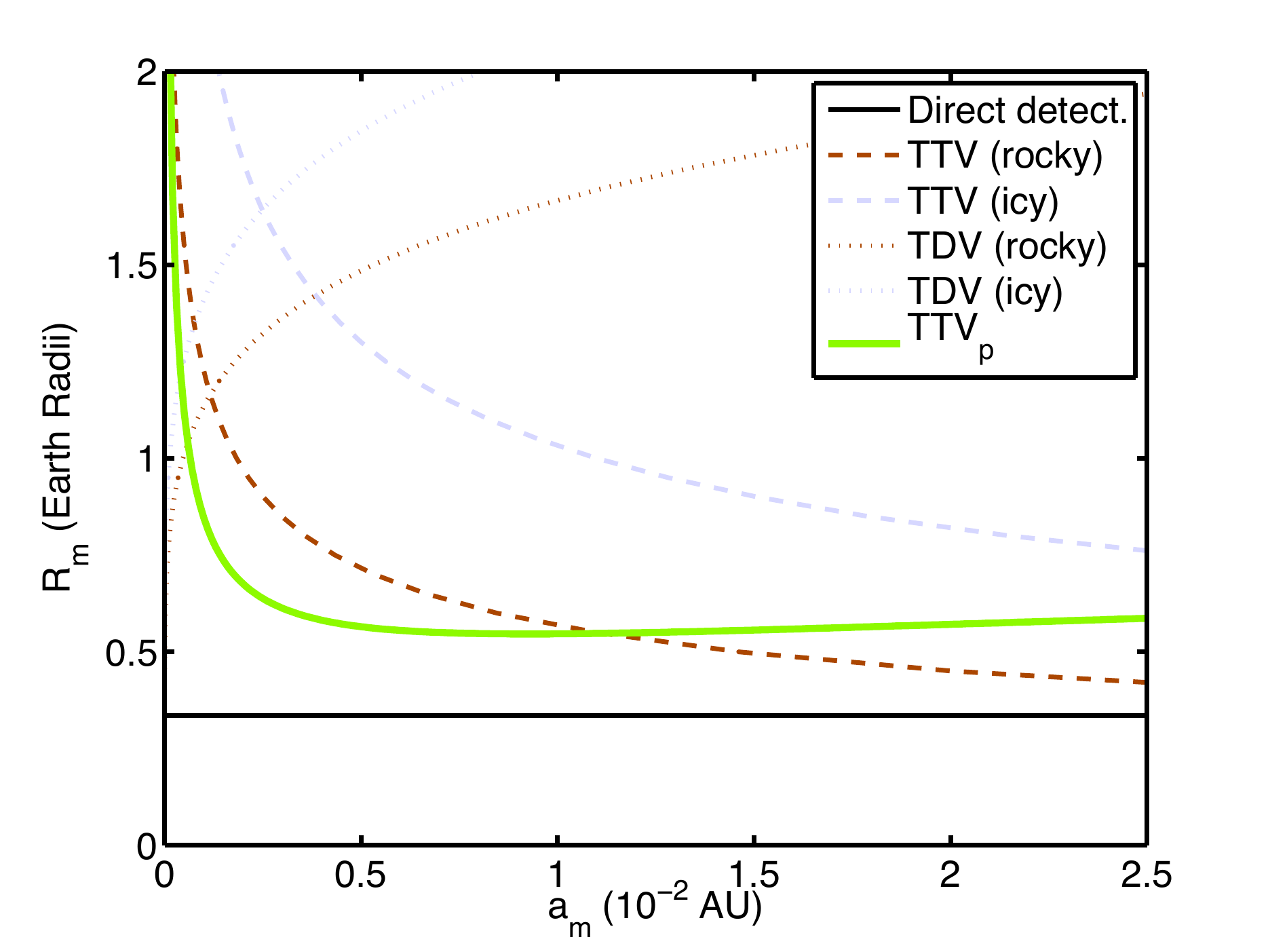}}
     \subfigure[$R_p$=$R_J$, $a_p=1$AU, $\delta_{min} = 0.5 R_{\sun}$.]{
          \label{Updated05NinfThresh}
          \includegraphics[width=.485\textwidth]{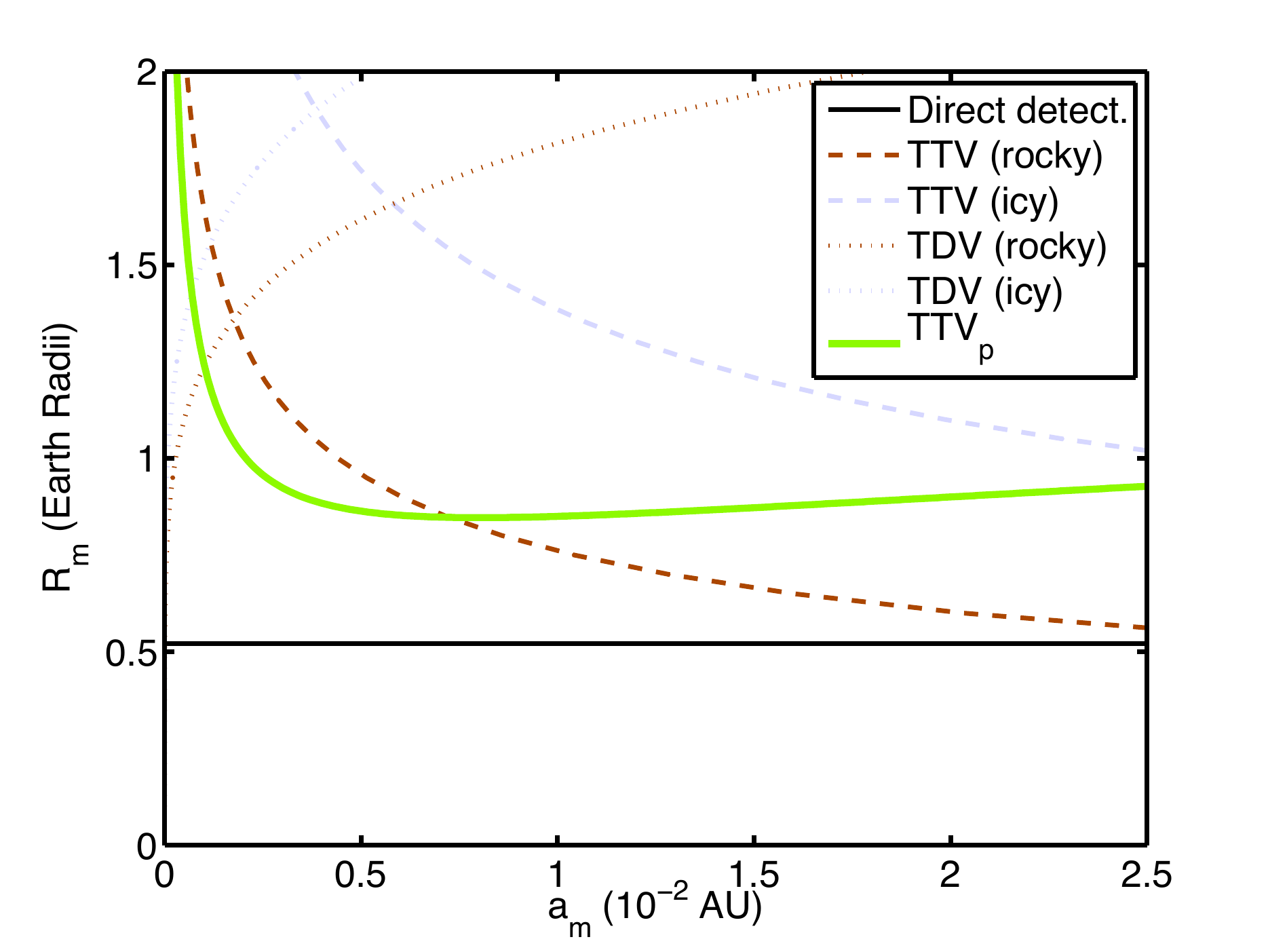}}
 \caption[Figure of the same form as figure \ref{ApproximateMoonDetThresh}, but also showing the TTV$_p$ threshold described by equation \eqref{transit_thresholds_expbehav_TTVpthresh} (green).]{Figure of the same form as figure \ref{ApproximateMoonDetThresh}, but also showing the TTV$_p$ threshold described by equation \eqref{transit_thresholds_expbehav_TTVpthresh}, plotted in green.}
     \label{ApproximateMoonDetThreshTTVp}
\end{figure}

One of the aims of this analysis is to compare the set of moons that can be detected by TTV$_p$ with the set of moons that can be detected using other transit moon detection methods, namely direct detection, barycentric transit timing and transit duration variation.  Approximate expressions for the three sigma detection thresholds associated with these techniques were derived for the case of large $N$ and white photometric noise in section~\ref{Intro_Dect_Moons_Transit} (equations~\eqref{intro_det_moon_DDtransthresh}, \eqref{intro_det_moon_TTVbtransthresh} and \eqref{intro_det_moon_TDVtransthresh}) and are restated below for convenience
\begin{equation*}
R_m = 0.0065 R_s  \frac{1}{N^{1/4}} \left[\frac{\sigma_{L}/L_0}{3.95 \times 10^{-4}} \left(\frac{\Delta t}{1\text{min}} \right)^{1/2}\right]^{1/2} \left(\frac{13\text{hrs}}{T_{tra}}\right)^{1/4},
\end{equation*}
\begin{multline*}
 R_m  = 0.0168R_s \frac{1}{N^{1/6}} \left[\frac{ \sigma_{L}/L_0}{3.95\times 10^{-4}} \left(\frac{\Delta t}{1 \text{min}}\right)^{1/2} \right]^{1/3} \left(\frac{13\text{hrs}}{T_{tra}}\right)^{1/6}, \\
 \times \left(\frac{R_s}{a_m}\right)^{1/3} \left(\frac{\rho_p}{\rho_m}\right)^{1/3} \left(\frac{R_p}{0.1R_s}\right)^{1/2}
\end{multline*}
\begin{multline*}
R_m = 0.0197 R_s  \frac{1}{N^{1/6}} \left[\frac{\sigma_{L}/L_0}{3.95\times10^{-4}} \left(\frac{\Delta t}{1\text{min}}\right)^{1/2} \right]^{1/3} \left(\frac{13 \text{hrs}}{T_{tra}}\right)^{1/2}, \\
\times \left(\frac{a_m}{R_s}\right)^{1/6} \left(\frac{\rho_p}{\rho_{ice}}\right)^{1/6} \left(\frac{\rho_{ice}}{\rho_m}\right)^{1/3}
\end{multline*}
and plotted as a function of $a_m$ in figure~\ref{ApproximateMoonDetThreshTTVp}, which is a recreation of figure~\ref{ApproximateMoonDetThresh}.

We now have the tools to derive an equivalent expression for the case of the photometric transit timing technique.  From equation~\eqref{transit_thresholds_expbehav_whitethresh}, we have that the three sigma detection threshold for the case of photometric transit timing is given by
\begin{multline}
R_m = 0.0066R_s \frac{1}{N^{1/4}} \left[\left(\frac{\sigma_L / L_0}{3.95 \times 10^{-4}}\right)\left(\frac{\Delta t}{1 \text{min}}\right)^{1/2}\right]^{1/2} \left(\frac{13 \text{hrs}}{T_{tra}} \right)^{1/4}  \\
\times  \left(\frac{a_m}{R_s}\right)^{-1/2} \left( 1 + \frac{a_m}{R_s} \right)^{3/4}.\label{transit_thresholds_expbehav_TTVpthresh}
\end{multline}
where the equation has been recast into physical variables for ease of comparison.  This threshold is also shown in green in figure~\ref{ApproximateMoonDetThreshTTVp}.

Through a comparison of equations~\eqref{intro_det_moon_DDtransthresh}, \eqref{intro_det_moon_TTVbtransthresh}, \eqref{intro_det_moon_TDVtransthresh} and \eqref{transit_thresholds_expbehav_TTVpthresh}, and a visual examination of figure~\ref{ApproximateMoonDetThreshTTVp}, a number of similarities and differences between the thresholds corresponding to the four methods can be seen.  In particular the dependance of the threshold on $N$ and $a_m$.  These aspects will be discussed, followed by a short discussion on the comparative effect of correlated noise on each of these methods.

We begin the comparison of the TTV$_p$ detection  threshold with the three thresholds associated with the other methods, by considering the dependance on $N$, the number of transits recorded.  Comparing equations~\eqref{intro_det_moon_DDtransthresh}, \eqref{intro_det_moon_TTVbtransthresh}, \eqref{intro_det_moon_TDVtransthresh} and \eqref{transit_thresholds_expbehav_TTVpthresh} we see that the TTV$_p$ threshold has the same dependance on $N$ as the direct detection threshold as opposed to the $1/N^{1/6}$ dependance shown by barycentric transit timing and transit duration variation (see equations~\eqref{intro_det_moon_TTVbtransthresh} and \eqref{intro_det_moon_TDVtransthresh}).  The physical origin of this dependance is that photometric transit timing and direct detection measure the cross-sectional area of the moon ($\propto R_m^2$) while barycentric transit timing and transit duration variation measure its mass ($\propto R_m^3$).  Consequently, as a result of this dependance on $N$, the thresholds for the case of TTV$_p$ will decrease at the same rate as those for direct detection as the number of recorded transits increases, but more rapidly than those for the two other timing methods.

In addition to the dependance on $N$, the dependance of these thresholds on moon semi-major axis is also of interest.  As can be seen in equations~\eqref{intro_det_moon_DDtransthresh}, \eqref{intro_det_moon_TTVbtransthresh} and \eqref{intro_det_moon_TDVtransthresh}, and as was discussed in section~\ref{Intro_Dect_Moons_Transit}, the moon detection thresholds for the case of direct detection, barycentric transit timing and transit duration variation are independent of $a_m$, decrease with increasing $a_m$ and increase with increasing $a_m$ respectively.  Physically this means that the barycentric transit timing and transit duration variation techniques are optimised to detect long and short period moons respectively, while direct detection is equally as good at detecting all moons.  In comparison, the TTV$_p$ technique is optimised to detect moons with $a_m \approx 2R_s$ (for the case of white noise), and, in addition, cannot be used to detect very close or distant moons.  As the optimal range probed by the other two timing techniques brackets the range of semi-major axis for which this technique works, and recalling that the direct detection threshold has not yet been extended to deal with red noise, it can be seen that the TTV$_p$ technique could be a complimentary technique.

Finally, in addition to having different dependancies on $N$ and $a_m$, the direct detection, barycentric transit timing, transit duration variation and photometric transit timing techniques are also affected to differing extents by correlated noise.  Prior to this thesis, the only work to investigate the effect of correlated noise on moon detection was that of \citet{Kippingetal2009}.  They suggested that the transit duration variation technique was relatively immune to the effects of correlated noise.  In comparison, in chapter~\ref{Trans_TTV_Noise} of this thesis it was found that correlated noise leads to a substantial decrease in moon detectability using the the TTV$_p$ technique.  Consequently, this aspect needs to be kept in mind before applying this technique to real data.

So, in summary, the direct detection, barycentric transit timing, transit duration variation and photometric transit timing techniques all probe different portions of parameter space.  In particular, the direct detection and photometric transit timing  thresholds will decrease more rapidly with increasing $N$ than those of barycentric transit timing or transit duration variation as a result of the different physical quantities that they measure.  In addition, the TTV$_p$ method has a range of semi-major axes ($a_m \approx 2R_s$) for which it is optimised to detect moons, as opposed to barycentric transit timing and transit duration variation which are optimised to detect long and short period moons respectively.  Finally, the effect of correlated noise may affect these results.  For example, as shown in chapter~\ref{Trans_TTV_Noise} the timing noise on $\tau$ increases dramatically if the light curve is contaminated with correlated timing noise.  Consequently for the case of a host star with a low number of transits and correlated photometric noise, barycentric transit timing and transit duration variation, which are less affected by red noise, may be more optimal to use, despite their unfavourable dependance on $N$.

Now that the TTV$_p$ threshold has been discussed in isolation and  put into the context of the other transit detection methods presented in the literature, we are in the position of being able to discuss and understand realistic detection thresholds.  With this in mind numerically calculated detection thresholds for the case where $N$ is small will be discussed in the next section.


\section{Numerically calculated TTV$_p$ moon detection thresholds}\label{Trans_Thresholds_MC}

\begin{figure}
     \centering
     \vspace{-0.2cm}
     \subfigure[$M_p$=$10 M_J$, $a_p=0.2$AU.]{
          \label{TransitThresh10MJ02AUcc}
          \includegraphics[width=.315\textwidth]{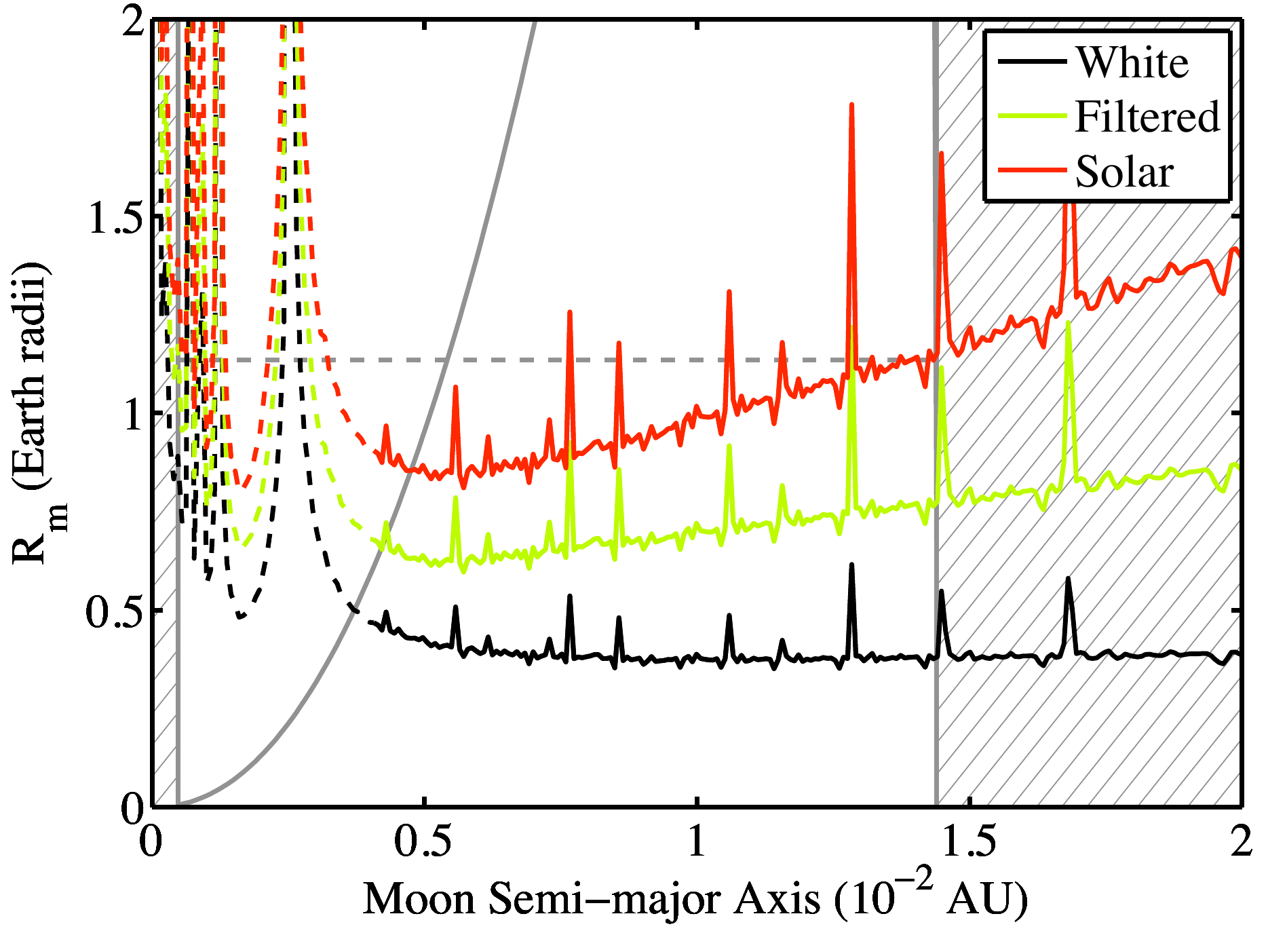}}
     \subfigure[$M_p$=$10 M_J$, $a_p=0.4$AU.]{
          \label{TransitThresh10MJ04AUcc}
          \includegraphics[width=.315\textwidth]{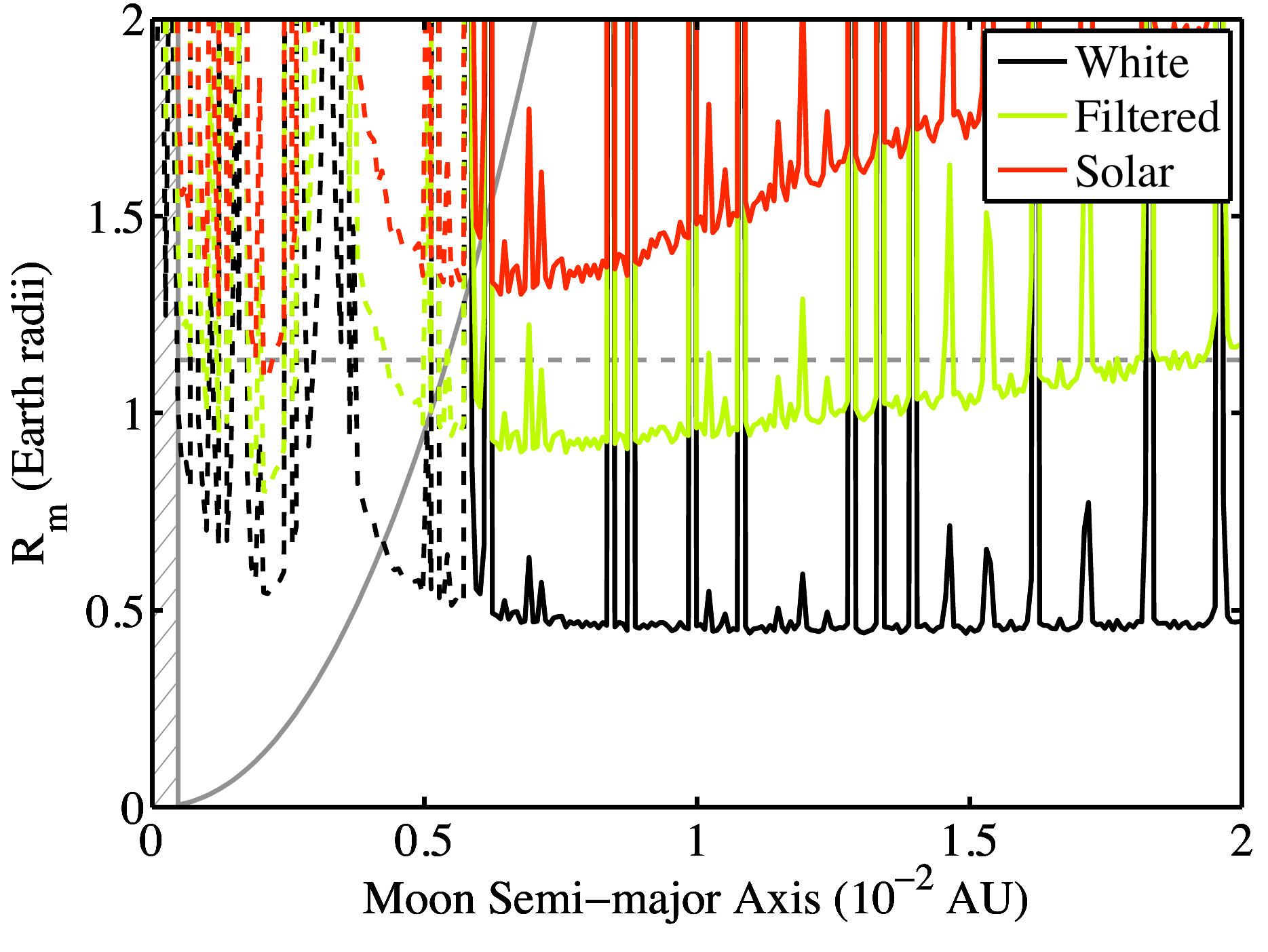}}
     \subfigure[$M_p$=$10 M_J$, $a_p=0.6$AU.]{
          \label{TransitThresh10MJ06AUcc}
          \includegraphics[width=.315\textwidth]{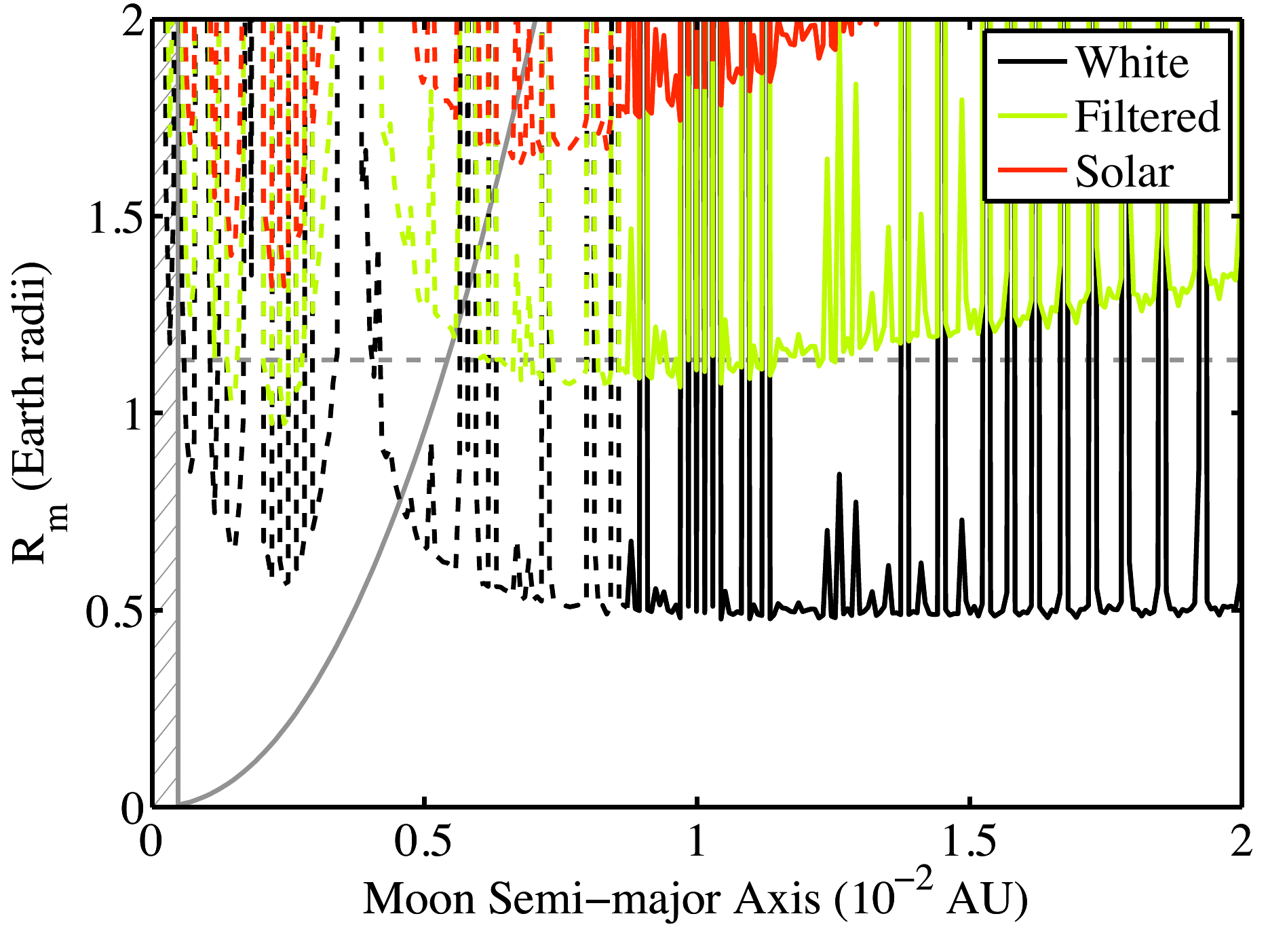}}\\ 
          \vspace{-0.2cm}
     \subfigure[$M_p = M_J$, $a_p=0.2$AU.]{
          \label{TransitThresh1MJ02AUcc}
          \includegraphics[width=.315\textwidth]{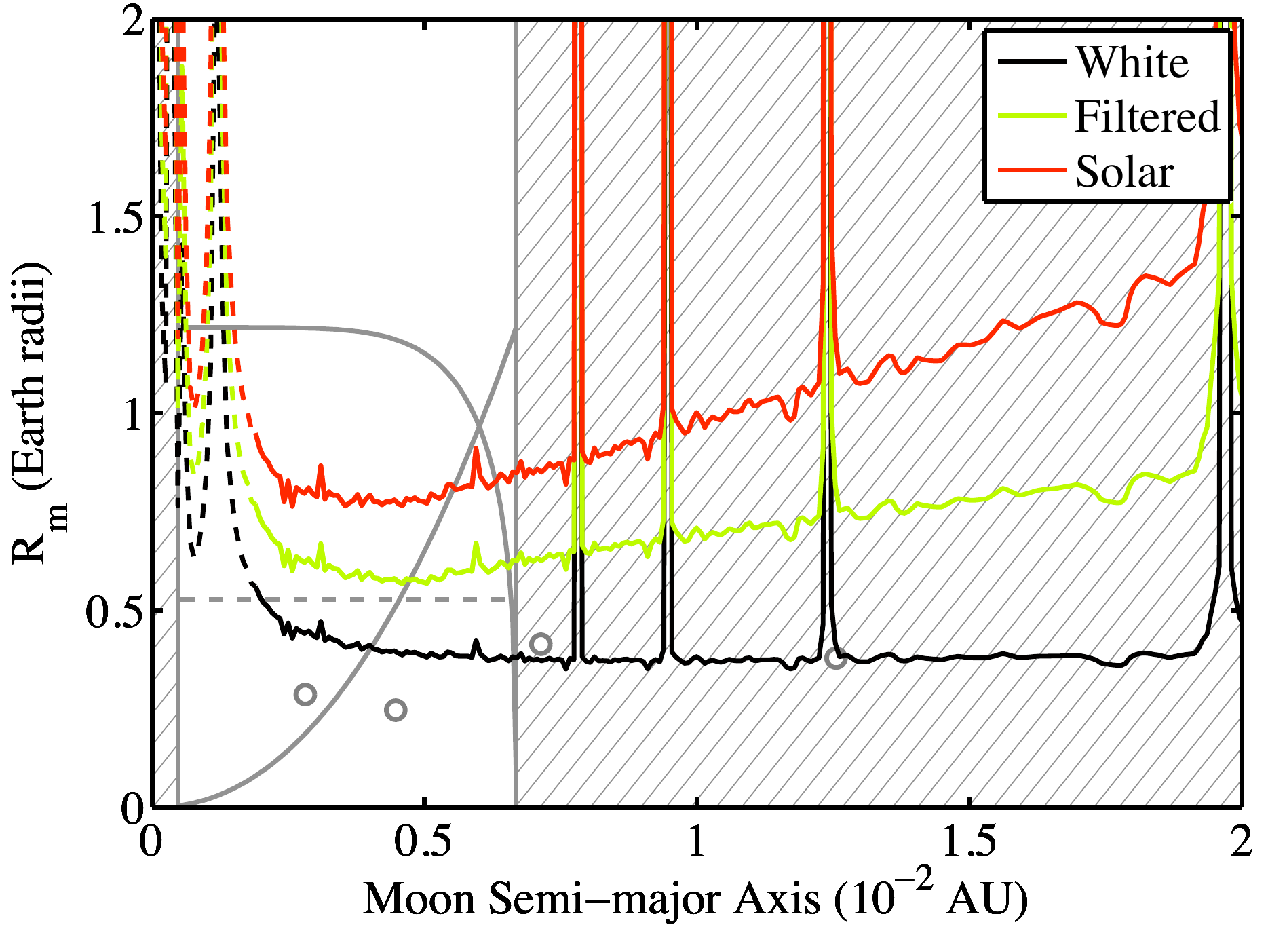}}
      \subfigure[$M_p = M_J$, $a_p=0.4$AU.]{
          \label{TransitThresh1MJ04AUcc}
          \includegraphics[width=.315\textwidth]{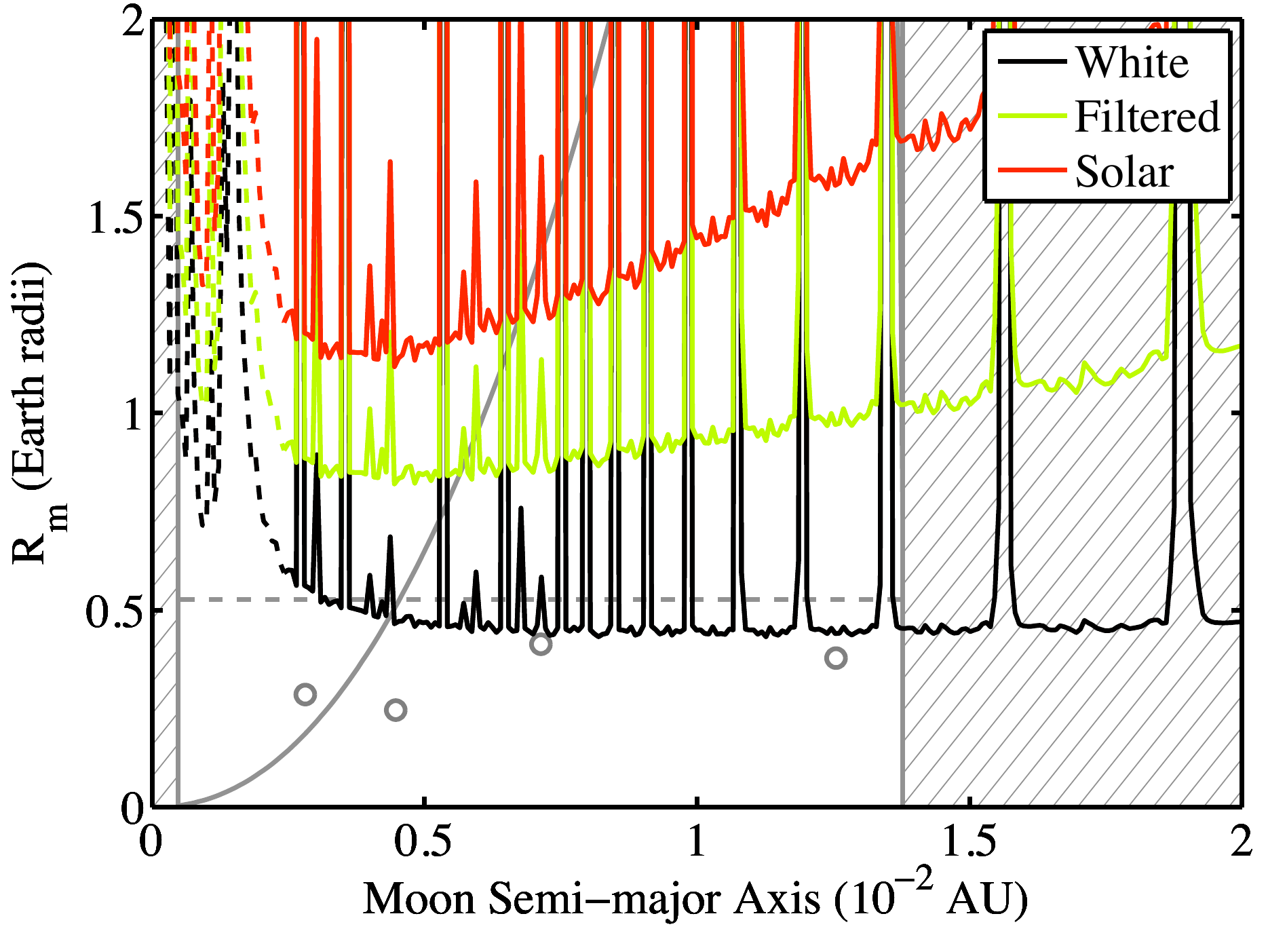}}
     \subfigure[$M_p = M_J$, $a_p=0.6$AU.]{
          \label{TransitThresh1MJ06AUcc}
          \includegraphics[width=.315\textwidth]{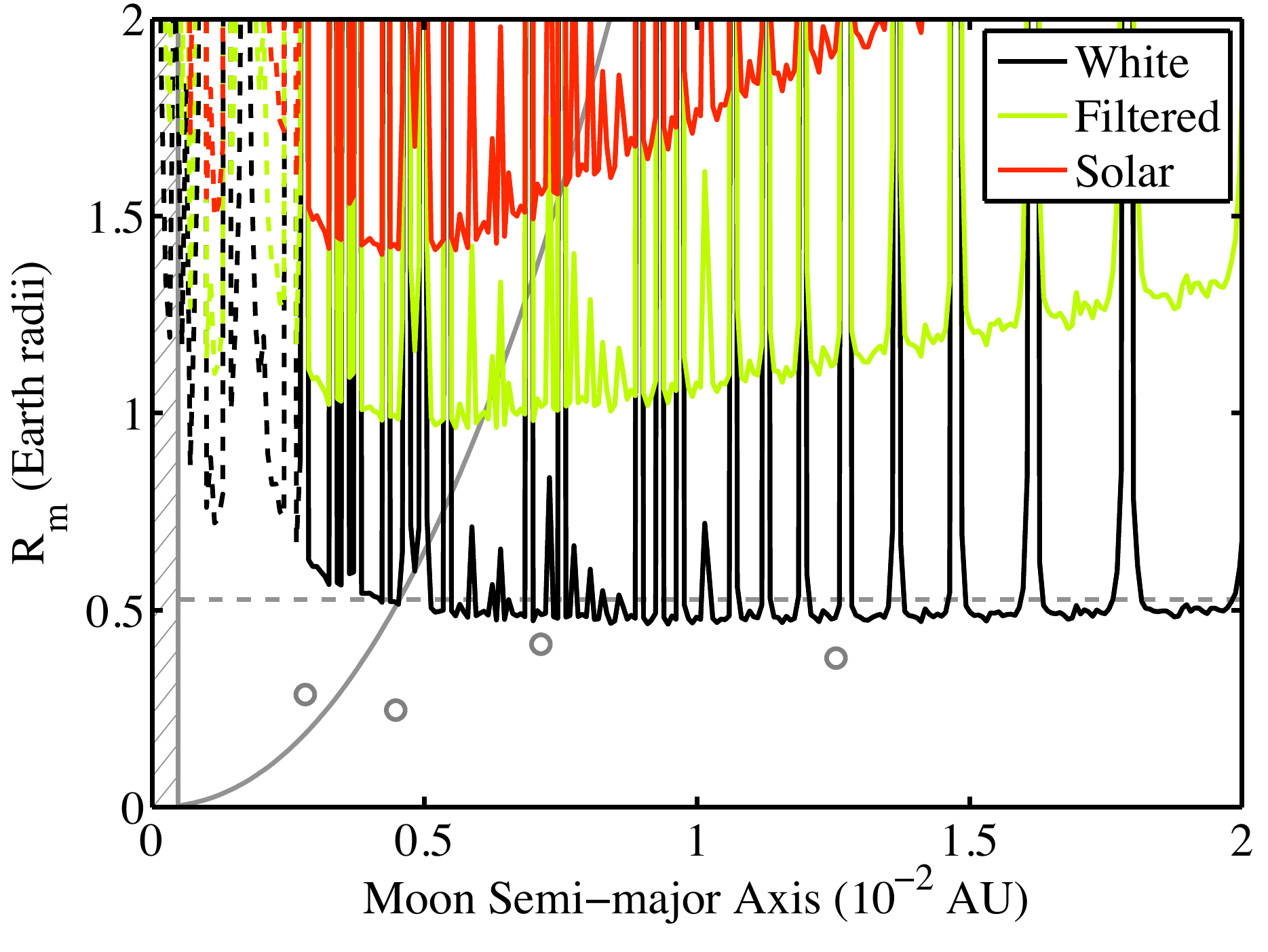}}\\ 
          \vspace{-0.2cm}
     \subfigure[$M_p = M_U$, $a_p=0.2$AU.]{
          \label{TransitThresh1MU02AUcc}
          \includegraphics[width=.315\textwidth]{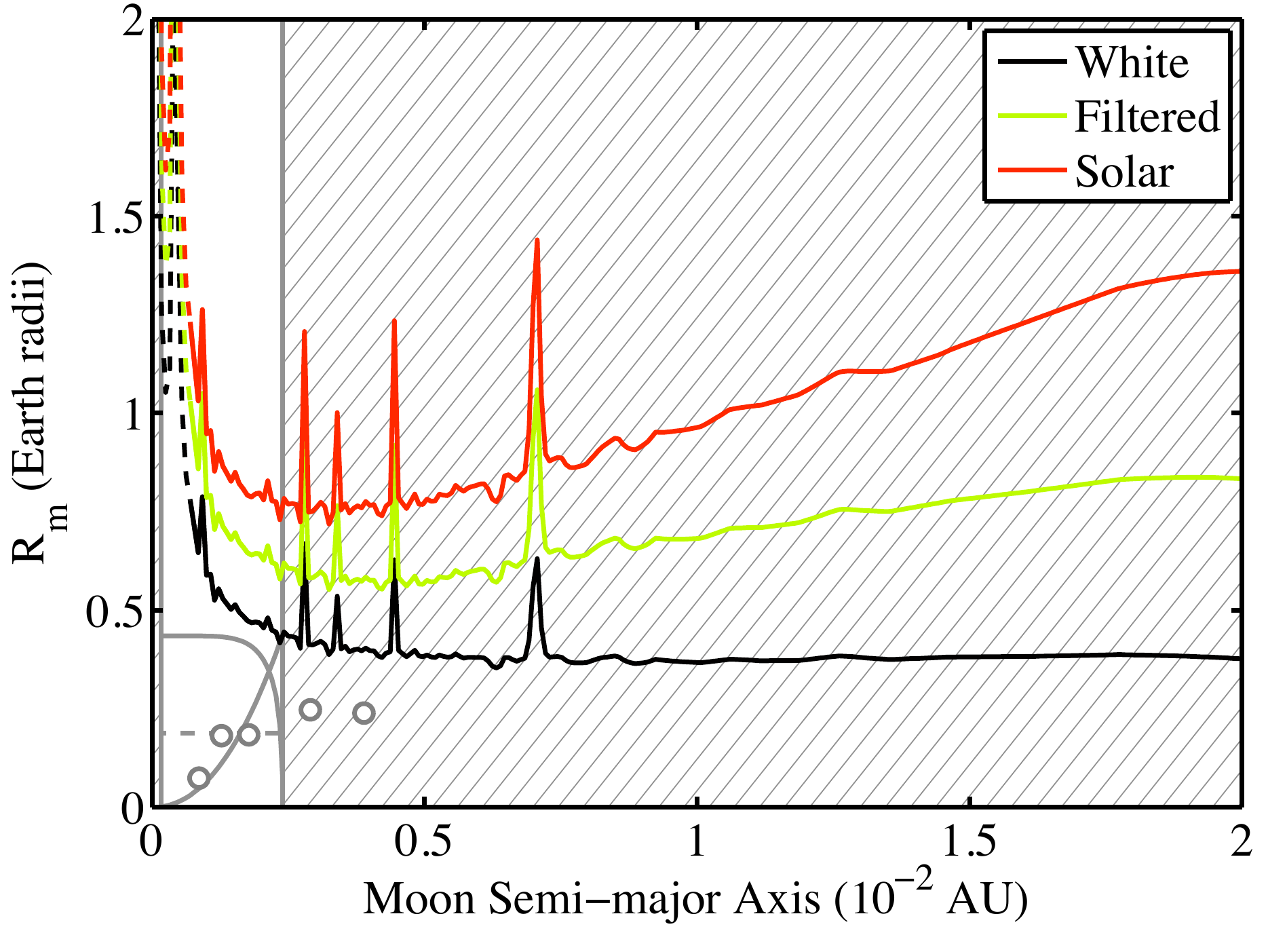}}
     \subfigure[$M_p = M_U$, $a_p=0.4$AU.]{
          \label{TransitThresh1MU04AUcc}
          \includegraphics[width=.315\textwidth]{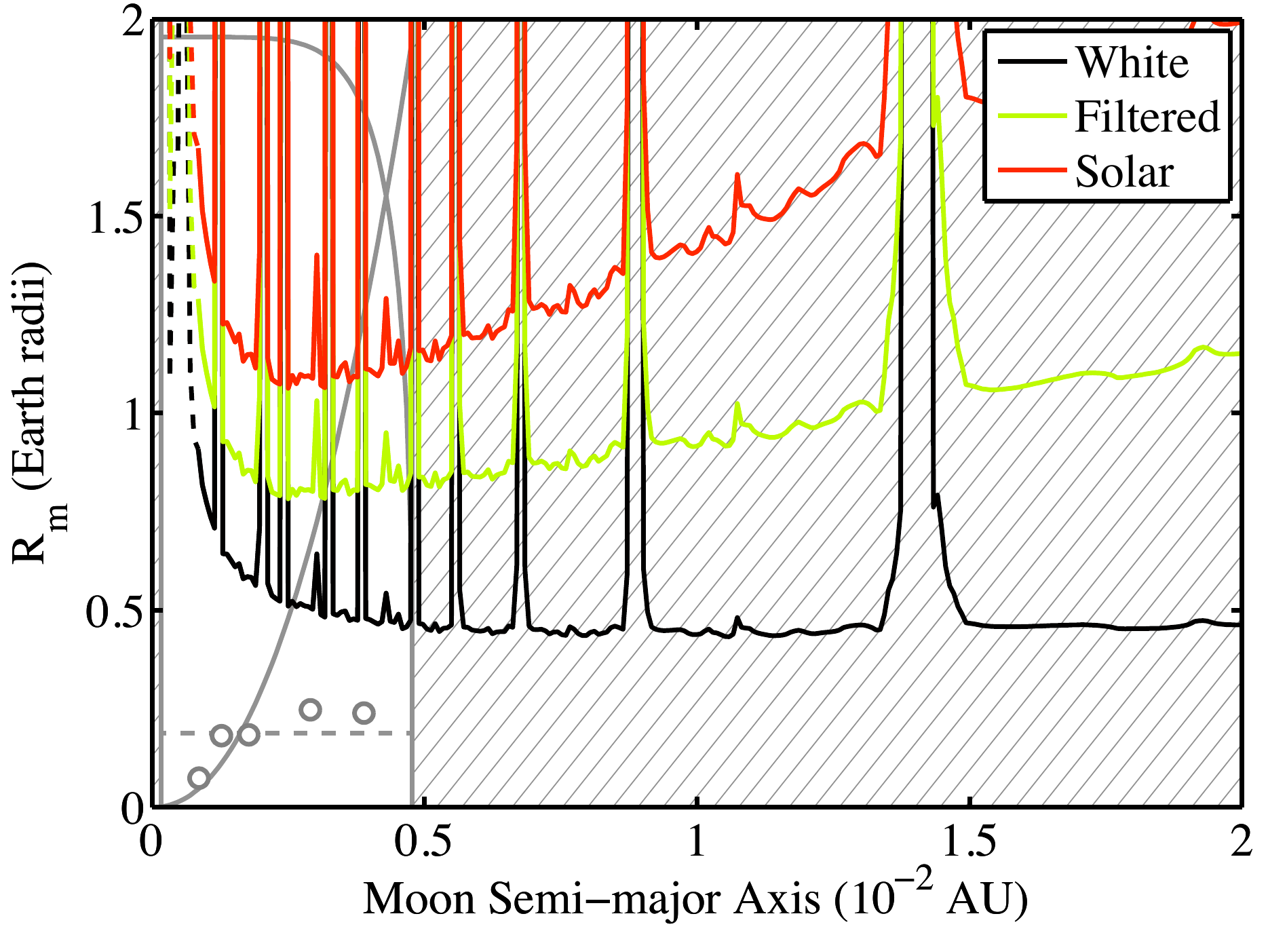}}
      \subfigure[$M_p = M_U$, $a_p=0.6$AU.]{
          \label{TransitThresh1MU06AUcc}
          \includegraphics[width=.315\textwidth]{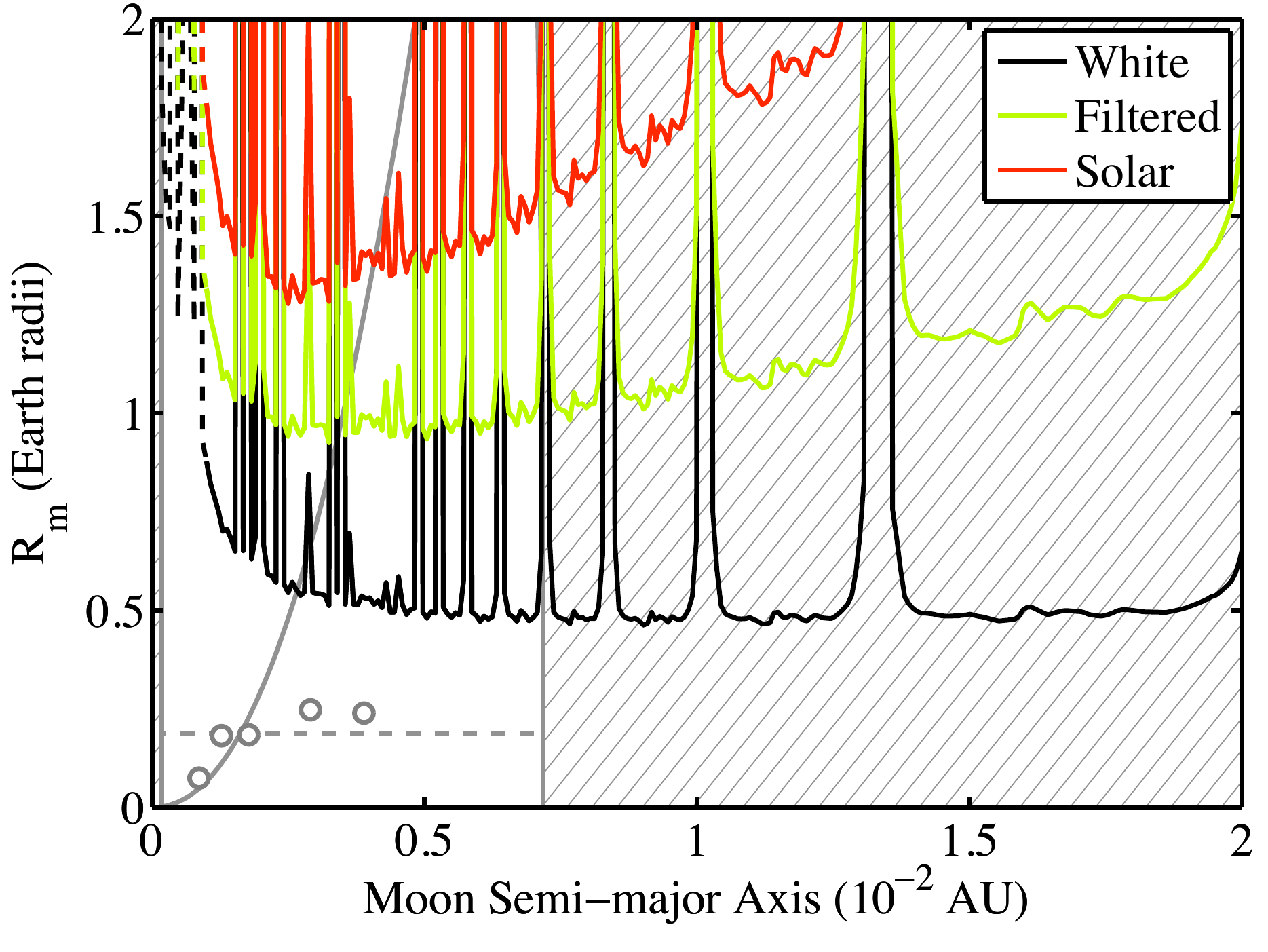}}\\ 
          \vspace{-0.2cm}
     \subfigure[$M_p = M_{\earth}$, $a_p=0.2$AU.]{
          \label{TransitThresh1ME02AUcc}
          \includegraphics[width=.315\textwidth]{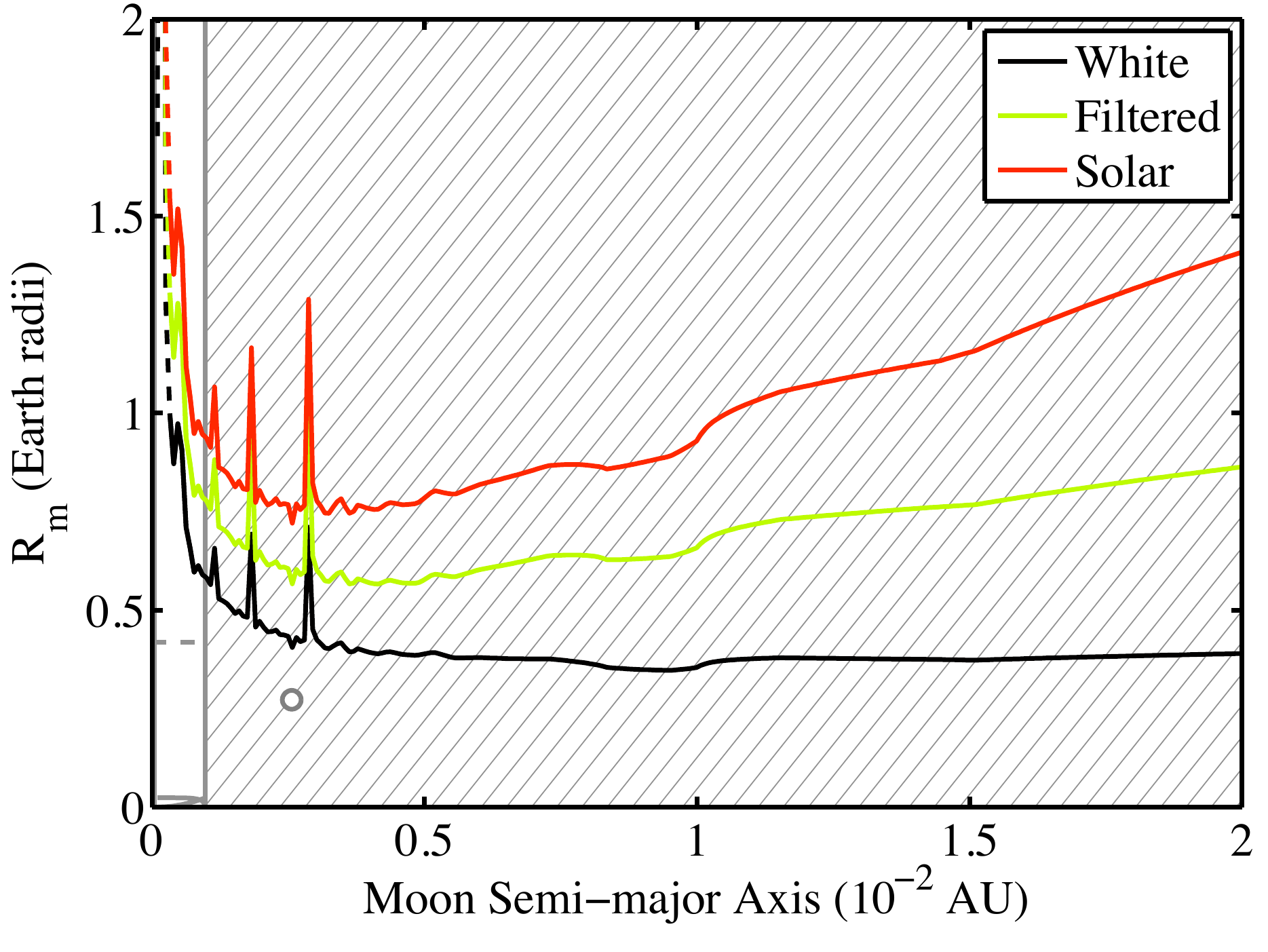}}
     \subfigure[$M_p = M_{\earth}$, $a_p=0.4$AU.]{
          \label{TransitThresh1ME04AUcc}
          \includegraphics[width=.315\textwidth]{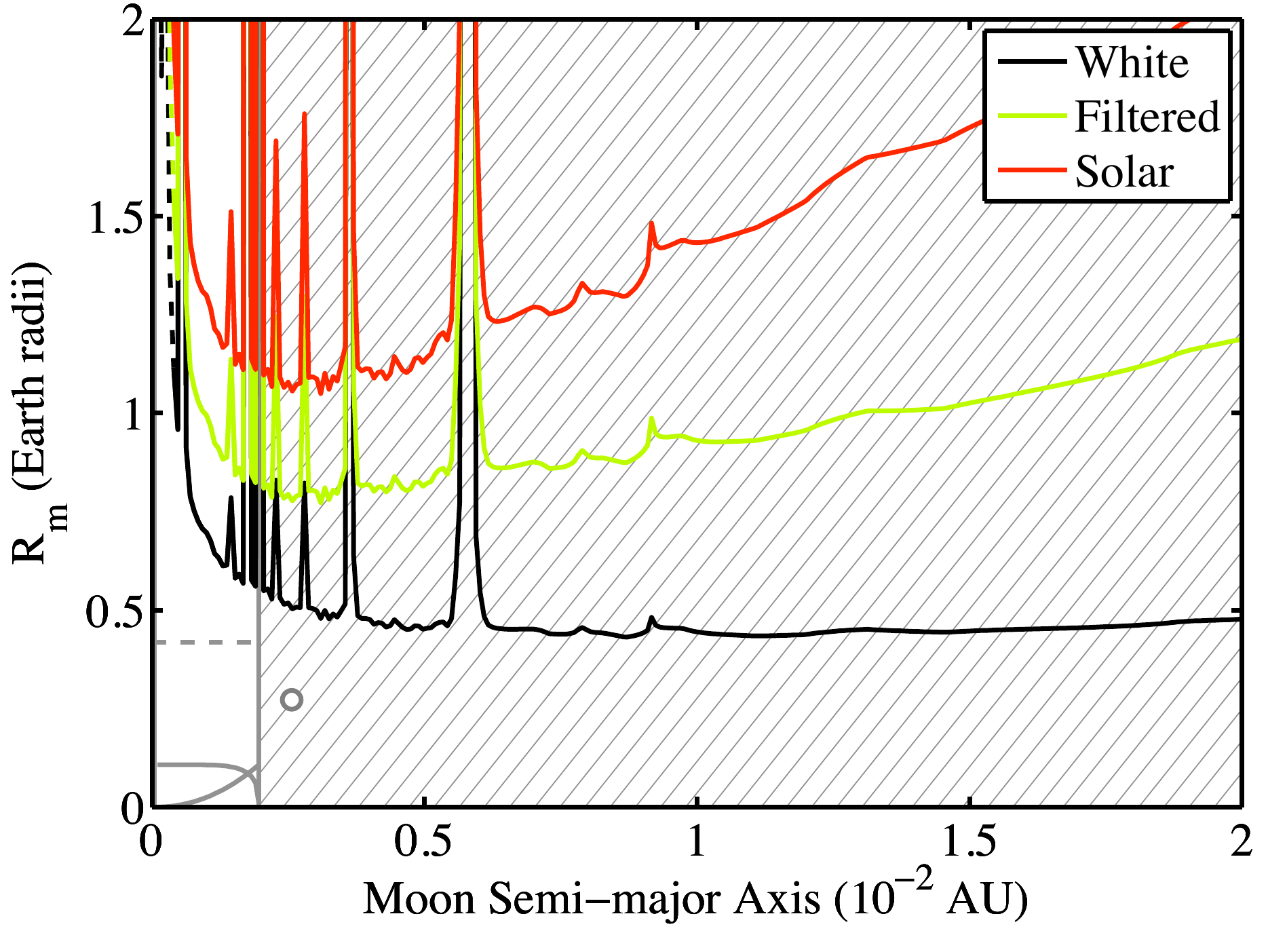}}
     \subfigure[$M_p = M_{\earth}$, $a_p=0.6$AU.]{
          \label{TransitThresh1ME06AUcc}
          \includegraphics[width=.315\textwidth]{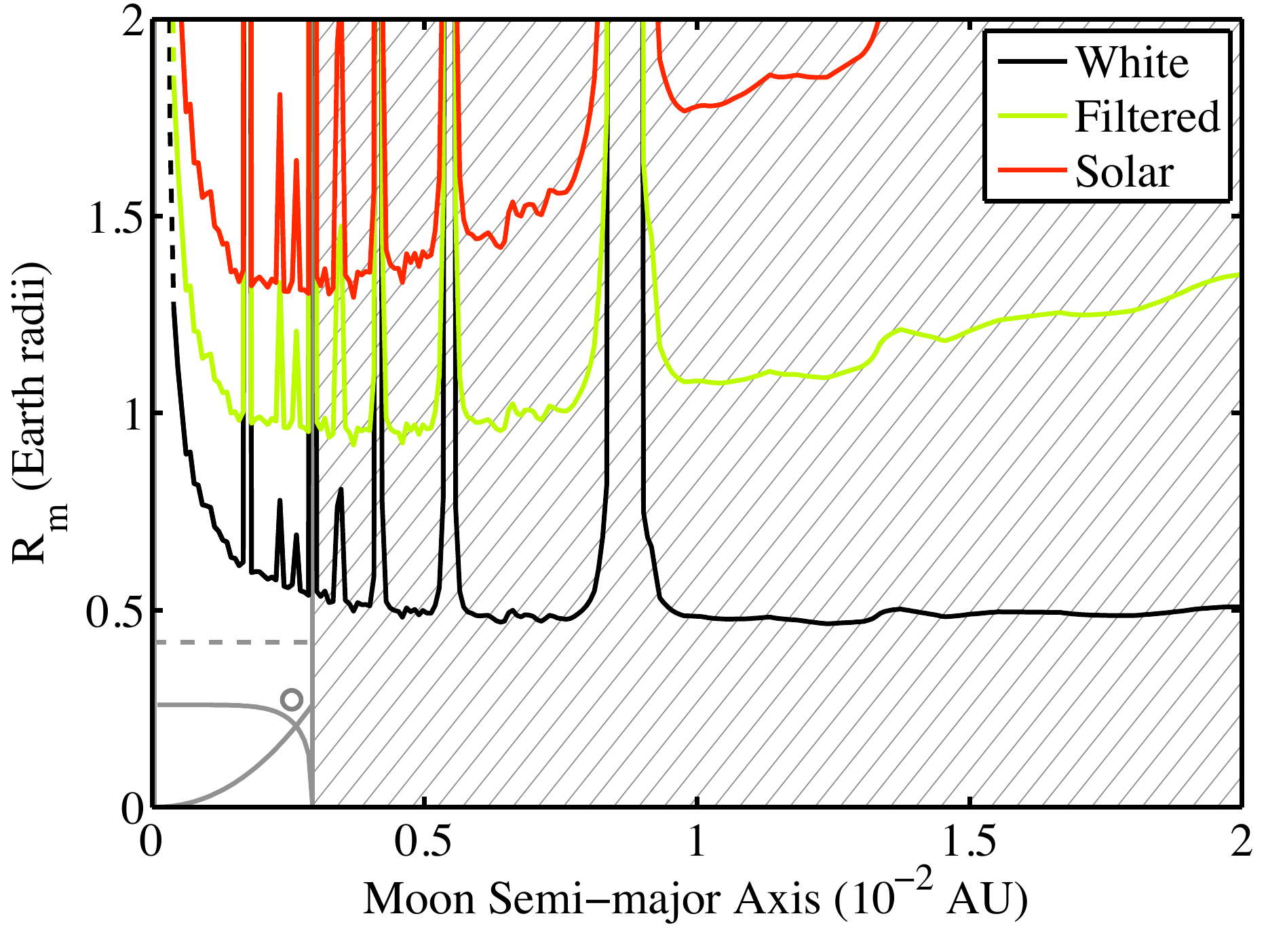}} 
          \vspace{-0.4cm}
     \caption[Plot of the 99.7\% detection threshold for the case where the orbit of the planet is circular and aligned to the line-of-sight for the 12 cases discussed in section~\ref{Trans_Thresholds_Parameters}.]{Plot of the 99.7\% detection threshold for the case where the orbit of the planet is circular and aligned to the line-of-sight for the 12 cases discussed in section~\ref{Trans_Thresholds_Parameters}.  The predicted maximum mass of a naturally formed moon is shown as a horizontal dashed line, while the minimum and maximum semi-major axis that a moon undergoing either inward or outward orbital evolution would have achieved, assuming a system lifetime of 5Gyr, are shown by solid grey lines.  For the case where the planet has the mass of Jupiter, Uranus or the Earth, the radii and semi-major axes of the satellites of Jupiter, Uranus or the Earth are also plotted (grey circles).   Finally, the interior of the planet and the three-body unstable region are hatched.}
     \label{MCThresholdsAligned}
\end{figure}

Now that we have the tools available to calculate and analyse detection thresholds for TTV$_p$, we are in a position to calculate thresholds for the more realistic case of a finite number of transits.  While we have equation~\eqref{transit_thresholds_method_thresholddef}, the expression for the statistic $2\log\Lambda$, defining the location of the thresholds, the process required to calculate the thresholds is not trivial and consequently we begin with a brief discussion of this aspect.  Then we move on to analysing and discussing the calculated thresholds.  As discussed in section~\ref{Trans_Thresholds_Parameters} we will perform this investigation in three stages.  First, we will investigate the simplest case, that of a planet on a circular orbit passing across the central chord of its host star (i.e. $\delta_{min} = 0$).  This case will then be used as a comparison case for the investigations into the effects of slight inclination of the planetary orbit ($\delta_{min} \ne 0$) and eccentricity in the planetary orbit on moon detectability.  We begin by summarising the method used for generating thresholds.

\subsection{Method for calculating Monte Carlo Thresholds}\label{Trans_Thresholds_MC_Method}

For this thesis we define the threshold as the line separating the region of parameter space where ``most" moons are ``detectable" from the region of parameter space where ``most" moons are not ``detectable", where ``detectable" is defined by whether or not the value of $2\log\Lambda$ calculated using equation~\eqref{transit_thresholds_method_thresholddef} lies above a critical value.  So, in order to practically calculate these thresholds, we need a definition of  ``most", as well as a way of determining if a system is ``detectable", which is not too computationally expensive.  These two choices will be discussed in turn.

We begin by discussing and defining the term ``most".  The selection of the definition of ``most" is not a trivial issue for two main reasons.  First, as a result of statistical fluctuations, there is a non-zero probability that a moon will be ``detected" when there is no moon, for example, for the thresholds presented in this chapter this probability is 0.3\%, so the limit cannot be set too low.  In addition, as mentioned in section~\ref{Trans_TTV_Signal_CC_PropSig} and in the footnote on page \pageref{phifootnote}, moon detectability depends on $\phi$, for example, the width of the non-detection spikes is a function of $\phi$.  As a result, selecting a definition of ``most" which is too high will effectively result in a measure of the detectability of the least detectable moons with that radius and semi-major axis (see the footnote on page \pageref{phifootnote} for an extreme example).  As a compromise between these two effects, it was decided that the point at which ``most" moons were detected corresponded to the point at which 50\% of the moons would have been detected.  Now that the term ``most" has been defined, the method used to determine $2\log\Lambda$, and thus if a moon has been ``detected" will be briefly outlined.

Following from the above discussion, the most intuitively obvious approach to determining $2\log\Lambda$, and thus  constructing thresholds would be to discretize the moon radius semi-major axis plane into a grid of points, simulate a statistically representative number of $\tau$ sequences for each point, determine the median $2\log\Lambda$ value for each point and draw the threshold where this value is equal to the critical value.  However, this approach is not feasible as it is computationally expensive, consequently a different approach was used.  First, three non-physical grids were set up corresponding to the cases of $N=9$ ($a_p = 0.6$AU), $N = 14$ ($a_p = 0.6$AU) and $N = 40$ ($a_p = 0.2$AU), with the aim of transforming the results to generate the thresholds.  Instead of discretizing on moon radius, the quantity $A/\sigma_\epsilon$ was used (ranging from 0 to 8) and instead of discretizing on moon semi-major axis, the quantity $\omega$ was used (ranging from 0 to $2\pi$).  In addition, for these models $t_0$ and $T_p$ were both set to 1 (such that they were the same order of magnitude as $A/\sigma$).  For each grid point a statistically significant number of realisations of $\tau$, defined as
\begin{equation}
\tau_j = t_0 + jT_p + \frac{A}{\sigma_\epsilon}\cos(\omega j + \phi) + \frac{\epsilon_j}{\sigma_\epsilon},
\end{equation}
were calculated.  In particular, note that $\epsilon_j/\sigma_\epsilon$ is a normally distributed variable with standard deviation 1 and mean 0, and $\phi$ is given by a random variable which is uniformly distributed between 0 and $2\pi$.  For this application 51 realisations were used.\footnote{The behaviour of moon detectability changes as $\phi$ changes by $\pi/2$ (see footnote on page \pageref{phifootnote} for an example).  Consequently, in order to categorise the behaviour, the number of simulations must be substantially larger than $(\pi/2)/2\pi = 4$.}

For each realisation, the value of $2\log\Lambda$ was calculated and recorded.  Finally the values were sorted by size and the 26$^{th}$ (the middle value) was selected.  This data was then transformed to a grid of 25 by 300 points in moon radius and semi-major axis space for realistic values of $t_0$ and $T_p$ using the method described in appendix~\ref{App_transformFitParams}.  While this method is advantageous in that it dramatically decreases the computation time required to produce each plot, it assumes that $\Delta \tau$ is well approximated by a sinusoid\footnote{Recall from sections~\ref{Transit_Signal_Method} and \ref{Trans_TTV_Signal_CC_Form_smallB}, that $\Delta \tau$ is only well-described by a sinusoid for the case where $v_m/v_{tr}$ is small and where the change in velocity during transit is not significant.  Consequently, the thresholds calculated using this method may be inaccurate for the case where $v_m/v_{tr}$ is large or the transit duration is comparable to, or larger than the moon's orbital period.  To give a feel for where this problem may arise, thresholds in the region where $v_m/v_{tr}$ is large ($v_m/v_{tr} > 0.66$), or the moon noticeably accelerates during transit ($T_{tra} > 1/4T_m $) a dashed line style as opposed to a solid line style is used.} for all values of moon semi-major axis.  Taking this approximation into account and continuing, to use this information to calculate the thresholds to a given significance, the probability distribution of $2\log\Lambda$ for the case where there is no moon is required.

For the case where the fitting model is linear, or, the number of transits, $N$, is large, the distribution of $2\log\Lambda$ for the case where there is no moon would be given by a $\chi^2$ distribution with three degrees of freedom.  However, as mentioned previously, the fitting model is not linear in the parameters $\omega$ and $\phi$.  Consequently, to determine the form of the distribution of $2\log\Lambda$, simulations were run for the case of $N = 9, 14$ and $40$ for the case where there was no moon.  For each simulation 10000 realizations of $\tau$ were generated for the case where $A=0$ and analysed to determine the shape of the probability distribution of $2\log\Lambda$.  This process was then repeated six times in order to gain an understanding of the variability of the calculated critical values.

For this work, the 99.7\% threshold is used (recall that for a standard normal distribution, the probability of being three sigma away from the mean is 100\% - 99.7\% = 0.3\%).  From these simulations it was found that 99.7\% of simulated $\tau$ sequences had $2\log\Lambda$ values less that $14.99\pm0.27$, $15.00\pm0.26$ and $15.33\pm0.36$ for the cases of $N = 9$, 14 and 40 respectively.  As the confidence limits on these values all overlap it was decided to use the value 15 to generate all the 99.7\% thresholds shown in this chapter.   However, while this is the value used for this work, different thresholds can be selected (see appendix~\ref{App_ExtraThresholds} for the 68.3\% and 95.4\% thresholds)

Now that the method for calculating the thresholds has been briefly described, the thresholds themselves can be calculated and analysed.  We begin with the case of circular planet and moon orbits, for the case where $\delta_{min} = 0$.

\subsection{Circular coplanar orbits}

We begin with the simplest configuration of planet and moon orbits, namely the case where both the planet and moon have circular orbits with the planet and moon transit the central chord of the star.  Using the method discussed above, the 99.7\% detection thresholds were calculated assuming white, realistic and filtered realistic photometric noise, for the 12 cases discussed in section~\ref{Trans_Thresholds_Parameters} (see figure~\ref{MCThresholdsAligned}).  These thresholds will be discussed in terms the dependance of the threshold on moon semi-major axis, dependance of the threshold on planet semi-major axis, and the location of the threshold with respect to moon formation and stability limits.

\subsubsection{Dependance of threshold on moon semi-major axis}

As discussed in section~\ref{Trans_Thresholds_ExpBehav}, we expect the detection threshold to have a lop-sided U-shape with a minimum at $a_m = 2 R_{\sun}$ for the case of white photometric noise, $a_m = R_{\sun}$ for the case of filtered noise, and between $1/2 R_{\sun}$ and $R_{\sun}$ for the case of realistic noise.  Noting that $R_{\sun} = 0.00465AU$, it can be seen from figure~\ref{MCThresholdsAligned}, that for the case where the minimum is well described by this analysis i.e. is not dashed, these approximate relations accurately describe the shape of the thresholds, especially for the case of filtered and solar photometric noise.  In addition, we expect that the threshold is decorated with a set of non-detection spikes with width and spacing proportional to $a_m^{2.5}$.  Such spikes can be clearly seen in the calculated thresholds for large values of $a_m$, for example for $a_m$ larger than $1.5\times10^{-2}$AU in figure~\ref{TransitThresh10MJ06AUcc}.  For smaller values of $a_m$ the regular pattern is broken as a result of finite numerical resolution, that is, the pattern is disrupted when the distance between neighbouring spikes becomes comparable or smaller than the discretisation used to construct the threshold.

\subsubsection{Dependance of threshold on planet semi-major axis}

As can be seen from figure~\ref{MCThresholdsAligned}, moons are more detectable around short period planets.  For example, for the case of filtered noise, the minimum radius of a detectable moon is a little over one Earth radius for the case of a planet at 0.6AU, compared to a little over half an Earth radius for a planet at 0.2AU.  This trend is a result of two main factors.  First, the shorter transit durations, and thus shorter observing durations, associated with short period planets result in a decrease in the amplitude of the timing noise.  This effect is particularly apparent for the case of realistic stellar photometric noise as a result of the superlinear relationship between $\epsilon_j$ and $T_{obs}$ (see figure~\ref{RedNoiseCompare}).  Second, planets with smaller semi-major axes undergo more transits ($N_{tra} \propto a_p^{-3/2}$) which counteracts the decrease in amplitude of $\Delta \tau$ ($\Delta \tau \propto a_p^{1/2}$) and results in a net increase in moon detectability.

\subsubsection{Comparison with formation and stability limits}

In addition to understanding the thresholds in isolation, it is also useful to view them in the context of the set of moons which are physically realistic, that is, the set of moons which are able to form, and are orbitally stable for the lifetime of the system (see chapter~\ref{Intro_Moons_Const} for an overview).  In each of figures~\ref{TransitThresh10MJ02AUcc} through to \ref{TransitThresh1ME06AUcc} the set of moons which are likely to form and are orbitally stable is indicated in a number of different ways.  First, for each planet where a Solar System analog exists, the moons of that analog are shown as grey circles, for example for the case of a one Jupiter mass host, the four Gallilean satellites are shown.  Second the maximum moon mass limit presented in the literature (see sections~\ref{Intro_Moons_Form_Impact} and \ref{Intro_Moons_Form_Disk}) is shown as a dashed grey line.  Third, to indicate the effect of orbital evolution, the two filled grey lines show the semi-major axis that the innermost and outermost stable moons would have achieved given a system lifetime of 5Gyr (see section~\ref{Intro_Moons_Stab_Lims}).  Finally, the region which is  three-body unstable is hatched.

As moon formation and stability processes depend strongly on the host planet mass, the cases of the four different host planet masses will be discussed in turn.  For the case a the ten Jupiter mass host planet, physically realistic moons are detectable the the case of white and filtered noise for all planet semi-major axes of investigated.  In addition, for the case where the planetary semi-major axis is 0.2AU, physically realistic moons could be detected even for the case of unfiltered noise.  This favourable situation is a result of the linear dependance of the maximum moon mass on planetary mass predicted by \citet{Canupetal2006}.  For the case of a one Jupiter mass host, physically realistic moons are again detectable, but only for the case of white photometric noise.  Interestingly, by comparing the threshold for the case of 0.2AU and 0.4AU, it can be seen that the detection of a Ganymede analog may be possible.  In comparison, for the case of a one Uranus mass host planet, while stable moons are detectable, no moons predicted to form according to the \citet{Canupetal2006} model are detectable at either the 99.7\% level or even the 68.3\% level (see appendix~\ref{App_ExtraThresholds}).  Finally, for the case of Earth-like planets, neither stable or physically realistic moons are detectable.  Consequently, it may be possible to detect moons of large gas giants, and unexpectedly large moons of smaller gas giants, but it is not possible to detect moons of terrestrial planets as a result of formation and stability constraints.

\begin{figure}
     \centering
     \vspace{-0.2cm}
     \subfigure[$M_p$=$10 M_J$, $a_p=0.2$AU.]{
          \label{TransitThresh10MJ02AUInc}
          \includegraphics[width=.315\textwidth]{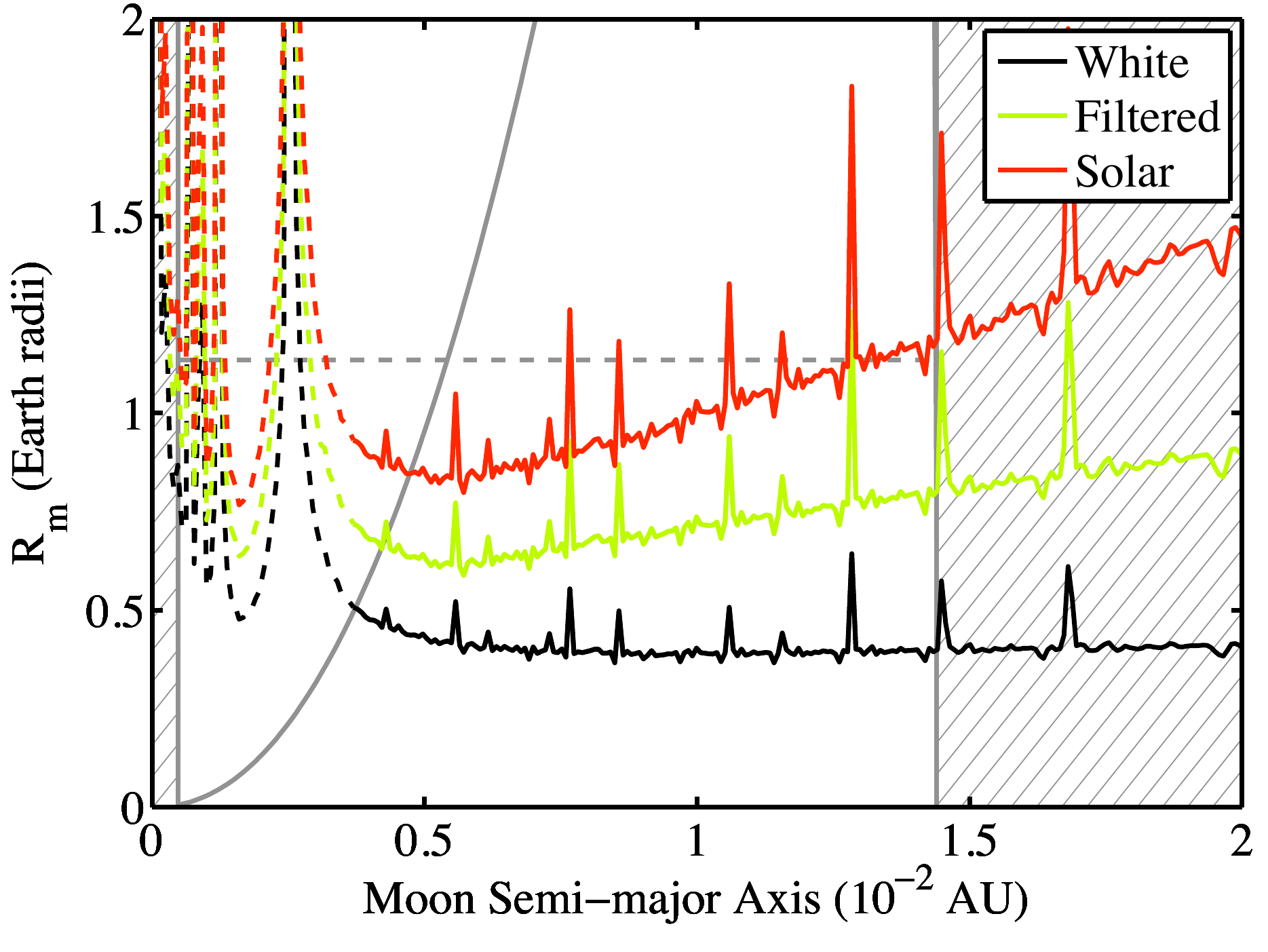}}
     \subfigure[$M_p$=$10 M_J$, $a_p=0.4$AU.]{
          \label{TransitThresh10MJ04AUInc}
          \includegraphics[width=.315\textwidth]{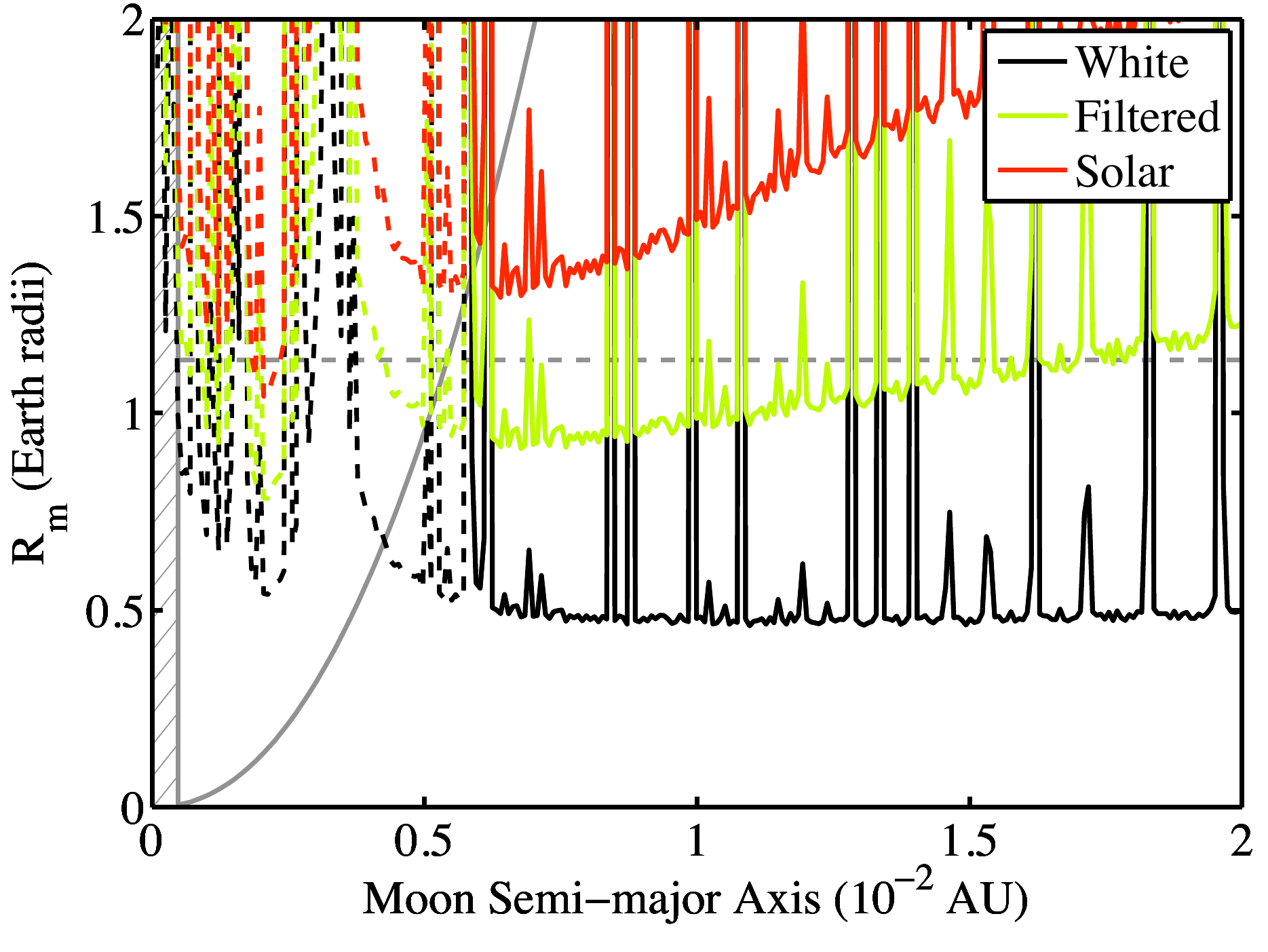}}
     \subfigure[$M_p$=$10 M_J$, $a_p=0.6$AU.]{
          \label{TransitThresh10MJ06AUInc}
          \includegraphics[width=.315\textwidth]{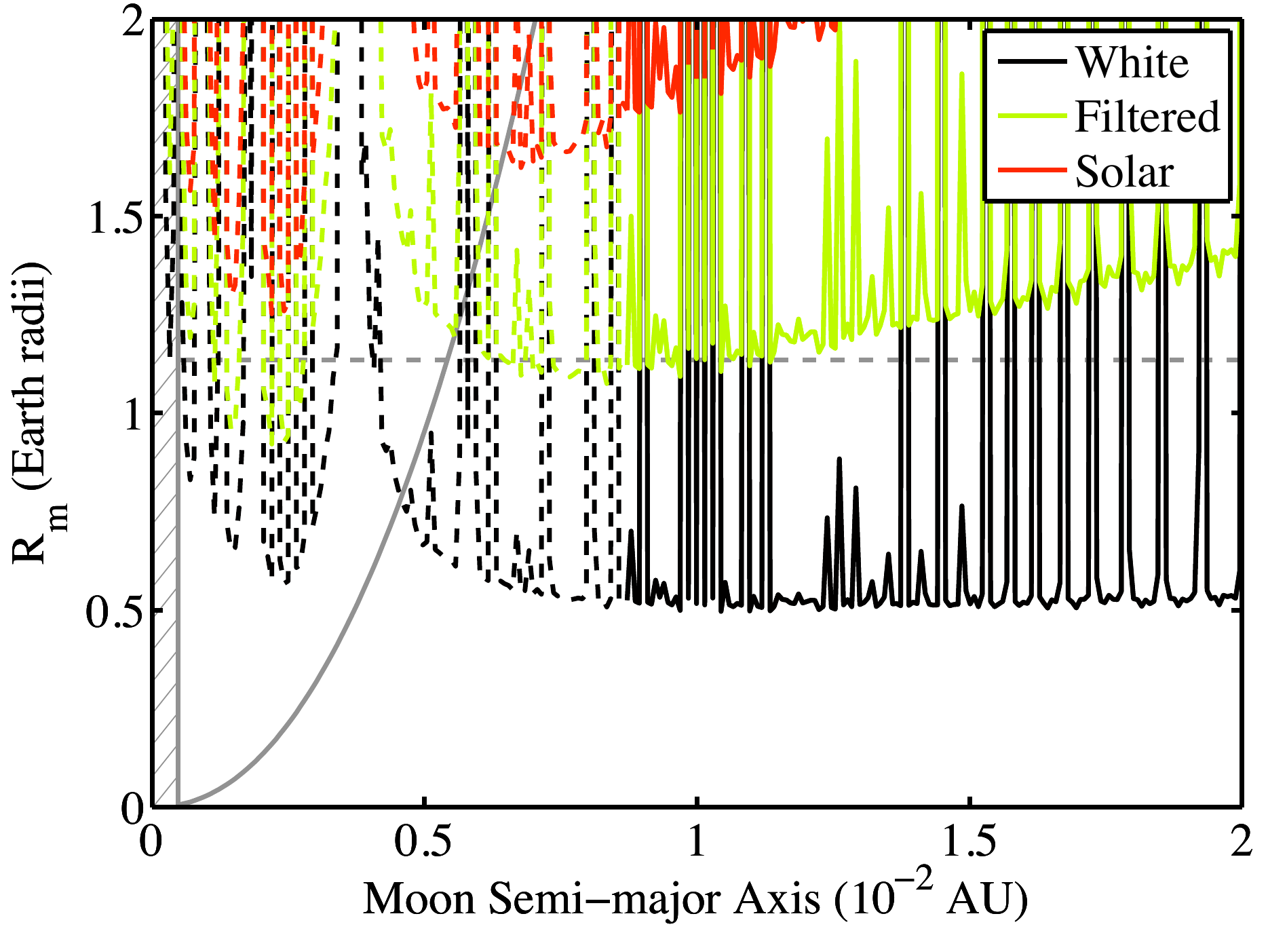}}\\ 
           \vspace{-0.2cm}
     \subfigure[$M_p = M_J$, $a_p=0.2$AU.]{
          \label{TransitThresh1MJ02AUInc}
          \includegraphics[width=.315\textwidth]{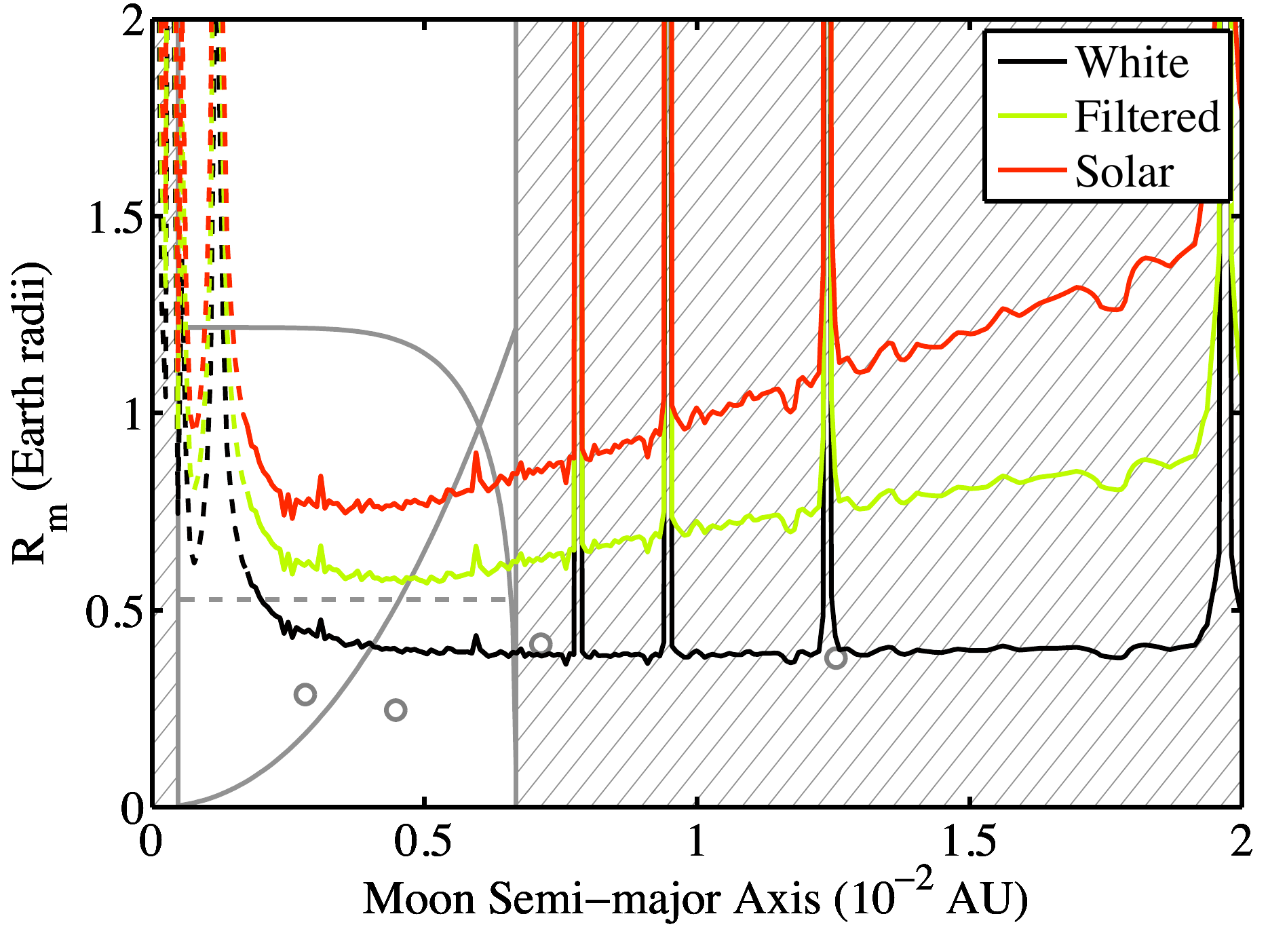}}
      \subfigure[$M_p = M_J$, $a_p=0.4$AU.]{
          \label{TransitThresh1MJ04AUInc}
          \includegraphics[width=.315\textwidth]{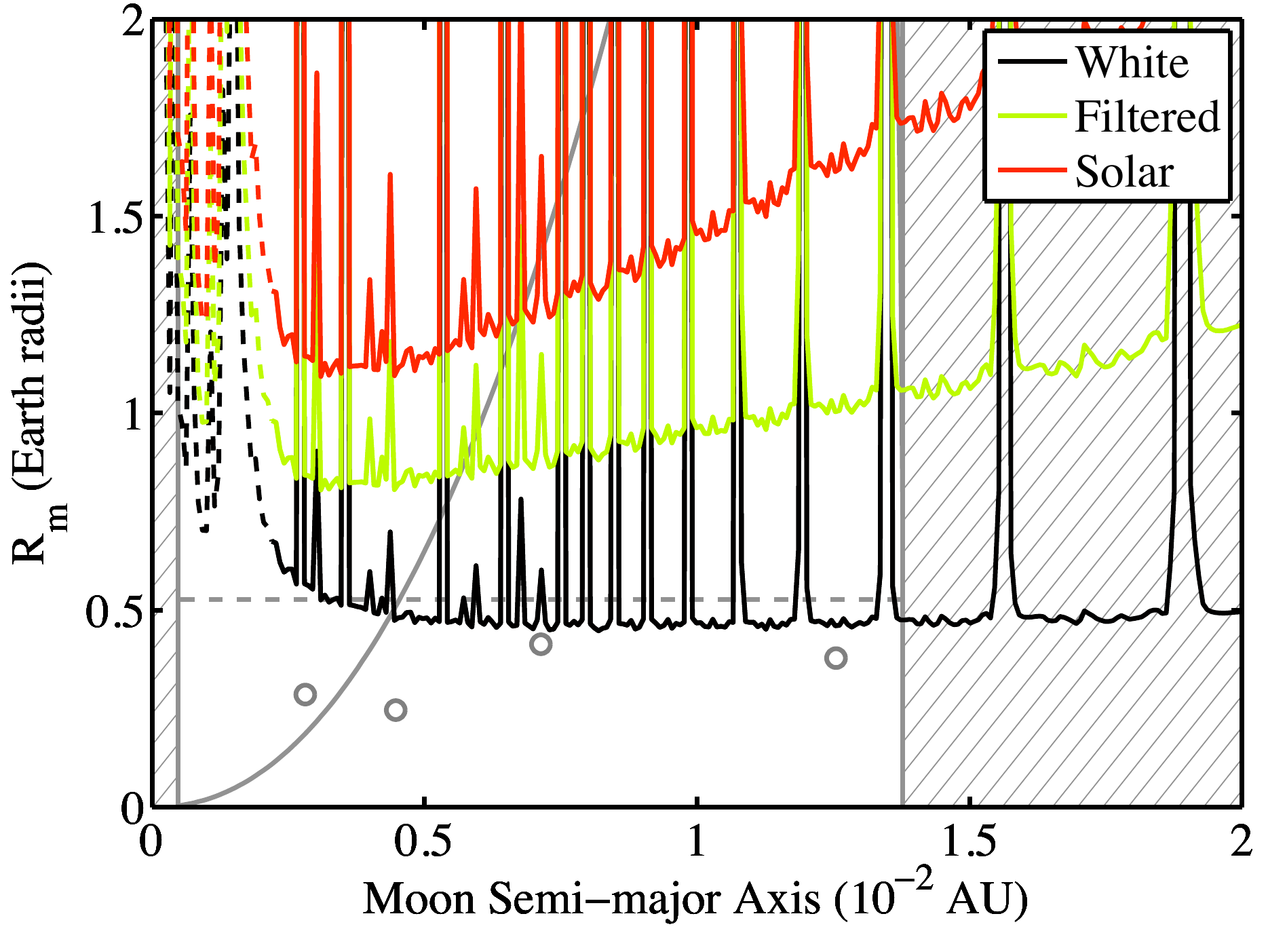}}
     \subfigure[$M_p = M_J$, $a_p=0.6$AU.]{
          \label{TransitThresh1MJ06AUInc}
          \includegraphics[width=.315\textwidth]{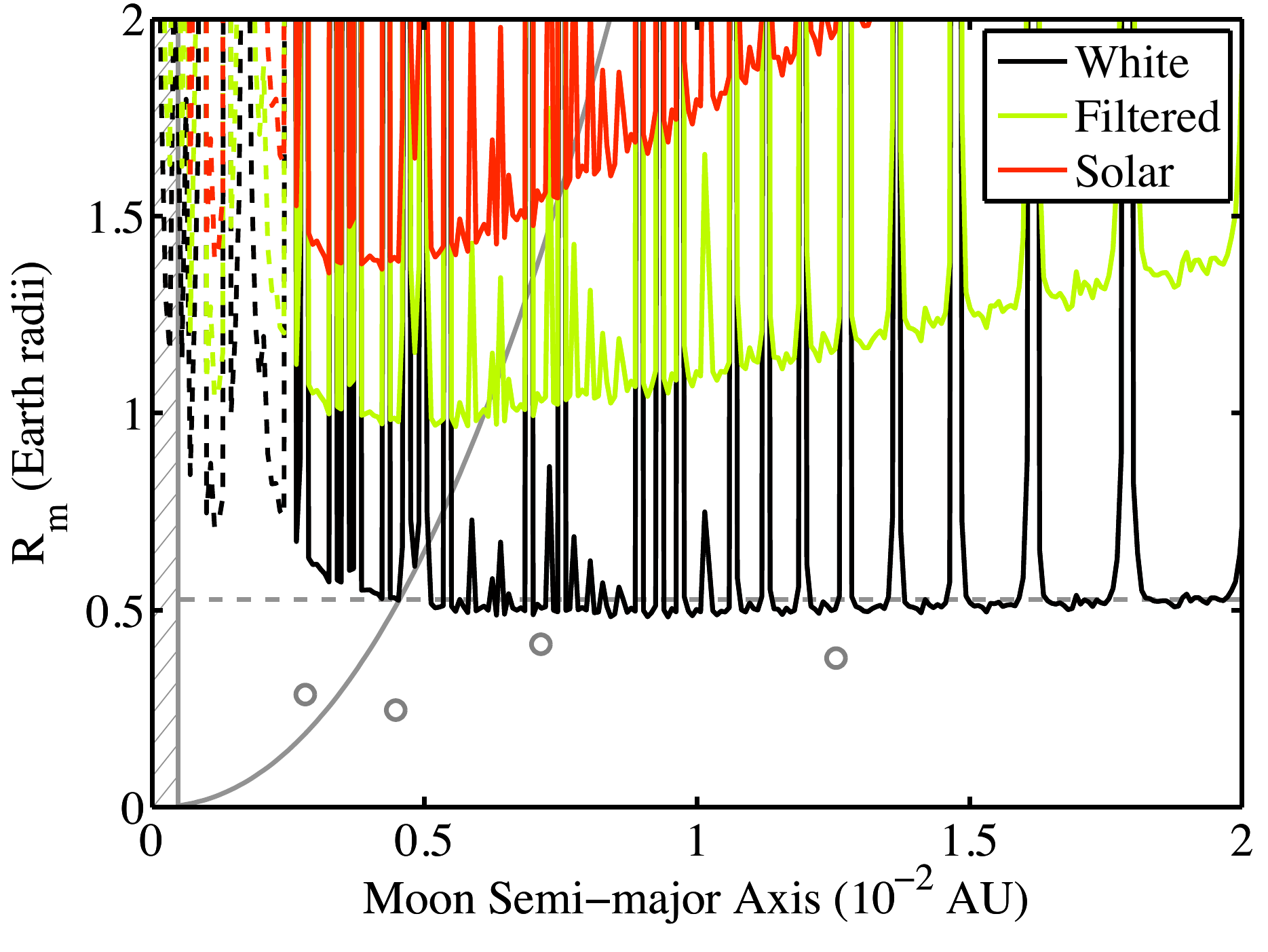}}\\ 
           \vspace{-0.2cm}
     \subfigure[$M_p = M_U$, $a_p=0.2$AU.]{
          \label{TransitThresh1MU02AUInc}
          \includegraphics[width=.315\textwidth]{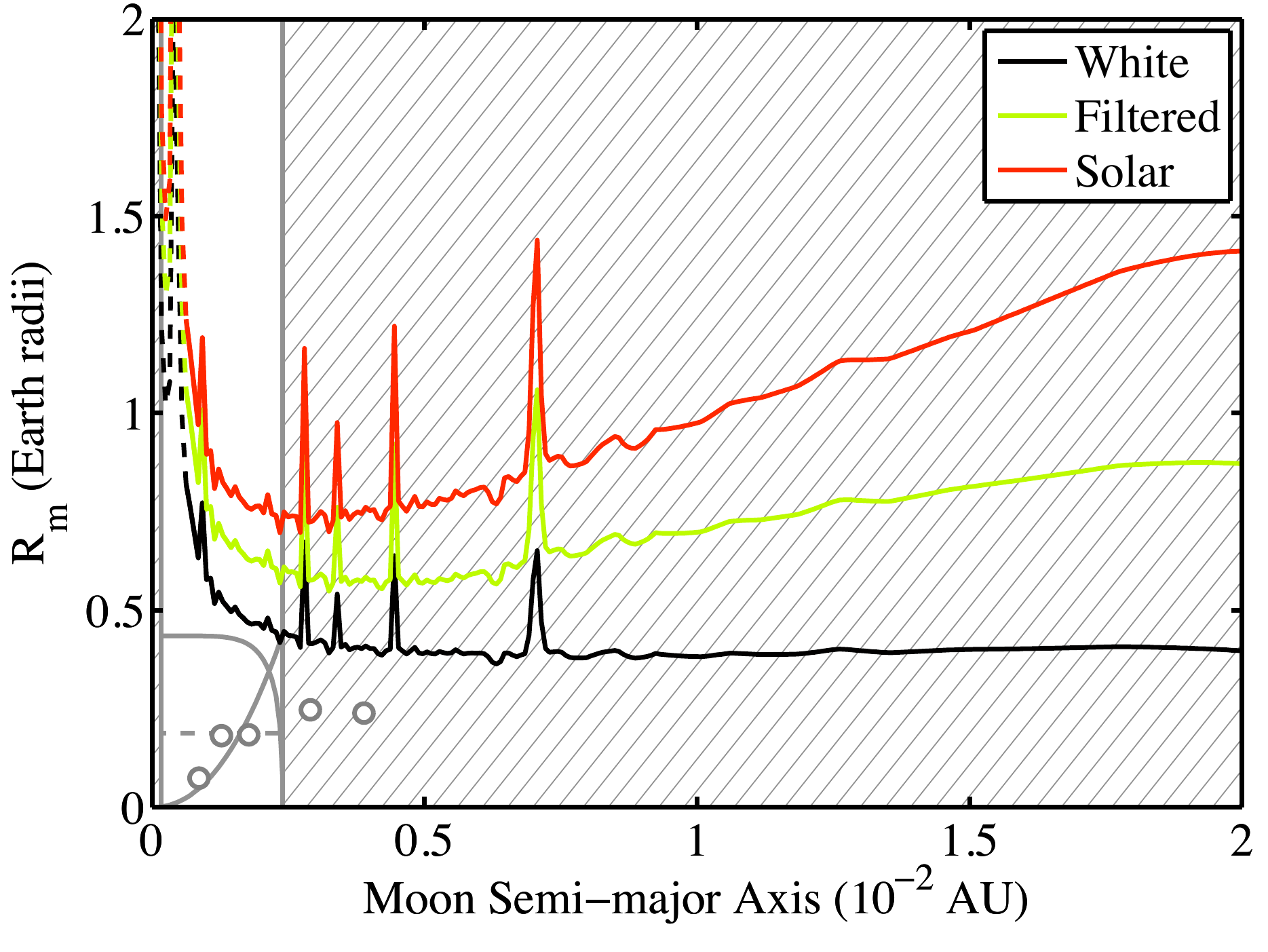}}
     \subfigure[$M_p = M_U$, $a_p=0.4$AU.]{
          \label{TransitThresh1MU04AUInc}
          \includegraphics[width=.315\textwidth]{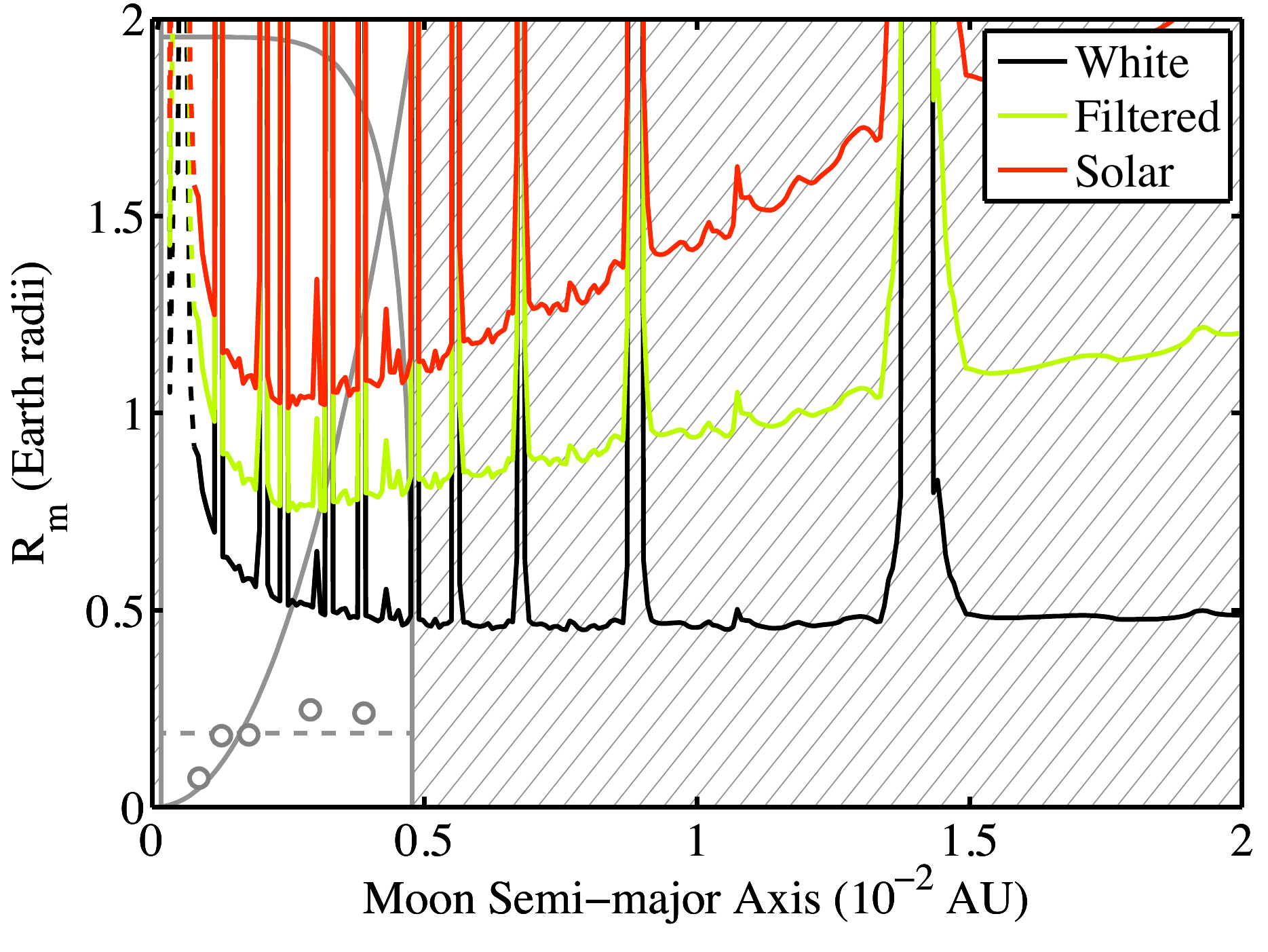}}
      \subfigure[$M_p = M_U$, $a_p=0.6$AU.]{
          \label{TransitThresh1MU06AUInc}
          \includegraphics[width=.315\textwidth]{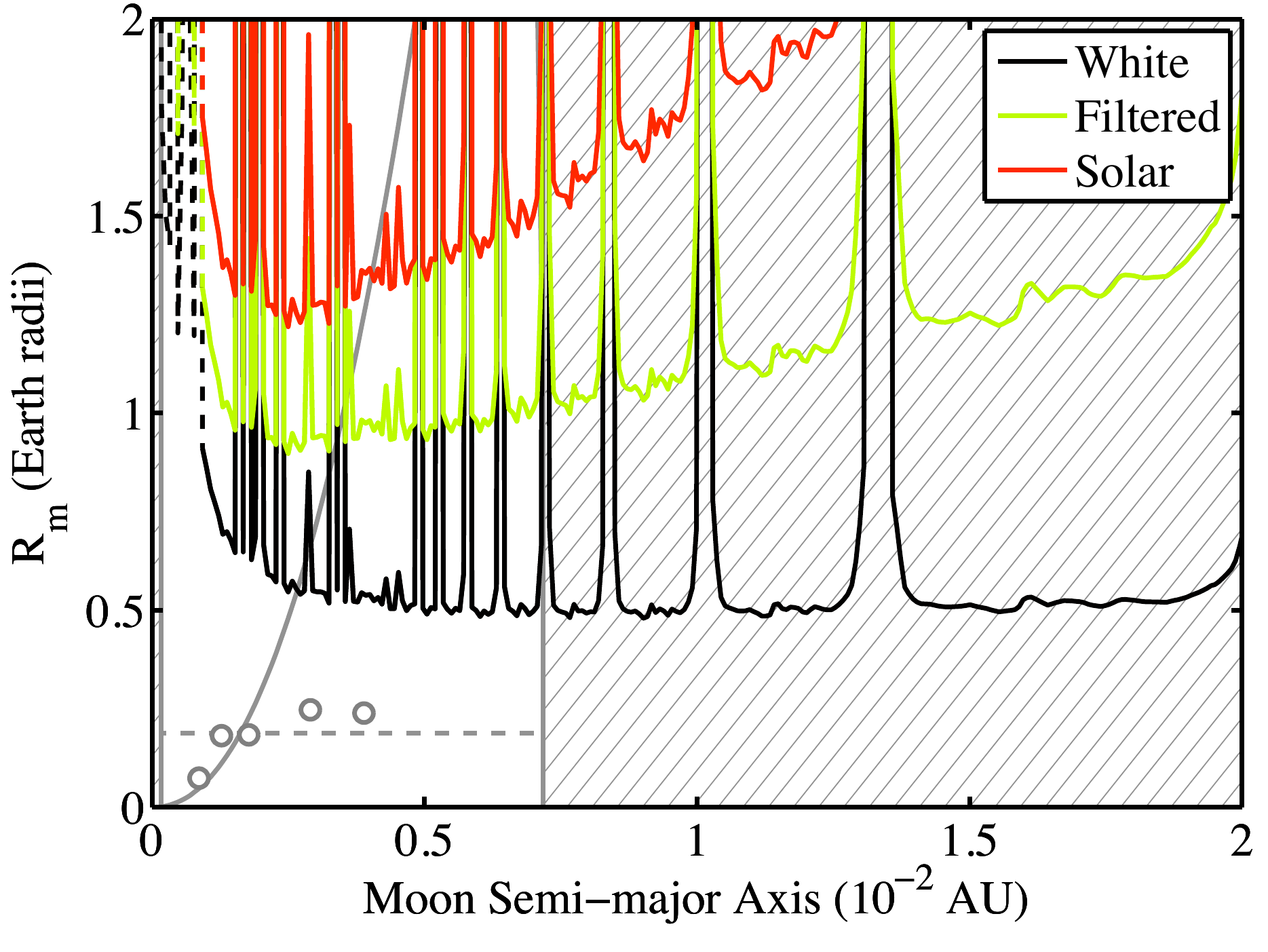}}\\ 
           \vspace{-0.2cm}
     \subfigure[$M_p = M_{\earth}$, $a_p=0.2$AU.]{
          \label{TransitThresh1ME02AUInc}
          \includegraphics[width=.315\textwidth]{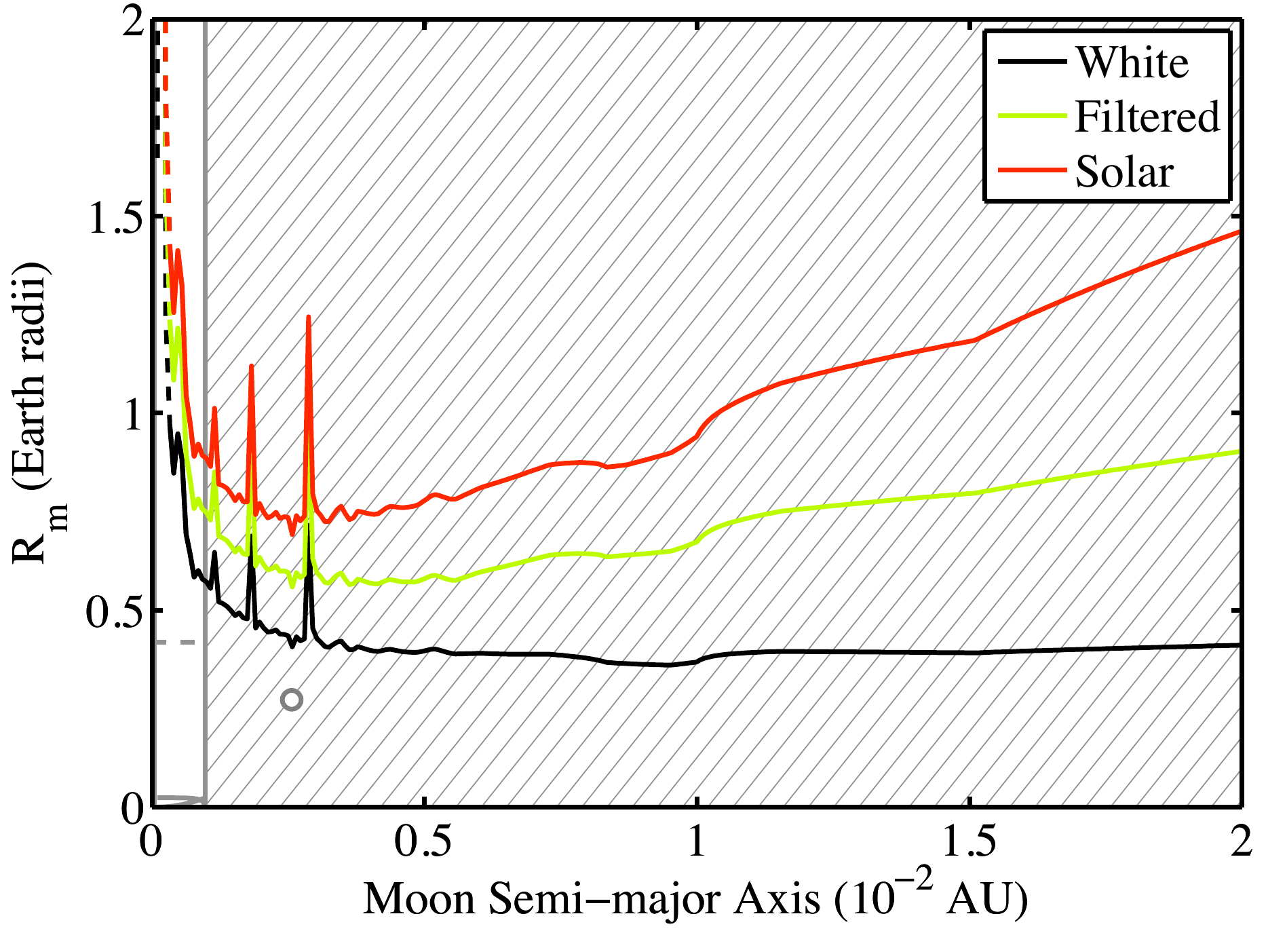}}
     \subfigure[$M_p = M_{\earth}$, $a_p=0.4$AU.]{
          \label{TransitThresh1ME04AUInc}
          \includegraphics[width=.315\textwidth]{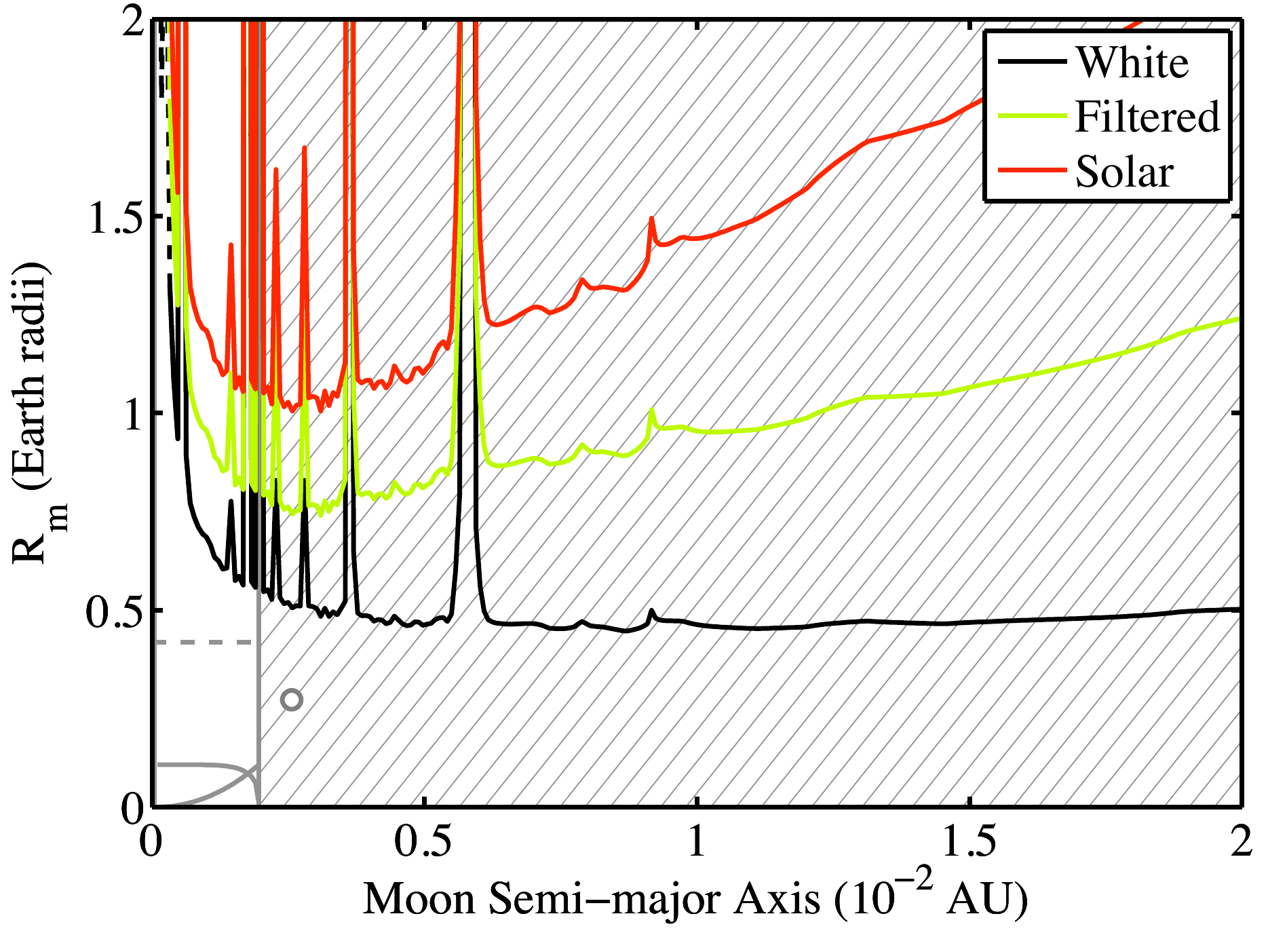}}
     \subfigure[$M_p = M_{\earth}$, $a_p=0.6$AU.]{
          \label{TransitThresh1ME06AUInc}
          \includegraphics[width=.315\textwidth]{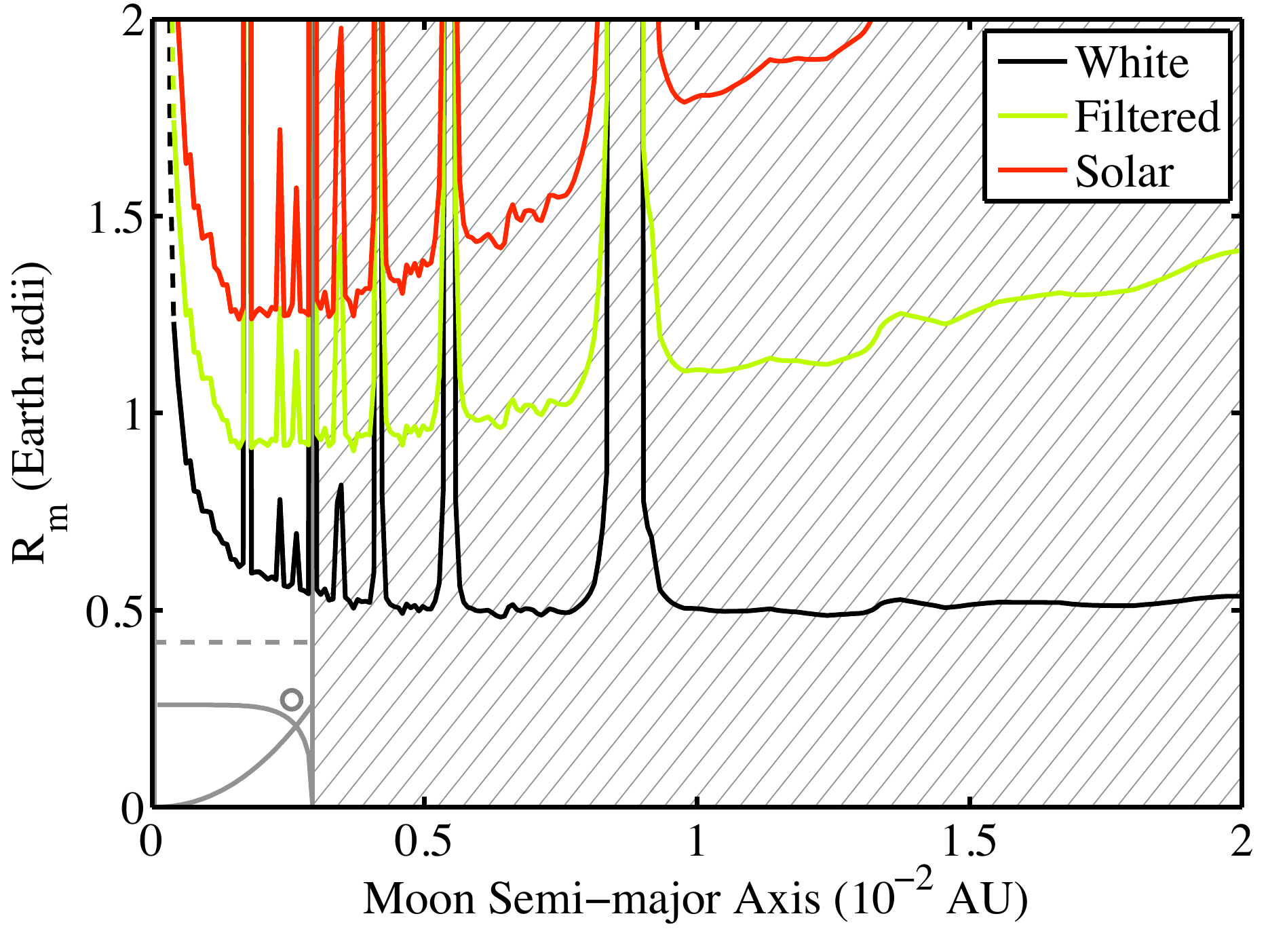}} 
           \vspace{-0.2cm}
     \caption[Figure of the same form as figure \ref{MCThresholdsAligned}, except calculated for the case of a slightly inclined ($\delta_{min} = 0.5 R_{\sun}$) planet orbit.]{Figure of the same form as figure \ref{MCThresholdsAligned}, except calculated for the case of a slightly inclined ($\delta_{min} = 0.5 R_{\sun}$) planet orbit.  As can be seen by comparing with figure \ref{MCThresholdsAligned}, the effect of slight inclination in the planet's orbit is two-fold.  First, the thresholds, and in particular, the threshold minima are shifted towards slightly lower values of moon semi-major axis.  Second, for the case of solar photometric noise, the relationship between large planetary semi-major axis and low moon detectability becomes less marked (see text).}
     \label{MCThresholdsInclined}
\end{figure}

\subsection{Slightly inclined orbits}\label{Trans_Thresholds_Thresh_Inclined}

For the case where the planet's orbit is circular, but slightly inclined (so that the planet no longer passes across the central chord of the star) the TTV$_p$ moon thresholds change.  This is because the shorter chord traveled by the planet across the face of the star results in a shorter transit and observing duration.  In particular, instead of being $2R_s/v_{tr}$, the transit duration is now $2R_s(1 - \delta_{min}^2)^{1/2}/v_{tr}$ and instead of being $2(R_s + a_m)/v_{tr}$ the observing duration is now $2(R_s(1 - \delta_{min}^2)^{1/2} + a_m)/v_{tr}$.  To investigate this effect, the thresholds in figure~\ref{MCThresholdsAligned} were recalculated assuming $\delta_{min} = 0.5 R_s$ and are presented in figure~\ref{MCThresholdsInclined}.  The effect of this change will be discussed by comparing with the thresholds calculated in the previous section.

\subsubsection{Dependance of threshold on moon semi-major axis}

For the case of slightly inclined orbits the dependance of the threshold on semi-major axis varies slightly from the dependance for the case of an aligned orbit.  As discussed in section~\ref{Trans_Thresholds_ExpBehav}, we expect that for the case of a transit with $\delta_{min} = 0.5$, the threshold minima would shift from $2R_{\sun}$, $R_{\sun}$ and between $R_{\sun}$ and $1/2R_{\sun}$ to $1.73R_{\sun}$, $0.87R_{\sun}$ and between $0.87R_{\sun}$ and $0.43R_{\sun}$ for the case of white, filtered and realistic stellar photometric noise respectively.  As the minima for the case of white noise are very shallow this difference is not obvious.  In addition, the minima for the case of filtered noise and red noise lie very close to the region where the expression for $\Delta \tau$ is inaccurate and inside the region which is the most affected by unresolved non-detection spikes, and consequently, this slight change is not immediately discernible.  To see this change, compare the red and filtered thresholds from figures~\ref{TransitThresh1ME06AUcc} and \ref{TransitThresh1ME06AUInc} between $a_m = 0.002$AU and $a_m = 0.008$AU.  In figure~\ref{TransitThresh1ME06AUcc}, the threshold for the aligned orbit, the filtered noise threshold is nearly horizontal in this region, while for the case of realistic noise it begins horizontal and then attains a slight positive slope.  Compare this to the case of the threshold for the analogous inclined system shown in figure~\ref{TransitThresh1ME06AUInc}.  In this case the threshold for filtered noise and red noise have a positive slope throughout this region.  This change in gradient indicates that the effect of changing $\delta_{min}$ from zero to 0.5 is to shift the minimum of the threshold to the left, as expected.

\subsubsection{Dependance of threshold on planet semi-major axis}

Unsurprisingly, the behaviour of the thresholds for the case of slightly inclined orbits as a function of planetary semi-major axis shows the same trend as for the case of aligned orbits.  In particular, moons of planets with smaller semi-major axes are more detectable than moons of planets with larger semi-major axes.  However, for the case where the orbit is inclined and the noise is red, this progression is not as marked.  To see this compare the moon radius associated with the red noise threshold minima between figures~\ref{TransitThresh1ME02AUcc} and \ref{TransitThresh1ME06AUcc}, and \ref{TransitThresh1ME02AUInc} and \ref{TransitThresh1ME06AUInc}.

To understand the origin of this effect, recall that, inclination of the planet's orbit modifies the transit duration, $T_{tra}$ (see equation~\eqref{transit_intro_dur_inc_Ddef}), and through that, the observing duration (see equation~\eqref{transit_thresholds_method_DobsDef}).  As the amplitude of $\Delta \tau$ depends on ($T_{obs}$ - $T_{tra}$), a quantity does not depend on $\delta_{min}$, this change in the position of the threshold is due to the dependance of $\sigma_\epsilon$ on $T_{tra}$ and $T_{obs}$.  In particular, recall from equation~\eqref{red_epsdef} that $\sigma_\epsilon$ is proportional to $\sigma_{\sun}T_{obs}/T_{tra}$.  As a result of the super-linear dependance of $\sigma_{\sun}$ on $T_{obs}$ for the case of red noise, for the case where $T_{obs}$ is large,  small decrease in the size of $T_{obs}$ causes a large decrease in the size of $\sigma_{\sun}$, while for the case where $T_{obs}$ is small, a small decrease in the size of $T_{obs}$ causes a small decrease in the size of $\sigma_{\sun}$.  This difference means that for large semi-major axes (large $T_{obs}$) the relative change in $\sigma_{\sun}$ can be larger than the relative change in $T_{tra}$, and thus the amplitude of $\sigma_\epsilon$ reduced, while for smaller semi-major axes (small $T_{obs}$) the relative change in $\sigma_{\sun}$ can be smaller than the relative change in $T_{tra}$, and thus, the amplitude of $\sigma_\epsilon$ increased.  This increase in the size of $\sigma_\epsilon$ for large planet semi-major axes and decrease in the size of $\sigma_\epsilon$ for small planet semi-major axes is the origin of the slight change in the dependance of moon detectability on planet semi-major axis for the case where the planet's orbit is slightly inclined.

\subsubsection{Comparison with formation and stability limits}

As slight inclination changes the position of both the threshold and the region of parameter space which is three-body stable only slightly, the overlap between the set of detectable moons and the set of moons which can form and be stable for the lifetime of the system is effectively equivalent to the overlap for the aligned case.  Now that the effect of inclination has been discussed, we can move onto a discussion of the effect of eccentricity of the planet's orbit on the TTV$_p$ thresholds and thus the set of realistic moons likely to be detected.

\begin{figure}
     \centering
     \vspace{-0.2cm}
     \subfigure[$M_p$=$10 M_J$, $a_p=0.2$AU.]{
          \label{TransitThresh10MJ02AUEccP}
          \includegraphics[width=.315\textwidth]{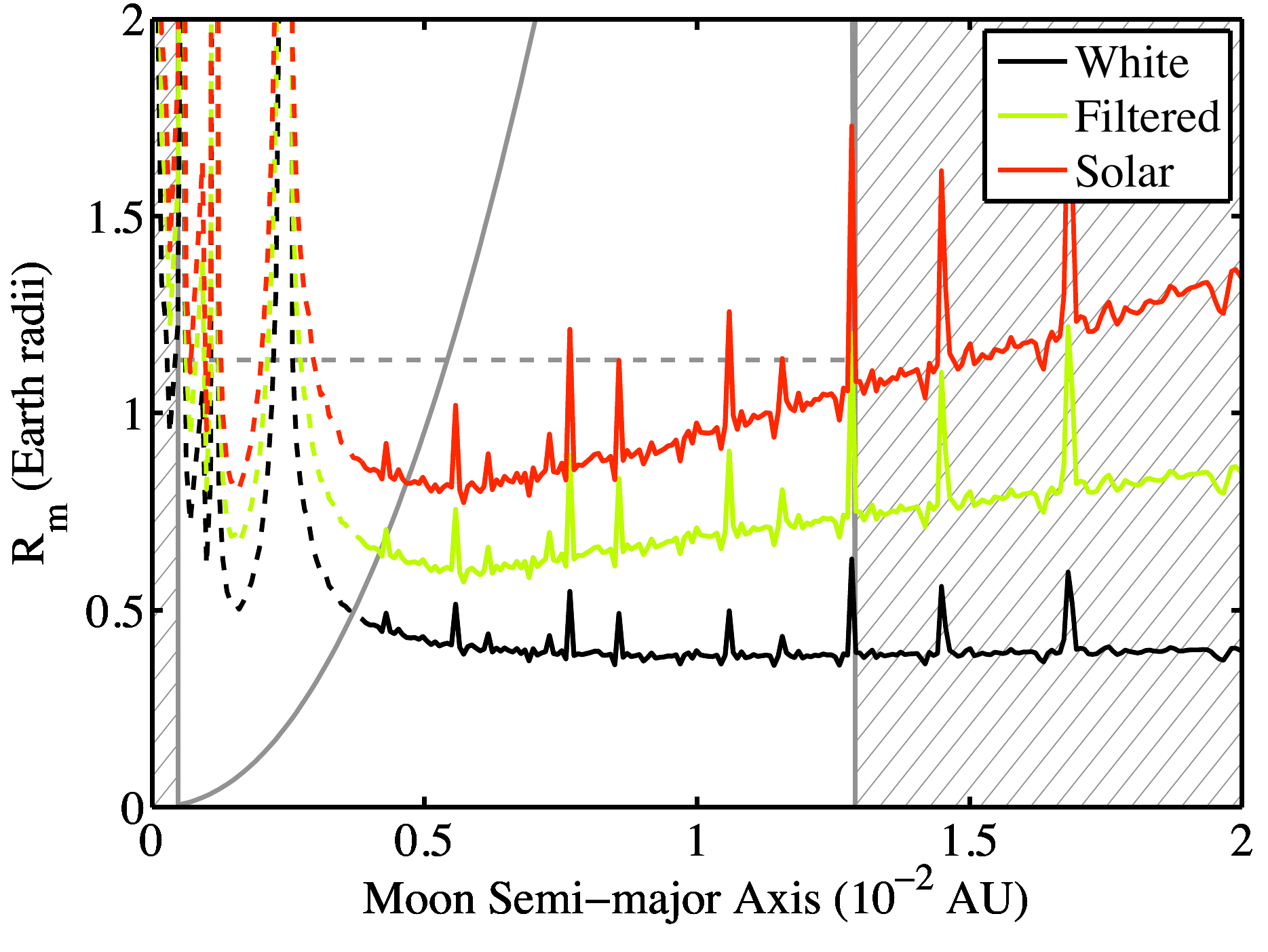}}
     \subfigure[$M_p$=$10 M_J$, $a_p=0.4$AU.]{
          \label{TransitThresh10MJ04AUEccP}
          \includegraphics[width=.315\textwidth]{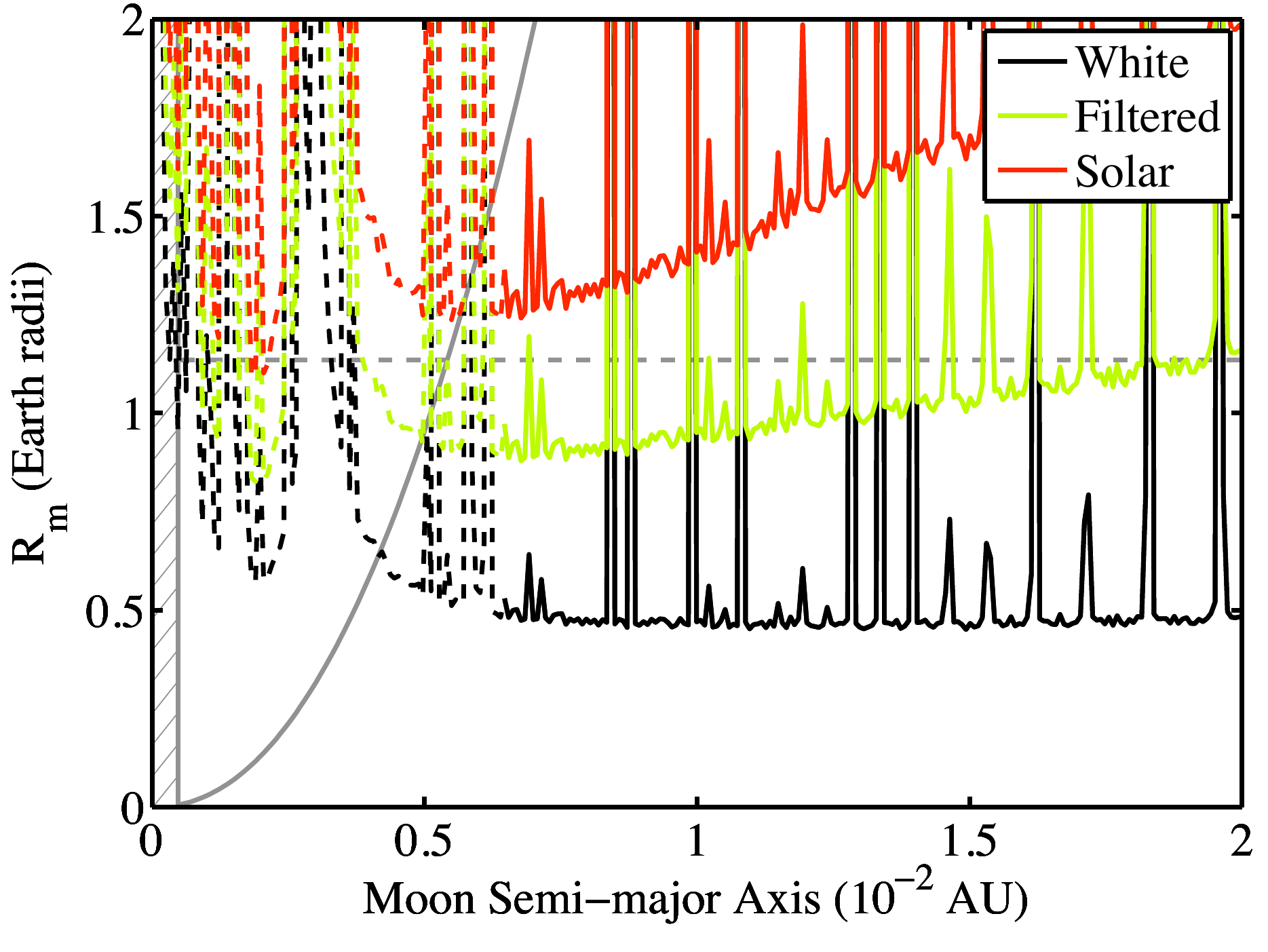}}
     \subfigure[$M_p$=$10 M_J$, $a_p=0.6$AU.]{
          \label{TransitThresh10MJ06AUEccP}
          \includegraphics[width=.315\textwidth]{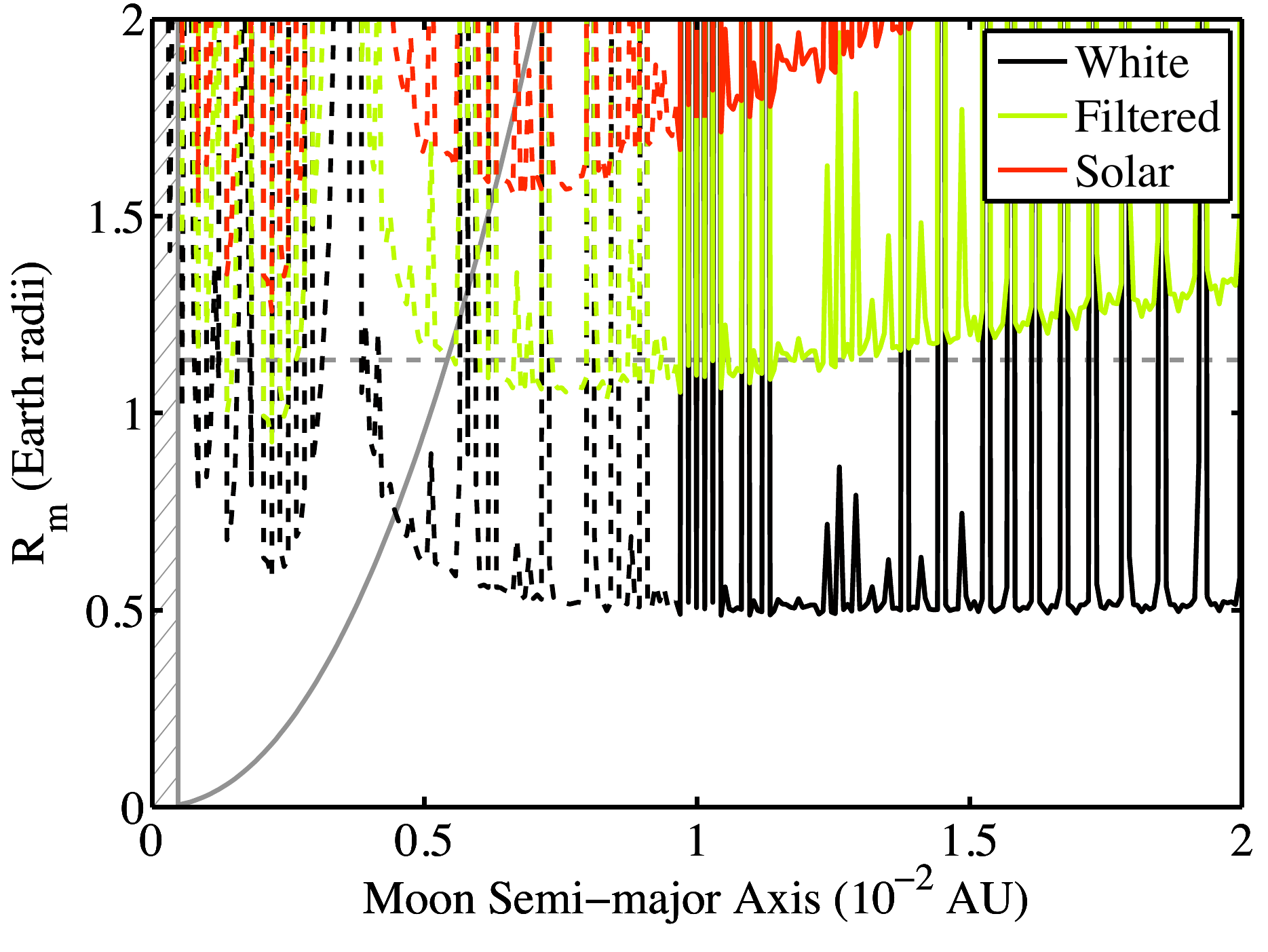}}\\ 
          \vspace{-0.2cm}
     \subfigure[$M_p = M_J$, $a_p=0.2$AU.]{
          \label{TransitThresh1MJ02AUEccP}
          \includegraphics[width=.315\textwidth]{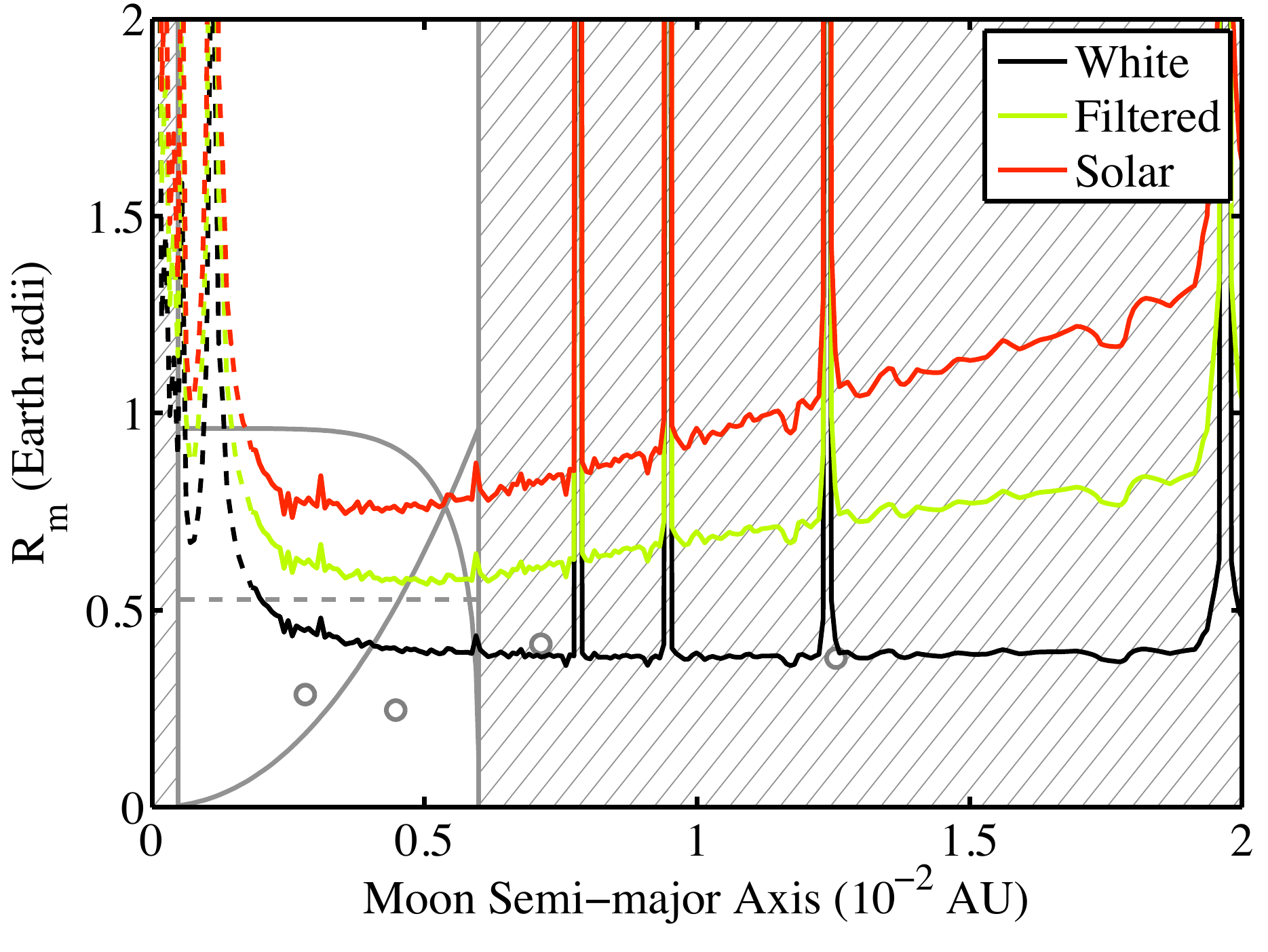}}
      \subfigure[$M_p = M_J$, $a_p=0.4$AU.]{
          \label{TransitThresh1MJ04AUEccP}
          \includegraphics[width=.315\textwidth]{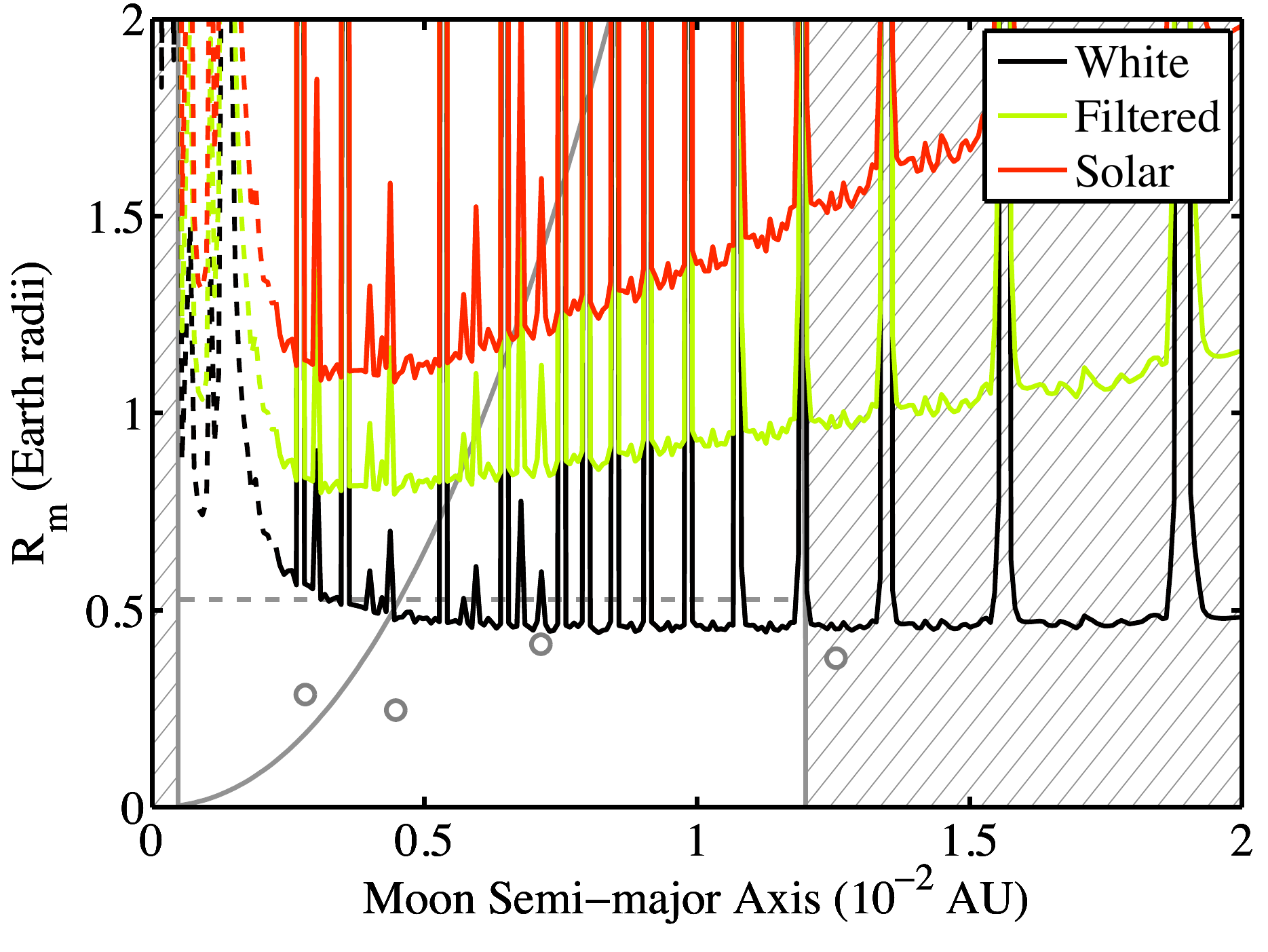}}
     \subfigure[$M_p = M_J$, $a_p=0.6$AU.]{
          \label{TransitThresh1MJ06AUEccP}
          \includegraphics[width=.315\textwidth]{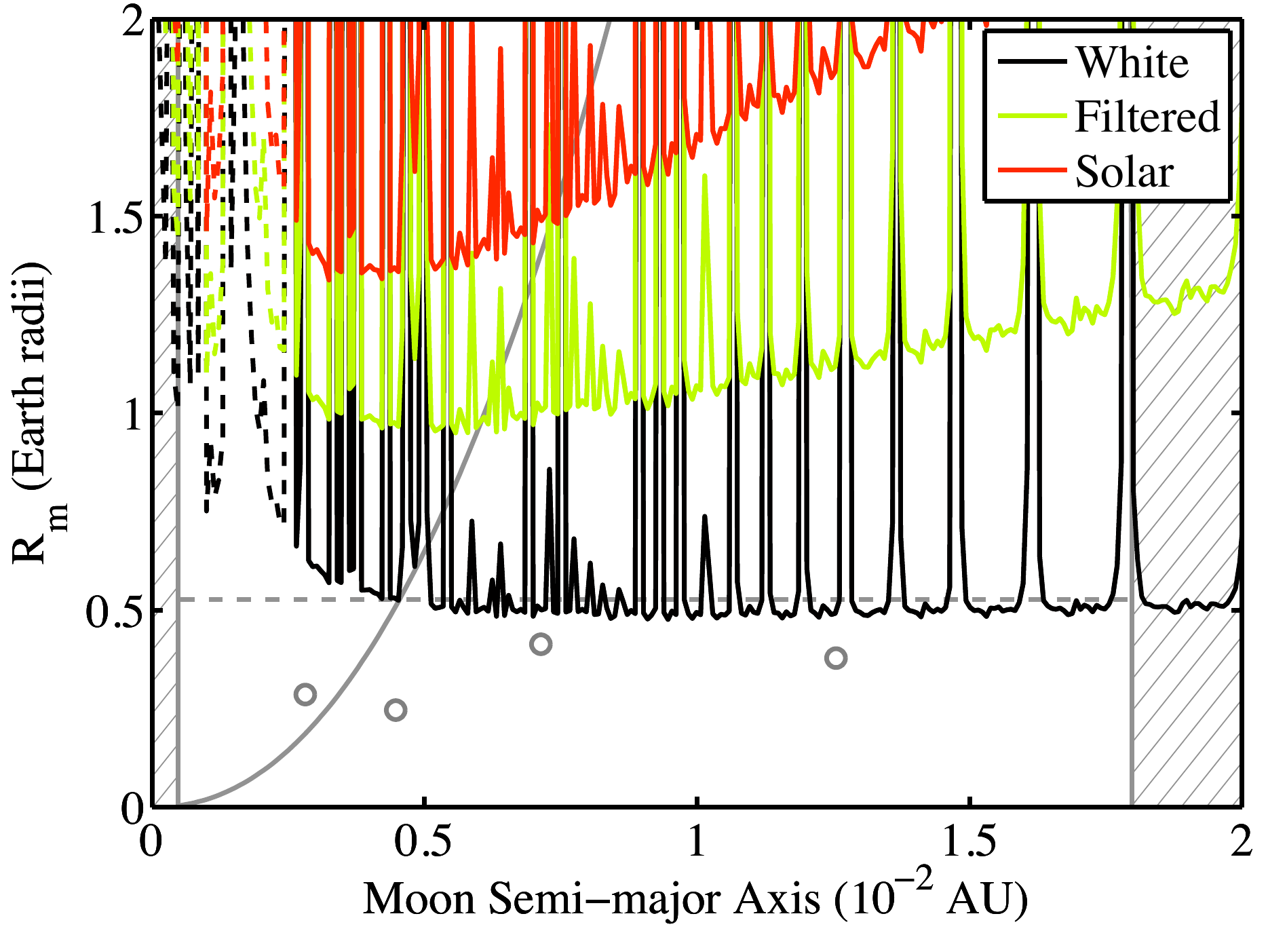}}\\ 
          \vspace{-0.2cm}
     \subfigure[$M_p = M_U$, $a_p=0.2$AU.]{
          \label{TransitThresh1MU02AUEccP}
          \includegraphics[width=.315\textwidth]{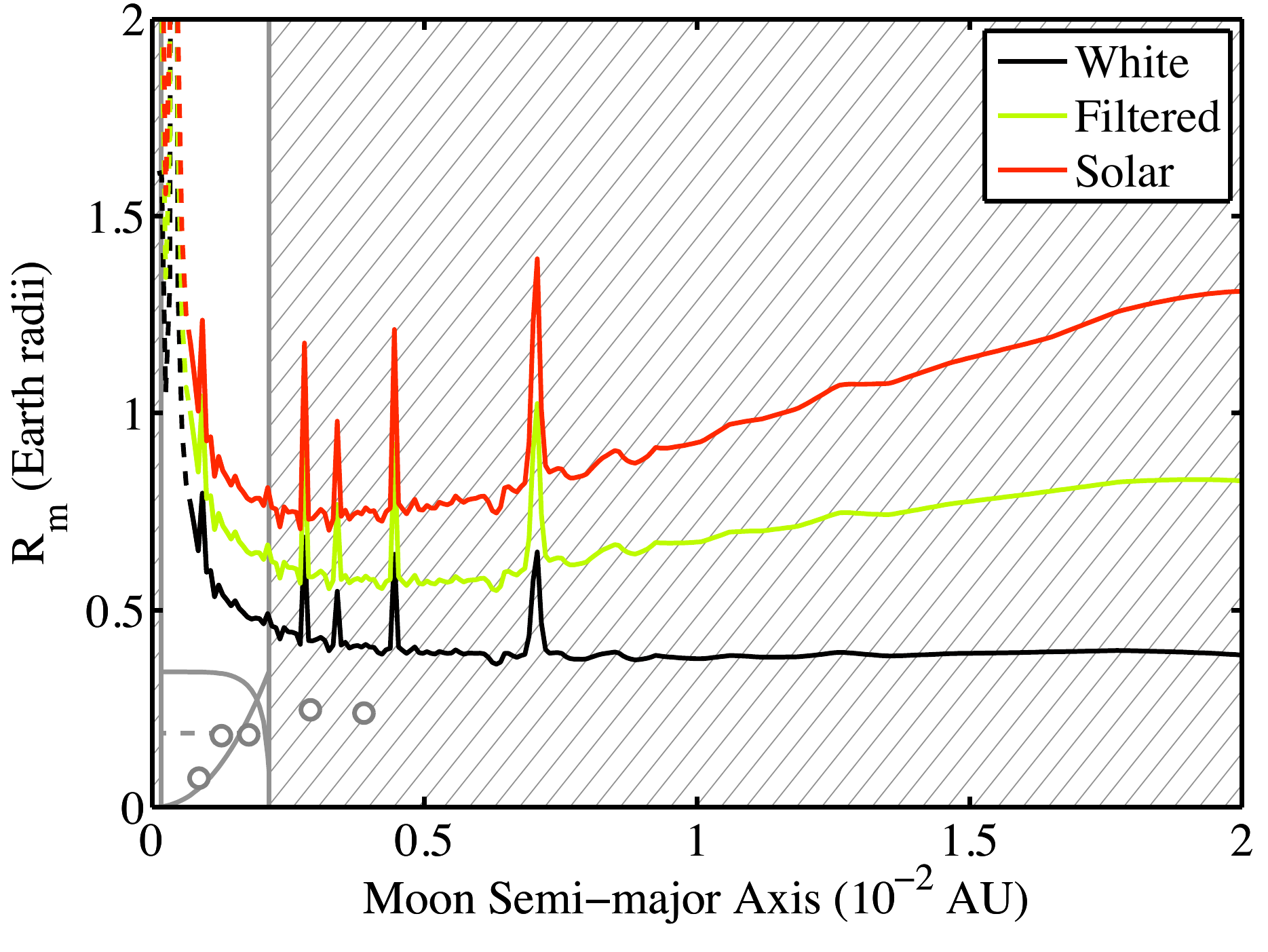}}
     \subfigure[$M_p = M_U$, $a_p=0.4$AU.]{
          \label{TransitThresh1MU04AUEccP}
          \includegraphics[width=.315\textwidth]{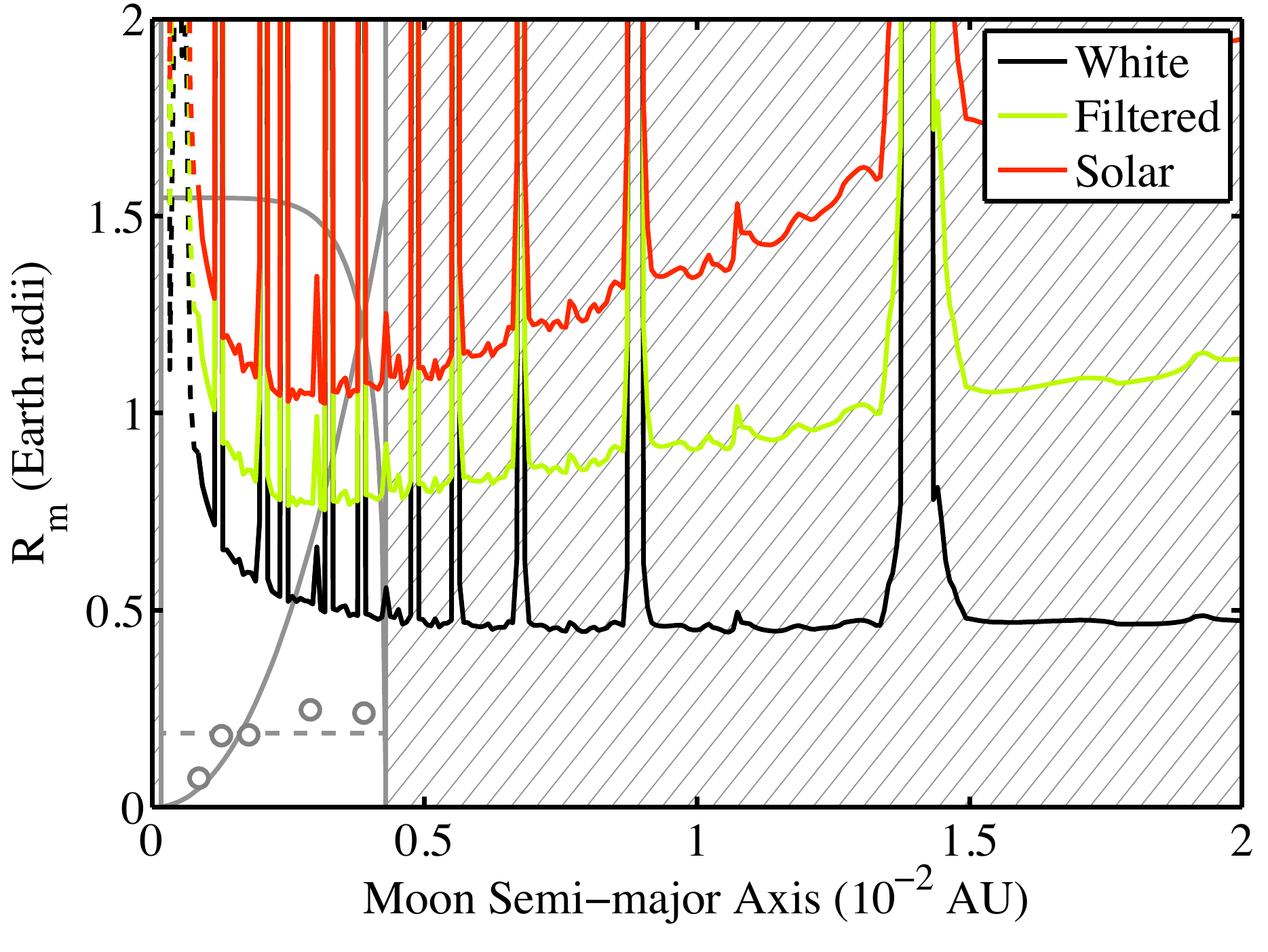}}
      \subfigure[$M_p = M_U$, $a_p=0.6$AU.]{
          \label{TransitThresh1MU06AUEccP}
          \includegraphics[width=.315\textwidth]{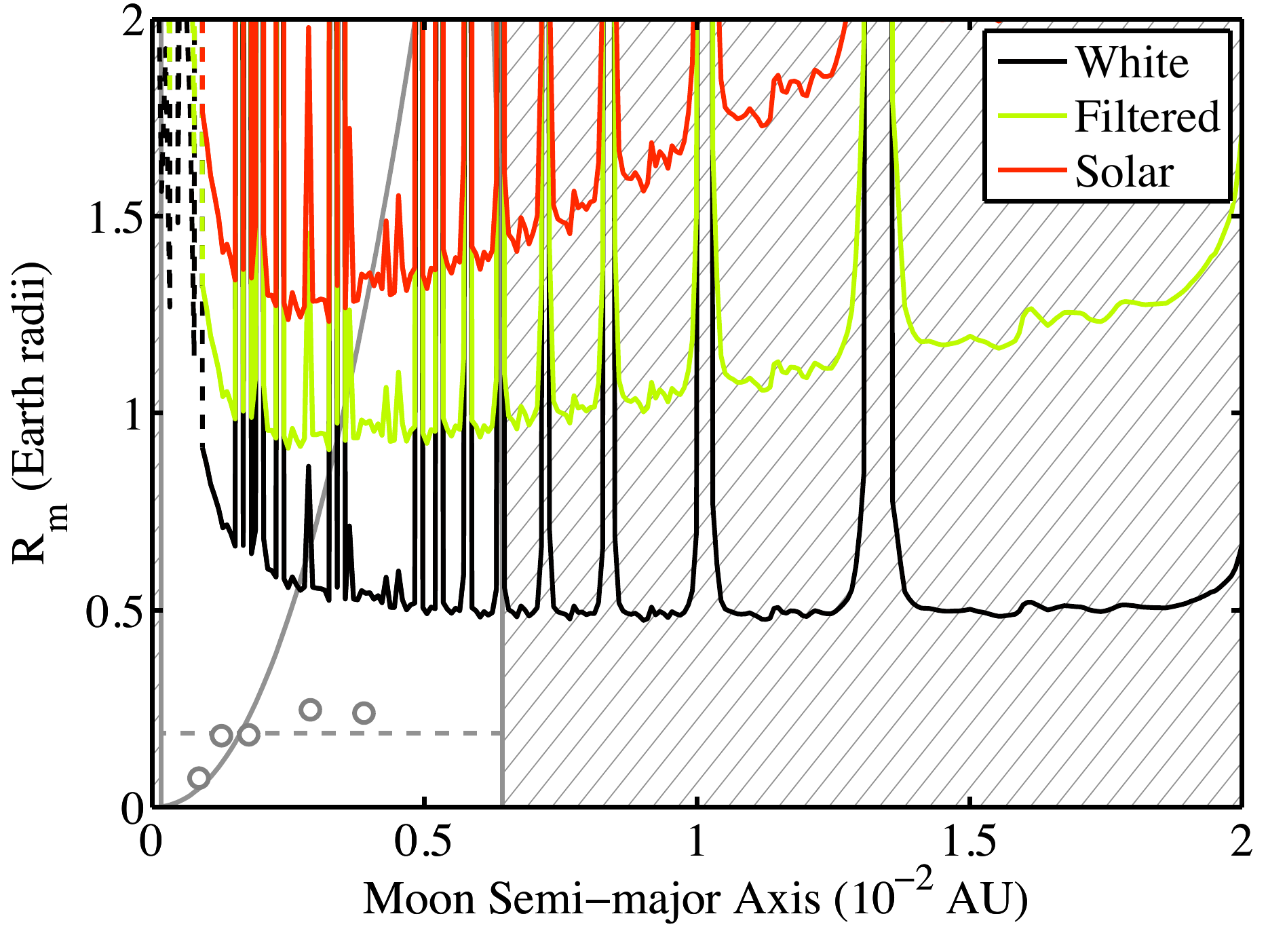}}\\ 
          \vspace{-0.2cm}
     \subfigure[$M_p = M_{\earth}$, $a_p=0.2$AU.]{
          \label{TransitThresh1ME02AUEccP}
          \includegraphics[width=.315\textwidth]{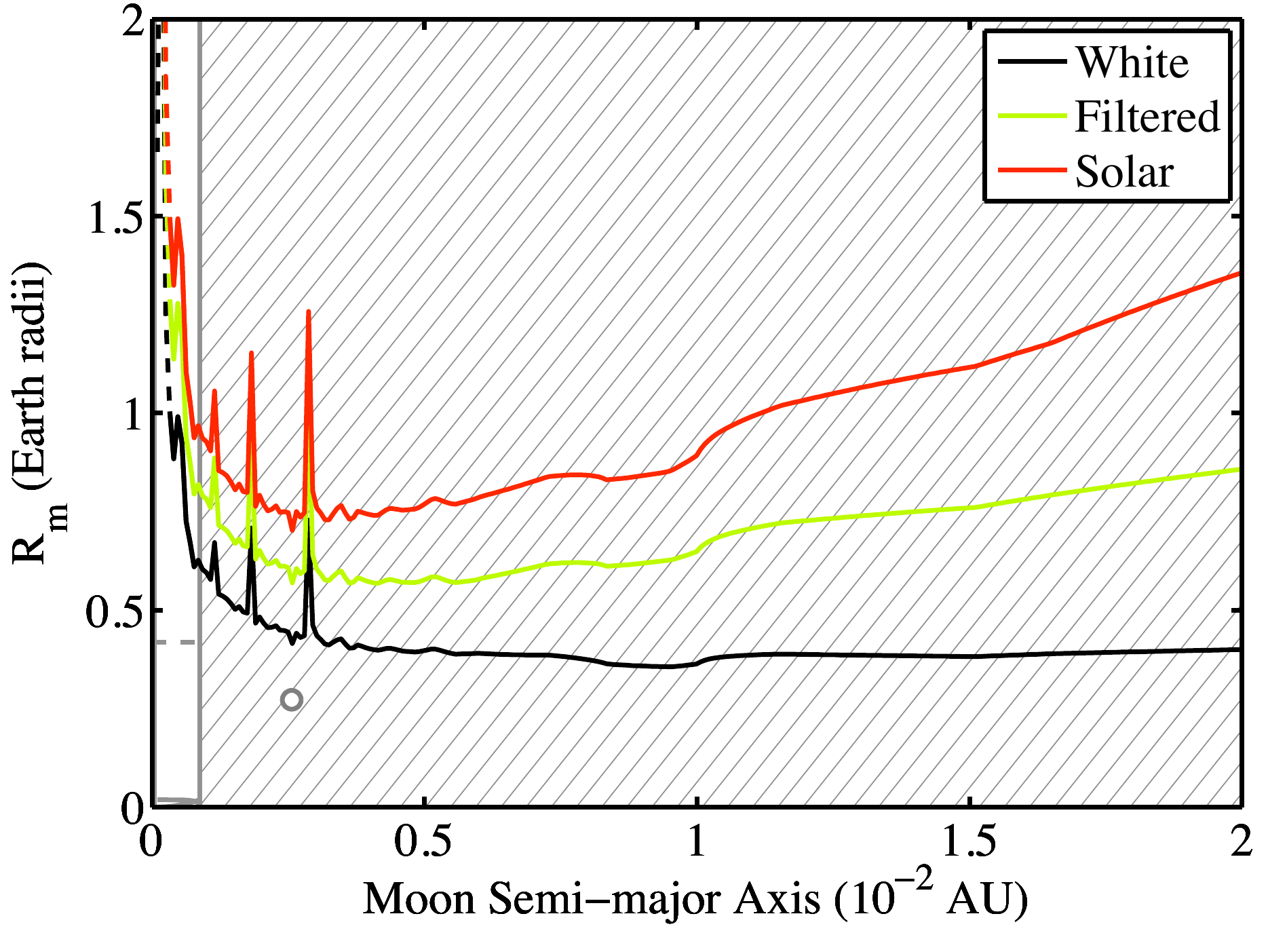}}
     \subfigure[$M_p = M_{\earth}$, $a_p=0.4$AU.]{
          \label{TransitThresh1ME04AUEccP}
          \includegraphics[width=.315\textwidth]{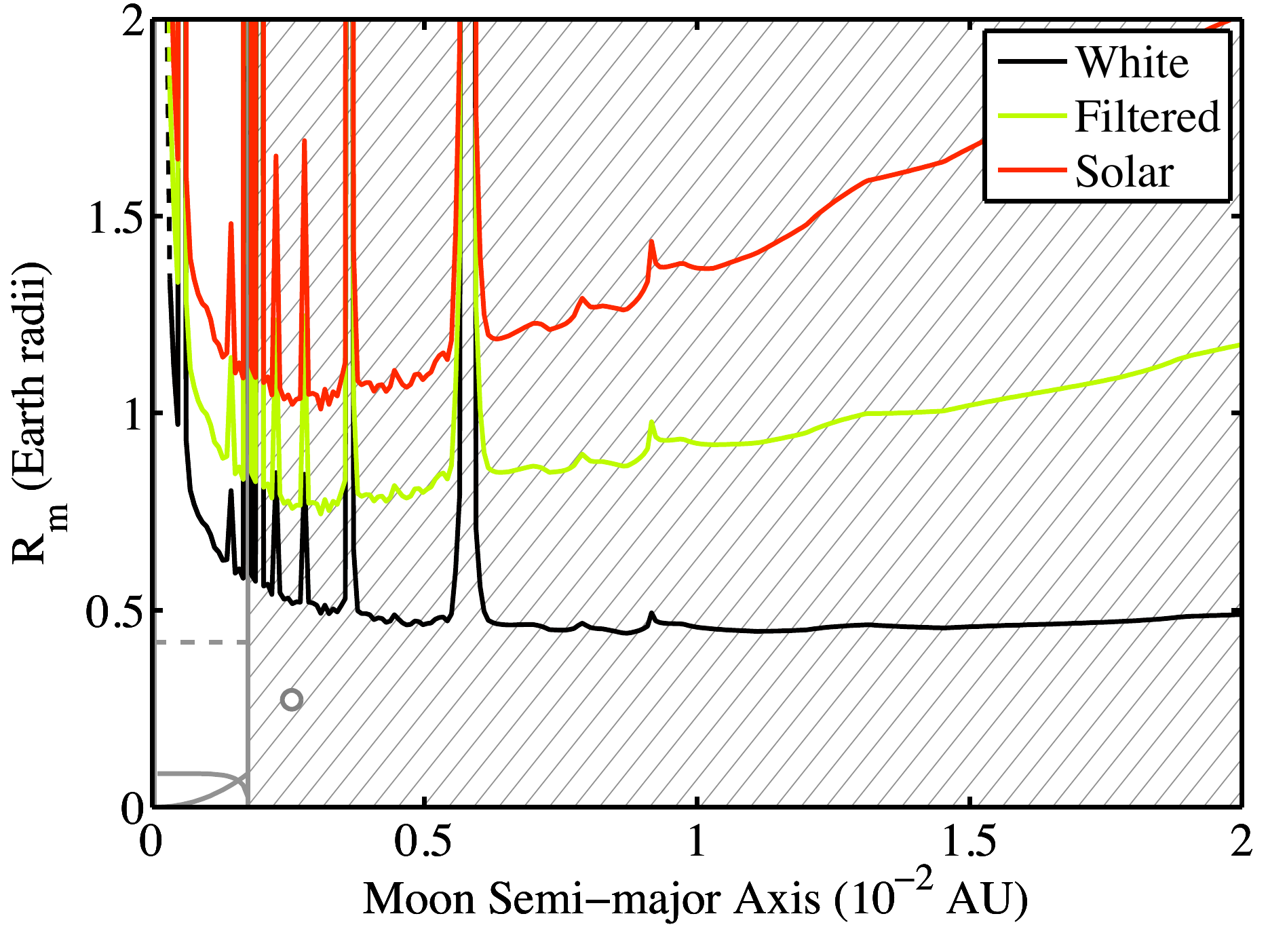}}
     \subfigure[$M_p = M_{\earth}$, $a_p=0.6$AU.]{
          \label{TransitThresh1ME06AUEccP}
          \includegraphics[width=.315\textwidth]{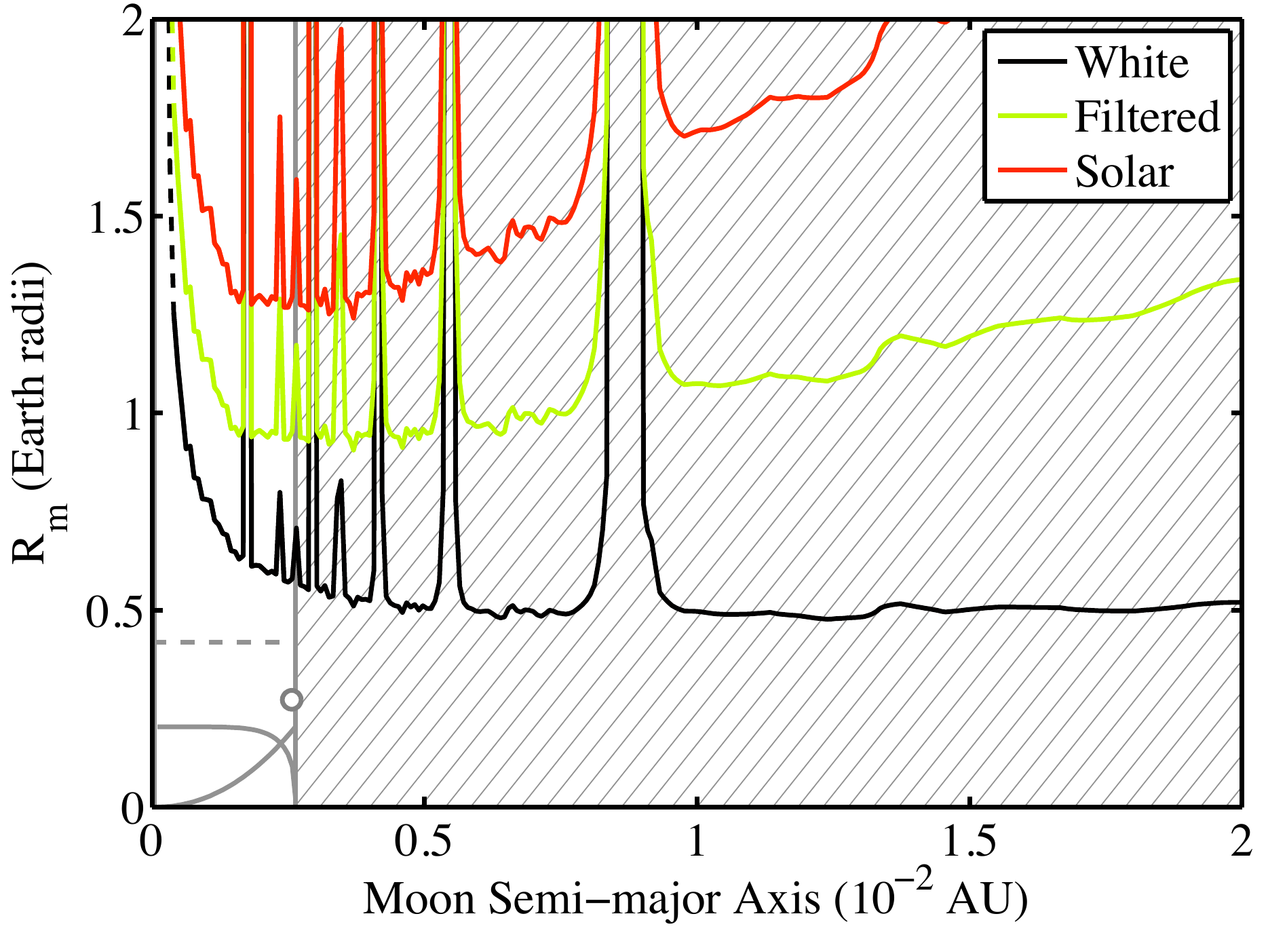}} 
          \vspace{-0.2cm}
     \caption[Figure of the same form as figure \ref{MCThresholdsAligned} except calculated for the case of an eccentric ($e_p = 0.1$) planet orbit, oriented such that the transit occurs at \emph{periastron}.]{Figure of the same form as figure \ref{MCThresholdsAligned} except calculated for the case of an eccentric ($e_p = 0.1$) planet orbit, oriented such that the transit occurs at \emph{periastron}.  As can be seen by comparing with figure \ref{MCThresholdsAligned}, the effect of eccentricity on the thresholds for this orbital orientation is to shift the white and solar noise thresholds vertically upward and downward respectively.  As a result, the three thresholds are closer to each other than for the circular case.  In addition, eccentricity reduces the size of the three-body stable region (hatched),  which in turn reduces the set of moons that are tidally stable for the lifetime of the system (compare figures \ref{TransitThresh1MJ02AUcc} and \ref{TransitThresh1MJ02AUEccP}).}
    \label{MCThresholdsEccentricPeri}
    \end{figure}
    
\begin{figure}
     \centering
     \vspace{-0.2cm}
     \subfigure[$M_p$=$10 M_J$, $a_p=0.2$AU.]{
          \label{TransitThresh10MJ02AUEccA}
          \includegraphics[width=.315\textwidth]{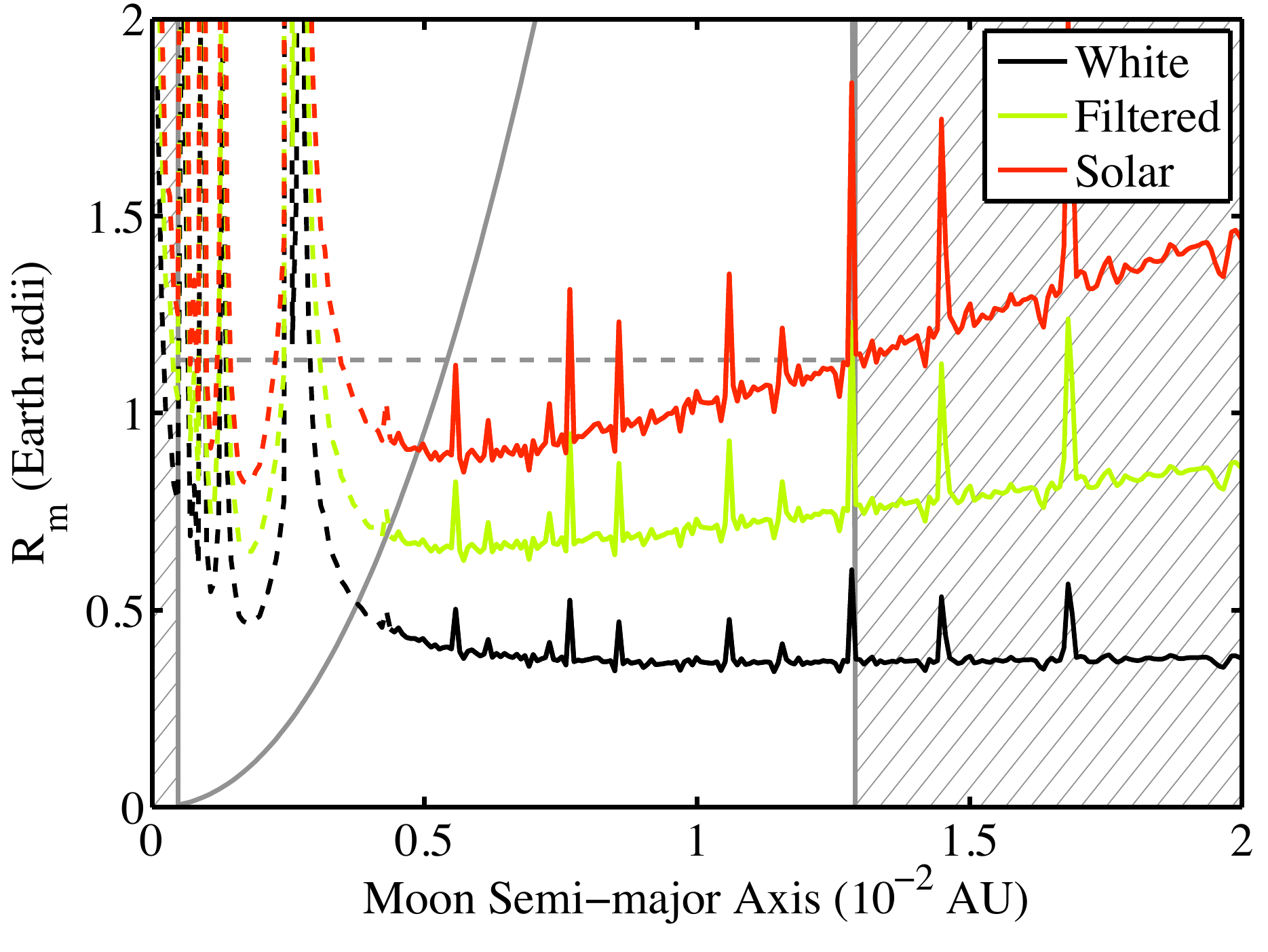}}
     \subfigure[$M_p$=$10 M_J$, $a_p=0.4$AU.]{
          \label{TransitThresh10MJ04AUEccA}
          \includegraphics[width=.315\textwidth]{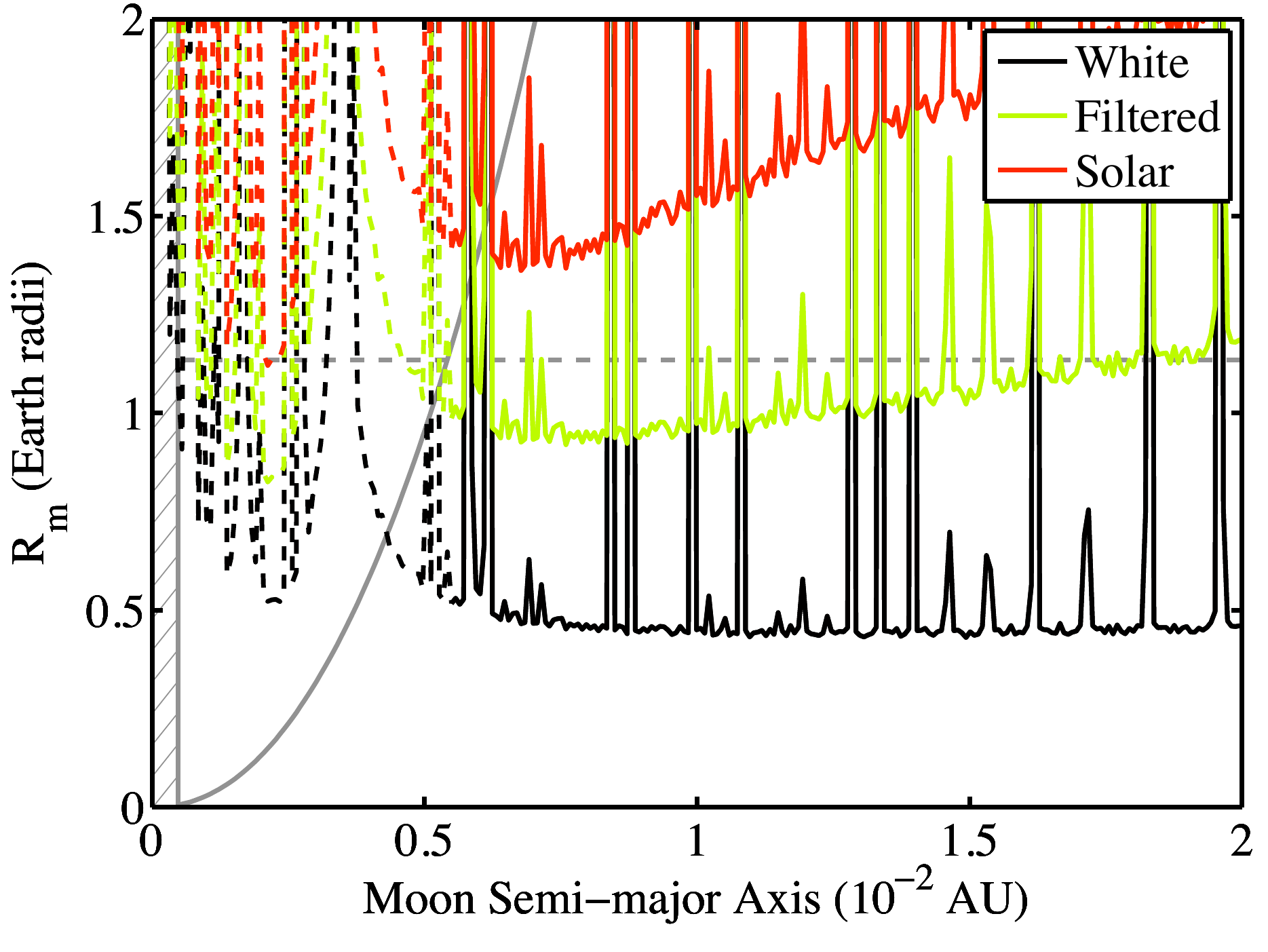}}
     \subfigure[$M_p$=$10 M_J$, $a_p=0.6$AU.]{
          \label{TransitThresh10MJ06AUEccA}
          \includegraphics[width=.315\textwidth]{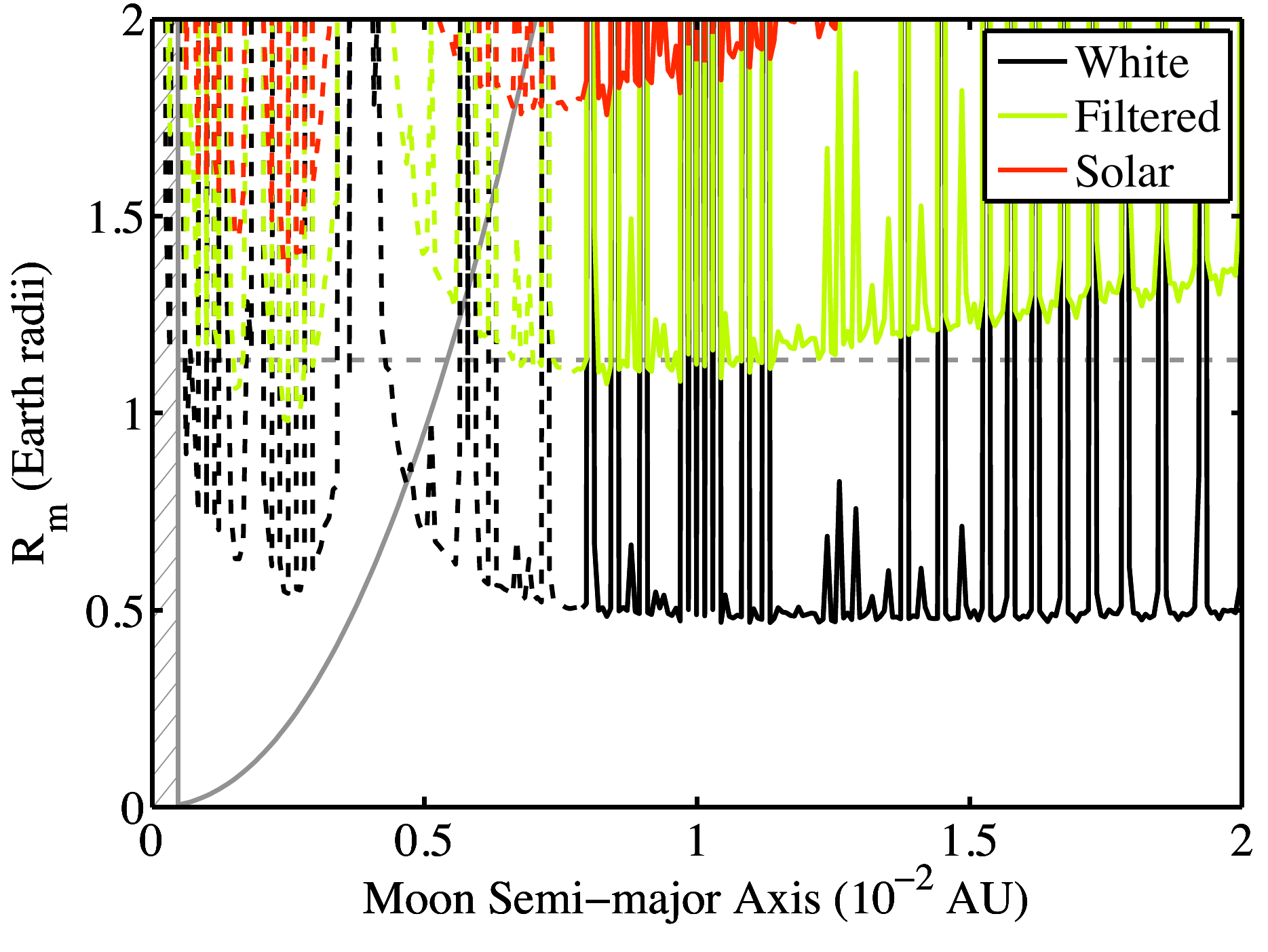}}\\ 
          \vspace{-0.2cm}
     \subfigure[$M_p = M_J$, $a_p=0.2$AU.]{
          \label{TransitThresh1MJ02AUEccA}
          \includegraphics[width=.315\textwidth]{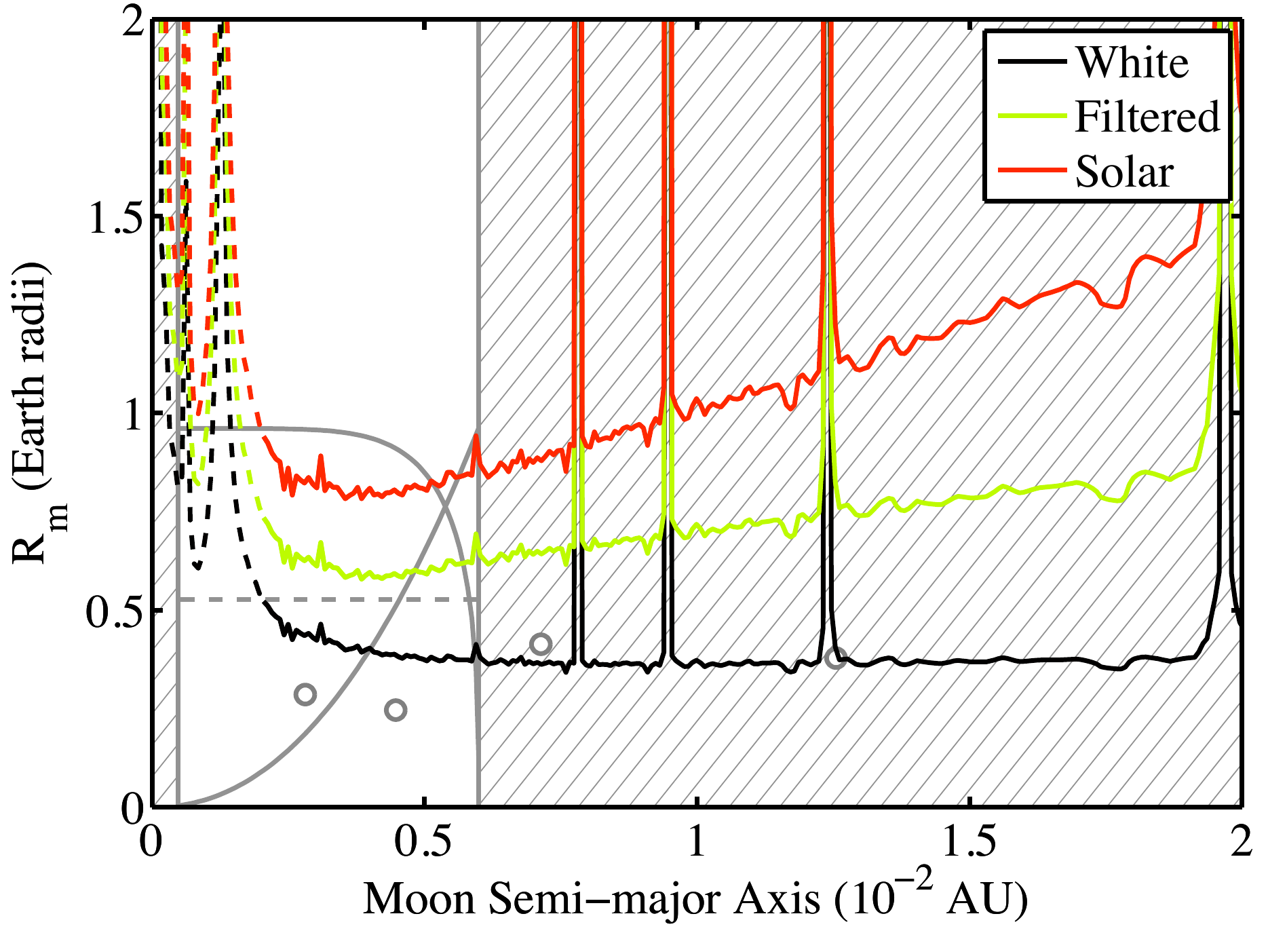}}
      \subfigure[$M_p = M_J$, $a_p=0.4$AU.]{
          \label{TransitThresh1MJ04AUEccA}
          \includegraphics[width=.315\textwidth]{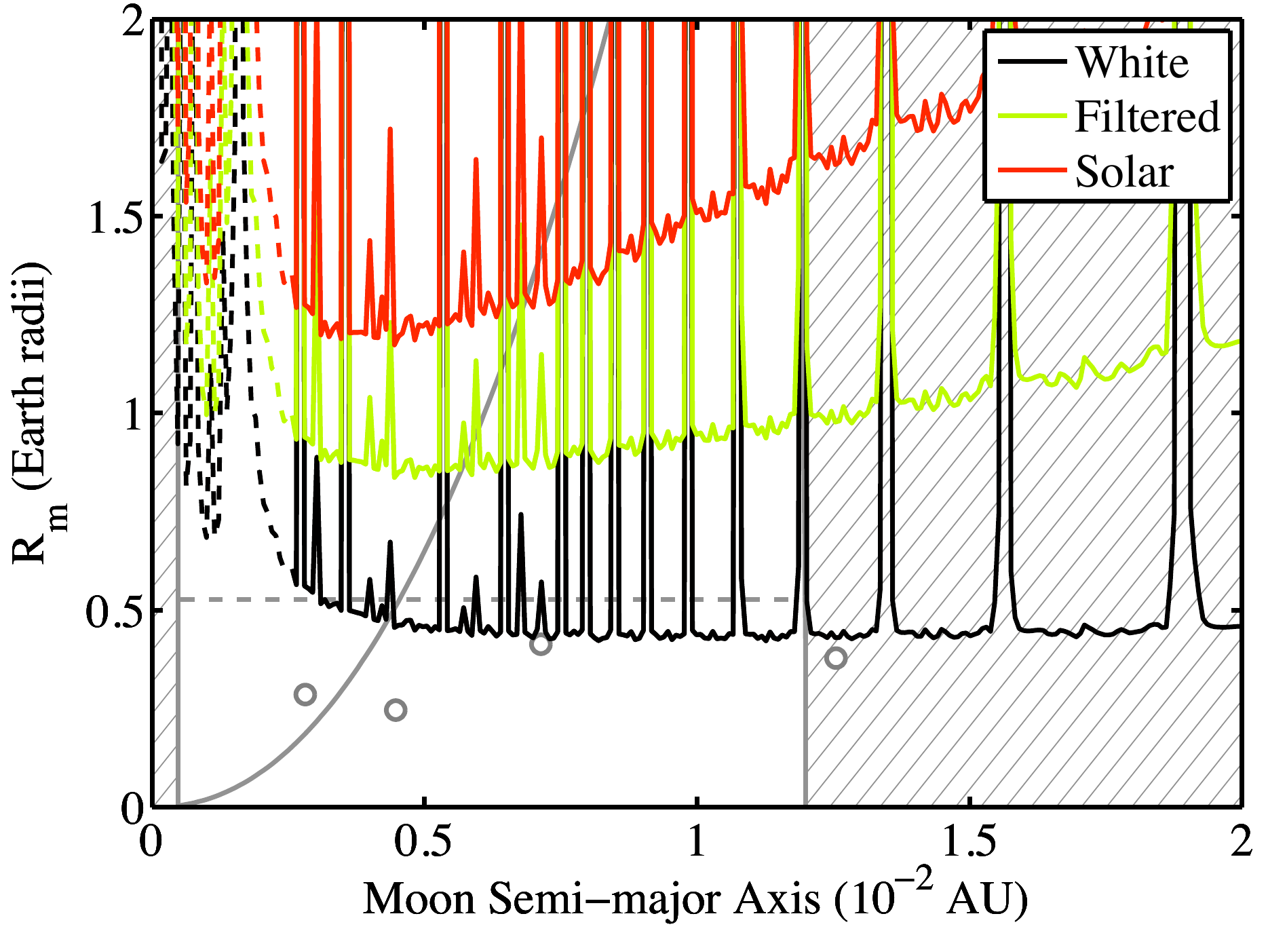}}
     \subfigure[$M_p = M_J$, $a_p=0.6$AU.]{
          \label{TransitThresh1MJ06AUEccA}
          \includegraphics[width=.315\textwidth]{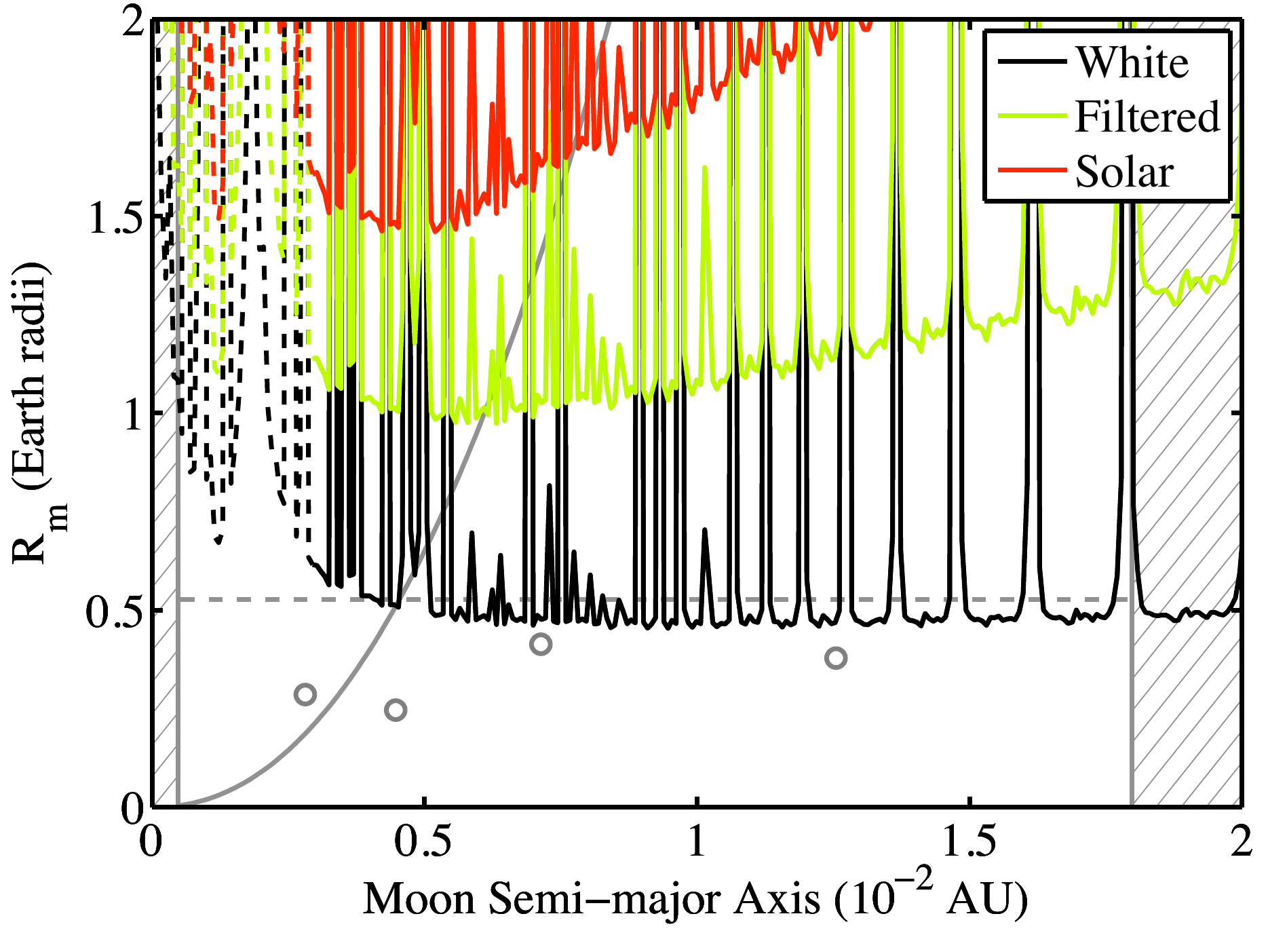}}\\ 
          \vspace{-0.2cm}
     \subfigure[$M_p = M_U$, $a_p=0.2$AU.]{
          \label{TransitThresh1MU02AUEccA}
          \includegraphics[width=.315\textwidth]{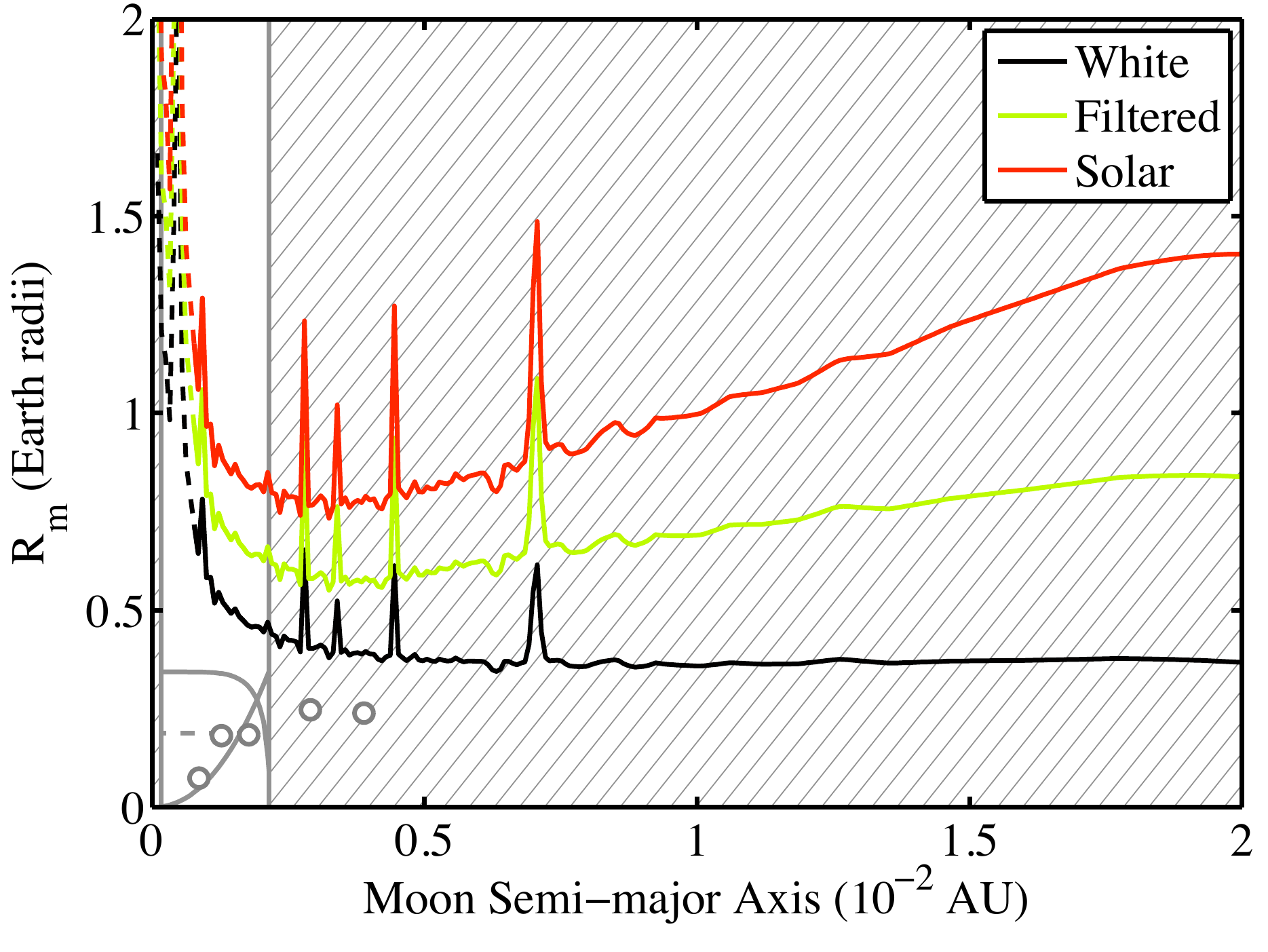}}
     \subfigure[$M_p = M_U$, $a_p=0.4$AU.]{
          \label{TransitThresh1MU04AUEccA}
          \includegraphics[width=.315\textwidth]{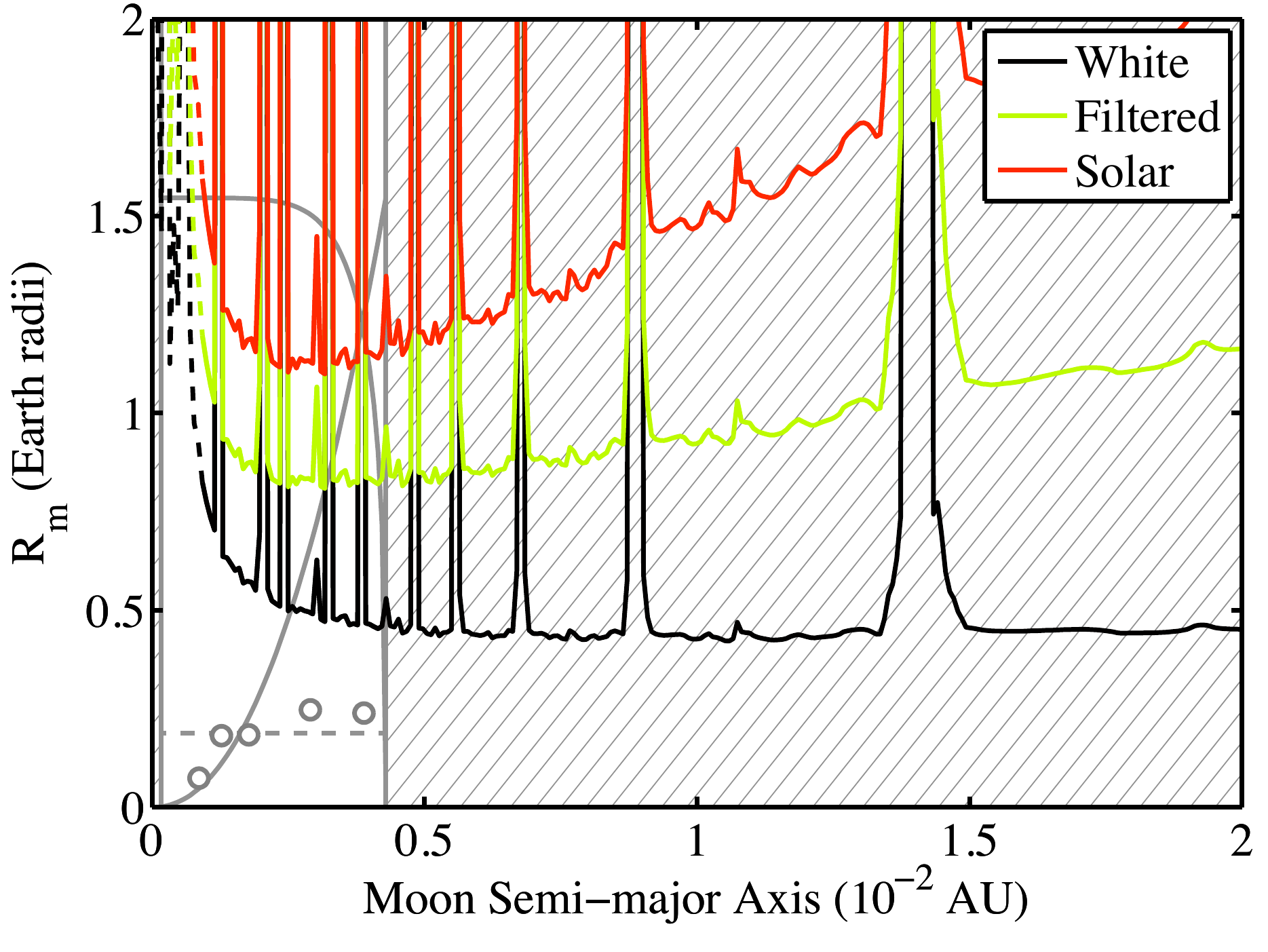}}
      \subfigure[$M_p = M_U$, $a_p=0.6$AU.]{
          \label{TransitThresh1MU06AUEccA}
          \includegraphics[width=.315\textwidth]{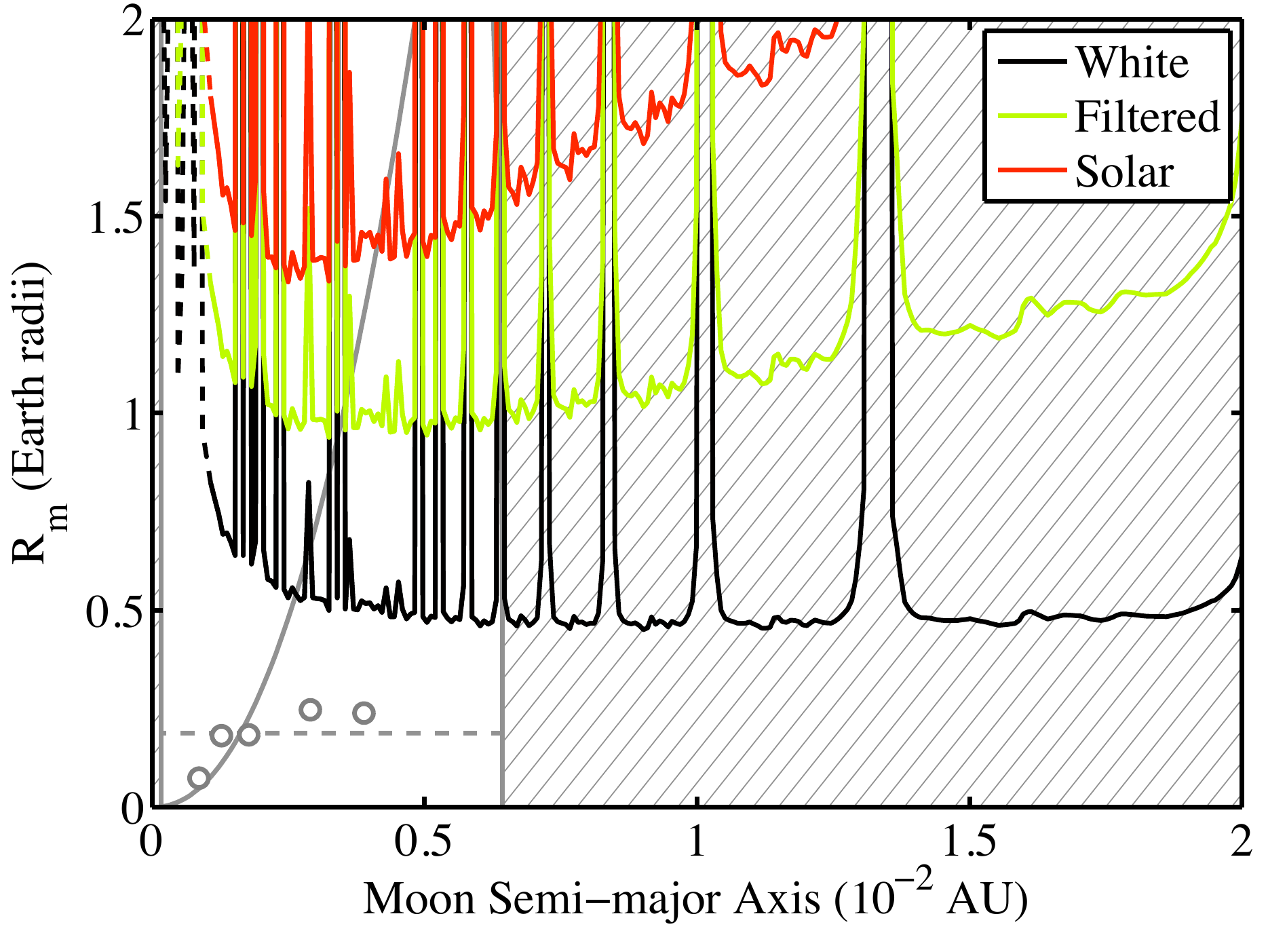}}\\ 
          \vspace{-0.2cm}
     \subfigure[$M_p = M_{\earth}$, $a_p=0.2$AU.]{
          \label{TransitThresh1ME02AUEccA}
          \includegraphics[width=.315\textwidth]{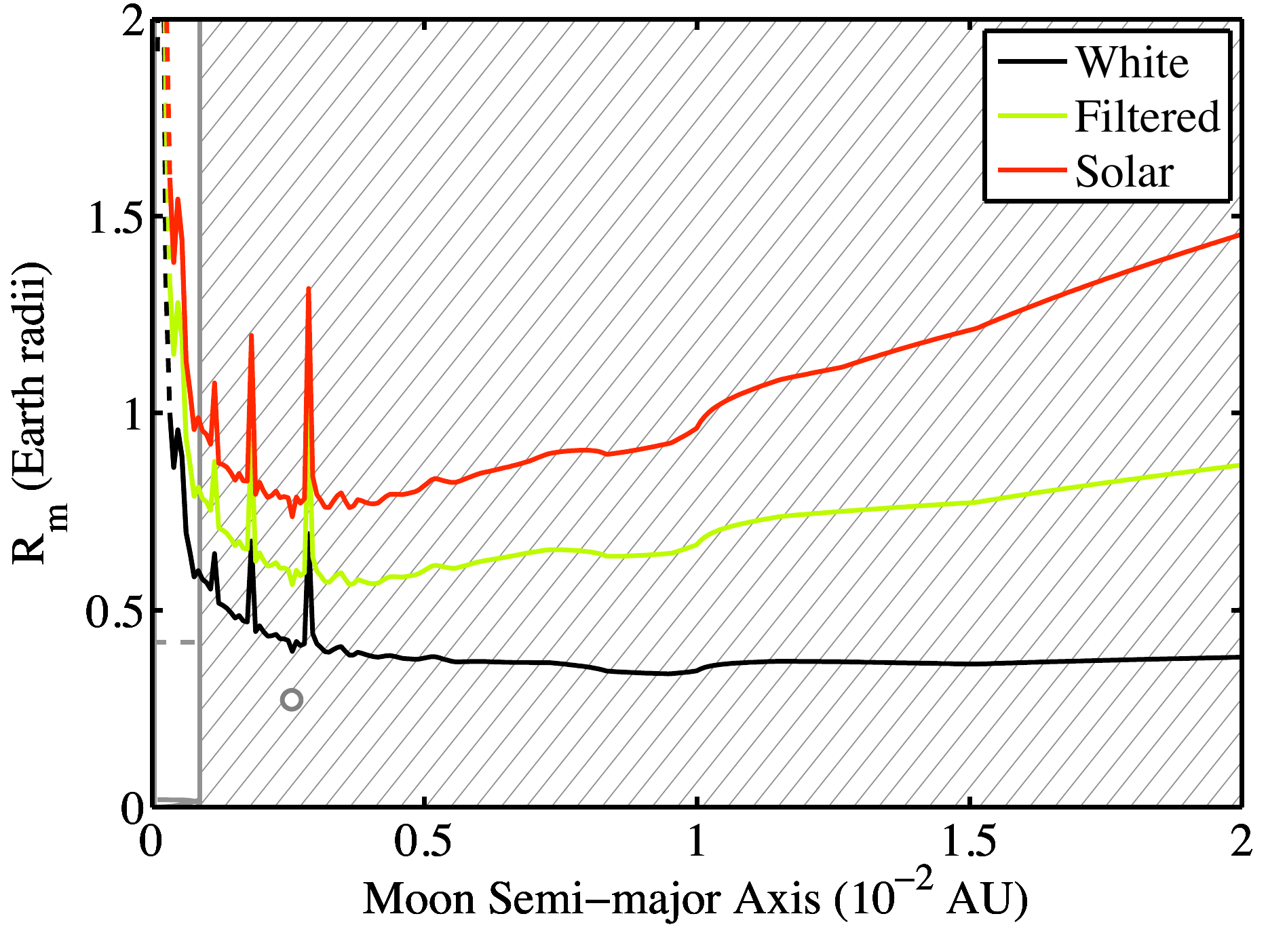}}
     \subfigure[$M_p = M_{\earth}$, $a_p=0.4$AU.]{
          \label{TransitThresh1ME04AUEccA}
          \includegraphics[width=.315\textwidth]{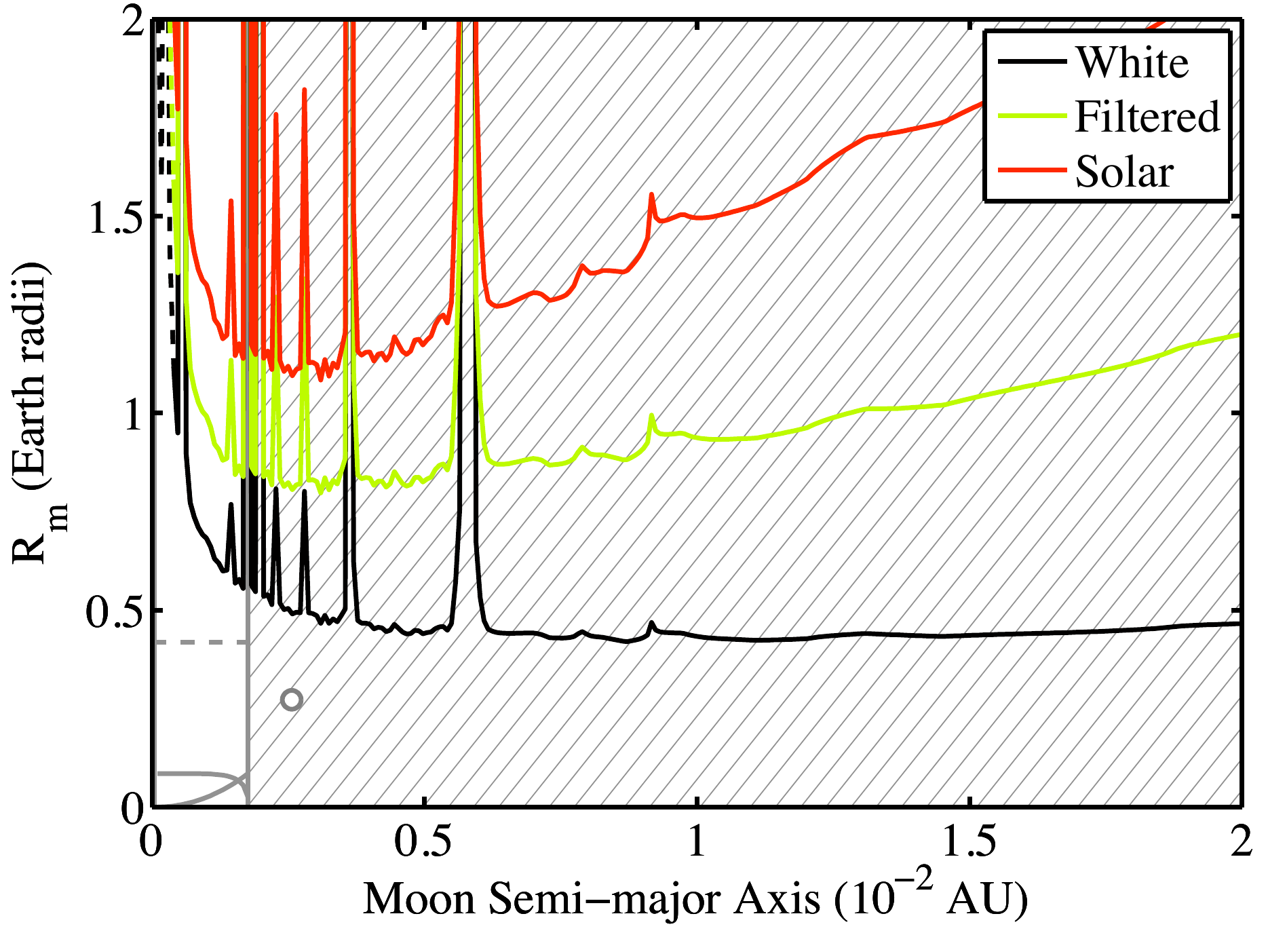}}
     \subfigure[$M_p = M_{\earth}$, $a_p=0.6$AU.]{
          \label{TransitThresh1ME06AUEccA}
          \includegraphics[width=.315\textwidth]{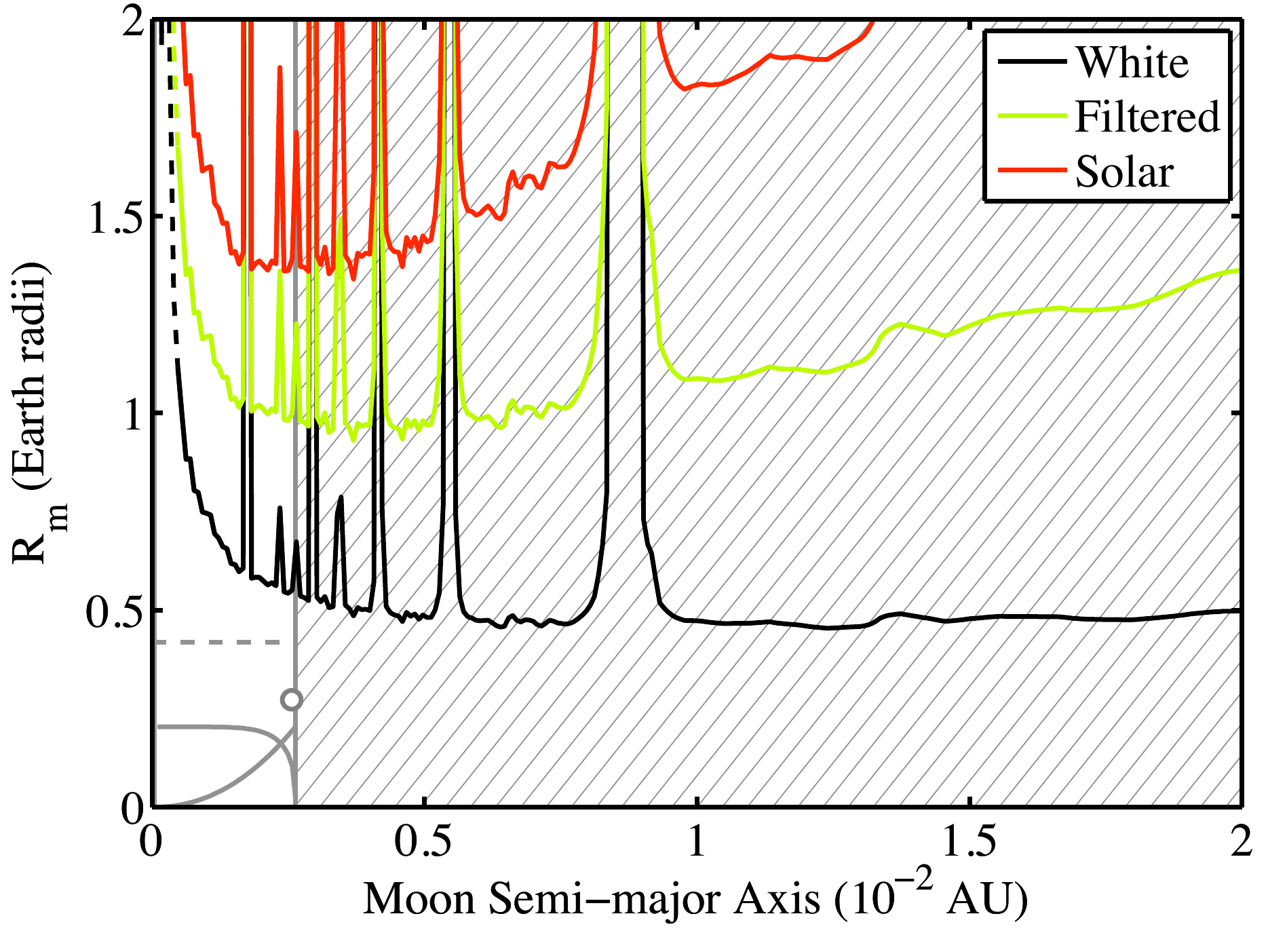}} 
          \vspace{-0.2cm}
     \caption[Figure of the same form as figure \ref{MCThresholdsAligned} except calculated for the case of an eccentric ($e_p = 0.1$) planet orbit, oriented such that the transit occurs at \emph{apastron}.]{Figure of the same form as figure \ref{MCThresholdsAligned} except calculated for the case of an eccentric ($e_p = 0.1$) planet orbit, oriented such that the transit occurs at \emph{apastron}.  As can be seen by comparing with figure \ref{MCThresholdsAligned}, the effect of eccentricity on the thresholds for this orbital orientation is to shift the white and solar noise thresholds vertically downward and upward respectively.  As a result, the three thresholds are more spread out than for the circular case.  In addition, eccentricity reduces the size of the three-body stable region (hatched),  which in turn reduces the set of moons that are tidally stable for the lifetime of the system (compare figures \ref{TransitThresh1MJ02AUcc} and \ref{TransitThresh1MJ02AUEccA}).}
     \label{MCThresholdsEccentricApo}
\end{figure}

\subsection{Eccentric orbits}\label{Trans_Thresholds_Thresh_Eccentric}

For the case where the orbit of the planet is eccentric, the transit light curve is either stretched or contracted as a result of the different value of $v_{tr}$, the velocity of the planet in the plane of the sky during transit.  In particular, there are two factors which could alter $v_{tr}$ and consequently affect the detection thresholds, the eccentricity of the orbit, and the orientation of the orbit.  To explore these two effects, two cases were analysed.  First the case of a planet with orbital eccentricity 0.1 which transits at pericenter is examined (see figure~\ref{MCThresholdsEccentricPeri}).  Second the case of a planet with orbital eccentricity 0.1 which transits at apocenter is examined (see figure~\ref{MCThresholdsEccentricApo}).  This particular eccentricity were selected as first, 0.1 is a representative value of eccentricity for extrasolar planets, second, as the size of the stable region decreases rapidly with increasing eccentricity, and third as it would be of interest to relate this work to the planet CoRoT-9~b as it has an orbital eccentricity of 0.11 and is capable of hosting large tidally stable moons \citep{Weidneretal2010}.  As pericenter and apocenter are the positions on an eccentric orbit where $v_{tr}$ is maximised and minimised respectively, these two cases should bracket the behaviour of the thresholds for any other orientation. These two cases are compared to the results for circular orbits, and then will be discussed in terms of the planet CoRoT-9~b.
 
 \subsubsection{Dependance of threshold on moon semi-major axis}
 
As can be seen in figures~\ref{MCThresholdsEccentricPeri} and \ref{MCThresholdsEccentricApo}, the threshold curves for the case of eccentric planet orbits have the same general shape as those for the case of circular orbits, in particular, have minima at $a_m = 2 R_{\sun}$, $a_m = R_{\sun}$, and between $1/2 R_{\sun}$ and $R_{\sun}$ for the case of white photometric noise, filtered photometric noise and realistic photometric noise respectively.  This is exactly what we would expect from the analysis in section~\ref{Trans_Thresholds_ExpBehav}.  However, while the thresholds have the same general shape they are vertically displaced from the equivalent aligned thresholds, with the degree of vertical displacement depending on the type of photometric noise and the orientation of the orbit.  As the amplitude of the photometric noise depends strongly on $T_{obs}$, which, in turn, depends on the semi-major axis of the planet's orbit, this effect will be discussed in greater detail in the following section.
 
 \subsubsection{Dependance of threshold on planet semi-major axis}
 
 This scaling of the transit light curve caused by the modified value of $v_{tr}$ has different affects for the different types of noise.  Consider the case at pericenter, where $v_{tr}$ is large.  In particular, $T_{tra}$ and $T_{obs}$ are 82\% of their original length.  As signal is proportional to ($T_{tra}-T_{obs}$), the thresholds will increase for the case of white noise ($\sigma_\epsilon \propto T_{obs}^{1/2}$), remain steady for the case of filtered noise ($\sigma_\epsilon \propto T_{obs}$) and decrease for the case of red noise ($\sigma_\epsilon$ is a superlinear function of $T_{obs}$).  Conversely, for the case of transit at apocenter where $v_{tr}$ is small and $T_{tra}$ and $T_{obs}$ are 122\% of their original length.  Now the thresholds will decrease for the case of white noise, remain steady for the case of filtered noise and increase for the case of red noise.  These effects can be seen in figures~\ref{MCThresholdsEccentricPeri} and \ref{MCThresholdsEccentricApo}.

\subsubsection{Comparison with formation and stability limits}

In addition to the effect that planetary orbital eccentricity has on the shape of the detection threshold, it also reduces the size of the stability region (compare the size of the hatched regions in figures~\ref{MCThresholdsAligned} and \ref{MCThresholdsEccentricPeri}),  which in turn alters moon's longevity and thus the set of physically realistic moons each planet could have.  Again, the comparison between the thresholds and the set of realistic moons will be conducted for each of the four different mass planets individually.  

For the case of a ten Jupiter mass host planet, physically realistic moons can again be detected for all investigated values of the planet semi-major axis for both white and filtered noise.  In addition, physically realistic moons can also be detected for the case of white photometric noise and a Jupiter mass host.  For the case of a one Uranus mass host and a one Earth mass host, it becomes even more difficult to detect moons as a result of the smaller stability region.  Consequently, for the case of planets in eccentric orbits, detection of physically realistic moons is only possible for larger, Jupiter-sized, planets.

\subsubsection{Comparison with the case of CoRoT-9~b}

This analysis is also of interest as a result of the similarities between the simulated cases and the detected transiting planet CoRoT-9~b.  This 0.84 Jupiter mass planet is on an orbit with semi-major axis 0.407AU and eccentricity 0.11 around a 0.99 solar mass star \citep{Deegetal2010}.  Consequently, as it transits near periastron, it is bracketed by the cases shown in figures~\ref{TransitThresh1MJ04AUEccP} and \ref{TransitThresh1MU04AUEccP}.

Unfortunately, these thresholds as they stand cannot be used to describe moon thresholds for this planet for two reasons.  First, to construct these thresholds it was assumed that the semi-major axis of the moon was known prior to detection.  Consequently, to create realistic moon detection thresholds, this assumption would have to be relaxed.  Second, the measured photometric error is approximately 4 times as large as that assumed to make these plots as first, the instrument used is CoRoT, not Kepler, and second, as the host star is a 13.7$^{th}$ magnitude star, not a 12$^{th}$ magnitude star.

However, while the thresholds shown in figure~\ref{MCThresholdsEccentricPeri} cannot be directly used to constrain moons around this planet they do indicate that such constraints would be scientifically interesting.  For example, while it is likely that a TTV$_p$ threshold could be used to place limits on moons which could be orbitally stable for the lifetime of this system, it is unlikely that it could be used to place limits on moons which we think should form, as first, the threshold and the dashed \citet{Canupetal2006} limit only barely intersect for the one Jupiter mass case analysed, and CoRoT-9 b is 0.84 Jupiter masses, and second, as the photometric noise is four times larger.  Consequently, a moon detection would be interesting as it would indicate that the formation model of \citet{Canupetal2006} is not correct, while a non-detection would also be scientifically interesting as it would increase the number of planets which follow the \citet{Canupetal2006} criterion from 4 to 5.

\section{Conclusion}

Detection thresholds were constructed for a range of realistic planet and moon parameters.  Informed by the results presented in chapters~\ref{Transit_Signal} and \ref{Trans_TTV_Noise} on the properties of $\Delta \tau$ and $\sigma_\epsilon$ the method of generalised likelihood ratio testing was used to derive an expression defining the position of the detection thresholds.  These thresholds were then investigated analytically for the case of a large number of transits and numerically for the case of a smaller, more realistic number of transits.  It was found that the TTV$_p$ moon detection threshold is given by a skewed U-shaped curve, superimposed with a comb of non-detection spikes, and that moons are more detectable for planets which are closer to their host star.  The minima of these curves is defined by the type of noise contaminating the transit light curves and the inclination of the orbit.  The depth of these curves is determined by the type of noise and is modified by the inclination, and the orbital eccentricity and orientation, with the sign of the modification depending on the orientation of the orbit and the type of photometric nose.  In addition, it may be possible to test moon formation theories with this method for the case of CoRoT-9~b.

Now that the second of the two moon detection methods investigated in this thesis has been analysed, we will summarise the work conducted and indicate directions for future research.












\cleardoublepage \pagestyle{empty} 
\part{Conclusion}
\pagestyle{plain} 
\chapter{Summary and future research directions}\label{Conclusion}

Over the course of this thesis, the question of the detectability of extra-solar moons has been addressed in various ways.  First, the pertinent issues were introduced by considering the types of moons likely to form and be retained around extra-solar planets (see chapter~\ref{Intro_Moons_Const}) and the suite of methods presented in the literature for detecting them (see chapter~\ref{Intro_Dect}).  Then, two of these methods, pulse time-of-arrival perturbation, for the case of moons of pulsar planets (see chapters~\ref{Pulsar_Paper} and \ref{Pulsar_Extension}) and photometric transit timing, for the case of moons of transiting planets (see chapters~\ref{Trans_Intro}, \ref{Transit_Signal}, \ref{Trans_TTV_Noise} and \ref{Trans_Thresholds}) were investigated in detail, in turn.  As these two sections of work were quite independent, the results and future research directions for each will be discussed in turn.

\section{Moon detection by pulse time-of-arrival perturbation}

The detectability of moons of pulsar planets was first considered using a simple model.  In particular, it was assumed that the orbit of the planet and moon were both circular and in the same plane.  This model was then applied to the case of the pulsar planet PSR B1620-26 b and used to exclude the existence of moons with mass greater than 0.125 Jupiter masses, a distance of greater than 0.46AU away from the planet.  This was the first published limit on the mass and orbital characteristics of a moon of a pulsar planet and, to my knowledge, the third\footnote{The first and second published limits were for the planets HD 209458 \citep{Brownetal2001} and OGLE-TR-113b \citep{Gillonetal2006}.  Both these planets are transiting planets.} published limit on the moon of an extra-solar planet.

This analysis was then extended by considering the case where the orbits of the planet and moon were no longer circular and in the same plane.  In particular, the effects of mutual inclination and eccentricity in both the planet's and moon's orbits on the time-of-arrival perturbation were investigated by deriving expressions correct to first order in the sine of the mutual inclination, correct for all values of mutual inclination, correct to first order in the eccentricity of the moon's orbit and correct to first order in the eccentricity of the planet's orbit.  The results of this analysis are summarised in figure~\ref{TOAFreqSplit}.

Both of these investigations constitute a preliminary analysis into this problem and could be built on in a variety of ways.  In particular:
\begin{itemize}
\item The moon threshold for the case of PSR B1620-26~b was derived assuming that the timing noise was white, with standard deviation given by a weighted average of the values presented in the literature.  However, the timing noise of pulsars is not necessarily white, in particular, it may be red, that is, have an over abundance of low frequency components.  As discussed in a different context in chapter~\ref{Trans_TTV_Noise}, red noise can affect detection thresholds.  Consequently thresholds derived directly from the observations would be much more realistic.
\item 
The expressions for the timing perturbation could be calculated using an inertial coordinate system.  The expressions for the timing perturbation presented in chapters~\ref{Pulsar_Paper} and \ref{Pulsar_Extension} were calculated using a coordinate system which was tied to the planet's orbit.  This assumption is not necessarily realistic as the orbits of the planet and moon may evolve as a result of external perturbations e.g. for the case of the PSR B1620-26 system, perturbation from the white dwarf.  As the timescale over which data has been taken is so long, over 20 years, this may be an issue.  The way to solve this is to use an inertial coordinate system and to rotate both the planet and moon's orbits using equation~\eqref{Int-Rev-RotDefn}.  While this approach would introduce more terms into the expansion, it would be more robust.
\item The moon detection thresholds calculated for PSR B1620-26 b could be extended to include the case of mutually inclined and eccentric orbits.  Using the methods presented in chapter~\ref{Pulsar_Extension}, and using the expressions given as test cases, it would be relative straightforward to implement the expansions presented in section~\ref{Pulsar_Ext_ExpansionIntro} in a computer program which produces the time of arrival perturbation as a function of the orbital elements of the planet and the moon.  Combining these results with the stability method presented in \citet{Mardling2008}, would result in a substantial improvement on the work conducted and thresholds presented in chapter~\ref{Pulsar_Paper}.
\item Finally, the PSR B1620-26 system contains a pulsar, white dwarf and planet.  Consequently, as mentioned in the paper, if the planet had a moon, the system would be a 4-body system.  Consequently, to provide more realistic constraints for this system, the full 4-body expansion would need to be used.
\end{itemize}
Now that the results and possible future research directions have been summarised for the case of this technique, the main results and and possible future research directions of the photometric transit timing technique will be discussed.

\section{Moon detection by photometric transit timing}

This thesis also addressed the detection of moons of transiting planets through an in-depth look at the photometric transit timing technique.  Initially proposed by \citet{Szaboetal2006}, this technique has only been investigated in two  works, \citet{Szaboetal2006} and \citet{Simonetal2007}.  In \citet{Szaboetal2006}, a preliminary investigation was conducted using a Monte Carlo simulation with 500 realisations to see if moons could be detected.  This was followed up with the work of \citet{Simonetal2007} who presented a formula for the amplitude of this perturbation.  In this context, the work presented in this thesis had five main objectives, which will be discussed individually.

First, in this work some of the assumptions made by \citet{Szaboetal2006} were relaxed.  In particular, the assumption that the position of the moon's transit was known prior to detection was relaxed to give the assumption that the quantity, $a_m$, was known prior to detection.  This is a substantial improvement as first, it reduces the number of variables which must be known a priori from three ($a_m$, $T_p/T_m$ and $f_m(0) + \varpi_m$) to one ($a_m$), and, second, as  $a_m$ can be estimated from formation models.  

Second, the definition of $\tau$ given by \citet{Szaboetal2006} was expanded to give explicit expressions for the perturbation due to the moon (named $\Delta \tau$) and the perturbation due to the photometric noise (named $\epsilon_j$) which are accurate for all transiting systems likely to be detected (see appendix~\ref{SecOrdNoise_App}).   Using this approach the $\Delta \tau$ signal and the noise on that signal could be investigated separately.

Third, explicit expressions for the form of $\Delta \tau$ were derived for the case where the moon's orbit was circular and aligned with the planet's orbit, and where the planet's orbit was circular and aligned to the line-of-sight, circular and inclined with respect to the line-of-sight and eccentric and aligned to the line-of-sight.  In addition, the case where the moon's orbit was slightly eccentric and the planet's orbit was circular and aligned to the line-of-sight was also investigated in appendix~\ref{EccMoon_App}.  In particular, these expressions were derived for the case where the motion of the moon during transit was approximately uniform, and the velocity of the moon around its orbit was much smaller than the velocity at which the planet-moon pair transited the star.  By comparison with $\Delta \tau$ values calculated by simulating the full light curve, it was found that these expressions described most detectable moons for low mass ($M_p < 1M_{J}$) extra-solar planets.  It was found that for the case where the moon's orbital velocity was small with respect to the transit velocity, $\Delta \tau$ was given by a sinusoid with coefficients tabulated in table~\ref{DelTauCoeffTab}.  The addition of eccentricity into the moon's orbit or a high value of the moon's orbital velocity resulted in the addition of higher order harmonics to $\Delta \tau$.  These results represent a substantial improvement over the amplitude measure of \citet{Simonetal2007} as they describe the behaviour of $\Delta \tau$ for a planet-moon pair at any position about on their mutual orbit, and as they were derived analytically, the regions for which they fail can and were documented.  

Fourth, the error in $\Delta \tau$ due to photometric noise was investigated.  To begin, a method was presented which allows the probability distribution of $\epsilon_j$ to be determined for any variety of photometric noise given that a sufficiently long sample of light curve containing this noise (e.g. out of transit light curves) is available beforehand.  Then, the case of white photometric noise was investigated analytically, followed by an investigation using this method of the effect of solar and filtered solar photometric noise.  For each of these three cases it was found that $\epsilon_j$ was Gaussian and uncorrelated for the case of planets capable of hosting substantial moons.  In addition expressions for the standard deviation of $\epsilon_j$ were also derived.  For the case of white noise the derivation of an expression for $\sigma_\epsilon$ was of particular interest as it allowed direct comparison with the qualitative results of \citet{Szaboetal2006}, who also assumed white noise.  In particular \citet{Szaboetal2006} found a relationship between decreased exposure time and increased moon detectability.  In this work, using the formula for $\sigma_\epsilon$ it was shown that this result depends strongly on the origin of the photometric noise, for example, if the photometric noise is shot noise dominated, $\sigma_\epsilon$ is independent of exposure time, if it is read noise dominated, then moon detectability increases with increasing exposure time.  Also, it was found that realistic solar noise resulted in a dramatic increase in the timing error and that while filtering reduced this, it did not completely reverse this effect.

Finally, the expressions derived for the form of $\Delta \tau$ and the behaviour of $\epsilon_j$ were combined using the technique of generalised likelihood ratio testing to give expressions for moon detection thresholds and to calculate moon detection thresholds for a number of characteristic cases.  In particular, analytic expressions were derived for the case where the number of transits tended to infinity.  For this case it was found that the TTV$_p$ technique was most sensitive to moons located at $a_m=2R_s$, $a_m=R_s$ and between $a_m=R_s$ and $a_m=1/2R_s$ for the case of white, filtered and solar noise respectively.  In addition, as a result of an analysis of the functional form of $\tau$, it was found that the threshold should be decorated with a comb of non-detection spikes with spacing dependent on $a_m$ and width dependent on $f_m(t_0)$, $a_m$ and $N$.  Both these results are in stark contrast with the result of \citet{Szaboetal2006} who qualitatively found that moons with larger semi-major axes were more detectable.  These results were then used to analyse a Monte Carlo simulation investigating the detectability of moons for small $N$.  It was found that for the test case selected, a 12$^{th}$ magnitude Sun-like host star detected by Kepler, that physically realistic moons could be detected for the case of large gas giant ($M_p \ge 1 M_{J}$) host planets for the case where the light curve is dominated by white noise and very large gas giant planets ($M_p \approx 10M_J$)  where it is dominated\footnote{Recall from chapter \ref{Trans_Thresholds} that for the light curve to be dominated by filtered noise, the amplitude of the intrinsic photometric variability of the host star would have to be approximately twice that of the Sun.} by filtered intrinsic photometric noise.  In addition, it is suggested that that following up planets such as CoRoT-9~b using this technique would be of scientific interest as it could place constraints on the moon formation model of \citet{Canupetal2006}.

\begin{figure}[tb]
\begin{center}
\includegraphics[width= 0.8\textwidth]{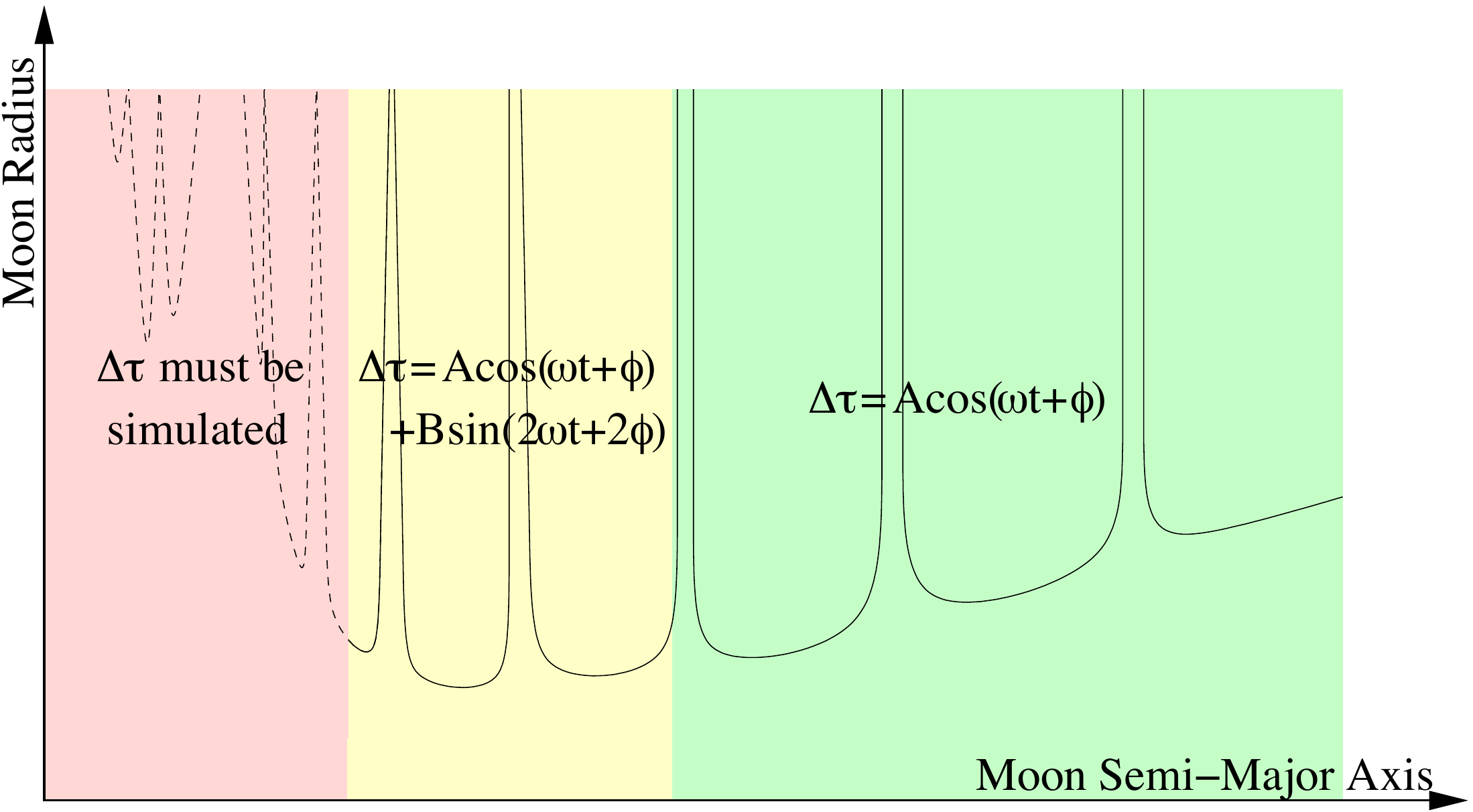}
\caption{Cartoon of a sample TTV$_p$ detection threshold showing the region where the threshold is accurate (green), and the two regions where it may not be accurate (yellow and red).} \label{TTVpThreshImprovement}
\end{center}
\end{figure}

While these results have substantially built and extended on those presented in the literature, there are many ways in which they could be improved, in particular:
\begin{itemize}
\item The assumption that $a_m$ is known a priori could be relaxed.  As this method currently stands, in order to practically search for moons using it, a semi-major axis would have to be selected (informed by the sensitivity of the method and our understanding of moon formation), e.g. $a_m = R_s$ for filtered noise, and that window using to search for moons with $a_m \le R_s$.  This is not an optimal approach, and the ability to search for the correct sized window would be a distinct improvement.  
\item Extend the method used in chapter~\ref{Transit_Signal} to describe inclined moon orbits, in particular, investigate the effect of allowing $\hat{A}_p$ and $\hat{A}_m$ to vary from transit to transit as a result of the different chords taken across the star by the planet and moon.
\item Investigate the effect of correlated noise from other sources on the TTV$_p$ moon detection threshold.  An example of a noise source of interest would be that of the CoRoT-9~b host star.
\item Investigate the behaviour of the non-detection spikes associated with $\phi$.  While these were only mentioned in this thesis, it would be useful to investigate them using an approach similar to the one used in section~\ref{Trans_TTV_Signal_CC_PropSig}.  This would be of use to first, get a handle on the number of undetectable systems as a function of $\phi$ and $N$, and second, to help explain some of the fine structure seen in the detection thresholds (for example the wiggles in gaps between the non-detection spikes on the right hand side of figure~\ref{TransitThresh1MU06AUcc}).
\item The TTV$_p$ detection thresholds could be investigated for the case where the the second harmonic of $\Delta \tau$ is important, for example, as the orbital velocity of the moon is not small or where $e_m \ne 0$.  For the case where additional harmonic is due to the large value of $v_m$, this extension results in two main challenges.  First, the $v_m/v_{tr}$ term in the expressions for $\sigma_\epsilon$ would have to be taken into account.  While generalised likelihood ratio testing can be used to investigate signals with standard deviations which depend on the signal, it is not a trivial extension.  Second, As the amplitude of the second harmonic of $\Delta \tau$ depends on $a_m$ (see equation~\eqref{transit_signal_cc_form_hB}), the transformation method presented in appendix~\ref{App_transformFitParams} can no longer be used and a simulation will have to be run for each semi-major axis of interest.  However, as $\Delta \tau$ can be accurately estimated as a function of $\phi$, it should not be computationally challenging to determining accurate thresholds in this region.  This region is shaded yellow in figure~\ref{TTVpThreshImprovement}.
\item The TTV$_p$ detection thresholds could be investigated for the case where the expressions derived in chapter~\ref{Transit_Signal} fail completely (the moon's transit is asymmetric or moon's orbital velocity is of the same order of magnitude or larger than the transit velocity).  As $\Delta \tau$ must now be numerically simulated for each realisation of $\phi$, $a_m$ and $R_m$, this is a very computationally intensive operation.  The computational load could be reduced by using information on the behaviour of the $\phi$ non-detection spikes  to select a much smaller set of representative values of $\phi$ (5 as opposed to 50), simulating the light curves for these values and then investigating 10 realisations for each $\phi$ value.  While this is not a trivial task it has the advantage that it will allow construction of realistic thresholds for the case of large ($M_p \approx 10 M_J$) planets,  which are the planets likely to have the largest detectable moons and thus opening up a scientifically interesting portion of parameter space.  This region is shaded red in figure~\ref{TTVpThreshImprovement}.
\item Extend the analysis to include additional significance levels for the detection thresholds.  One way that this could be done would be by running additional Monte Carlo simulations to determine the form of the null distribution of $2\log\Lambda$.  To give a context, recall from section~\ref{Trans_Thresholds_MC_Method} that 60,000 simulations were required to calculate the 99.7\% confidence limit (3 sigma threshold) for the null distribution.  Using this method, approximately 2,500,000 simulations would have to be run to obtain the 99.99994\% confidence limit (4 sigma threshold) with comparable accuracy.  Another approach would be to assume that $N$ is large and approximate the null distribution with a $\chi^2$ distribution with three degrees of freedom.
\end{itemize}

\section{Conclusion}

Over the course of this thesis the detectability of moons of extra-solar planets has been investigated with particular reference to the pulse time-of-arrival perturbation and photometric transit timing techniques.  Approximate expressions for the moon detection thresholds were derived analytically for both cases and compared with moon formation and stability predictions.  This culminated in a limit being placed on the possible moons of PSR~B1620-26~b for the case of the pulse time-of-arrival perturbation technique, and a sequence of generic thresholds being calculated for moons of transiting planets for the case of the photometric transit timing technique.  While this represents a substantial advance in our understanding of the detectability of moons of extra-solar planets, there is still much more to be done!


\cleardoublepage \pagestyle{empty} 
\part*{Appendicies}
\pagestyle{plain} 
\appendix

\chapter{Definition of variables}\label{VarDef_App}

\begin{longtable}{lp{9cm}}
Variable Name       	& Definition\\
\hline
$a_m$ 			& Semi-major axis of the moon's orbit.\\
$a_p$ 			& Semi-major axis of the planet's orbit.\\
$A$ 				& Amplitude of $\Delta \tau$.\\
$\hat{A}$ 			& Most likely value of $A$, assuming that there is a moon.\\
$A_1$ 			& Relative area of the $1^{st}$ sunspot.\\
$A_2$ 			& Relative area of the $2^{nd}$ sunspot.\\
$A_3$ 			& Relative area of the $3^{rd}$ sunspot.\\
$A_m$ 			& Sum of $\alpha_m$ over a given transit.\\
$\hat{A}_m$ 		& The value of $A_m$ for the case where the moon travels with uniform velocity $v_{tr}$ across the face of the star during transit.\\
$A_p$  			& Sum of $\alpha_p$ over a given transit.\\
$\hat{A}_p$ 		& The value of $A_p$ for the case where the planet travels with uniform velocity $v_{tr}$ across the face of the star during transit.\\
$\alpha(t)$		& Total photometric deficit during a transit.\\
$\alpha_m(t)$ 		& Photometric deficit resulting from the transit of the moon.\\
$\alpha_n(t)$ 		& Photometric deficit due to photometric noise.\\
$\alpha_p(t)$		& Photometric deficit resulting from the transit of the planet.\\
$B$ 				& Variable representing $v_m/v_{tr}$, $-v_m/v_{tr}$, $-v_p/v_{tr}$ or $v_p/v_{tr}$ depending on whether the equation describes the ingress of the moon's transit, the egress of the moon's transit, the ingress of the planet's transit or the egress of the planet's transit.\\
$\beta$ 			& Factor by which the photometric noise of the Sun is scaled.  In chapter~\ref{Trans_Thresholds}, $\beta$ is set to 1.9.\\
$c$  				& The speed of light.\\
$D_{lmm'}\left(I,\omega,\Omega \right)$ & A Wigner D-function.  See equation \eqref{Int-Rev-DfunDefn} for the definition.\\
$\delta_m(t)$ 		& Projected distance between the center of the star and the moon.\\
$\delta_{min}$ 		& Minimum value of $\delta_m(t)$ or $\delta_p(t)$.\\
$\delta_p(t)$ 		& Projected distance between the center of the star and the planet.\\
$e_m$ 			& Eccentricity of the moon's orbit.\\
$e_p$ 			& Eccentricity of the planet's orbit.\\
$E_m(t)$ 			&Eccentric anomaly of the moon.  See equation \eqref{transit_signal_coord_Eidef} for the definition.\\
$E_p(t)$ 			&Eccentric anomaly of the planet.  See equation \eqref{transit_signal_coord_Eodef} for the definition.\\
$\epsilon_j$ 		& TTV$_p$ timing perturbation due to photometric noise.\\
$\epsilon^*$  		& The error in $\tau^*$ due to photometric noise.\\
$\epsilon_{\sun}$ 	& The error in $\tau^*$ due to solar photometric noise.\\
$f_m(t)$ 			& True anomaly of the moon's orbit.\\
$f_p(t)$ 			& True anomaly of the planet's orbit.\\
$F(e,\omega)$  	& Function describing the effect of orbital eccentricity and orientation on transit duration (see equation~\eqref{transit_intro_dur_ecc_Fdef}).\\
$F^{(lm)}_n(e_p)$  	& Moon eccentricity function.  See appendix \ref{App_Ecc_Fun} for more information.\\
$\phi$ 			& Phase of $\Delta \tau$.\\
$\Phi$ 			& Variable representing the position of the moon during the $j^{th}$ transit.  The exact definitions are given in table~\ref{ABTable} or \ref{ABTableInEc} depending on whether the moon's orbit is circular or eccentric. \\
$\hat{\phi}$ 		& Most likely value of $\phi$, assuming that there is a moon.\\
$G$ 				& The universal gravitational constant.\\
$\gamma_{lmm'}(I_m)$ & Inclination functions.  See appendix \ref{Incl_App} for more information.\\
$I_m$ 			& Inclination of the moon's orbit.\\
$I_p$ 			& Inclination of the planet's orbit.\\
$j$ 				& Transit number.\\
$J_k(x)$ 			& A Bessel function of the first kind.  See equation \eqref{transit_signal_cc_Besseldef} for the definition.\\
$k_{2p}$  			& Love number of the planet.\\
$l$ 				& Summation index.\\
$L(t)$  				& Luminosity of a star.\\
$L^*(t)$ 			& Luminosity of a star where the amplitude of the photometric noise has been increased by a factor of $\beta$.\\
$L_0$  			& Average luminosity of a star.\\
$L_{fit}(t)$  			& Best fit curve to $L$.\\
$L_n(t)$  			& Zero mean photometric noise of a star.\\
$L_r$ 			& Background intensity due to the solar network. \\
$\lambda_1$ 		& Longitude of the $1^{st}$ sunspot.\\
$\lambda_2$ 		& Longitude of the $2^{nd}$ sunspot.\\
$\lambda_3$ 		& Longitude of the $3^{rd}$ sunspot.\\
$\Lambda$  		& The ratio of the probability that data was produced under the null hypothesis to the probability that it was produced under the alternative hypothesis.\\
$m$ 				& Summation index.\\
$m'$ 			& Summation index.\\
$M_m$ 			& Mass of the moon.\\
$M_m(t)$ 			& Mean anomaly of the moon's orbit. Note that the time dependance differentiates this from $M_m$, the mass of the moon.\\
$M_p$ 			& Mass of the planet.\\
$M_p(t)$ 			& Mean anomaly of the planet's orbit.  Note that the time dependance differentiates this from $M_p$, the mass of the planet.\\
$M_s$ 			& Mass of the star.\\
$M_J$  			& Mass of Jupiter, $1.90\times 10^{27}$kg.\\
$M_U$  			& Mass of Uranus, $8.68\times 10^{25}$kg.\\
$M_{\earth}$  		& Mass of the Earth, $5.97\times 10^{24}$kg.\\
$M_{\sun}$  		& Mass of the Sun, $1.99\times 10^{30}$kg.\\
$\mu_\epsilon$  	& The mean of the distribution of $\epsilon_j$.  This is equal to zero by definition.\\
$n$ 				& Summation index.\\
$n_m$ 			& Mean motion of the planet and moon about their common barycenter.\\
$n_p$ 			& Mean motion of the planet-moon system around the host star.\\
$\mathbf{n}$ 		& A unit vector directed along the line-of-sight. \\
$\mathbf{n}_m$ 	& A unit vector normal to the plane of the moon's orbit.\\
$N$ 				& Number of recorded transits.\\
$N_{obs}$ 		& Number of exposures taken during the observing window.\\
$N_{tra}$ 			& Number of exposures taken during the transit.\\
$P_l^m(\cos \theta)$ & Associated Legendre polynomial of degree $l$ and order $m$.  Defined in equation \eqref{Int-Rev-PlmDef}.\\
$\rho_m$ 			& Density of the moon.\\
$\rho_p$ 			& Density of the planet.\\
$Q_p$ 			& $Q$-value of the planet.\\
$\theta_1$ 		& Latitude of the $1^{st}$ sunspot.\\
$\theta_2$ 		& Latitude of the $2^{nd}$ sunspot.\\
$\theta_3$ 		& Latitude of the $3^{rd}$ sunspot.\\
$\theta_{eg,m}$ 	& $n_m t_{eg,m} + f_m(0)+\omega_m + \pi/2$.\\
$\theta_{eg,p}$ 	& $n_m t_{eg,p} + f_m(0)+\omega_m + \pi/2$.\\
$\theta_{in,m}$ 	& $n_m t_{in,m} + f_m(0)+\omega_m + \pi/2$.\\ 
$\theta_{in,p}$ 		& $n_m t_{in,p} + f_m(0) +\omega_m + \pi/2$.\\
$\theta_m$ 		& Spherical polar angle describing the angular orientation of $\mathbf{r}_m$ (see figure \ref{RpRmOrientDia}).\\
$\theta_p$ 		& Spherical polar angle describing the angular orientation of $\mathbf{r}_p$ (see figure \ref{RpRmOrientDia}).\\
$\mathbf{r}$ 		& Jacobian coordinate defined in figure \ref{ThreeBodySchematic}.\\
$\mathbf{r}_m$ 	& Jacobian coordinate directed from the planet to the moon.  See figure \ref{StarPlanetMoonThreeBodySchematic}.\\
$\mathbf{r}_p$ 		& Jacobian coordinate directed from the planet-moon barycenter to the star.  See figure \ref{StarPlanetMoonThreeBodySchematic}.\\
$R_c$ 			& The centrifugal radius for the accretion disk around a planet.  For this thesis it is taken to be equal to $R_H/48$.\\
$R_m$ 			& Radius of the moon.\\
$R_p$ 			& Radius of the planet.\\
$R_s$ 			& Radius of the star.\\
$R_H$  			& The Hill radius of the planet.  Defined in equation~\eqref{intro_limits_stab_RHdef}.\\
$R_R$  			& The Roche tidal radius of the planet.  Defined in equation~\eqref{intro_limits_stab_RRdef}.\\
$R_{\sun}$  		& Radius of the Sun, $6.96\times10^{8}$m.\\
$\mathbf{R}$ 		& Jacobian coordinate defined in figure \ref{ThreeBodySchematic}.\\
$\mathbf{R}_s$ 	& The vector from the system barycenter to the star.\\
$\mathcal{R}$ 		& The disturbing function.  See equation \eqref{intro_not_spm_DistFunDef}.\\
$s^{(lm)}_n(e_m)$  	& Moon eccentricity function.  See appendix \ref{App_Ecc_Fun} for more information.\\
$\sigma_L$ 		& The standard deviation of white photometric noise.\\
$\sigma_\epsilon$ 	& The standard deviation of $\epsilon_j$.\\
$\sigma_{\sun}$  	& The standard deviation of $\epsilon_{\sun}$.\\
$t$ 				& Time.\\
$t_0$			& Mid-time of the zeroth transit assuming that there is no moon.\\
$\overline{t}_0$ 	& Most likely value of $t_0$, assuming that there is no moon.\\
$\hat{t}_0$ 		& Most likely value of $t_0$, assuming that there is a moon.\\
$t_{eg,m}$ 		& Time of moon egress, that is, the time at which the center of the moon passes off the limb of the star.\\
$t_{eg,p}$ 		& Time of planetary egress, that is, the time at which the center of the planet passes off the limb of the star.\\
$t_{in,m}$ 		& Time of moon ingress, that is, the time at which the center of the moon passes onto the limb of the star.\\
$t_{in,p}$ 			&Time of planetary ingress, that is, the time at which the center of the planet passes onto the limb of the star.\\
$t_{mid}$ 			& Mid-time of the planetary transit for the case where the planet has no moon.\\
$t_{mid,m}$  		& Mid-time of the moon's transit for the case where the planet has a moon.\\
$t_{mid,p}$  		& Mid-time of the planetary transit for the case where the planet has a moon.\\
$T_{in}$ 			& The duration of ingress for the case of a planet with no moon.\\
$T_{in,m}$ 		& The duration of ingress for the moon for the case of a planet with a moon.\\
$T_{in,p}$ 		& The duration of ingress for the planet for the case of a planet with a moon.\\
$T_m$  			& Orbital period of moon.\\
$T_p$  			& Orbital period of planet.\\
$\overline{T}_p$ 	& Most likely value of $T_p$, assuming that there is no moon.\\
$\hat{T}_p$ 		& Most likely value of $T_p$, assuming that there is a moon.\\
$T_{tra}$ 			& The transit duration of a planet with no moon.\\
$T_{tra,m}$ 		& The transit duration of the moon for the case of a planet with a moon.\\
$T_{tra,p}$ 		& The transit duration of the planet for the case of a planet with a moon.\\
$TOA_{pert,p}(t)$ 	& Time-of-arrival perturbation due to orbit of the the planet-moon system about the pulsar.\\
$TOA_{pert,pm}(t)$ 	& Time-of-arrival perturbation due to planet-moon binarity.\\
$\Delta t$  		& Time between consecutive exposures.\\
$\Delta t_p$ 		& Time delay in the center of the planetary transit due to the motion of the planet about the planet-moon barycenter.\\
$\tau_m$ 			& First moment of $\alpha_m$ over a given transit.\\
$\tau_j$ 			& The $\tau$ value calculated for the $j^{th}$ transit.  See equation~\eqref{TraM-TTV-taudef} for the definition of $\tau$.\\
$\tau_p$ 			& First moment of $\alpha_p$ over a given transit.\\
$\tau^*$ 			& Defined in equation \eqref{transit_noise_method_taustardef}.\\
$\Delta \tau(j)$ 		& TTV$_p$ timing perturbation due to a moon.\\
$v_m$ 			& Velocity of the moon about the planet-moon barycenter.\\
$v_p$ 			& Velocity of the planet about the planet-moon barycenter.\\
$v_{tr}$ 			& Velocity at which the planet-moon barycenter transits the face of a star.\\
$\omega$  		& Angular frequency of $\Delta \tau$.\\
$\hat{\omega}$ 	& Most likely value of $\omega$, assuming that there is a moon.\\
$\omega_m$ 		& Argument of pericenter of the moon's orbit.\\
$\omega_p$ 		& Argument of pericenter of the planet's orbit.\\
$\Omega$  		& Angular velocity associated with the rotation of the Sun.\\
$\Omega_m$		& Longitude of the ascending node of the moon's orbit.\\
$\Omega_p$ 		& Longitude of the ascending node of the planet's orbit.\\
$x_m(t)$ 			& $x$-coordinate of moon.\\
$x_p(t)$ 			& $x$-coordinate of planet.\\
$y_m(t)$ 			& $y$-coordinate of moon.\\
$y_p(t)$ 			& $y$-coordinate of planet.\\
$\varpi_m$ 		& $\omega_m + \Omega_m$.\\ 
$\varpi_p$ 		& $\omega_p + \Omega_p$.\\
$Y_{lm}(\theta,\psi)$ & Spherical harmonic of degree $l$ and order $m$. Defined in equation \eqref{Int-Rev-YlmDef}.\\
$\psi_m$ 			& Spherical polar angle describe the angular orientation of $\mathbf{r}_m$ (see figure \ref{RpRmOrientDia}).\\
$\psi_p$ 			& Spherical polar angle describe the angular orientation of $\mathbf{r}_p$ (see figure \ref{RpRmOrientDia}).\\
\end{longtable}
 
\chapter{Important equations}

\section{Detecting moons of pulsar planets}

Definition of $TOA_{pert,pm}$:
\begin{equation}\tag{\ref{PulM-GovEq-GovEq2}}
TOA_{pert,pm} = -\frac{1}{c} \frac{1}{M_s} \int_0^t \int_0^{t'}
\frac{\partial \mathcal{R}}{\partial \mathbf{r}_p} 
\cdot \mathbf{n} dt' dt.
\end{equation}
Expansion of the disturbing function using spherical harmonics:
\begin{equation}\tag{\ref{pulsar_signal_DistFunExpan}}
\mathcal{R} = - \frac{G M_m M_p M_s}{M_m + M_p}\sum_{l=2}^\infty
\sum_{m=-l}^l \frac{4 \pi}{2l+1} M_l \frac{
r_m^l}{r_p^{l+1}}Y_{lm}(\theta_m,\psi_m)Y_{lm}^*(\theta_p,\psi_p).
\end{equation}
Expressions for $TOA_{pert,pm}$ are presented in equations~\eqref{PulM-CC-Eq5}, \eqref{PulM-Incl-GenI3}, \eqref{PulM-Ecc-Inn-Small5} and  \eqref{PulM-Ecc-Out-Eq5} for the cases of circular coplanar orbits, circular mutually inclined orbits, slightly eccentric moon orbits, and slightly eccentric planet orbits respectively.

\section{Detecting moons of transiting planets}

Definition of $\tau$:
\begin{equation}\tag{\ref{TraM-TTV-taudef}}
\tau = \frac{\sum_i t_i \alpha(t_i)}{\sum_i \alpha(t_i)},
\end{equation}
\begin{equation}\tag{\ref{TraM-TTV-taufitdef}}
\tau_j = t_0 + jT_p + \Delta \tau + \epsilon_j.
\end{equation}
Definition of $\Delta \tau$:
\begin{equation}\tag{\ref{transit_intro_ground_deltaudef}}
jT_p + t_0 + \Delta \tau = \frac{A_p\tau_p + A_m\tau_m}{A_p + A_m}
\end{equation}
Approximate expression for $\Delta \tau$ for the case where the moon's orbit is circular and coplanar with the planet's orbit:
\begin{equation}
\Delta \tau = A \cos (\omega j + \phi),
\end{equation}
where expressions for the coefficients $A$, $\omega$ and $\phi$ are given in table~\ref{DelTauCoeffTab}.
Definition of $\epsilon_j$:
\begin{equation}\tag{\ref{transit_intro_ground_noisedef}}
\epsilon_i = \frac{1}{A_p + A_m}\sum_i\left[t_i -  (jT_p + t_0 +\Delta \tau) \right]\alpha_n(t_i).
\end{equation}
Expressions for standard deviation of $\epsilon_j$ for the cases of white, realistic and filtered realistic noise:
\begin{multline}\tag{\ref{transit_noise_white_sigdefphys}}
\sigma_\epsilon =  47.9\text{s} \left[\left(\frac{\sigma / L_0}{3.95 \times 10^{-4}}\right)\left(\frac{\Delta t}{1 \text{min}}\right)^{1/2}\right] \left[\frac{100(A_p + A_m)}{N L_0} \right]^{-1} \\
\times \left[\left(\frac{D_{obs}}{24 \text{hrs}}\right)^{3/2} \left(\frac{D_{tra}}{13 \text{hrs}}\right)^{-1}\right],
\end{multline}
\begin{multline}\tag{\ref{transit_noise_red_sigdefphys}}
\sigma_\epsilon = 103.7\text{s} \left[\beta \right]\left[\frac{L_{0}N_{tra}}{100(A_p + A_m)}\right] \left[ \left(\frac{D_{tra}}{13 hr}\right)^{-1} \right. \\ \left.
\times \left(0.010 \left(\frac{D_{obs}}{24 \text{hrs}}\right) + 0.277 \left(\frac{D_{obs}}{24 \text{hrs}}\right)^2 + 0.714 \left(\frac{D_{obs}}{24 \text{hrs}}\right)^3\right)\right], 
\end{multline}
\begin{multline}\tag{\ref{transit_noise_filt_sigdefphys}}
\sigma_\epsilon = 53.2\text{s} \left[\beta\right] \left[\frac{100(A_p + A_m)}{L_{0}N_{tra}}\right]^{-1} \\ \times\left[ \left(\frac{D_{tra}}{13 \text{hrs}}\right)^{-1} \left(-8 \times 10^{-3} \left(\frac{D_{obs}}{24 \text{hrs}}\right) + 1.008 \left(\frac{D_{obs}}{24 \text{hrs}}\right)^2 \right)\right]. 
\end{multline}
Definition of threshold test statistic $\Lambda$:
\begin{multline}\tag{\ref{transit_thresholds_method_thresholddef}}
2\log(\Lambda) = \sum_{j = 1}^N\frac{(\tau_j - (\overline{t}_0+j\overline{T}_p))^2}{\sigma_\epsilon^2} \\- \sum_{j = 1}^N\frac{(\tau_j - (\hat{t}_0+j\hat{T}_p+ \hat{A}\cos(\hat{\omega} j + \hat{\phi})))^2}{\sigma_\epsilon^2}.
\end{multline}

\chapter{Inclination functions}
\label{Incl_App}

A spherical harmonic $Y_{lm}$ can be written as a sum of spherical harmonics of the same degree, which have been rotated by the three Euler angles $I$, $\omega$ and $\Omega$.  This process is
described in the following equation:
\begin{equation}
Y_{lm}(\theta,\psi) = \sum_{m'=-l,2}^l
\mathcal{D}_{mm'}^{(l)}(I,\omega,\Omega)Y_{lm'}(\pi/2,f),
\label{AppIncInt1}
\end{equation}
where the $\mathcal{D}_{mm'}^{(l)}(I,\omega,\Omega)$ are the Wigner-D functions
used in quantum mechanics. For a more complete
review of the use of Wigner-D functions to describe inclined
systems, please see \citet{MardlingPrep}.

\section{Derivation of inclination functions}

Following \citet{MardlingPrep}, the Wigner-D functions can be written as
\begin{equation}
\mathcal{D}_{mm'}^{(l)}(I,\omega,\Omega) =
\left(-i\right)^{2l+m+m'}\gamma_{lmm'}(I)e^{i(m'\omega + m\Omega)},
\label{AppIncDer1}
\end{equation}
where
\begin{equation}
\gamma_{lmm'}(I) = \sum_{n = n_{min}}^{n_{max}} \beta_{lmm'}^{(n)}
\left(\cos \frac{I}{2}\right)^{2n - m - m'} \left(\sin
\frac{I}{2}\right)^{2l - 2n + m + m'}, \label{AppIncDer2}
\end{equation}
and
\begin{equation}
\beta_{lmm'}^{(n)} = (-1)^n \left(
\begin{array}{c}
    l+m \\
    n   \\
\end{array}
\right)\left(
\begin{array}{c}
    l-m    \\
    n-m-m' \\
\end{array}
\right) \left[\frac{(l+m')!(l-m')!}{(l+m)!(l-m)!}\right]^{1/2},
\label{AppIncDer3}
\end{equation}
where $n_{min}=\text{max}\left[0,m+m'\right]$ and
$n_{max}=\text{min}\left[l+m,l+m'\right]$. For convenience, these
$\gamma_{lmm'}(I)$ have been tabulated in table~\ref{App-Inc-GamTab}
for the case of $l=2$.

\begin{table}[tb]
\begin{center}
  \begin{tabular}{llll}
  \hline
  $l$ 	& $m$ 	& $m'$ 	& $\gamma_{lmm'}(I)$\\
  \hline
  2   	& 2   		& 2    	& $\frac{1}{4}\left(1 + \cos I\right)^2$\\
      	&     		& 0    	& $\frac{1}{2}\sqrt{\frac{3}{2}}\sin^2 I$\\
      	&     		& -2   	& $\frac{1}{4}\left(1 - \cos I\right)^2$\\
  2   	& 1   		& 2    	& -$\frac{1}{2}\sin I\left(1 + \cos I\right)$\\
      	&     		& 0    	& $\frac{1}{2}\sqrt{\frac{3}{2}}\sin(2I)$\\
      	&     		& -2   	& $\frac{1}{2}\sin I\left(1 - \cos I\right)$\\
  2   	& 0   		& 2    	& $\frac{1}{2}\sqrt{\frac{3}{2}}\sin^2I$\\
      	&     		& 0    	& $P_2(\cos I)$\\
      	&     		& -2   	& $\frac{1}{2}\sqrt{\frac{3}{2}}\sin^2I$\\
  \hline
  \end{tabular}\\
\caption[Table of inclination functions $\gamma_{lmm'}(I)$. ]{Table of inclination functions $\gamma_{lmm'}(I)$.  While $\gamma_{lmm'}(I)$
is also defined for negative $m$, these have not been tabulated as
$\gamma_{l-m m'}(I)$ can easily be generated from $\gamma_{lm -m'}(I)$
using equation~\eqref{AppIncProp1}.}\label{App-Inc-GamTab}
\end{center}
\end{table}

\section{Properties of inclination functions}

From equations~\eqref{Int-Rev-YlmDef} and \eqref{Int-Rev-PlmProp},
it can be seen that $Y_{l-m} = (-1)^m Y_{lm}^*$.  Using this and
equation~\eqref{AppIncDer1} it can be shown that
\begin{equation}
\gamma_{l-m-m'}(I) = (-1)^{m + m'}\gamma_{lmm'}(I). \label{AppIncProp1}
\end{equation}

\chapter{Eccentricity functions} \label{App_Ecc_Fun}

Following \citet{Mardling2008}, the coefficients $s_n^{lm}(e_m)$ and $F_n^{lm}(e_p)$ are defined as
\begin{align}
s_n^{lm}(e_m) &= \frac{1}{2\pi}\int_0^{2\pi}\frac{r_m^l}{a_m^l}e^{imf_m}
e^{-inM_m(t)}dM_m(t),\label{AppEccIntDefGens}\\
F_n^{lm}(e_p) &=
\frac{1}{2\pi}\int_0^{2\pi}\frac{a_p^{(l+1)}}{r_p^{(l+1)}}e^{-imf_p}
e^{inM_p(t)}dM_p(t),\label{AppEccIntDefGenF}
\end{align}
such that
\begin{align}
\frac{r_m^l}{a_m^l}e^{imf_m} &= \sum_{n = -\infty}^\infty s_n^{lm}(e_m)
e^{inM_m(t)},
\label{AppEccIntSums}\\
\frac{a_p^{(l+1)}}{r_p^{(l+1)}}e^{-imf_p} &= \sum_{n = -\infty}^\infty
F_n^{lm}(e_p) e^{-inM_p(t)},  \label{AppEccIntSumF}
\end{align}
where $r_m$, $a_m$, $f_m$ and $M_m(t)$ are the radius, semi-major axis, true
anomaly and mean anomaly of the moon's orbit and $r_p$, $a_p$, $f_p$ and
$M_p(t)$ are the radius, semi-major axis, true anomaly and mean anomaly
of the planet's orbit

For the applications investigated in this thesis only the $l=2$ terms for $s_n^{lm}(e_m)$ and the $l=3$ terms for $F_n^{lm}(e_p)$ are used.  Consequently equations~\eqref{AppEccIntDefGens} and
\eqref{AppEccIntDefGenF} can be written as
\begin{align}
s_n^{2m}(e_m) &= \frac{1}{2\pi}\int_0^{2\pi}\frac{r_m^2}{a_m^2}e^{imf_m}
e^{-inM_m(t)}dM_m(t),\label{AppEccIntDefs}\\
F_n^{3m}(e_p) &= \frac{1}{2\pi}\int_0^{2\pi}\frac{a_p^4}{r_p^4}e^{-imf_p}
e^{inM_p(t)}dM_p(t).\label{AppEccIntDefF}
\end{align}

\section{Derivation of moon eccentricity functions}
\label{App_Ecc_Inn_Der}

In order for equation~\eqref{AppEccIntDefs} to be evaluated, $f_m$
in the complex exponential must be written in terms of $M_m(t)$, the
mean anomaly. While $f_m$ cannot be written as an explicit function
of $M_m$, both $M_m$ and $f_m$ can be written in terms
of the eccentric anomaly of the moon's orbit, $E_m$.  Using the
relation between the mean and eccentric anomalies \citep[][p. 34]{Murrayetal1999},
\begin{equation}
M = E - e\sin E,\label{AppEccInnEq1}
\end{equation}
it can be seen that equation~\eqref{AppEccIntDefs} can be rewritten
as
\begin{equation}
s_n^{2m}(e_m) = \frac{1}{2\pi}\int_0^{2\pi}\frac{r_m^2}{a_m^2}e^{imf_m}
e^{-in(E_m - e_m\sin E_m)}\left(1 - e_m\cos E_m\right)
dE_m\label{AppEccInnEq2}
\end{equation}
Using the relation between eccentric anomaly and $r_m$ and $f_m$ \citep[][p. 33]{Murrayetal1999} given
by 
\begin{align}
r_m &= a_m\left(1 - e_m \cos E_m\right), \label{AppEccInnEq3}\\
\cos f_m &= \frac{\cos E_m - e_m}{1 - e_m \cos E_m},
\label{AppEccInnEq4}
\end{align}
it can be seen that $r_m^2 e^{imf_m}$ can be entirely specified by $E_m$.  While analytic solutions to equation~\eqref{AppEccInnEq2}, using equations~\eqref{AppEccInnEq3} and \eqref{AppEccInnEq4} exist for $n=0$, for $n \ne 0$, the dependence of the Fourier coefficients $s^{lm}_n(e_m)$ on $e_m$ must
be determined numerically. This dependence up to order $e_m^3$ is given in table~\ref{App-Ecc-Inn-sTab}.

\begin{table}[tb]
\begin{center}
  \begin{tabular}{llll}
  \hline
  $l$ 	& $m$ 	& $n$ 	& $s_n^{(lm)}$\\
  \hline
 2	& 2  		& 5   		& $\frac{25}{24}e_m^3$\\
 	&     		& 4   		& $e_m^2$\\
	&     		& 3   		& $e_m-\frac{19}{8}e_m^3$\\
	&     		& 2   		& $1 - \frac{5}{2}e_m^2$\\
	&		& 1   		& $-3e_m$\\
	&   		& 0   		& $\frac{5}{2}e_m^2$\\
   	&    		& -1   	& $-\frac{7}{24}e_m^3$\\
2	& 0    	& 3   		& $-\frac{1}{8}e_m^3$\\
	&     		& 2   		& $-\frac{1}{4}e_m^2$\\
      	&     		& 1   		& $-e_m + \frac{1}{8}e_m^3$\\
      	&    		& 0   		& $1 + \frac{3}{2}e_m^2$\\
  \hline
  \end{tabular}\\
 \caption[Table of the dependence of the Fourier coefficients
$s_n^{lm}(e_m)$ on the moon's orbital eccentricity to order $e_m^{3}$.]{Table of the dependence of the Fourier coefficients $s_n^{lm}(e_m)$ on the moon's orbital eccentricity to order $e_m^{3}$.
 While the coefficients are defined for negative $m$, they can be
derived from coefficients with positive $m$ using equation~\eqref{AppEccInnProp4}.}\label{App-Ecc-Inn-sTab}
\end{center}
 \end{table}

\section{Properties of moon eccentricity functions}

There is some redundancy in the coefficients $s_n^{lm}(e_m)$. This can be
seen by taking the complex conjugate of equation~\eqref{AppEccIntSums},
\begin{equation}
\frac{r_m^l}{a_m^l}e^{-imf_m} = \sum_{n = -\infty}^\infty s_n^{lm*}(e_m)
e^{-inM_m(t)}, \label{AppEccInnProp1}
\end{equation}
where $^*$ denotes complex conjugation, and comparing it to the
equation where $m$ and $n$ have been replaced with their negative
values,
\begin{equation}
\frac{r_m^l}{a_m^l}e^{-imf_m} = \sum_{n = \infty}^{-\infty} s_{-n}^{l-m}(e_m)
e^{-inM_m(t)}. \label{AppEccInnProp2}
\end{equation}
Reversing the order of summation gives
\begin{equation}
\frac{r_m^l}{a_m^l}e^{-imf_m} = \sum_{n = -\infty}^{\infty} s_{-n}^{l-m}(e_m)
e^{-inM_m(t)}. \label{AppEccInnProp3}
\end{equation}
By comparing coefficients of like exponentials in equations~\eqref{AppEccInnProp1} and \eqref{AppEccInnProp3}, we have that
\begin{equation}
s_n^{lm*}(e_m) = s_{-n}^{l-m}(e_m). \label{AppEccInnProp4}
\end{equation}

\section{Derivation of planet eccentricity functions}

As in the case of $s_n^{lm}(e_m)$, in order to evaluate equation~\eqref{AppEccIntDefF}, both $M_p(t)$ and $f_p$ need to be written in
terms of the eccentric anomaly $E_p$. Again this can be done by noting that
\begin{align}
r_p &= a_p\left(1 - e_p \cos E_p\right), \label{AppEccOutEq1}\\
\cos f_p &= \frac{\cos E_p - e_p}{1 - e_p \cos E_p}.
\label{AppEccOutEq2}
\end{align}
Again, while analytic solutions exist to equation~\eqref{AppEccIntDefF} using equations~\eqref{AppEccOutEq1} and \eqref{AppEccOutEq2} for $n=0$, for $n \ne
0$ the dependence of $F_{n}^{lm}(e_p)$ on $e_p$ must be numerically
determined.  This dependence up to order $e_p^3$ is given in
table~\ref{App-Ecc-Out-FTab}.

\begin{table}[tb]
\begin{center}
  \begin{tabular}{lllllllll}
  \hline
  $l$ 	& $m$ 	& $n$ 	& $F_n^{(lm)}$ 				& & $l$ 	& $m$ 	& $n$ 	& $F_n^{(lm)}$\\
  \hline
 3 	& 3 		& 6   		& $\frac{163}{4}e_p^3$ 		& & 3          & 1            & 4             & $\frac{77}{6}e_p^3$\\
 	&		& 5   		&  $\frac{127}{8}e_p^2$ 		& &	          &               & 3             & $\frac{53}{8}e_p^2$\\
	& 		& 4   		& $5e_p - 22e_p^3$			& &             &               & 2             & $2e_p + \frac{11}{4}e_p^3$\\
	&		& 3   		&  $1 - 6e_p^2$       	                  & &     	 &               & 1   		& $1 + 2e_p^2$\\
	&		& 2   		& $-e_p + \frac{5}{4}e_p^3$ 	& &		&               & 0   		& $e_p + \frac{5}{2}e_p^3$\\
	&		& 1   		&  $\frac{1}{8}e_p^2$ 	         & &  		&               & -1   	        & $\frac{11}{8}e_p^2$\\
     	&    		& 0   		& 0 						& &     	&     		& -2   	& $\frac{23}{12}e_p^3$\\
3	&  2 		& 5   		& $\frac{145}{6}e_p^3$	         & & 3          &  0           & 3            & $\frac{23}{4}e_p^3$\\
	&    		& -1   	& $\frac{1}{3}e_p^3$	         & &             &                & 2           & $\frac{7}{2}e_p^2$\\
	&   		& 0   		& $\frac{1}{4}e_p^2$      	         & &             &     		& 1   		& $2e_p + \frac{17}{4}e_p^3$\\
	&		& 1   		& $\frac{1}{2}e_p^3$			& &   		&     		& 0   		& $1 + 3e_p^2$\\
      	&     		& 2   		& $1 - e_p^2$\\
      	&     		& 3   		& $4e_p - 5e_p^3$\\
	&     		& 4   		& $\frac{43}{4}e_p^2$\\
  \hline
  \end{tabular}\\
 \caption[Table of the dependence of the Fourier coefficients $F_n^{lm}(e_p)$ on the planet's orbital eccentricity to order $e_p^{3}$.]{Table of the dependence of the Fourier coefficients
$F_n^{lm}(e_p)$ on the planet's orbital eccentricity to order $e_p^{3}$.
 While the coefficients are defined for negative $m$, they can be
derived from coefficients with positive $m$ using equation~\eqref{AppEccOutProp4}.}\label{App-Ecc-Out-FTab}
\end{center}
 \end{table}

\section{Properties of planet eccentricity functions}

As with the case of the $s_n^{lm}(e_m)$, the coefficients $F_{n}^{lm}(e_p)$
are not independent.  Taking the complex conjugate of equation~\eqref{AppEccIntSumF} gives
\begin{equation}
\frac{a_p^{(l+1)}}{r_p^{(l+1)}}e^{imf_p} = \sum_{n = -\infty}^\infty
F_n^{lm*}(e_p) e^{inM_p(t)}. \label{AppEccOutProp1}
\end{equation}
Considering the expression where $m$ and $n$ have been replaced by
their negative values
\begin{equation}
\frac{a_p^{(l+1)}}{r_p^{(l+1)}}e^{imf_p} = \sum_{n = \infty}^{-\infty}
F_{-n}^{l-m}(e_p) e^{inM_p(t)}, \label{AppEccOutProp2}
\end{equation}
and the order of summation has been reversed, gives
\begin{equation}
\frac{a_p^{(l+1)}}{r_p^{(l+1)}}e^{imf_p} = \sum_{n = -\infty}^{\infty}
F_{-n}^{l-m}(e_p) e^{inM_p(t)}. \label{AppEccOutProp3}
\end{equation}
It can be seen by comparing like coefficients in equations~\eqref{AppEccOutProp1} and \eqref{AppEccOutProp3} that
\begin{equation}
F_n^{lm*}(e_p) = F_{-n}^{l-m}(e_p). \label{AppEccOutProp4}
\end{equation}

\chapter[Estimate for transit ingress duration]{Analytic estimate for the duration of transit ingress and egress}\label{IngressDur_App}

\begin{figure}
     \centering
     \subfigure[Exact geometry.]{
          \label{App_ingress_figa}
          \includegraphics[width=.48\textwidth]{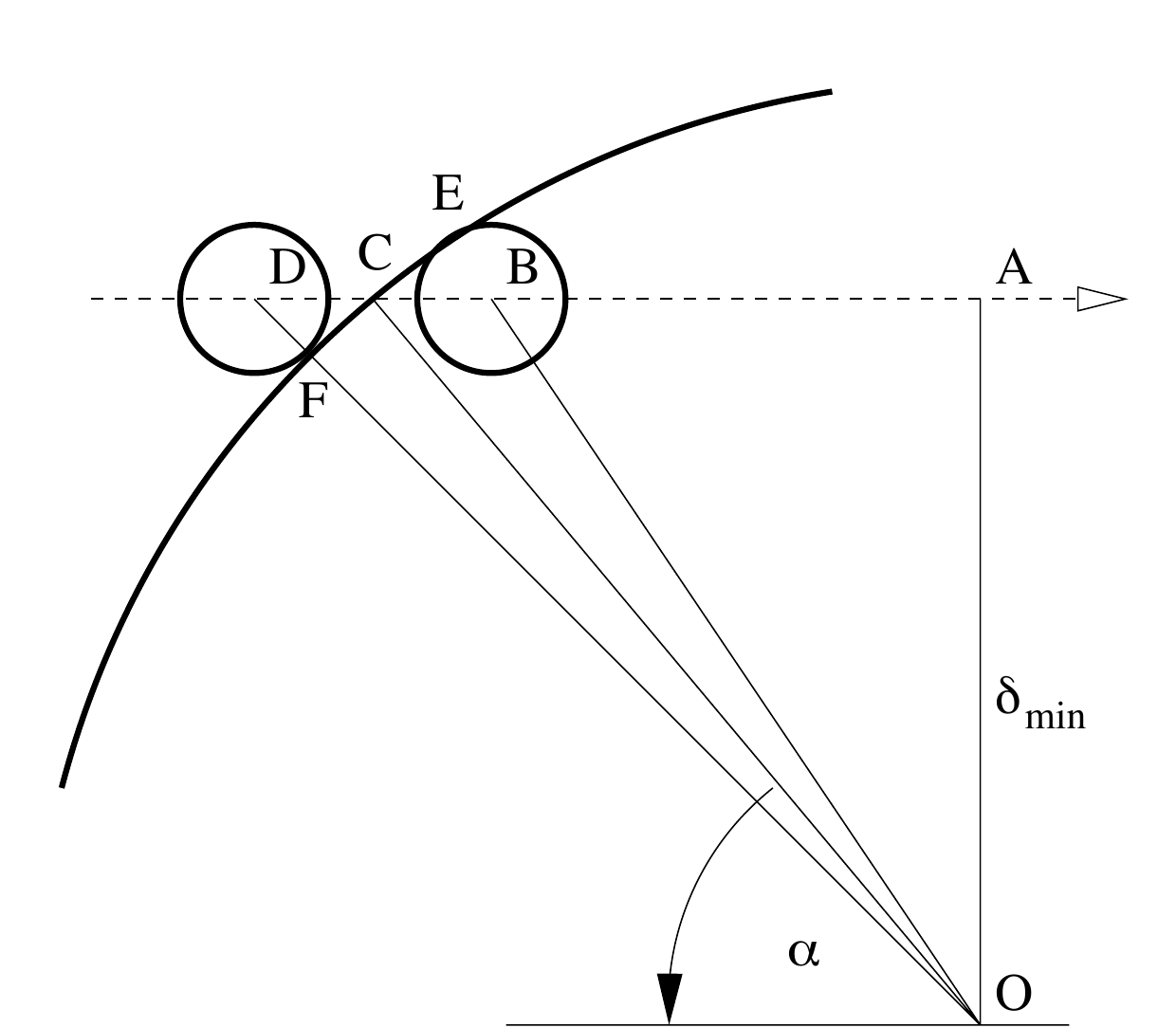}}
     \subfigure[Approximated geometry.]{
          \label{App_ingress_figb}
          \includegraphics[width=.48\textwidth]{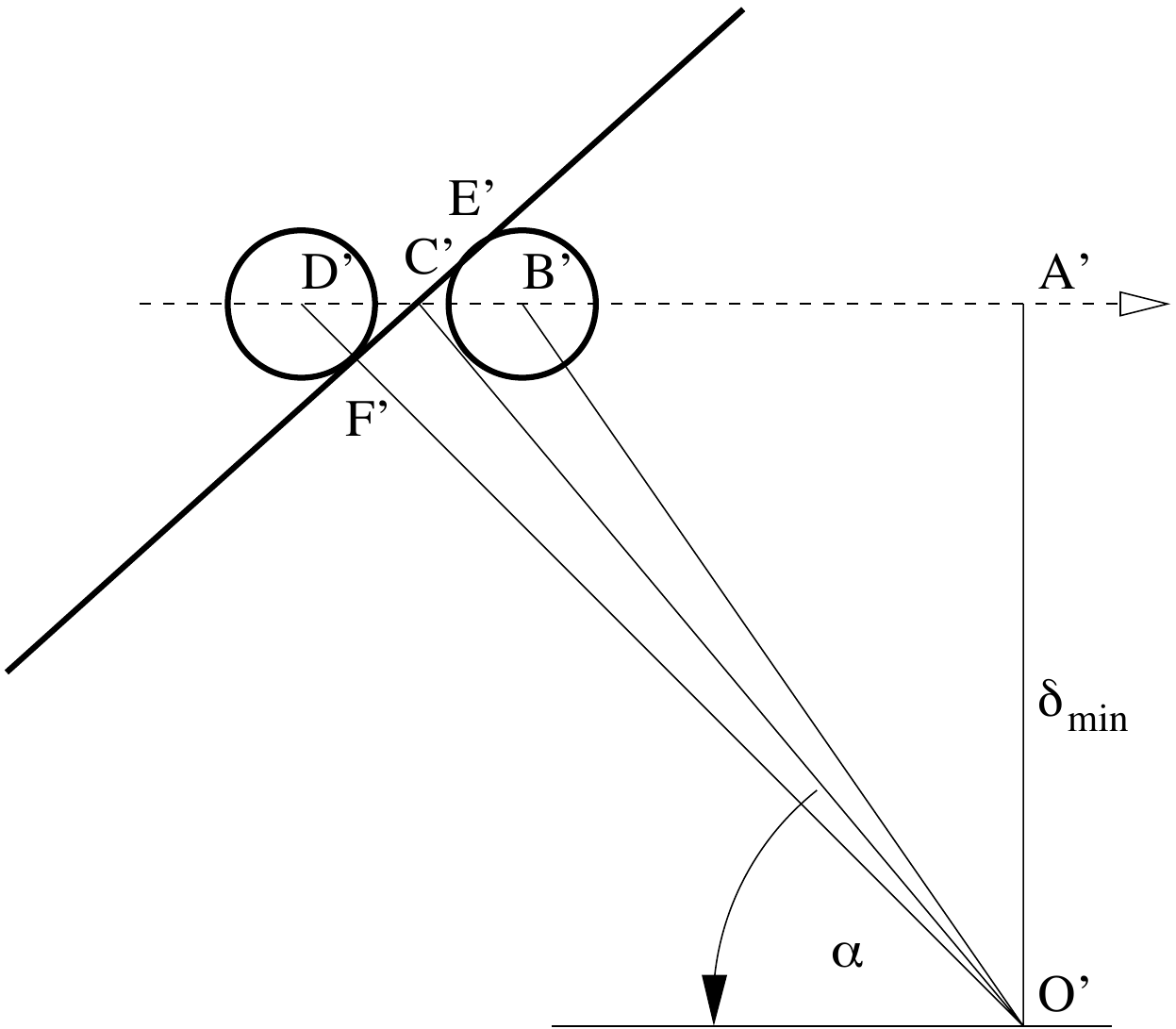}}\\
     \caption[Schematic of the position of a planet on the face of its star, at the beginning and end of transit ingress.]{Schematic of the position of a planet on the face of its star, at the beginning and end of transit ingress.  The limb of the star is shown as a thick black line while the position of the planet at the beginning and end of ingress is shown by two circles.  The path taken by the center of the planet is shown as a dashed line, and the center of the star is treated as the origin, and labeled $O$.}
     \label{IngressGeom}
\end{figure}

For this thesis, we would like to be able to compare detection thresholds for the four methods proposed in the literature for detecting moons of transiting planets.  However, the thresholds corresponding to two of these methods, barycentric transit timing and transit duration variation, depend on the duration of the ingress and egress of transit, that is, the time it takes for a planet not overlapping with the disk of its host star to completely overlap with it, and vice versa.  Consequently, in order to perform this comparison, we need an expression for the duration of transit ingress and egress for the case of circular orbits, the comparison case considered.  In particular, to fit with the other expressions derived, we would like this expression to be formulated in terms of the planetary radius, the stellar radius, the transit duration and the impact parameter, and also be simple enough to allow insight into the physics.

For the case of circular orbits, the transit light curve is symmetric, and thus the duration of ingress and egress are equal.  Consequently, for this appendix, we will concentrate on deriving the duration of ingress only, but note that the expressions derived can be just as equally applied to the egress. First, we note that the duration of ingress can be written in terms of the distance travelled by the planet during ingress and the velocity at which the planet travels, that is,
\begin{equation}
T_{in} = \frac{DB}{v_{tr}},\label{AppIngDurDef}
\end{equation}
where the distance $DB$ is defined in figure~\ref{App_ingress_figa} and where $v_{tr}$ is the velocity of the planet along the plane of the sky during transit.  Noting that the duration of the transit is given by 
\begin{align}
T_{tra} &= \frac{2CA}{v_{tr}},\\
 &= \frac{2\sqrt{R_s^2 - \delta_{min}^2}}{v_{tr}},
 \label{AppIngDtraDef}
\end{align}
we can write equation \eqref{AppIngDurDef} as
\begin{equation}
T_{in} = T_{tra} \frac{DB}{2\sqrt{R_s^2 - \delta_{min}^2}}.
\label{AppIngDurDtraDef}
\end{equation}
Thus, in order to determine $T_{in}$, we must determine the distance $DB$.

From figure~\ref{App_ingress_figa} we have that the distance $DB$ is given by $DA - BA$.  Using Pythagorus, we have that 
\begin{equation}
DA = \sqrt{(R_s + R_p)^2 - \delta_{min}^2},
\label{AppIngDurDADef}
\end{equation} 
and 
\begin{equation}
BA = \sqrt{(R_s - R_p)^2 - \delta_{min}^2}.
\label{AppIngDurBADef}
\end{equation}
Consequently we have that
\begin{equation}
DB = \sqrt{(R_s + R_p)^2 - \delta_{min}^2} - \sqrt{(R_s - R_p)^2 - \delta_{min}^2},
\label{AppIngDurDBDef}
\end{equation}
and thus from equation \eqref{AppIngDurDtraDef}
\begin{equation}
T_{in} = T_{tra} \frac{\sqrt{(R_s + R_p)^2 - \delta_{min}^2} - \sqrt{(R_s - R_p)^2 - \delta_{min}^2}}{2\sqrt{R_s^2 - \delta_{min}^2}}.
\label{AppIngDurDtraExact}
\end{equation}

While this expression is exact, it is functionally complicated.  As we are deriving this formula with the final aim of gaining insight into the effect that planet size, transit duration and impact parameter have on moon detection thresholds, it would be useful to use an approximation to this equation which describes its general behaviour, but is more physically insightful.  In particular, as planets are in general much smaller than their host stars ($R_p = 0.103R_{\sun}$ for Jupiter and $R_p = 0.0092R_{\sun}$ for Earth), we can use this to obtain such an approximation of equation \eqref{AppIngDurDBDef} and consequently a more useful expression for $T_{in}$. 

This approximation can be derived either by expanding equation \eqref{AppIngDurDBDef} and only retaining terms of order $R_p/R_s$, or by using an approximation to figure~\ref{App_ingress_figa}.  As both these approaches give the same result, we will derive an approximation to equation \eqref{AppIngDurDBDef} by approximating figure~\ref{App_ingress_figa}, as this approach involves less algebra and is easier to follow.  We begin by replacing the curved limb of the star locally with a straight line (see figure~\ref{App_ingress_figb}).  For ease of comparison, the points which are equivalent to those in figure~\ref{App_ingress_figa} have been marked with a dashed version of the same letter.  Consequently, by determining $D'B'$, we obtain an estimate, for $DB$ and consequently, can calculate $T_{in}$.

We begin by considering the angle $\angle C'B'E'$. From geometry, we have that $\angle O'C'A' = \alpha$, and consequently, that $\angle C'B'E' = \alpha$ as the line $\overline{O'C'}$ and $\overline{B'E'}$ are parallel.  From trigonometry we have that the distance $C'B'$ is given by 
\begin{equation}
C'B' = \frac{R_p}{\sin \alpha}.  
\end{equation}
Using the same arguments to determine $D'C'$, we have that $D'C'$ is also $R_p/\sin \alpha$ and consequently that 
\begin{equation}
D'B' = \frac{2R_p}{\sin \alpha}.
\end{equation}
or 
\begin{equation}
\sin \alpha = \frac{2R_p}{D'B'}.
\end{equation}

From $\triangle O'A'C'$ we have that 
\begin{equation}
\cos \alpha = \frac{\delta_{min}}{R_s}
\end{equation}
Squaring both expressions, and adding them gives
\begin{equation}
1 = \left(\frac{2R_p}{D'B'}\right)^2 + \left(\frac{\delta_{min}}{R_s}\right)^2,
\end{equation}
which, upon simplification yields
\begin{equation}
D'B'  = 2R_p \left(1 - \left(\frac{\delta_{min}}{R_s}\right)^2\right)^{-1/2}.
\label{AppIngDurDBAprox}
\end{equation}

Recall that equation \eqref{AppIngDurDBAprox} is the same expression as the one that would have been obtained by expanding equation \eqref{AppIngDurDBDef} and retaining terms up to order $R_p/R_s$.  In this context, consider equation \eqref{AppIngDurDtraDef}, restated below
\begin{equation}
T_{in} = T_{tra} \left[\frac{1}{\sqrt{R_s^2 - \delta_{min}^2}}\right] \times DB.
\end{equation}
As our approximation for $DB$ is only correct to first order in $R_p/R_s$, and the term in square brackets is zeroth order in $R_p/R_s$, $T_{in}$ can only be correct to first order in $R_p/R_s$.  Completing this expansion and substituting in equation~\eqref{AppIngDurDBAprox} for $DB$, we have that the duration of ingress is approximated by
\begin{align}
T_{in} &\approx T_{tra} \frac{1}{\sqrt{R_s^2  - \delta_{min}^2}} 2R_p \left(1 - \left(\frac{\delta_{min}}{R_s}\right)^2\right)^{-1/2},\\
&= 2\frac{R_p}{R_s} T_{tra} \left(1  - \left(\frac{\delta_{min}}{R_s}\right)^2\right)^{-1}.
\label{AppIngDurDtraApprox}
\end{align}

This formula is much more appropriate for the purposes of this thesis than equation \eqref{AppIngDurDtraExact}.  To see this, consider how the behaviour described by equation \eqref{AppIngDurDtraApprox} compares to the relationships that one would naively expect between the ingress duration and the planetary radius, the transit duration and the impact parameter.  For the case of planetary radius, we would expect that larger planets would have larger ingress durations as a result of the longer distance that they have to travel to pass over the limb of the star, and from equation \eqref{AppIngDurDtraApprox} we see that this is the case as $T_{in}$ is proportional to $R_p$.  In addition, the fact that $T_{in}$ is proportional to $T_{tra}$ is also expected as the shape of the transit is affected by the geometry of the transit and the scale of the transit is affected by the transit duration.  So, any of the factors which affect the transit duration, either by altering the velocity of the planet during transit (e.g. the eccentricity or semi-major axis of the planet's orbit) or by altering the distance that the planet has to travel during transit (e.g. the impact parameter of the transit) can affect $T_{in}$ by way of this term.  Finally we would also expect that the ingress duration would increase as $\delta_{min}$ increases as the slope of the segment of stellar limb that the planet must traverse becomes increasingly shallow.  Noting that to zeroth order in $R_p/R_s$, $T_{tra} \propto \left(1  - (\delta_{min}/R_s)^2\right)^{1/2}$ (see equation \eqref{AppIngDtraDef}), we find that $T_{in} \propto \left(1  - (\delta_{min}/R_s)^2\right)^{-1/2}$.  Thus equation \eqref{AppIngDurDtraApprox} does display the expected increase in $T_{in}$ with increasing $\delta_{min}$.\footnote{Note that this formula fails for $\delta_{min} \approx R_s$ as a result of approximating the circular stellar limb with a straight line.  In particular for the case of $\delta_{min} =R_s$, the stellar limb would be approximated by a horizontal line and the infinite transit duration implied by equation \eqref{AppIngDurDtraApprox} physically corresponds to a planet traveling along this horizontal line.  The discrepancy between the true ingress duration described by equation \eqref{AppIngDurDtraExact} and the approximate ingress duration described by equation \eqref{AppIngDurDtraApprox} is only a problem for planets with very grazing transits.}  Consequently equation~\eqref{AppIngDurDtraApprox} is a suitable equation to derive the detection thresholds shown in chapter~\ref{Intro_Dect}.

\chapter[Effect of incorrect unobscured flux]{Effect of incorrect value of unobscured flux on $\tau$}
\label{FluxErr_App}

\begin{figure}[tb]
\begin{center}
\includegraphics[width=.95\textwidth]{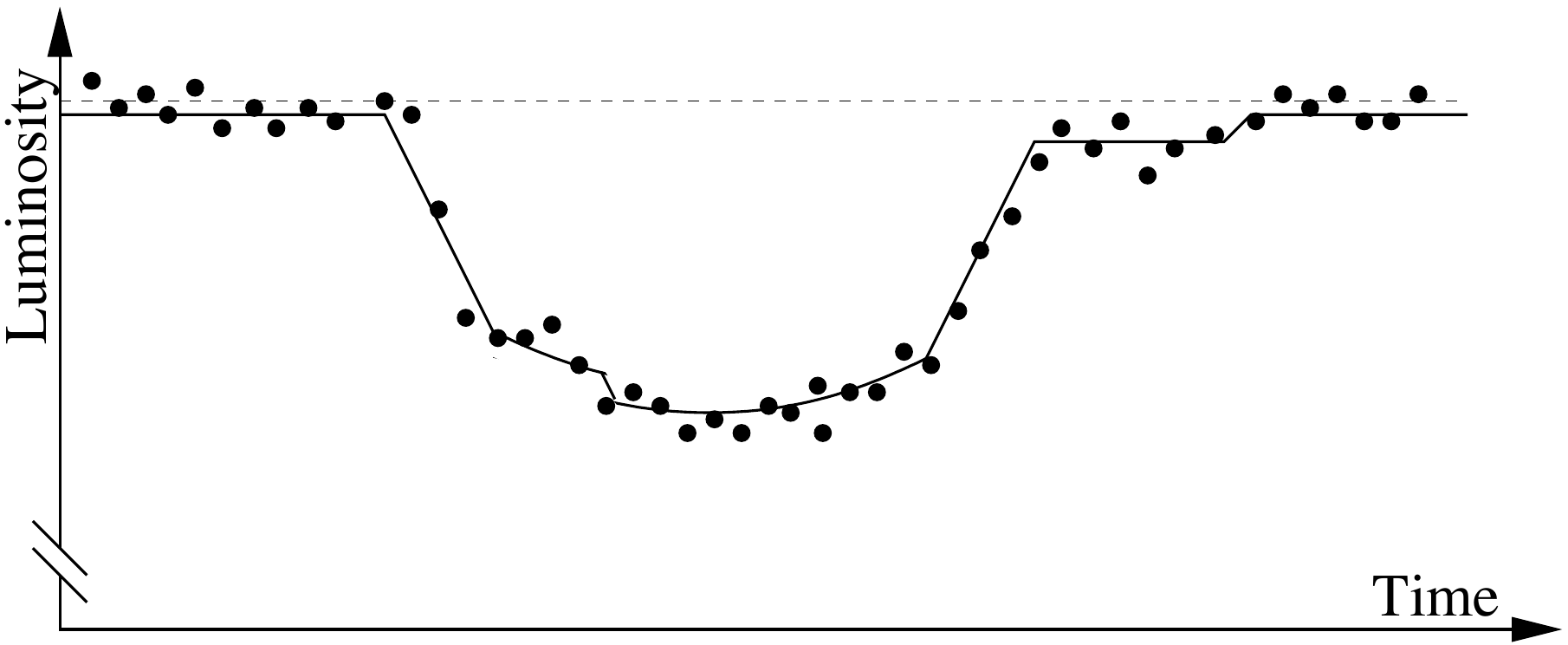}
\caption{A cartoon of a possible light curve which could be produced by a planet-moon pair, showing both the theoretically predicted (black line) and observationally measured (black dots) luminosity.  In addition, the observationally estimated unobscured flux is also shown (dashed line).}
\label{App_TransBase}
\end{center}
\end{figure}

Light curve quantities such as $\tau$ depend on the value assumed for the unobscured flux of the host star.  As a result of factors such as inherent photometric variability and small number statistics, the value of this flux cannot be exactly observationally determined. In this appendix the effect of this error on $\tau$ will be investigated, with the aim of showing that this source of error can be effectively neglected.

\begin{figure}
     \centering
     \subfigure[Planet transit and flux error separate.]{
          \label{App_TransBase_Break1}
          \includegraphics[width=.48\textwidth]{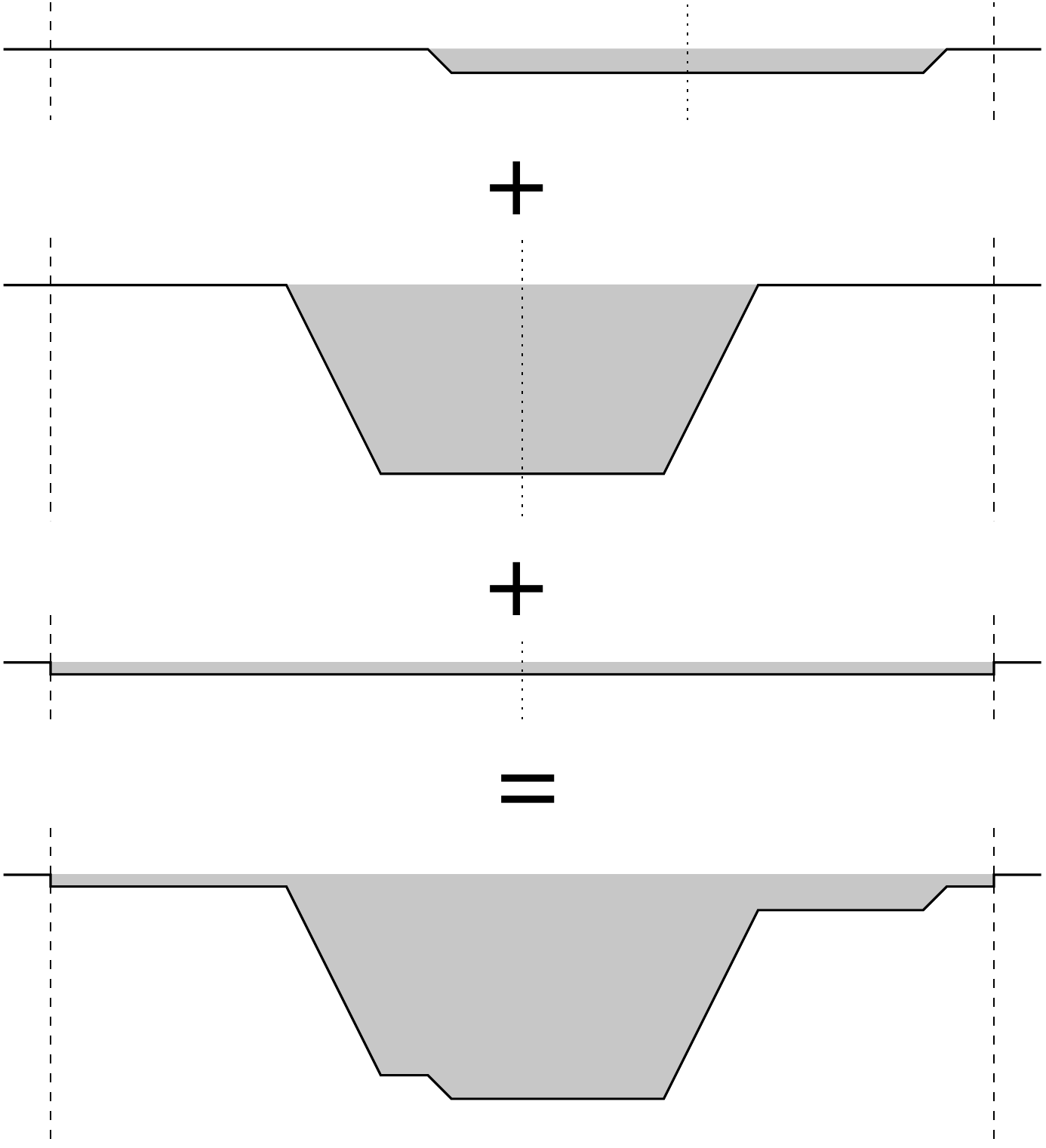}}
     \subfigure[Planet transit and flux error combined.]{
          \label{App_TransBase_Break2}
          \includegraphics[width=.48\textwidth]{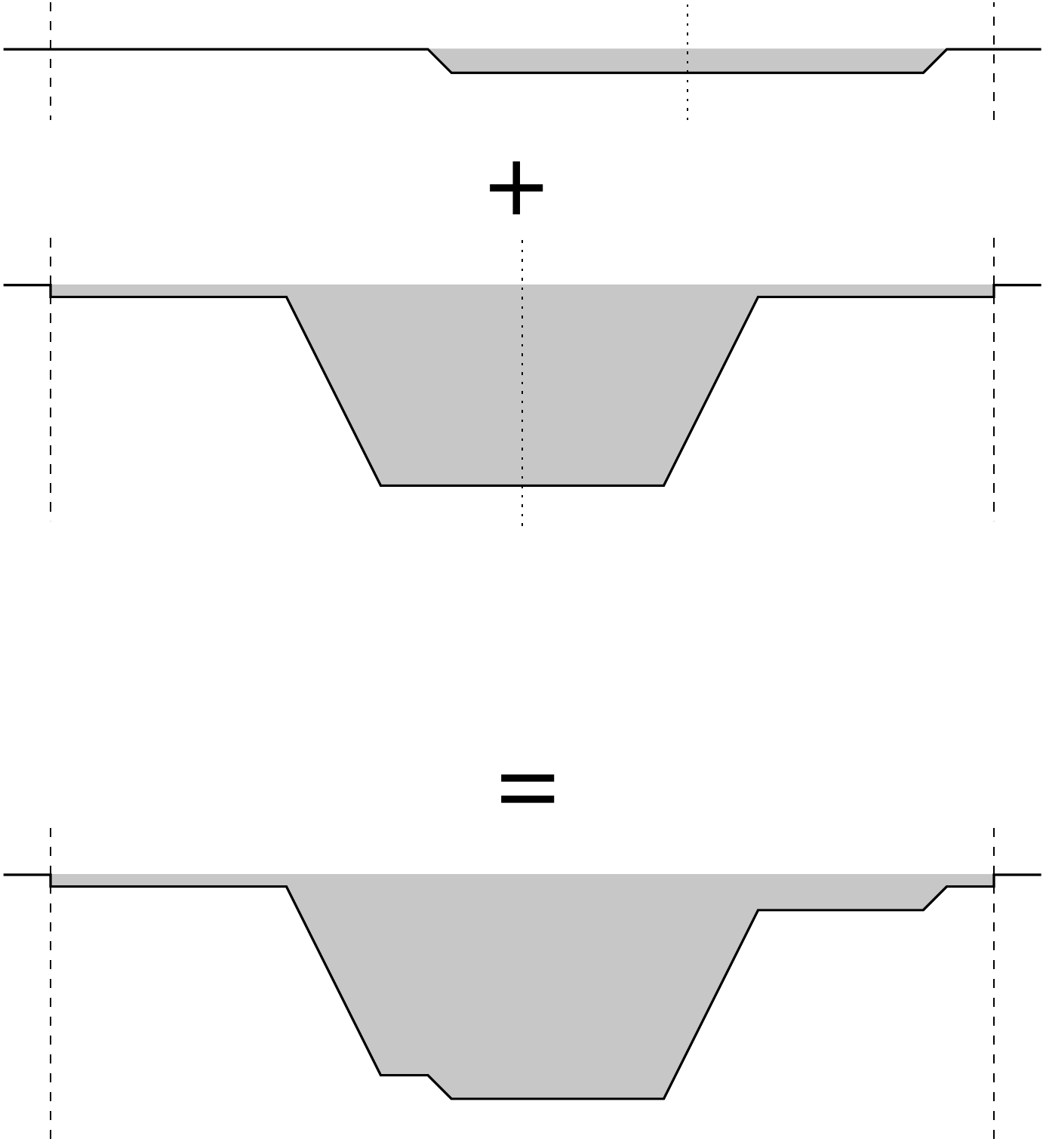}}
     \caption[Figure showing two different ways of dividing the composition of a transit light curve for the case where there is an incorrect assumed value of the unoccluded flux.]{Figure showing two different ways of dividing the composition of a transit light curve.  The constituent components (moon, planet and error due to incorrect assumed value of the unoccluded flux) which go into a transit light curve are shown above, and the full light curve is shown below.  In particular, any region in which there is a photon deficit is shaded grey.  In addition, the beginning and end of the observation window are shown by vertical dashed lines, while the line of symmetry of each component is shown by a vertical dotted line.}
     \label{App_TransBase_Break}
\end{figure}

To begin, consider the light curve shown in figure~\ref{App_TransBase},  In particular note that the observationally selected unobscured flux (dashed) is not correct, and in this case, is slightly above the actual unobscured flux.  In order to analyse the effect of this error on $\tau$, the definition of $\alpha$ given by equation~\eqref{TraM-TTV-alphasplit} must be updated.  With this in mind we introduce a photon deficit $\alpha_f$, associated with selecting the wrong unobscured flux, such that $\alpha$ is now
\begin{equation}
\alpha = \alpha_p + \alpha_m + \alpha_f + \alpha_n \label{App_Baseline_alphadef}
\end{equation}
where we recall that $\alpha_p$, $\alpha_m$ and $\alpha_n$ are the photon deficits associated with the transit of the planet, the transit of the moon and the photometric noise respectively.  This breakdown is shown in figure~\ref{App_TransBase_Break1}.

To investigate the effect of this additional term we will use an approach equivalent to that used in section~\ref{Transit_Intro_Deriv} to write $\Delta \tau$ in terms of $A_p$, $A_m$, $\tau_p$ and $\tau_m$.  To begin, we group $\alpha_p$ and  $\alpha_f$ into a single term (see figure~\ref{App_TransBase_Break2}), such that equation \eqref{App_Baseline_alphadef} becomes
\begin{equation}
\alpha = (\alpha_p + \alpha_f) + \alpha_m + \alpha_n.
\end{equation}
Using the same method as used in section~\ref{Transit_Intro_Deriv}, it can be shown that
\begin{align}
\Delta \tau + t_0 + jT_p &= \frac{\frac{\sum_i t (\alpha_p + \alpha_f)}{\sum_i (\alpha_p + \alpha_f)}\sum_i (\alpha_p + \alpha_f) + \frac{\sum_i t \alpha_m}{\sum_i \alpha_m} \sum_i \alpha_m }{\sum_i (\alpha_p + \alpha_f) + \alpha_m},\\
&= \frac{\frac{\sum_i t \alpha_p}{\sum_i \alpha_p}\sum_i \alpha_p + \frac{\sum_i t \alpha_f}{\sum_i \alpha_f}\sum_i  \alpha_f + \frac{\sum_i t \alpha_m}{\sum_i \alpha_m} \sum_i \alpha_m }{\sum_i \alpha_p + \alpha_f + \alpha_m}.
\end{align}
Noting that as the $\alpha_p$ and $\alpha_f$ are symmetric and centered on the transit window, $\sum_i (t \alpha_p)/\sum_i \alpha_p$ and $\sum_i (t \alpha_f)/\sum_i \alpha_f$ are both equal and given by $\tau_p$.  Consequently
\begin{equation}
\Delta \tau + t_0 + jT_p = \frac{\tau_p (A_p + \Delta A_p) + \tau_m A_m}{A_p + \Delta A_p + A_m}\\
\end{equation}
where the definition of $\tau_p$, $\tau_m$, $A_p$ and $A_m$ given in section~\ref{Transit_Intro_Deriv} have been used, and where we define $\Delta A_p = \sum_i \alpha_f$.

Consequently, the effect of a small error in the baseline is equivalent to a small error in $A_p$, $\Delta A_p$.  Now, recall from the discussion in section~\ref{Trans_TTV_Signal_CC_Form_smallB} that $\Delta \tau$ is approximately proportional to $A_m/(A_p+A_m)$.  Consequently, using the binomial expansion we expect the error in $\Delta \tau$ caused by selecting the wrong baseline to be a factor of  $\Delta A_p/A_p$ smaller than $\Delta \tau$.  As a statistically significant difference between the base of the transit and the unobscured flux must be detected to confirm a planet, we would expect the error in the unobscured flux to be much smaller than the dip depth, and consequently, the error in $A_p$ to be much smaller than $A_p$, and thus, the error in $\Delta \tau$ to be much smaller than $\Delta \tau$.  Thus, the small changes in $A_p$ caused by an incorrect baseline should not strongly affect the measured value of $\Delta \tau$.  Consequently, this error can be ignored.

\chapter[Shape of transit light curve does not depend on $\Omega_p$]{Proof that the shape of a transit light curve does not depend on $\Omega_p$}\label{App_Omega_Dep}

In chapter~\ref{Transit_Signal}, the coordinate system is rotated by $\Omega_p$ for convenience.  However, in order to justify this approach it must be shown that the shape and mid-time of the transit light curve are unchanged by such a transformation.  This will be done by considering the dependance of $L(t)$, the measured luminosity of the star on $\delta_p(t)$, the projected distance between the planet and the center of the star, and then showing that $\delta_p(t)$ is not dependent on $\Omega_p$.

To begin, consider equation~\eqref{TraM-DescT-Ldef}, the expression linking the measured luminosity and $\delta_p(t)$
\begin{equation*}
L(t) = L_0 - \alpha_p(\delta_p(t)).
\end{equation*}
As $L_0$ is a constant related to the star (and thus doesn't depend on $\Omega_p$) any effect that $\Omega_p$ has on the light curve will be through the $\alpha_p$ term.  Recalling from equation~\eqref{TraM-DescT-alphadef} that $\alpha_p$ depends only on the geometry, that is, $R_s$, $R_p$ and $\delta_p(t)$, so any dependance on $\Omega_p$ must be through the $\delta_p(t)$ term.

Recalling that for Part~\ref{TransitPart} of this thesis, the $xy$ plane is defined to lie in the plane of the sky, we have that
\begin{equation}
\delta_p(t)^2 = x_p(t)^2 + y_p(t)^2.
\end{equation}
Substituting in equations~\eqref{transit_intro_dur_xpdef} and \eqref{transit_intro_dur_ypdef} for $x_p(t)$ and $y_p(t)$ and simplifying gives
\begin{multline}
\delta_p(t)^2 = r_p(t)^2 \cos^2(f_p(t) + \omega_p) + r_p(t)^2 \cos^2 I_p \sin^2(f_p(t) + \omega_p),
\end{multline}
where the time dependance has been explicitly written.  Note that while the expressions for $x_p(t)$ and $y_p(t)$ depended on $\Omega_p$, this expression is independent of $\Omega_p$.  As both $L_0$ and $\delta_p(t)$ and thus $\alpha_p(\delta_p(t))$ are independent of $\Omega_p$, $L(t)$ must also be independent of $\Omega_p$.

\begin{figure}[tb]
\begin{center}
\includegraphics[height=2.5in,width=2.5in]{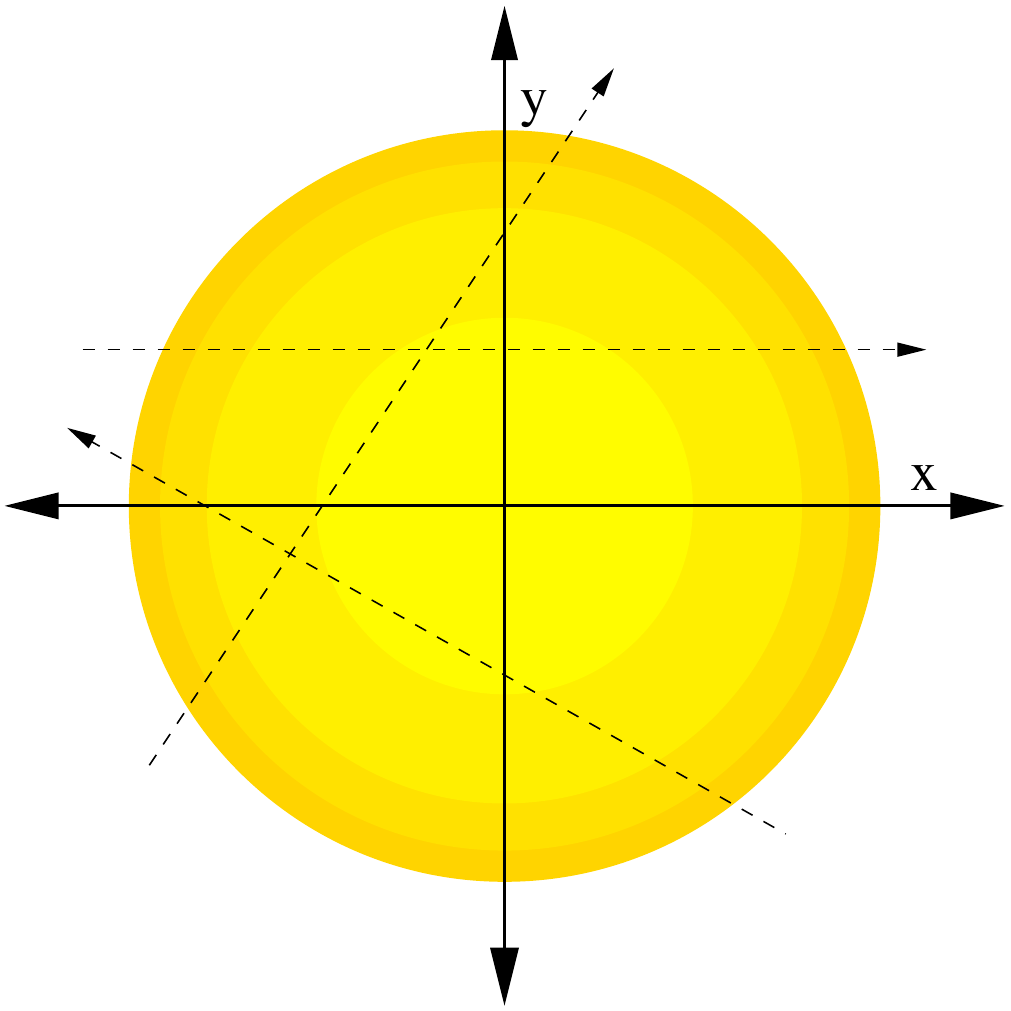}
\caption{A cartoon showing the path taken by the planet across the face of the star (dashed line) for three orbits, which vary only in their $\Omega_p$ values.  Note that while the path changes, the intensity along it does not.}
\label{App_Om_Sch}
\end{center}
\end{figure}

This is exactly what we would physically expect, as $\Omega_p$ merely rotates the chord taken by the planet across the face of the star about the center of the star (see figure~\ref{App_Om_Sch}).  As the measured intensity depends on the projected distance from the center of the star only, this should not modify the intensity along the chord and thus not modify the transit light curve.

\chapter[Estimate of probability of high transit velocity]{Estimate of probability that a planet on an eccentric orbit has a higher transit velocity than a planet on an equivalent circular orbit}\label{EccProb_App}

For the case of a planet on an eccentric orbit, the velocity at which the planet passes across the star may no longer be given by $a_p n_p$, the value for the case of a circular orbit.  This difference is of particular importance as, in chapter~\ref{Transit_Signal} an expansion is used to derive expressions for $\Delta \tau$ which assumes that $v_m/v_{tr} < 0.66$.  As $v_{tr}$ defines where this expression breaks down, discovering the effect of orbital eccentricity and orientation on $v_{tr}$, and in particular the set of orbits for which $v_{tr} > a_p n_p$, is of interest.  Consequently, the set of orbits (parameterised by $f_{tr}$, the value of $f$ during transit) for which this occurs, and in particular the values of $f_{tr}$ for which the transit velocity is equal to $a_p n_p$, will be investigated, and used to calculate the probability that a given transiting planet will have $v_{tr} > a_p n_p$.

In order to determine which orbits have $v_{tr} > a_p n_p$, we need an equation describing the transit velocity.  While the expression given in section~\eqref{Trans_Intro_Transtech} is exact, it is not very analytically tractable.  Consequently we use a simplified version derived from angular momentum constraints.  From \citet{Murrayetal1999} we have that
\begin{equation}
\mathbf{r}_p\times\mathbf{v}_p = r_p v_{perp} = n_pa_p^2\sqrt{1 - e_p^2},\label{TraM-TTV-Oecc-angmomdef}
\end{equation}
where $\mathbf{r}_p$ and $\mathbf{v}_p$ are position and velocity vectors of the planet, and where $v_{perp}$ is the component of the planet's velocity perpendicular to the position vector $\mathbf{r}_p$.  As equation~\eqref{TraM-TTV-Oecc-angmomdef} is true for all points on the orbit, equation~\eqref{TraM-TTV-Oecc-angmomdef} can be evaluated for the case where the planet is transiting.  In this case $r_p=r_p(f_{tr})$ and $v_{perp} =v_{tr}$.   Substituting these identities into equation~\eqref{TraM-TTV-Oecc-angmomdef}, rearranging, and expanding $r_p(f_{tr})$ using equation~\eqref{transit_signal_coord_Rdef} gives
\begin{equation}
v_{tr} = (n_pa_p)\frac{1 + e_p \cos(f_{tr})}{\sqrt{1 - e_p^2}}.\label{TraM-TTV-Oecc-vdef2}
\end{equation}
This is the expression for transit velocity which will be used in this appendix.

In order to determine whether it is more or less likely for planets on eccentric orbits to transit while their transit velocity is higher than that of a planet on a circular orbit with the same semi-major axis, we need to determine the values of $f_{tr}$ at which this transition occurs.  Setting the transit velocity of an eccentric orbit equal to that of a circular orbit with the same semi-major axis gives
\begin{equation}
n_pa_p = n_pa_p\frac{1 + e_p\cos(f_{tr})}{\sqrt{1-e_p^2}},
\end{equation}
which gives, after rearrangement
\begin{align}
f_{tr} &= \pm \cos^{-1}\left(\frac{\sqrt{1-e_p^2} - 1}{e_p}\right),\label{TraM-TTV-Oecc-folimdef}\\
 &= \pm f_{p,lim}.
\end{align}

As $\sqrt{1-e_p^2} - 1 < 0$, the argument of the inverse cosine is always negative.  Consequently, the range of angles for which the transit velocity in the eccentric case is always described by $f_{p,lim} \ge \pi/2$.  Also, as $e_p$ tends to 1, equation~\eqref{TraM-TTV-Oecc-folimdef} simplifies to give $f_p = \pm f_{p,lim} = \pm \pi$.

\begin{figure}[tb]
\begin{center}
\includegraphics[width=4.5in]{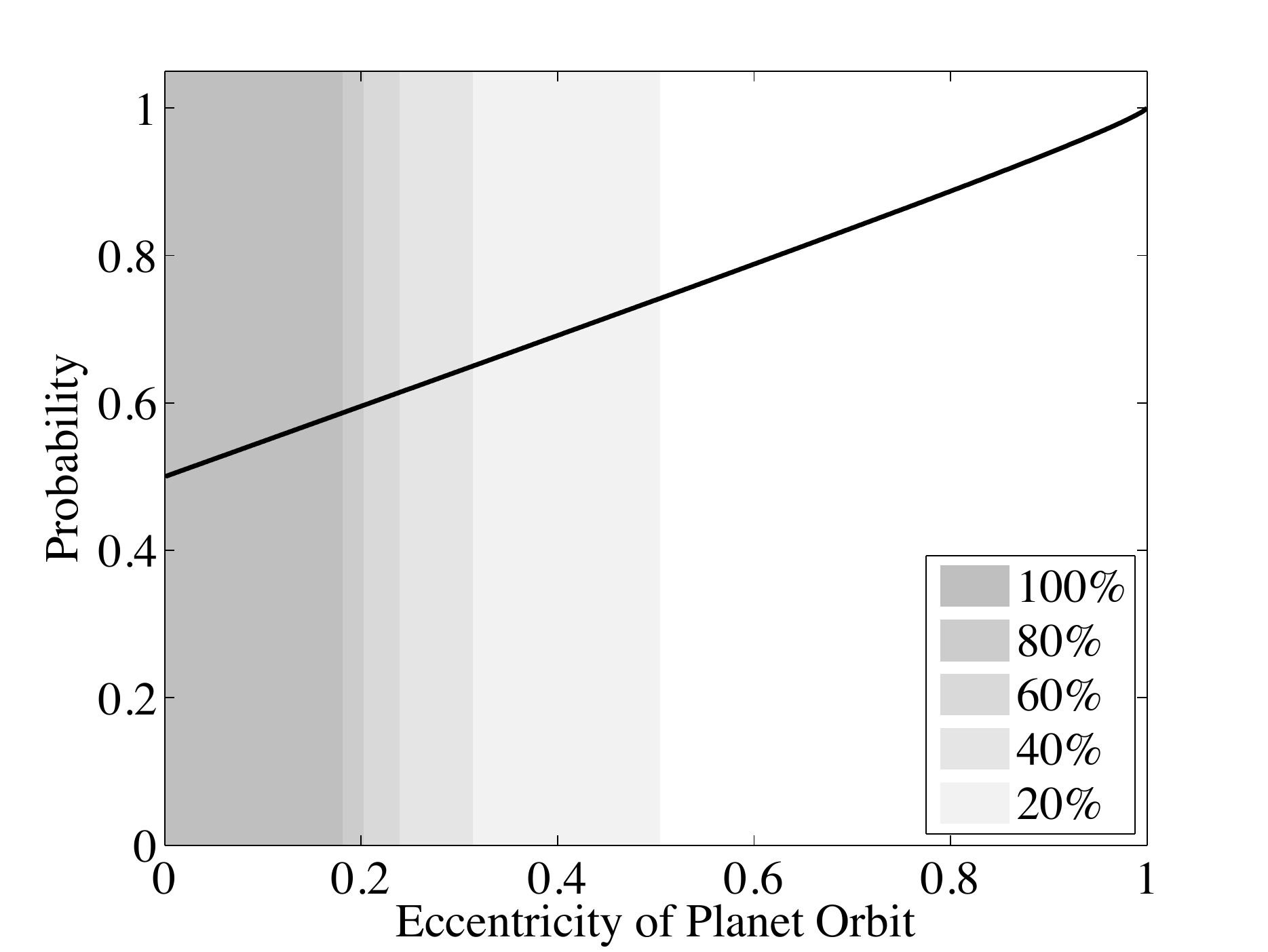}
\caption[Plot of the probability that a planet on an orbit will transit its host star with a velocity higher than that for a planet on a circular orbit with the same semi-major axis.]{Plot of the probability that a planet on an orbit will transit its host star with a velocity higher than that for a planet on a circular orbit with the same semi-major axis.  The shaded regions indicated the probability that a planet with that eccentricity will be observed with transit velocity within 20 percent of that for the equivalent circular orbit.}
\label{AppEccProb}
\end{center}
\end{figure}

As the probability, for a planet to transit is inversely proportional to the distance between the planet and star at the time of transit (see section~\ref{Trans_Intro_Transtech}), the probability that a planet on an eccentric orbit transits is described by
\begin{align}
P_{tot} &\propto \int_{-\pi}^\pi \frac{1}{r_p(f_p)}df_p,\\
&= \int_{-\pi}^\pi \frac{1 + e_p\cos(f_p)}{a_p(1-e_p^2)}df_p,\\
&= \left[\frac{f_p + e_p\sin(f_p)}{a_p(1-e_p^2)}\right]_{-\pi}^\pi,\\
&= \frac{2\pi}{a_p(1-e_p^2)}.
\end{align}
In addition, the probability that a planet will transit only when its transit velocity is larger than that of the equivalent circular orbit is described by
\begin{align}
P_{>} &\propto \int_{-f_{p,lim}}^{f_{p,lim}} \frac{1}{r_p(f_p)}df_p,\\
&= \int_{-f_{p,lim}}^{f_{p,lim}} \frac{1 + e_p\cos(f_p)}{a_p(1-e_p^2)}df_p,\\
&= \left[\frac{f_p + e_p\sin(f_p)}{a_p(1-e_p^2)}\right]_{-f_{p,lim}}^{f_{p,lim}},\\
&= \frac{2f_{p,lim} + 2e_p\sin(f_{p,lim})}{a_p(1-e_p^2)}.
\end{align}
As the proportionality constant for both equations is the same, we have that the probability that a planet on an eccentric orbit will transit while its orbital velocity is larger than that for the equivalent circular orbit, given that it transits, is equal to
\begin{equation}
\frac{P_{>}}{P_{tot}} = \frac{f_{p,lim} + e_p\sin(f_{p,lim})}{\pi}.\label{TraM-TTV-Oecc-probrat1}
\end{equation}
As $e_p\sin(f_{p,lim}) > 0 $ for $\pi/2 \le f_{p,lim} \le \pi$, the range for $f_{p,lim}$ identified above, equation~\eqref{TraM-TTV-Oecc-probrat1} can be written as
\begin{equation}
\frac{P_{>}}{P_{tot}} > \frac{f_{p,lim}}{\pi}.\label{TraM-TTV-Oecc-probrat2}
\end{equation}
Substituting in the lowest possible value for $f_{p,lim}$, $\pi/2$, we obtain
\begin{equation}
\frac{P_{>}}{P_{tot}} > \frac{1}{2}.\label{TraM-TTV-Oecc-probrat3}
\end{equation}
Consequently, planets on eccentric orbits have a higher probability of transiting near their periastron passages while their orbital velocity is higher than that of the equivalent circular orbit, than near apastron when their velocity is lower.  In addition, this probability increases nearly linearly as the eccentricity increases (see figure~\ref{AppEccProb}).

\chapter[$\Delta \tau$ for eccentric moon orbit aligned to the line-of-sight]{Form of TTV$_p$ perturbation for the case of slightly eccentric moon orbits}\label{EccMoon_App}

\begin{figure}[tb]
\begin{center}
\includegraphics[height=2.55in,width=4.5in]{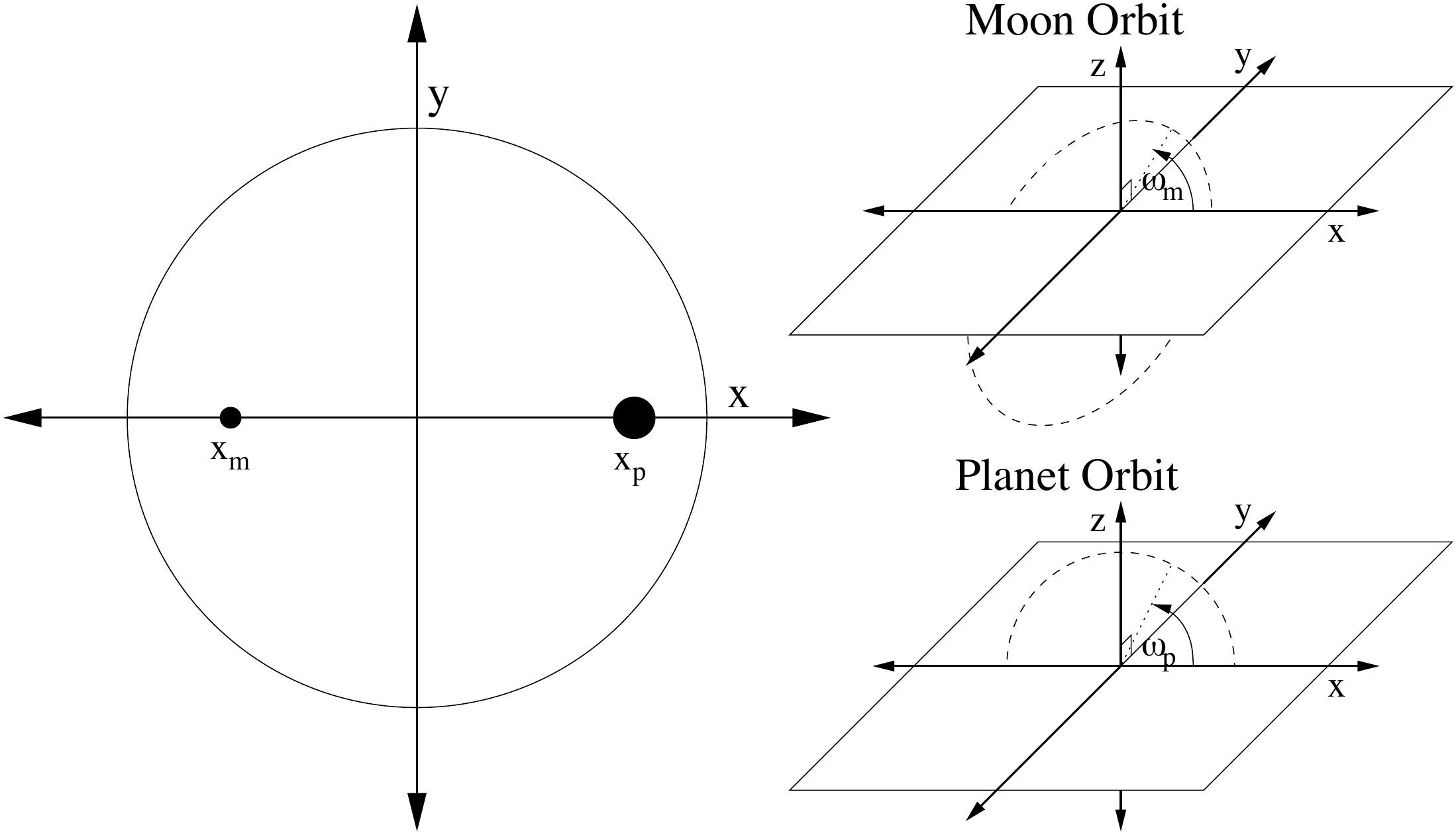}
\caption[Schematic diagram of the same form as figure~\ref{TransitSignalGenCoordSysRot} of the coordinate system for the case of eccentric moon orbits.]{Schematic diagram of the same form as figure~\ref{TransitSignalGenCoordSysRot} of the coordinate system for the case of eccentric moon orbits. In particular, it is assumed that $I_p = \pi/2$, $I_m = \pi/2$ and $\Omega_m = \Omega_p$.}
\label{TransitSignalCoordSysIe}
\end{center}
\end{figure}

Eccentricity in the moon's orbit can affect the form of $\Delta \tau$ through altering the time dependence of the planet and moon's position on the face of the star.  To investigate this effect, the case where the moon's orbit was slightly eccentric ($e_m \ll 1$), but aligned to the line-of-sight ($I_m = 0$ and $\Omega_p = \Omega_m$), and the planet's orbit was circular ($r_p = a_p$) and also aligned to the line-of-sight ($I_p = 0$), was examined (see figure~\ref{TransitSignalCoordSysIe}).  Again expanding the first term of equations~\eqref{transit_signal_coord_xpdef} and \eqref{transit_signal_coord_xmdef} about $t=jT_p+t_0$, we have that
\begin{align}
x_p &= v_{tr}(t - (jT_p+t_0)) - r_m(t)\frac{M_m}{M_{pm}}\cos(f_m(t)+ \omega_m), \label{transit_signal_inec_xp}\\
x_m &= v_{tr}(t - (jT_p+t_0)) + r_m(t)\frac{M_p}{M_{pm}}\cos(f_m(t)+ \omega_m), \label{transit_signal_inec_xm}
\end{align}
where $r_m(t)$ is described by equation~\eqref{transit_signal_coord_rdef} and $f_m(t)$ is described by equations~\eqref{transit_signal_coord_fidef} and \eqref{transit_signal_coord_Eidef}.  To allow easy comparison with the equations derived for the case of circular moon orbits, it would be useful if the term $r_m(t)\cos(f_m(t) + \omega_m)$ could be written in terms of $n_m$ and $t$.  

As we are investigating the signal $\Delta \tau$ for the case of small $e_m$, we can approximate the $r_m(t)\cos(f_m(t) + \omega_m)$ term using an expansion valid for small $e_m$.  From \citet[][p. 39-40]{Murrayetal1999} have that
\begin{align}
r_m &= a_m  - e_m a_m \cos (M_m(t)) + O(e_m^2),\\
\cos (f_m(t)) &= \cos (M_m(t)) + e_m(\cos (2M_m(t)) -1) + O(e_m^2),\\
\sin (f_m(t)) &= \sin (M_m(t)) + e_m \sin(2M_m(t)) + O(e_m^2),
\end{align}
where $M_m(t)$, the mean anomaly of the moon's orbit is equal to $n_m t + M_m(0)$, and where we note that the explicitly stated time dependance differentiates between $M_m(t)$, the mean anomaly of the moon and $M_m$, the mass of the moon.  Expanding the cosine function and using these expressions gives
\begin{multline}
r_m(t) \cos(f_m(t) + \omega_m) = a_m\cos(M_m(t) + \omega_m) \\+ e_m a_m \frac{1}{2} \cos(2 M_m(t) + \omega_m) - e_m a_m \frac{3}{2}\cos \omega_m + O(e_m^2).
\end{multline}
Thus, to first order in $e_m$, equations~\eqref{transit_signal_inec_xp} and \eqref{transit_signal_inec_xm} become
\begin{multline}
x_p = v_{tr}(t - (jT_p+t_0)) - a_m\frac{M_m}{M_{pm}}\cos(M_m(t) + \omega_m) \\-  e_ma_m\frac{1}{2} \frac{M_m}{M_{pm}}\cos(2 M_m(t) + \omega_m) +  e_m a_m\frac{3}{2} \frac{M_m}{M_{pm}} \cos \omega_m,
\end{multline}
\begin{multline}
x_m = v_{tr}(t - (jT_p+t_0)) + a_m \frac{M_p}{M_{pm}}\cos(M_m(t) + \omega_m) \\+ e_m a_m \frac{1}{2} \frac{M_p}{M_{pm}} \cos(2 M_m(t) + \omega_m) - e_m a_m \frac{3}{2} \frac{M_p}{M_{pm}} \cos \omega_m.
\end{multline}

Rearranging, writing $M_m(t)$ out in full, and setting $x_p$ and $x_m$ to the location of the star's limb gives
\begin{multline}
-R_{s} = v_{tr}(t_{in,p} - (jT_p+t_0)) - a_m\frac{M_m}{M_{pm}}\cos(n_m t_{in,p} + M_m(0) + \omega_m) \\- e_m  a_m \frac{M_m}{M_{pm}}\left[\frac{1}{2}\cos(2(n_m t_{in,p} + M_m(0)) + \omega_m) - \frac{3}{2}\cos \omega_m\right],\label{transit_signal_inec_tinp}
\end{multline}
\begin{multline}
-R_{s} = v_{tr}(t_{in,m} - (jT_p+t_0)) + a_m \frac{M_p}{M_{pm}}\cos(n_m t_{in,m} + M_m(0) + \omega_m) \\+ e_m a_m \frac{M_p}{M_{pm}} \left[\frac{1}{2}\cos(2 (n_m t_{in,m} + M_m(0)) + \omega_m) - \frac{3}{2}\cos \omega_m\right], \label{transit_signal_inec_tinm}
\end{multline}
\begin{multline}
R_{s} = v_{tr}(t_{eg,p} - (jT_p+t_0)) - a_m\frac{M_m}{M_{pm}}\cos(n_m t_{eg,p} + M_m(0) + \omega_m) \\- e_m a_m  \frac{M_m}{M_{pm}} \left[\frac{1}{2}\cos(2(n_m t_{eg,p} + M_m(0)) + \omega_m) -  \frac{3}{2}\cos \omega_m\right],\label{transit_signal_inec_tegp}
\end{multline}
\begin{multline}
R_{s} = v_{tr}(t_{eg,m} - (jT_p+t_0)) + a_m \frac{M_p}{M_{pm}}\cos(n_m t_{eg,m} + M_m(0) + \omega_m) \\+ e_m a_m \frac{M_p}{M_{pm}} \left[\frac{1}{2}\cos(2 (n_m t_{eg,m} + M_m(0)) + \omega_m) - \frac{3}{2}\cos \omega_m\right].\label{transit_signal_inec_tegm}
\end{multline}
Comparing equations~\eqref{transit_signal_inec_tinp} to \eqref{transit_signal_inec_tegm}, with equations~\eqref{transit_signal_cc_tinp} to \eqref{transit_signal_cc_tegm}, the equivalent equations for the case of circular and coplanar orbits, it can be seen that they are very similar.  In particular, noting that for circular orbits $f_m(t) = M_m(t)$ (and thus $f_m(0) = M_m(0)$), the first line of each of equations~\eqref{transit_signal_inec_tinp} to \eqref{transit_signal_inec_tegm} are exactly the same as the corresponding equations for circular coplanar orbits, while the second line is proportional to the eccentricity of the moon's orbit (and would consequently vanish for the circular case).

Again substituting $\theta_{in,p} = n_m t_{in,p}+ M_m(0) +\pi/2 + \omega_m$, $\theta_{eg,p} = n_m t_{eg,p}+ M_m(0) +\pi/2+ \omega_m$, $\theta_{in,m} = n_m t_{in,m}+ M_m(0) +\pi/2+ \omega_m$ and $\theta_{eg,m} = n_m t_{eg,m}+ M_m(0) +\pi/2+ \omega_m$, we obtain,
\begin{multline}
-\frac{n_mR_{s}}{v_{tr}} + n_m(jT_p+t_0) + M_m(0) +\frac{\pi}{2} + \omega_m = \theta_{in,p}  - \frac{v_p}{v_{tr}}\sin \theta_{in,p} \\+ e_m \frac{v_p}{v_{tr}}\left(\frac{1}{2}\cos(2\theta_{in,p} - \omega_m) +  \frac{3}{2}\cos \omega_m\right),\label{transit_signal_inec_tinp2}
\end{multline}
\begin{multline}
-\frac{n_mR_{s}}{v_{tr}} + M_m(0) +\frac{\pi}{2} + \omega_m + n_m(jT_p+t_0) = \theta_{in,m} +  \frac{v_m}{v_{tr}}\sin \theta_{in,m} \\-  e_m\frac{v_m}{v_{tr}}\left(\frac{1}{2}\cos(2\theta_{in,m} - \omega_m) + \frac{3}{2}\cos \omega_m\right),\label{transit_signal_inec_tinm2}
\end{multline}
\begin{multline}
\frac{n_mR_{s}}{v_{tr}} + M_m(0) +\frac{\pi}{2} + \omega_m + n_m(jT_p+t_0) = \theta_{eg,p} - \frac{v_p}{v_{tr}}\sin\theta_{eg,p} \\+ e_m \frac{v_p}{v_{tr}}\left(\frac{1}{2}\cos(2\theta_{eg,p} - \omega_m) + \frac{3}{2}\cos \omega_m\right),\label{transit_signal_inec_tegp2}
\end{multline}
\begin{multline}
\frac{n_mR_{s}}{v_{tr}} + M_m(0) +\frac{\pi}{2} + \omega_m + n_m(jT_p+t_0) = \theta_{eg,m} + \frac{v_m}{v_{tr}}\sin \theta_{eg,m} \\- e_m\frac{v_m}{v_{tr}} \left(\frac{1}{2}\cos(2\theta_{eg,m} - \omega_m) + \frac{3}{2}\cos \omega_m\right).\label{transit_signal_inec_tegm2}
\end{multline}

Each of these four equations is mathematically equivalent to
\begin{equation}
\Phi= \theta_{ec} + B\left[\sin \theta_{ec} - e_m \left(\frac{1}{2}\cos(2\theta_{ec} - \omega_m) + \frac{3}{2}\cos \omega_m)\right)\right],  \label{transit_signal_inec_AB}
\end{equation}
where $\Phi$ and $B$ are equivalent to their definitions in section~\ref{Trans_TTV_Signal_CC} (noting that $f_m(0) = M_m(0)$ for circular orbits) and are explicitly given in table~\ref{ABTableInEc}.  

\begin{table}[tb]
	\begin{center}
  \begin{tabular}{l|c|c}
 $X$ & $\Phi_X$  & $B_X$ \\
  \hline
  ${in,p}$  & $M_m(0) + \omega_m + \frac{\pi}{2} +  n_m(jT_p + t_0) -\frac{n_m R_s}{v_{tr}}$ & $ - \frac{v_p}{v_{tr}}$ \\
  ${in,m}$  & $M_m(0) + \omega_m + \frac{\pi}{2}+  n_m(jT_p + t_0) -\frac{n_m R_s}{v_{tr}}$ & $\frac{v_m}{v_{tr}}$ \\
  ${eg,p}$  & $M_m(0) + \omega_m + \frac{\pi}{2}+  n_m(jT_p + t_0) + \frac{n_m R_s}{v_{tr}}$ & $- \frac{v_p}{v_{tr}}$ \\
  ${eg,m}$  & $M_m(0) + \omega_m + \frac{\pi}{2}+  n_m(jT_p + t_0) + \frac{n_m R_s}{v_{tr}}$ & $\frac{v_m}{v_{tr}}$\\
  \end{tabular}\\
 \caption{The values of $\Phi$ and $B$ corresponding to equations~\eqref{transit_signal_inec_tinp2} to \eqref{transit_signal_inec_tegm2}.}
 \label{ABTableInEc}
 \end{center}
 \end{table}

While this expression is not analytically soluble, it can be explored using a perturbation expansion for the case where $e_m$ is small.  As $\theta_{ec}$ is likely to be similar to the value of $\theta_{cc}$ calculated for the case of circular coplanar orbits, especially for the case of small $e_m$, we can write
\begin{equation}
\theta_{ec} = \theta_{cc} + \Delta \theta_{ec}.
\end{equation}
Substituting this into equation~\eqref{transit_signal_inec_AB}, we have that
\begin{multline}
0 = \Delta \theta_{ec} - B\sin \theta_{cc} + B\Bigg[\sin (\theta_{cc} + \Delta \theta_{ec}) \\- e_m \left(\frac{1}{2}\cos(2(\theta_{cc} + \Delta \theta_{ec}) - \omega_m) + \frac{3}{2}\cos \omega_m \right)\Bigg],  \label{transit_signal_inec_ABseDel}
\end{multline}
where equation~\eqref{transit_signal_cc_AB} has been used to cancel terms.

Expanding $\Delta \theta_{ec}$ using
\begin{equation}
\Delta \theta_{ec} = e_m g(\theta_{cc}) + \ldots
\end{equation}
and substituting into equation~\eqref{transit_signal_inec_ABseDel} gives
\begin{multline}
0 = e_m g(\theta_{cc}) - B\sin \theta_{cc} + B\Bigg[\sin (\theta_{cc} + e_m g(\theta_{cc})) \\- e_m \left(\frac{1}{2}\cos(2(\theta_{cc} + e_m g(\theta_{cc})) - \omega_m) + \frac{3}{2}\cos \omega_m \right)\Bigg].  
\end{multline}
Expanding the sinusoids and gathering first order terms in $e_m$ gives, after rearrangement
\begin{equation}
g(\theta_{cc}) =  \frac{1}{2} \frac{B (\cos(2 \theta_{cc} - \omega_m) + 3\cos \omega_m))  }{1 + B\cos \theta_{cc}}.
\end{equation}
Consequently, we have that
\begin{multline}
\theta_{ec} = \Phi + \sum_{k =1}^\infty \frac{2}{k} J_k(kB) \sin(k\Phi) \\+ \frac{e_m}{2} B \frac{3\cos \omega_m + \cos(2\Phi + \sum_{k =1}^\infty \frac{4}{k} J_k(kB) \sin(k\Phi) - \omega_m)}{1 + B \cos (A + \sum_{k =1}^\infty \frac{2}{k} J_k(kB) \sin(k\Phi))}.\label{transit_signal_inec_tecdef}
\end{multline}

As this expression is structurally complex, while an expression for $\Delta \tau$ could be derived using this equation, it would provide little physical insight into the system.  Consequently, expressions for $\Delta \tau$ will only be derived for the special case, $v_m/v_{tr} \ll 1$, considered in the next section.

\section{Form of $\Delta \tau$}

To investigate the form of $\Delta \tau$ for the case of non-zero values of $e_m$, it was decided to consider only the simplest case, that is, where $v_m/v_{tr} \ll 1$.  While more general expressions for $\Delta \tau$ can be derived (e.g. by retaining second order terms in $v_m/v_{tr}$ and $v_p/v_{tr}$, as in section~\ref{Trans_TTV_Signal_CC}), they are much more complicated and this case will not be considered in this thesis.

\subsection{Case where $v_m/v_{tr} \ll 1$}

For the case where $v_m/v_{tr} \ll 1$ and $v_p/v_{tr} \ll 1$, equation~\eqref{transit_signal_inec_tecdef} simplifies to 
\begin{equation}
\theta_{ec} = \Phi + B \sin(\Phi) + \frac{e_m}{2} B (3\cos \omega_m + \cos(2\Phi - \omega_m)).\label{transit_signal_inec_tecdefsB}
\end{equation}

Using the definitions of $\theta_{in,p}$, $\theta_{in,m}$, $\theta_{eg,p}$ and $\theta_{eg,m}$ and the expressions for $\Phi$ and $B$ in table~\ref{ABTableInEc}, expressions for $t_{in,p}$, $t_{in,m}$, $t_{eg,p}$ and $t_{eg,m}$ can be derived.  Retaining only first order terms in $v_m/v_{tr}$ and $v_p/v_{tr}$ gives
\begin{multline}
 t_{in,p} = jT_p + t_0 -\frac{R_s}{v_{tr}} - \frac{1}{n_m}\frac{v_p}{v_{tr}} \cos\left(M_m(0) + \omega_m +  n_m(jT_p + t_0) -\frac{n_m R_s}{v_{tr}}\right) \\- \frac{e_m}{2}  \frac{1}{n_m}\frac{v_p}{v_{tr}} \Bigg(3\cos \omega_m \\- \cos\left(2M_m(0) +  2n_m(jT_p + t_0) - 2\frac{n_m R_s}{v_{tr}} + \omega_m\right)\Bigg),\label{transit_signal_inec_tecdef}
\end{multline}
\begin{multline}
 t_{in,m} = jT_p + t_0 -\frac{R_s}{v_{tr}} + \frac{1}{n_m}\frac{v_m}{v_{tr}} \cos\left(M_m(0) + \omega_m +  n_m(jT_p + t_0) -\frac{n_m R_s}{v_{tr}}\right) \\+   \frac{e_m}{2} \frac{1}{n_m}\frac{v_m}{v_{tr}} \Bigg(3\cos \omega_m \\ - \cos\left(2M_m(0) +  2n_m(jT_p + t_0) - 2\frac{n_m R_s}{v_{tr}} + \omega_m\right)\Bigg),\label{transit_signal_inec_tecdef}
\end{multline}
\begin{multline}
t_{eg,p} = jT_p + t_0 + \frac{R_s}{v_{tr}} - \frac{1}{n_m}\frac{v_p}{v_{tr}} \cos\left(M_m(0) + \omega_m +  n_m(jT_p + t_0) + \frac{n_m R_s}{v_{tr}}\right) \\- \frac{e_m}{2} \frac{1}{n_m} \frac{v_p}{v_{tr}} \Bigg(3\cos \omega_m \\ - \cos\left(2 M_m(0) +  2n_m(jT_p + t_0) + 2\frac{n_m R_s}{v_{tr}}) + \omega_m\right)\Bigg),\label{transit_signal_inec_tecdef}
\end{multline}
\begin{multline}
t_{eg,m} = jT_p + t_0 + \frac{R_s}{v_{tr}} + \frac{1}{n_m}\frac{v_m}{v_{tr}} \cos\left(M_m(0) + \omega_m +  n_m(jT_p + t_0) + \frac{n_m R_s}{v_{tr}}\right) \\+ \frac{e_m}{2} \frac{1}{n_m} \frac{v_m}{v_{tr}} \Bigg(3\cos \omega_m \\- \cos\left(2 M_m(0) +  2n_m(jT_p + t_0) + 2\frac{n_m R_s}{v_{tr}}) + \omega_m\right)\Bigg).\label{transit_signal_inec_tecdef}
\end{multline}

These expressions can now be used to construct expressions for $\tau_p$, $\tau_m$, $A_p$ and $A_m$.  As $e_m$ is small, the eccentricity does not strongly affect the moon's velocity.  Consequently, the assumption of uniform velocities made in section~\ref{Transit_Signal_Method_Implementation} is still valid, and consequently equations~\eqref{transit_signal_method_taupdef}, \eqref{transit_signal_method_taumdef}, \eqref{transit_signal_method_Apdef} and \eqref{transit_signal_method_Amdef}, the equations defining $\tau_p$, $\tau_m$, $A_p$ and $A_m$ can be used.  Substituting these expressions into equations~\eqref{transit_signal_method_taupdef}, \eqref{transit_signal_method_taumdef}, \eqref{transit_signal_method_Apdef} and \eqref{transit_signal_method_Amdef} and again retaining only first order terms yields
\begin{multline}
\tau_p = jT_p + t_0 - \frac{1}{n_m}\frac{v_p}{v_{tr}} \cos \left(\frac{n_m R_s}{v_{tr}}\right) \cos\left(M_m(0) + \omega_m +  n_m(jT_p + t_0)\right)  \\-   \frac{e_m}{2}\frac{1}{n_m}\frac{v_p}{v_{tr}} \Bigg(3\cos \omega_m \\- \cos\left(2\frac{n_m R_s}{v_{tr}}\right) \cos(2M_m(0) +  2n_m(jT_p + t_0) + \omega_m) \Bigg), \label{transit_signal_inec_taupdef}
\end{multline}
\begin{multline}
\tau_m = jT_p + t_0 + \frac{1}{n_m}\frac{v_m}{v_{tr}} \cos \left(\frac{n_m R_s}{v_{tr}}\right) \cos\left(M_m(0) + \omega_m +  n_m(jT_p + t_0)\right)  \\+ \frac{e_m}{2} \frac{1}{n_m}\frac{v_m}{v_{tr}} \Bigg(3\cos \omega_m \\- \cos\left(2\frac{n_m R_s}{v_{tr}}\right)\cos(2M_m(0) +  2n_m(jT_p + t_0) + \omega_m) \Bigg),\label{transit_signal_inec_taumdef}
\end{multline}
\begin{multline}
A_p = \hat{A}_p  + \hat{A}_p\frac{1}{n_m} \frac{v_{tr}}{R_{s}}\frac{v_p}{v_{tr}} \sin\left( \frac{n_m R_s}{v_{tr}}\right) \sin\left(M_m(0) + \omega_m +  n_m(jT_p + t_0)\right)  \\- \frac{e_m}{2} \frac{\hat{A}_p}{n_m}  \frac{v_{tr}}{R_{s}} \frac{v_p}{v_{tr}} \sin\left(2\frac{n_m R_s}{v_{tr}}\right) \\
\times \sin(2 M_m(0) +  2n_m(jT_p + t_0) + \omega_m),\label{transit_signal_inec_Apdef}
\end{multline}
\begin{multline}
A_m = \hat{A}_m  - \hat{A}_m\frac{1}{n_m} \frac{v_{tr}}{R_{s}}\frac{v_m}{v_{tr}} \sin\left( \frac{n_m R_s}{v_{tr}}\right) \sin\left(M_m(0) + \omega_m +  n_m(jT_p + t_0)\right)  \\+ \frac{e_m}{2} \frac{\hat{A}_m}{n_m}  \frac{v_{tr}}{R_{s}} \frac{v_m}{v_{tr}} \sin\left(2\frac{n_m R_s}{v_{tr}}\right) \\
\times \sin(2 M_m(0) +  2n_m(jT_p + t_0) + \omega_m).\label{transit_signal_inec_Amdef}
\end{multline}

Combining equations~\eqref{transit_signal_inec_taupdef} to \eqref{transit_signal_inec_Amdef} using equation~\eqref{transit_intro_ground_deltaudef} and neglecting any terms of order $v_m^2/v_{tr}^2$,  $v_m v_p/v_{tr}^2$ and $v_p^2/v_{tr}^2$ or greater gives
\begin{multline}
\Delta \tau = \frac{\hat{A}_m M_p - \hat{A}_p M_m}{\hat{A}_{pm} M_{pm}} \frac{a_m}{v_{tr}} \cos \left(\frac{n_m R_s}{v_{tr}}\right) \\
\times \cos\left(M_m(0) + \omega_m +  n_m(jT_p + t_0)\right) \\+  
\frac{e_m}{2} \frac{\hat{A}_m M_p - \hat{A}_p M_m}{\hat{A}_{pm} M_{pm}} \frac{a_m}{v_{tr}} \Bigg(3\cos \omega_m \\ - \cos\left(2\frac{n_m R_s}{v_{tr}}\right) \cos(2M_m(0) +  2n_m(jT_p + t_0) + \omega_m)\Bigg).\label{transit_signal_inec_form_lB}
\end{multline}

This expression can also be approximated by comparing the relative sizes of $\hat{A}_p M_m$ and $\hat{A}_m M_p$ to give
\begin{multline}
\Delta \tau = \frac{M_p}{M_p + M_m} \frac{\hat{A}_m}{\hat{A}_p + \hat{A}_m} \frac{a_m}{v_{tr}} \cos \left(\frac{n_m R_s}{v_{tr}}\right) \\
\times \cos\left(M_m(0) + \omega_m +  n_m(jT_p + t_0)\right) \\+  
\frac{e_m}{2} \frac{M_p}{M_p + M_m} \frac{\hat{A}_m}{\hat{A}_p + \hat{A}_m} \frac{a_m}{v_{tr}} \Bigg(3\cos \omega_m \\ - \cos\left(2\frac{n_m R_s}{v_{tr}}\right) \cos(2M_m(0) +  2n_m(jT_p + t_0) + \omega_m)\Bigg).\label{transit_signal_inec_form_lBsimp}
\end{multline}

Now that expressions for $\Delta \tau$ have been calculated for the case of eccentric moon orbits, the effect of this eccentricity on the form of $\Delta \tau$ will be discussed.

\section{Effect of eccentricity in the moon's orbit on $\Delta \tau$}

As can be seen from equations~\eqref{transit_signal_inec_form_lB} and \eqref{transit_signal_inec_form_lB}, eccentricity in the orbit of the moon affects $\Delta \tau$ in one very important way.  It leads to the distortion of the shape of $\Delta \tau$ as a function of transit number, so that it is no longer a sinusoid, by introducing higher order harmonics.  In particular, for low values of $e_m$ and  $v_m/v_{tr}$, the amplitude of these harmonics is proportional to the eccentricity of the moon's orbit, while the phase depends on the orientation of the moon's orbit.

\begin{figure}
     \centering
     \subfigure[$e_m = 0.1$, $\omega_m = \pi/2$.]{
          \label{TauAgreementEcc2B066}
          \includegraphics[width=.46\textwidth]{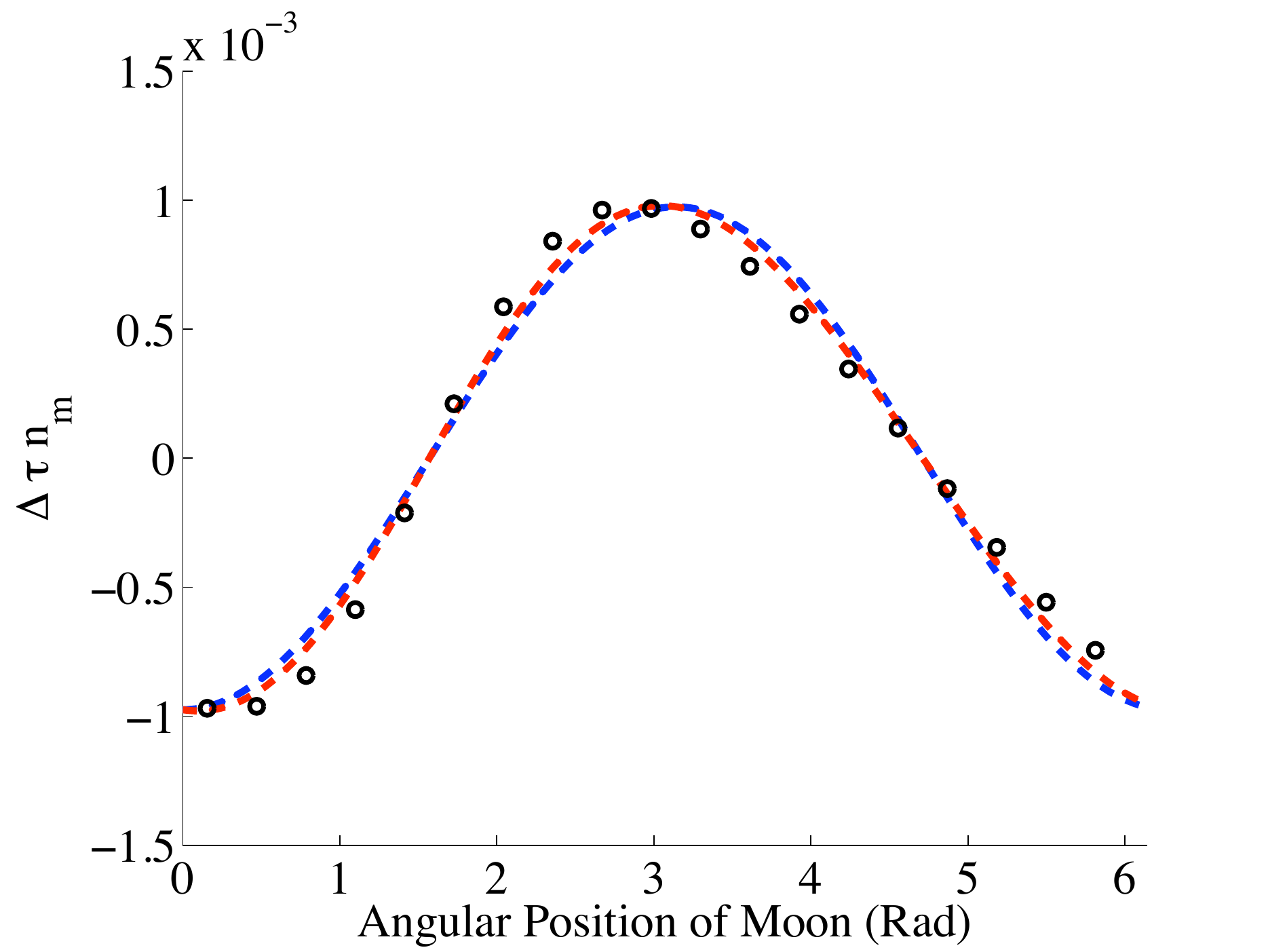}}
     \subfigure[$e_m = 0.4$, $\omega_m = \pi/2$.]{
          \label{TauAgreementEcc1B066}
          \includegraphics[width=.46\textwidth]{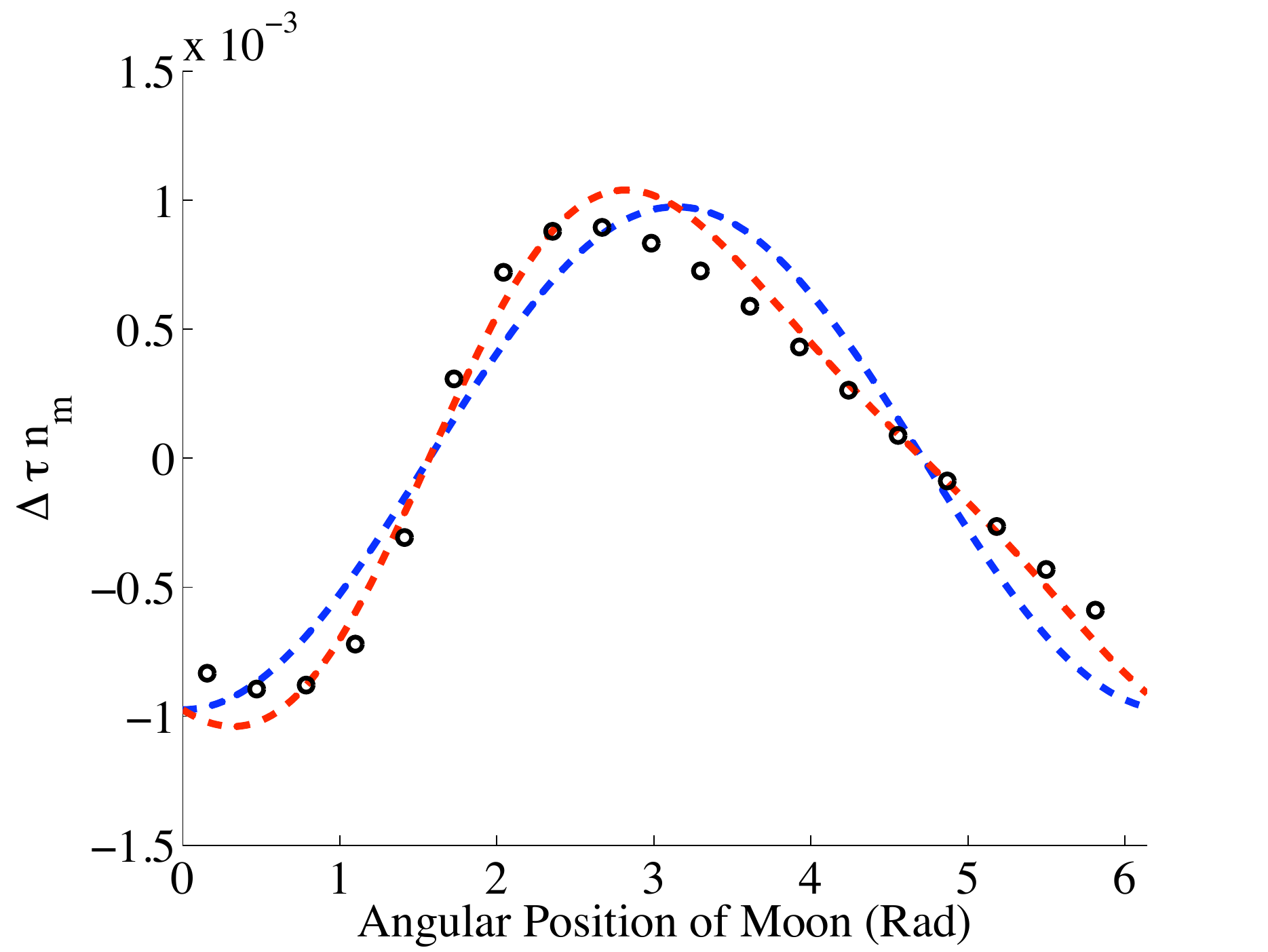}}\\
          \vspace{-0.2cm}
     \subfigure[$e_m = 0.1$, $\omega_m = \pi$.]{
           \label{TauAgreementEcc05B066}
           \includegraphics[width=.46\textwidth]{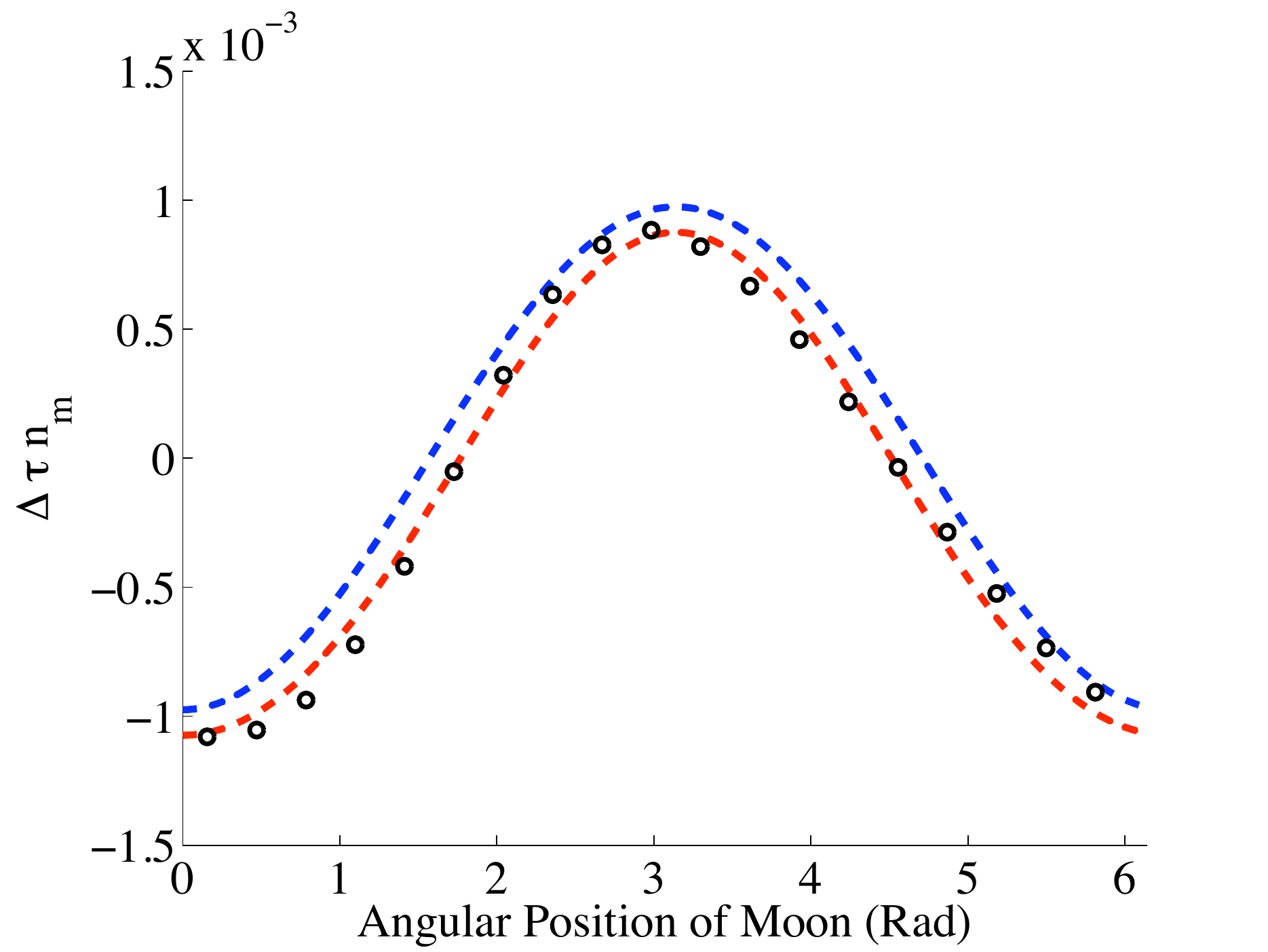}}
           \subfigure[$e_m = 0.4$, $\omega_m = \pi/2$.]{
           \label{TauAgreementEcc1B033}
           \includegraphics[width=.46\textwidth]{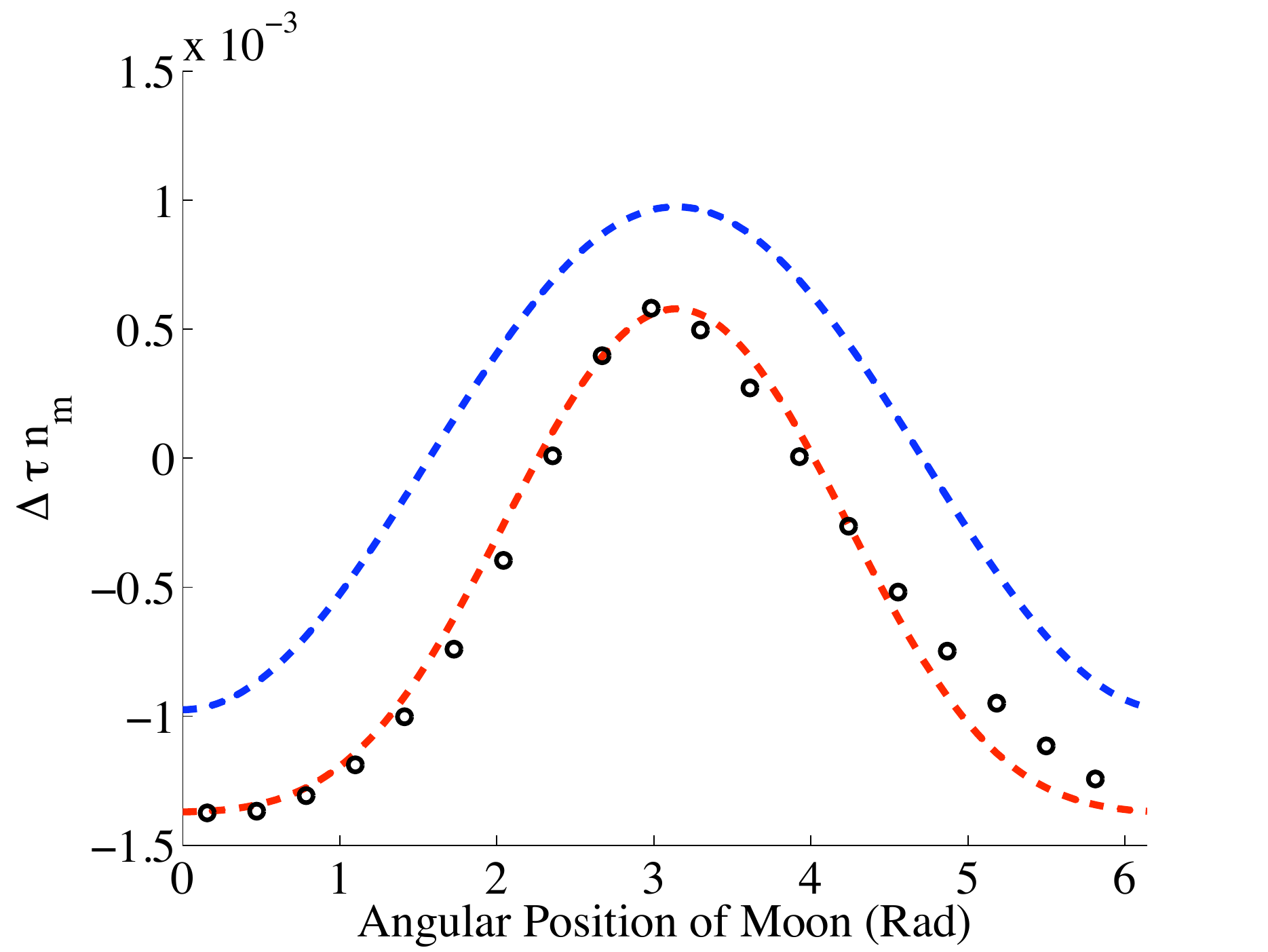}}\\
               \vspace{-0.2cm}
           \subfigure[$e_m = 0.1$, $\omega_m = 3\pi/2$.]{
          \label{TauAgreementEcc2B066}
          \includegraphics[width=.46\textwidth]{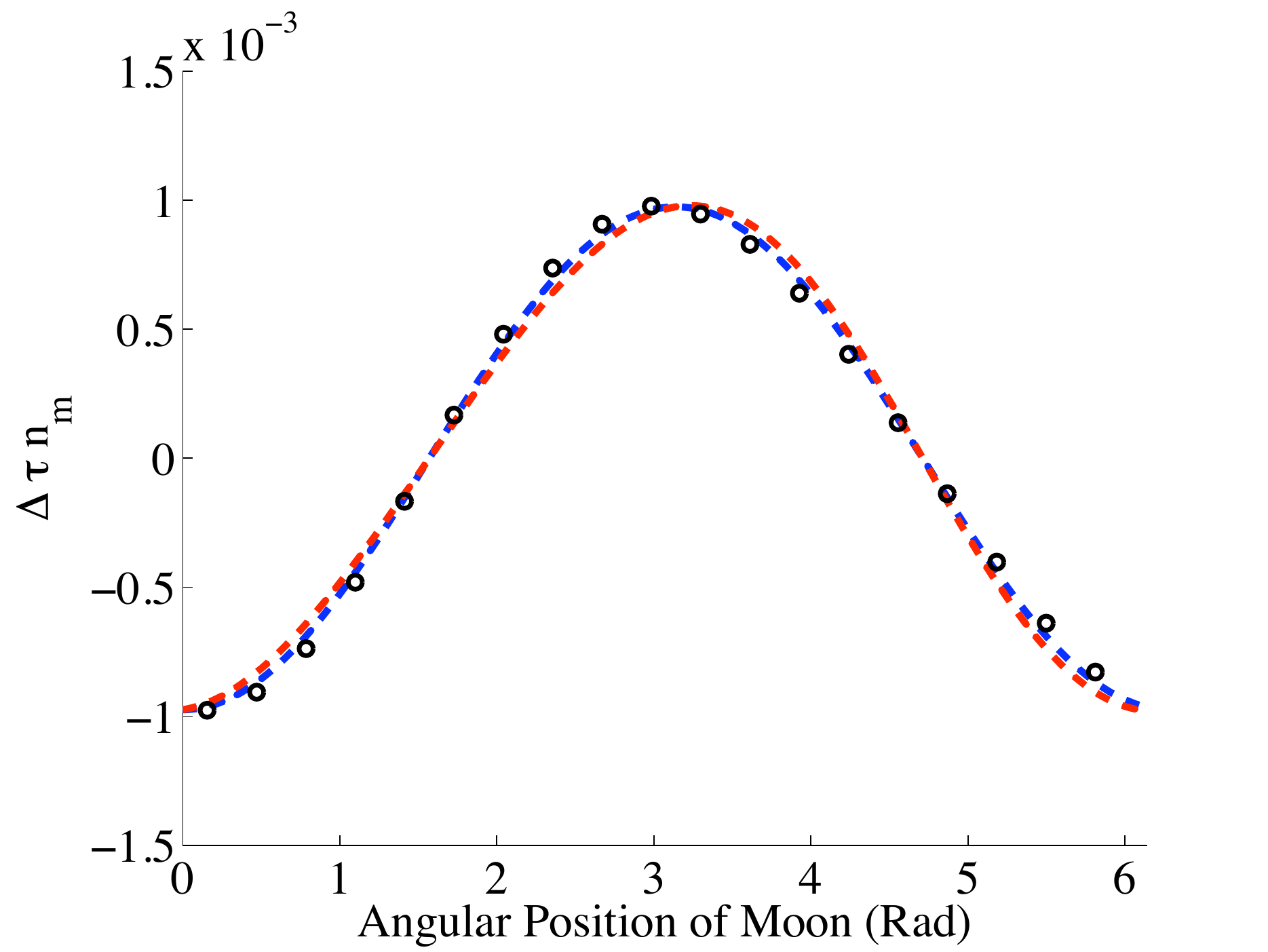}}
     \subfigure[$e_m = 0.4$, $\omega_m = \pi/2$.]{
          \label{TauAgreementEcc1B066}
          \includegraphics[width=.46\textwidth]{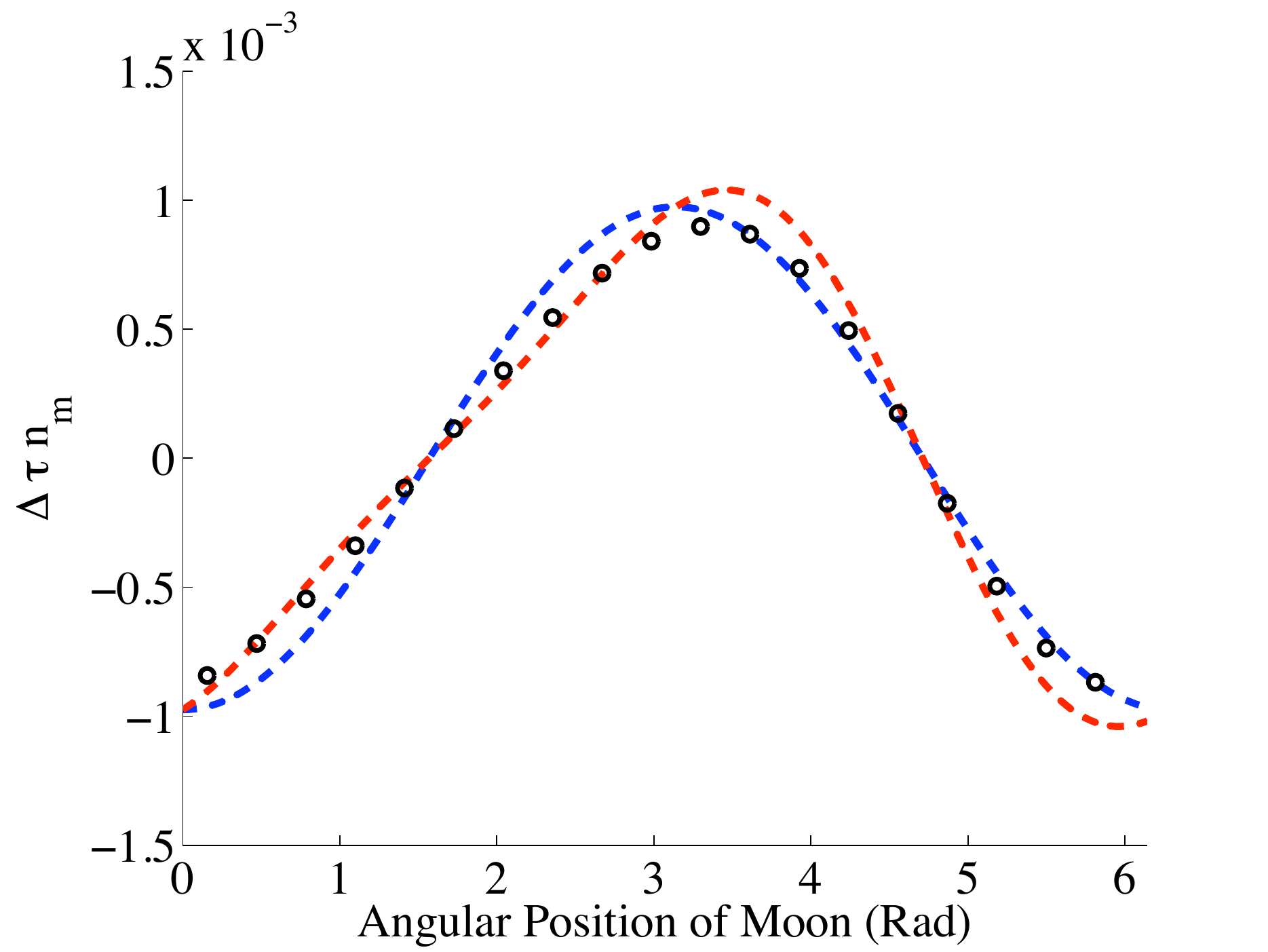}}
     \caption[Comparison of the value of $\Delta \tau$ calculated directly from simulated transit light curves (black), with that of equations~\eqref{transit_signal_cc_form_lB} and \eqref{transit_signal_inec_form_lB}, the analytic approximations to $\Delta \tau$ assuming circular (blue) and eccentric (red) moon orbits.]{Comparison of the value of $\Delta \tau$ calculated directly from simulated transit light curves (black), with that of equations~\eqref{transit_signal_cc_form_lB} and \eqref{transit_signal_inec_form_lB}, the analytic approximations to $\Delta \tau$ assuming circular (blue) and eccentric (red) moon orbits.  The orientations selected ($\omega_m = \pi/2$, $\pi$ and $3 \pi/2$) physically represent orbits with pericenter directions pointing toward the observer, along the plane of the sky, and away from the observer, respectively.  Finally, these plots were constructed for the case of a large gas giant moon, in particular, it was assumed that $R_p = 0.1 R_s$, $R_m = 0.01 R_s$ and $a_m = R_s$.}
     \label{TauAgreementie}
\end{figure}

To investigate this effect, a simulation was run for the case of low (0.1) $v_m/v_{tr}$, comparing $\Delta \tau$ values calculated from equation~\eqref{transit_signal_cc_form_lB}, assuming a circular moon orbit, calculated from equation~\eqref{transit_signal_inec_form_lB} assuming an eccentric moon orbit, and calculated directly from the simulated light curve.  These simulations were conducted for the case of ``low" eccentricity ($e_m = 0.1$) and ``moderate" eccentricity ($e_m = 0.4$) and a range of different orbital orientations ($\omega = \pi/2$, $\pi$ and $3\pi/2$) and are shown in figure~\ref{TauAgreementie}.  Recalling from table~\ref{SSMoonsTable}  that the regular satellites in the Solar System all have eccentricities less than 0.06, this simulation indicates that, equation~\eqref{transit_signal_inec_form_lB} accurately describes $\Delta \tau$ for a range of realistic, non-negligible eccentricities and a typical range of orientations.

\chapter[The effect of second order terms on $\epsilon_j$]{The effect of second order terms on $\epsilon_j$ for the case of white noise}
\label{SecOrdNoise_App}

In chapter~\ref{Trans_Intro},  expressions for $\Delta \tau$ and $\epsilon_j$ were derived by performing a binomial expansion on equation~\eqref{TraM-TTV-tauderiv1}, the equation defining $\tau$, and retaining first order terms in $\sum_i \alpha_n/\sum_i (\alpha_p + \alpha_m)$.  However, for the case where $\sum_i \alpha_n$ is large, due to, for example, a dim host star or a bad pixel, or where $\sum_i (\alpha_p + \alpha_m)$ is small due to e.g. a small planet or a short transit duration, $\sum_i \alpha_n \ll \sum_i (\alpha_p + \alpha_m)$ may no longer hold.  

To explore the effect of neglecting higher order terms in $\sum_i \alpha_n/\sum_i (\alpha_p + \alpha_m)$ on $\epsilon_j$, the next term in the binomial expansion of equation~\eqref{TraM-TTV-tauderiv1} will be retained and investigated for the case of white noise.  Expanding equation~\eqref{TraM-TTV-tauderiv1} and retaining all terms to second order in $\alpha_n$ gives
\begin{multline}
\tau = \frac{\sum_i t_i (\alpha_p(t_i) + \alpha_m(t_i))}{\sum_i \alpha_p(t_i) + \alpha_m(t_i) } 
+ \frac{\sum_i \left( t_i - \left(t_0 + jT_p + \Delta \tau\right) \right)\alpha_n(t_i)}{A_p + A_m }  \\
+ \frac{\sum_i \sum_k  \left(-t_i + t_0 + jT_p + \Delta \tau \right)\alpha_n(t_i) \alpha_n(t_k)}{\left(A_p + A_m \right)^2 } .\label{TraM-Noi-White-taun2ndord}
\end{multline}
Consequently, to second order,
\begin{multline}
\epsilon_j =  \frac{\sum_i \left( t_i - \left(t_0 + jT_p + \Delta \tau\right) \right)\alpha_n(t_i)}{A_p + A_m }  \\
+ \frac{\sum_i \sum_k  \left(-t_i + t_0 + jT_p + \Delta \tau \right)\alpha_n(t_i) \alpha_n(t_k)}{\left(A_p + A_m \right)^2 } .\label{TraM-Noi-White-muderive1}
\end{multline}

In chapter~\ref{Trans_TTV_Noise} it was shown that the the first term in equation~\eqref{TraM-Noi-White-muderive1} is normally distributed.  Consequently, any non-normal behaviour exhibited by $\epsilon_j$ must be due to the effect of the second term.  As this term is comprised of weighted sums of pairs of $\alpha_n$ multiplied together, the simple normal formulas can no longer be used to analyse this term.  In particular, the author is unaware of any method which will provide an analytic description of the distribution of $\epsilon_j$ using equation~\eqref{TraM-Noi-White-muderive1}.  However, while the shape of the distribution cannot be calculated, the mean of the distribution can.

This can be achieved by noting that the mean of a weighted sum of random variables is equal to the sum of the means of the variables multiplied by their associated weights, that is, if
\begin{equation}
Y = \sum_{i = 1}^N \beta_i X_i,
\end{equation}
where $X_1$ to $X_N$ are random variables, then
\begin{equation}
\mu_Y = \sum_{i = 1}^N \beta_i \mu_i,
\end{equation}
where $\mu_Y$ is the mean of $Y$ and $\mu_i$ is the mean of $X_i$ for $i = 1 \ldots N$.
Consequently, to calculate $\mu_\epsilon$ using equation~\eqref{TraM-Noi-White-muderive1}, the means of all the individual terms need to be calculated and summed.

There are three types of terms in equation~\eqref{TraM-Noi-White-muderive1}, terms which are proportional to $\alpha_n(t_i)$, terms which are proportional to $\alpha_n(t_i)\alpha_n(t_k)$ with $i \ne k$, and terms which are proportional to $(\alpha_n(t_i))^2$.  The contribution of each of these types of terms to the mean will be investigated in turn.  From the definition of $\alpha_n(t_i)$,\footnote{Recall that for the case of white noise $\alpha_n(t_i)$ is normally distributed with mean zero.} we have that the mean of all terms proportional to $\alpha_n(t_i)$ is zero.  In addition, as $\alpha_n(t_i)$ is symmetric and centered on 0, the distribution of $\alpha_n(t_i)\alpha_n(t_k)$ for $i \ne k$ is also symmetric and centered on 0.  Consequently these terms also do not contribute to $\mu_\epsilon$.  However, the terms proportional to $(\alpha_n(t_i))^2$ are always greater than or equal to zero and consequently have non-zero mean.  Neglecting the first order terms and the terms with $i \ne k$ gives
\begin{equation}
\mu_\epsilon =   \frac{\sum_i  \left(-t_i + t_0 + jT_p + \Delta \tau \right)\overline{\alpha_n(t_i)^2 }}{\left(A_p + A_m \right)^2 } .\label{TraM-Noi-White-muderive2}
\end{equation}
where $\overline{\alpha_n(t_i)^2 }$ represents the mean of $\alpha_n(t_i)^2 $.

To determine $\overline{\alpha_n(t_i)^2 }$ we begin by considering a random variable $X$ which is distributed according to a normal distribution with mean zero and standard deviation 1.   From the definition of the chi squared distribution we then have that $X^2$ is distributed according to a chi squared distribution with one degree of freedom.  Consequently it follows that $\alpha_n(t_i)^2/ \sigma_L^2$ is also distributed according to a chi squared distribution with one degree of freedom.

As the mean of a chi squared distribution is equal to the number of degrees of freedom, we have that $\overline{\alpha_n(t_i)^2 } = \sigma_L^2$.  Substituting this into equation~\eqref{TraM-Noi-White-muderive2} gives
\begin{equation}
\mu_\epsilon =  \frac{\sum_i  \left(-t_i + t_0 + jT_p + \Delta \tau \right)\sigma_L^2 }{\left(A_p + A_m \right)^2 } .\label{TraM-Noi-White-muderive3}
\end{equation}
Substituting equation~\eqref{TraM-Noi-White-sumi} for $\sum_{i = 0}^N i$ gives:
\begin{align}
\mu_\epsilon& =  \frac{\sum_{i = 0}^{N_{obs}}  \left(\Delta \tau - \Delta t_p - \left(i - \frac{N_{obs}}{2}\right) \right)\sigma_L^2 }{\left(A_p + A_m \right)^2 }\label{TraM-Noi-White-muderive4}\\
& =  \frac{\left((N_{obs}+1)(\Delta \tau - \Delta t_p) - \left(\frac{N_{obs}(N_{obs}+1)}{2} - \frac{N_{obs}(N_{obs}+1)}{2}\right) \right)\sigma_L^2 }{\left(A_p + A_m \right)^2 }\label{TraM-Noi-White-muderive5}\\
& =  \frac{ (N_{obs}+1)\left(\Delta \tau - \Delta t_p  \right)\sigma_L^2 }{\left(A_p + A_m \right)^2 }
 .\label{TraM-Noise-White-muderive6}
\end{align}

As $\Delta \tau - \Delta t_p $ is the time difference between the photocenter of the dip and the center of the planetary transit, it can be seen that this error only occurs when the transit is asymmetric, for example, due to an orbiting moon.

\begin{figure}
     \centering
     \subfigure[$\Delta \tau = 0$s, $N_{obs} = 280$.]{
          \label{fig:dl2858}
          \includegraphics[width=.48\textwidth]{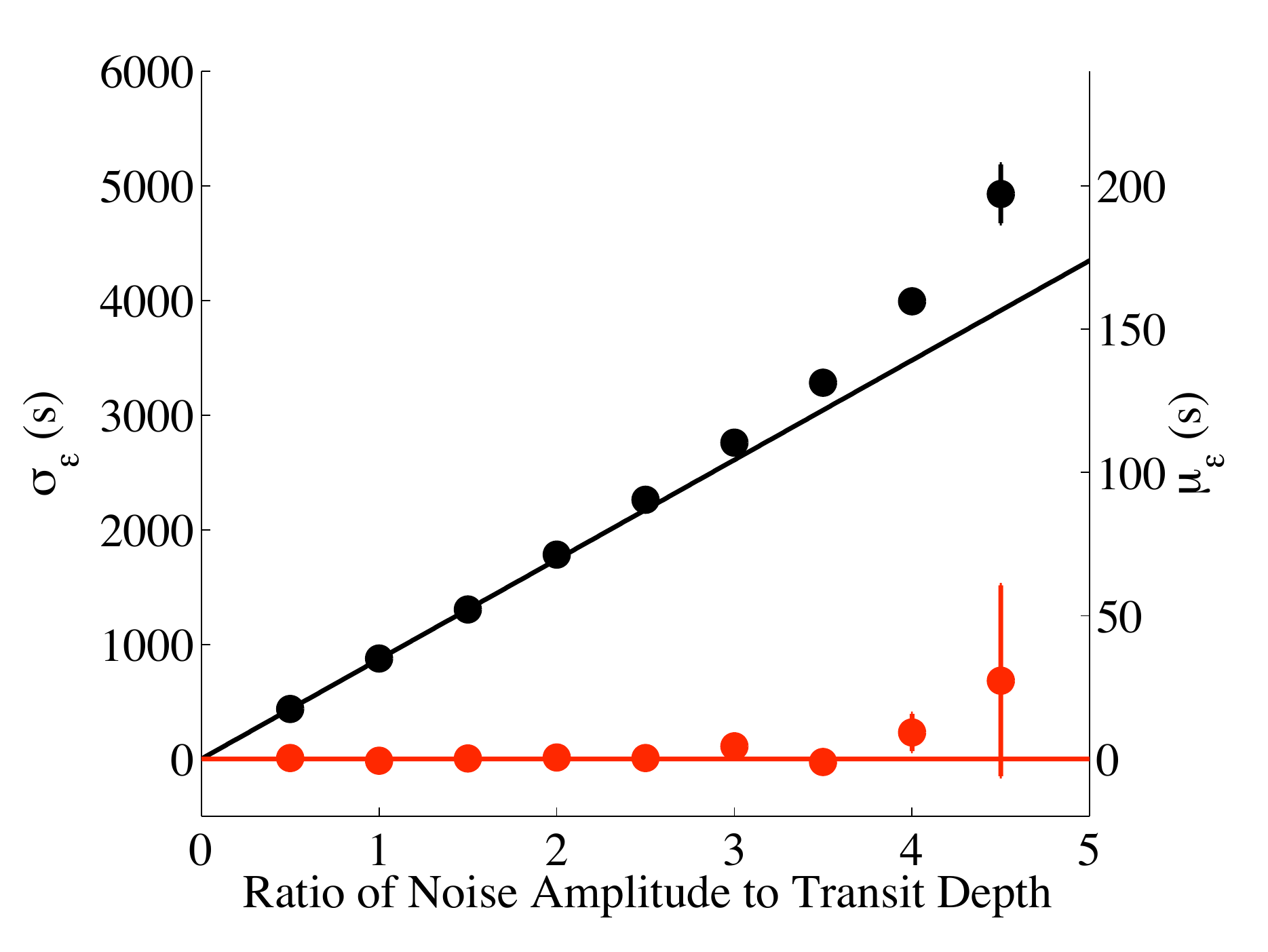}}
     \subfigure[$\Delta \tau = 1000$s, $N_{obs} = 280$.]{
          \label{fig:er2858}
          \includegraphics[width=.48\textwidth]{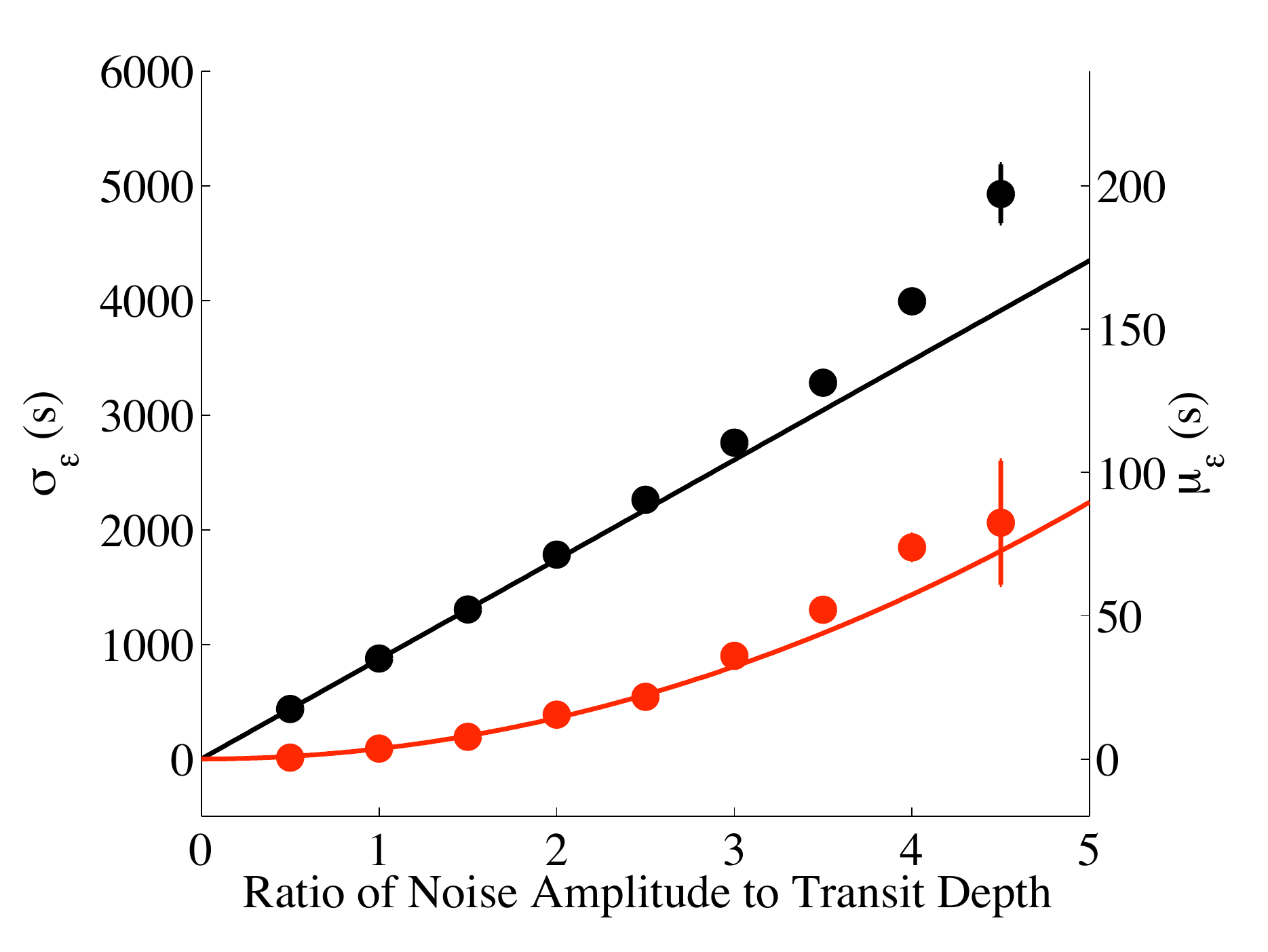}}\\
           \subfigure[$\Delta \tau = 0$s, $N_{obs} = 140$.]{
           \label{fig:cminusscalar2858}
           \includegraphics[width=.48\textwidth]{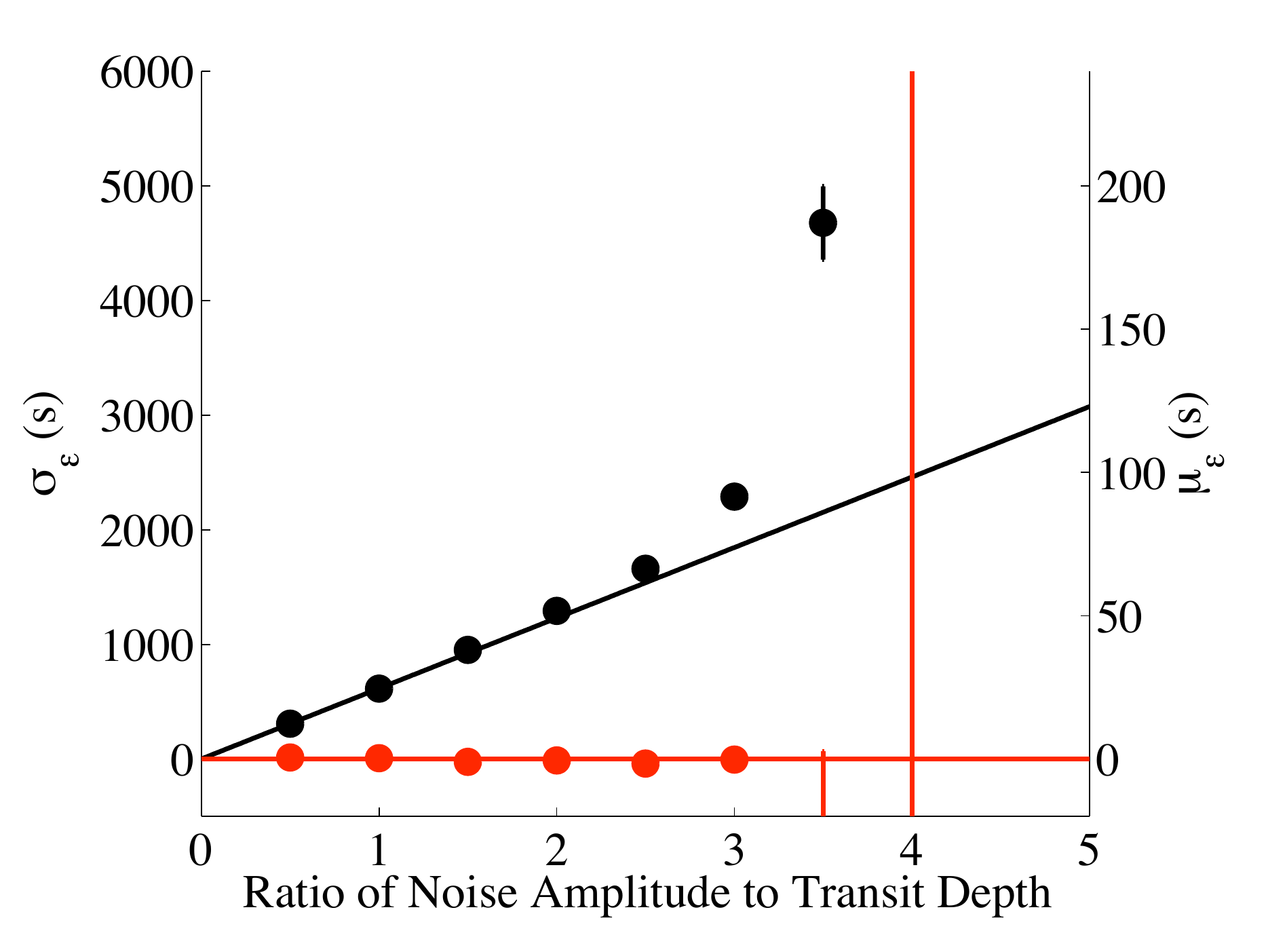}}
           \subfigure[$\Delta \tau = 1000$s, $N_{obs} = 140$.]{
           \label{fig:cminusscalar2858}
           \includegraphics[width=.48\textwidth]{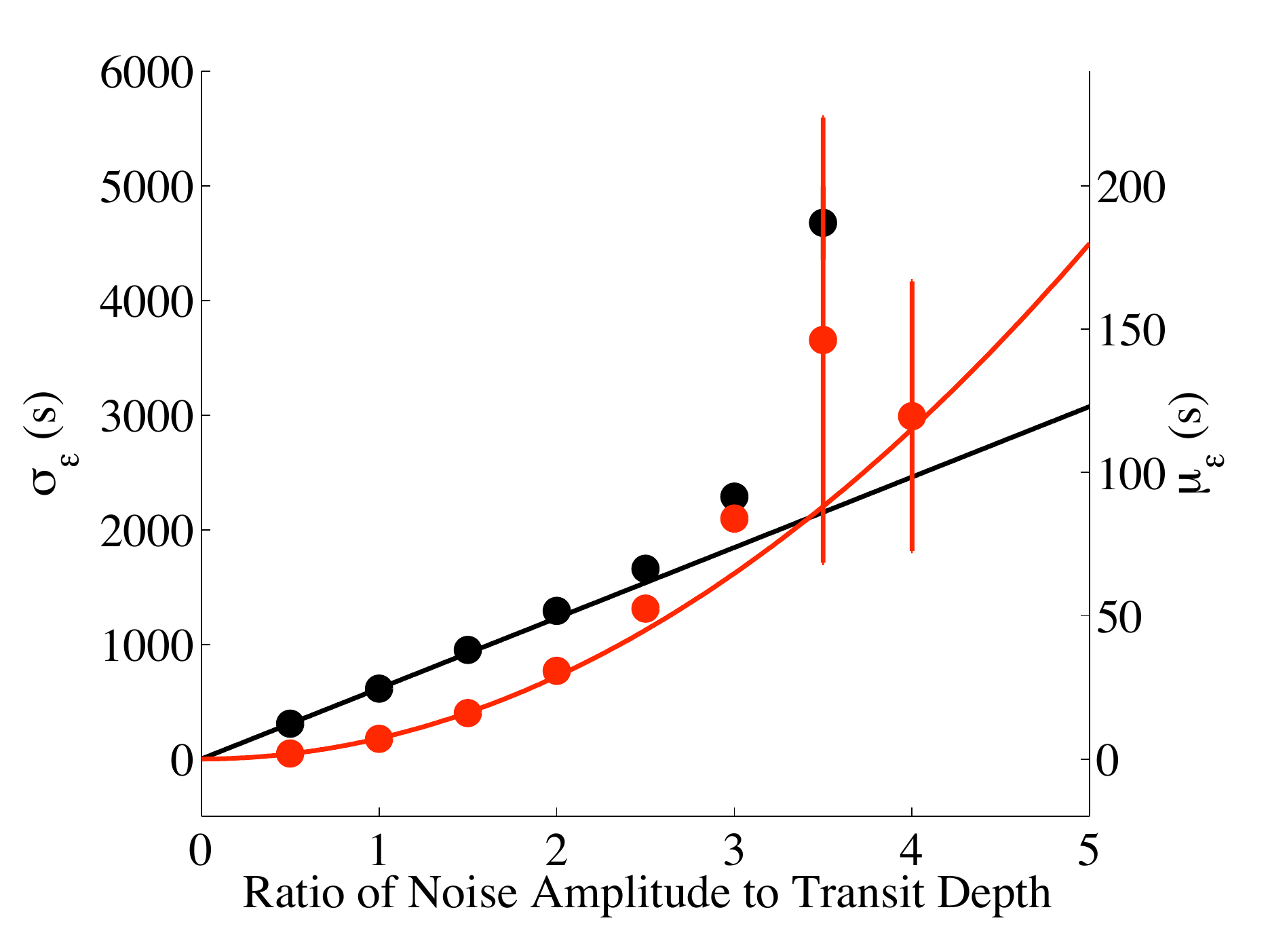}}
     \caption[Comparison between equations~\eqref{transit_noise_white_sigdef} and \eqref{TraM-Noise-White-muderive6}, the theoretical predictions for the behaviour of $\sigma_\epsilon$ (black) and $\mu_\epsilon$ (red) as a function of the ratio between the $\sigma_L$, the photometric noise, and transit depth, and the values obtained from a Monte Carlo simulation.]{Comparison between equations~\eqref{transit_noise_white_sigdef} and \eqref{TraM-Noise-White-muderive6}, the theoretical predictions for the behaviour of $\sigma_\epsilon$ (black) and $\mu_\epsilon$ (red) as a function of the ratio between the $\sigma_L$, the photometric noise, and transit depth, and the values obtained from a Monte Carlo simulation.  Exposure times of three minutes were used for all plots.  $N_{obs}$, the number of exposures was selected so that the two cases explored corresponded to a central transit at 1AU and 0.2AU respectively.  $10^6$ simulated transits were used to construct each mean data point while $5 \times 10^5$ simulated transits were used to construct the standard deviation data points.   As $\sum \alpha_n$ became more comparable to $\sum (\alpha_p + \alpha_m)$, the probability of $\sum(\alpha_p + \alpha_m + \alpha_n) \approx 0$, and consequently small denominators increased.  Thus, in this region there was no appreciable tightening of error bars as the number of simulations increased.  Note that error bars are only shown for points where the bars are larger than the marker.}
     \label{WhiteNoiseShift}
\end{figure}

To demonstrate the effect of neglecting the higher order $\alpha_n$ terms, equation~\eqref{TraM-Noi-White-muderive1} was numerically investigated.  The mean and the standard deviation of $\epsilon_j$ was calculated for a selection of values of $\sigma_L/(A_p + A_m)$, ranging from 0 to 4.5.  These numerical results were then compared with equations~\eqref{transit_noise_white_sigdef} and \eqref{TraM-Noise-White-muderive6} the first order formula for $\sigma_\epsilon$ and the second order formula for $\mu_\epsilon$ (see figure~\ref{WhiteNoiseShift}).

For reference, the point at which the numerical results start to diverge from the predicted curves corresponds  to an $0.8R_{\earth}$ planet orbiting a Sun-like star, observed with a relative photometric error of $2.2 \times 10^{-4}$.  As sub-Earth mass planets capable of hosting stable moons are unlikely to be detected by COROT and Kepler, the higher order terms in equation~\eqref{TraM-Noi-White-muderive1} can be safely neglected.

\chapter{Derivation of equivalent white noise amplitude}\label{App_WhiteNoiseAmp}

In order to compare the calculated values for $\sigma_\epsilon$ for the case raw and filtered solar photometric noise with those for white noise, a method is needed to select the white noise which is the \emph{same} as the solar photometric noise.  For this work it was decided to compare white and solar photometric noise with the same power.  To ensure that photometric variation on timescales relevant to transiting systems was included but variation due to longer term phenomenon such as the solar cycle were not, a three month span of data was used to determine the amount of power per unit time in the raw and filtered solar data sets.  
Using the definition of power
\begin{equation}
P_n = \sum_{i=1}^{N} \left| \alpha_n(t_i) \right|^2,
\end{equation}
this was found to be 6491 W/m$^2$ and 5644 W/m$^2$ respectively.  Consequently, a three month segment of white noise which is equivalent to the raw and filtered solar data should show a power of 6491 W/m$^2$ and 5644 W/m$^2$ respectively.

Noting that for the case of white noise,
\begin{equation}
\sum_{i=1}^{N} \left| \alpha_n(t_i) \right|^2 = N \left< \alpha_n^2 \right>,
\end{equation}
where $N$ is large and $\left< \alpha_n^2 \right>$ is the expectation value of $\left(\alpha_n\right)^2$. From equation~\eqref{white_intro_gaudef}, the definition of a Gaussian distribution, and the definition of expectation value we have that
\begin{align}
\left<\alpha_n^2\right> &= \frac{1}{\sigma_L \sqrt{2\pi}} \int_{-\infty}^{\infty} x^2 e^{-\frac{x^2}{2\sigma_L^2}} dx,\\
&=\sigma_L^2.
\end{align}
Consequently
\begin{equation}
\sigma_L = \sqrt{\frac{P_n}{N}}.
\label{noisecomp_whtnoiseamp}
\end{equation}
Thus $\sigma_L$ is equal to 0.0518 W/m$^2$ for the case of raw solar photometric noise and 0.0483 W/m$^2$ for the case of filtered solar photometric noise.  Noting that $L_0$ is equal to 1367.3 W/m$^2$, this give a relative photometric variability of $3.79 \times 10^{-5}$ and $3.53 \times 10^{-5}$ respectively.

\chapter{Proof that $\overline{t}_0$ and $\overline{T}_p$ tend to $t_0$ and $T_p$ as $N \to \infty$}\label{App_t0Tp}

To begin, consider a sequence of $\tau_j$ values described by 
\begin{equation}
\tau_j = t_0 + jT_p + A\cos(\omega j + \phi) + \epsilon_j,
\label{app_t0Tp_taudef}
\end{equation}
where each of the $\epsilon_j$ are normally distributed, independent, random variables with mean zero and standard deviation $\sigma_\epsilon$.  As discussed in chapter~\ref{Trans_Thresholds}, the process of determining if the sinusoidal perturbation in the recorded $\tau_j$ values is detectable, involves least-squares fitting a linear model ($\overline{t}_0 + j\overline{T}_p$) and a linear model plus a sinusoid ($\hat{t}_0 + j\hat{T}_p + \hat{A}\cos(\hat{\omega}j + \hat{\phi})$) to these values, and comparing the residuals.  While we expect the fitting parameters of the line plus a sinusoid model to tend towards the true values as $N$ tends to infinity, it is less obvious what $\overline{t}_0$ and $\overline{T}_p$ tend to as $N$ becomes large.  To investigate this, expressions for $\overline{t}_0$ and $\overline{T}_p$ will be derived in this appendix.  Then, through a consideration of the error on these quantities and their limit as $N$ tends to infinity, it will be shown that for the case where there is a detectable sinusoidal perturbation, $\overline{t}_0$ and $\overline{T}_p$ tend to $t_0$ and $T_p$ as $N$ becomes large.

\section{Expressions for $\overline{t}_0$ and $\overline{T}_p$}

From \citet[][p. 656]{Pressetal1992}, we have that the coefficients for a least-squares linear fit are given by
\begin{align}
\overline{t}_0 &= \frac{\sum_{j = 1}^N j^2 \sum_{j = 1}^N \tau_j - \sum_{j = 1}^N j \sum_{j = 1}^N j \tau_j}{N \sum_{j = 1}^N j^2 - \left(\sum_{j = 1}^N j\right)^2},\label{app_t0Tp_t0fit}\\
\overline{T}_p &= \frac{N \sum_{j = 1}^N j \tau_j - \sum_{j = 1}^N j \sum_{j = 1}^N \tau_j}{N \sum_{j = 1}^N j^2 - \left(\sum_{j = 1}^N j\right)^2},\label{app_t0Tp_Tpfit}
\end{align}
where we note that there is no dependance on $\sigma_\epsilon$ as it is the same for all $\tau_j$ values.  Substituting equation~\eqref{app_t0Tp_taudef} into equations~\eqref{app_t0Tp_t0fit} and \eqref{app_t0Tp_Tpfit} and separating the $\epsilon_j$ terms from the other terms gives
\begin{multline}
\overline{t}_0 = \frac{\sum_{j = 1}^N j^2 \sum_{j = 1}^N (t_0 + jT_p + A\cos(\omega j + \phi)) }{N \sum_{j = 1}^N j^2 - \left(\sum_{j = 1}^N j\right)^2} \\
- \frac{\sum_{j = 1}^N j \sum_{j = 1}^N j (t_0 + jT_p + A\cos(\omega j + \phi))}{N \sum_{j = 1}^N j^2 - \left(\sum_{j = 1}^N j\right)^2}  \\
+ \frac{\sum_{j = 1}^N (\sum_{i = 1}^N i^2 - j\sum_{i = 1}^N i) \epsilon_j}{N \sum_{j = 1}^N j^2 - \left(\sum_{j = 1}^N j\right)^2},\label{app_t0Tp_t0def}
\end{multline}
\begin{multline}
\overline{T}_p = \frac{N \sum_{j = 1}^N j (t_0 + jT_p + A\cos(\omega j + \phi))}{N \sum_{j = 1}^N j^2 - \left(\sum_{j = 1}^N j\right)^2} \\
- \frac{\sum_{j = 1}^N j \sum_{j = 1}^N (t_0 + jT_p + A\cos(\omega j + \phi))}{N \sum_{j = 1}^N j^2 - \left(\sum_{j = 1}^N j\right)^2} \\
+ \frac{ \sum_{j = 1}^N (Nj - \sum_{i = 1}^N i) \epsilon_j}{N \sum_{j = 1}^N j^2 - \left(\sum_{j = 1}^N j\right)^2}.\label{app_t0Tp_Tpdef}
\end{multline}
The first two terms in each of these expressions do not depend on any of the $\epsilon_j$, and consequently represent the fit that would occur if there were no noise on the $\tau_j$ values.  The third term in each of these two expressions consists of a weighted sum of $\epsilon_j$ values and determines the degree to which the fit is perturbed by the presence of timing noise.  We will consider these two aspects in turn for the case of large $N$, and combine them to give a full description of the behaviour of  $\overline{t}_0$ and $\overline{T}_p$ as $N$ increases.

\section{Limiting behaviour of $\overline{t}_0$ and $\overline{T}_p$ as $N \to \infty$ for the case where $\sigma_\epsilon = 0$}

For the case where there is no timing noise, i.e., $\sigma_\epsilon = 0$, the third term of equations~\eqref{app_t0Tp_t0def} and \eqref{app_t0Tp_Tpdef} vanishes, and the equations become
\begin{multline}
\overline{t}_0 = \frac{\sum_{j = 1}^N j^2 \sum_{j = 1}^N (t_0 + jT_p + A\cos(\omega j + \phi)) }{N \sum_{j = 1}^N j^2 - \left(\sum_{j = 1}^N j\right)^2} \\
- \frac{\sum_{j = 1}^N j \sum_{j = 1}^N j (t_0 + jT_p + A\cos(\omega j + \phi))}{N \sum_{j = 1}^N j^2 - \left(\sum_{j = 1}^N j\right)^2},\label{app_t0Tp_t0def_sig0}
\end{multline}
\begin{multline}
\overline{T}_p = \frac{N \sum_{j = 1}^N j (t_0 + jT_p + A\cos(\omega j + \phi))}{N \sum_{j = 1}^N j^2 - \left(\sum_{j = 1}^N j\right)^2} \\
- \frac{\sum_{j = 1}^N j \sum_{j = 1}^N (t_0 + jT_p + A\cos(\omega j + \phi))}{N \sum_{j = 1}^N j^2 - \left(\sum_{j = 1}^N j\right)^2}.\label{app_t0Tp_Tpdef_sig0}
\end{multline}
We begin by considering equation~\eqref{app_t0Tp_t0def_sig0}.

From section~\ref{Trans_TTV_Noise_White_epj} and equations~\eqref{TraM-Noi-White-sumi} and \eqref{TraM-Noi-White-sumi2}, we have that $\sum_{j = 1}^N j = N(N+1)/2$ and that  $\sum_{j = 1}^N j^2 = N(N+1)(2N+1)/6$.  However the behaviour of the $\sum_{j = 1}^N \cos(\omega j + \phi)$ and $\sum_{j = 1}^N j\cos(\omega j + \phi)$ terms is less obvious.  As a result we will derive expressions for these terms.  Consider the $\sum_{j = 1}^N \cos(\omega j + \phi)$ term first.  Writing the sinusoid as a complex exponential we have
\begin{align}
\sum_{j = 1}^N \cos(\omega j + \phi) &= \sum_{j = 1}^N \frac{1}{2} \left(e^{i(\omega j + \phi)} + e^{-i(\omega j + \phi)} \right),\\
 &= \sum_{j = 1}^N \frac{e^{i\phi}}{2} \left(e^{i\omega}\right)^j + \sum_{j = 1}^N \frac{e^{-i\phi}}{2} \left(e^{-i\omega}\right)^j.
\end{align}
Both of these terms are geometric series, so they can be analytically evaluated.  Recalling that
\begin{equation}
\sum_{n= 0}^N a r^n = a\frac{1 - r^{N+1}}{1 - r},
\end{equation}
and thus that 
\begin{equation}
\sum_{n= 1}^N a r^n = a\frac{1 - r^{N+1}}{1 - r} - a,
\end{equation}
we have that
\begin{multline}
\sum_{j = 1}^N \cos(\omega j + \phi) =  \frac{e^{i\phi}}{2} \frac{1 - e^{i\omega(N+1)}}{1 - e^{i\omega}} - \frac{e^{i\phi}}{2} \\
+ \frac{e^{-i\phi}}{2} \frac{1 - e^{-i\omega(N+1)}}{1 - e^{-i\omega}} - \frac{e^{-i\phi}}{2},
\end{multline}
which simplifies to
\begin{multline}
\sum_{j = 1}^N \cos(\omega j + \phi) =  \\
\frac{\cos \phi  - \cos(\omega(N+1) + \phi) - \cos(\omega - \phi) + \cos(\omega N + \phi)}{2 - 2\cos \omega} - \cos \phi.
\end{multline}
So unless $\omega = k\times 2 \pi$, where $k$ is an integer, (corresponding to the case where the moon completes an integer number of orbits per planetary orbit and is thus undetectable), this term is of order 1.

The equivalent equation for the case of $\sum_{j = 1}^N j\cos(\omega j + \phi)$ can be constructed from the one derived for $\sum_{j = 1}^N \cos(\omega j + \phi)$, by noting that
\begin{align}
\frac{d}{d\omega} \sum_{j = 1}^N \cos(\omega j + (\phi - \pi/2)) &= \frac{d}{d\omega} \sum_{j = 1}^N \sin(\omega j + \phi),\\
 &=  \sum_{j = 1}^N j \cos(\omega j + \phi).
\end{align}
Thus
\begin{multline}
\sum_{j = 1}^N j \cos(\omega j + \phi) = \frac{d}{d\omega} \left[\frac{\cos (\phi - \pi/2)  - \cos(\omega(N+1) + \phi - \pi/2) }{2 - 2\cos \omega} \right. \\
\left. + \frac{- \cos(\omega - \phi + \pi/2) + \cos(\omega N + \phi - \pi/2)}{2 - 2\cos \omega} - \cos (\phi - \pi/2).
\right],
\end{multline}
which evaluates to
\begin{multline}
\sum_{j = 1}^N j \cos(\omega j + \phi) = \left[\frac{ - (N+1)\cos(\omega(N+1) + \phi) + \cos(\omega - \phi)}{2 - 2\cos \omega} \right. \\
\left. + \frac{N \cos(\omega N + \phi)}{2 - 2\cos \omega} \right] - 2 \sin \omega \left[\frac{\sin \phi  - \sin(\omega(N+1) + \phi) }{(2 - 2\cos \omega)^2} \right. \\
\left. + \frac{\sin(\omega - \phi) + \sin(\omega N + \phi)}{(2 - 2\cos \omega)^2} \right].
\end{multline}
Consequently $\sum_{j = 1}^N j\cos(\omega j + \phi)$ is of order $N$.

Now that we have expressions for $\sum_{j = 1}^N j$, $\sum_{j = 1}^N j^2$, $\sum_{j = 1}^N \cos(\omega j + \phi)$ and $\sum_{j = 1}^N j\cos(\omega j + \phi)$ we can evaluate equation~\eqref{app_t0Tp_t0def_sig0}.  Substituting these expressions gives
\begin{multline}
\overline{t}_0 = \frac{\frac{N(N+1)(2N+1)}{6}\left( N t_0 + \frac{N(N+1)}{2}T_p + O(1) \right)}{N \frac{N(N+1)(2N+1)}{6} - \left(\frac{N(N+1)}{2}\right)^2} \\
- \frac{\frac{N(N+1)}{2} \left(\frac{N(N+1)}{2}t_0 + \frac{N(N+1)(2N+1)}{6}T_p + O(N)\right)}{N \frac{N(N+1)(2N+1)}{6} - \left(\frac{N(N+1)}{2}\right)^2},
\end{multline}
where the sums involving sinusoid terms have been left in order of magnitude notation for simplicity.  Canceling like terms and retaining only the highest order terms in $N$ in the numerator and denominator, gives
\begin{align}
\lim_{N \to \infty} \overline{t}_0 &= \frac{\frac{N^4}{3} t_0 - \frac{N^4}{4} t_0}{N \frac{N^3}{3} - \left(\frac{N(N+1)}{2}\right)^2}, \\
&=t_0.
\end{align}
Similarly, we can now also evaluate equation~\eqref{app_t0Tp_Tpdef_sig0} describing $\overline{T}_p$.  
\begin{multline}
\overline{T}_p = \frac{N \left(\frac{N(N+1)}{2}t_0 + \frac{N(N+1)(2N+1)}{6} T_p + O(N)\right) }{N \frac{N(N+1)(2N+1)}{6} - \left(\frac{N(N+1)}{2}\right)^2} \\
- \frac{\frac{N(N+1)}{2}\left(N t_0 + \frac{N(N+1)}{2}T_p + O(1)\right)}{N \frac{N(N+1)(2N+1)}{6} - \left(\frac{N(N+1)}{2}\right)^2},
\end{multline}
where again, the sums involving sinusoid terms have been left in order of magnitude notation for simplicity.  Canceling like terms and retaining only the highest order terms in $N$ in the numerator and denominator, gives
\begin{align}
\lim_{N \to \infty} \overline{T}_p &= \frac{\frac{N^4}{3} T_p - \frac{N^4}{4}T_p}{\frac{N^4}{3} - \frac{N^4}{4}}, \\
& = T_p.
\end{align}
Consequently, for the case where a sinusoidal perturbation would be detectable, that is $\omega \ne 2k\pi$ where $k$ is an integer, and the standard deviation $\sigma_\epsilon$ is zero, $\overline{t}_0$ converges to $t_0$ and $\overline{T}_p$ converges to $T_p$ as $N$ tends to infinity.

\section{Limiting behaviour of the error in $\overline{t}_0$ and $\overline{T}_p$ as $N \to \infty$}

From equations~\eqref{app_t0Tp_t0def} and \eqref{app_t0Tp_Tpdef} we have that $\epsilon_{t_0}$ and $\epsilon_{T_p}$, the error in $\overline{t}_0$ and $\overline{T}_p$ due to timing noise is given by 
\begin{equation}
\epsilon_{t_0} =  \frac{\sum_{j = 1}^N (\sum_{i = 1}^N i^2 - j\sum_{i = 1}^N i) \epsilon_j}{N \sum_{j = 1}^N j^2 - \left(\sum_{j = 1}^N j\right)^2},\label{app_t0Tp_epst0def}
\end{equation}
and
\begin{equation}
\epsilon_{T_p} =  \frac{ \sum_{j = 1}^N (Nj - \sum_{i = 1}^N i) \epsilon_j}{N \sum_{j = 1}^N j^2 - \left(\sum_{j = 1}^N j\right)^2}.\label{app_t0Tp_epsTpdef}
\end{equation}
Using equation~\eqref{TraM-Noi-White-Prop-sumstd} from section~\ref{Trans_TTV_Noise_White}, these expressions can be transformed to give the standard deviations $\sigma_{t_0}$ and $\sigma_{T_p}$, of $\epsilon_{t_0}$ and $\epsilon_{T_p}$,
\begin{equation}
\sigma_{t_0} =  \frac{\sqrt{\sum_{j = 1}^N (\sum_{i = 1}^N i^2 - j\sum_{i = 1}^N i)^2 \sigma_\epsilon^2}}{N \sum_{j = 1}^N j^2 - \left(\sum_{j = 1}^N j\right)^2},\label{app_t0Tp_sigmat0def}
\end{equation}
and
\begin{equation}
\sigma_{T_p} =  \frac{\sqrt{ \sum_{j = 1}^N (Nj - \sum_{i = 1}^N i)^2 \sigma_\epsilon^2}}{N \sum_{j = 1}^N j^2 - \left(\sum_{j = 1}^N j\right)^2}.\label{app_t0Tp_sigmaTpdef}
\end{equation}
where we recall that $\sigma_\epsilon$ is the standard deviation of $\epsilon_j$.

Consider equation~\eqref{app_t0Tp_sigmat0def}, the equation for $\sigma_{t_0}$.  We have that
\begin{align}
\sigma_{t_0} &=  \frac{\sqrt{\sum_{j = 1}^N (\sum_{i = 1}^N i^2 - j\sum_{i = 1}^N i)^2 \sigma_\epsilon^2}}{N \sum_{j = 1}^N j^2 - \left(\sum_{j = 1}^N j\right)^2},\\
\sigma_{t_0} &= \sigma_\epsilon \frac{\sqrt{\sum_{j = 1}^N \left(\frac{N(N+1)(2N+1)}{6} - j\frac{N(N+1)}{2}\right)^2}}{N \sum_{j = 1}^N j^2 - \left(\sum_{j = 1}^N j\right)^2},\\
\sigma_{t_0} &= \sigma_\epsilon \frac{\sqrt{\frac{N^3(N+1)^2(2N+1)^2}{36} - \frac{N^3(N+1)^3(2N+1)}{12} + \frac{N^3(N+1)^3(2N+1)}{24} }}{\frac{N^2(N+1)(2N+1)}{6} - \frac{N^2(N+1)^2}{4}}.
\end{align}
Consequently,
\begin{align}
\lim_{N \to \infty} \sigma_{t_0} &= \sigma_\epsilon \frac{\sqrt{\frac{N^7}{9} - \frac{N^7}{6} + \frac{N^7}{12} }}{\frac{N^4}{3} - \frac{N^4}{4}},\\
&= \sigma_\epsilon \frac{\sqrt{\frac{N^7}{36}  }}{\frac{N^4}{12} },\\
&= \sigma_\epsilon \frac{2}{\sqrt{N}},\\
&= 0.
\end{align}

Similarly, for the case of $\sigma_{T_p}$ we have that
\begin{align}
\sigma_{T_p} &=  \frac{\sqrt{ \sum_{j = 1}^N (Nj - \sum_{i = 1}^N i)^2 \sigma_\epsilon^2}}{N \sum_{j = 1}^N j^2 - \left(\sum_{j = 1}^N j\right)^2},\\
 &=  \sigma_\epsilon \frac{\sqrt{ \sum_{j = 1}^N \left(Nj - \frac{N(N+1)(2N + 1)}{6} \right)^2 }}{N \sum_{j = 1}^N j^2 - \left(\sum_{j = 1}^N j\right)^2},\\
 &=  \sigma_\epsilon \frac{\sqrt{ \frac{N^3(N+1)(2N + 1)}{6} - \frac{N^3(N+1)^2(2N + 1)}{6} + \frac{N^3(N+1)^2(2N + 1)^2}{36}  }}{\frac{N^2(N+1)(2N+1)}{6} - \frac{N^2(N+1)^2}{4}},
\end{align}
and thus,
\begin{align}
\lim_{N \to \infty} \sigma_{T_p} &= \sigma_\epsilon \frac{\sqrt{ \frac{N^5}{3} - \frac{N^6}{3} + \frac{N^7}{9}  }}{\frac{N^4}{3} - \frac{N^4}{4}},\\
&= \sigma_\epsilon \frac{\sqrt{\frac{N^7}{9}  }}{\frac{N^4}{12} },\\
&= \sigma_\epsilon \frac{4}{\sqrt{N}}, \\
&= 0.
\end{align}
Consequently, the errors in $\overline{t}_0$ and $\overline{T}_p$ tend to 0 as $N$ tends to infinity.

\section{Summary of behaviour of $\overline{t}_0$ and $\overline{T}_p$ as $N \to \infty$}

The behaviour of the fitting parameters $\overline{t}_0$ and $\overline{T}_p$ as $N$ tended to infinity was considered using a two-pronged approach.  First, the behaviour of $\overline{t}_0$ and $\overline{T}_p$ for the case where $\sigma_\epsilon = 0$ was investigated.  For this case it was found that $\overline{t}_0$ and $\overline{T}_p$ tended to $t_0$ and $T_p$.  Then the errors on $\overline{t}_0$ and $\overline{T}_p$ were considered for the case of non-zero $\sigma_\epsilon$, and it was found that the size of these errors approach 0 as $N$ tends to infinity.  Thus, as the average values of $\overline{t}_0$ and $\overline{T}_p$ (equations~\eqref{app_t0Tp_t0def_sig0} and \eqref{app_t0Tp_Tpdef_sig0}) do not depend on $\epsilon_j$, this implies that $\overline{t}_0$ and $\overline{T}_p$ tend to $t_0$ and $T_p$ as $N$ tends to infinity.

\chapter{Method for transforming fitting parameters derived from a general $\tau$ model to a specific $\tau$ model}\label{App_transformFitParams}

In chapter~\ref{Trans_Thresholds}, a number of Monte Carlo simulations are done for the case of small $N$ in order to derive detection thresholds.  Practically this involves simulating a large number of realisations of $\tau$, performing a linear and a non-linear least-squares fit to each of them and determining the difference in the sum of the squares of the residuals of the two fits.  As a result of the sheer number of fits required to make one plot,\footnote{Each of the plots in figures~\ref{MCThresholdsAligned}, \ref{MCThresholdsInclined}, \ref{MCThresholdsEccentricPeri} and \ref{MCThresholdsEccentricApo} were constructed using 300 different values of semi major axis and 25 different values of $A/\sigma_\epsilon$.  In addition 51 models were used to estimate the percentage of systems which would be detectable at the level of 99.7\% for each point.  This would correspond to running $300\times25\times51 = 382500$ models for each of the these 48 plots as opposed to the $25\times25\times51 = 31875$ models run in total using this method.} and the artificially large value of the condition number of the covariance matrix (and its inverse) when realistic values for the coefficients are used,\footnote{As $T_p \gg A$, some columns of the design matrix are many orders of magnitude larger than others.  This makes it numerically difficult to perform the non-linear least squares fitting.  In particular, for the case of the MATLAB function \texttt{nlinfit}, the computation time for a model with $T_p \gg A$, compared to a more general model with $T_p \approx A$ was approximately factor of 10 larger as a result of the additional function calls.} it would be useful to perform these fits on more general models, and then transform the derived fitting parameters into those corresponding to the models of interest.  In particular, using such a method, would mean that fewer models are run as one general model can describe many simpler models, and each model takes a shorter time to fit as a result of the smaller condition number.  Such a method exists.  In particular, it can be used to transform the fitting parameters for model with any value of $t_0$ and $T_p$ to that for any other value of $t_0$ and $T_p$, given that $A/\sigma_\epsilon$, $\omega$ and $\phi$ are the same.  Then, using these expressions, the difference in the sum of the residuals squared for the two models can be easily calculated.

To begin, consider a sequence of $\tau$ values described by
\begin{equation}
\tau_j = t_0 + T_p j + A\cos(\omega j + \phi) + \epsilon_j,
\label{App_FitMethod_Intro_Par_taudef}
\end{equation}
where $\epsilon_j$ is a normally distributed variable with mean 0 and standard deviation $\sigma_\epsilon$.  Performing the least-squares linear (no moon) and least-squares non-linear (moon) fits detailed in section~\ref{Trans_Thresholds_Method_GenLikelihood} gives the fitting parameters $\overline{t}_0$ and $\overline{T}_p$, and $\hat{t}_0$, $\hat{T}_p$, $\hat{A}$, $\hat{\omega}$ and $\hat{\phi}$ respectively.  In particular, recalling that at a minimum the derivative equals zero, these parameters are defined by
\begin{equation}
0 = \frac{\partial}{\partial\overline{x}} \sum_{j = 1}^N \left(t_0 + T_p j + A\cos(\omega j + \phi) + \epsilon_j - (\overline{t}_0 + \overline{T}_p j)\right)^2,
\label{App_FitMethod_Intro_Par_lincoeffdef}
\end{equation}
and
\begin{multline}
0 = \frac{\partial}{\partial\hat{x}} \sum_{j = 1}^N \left(t_0 + T_p j + A\cos(\omega j + \phi) + \epsilon_j \right. \\
\left. - (\hat{t}_0 + \hat{T}_p j + \hat{A}\cos(\hat{\omega}j + \hat{\phi}))\right)^2,\label{App_FitMethod_Intro_Par_nonlincoeffdef}
\end{multline}
where $\overline{x}$ could be either one of $\overline{t}_0$ or $\overline{T}_p$ and where $\hat{x}$ could be any of $\hat{t}_0$, $\hat{T}_p$, $\hat{A}$, $\hat{\omega}$ or $\hat{\phi}$.  As will be shown in this appendix, these model parameters can be directly derived from a more general model given by
\begin{equation}
\tau_j = a + b j + \frac{A}{\sigma_\epsilon}\cos(\omega j + \phi) + \frac{\epsilon_j}{\sigma_\epsilon}.
\label{App_FitMethod_Intro_Gen_taudef}
\end{equation}
where we note that the $\epsilon_j$ in equation~\eqref{App_FitMethod_Intro_Par_taudef} and the $\epsilon_j$ in this equation are the same.  By analogy, fitting parameters for this model can also be defined by equations similar to equations~\eqref{App_FitMethod_Intro_Par_lincoeffdef} and \eqref{App_FitMethod_Intro_Par_nonlincoeffdef}.  In particular
\begin{equation}
0 = \frac{\partial}{\partial\overline{x}} \sum_{j = 1}^N \left(a + b j + \frac{A}{\sigma_\epsilon}\cos(\omega j + \phi) + \frac{\epsilon_j}{\sigma_\epsilon} - (\overline{a} + \overline{b} j)\right)^2,\label{App_FitMethod_Intro_Gen_lincoeffdef}
\end{equation}
and
\begin{multline}
0 = \frac{\partial}{\partial\hat{x}} \sum_{j = 1}^N \left(a + b j + \frac{A}{\sigma_\epsilon}\cos(\omega j + \phi) + \frac{\epsilon_j}{\sigma_\epsilon} \right. \\
\left. - (\hat{a} + \hat{b} j + \frac{\hat{A}}{\sigma_\epsilon} \cos(\hat{\omega}j + \hat{\phi}))\right)^2,\label{App_FitMethod_Intro_Gen_nonlincoeffdef}
\end{multline}
where again we note that $\overline{x}$ could be either one of $\overline{a}$ or $\overline{b}$ and where $\hat{x}$ could be any of $\hat{a}$, $\hat{b}$, $\hat{A}$, $\hat{\omega}$ and $\hat{\phi}$. 

By transforming equation~\eqref{App_FitMethod_Intro_Gen_lincoeffdef} into an equation equivalent to equation~\eqref{App_FitMethod_Intro_Par_lincoeffdef}, and similarly, transforming equation~\eqref{App_FitMethod_Intro_Gen_nonlincoeffdef} into an equation equivalent to equation~\eqref{App_FitMethod_Intro_Par_nonlincoeffdef}, we will derive equations for transforming the fitting coefficients.  In addition it will also be shown that the difference of the sums of residuals squared can also be easily transformed.

\section{Expressions for $\overline{t}_0$ and $\overline{T}_p$ in terms of $\overline{a}$ and $\overline{b}$}

Consider equation~\eqref{App_FitMethod_Intro_Gen_lincoeffdef},
\begin{equation*}
0 = \frac{\partial}{\partial\overline{x}} \sum_{j = 1}^N \left(a + b j + \frac{A}{\sigma_\epsilon}\cos(\omega j + \phi) + \frac{\epsilon_j}{\sigma_\epsilon} - (\overline{a} + \overline{b} j)\right)^2.
\end{equation*}
Multiplying this equation by $\sigma_\epsilon$ gives
\begin{equation}
0 = \frac{\partial}{\partial\overline{x}} \sum_{j = 1}^N \left(a \sigma_\epsilon + b \sigma_\epsilon j + A \cos(\omega j + \phi) + \epsilon_j - (\overline{a} \sigma_\epsilon + \overline{b} \sigma_\epsilon j)\right)^2.
\end{equation}
Adding and taking away $t_0 + jT_p$ to each term gives
\begin{multline}
0 = \frac{\partial}{\partial\overline{x}} \sum_{j = 1}^N \left(t_0 +  T_p j + A \cos(\omega j + \phi) + \epsilon_j \right. \\
\left. - ([t_0 + (\overline{a} - a) \sigma_\epsilon] + [T_p + (\overline{b} - b) \sigma_\epsilon] j)\right)^2.
\end{multline}

This equation is equivalent to equation~\eqref{App_FitMethod_Intro_Par_lincoeffdef} in that if $\overline{t}_0$ and $\overline{T}_p$ are the solutions to equation~\eqref{App_FitMethod_Intro_Par_lincoeffdef}, then they must also be the solutions to this equation.  In other words,
\begin{align}
\overline{t}_0 &= t_0 + (\overline{a} - a) \sigma_\epsilon,\label{App_FitMethod_Lin_adef}\\
\overline{T}_p &= T_p + (\overline{b} - b) \sigma_\epsilon.\label{App_FitMethod_Lin_bdef}
\end{align}
These expressions allow the fitting coefficients defined for the general model to be transformed to a particular model.  We now consider the case of the non-linear fit.

\section{Expressions for $\hat{t}_0$ and $\hat{T}_p$ in terms of $\hat{a}$ and $\hat{b}$}

Again we begin by considering equation~\eqref{App_FitMethod_Intro_Par_nonlincoeffdef},
\begin{multline*}
0 = \frac{\partial}{\partial\hat{x}} \sum_{j = 1}^N \left(a + b j + \frac{A}{\sigma_\epsilon}\cos(\omega j + \phi) + \frac{\epsilon_j}{\sigma_\epsilon} \right. \\
\left. - (\hat{a} + \hat{b} j + \frac{\hat{A}}{\sigma_\epsilon} \cos(\hat{\omega}j + \hat{\phi}))\right)^2.
\end{multline*}
Multiplying through by $\sigma_\epsilon$ gives
\begin{multline}
0 = \frac{\partial}{\partial\hat{x}} \sum_{j = 1}^N \left(a \sigma_\epsilon + b \sigma_\epsilon j + A \cos(\omega j + \phi) + \epsilon_j \right. \\
\left. - (\hat{a} \sigma_\epsilon + \hat{b} \sigma_\epsilon j + \hat{A} \cos(\hat{\omega}j + \hat{\phi}))\right)^2.
\end{multline}
Again, adding and taking away $t_0 + jT_p$ to each term gives
\begin{multline}
0 = \frac{\partial}{\partial\hat{x}} \sum_{j = 1}^N \left(t_0 + jT_p + A \cos(\omega j + \phi) + \epsilon_j \right. \\
\left. - ([t_0 + (\hat{a} - a) \sigma_\epsilon] + [T_p + (\hat{b} - b) \sigma_\epsilon] j + \hat{A} \cos(\hat{\omega}j + \hat{\phi}))\right)^2
\end{multline}
This equation is again equivalent to equation~\eqref{App_FitMethod_Intro_Par_nonlincoeffdef} in that if $\hat{t}_0$, $\hat{T}_p$, $\hat{A}$, $\hat{\omega}$ and $\hat{\phi}$ are the solutions to equation~\eqref{App_FitMethod_Intro_Par_nonlincoeffdef}, then they must also be the solutions to this equation.  In particular, $\hat{t}_0$ should be equal to the first term in square brackets, $\hat{T}_p$ should be equal to the second term in square brackets, and $\hat{A}$, $\hat{\omega}$ and $\hat{\phi}$ should be the same in both cases.  Mathematically this means that
\begin{align}
\hat{t}_0 &= t_0 + (\hat{a} - a) \sigma_\epsilon,\label{App_FitMethod_NonLin_adef}\\
\hat{T}_p &= T_p + (\hat{b} - b) \sigma_\epsilon\label{App_FitMethod_NonLin_bdef}.
\end{align}
Now we will consider the transformation of the difference of the sum of the residuals squared, $2\log (\Lambda)$.

\section[Expression for the difference of the sum of residuals squared]{Expression for the difference of the sum of residuals squared in terms of the difference of the sum of residuals squared for the general model}

Now that we have expressions for $\overline{t}_0$, $\overline{T}_p$ $\hat{t}_0$ and $\hat{T}_p$ in terms of $\overline{a}$, $\overline{b}$ $\hat{a}$, $\hat{b}$ and $\sigma_\epsilon$ we can show that the difference in the sum of the residuals squared for the case of the general model can be transformed to that for a physically realistic model.  In other words, given a set of these differences calculated for set of general models, the set of differences which would have been observed for the equivalent (the same values of $A/\sigma_\epsilon$, $\omega$ and $\phi$) realistic models can be calculated.

To begin, consider the difference of the sum of squares of the residuals for the two models, for the realistic model
\begin{multline}
2\sigma_\epsilon^2\log \Lambda_{real} = \sum_{j = 1}^N \left(t_0 + T_p j + A\cos(\omega j + \phi) + \epsilon_j - (\overline{t}_0 + \overline{T}_p j)\right)^2 
\\
-  \sum_{j = 1}^N \left(t_0 + T_p j + A\cos(\omega j + \phi) + \epsilon_j \right. \\
\left. - (\hat{t}_0 + \hat{T}_p j +\hat{A}\cos(\hat{\omega}j + \hat{\phi}))\right)^2,\label{App_FitMethod_Lambda_real}
\end{multline}
and the general model
\begin{multline}
2\log \Lambda_{gen} = \sum_{j = 1}^N \left(a + b j + \frac{A}{\sigma_\epsilon}\cos(\omega j + \phi) + \frac{\epsilon_j}{\sigma_\epsilon} - (\overline{a} + \overline{b} j)\right)^2 
\\
-  \sum_{j = 1}^N \Bigl(a + b j + \frac{A}{\sigma_\epsilon}\cos(\omega j + \phi) + \frac{\epsilon_j}{\sigma_\epsilon} \\
\left. - \left(\hat{a} + \hat{b} j + \frac{\hat{A}}{\sigma_\epsilon}\cos(\hat{\omega}j + \hat{\phi})\right)\right)^2.\label{App_FitMethod_Lambda_gen}
\end{multline}
Using equations~\eqref{App_FitMethod_Lin_adef}, \eqref{App_FitMethod_Lin_bdef}, \eqref{App_FitMethod_NonLin_adef} and \eqref{App_FitMethod_NonLin_bdef} to substitute for $\hat{a}$, $\hat{b}$, $\overline{a}$ and $\overline{b}$, equation~\eqref{App_FitMethod_Lambda_real} becomes
\begin{multline}
2\sigma_\epsilon^2\log \Lambda_{real} =  \sum_{j = 1}^N \left(t_0 + T_p j + A\cos(\omega j + \phi) + \epsilon_j  \right. \\
\left. - (t_0 + (\overline{a} - a)\sigma_\epsilon + (T_p + (\overline{b} - b)\sigma_\epsilon) j)\right)^2 
\\
-  \sum_{j = 1}^N \Bigl(t_0 + T_p j + A\cos(\omega j + \phi) + \epsilon_j  \\
\left. - (t_0 + (\hat{a} - a) \sigma_\epsilon + (T_p + (\hat{b} - b) \sigma_\epsilon) j +\hat{A}\cos(\hat{\omega}j + \hat{\phi}))\right)^2,
\end{multline}
which simplifies to 
\begin{multline}
2 \log \Lambda_{real} = \sum_{j = 1}^N \left(a + b j + \frac{A}{\sigma_\epsilon}\cos(\omega j + \phi) + \frac{\epsilon_j}{\sigma_\epsilon} - (\overline{a} + \overline{b} j)\right)^2 
\\
-  \sigma_\epsilon^2 \sum_{j = 1}^N \left(a + b j + \frac{A}{\sigma_\epsilon}\cos(\omega j + \phi) + \frac{\epsilon_j}{\sigma_\epsilon}  \right. \\
\left. - \left(\hat{a} + \hat{b} j +\frac{\hat{A}}{\sigma_\epsilon}\cos(\hat{\omega}j + \hat{\phi})\right)\right)^2.\label{App_FitMethod_Lambda_realeq3}
\end{multline}
Comparing equations~\eqref{App_FitMethod_Lambda_realeq3} and \eqref{App_FitMethod_Lambda_gen}, it can be seen that
\begin{equation}
2\log \Lambda_{real} = 2\log \Lambda_{gen}.
\end{equation}
Consequently, a set of general models can be simulated, and have their $2\log \Lambda$ values calculated and recorded.  Then, these values can be directly applied to any equivalent realistic model.

\chapter{68.3\% and 95.4\% TTV$_p$ thresholds}\label{App_ExtraThresholds}

\begin{figure}
     \centering
     \subfigure[$M_p$=$10 M_J$, $a_p=0.2$AU.]{
          \label{TransitThresh1s10MJ02AUcc}
          \includegraphics[width=.315\textwidth]{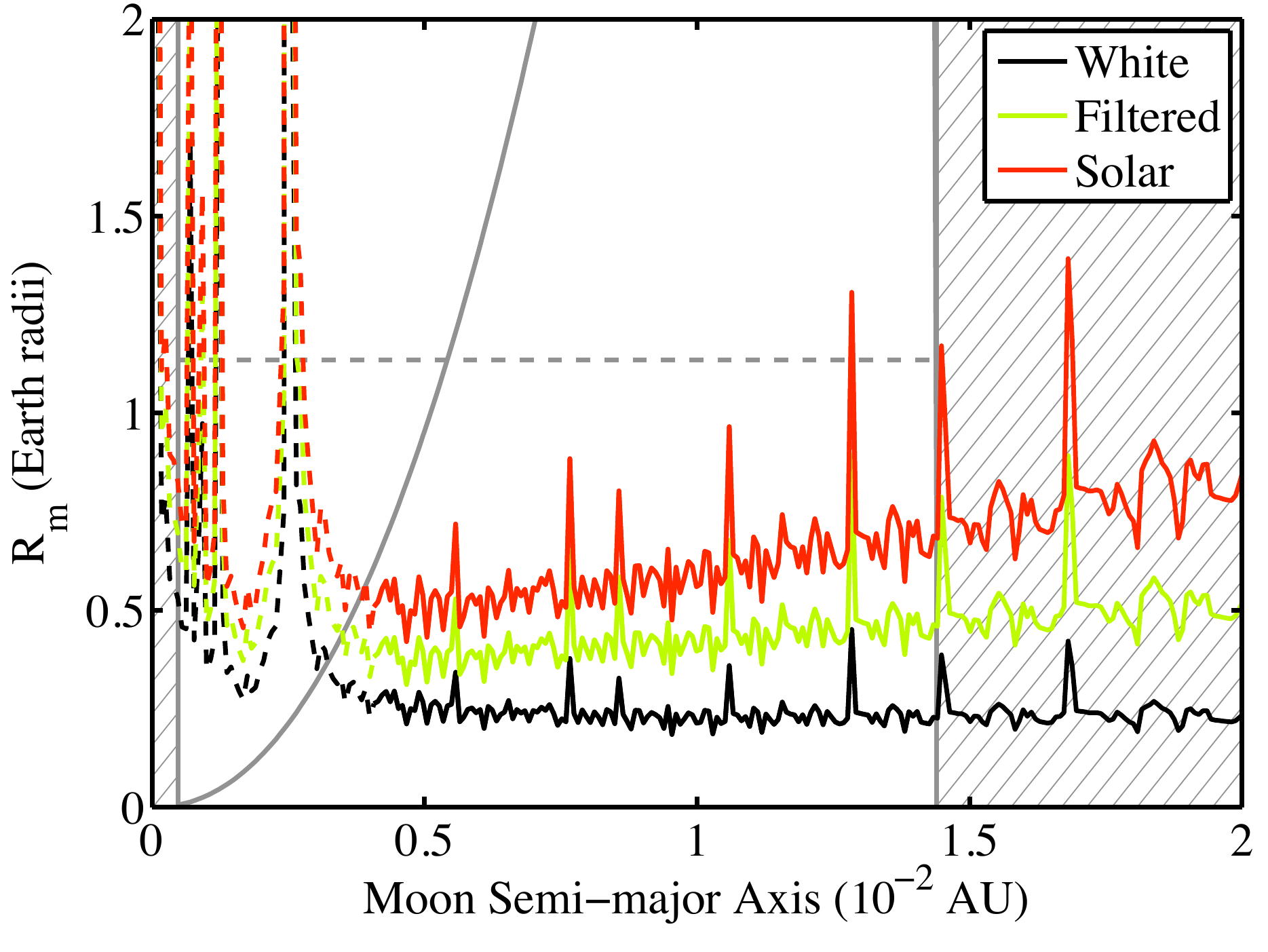}}
     \subfigure[$M_p$=$10 M_J$, $a_p=0.4$AU.]{
          \label{TransitThresh1s10MJ04AUcc}
          \includegraphics[width=.315\textwidth]{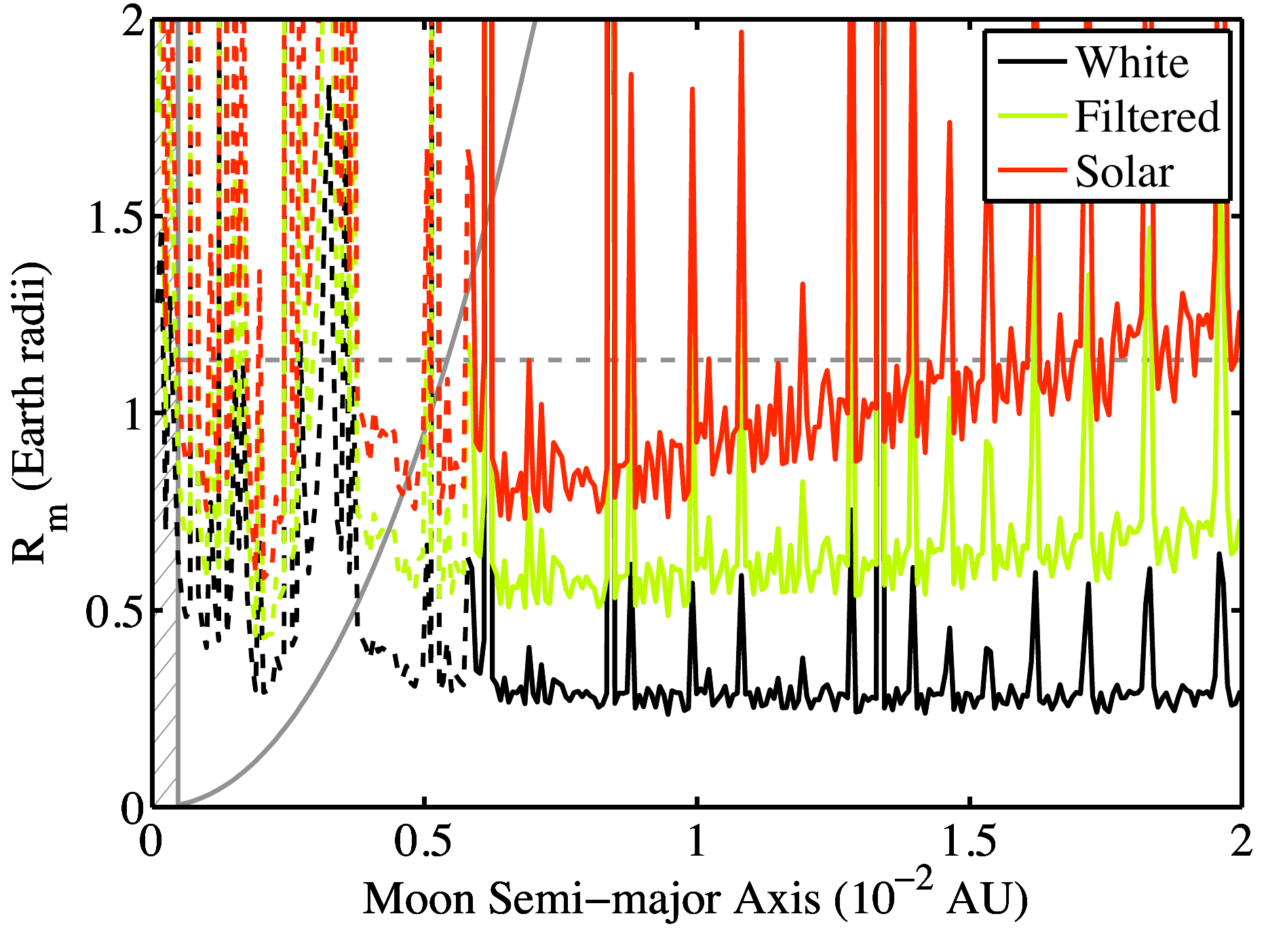}}
     \subfigure[$M_p$=$10 M_J$, $a_p=0.6$AU.]{
          \label{TransitThresh1s10MJ06AUcc}
          \includegraphics[width=.315\textwidth]{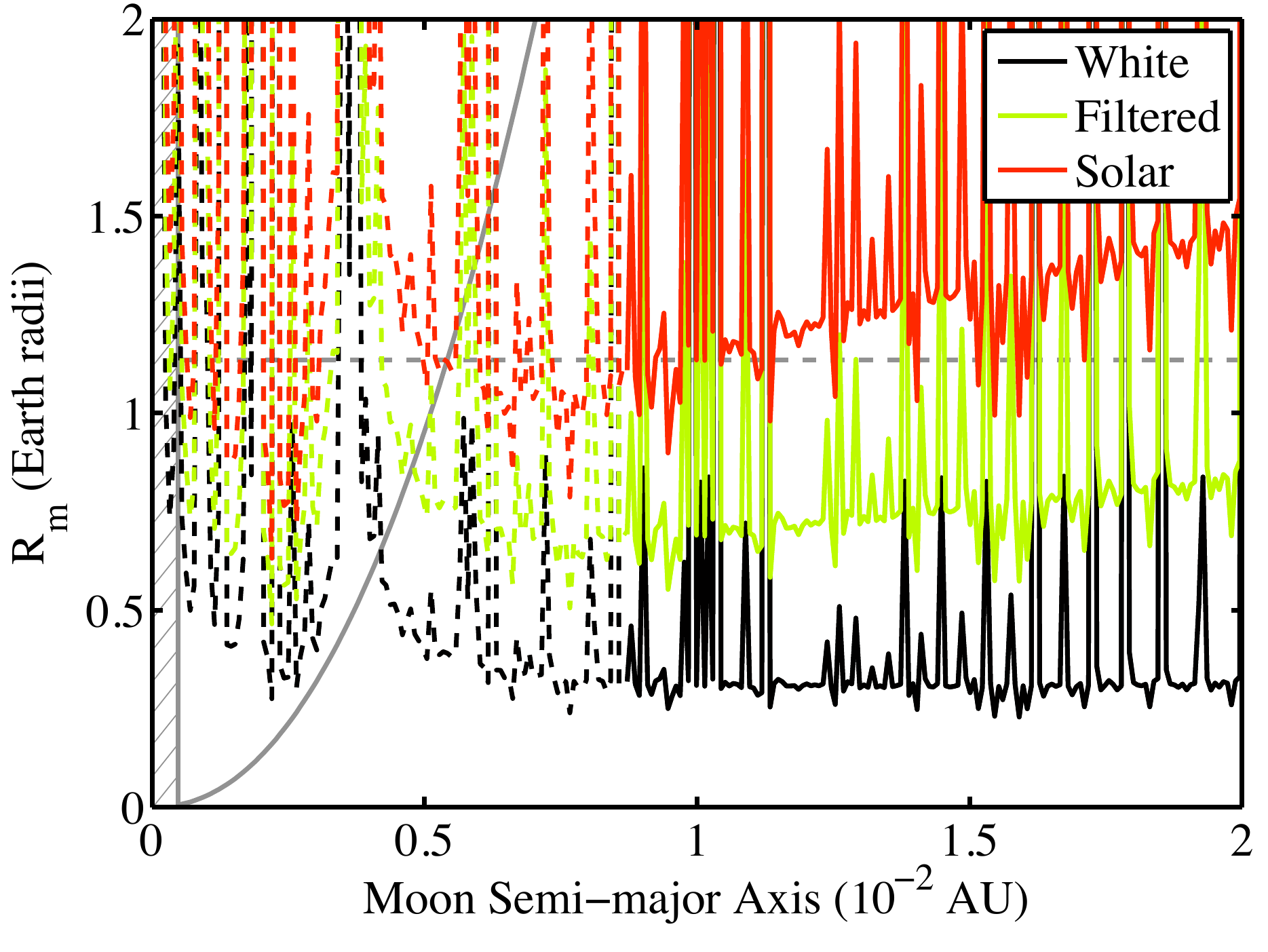}}\\ 
     \subfigure[$M_p = M_J$, $a_p=0.2$AU.]{
          \label{TransitThresh1s1MJ02AUcc}
          \includegraphics[width=.315\textwidth]{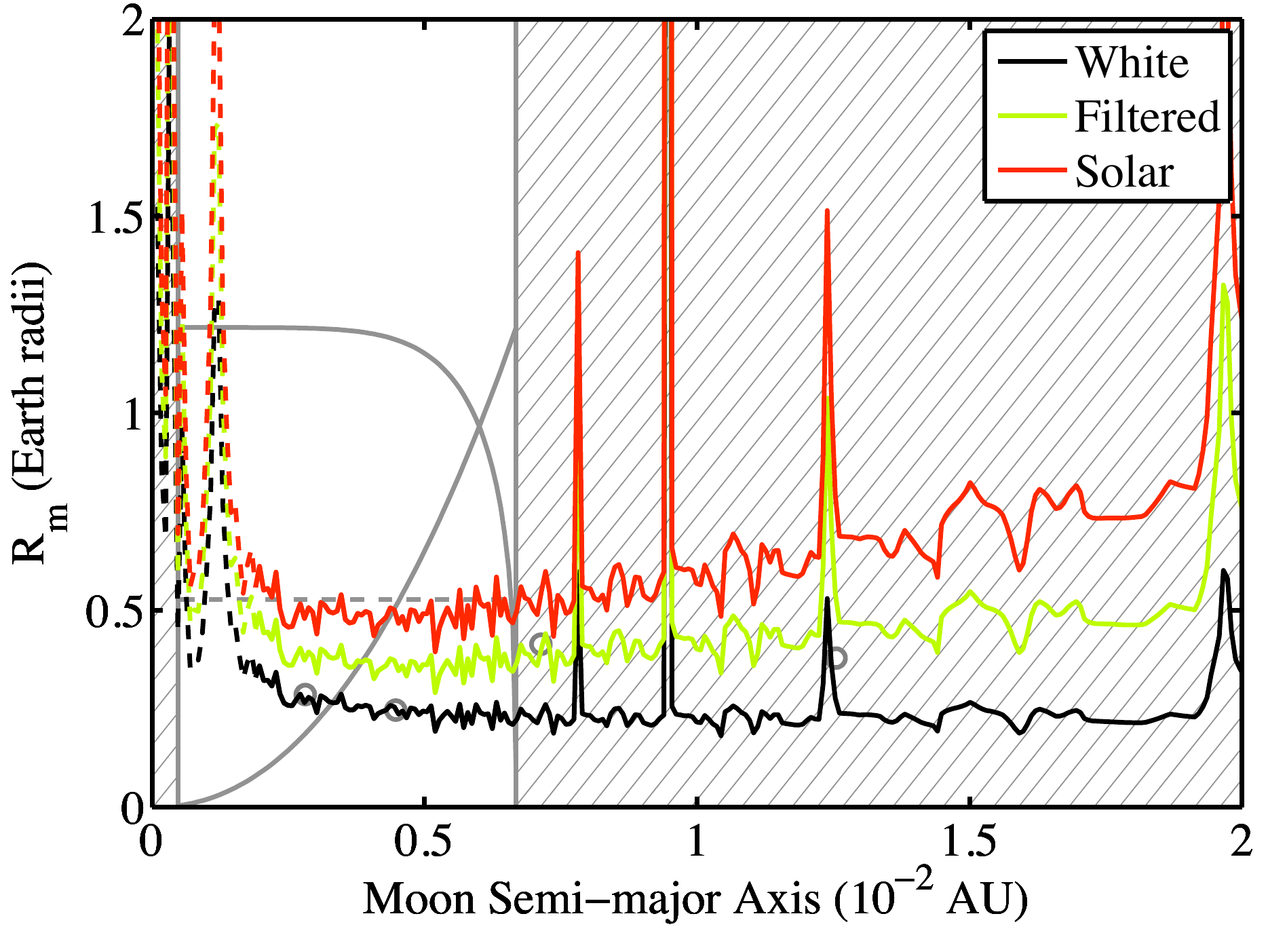}}
      \subfigure[$M_p = M_J$, $a_p=0.4$AU.]{
          \label{TransitThresh1s1MJ04AUcc}
          \includegraphics[width=.315\textwidth]{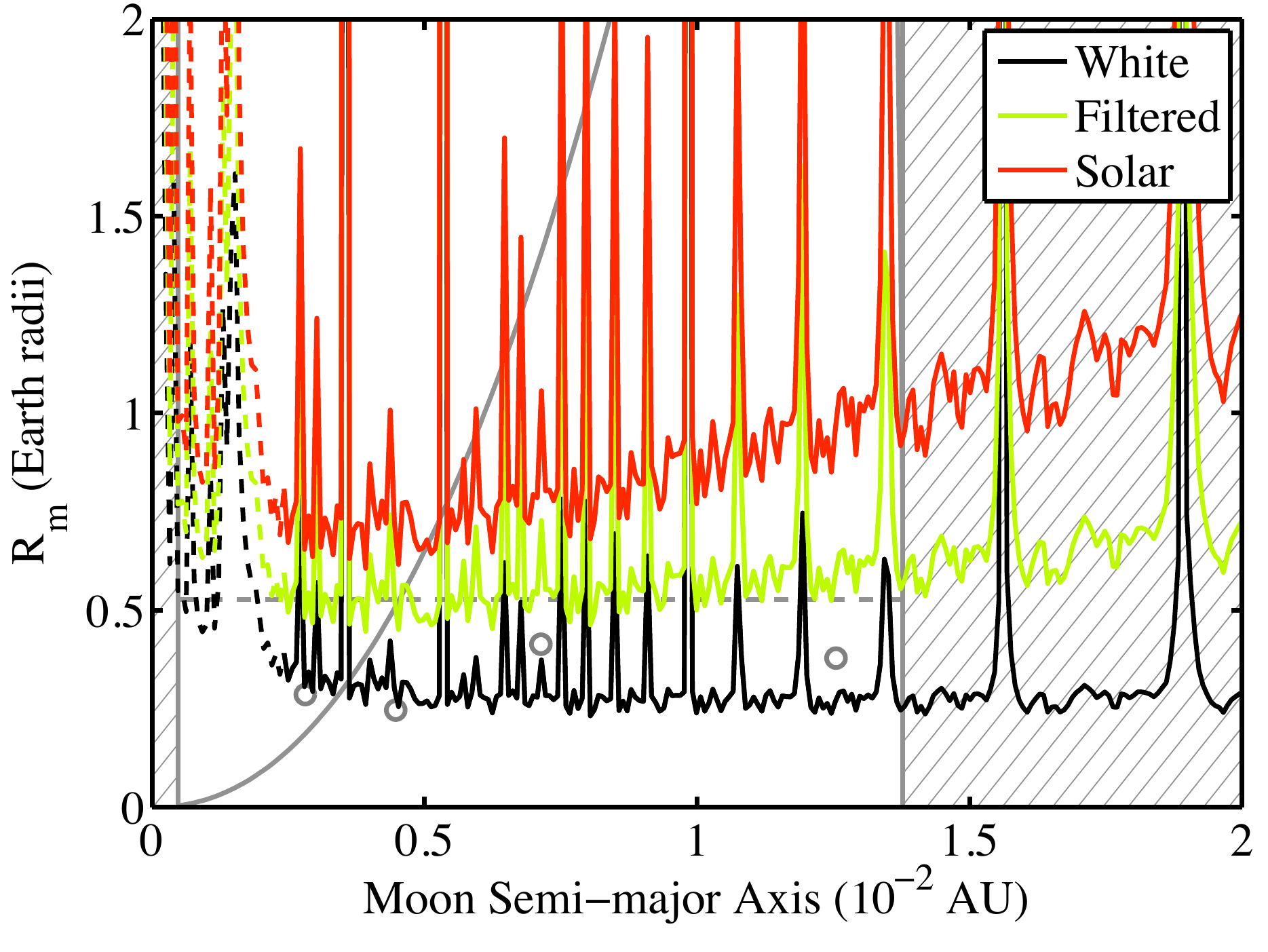}}
     \subfigure[$M_p = M_J$, $a_p=0.6$AU.]{
          \label{TransitThresh1s1MJ06AUcc}
          \includegraphics[width=.315\textwidth]{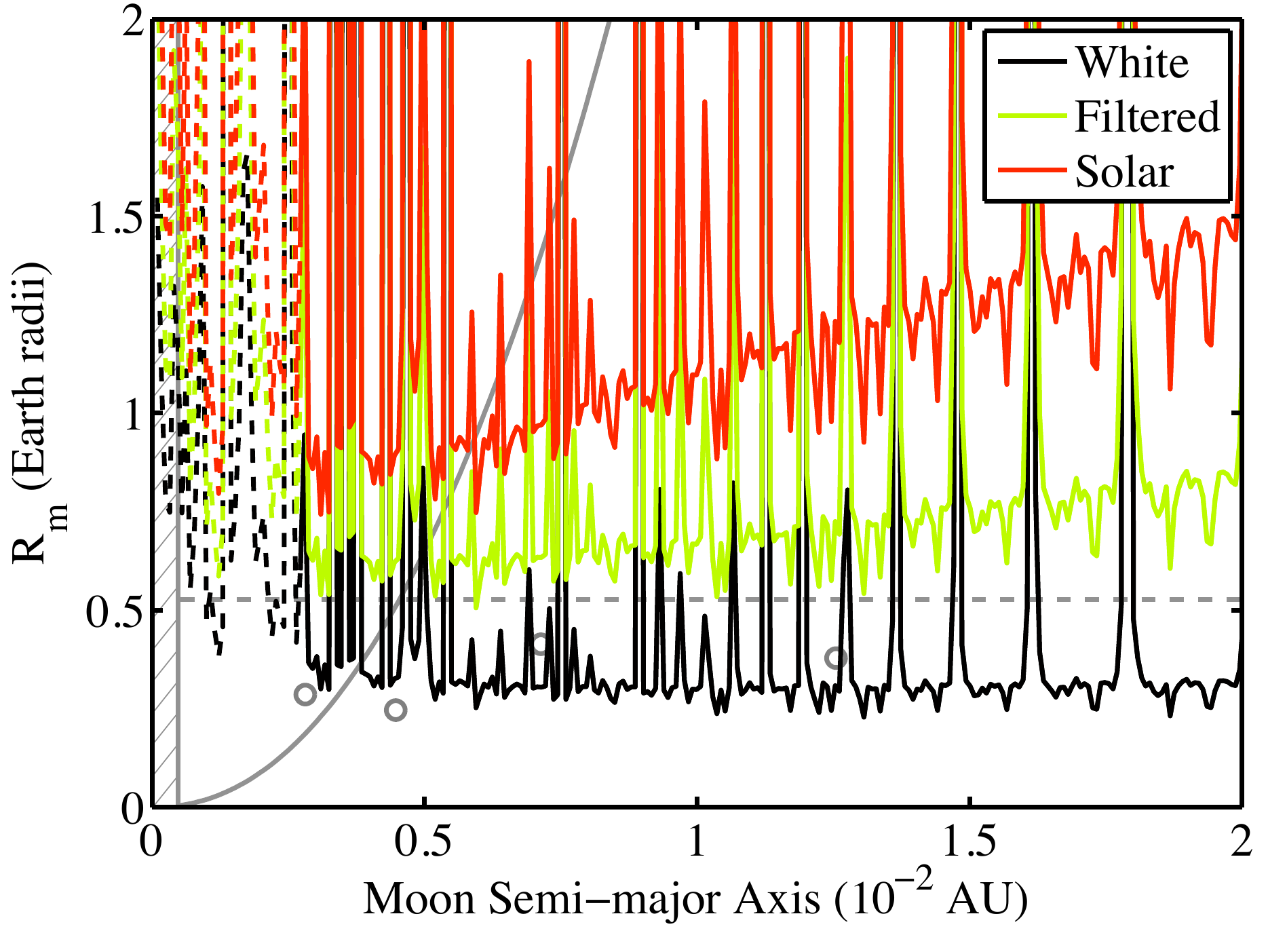}}\\ 
     \subfigure[$M_p = M_U$, $a_p=0.2$AU.]{
          \label{TransitThresh1s1MU02AUcc}
          \includegraphics[width=.315\textwidth]{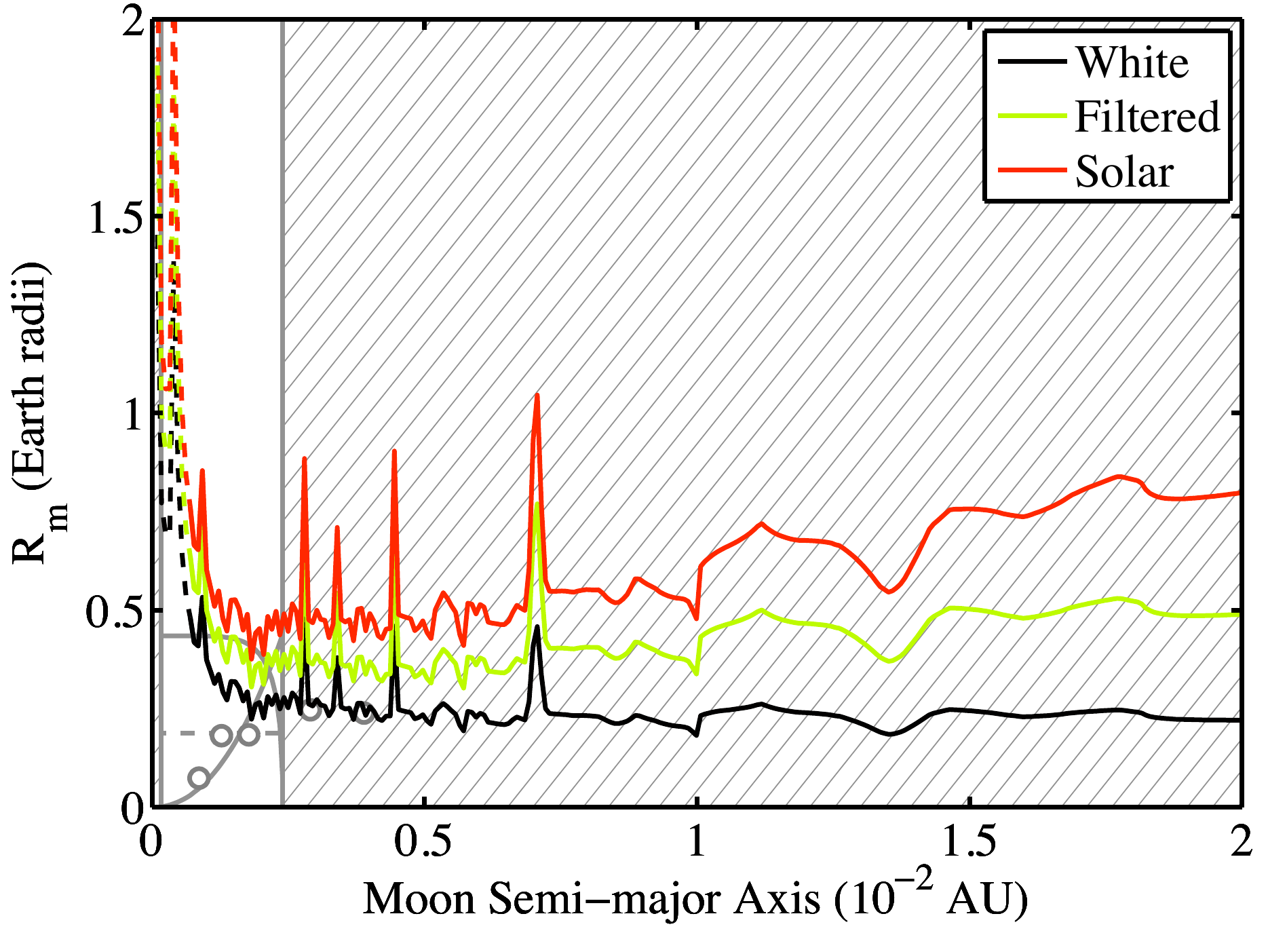}}
     \subfigure[$M_p = M_U$, $a_p=0.4$AU.]{
          \label{TransitThresh1s1MU04AUcc}
          \includegraphics[width=.315\textwidth]{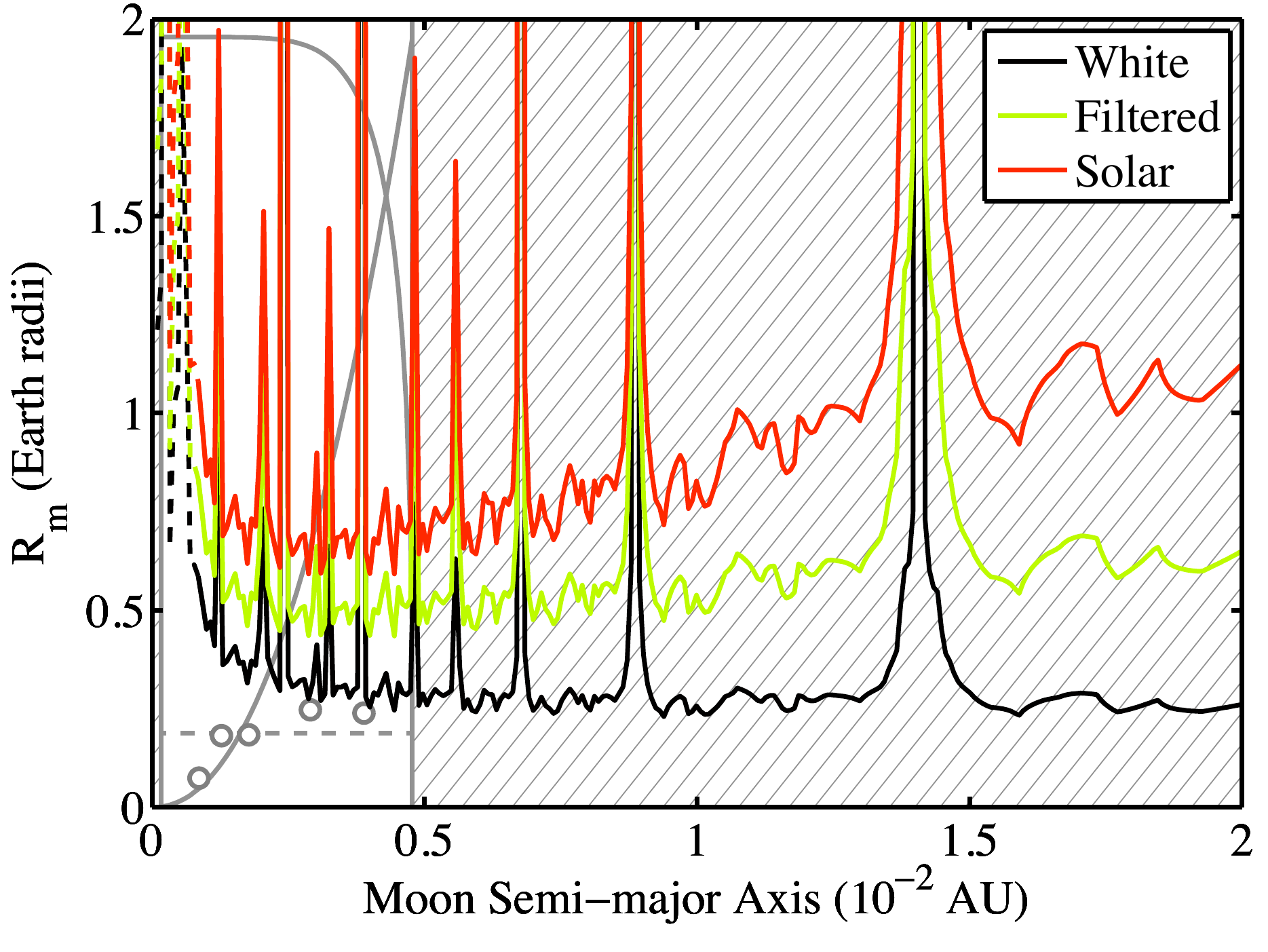}}
      \subfigure[$M_p = M_U$, $a_p=0.6$AU.]{
          \label{TransitThresh1s1MU06AUcc}
          \includegraphics[width=.315\textwidth]{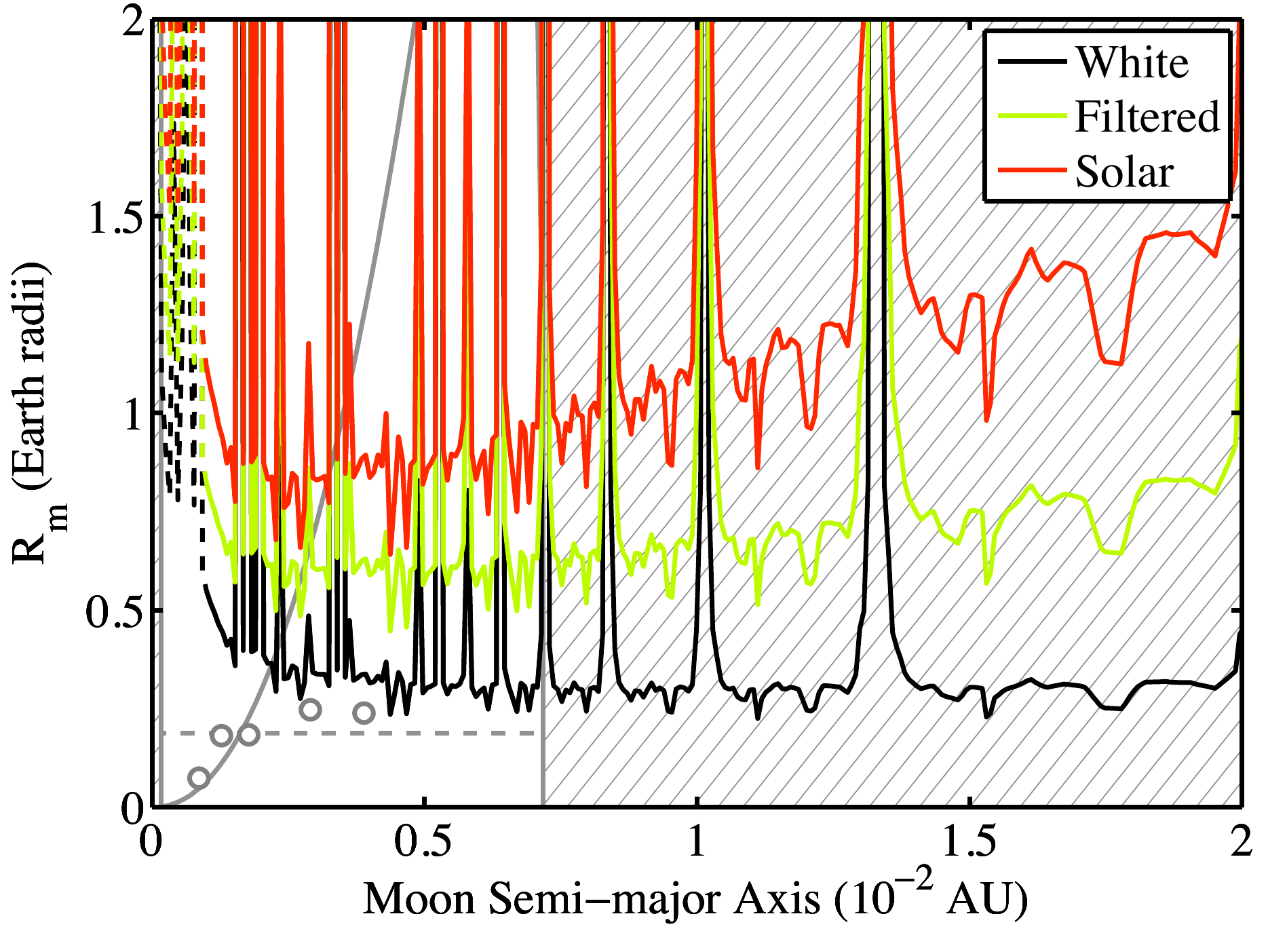}}\\ 
     \subfigure[$M_p = M_{\earth}$, $a_p=0.2$AU.]{
          \label{TransitThresh1s1ME02AUcc}
          \includegraphics[width=.315\textwidth]{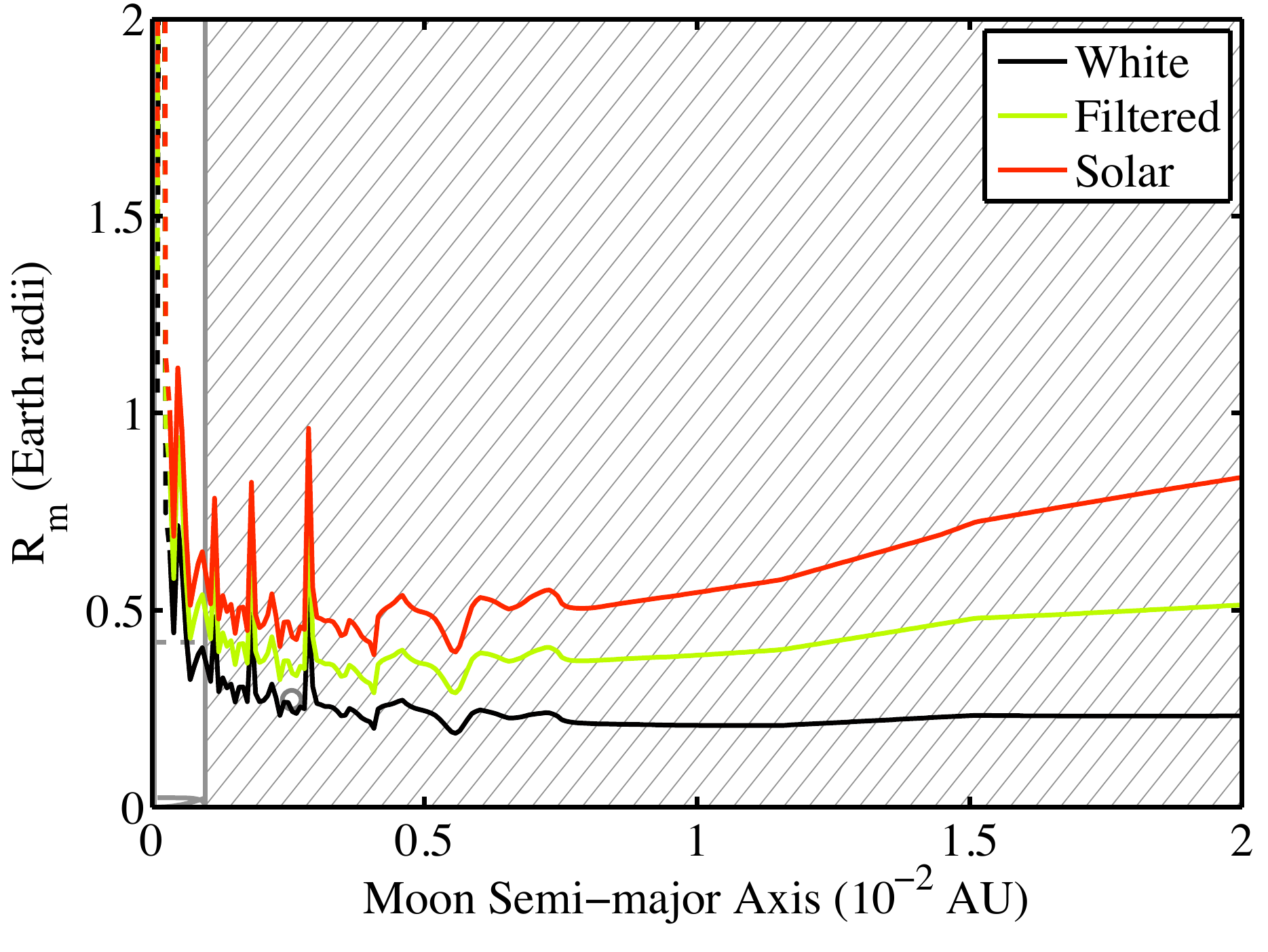}}
     \subfigure[$M_p = M_{\earth}$, $a_p=0.4$AU.]{
          \label{TransitThresh1s1ME04AUcc}
          \includegraphics[width=.315\textwidth]{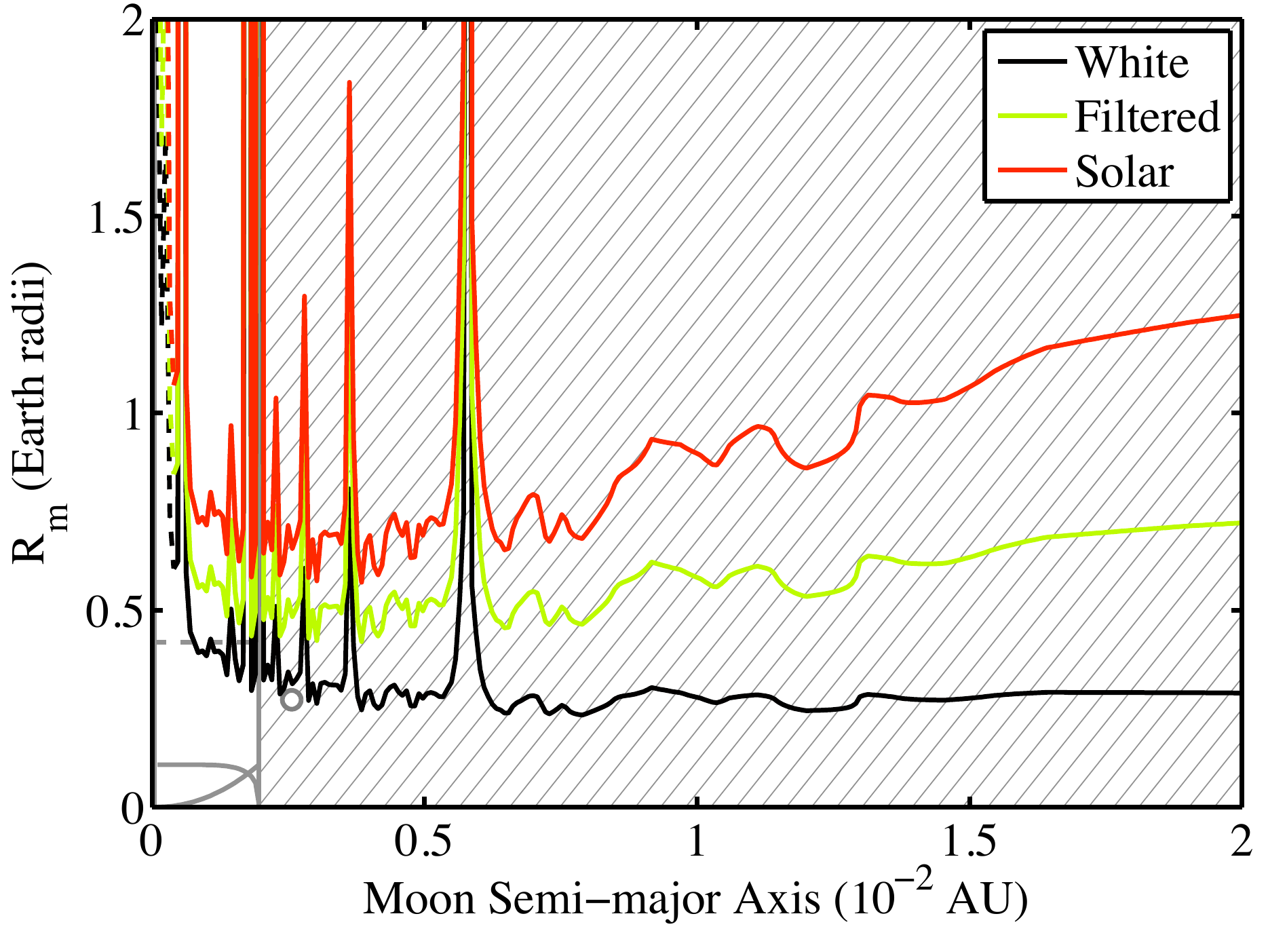}}
     \subfigure[$M_p = M_{\earth}$, $a_p=0.6$AU.]{
          \label{TransitThresh1s1ME06AUcc}
          \includegraphics[width=.315\textwidth]{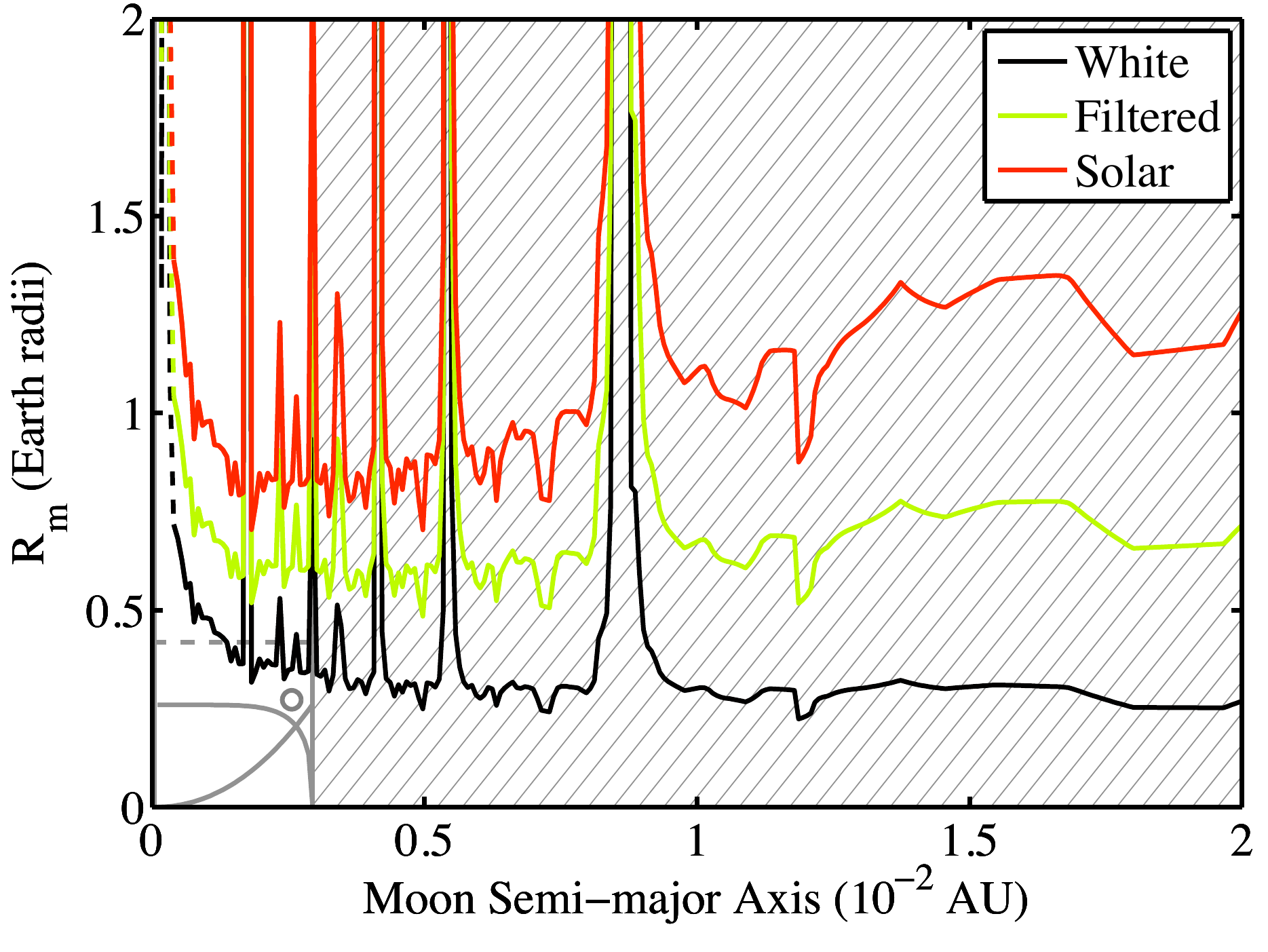}} 
     \caption{Figure of the same form as figure~\ref{MCThresholdsAligned}, but showing the 68.3\% thresholds.}
     \label{MCThresholdsAligned1S}
\end{figure}

\begin{figure}
     \centering
     \subfigure[$M_p$=$10 M_J$, $a_p=0.2$AU.]{
          \label{TransitThresh1s10MJ02AUInc}
          \includegraphics[width=.315\textwidth]{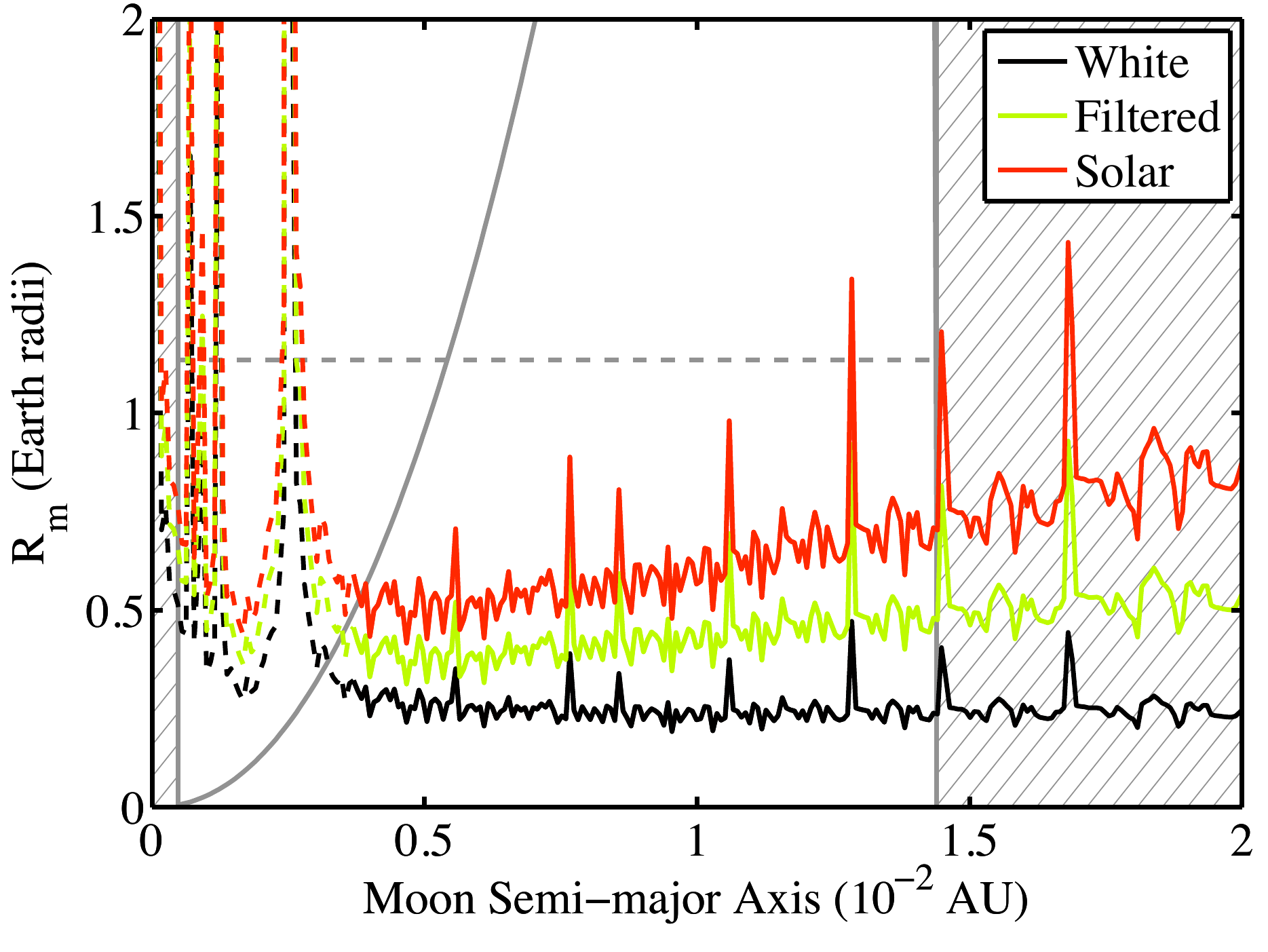}}
     \subfigure[$M_p$=$10 M_J$, $a_p=0.4$AU.]{
          \label{TransitThresh1s10MJ04AUInc}
          \includegraphics[width=.315\textwidth]{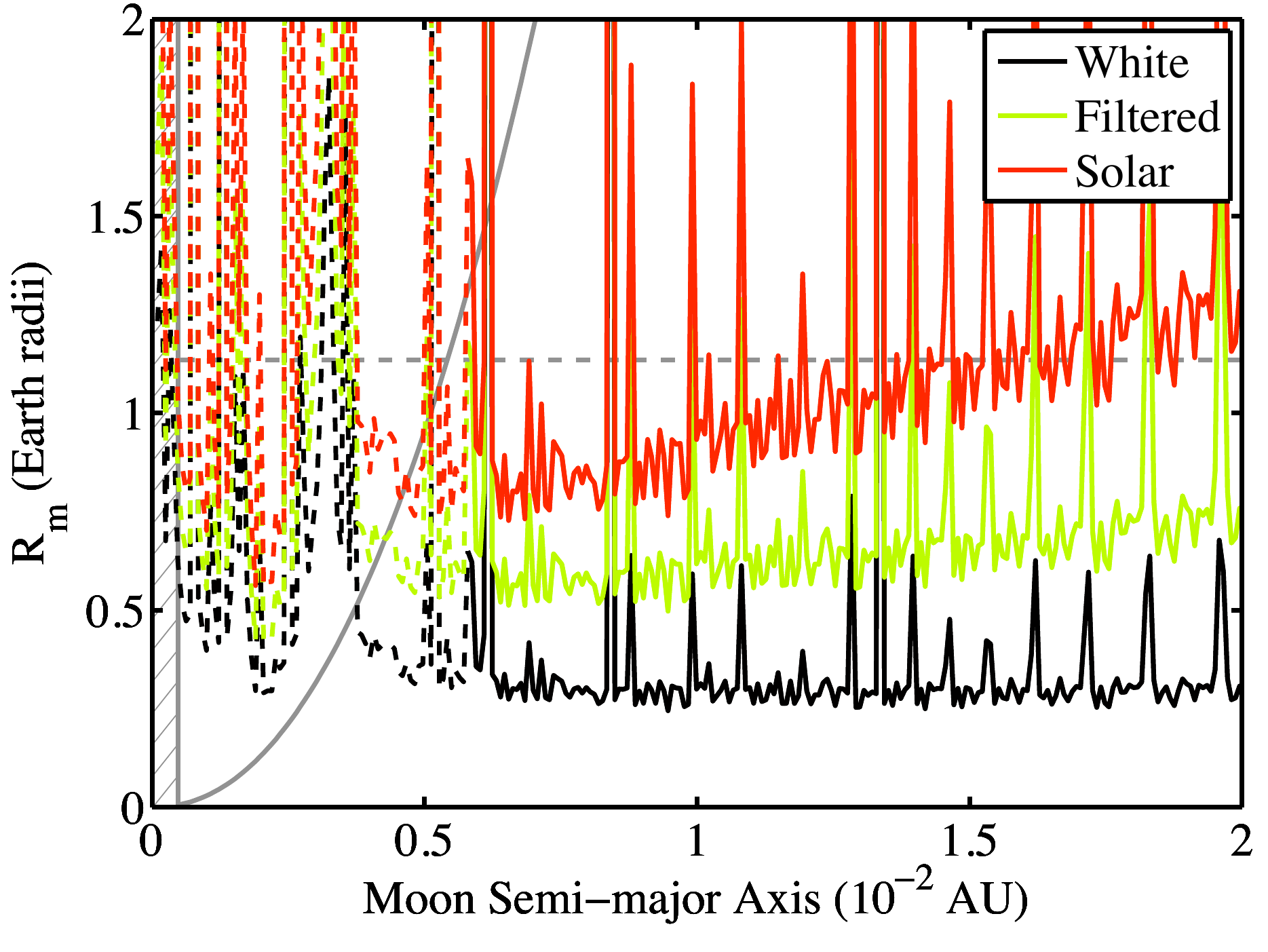}}
     \subfigure[$M_p$=$10 M_J$, $a_p=0.6$AU.]{
          \label{TransitThresh1s10MJ06AUInc}
          \includegraphics[width=.315\textwidth]{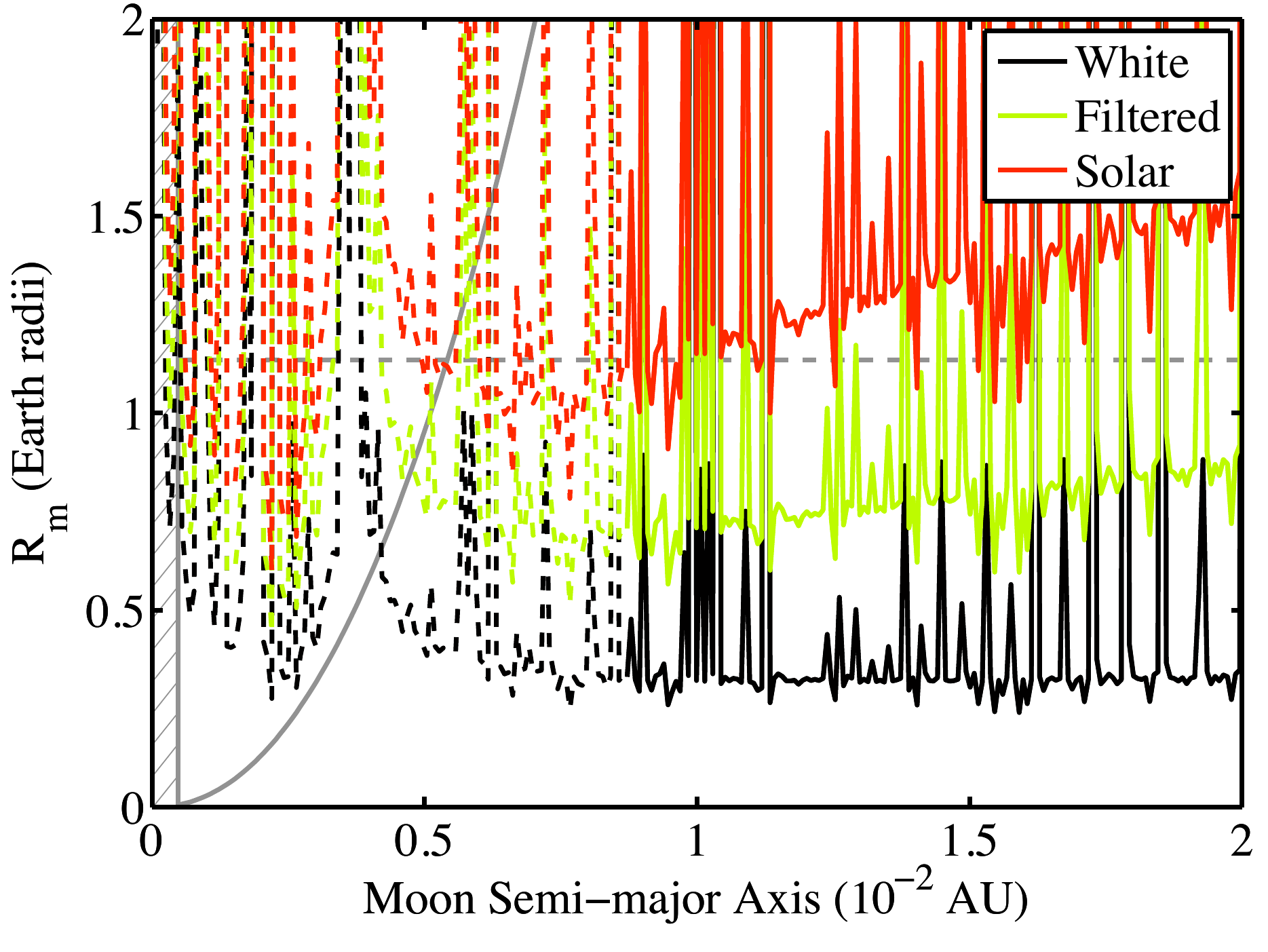}}\\ 
     \subfigure[$M_p = M_J$, $a_p=0.2$AU.]{
          \label{TransitThresh1s1MJ02AUInc}
          \includegraphics[width=.315\textwidth]{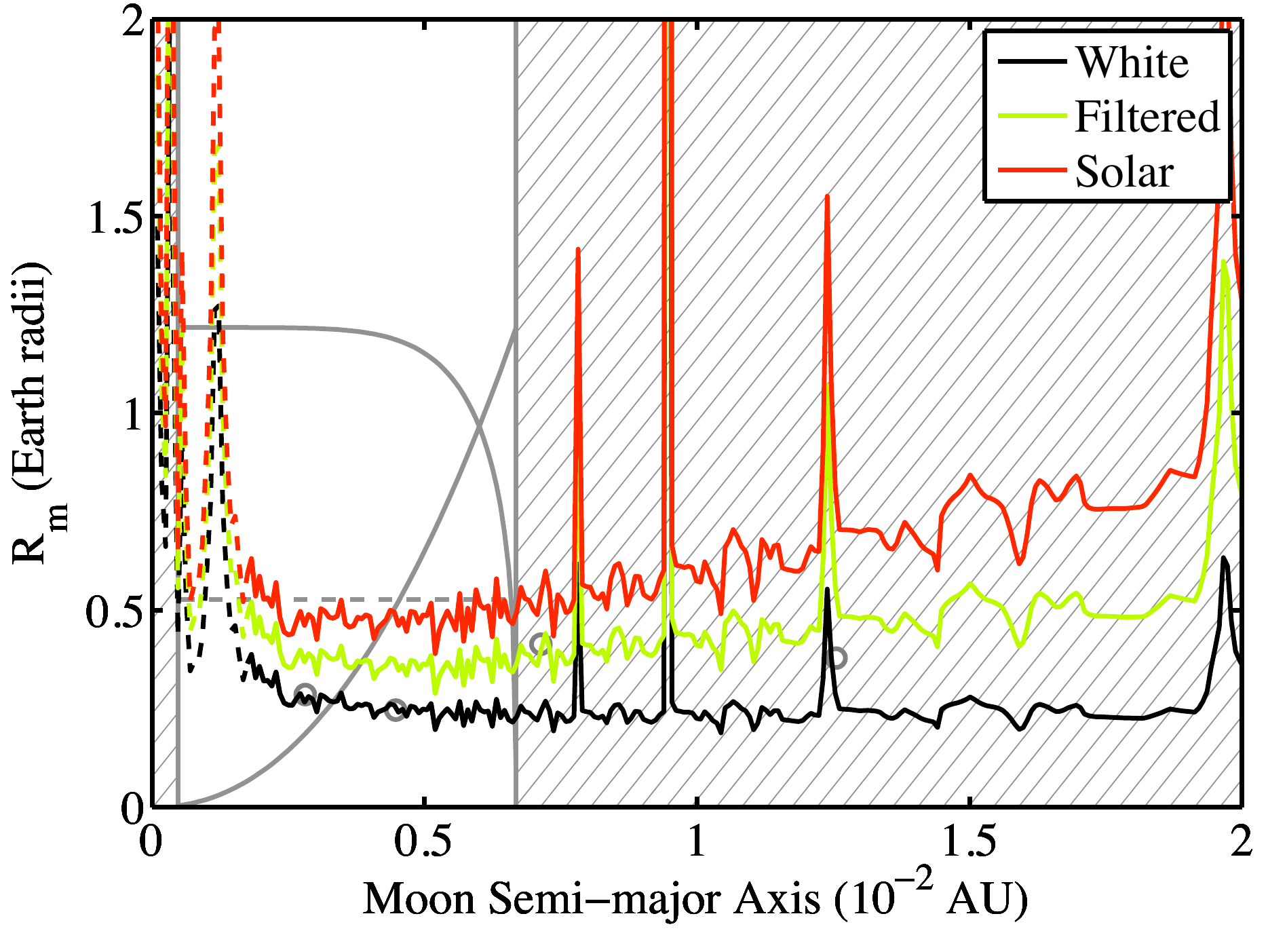}}
      \subfigure[$M_p = M_J$, $a_p=0.4$AU.]{
          \label{TransitThresh1s1MJ04AUInc}
          \includegraphics[width=.315\textwidth]{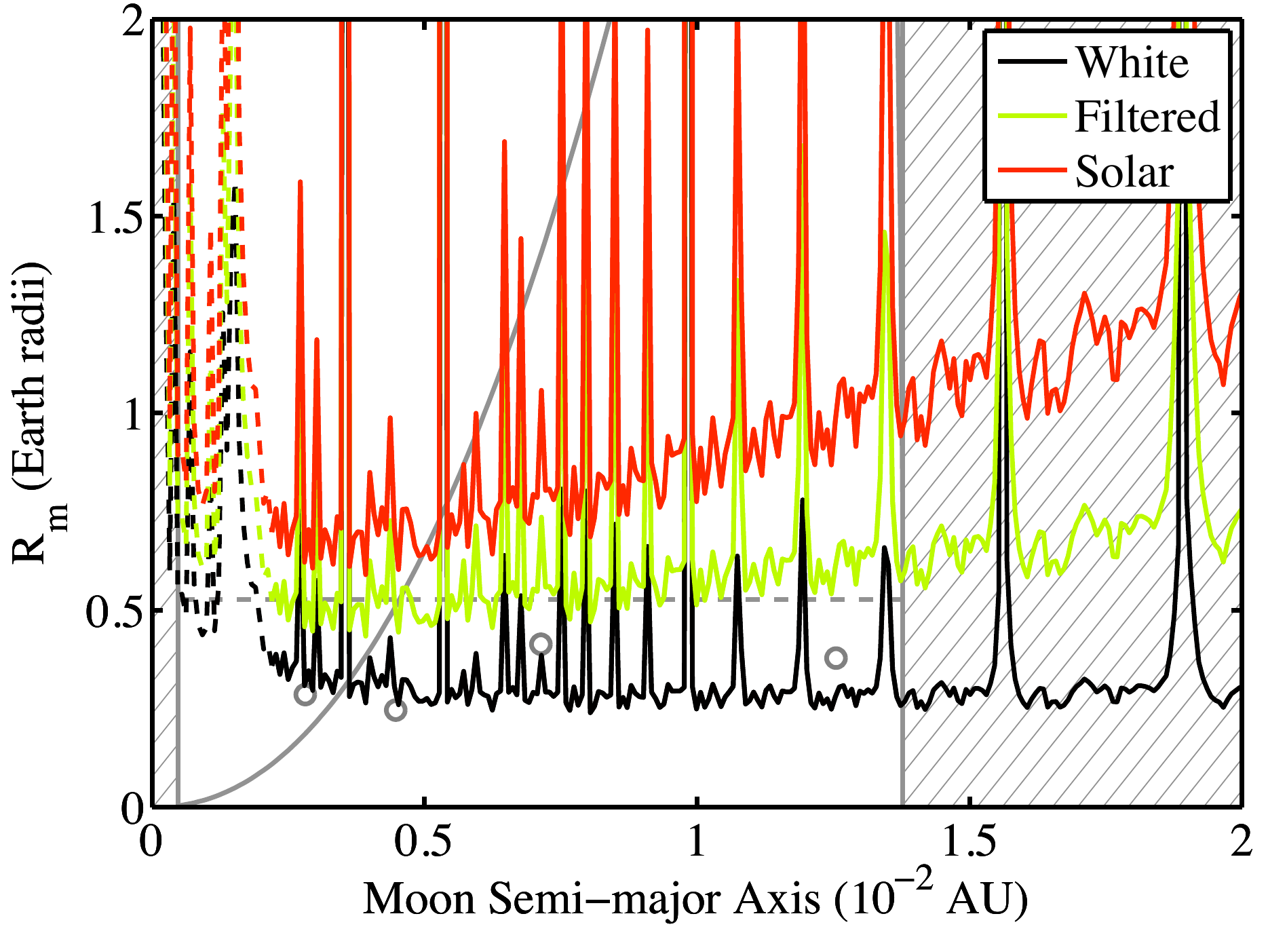}}
     \subfigure[$M_p = M_J$, $a_p=0.6$AU.]{
          \label{TransitThresh1s1MJ06AUInc}
          \includegraphics[width=.315\textwidth]{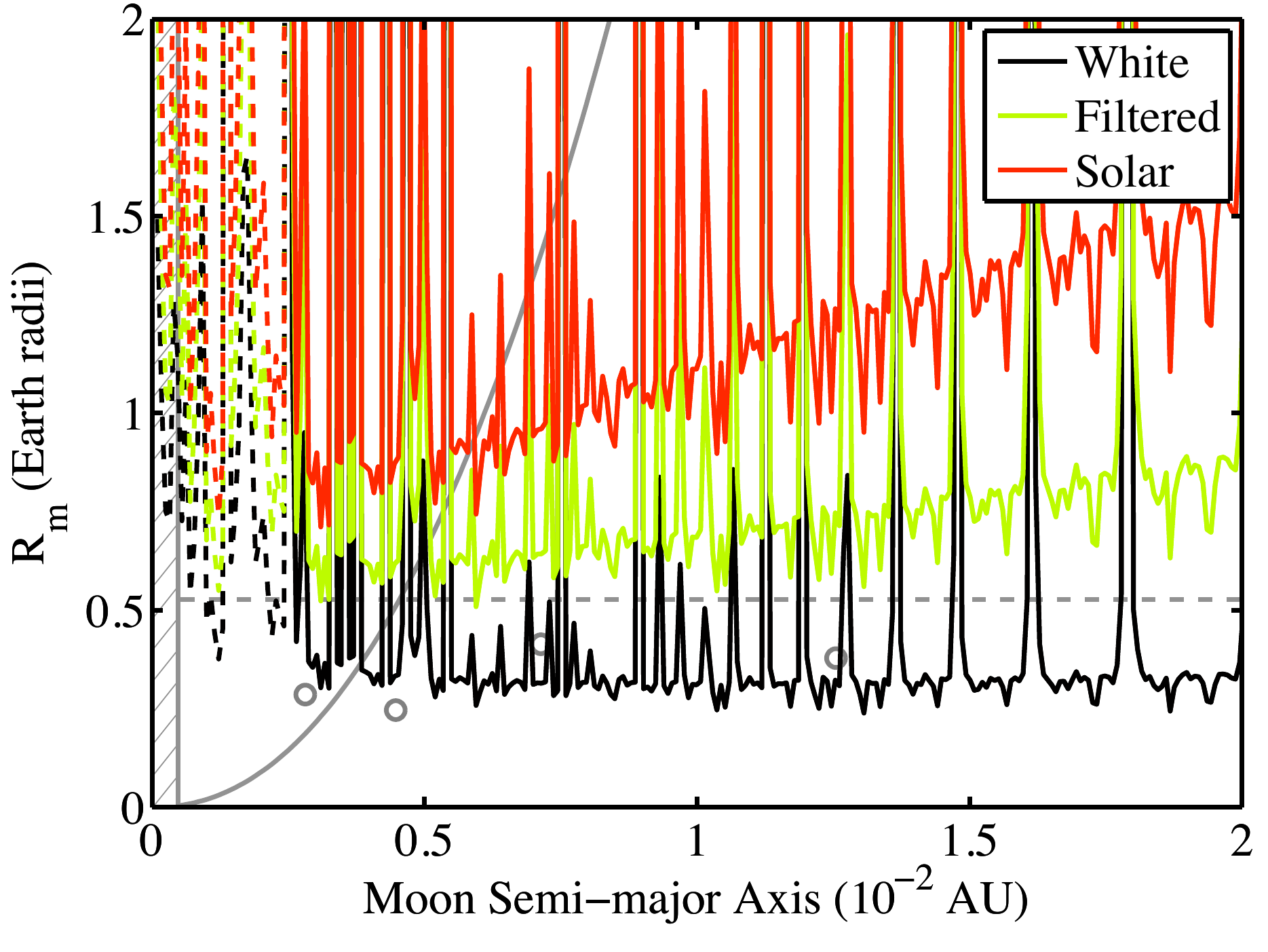}}\\ 
     \subfigure[$M_p = M_U$, $a_p=0.2$AU.]{
          \label{TransitThresh1s1MU02AUInc}
          \includegraphics[width=.315\textwidth]{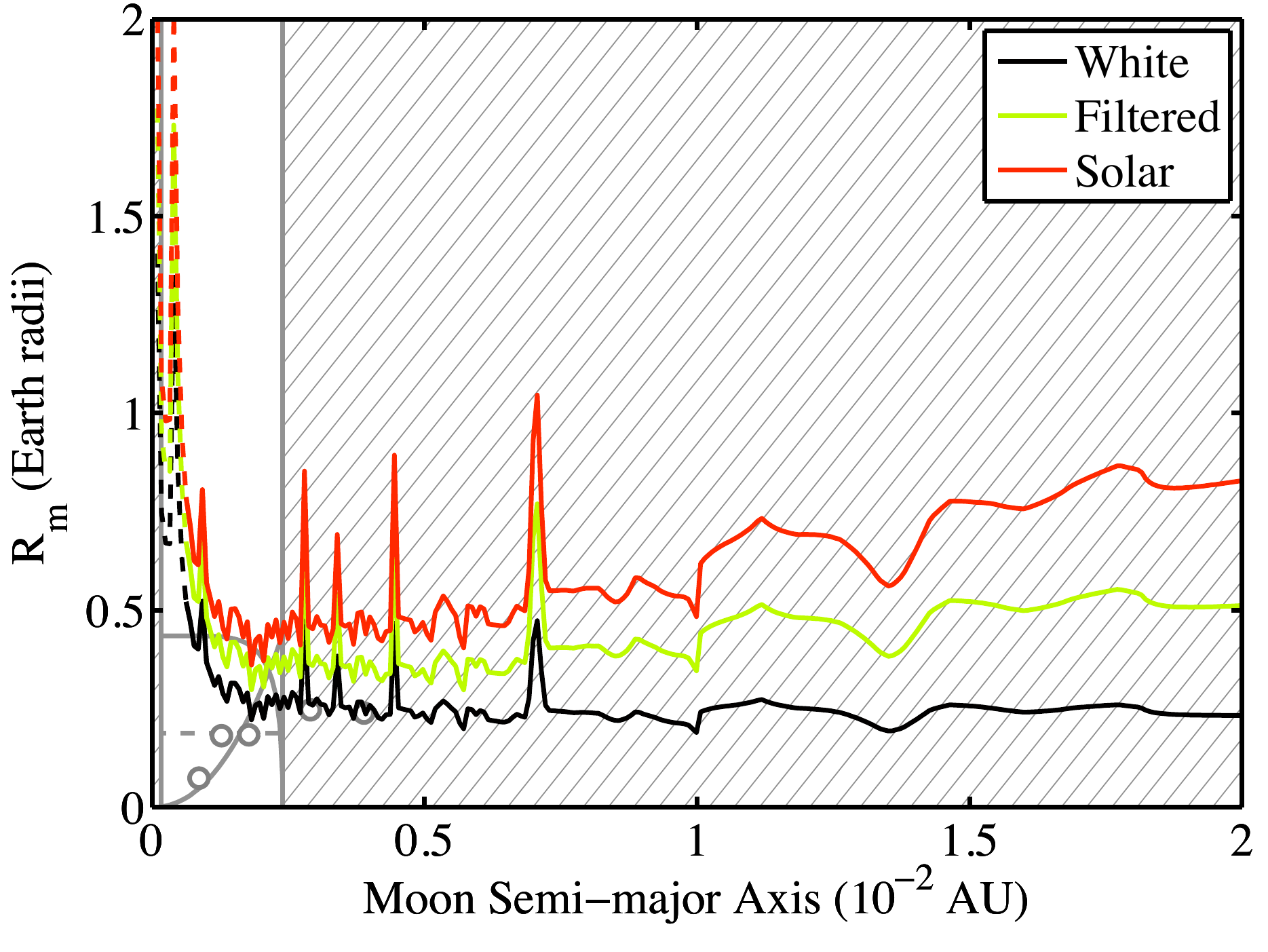}}
     \subfigure[$M_p = M_U$, $a_p=0.4$AU.]{
          \label{TransitThresh1s1MU04AUInc}
          \includegraphics[width=.315\textwidth]{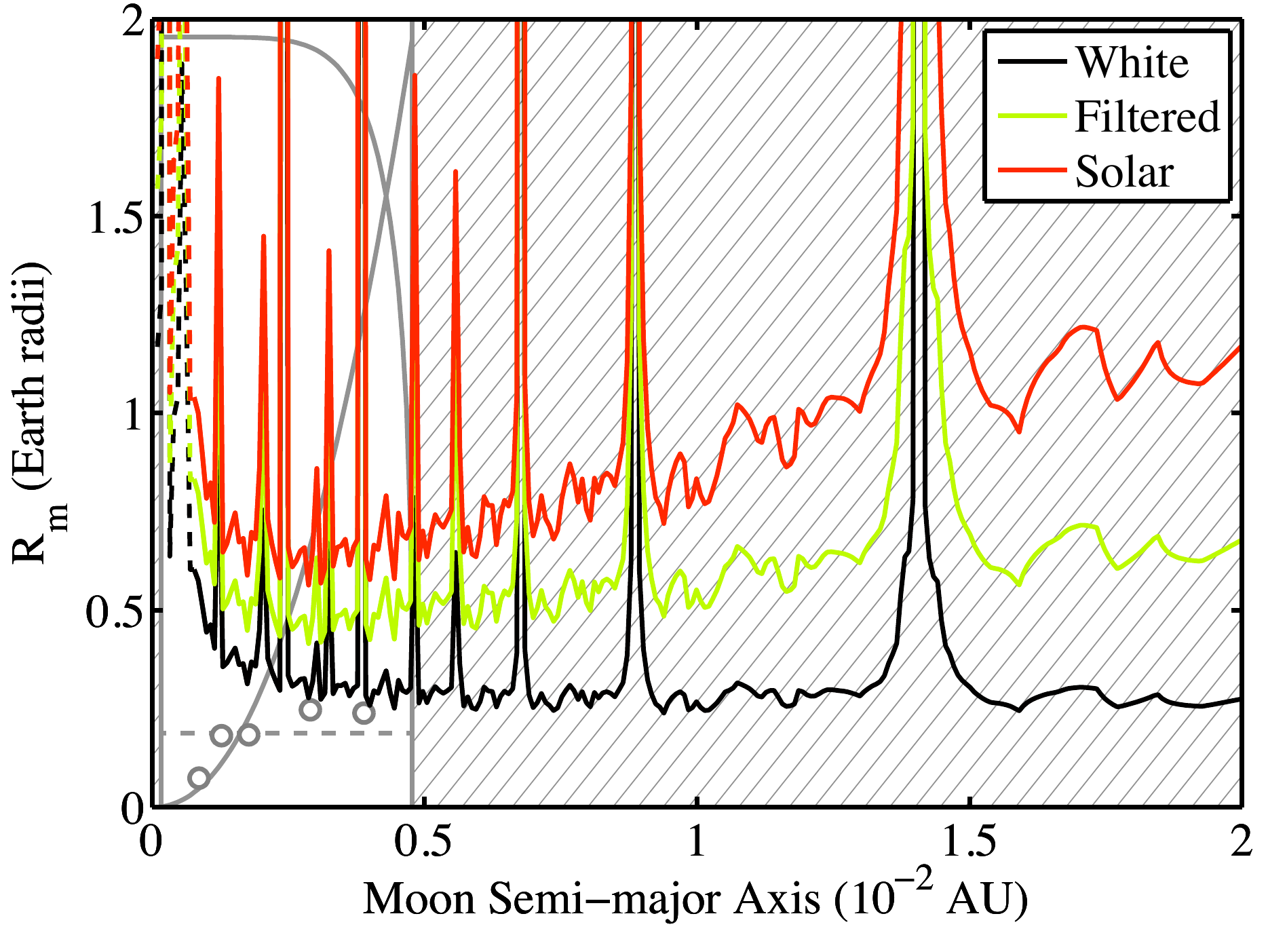}}
      \subfigure[$M_p = M_U$, $a_p=0.6$AU.]{
          \label{TransitThresh1s1MU06AUInc}
          \includegraphics[width=.315\textwidth]{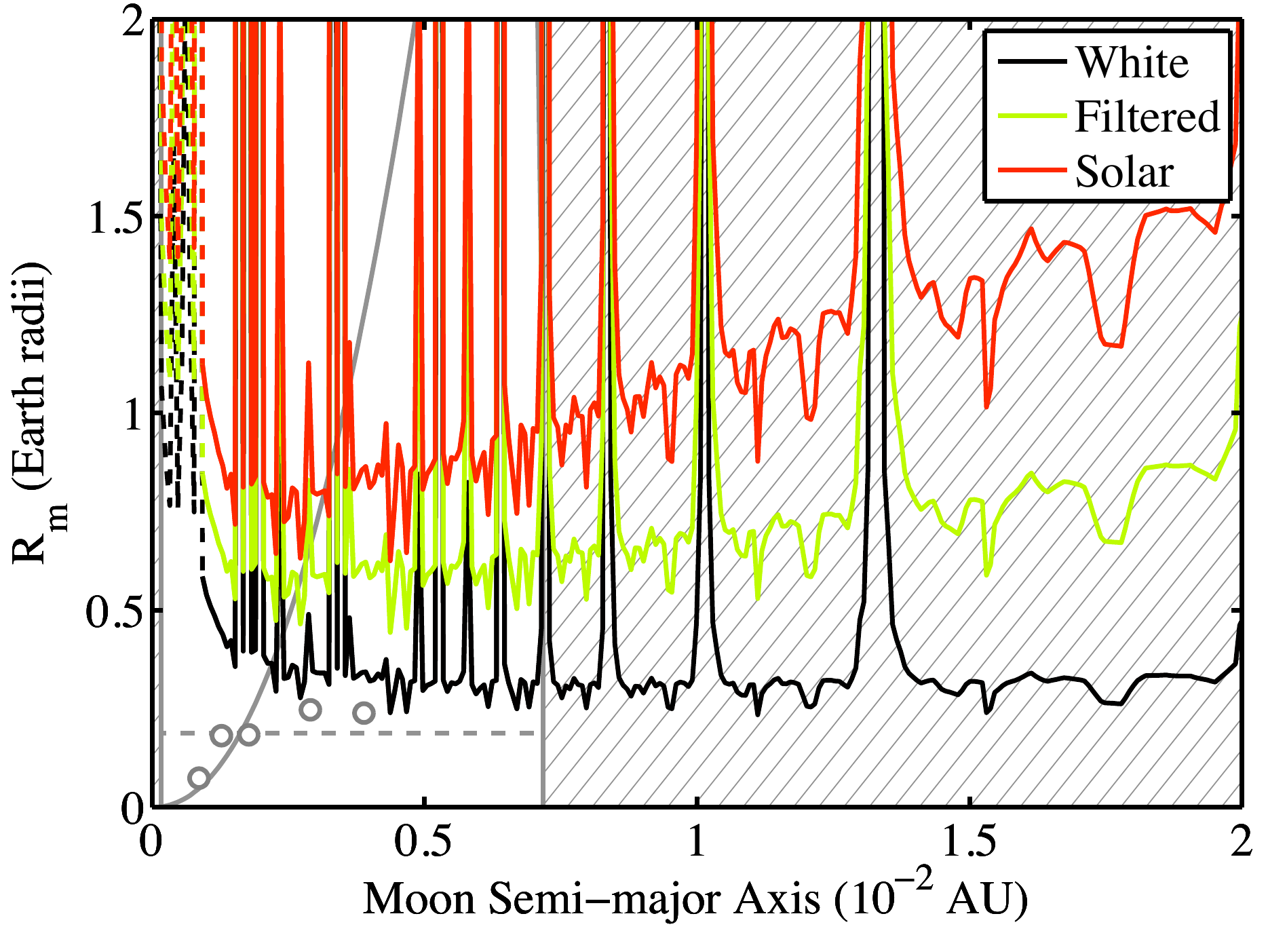}}\\ 
     \subfigure[$M_p = M_{\earth}$, $a_p=0.2$AU.]{
          \label{TransitThresh1s1ME02AUInc}
          \includegraphics[width=.315\textwidth]{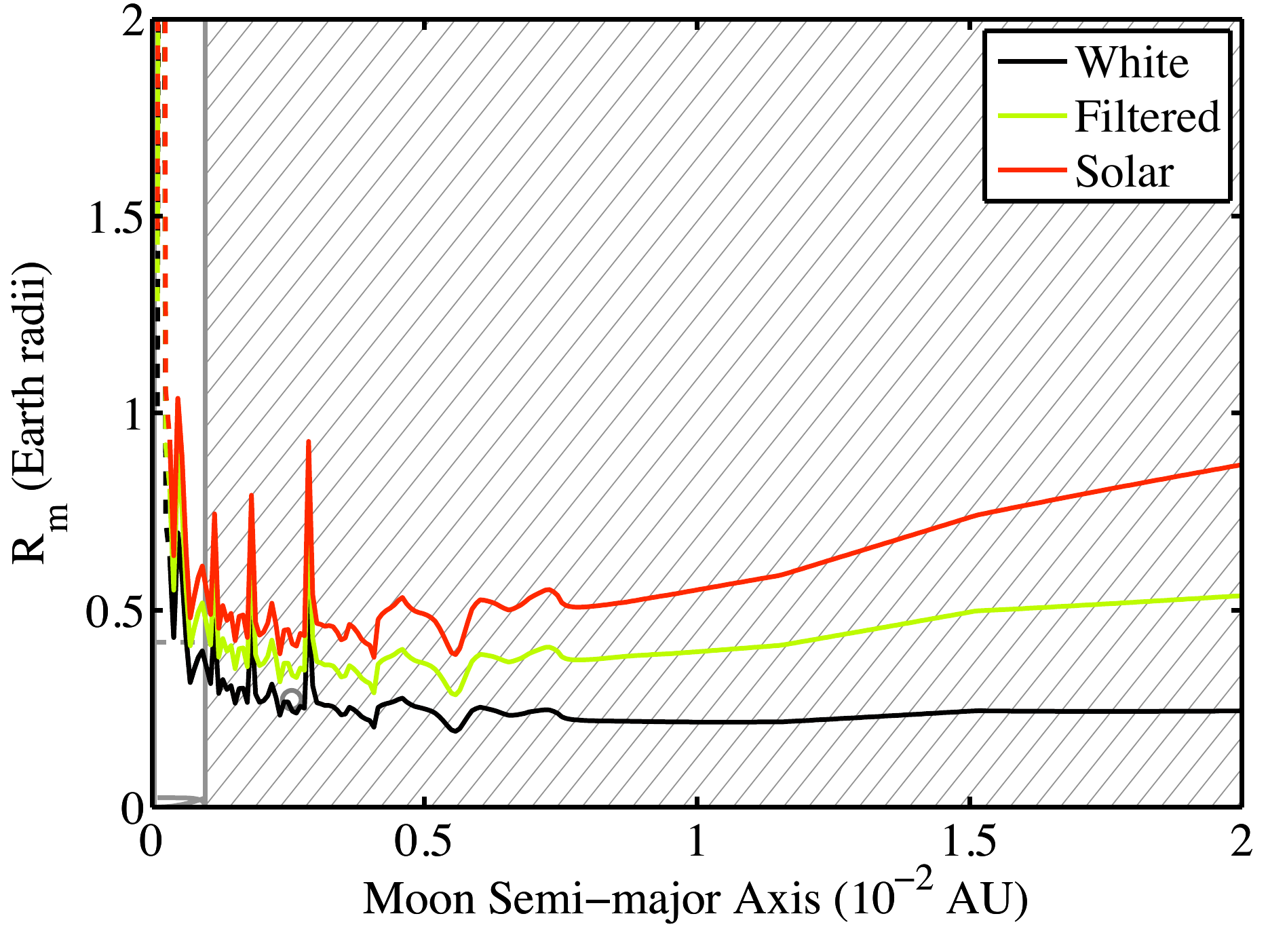}}
     \subfigure[$M_p = M_{\earth}$, $a_p=0.4$AU.]{
          \label{TransitThresh1s1ME04AUInc}
          \includegraphics[width=.315\textwidth]{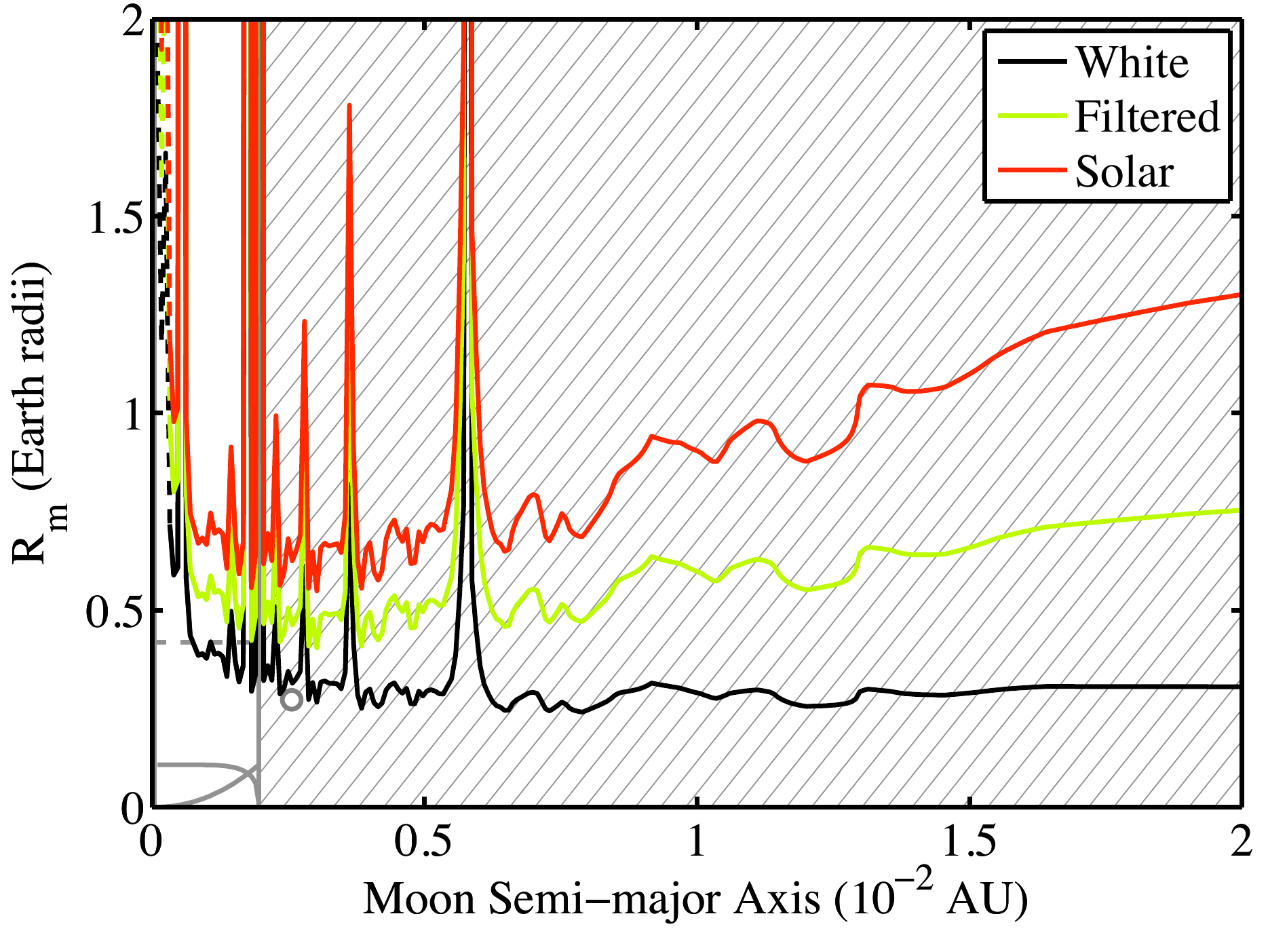}}
     \subfigure[$M_p = M_{\earth}$, $a_p=0.6$AU.]{
          \label{TransitThresh1s1ME06AUInc}
          \includegraphics[width=.315\textwidth]{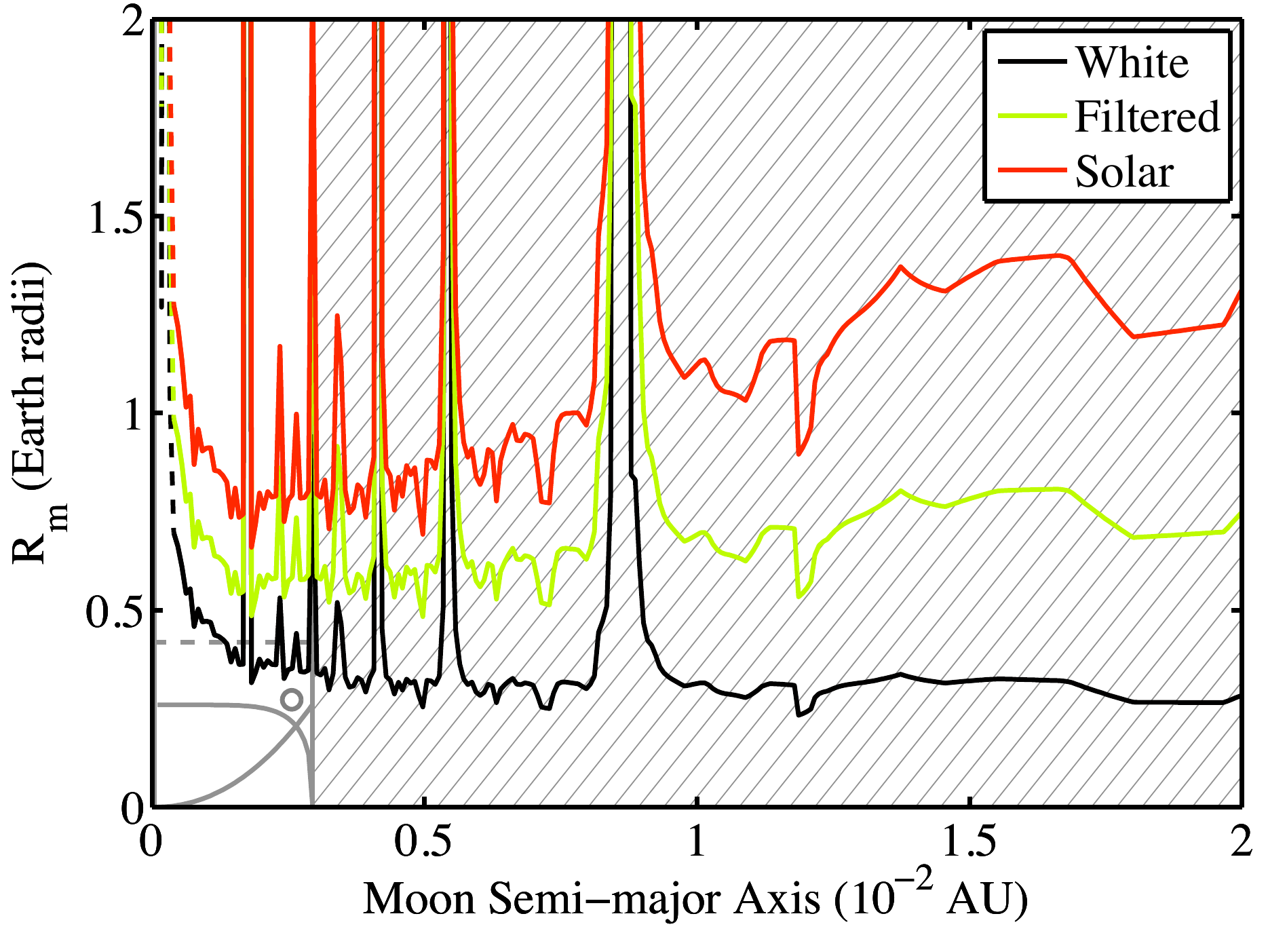}} 
     \caption{Figure of the same form as figure~\ref{MCThresholdsInclined}, but showing the 68.3\% thresholds.}
     \label{MCThresholdsInclined1S}
\end{figure}

\begin{figure}
     \centering
     \subfigure[$M_p$=$10 M_J$, $a_p=0.2$AU.]{
          \label{TransitThresh1s10MJ02AUEccP}
          \includegraphics[width=.315\textwidth]{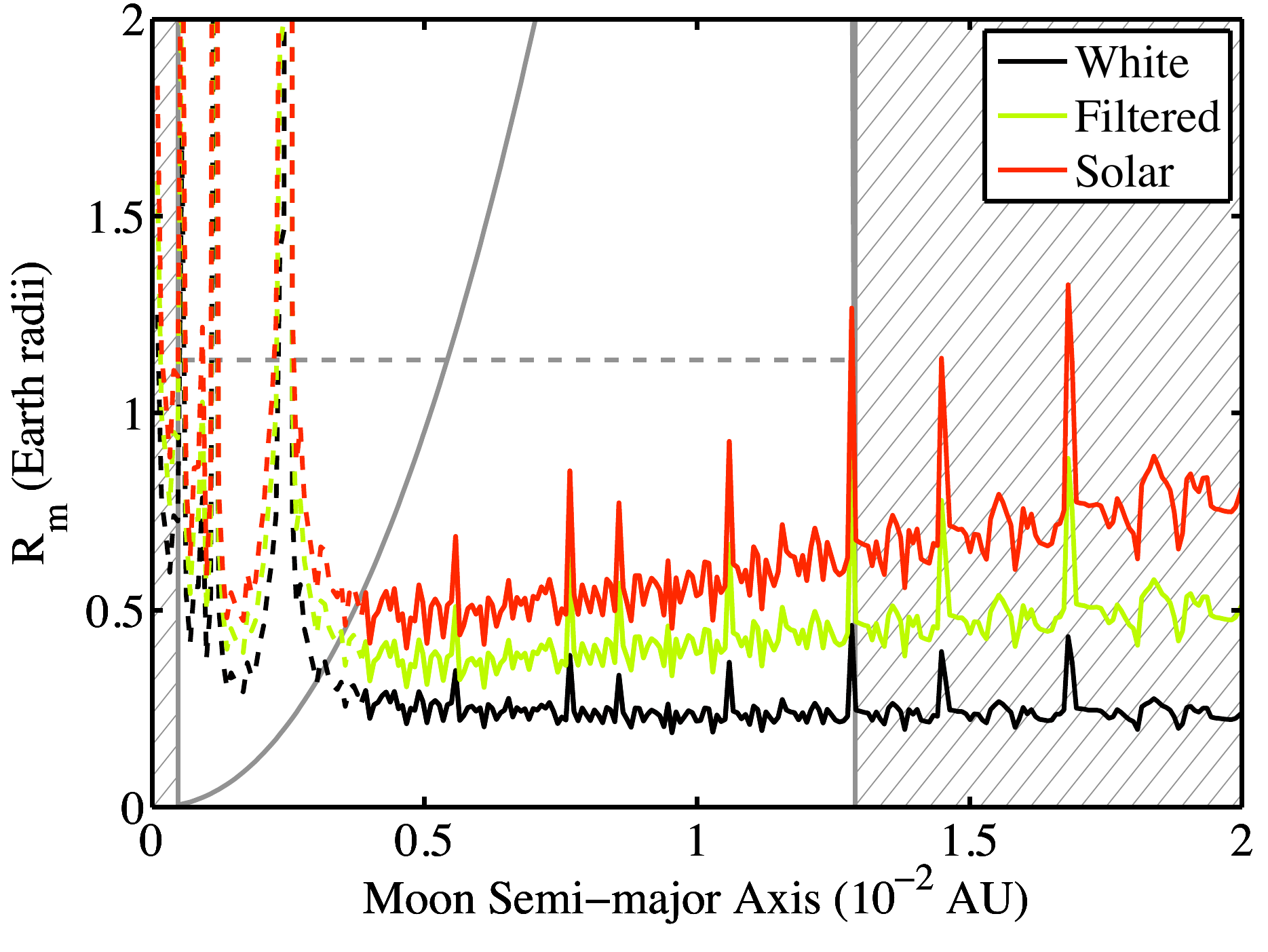}}
     \subfigure[$M_p$=$10 M_J$, $a_p=0.4$AU.]{
          \label{TransitThresh1s10MJ04AUEccP}
          \includegraphics[width=.315\textwidth]{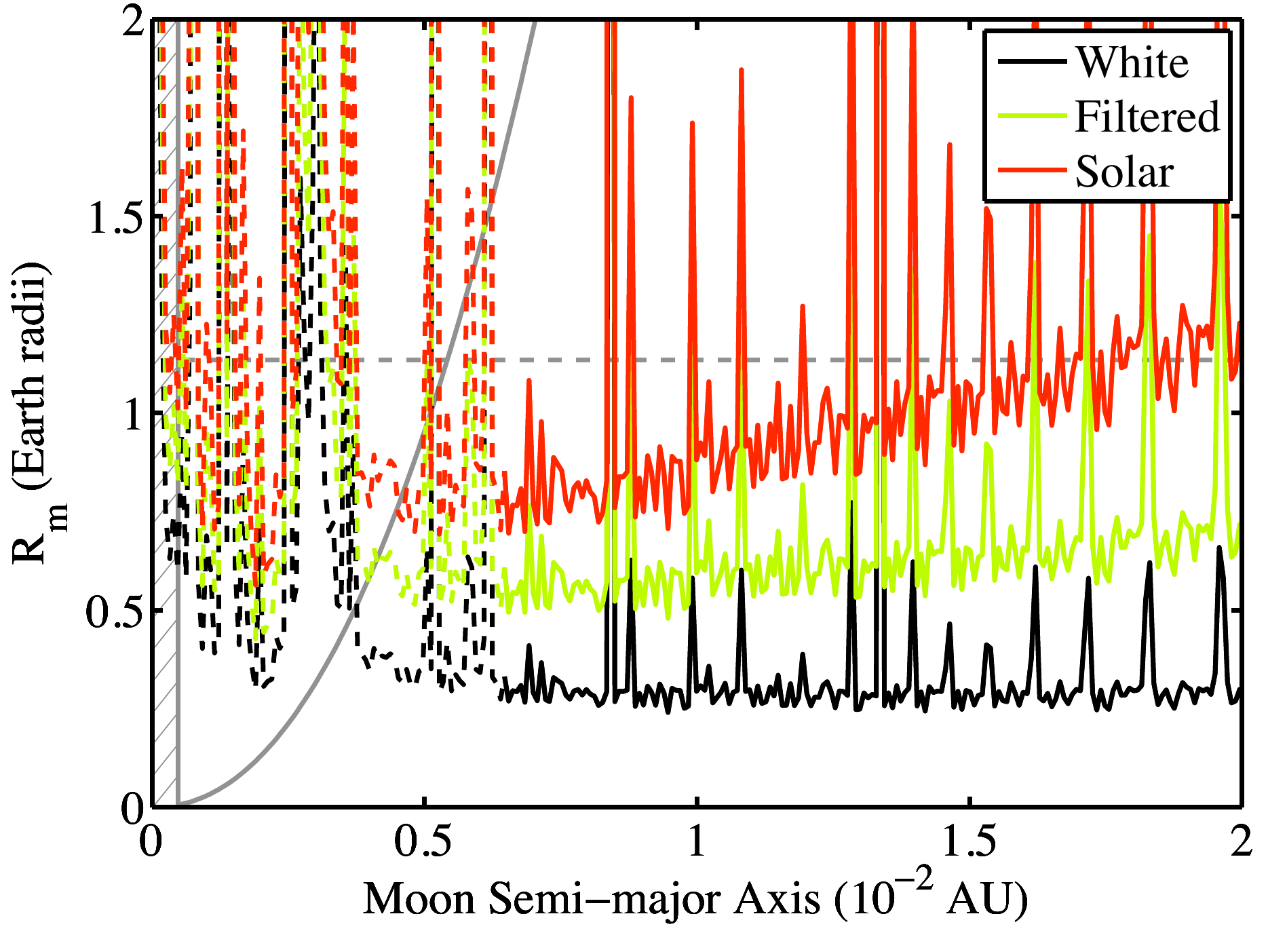}}
     \subfigure[$M_p$=$10 M_J$, $a_p=0.6$AU.]{
          \label{TransitThresh1s10MJ06AUEccP}
          \includegraphics[width=.315\textwidth]{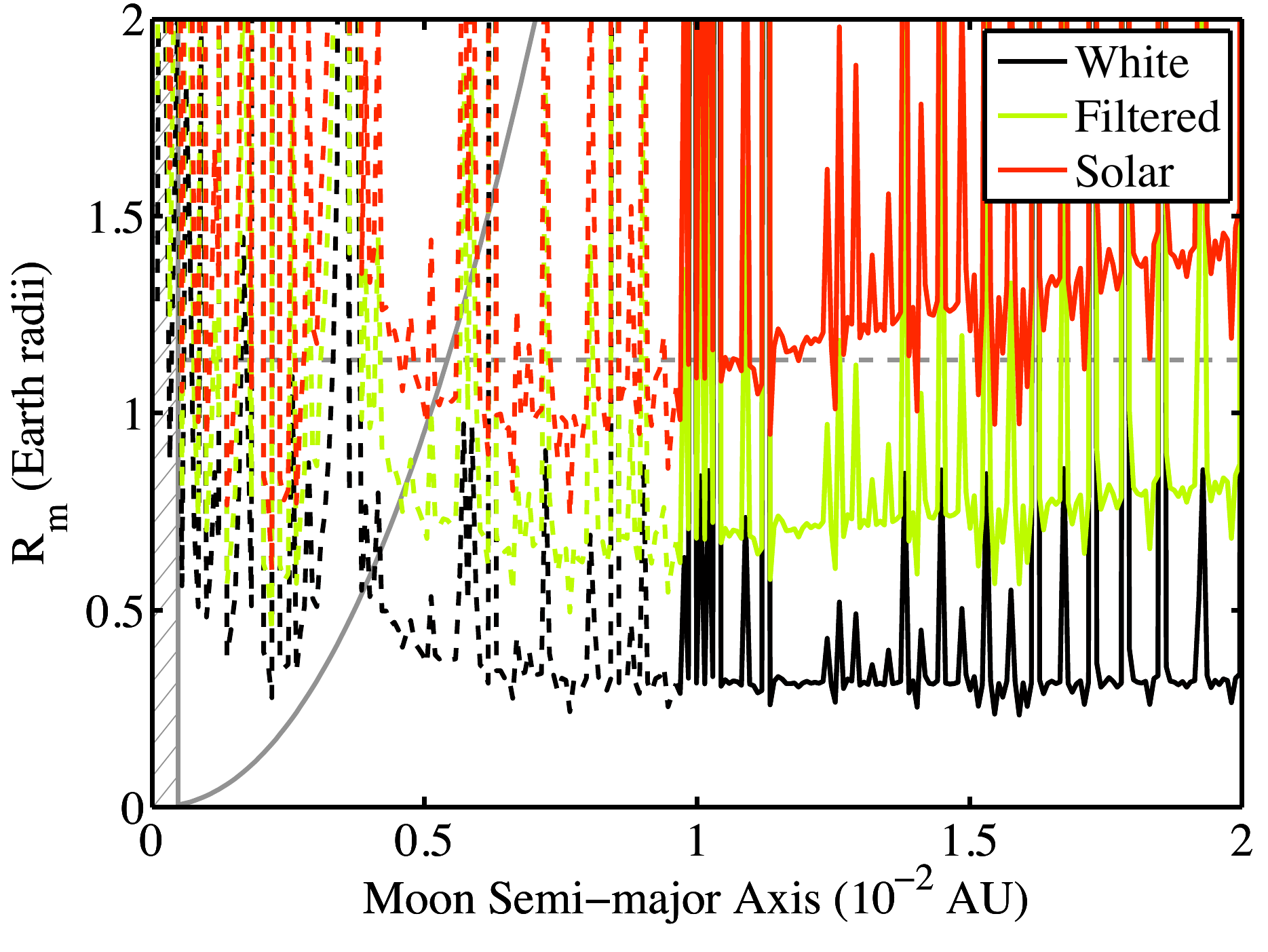}}\\ 
     \subfigure[$M_p = M_J$, $a_p=0.2$AU.]{
          \label{TransitThresh1s1MJ02AUEccP}
          \includegraphics[width=.315\textwidth]{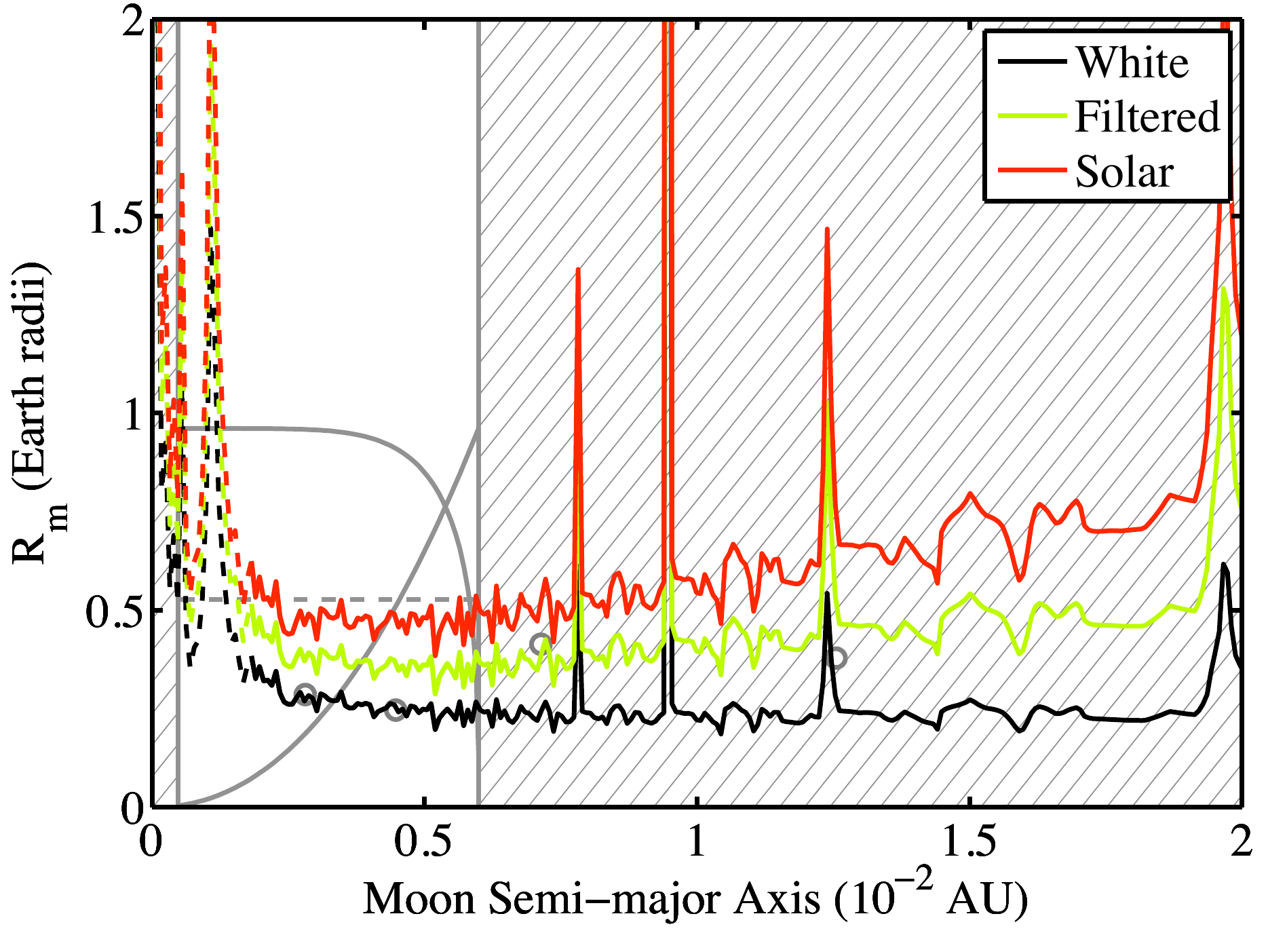}}
      \subfigure[$M_p = M_J$, $a_p=0.4$AU.]{
          \label{TransitThresh1s1MJ04AUEccP}
          \includegraphics[width=.315\textwidth]{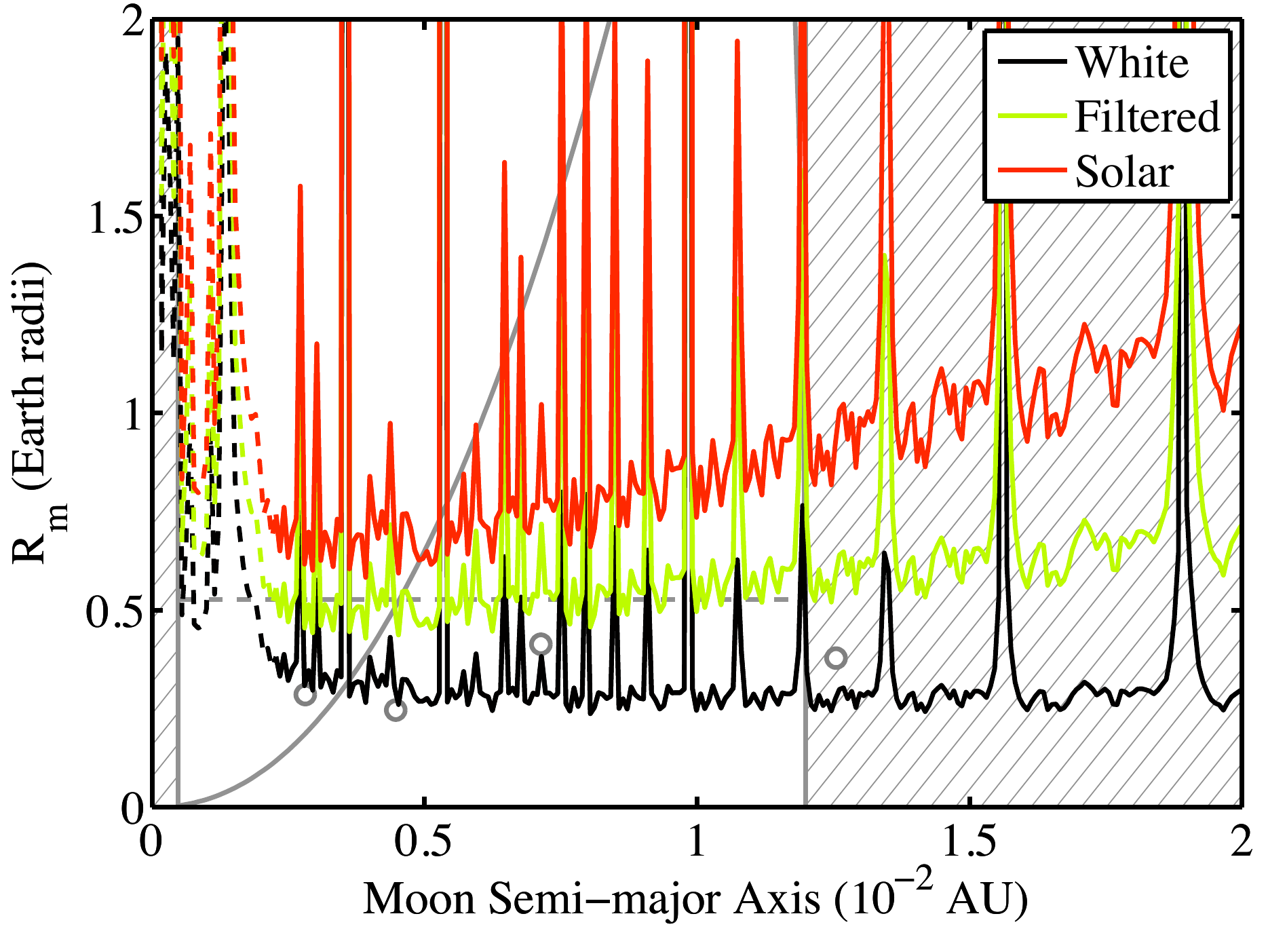}}
     \subfigure[$M_p = M_J$, $a_p=0.6$AU.]{
          \label{TransitThresh1s1MJ06AUEccP}
          \includegraphics[width=.315\textwidth]{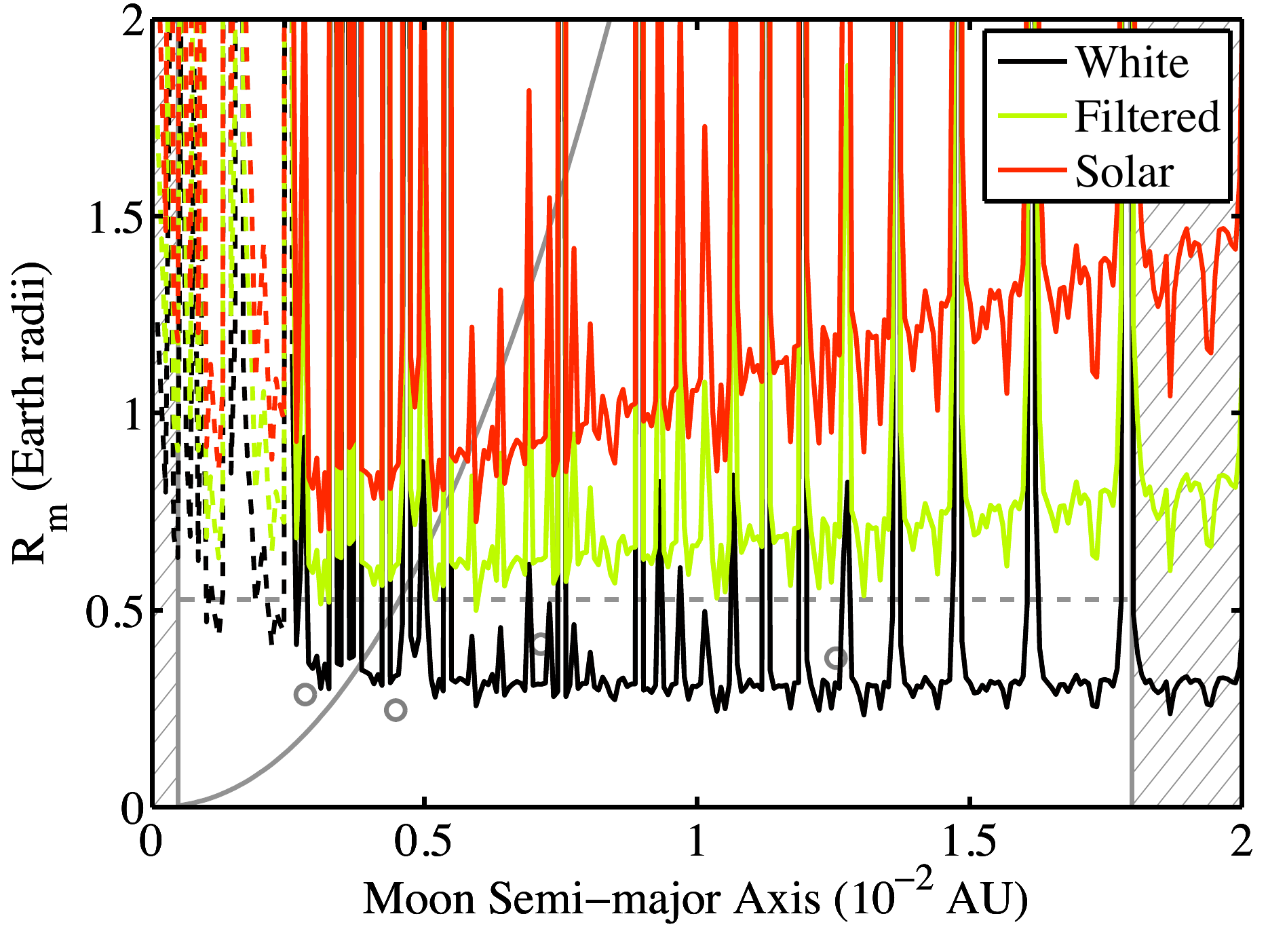}}\\ 
     \subfigure[$M_p = M_U$, $a_p=0.2$AU.]{
          \label{TransitThresh1s1MU02AUEccP}
          \includegraphics[width=.315\textwidth]{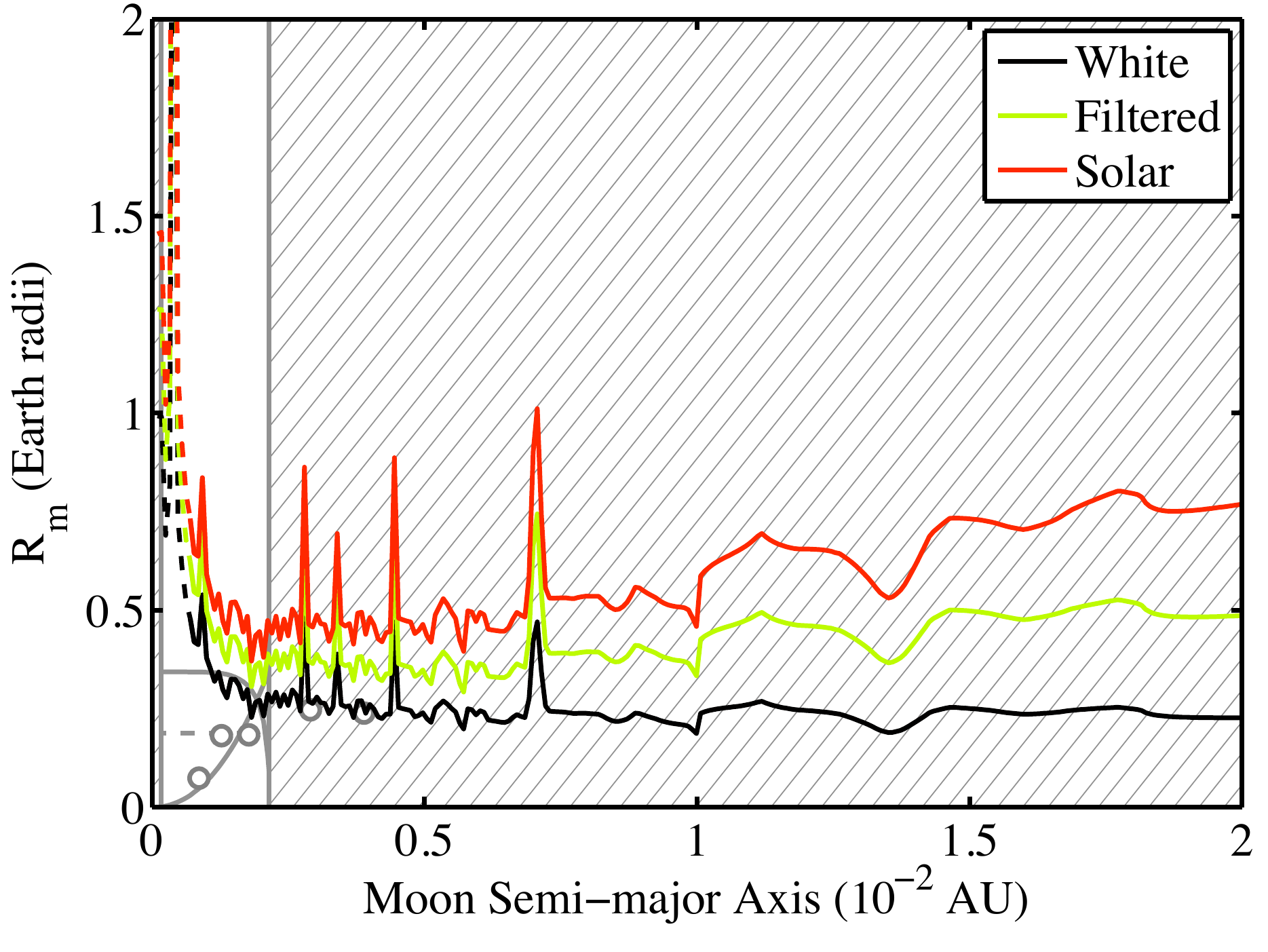}}
     \subfigure[$M_p = M_U$, $a_p=0.4$AU.]{
          \label{TransitThresh1s1MU04AUEccP}
          \includegraphics[width=.315\textwidth]{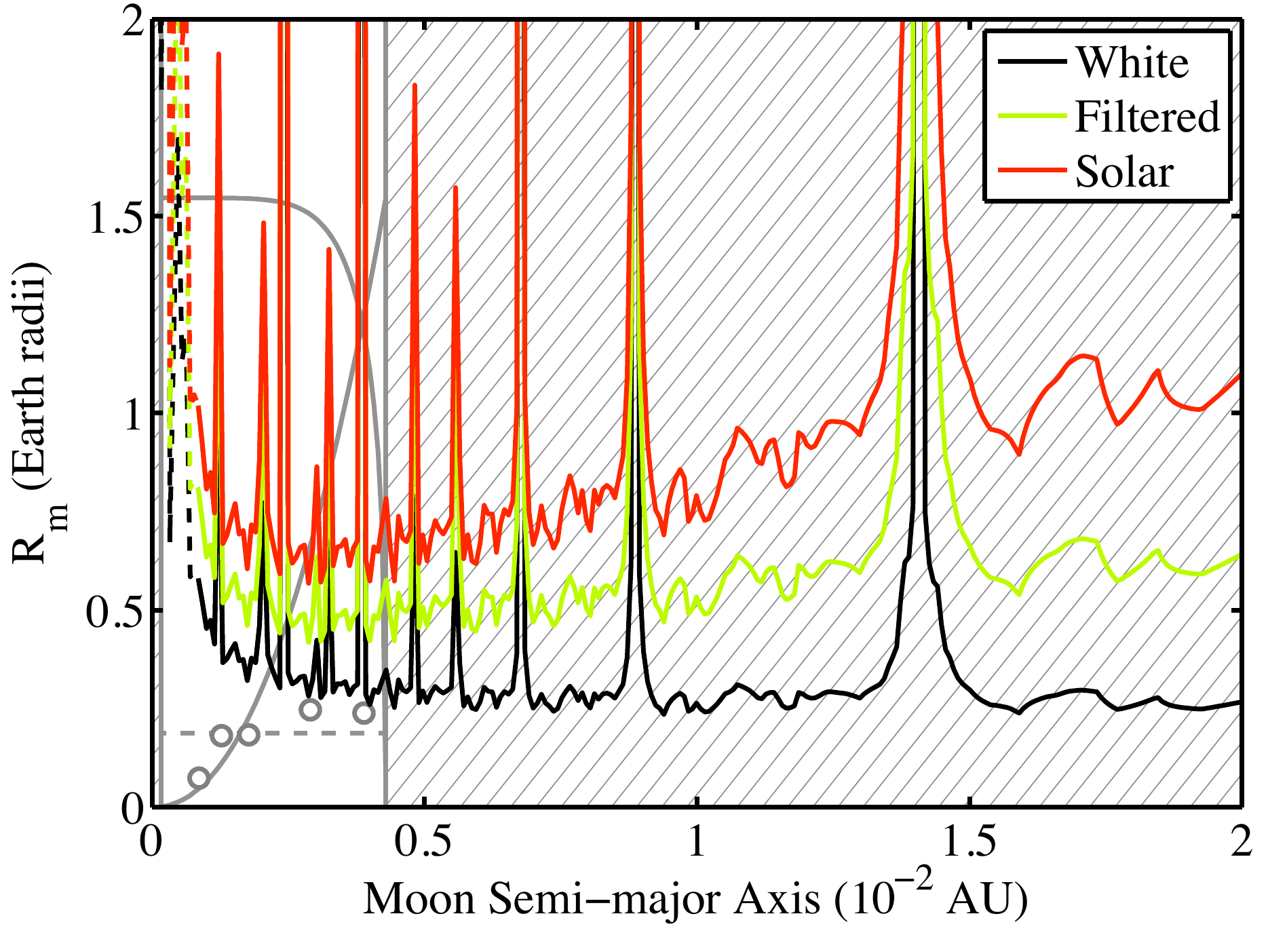}}
      \subfigure[$M_p = M_U$, $a_p=0.6$AU.]{
          \label{TransitThresh1s1MU06AUEccP}
          \includegraphics[width=.315\textwidth]{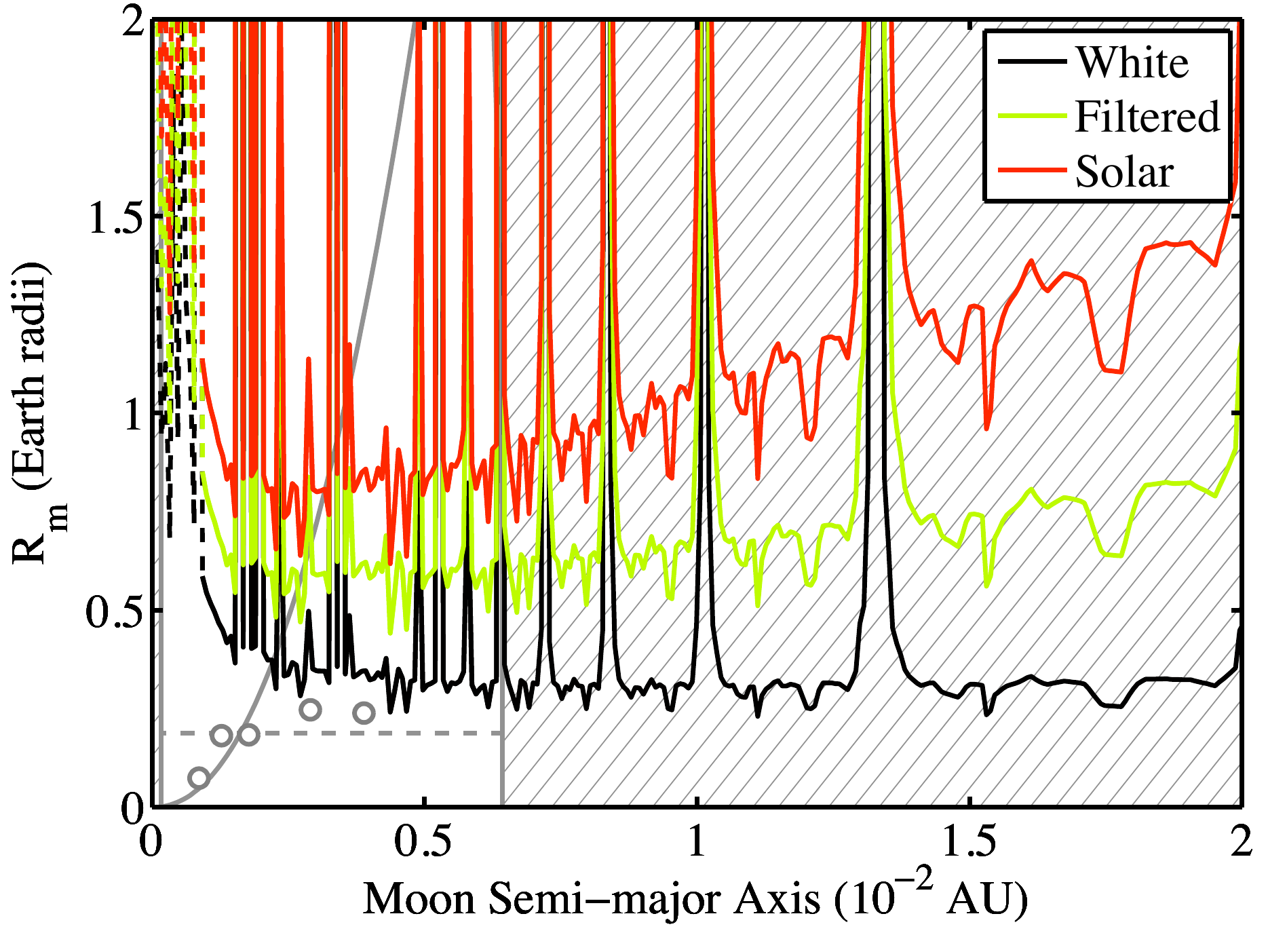}}\\ 
     \subfigure[$M_p = M_{\earth}$, $a_p=0.2$AU.]{
          \label{TransitThresh1s1ME02AUEccP}
          \includegraphics[width=.315\textwidth]{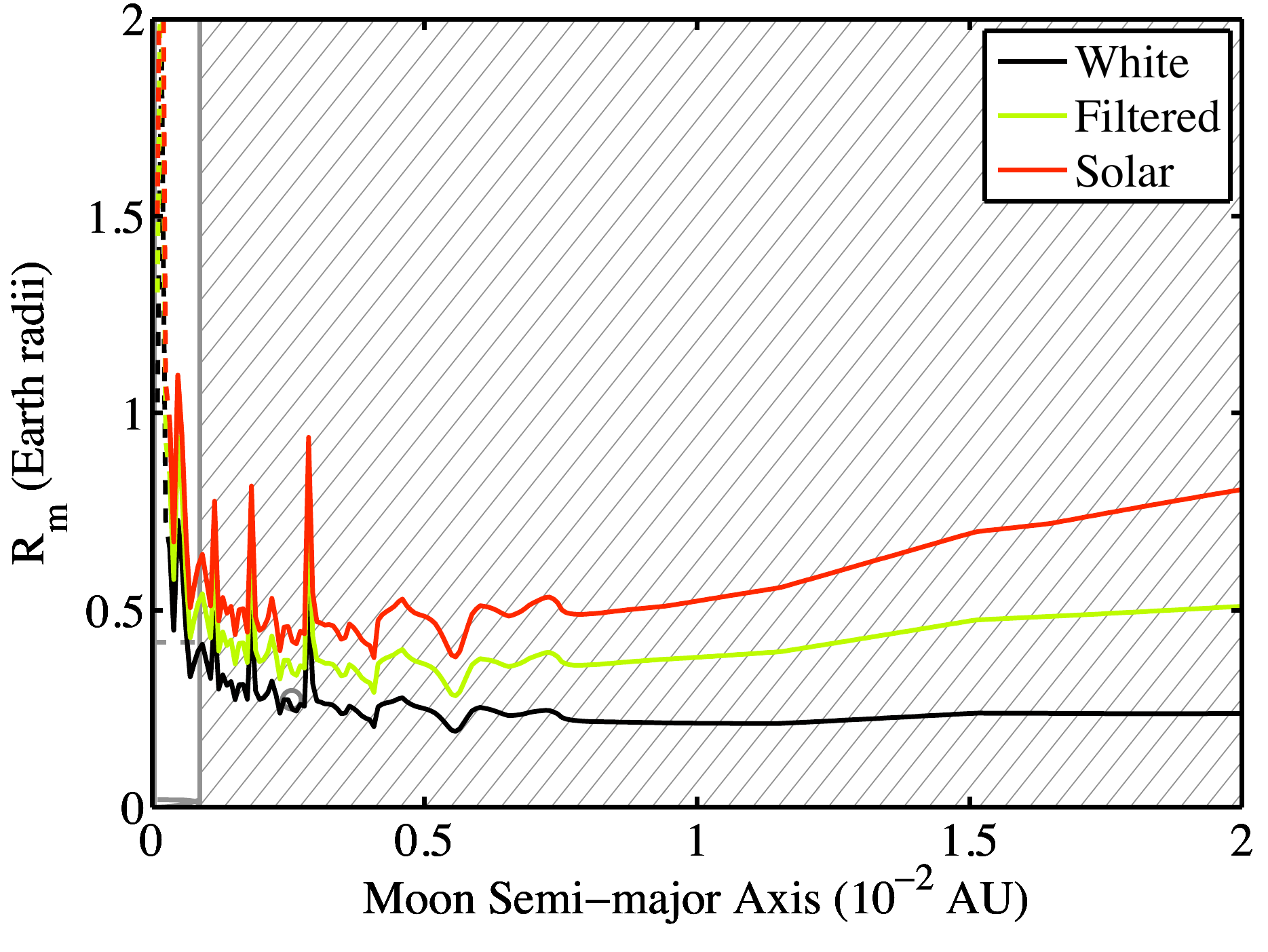}}
     \subfigure[$M_p = M_{\earth}$, $a_p=0.4$AU.]{
          \label{TransitThresh1s1ME04AUEccP}
          \includegraphics[width=.315\textwidth]{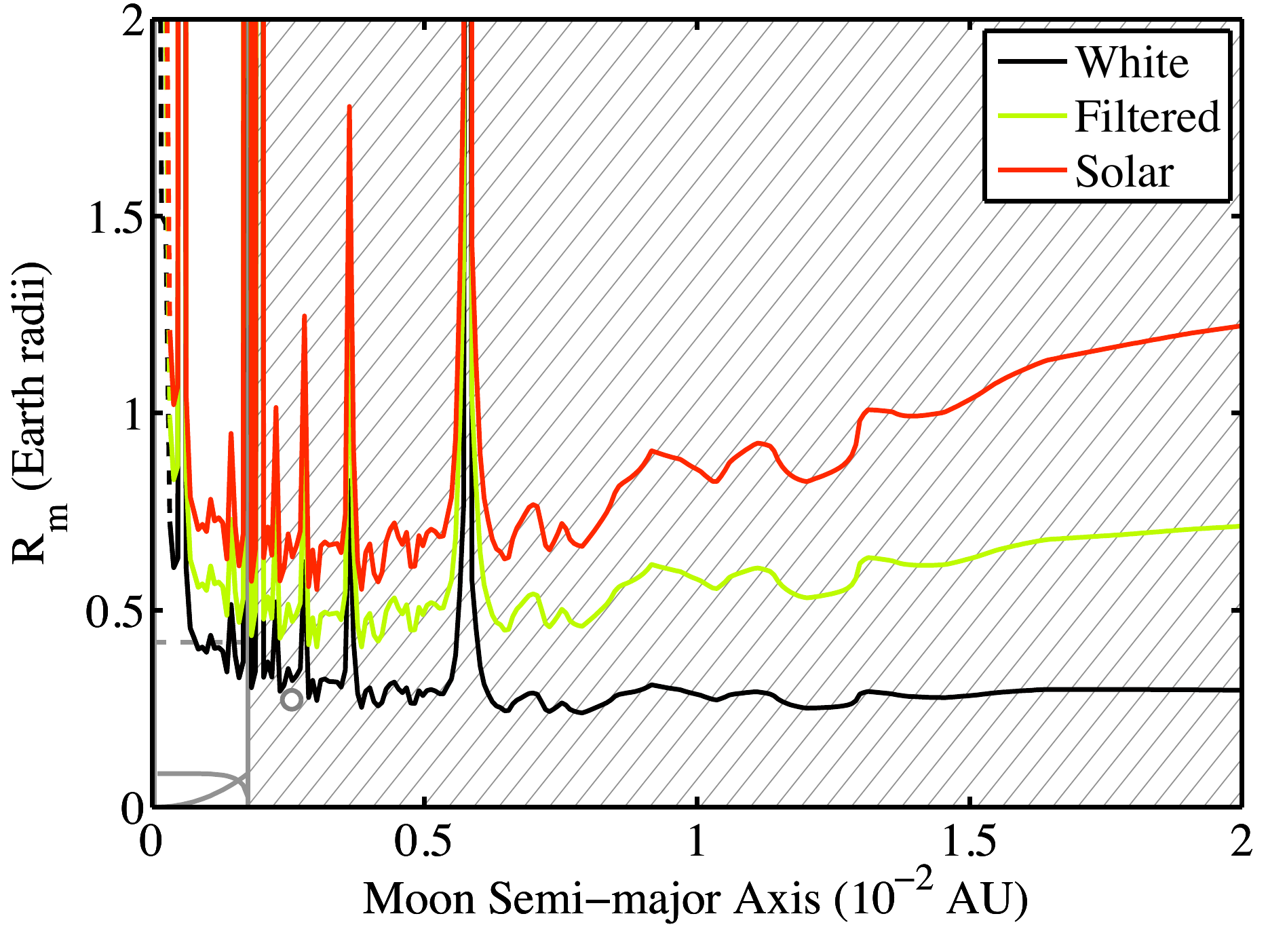}}
     \subfigure[$M_p = M_{\earth}$, $a_p=0.6$AU.]{
          \label{TransitThresh1s1ME06AUEccP}
          \includegraphics[width=.315\textwidth]{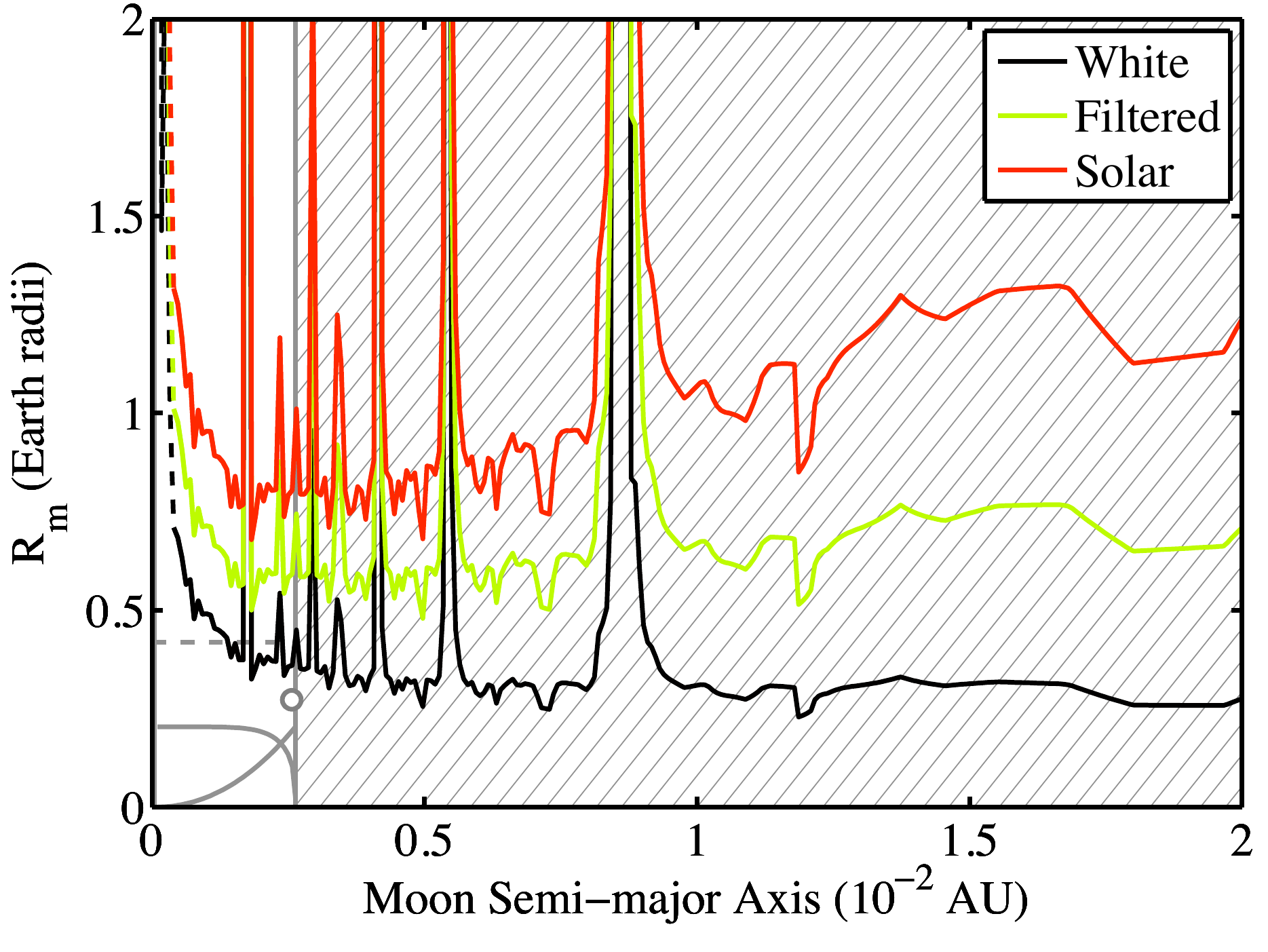}} 
     \caption{Figure of the same form as figure~\ref{MCThresholdsEccentricPeri}, but showing the 68.3\% thresholds.}
    \label{MCThresholdsEccentricPeri1S}
    \end{figure}

\begin{figure}
     \centering
     \subfigure[$M_p$=$10 M_J$, $a_p=0.2$AU.]{
          \label{TransitThresh1s10MJ02AUEccA}
          \includegraphics[width=.315\textwidth]{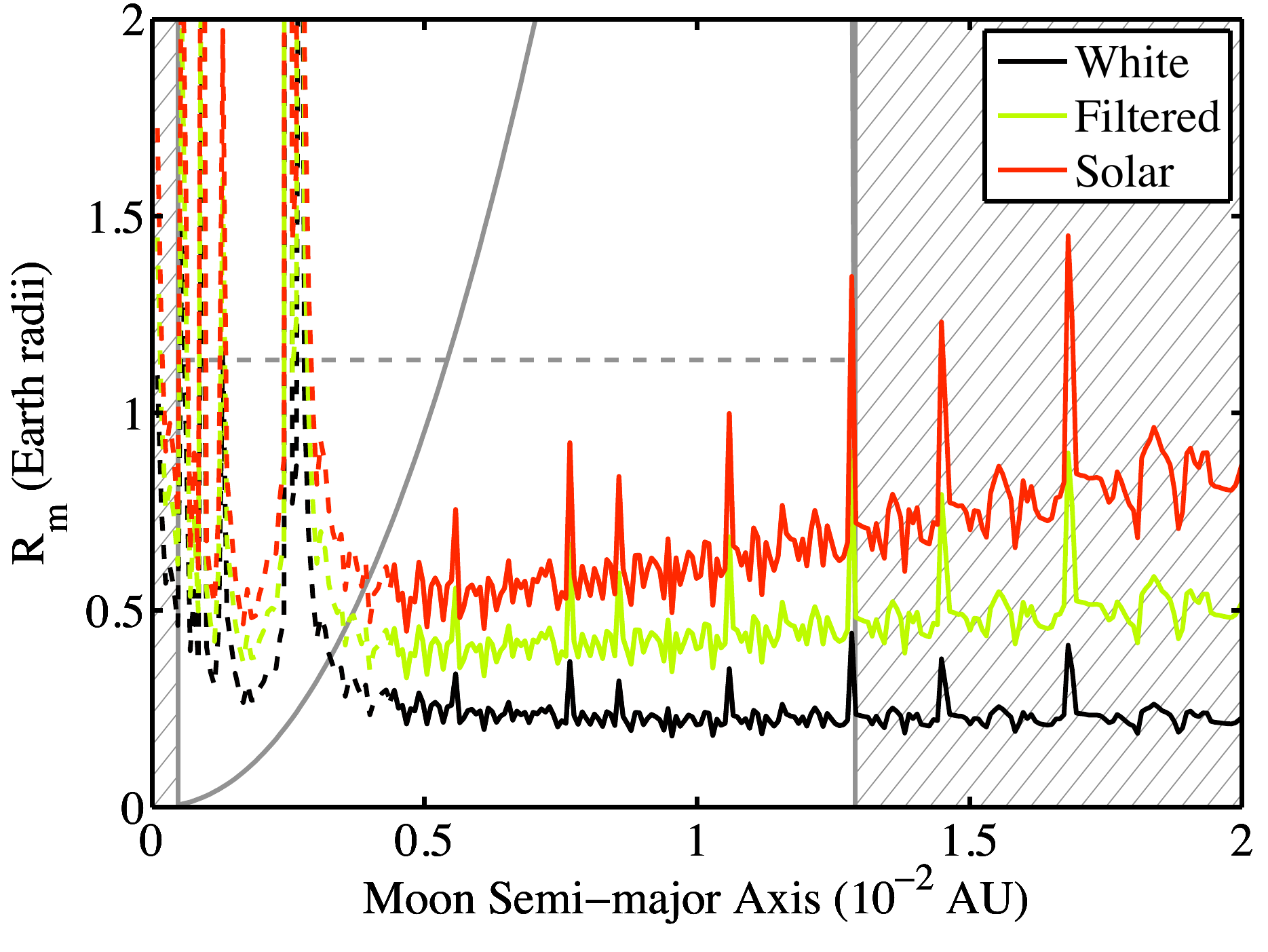}}
     \subfigure[$M_p$=$10 M_J$, $a_p=0.4$AU.]{
          \label{TransitThresh1s10MJ04AUEccA}
          \includegraphics[width=.315\textwidth]{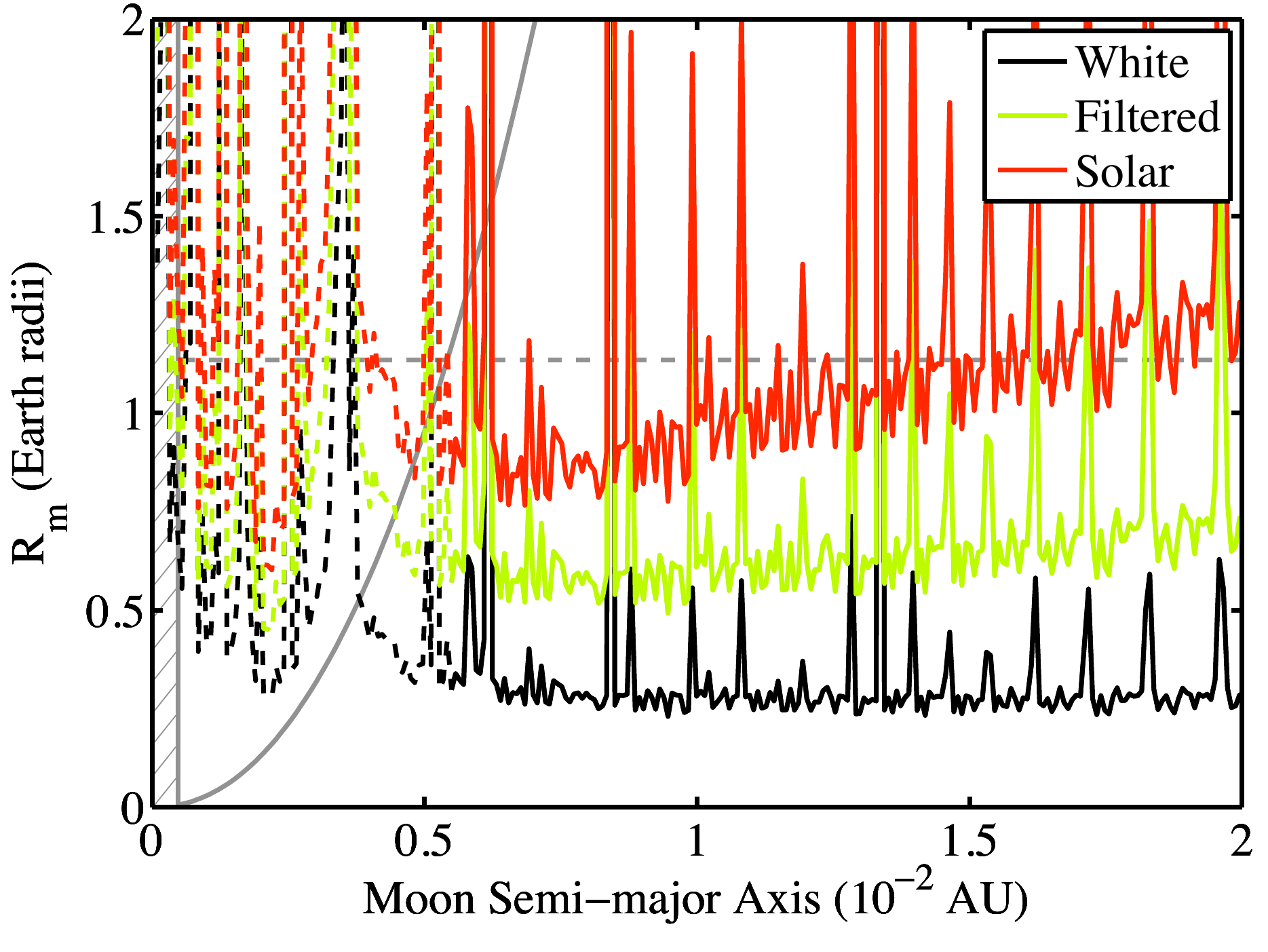}}
     \subfigure[$M_p$=$10 M_J$, $a_p=0.6$AU.]{
          \label{TransitThresh1s10MJ06AUEccA}
          \includegraphics[width=.315\textwidth]{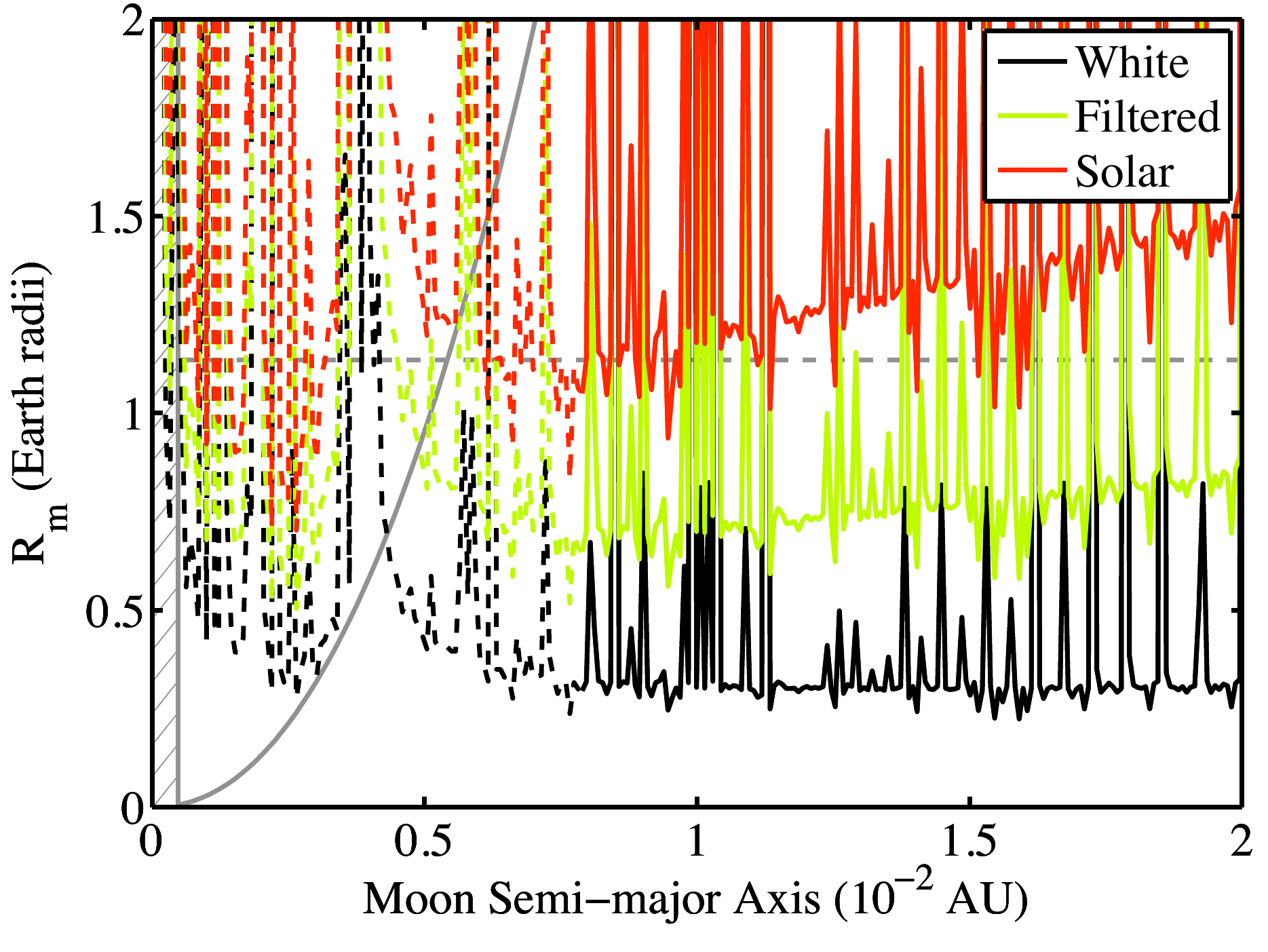}}\\ 
     \subfigure[$M_p = M_J$, $a_p=0.2$AU.]{
          \label{TransitThresh1s1MJ02AUEccA}
          \includegraphics[width=.315\textwidth]{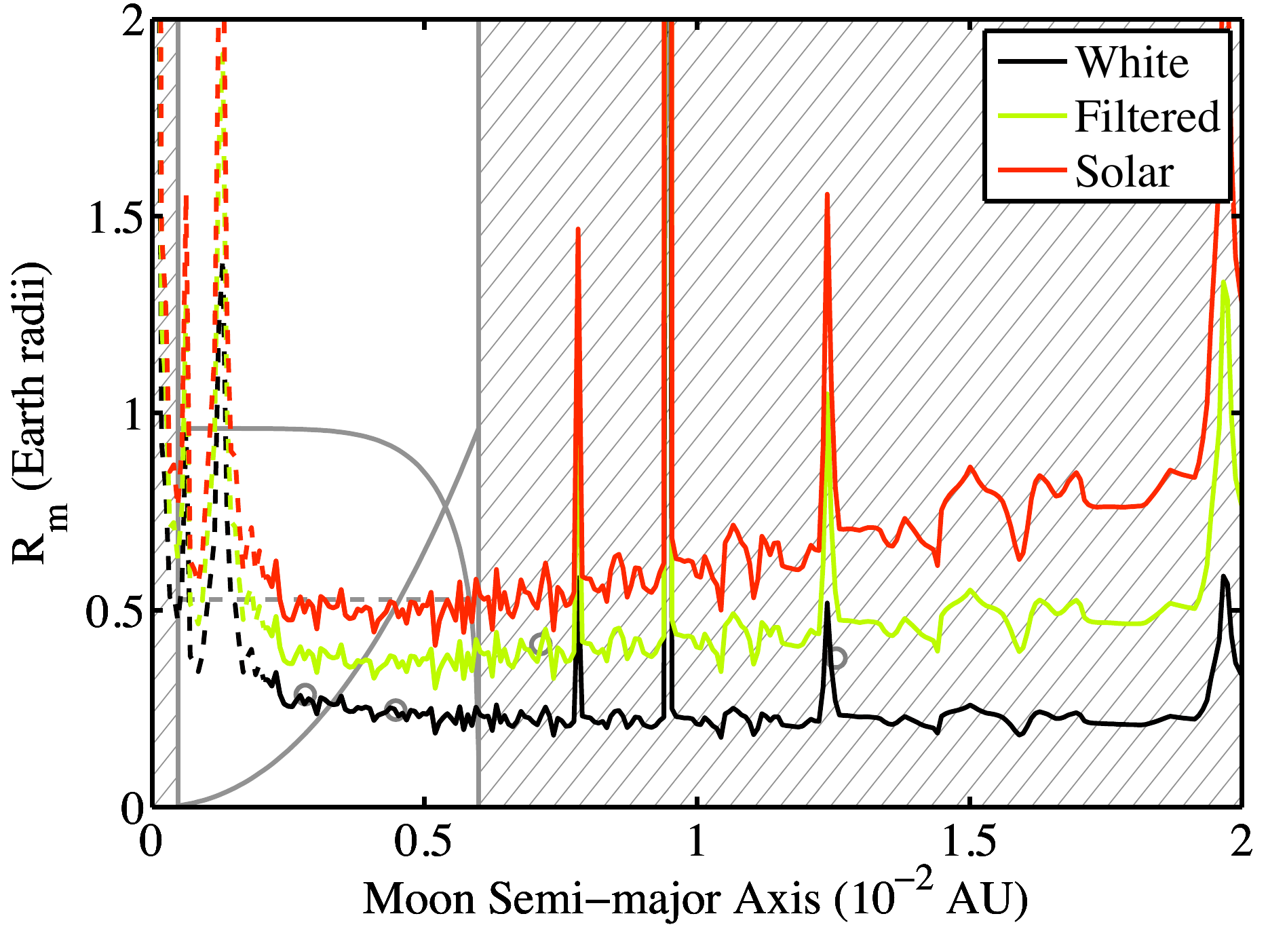}}
      \subfigure[$M_p = M_J$, $a_p=0.4$AU.]{
          \label{TransitThresh1s1MJ04AUEccA}
          \includegraphics[width=.315\textwidth]{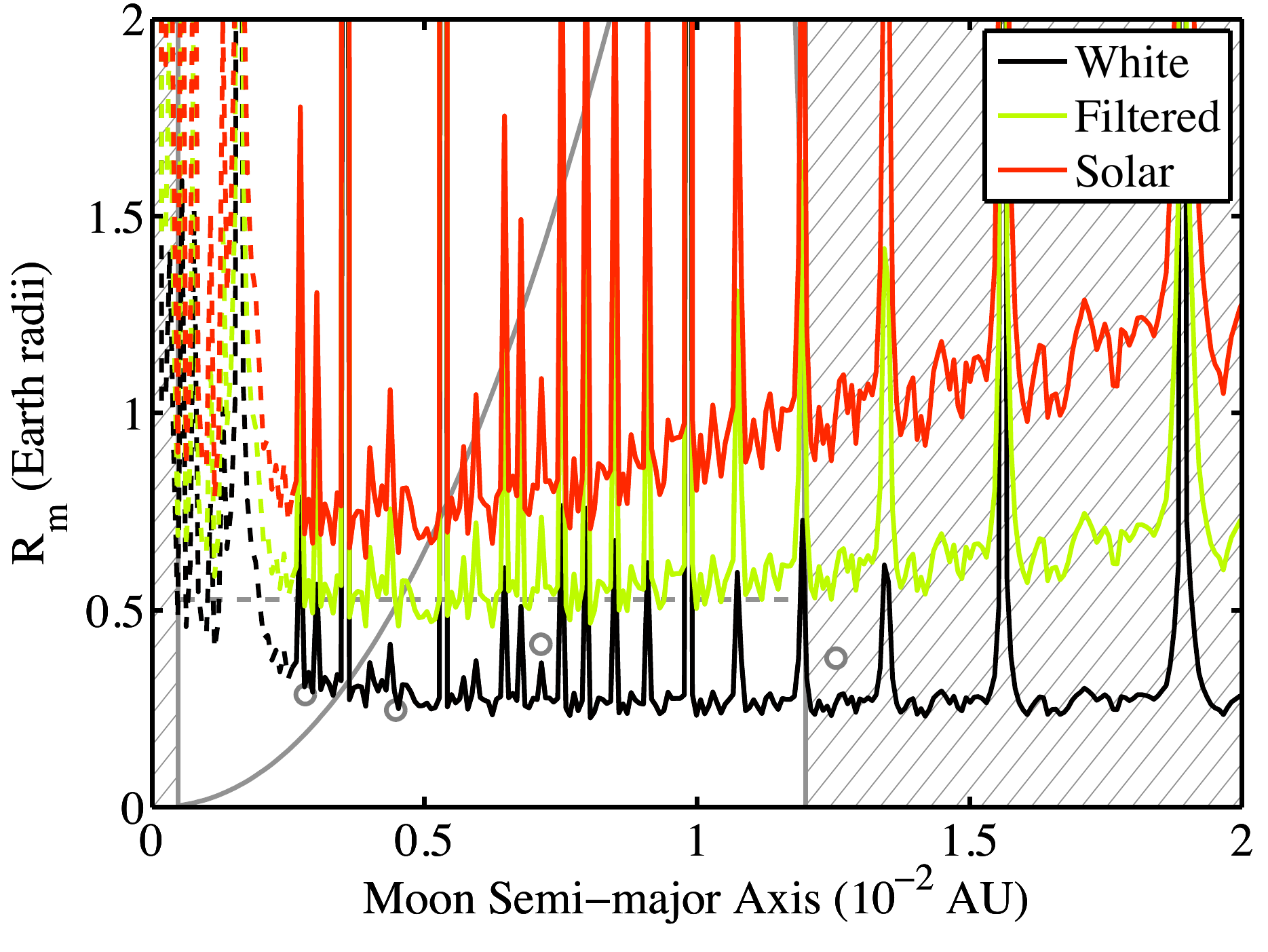}}
     \subfigure[$M_p = M_J$, $a_p=0.6$AU.]{
          \label{TransitThresh1s1MJ06AUEccA}
          \includegraphics[width=.315\textwidth]{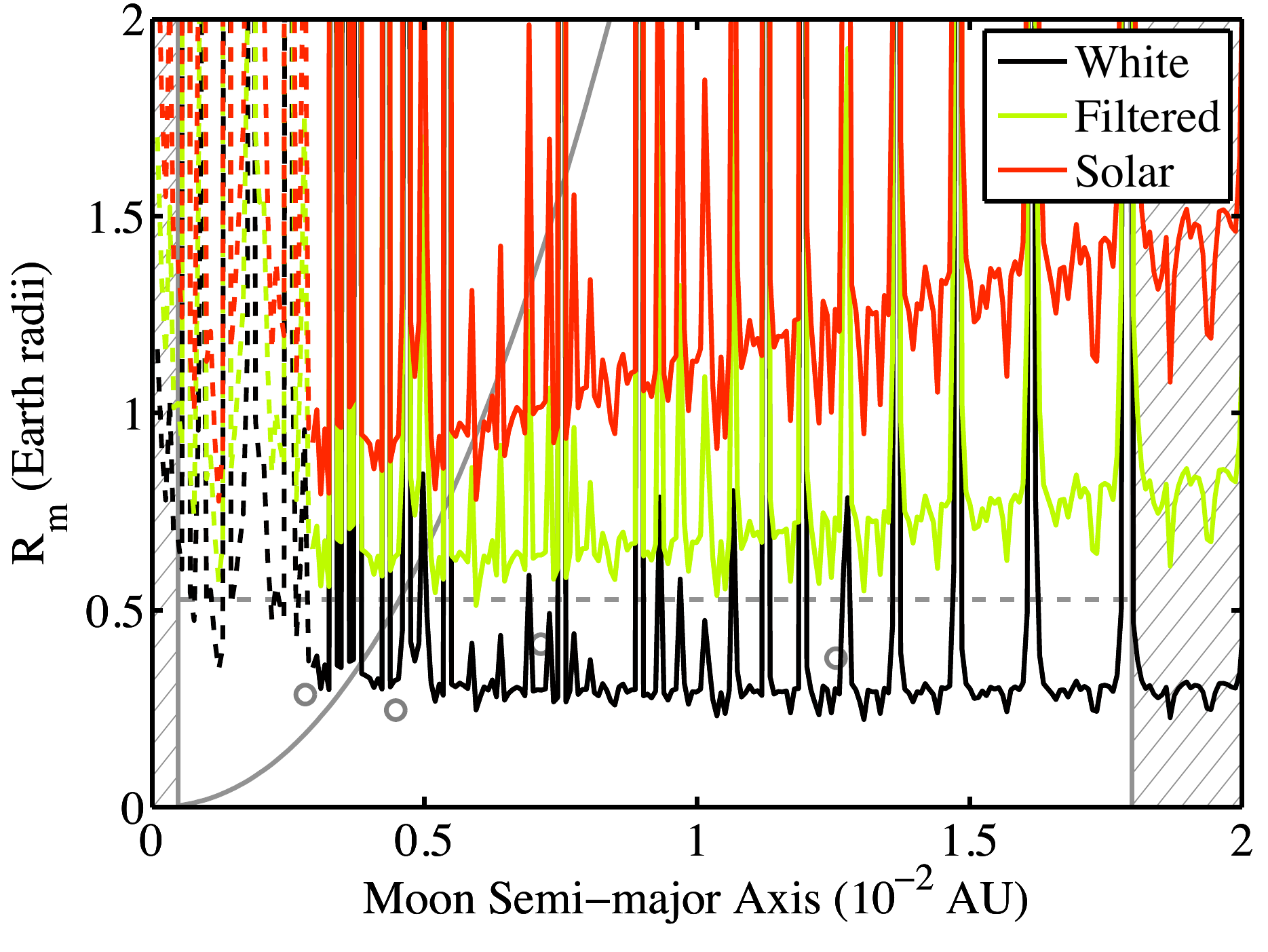}}\\ 
     \subfigure[$M_p = M_U$, $a_p=0.2$AU.]{
          \label{TransitThresh1s1MU02AUEccA}
          \includegraphics[width=.315\textwidth]{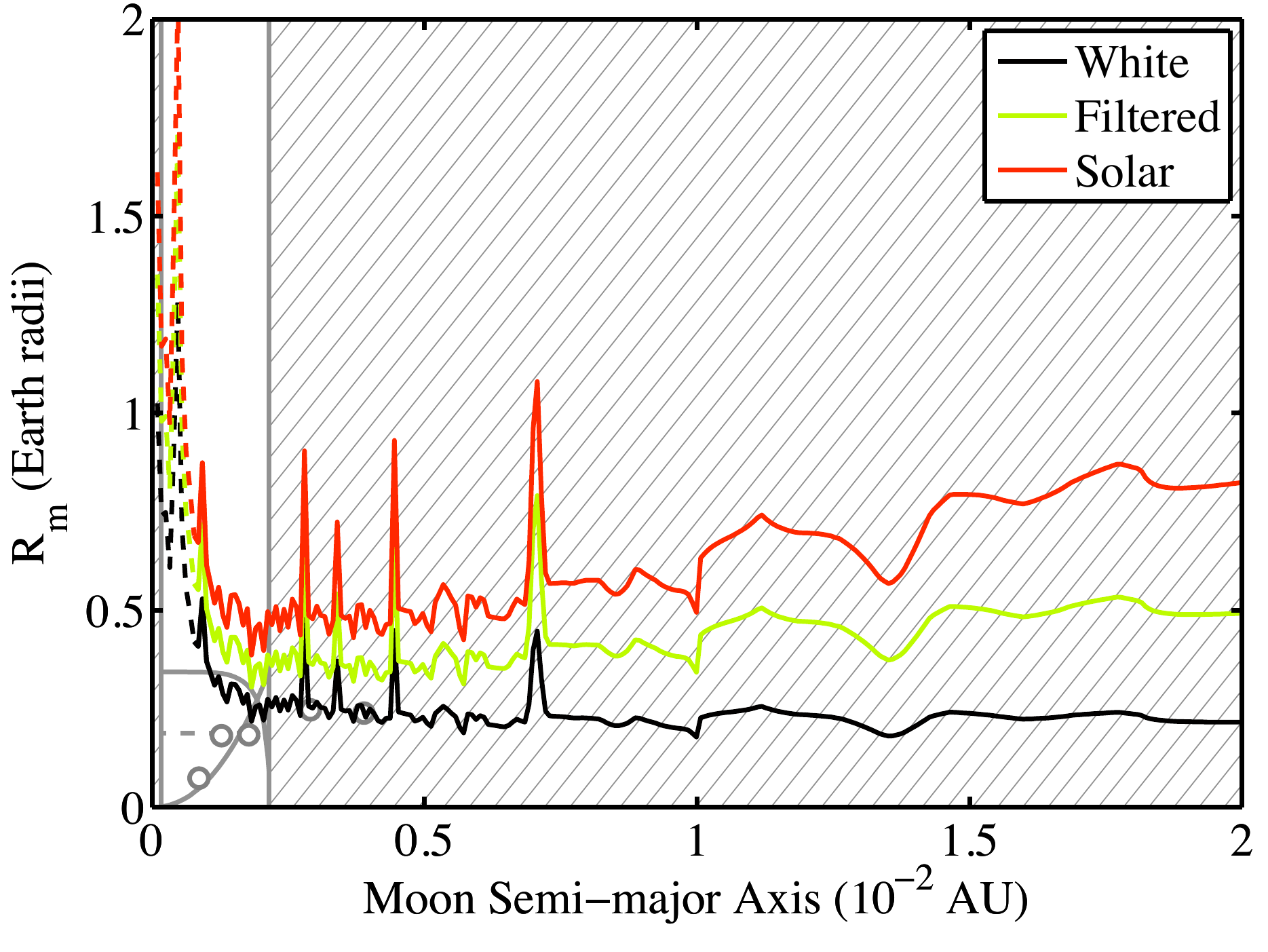}}
     \subfigure[$M_p = M_U$, $a_p=0.4$AU.]{
          \label{TransitThresh1s1MU04AUEccA}
          \includegraphics[width=.315\textwidth]{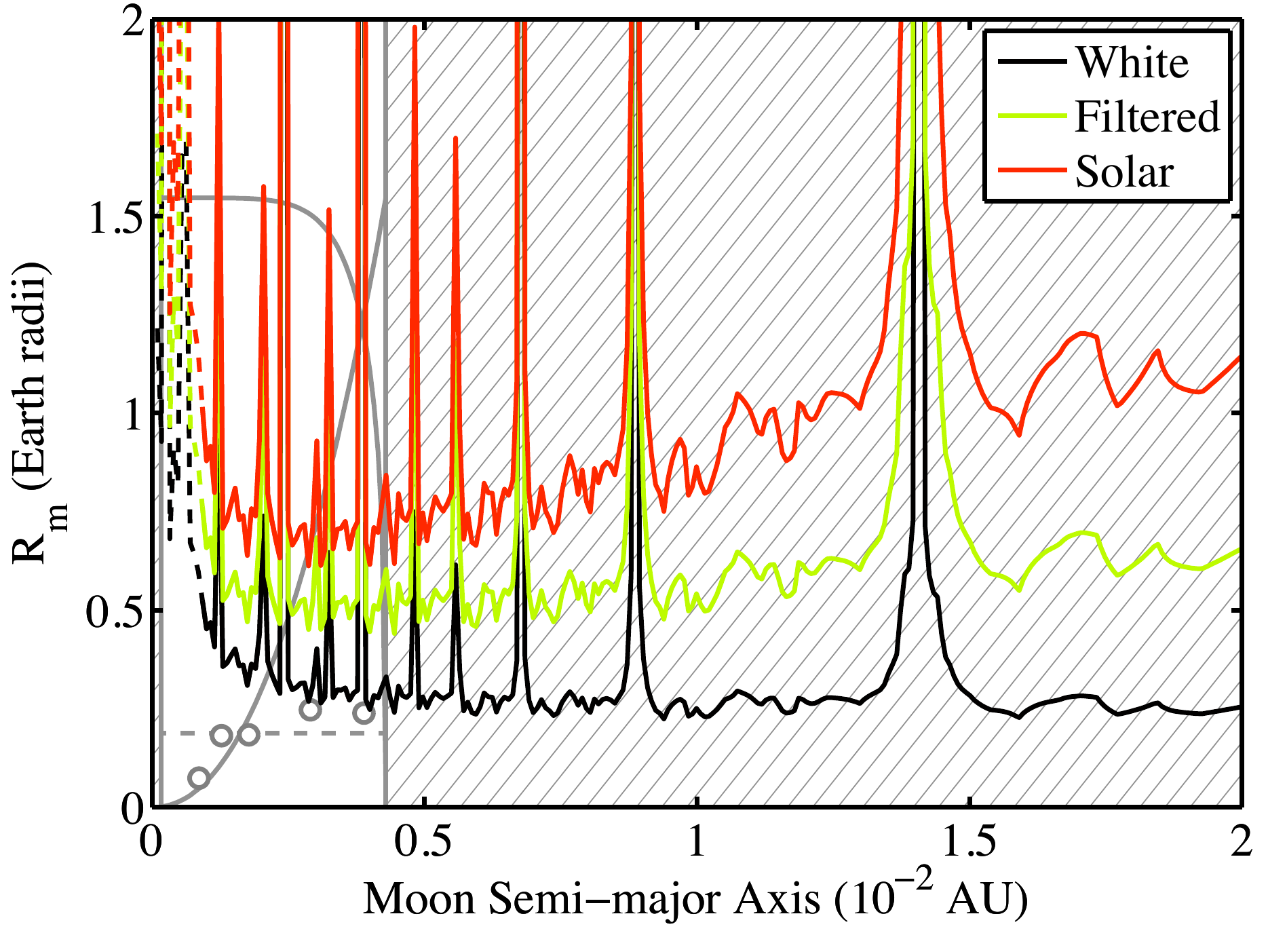}}
      \subfigure[$M_p = M_U$, $a_p=0.6$AU.]{
          \label{TransitThresh1s1MU06AUEccA}
          \includegraphics[width=.315\textwidth]{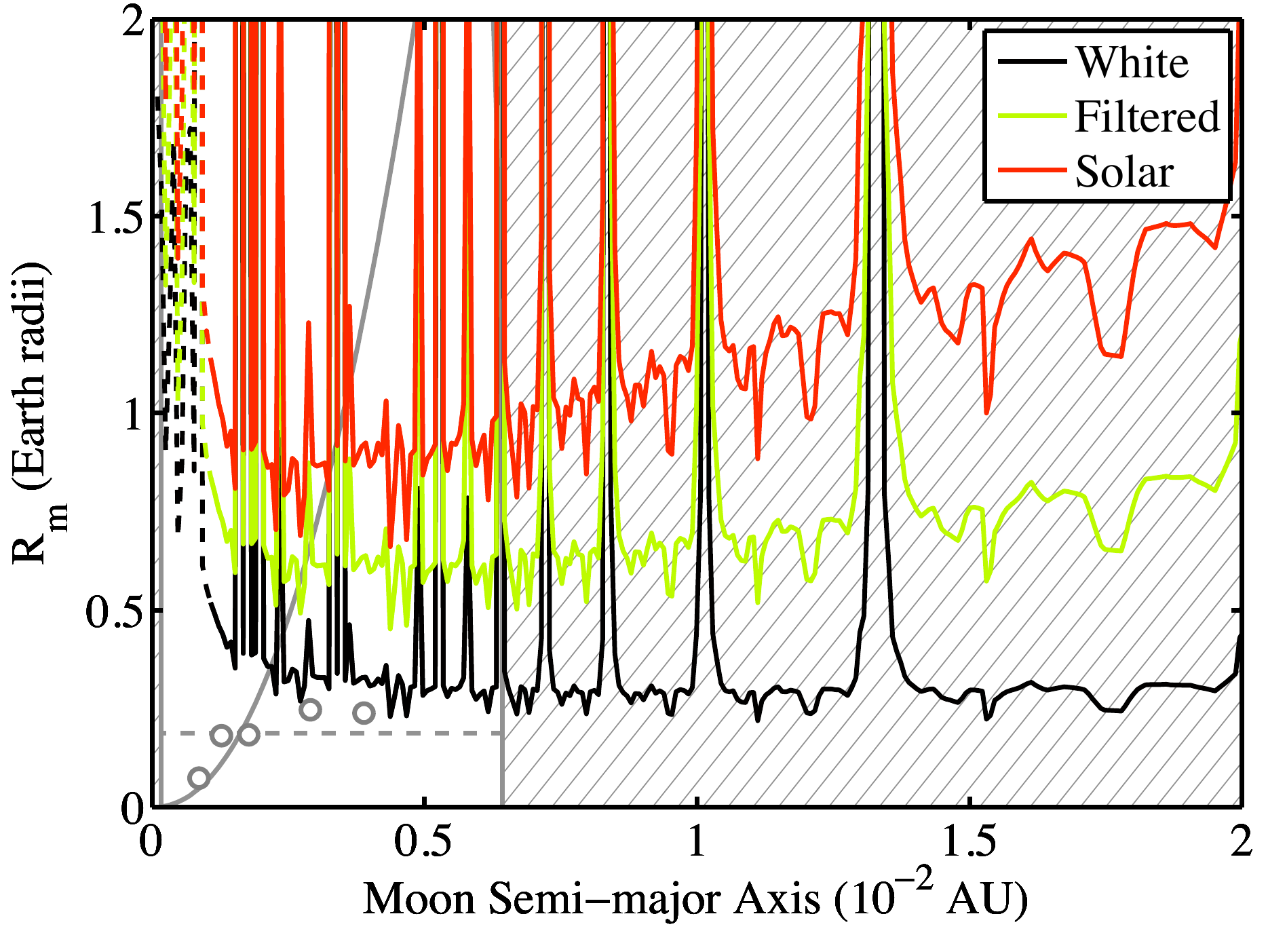}}\\ 
     \subfigure[$M_p = M_{\earth}$, $a_p=0.2$AU.]{
          \label{TransitThresh1s1ME02AUEccA}
          \includegraphics[width=.315\textwidth]{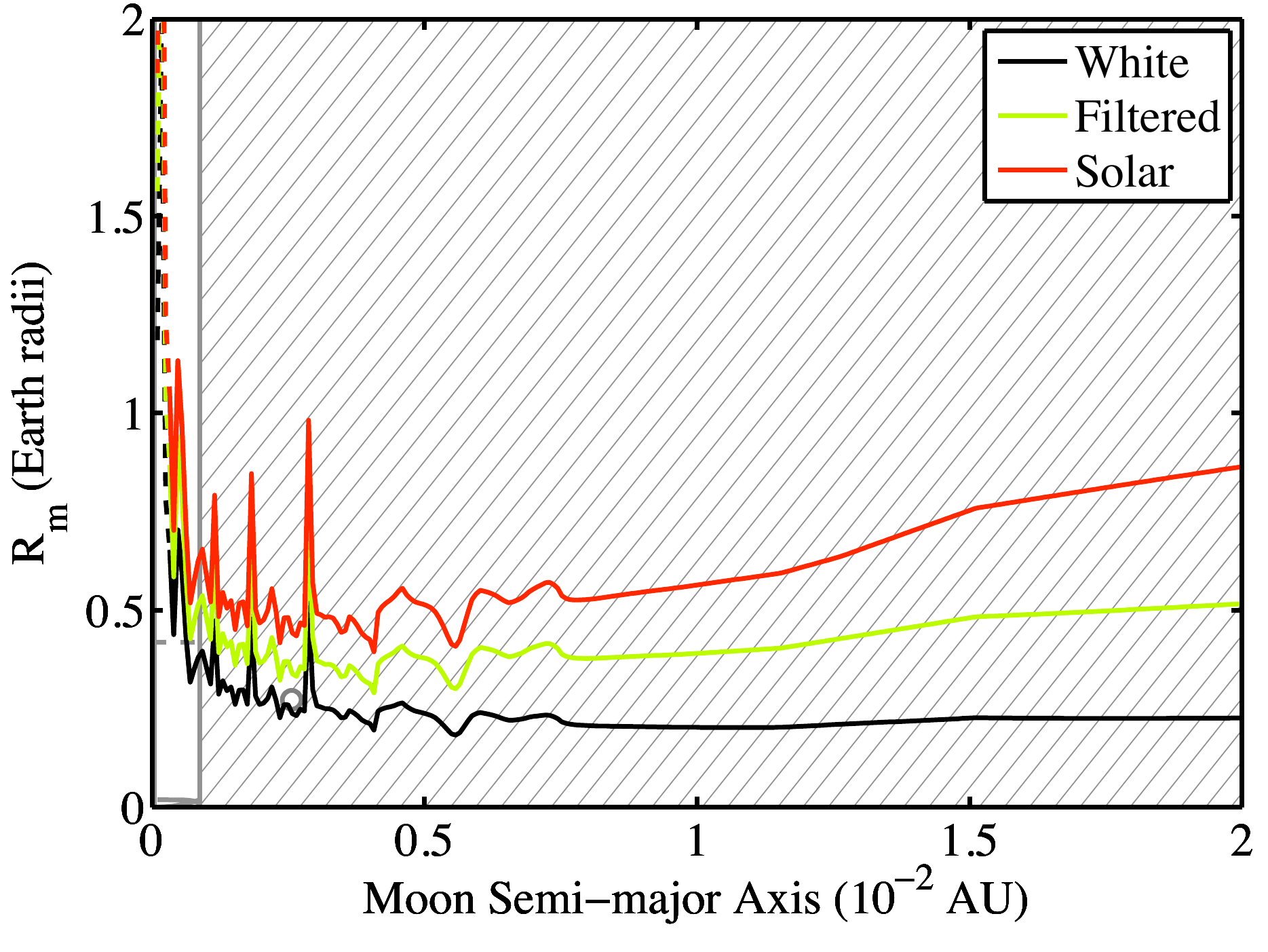}}
     \subfigure[$M_p = M_{\earth}$, $a_p=0.4$AU.]{
          \label{TransitThresh1s1ME04AUEccA}
          \includegraphics[width=.315\textwidth]{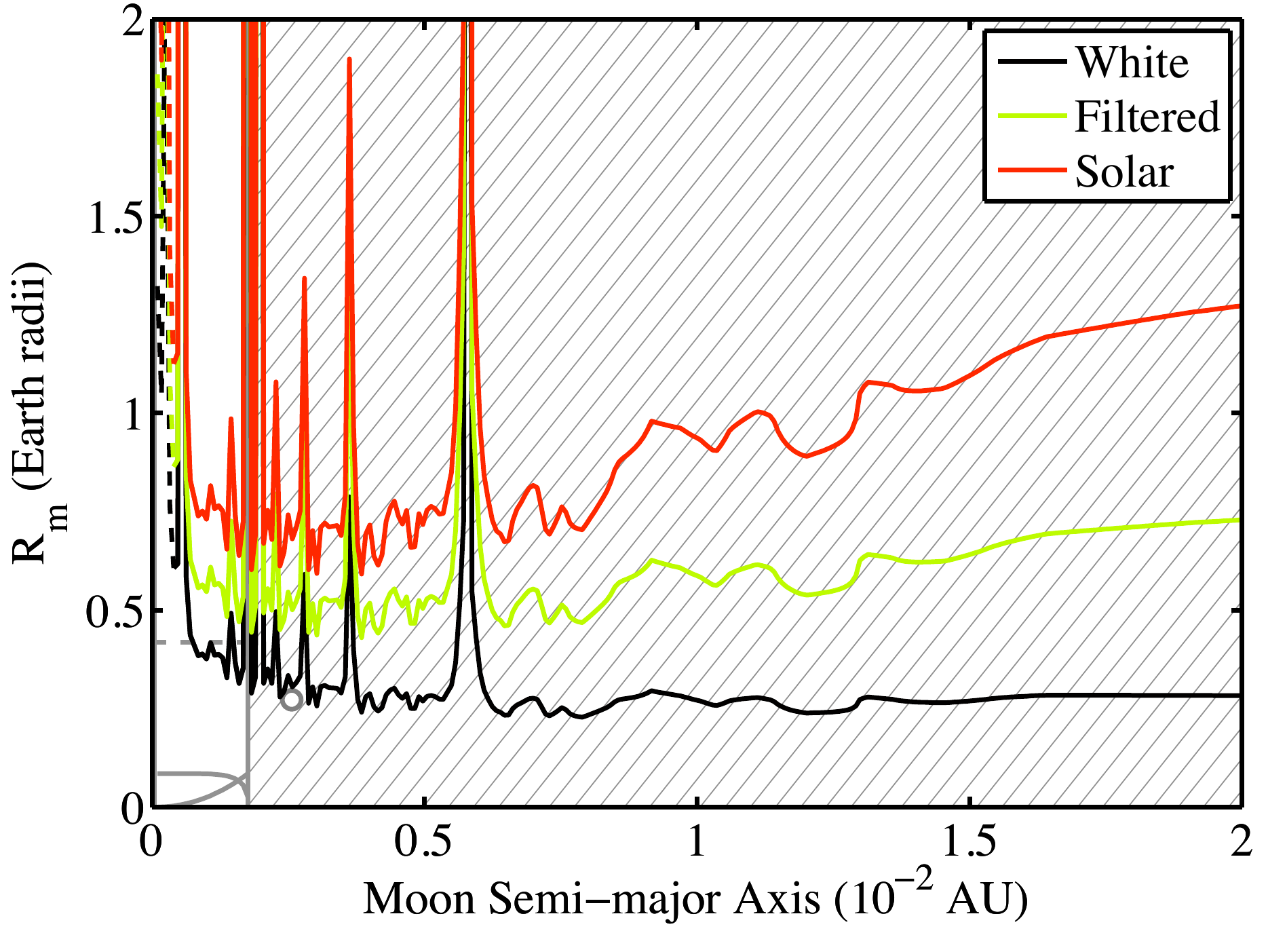}}
     \subfigure[$M_p = M_{\earth}$, $a_p=0.6$AU.]{
          \label{TransitThresh1s1ME06AUEccA}
          \includegraphics[width=.315\textwidth]{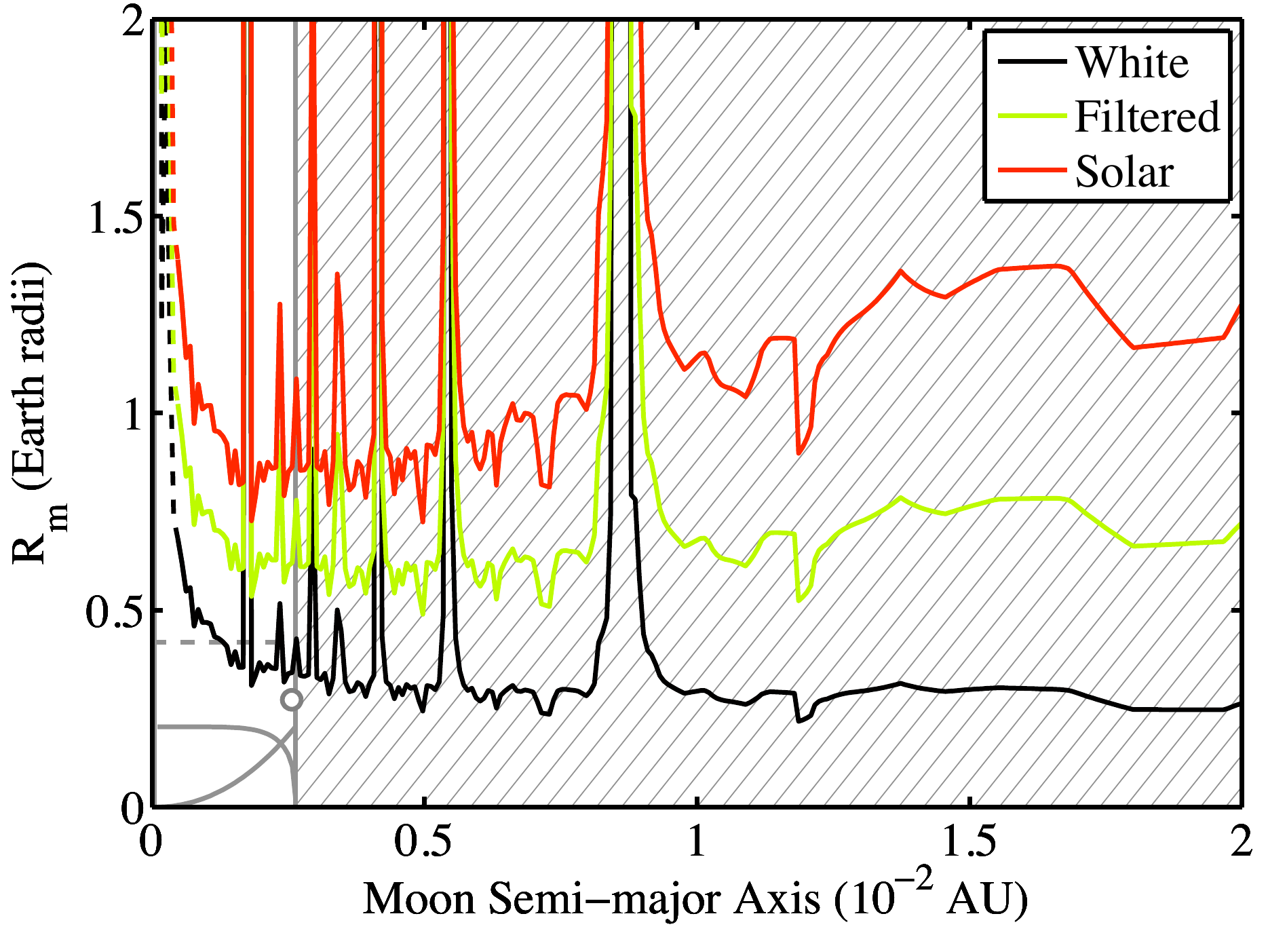}} 
     \caption{Figure of the same form as figure~\ref{MCThresholdsEccentricApo}, but showing the 68.3\% thresholds.}
     \label{MCThresholdsEccentricApo1S}
\end{figure}

\begin{figure}
     \centering
     \subfigure[$M_p$=$10 M_J$, $a_p=0.2$AU.]{
          \label{TransitThresh2s10MJ02AUcc}
          \includegraphics[width=.315\textwidth]{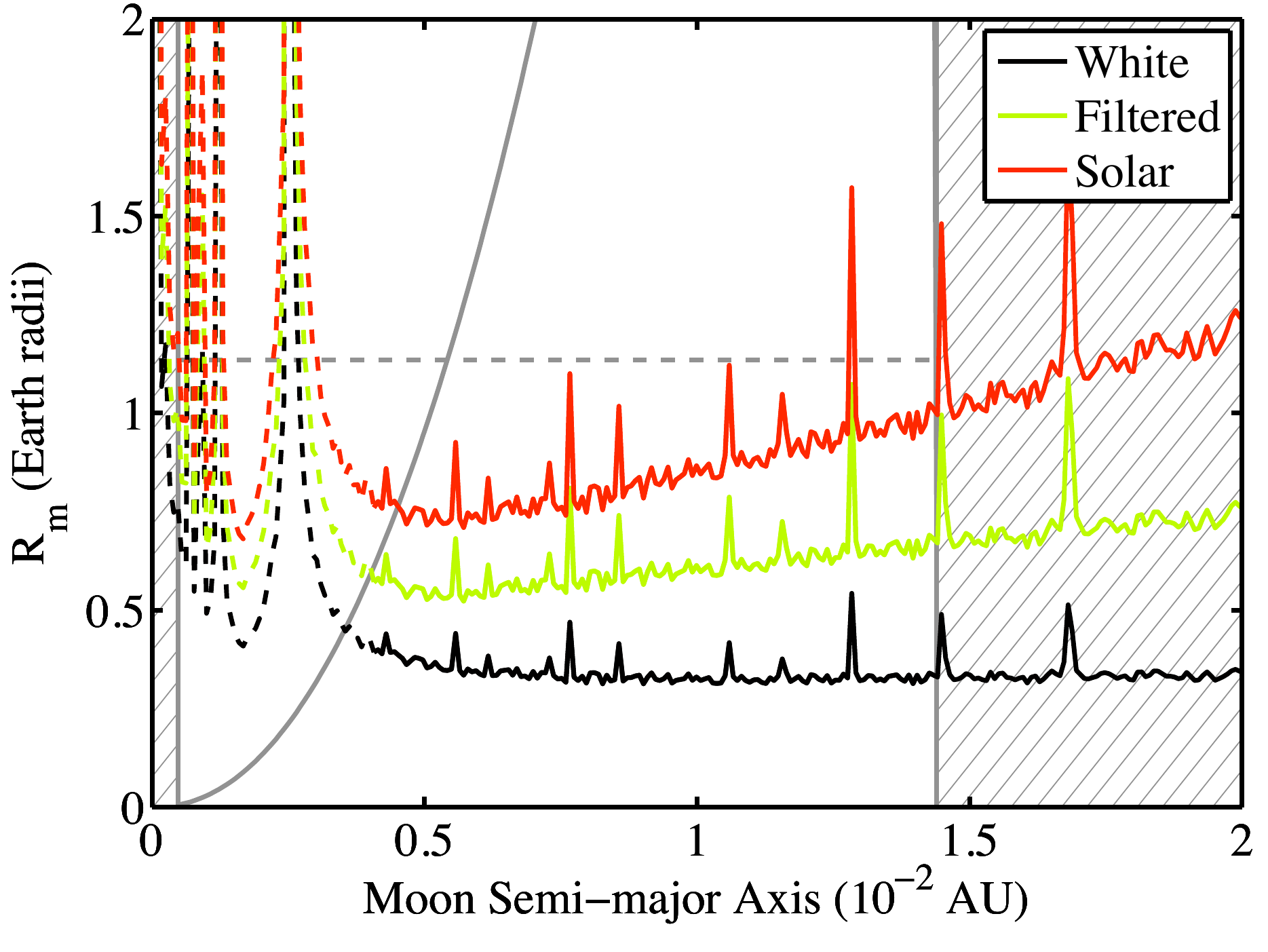}}
     \subfigure[$M_p$=$10 M_J$, $a_p=0.4$AU.]{
          \label{TransitThresh2s10MJ04AUcc}
          \includegraphics[width=.315\textwidth]{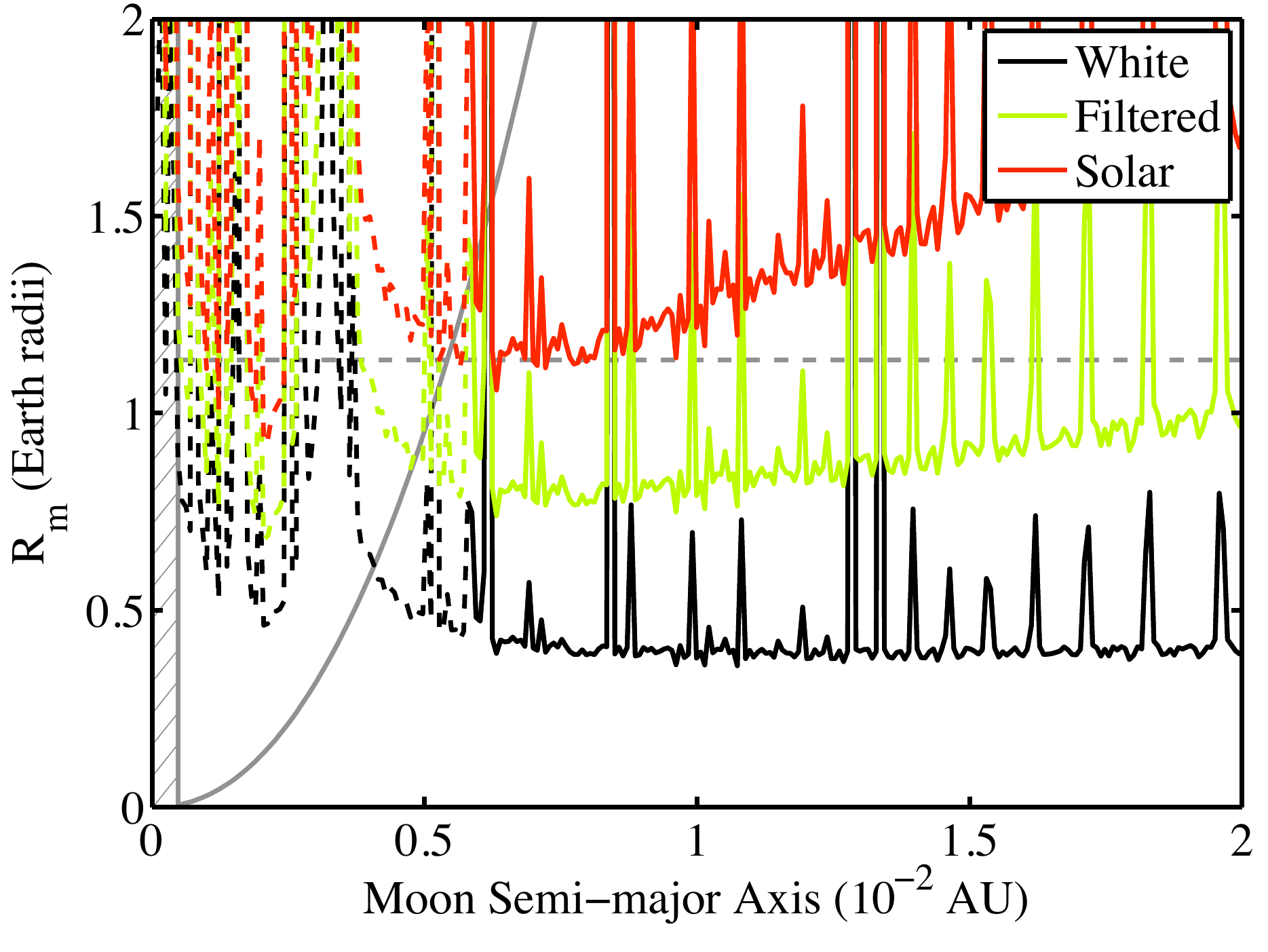}}
     \subfigure[$M_p$=$10 M_J$, $a_p=0.6$AU.]{
          \label{TransitThresh2s10MJ06AUcc}
          \includegraphics[width=.315\textwidth]{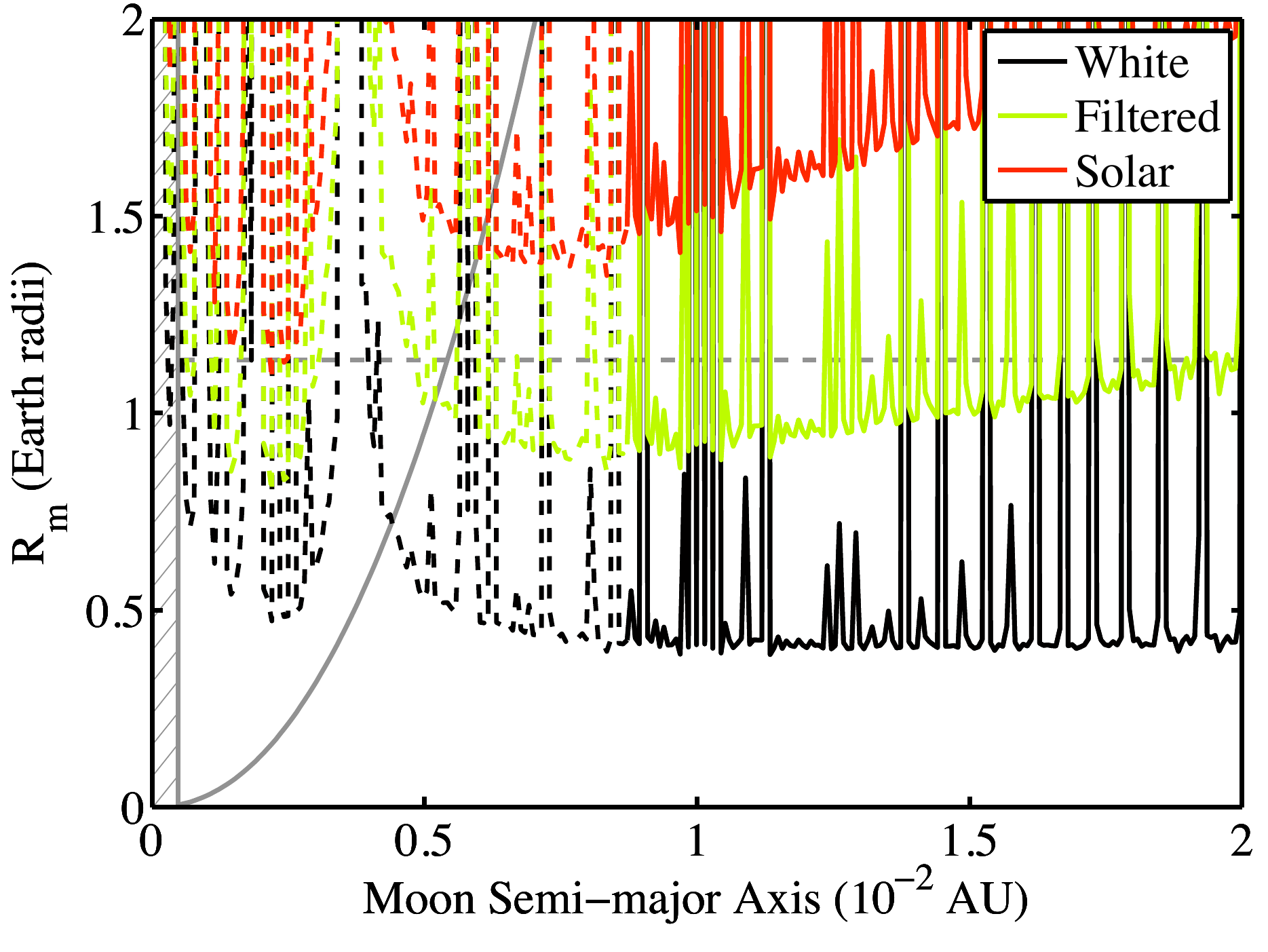}}\\ 
     \subfigure[$M_p = M_J$, $a_p=0.2$AU.]{
          \label{TransitThresh2s1MJ02AUcc}
          \includegraphics[width=.315\textwidth]{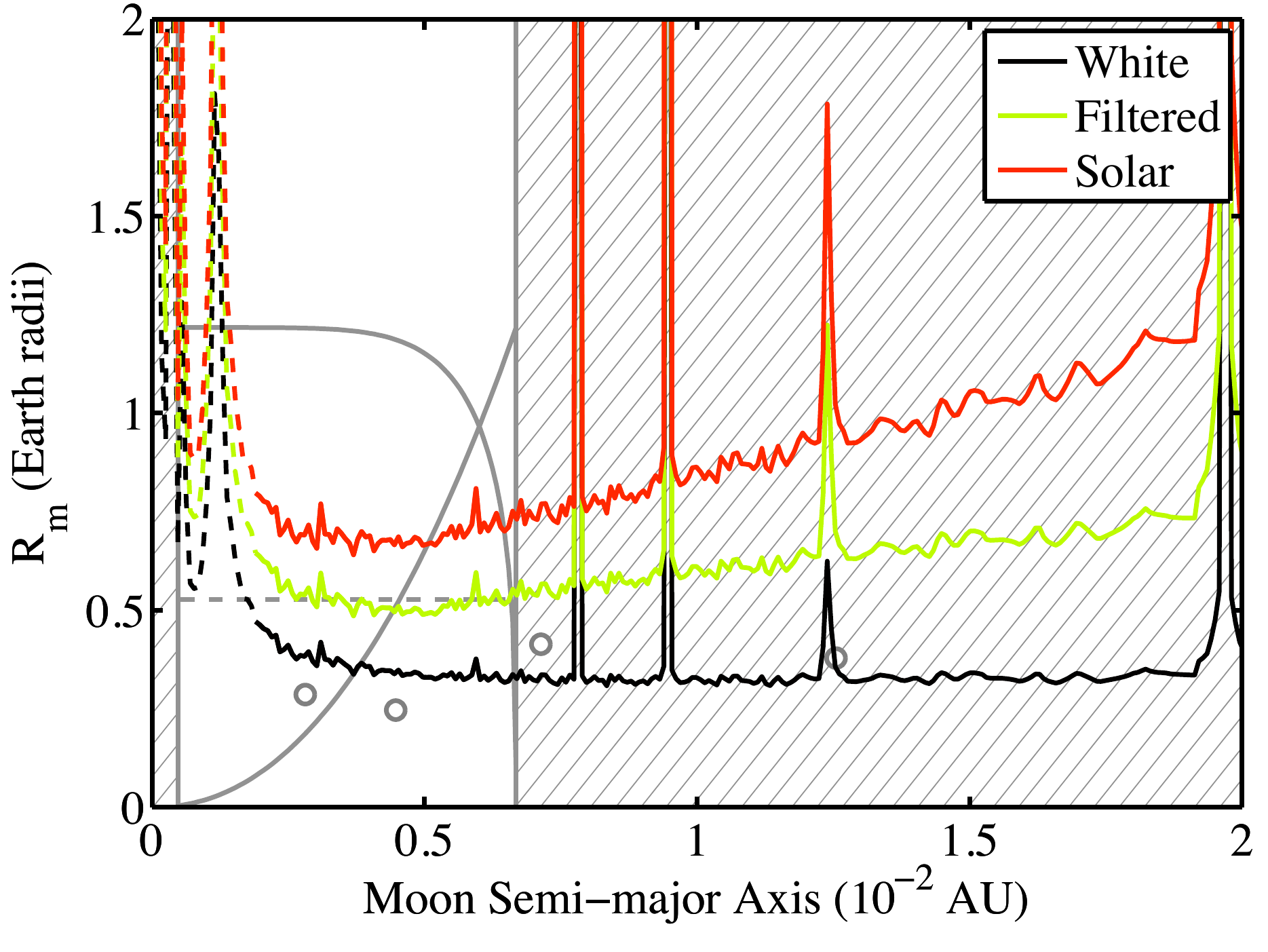}}
      \subfigure[$M_p = M_J$, $a_p=0.4$AU.]{
          \label{TransitThresh2s1MJ04AUcc}
          \includegraphics[width=.315\textwidth]{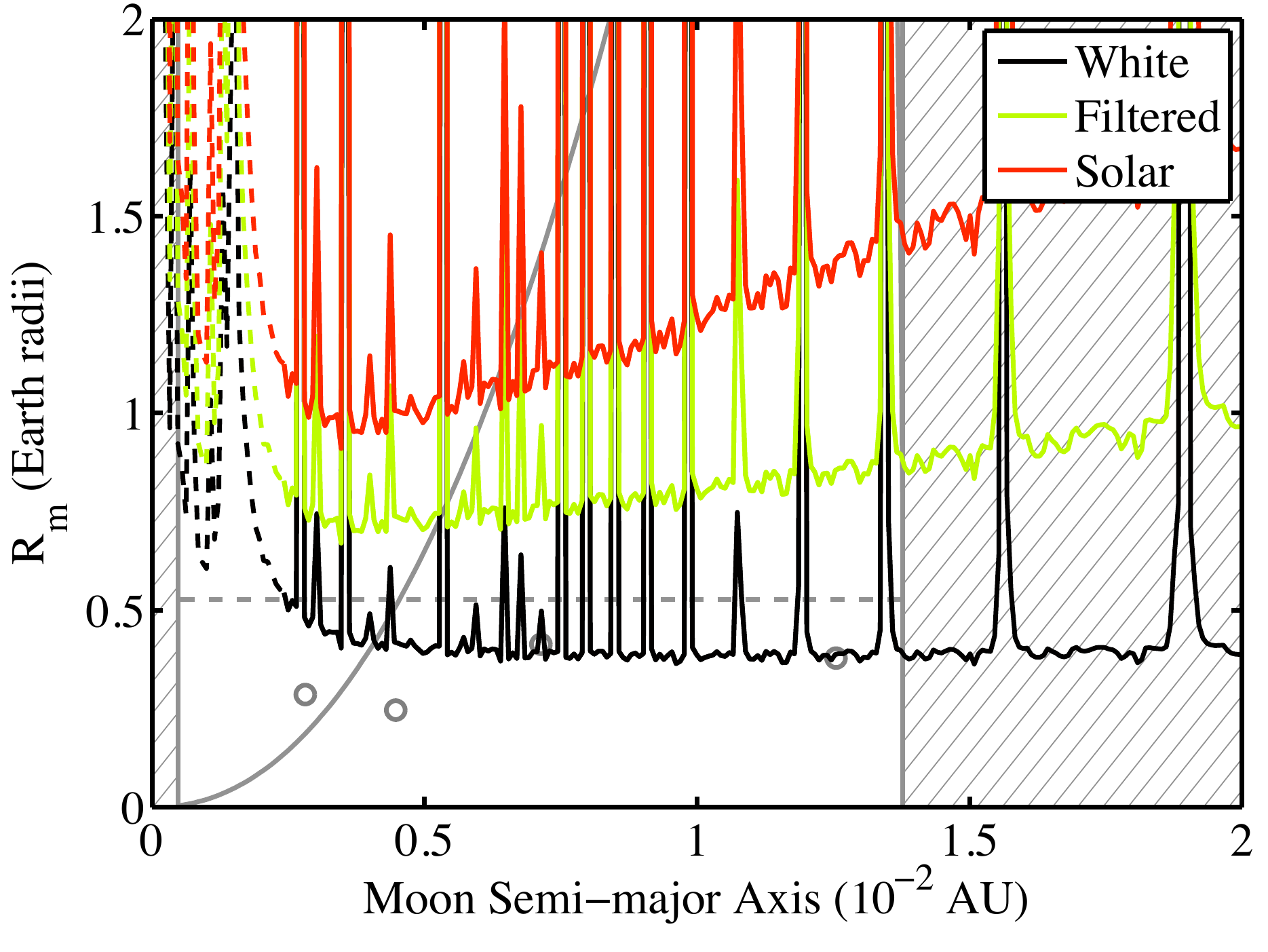}}
     \subfigure[$M_p = M_J$, $a_p=0.6$AU.]{
          \label{TransitThresh2s1MJ06AUcc}
          \includegraphics[width=.315\textwidth]{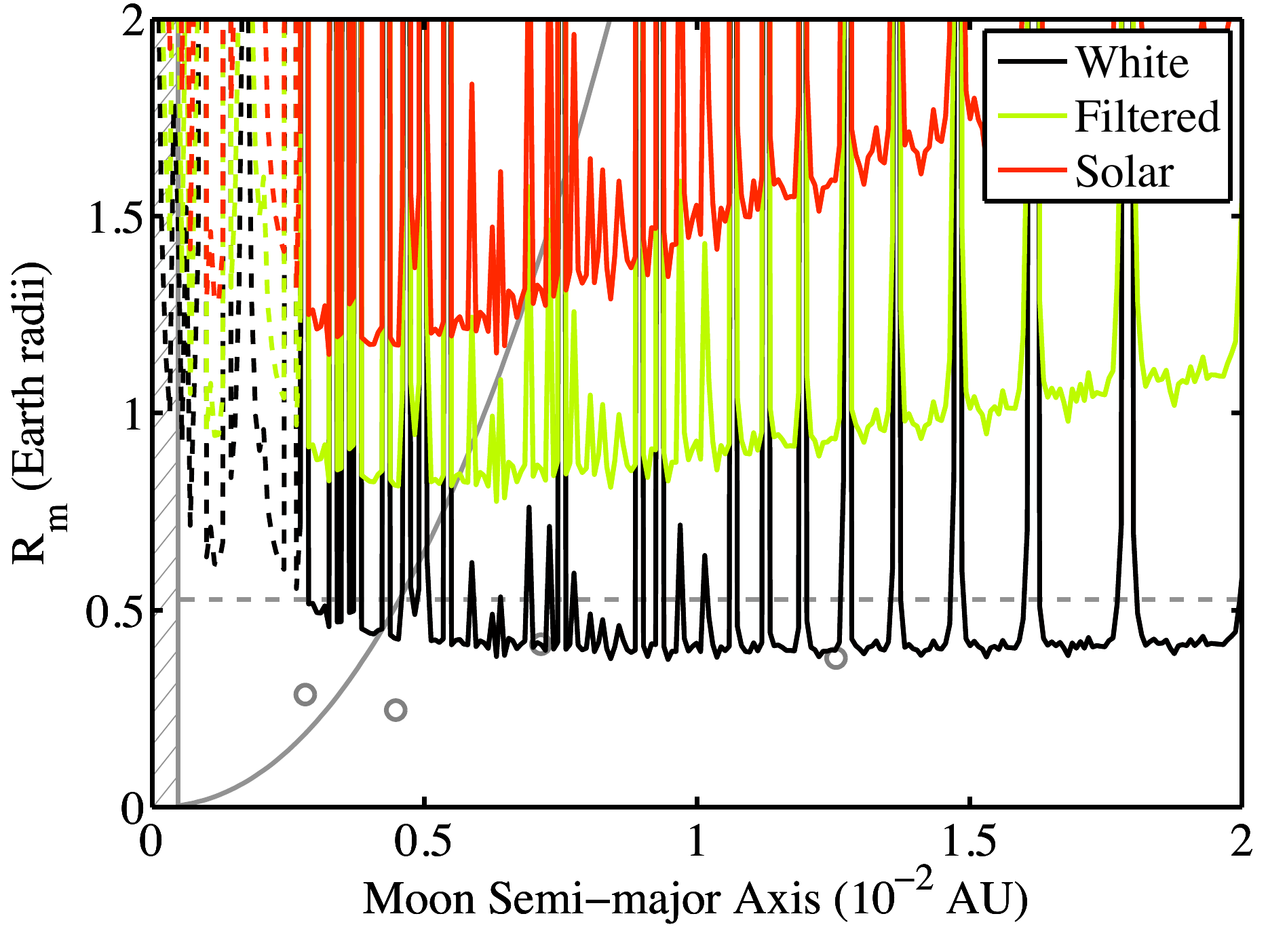}}\\ 
     \subfigure[$M_p = M_U$, $a_p=0.2$AU.]{
          \label{TransitThresh2s1MU02AUcc}
          \includegraphics[width=.315\textwidth]{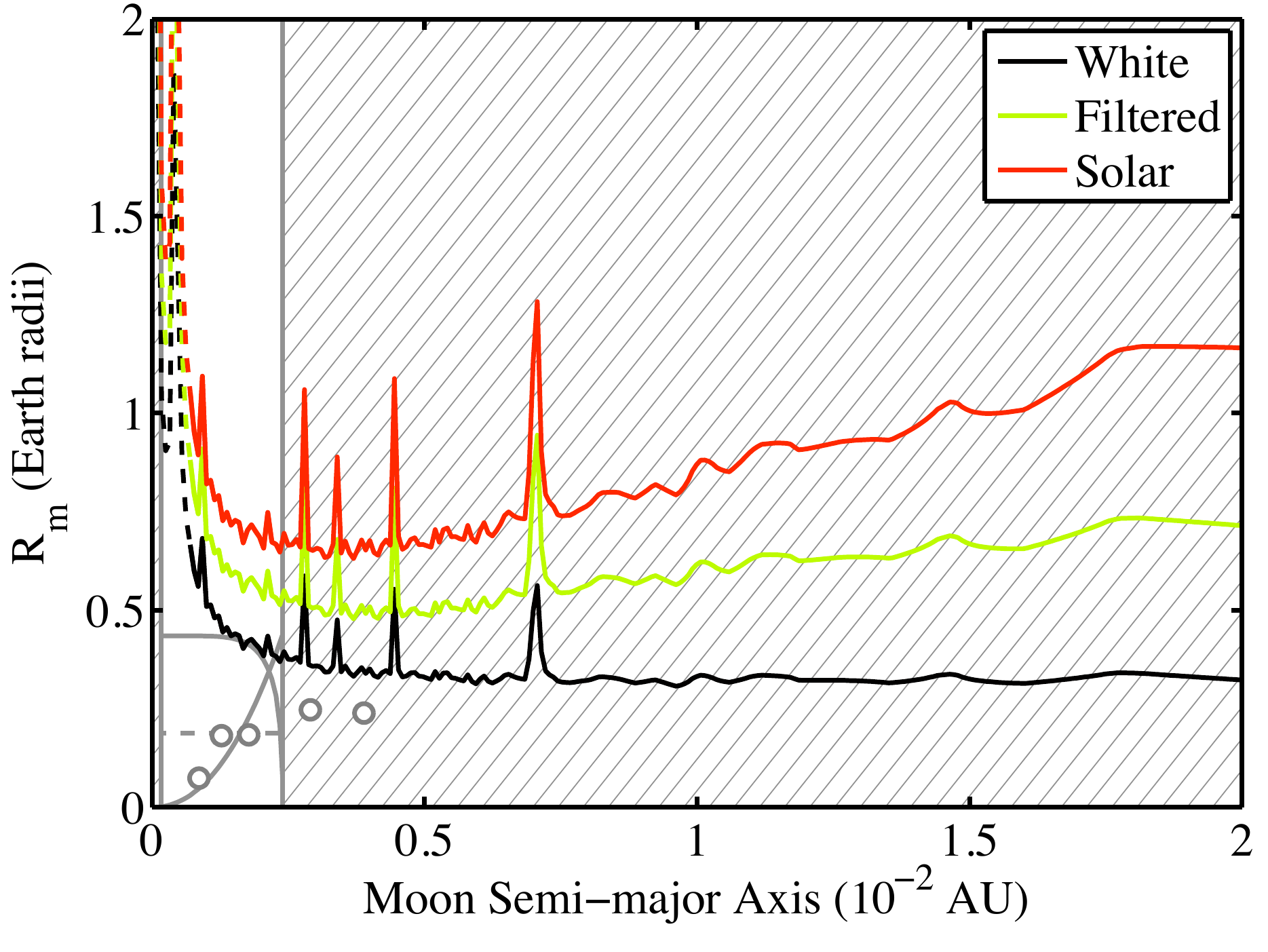}}
     \subfigure[$M_p = M_U$, $a_p=0.4$AU.]{
          \label{TransitThresh2s1MU04AUcc}
          \includegraphics[width=.315\textwidth]{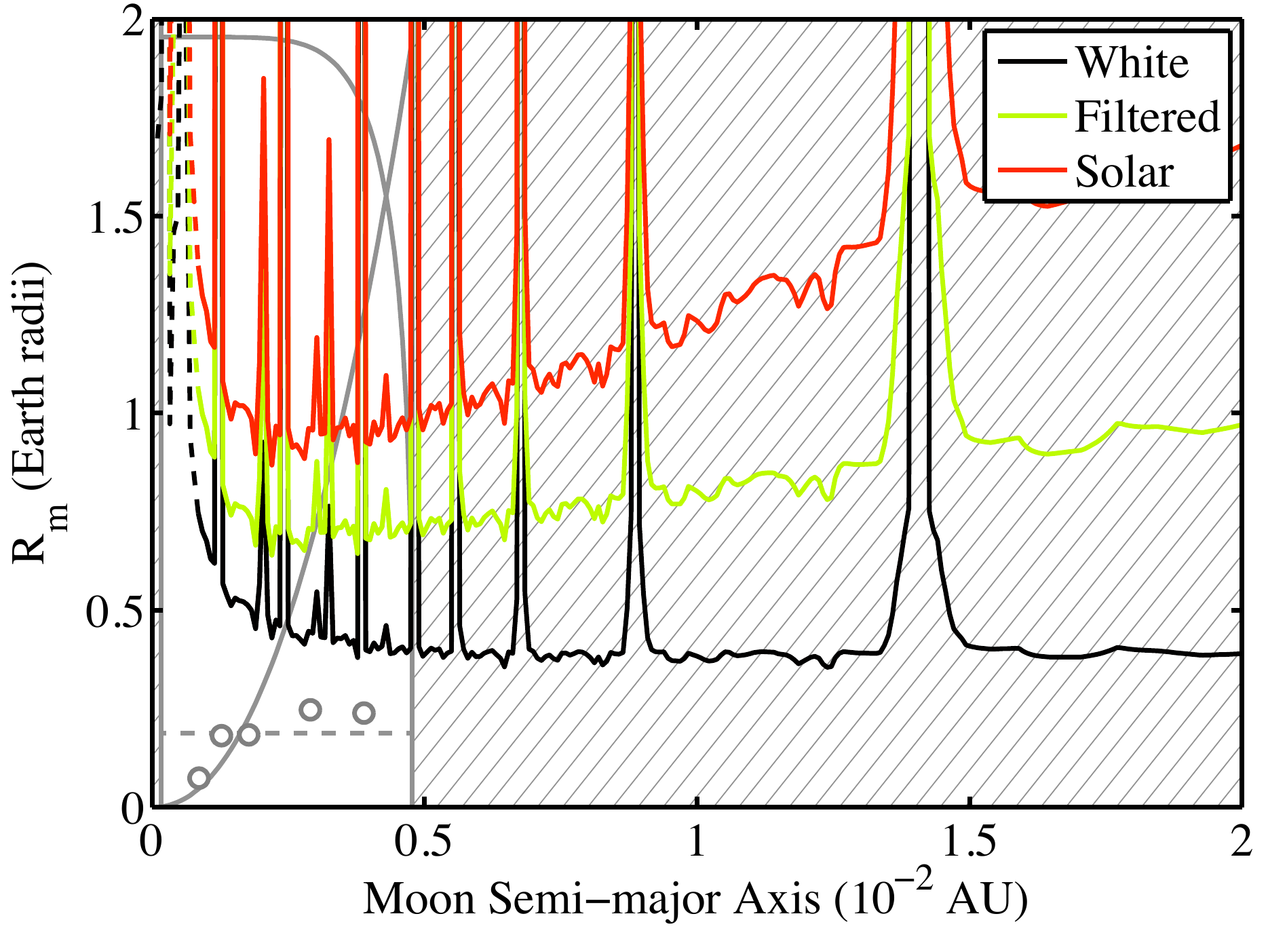}}
      \subfigure[$M_p = M_U$, $a_p=0.6$AU.]{
          \label{TransitThresh2s1MU06AUcc}
          \includegraphics[width=.315\textwidth]{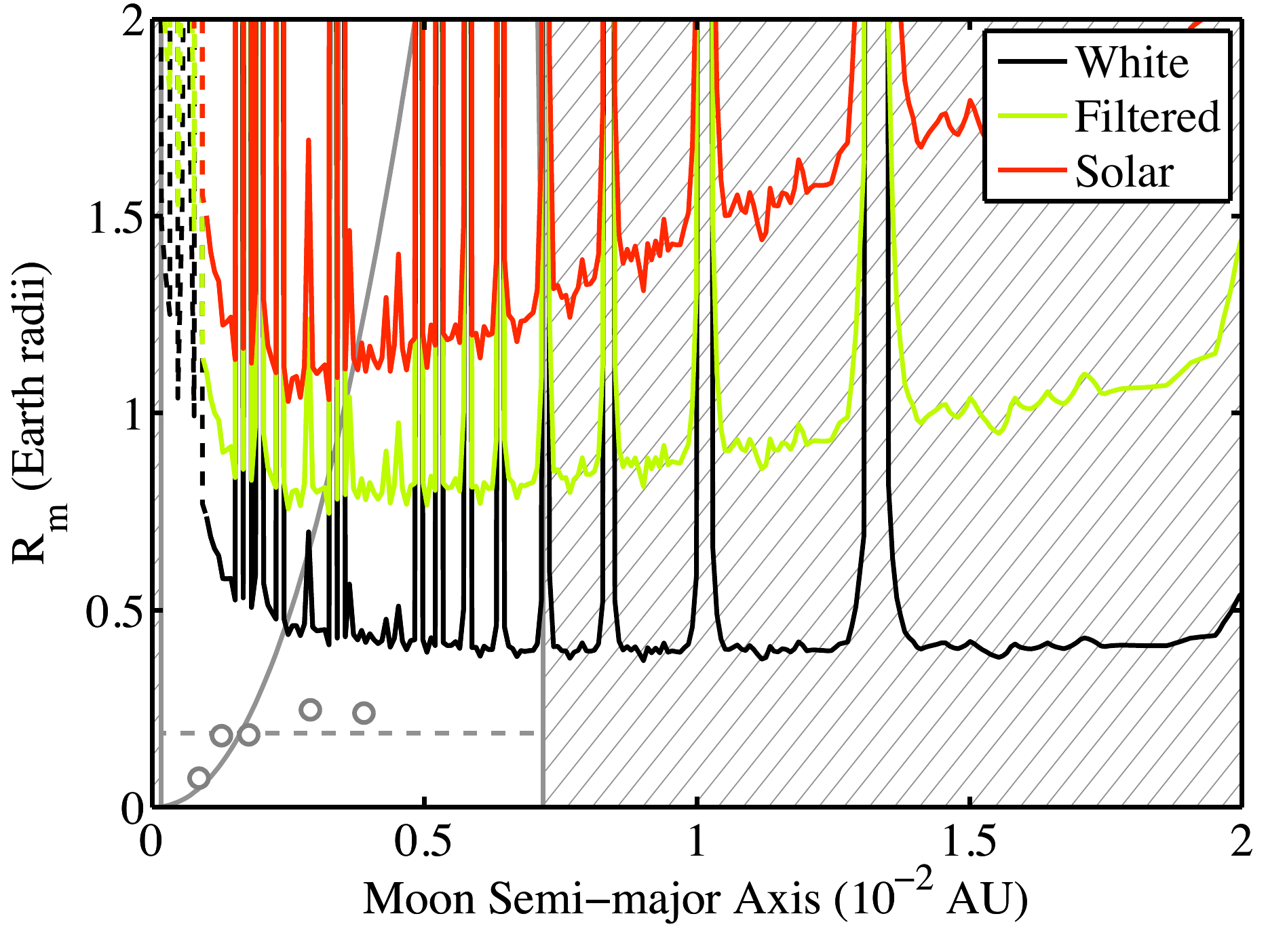}}\\ 
     \subfigure[$M_p = M_{\earth}$, $a_p=0.2$AU.]{
          \label{TransitThresh2s1ME02AUcc}
          \includegraphics[width=.315\textwidth]{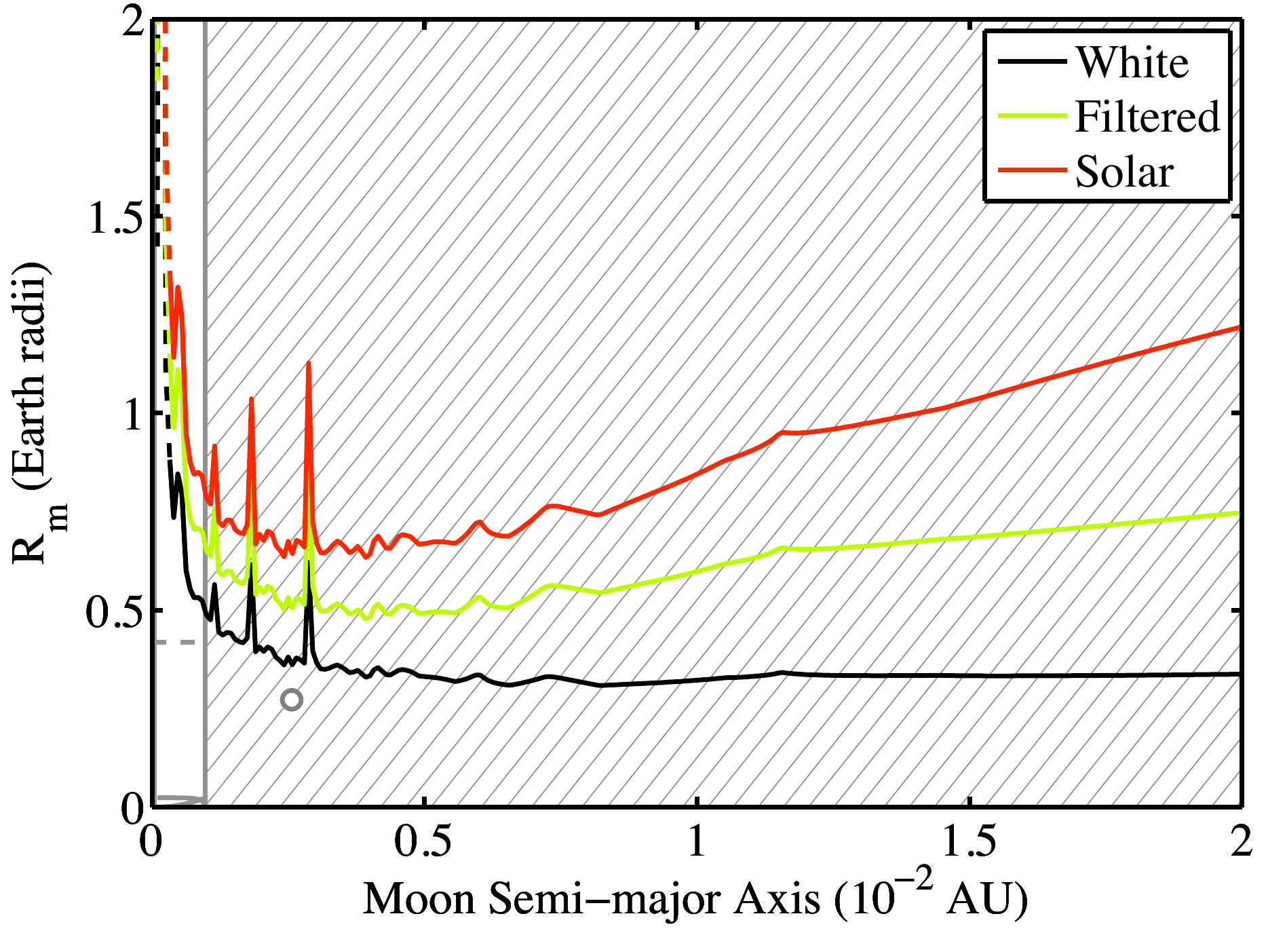}}
     \subfigure[$M_p = M_{\earth}$, $a_p=0.4$AU.]{
          \label{TransitThresh2s1ME04AUcc}
          \includegraphics[width=.315\textwidth]{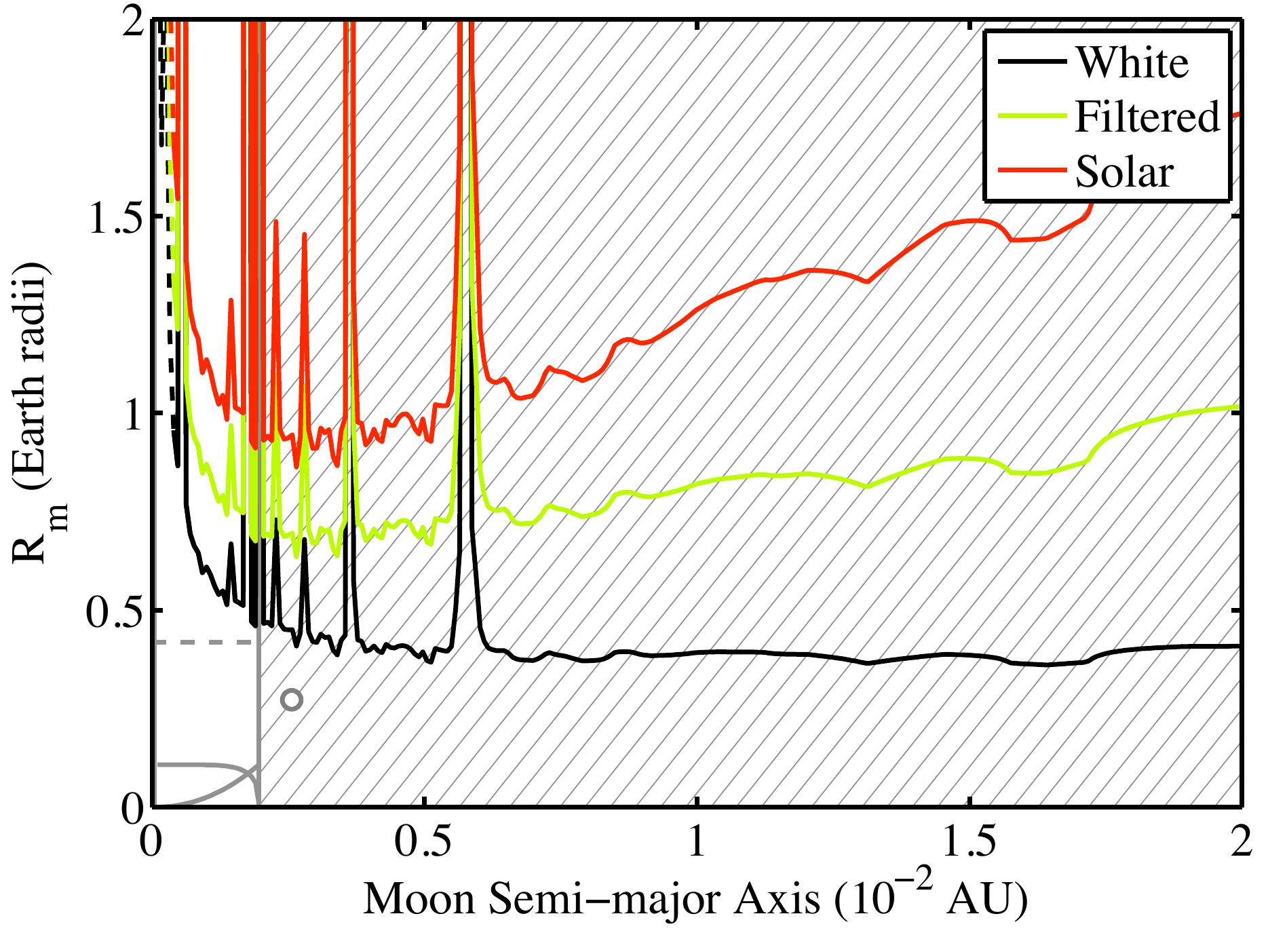}}
     \subfigure[$M_p = M_{\earth}$, $a_p=0.6$AU.]{
          \label{TransitThresh2s1ME06AUcc}
          \includegraphics[width=.315\textwidth]{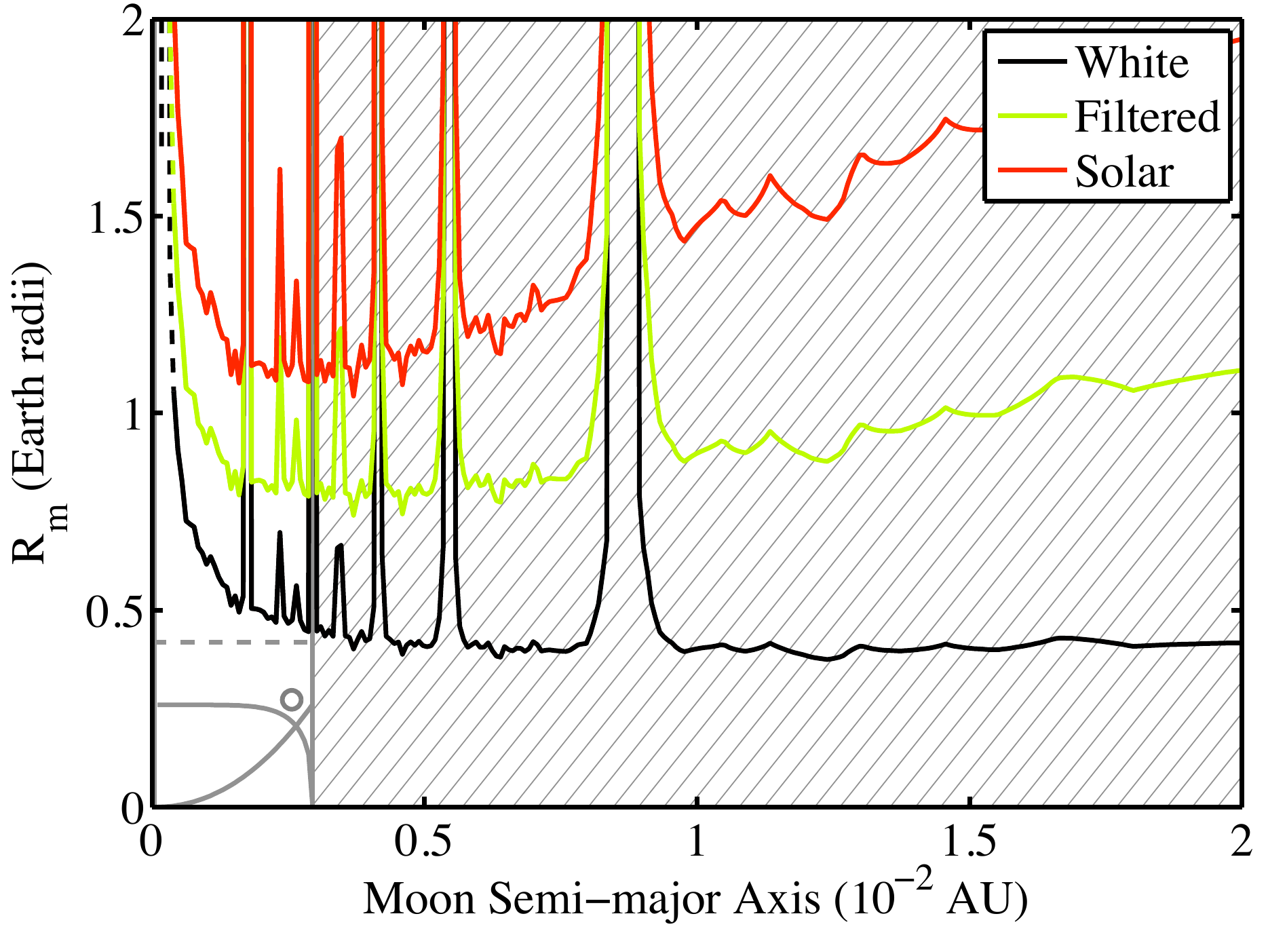}} 
     \caption{Figure of the same form as figure~\ref{MCThresholdsAligned}, but showing the 95.4\% thresholds.}
     \label{MCThresholdsAligned2S}
\end{figure}

\begin{figure}
     \centering
     \subfigure[$M_p$=$10 M_J$, $a_p=0.2$AU.]{
          \label{TransitThresh2s10MJ02AUInc}
          \includegraphics[width=.315\textwidth]{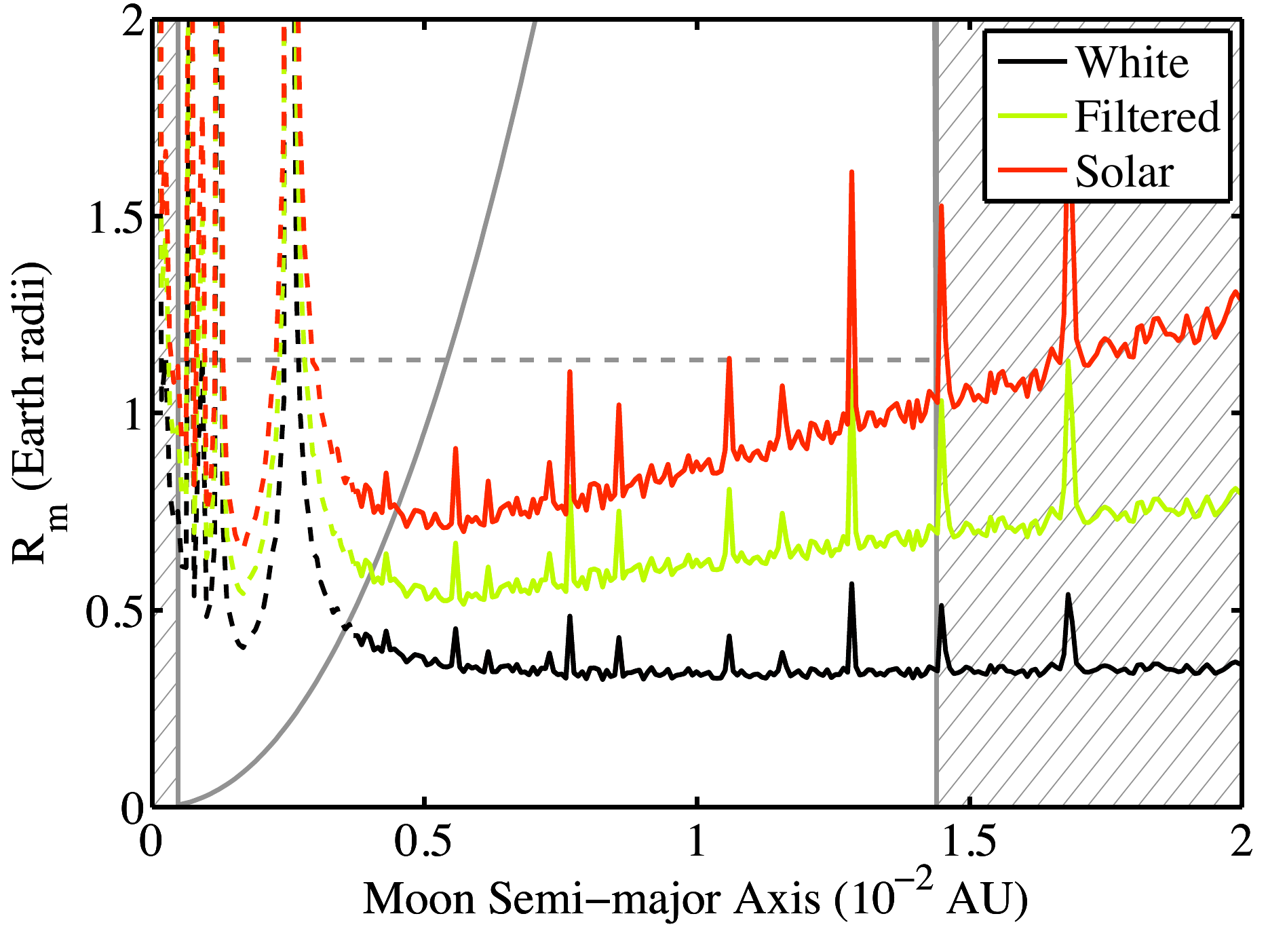}}
     \subfigure[$M_p$=$10 M_J$, $a_p=0.4$AU.]{
          \label{TransitThresh2s10MJ04AUInc}
          \includegraphics[width=.315\textwidth]{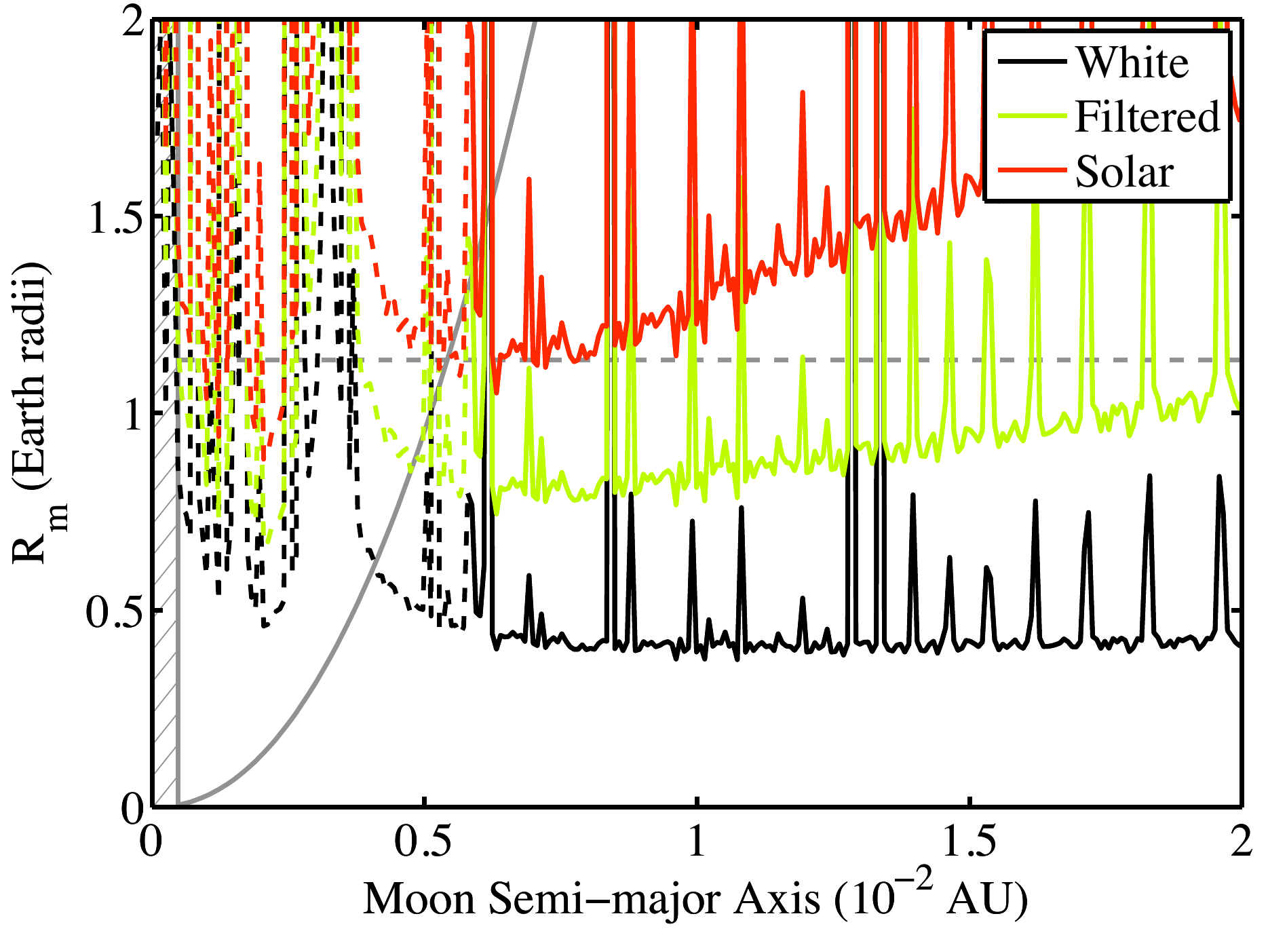}}
     \subfigure[$M_p$=$10 M_J$, $a_p=0.6$AU.]{
          \label{TransitThresh2s10MJ06AUInc}
          \includegraphics[width=.315\textwidth]{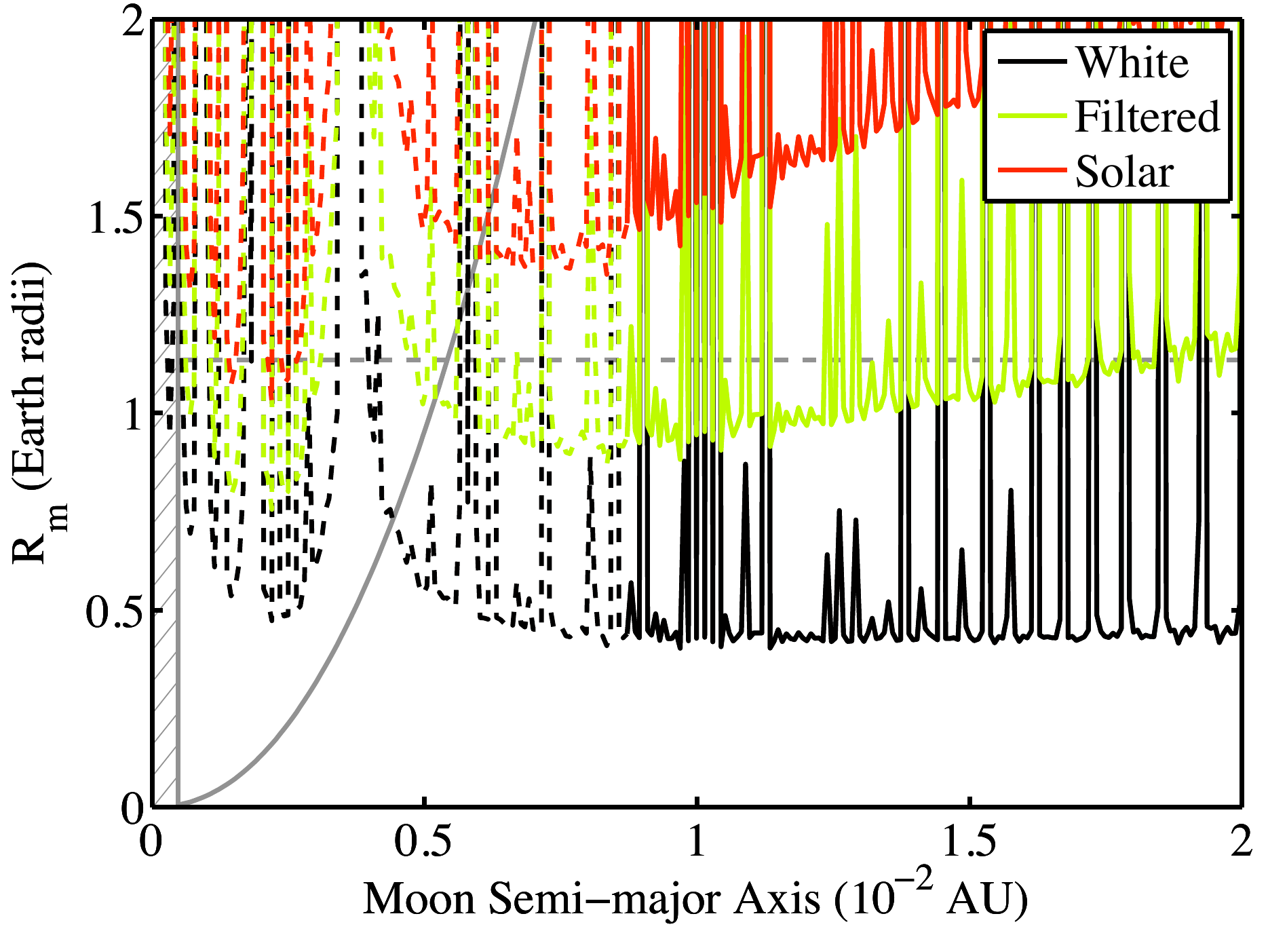}}\\ 
     \subfigure[$M_p = M_J$, $a_p=0.2$AU.]{
          \label{TransitThresh2s1MJ02AUInc}
          \includegraphics[width=.315\textwidth]{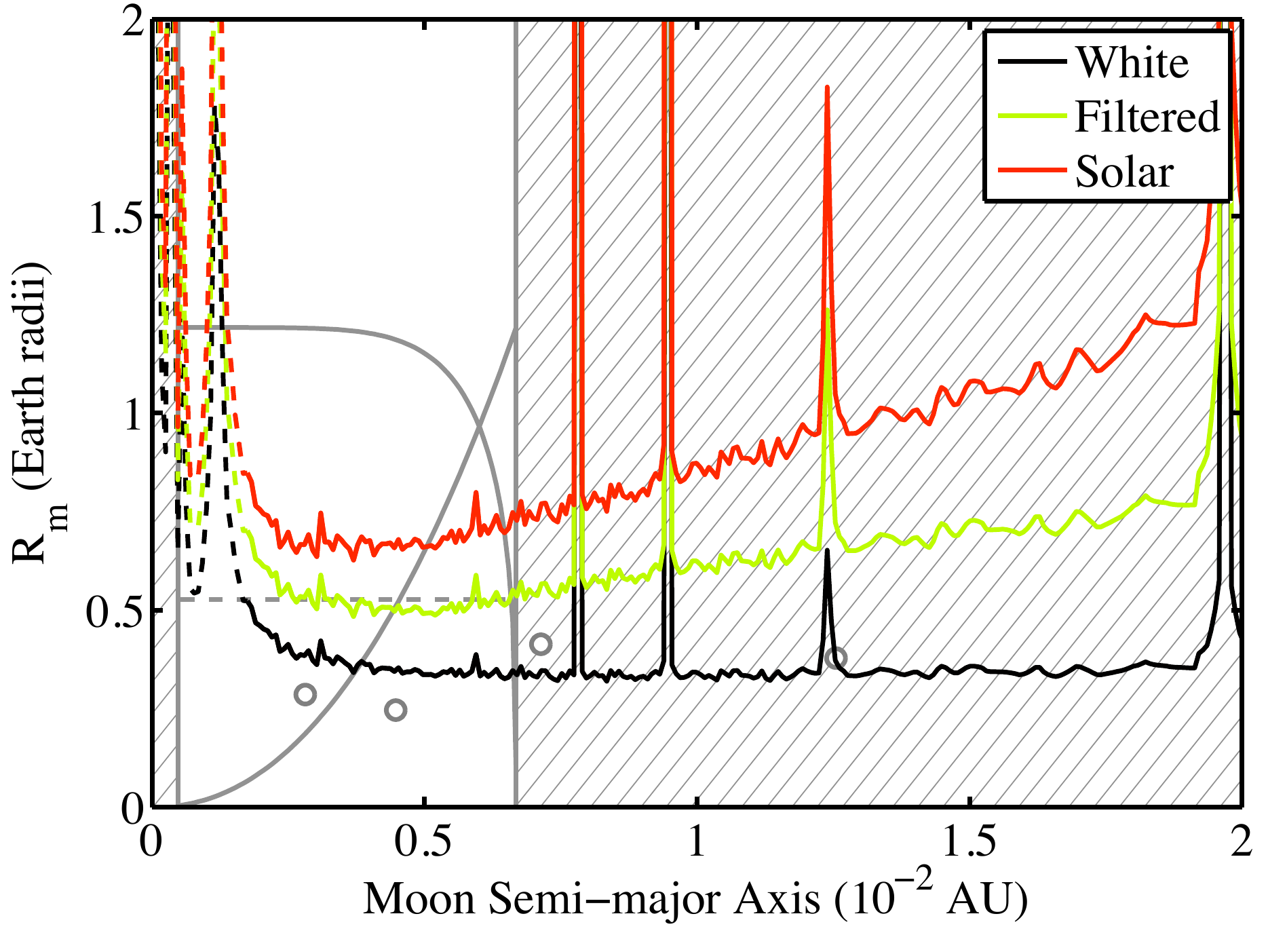}}
      \subfigure[$M_p = M_J$, $a_p=0.4$AU.]{
          \label{TransitThresh2s1MJ04AUInc}
          \includegraphics[width=.315\textwidth]{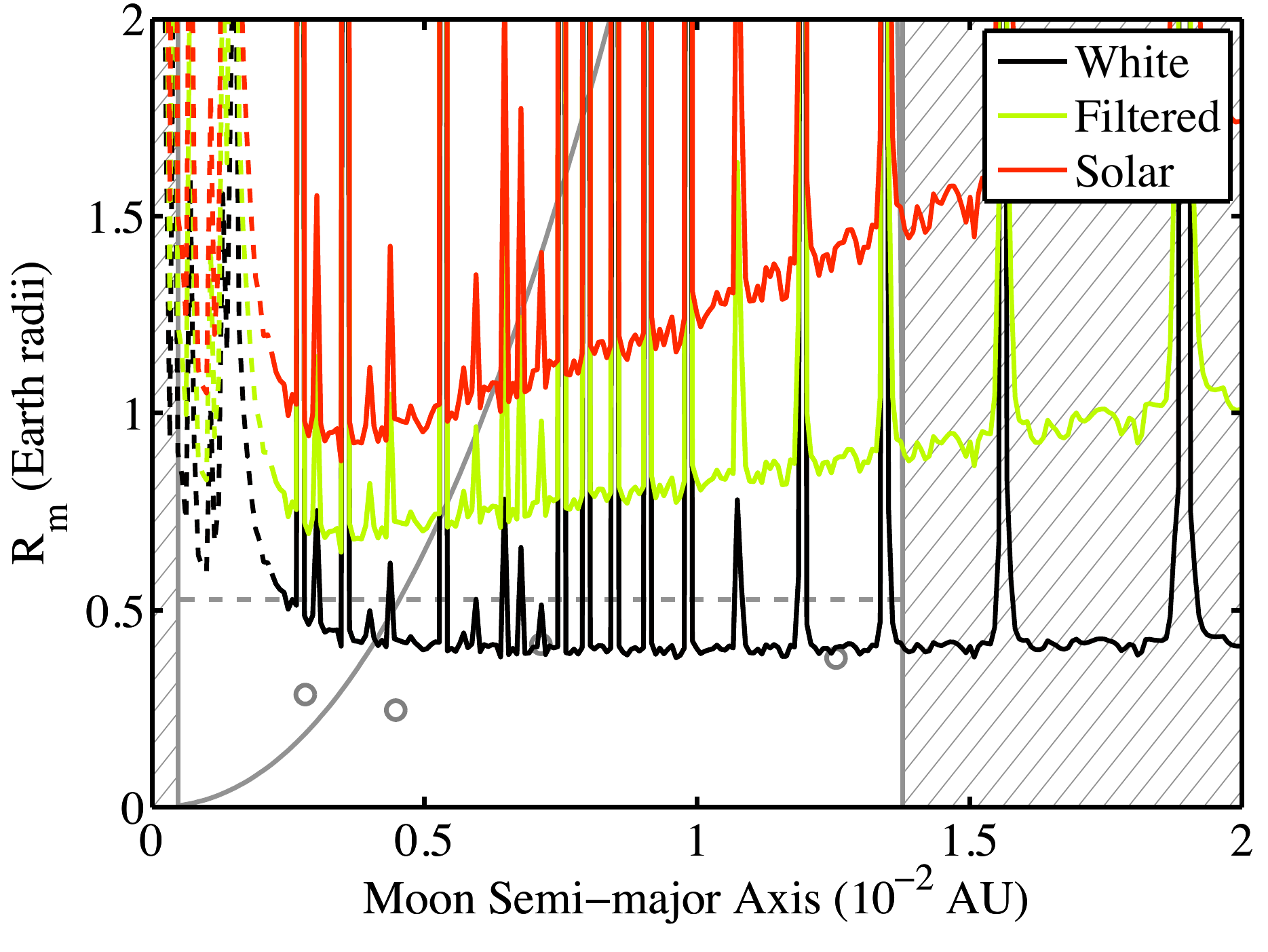}}
     \subfigure[$M_p = M_J$, $a_p=0.6$AU.]{
          \label{TransitThresh2s1MJ06AUInc}
          \includegraphics[width=.315\textwidth]{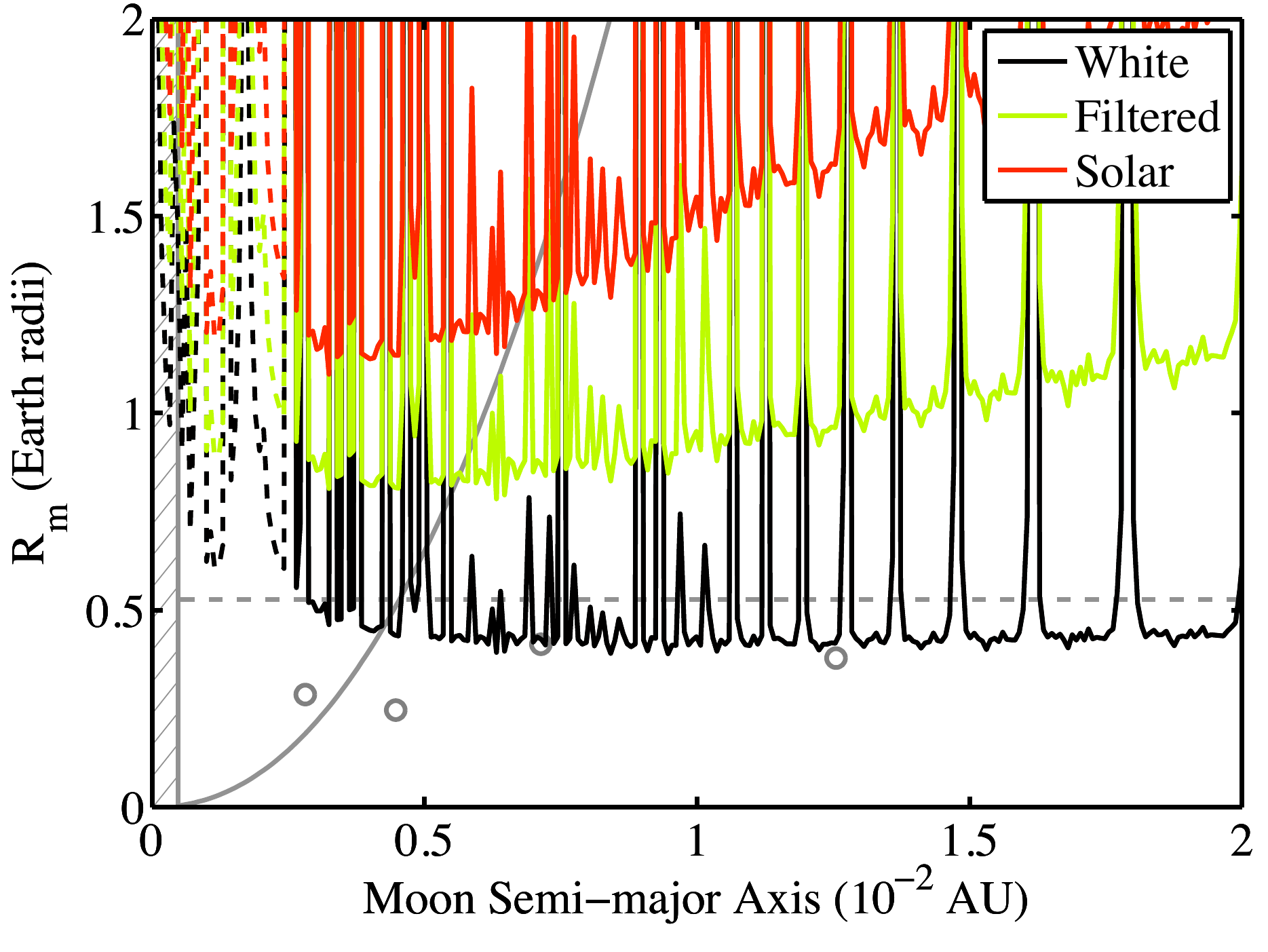}}\\ 
     \subfigure[$M_p = M_U$, $a_p=0.2$AU.]{
          \label{TransitThresh2s1MU02AUInc}
          \includegraphics[width=.315\textwidth]{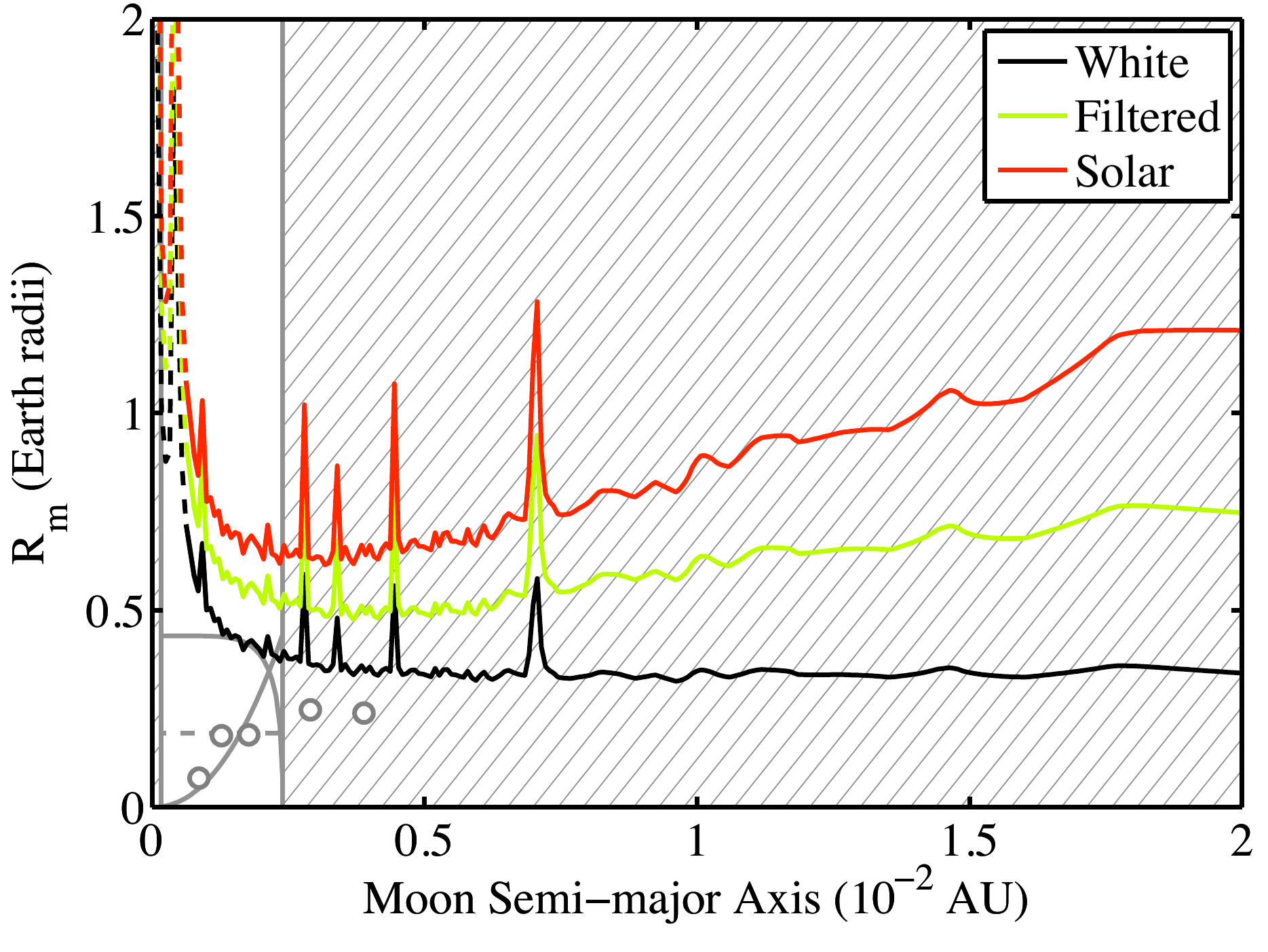}}
     \subfigure[$M_p = M_U$, $a_p=0.4$AU.]{
          \label{TransitThresh2s1MU04AUInc}
          \includegraphics[width=.315\textwidth]{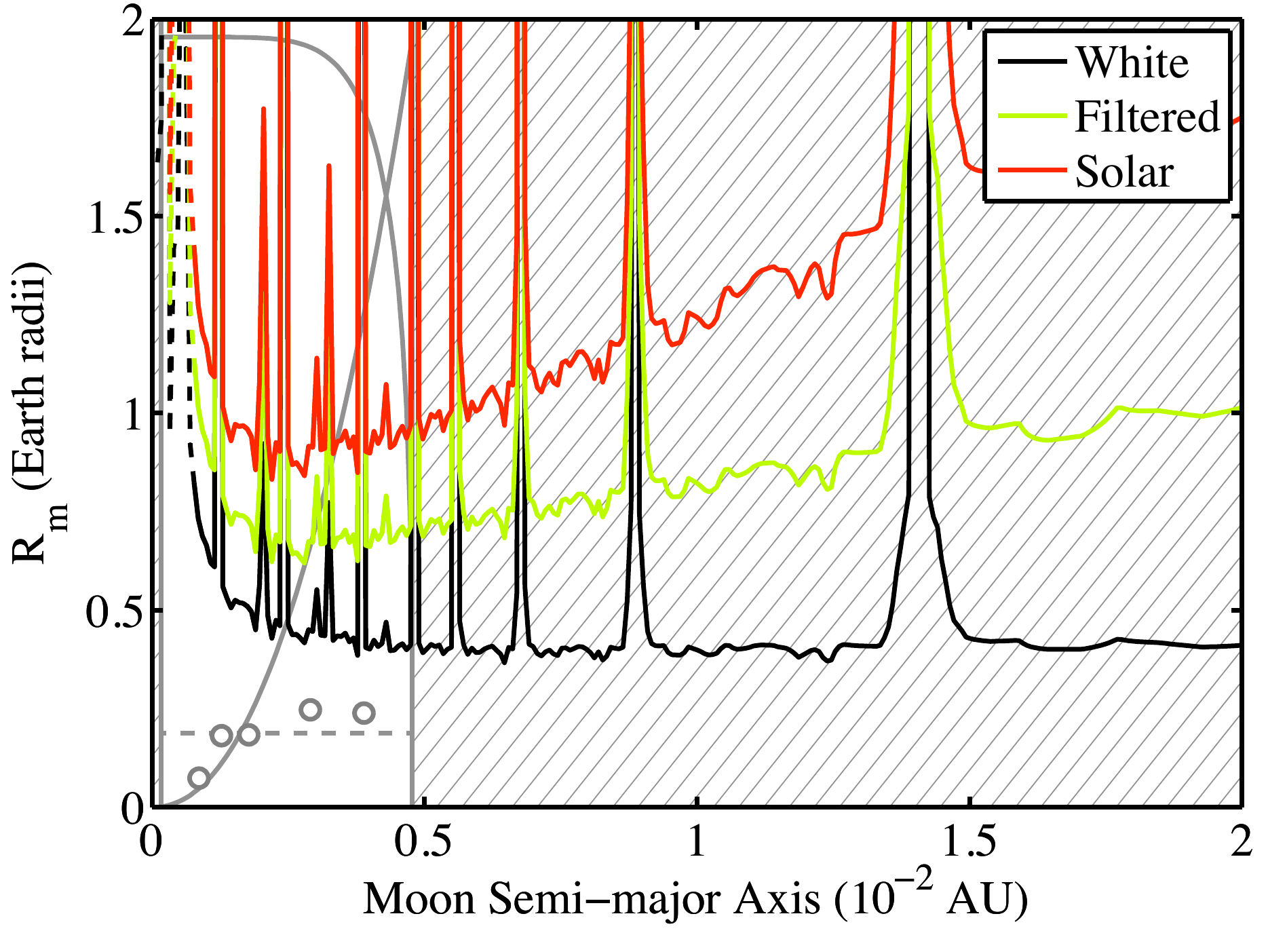}}
      \subfigure[$M_p = M_U$, $a_p=0.6$AU.]{
          \label{TransitThresh2s1MU06AUInc}
          \includegraphics[width=.315\textwidth]{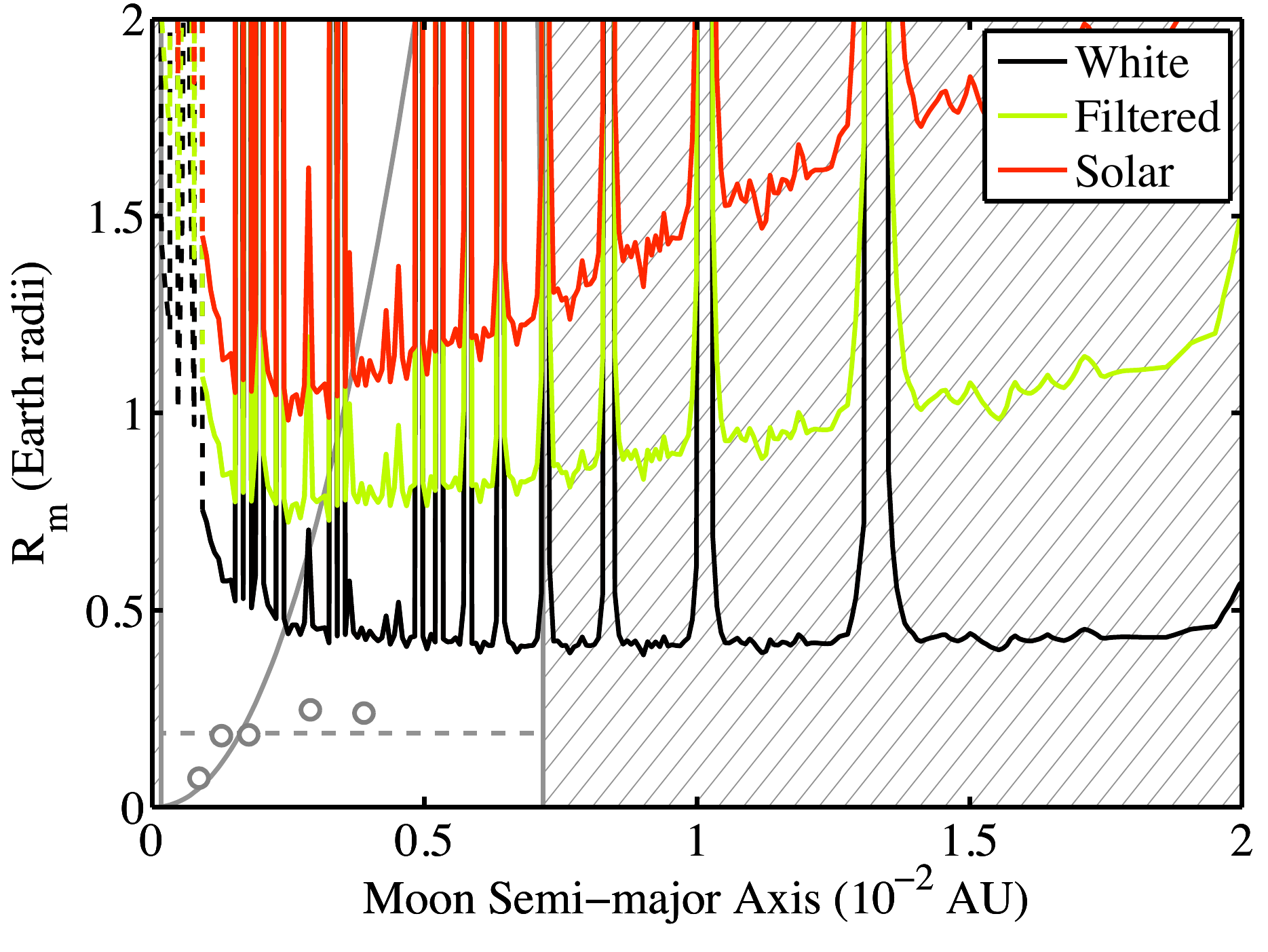}}\\ 
     \subfigure[$M_p = M_{\earth}$, $a_p=0.2$AU.]{
          \label{TransitThresh2s1ME02AUInc}
          \includegraphics[width=.315\textwidth]{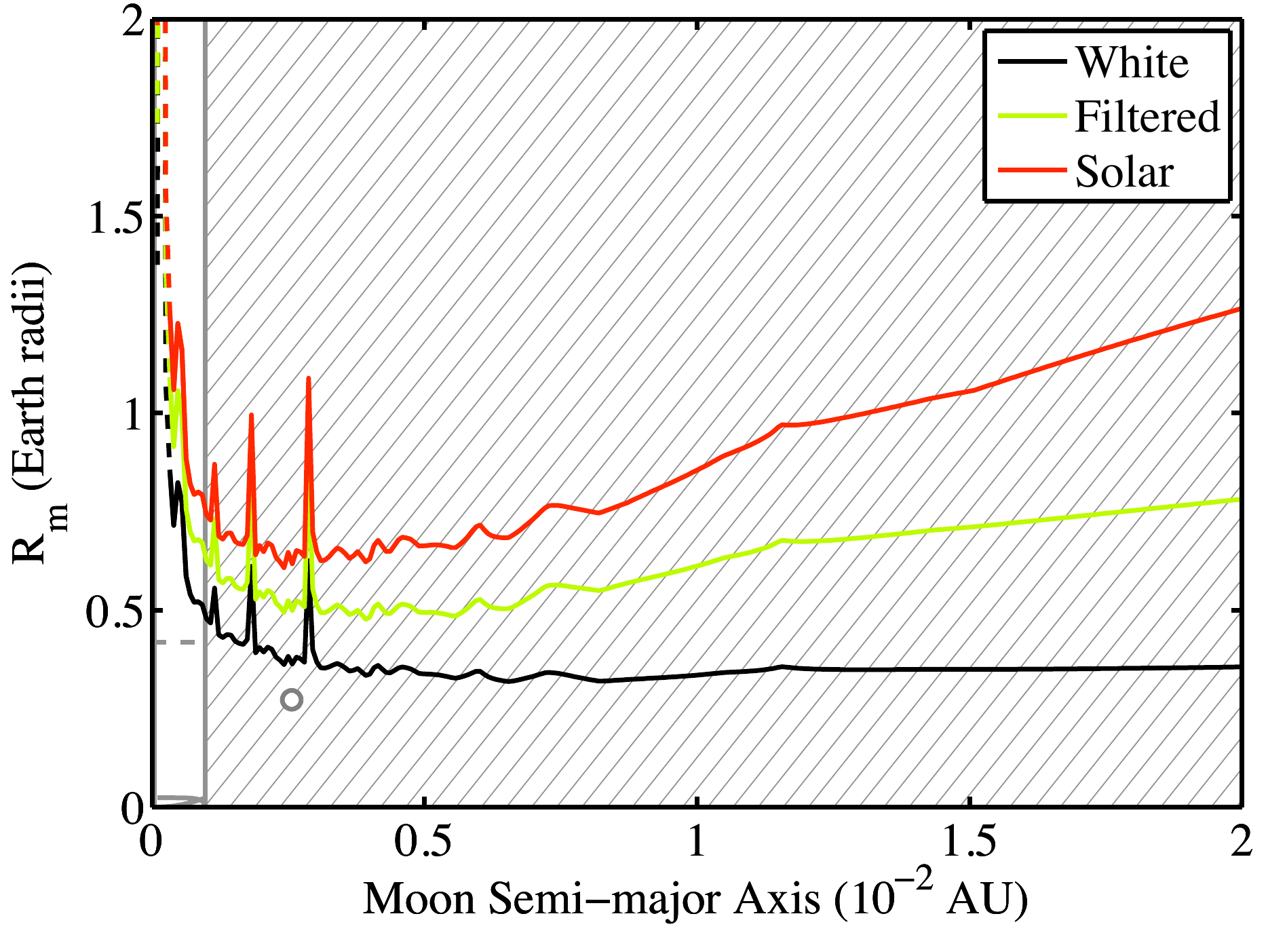}}
     \subfigure[$M_p = M_{\earth}$, $a_p=0.4$AU.]{
          \label{TransitThresh2s1ME04AUInc}
          \includegraphics[width=.315\textwidth]{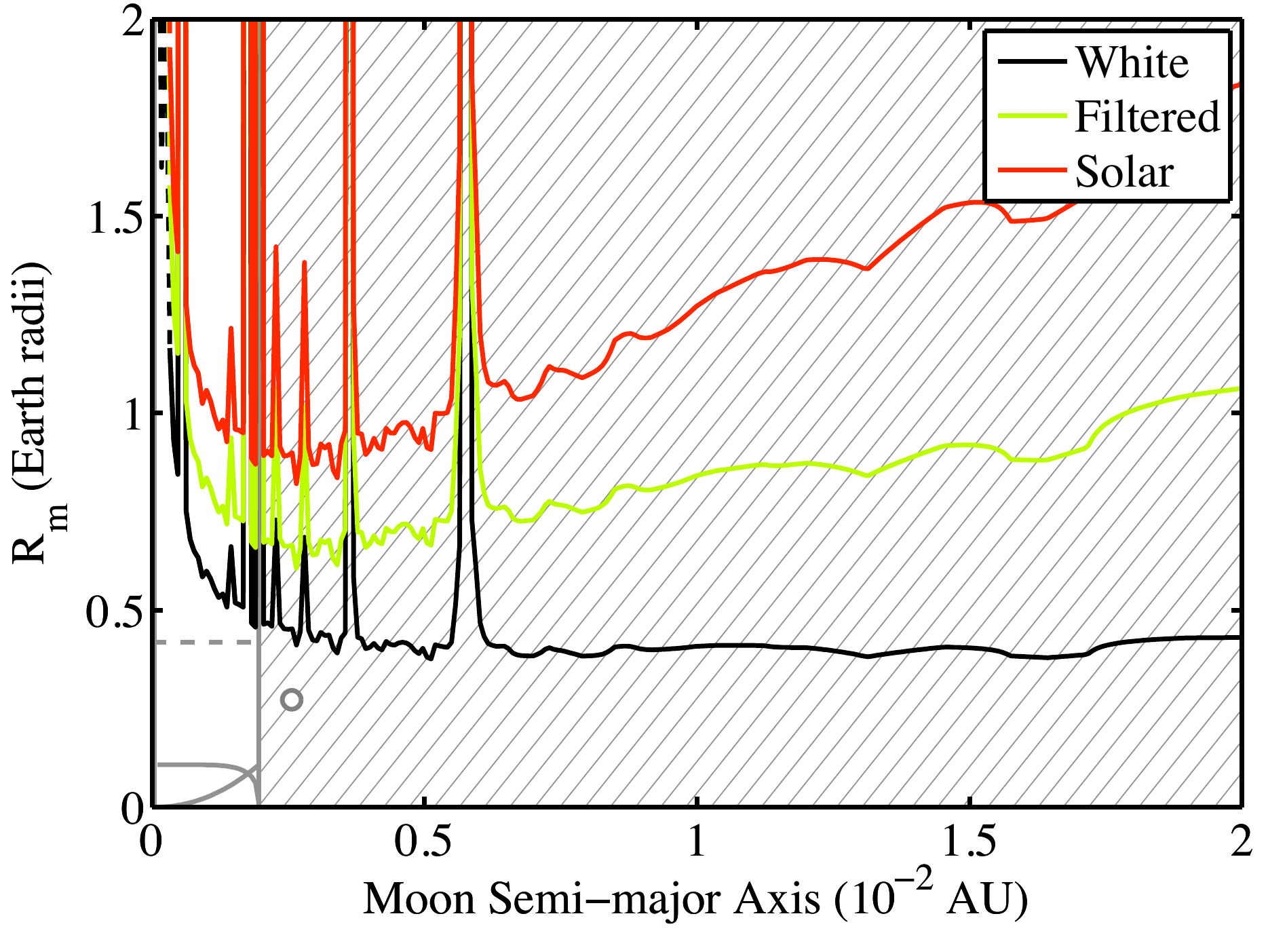}}
     \subfigure[$M_p = M_{\earth}$, $a_p=0.6$AU.]{
          \label{TransitThresh2s1ME06AUInc}
          \includegraphics[width=.315\textwidth]{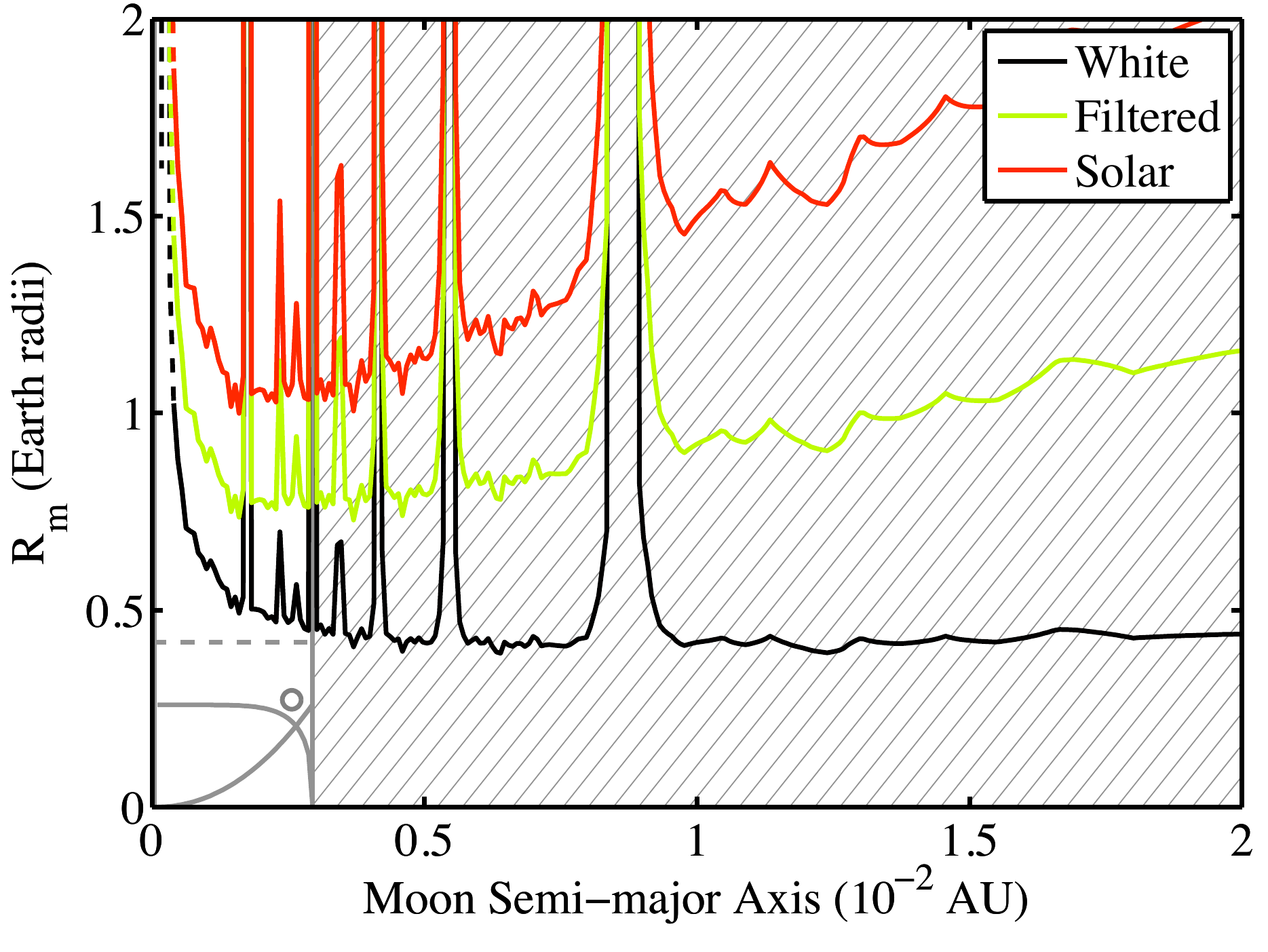}} 
     \caption{Figure of the same form as figure~\ref{MCThresholdsInclined}, but showing the 95.4\% thresholds.}
     \label{MCThresholdsInclined2S}
\end{figure}

\begin{figure}
     \centering
     \subfigure[$M_p$=$10 M_J$, $a_p=0.2$AU.]{
          \label{TransitThresh2s10MJ02AUEccP}
          \includegraphics[width=.315\textwidth]{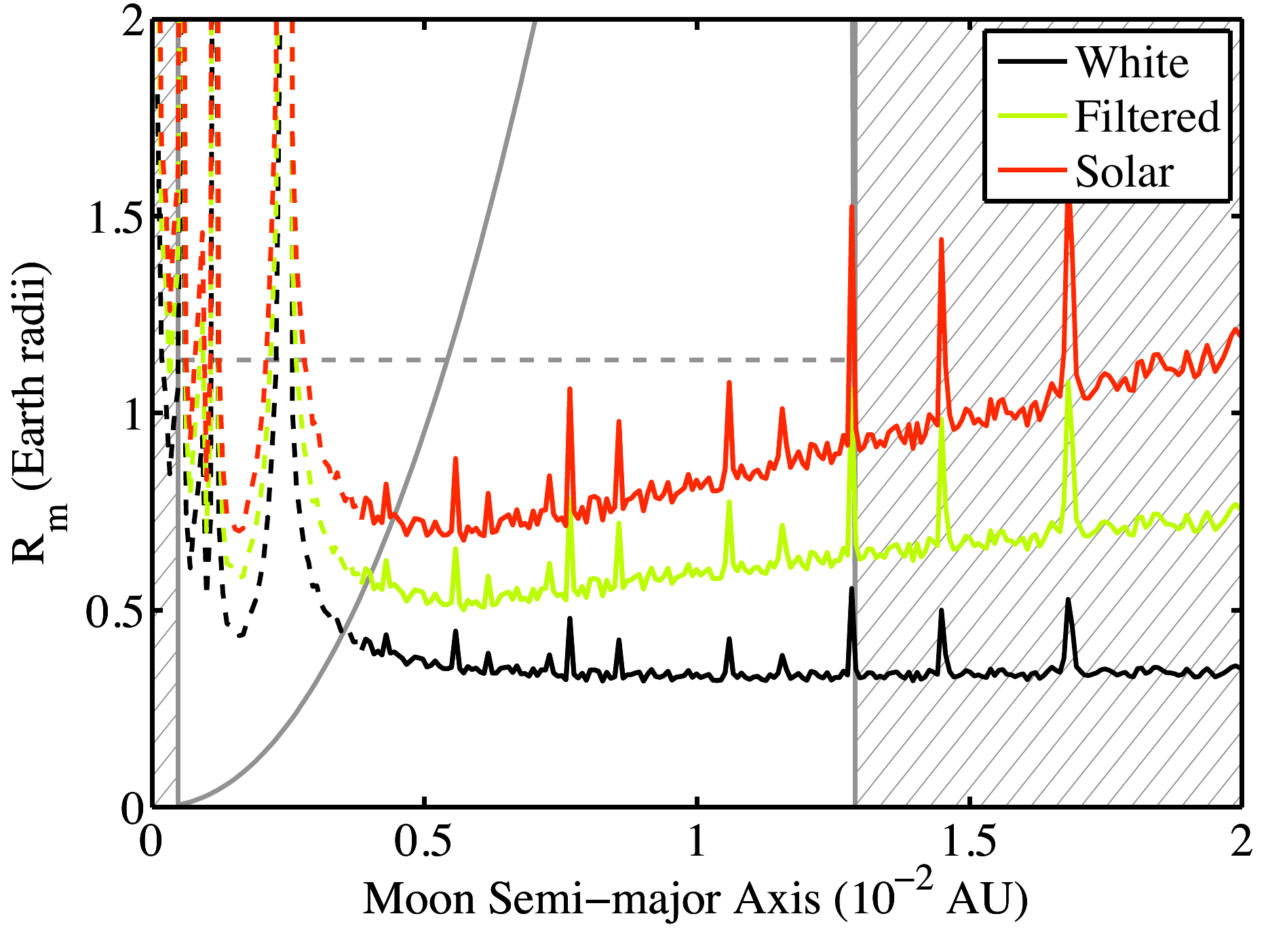}}
     \subfigure[$M_p$=$10 M_J$, $a_p=0.4$AU.]{
          \label{TransitThresh2s10MJ04AUEccP}
          \includegraphics[width=.315\textwidth]{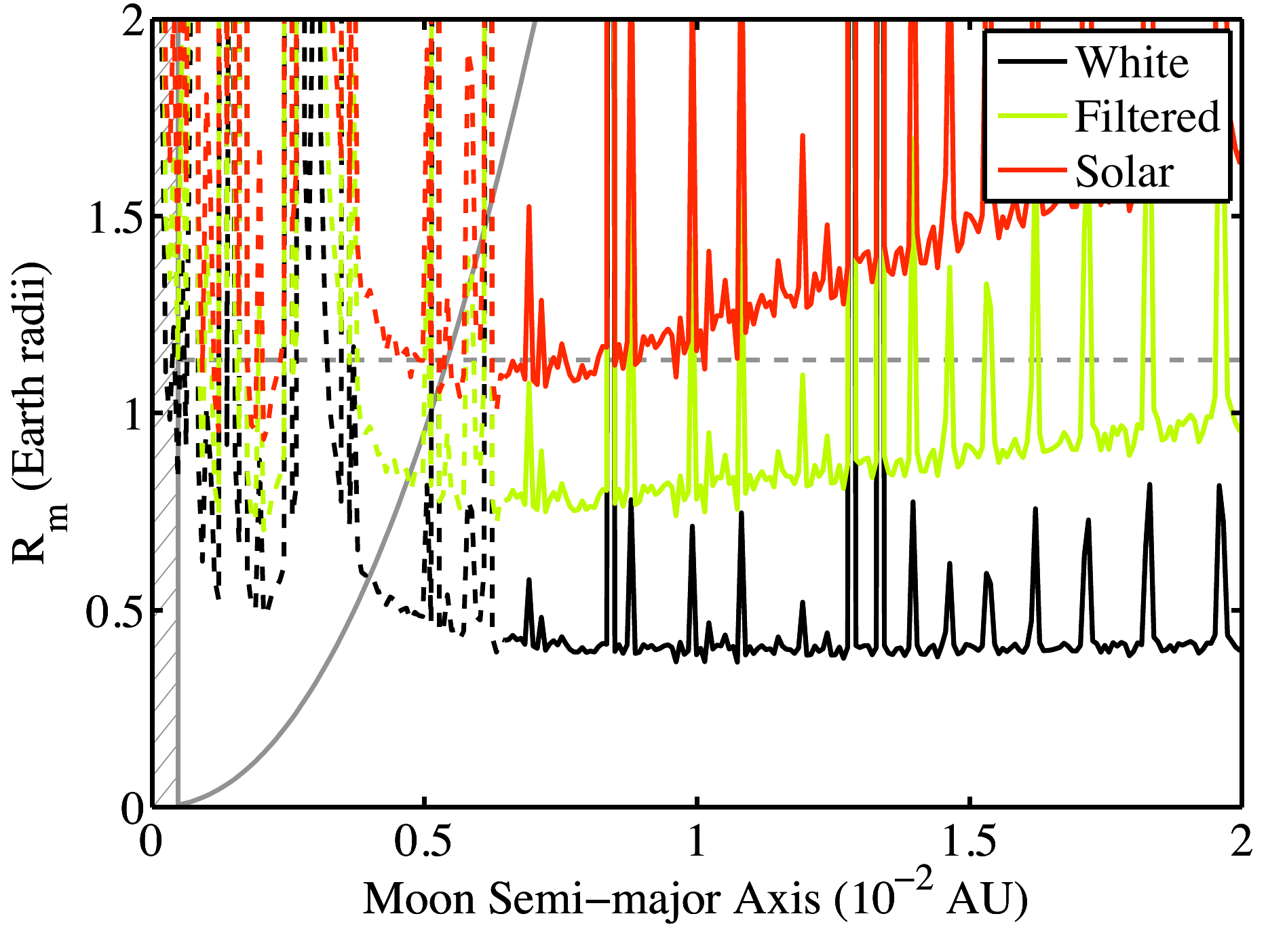}}
     \subfigure[$M_p$=$10 M_J$, $a_p=0.6$AU.]{
          \label{TransitThresh2s10MJ06AUEccP}
          \includegraphics[width=.315\textwidth]{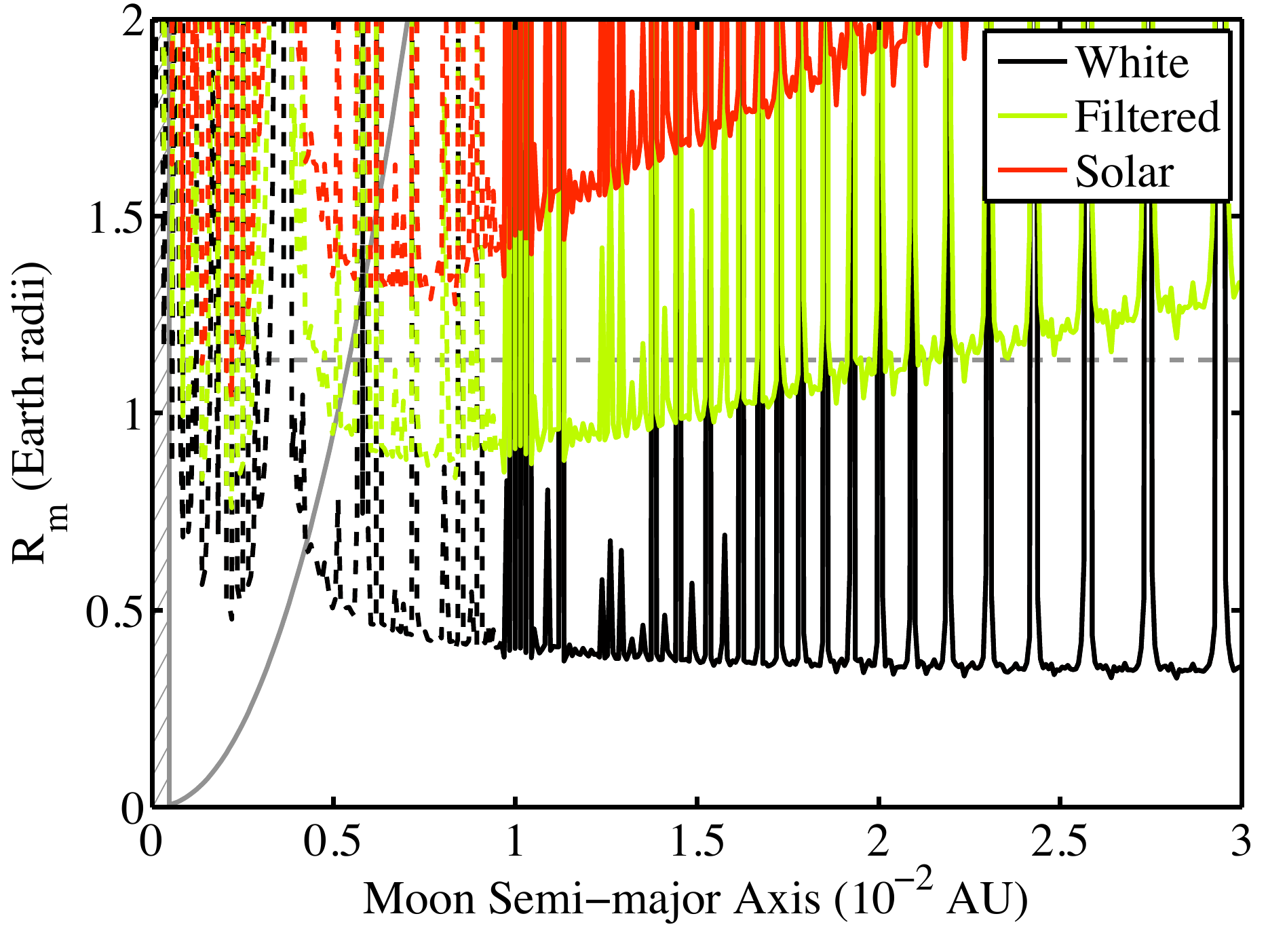}}\\ 
     \subfigure[$M_p = M_J$, $a_p=0.2$AU.]{
          \label{TransitThresh2s1MJ02AUEccP}
          \includegraphics[width=.315\textwidth]{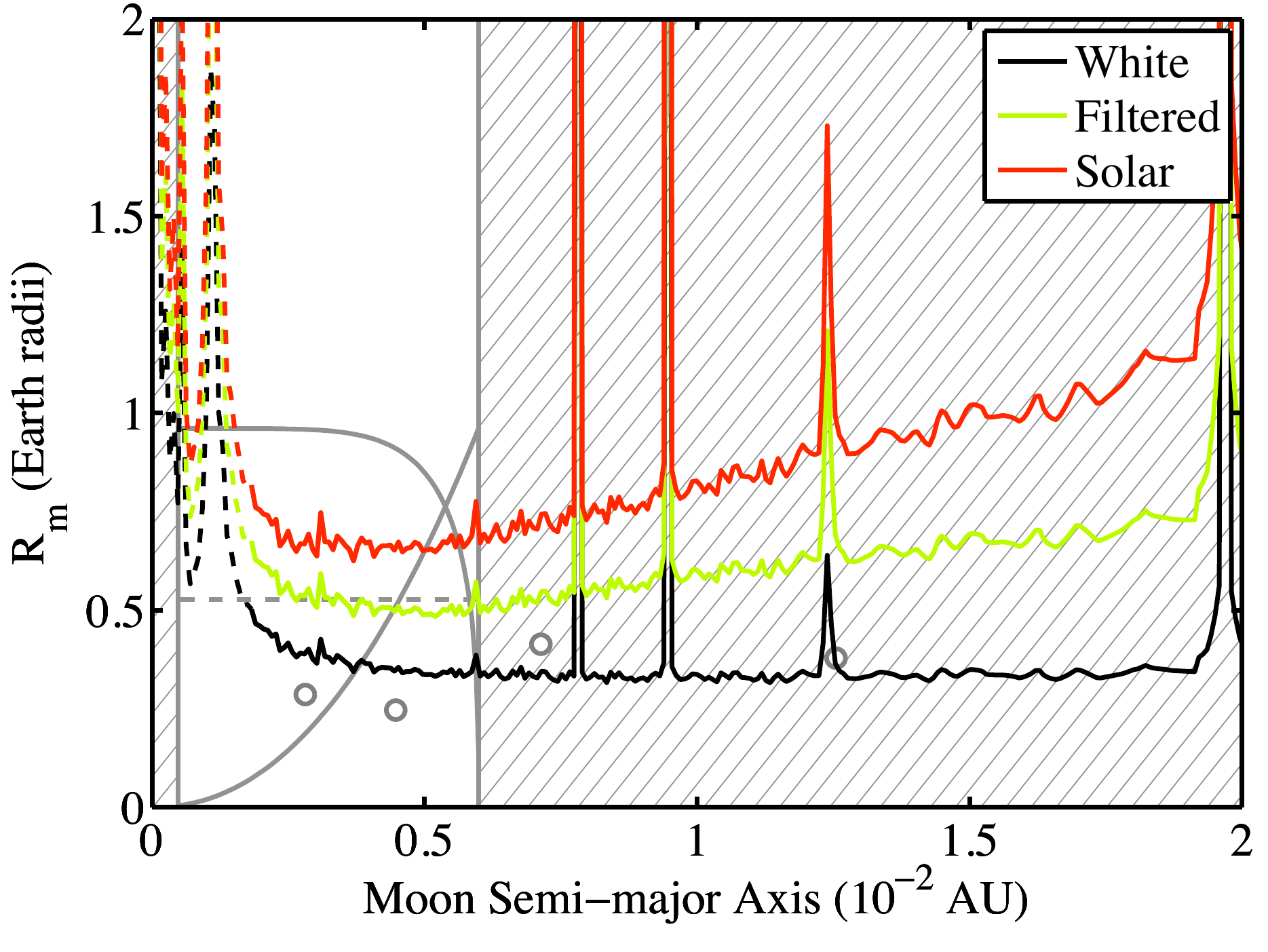}}
      \subfigure[$M_p = M_J$, $a_p=0.4$AU.]{
          \label{TransitThresh2s1MJ04AUEccP}
          \includegraphics[width=.315\textwidth]{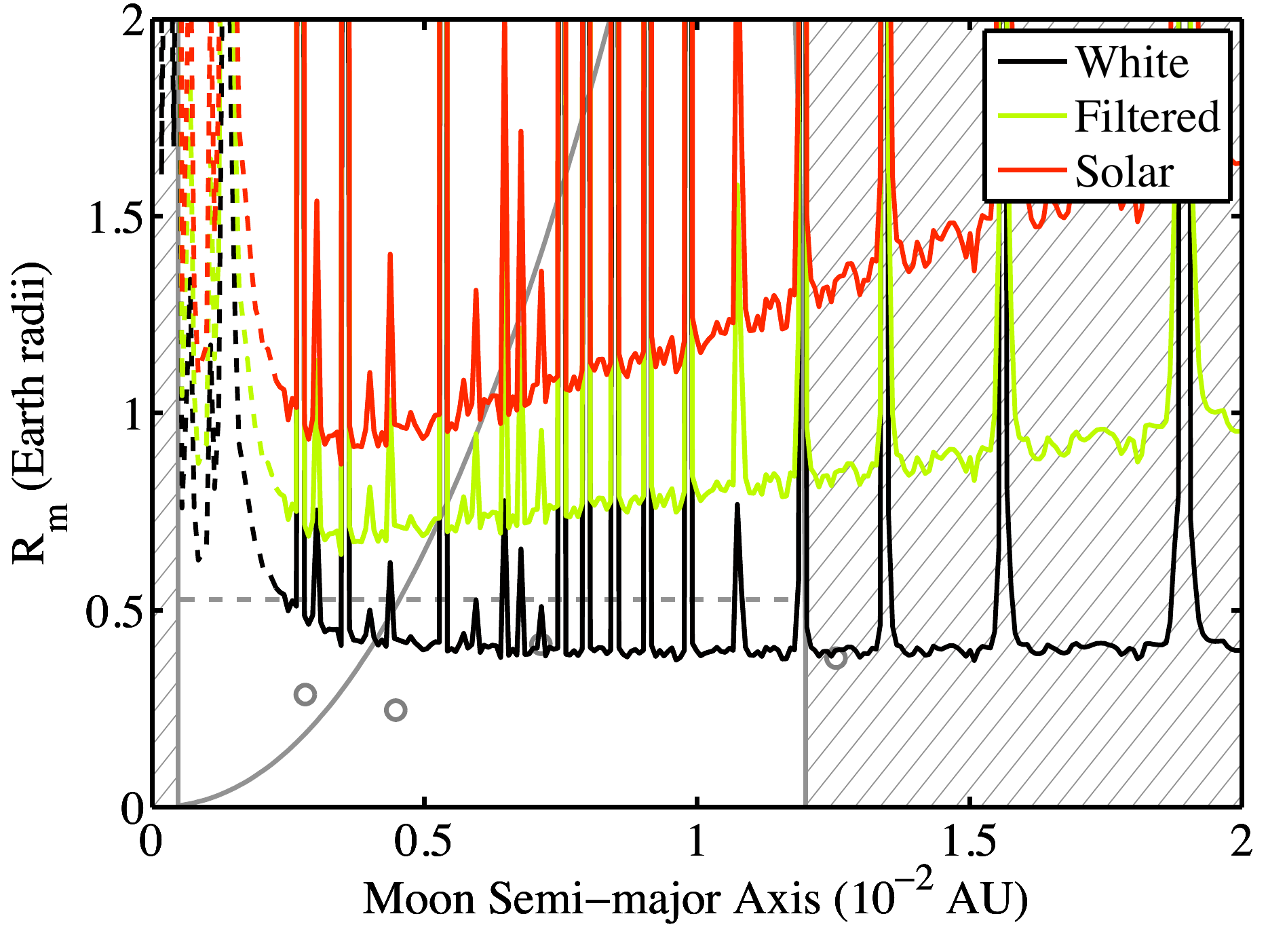}}
     \subfigure[$M_p = M_J$, $a_p=0.6$AU.]{
          \label{TransitThresh2s1MJ06AUEccP}
          \includegraphics[width=.315\textwidth]{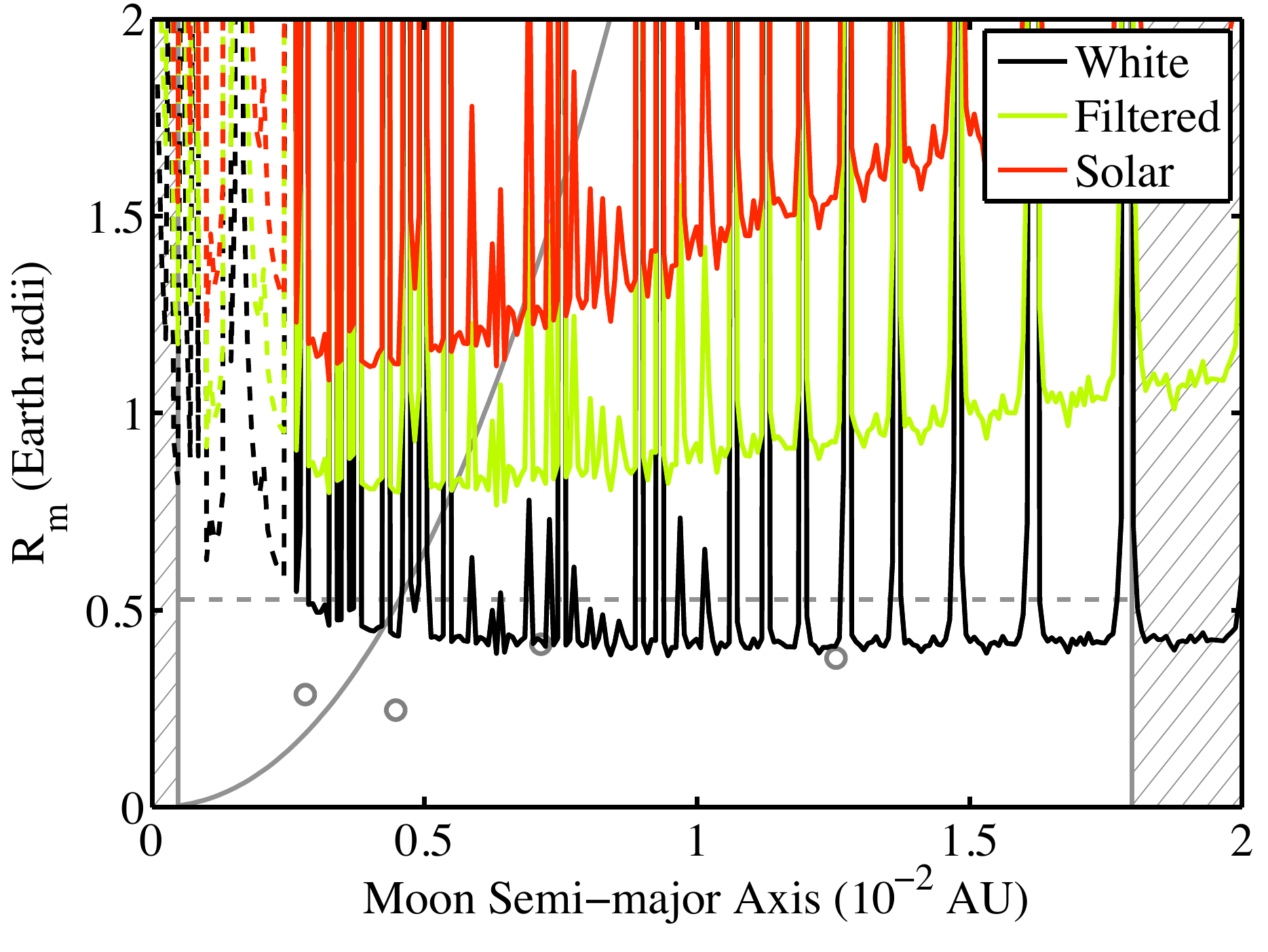}}\\ 
     \subfigure[$M_p = M_U$, $a_p=0.2$AU.]{
          \label{TransitThresh2s1MU02AUEccP}
          \includegraphics[width=.315\textwidth]{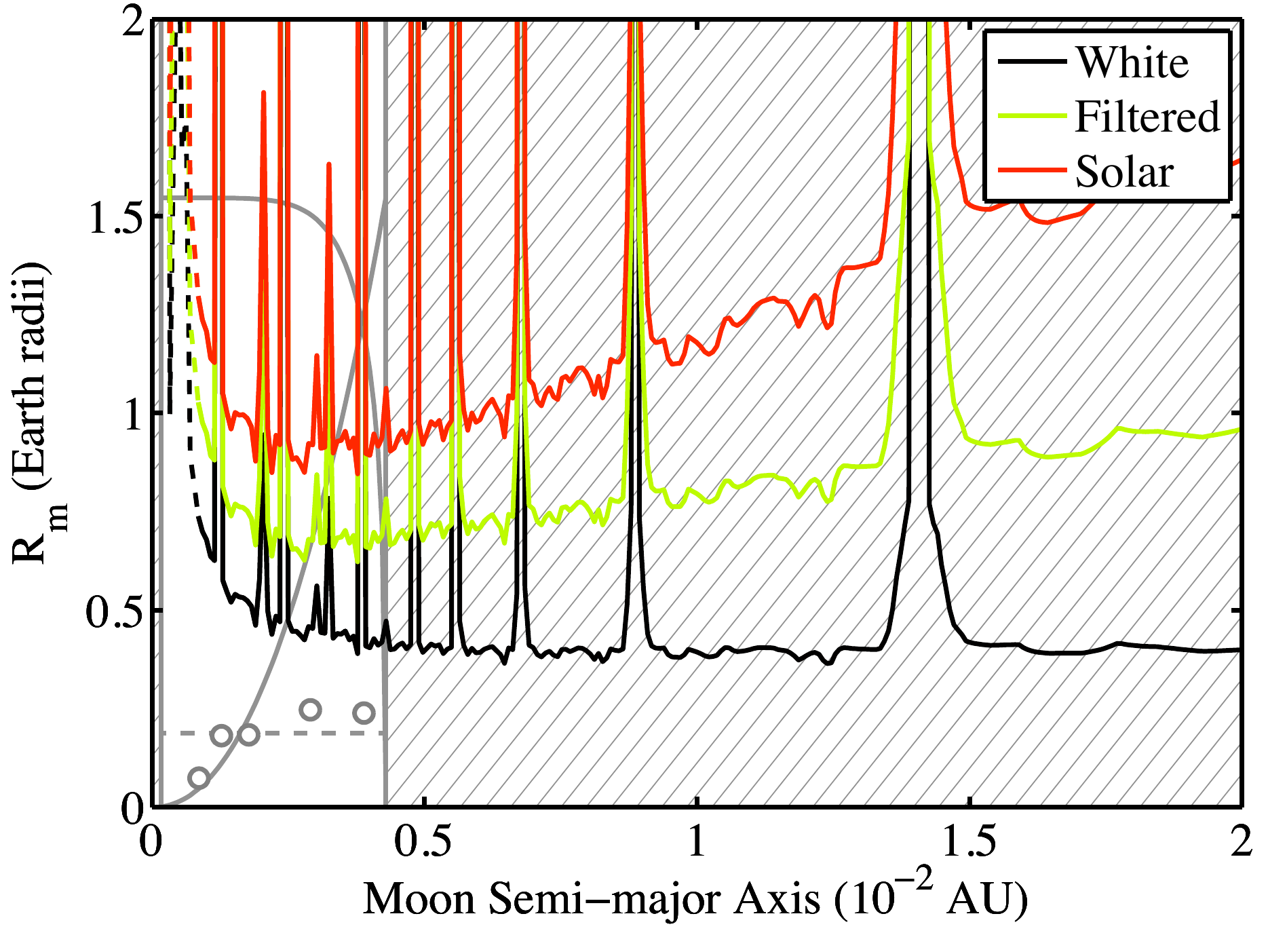}}
     \subfigure[$M_p = M_U$, $a_p=0.4$AU.]{
          \label{TransitThresh2s1MU04AUEccP}
          \includegraphics[width=.315\textwidth]{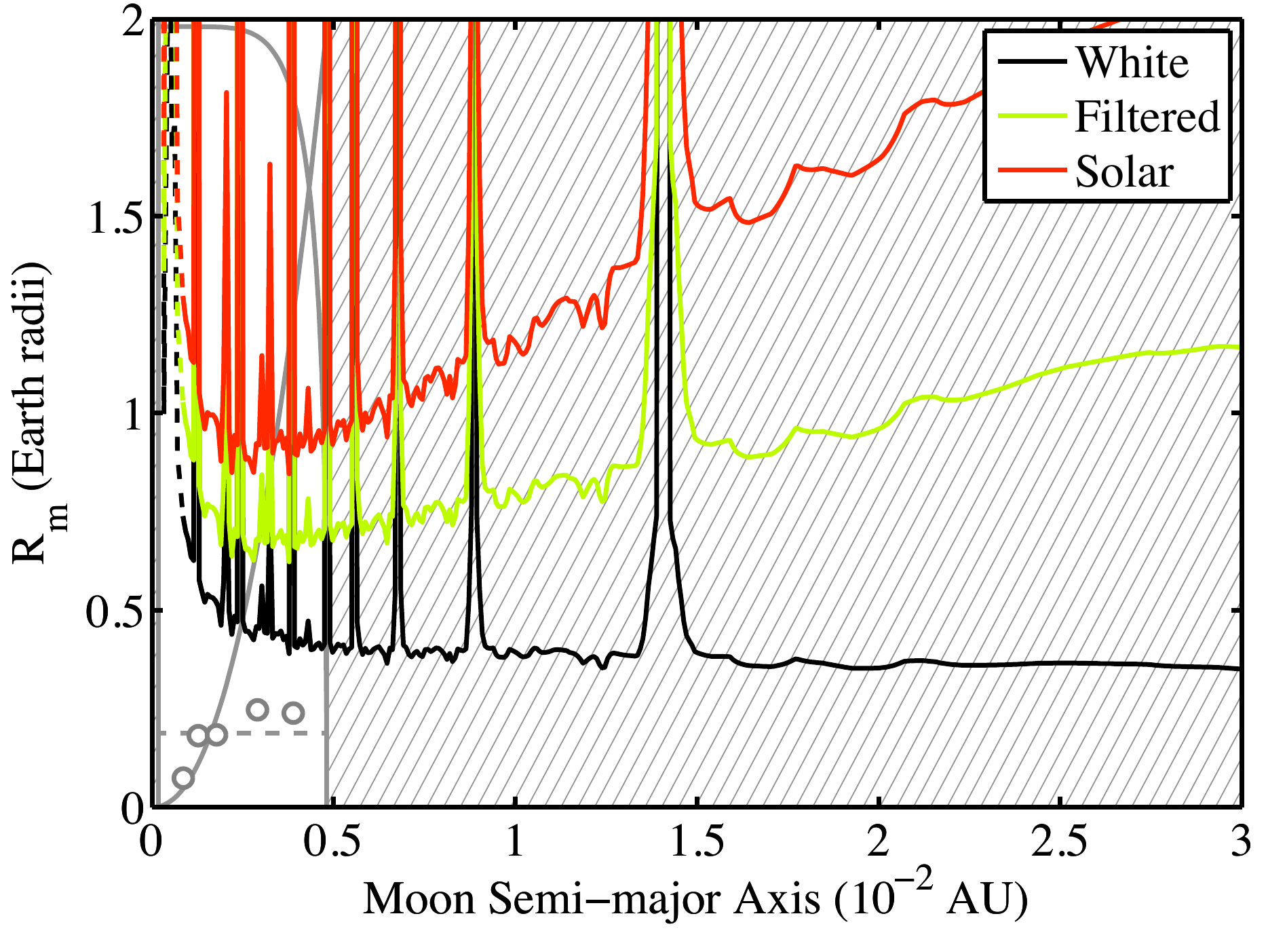}}
      \subfigure[$M_p = M_U$, $a_p=0.6$AU.]{
          \label{TransitThresh2s1MU06AUEccP}
          \includegraphics[width=.315\textwidth]{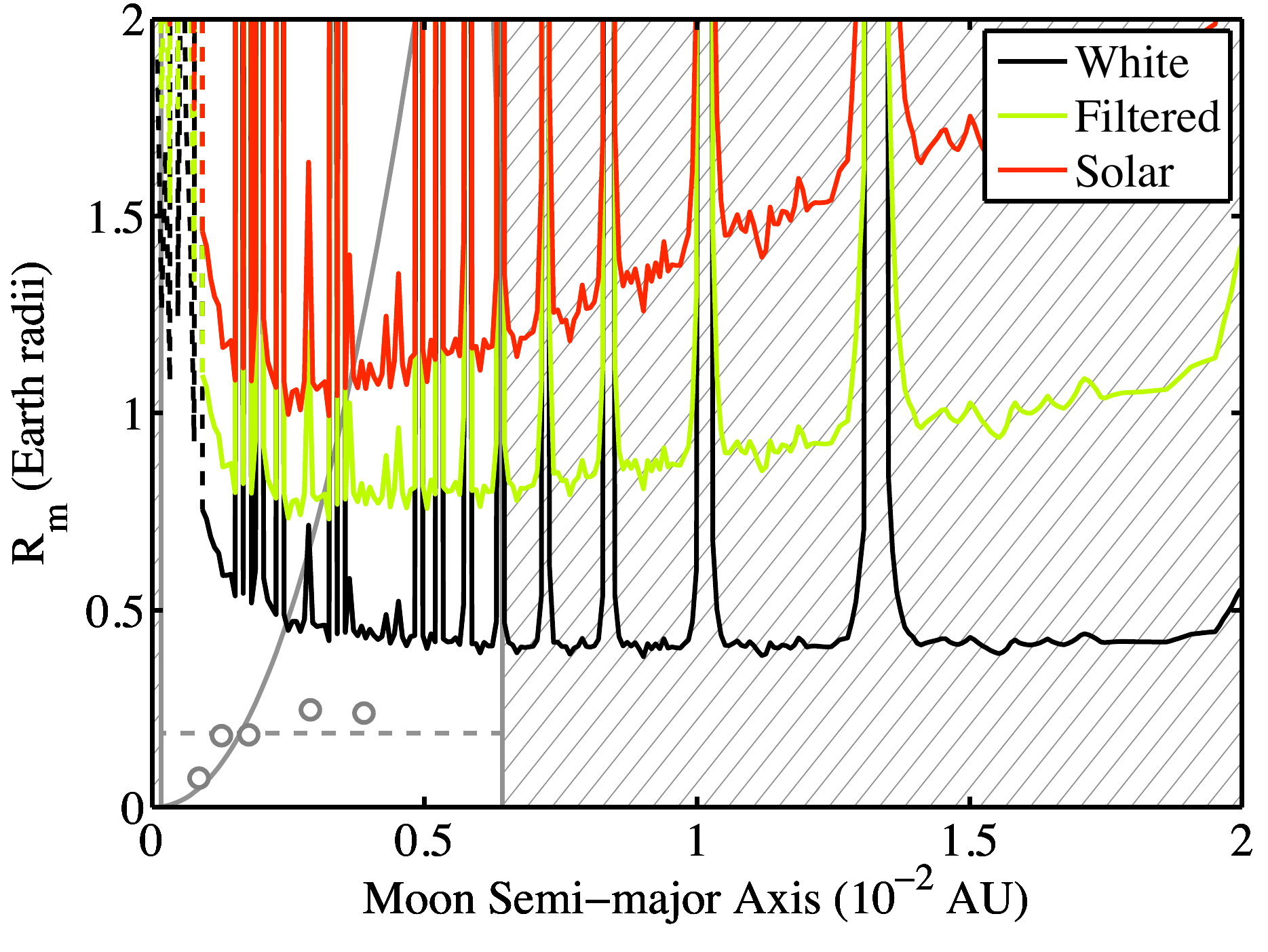}}\\ 
     \subfigure[$M_p = M_{\earth}$, $a_p=0.2$AU.]{
          \label{TransitThresh2s1ME02AUEccP}
          \includegraphics[width=.315\textwidth]{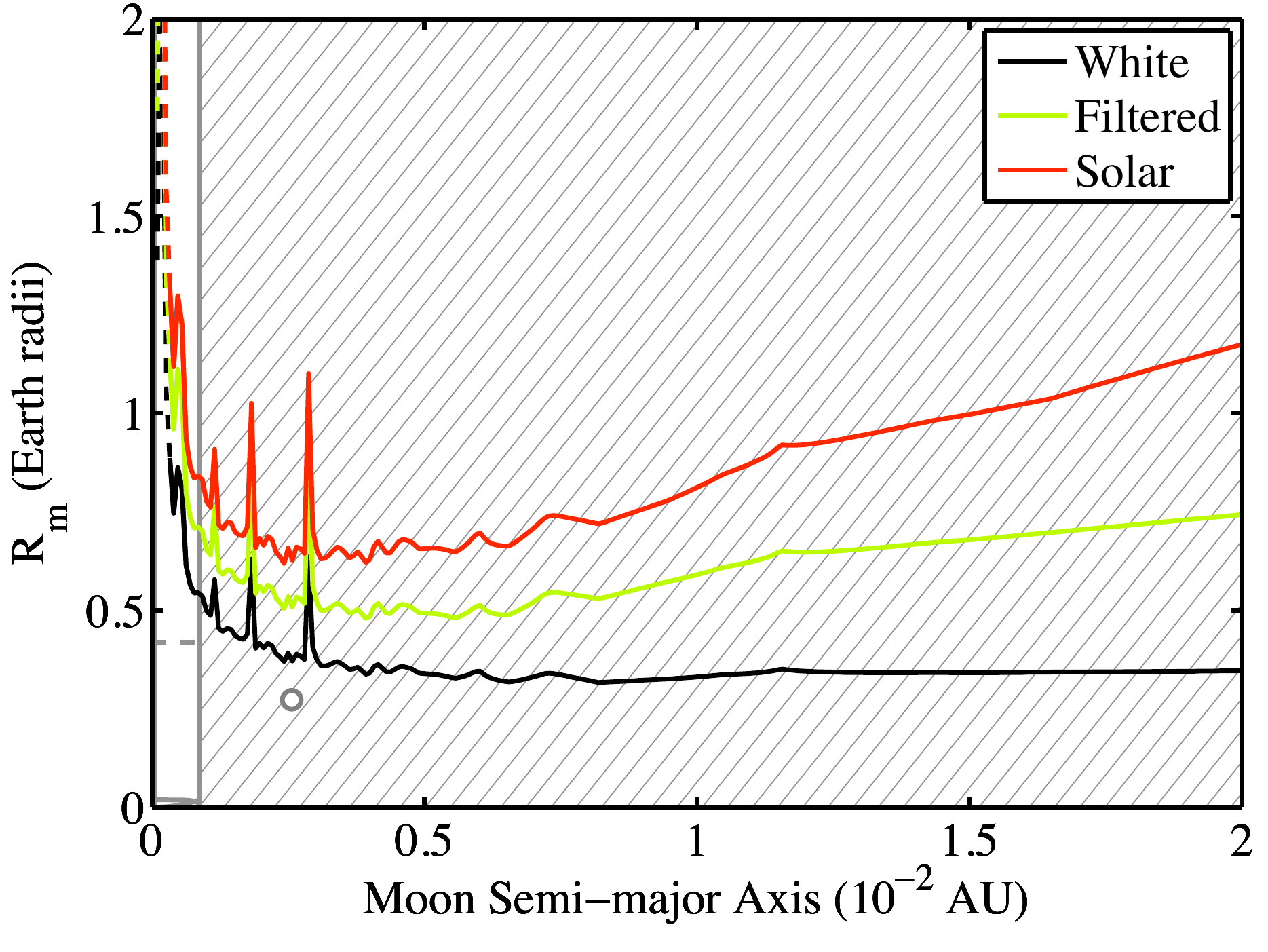}}
     \subfigure[$M_p = M_{\earth}$, $a_p=0.4$AU.]{
          \label{TransitThresh2s1ME04AUEccP}
          \includegraphics[width=.315\textwidth]{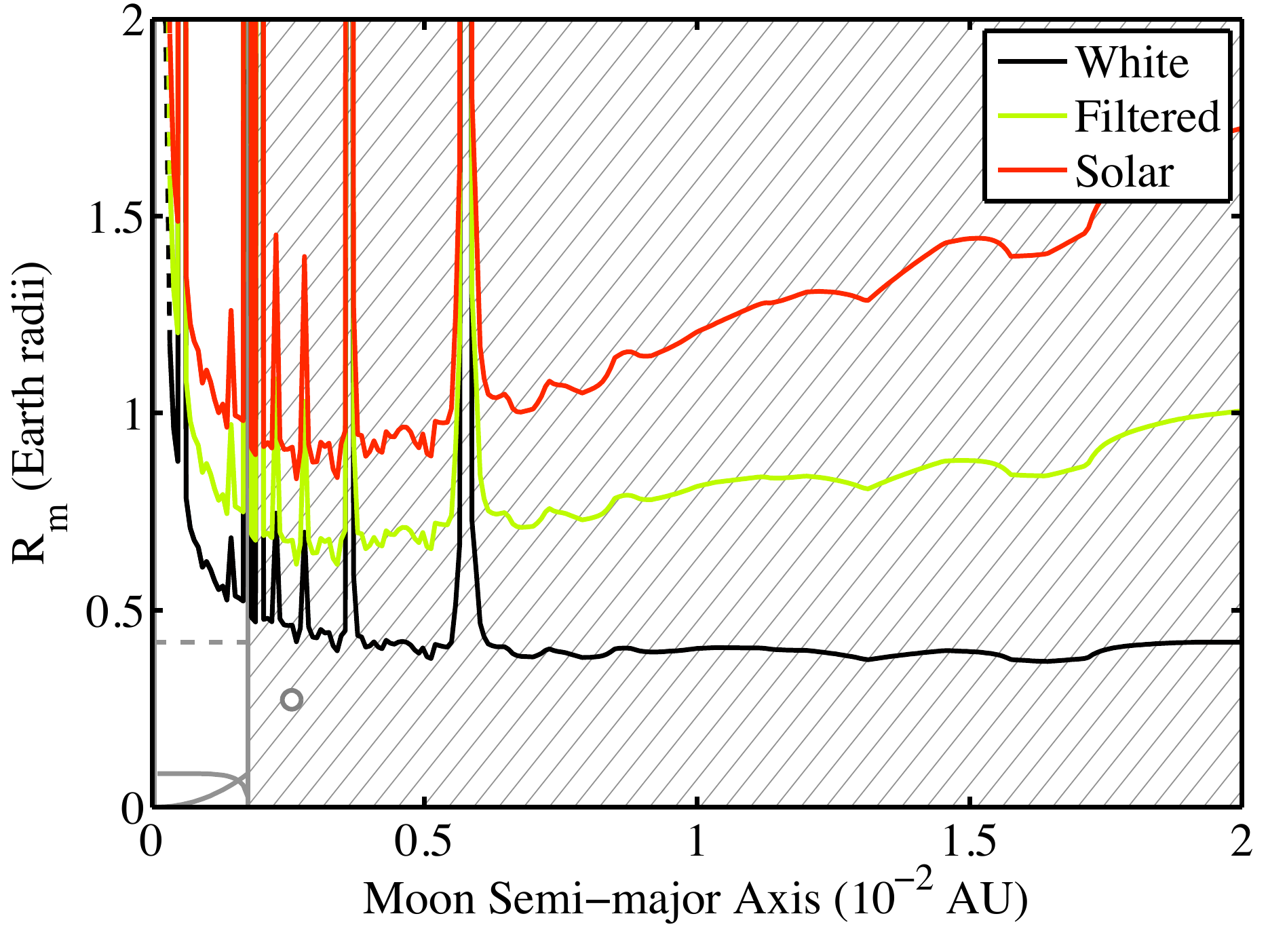}}
     \subfigure[$M_p = M_{\earth}$, $a_p=0.6$AU.]{
          \label{TransitThresh2s1ME06AUEccP}
          \includegraphics[width=.315\textwidth]{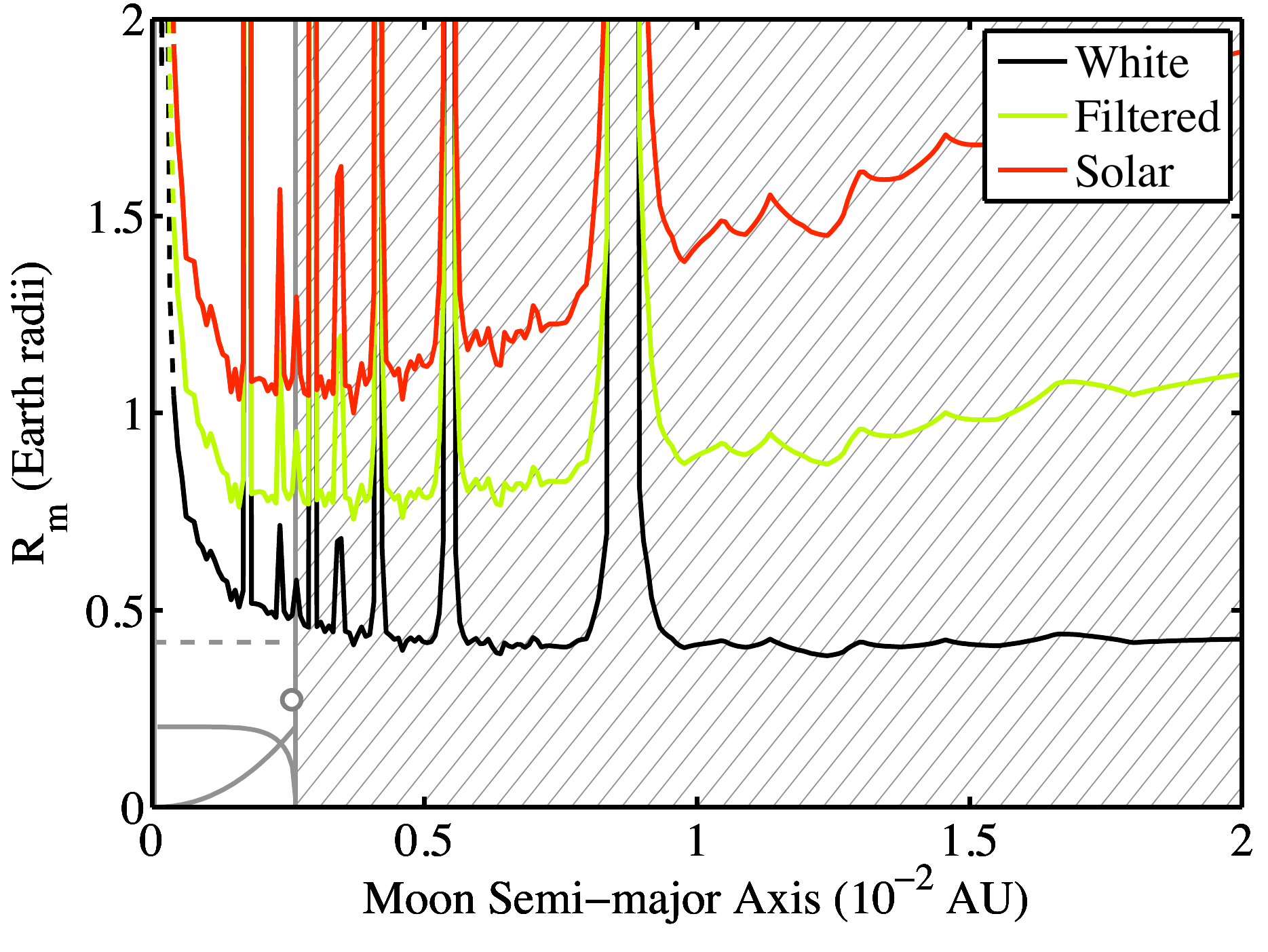}} 
     \caption{Figure of the same form as figure~\ref{MCThresholdsEccentricPeri}, but showing the 95.4\% thresholds.}
    \label{MCThresholdsEccentricPeri2S}
    \end{figure}

\begin{figure}
     \centering
     \subfigure[$M_p$=$10 M_J$, $a_p=0.2$AU.]{
          \label{TransitThresh2s10MJ02AUEccA}
          \includegraphics[width=.315\textwidth]{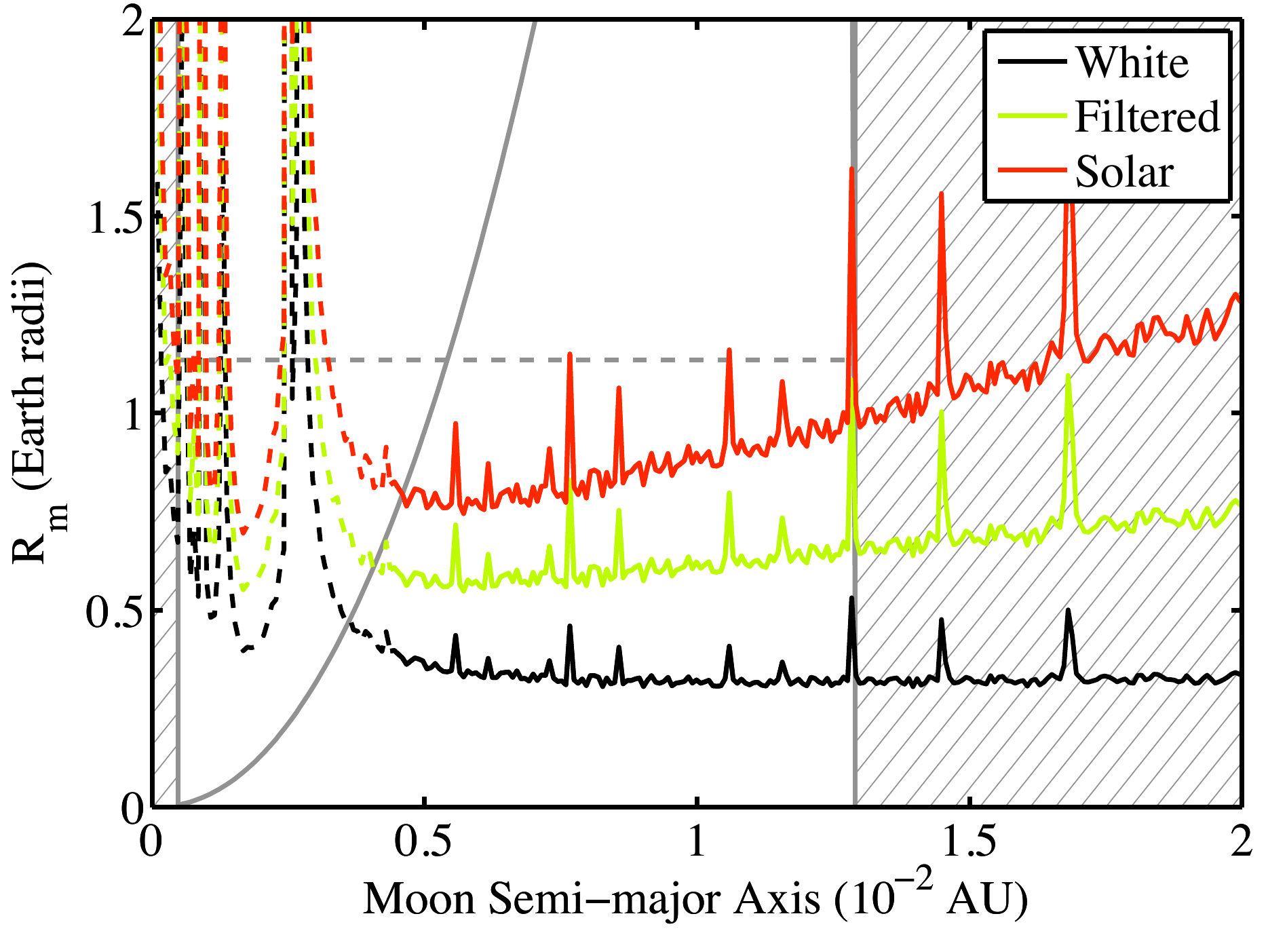}}
     \subfigure[$M_p$=$10 M_J$, $a_p=0.4$AU.]{
          \label{TransitThresh2s10MJ04AUEccA}
          \includegraphics[width=.315\textwidth]{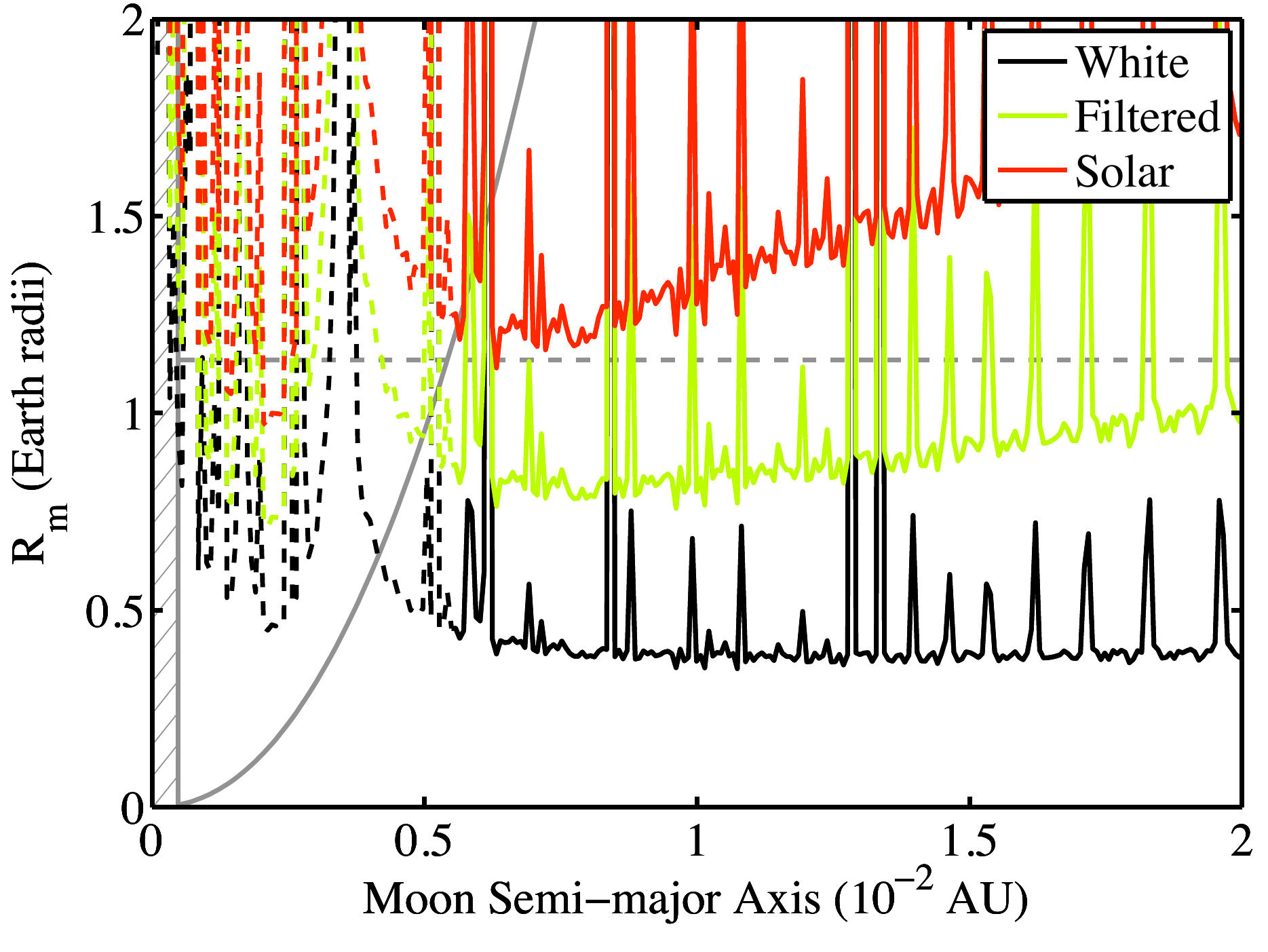}}
     \subfigure[$M_p$=$10 M_J$, $a_p=0.6$AU.]{
          \label{TransitThresh2s10MJ06AUEccA}
          \includegraphics[width=.315\textwidth]{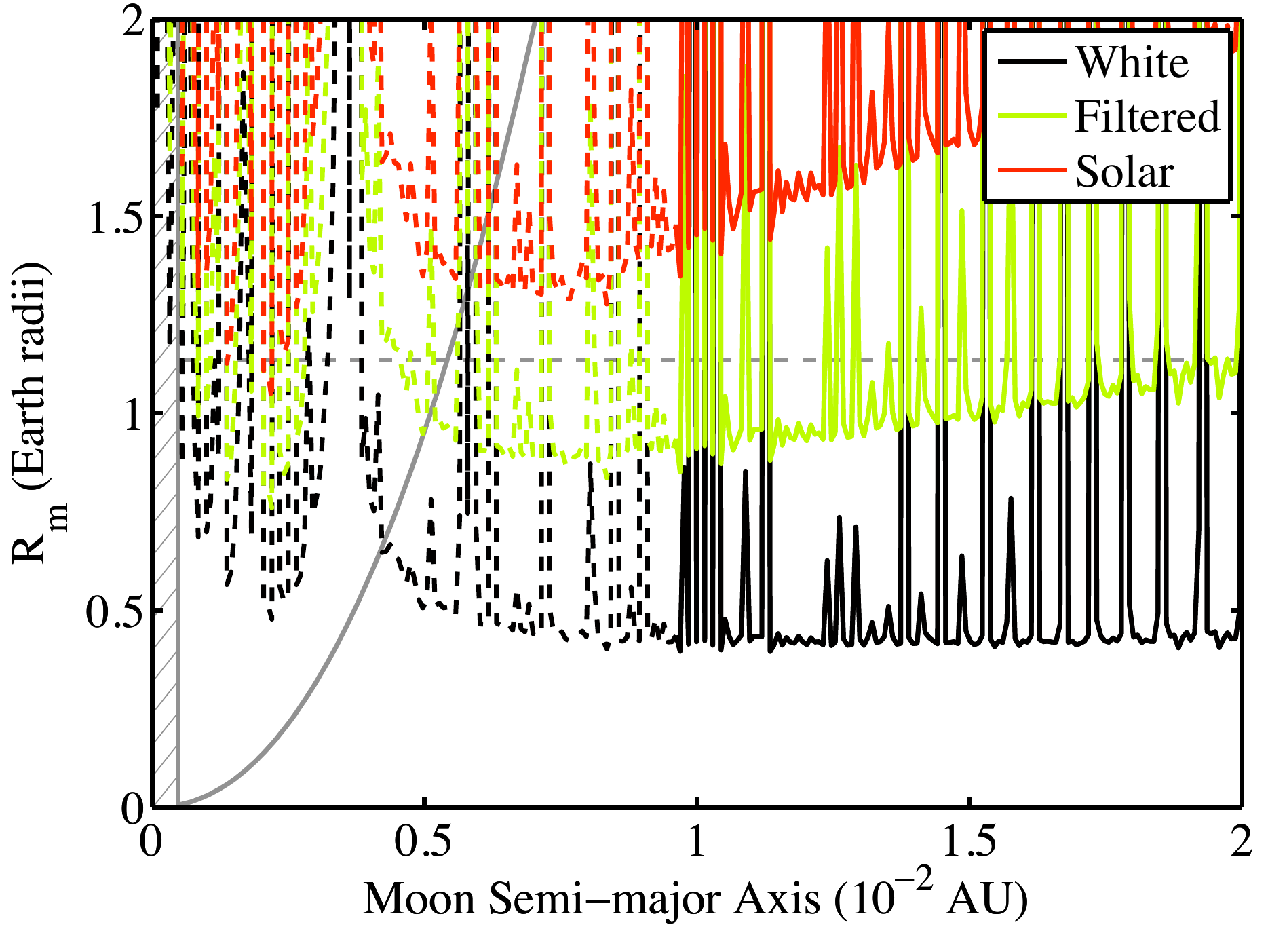}}\\ 
     \subfigure[$M_p = M_J$, $a_p=0.2$AU.]{
          \label{TransitThresh2s1MJ02AUEccA}
          \includegraphics[width=.315\textwidth]{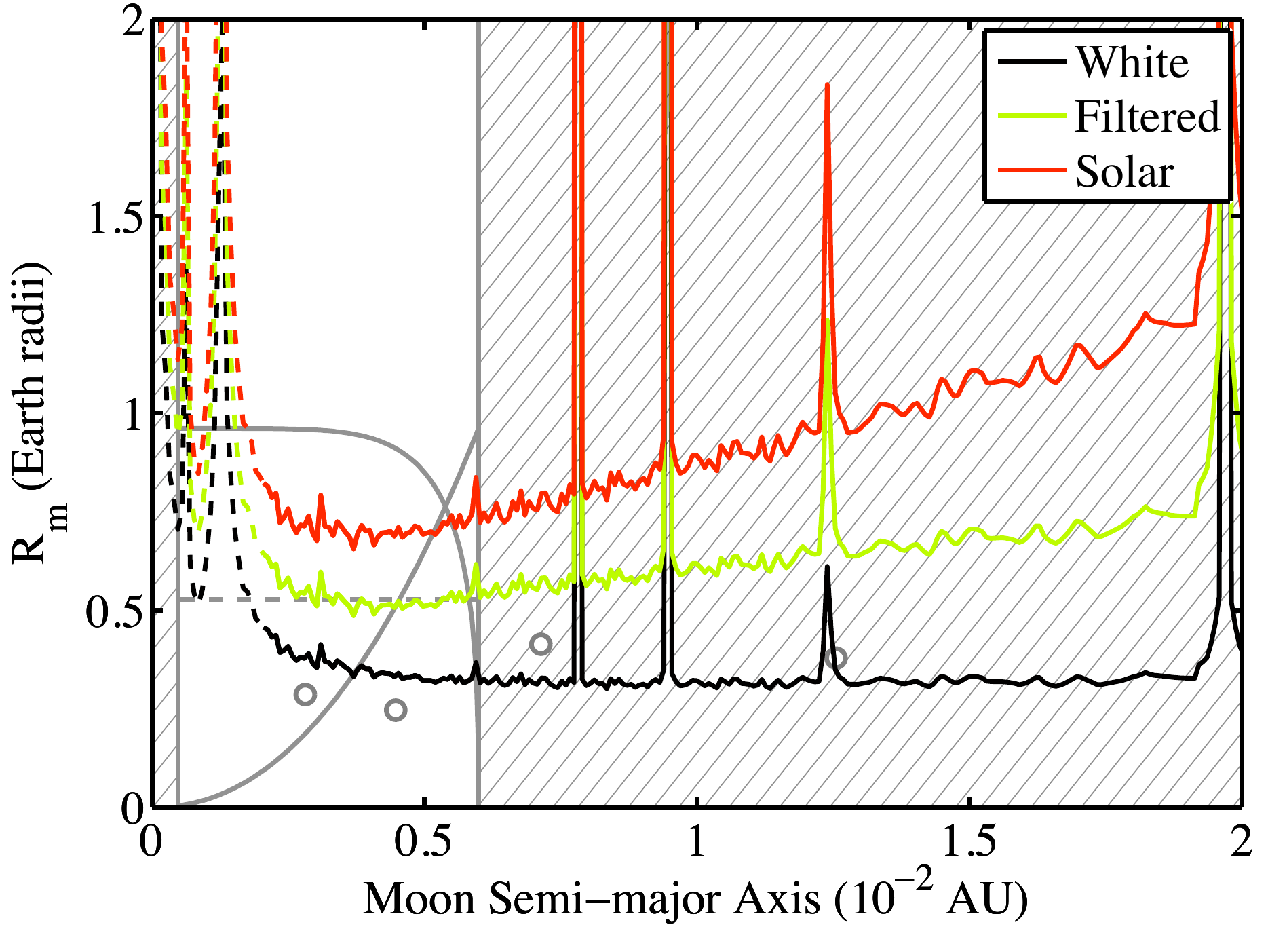}}
      \subfigure[$M_p = M_J$, $a_p=0.4$AU.]{
          \label{TransitThresh2s1MJ04AUEccA}
          \includegraphics[width=.315\textwidth]{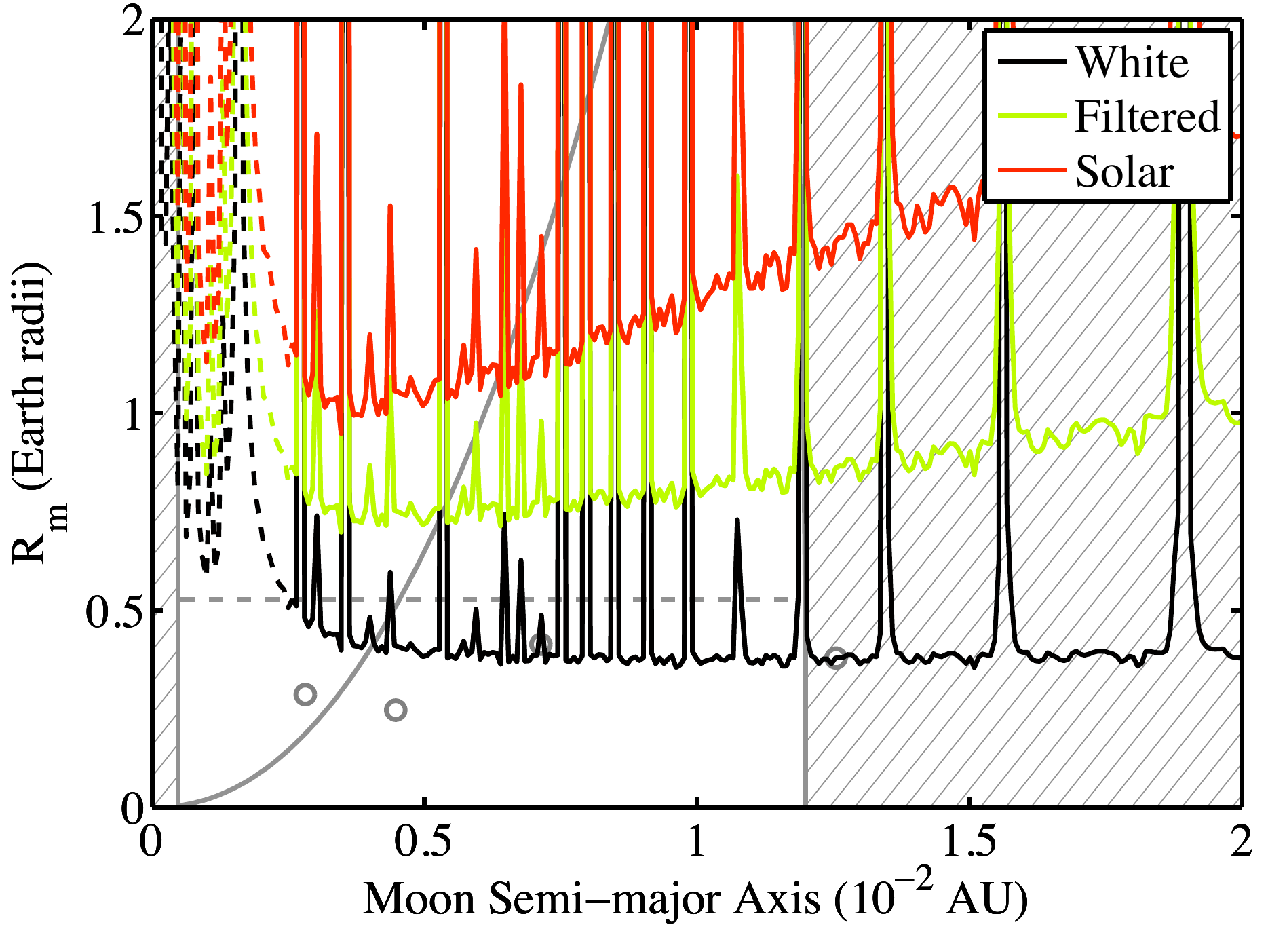}}
     \subfigure[$M_p = M_J$, $a_p=0.6$AU.]{
          \label{TransitThresh2s1MJ06AUEccA}
          \includegraphics[width=.315\textwidth]{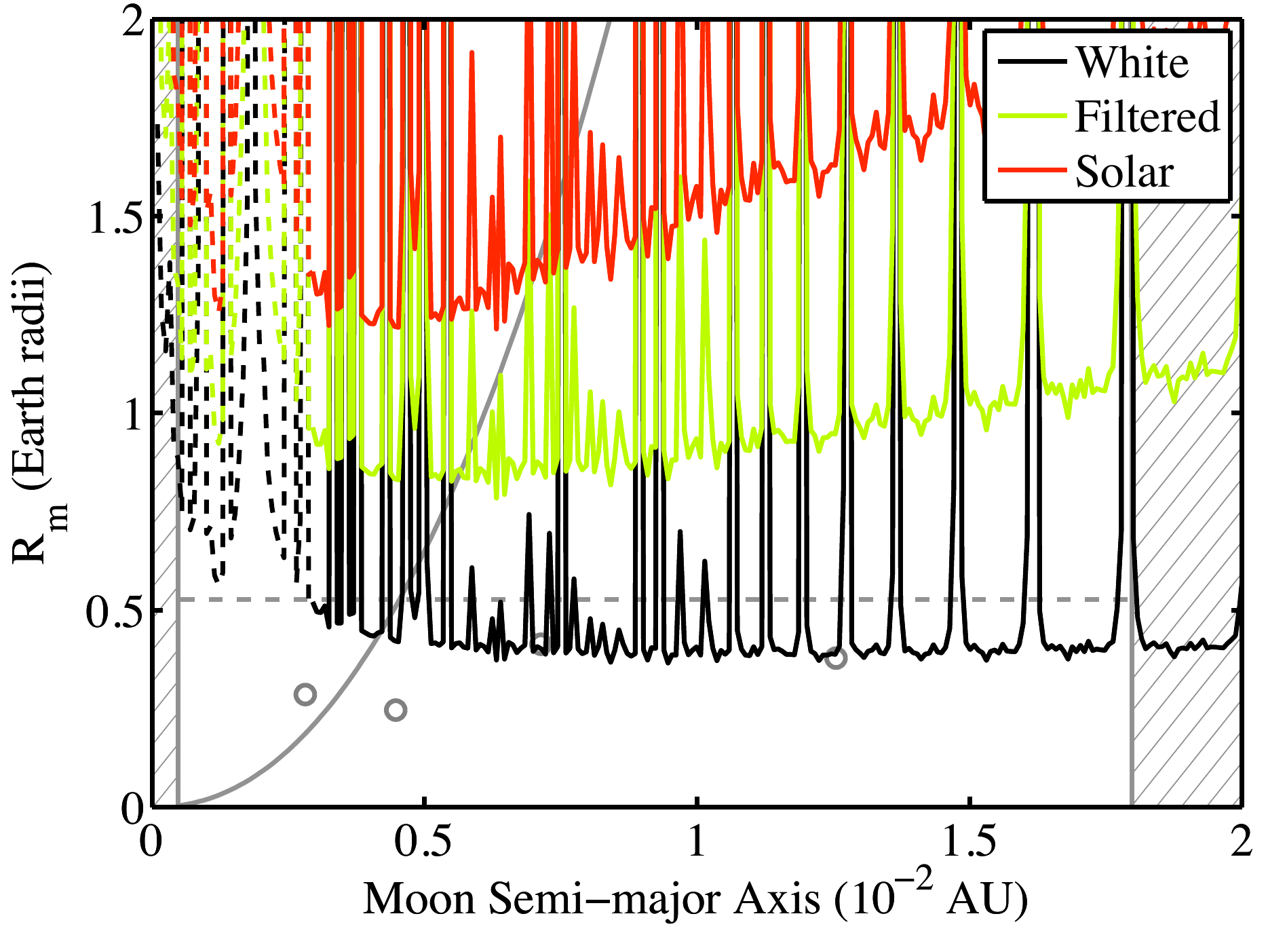}}\\ 
     \subfigure[$M_p = M_U$, $a_p=0.2$AU.]{
          \label{TransitThresh2s1MU02AUEccA}
          \includegraphics[width=.315\textwidth]{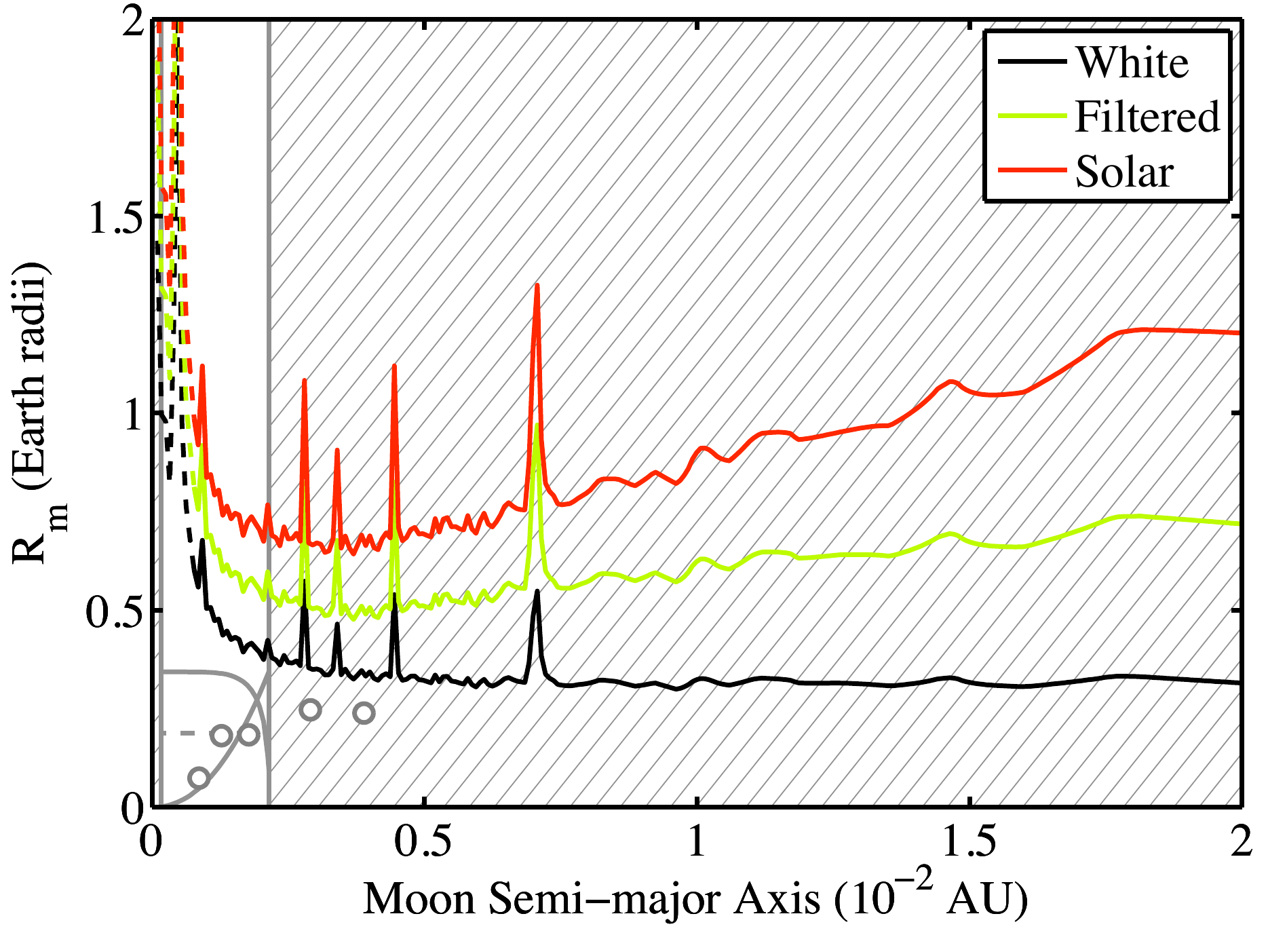}}
     \subfigure[$M_p = M_U$, $a_p=0.4$AU.]{
          \label{TransitThresh2s1MU04AUEccA}
          \includegraphics[width=.315\textwidth]{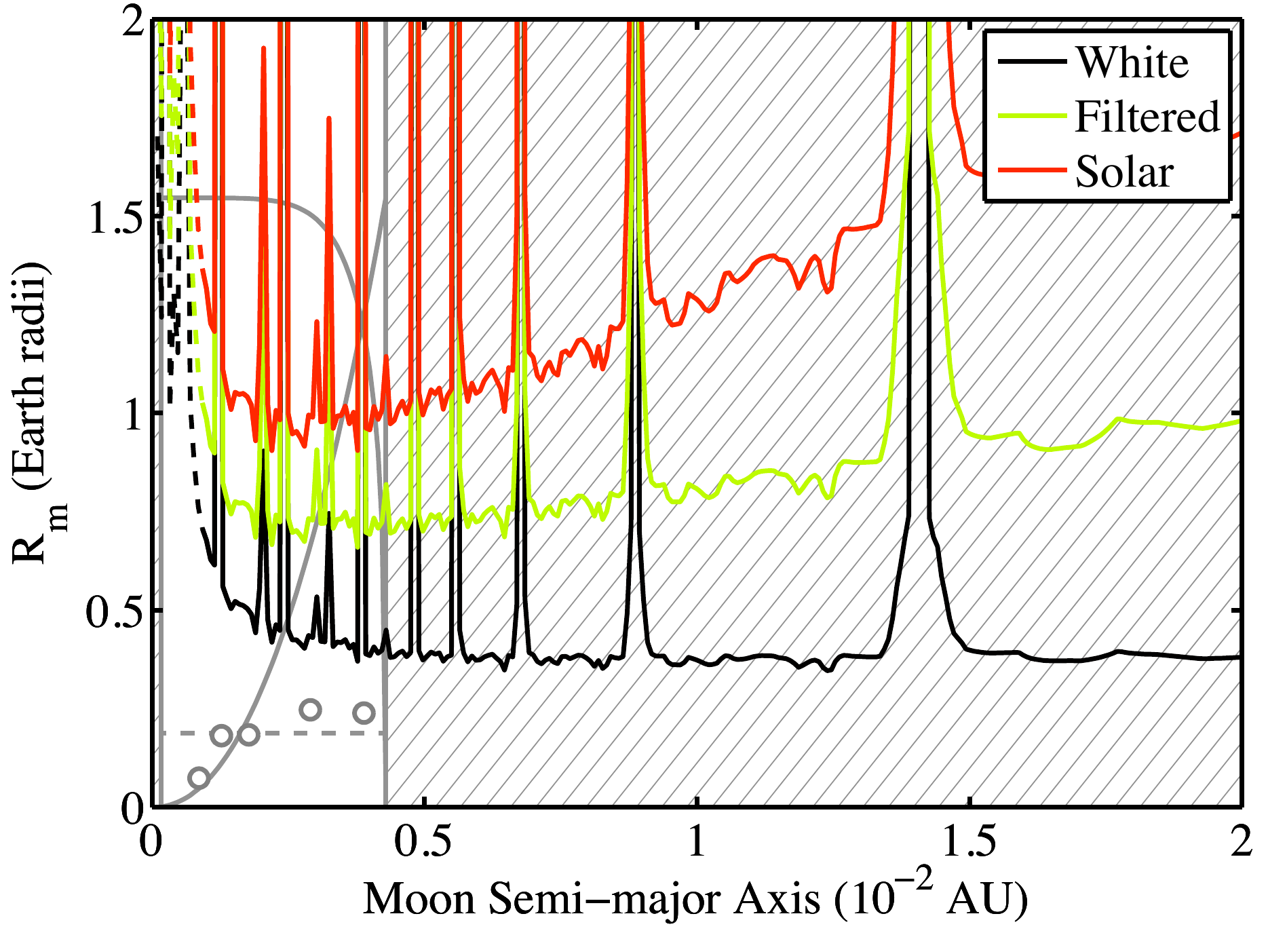}}
      \subfigure[$M_p = M_U$, $a_p=0.6$AU.]{
          \label{TransitThresh2s1MU06AUEccA}
          \includegraphics[width=.315\textwidth]{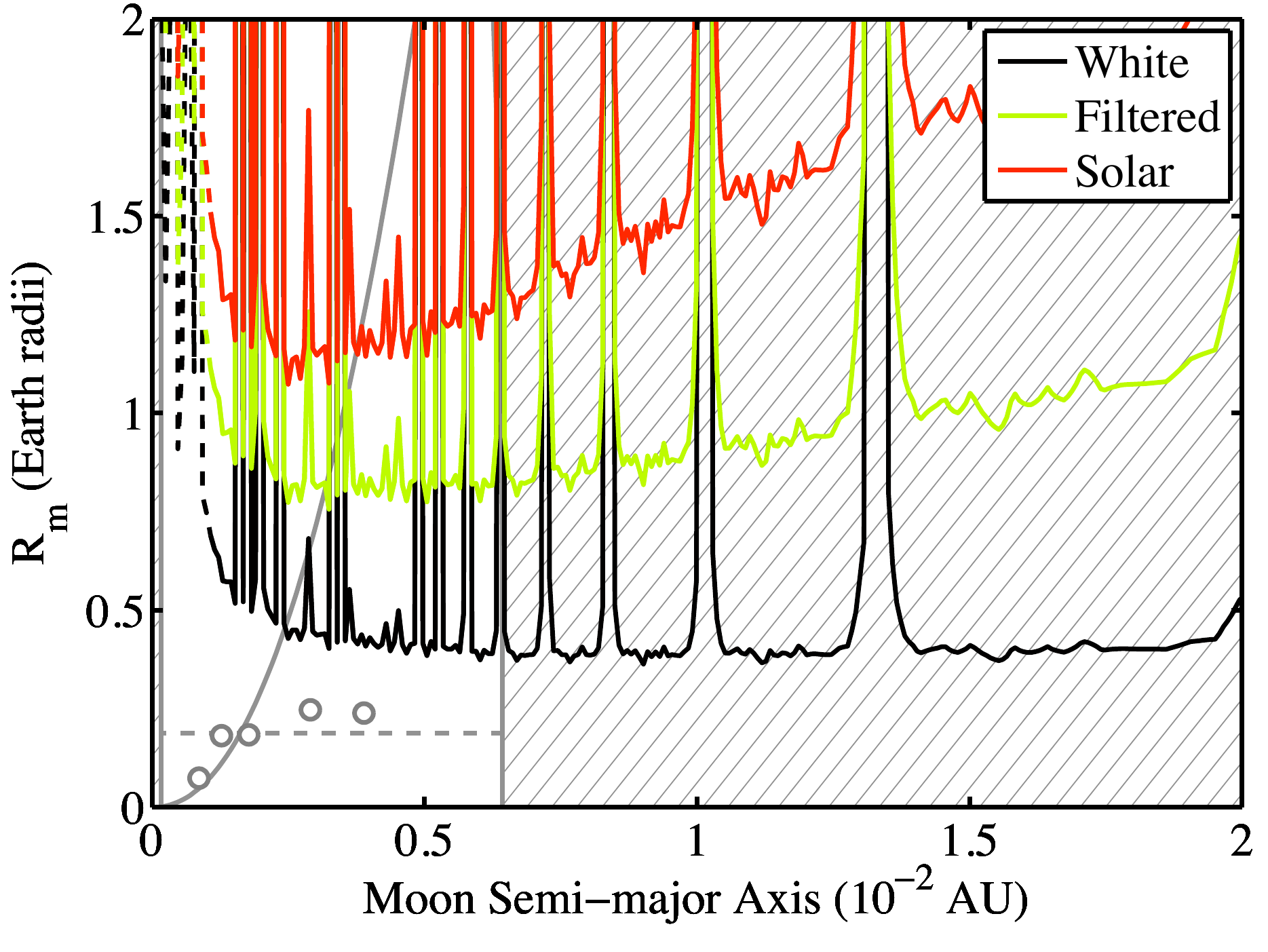}}\\ 
     \subfigure[$M_p = M_{\earth}$, $a_p=0.2$AU.]{
          \label{TransitThresh2s1ME02AUEccA}
          \includegraphics[width=.315\textwidth]{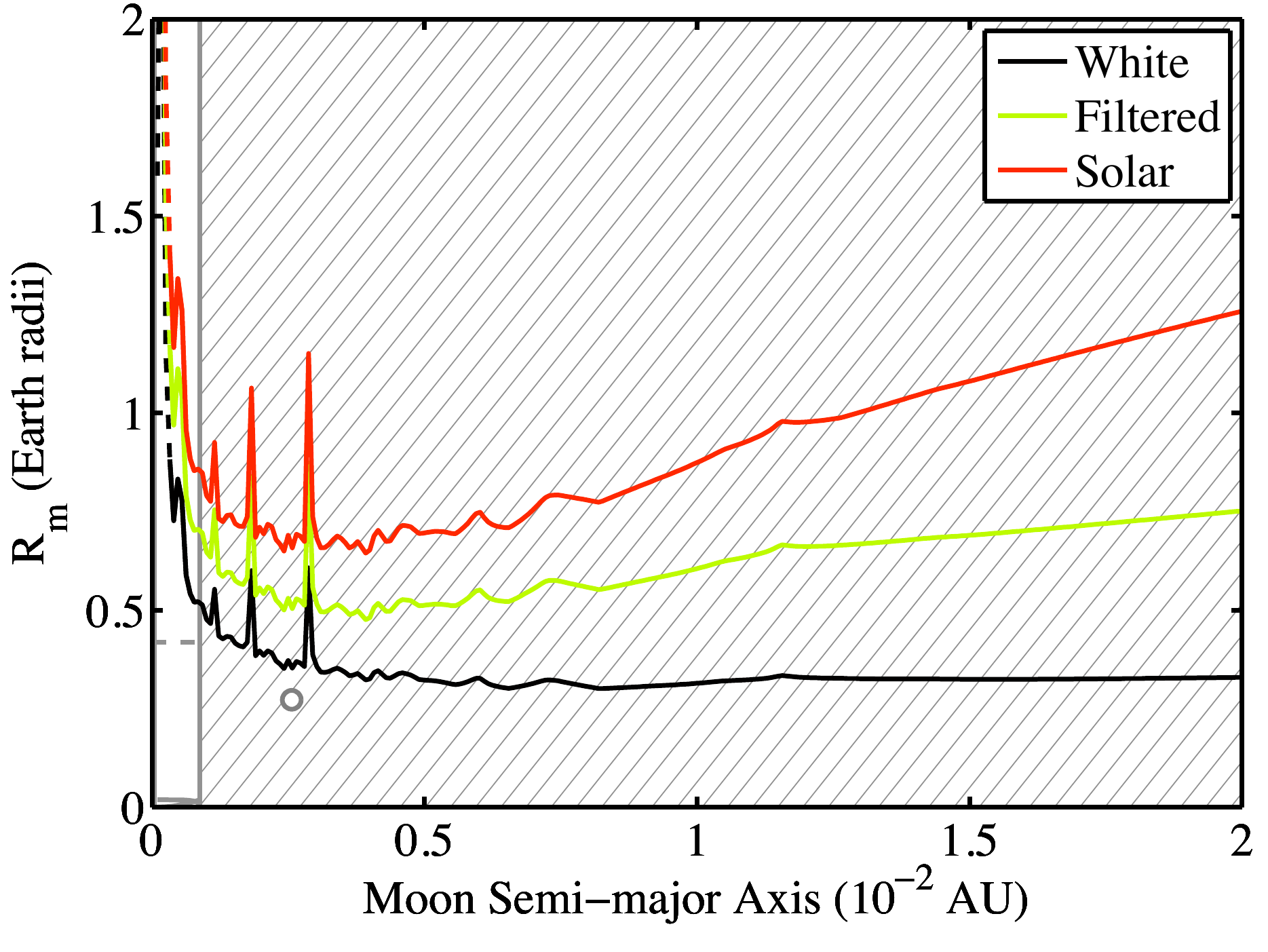}}
     \subfigure[$M_p = M_{\earth}$, $a_p=0.4$AU.]{
          \label{TransitThresh2s1ME04AUEccA}
          \includegraphics[width=.315\textwidth]{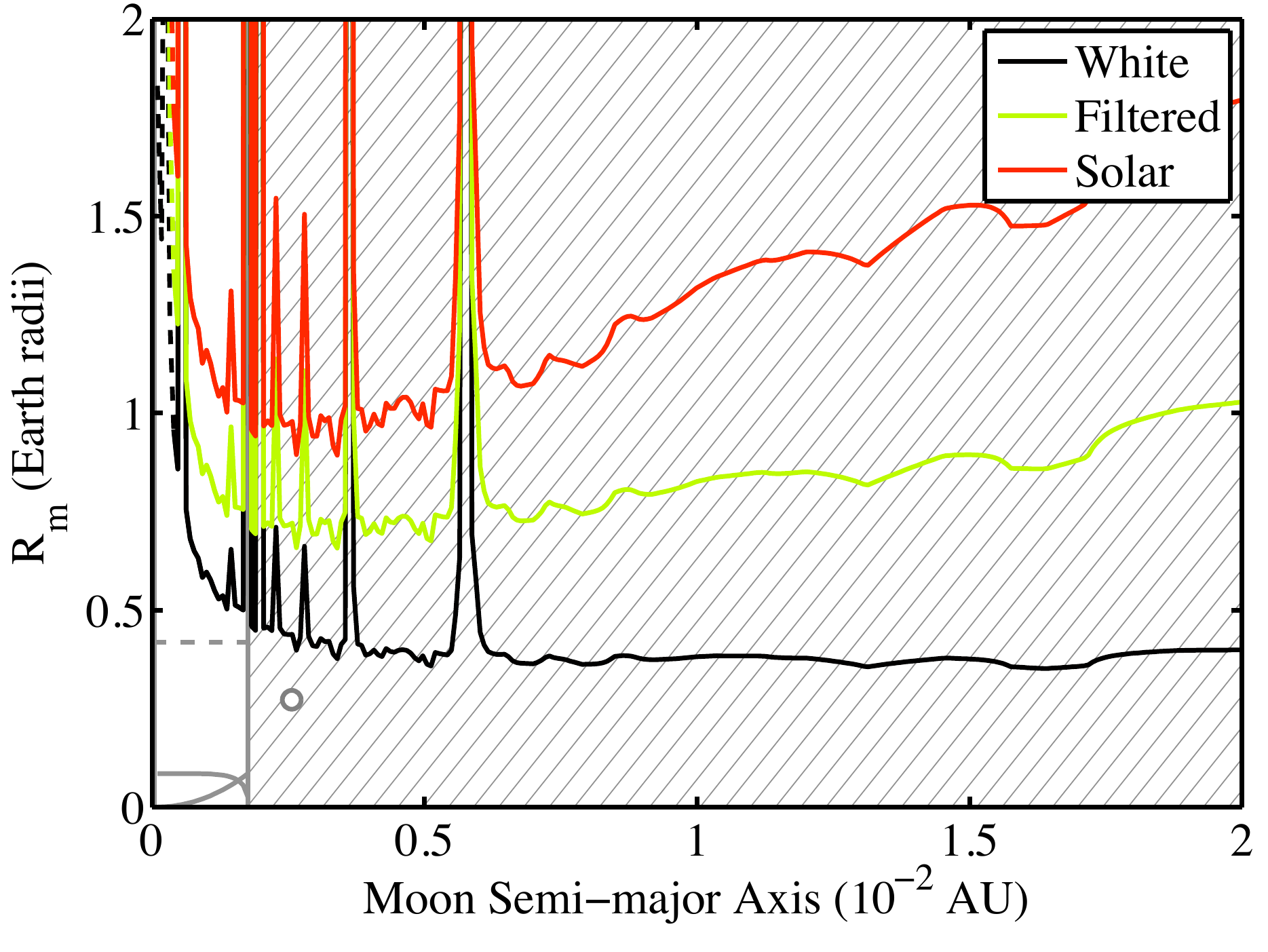}}
     \subfigure[$M_p = M_{\earth}$, $a_p=0.6$AU.]{
          \label{TransitThresh2s1ME06AUEccA}
          \includegraphics[width=.315\textwidth]{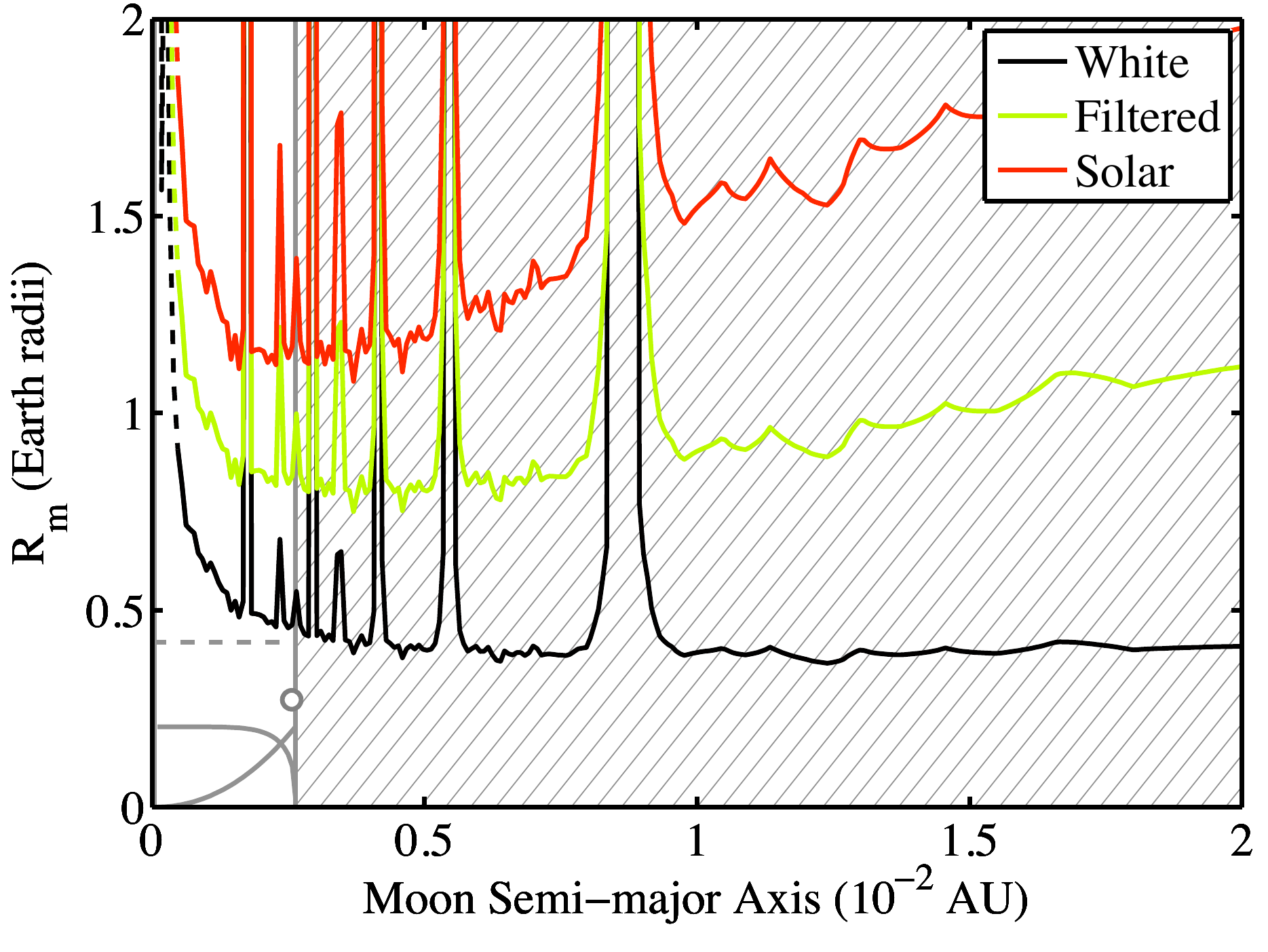}} 
     \caption{Figure of the same form as figure~\ref{MCThresholdsEccentricApo}, but showing the 95.4\% thresholds.}
     \label{MCThresholdsEccentricApo2S}
\end{figure}

In figures~\ref{MCThresholdsAligned1S} to \ref{MCThresholdsEccentricApo1S}, and \ref{MCThresholdsAligned2S} to \ref{MCThresholdsEccentricApo2S} we show the 68.3\% and 95.4\% moon detection thresholds equivalent to the 99.7\% thresholds discussed in chapter~\ref{Trans_Thresholds} and shown in figures~\ref{MCThresholdsAligned} to \ref{MCThresholdsEccentricApo}.  For reference, using the Monte Carlo method detailed in section \ref{Trans_Thresholds_MC_Method} it was found that the 68.3\% and 95.4\% limits on $2 \log \Lambda$ were $3.96 \pm 0.04$ and $8.63 \pm 0.09$ for the case where $N=9$ (0.6AU), $4.22 \pm 0.04$ and $8.99 \pm 0.07$ for the case where $N=14$ (0.4AU) and $4.49 \pm 0.04$ and $9.26 \pm 0.11$ for the case where $N = 40$ (0.2AU).

\bibliography{../References/MoonRef,../References/PlanetRef,../References/PlanetTransRef,../References/ReferenceLibrary,../References/FormationRef,../References/PulsarRef,../References/MathRef}


\end{document}